\documentclass{aa}  

\newcommand{\kms}{km\,s$^{-1}$\,}

\usepackage{graphicx,natbib}
\begin{document}
   \title{Study of the stellar line-strength indices and kinematics along bars\thanks{Based on observations
 obtained at Siding Spring Observatory (RSAA, ANU, Australia) and the INT telescope at the ING, La Palma, Spain}}
   
   \subtitle{I. Bar age and metallicity gradients}

     \author{I. P\'erez \inst{1}\thanks{Veni
          Fellow}\fnmsep\inst{2}\thanks{Associate Researcher}\and
          P. S\'anchez-Bl\'azquez\inst{3}\thanks{Marie Curie
          Fellow}\and A. Zurita\inst{2}\thanks{Retorno J.A. Fellow}}
  
 \institute{Kapteyn Astronomical Institute, University of Groningen,
   Postbus 800, Groningen 9700 AV, the Netherlands\\ email:isa@astro.rug.nl \and Departamento de F\'isica te\'orica  y del Cosmos, Campus de Fuentenueva, Universidad de Granada, 18071 Granada, Spain\\ email:azurita@ugr.es \and Jeremiah Horrocks Institute for Astrophysics \& Supercomputing, University of Central Lancashire, PR1 2HE Preston, UK\\email:psanchez-blazquez@uclan.ac.uk }

   \offprints{I. P\'erez}

   \date{Accepted December 08, 2008}

 
  \abstract
   {}
   {This is the first paper of a series aimed to understand the formation and evolution of bars in early-type spirals and their 
    influence in the evolution of the galaxy.}
   {Optical long-slit spectra along the major axis of the bar of a sample of 20 galaxies are analyzed. velocity and 
   velocity dispersion profiles along the bar are presented.  Line-strength indices in the bar region are also measured to derive stellar mean-age and metallicity distributions along the bars using stellar population models.}
   {We obtain mean ages, metallicities and chemical abundances along the bar of 20  galaxies with morphological types from SB0 to SBbc. The main result is that we find large variation in age and metallicity along the bar in $~$45\% of our sample. We find three different types of bars according to their metallicity and age distribution along the radius:  1) Bars with negative metallicity gradients. They show mean young/intermediate population ($<$~2~Gyr),  and have amongst the lowest stellar maximum central velocity dispersion of the sample. 2) Bars with null metallicity gradients. These galaxies that do not show any gradient in their metallicity distribution along the bar and have negative age gradients (i.e younger populations at the bar end).  3)  Bars with positive metallicity gradients, i.e. more metal rich at the bar ends. These galaxies are predominantly those with higher velocity dispersion and older mean population.  We found no significant correlation between the age and metallicity distribution, and bar/galaxy parameters such as the AGN presence, size  or the bar strength. From the kinematics, we find that all the galaxies show a disk--like central component.}
   {The results from the metallicity and age gradients indicate that most galaxies with high central stellar velocity dispersion host bars that could have been formed more than 3~Gyrs ago, while galaxies with lower central velocity dispersions show a wider distribution in their population and age gradients.  A few bars show characteristics compatible with having been formed less than $<$2~Gy ago. However, we do not have a definite answer to explain the observed gradients and these results place strong constrains to models of bar formation and evolution. The distribution of mean stellar population parameters in the bar with respect to $\sigma$ is similar to that found in bulges, indicating a close link in the evolution of both components. The disk-like central components also show the important role played by bars in the secular evolution of the central structure.}
   \keywords{ Galaxies: spirals-- Galaxies: abundances -- Galaxies: evolution -- Galaxies: formation --  Galaxies: structure -- Galaxies: kinematics
               }

   \maketitle
%

\section{Introduction}

Bars are known to be efficient mechanisms to redistribute angular momentum and matter in the galaxies up to large distances 
and, therefore, they are expected to play an important role in the secular evolution of disk galaxies. The bar gravity torques
make the gas lose angular momentum which, in turn, provokes an inflow of gas towards the central parts. This inflow of gas accumulates in central
mass concentrations (CMC), fueling the central black hole (e.g Sholsman et al. 1989) and, possibly, forming a
stellar bulge. Bars are found in $\sim$60\% of spiral galaxies in the local Universe (e.g. Eskridge et al.\ 2000; Whyte et al.\ 2002;
Men\'endez-Delmestre et al. 2007; Marinova \& Jogee 2007)\nocite{menendezdelmestre2007,marinova2007}\nocite{eskridge2000,whyte2002}, or 30\% if only strong bars are considered.
Despite their obvious importance in disk galaxy evolution, their own fate is still a matter of debate. Several 
mechanisms are able to destroy the bar: for example; 1) the bar may dissolve in the presence of a sufficiently
massive central component (e.g. a black hole) (Friedli \& Pfenniger 1991; Friedli \& Benz 1993, Norman, Sellwood \& Hasan 1996; Berentzen et al. 1998; 
Fukuda et al.\ 2000; Bornaud \& Combes 2002; Shen \& Sellwood 2004; Athanassoula, Lambert \& Dehnen 2005;
Bornaud, Combes \& Semelin 2005)\nocite{shen,friedli91,friedli93,norman96,berentzen98}.
However, the masses of the CMC needed to destroy the bar are much larger than those of the most super massive black 
holes found so far in disk galaxies (Shen \& Selwood 2004; Athanassoula, Lambert \& Dehnen 2005; 
Ferrarese \& Ford 2005)\nocite{ferrarese}. 2) Transfer of angular momentum from infalling gas to the bar has also been proposed as a mechanism that 
could strongly weaken the bar. The combined effect of this angular momentum transfer and the CMC can destroy the bar in Sb-Sc galaxies
in 1-2 Gyr  (Bournaud \& Combes 2002)\nocite{bournaud}. Then, if gas is present, a new bar could form with characteristics different from those of the parent bar.
However, simulations of the effect of gas on the stellar evolution of a bar embedded in a live halo (Berentzen et al. 2007)\nocite{berentzen} have shown that, althogh there are some structural  differences in bars evolving with and without gas, in both cases the stellar bar can survives, at least, 5 Gyr (the whole computation time of these simulations

Do all bars go trough these processes? Do some bars die while others are robust over many Hubble times?
Previous studies have tried to put some limits to the age of the bars by analyzing the evolution with redshift of the bar's fraction in galaxies. 
These studies have also  obtained  discrepant results: 
Some authors have found a fraction of barred galaxies at z$>$0.5 lower
than the local fraction (Abraham et al.\ 1999)\nocite{abraham}, although some authors claim that this may be 
the consequence of  selection effects, due to the high angular resolution
needed to find bars (Elmegreen, Elmegreen \& Hirst.\ 2004; although see van den Bergh et~al.\ 2002)\nocite{elmegreen2004,vandenbergh}.
In order to avoid the problem of the angular resolution, several studies have carried out this analysis using
the ACS camera on the HST: 
Elmegreen et al. (2004) and Jogee et al.\ (2004)\nocite{jogee2004}
found the same bar fraction ($\sim$0.4) at redshift z=1.1
than in the local Universe, suggesting that the bar dissolution cannot be common during a Hubble time unless of course, the bar formation 
rate is comparable to the bar destruction rate. 
On the contrary, Sheth et al.\ (2008)\nocite{sheth2008}, in a recent study, using images from the COSMOS survey (Scoville et al.\ 2007)\nocite{scoville2007} and using
a larger sample than previous studies, found that the bar fraction at z=0.84 is one-third of the present-day value. They also 
found a much stronger evolution for low mass galaxies and late-type spectral types. Part of the discrepant results may be due 
to the selection effects and other systematic effects that need to be still further investigated. 

 Detailed analysis of the stellar populations in the bar region of local galaxies can shed some light to the formation and
 evolution of bars.
 Gadotti \& de Souza (2006)\nocite{gadotti} obtained the color and color gradients in the bar region of a sample 
 of 18 barred galaxies. They interpreted
 the color differences as differences in stellar ages. They concluded that younger bars were hosted by galaxies of later types.  However, the conclusions are hampered by the assumption that the color traces the age, and that the metallicity effect or dust extinction are not important. Clearly, a study attempting to break the age--metallicity degeneracy and taking into account the effect of dust is needed. In addition, the influence of dust in the 
line-strenght indices have been shown to be minimum (MacArthur 2005)\nocite{macarthur} while ages and metallicities from broad--band
colors are heavily affected by dust as well as by the age--metallicity degeneracy. 

To date, it has been difficult to obtain stellar population parameters along the bar because a high 
signal-to-noise is required for this analysis. Although bars are high surface brightness structures, this still implies long 
integration times on medium-size telescopes. Recent work has presented the radial distribution of line-strength indices along bars 
(P\'erez, S\'anchez-Bl\'azquez \& Zurita 2007) \nocite{perez07} for 6 galaxies.

The theoretical model of abundance gradients in bars developed by Friedli (1994) \nocite{friedli94} predicted, using N-body 
simulations of bars with pre-existing exponential abundances, a null evolution of the stellar abundance profile along the bar  
(although with a decrease in the mean metallicity)
 while the gas abundance profile flattened rapidly.  Studies of the radial gas-phase abundance distribution have shown that there is, indeed, a flattening in the gas 
abundance  along the bar (Martin \& Friedli 1997, 1999)\nocite{martin97,martin99}.  However, recent work (P\'erez, S\'anchez-Bl\'azquez \& Zurita 2007) has shown, from stellar absorption line-strengths,  that  some bars present stellar metallicity gradients 
opposite to those found in the disks (i.e. more metal rich population at the end of the bar than in the regions of the bar closer to the center) , contradicting the numerical
results. This surprising result needed a more systematic study of the stellar population along the bars, properly comparing the line-strength indices of a larger sample of galaxies with the state-of-the art stellar populaton models to derive and quantify the variations of age and metallicity along the bar. 

In this paper, we present a  detailed stellar population analysis of the bar region 
of  a sample of 20 early-type spirals (from S0 to Sbc) with and without nuclear activity in order to derive ages, metallicities and relative chemical abundances ratios.  The sample selection is described in
Sect.~\ref{sample}. The observations and data reduction are presented in Sect.~\ref{observation}. The velocity dispersion measurements and the emission correction applied to the data are explained in Section~\ref{velocity}. The line-strength derivation is described in Sect.~\ref{line}.  The morphological characterization of the sample is described in Sect.~\ref{morphology}. Section~\ref{deri.para} gives a description of the methodology followed to derived the stellar population parameters.The results of the line strength-indices and the kinematics are presented in Sect.~\ref{results}. In this work, we concentrate only in the stellar populations of the bar region. 
The results of the central indices, bulge and bulge gradients will be presented in the second series of this paper.

 \section{Sample characterization
 \label{sample}}
 
 We have selected barred galaxies from the Third Reference Catalogue of bright galaxies (RC3), de Vaucouleurs 1948, with 
 the following criteria; to be  classified as barred and with inclinations between 
 10$^{\circ}$ and 70$^{\circ}$, and nearby ($cz$~$\le$~4000~kms$^{-1}$) to be able to properly resolve the bar.
 The sample is biased towards early-types barred galaxies, which have higher surface brightness, to ensure a high signal-to-noise ratio (S/N),
 crucial for a reliable determination of line-strength indices.
 Since we also aim to relate the mean stellar ages and metallicities of the bar 
to the nuclear  activity in the center of the galaxy, half of the galaxies were selected to 
have nuclear activity. Eight of the galaxies present 
nuclear bars. 
 Our final sample comprises 20 galaxies. 
 It is, by no means, a 
statistically complete sample, but it is a representative sample of early-type galaxies bar population. 
 Table~\ref{sample.tab} shows the main properties of the sample  as taken from the Hyperleda galaxy catalog (Paturel et al.\ 2003)~\footnote{http://leda.univ-lyon1.fr}. The bar strengths (shown in Tab.~\ref{sample.tab} as bar--class; for assigment of bar--class to a certain bar strength, see Buta \& Block 2001, is a scale from 0 to 6, being 6 the strongest bar)\nocite{buta2001} have been taken from the literature, where strength is defined as the torque of a 
bar embedded in its disk (Combes \& Sanders 1981\nocite{combes81}), see Table~\ref{sample.tab} for the references for the individual galaxies. The nuclear types have been obtained  from~\cite{veron}. The sample shows a wide distribution of maximum rotational velocities (80 - 260 kms$^{-1}$).
\begin{table*}
\begin{center}
\caption{General properties of the sample
\label{sample.tab}}
\begin{tabular}{l l l l l l l l l }     
\hline\hline
Object          & v(km~s$^{-1}$)    & Type & Bar-class$^{1}$& Nuclear type & Inner morph.$^{2}$&$B$& V$_{max,gas}$  (km\,s$^{-1}$\ )$^{3}$& $i$(deg)\\
\hline
NGC~1169$^{a}$   &  2387  & SABb    &3          & --             &   --         &12.35        & 259.1 $\pm$ 7.3              & 57.1 \\
NGC~1358        &  4028      & SAB(R)0    &    --       &Sy2           &   --         &13.19        & 136.1 $\pm$ 10.6              & 62.8 \\
NGC~1433$^{b}$   &1075        &(R)SB(rs)ab&4&Sy2&Double--bar$^{a}$&10.81& 85.1 $\pm$ 2.4 & 68.1\\
NGC~1530$^{b} $  &2461&SBb&6&--&--&12.50&169.1$\pm$3.5&58.3\\
NGC~1832$^{d}$   &1939&SB(r)bc&2&--&--&12.50&129.9$\pm$2.0& 71.8\\
NGC~2217        &1619  &(R)SB(rs)0/a&--&LINER?&Double--bar$^{b}$&11.36& 183.4$\pm$ 9.2 & 30.7\\
NGC~2273$^{c}$  &1840&SB(r)a&2&Sy2&--&12.62&192.2$\pm$5.5&57.3\\
NGC~2523        &3471&SBbc&--&--&--&12.64 &211.4 $\pm$10.9 &61.3\\
NGC~2665        &1734&(R)SB(r)a&--&--&--&12.47& 130.9$\pm$7.1&32.8\\
NGC~2681$^{c}$  &692&(R)SAB(rs)0/a&1&Sy3&Triple--bar$^{c}$&11.15&87.5$\pm$6.7&15.9\\
NGC~2859$^{c}$  &1687&(R)SB(r)0~\^&1&Sy&Double--bar$^{d}$&11.86& 238.5$\pm$13.3& 33.0\\
NGC~2935 & 2271 & (R)SAB(s)b  & --& -- &-- &12.26 &188.3$\pm$2.0 &42.7\\
NGC~2950       &1337&(R)SB(r)0\^~0&--&--&Double--bar$^{e}$&11.93& --& 62.0\\
NGC~2962       &1966&(R)SAB(rs)0&--&--&Double--bar$^{c}$&12.91&202.9$\pm$9.9& 72.7\\
NGC~3081$^{c}$  &2391&(R)SAB(r)0/a&3&Sy2&Double--bar$^{d}$&12.89 &99.9$\pm$4.0&60.1\\
NGC~4245$^{c}$  &815&SB(r)0/a&2&&--&12.33&113.5$\pm$5.4&56.1\\
NGC~4314$^{a}$  &963&SB(rs)a&3&LINER&Double--bar$^{c}$&11.42&253.3$\pm$24.6&16.2\\
NGC~4394$^{d}$  &922&(R)SB(r)b&3&LINER&--&11.59& 212.5$\pm$16.0&20.0\\
NGC~4643$^{c}$  &1335&SB(rs)0/a&3&LINER&--&11.68&171.4$\pm$7.2&42.9\\
NGC~5101$^{d}$  &1868&(R)SB(r)0/a&2&--&--&11.59&195.7$\pm$9.0&23.2\\
\hline
\end{tabular}
\end{center}
{\footnotesize 
$(1):$\\
$(a)$ Bar class derived from the $K-band$ light distribution, ~\cite{block}\\
$(b)$ Bar class derived from the $K-band$ light distribution,~\cite{block04}\\
$(c)$ Bar class derived from the $K-band$ light distribution,~\cite{buta} \\
$(d)$ Bar class derived from the $H-band$ light distribution,~\cite{laurikainen}\\
$(2):$
$(a)$~\cite{buta86}
$(b)$~\cite{jungwiert97}
$(c)$~\cite{erwin04}
$(d)$~\cite{wozniak95}\\
$(3)$ Rotational velocity corrected for inclination}
\end{table*}

\section{Observations and data reduction
\label{observation}}
 
We obtained long-slit spectra along the bar in our sample of  20 barred galaxies.
The bar position  angles were derived using the Digital Sky Survey (DSS) images.

The observations were performed in two  different runs. 
In the first run (run1 hereafter), spectroscopy of 6 galaxies  was obtained at   
the 2.3m telescope at Siding Spring Observatory (RSAA, ANU, Australia) during February 2006 with the Double 
Beam Spectrograph (DBS, Rodgers, Conroy \& Bloxham 1998)\nocite{rodgers88}.
The gratings  
 600B and the 600R were used for the blue and red arms, respectively, with a slit-width of 2~arcsec. 
This set-up gives a dispersion of 1.1~\AA~pixel$^{-1}$ for the blue-arm and 1.09~\AA~pixel$^{-1}$ for the red arm 
in the wavelength intervals 3892--5814 \AA\ and 5390--7314 \AA\ respectively, and  a final spectral 
resolution of FWHM$\sim2.2$~\AA. 
One of the main problems in the derivation of accurate line-strength indices in regions of 
low surface brightness comes from the systematic effects due to sky subtraction. 
One of the advantages of the DBS is the length of the slit  (6.7~arcmin) which allows the selection 
of sky regions at large radii,  not contaminated by the  light of the galaxy.

The second set of observations (run2 hereafter) were carried out in February 2007 at El Roque de Los Muchachos Observatory (La Palma, Spain).
 We used the IDS  mounted on the  Isaac Newton telescope (2.5m).
The R632V grating  combined with a 1.5~arcsec slit--width gives a resolution of 3\AA~(FWHM).
The full unvignetted
slit length in IDS is 3.3 arcmin. In order to obtain enough sky coverage in one of the galaxy sides, we displaced the center of the slit from the galaxy center during the observations.

 Comparison arc lamp exposures were obtained in both runs
to wavelength calibrate the frames. Spectrophotometric standards were observed with a slit-width of 6 arcsec in 
order to avoid differential flux-loses due to atmospheric diffraction.
Additionally, we observed  11 and 35 G-K stars in the first and second run respectively 
from the Lick/IDS (Gorgas et al.\ 1993; Worthey et al.\ 1994) and MILES libraries (S\'anchez-Bl\'azquez et al. 2006)
in order to calibrate our data into the Lick/IDS spectrophotometric system.
The total integration time for each galaxy is listed in  Table~\ref{observations}.
Only the blue part of the spectra in the case of the DBS observations is analyzed in this paper.  The red part of the 
spectra, i.e. the nebular emission properties, will be presented elsewhere.

The reduction of the two runs has been carried out with the package REDUCEME (Cardiel 1999)\nocite{cardiel}. Standard data reduction procedures (flat-fielding, cosmic ray removal, wavelength calibration, sky 
subtraction and fluxing) were
performed. Error images were created at the beginning of the reduction and were processed in parallel
with the science images.
Initial reduction of the CCD frames involved bias subtraction and removal of pixel-to-pixel sensitivity variations (using flat-field
exposures of a tungsten calibration lamp). Correction of a two-dimensional low-frequency scale sensitivity variations was done using 
twilight sky exposures.
Arc frames were used to convert the spectra into a linear
wavelength scale. We use typically $\sim$ 60 lines. The lines were fitted with a 1st order polynomia, 
with r.m.s. errors lower than 0.1 \AA. Atmospheric extinction was calculated using the extinction curve of King (1985)\nocite{king85}. To correct
the effect of interstellar extinction, we used the curve of Cardelli, Clayton \& Mathis (1989)\nocite{cardelli1989}. The reddenings applied were
extracted from the RC3 catalogue of galaxies ~\citep{vaucouleurs}.
Relative flux calibration was achieved using the spectra  of the observed standard stars.
All the flux-calibration curves of each night  were averaged and the flux calibration 
errors were estimated by the differences between the indices measured with different 
curves. 
After correction for geometrical distortions (both spatial and spectral), a sky image was generated for each 
observation by fitting, for each channel, a low order polynomial using regions 
selected at both sides of the galaxy frame in the first run, and only in one of the sides of the frame for the galaxies in the second run.

In order to derive the index spatial distribution, for the fully reduced galaxy frames, a final frame was created by extracting spectra
along the slit, binning in the spatial direction to guarantee a minimum signal-to-noise
ratio of 20 per \AA~ in the spectral region of Mgb. This minimum S/N ensures errors lower than 15\% 
in most of the Lick/IDS indices (Cardiel et~al.\ 1998)\nocite{cardiel1998}.

\section{Analysis}
\subsection{Velocity dispersion measurements and emission correction
\label{velocity}}

Whenever ionized gas is present in the galaxy, emission lines can affect 
the measurement of the absorption indices. In particular, all the Balmer
lines (H$\delta$, H$\gamma$ and H$\beta$), Fe5015, 
and Mgb indices are affected by the presence of H$\delta\lambda 4101$ ,  H$\gamma\lambda 4340$,  H$\beta\lambda4861$,
 [OIII]$\lambda\lambda$ 4959,5007, and [NII]$\lambda\lambda$5198,5200 lines in emission.
The presence of emission in the Balmer lines tend to fill the absorption profiles making the indices weaker. If this is not corrected properly, 
older ages can be artificially measured.
 In order to correct from the  emission lines contamination we have used the routine GANDALF  
(Sarzi et~al.\ 2005, S05 hereafter). This routine  fits,  simultaneously, the stellar and the emission line spectra, by treating the  emission
 lines as additional Gaussian templates and iteratively searching for the best radial velocity and velocity dispersion. To model the stellar 
spectra, we use the stellar population models by Vazdekis et al. (2008) based on the 
MILES stellar library  (S\'anchez-Bl\'azquez et al. 2006)\nocite{patri06}, 
previously  degraded to our instrumental resolution. 
The continuum shape of the stellar templates and that of the galaxies is
fitted and removed using a multiplicative Legrende polynomial and subtracted
from all spectra.

As it is explained in S05, despite that the kinematic of the different emission lines can be fitted independently, 
in some cases, as in H$\beta$, degeneracies between the stellar and emission 
templates can cause spurious detections.  Furthermore, it is very difficult 
to constrain the kinematic of the weak emission lines. For these reasons, we follow
the same procedure as in S05 and impose that all the emission lines have the same  kinematic
as the strong line [OIII]$\lambda 5007$, except for the cases when H$\beta$ line is very strong.
The steps followed are:
(1) Masking the emission lines and calculating a first estimate of the  stellar velocity 
and velocity dispersion. 
(2) Then, convolving our templates with the calculated velocity dispersion and shifting them according
to the recessional velocity, and then, calculating the kinematics of the [OIII]$\lambda 5007$ lines.
(3) We force all the emission lines to have the same kinematic as the one calculated
with [OIII]$\lambda 5007$ and fit, simultaneously, emission and absorption spectra.

\begin{figure*}
\begin{center}
\resizebox{0.7\textwidth}{!}{\includegraphics[angle=-90]{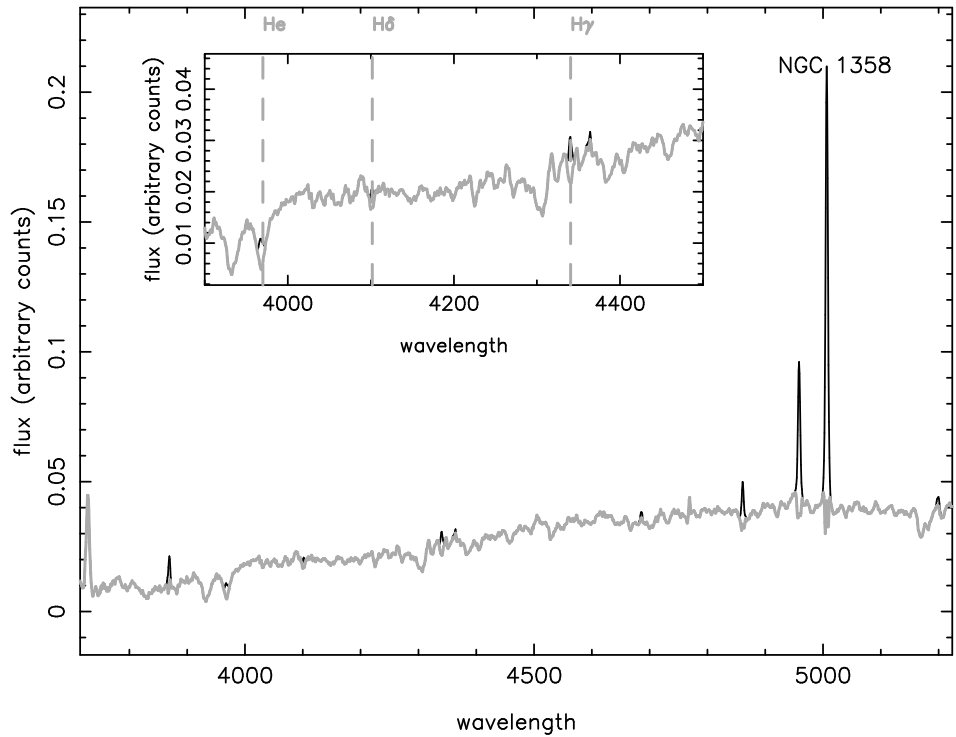}}
\caption{Example of the emission correction with GANDALF for the central spectrum of NGC1358. Grey 
line shows the spectrum after being cleaned from emission lines, while the black spectra show the 
spectrum before.  NGC~1358 is a galaxy from the sample with particularly strong emission lines, the fitting process applied to correct for the emission does a very good job even in this extreme case (Sect.\ref{velocity} for details).\label{n1358.gandalf.13.ps}}
\end{center}
\end{figure*}

Figure~\ref{n1358.gandalf.13.ps} shows the central spectrum of NGC1358, a galaxy with 
particularly strong emission, before and after cleaning the emission using the method described above.
In some cases (i.e in the nuclei of active galaxies), the emission lines cannot be fitted with a single Gaussian, 
as the emission line profiles show asymmetries.
In those cases, double Gaussians are fitted in order to reproduce the observed profile. This 
process will be explained in more detail in a future paper, when we analyze the properties
of the bulges and central parts in our sample (P\'erez, S\'anchez-Bl\'azquez, \& Zurita 2009, in preparation).

GANDALF makes use of PPXF (penalized pixel-fitting) to calculate the stellar kinematics. PPXF is a software described
in Capellari \& Emsellem (2004). It is specially appropriate for low signal-to-noise data or 
when the data is not well sampled. 
The errors in the radial velocity and velocity dispersion were computed through numerical simulations. In each 
simulation, a bootstrapped galaxy spectrum, obtained using the error spectrum provided by the reduction with 
REDUCEME, is fed into the algorithm. Errors in the parameters are then calculated as the unbiased
deviation of the different solutions.
The kinematic of the gas and the diagnostic emission lines will be subject of a future paper (Florido et al. 2009, in preparation).

\subsection{Line-strength indices
\label{line}}

Lick/IDS line-strength indices using the definition in Trager et al. (1998)\nocite{trager}
are measured in all the --cleaned from emission lines-- binned spectra along the radius. The 
errors are calculated by the uncertainties caused by photon noise, wavelength
calibration and flux calibration. 

Line-strength indices depend on the broadening of the lines caused by 
instrumental resolution and by the internal motions of the stars.
Therefore, when comparing with the stellar population models, the synthetic and observed indices
need to be measured at the same resolution.

We are using two different sets of models to derive the synthetic indices. The first one, Vazdekis et al. (2008)
is built using a library (MILES) with constant resolution along the wavelength direction (2.3~\AA) which is also
flux calibrated (S\'anchez-Bl\'azquez et al. 2006). 
The main advantage of these models is that they predict the whole spectrum for a population of a given 
age and metallicity and, therefore, it can be convolved with a broadening function to mimic 
the desired resolution.  To compare with these models we degrade all the spectra to the resolution of the run1 spectra (FWHM$\approx$5.2~\AA).

The second set of models we are using are those of Thomas, Maraston \& Bender (2003, TMB03 hereafter)\nocite{thomas2003}.
These models are based on the Lick/IDS fitting functions (Gorgas et al.\ 1993; Worthey et al.\ 1994)\nocite{worthey,gorgas}
and, therefore, their resolution cannot be modified. To compare our measurements with this model 
we degrade our spectra to the 
wavelength-dependent resolution of the Lick stars 
following the prescriptions given in Worthey \& Ottaviani (1997)\nocite{worthey97}.

Absorption lines are not only broadened by the line-spread function of the 
instrument, but also by Doppler broadening due to the velocity dispersion of the stars.
To compare with the models and with galaxies with different velocity dispersions, 
this needs to be corrected. Because the variation of the indices
with the velocity dispersion depends on the strength of the index itself (e.g. Kuntschner 2000)\nocite{kuntschner2000}, one should 
perform a different correction for each individual spectrum along the radius. 
To calculate these corrections we use
the template obtained in GANDALF when fitting the spectra in the first bin, i.e. the bar--bin closest to the galaxy center. We degrade then this spectrum at the Lick/IDS resolution 
and broaden it  to different velocity dispersions in steps of 10~kms$^{-1}$. We then 
build polynomia (for each individual spectrum) of the form:
C($\sigma$)=I($\sigma$=0)/I($\sigma$) for atomic indices, and C($\sigma$)=I($\sigma=0$)-I($\sigma$)
for molecular indice, where I($\sigma$=0) represents the 
index at the Lick resolution and I($\sigma$), the index at Lick resolution broadened with a velocity dispersion $\sigma$. Polynomia are built as well with the initial broadening equal to the 
instrumental resolution of run1, in order
to compare with the V08 models at the resolution of the data.
We  finally apply a single correction to each  galaxy using the polynomia derived 
for the spectra at the beginning of the bar, and use the rest of the templates along 
the bar to calculate the error for this approximation. This error is added in quadrature 
to the final index error. 

Once the indices in the galaxy spectra and models are measured at the same resolution 
and corrected by the velocity dispersion broadening,  a further correction is needed in order to fully 
transform the measurements into the right spectrophotometric system. These small corrections 
try to compensate for the differences between the models and the data that appear due to a  deficient
calibration of the Lick/IDS stellar spectra. To obtain these offsets, we compare the index measurements
in our sample of stars in common with the Lick/IDS library observed for this purpose. The final offsets are very small and listed
in Table~\ref{table.offsets}. Figures with the comparison of individual indices can be found in appendix~\ref{appen.compara.lick}.
 In principle, because our data are relatively flux calibrated, 
a further offset is not necessary to transform our measurements into the MILES spectrophotometric system (that is also flux calibrated).
However, we also compare the indices measured in stars in common with this library. As expected, 
for most of the indices the offsets are null (see Table~\ref{table.offsets}), although small offsets were found for some of them.

\begin{table}
\begin{center}
\caption{Log of observations. \label{observations}}
\begin{tabular}{c l l l l }     
\hline\hline
Object & Telescope & Date       & $PA$(deg) & Exp. Time \\
\hline
NGC~1169& INT       &09/02/2007   &95  &3h\\  
NGC~1358& INT       &10,11/02/2007&127 &3h\\
NGC~1433& 2.3m,SSO  & 03,06/02/2006& 96&2.5h    \\
NGC~1530& INT        &08/02/2007   &121 &3h\\
NGC~1832& INT        &12/02/2007&171&3h\\
NGC~2217& 2.3m, SSO  &6/02/2006&290&2h\\
NGC~2273& INT        &07/02/2007&116&3h\\
NGC~2523& INT        &09/02/2007&118&3h\\
NGC~2665& 2.3m,SSO &01,02/02/2006&54&3.3h\\
NGC~2681& INT        &02/02/2007&28&3h\\
NGC~2859& INT        &08/02/2007&162&3h\\
       &            &12/02/2007&72&3h\\
NGC~2935&2.3m,SSO& 5,6/02/2006& 135 & 3h\\       
NGC~2950&INT         &11/02/2007&162&3h\\
NGC~2962&INT         &10/02/2007&175&3h\\
NGC~3081&2.3m,SSO &01/02/2006&64&2.5h\\
NGC~4245&INT         &07,08/02/2007&135&3h\\
NGC~4314&INT         &09/02/2007&147&2.5h\\
       &            &12/02/2007&237&3h\\
NGC~4394&INT         &11/02/2007&143&3h\\
NGC~4643&INT &10/02/2007&134&3h\\
		 & 2.3m,SSO &1,5,6/02/2006& 134&3.3h\\	
NGC~5101&2.3m, SSO   &  3,5,6/02/2006& 122 & 3h  \\
\hline
\end{tabular}
\end{center}
\end{table}

\subsection{Morphological characterization
\label{morphology}}
 In order to relate the stellar population properties with the morphological features, we  perform ellipse fitting analysis on available archived $R-band$ images of the galaxy sample. For nine out of the 20 galaxies, Sloan 
 Digital Sky Survey~\footnote{http://www.sdss.org} (SDSS) data was available.  For NGC~1530 we use the data published in Zurita \& P\'erez (2008)\nocite{zurita2008}, for NGC~2273  the $R-band$
 image from Hameed \& Devereux (1999)\nocite{hameed} and for NGC~1169 the $R-band$ from Knapen et al. (2004)\nocite{knapen04}. 
 Ellipse fitting to the light distribution is performed using the ELLIPSE task within IRAF.  As all the galaxies showed
 point-like nuclear regions, the center is fixed using the coordinates obtained by fitting a
 Gaussian to the nucleus. The position angle ($PA$) and the ellipticity ($\epsilon$) are left as free parameters in the fitting.
 The minimum ellipticity  in the bar region has been 
 adopted as the definition for the  size of the bar. It has been shown (Michel-Dansac \& Wozniak 2006)\nocite{micheldansac06} that 
this measurement is correlated with the corotation radius and, therefore, gives a bar structural size, it is also the method to calculate the bar size that is the least sensitive to dust absorption.  Furthermore, it is a very simple criteria to reproduce
 and can be compared to other results in the
 literature. For a discussion about the differences in the  bar size obtained using different methods  see for example~\cite{erwin}. The bulge
 radius is considered as to the radius
 where the $PA$ and ellipticity $\epsilon$ starts showing the typical bar profile, i.e. increasing $\epsilon$ and constant $PA$. 
 Figure~\ref{morf1} and Fig.~\ref{morf2} show the results of the ellipse
 fitting for all the analyzed galaxies together with a $R-band$ image, and an image of the residuals obtained subtracting
 the  galaxy model  to the $R-band$ galaxy image. The radius of the bar, bulge and the extent of our spectrocopic data are overplotted on the two 
 figures. For the galaxies for which no SDSS 
  data was available, Fig.~\ref{2mass} shows 2MASS data  together with the spectroscopic data extent.
 Table~\ref{barsize} shows the derived bar size for the galaxies, with good imaging data from the SDSS and the literature, in comparison with the bar sizes obtained by other authors.  There is good agreement 
between the values. The beginning of the bar coincides, in many cases, with a change on the trends of the line strength indices with radius
 (see Sec.~\ref{deri.para}), giving confidence  in our size estimations. 
The general morphology of the galaxies in the sample shows a large range of properties. Table~\ref{sample.tab} shows the detailed
morphological types of all the galaxies. 
Some galaxies, independently of their type, have outer rings, some host 
strong bars and some are of intermediate $AB$ type. We, therefore, cover a large range in general galaxy morphology. This allows us to test  
systematic behaviors associated to different morphologies.  
\begin{figure*}[H]
\resizebox{1.0\textwidth}{!}{\includegraphics[angle=0]{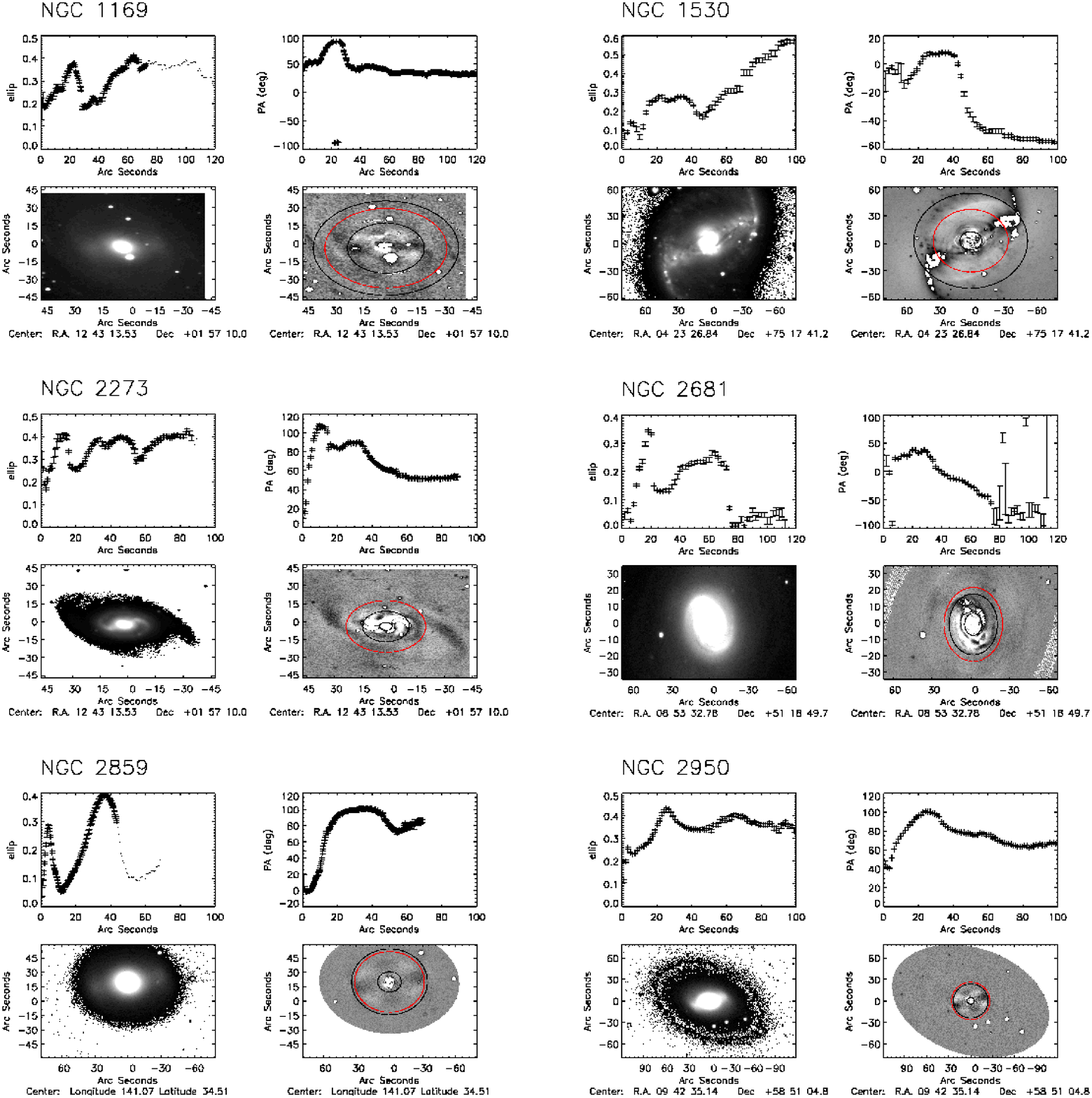}}
\caption{\label{morf1} Morphological analysis of the sample galaxies from SDSS images. For every galaxy it is shown the ellipticity distribution, position angle distribution the $R$-band image, the red circle shows the extent of the data and the black circle the position of the bar end, and the residual image.}
\end{figure*}

\begin{figure*}[H]
\resizebox{1.0\textwidth}{!}{\includegraphics[angle=0]{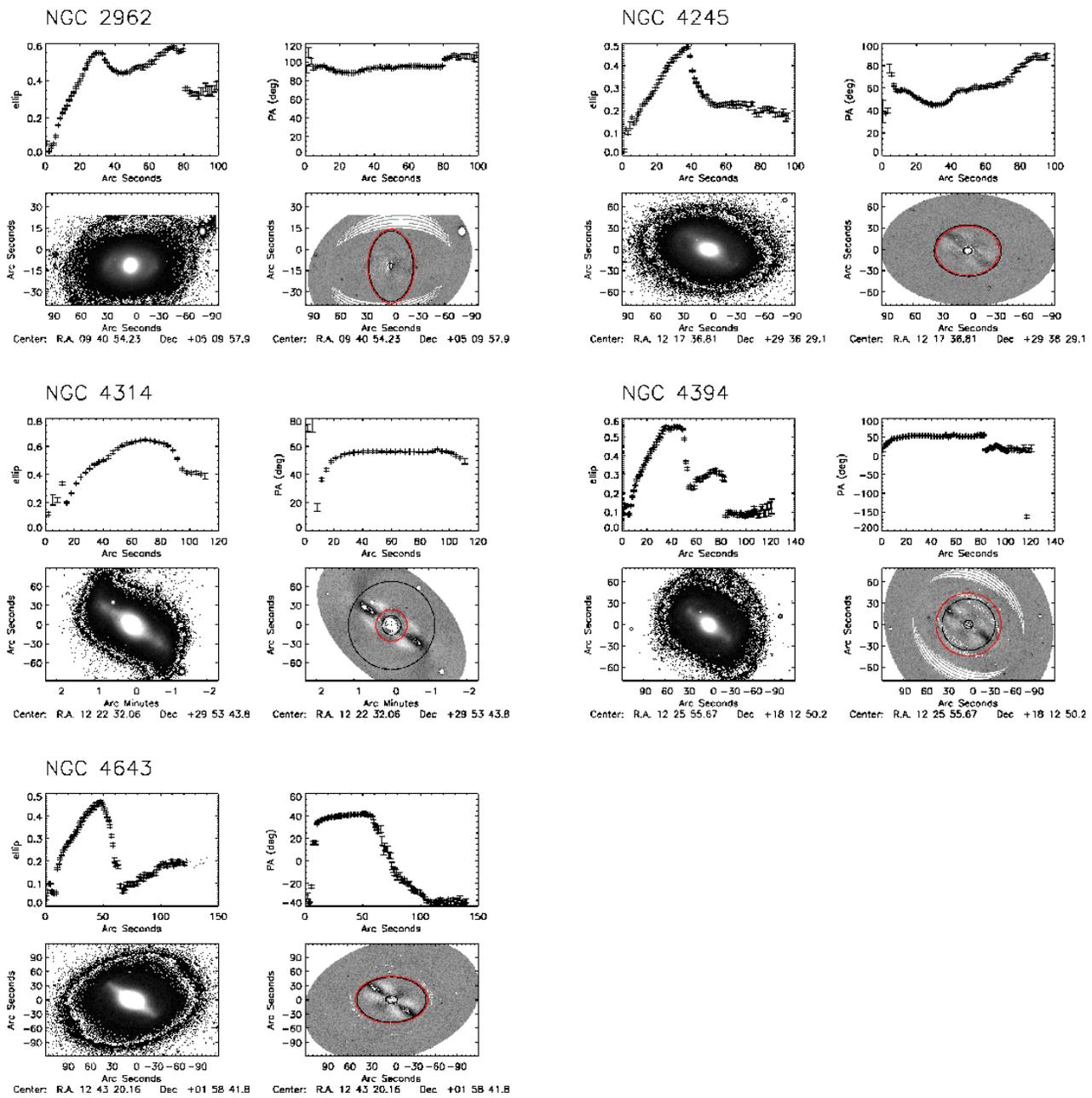}}
\caption{Continuation of Fig~\ref{morf1} \label{morf2}}
\end{figure*}  

\begin{figure*}
\resizebox{1.0\textwidth}{!}{\includegraphics[angle=0]{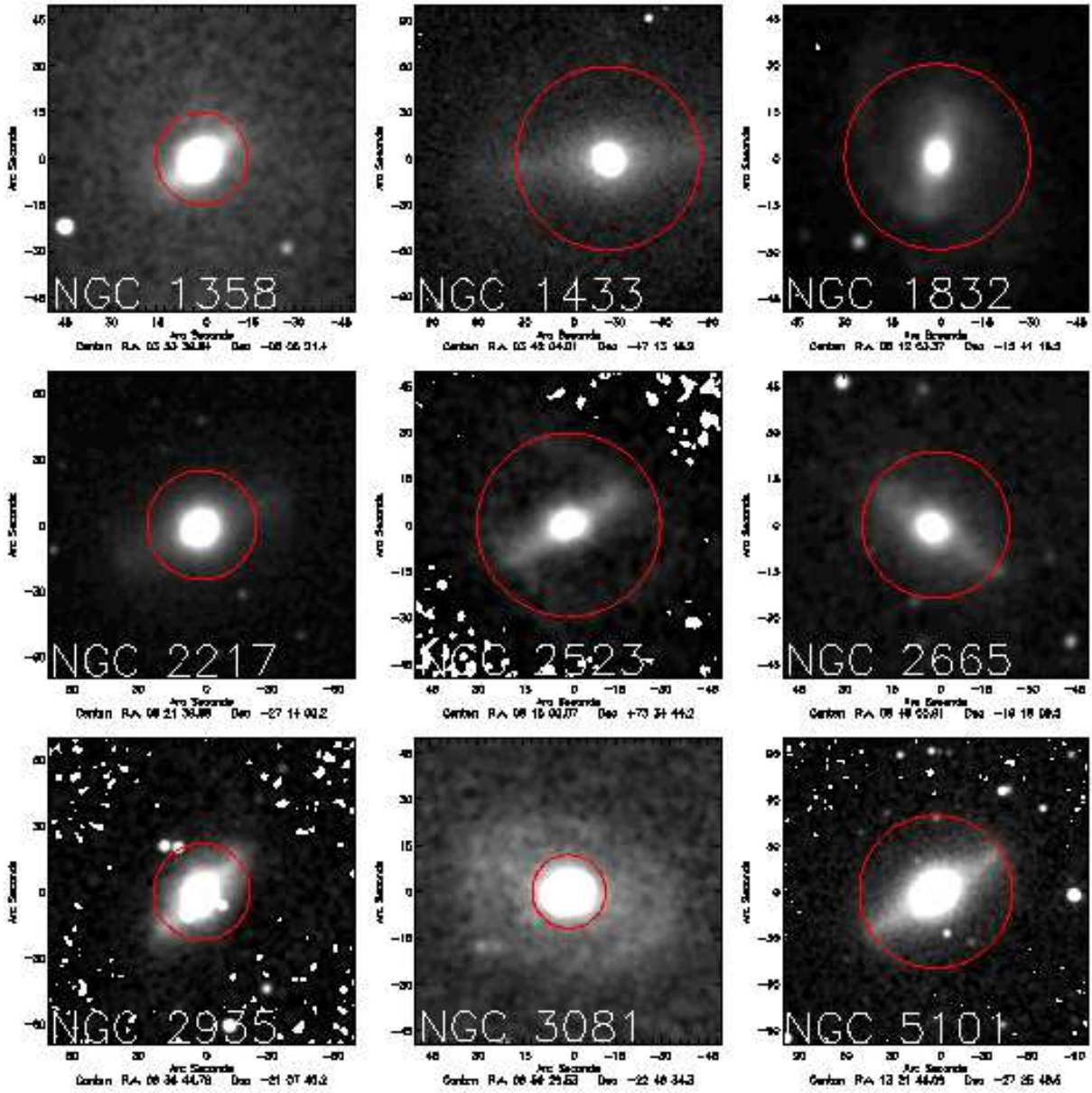}}
\caption{\label{2mass} 2MASS images of the galaxies without SDSS imaging. The red circles show the extent of the spectroscopic data}
\end{figure*}

\begin{table}
\caption{Bar sizes.  The bar size derivation is described in Section~\ref{morphology}
\label{barsize}}
\begin{center}
\begin{tabular}{l l l }     
\hline \hline
 Galaxy name & Bar size (arcsec) & Bar size (arcsec)  \\ 
        & (other studies)   & (our study)  \\
\hline
NGC~1169 &   --                &  29   \\
NGC~1358$^{d}$ &   25    &   --    \\
NGC~1433$^{b, i}$&   90, 66 &   --               \\
NGC~1530$^{e}$ &  50                 & 69  \\
NGC~1832$^{f}$& 17 &  --     \\
NGC~2217$^{b; f; i}$ &50, 37, 43  &   -- \\        
NGC~2273 $^{a}$& 14  &  21           \\     
NGC~2523 &   --       &  --    \\      
NGC~2665 &    --                &  --    \\    
NGC~2681$^{a, 1; c; g}$& 50, 25, 29  &  23          \\       
NGC~2859$^{b; c}$ &34, 46  &  48            \\       
NGC~2935$^{f; i}$&25, 39 &--                          \\
NGC~2950 $^{a; g}$& 24, 38 &  44            \\       
NGC~2962 $^{a}$&  29 &  45            \\   
NGC~3081$^{b; g; c}$ &35, 41, 38  &  35            \\      
NGC~4245$^{a}$ &  37&  59          \\        
NGC~4314 $^{a, g}$& 67, 75  &  92    \\      
NGC~4394$^{h;i}$ & 52,  35    & 56\\      
NGC~4643$^{a}$ & 50   & 67    \\  
NGC~5101$^{f}$& 50 & --  \\ 
 \hline
\end{tabular}
\end{center}
{\footnotesize 
(1) NGC~2681 is a triple barred system. Erwin (2005) gives the value of the primary bar, while we give here the value of the secondary bar
(a)~\cite{erwin} bar size given as the radius of maximum ellipticity in the bar region\\
(b)~\cite{erwin04} bar size given at the radius of maximum ellipticity in the bar region\\
(c)~\cite{wozniak95}, they give the projected bar size at the radius of minimum ellipticity \\
(d)~\cite{mulchaey97}, they give the bar size as the radius at the minimum of ellipticity\\
(e)~\cite{regan96}\\
(f)~\cite{jungwiert97}, bar size given as the radius of maximum ellipticity in the bar region\\
(g)~\cite{friedli96}, they give the projected bar size as the radius at the ellipticity minimum\\
(h)~\cite{gadotti06}, bar size given as the radius of maximum ellipticity in the bar region\\
(i)~\cite{martin95}, bar size given by visual inspection of photographic plates
}
\end{table}
\subsection{Derivation of stellar population parameters
\label{deri.para}}
As we mentioned before (Section~\ref{line}), to derive stellar population parameters we compare our line-strength
indices with two different sets of stellar population models: TMB03 and V08. 
TMB03 use the fuel consumption theorem (Renzini \& Buzzoni 1986)\nocite{renzini86} 
to evaluate the energetics
of the post-main sequence phases. They use the stellar tracks of Cassisi, Castellani \& Castellani (1997)\nocite{cassisi97}, 
Bono et al.\ (1997) and Cassisi et al.\ (1999), and for the more metal rich models the ones by Salasnich et al. (2000). V08 
uses the isochrones of Girardi et al.\ (2000)\nocite{salasnich00,cassisi99,bono97,girardi2000} for all metallicities, and  an empirical calibration to transform the
models from the  theoretical to the observational plane.
By comparing with two different sets of models we take into account some of  the uncertainties coming 
from the variety of ingredients chosen by different authors. Below, we give a brief explanation of such possible uncertainties.

V08 models provide synthetic spectra of single stellar populations with a given age, metallicity and 
solar chemical abundances ratios (the ones of the empirical library).
We know that, in some systems, the ratio between different elements do not match the solar values. For example, 
in giant elliptical galaxies,  [Mg/Fe]$>$0.0 (e.g. O'Connell 1976; Faber et al. 1992)\nocite{oconnell76,faber92}.
It is important to take this effect into account because the presence of differences in the chemical abundance partitions
affects the age measurements (e.g. TMB03). Several groups have  characterized the variation of line-strength 
indices for single stellar populations to variations of individual chemical abundances (e.g. Tantalo et al. 1998; Trager et al. 2000;
Proctor \& Sansom 2002; TMB03)\nocite{proctor2002,tantalo98}. 
 In order to do this, we follow the prescriptions given in Trager et al.\ (2000) with the corrections for negative indices given 
in Thomas et al.\ (2003) using the response functions by Korn et al.\ (2004)\nocite{korn2004}.
We make the assumption that all the $\alpha$-elements are equally enhanced with the exception of Ca, which
is assumed to follow Fe-peak element abundances. C is also assumed to be enhanced. Therefore, we define two groups: The enhanced group (C, N, O, Ne, Na, Mg, Al, Si, S and Ar) and the depressed group (Fe, Ca, Cr and Ni).
This matches the model 4 in Trager et al. (2000)\nocite{trager2000}.
We assume a linear relationship between the [Z/H], [Fe/H] and [E/Fe] (where E represents the mass of all elements with enhanced 
abundance ratios) of the form  [Z/H]=[Fe/H]+ 0.929 [E/Fe] (see Trager et al.\ 2000) and 
all the variations of [E/Fe] are compensated by variations of the Fe-peak group in such a way that [Z/H] 
stay constant.

 The changes of the indices to different chemical composition is calculated as (see Trager et al.\ 2000):
\begin{equation}
\frac{\delta I}{I_0}=\prod_i (1+R_{0.3}(X_i))^{[X_i/H]/0.3} - 1 ;
\end{equation}
where $R_{0.3}(X_i)$ is the  metallicity dependent response function by Korn et al.\ (2004) to
a variation of [X$_i$/H] of the element $i$.

The models of TMB03 already incorporate a similar treatment to take into account the differences in the 
chemical abundances ratios. We use the original version of the models with [C/Fe] and [N/Fe]=0 and [Ca/$\alpha$]=0, 
but this choice does not affect our results.

To obtain stellar population parameters 
we use 4 indices, namely; H$\delta_A$, H$\gamma_A$, Fe4383 and Mgb and follow a multi-index approach
as described in Proctor \& Samson (2002).  
We avoid using H$\beta$ because this is the Balmer index most affected by the presence of emission. 
The errors are estimated by performing 100 Monte Carlo simulations where each index was perturbed
by a gaussian distribution with $\sigma$ equal to the index error. 

The average deviations of the measured indices from  the indices predicted by the model
for the derived  values of age, metallicity and [E/Fe],  are shown in Fig.~\ref{fig.chi}. 
Despite the fact that we have only used 4 indices for the fit, it can be seen that most of the indices are 
reproduced within the errors by the stellar population parameters derived with this method. 
Some of the indices that are worse reproduced for some galaxies are the broad indices Mg$_1$, Mg$_2$, CN$_1$ and CN$_2$. This is understandable as those indices are very sensitive to the flux calibration. As we have not corrected from internal extinction in the galaxy, the shape of the continuum in our spectra is affected by reddening. For the rest of the indices, the effect of dust is almost negligible (see, e.g. MacArthur 2005). In order to analyze the possible correlations between the results from the stellar population analysis and other parameters,  we fit a slope to the derived age, metallicity and [E/Fe] distributions with a simple linear regression weighted with the errors in the $y$-direction. Table~\ref{gradtab} shows the fitted slope together with the errors. For four galaxies out of the 20, the obtained gradients are not reliable due to large fitting errors and  these galaxies are discarded in the plots and the analysis regarding the gradient values (NGC~1530,NGC~3081,NGC~4314,NGC~2935), we include them in Tab.~\ref{gradtab}.

\begin{figure*}
\resizebox{0.3\textwidth}{!}{\includegraphics[angle=-90]{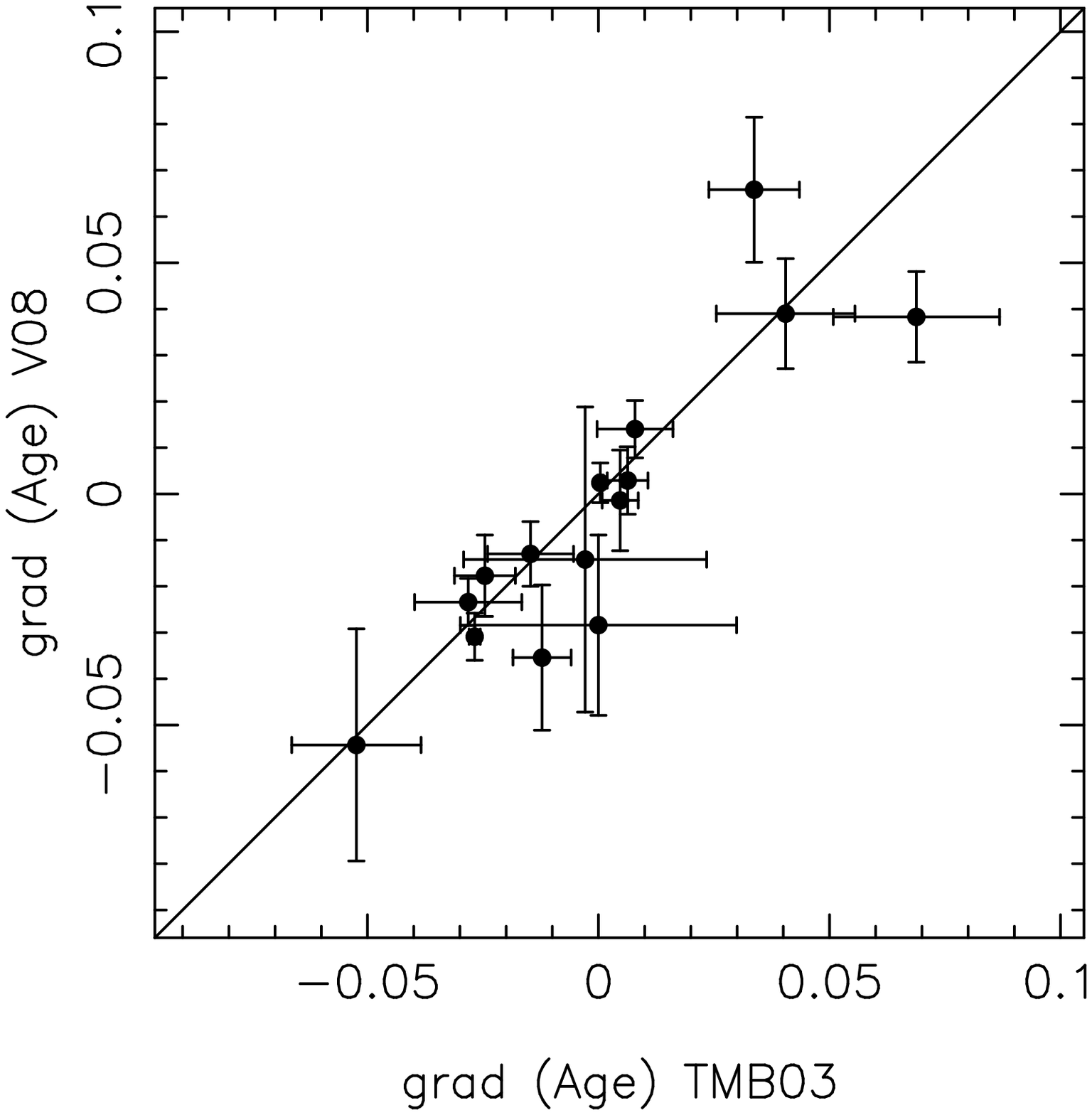}}
\resizebox{0.3\textwidth}{!}{\includegraphics[angle=-90]{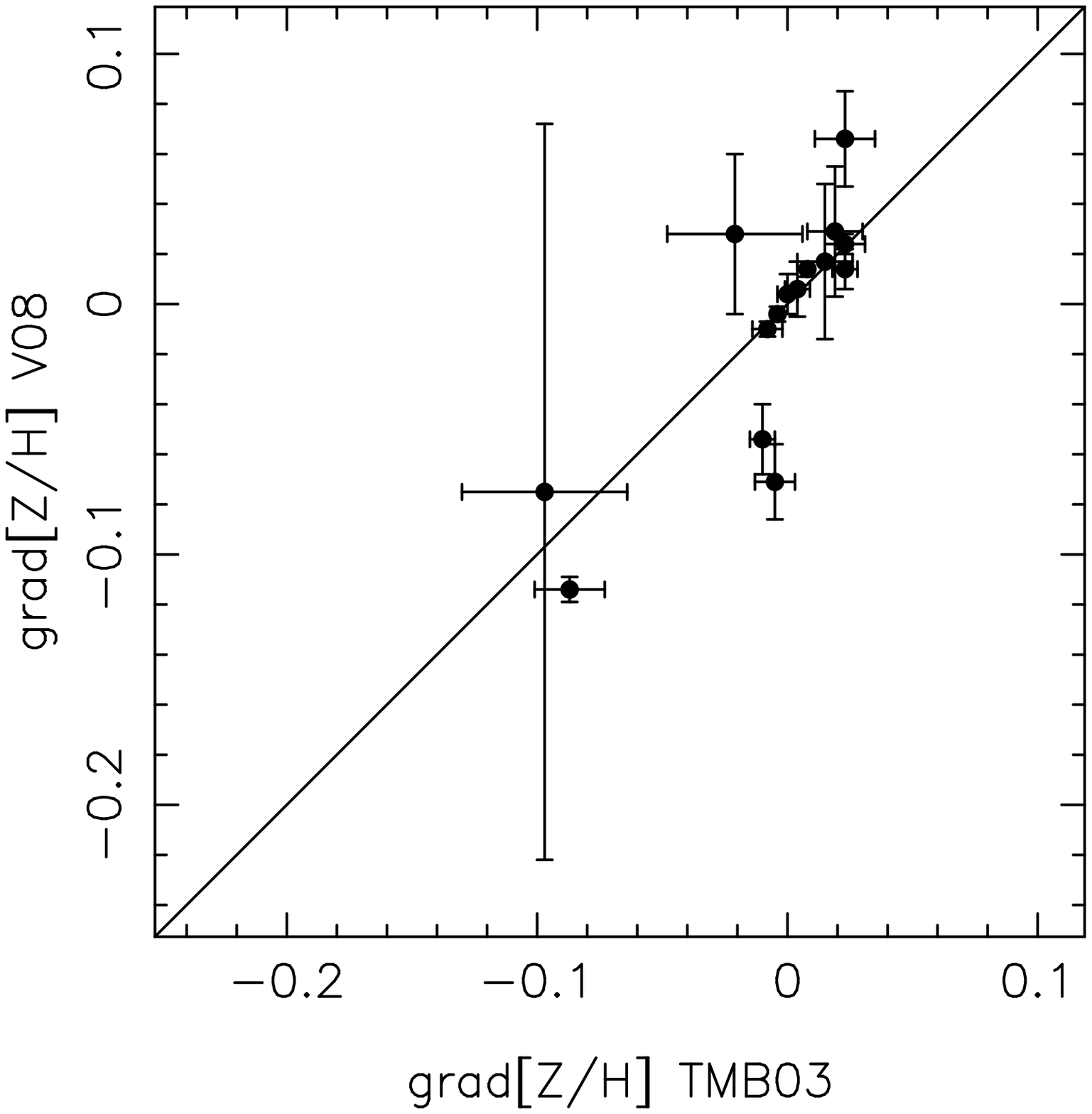}}
\resizebox{0.3\textwidth}{!}{\includegraphics[angle=-90]{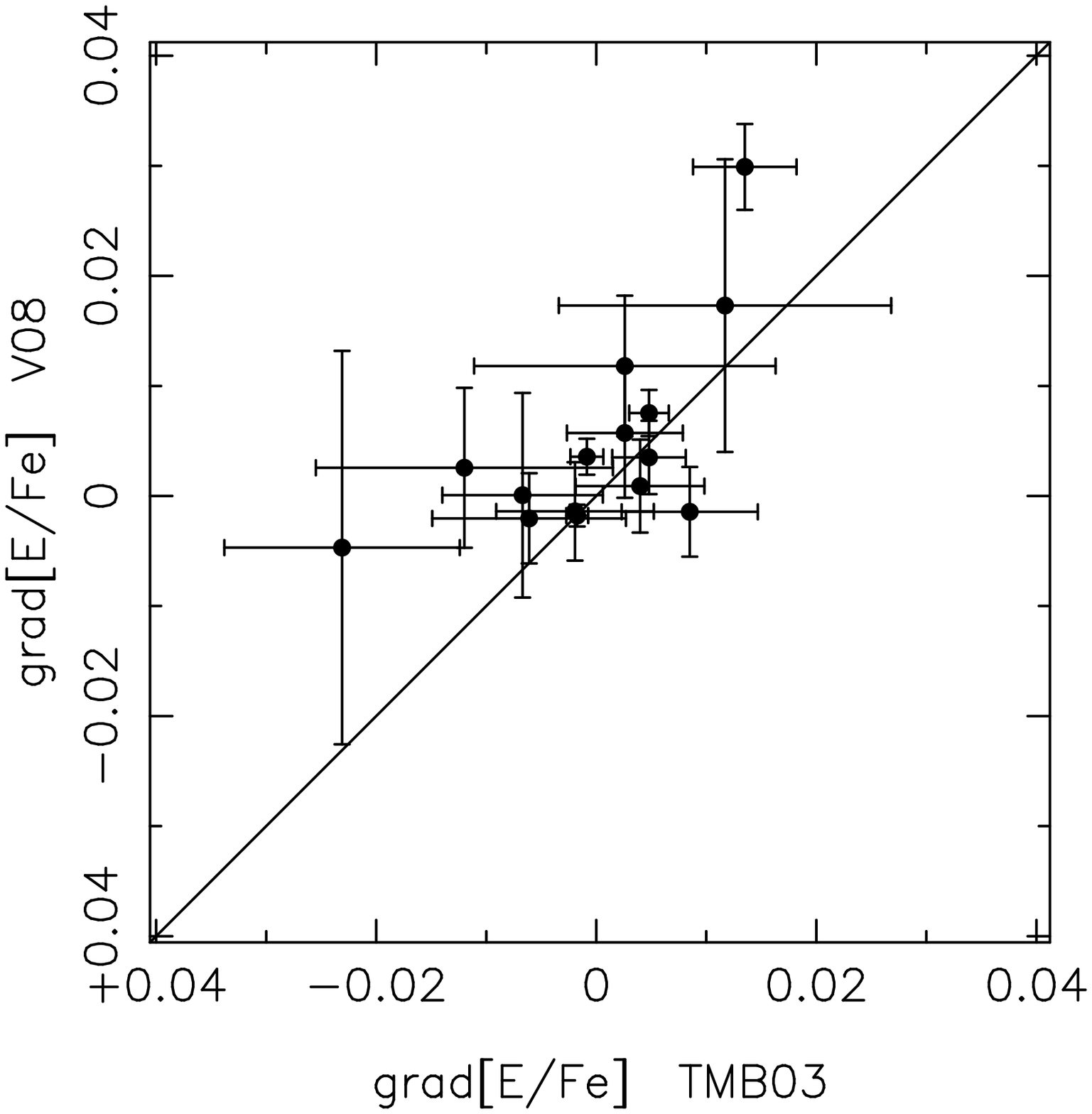}}
\caption{Comparison between the mean gradients measured using TMB03 and V08 models. We have carried out a comparison of the metallicity and age gradients using two different models to asses the dependance of the results with the model choice. As it can be seen, the comparison is very good in general.\label{comparison.models}}
\end{figure*}

To check the dependence of our results to the choice of stellar population model, we repeat the calculations 
using the new models by V08. 
Fig.~\ref{comparison.models} shows the comparison of the age, metallicity and [E/Fe] gradients (the slope of the fit) using different models. As can be seen, 
the comparison is very good in general. We have checked that none of our results is  affected by our choice of model.

An important concern about the obtained results is that a non-perfect emission correction could still be affecting 
our derived indices.  The effect of a poor emission correction is to make the age older and, therefore, because
the errors in age and metallicity are correlated, the metallicity is lowered.
It is difficult to estimate an error due to emission correction that takes into account the systematic
effects as, for example, a template mismatch. Therefore, to test if our results are biased 
due to a poor emission subtraction, we re-calculate the ages and metallicities
 using 8  indices non-affected by emission:
 G4300, Fe4383, Ca4455, Fe4531, C4668, Mg1, Fe5270, Fe5335. Basically we use all the indices
except for the ones that may be affected  by emission.
It has been shown that even without Balmer lines, it is possible to obtain a reliable age (although 
with larger errors) (e.g. Chillingarian et al. 2008)\nocite{chilingarian08}. 

\begin{figure*}
\resizebox{0.3\textwidth}{!}{\includegraphics[angle=-90]{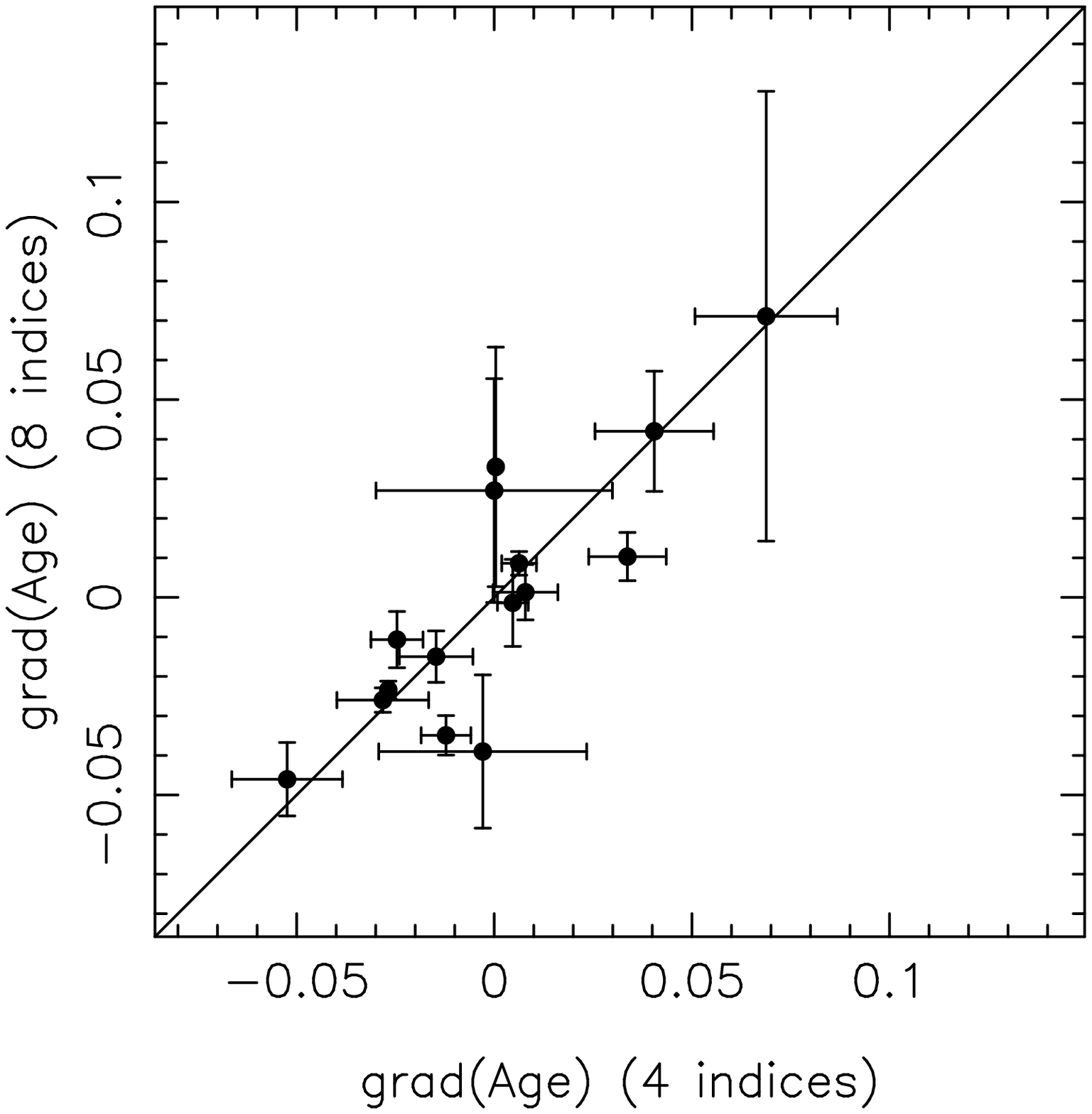}}
\resizebox{0.3\textwidth}{!}{\includegraphics[angle=-90]{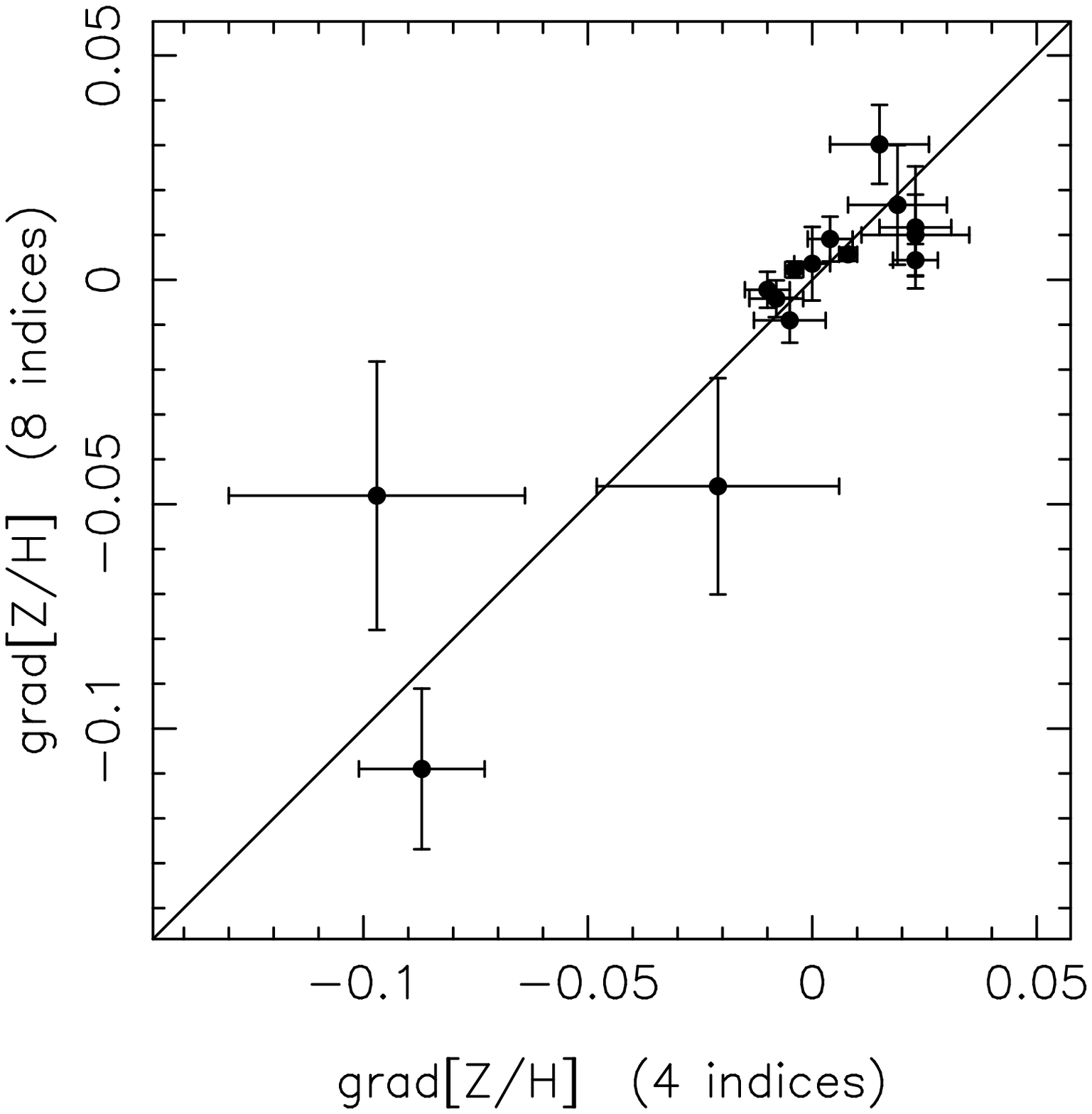}}
\resizebox{0.3\textwidth}{!}{\includegraphics[angle=-90]{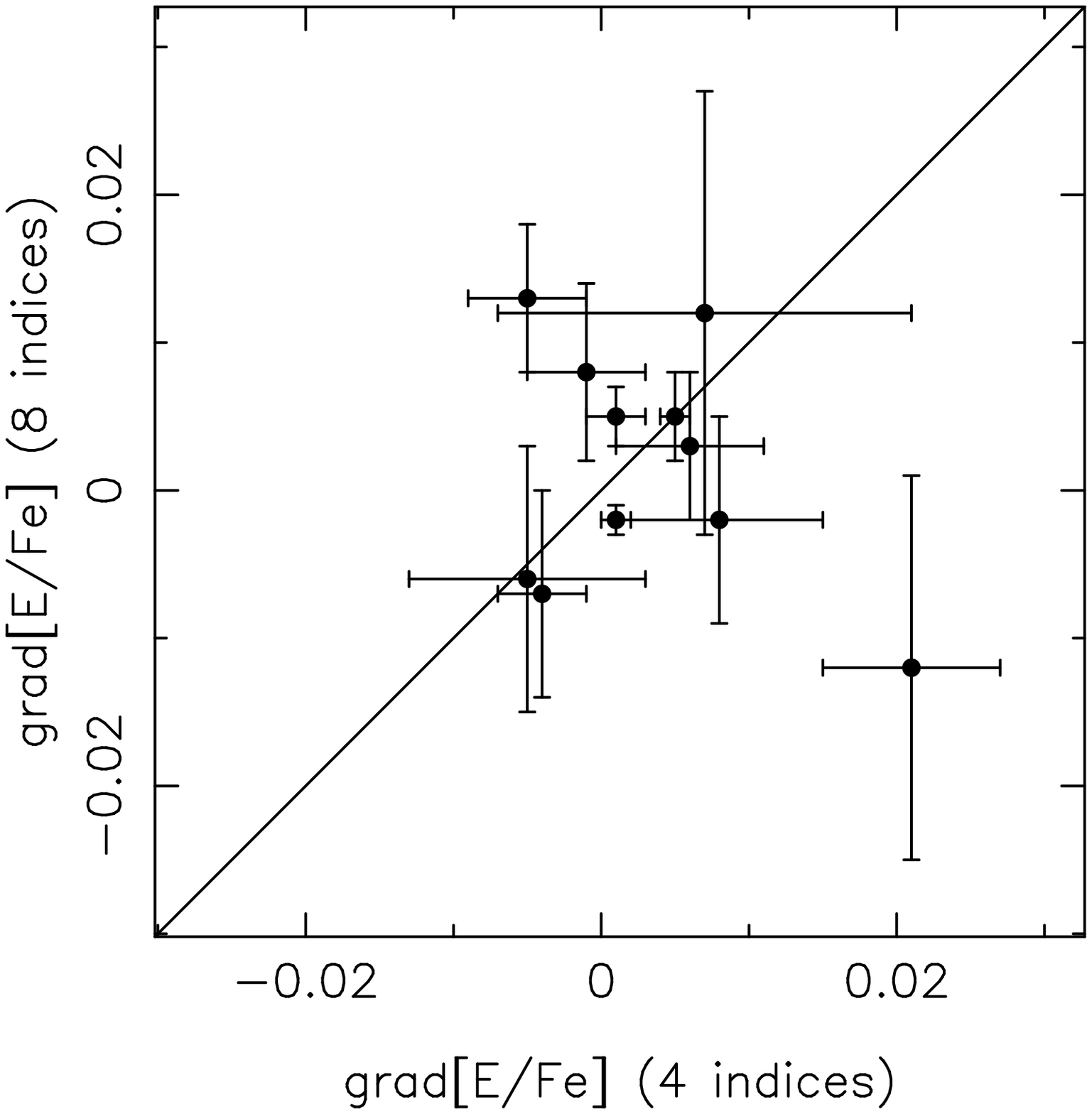}}
\caption{Comparison between the age, metallicity  gradients obtained using 4 indices or 
a combination of  8 indices not affected by emission (see text for details). There is very good agreement between the results obtained using different sets of index combination.\label{fig.comparaemision}}
\end{figure*}

The differences between the mean gradients using these set of indices and using 
the previous set of 4 indices are shown in Fig.~\ref{fig.comparaemision}. As can be seen from this figure, there is very good agreement between the results obtained using a different set of index combination giving confidence in the derived results. However, for the [E/Fe] gradients there are three galaxies for which the values seem to slightly differ from the two methods.


Before continuing with the presentation of the results, 
we want to remind the reader that
the stellar population parameters derived here are  single stellar population (SSP)-equivalent parameters
and can be interpreted as mean values weighted with the light of individual stars. In the case of the age, 
the mean values are going to be strongly biased towards the age of the youngest components.

\section{Results
\label{results}}

\subsection{Kinematics
\label{kinematics}}

Figure~\ref{kinematics.movel} shows the line-of-sight position-velocity diagrams and velocity dispersion along the bar major-axis
for the galaxies in our sample.
Close examination of these diagrams can help to reveal the nature of the structures in the bar region.  All the galaxies present disk like structures from their kinematics. These structures are characterized by a central velocity dispersion minima ($\sigma$ drops),
central $\sigma$
plateaus and/or a clear rotating inner disks or ring as derived from the  line-of-sight position-velocity  diagrams. We define the central $\sigma$ as the maximum velocity dispersion in the bulge region. We have chosen this dispersion value to avoid being dominated in some galaxies by the low central velocity dispersion that could be indicative of nuclear star formation. Therefore, the maximum velocity dispersion should be closer to the velocity dispersion of the bulge.

Most of our galaxies show  $"$double-hump$"$ rotation curve (Bureau \& Athanassoula 2005; Chung \& Bureau 2004), that is, the 
rotation curve first rises rapidly and reaches a local maximum, then drops slightly and then raises again slowly. These profiles
were detected in N-body simulations of edge-on barred disks by Bureau \& Athanassoula (2005). They appear in strong bars and are detected in small 
viewing angles. The fact that these features are detected in pure N-body simulations indicate that they could arise naturally from the orbital structure of a barred disk and therefore gas dissipation is not indispensable.
Central velocity dispersion minima occur in 70\% of our sample. Galaxies 
not showing a $\sigma$-drop are : NGC~2859, NGC~2935, NGC~2950, NGC~2681, NGC~2217, NGC~1169 and NGC~1358. Although, the NGC~2950, NGC~2217 and NGC~2859 shallow and wide central dip could be considered as well as a $\sigma$-drop.
Chung \& Bureau (2004)\nocite{chung04} found $\sigma$-drops in 40\% of their barred sample, while Peletier et al. (2007)\nocite{peletier07}
found it in 50\% of their sample of 24 barred and unbarred galaxies. M\'arquez et al. (2003) \nocite{marquez03}reported that 12 out of 14 of the analyzed  
galaxies with $\sigma$-drops showed an inner disk from their HST images. The standard explanation for these drops is the presence
of cold central stellar disks originating from gas inflow (Emsellem et al. 2001; Wozniak et~al.\ 2003).
Wozniak \& Champavert (2006)\nocite{wozniak06} predicted from dynamical simulations of drops in barred
galaxies that these features can last for more than 1 Gyr if the central region is continuously fed by fresh gas leading to a continuos
star formation activity. In their simulations, continuous star formation is needed to replace the heated particles by new low-$\sigma$ ones.
They predict that  as the star formation disappears, the drop will disappear as well. However, this disappearance is not instantaneus, they find that a $\sigma$-drop dissapears at a linear rate of 10~\kms in 500~Myr, when star formation is switched off.
On the contrary, Athanassoula \& Misiriotis (2002) found these features in N-body dissipationless simulations, and Bureau \& Athanassoula (2005) showed
that they can arise from the orbital structure of strongly barred galaxies. These authors noted that, in their simulations of strong barred
galaxies, the width of the central flat section of the velocity dispersion profiles was roughly the same as that of the rapidly rising 
part of the rotation curve, and suggest a common origin ($x1$ orbits). However, these N-body simulations have no stellar populations predictions and, therefore, it is difficult to compare to our observations.
These two possible $\sigma$-drop origins (cold dissipative disk vs. bar orbital stellar structure) can be separated by studying the age in the central parts of the galaxies
with $\sigma$-drops.  If the $\sigma$-drop has to be maintained by continuous gas replenishment and star formation, we would expect a very young population dominating the central regions. On the other hand, if the drops are due to an old stellar structure, we would expect the central parts to be dominated by an old stellar component.
We will present the stellar population of the central parts of these galaxies in the second series of the paper, but 
we can already advance that, while most of these galaxies have, indeed, young components and star formation in their centers, eight galaxies (NGC~1358, NGC~2523, NGC~2859, NGC~2962, NGC~4245, NGC~4314, NGC~4643 and NGC~5101) have old central luminosity weighted ages ($>$ 3 Gyr) despite having central dips or plateaus.  However, this fact does not, necessarily, imply that the ages of the cold component are old, since we are measuring luminosity weighted ages and a bulge component could be contributing to the derived old age. It would be necessary to disentangle the populations coming from both components. This will be done in detail in a forthcoming paper addressing the central regions of barred galaxies.

Eight out of the 20 galaxies are classified as double barred (Erwin 2004) NGC~1433, NGC~2217, NGC~2681, NGC~2859, NGC~2950, NGC~2962, 
NGC~3081, NGC~4314.  They do not show any difference in their kinematic profiles, although all of them show a plateau in their central stellar
 velocity dispersion. 
\begin{figure*}
 \begin{center}
\resizebox{0.48\textwidth}{!}{\includegraphics[angle=-90]{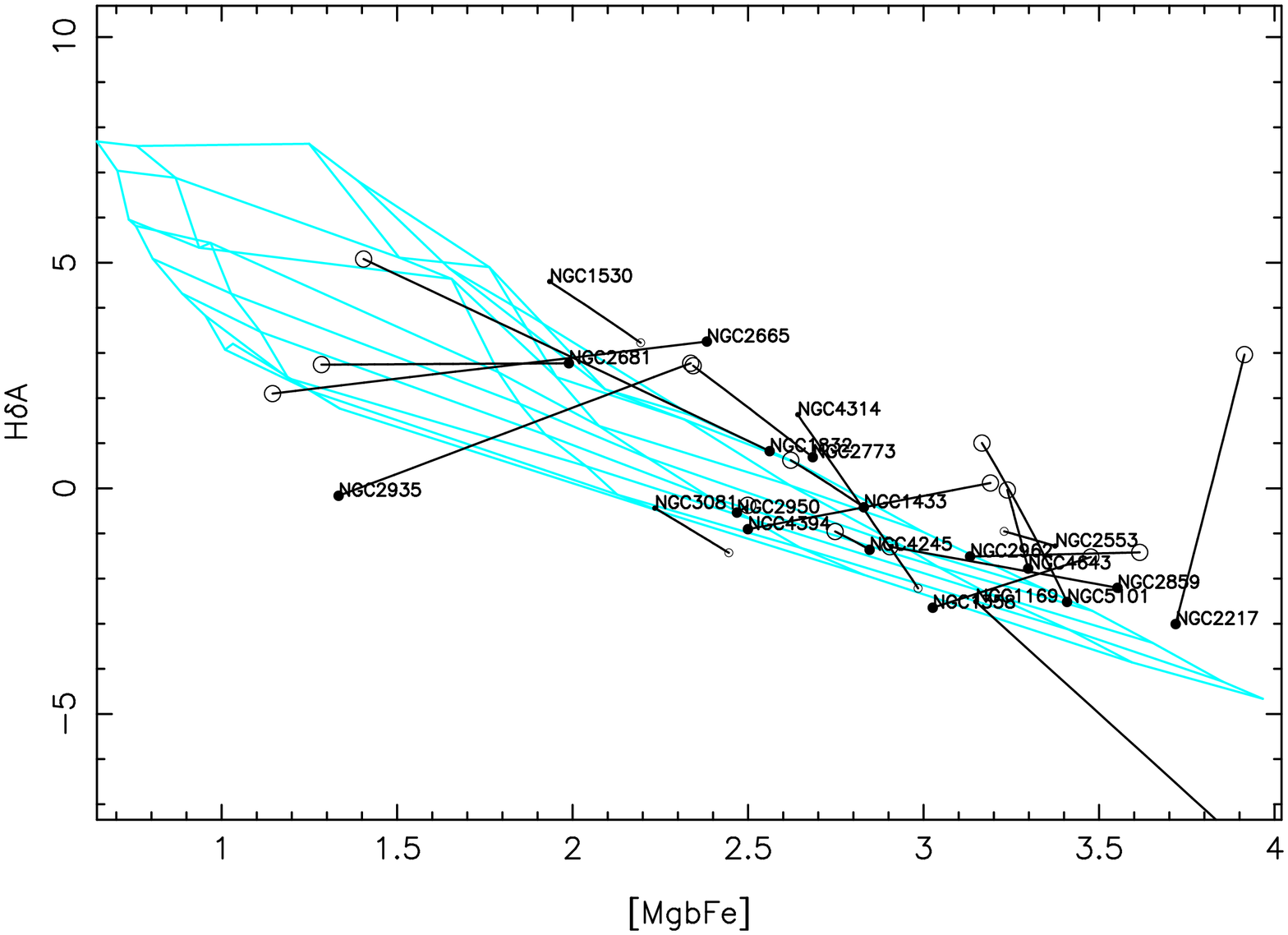}}
\resizebox{0.48\textwidth}{!}{\includegraphics[angle=-90]{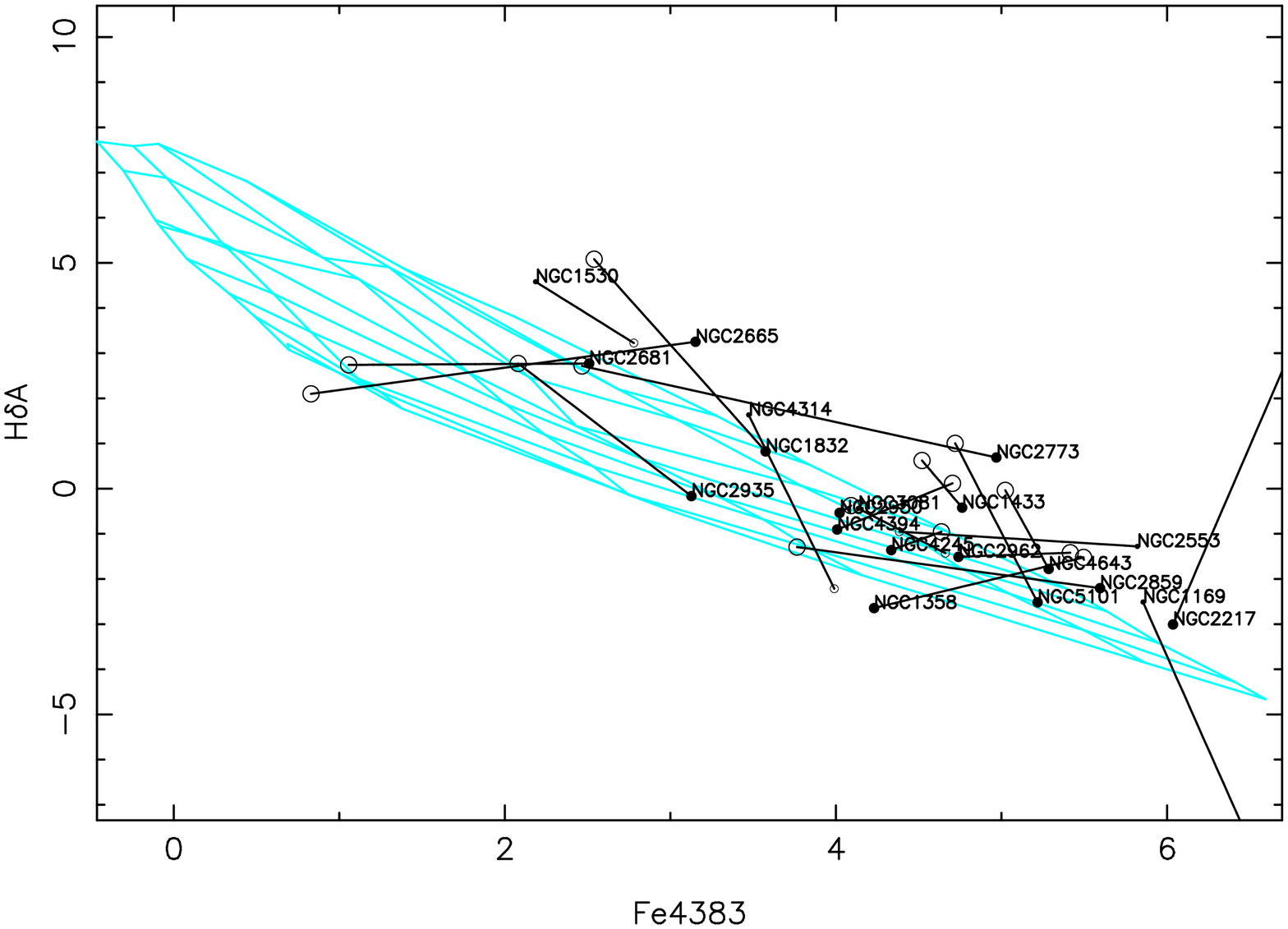}}
\caption{ \label{bargrad} [MgFe] and Fe4383 indices vs. H$\delta_A$ measured in the inner (filled symbols) and 
outer (open symbols) parts of the bar. Overplotted are the models by Vazdekis. Models are shown with ages~=~ 1,1.41,2.00,2.82,3.98,5.62,7.98,11.22,15.85 (Gyr) from top to bottom, and metallicities, [Z/H]~=~-1.68,-1.28,-0.68,-0.38,0.0,0.2 from left to right.}
\end{center}
\end{figure*}

\begin{figure*}[h]
\resizebox{0.3\textwidth}{!}{\includegraphics[angle=-90]{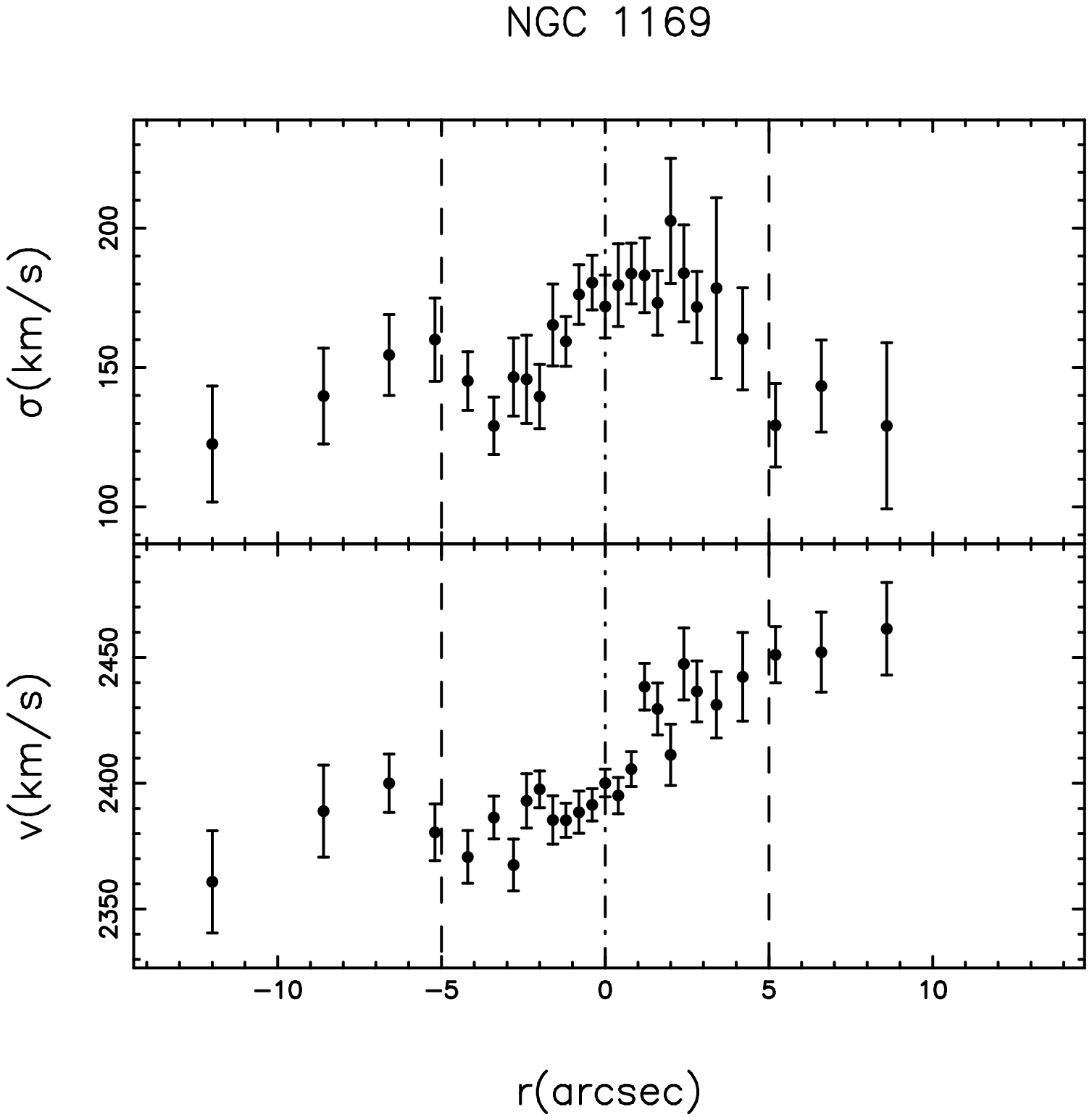}}
\resizebox{0.3\textwidth}{!}{\includegraphics[angle=-90]{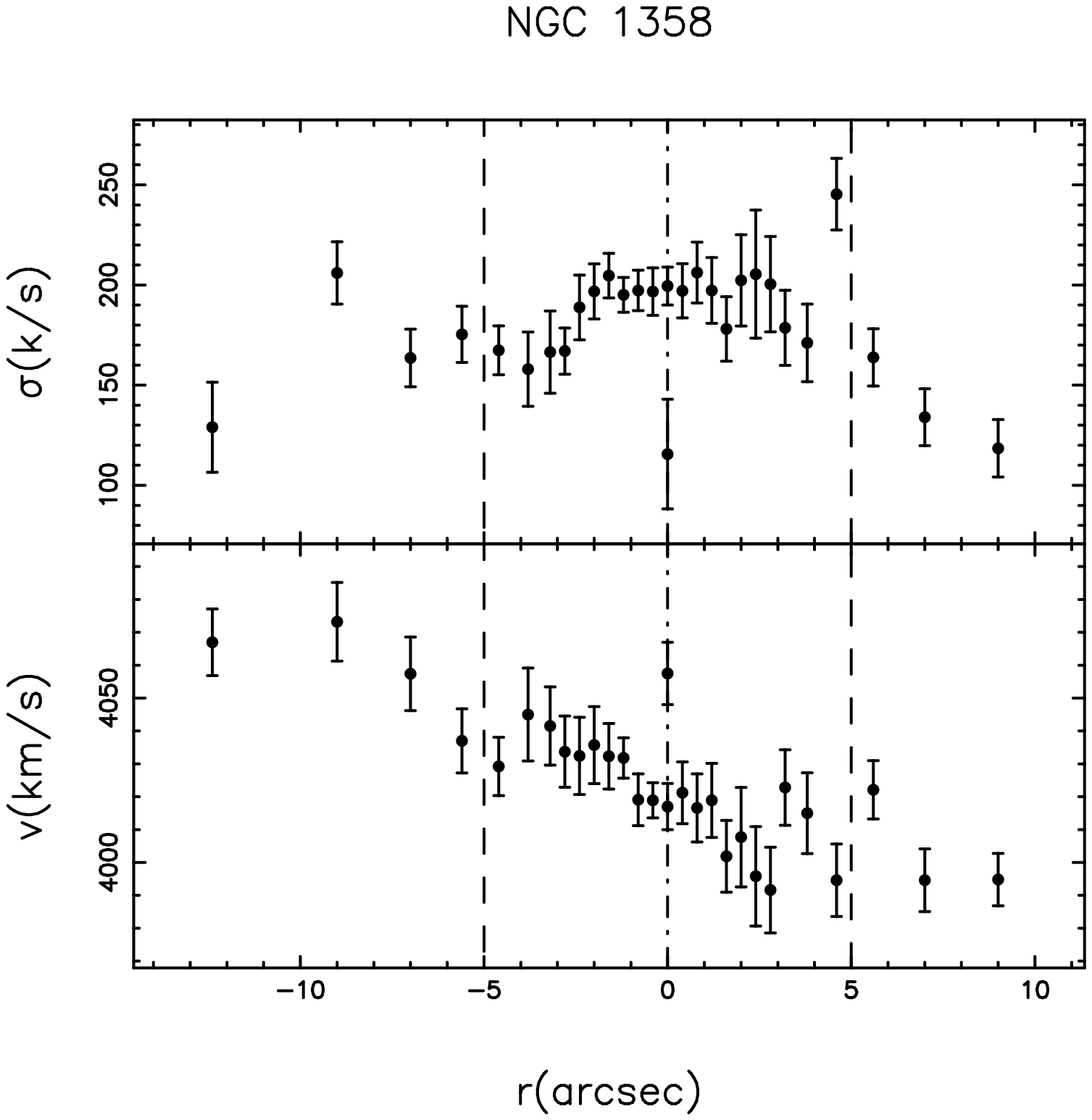}}
\resizebox{0.3\textwidth}{!}{\includegraphics[angle=-90]{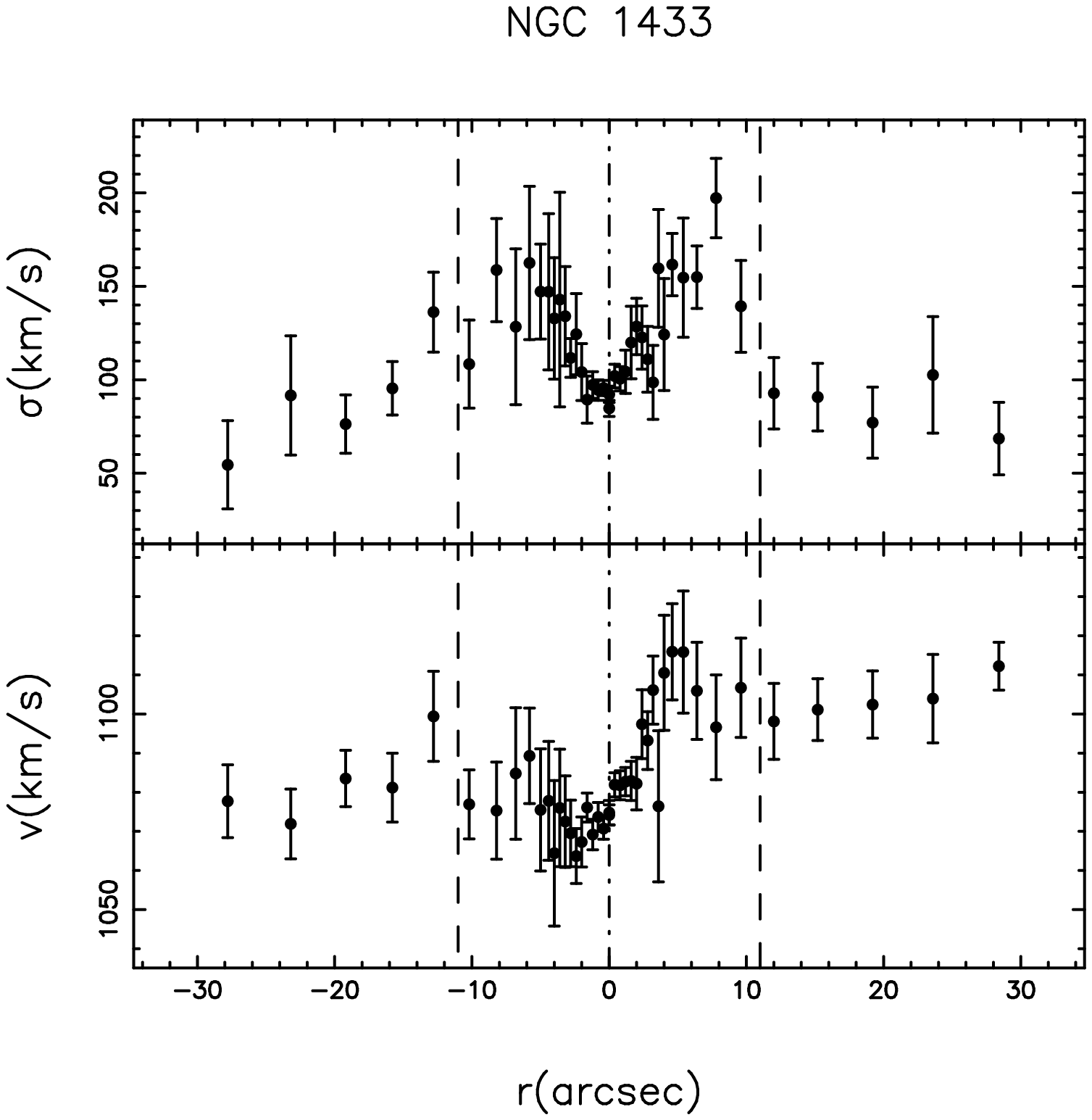}}
\resizebox{0.3\textwidth}{!}{\includegraphics[angle=-90]{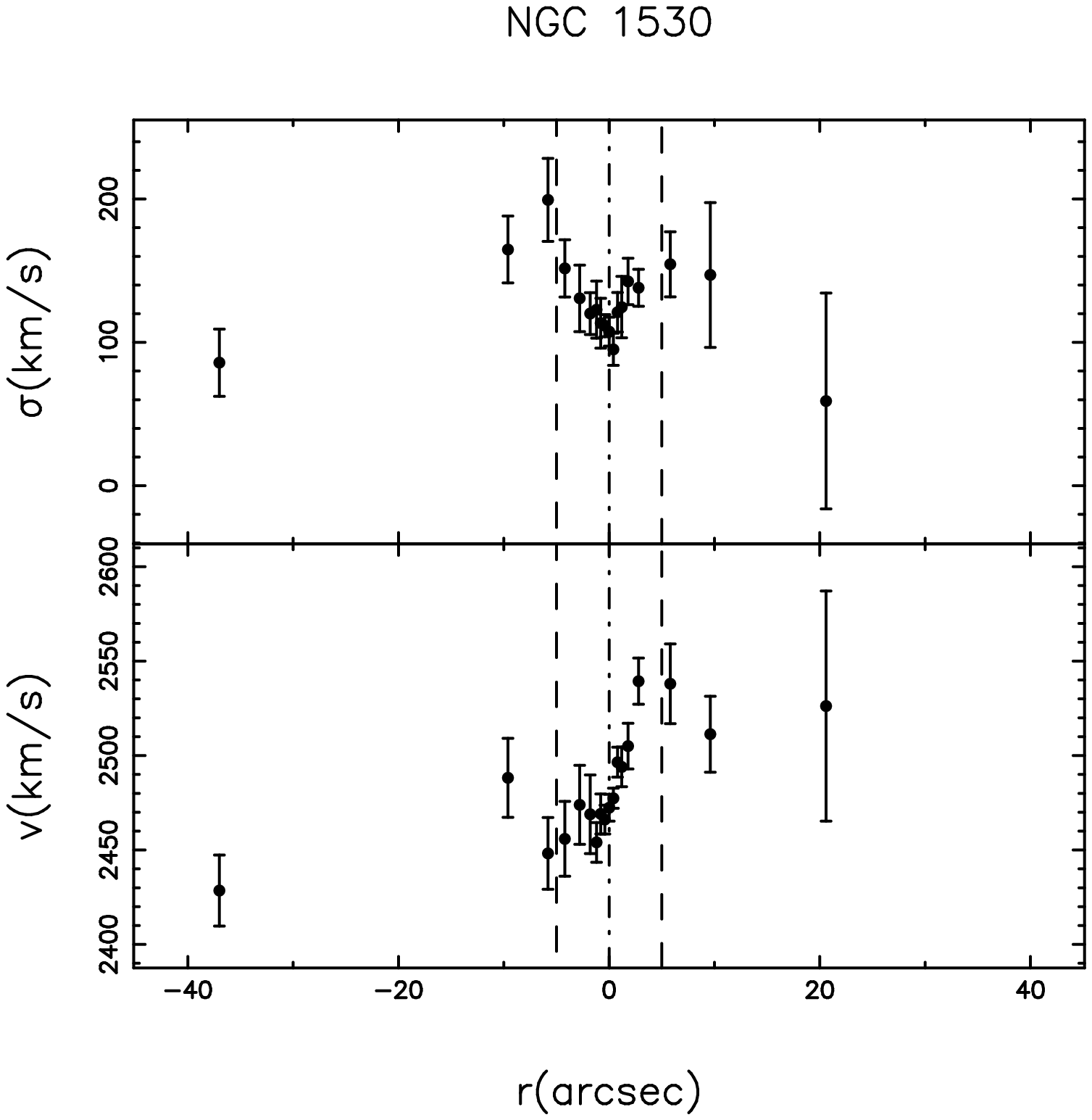}}
\resizebox{0.3\textwidth}{!}{\includegraphics[angle=-90]{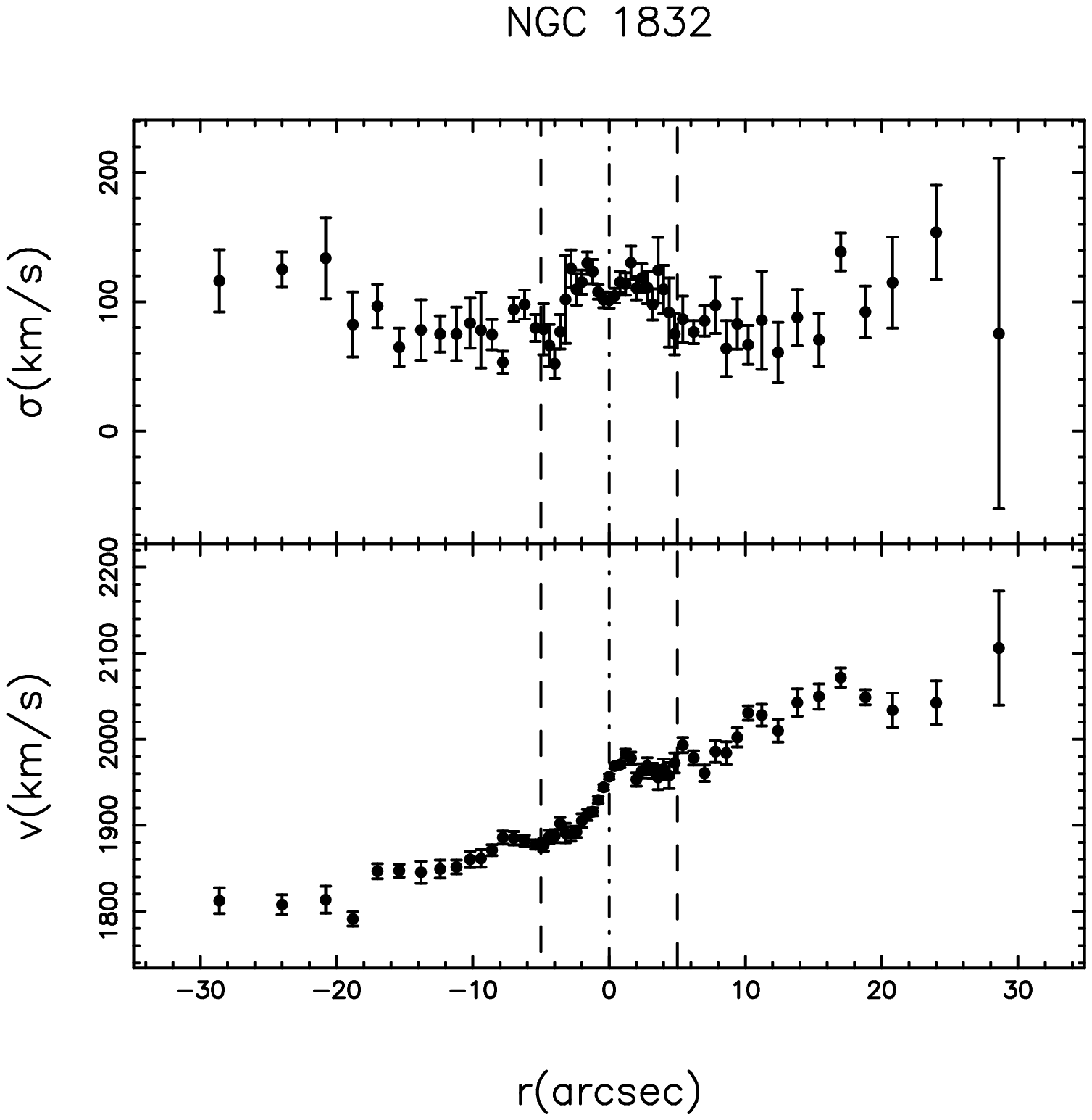}}
\resizebox{0.3\textwidth}{!}{\includegraphics[angle=-90]{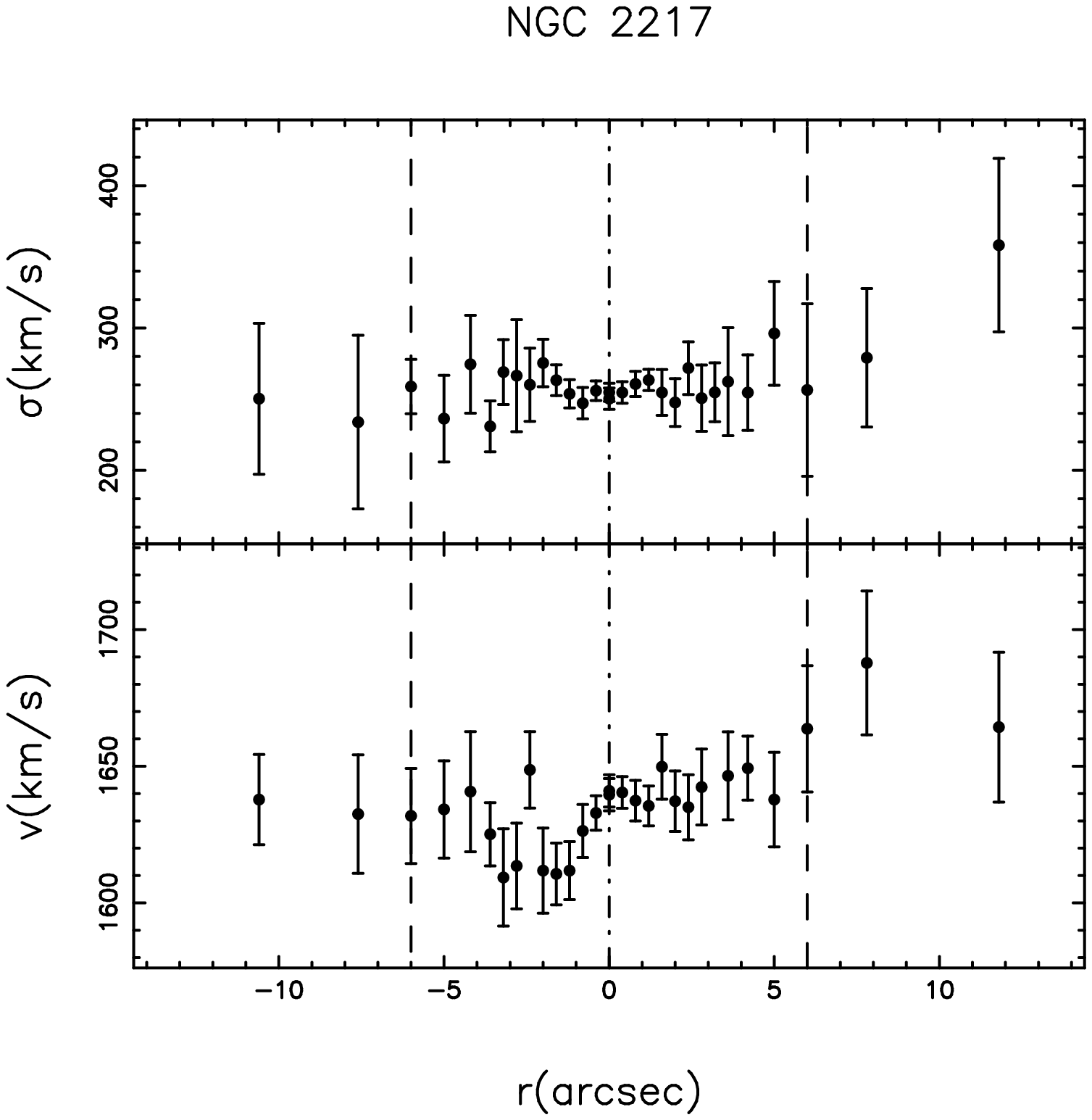}}
\resizebox{0.3\textwidth}{!}{\includegraphics[angle=-90]{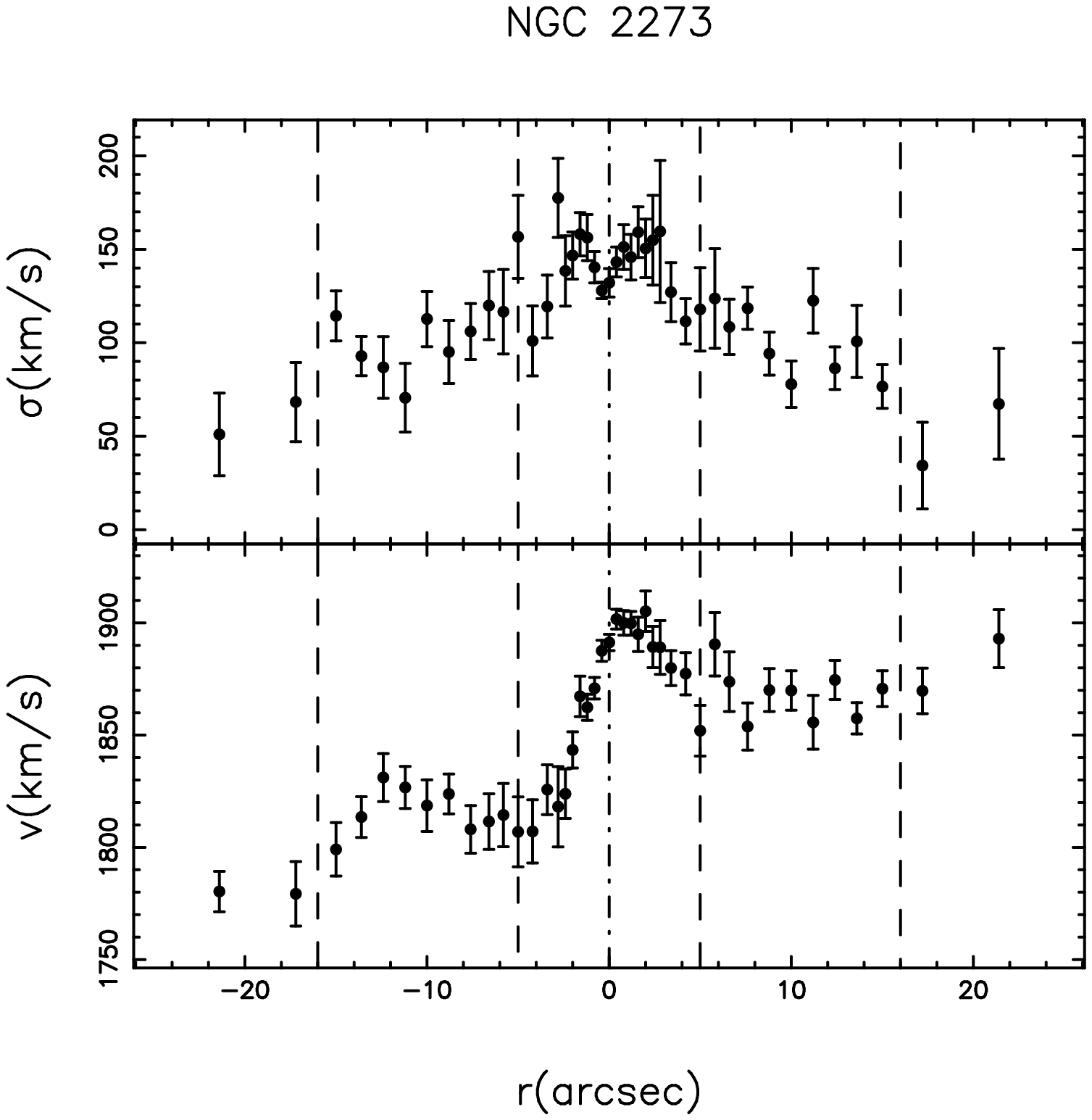}}\hspace{0.8cm}
\resizebox{0.3\textwidth}{!}{\includegraphics[angle=-90]{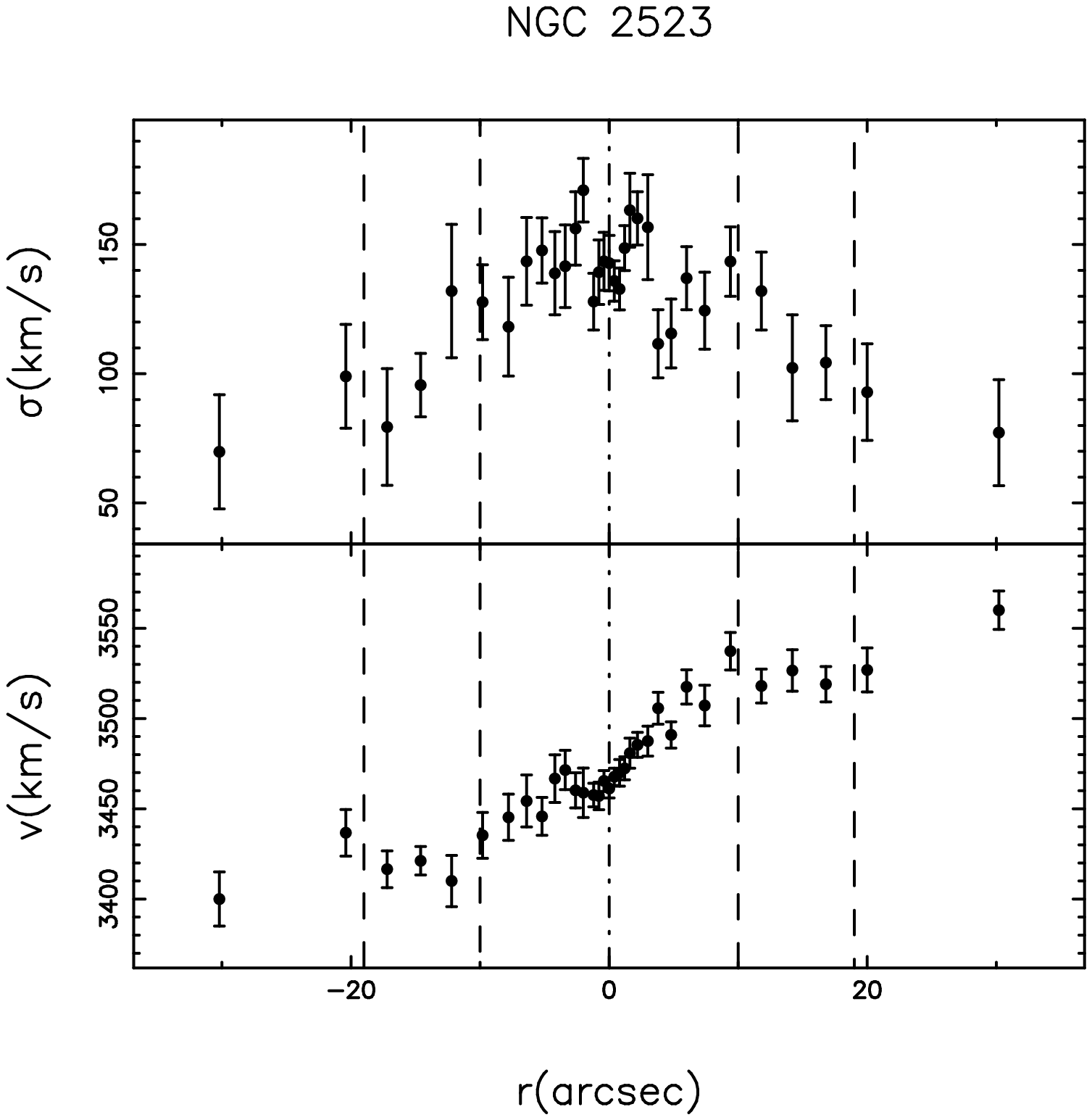}}\hspace{0.8cm}
\resizebox{0.3\textwidth}{!}{\includegraphics[angle=-90]{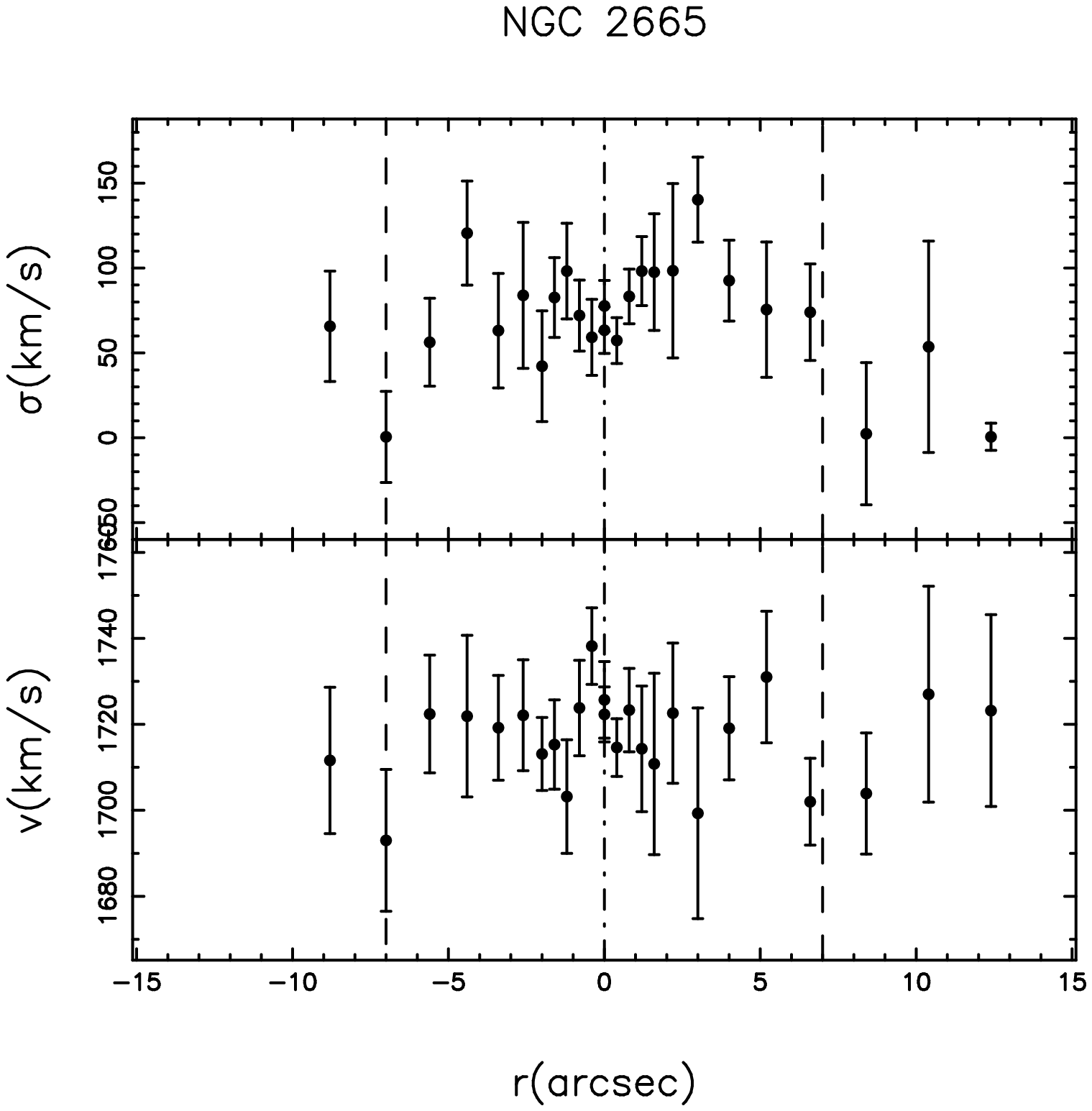}}
\caption{Position-velocity diagrams and velocity dispersion profiles along the bar major-axis for our sample of galaxies. Dashed lines 
indicate the beginning of  the region dominated by the bar.\label{kinematics.movel}}
\end{figure*}
\addtocounter{figure}{-1}
\begin{figure*}[h]
\resizebox{0.3\textwidth}{!}{\includegraphics[angle=-90]{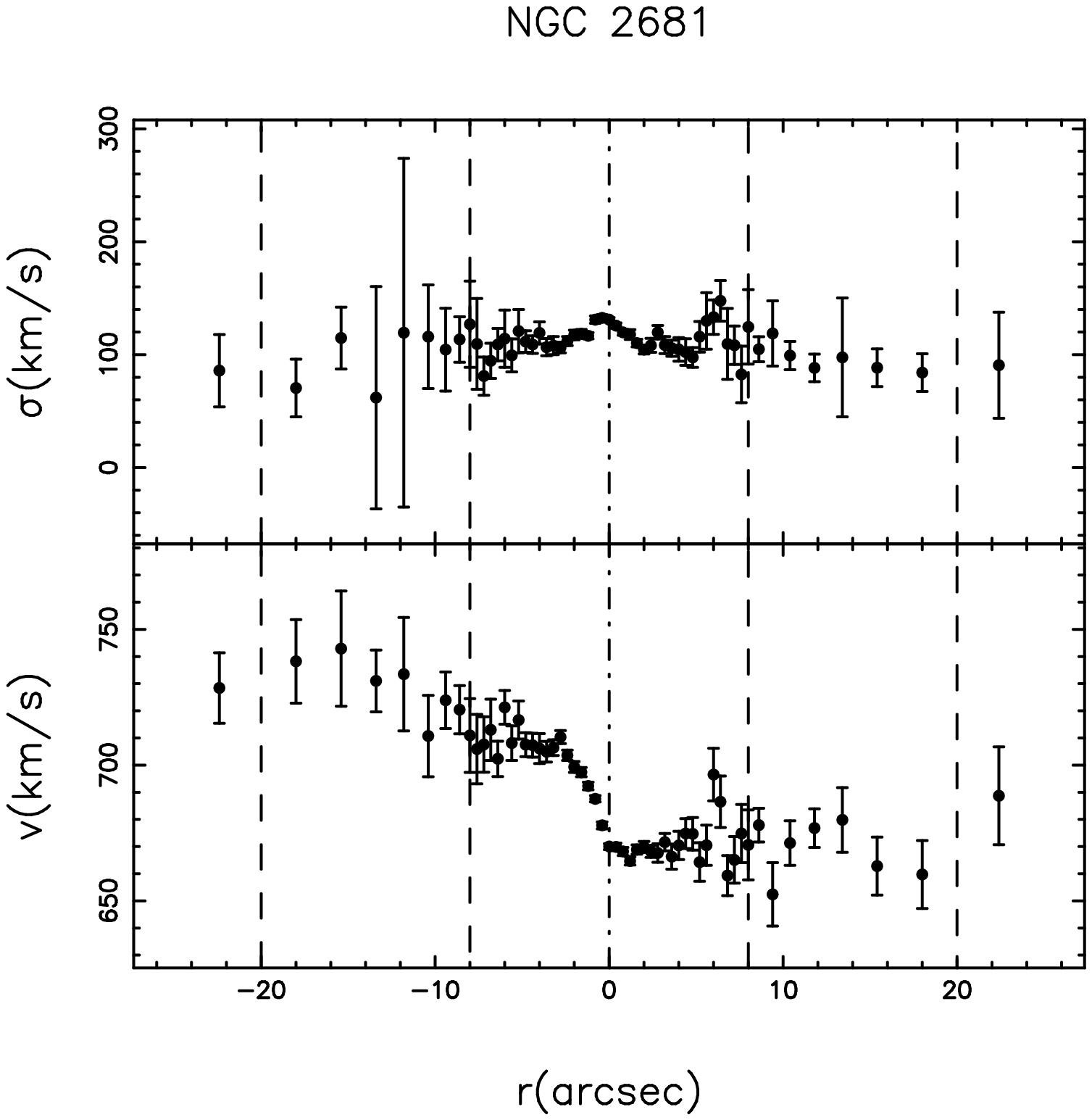}}
\resizebox{0.3\textwidth}{!}{\includegraphics[angle=-90]{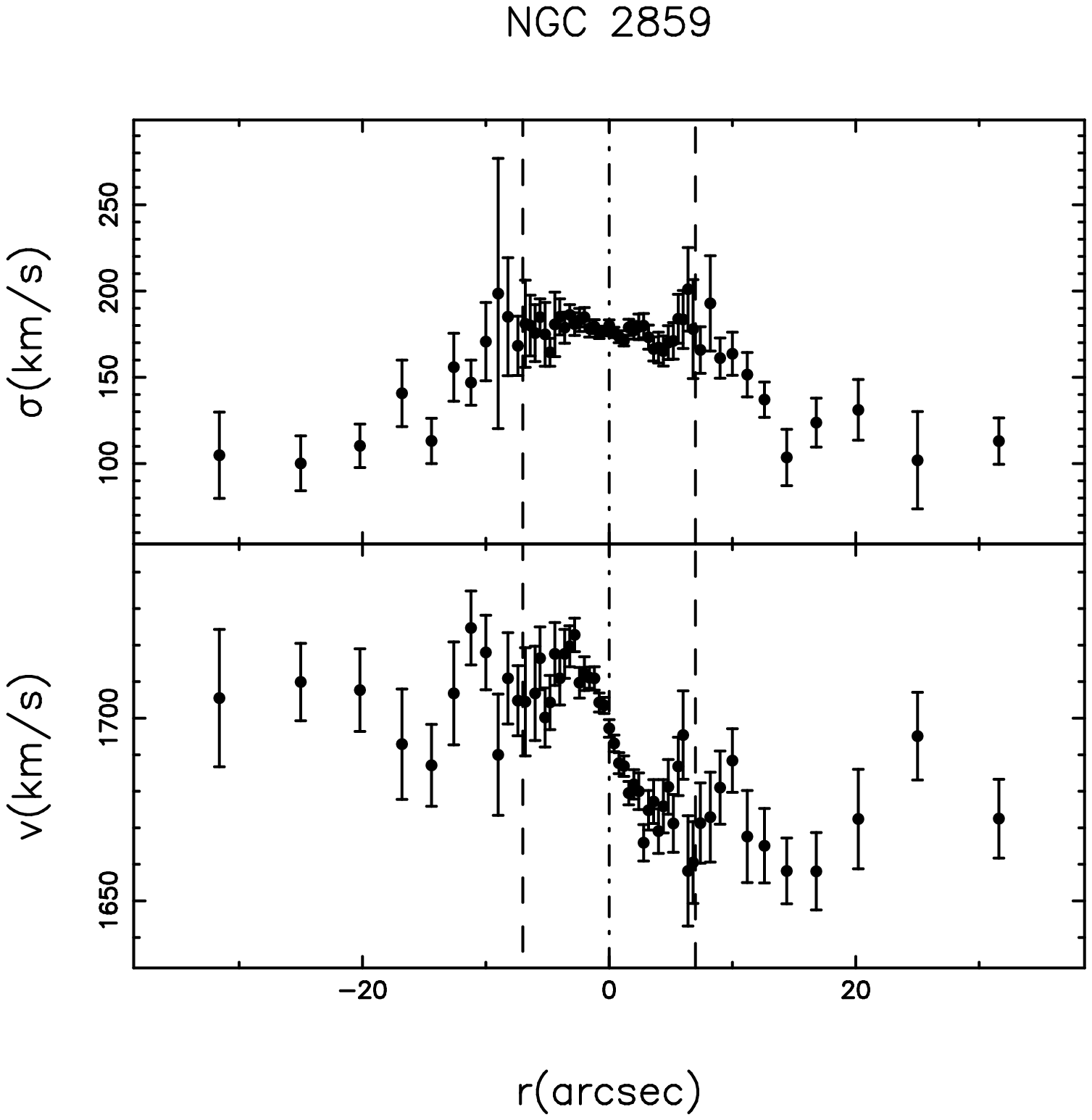}} 
\resizebox{0.3\textwidth}{!}{\includegraphics[angle=-90]{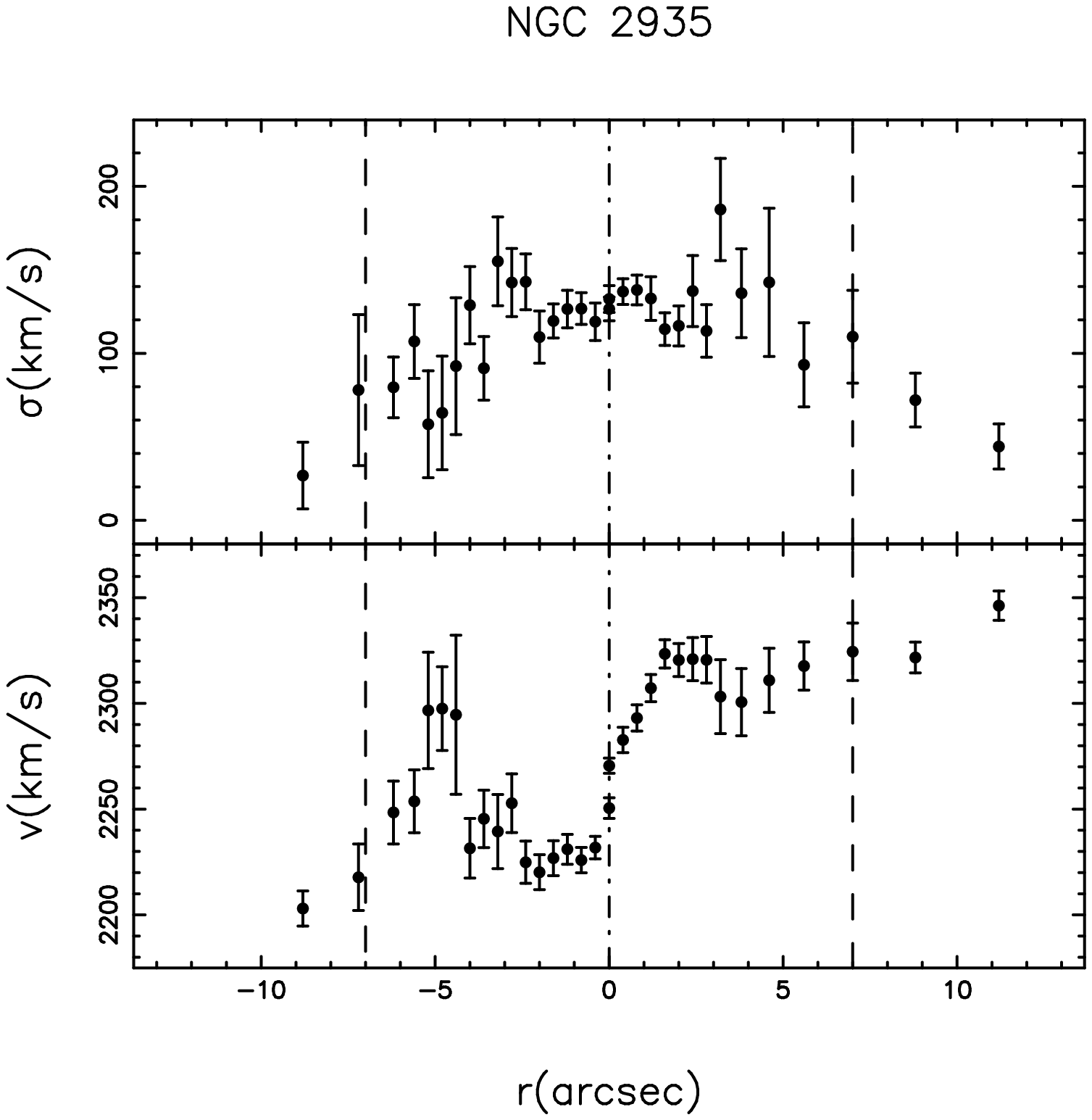}} 
\resizebox{0.3\textwidth}{!}{\includegraphics[angle=-90]{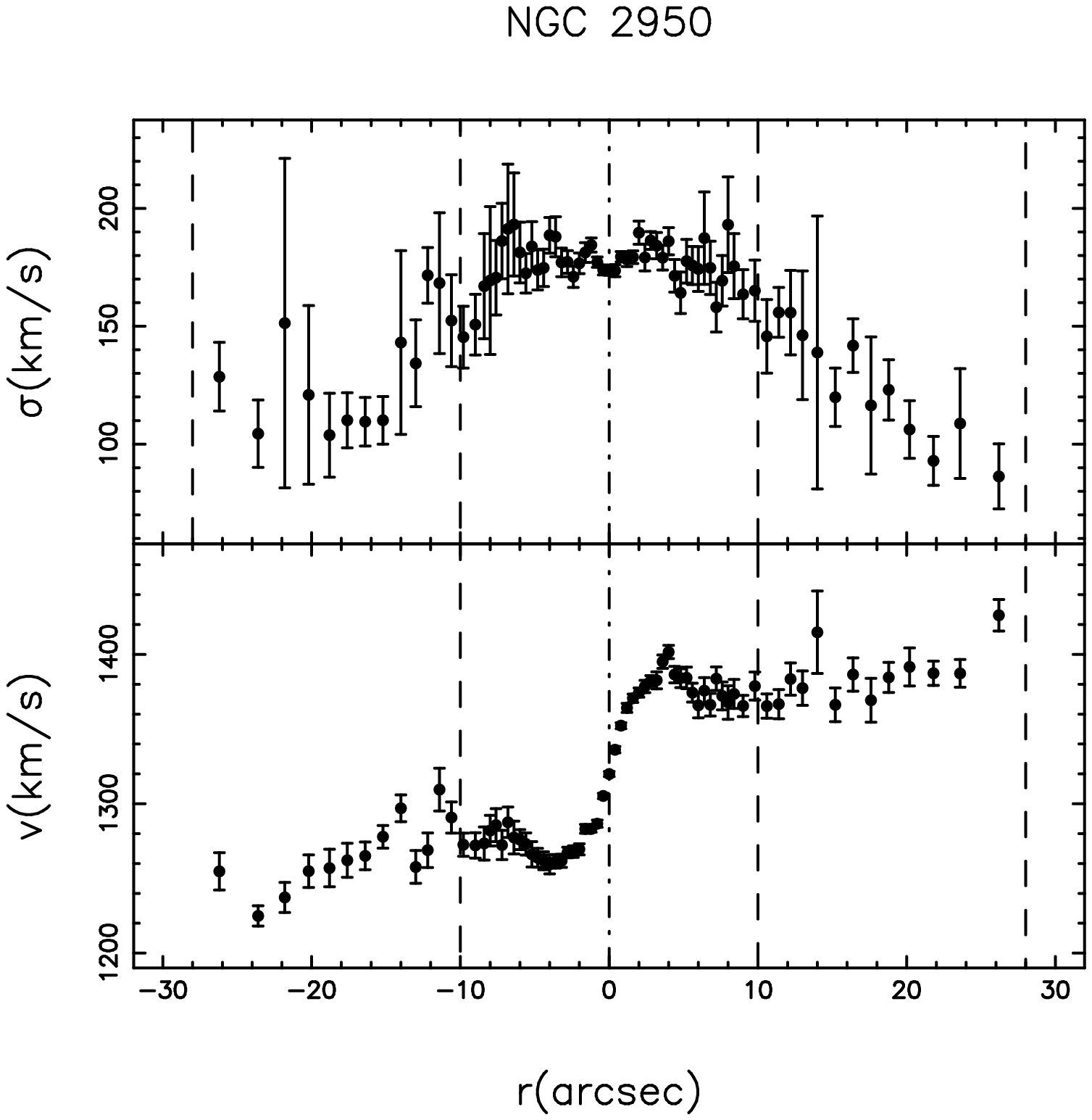}}
\resizebox{0.3\textwidth}{!}{\includegraphics[angle=-90]{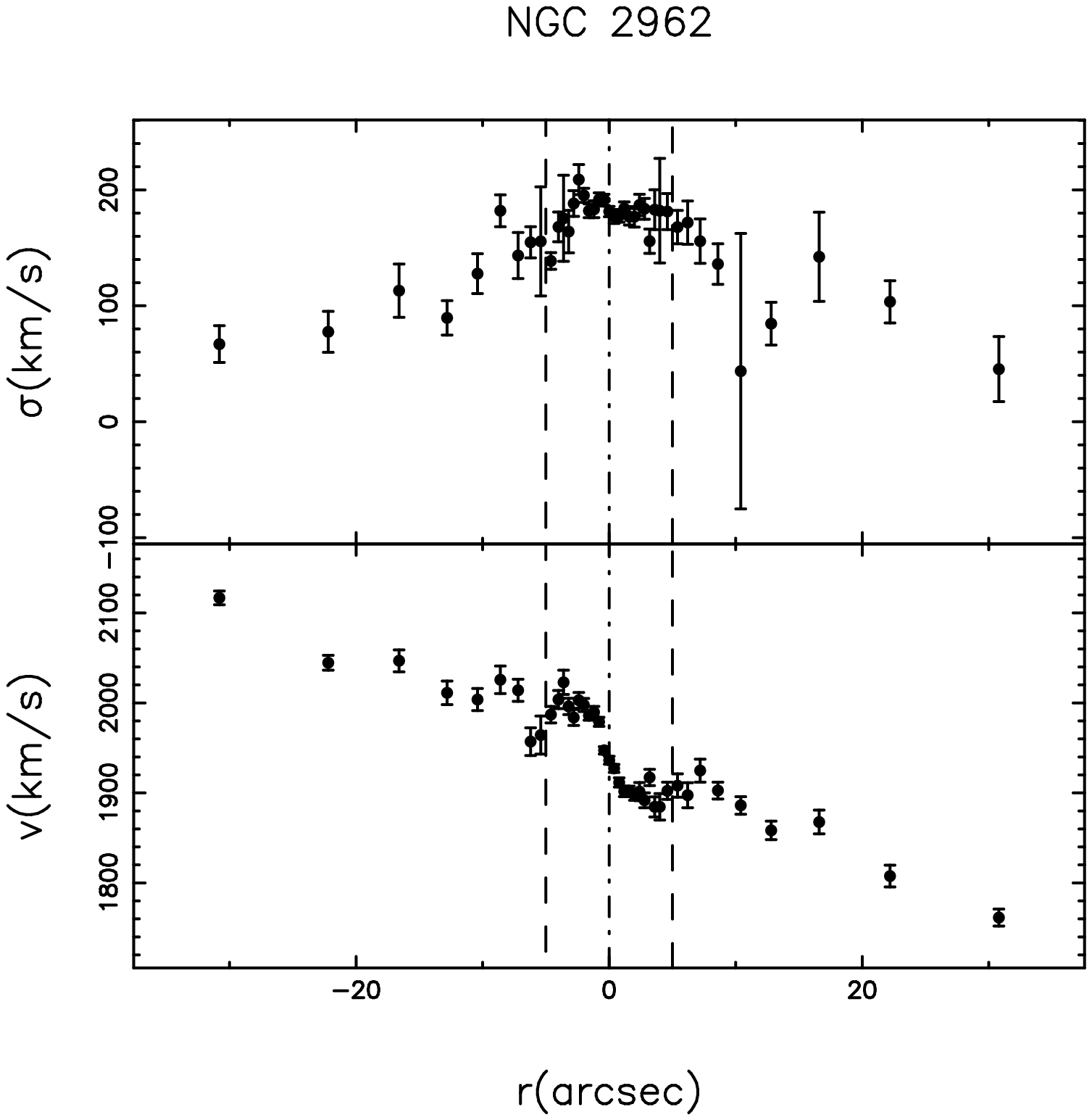}}
\resizebox{0.3\textwidth}{!}{\includegraphics[angle=-90]{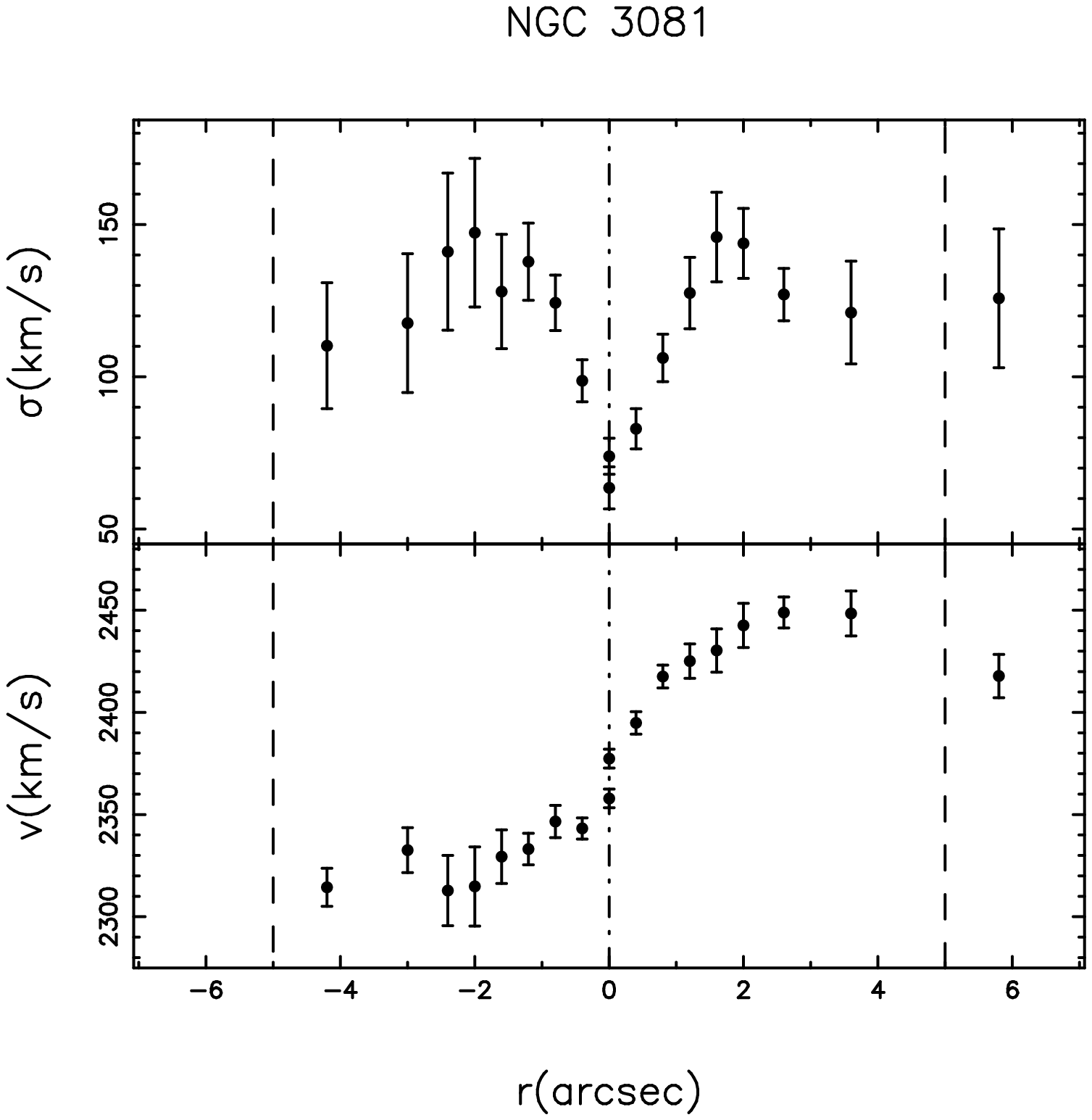}}
\resizebox{0.3\textwidth}{!}{\includegraphics[angle=-90]{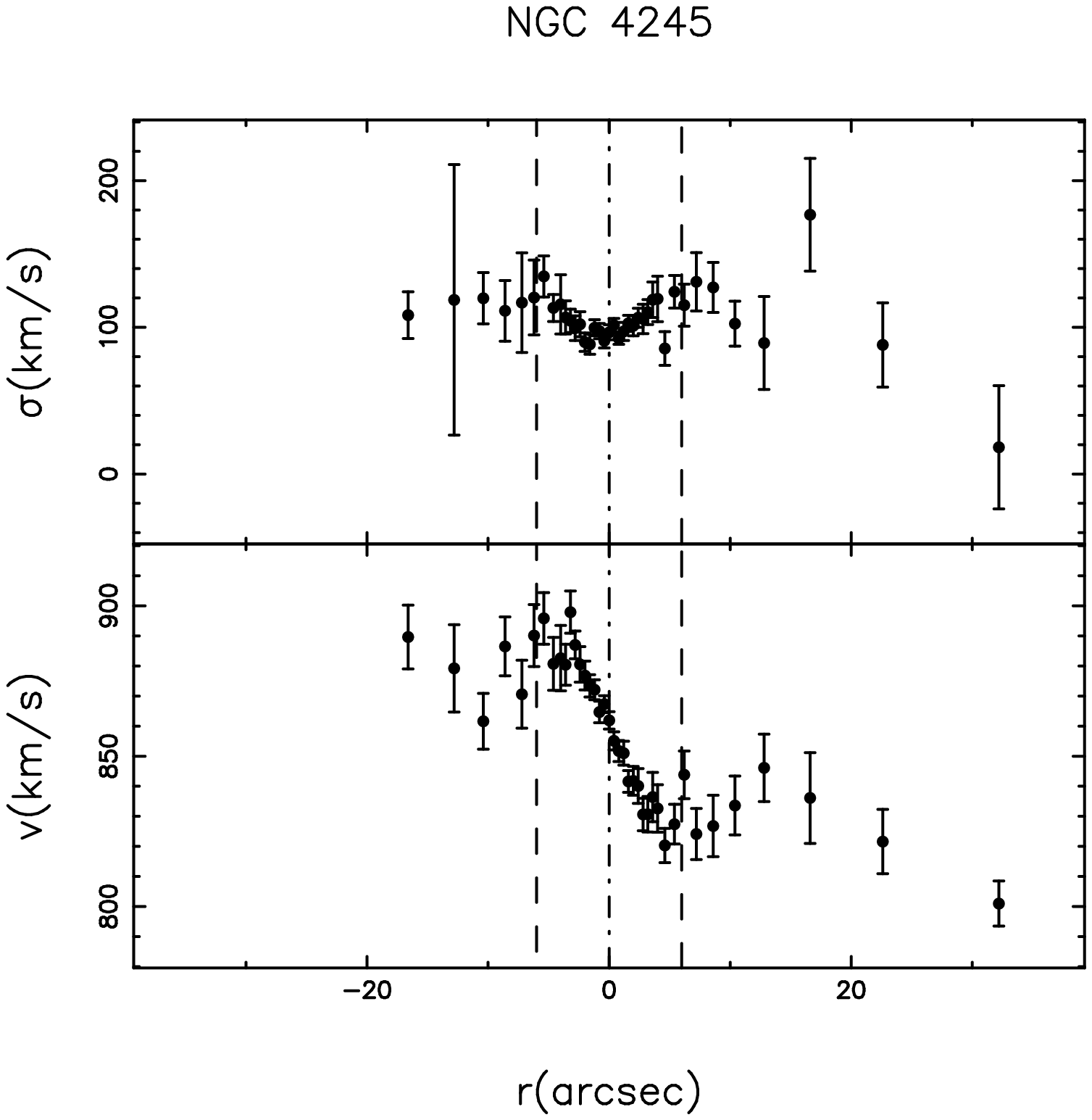}}
\resizebox{0.3\textwidth}{!}{\includegraphics[angle=-90]{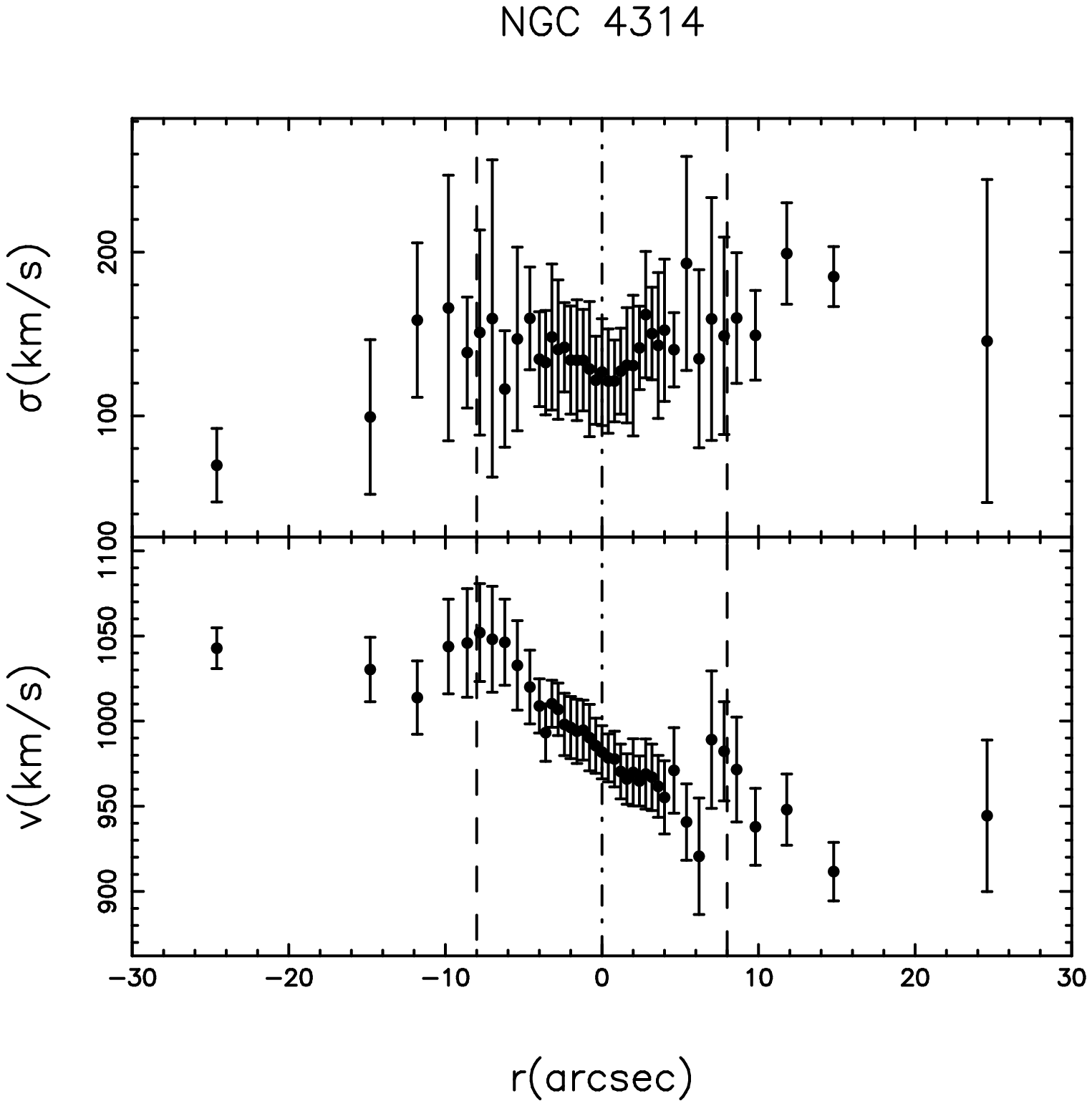}}
\resizebox{0.3\textwidth}{!}{\includegraphics[angle=-90]{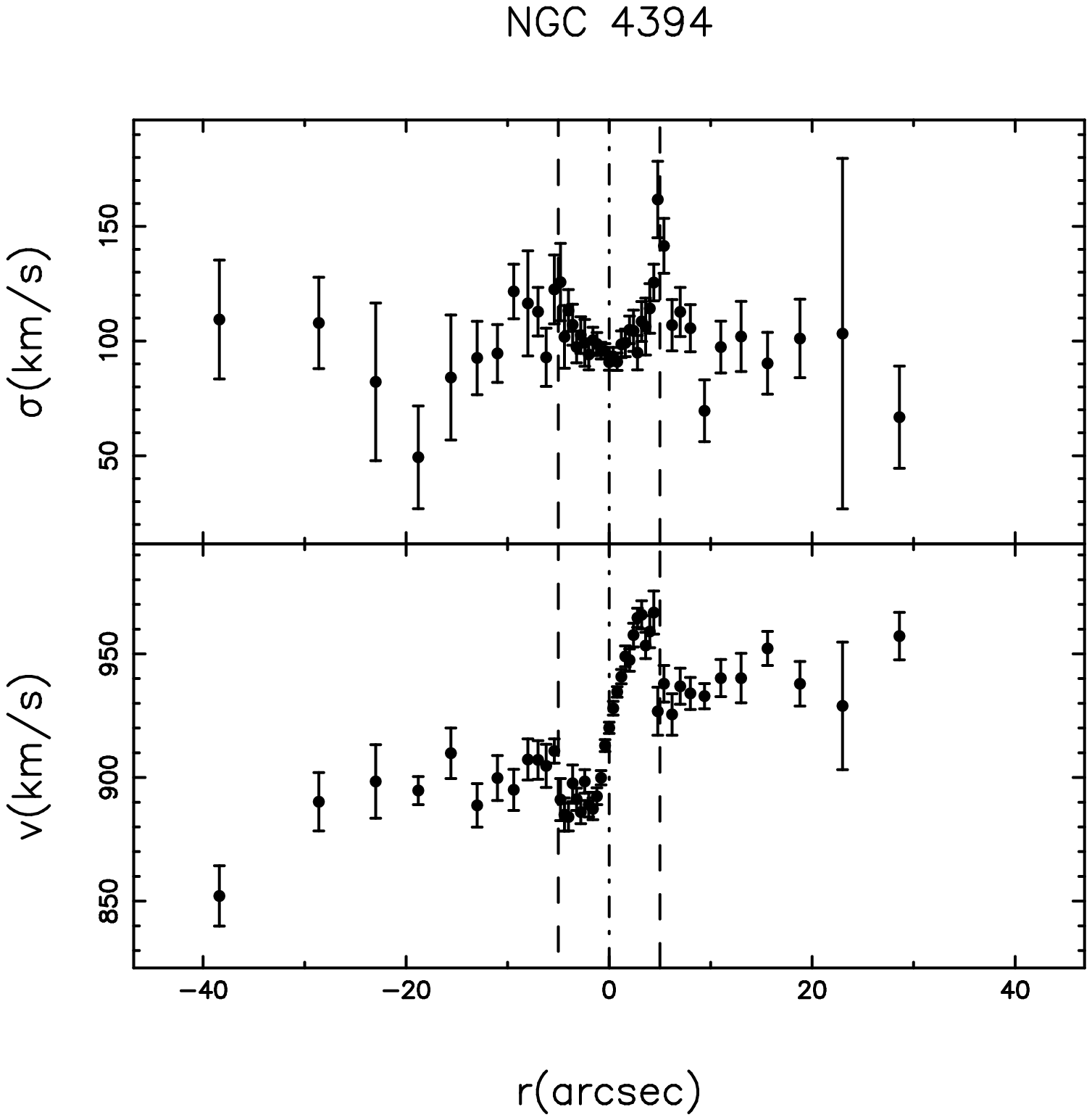}} 
\resizebox{0.3\textwidth}{!}{\includegraphics[angle=-90]{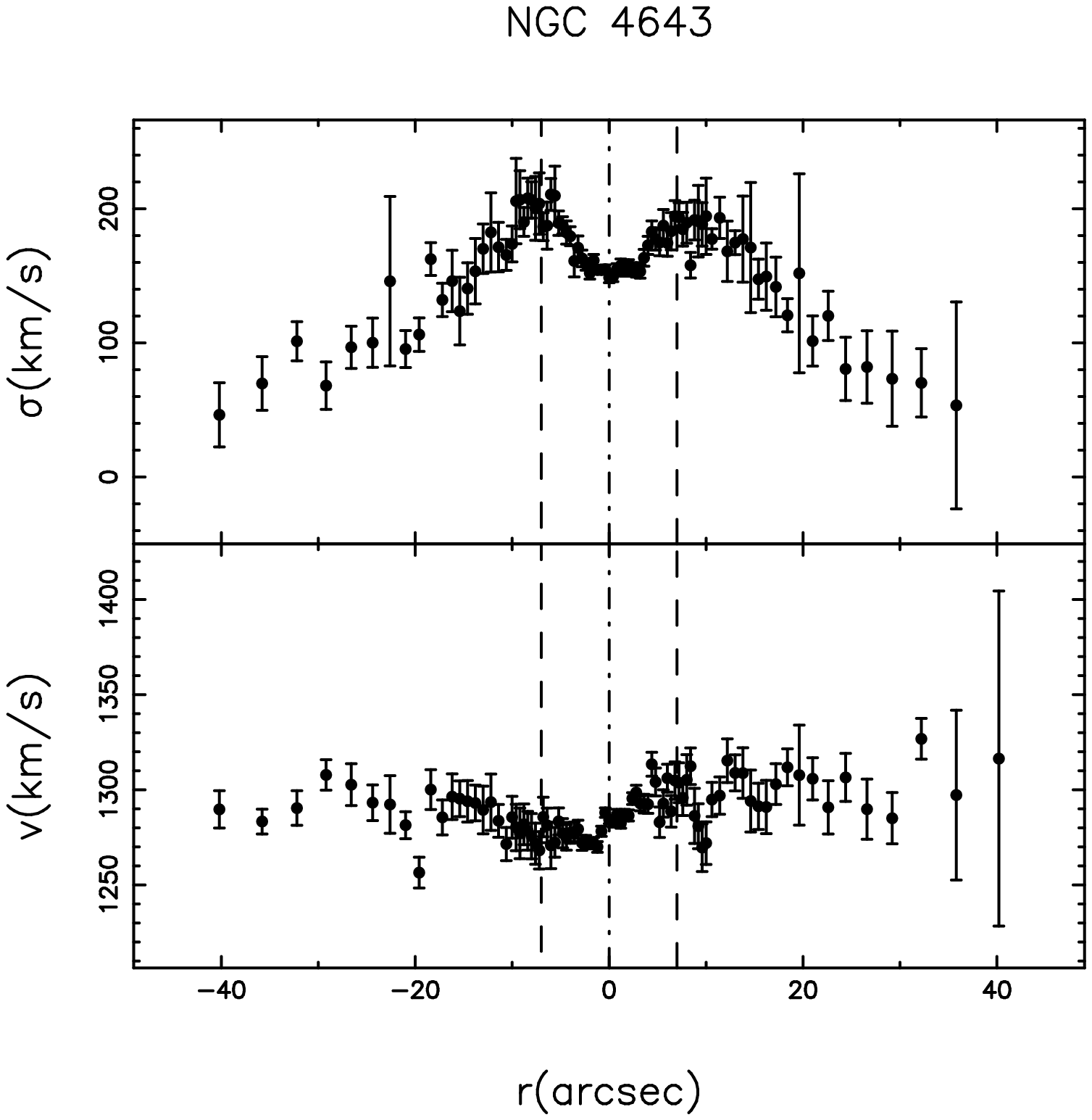}}\hspace{0.8cm}
\resizebox{0.3\textwidth}{!}{\includegraphics[angle=-90]{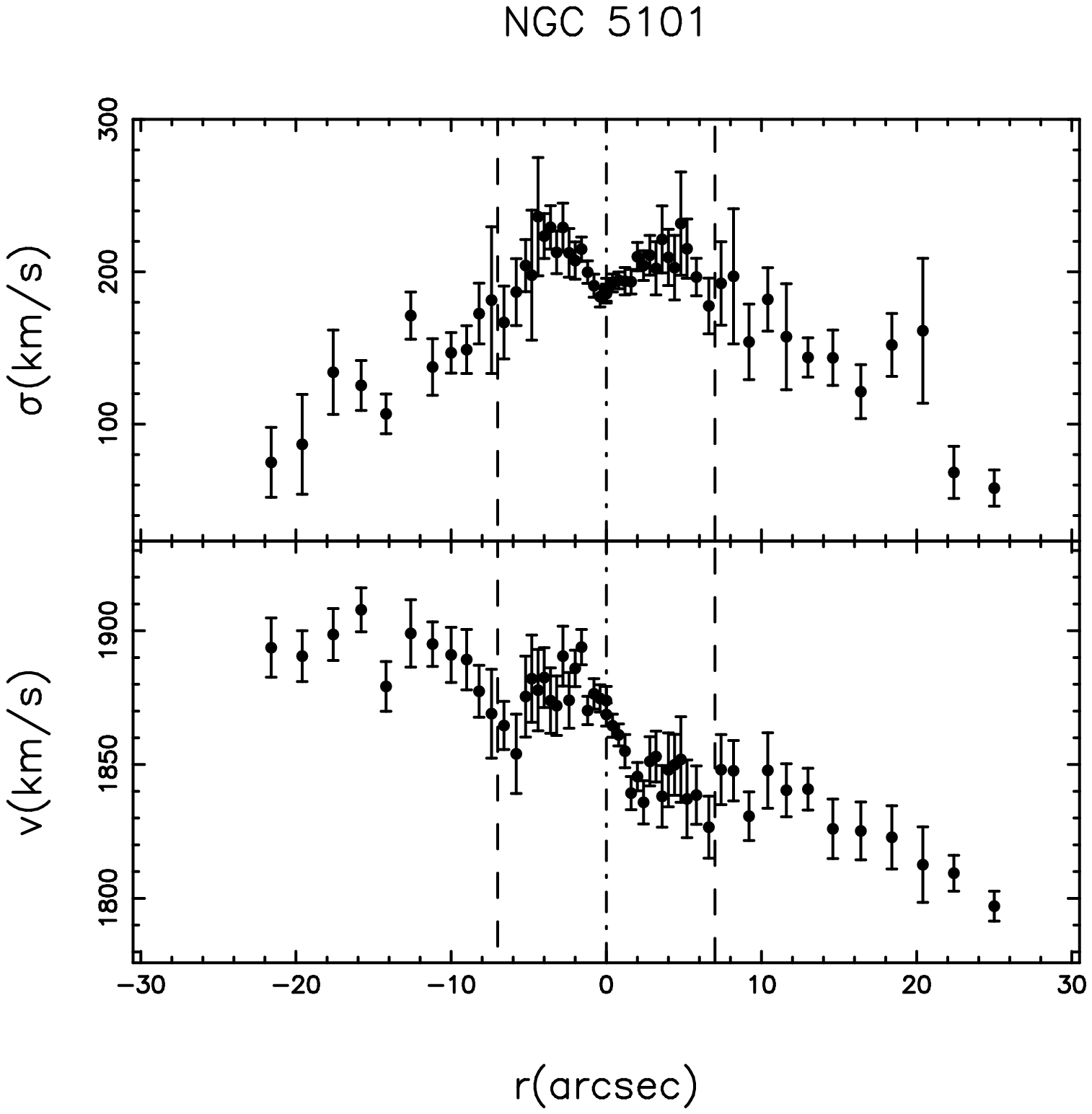}}
\caption{ Position-velocity diagrams and velocity dispersion profiles along the bar major-axis for our sample of galaxies. Dashed lines indicate the beginning of  the region dominated by the bar}
\end{figure*}

\subsection{Mean stellar population parameters in the bar region
\label{mean.stellar}}
 
\begin{figure*}
\hspace{0.5cm}
\resizebox{0.33\textwidth}{!}{\includegraphics[angle=0]{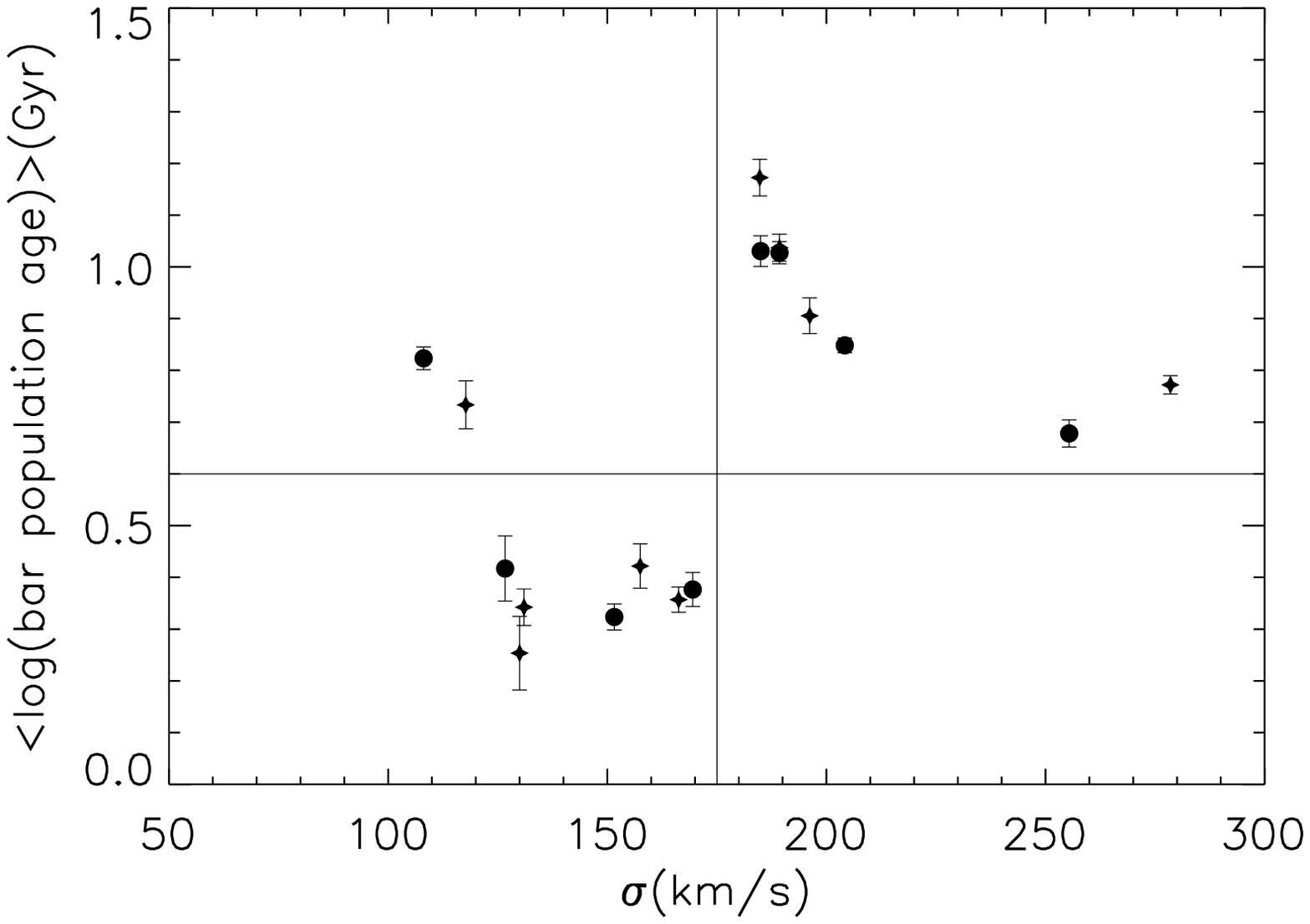}}
\resizebox{0.33\textwidth}{!}{\includegraphics[angle=0]{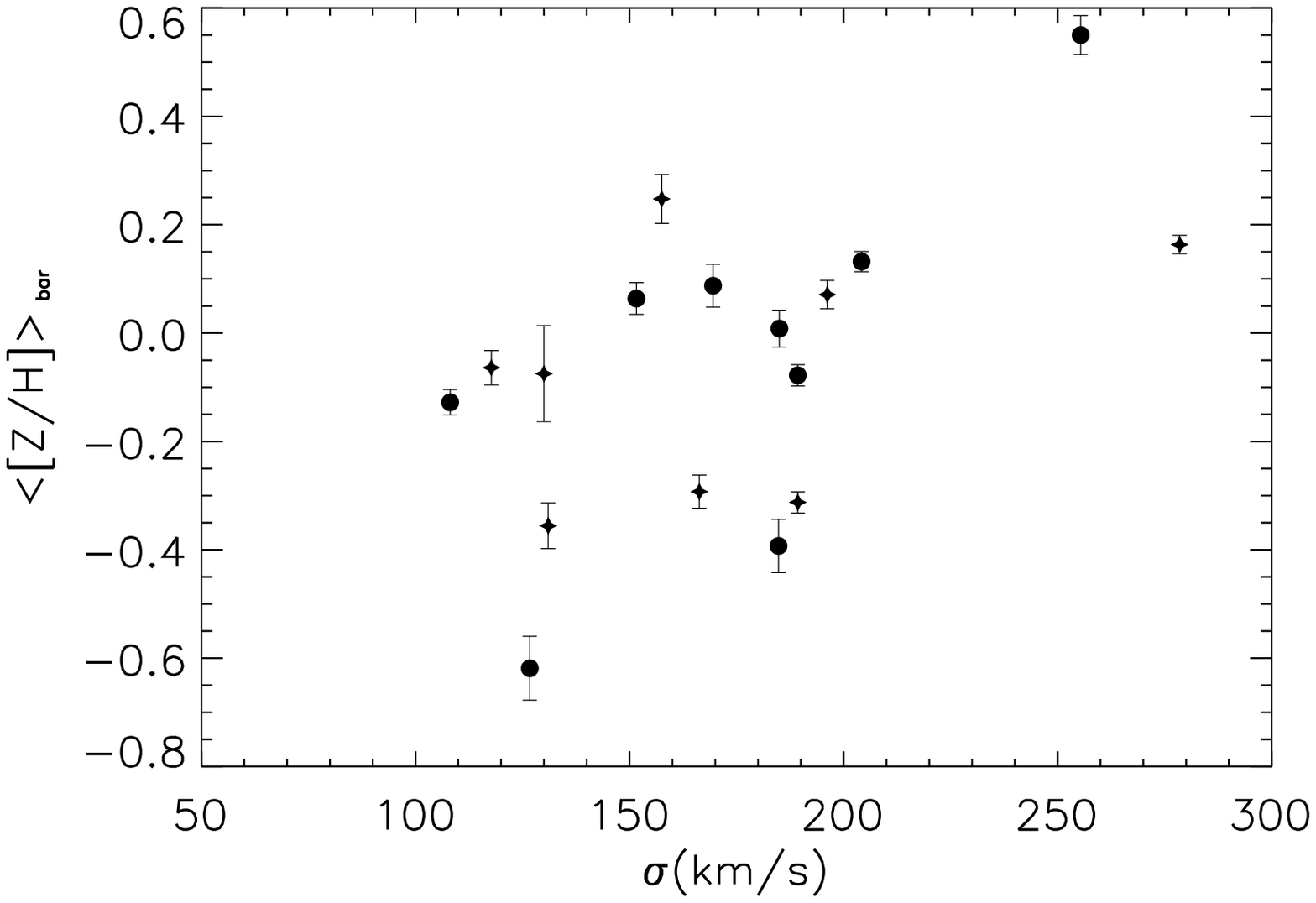}}
\resizebox{0.33\textwidth}{!}{\includegraphics[angle=0]{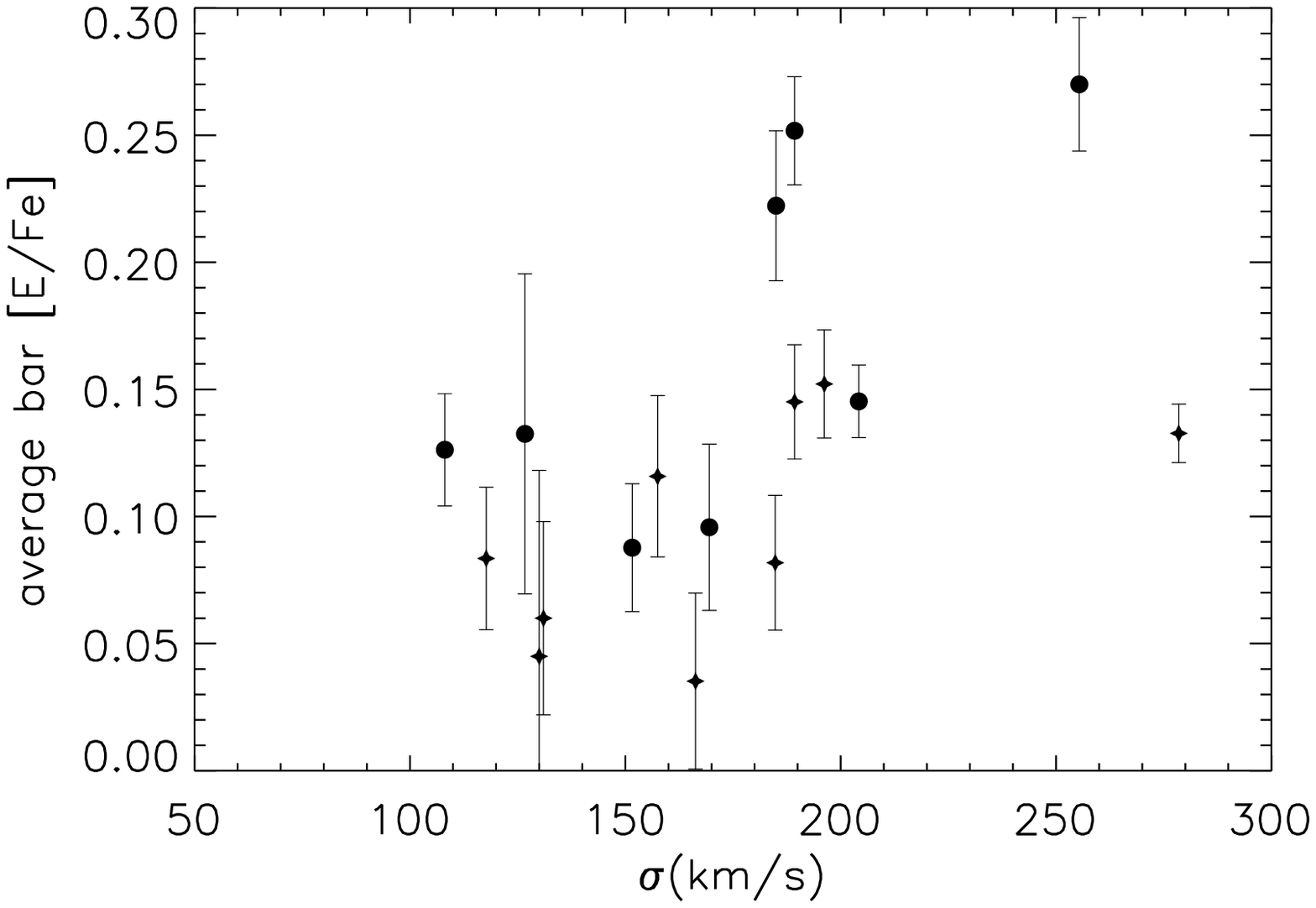}}
\caption{Mean age, metallicity and [E/Fe] in the bar region against maximum central velocity dispersion. The diamonds correspond to the galaxies classified as hosting and AGN, the circles are the galaxies with non-active nuclei\label{mean.age.sig}. Notice that galaxies with lower maximum central velocity dispersions are also the ones with younger ages and lower Z and [E/Fe] values, these trends are similar to those found for the bulge region}
\end{figure*}

In order to relate the bar properties with their stellar population parameters, we first calculate the mean stellar age, [E/Fe], and metallicity of the bar by averaging the derived values 
along the bar and weighting with the errors in these parameters.  At this point, we would like to clarify  the difference between what we mean by the age of the bar and the age of the bar stellar population, the later is the mean age of the stars as derived from the stellar populations, and the age of the bar is, for us, the age constrains that we obtain from the stellar population analysis, mean values and gradients of the ages and metallicities, to the time of bar formation.
The mean values are reported in Table~\ref{mean.ages}. The first thing that can be seen 
is that our sample of bars exhibits a large dispersion in the mean ages of their stellar 
populations. The differences in the mean ages in the bars between the different galaxies can be as large as 10 Gyr, 
with some galaxies showing a mean bar age larger than 10 Gyr and others
younger than $<$1 Gyr.  
Figure~\ref{mean.age.sig} shows the relationship between the mean stellar age, the mean Z and the average [E/Fe]  in the region 
of the bar against the maximum central velocity dispersion.
As can be seen, there is a trend in which galaxies with smaller  central 
velocity dispersion are also the ones showing younger mean-ages. In the same way, galaxies with lower maximum central velocity dispersion tend to have lower values of [E/Fe] and metallicity. In this Figure, we have separated the galaxies hosting a nuclear AGN from those without, there is no significant difference in the distribution of the two samples.  

This is the first time that stellar population parameters are derived for the bar region in galaxies. It is interesting to notice that the trends shown in Fig.~\ref{mean.age.sig} are similar to those found for galaxy bulges (Proctor \& Sansom 2002, Morthy \& Holtzman 2006)\nocite{moorthy} with values of the metallicity and [E/Fe] slightly lower than those found in the bulges. This result points out to an intimate link between the evolution of the bar and the bulge. 

Gadotti \& de Souza (2006) also found, using color as a proxy for ages, a large
difference in the ages of the stars in a sample of 18 barred galaxies.
These authors found a relation between the color of the bar and the morphological
type of spiral, being more likely that a later type spiral host a younger bar. We did not find any correlation between the mean age of the stars in the bars and the morphological type of the galaxies, although we do not have late spirals in our sample.

These authors also found a weak correlation between the mean age of the bar and the 
length of the bar, with older bars being also longer. We did not find any correlation 
between these two parameters. 
As a cautionary note, it is not a good assumption to consider broad--band colors as a probe for age, assuming a constant metallicity for all the 
galaxies, as we have shown here that bars in different galaxies show a wide metallicity distribution.

\begin{table}
\caption{Mean stellar population parameters. Mean ages (2), mean metallicities (3) and [E/Fe] (4) (see Section~\ref{deri.para} for an explanation on the parameter derivation).\label{mean.ages}}
\begin{tabular}{l ccc}
\hline\hline
Galaxy name & $<\log {\rm age}>$ & $<$[Z/H]$>$ & $<$[E/Fe]$>$ \\ 
\hline
NGC~1169  &   $1.172\pm  0.038$ &$  0.069\pm 0.032$&$ 0.112\pm 0.029$\\
NGC~1358  &   $1.043\pm  0.048$ &$  0.000\pm 0.034$&$ 0.225\pm 0.029$\\
NGC~1433  &   $0.266\pm  0.031$ &$  0.005\pm 0.041$&$ 0.138\pm 0.038$\\
NGC~1530  &   $0.078\pm  0.027$ &$ -0.023\pm 0.050$&$ 0.267\pm 0.073$\\
NGC~1832  &   $0.153\pm  0.015$ &$ -0.184\pm 0.028$&$ 0.002\pm 0.024$\\
NGC~2217  &   $0.344\pm  0.045$ &$  0.644\pm 0.046$&$ 0.336\pm 0.030$\\
NGC~2273  &   $0.241\pm  0.020$ &$  0.064\pm 0.030$&$ 0.111\pm 0.025$\\
NGC~2523  &   $0.530\pm  0.062$ &$  0.313\pm 0.054$&$ 0.136\pm 0.031$\\ 
NGC~2665  &   $0.334\pm  0.076$ &$ -0.606\pm 0.092$&$-0.010\pm 0.076$\\
NGC~2681  &   $0.439\pm  0.029$ &$ -0.602\pm 0.039$&$ 0.015\pm 0.036$\\
NGC~2859  &   $1.022\pm  0.029$ &$ -0.064\pm 0.021$&$ 0.238\pm 0.021$\\
NGC~2935  &   $0.374\pm  0.056$ &$ -0.192\pm 0.059$&$ 0.114\pm 0.052$\\
NGC~2950  &   $1.053\pm  0.025$ &$ -0.311\pm 0.019$&$ 0.114\pm 0.052$\\
NGC~2962  &   $0.920\pm  0.035$ &$  0.029\pm 0.026$&$ 0.135\pm 0.022$\\
NGC~3081  &   $0.950\pm  0.157$ &$ -0.225\pm 0.124$&$-0.060\pm 0.096$\\
NGC~4245  &   $0.774\pm  0.045$ &$ -0.095\pm 0.031$&$ 0.085\pm 0.028$\\
NGC~4314  &   $0.297\pm  0.031$ &$  0.066\pm 0.038$&$ 0.105\pm 0.032$\\
NGC~4394  &   $0.956\pm  0.033$ &$ -0.202\pm 0.024$&$ 0.131\pm 0.025$\\
NGC~4643  &   $0.741\pm  0.013$ &$  0.321\pm 0.009$&$ 0.126\pm 0.006$\\
NGC~5101  &   $0.477\pm  0.021$ &$  0.182\pm 0.213$&$ 0.113\pm 0.015$\\
\hline
\end{tabular}
\end{table}
 
 \subsection{Stellar populations along the bar
\label{stellar.populations.gradients}}

As shown in Sec.~\ref{deri.para}, we fit a slope to the derived age, metallicity and [E/Fe] distributions with a simple linear regression weighted with the errors in the $y$-direction (Table~\ref{gradtab} ).  We find that most galaxies show a gradient different from zero in both age and metallicity. 
Fig.~\ref{bargrad} shows the indices at the beginning and end of the bar (Sec.\ref{morphology}) together
 with the predictions of the models by V08.
The galaxies with the lowest mean metallicity values have, systematically, younger and more metal poor populations at the end of the bar  (these are the data points on the left-hand side  of the grid in Fig.~\ref{bargrad}). The more metal rich bars (data points on the right-hand side of the grid in Fig.~\ref{bargrad}) show more complex behavior with most of the galaxies showing younger and more metal rich populations and the end of the bar.  

\begin{figure*}
\resizebox{0.3\textwidth}{!}{\includegraphics[angle=-90]{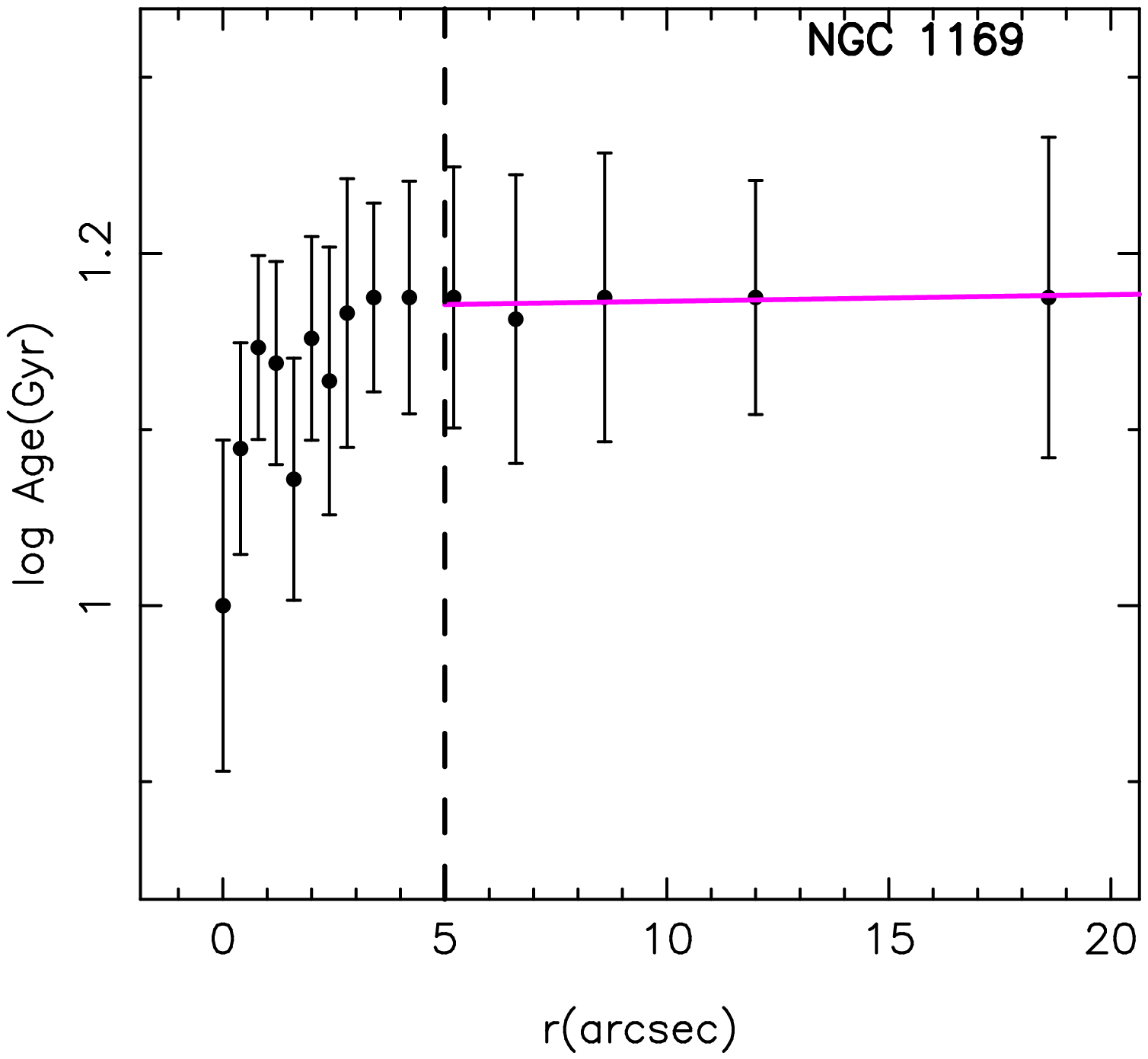}}
\resizebox{0.3\textwidth}{!}{\includegraphics[angle=-90]{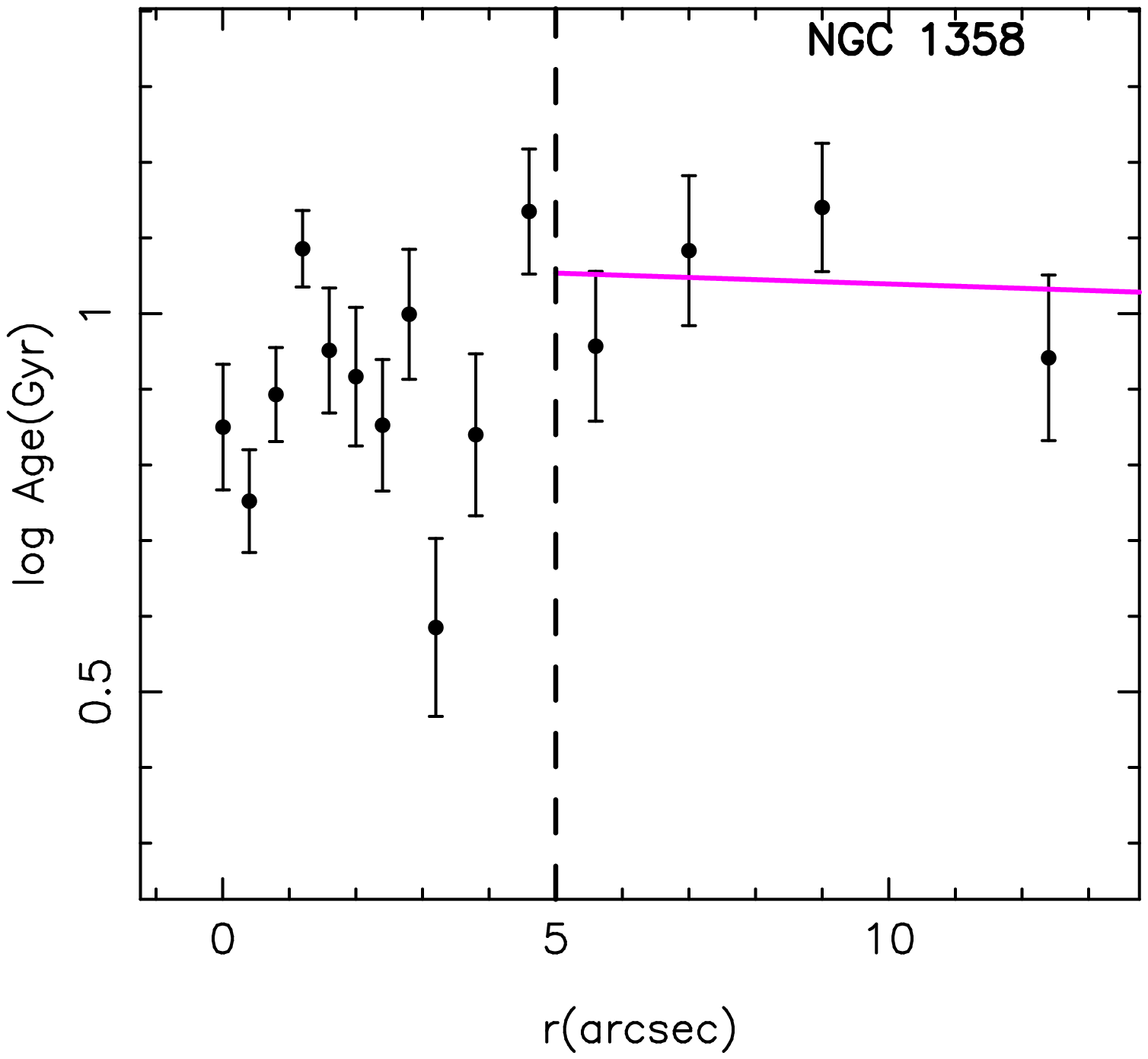}}
\resizebox{0.3\textwidth}{!}{\includegraphics[angle=-90]{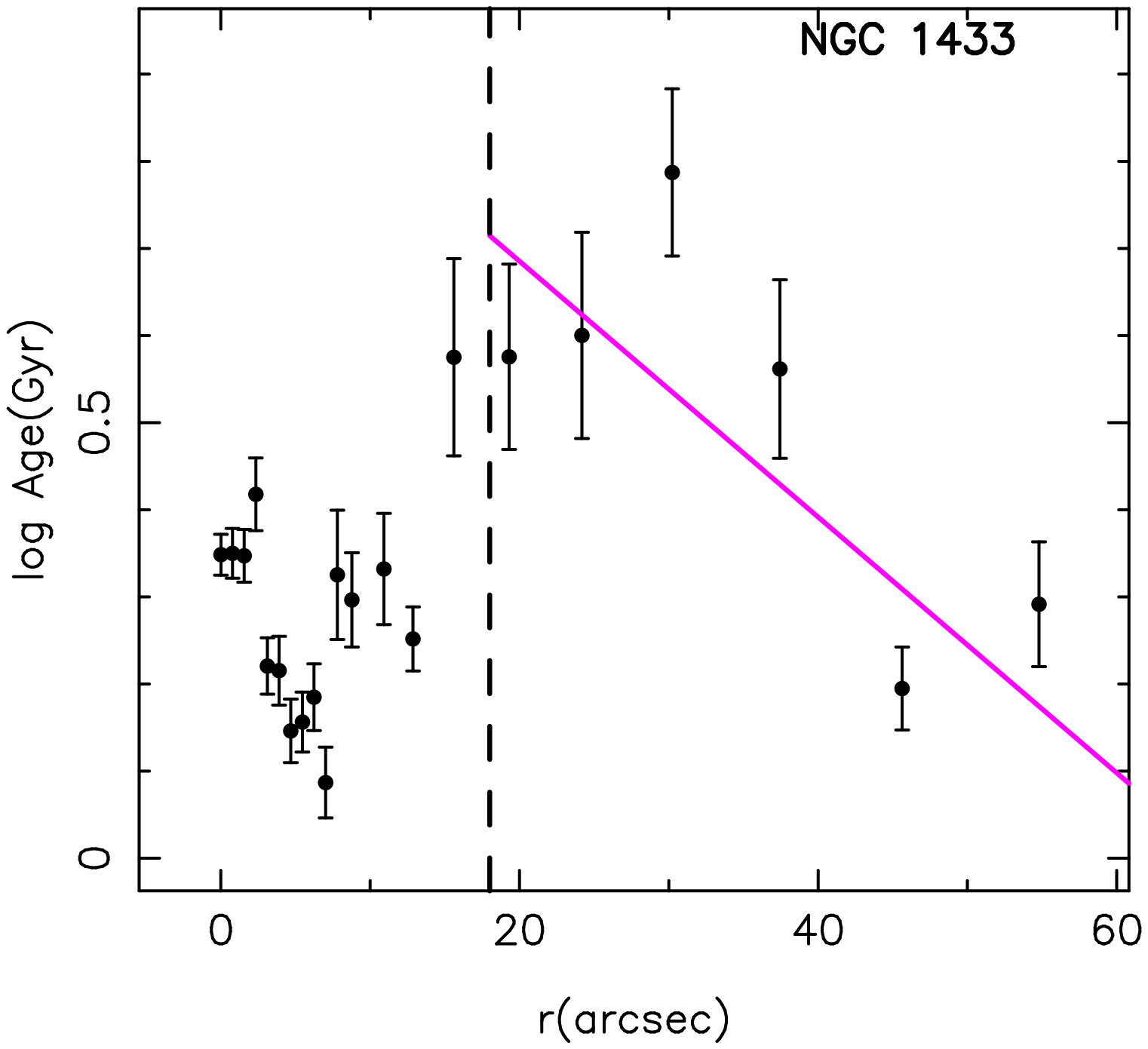}}
\resizebox{0.3\textwidth}{!}{\includegraphics[angle=-90]{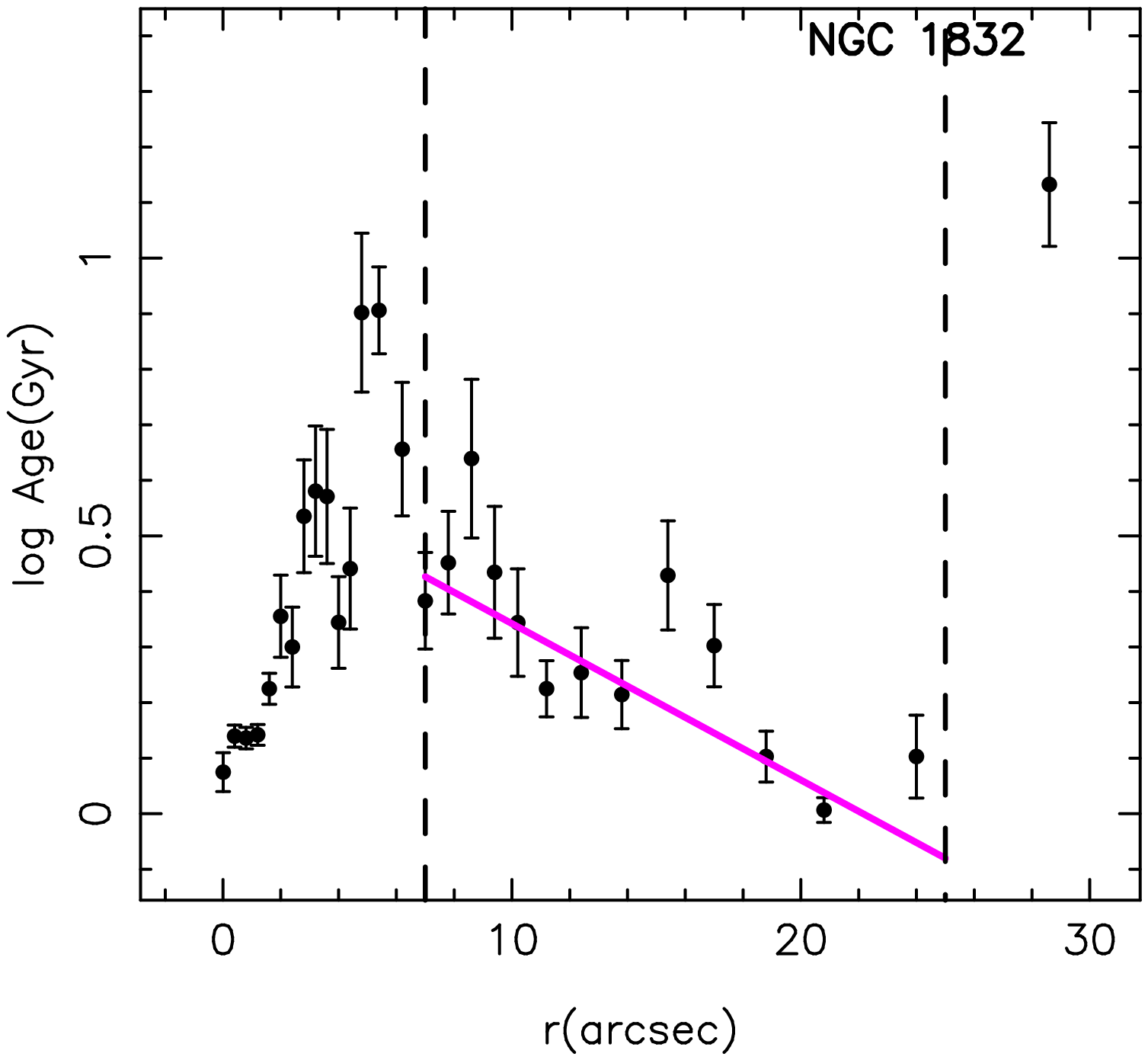}}
\resizebox{0.3\textwidth}{!}{\includegraphics[angle=-90]{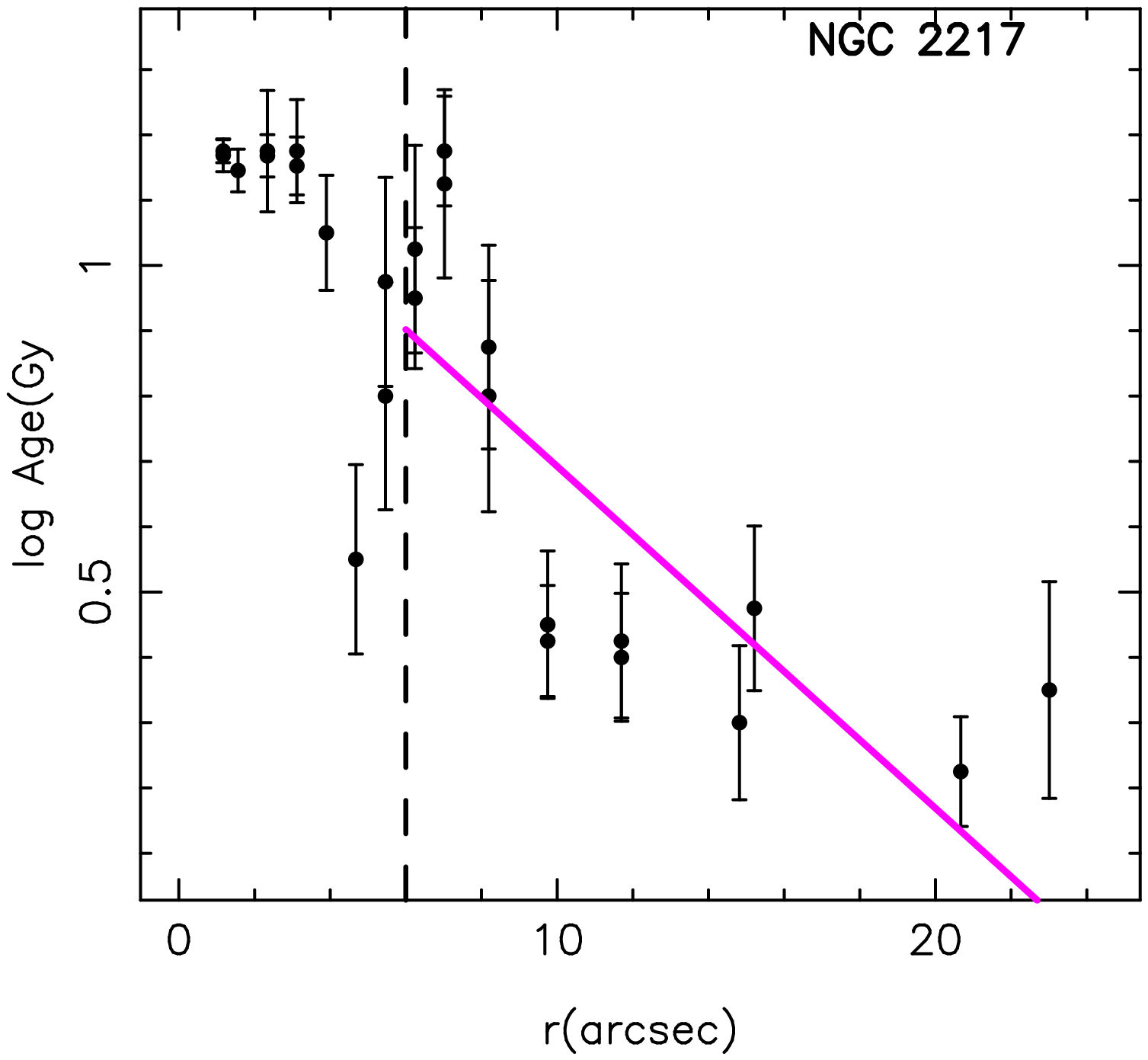}}
\resizebox{0.3\textwidth}{!}{\includegraphics[angle=-90]{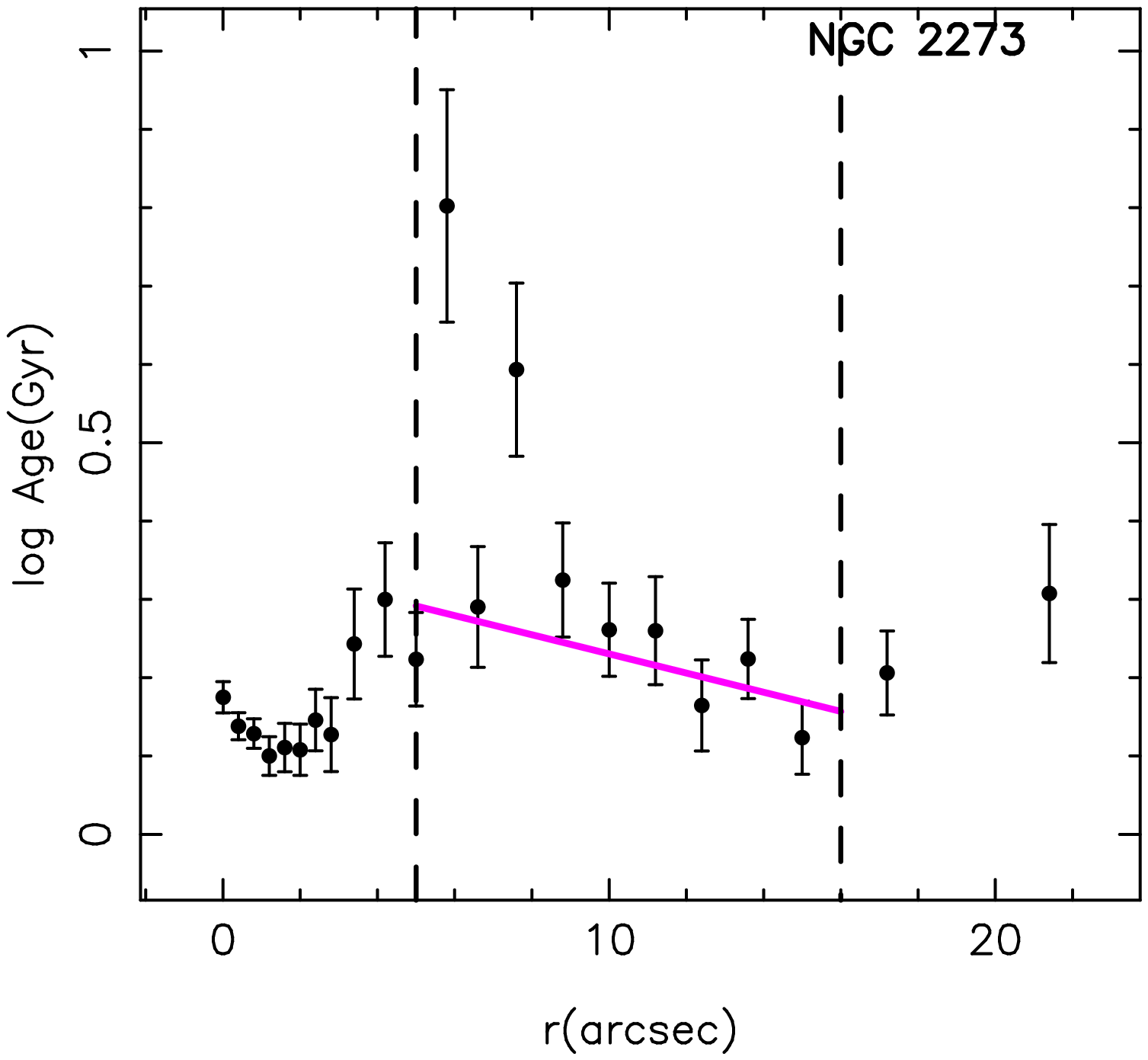}}
\resizebox{0.3\textwidth}{!}{\includegraphics[angle=-90]{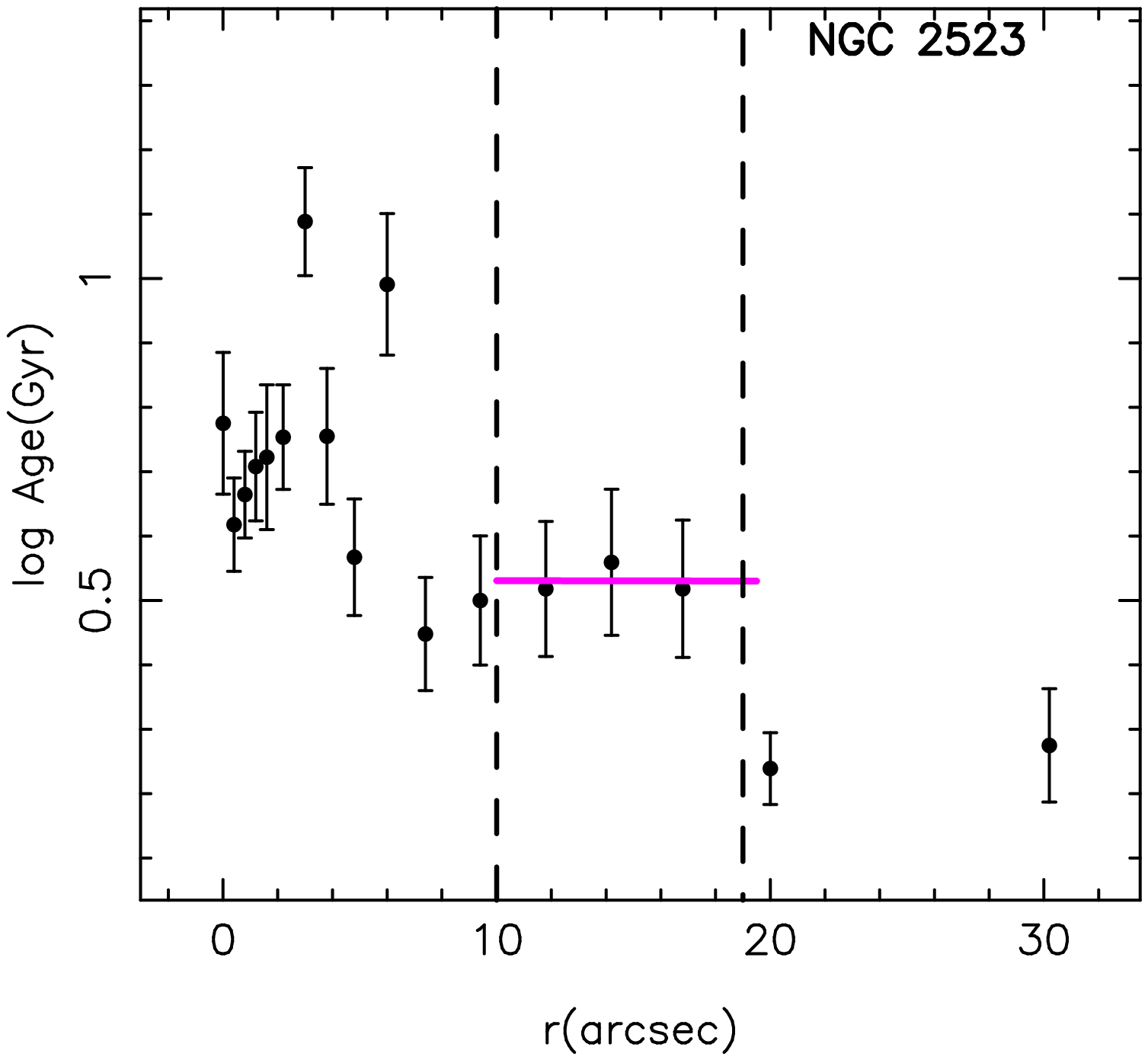}}\hspace{0.8cm}
\resizebox{0.3\textwidth}{!}{\includegraphics[angle=-90]{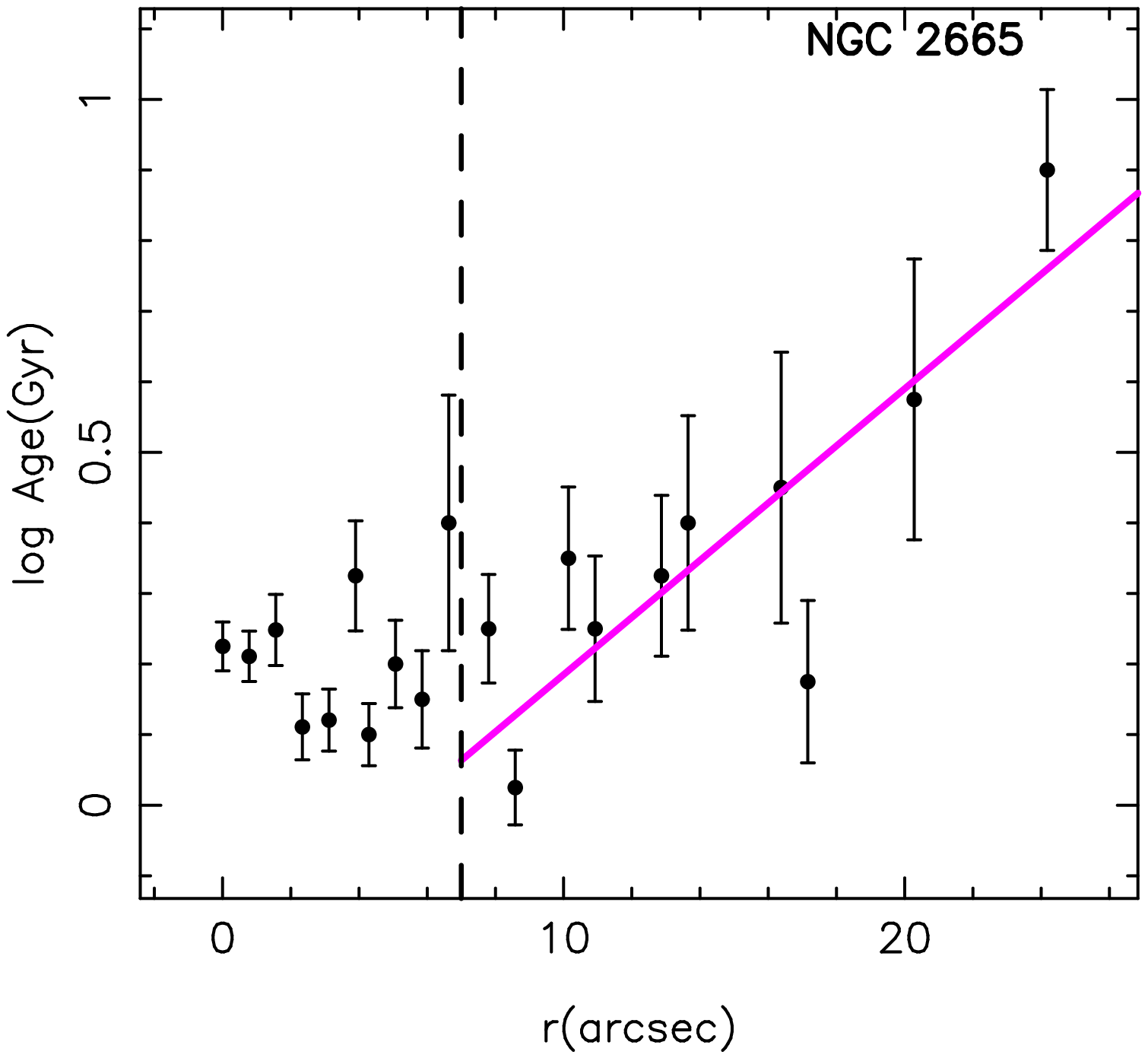}}\hspace{0.8cm}
\resizebox{0.3\textwidth}{!}{\includegraphics[angle=-90]{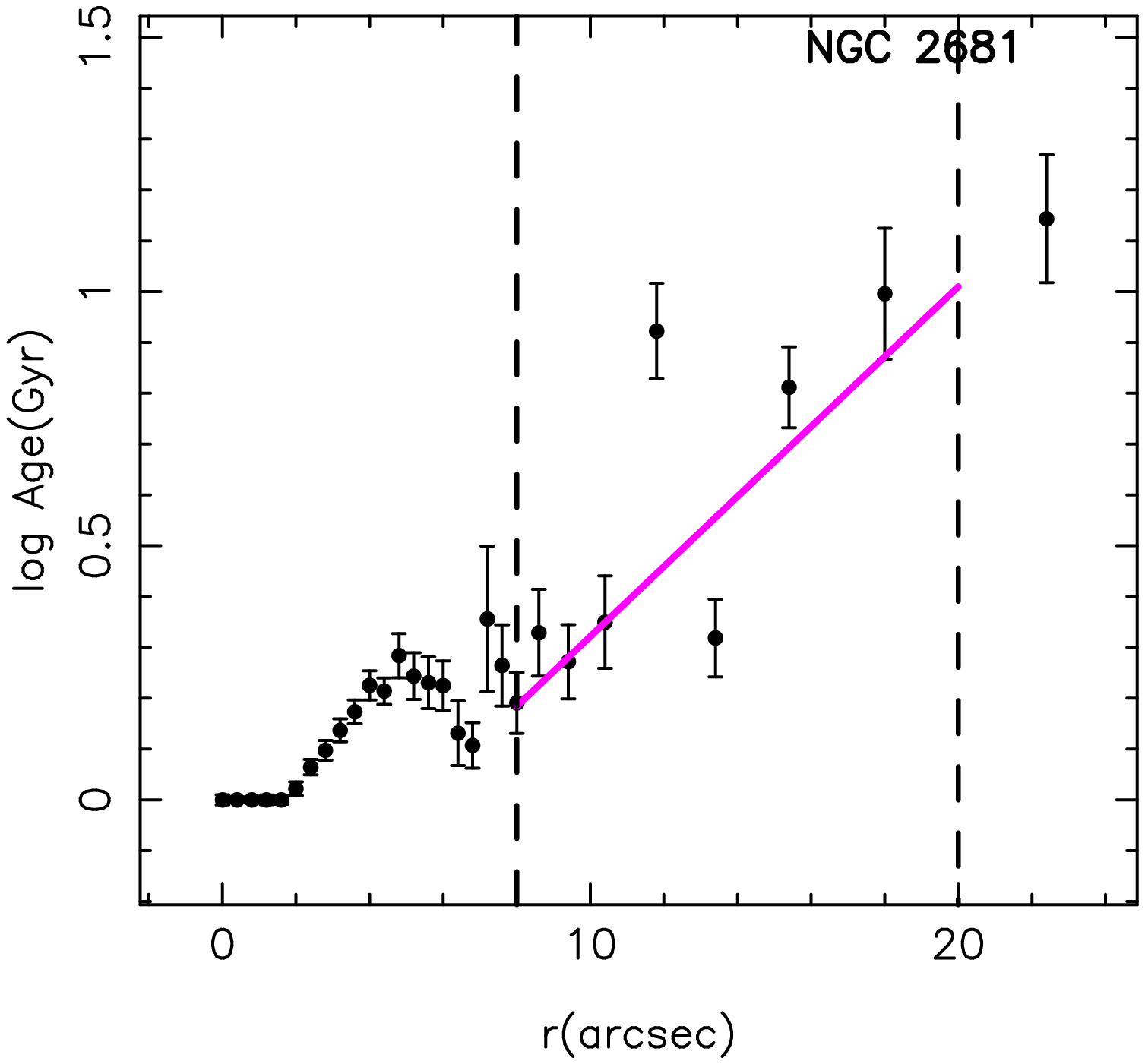}} 
\caption{SSP-equivalent age and metallicity along the radius. Dashed lines indicate 
 the beginning and the end of the bar region. A linear fit to the 
 bar region is also plotted. \label{fig.age.grad}}
\end{figure*}

Figure~\ref{fig.age.grad} shows the age and  metallicity distribution profiles for the galaxies in 
our sample. We have not included in this analysis NGC~1530, NGC~3081,NGC~4314 and NGC~2935 due to the large  fitting errors.
Seven galaxies have metallicity gradients significantly (more than 3-$\sigma$ significance) different from zero:
NGC~1169, NGC~2217, NGC~4394,  and NGC~5101 (positive) NGC~2665, NGC~2681, NGC~4245 (negative).

\addtocounter{figure}{-1}
\begin{figure*}
\resizebox{0.3\textwidth}{!}{\includegraphics[angle=-90]{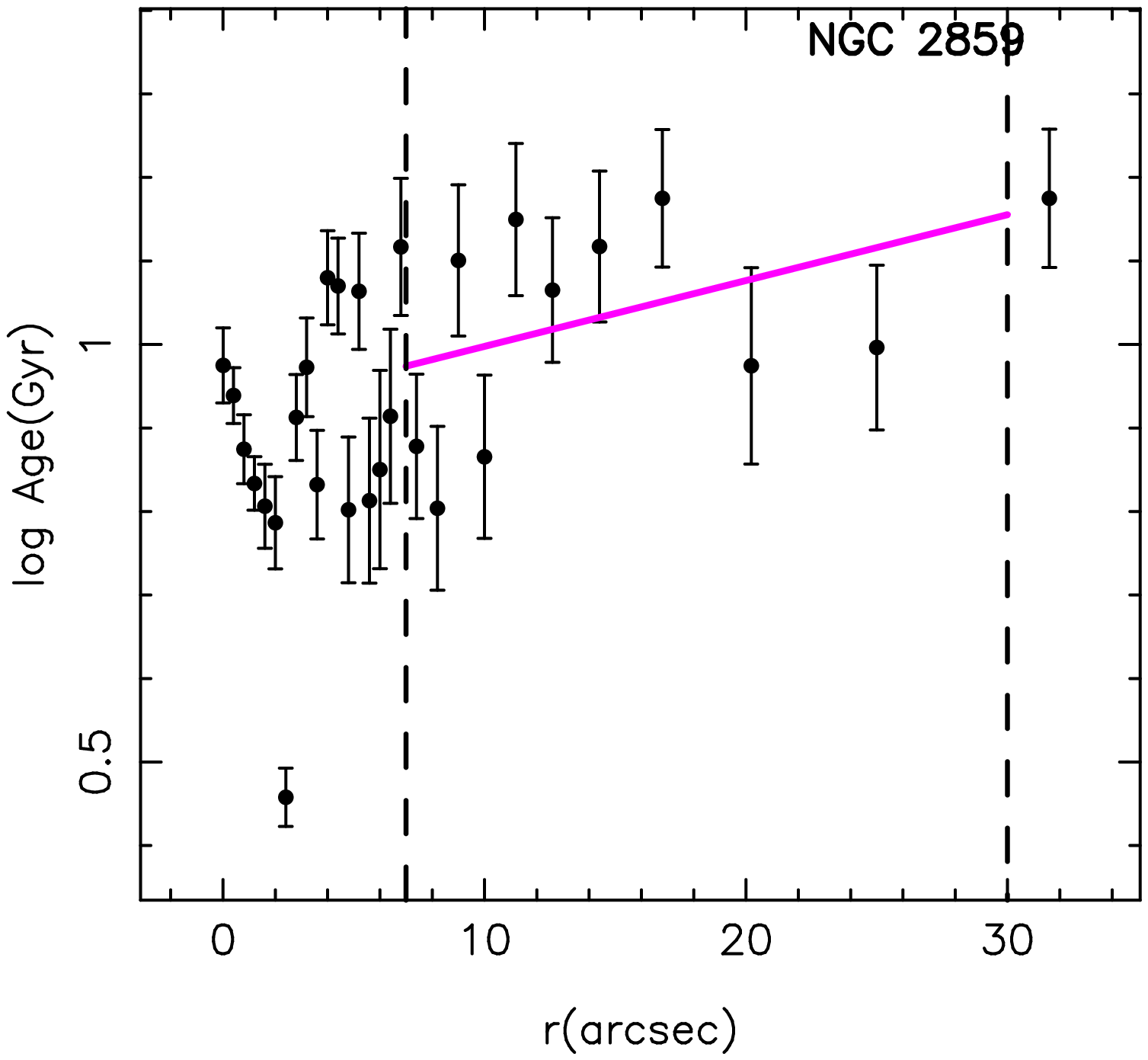}}
\resizebox{0.3\textwidth}{!}{\includegraphics[angle=-90]{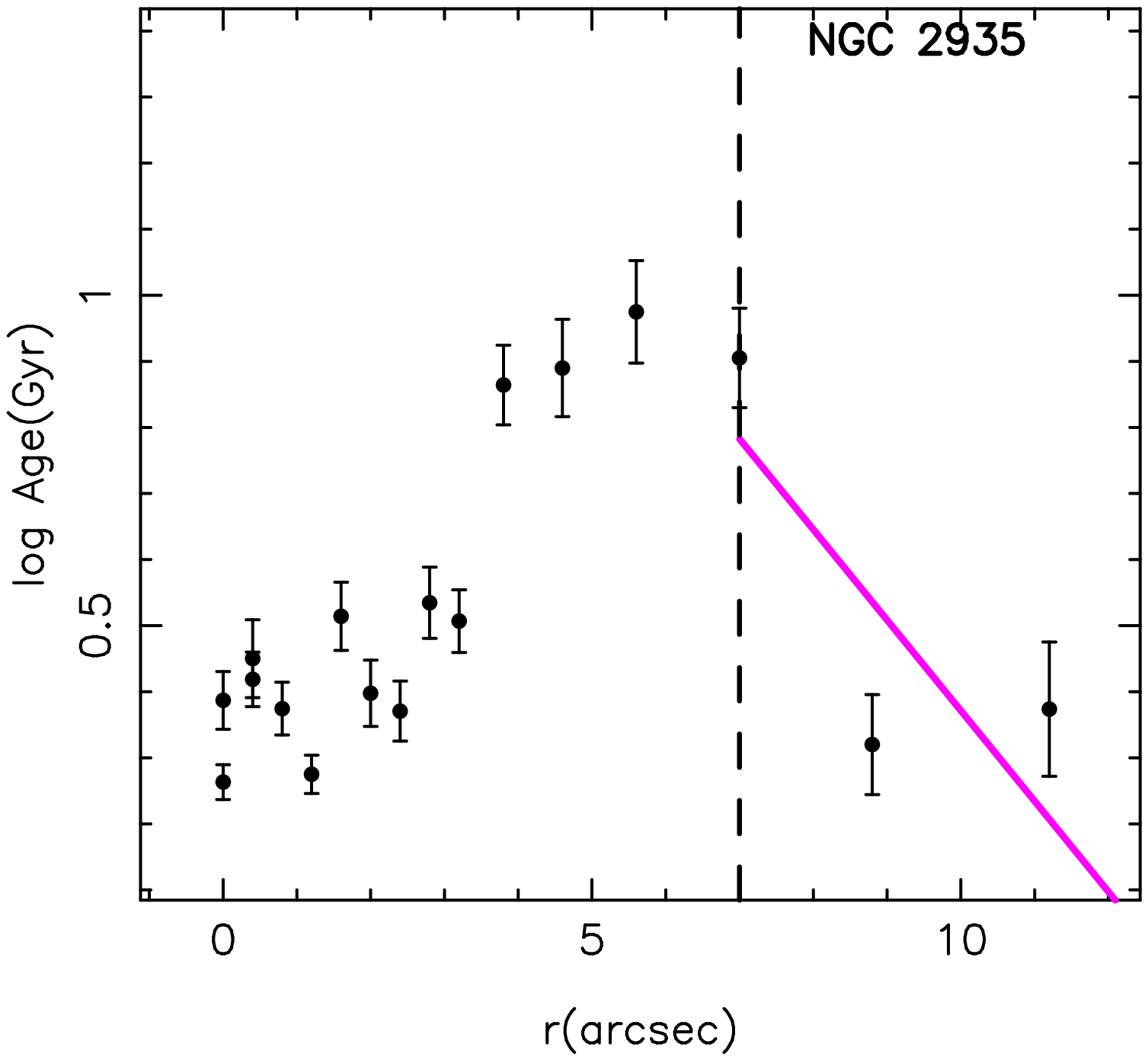}}
\resizebox{0.3\textwidth}{!}{\includegraphics[angle=-90]{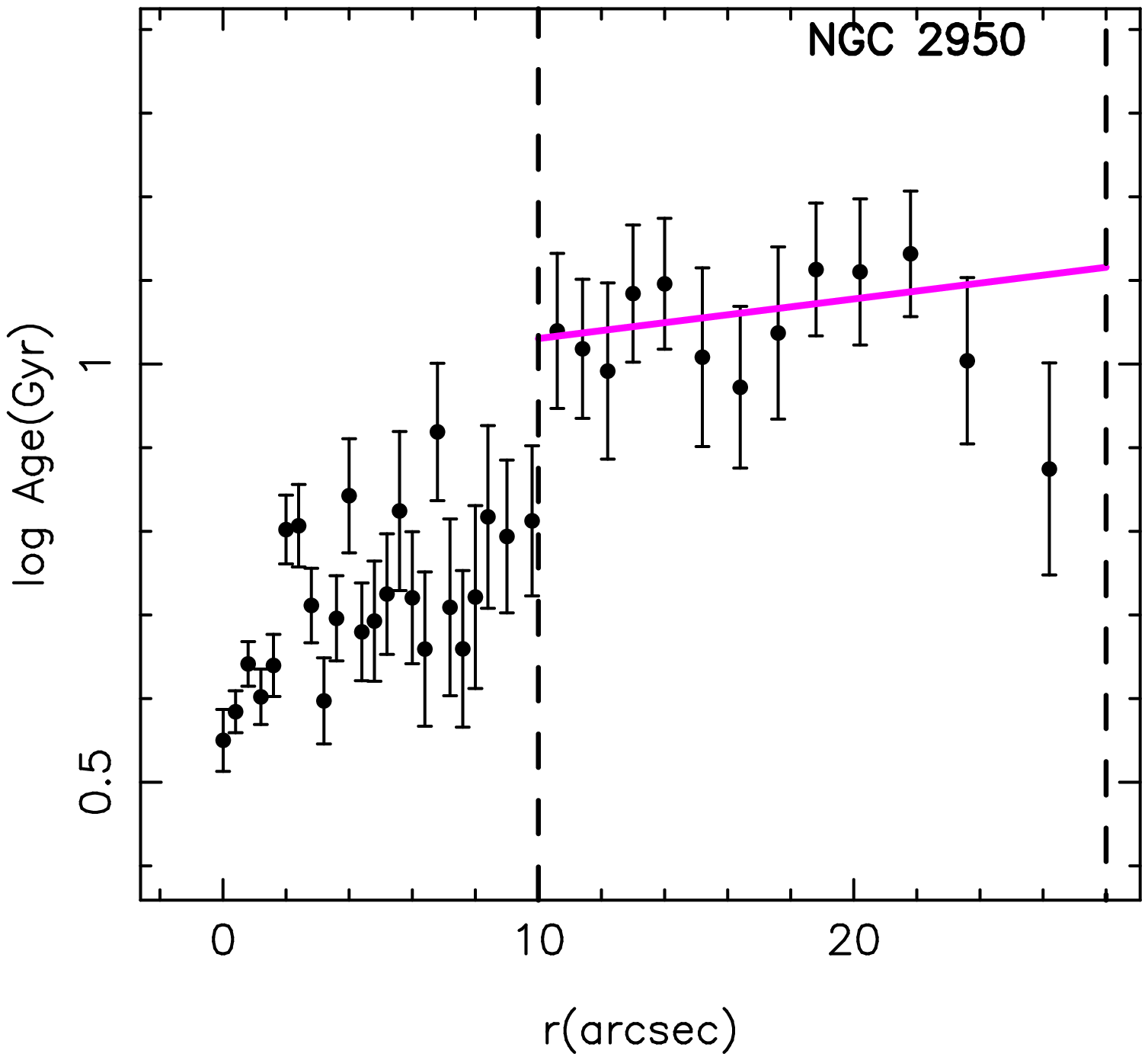}}
\resizebox{0.3\textwidth}{!}{\includegraphics[angle=-90]{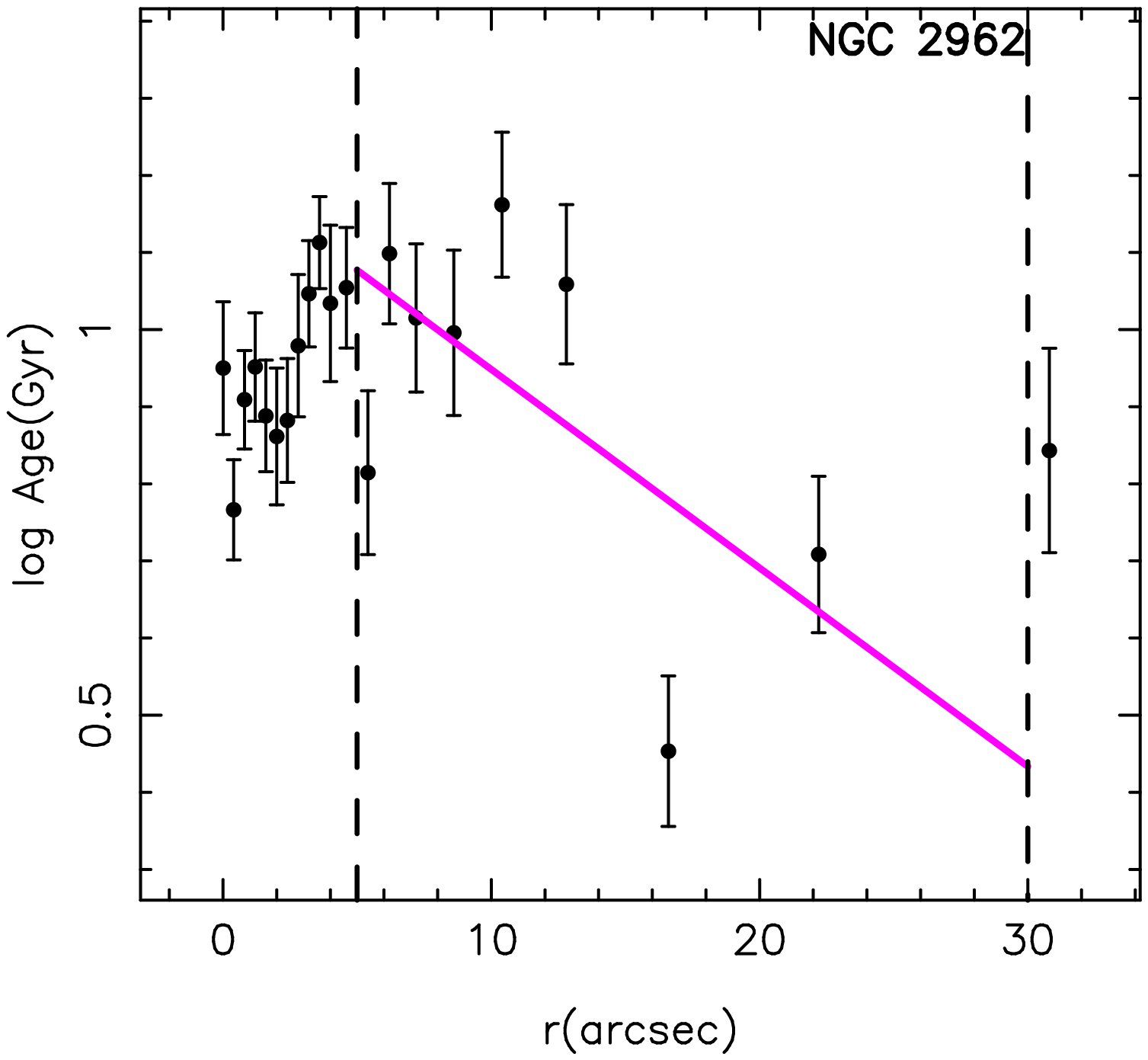}}
\resizebox{0.3\textwidth}{!}{\includegraphics[angle=-90]{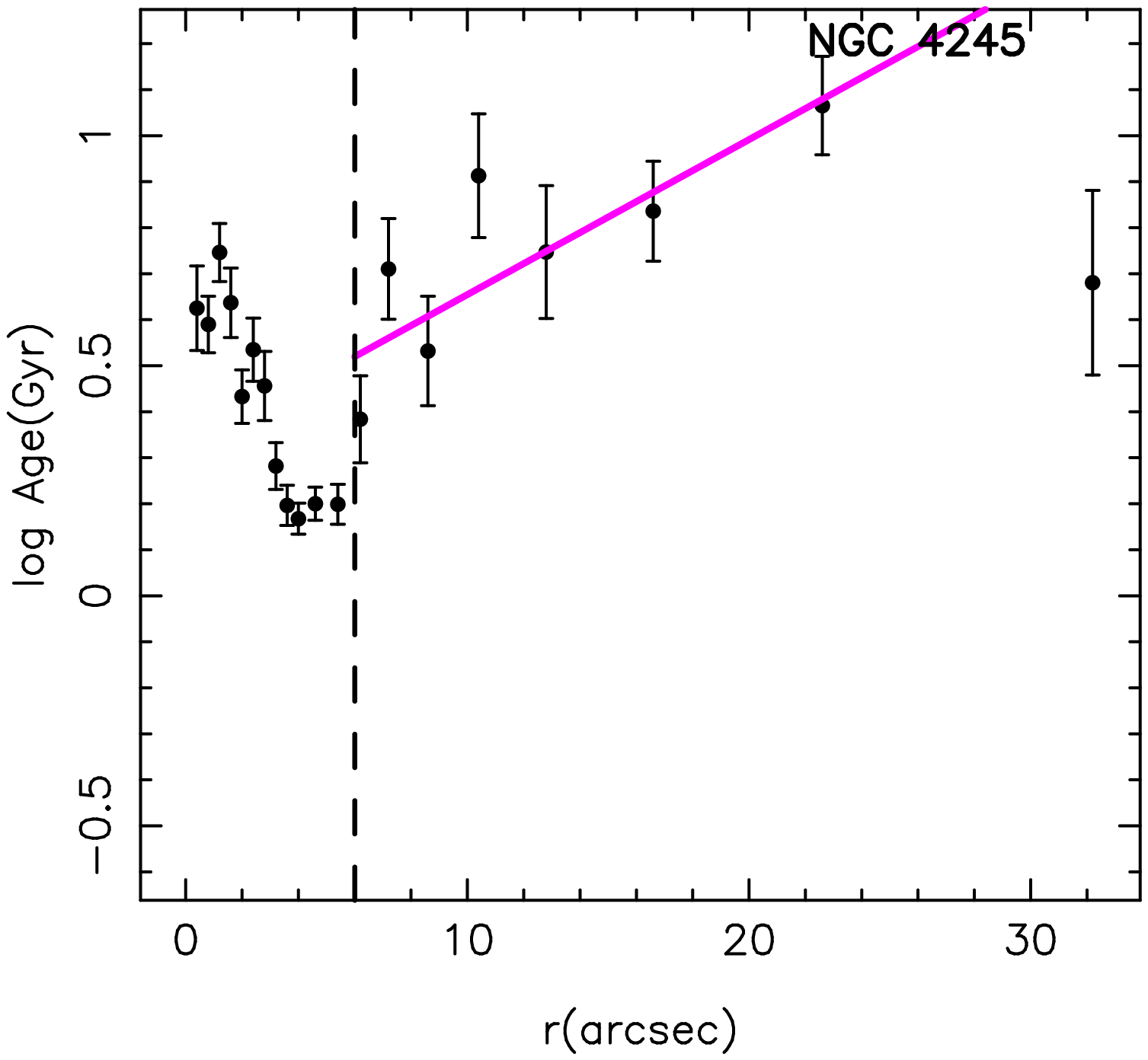}}
\resizebox{0.3\textwidth}{!}{\includegraphics[angle=-90]{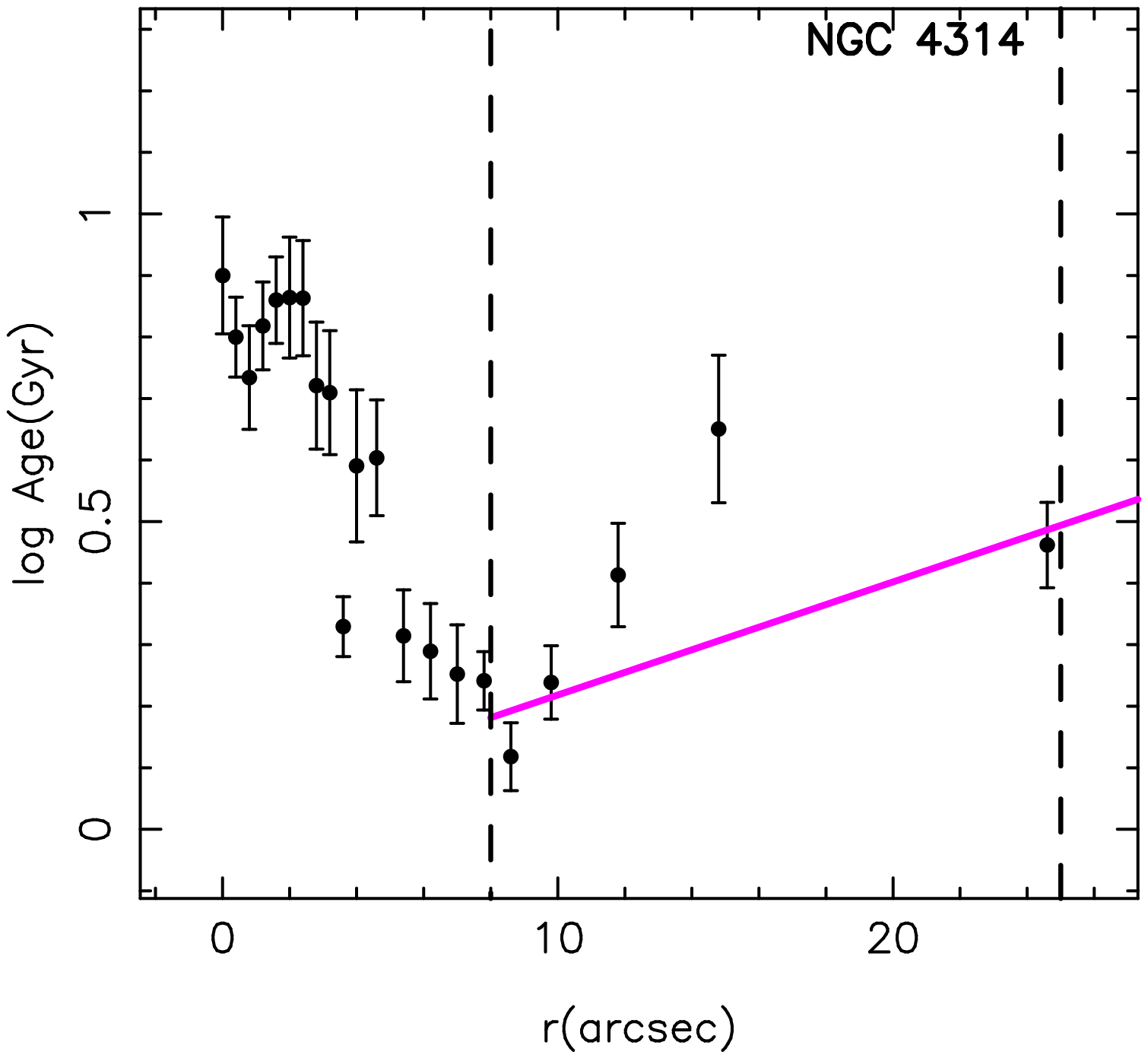}}
\resizebox{0.3\textwidth}{!}{\includegraphics[angle=-90]{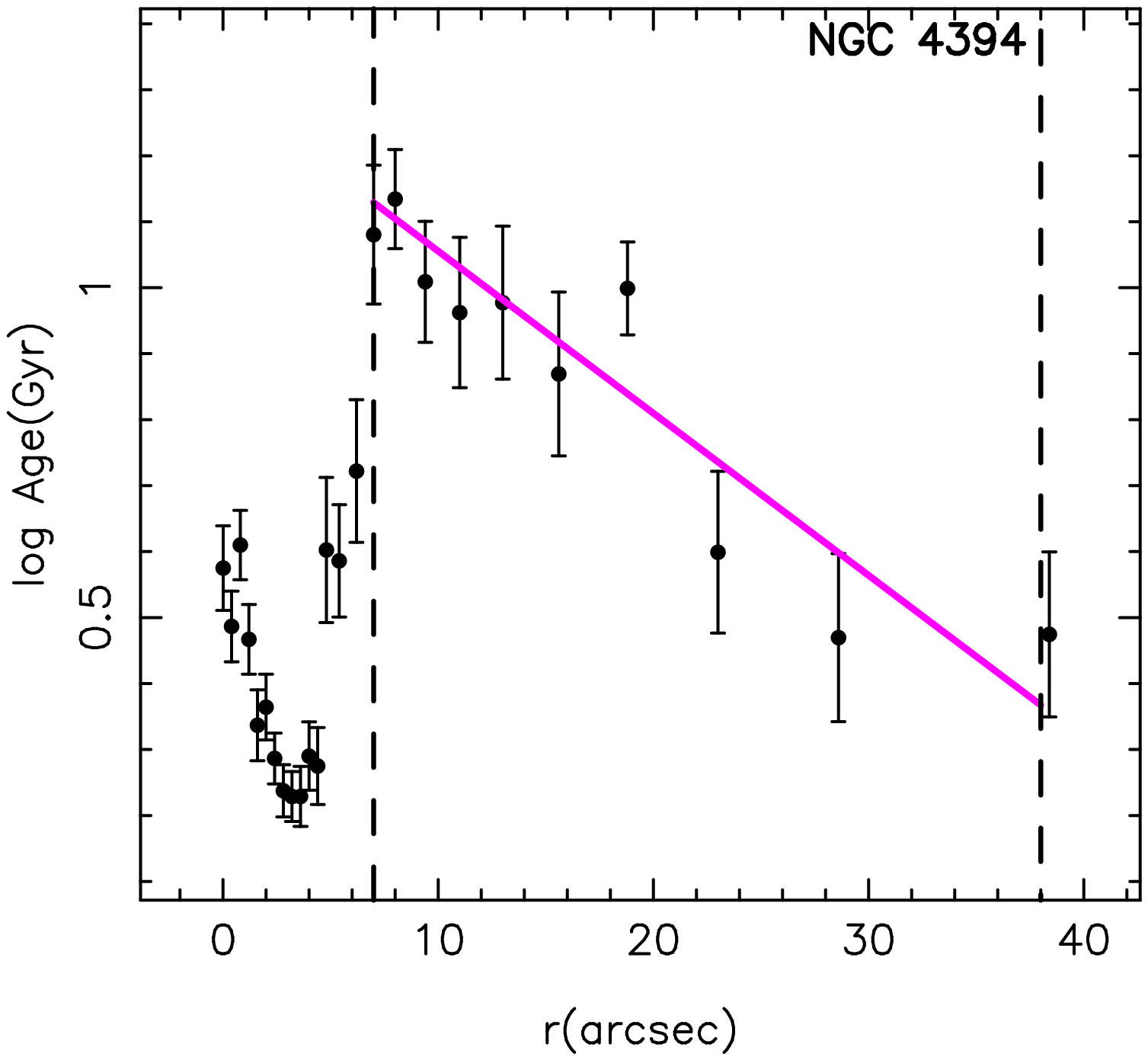}}\hspace{0.8cm}
\resizebox{0.3\textwidth}{!}{\includegraphics[angle=-90]{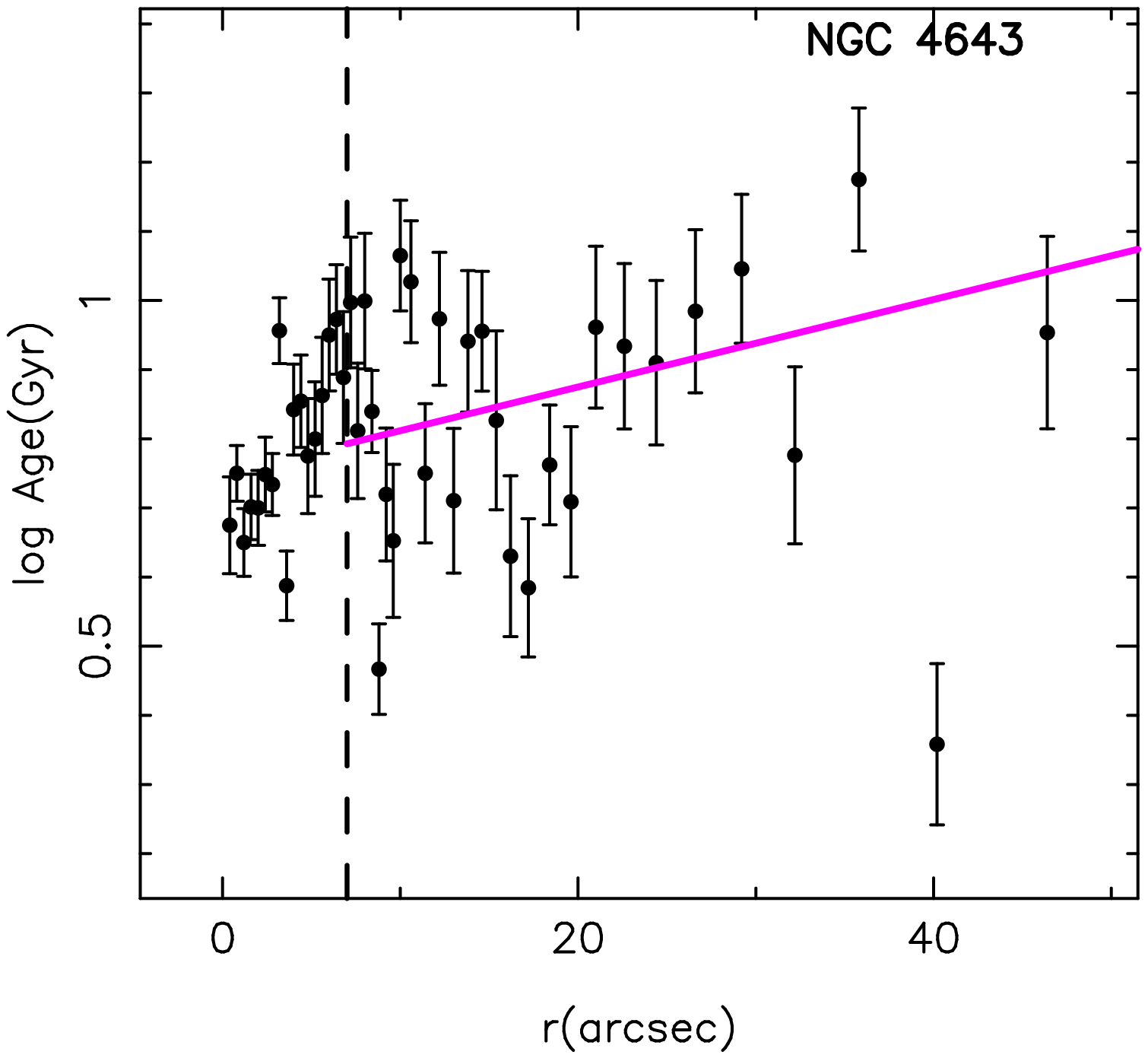}}\hspace{0.8cm}
\resizebox{0.3\textwidth}{!}{\includegraphics[angle=-90]{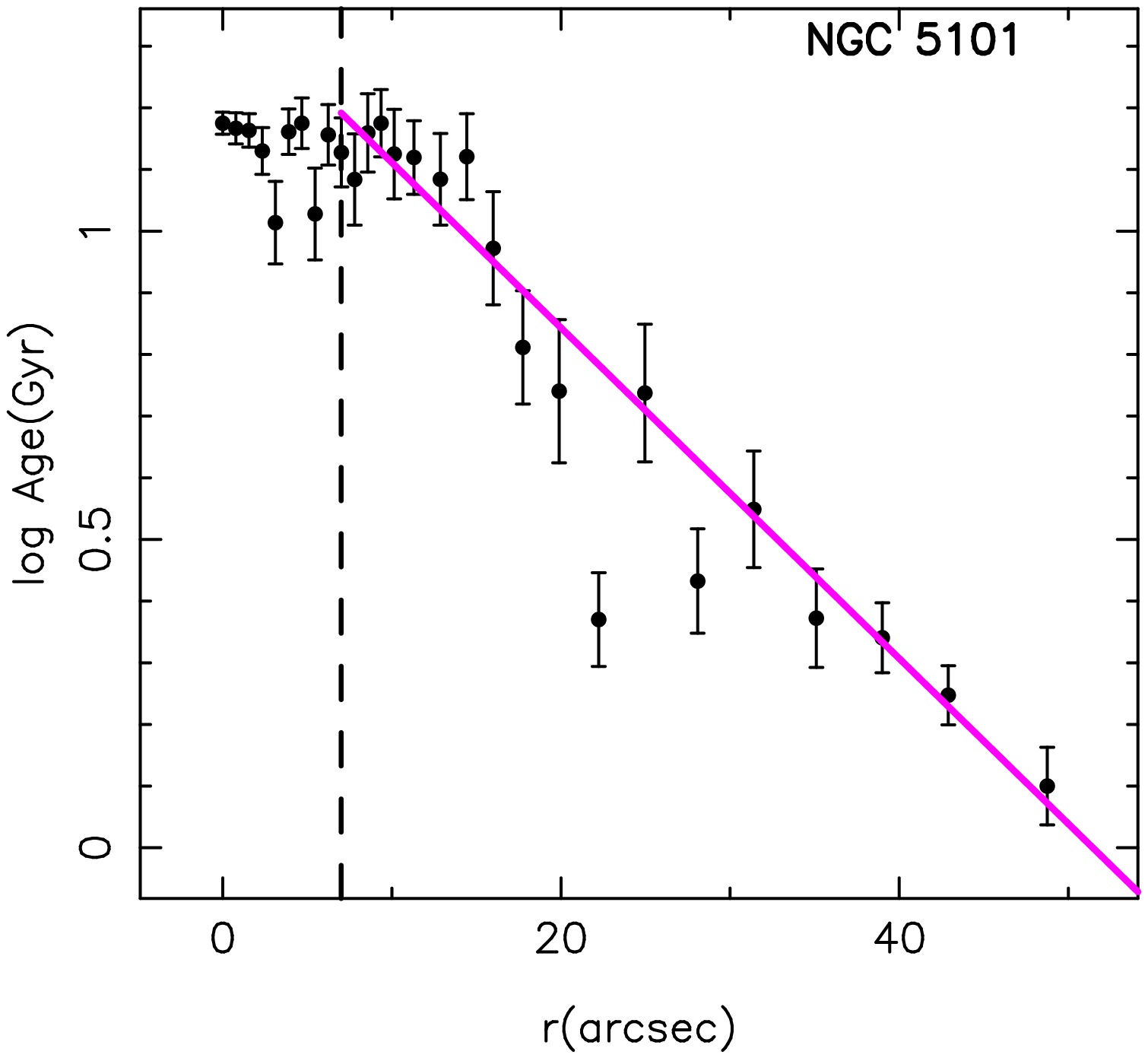}}
\caption{SSP-equivalent age and metallicity along the radius. Dashed lines indicate 
 the beginning and the end of the bar region. A linear fit to the 
 bar region is also plotted.}
 \end{figure*}
\addtocounter{figure}{-1}
\begin{figure*}
\resizebox{0.3\textwidth}{!}{\includegraphics[angle=-90]{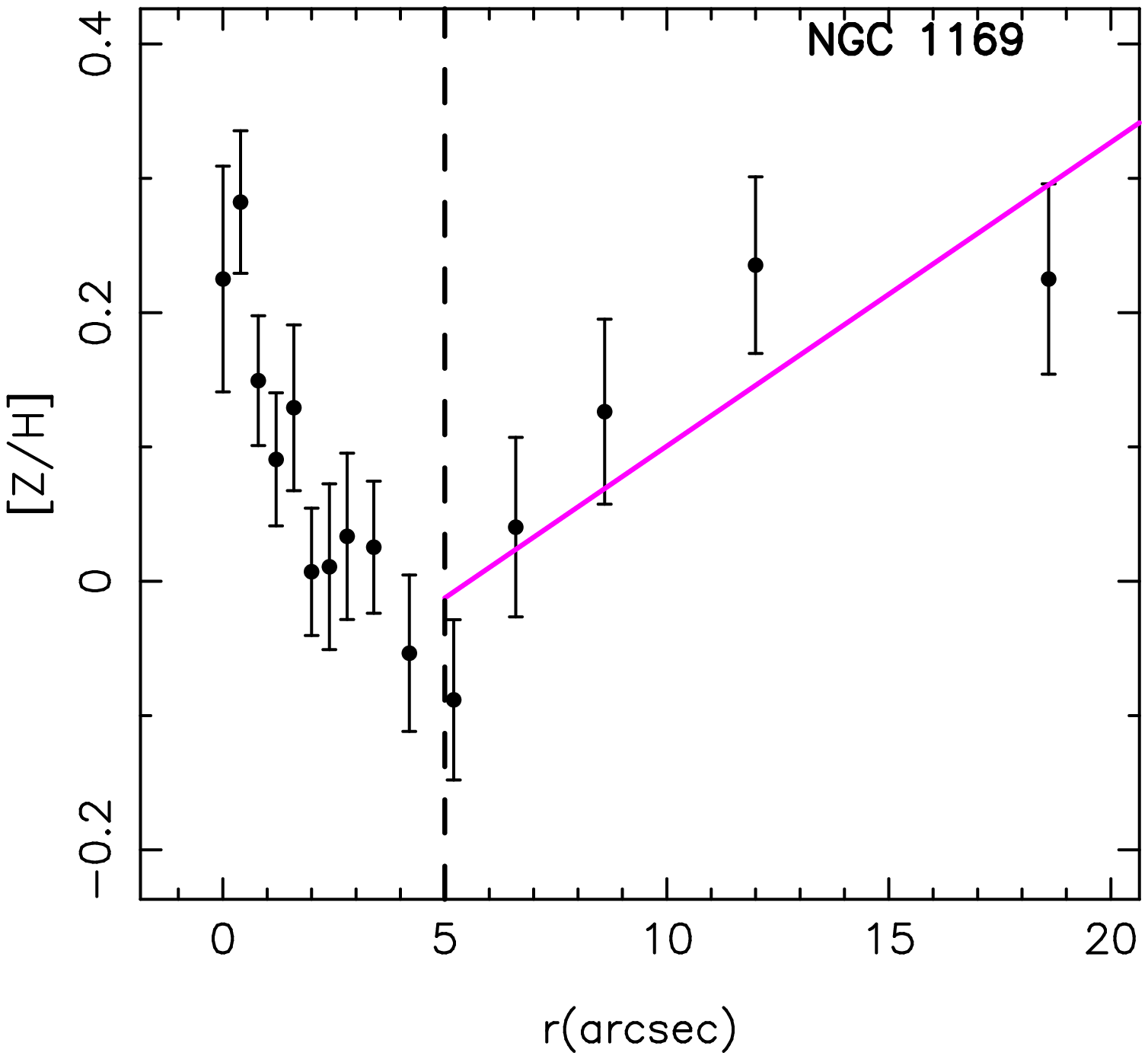}}
\resizebox{0.3\textwidth}{!}{\includegraphics[angle=-90]{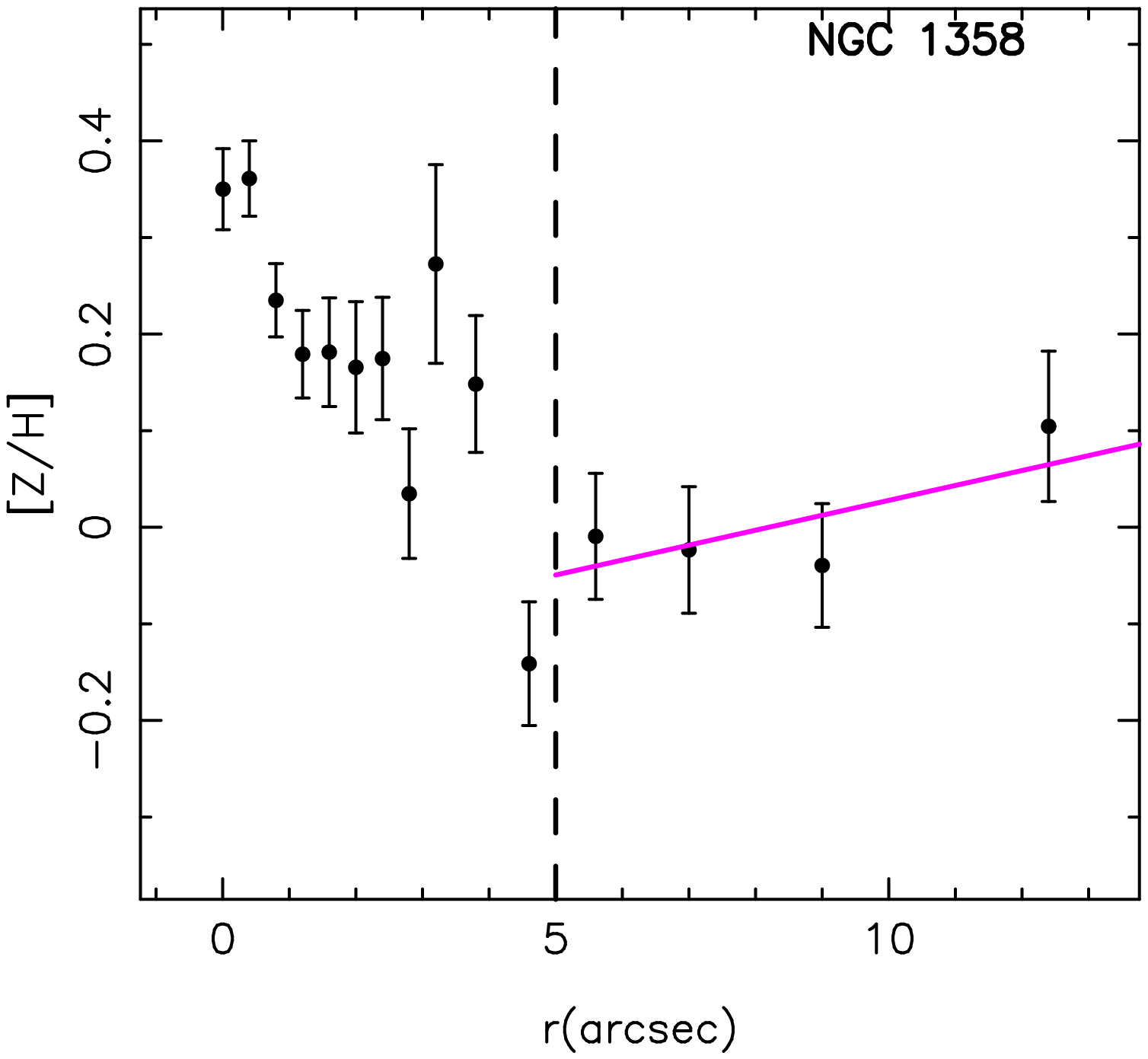}}
\resizebox{0.3\textwidth}{!}{\includegraphics[angle=-90]{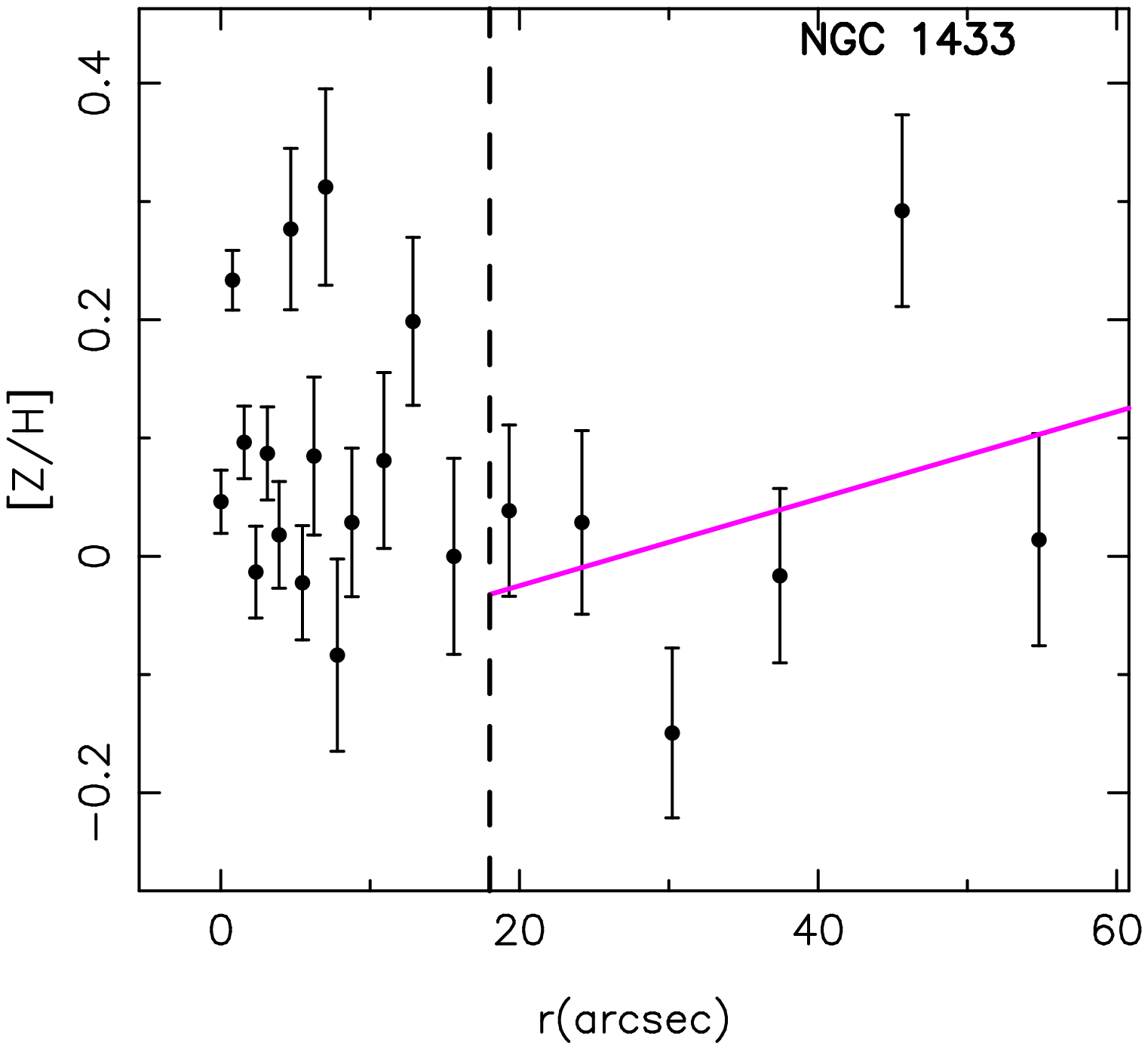}}
\resizebox{0.3\textwidth}{!}{\includegraphics[angle=-90]{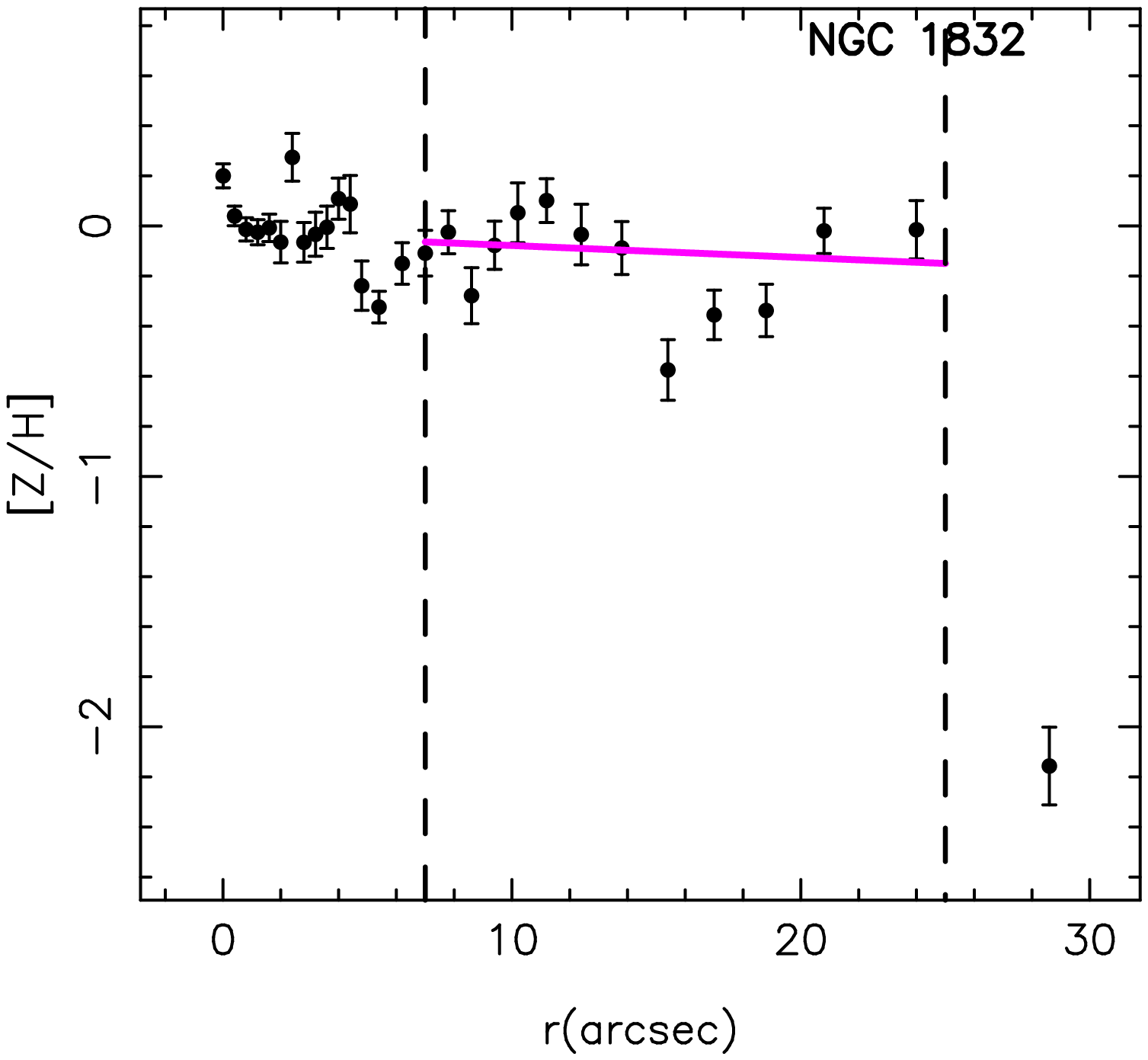}}
\resizebox{0.3\textwidth}{!}{\includegraphics[angle=-90]{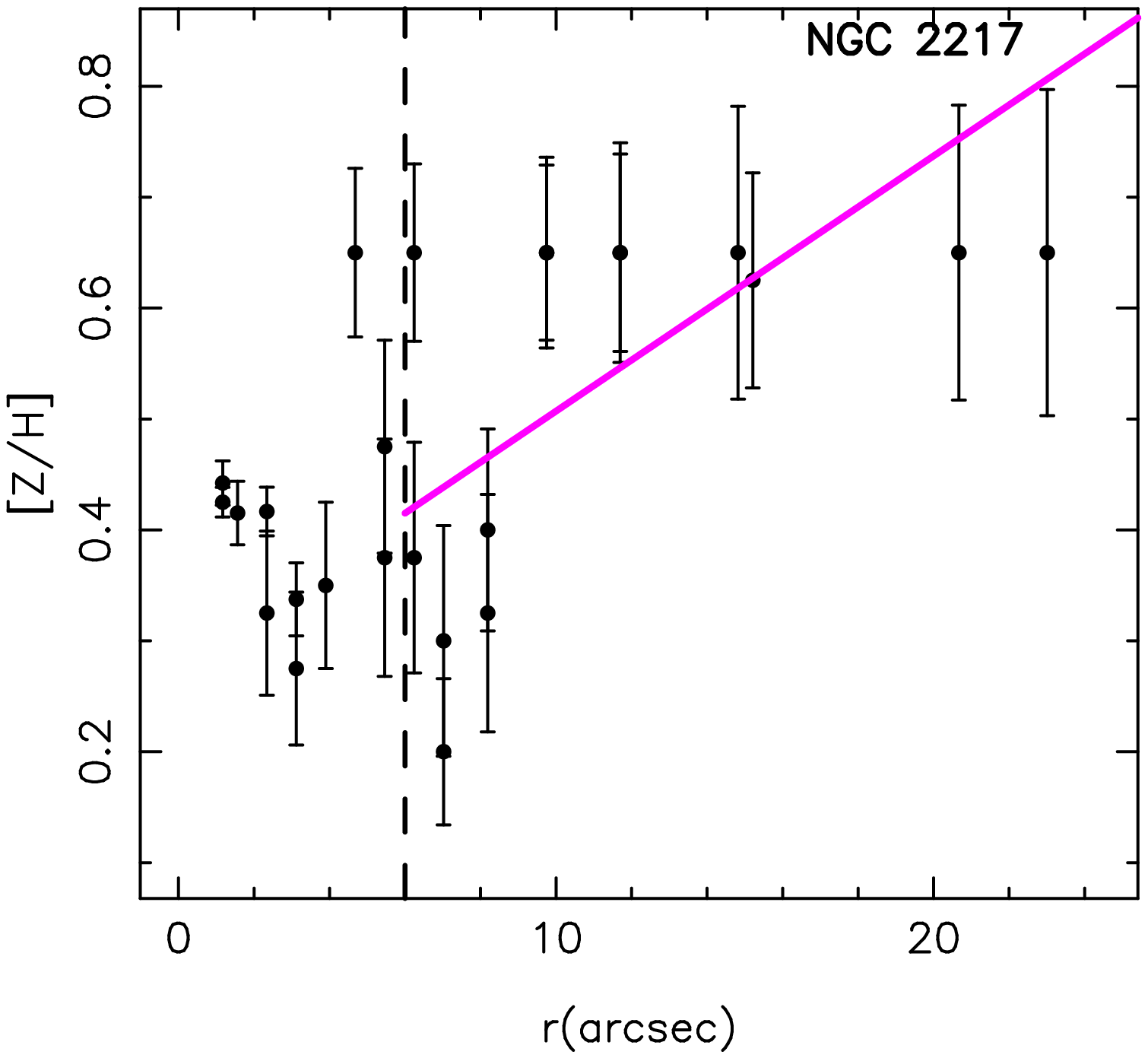}}
\resizebox{0.3\textwidth}{!}{\includegraphics[angle=-90]{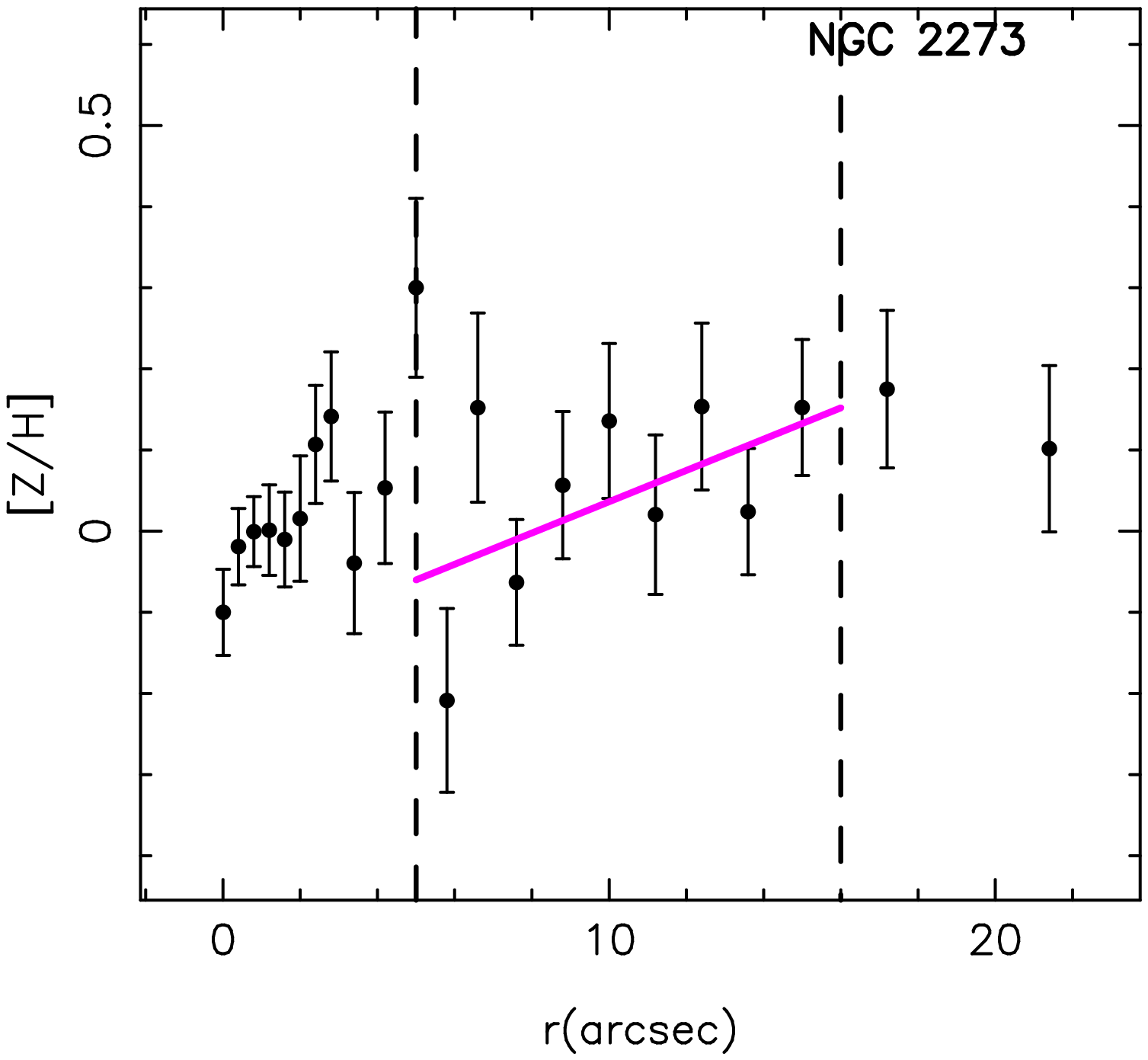}}
\resizebox{0.3\textwidth}{!}{\includegraphics[angle=-90]{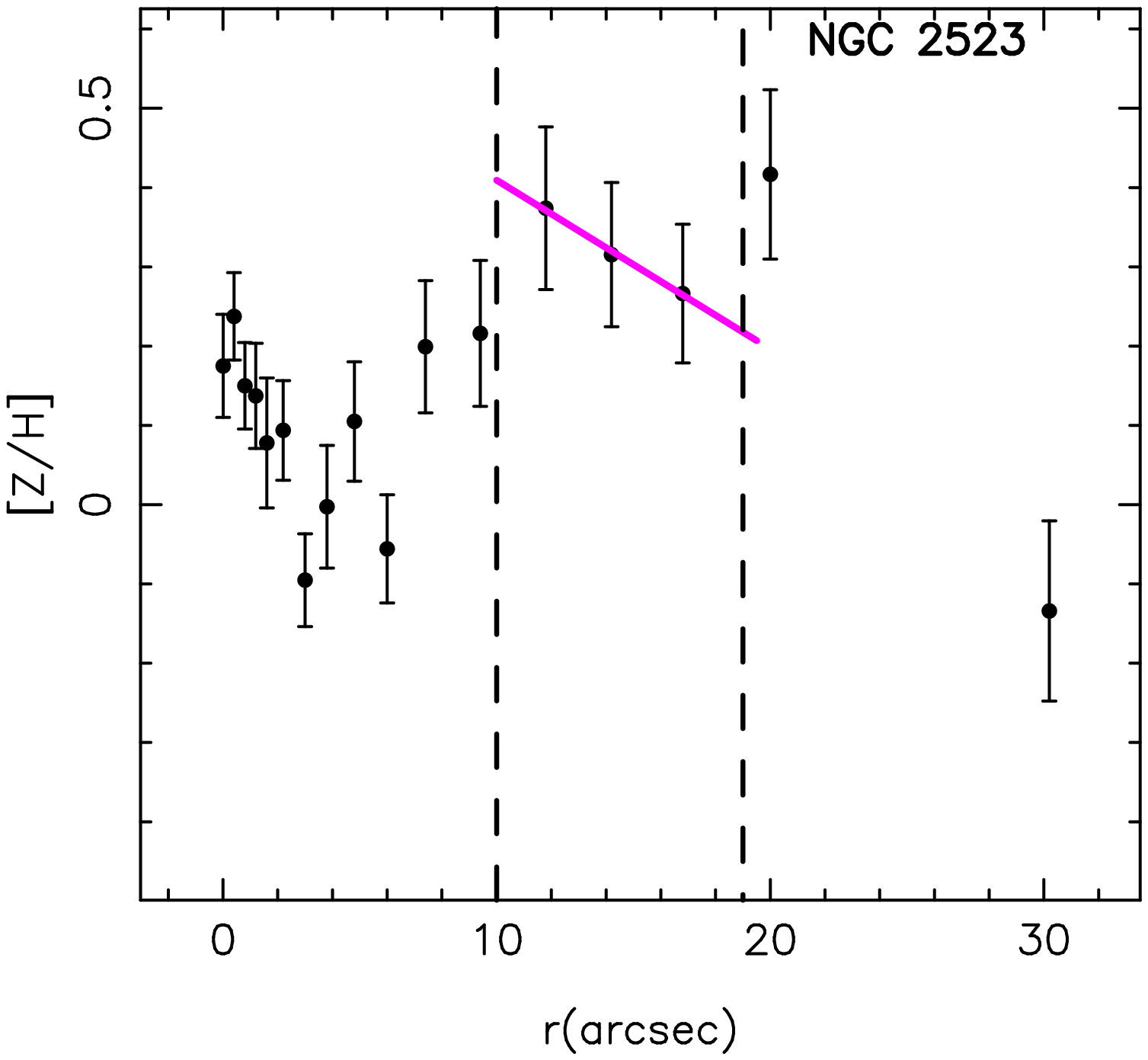}}\hspace{0.8cm}
\resizebox{0.3\textwidth}{!}{\includegraphics[angle=-90]{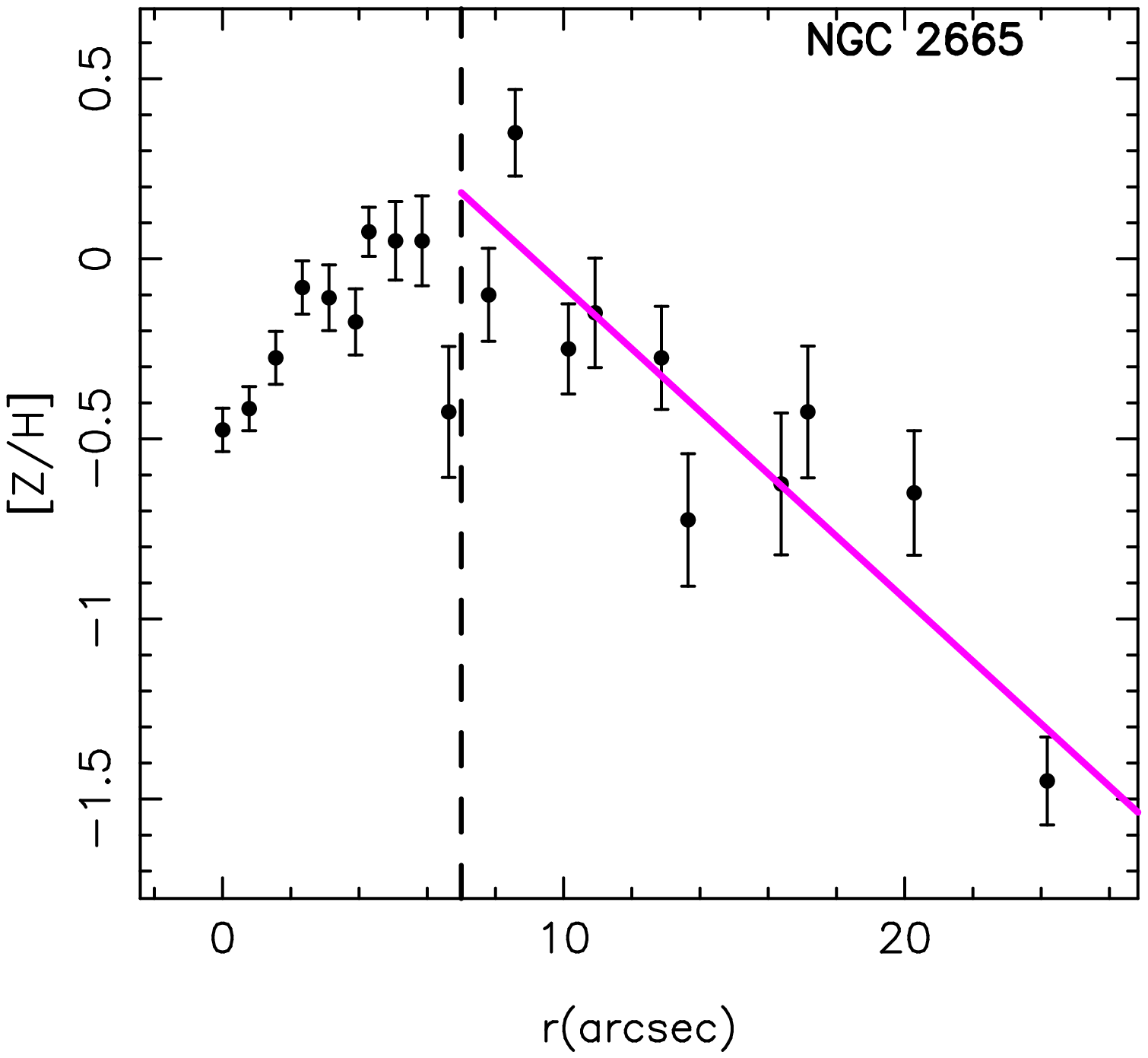}}\hspace{0.8cm}
\resizebox{0.3\textwidth}{!}{\includegraphics[angle=-90]{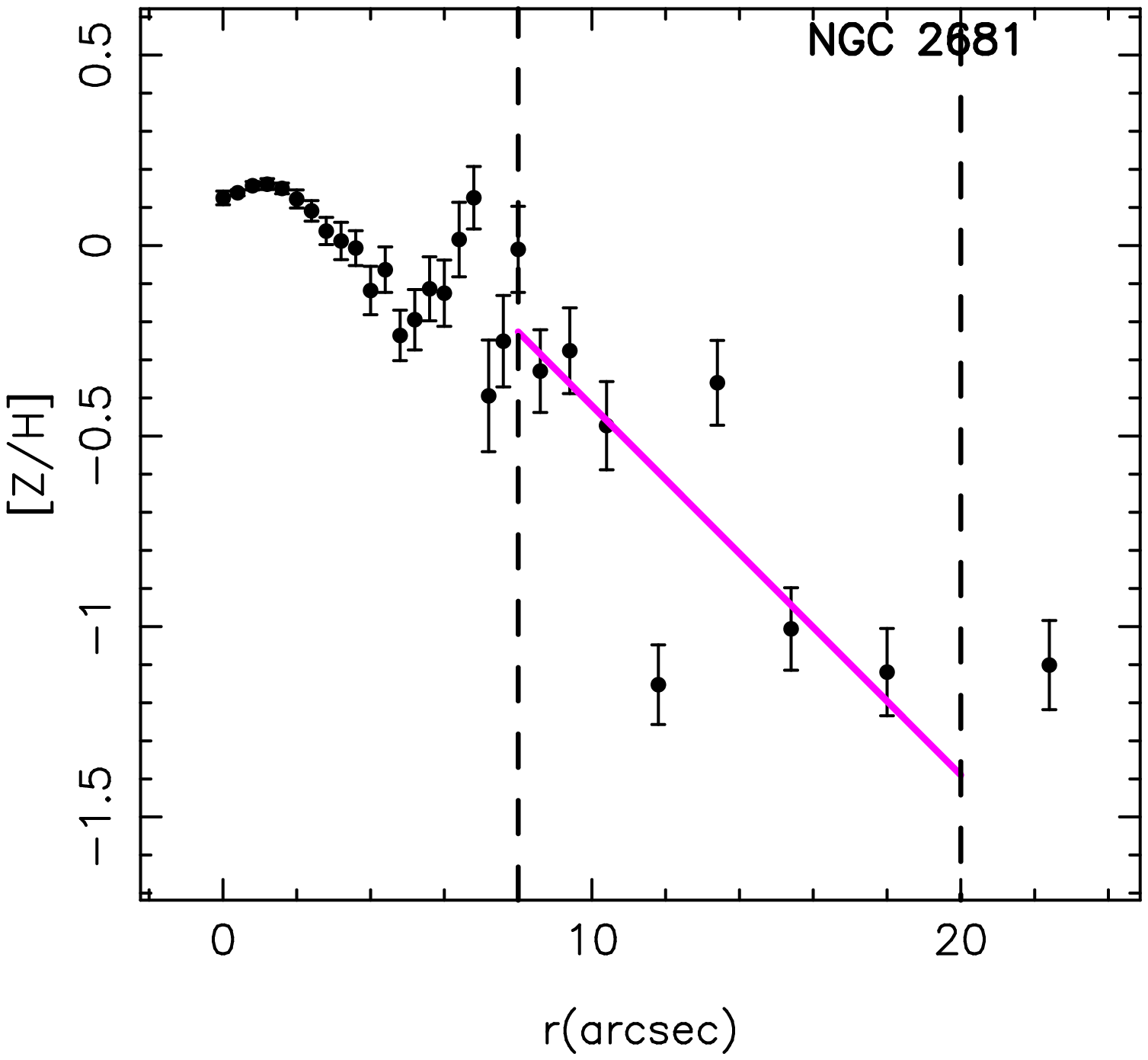}}
\caption{SSP-equivalent age and metallicity along the radius. Dashed lines indicate 
 the beginning and the end of the bar region. A linear fit to the 
 bar region is also plotted.}
\end{figure*}
\addtocounter{figure}{-1}
\begin{figure*}
\resizebox{0.3\textwidth}{!}{\includegraphics[angle=-90]{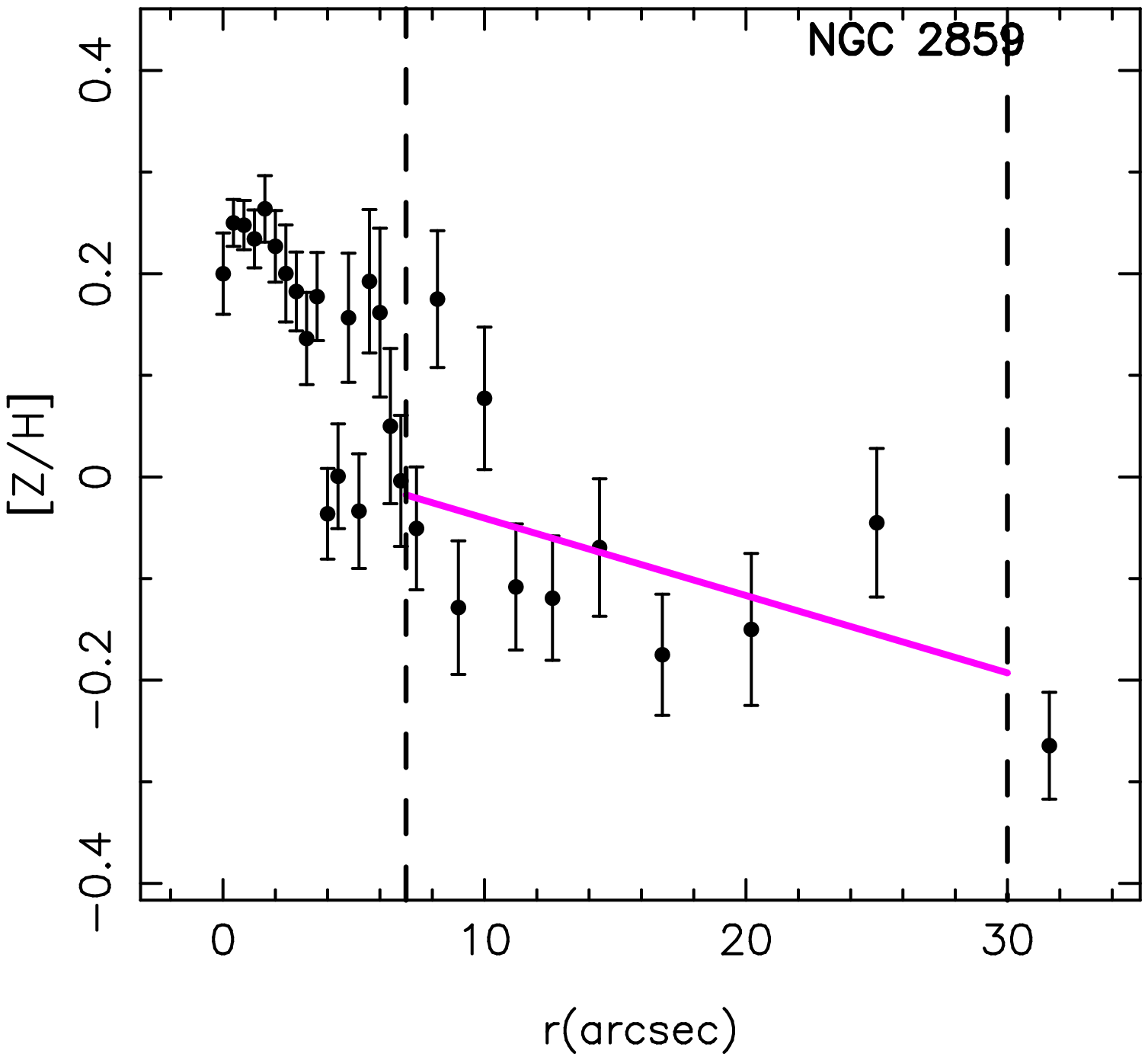}}
\resizebox{0.3\textwidth}{!}{\includegraphics[angle=-90]{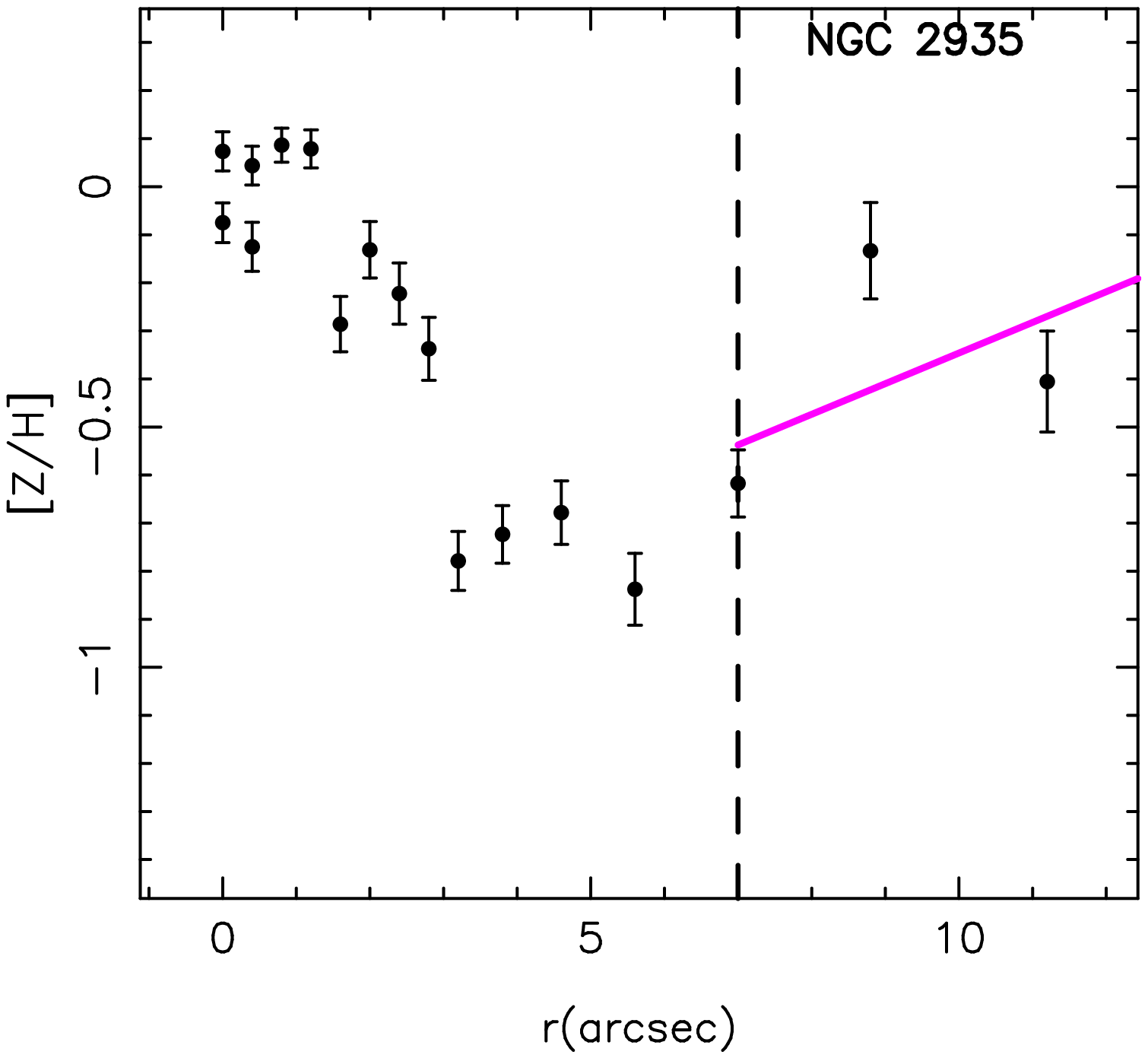}}
\resizebox{0.3\textwidth}{!}{\includegraphics[angle=-90]{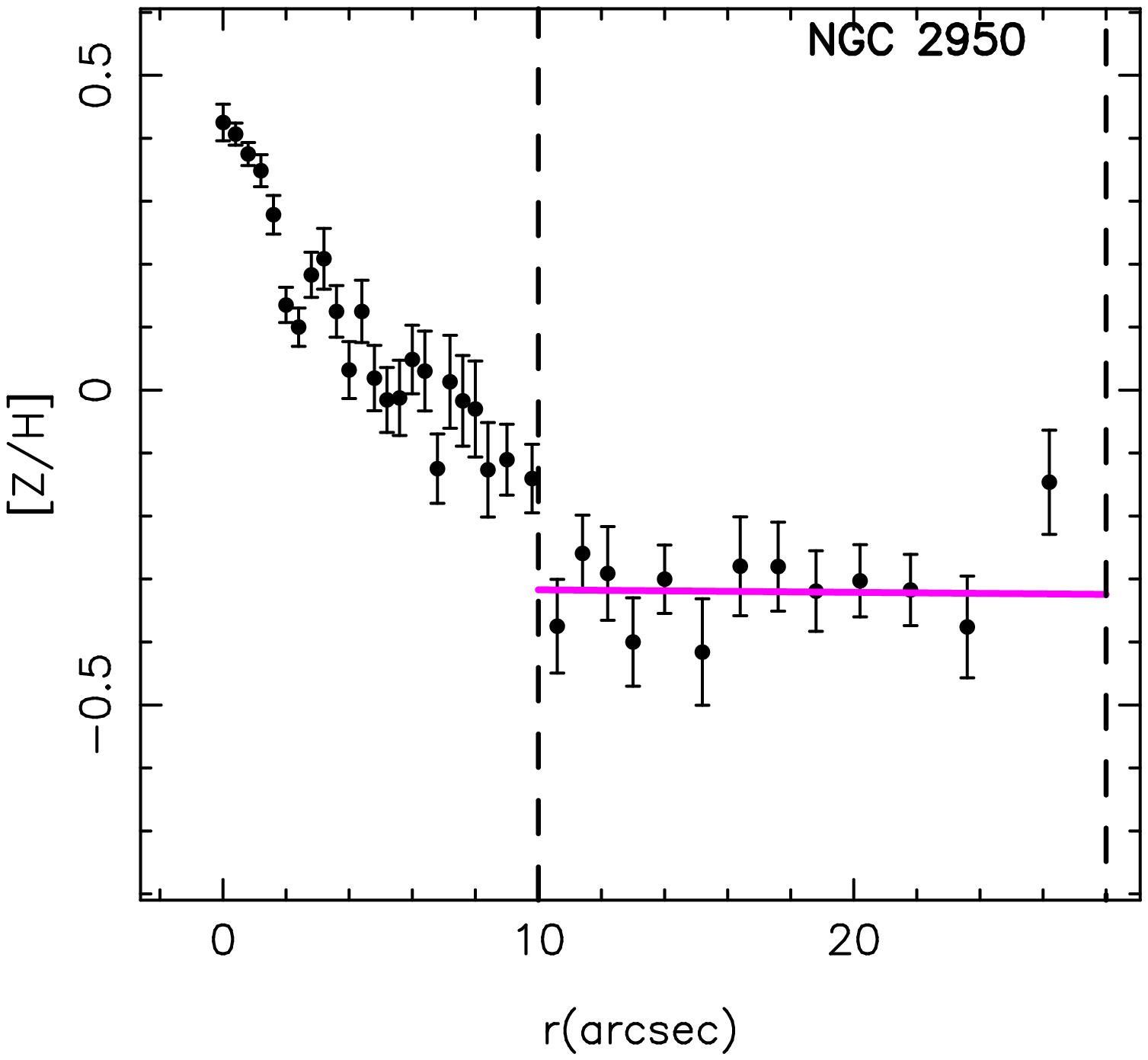}}
\resizebox{0.3\textwidth}{!}{\includegraphics[angle=-90]{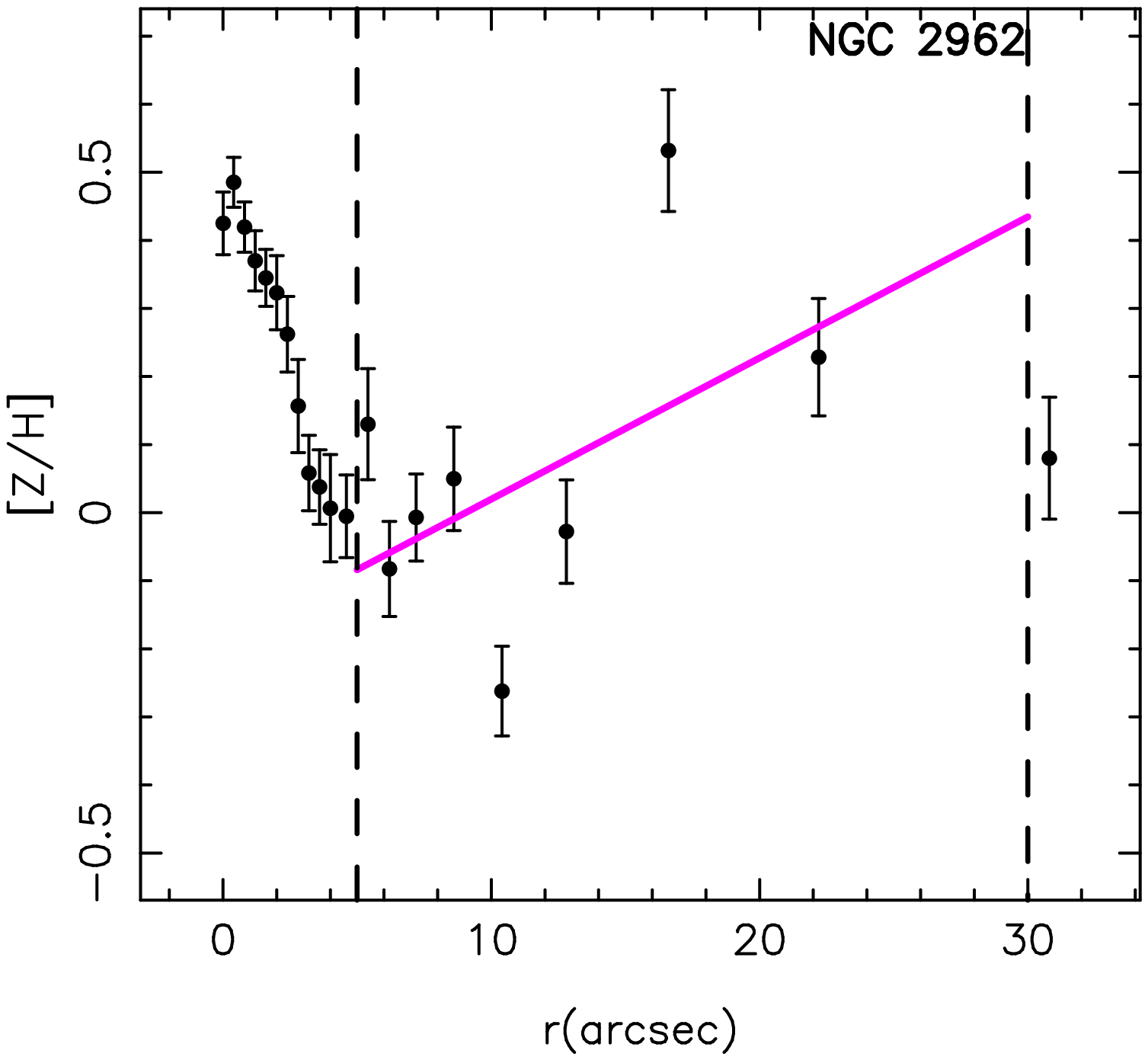}}
\resizebox{0.3\textwidth}{!}{\includegraphics[angle=-90]{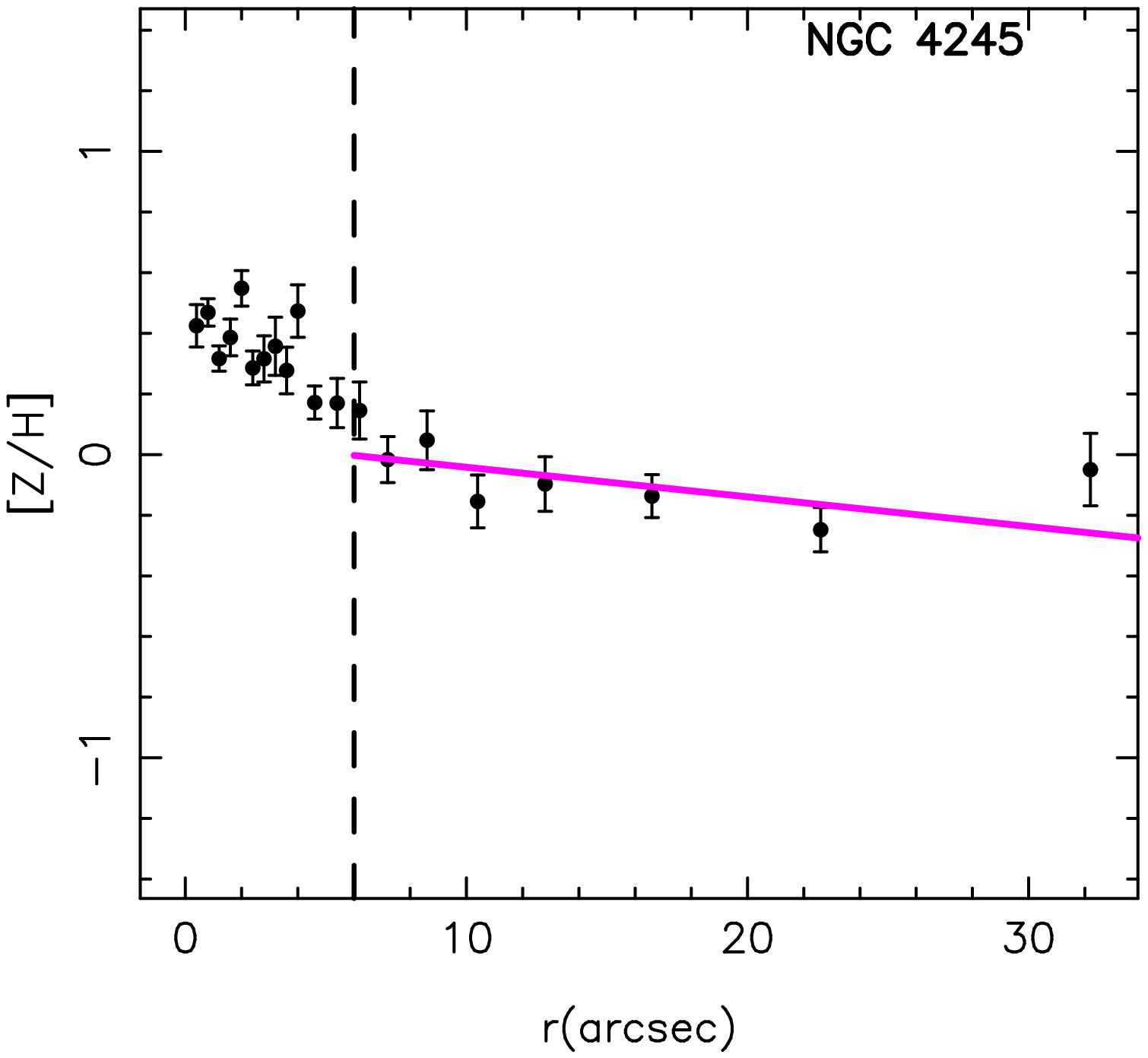}}
\resizebox{0.3\textwidth}{!}{\includegraphics[angle=-90]{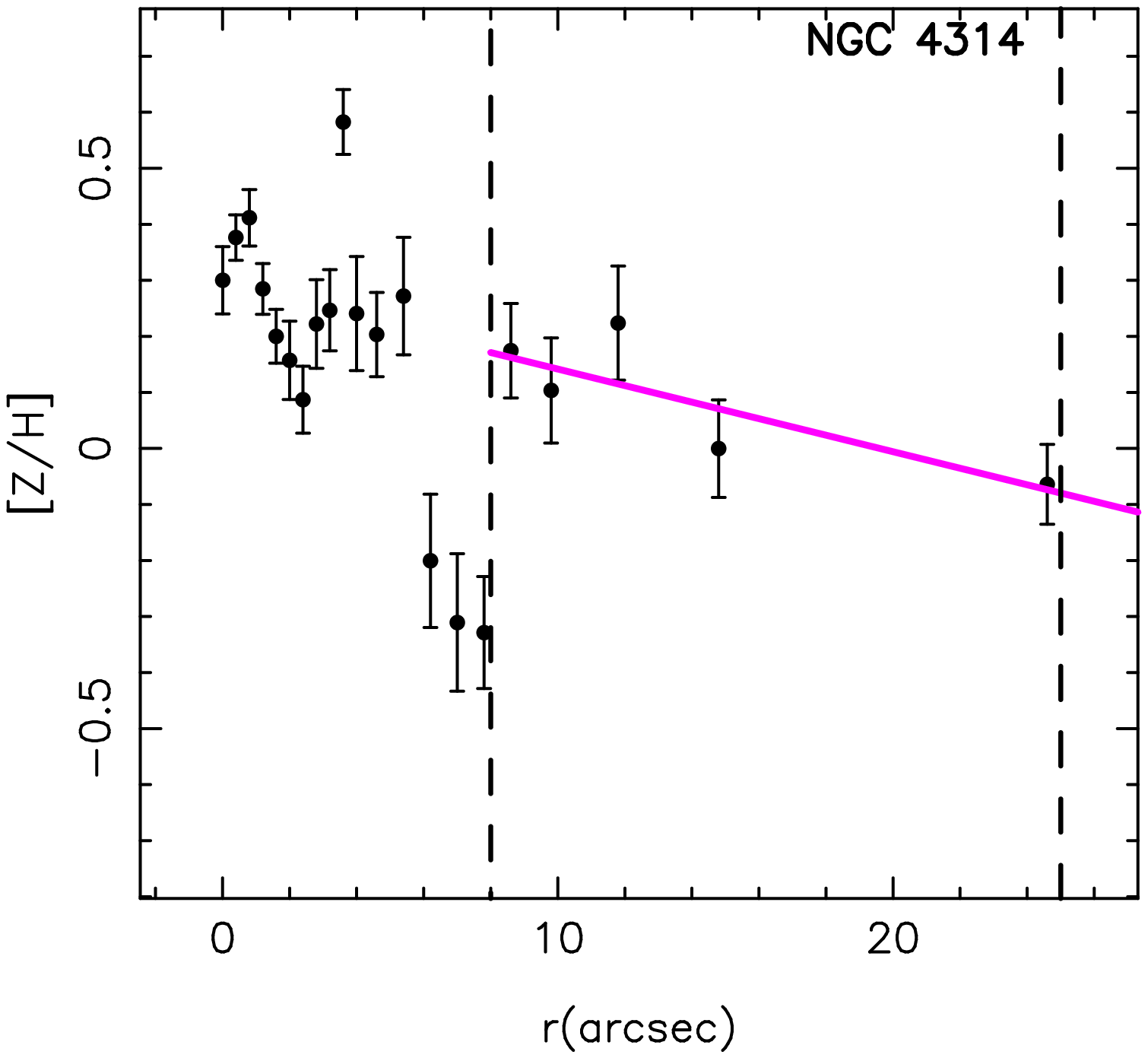}}
\resizebox{0.3\textwidth}{!}{\includegraphics[angle=-90]{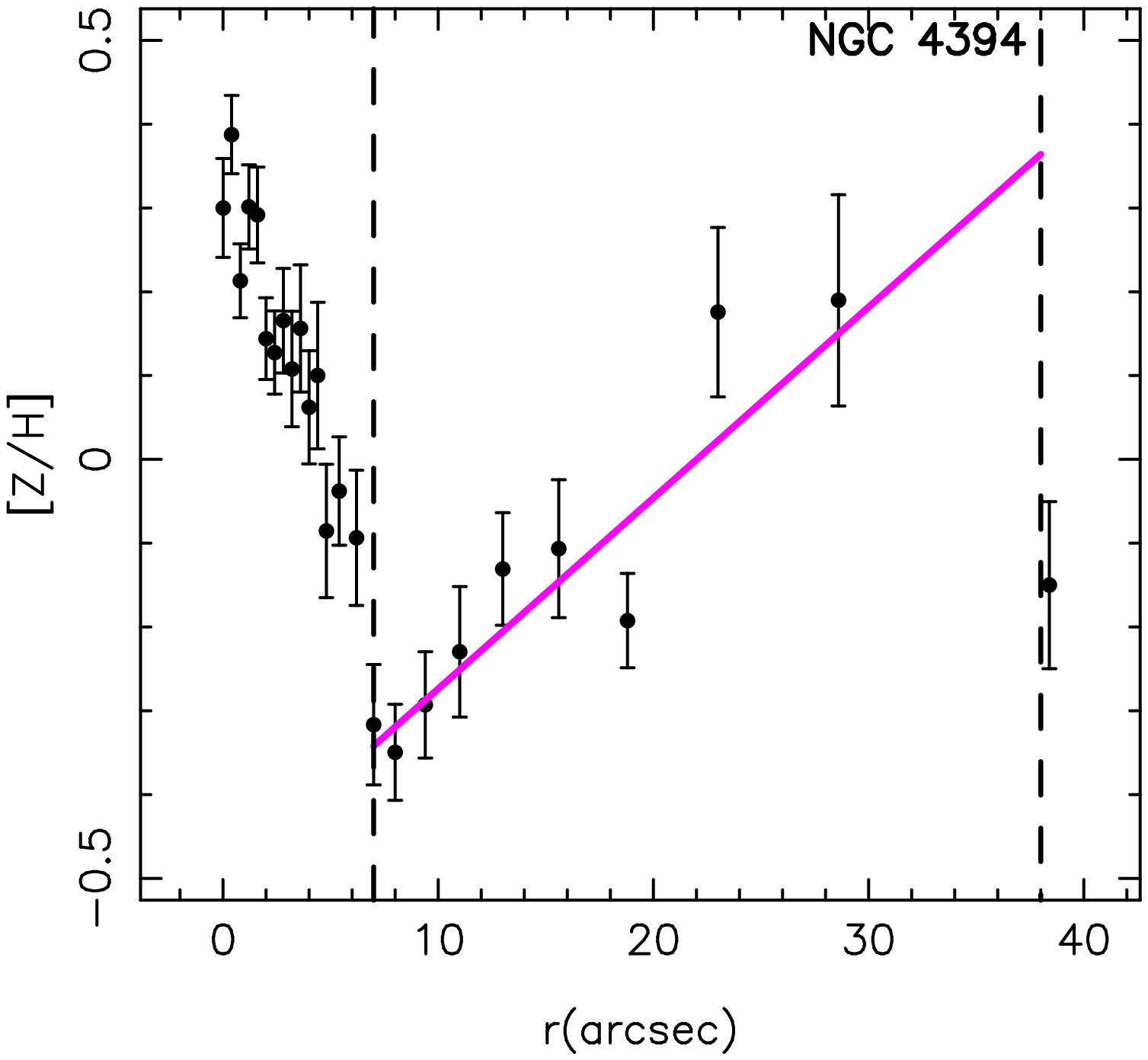}}\hspace{0.8cm}
\resizebox{0.3\textwidth}{!}{\includegraphics[angle=-90]{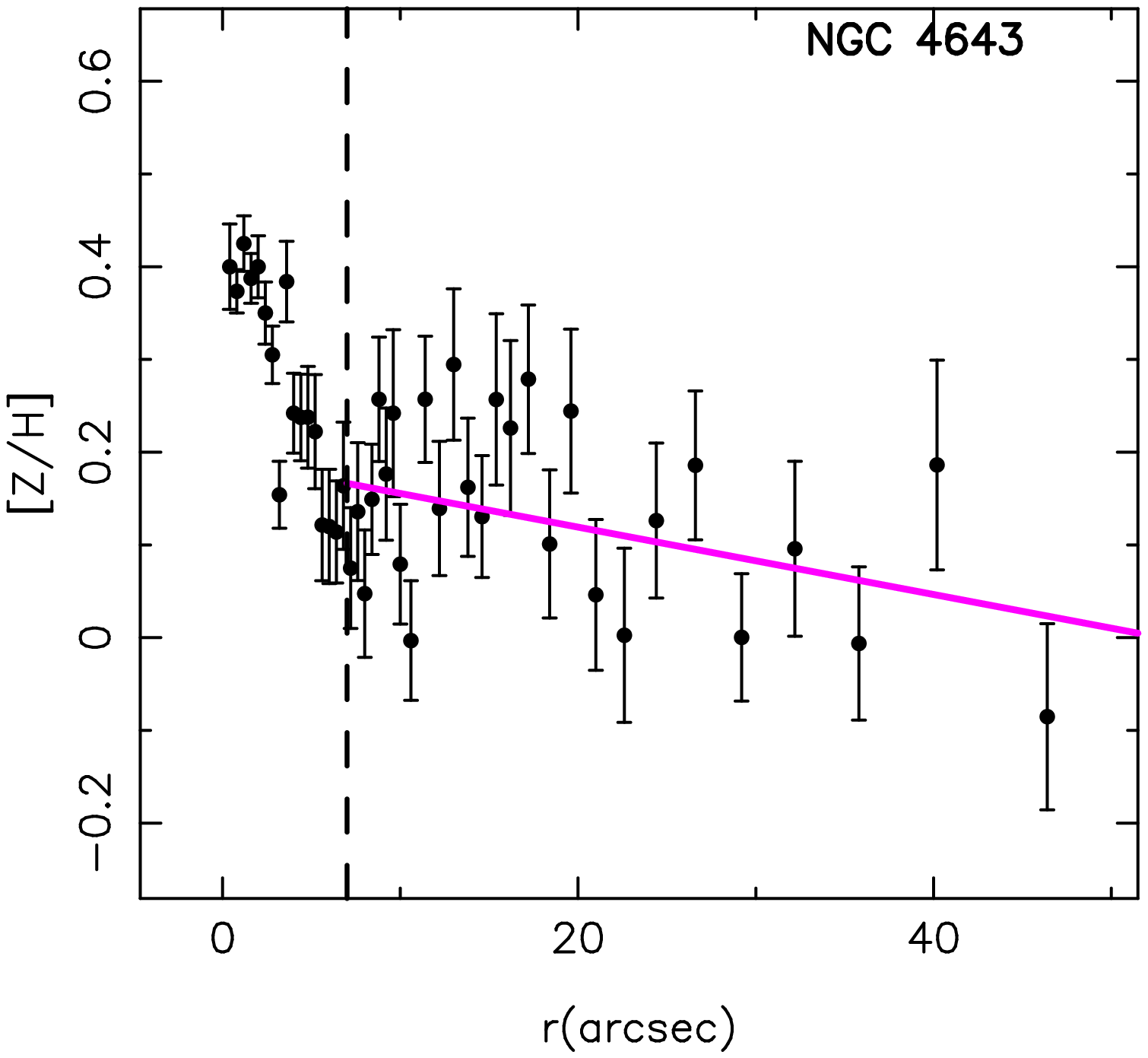}}\hspace{0.8cm}
\resizebox{0.3\textwidth}{!}{\includegraphics[angle=-90]{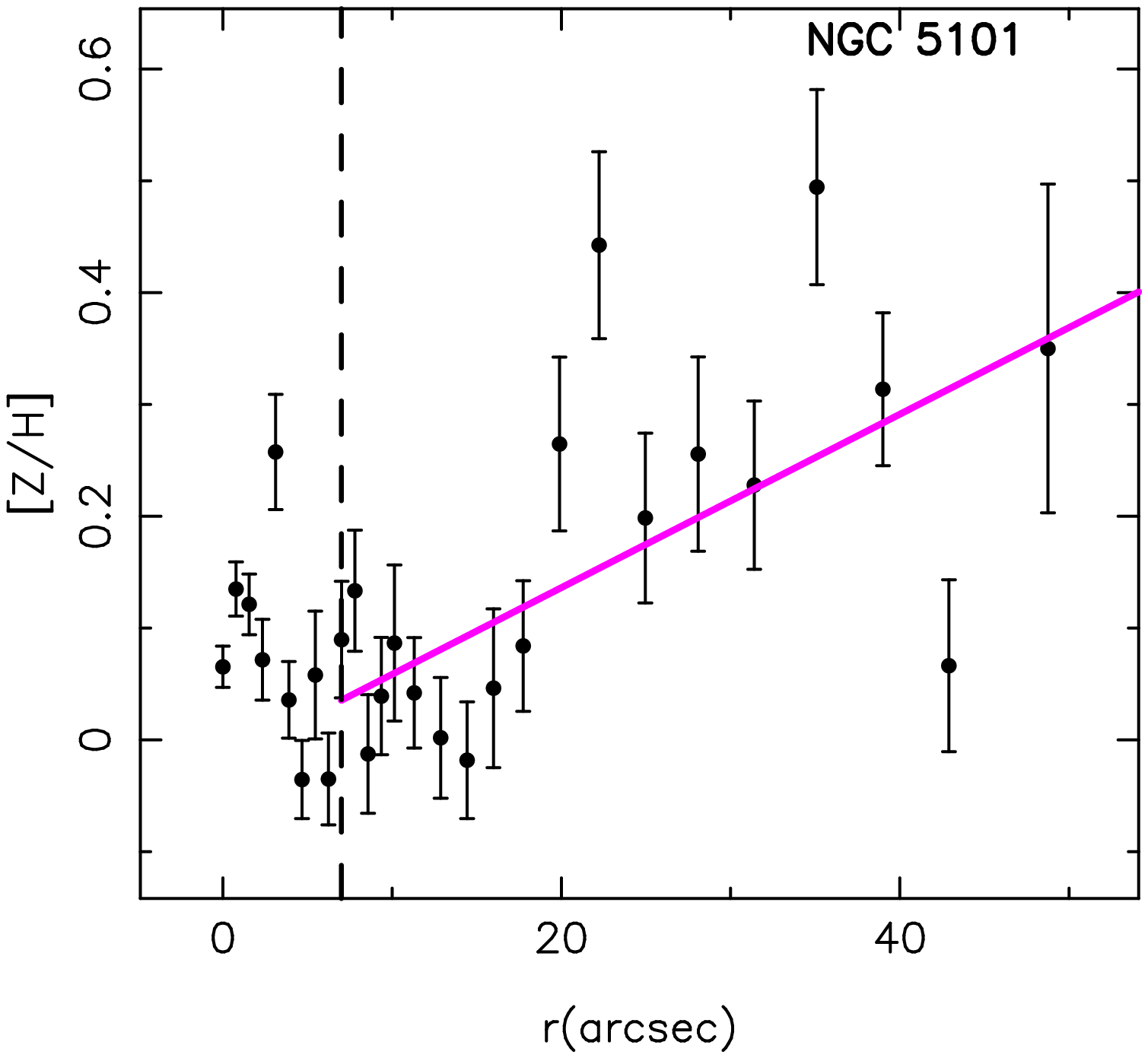}}
\caption{SSP-equivalent age and metallicity along the radius. Dashed lines indicate 
 the beginning and the end of the bar region. A linear fit to the 
 bar region is also plotted.}
 \end{figure*}



\begin{table}
\caption{Metallicity and age gradients
\label{gradtab}}
\begin{center}
\begin{tabular}{ l r  r}     
\hline \hline
Galaxy name & d[Z/H]/dr    & d$\log$(age)/dr \\
\hline
NGC~1169& $0.023\pm 0.008$  & $0.0004\pm  0.0006 $\\
NGC~1358 &  $0.015\pm  0.011$  &$-0.0029\pm  0.0263$\\
NGC~1433 &   $0.004\pm  0.005$ & $-0.0147\pm  0.0093$\\
NGC~1530 &    $0.052\pm  0.012$ & $-0.0143\pm  0.0070$\\
NGC~1832 &    $-0.005\pm  0.008$ & $-0.0282\pm  0.0116$\\
NGC~2217 &     $0.023\pm  0.012$& $-0.0524\pm  0.0140$\\
NGC~2273 &     $0.019\pm  0.011$ & $-0.0122\pm  0.0063$\\
NGC~2523 &    $-0.021\pm  0.027$  & $0.0000\pm  0.0299$\\
NGC~2665  &   $-0.087\pm  0.014$  & $0.0405\pm  0.0150$\\
NGC~2681  &   $-0.097\pm  0.033$  & $0.0688\pm  0.0180$\\
NGC~2859  &   $-0.008\pm  0.006 $ & $0.0079\pm  0.0082$\\
NGC~2935 &     $0.064\pm  0.040 $& $-0.1371\pm  0.0300$\\
NGC~2950 &    $-0.000\pm  0.004 $ & $0.0047\pm  0.0039$\\
NGC~2962 &     $0.021\pm  0.015 $& $-0.0257\pm  0.0138$\\
NGC~3081 &    $-0.587\pm  0.556 $& $-0.0661\pm  0.6848$\\
NGC~4245 &    $-0.010\pm  0.005 $ & $0.0337\pm  0.0098$\\
NGC~4314 &    $-0.014\pm  0.005 $ & $0.0184\pm  0.0096$\\
NGC~4394 &     $0.023\pm  0.005 $& $-0.0246\pm  0.0066$\\
NGC~4643 &    $-0.004\pm  0.002 $ & $0.0063\pm  0.0044$\\
NGC~5101 &     $0.008\pm 0.002 $&  $-0.0268\pm  0.0012$\\ 
\hline
\end{tabular}
\end{center}
\end{table}

\begin{figure}
\begin{center}
\hspace{-1.0cm}\includegraphics[width=6.50cm,angle=-90]{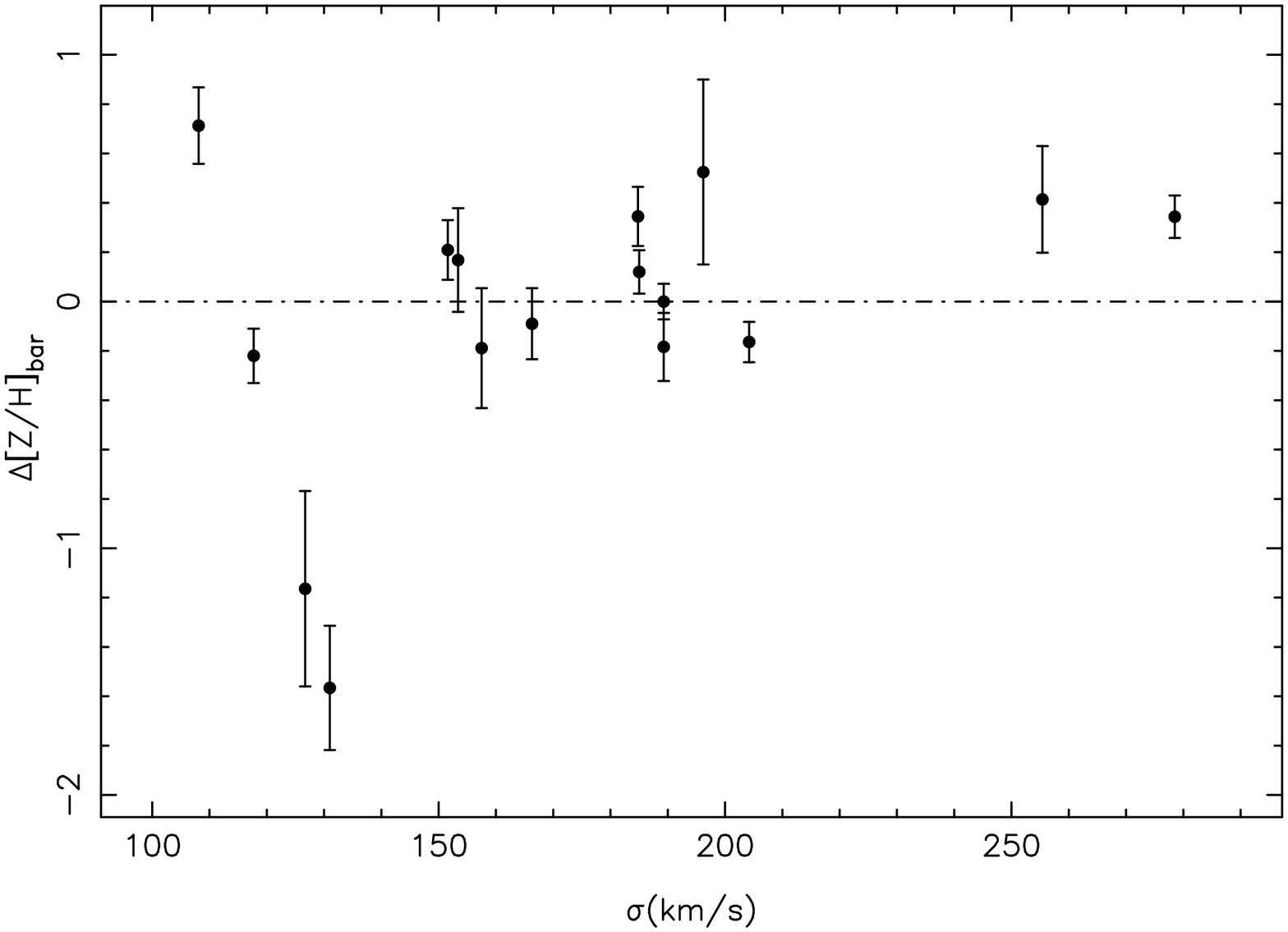}
\caption{ \label{sigma-gradz}  Metallicity difference between the inner and the outer bar region as a function of maximum central velocity dispersion. Notice that galaxies with $\sigma$$~<$~170~km~s$^{-1}$ tend to have negative gradient values.}
\end{center}
\end{figure}

We therefore, confirm the results from P\'erez et~al.\ (2007), where 
we found, in a much more reduced sample,  that some of the bars had positive metallicity gradients
and negative age gradients. These negative age gradients are expected since, as  it has been shown in numerical simulations, 
there is an accumulation 
of young stellar population trapped on elliptical-like orbits along the bar, near the ultra-harmonic resonance 
(Wozniak 2007)\nocite{wozniak07}. Although expected, this negative age gradient should be further investigated by the models to check whether the gradient can stay for longer than 3 Gyr, time at which the simulations where stopped and shorter than the age of the last star formation bursts found for our galaxies. 

What we found more striking is the presence of positive metallicity gradient in four galaxies, i.e. the younger stars at the end of the bar are also more metal rich. These gradients are difficult to understand in the context of our current understanding of bar formation (see Section 6), if they are a consequence of  the star formation ocurred during its formation, this would imply that there has been chemical enrichment and, therefore,  that the star formation lasted for a long period of time or that it has been very efficient. 

The galaxies showing a negative metallicity gradient (i.e. less metal rich in the outer parts of the bar) might be reflecting the original disk gradient. Only for a 
few galaxies the disk radial stellar gradients have been derived. Moll\'a, Hardy \& Beauchamp (1999) \nocite{molla} derived shallow 
negative gradients for the Mg$_{2}$ and Fe5270 indices in the discs of NGC~4303 (SABbc), NGC~4321 (SABbc) and 
NGC~4535 (SABc).  Ryder, Fenner \& Gibson (2005) \nocite{ryder05} obtained the radial disk profiles of the Mg$_{2}$ and Fe5270 indices 
for a sample of 8 late-type galaxies with similar results. In a recent work, Yoachim \& Dalcanton (2008) have derived luminosity weighted ages an metallicities in the disk region of nine edge-on galaxies, finding radial negative age gradients and almost no metallicity gradient. In order to follow up this issue, we have collected deep spectroscopic information in the disk of our sample (S\'anchez-Bl\'azquez et al. 2009, in preparation).

Figure \ref{sigma-gradz} shows the relation between the metallicity gradient along the bar and the velocity dispersion.
It can be seen that the galaxies with velocity dispersions higher than 170~kms$^{-1}$ tend to have positive metallicity gradients, while the 
lower velocity dispersion galaxies show a larger dispersion in their metallicity gradient value. Similar behavior is found for the age gradient, 
where galaxies with large velocity dispersion tend to have a shallow positive  age gradient while for the lower velocity dispersion galaxies the absolute
 value of the age gradient is larger, both negative and positive gradients. This seems to imply that the stellar population gradient is, somehow, related to the mass of the galaxy's bulge. To test whether there is a link between the mass of the bulge and the total galaxy mass, we have investigated the correlation between the bulge dispersion velocity and the total $H$-band 
luminosity (as a tracer of the stellar galaxy mass). There is, indeed, a correlation between the maximum central velocity dispersion and the total $H$-band 
luminosity, which might link the observed metallicity gradients to the stellar mass of the galaxy. Further investigation needs to be done to address this issue. The galaxies with negative metallicity gradients have systematically, positive age gradients and have average  bar--age values  of the order of 2 Gyr (except for NGC~4245 with an average of 5 Gyr). 

\subsection{Correlation with bar/galaxy parameters
\label{correlation.bar}}
We want to check possible connexions between observational structural bar parameters, that are thought to be important in the bar formation and evolution, with the stellar population parameters derived in this work.  The bar--strength is determinant in the orbital structure of bars and it is thought to evolve as the bar ages (e.g. Athanassoula 2005)\nocite{athanassoula05}.  For this purpose, we have used the available bar--strengths obtained as the ratio of the tangential force to the mean radial force derived from the NIR images (Combes \& Sanders 1981, 
Block et al. 2001, 2004)\nocite{combes81}.  For 13 out of the 20 galaxies there are values for the bar--strength from the literature (Table~\ref{sample.tab}). No clear relation is found between the
 bar--strength and the metallicity gradients.
 
It has been suggested that older bars should get longer and thinner in time as their pattern speed decreases, in order to check if there is any 
systematic behavior between the bar sizes and the metallicity gradients,  we have checked the absolute bar sizes and the relative sizes of the bar 
with respect to their disk  for the galaxies in our sample. We find, in agreement with previous studies (Erwin, 2005)\nocite{erwin}, that the bar sizes 
are amongst the typical sizes of S0--Sab (between  1 and 8 kpc in our sample) and that the sizes correlate well with the disk size (P\'erez et al. 2005,
 Erwin 2005)\nocite{perez05}. There is no significant trend between the size of the bar and the metallicity and age gradient.   However, there  might be  a hint of correlation
pointing to larger bars having positive age gradients, negative metallicity gradients and an average young population. But, due to low number statistics and the  large spread in the values, nothing conclusive can be said to this respect. However, it would be interesting to follow-up this point calculating the bar--size for all the galaxies with new broad-band data to test if there is a significant trend of the bar stellar population parameters with bar--size.
 Rings are located near resonances induced by the bar component, from numerical simulations outer rings form in time-scales longer than 1~Gyr and
 therefore, they have been associated to and relatively aged system. However, in some simulations, rings appear very quickly, about 0.2~Gyr after 
the bar formation (Rautiainen \& Salo, 2000). We do not see any correlation between the presence of an outer ring and the bar age. Some of the bars
 with young populations and negative metallicity gradients present outer rings, therefore, the presence or absence of an outer ring is not a good discriminator of
 the bar age. 

There  seems  to be  no  clear  correlation  with morphological  type, although we do not cover a large
 range in morphological types (S0--Sbc), the galaxies with positive metallicity gradients in the bar region are not particularly different in
 morphology to the rest of the sample. 
  
\subsubsection{AGN connection 
\label{agn.connection}}
Since the bar is an efficient mechanism to drive material to the central parts of the galaxy it has been very much discussed its role in 
feeding the central massive black hole and the bar--AGN connection.
Recent statistical studies (e.g. Knapen, 2004) have concluded that there is not a statistically significant relation between the presence of an AGN 
in barred galaxies. Part of the problem in the lack of  correlation between the AGN and the bar presence could be the different time--scales of the
 different processes. There have been recent claims, using broad--band photometry (Gadotti \& Souza, 2004) that active galactic nuclei tend to have 
young bars, while galaxies with old bars do not generally host AGNs. These studies are based on the assumption that the change in broad--band galaxy 
color is due to age and not to metallicity as well. Because spectra provide a better way to discern between both age and metallicity, and since the 
indices are not affected by dust, we have carried out this project also to see whether  there is any correlation between the bar ages and metallicities 
with the presence of an AGN.  
In Fig.~\ref{mean.age.sig} we have separated, with different symbols, galaxies
hosting an AGN. We do not see any significant correlation with mean bar age and metallicity nor with the derived gradients. It might only 
be reflecting the small number statistics and a larger sample would be needed to properly cover the parameter space.
Therefore, we do not reproduce
the results of Gadotti \& de Souza (2006). However, the correlation 
that these authors
found is weak and they were using colors as a proxy for age.

\section{ Discussion and Summary \label{discussion}}


We have derived the kinematics and the distribution of ages and metallicities along the bar in a sample of 20 barred galaxies to study the question
 whether bars are
 long--lived or weather they are destroyed and reformed in a few Gyrs. We also want to explore the relation between the age, and stellar populations
 of the bar and
 other global physical properties of the host galaxies. 

From the study of the kinematics along the bar, we have found that all the galaxies in the sample show a disk--like component in their centers, showing 
as dips or plateaus
in their central stellar velocity dispersion and as stellar rotating disks. Therefore, this implies that bars, as expected, have a strong influence
in the building up and
later evolution of the central component. This aspect will be studied in more detail in S\'anchez-Bl\'azquez, P\'erez and Zurita (2009, in prep) 
where we compare the bulges
ages and metallicities of barred and unbarred bulges of otherwise similar galaxies.

 We have derived ages and metallicities from Lick/IDS line-strength indices of single stellar population (SSP). It is clear that the true star 
formation history in these 
objects will be more complicated than a simple SSP. However, a lot of information can be derived from the SSP analysis. The SSP-equivalent age of 
a composite stellar
population (CSP) is dominated primarily by the age of the youngest component and the mass fraction of the different populations. The SSP-equivalent
metallicity
reflects a $V-band$  luminosity weighted chemical composition (Serra and Trager 2007)\nocite{serra}. 

 As bars form from stars already in the disc, we have to distinguish
   between stellar population gradients  that may have already been present
   when the bar formed
   and those  that formed during the bar formation or afterwards.
   There are not many studies analysing the stellar population
   of discs outside the MW. In particular, for early-type spirals,
  only broadband colours have been traditionally used to estimate
  the ages and metallicities of the stellar discs. However,
   stellar parameters are very difficult to obtain with broad band
   colours. Recently, Yoachim \& Dalcanton (2008)\nocite{yoachim08} have measured Lick indices
   in the thin and thick disc of edge-on galaxies. They found
   a  very strong age gradient and a null or negative metallicity
   gradient dependending on the index they were using. In the MW
   a negative age and metallicity gradient are measured. We
   will assume, for our discussion,  that  this is the 
general trend on the discs
  before the formation of the bar. 

We have found three different types of bars according to their metallicity and age distribution along the bar:  
 
1) Bar with negative metallicity gradient. These bars show a mean young/intermediate population ($<$~2Gyr) and have amongst the 
lowest central stellar velocity dispersion of the sample. They also tend to have positive age gradients (i.e. younger populations at the beginning of the bar). In fact, all the  galaxies  showing these trends posses circumnuclear regions with young population ($~$1.5~Gyr). The negative gradient can be understood as the original gradient
   of the disc. This is because a small fraction of young stars can bias
  the mean-age measured
   with SSP towards very small values. However, the mean
   metallicity is more sensitive to the more massive component.
   In this picture, the circumnuclear region would be strongly affecting
   the mean age at the beggining of the bar (lowering it)  but its
   influence on the mean metallicity would not be so strong.
   Since one would expect the original metallicity gradients  to flatten by the  velocity dispersion of the bar (Friedli et al. 1994)\nocite{friedli94}, the presence of metallicity gradients may be indicating that they formed recently.

2) Bars with null metallicity gradients. The galaxies that do not show any gradient in their metallicity distribution along the bar have negative
 age gradients, that is, their ends of the bar are younger. These galaxies present gradients in both metallicity and age compatible with the models that studied mixing of metals and the distribution of stellar population ages in bars using numerical simulations (e.g Wozniak 2007\nocite{wozniak07}; Friedli et al. 1994). They predict low age regions at the end of bars due to the accumulation of young stellar populations trapped in orbits near the ultra-harmonic resonance. During the first Gyr, in the simulations, most of the star formation happens. We observe that, although, the population at the end of the bar is younger than the average bar value, it is still larger than 3 ~Gyr. This fact might be indicating that the bar in these galaxies is at least 3~Gyrs old.

 3) Bars with positive metallicity gradients. These galaxies
          are the ones with older mean ages in their stellar populations.
          They are also the ones with more massive bulges (larger
          $\sigma$).
          In principle, this positive metallicity gradients are
          difficult to understand in light of our current knowledge
          of formation and evolution of bars. Hydrodynamical simulations
          indicate that, even if metal rich population can be formed
          at the end of the bar, the diffusion timescale in this
          region is very short, of the order of 50 Myr (see figure 3
          in Wozniak 2007).
          The metallicity gradient in disc galaxies is believed
          to be negative (see references above) and, therefore,
          it is not likely that this gradient is a relic of the
          original metallicity gradient on the disc.
          We do not have a definitive answer to the origin
          of this metallicity gradients.
           One posibility is the presence of an inner
         ring at the beggining of the bar with enriched SF that could be populating the extremes of the bar in the same fashion as we explained for the presence of young stellar populations in the outer parts of the bar (case 2). However, we do not see this effect in all the galaxies hosting a nuclear young ring. Furthermore, the luminosity weighted mean age of the stars at the beggining of the bar is older than that in the external parts which is not expected if  the bar ends  are being populated with an enriched population coming from the inner ring. Projection effect of the old metal-rich population of the bulge on the bar region is not a likely scenario, since, at it can see from the kinematics, the outer part of the bars are dominated by a cooler stellar component.

 It has been claimed that the star formation in bars follows a sequential pattern, where younger bars ($<$~1Gyr) would show star formation along the entire bar and later the star formation would be concentrated
 on the bar--ends and the nuclear region (Phillips 1993; Martin \& Friedli 1997; Martin \& Roy 1995; 
 Friedli \& Benz 1995; Verley et al. 2007)\nocite{phillips,martin95,friedli95,martin97,verley}. We have checked the galaxies for emission lines coming from H{\small II} regions to see if the ones showing negative metallicity gradients present systematically H{\small II} along their bars.
   
 NGC~2665 and NGC~1530 show nebular emission along their bars belonging to H{\small II} regions, which reflects very recent
 star formation events. Unfortunately, we do not have high S/N information on the stellar indices for the whole bar in NGC~1530 that would allow us 
 to establish the stellar population age and metallicity. NGC~2665, on the other hand is one of the galaxies that show a negative metallicity and a  positive age gradient along its bar. This is compatible with the idea of this sequential 
 star formation pattern if we assume that these bar have formed less than 3 Gy ago. However, other bars such as, NGC~4245 and NGC~2681, although belonging to the same case as NGC~2665, do not present any  H{\small II} regions along their bars.
 
The lowest average bar value of the $\alpha/$Fe, here expressed as [E/Fe] (Trager et al. 2000),  tends to have lower central velocity 
dispersion values (Fig.~\ref{mean.age.sig}), these galaxies have also in general a lower average age (Fig.~\ref{mean.age.sig}).  The galaxies 
with higher central velocity dispersion stopped forming star earlier than galaxies with lower velocity dispersion. However, this is not an
indication that the galaxies with lower central velocity dispersion host a younger (i.e. a more recently formed) bar.

 \section {Conclusions \label{conclusions}}

We have  measured, for the first time, line-strength indices along the bar of a sample of 20 early-type spiral galaxies and S0.
We find that the mean values of age, metallicity and [E/Fe] correlate with central velocity dispersion in a similar way to that of bulges, pointing to a intimate evolution of both components. Galaxies with high stellar velocity dispersions ($>$170~kms$^{-1}$) host bars with old stars while galaxies with lower central velocity dispersion show  stars with a large dispersion in their ages.

We find, for the first time, gradients in age and metallicity along the bars of all the galaxies. We have found three types of bars according to their metallicity and age distribution along the radius. Most galaxies could be compatible with having formed more than 3 Gyr ago, although
a few of them show characteristics compatible with having been formed less than $<$ 2 Gyr ago. These derived gradients place strong constrains on models of bar evolution.  

We have explored the possible correlation of the bar age with other bar/galaxy parameters, such as
 bar--strength, bar size,  morphological features, and AGN presence without finding any significant trend. All the galaxies show disk--like central components, implying a 
 strong role played by bars in the bulge secular evolution. Among these disk--like central components we find some galaxies hosting an old, luminosity weighted, stellar component  ($>$ 3 Gyr) and others with a recent ($<$ 1 Gyr) burst. Therefore, it would interesting to separate the bulge an inner disk populations to analyse their stellar population parameters in order to investigate the origin of the $\sigma$-drops. This will be presented in a forthcoming paper.
 
 \begin{acknowledgements}
  We are indebted to Dr. Marc Sarzi and Dr. Jesus Falc\'on-Barroso for their help and advises in the use of GANDALF. This work would not have
been possible without it. We are also very thankful to Dr. Jes\'us Gonz\'alez for sharing his wisdom with us. We are thankful to Dr. H. Wozniak for his useful comments which have greatly improved the manuscript.
  This research has made use of the NASA/IPAC Extragalactic Database (NED) which is operated by the Jet Propulsion Laboratory, California
 Institute of Technology, under contract with the National Aeronautics and Space Administration.  We have also made use of the Digital Sky Survey. The Digitized Sky Surveys were produced at the Space Telescope Science Institute under U.S. Government grant NAG W-2166. The images of these surveys are based on photographic data obtained using the Oschin Schmidt Telescope on Palomar Mountain and the UK Schmidt Telescope. I. P\'erez is supported by a postdoctoral
 fellowship from
 the Netherlands Organisation for Scientific Research (NWO, Veni-Grant 639.041.511) and the Spanish Plan Nacional del Espacio del Ministerio de
 Educaci\'on y Ciencia. 
 P.S.B is  supported by a Marie Curie Intra-European
 Fellowship within the 6th European Community Framework Programme. I.P. thanks support for a visit to the University of Central Lancashire from
  a Livesey Award. A.Z acknowledges support from the Consejer\'{\i}a de Eduaci\'{o}n y Ciencia de la Junta de Andaluc\'{\i}a
   \end{acknowledgements}
 
\bibliography{abundance}
\bibliographystyle{natbib}
\appendix
\section{Line-strength distribution}
Fig.~\ref{line-strength} shows the line-strength gradients in the bar region
for all the galaxies.

\begin{figure*}
\resizebox{0.3\textwidth}{!}{\includegraphics[angle=-90]{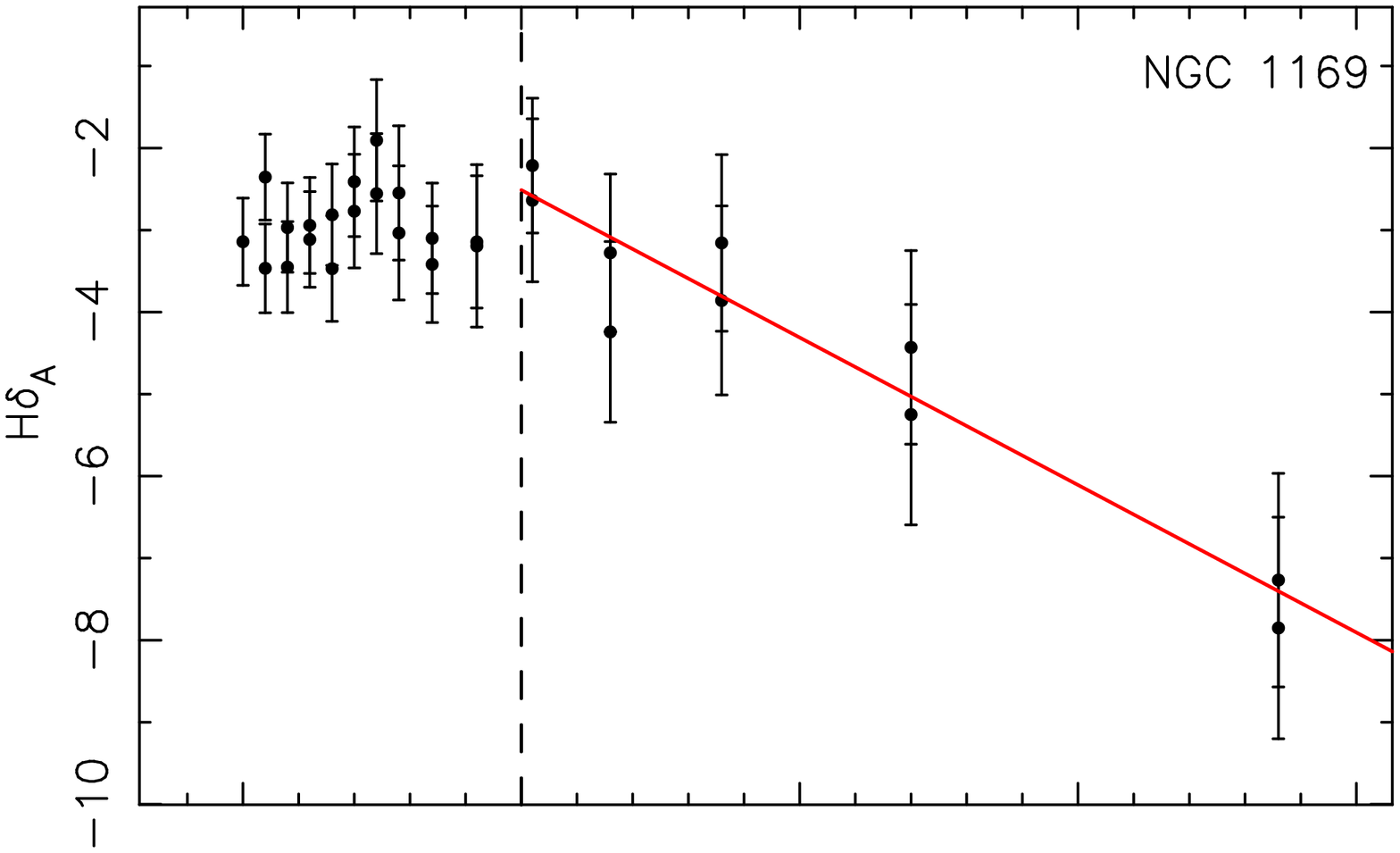}}
\resizebox{0.3\textwidth}{!}{\includegraphics[angle=-90]{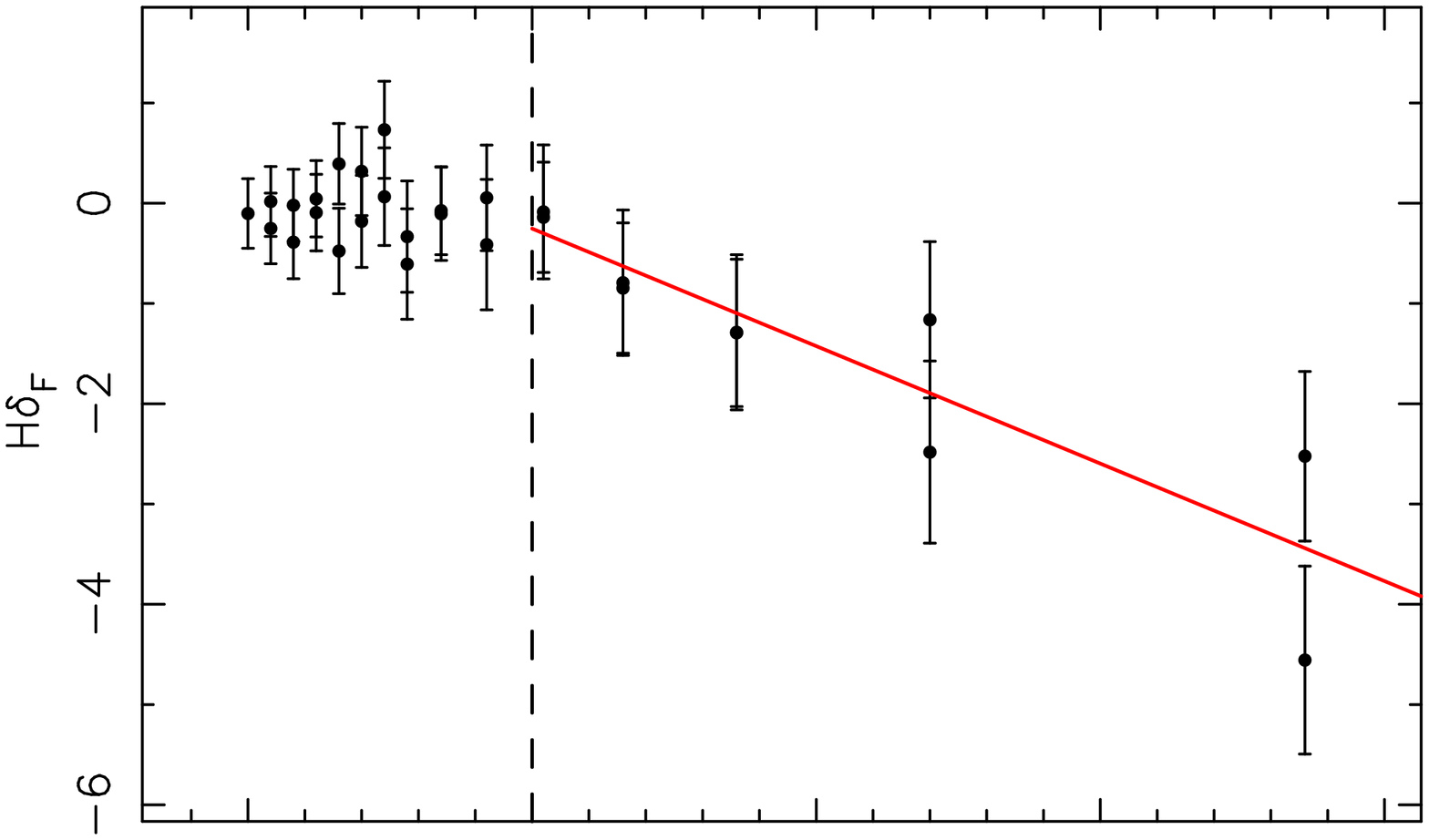}}
\resizebox{0.3\textwidth}{!}{\includegraphics[angle=-90]{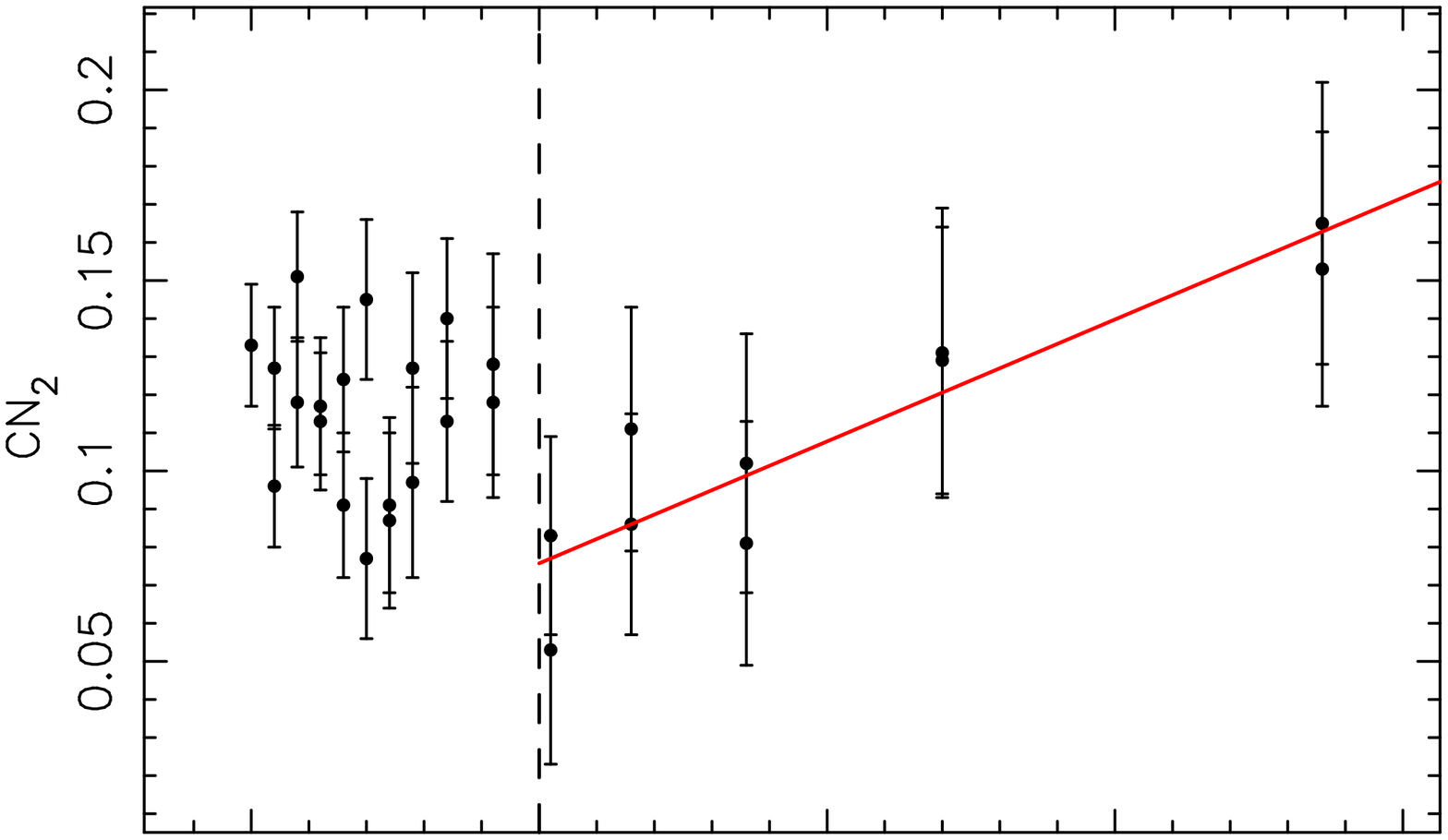}}
\resizebox{0.3\textwidth}{!}{\includegraphics[angle=-90]{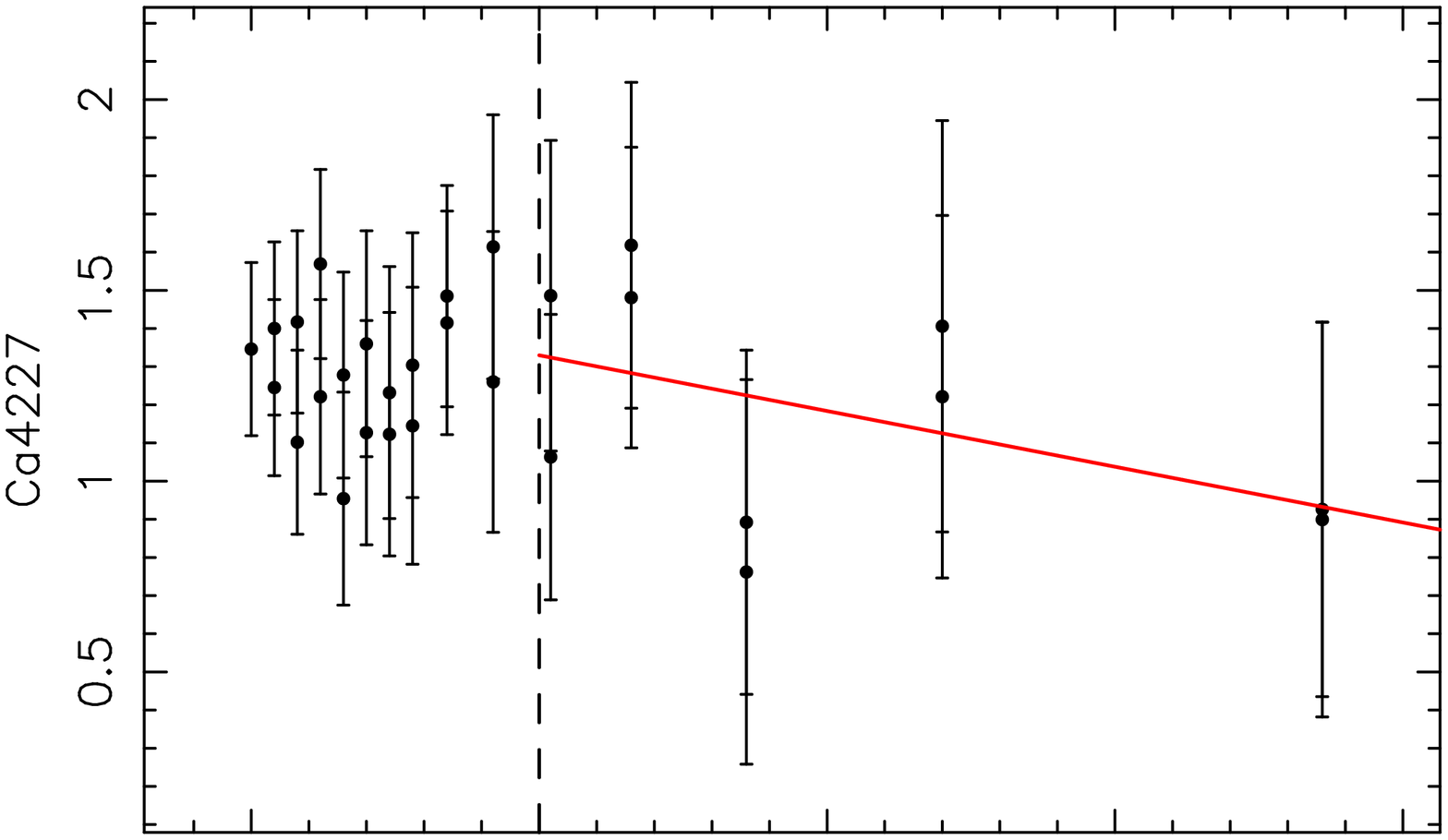}}
\resizebox{0.3\textwidth}{!}{\includegraphics[angle=-90]{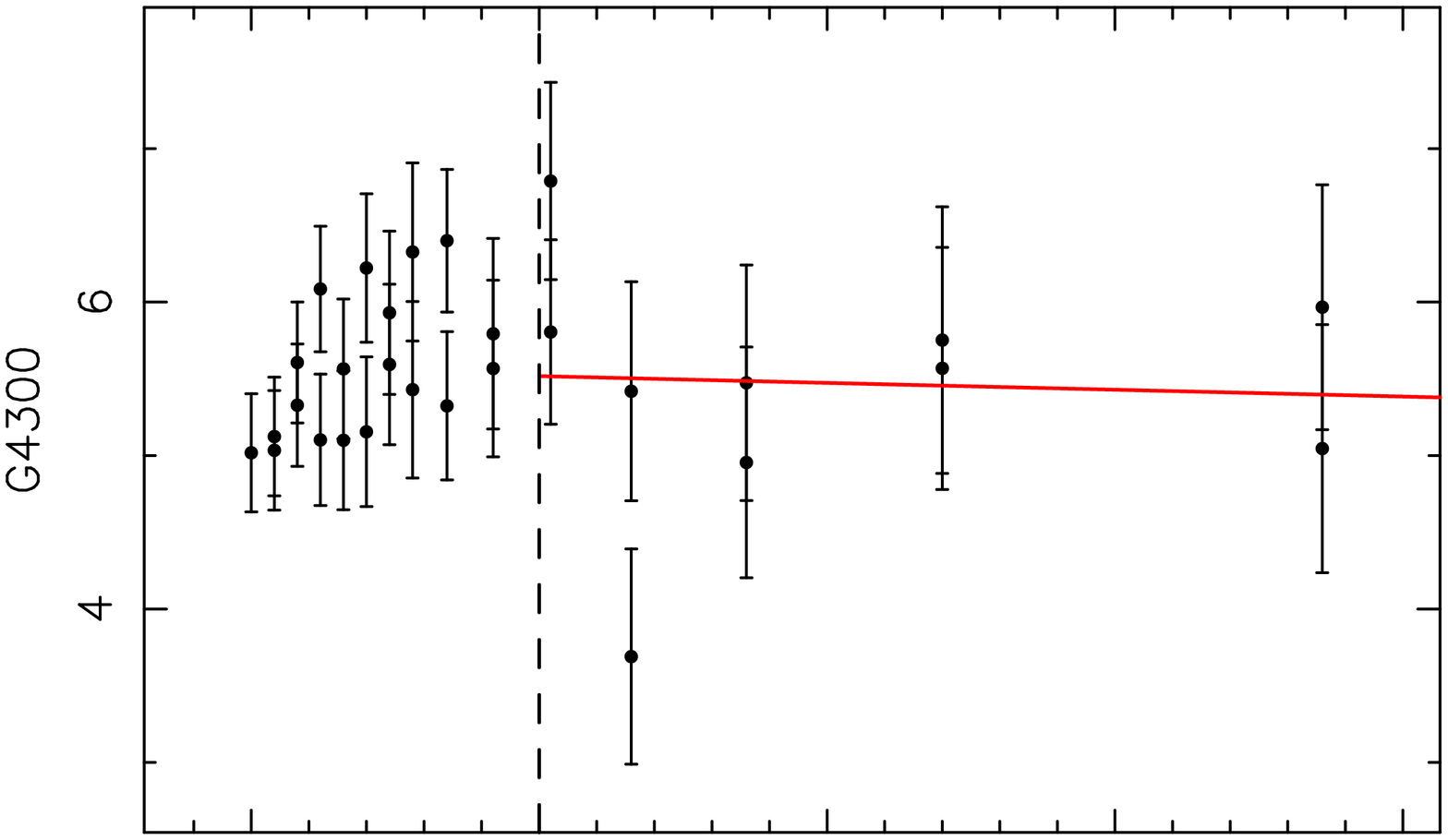}}
\resizebox{0.3\textwidth}{!}{\includegraphics[angle=-90]{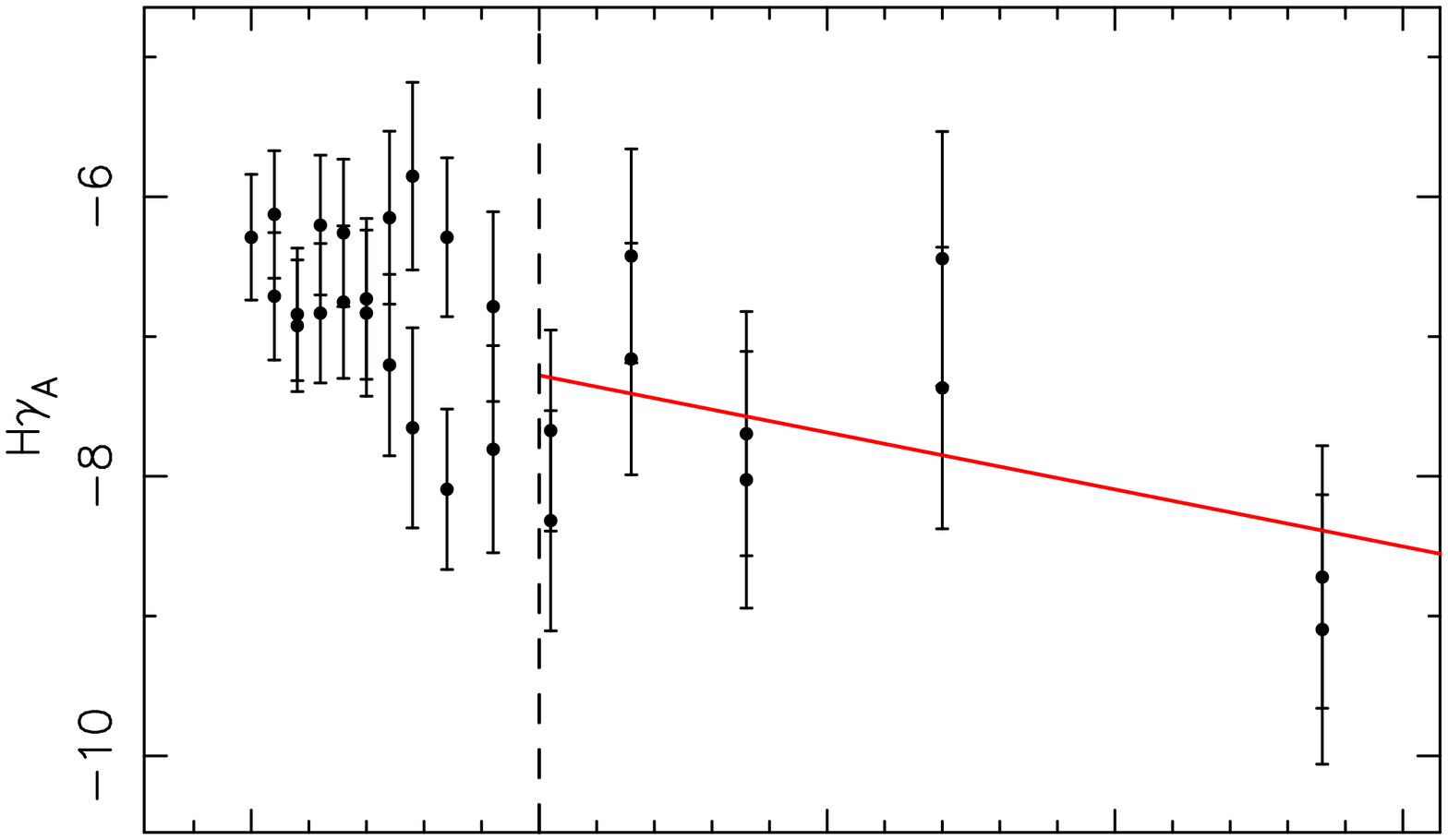}}
\resizebox{0.3\textwidth}{!}{\includegraphics[angle=-90]{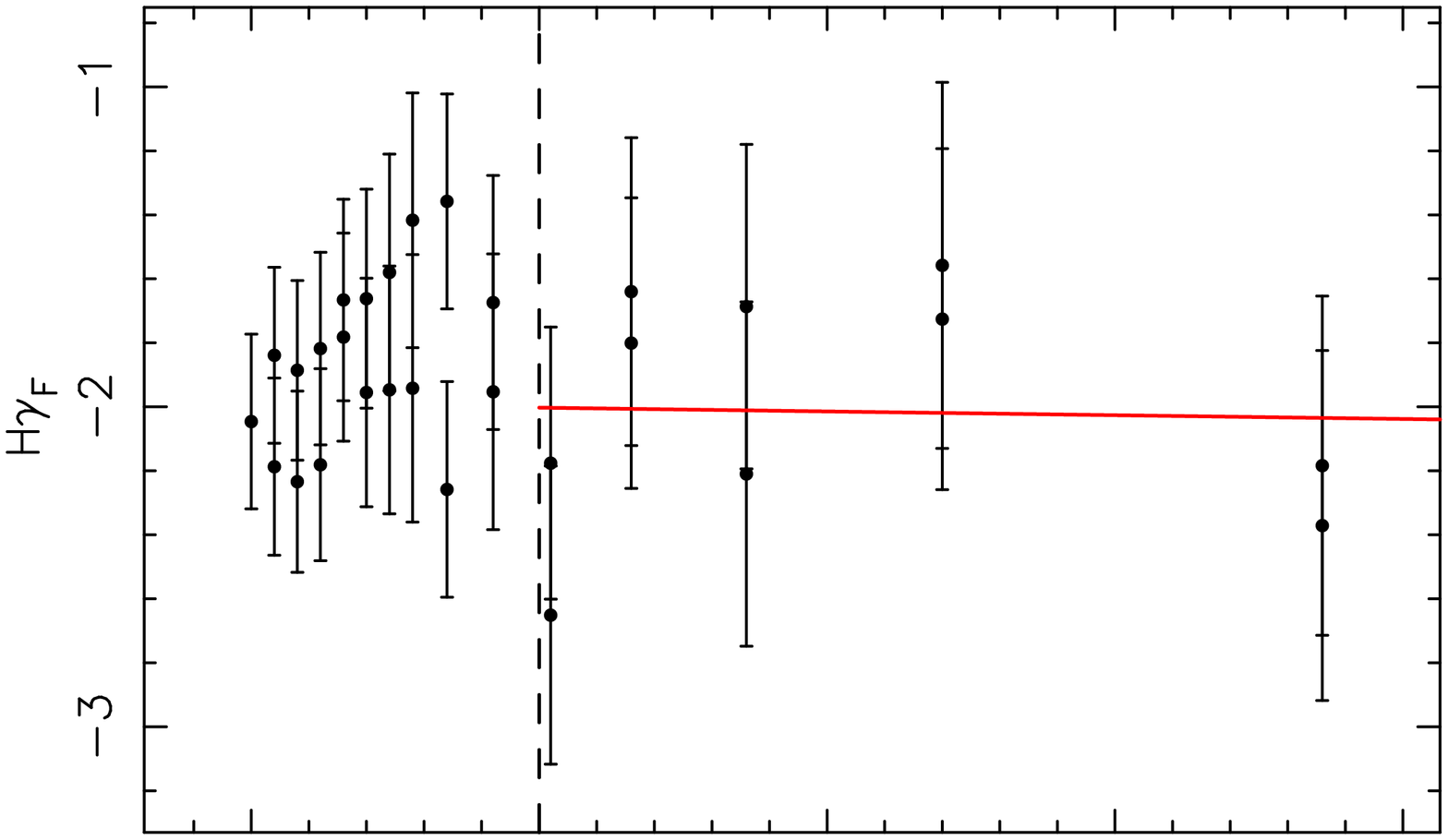}}
\resizebox{0.3\textwidth}{!}{\includegraphics[angle=-90]{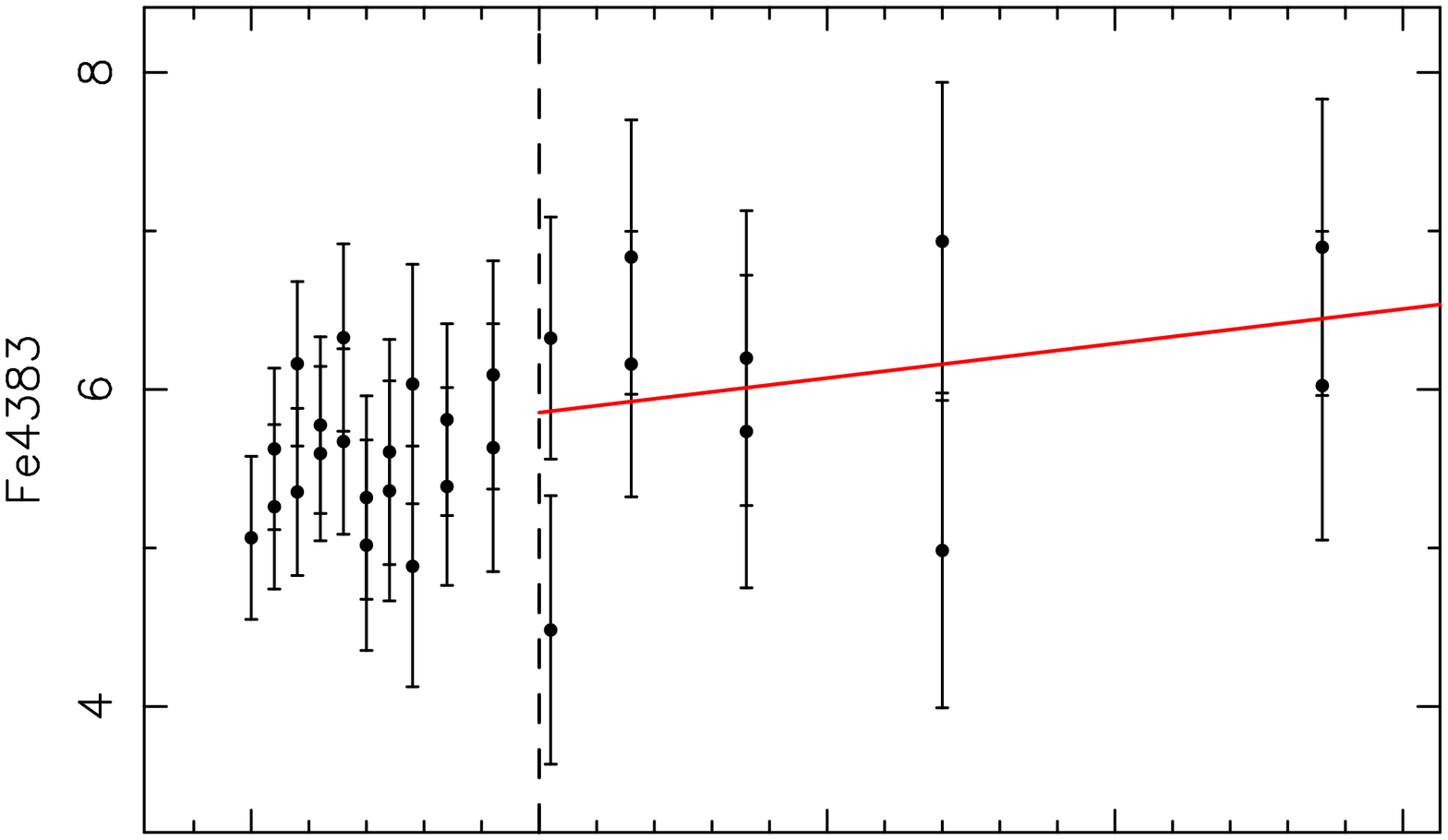}}
\resizebox{0.3\textwidth}{!}{\includegraphics[angle=-90]{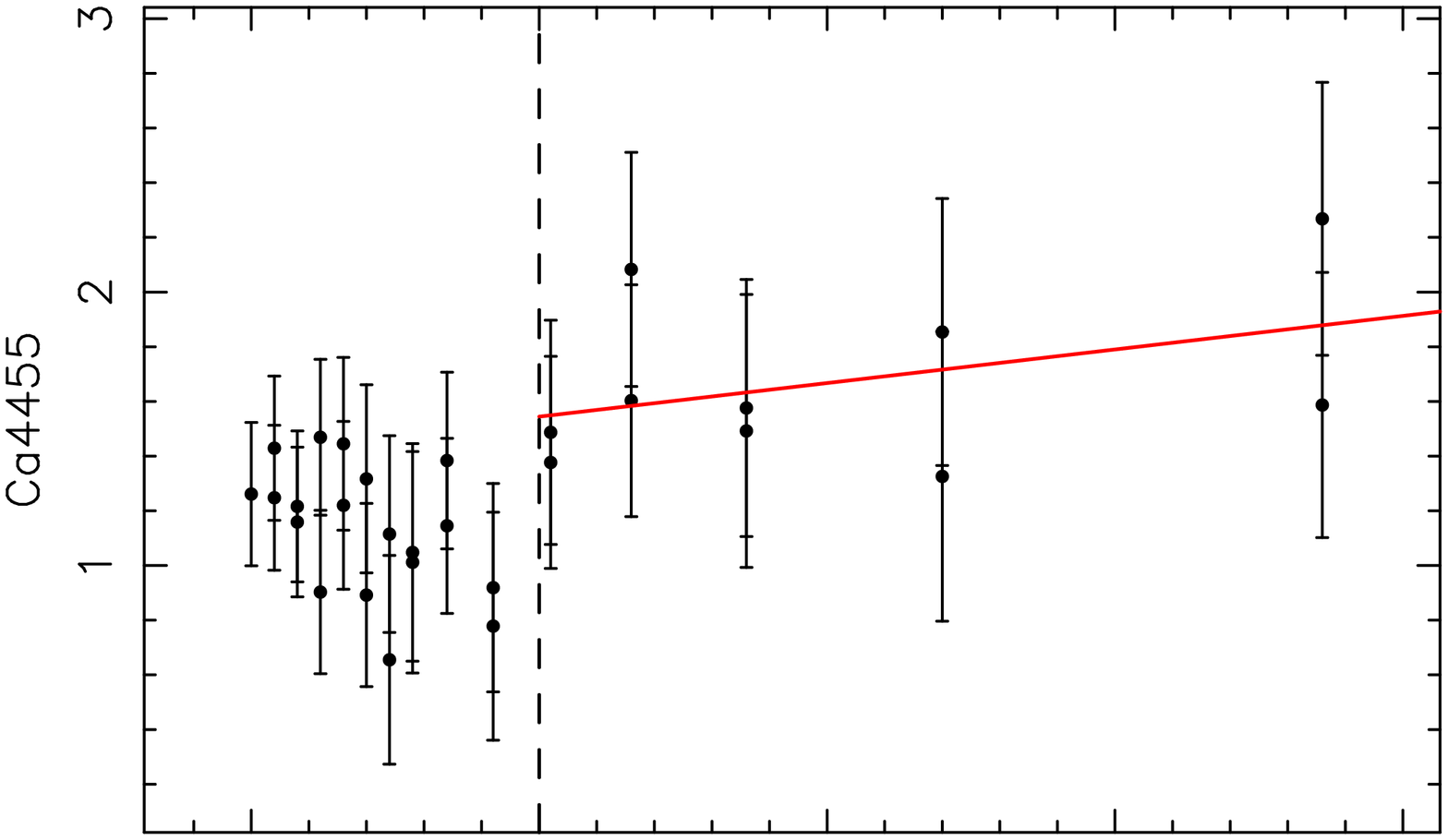}}
\resizebox{0.3\textwidth}{!}{\includegraphics[angle=-90]{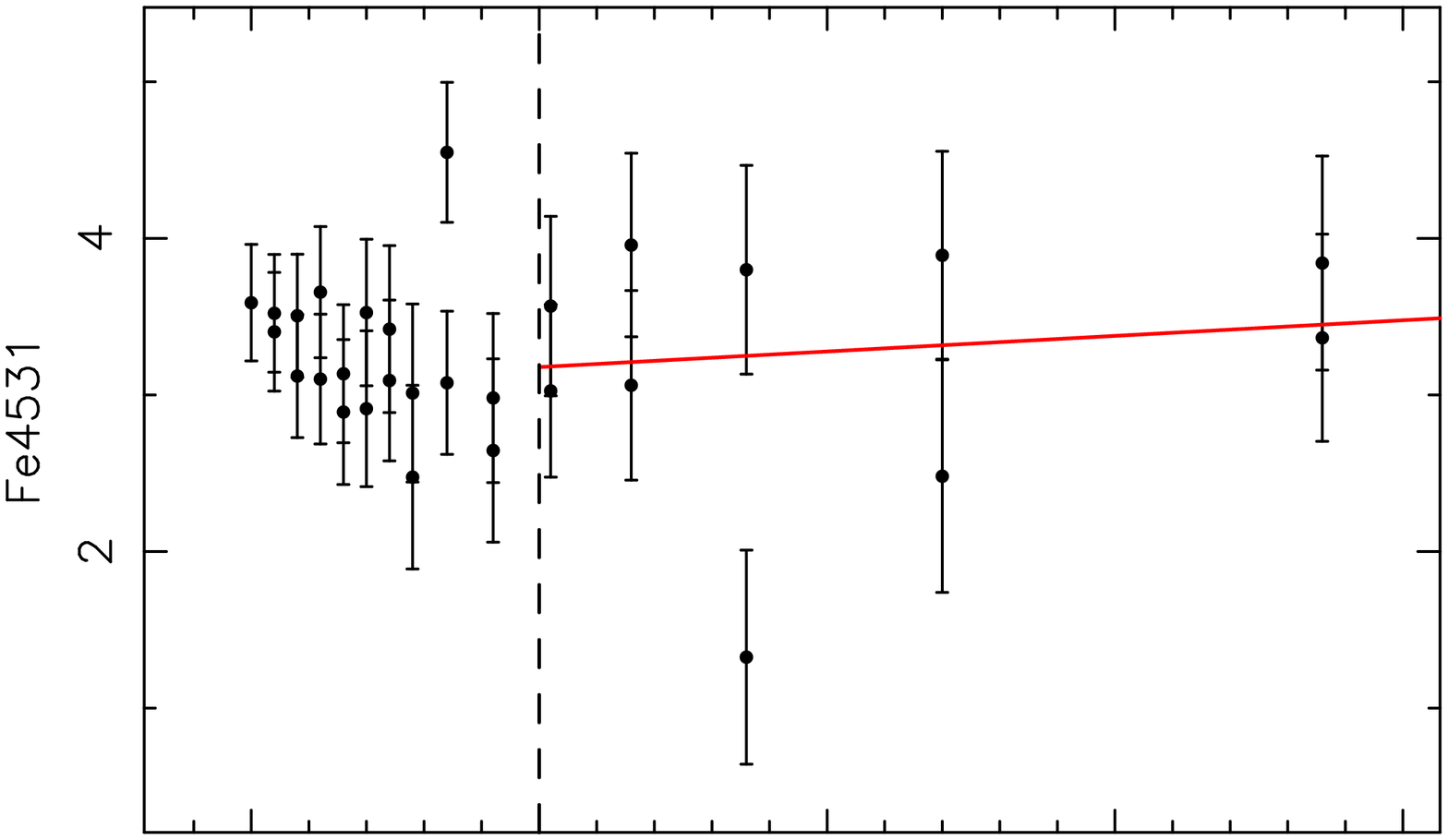}}
\resizebox{0.3\textwidth}{!}{\includegraphics[angle=-90]{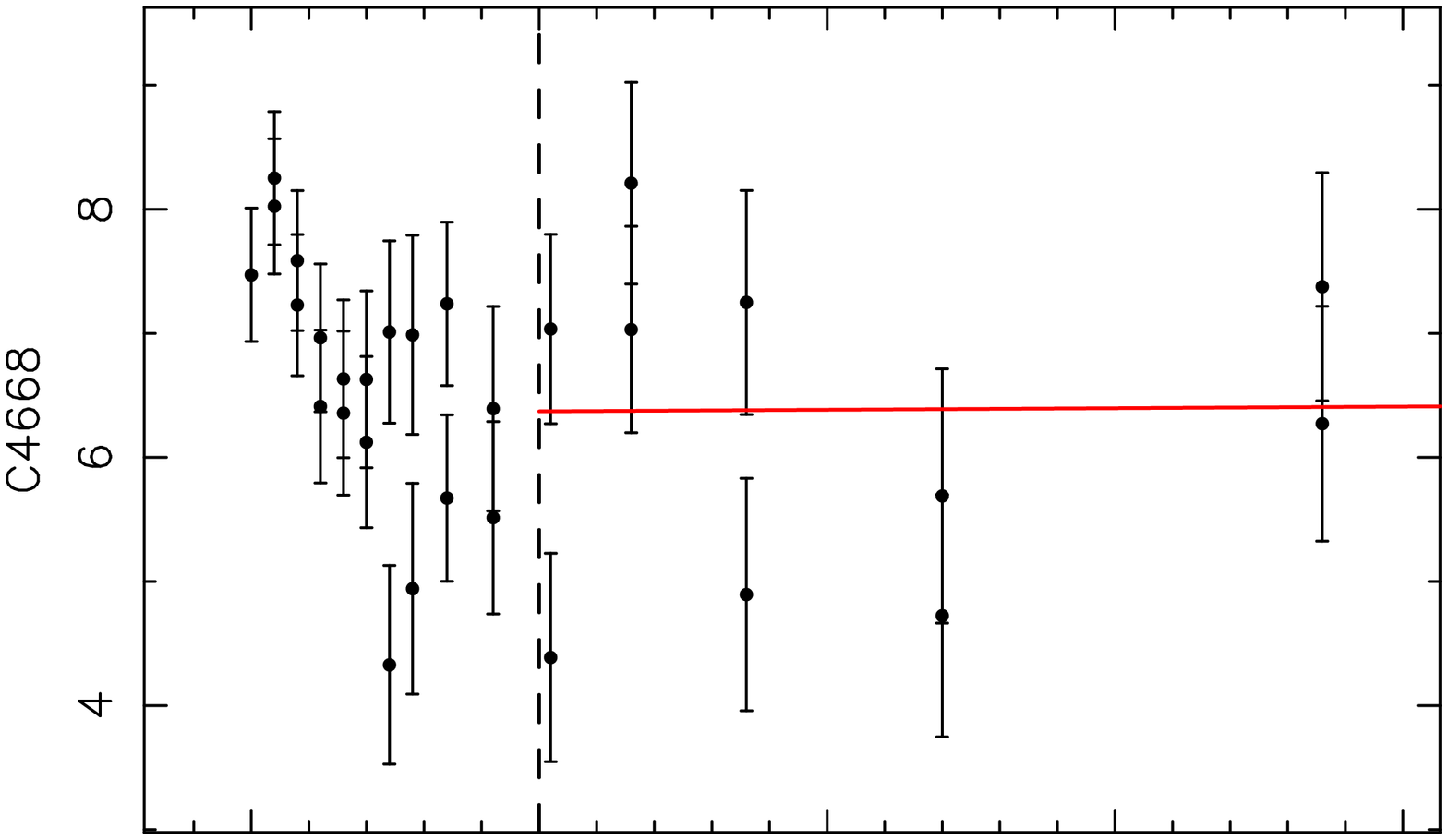}}
\resizebox{0.3\textwidth}{!}{\includegraphics[angle=-90]{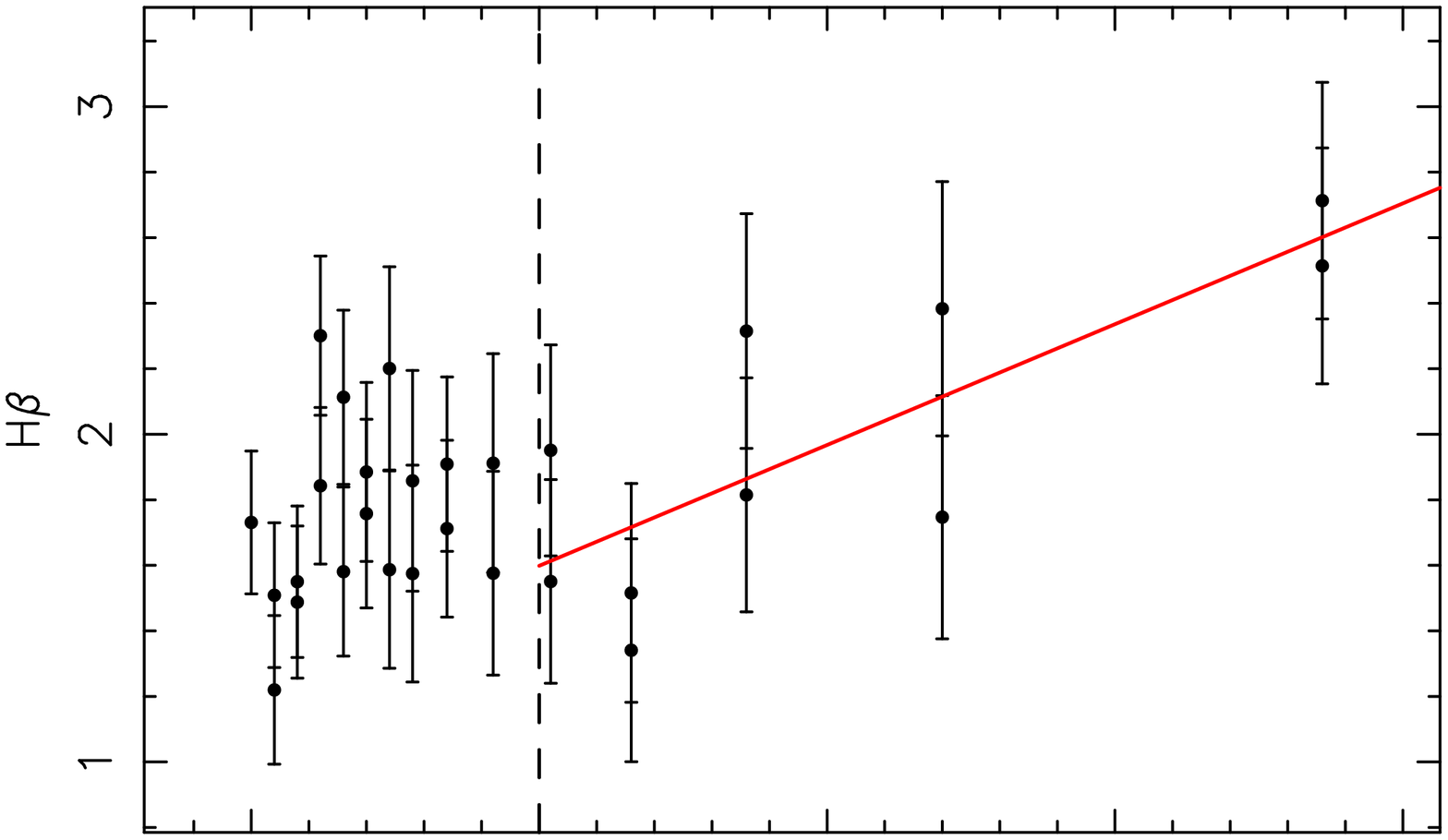}}
\resizebox{0.3\textwidth}{!}{\includegraphics[angle=-90]{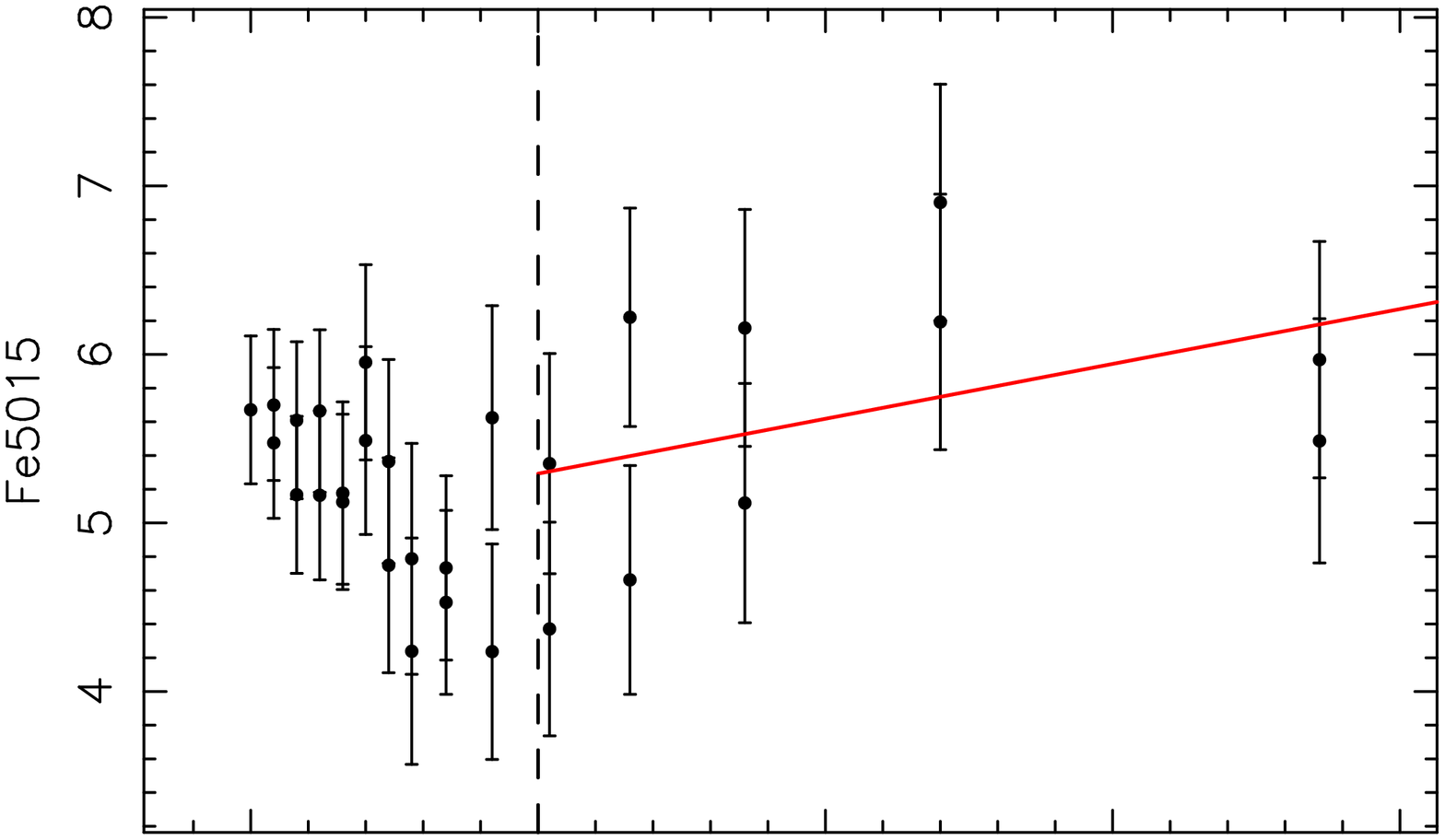}}
\resizebox{0.3\textwidth}{!}{\includegraphics[angle=-90]{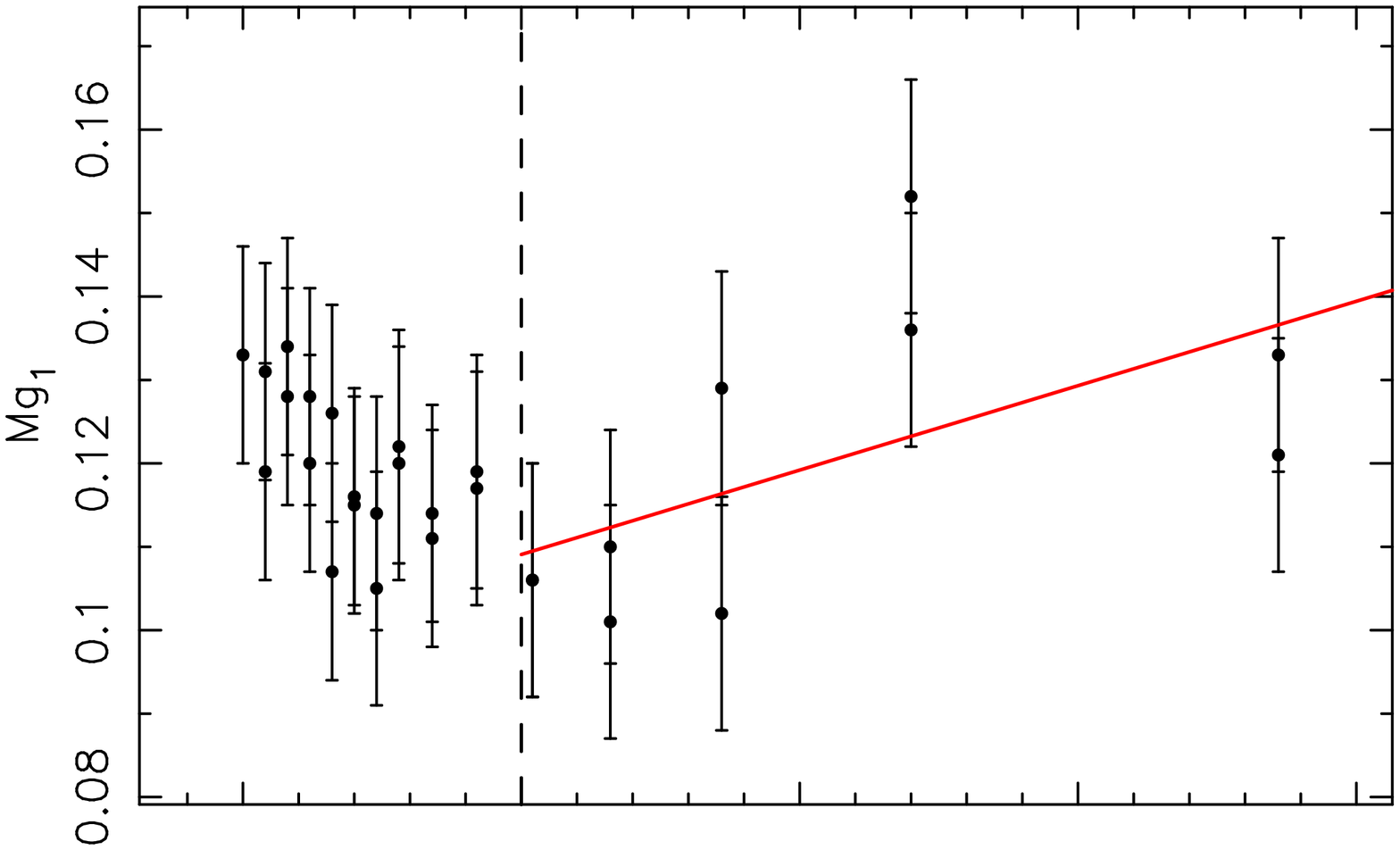}}
\resizebox{0.3\textwidth}{!}{\includegraphics[angle=-90]{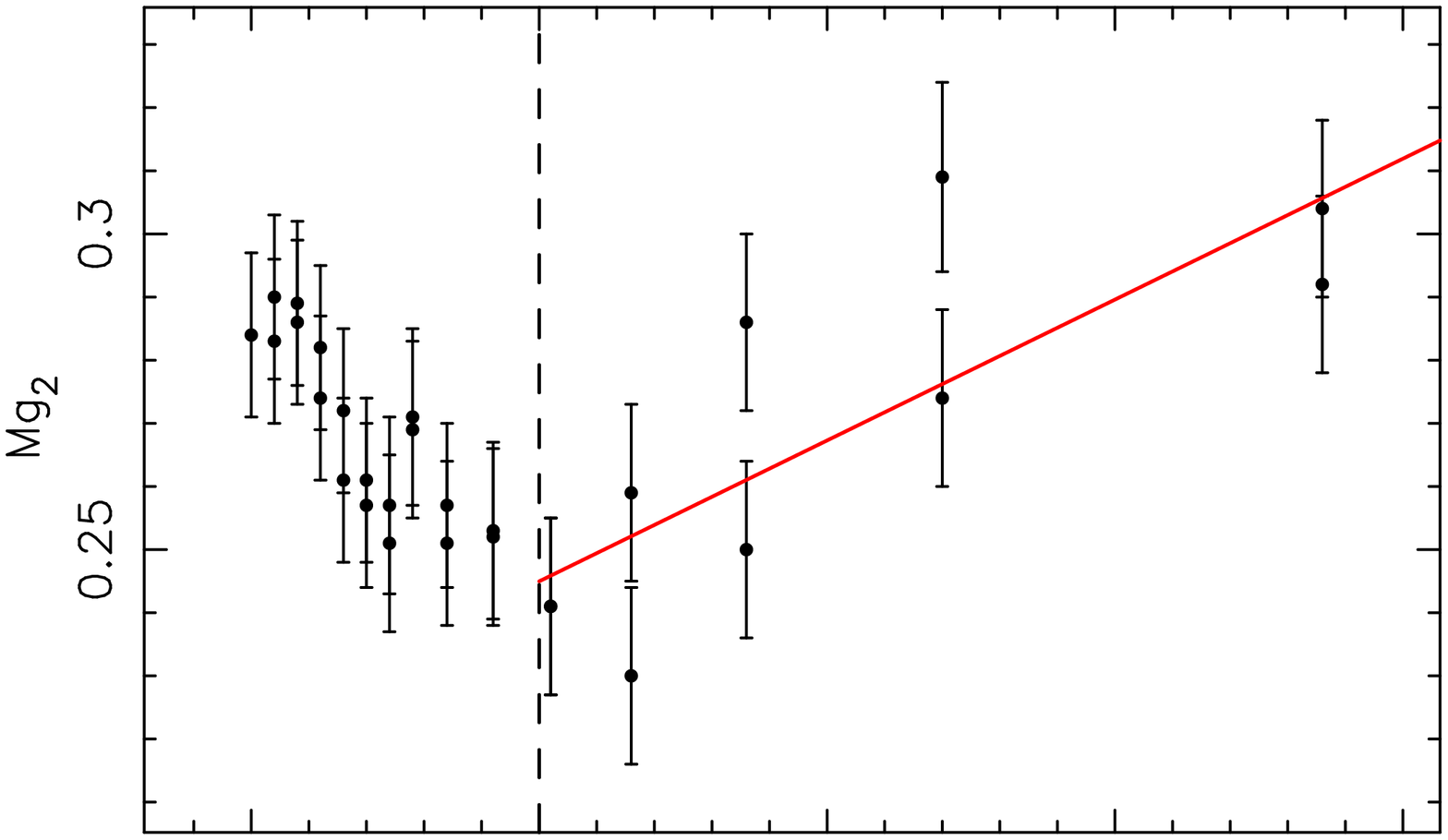}}
\resizebox{0.3\textwidth}{!}{\includegraphics[angle=-90]{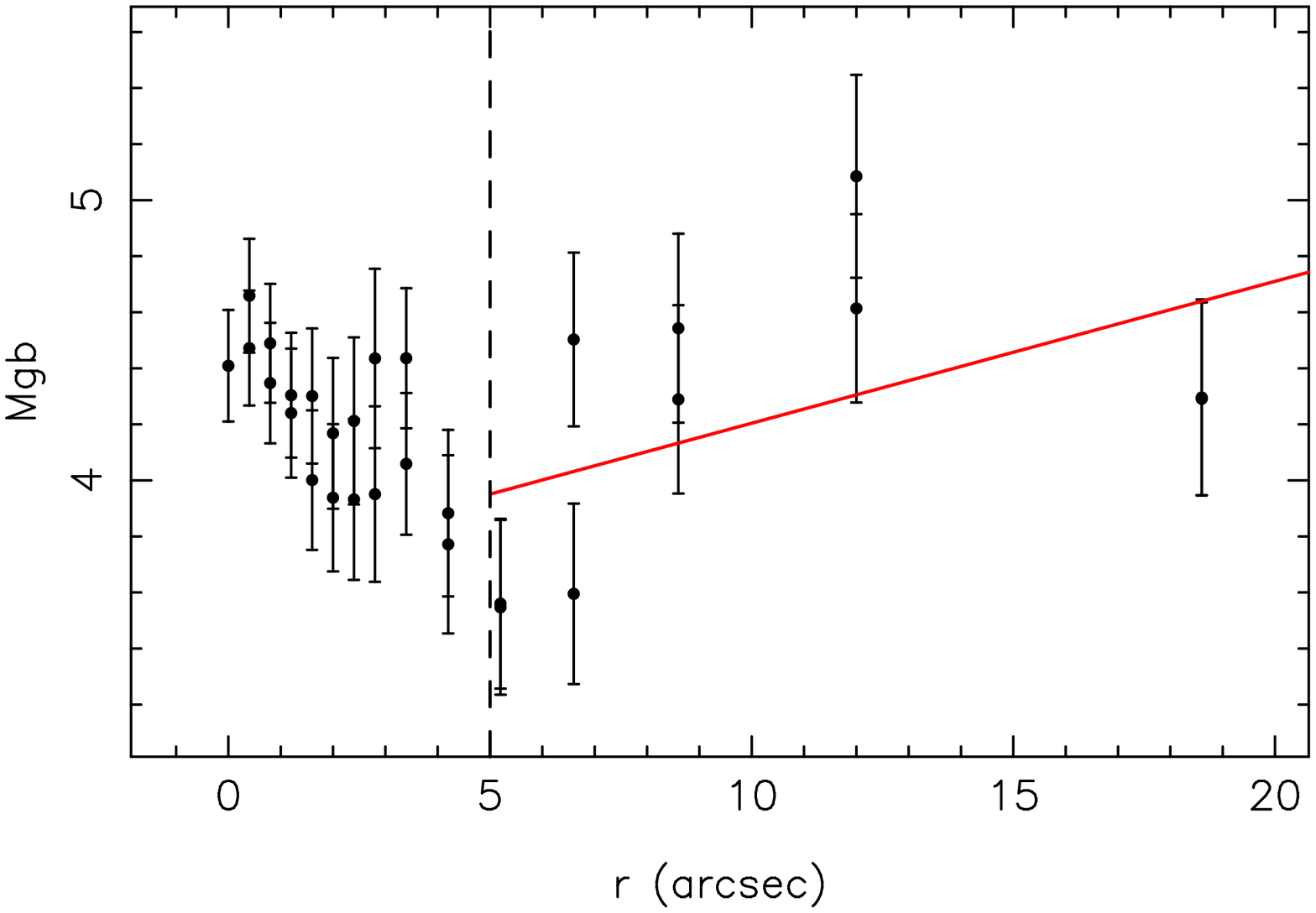}}\hspace{0.85cm}
\resizebox{0.3\textwidth}{!}{\includegraphics[angle=-90]{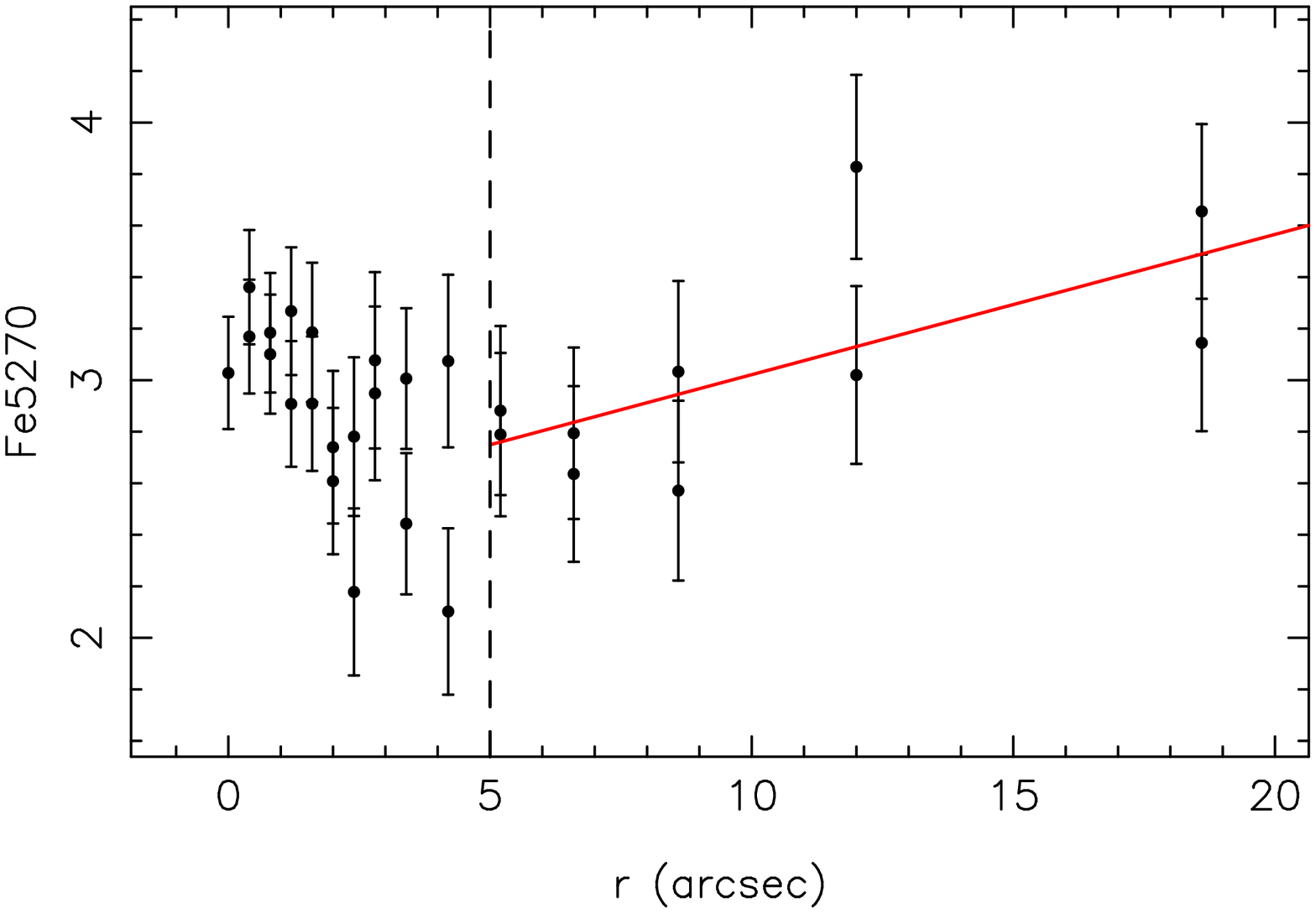}}\hspace{0.85cm}
\resizebox{0.3\textwidth}{!}{\includegraphics[angle=-90]{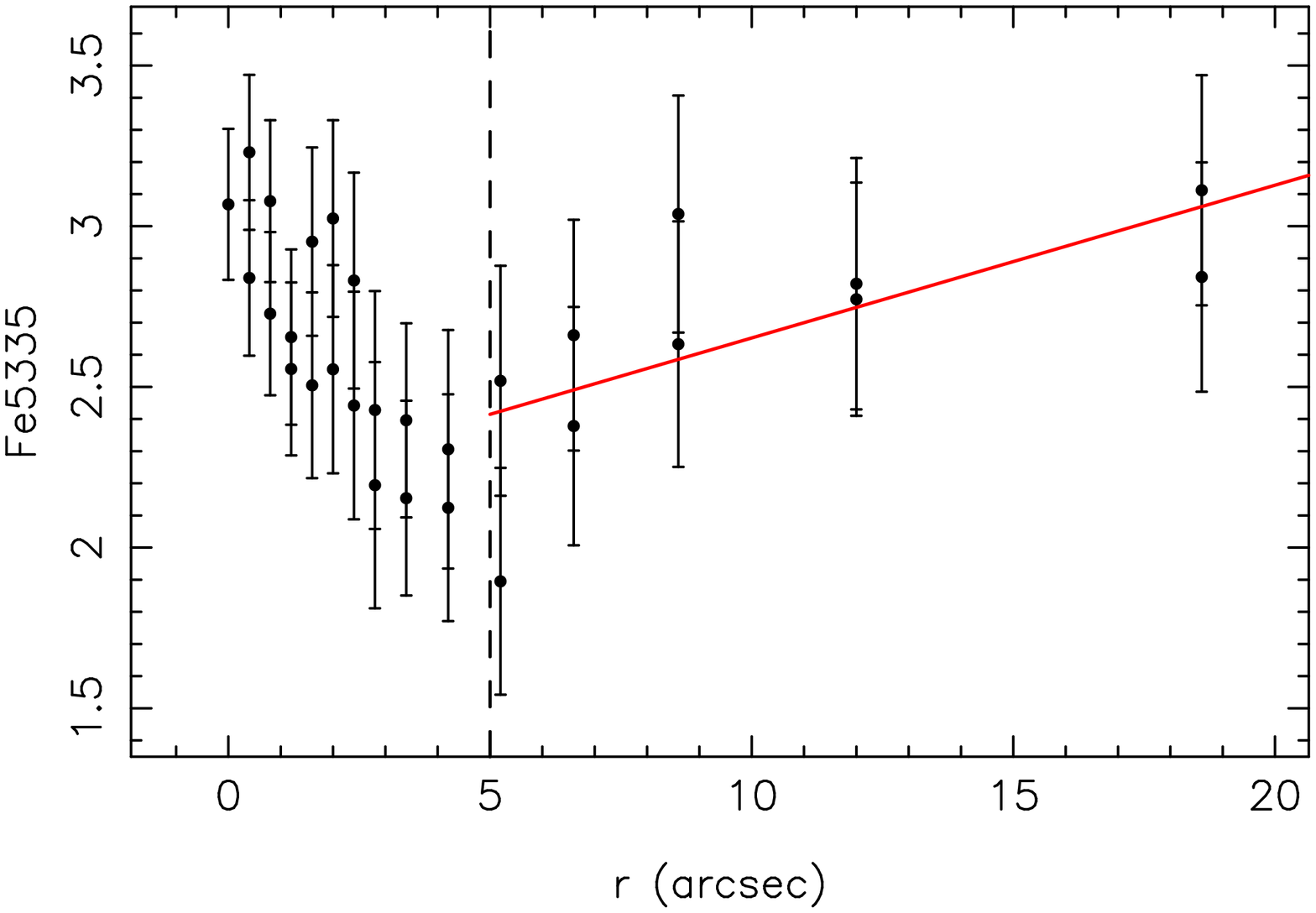}}
\caption{Line-strength distribution in the bar region for all the galaxies \label{line-strength}}
\end{figure*}
\begin{figure*}
\addtocounter{figure}{-1}
\resizebox{0.3\textwidth}{!}{\includegraphics[angle=-90]{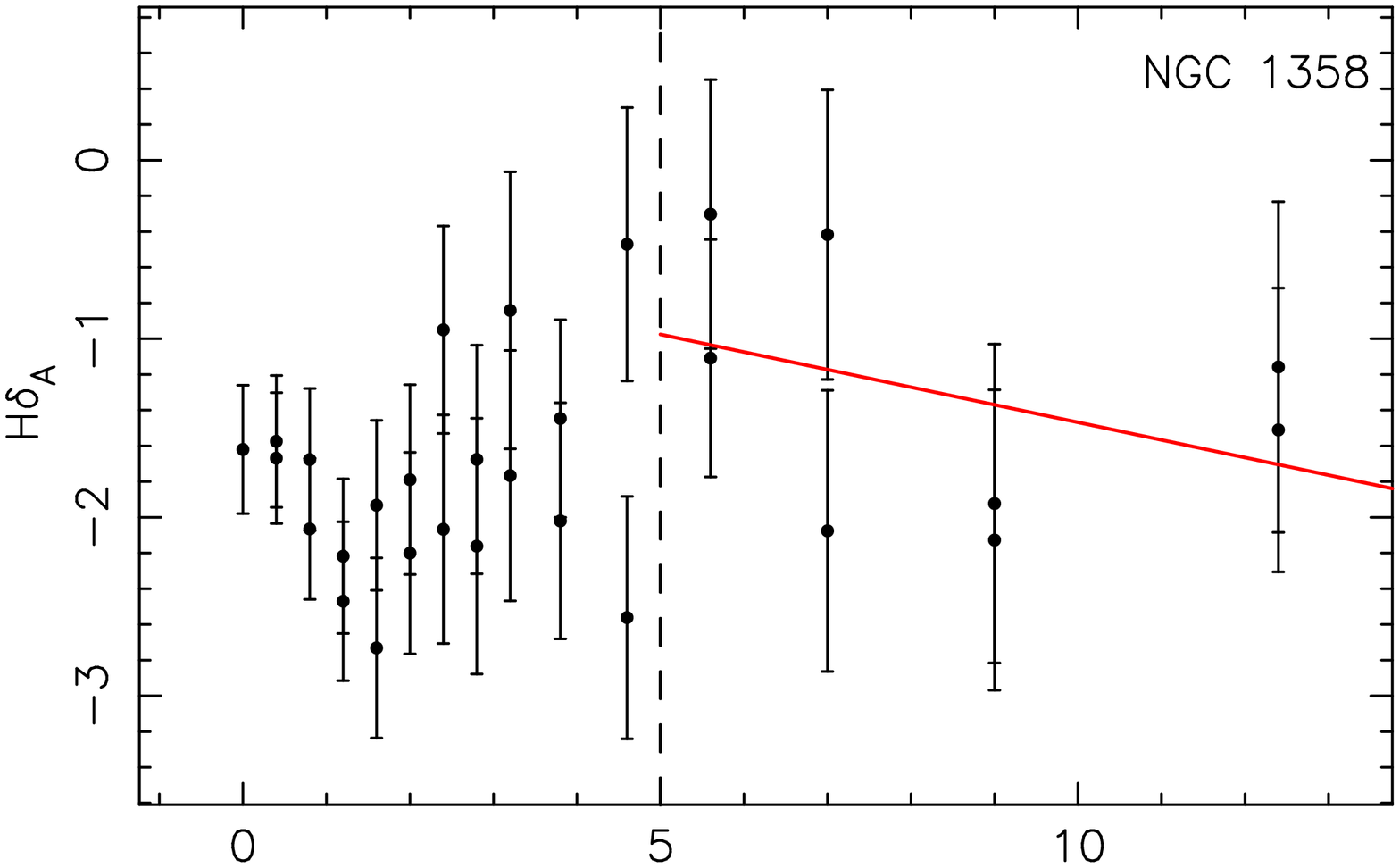}}
\resizebox{0.3\textwidth}{!}{\includegraphics[angle=-90]{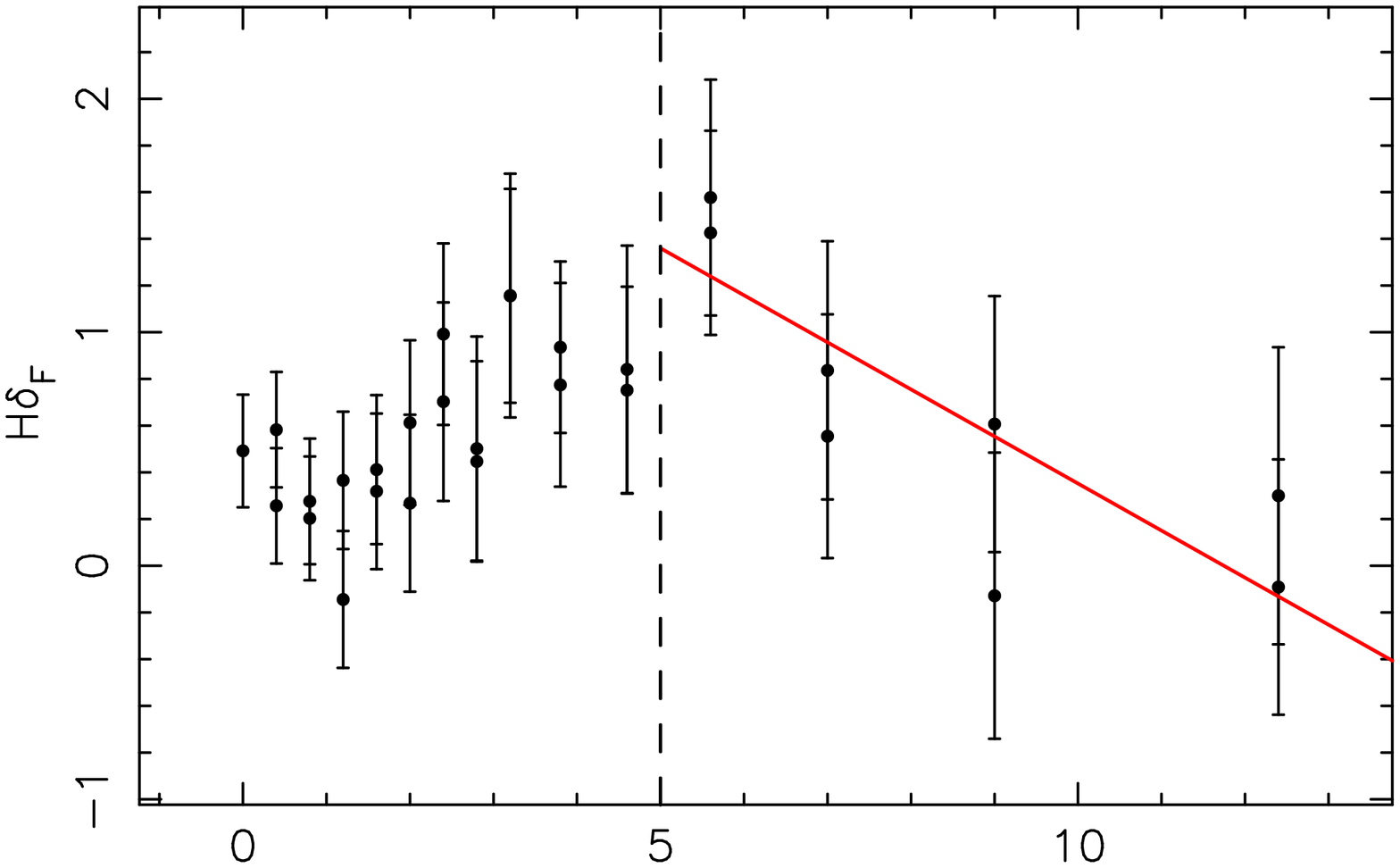}}
\resizebox{0.3\textwidth}{!}{\includegraphics[angle=-90]{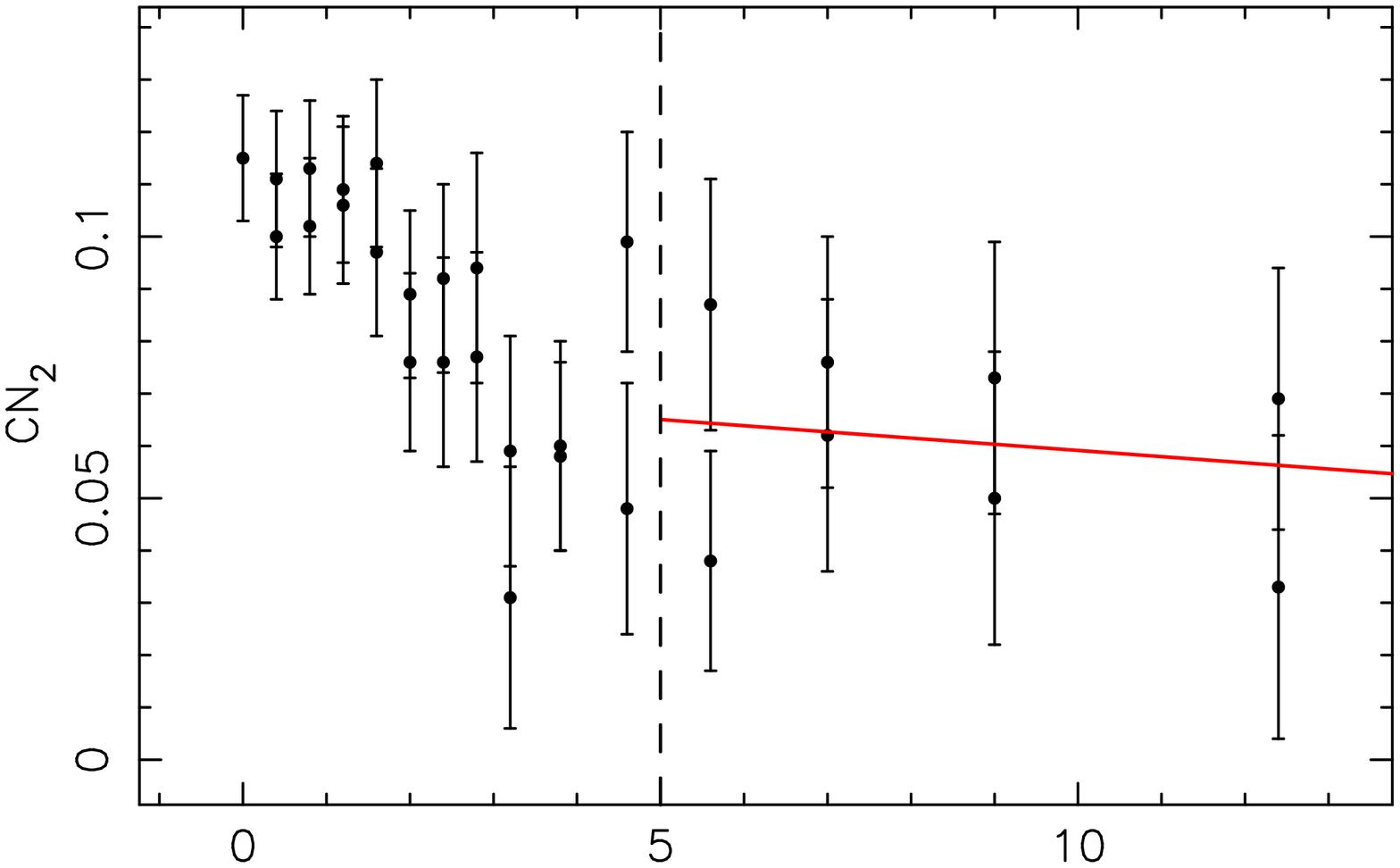}}
\resizebox{0.3\textwidth}{!}{\includegraphics[angle=-90]{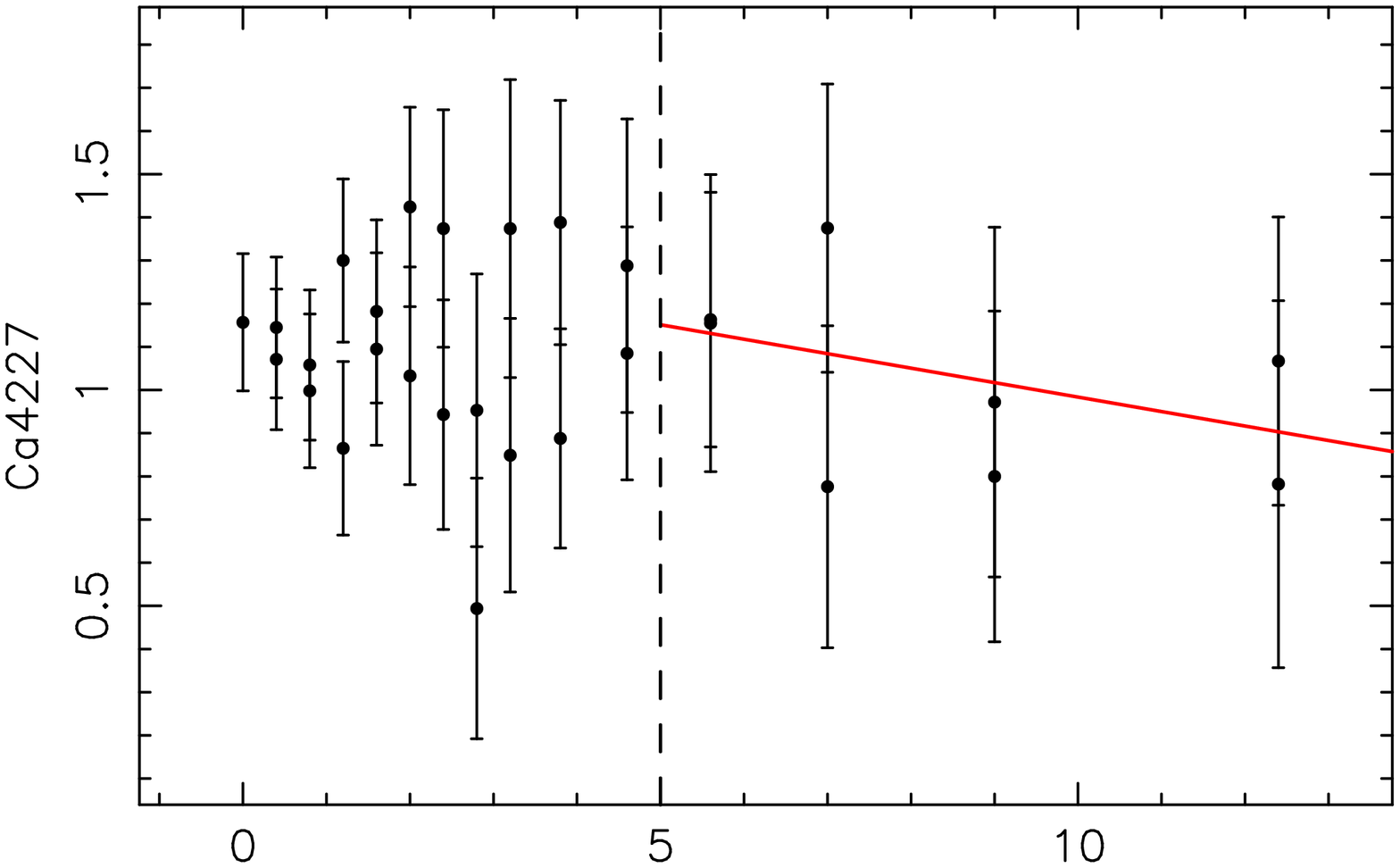}}
\resizebox{0.3\textwidth}{!}{\includegraphics[angle=-90]{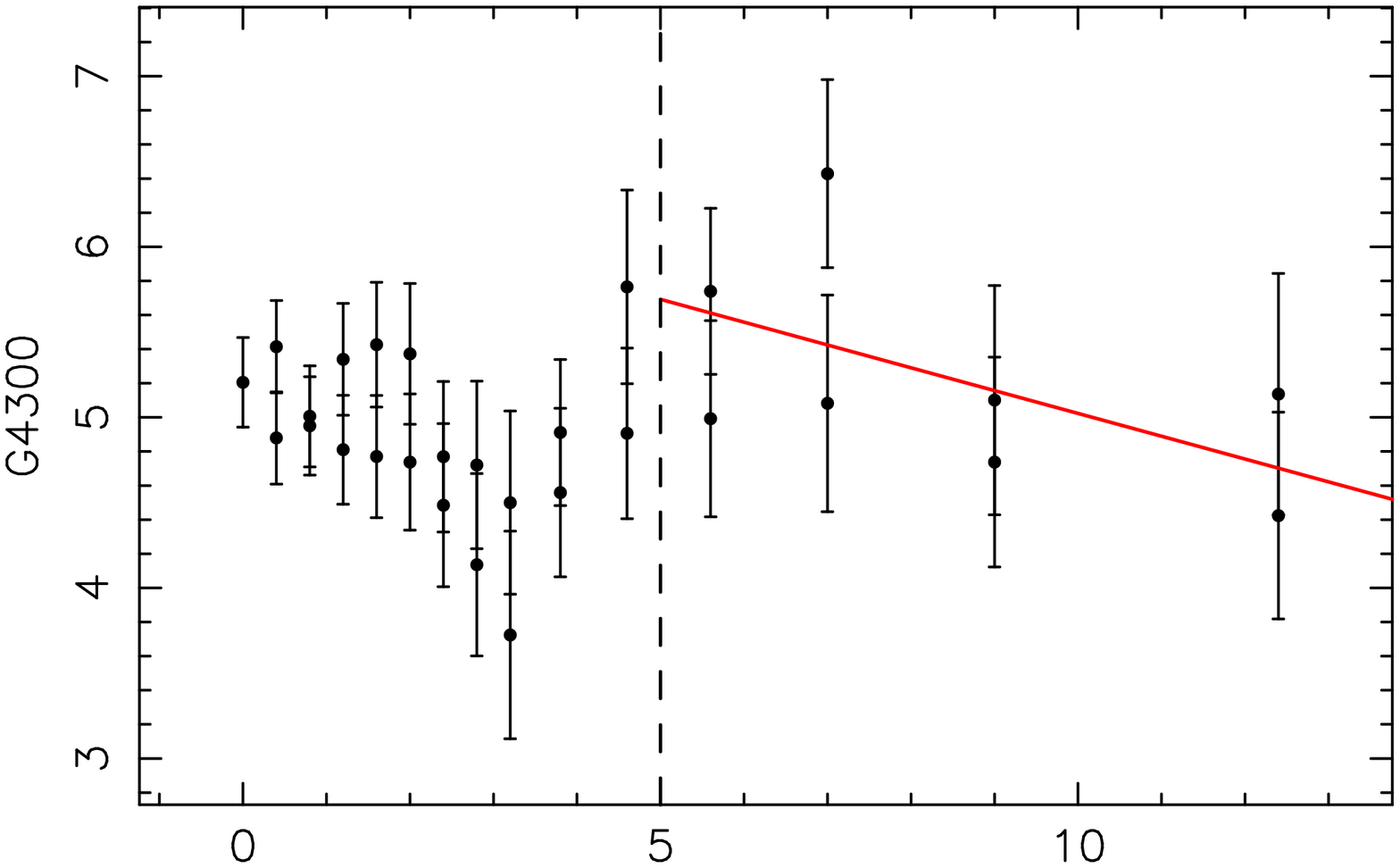}}
\resizebox{0.3\textwidth}{!}{\includegraphics[angle=-90]{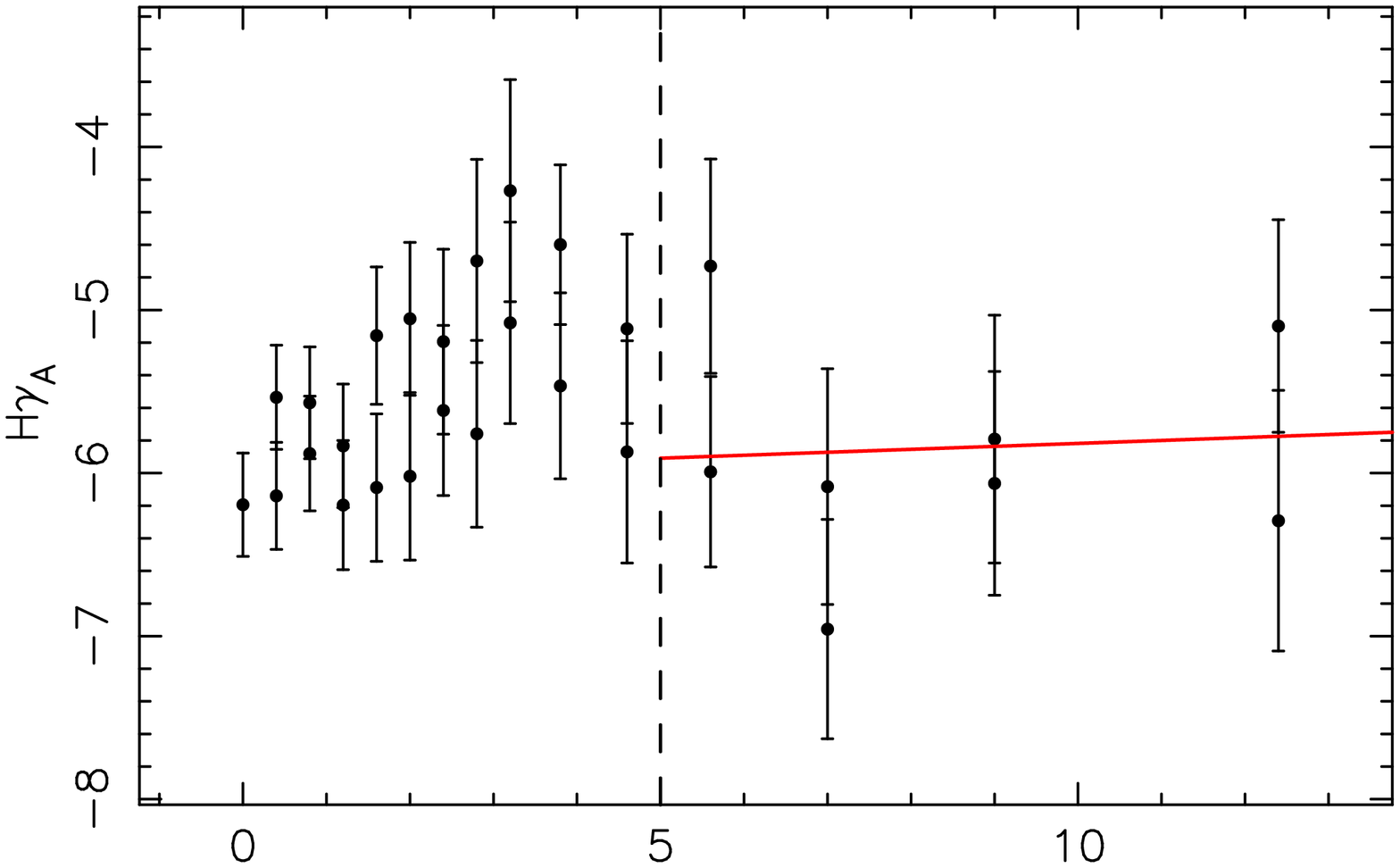}}
\resizebox{0.3\textwidth}{!}{\includegraphics[angle=-90]{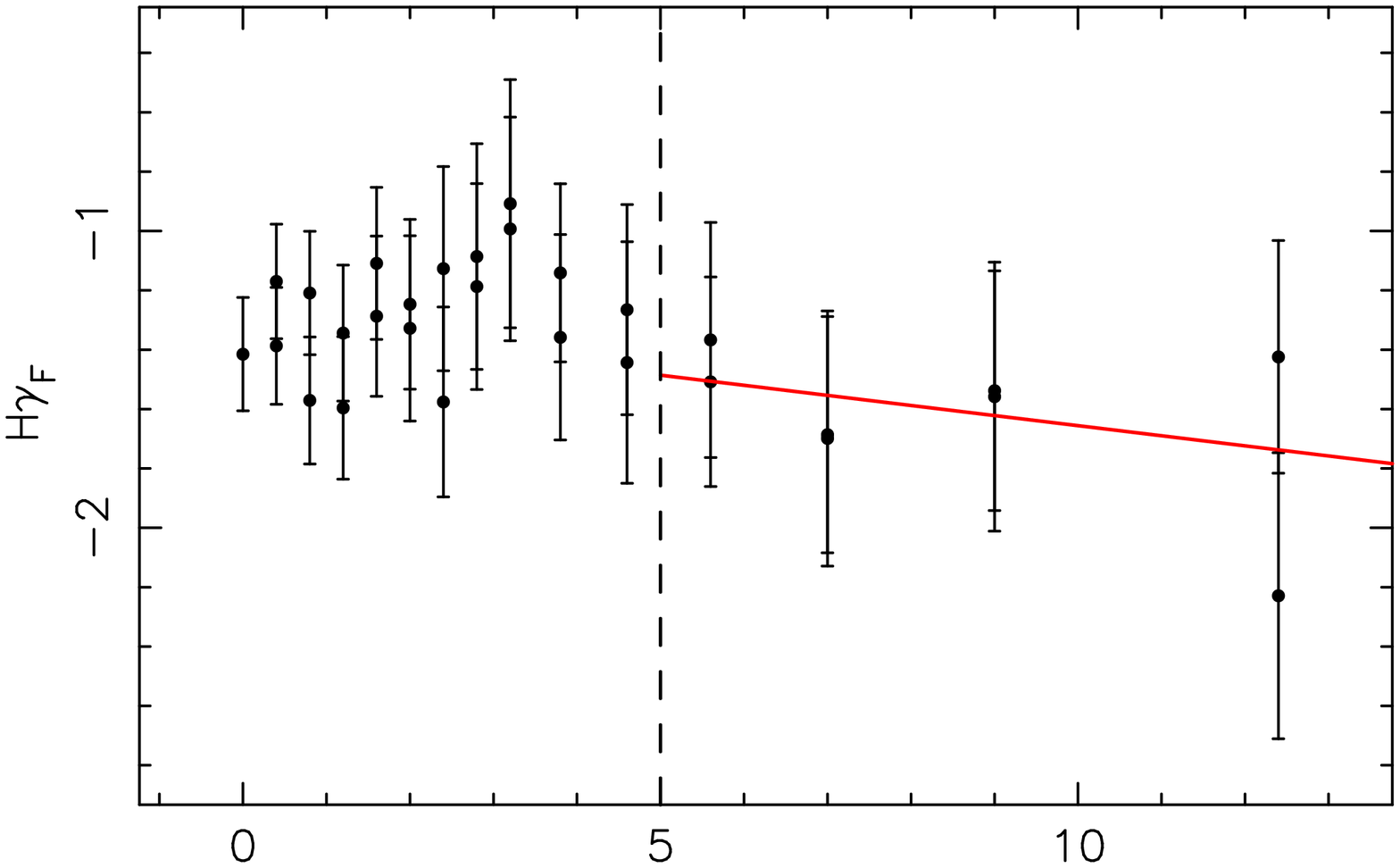}}
\resizebox{0.3\textwidth}{!}{\includegraphics[angle=-90]{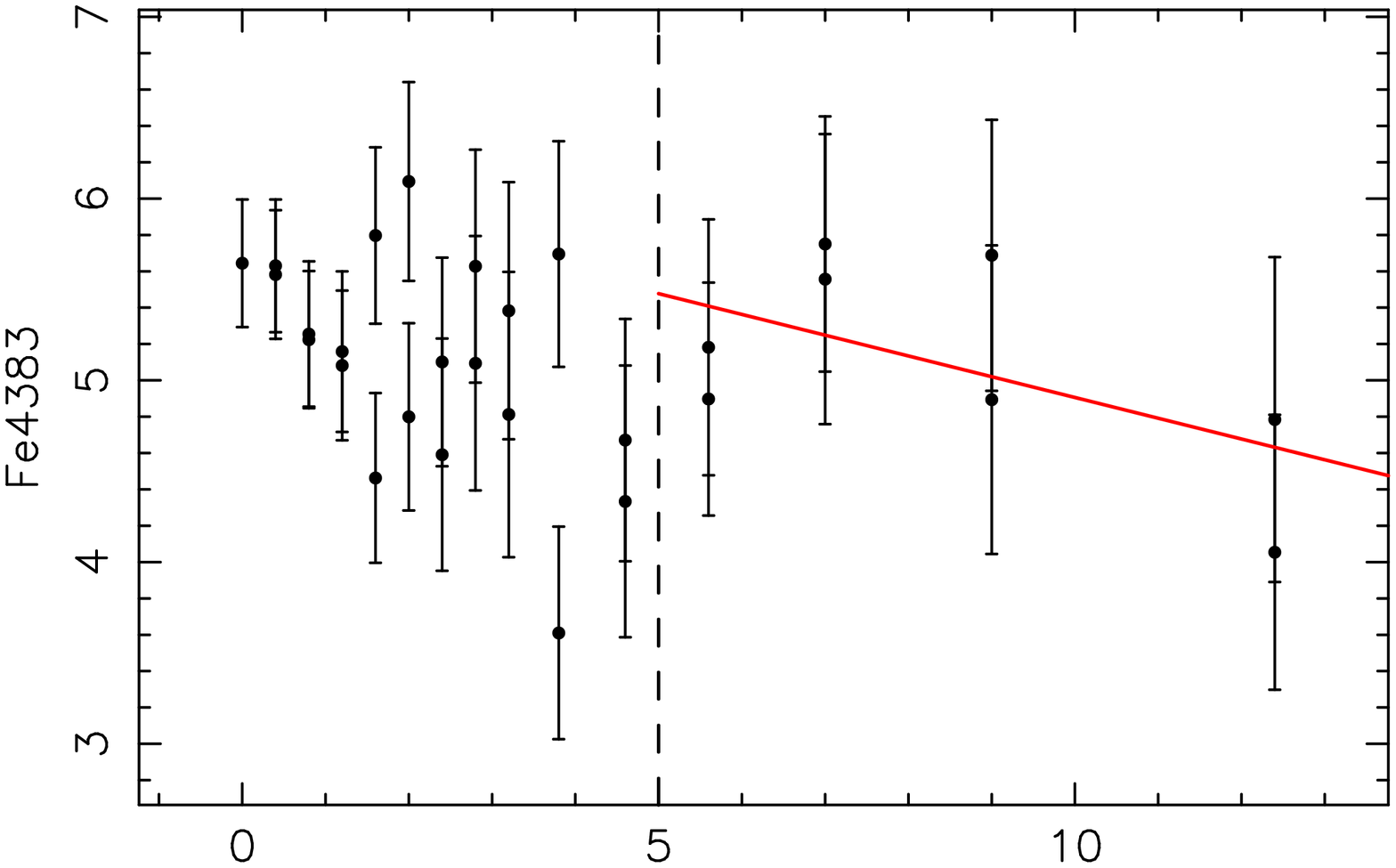}}
\resizebox{0.3\textwidth}{!}{\includegraphics[angle=-90]{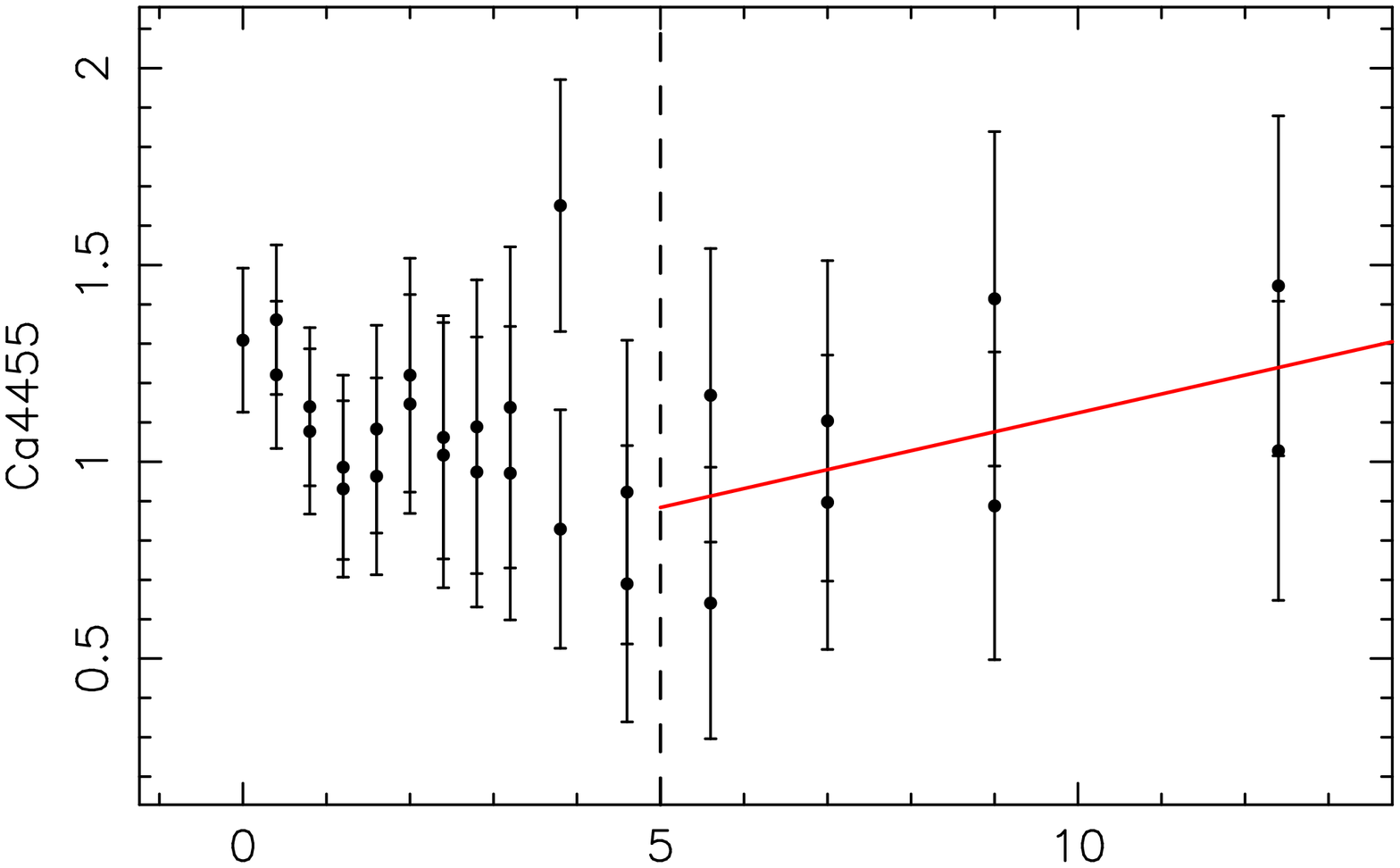}}
\resizebox{0.3\textwidth}{!}{\includegraphics[angle=-90]{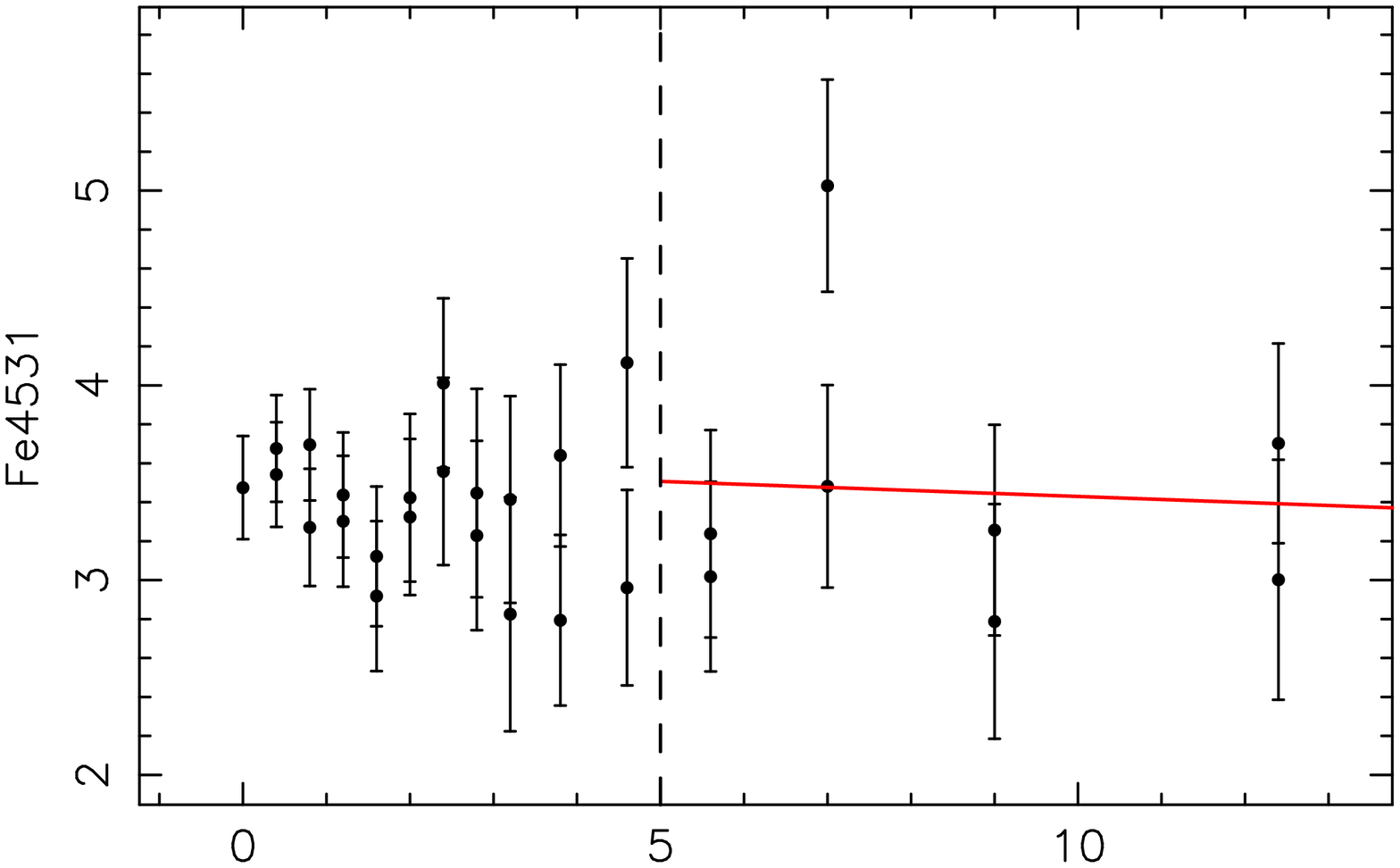}}
\resizebox{0.3\textwidth}{!}{\includegraphics[angle=-90]{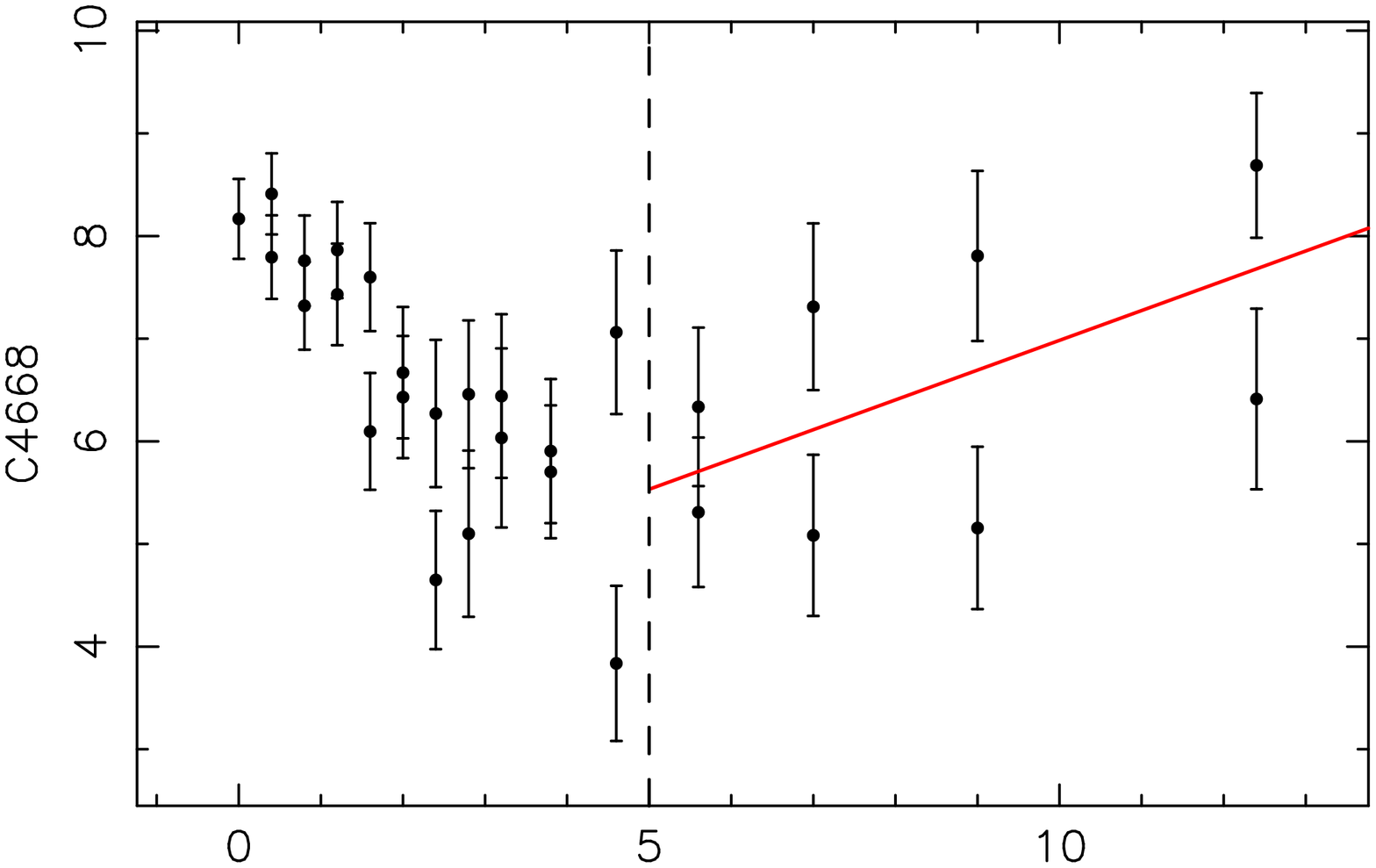}}
\resizebox{0.3\textwidth}{!}{\includegraphics[angle=-90]{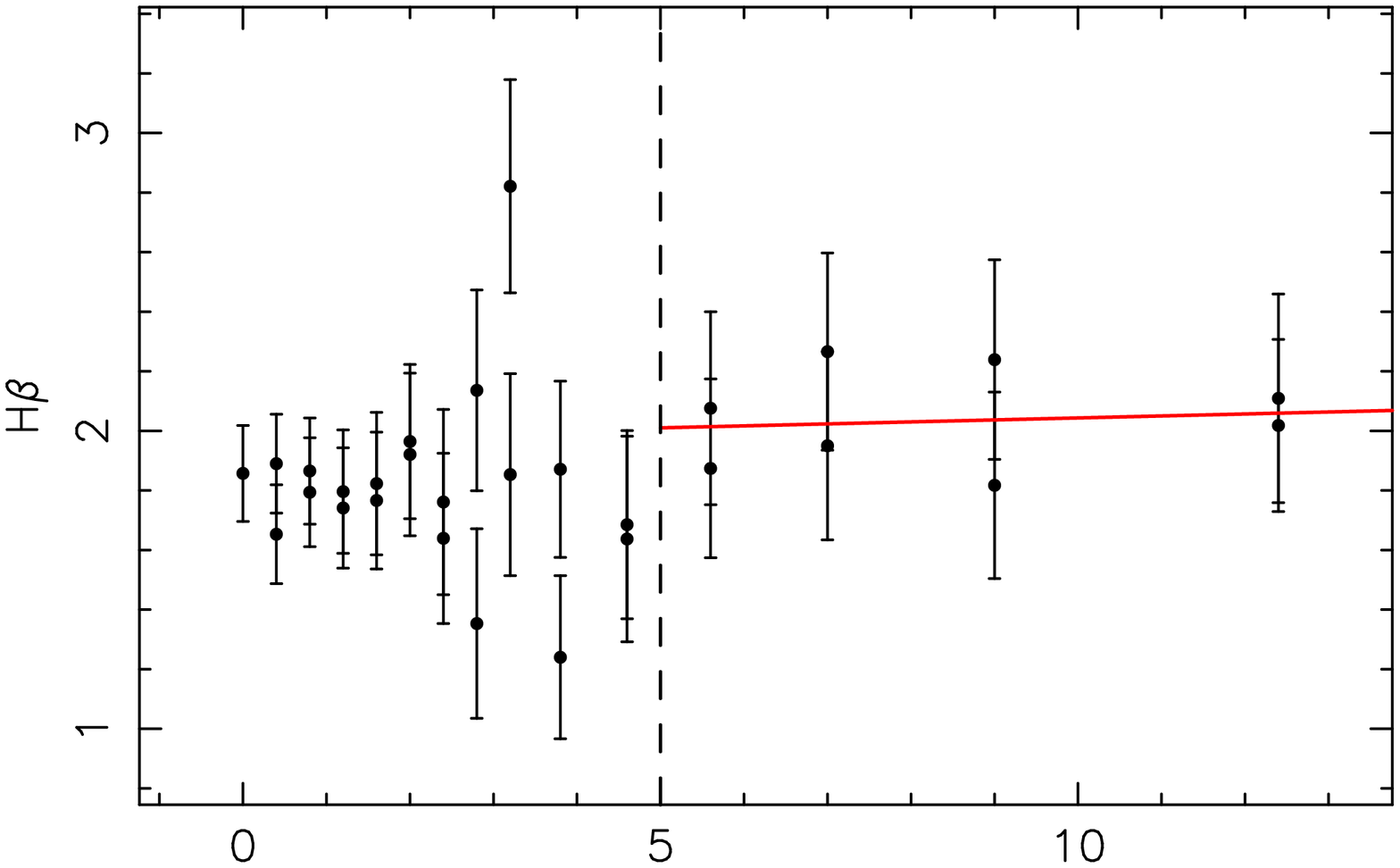}}
\resizebox{0.3\textwidth}{!}{\includegraphics[angle=-90]{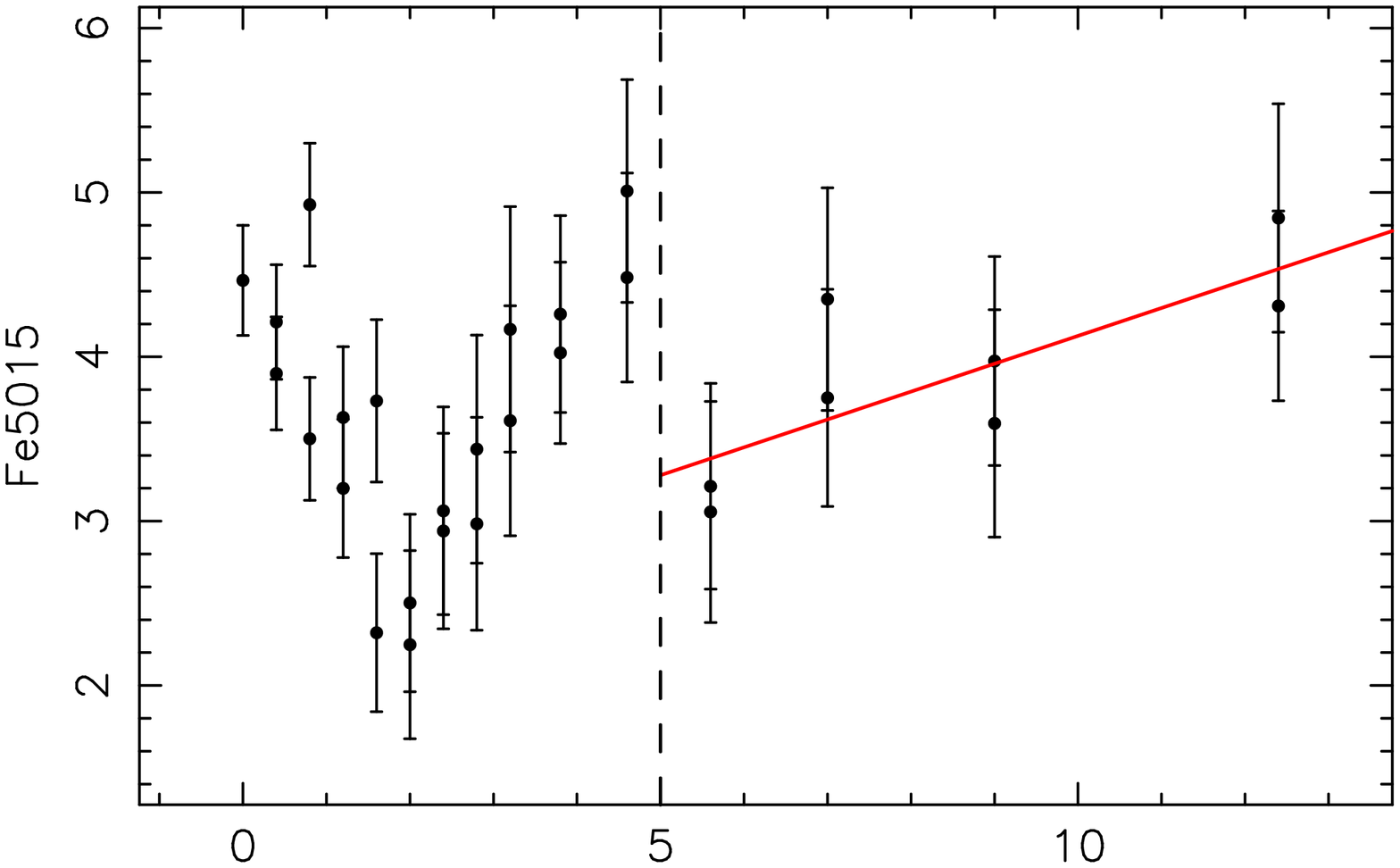}}
\resizebox{0.3\textwidth}{!}{\includegraphics[angle=-90]{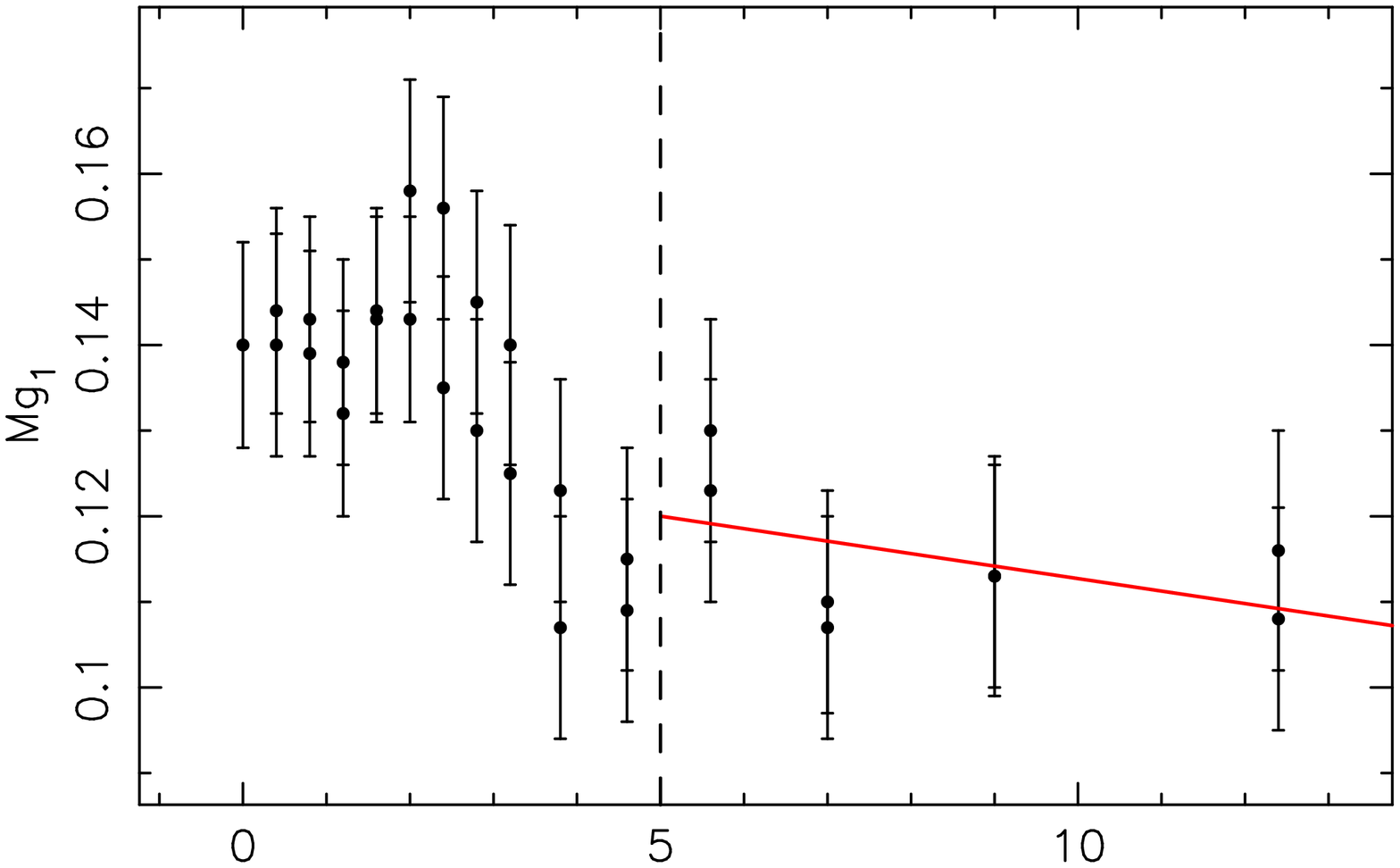}}
\resizebox{0.3\textwidth}{!}{\includegraphics[angle=-90]{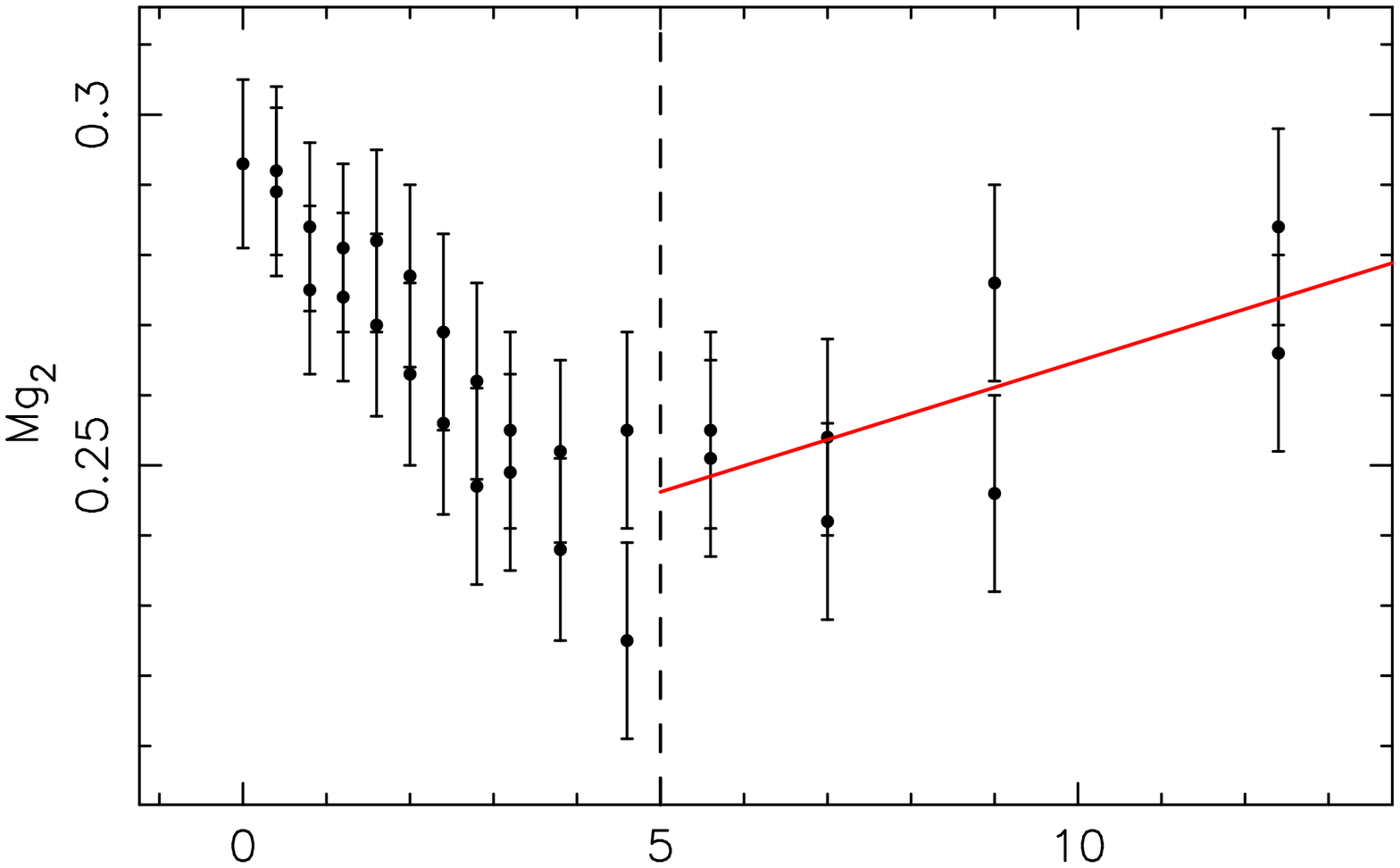}}
\resizebox{0.3\textwidth}{!}{\includegraphics[angle=-90]{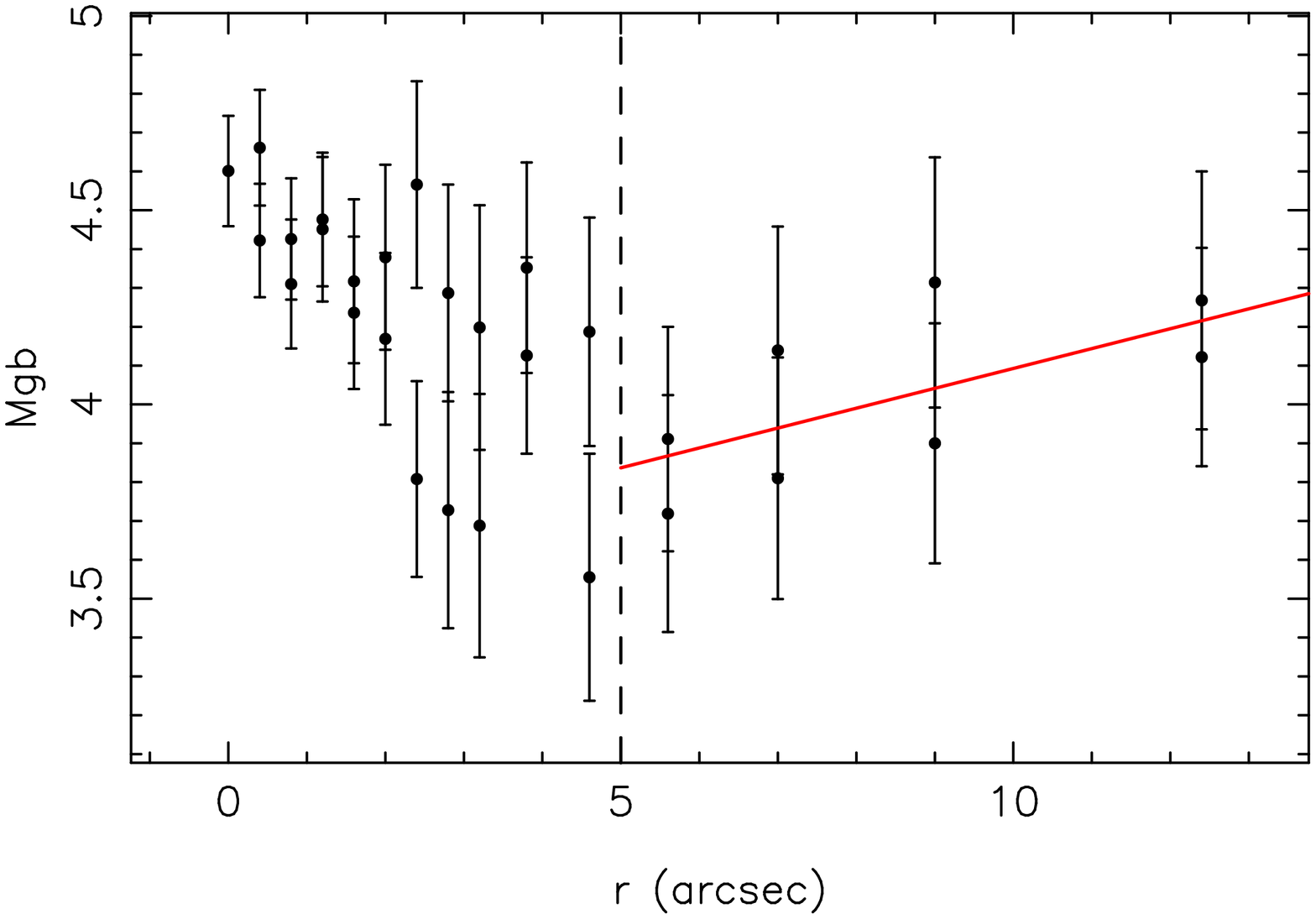}}\hspace{0.85cm}
\resizebox{0.3\textwidth}{!}{\includegraphics[angle=-90]{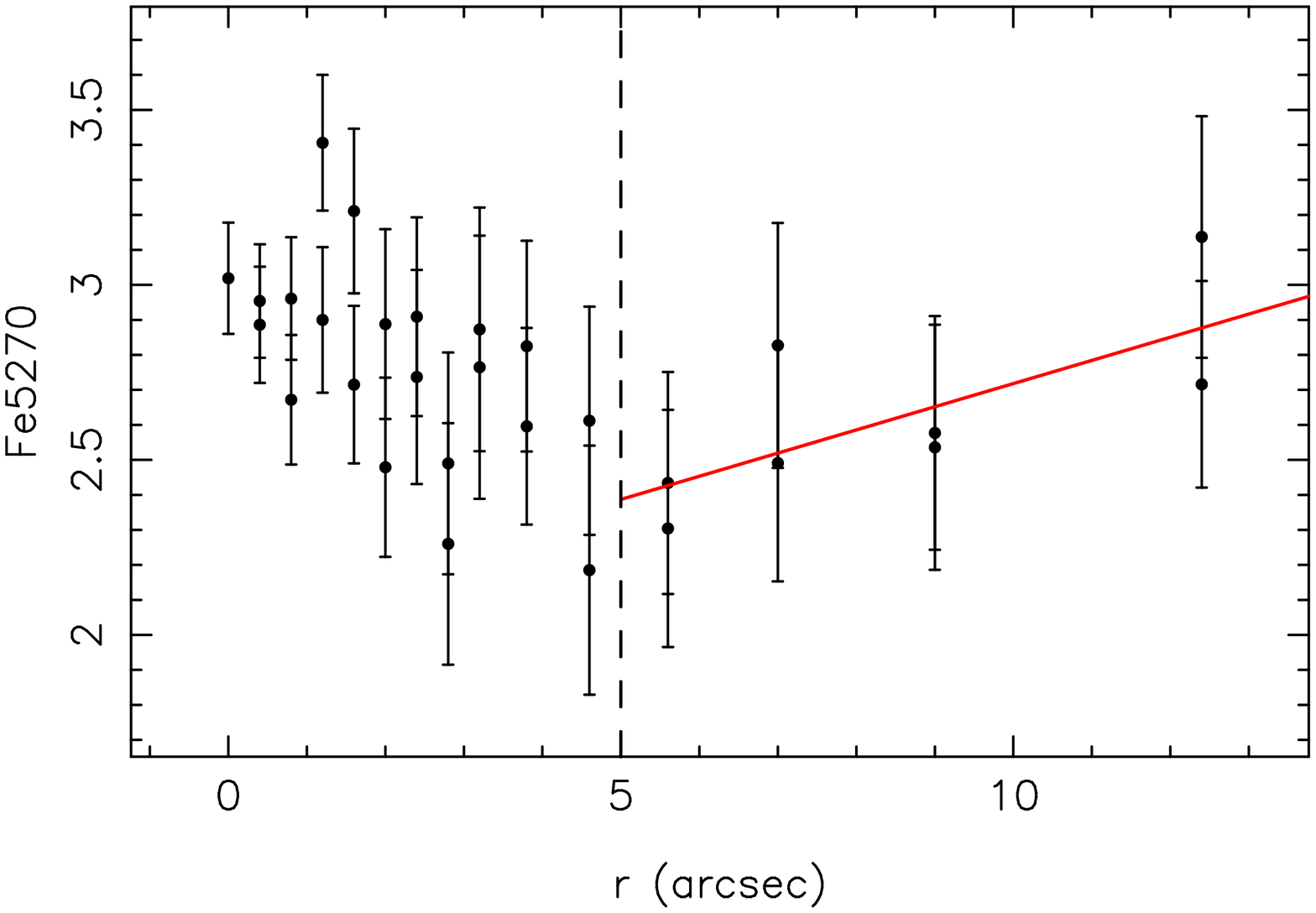}}\hspace{0.85cm}
\resizebox{0.3\textwidth}{!}{\includegraphics[angle=-90]{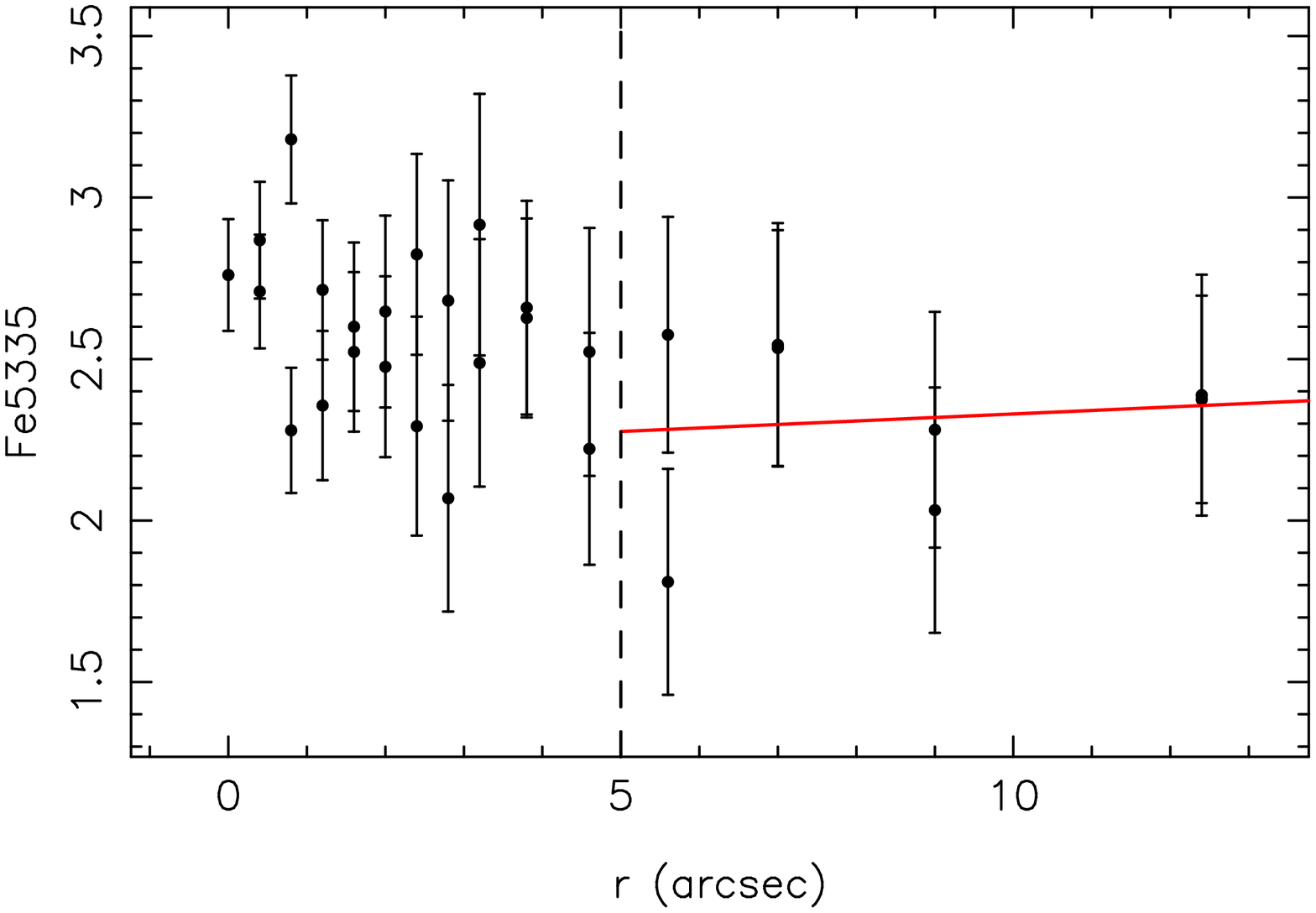}}
\caption{Line-strength distribution in the bar region for all the galaxies}
\end{figure*}
\begin{figure*}
\addtocounter{figure}{-1}
\resizebox{0.3\textwidth}{!}{\includegraphics[angle=-90]{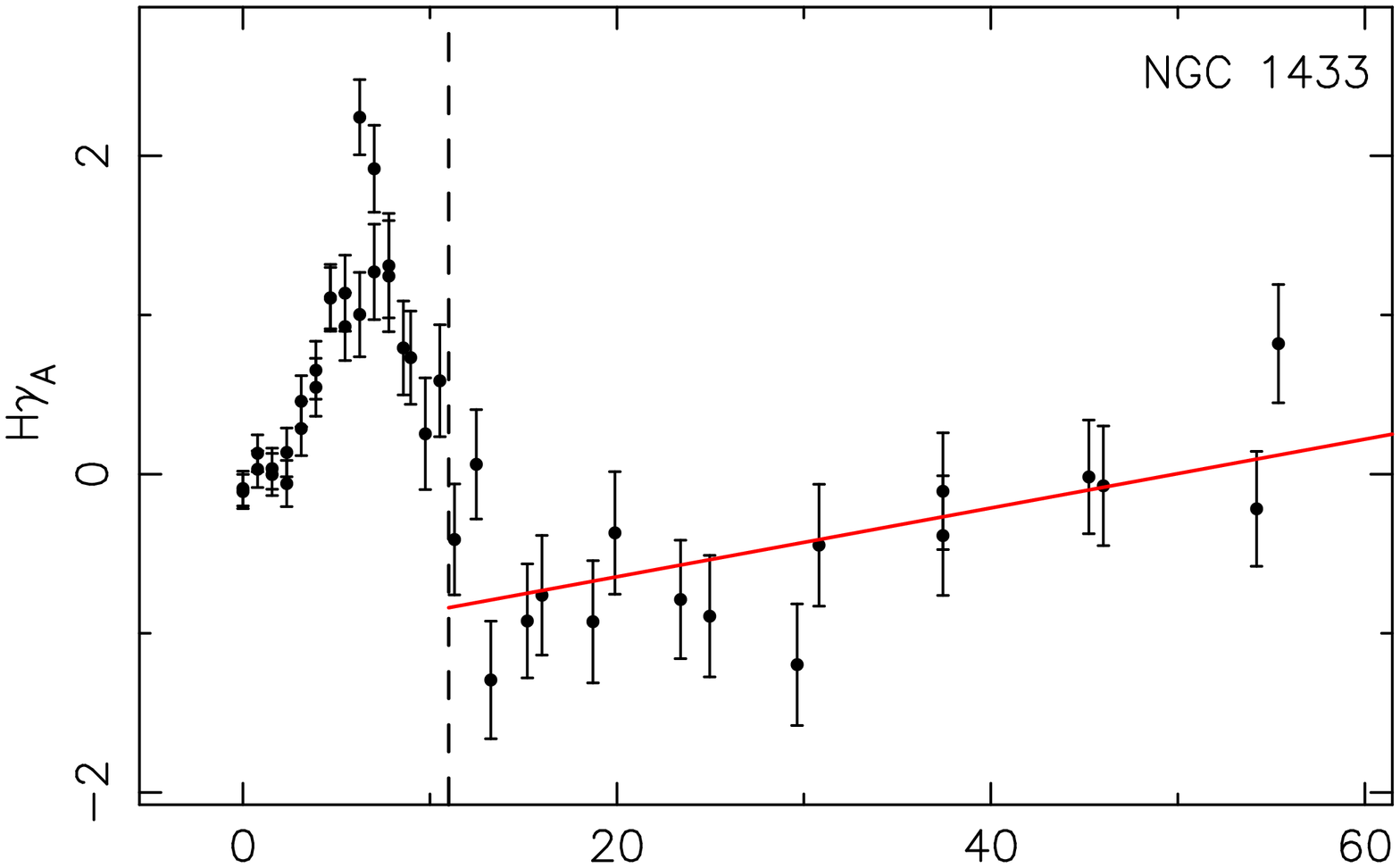}}
\resizebox{0.3\textwidth}{!}{\includegraphics[angle=-90]{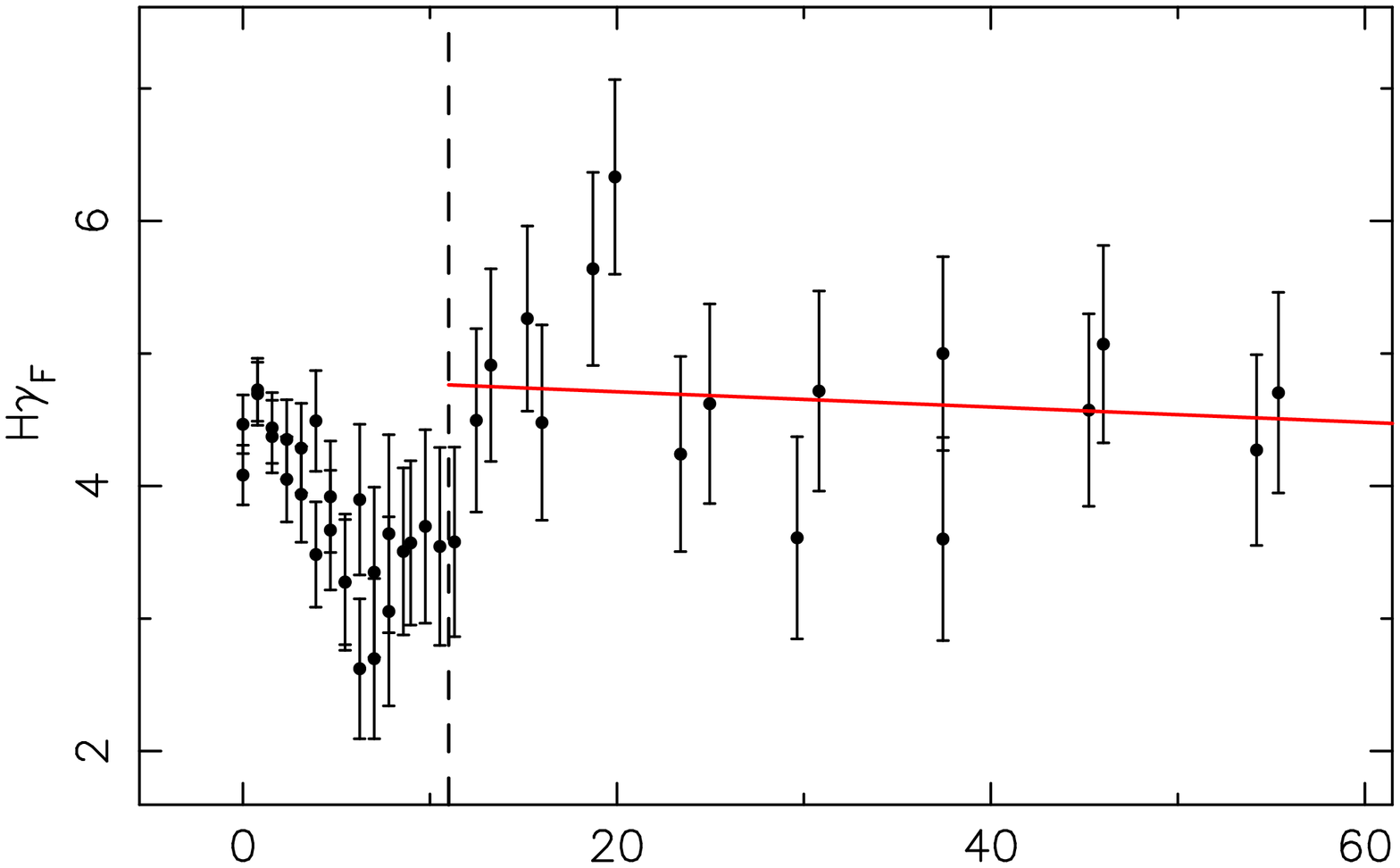}}
\resizebox{0.3\textwidth}{!}{\includegraphics[angle=-90]{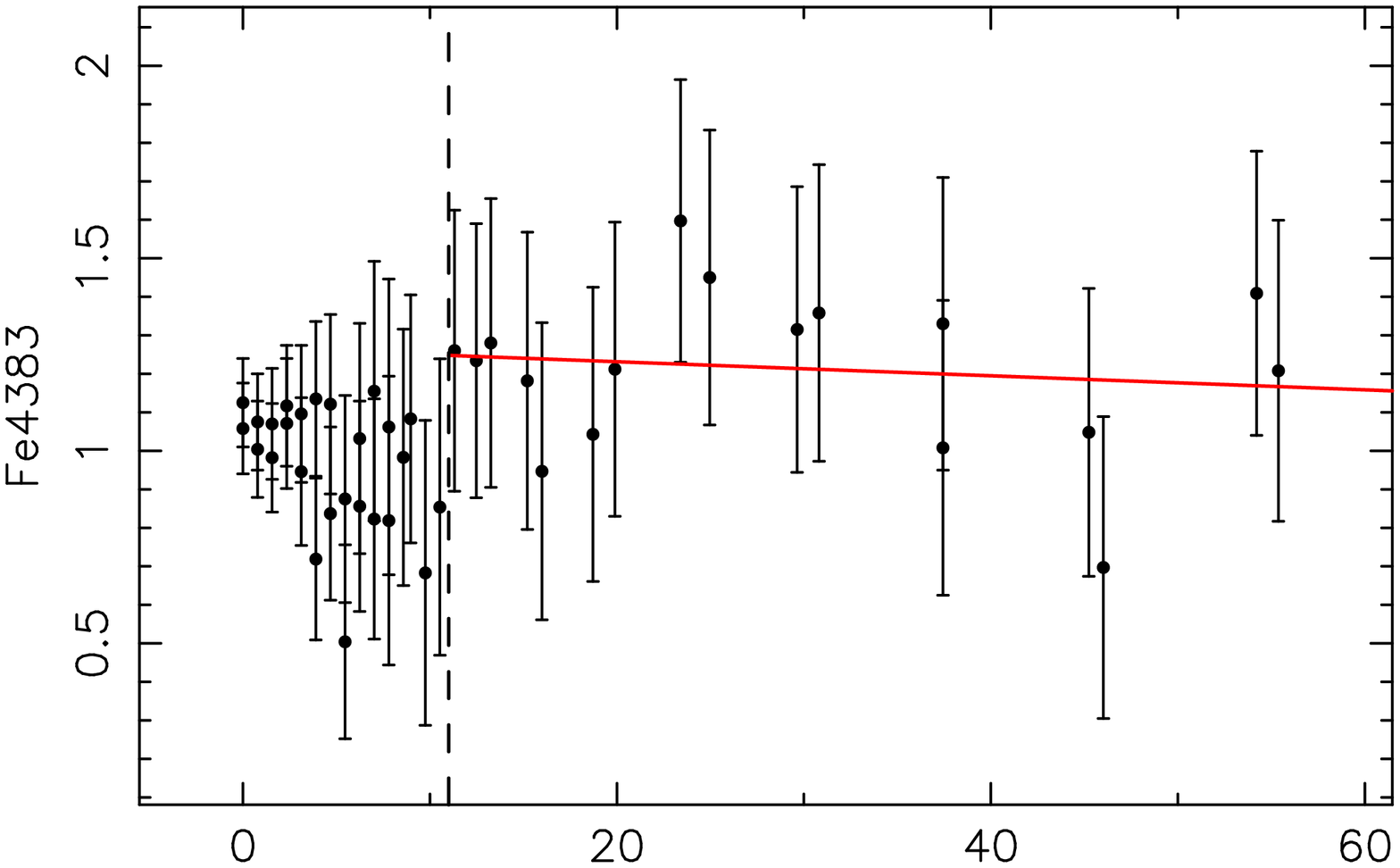}}
\resizebox{0.3\textwidth}{!}{\includegraphics[angle=-90]{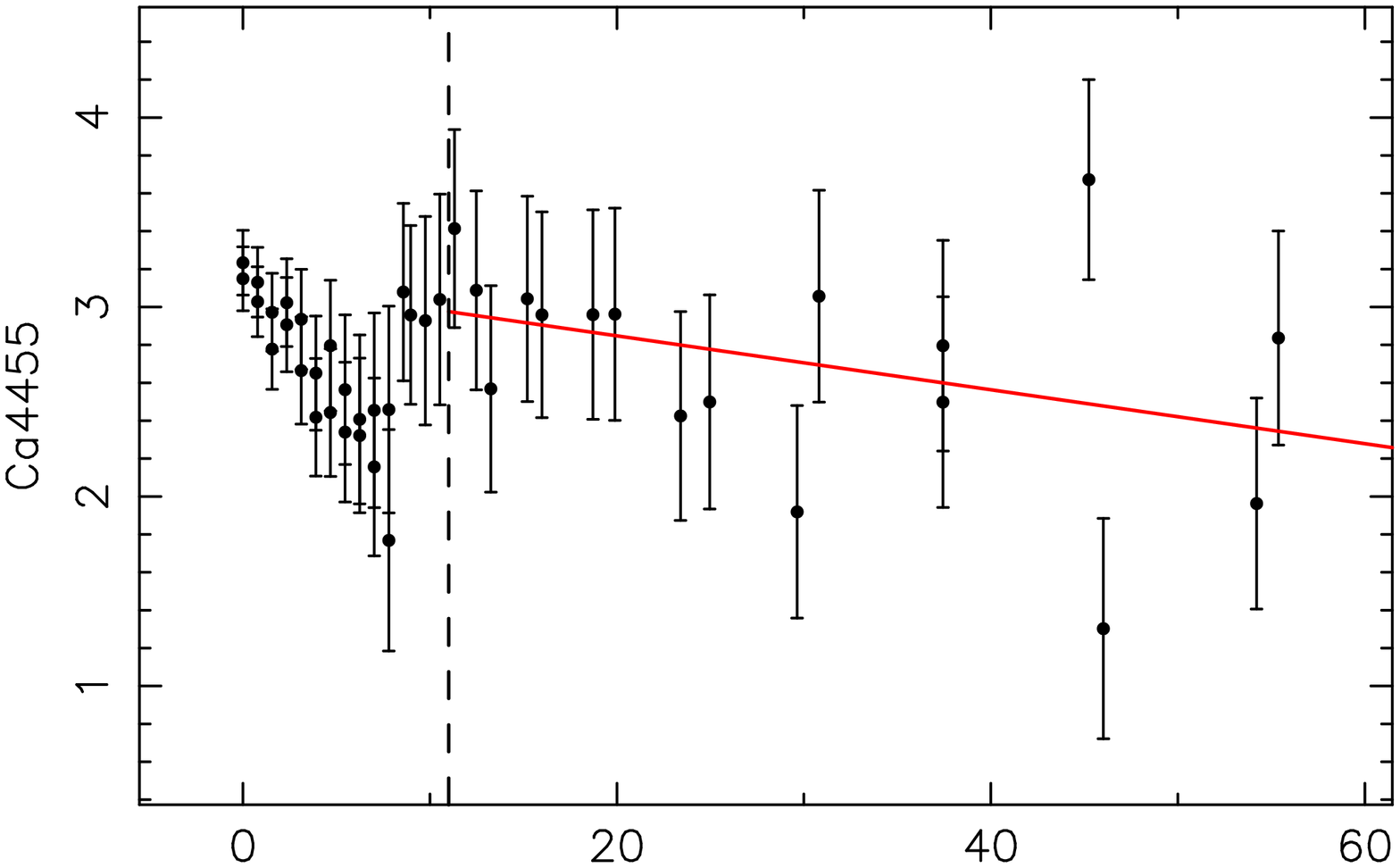}}
\resizebox{0.3\textwidth}{!}{\includegraphics[angle=-90]{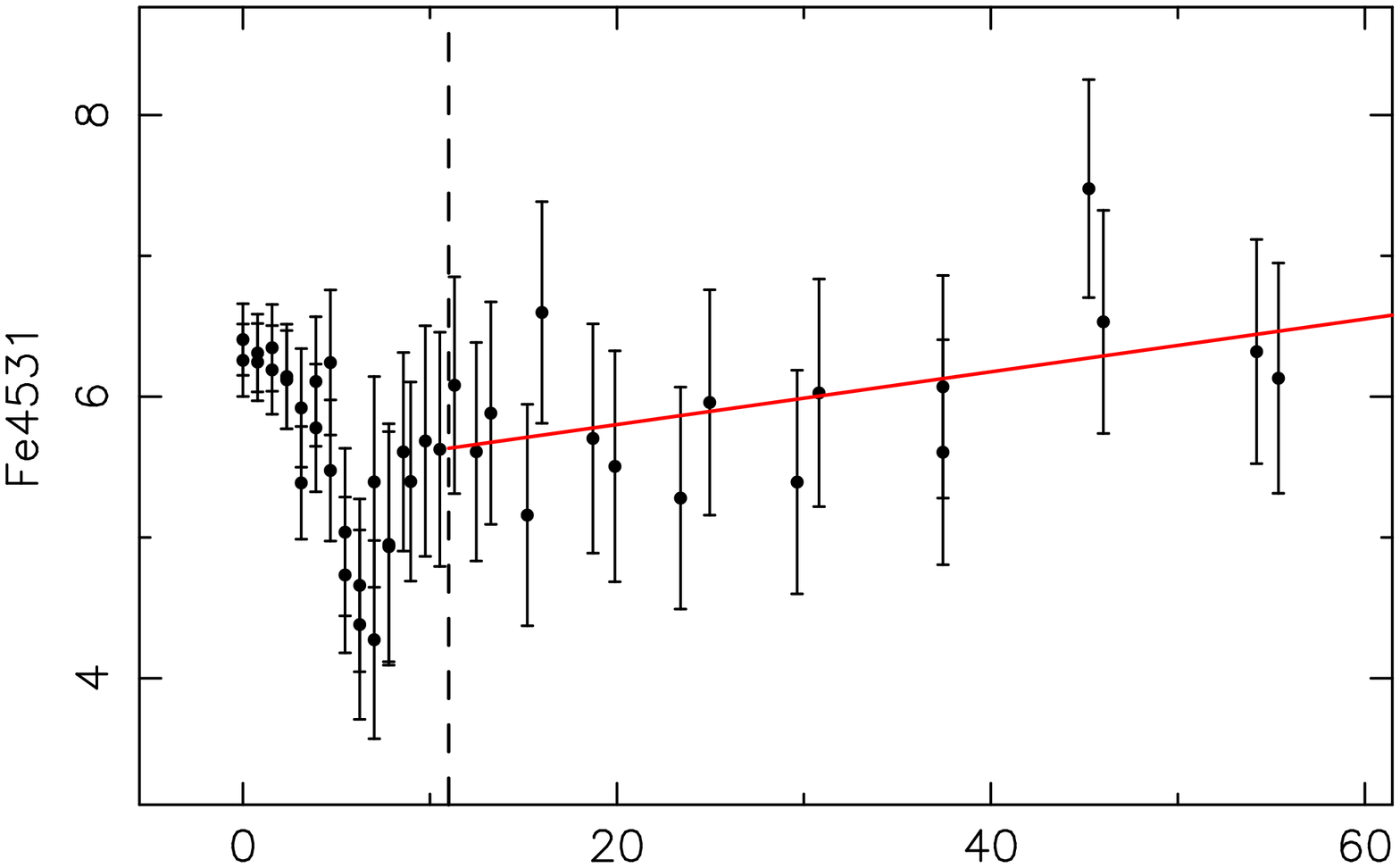}}
\resizebox{0.3\textwidth}{!}{\includegraphics[angle=-90]{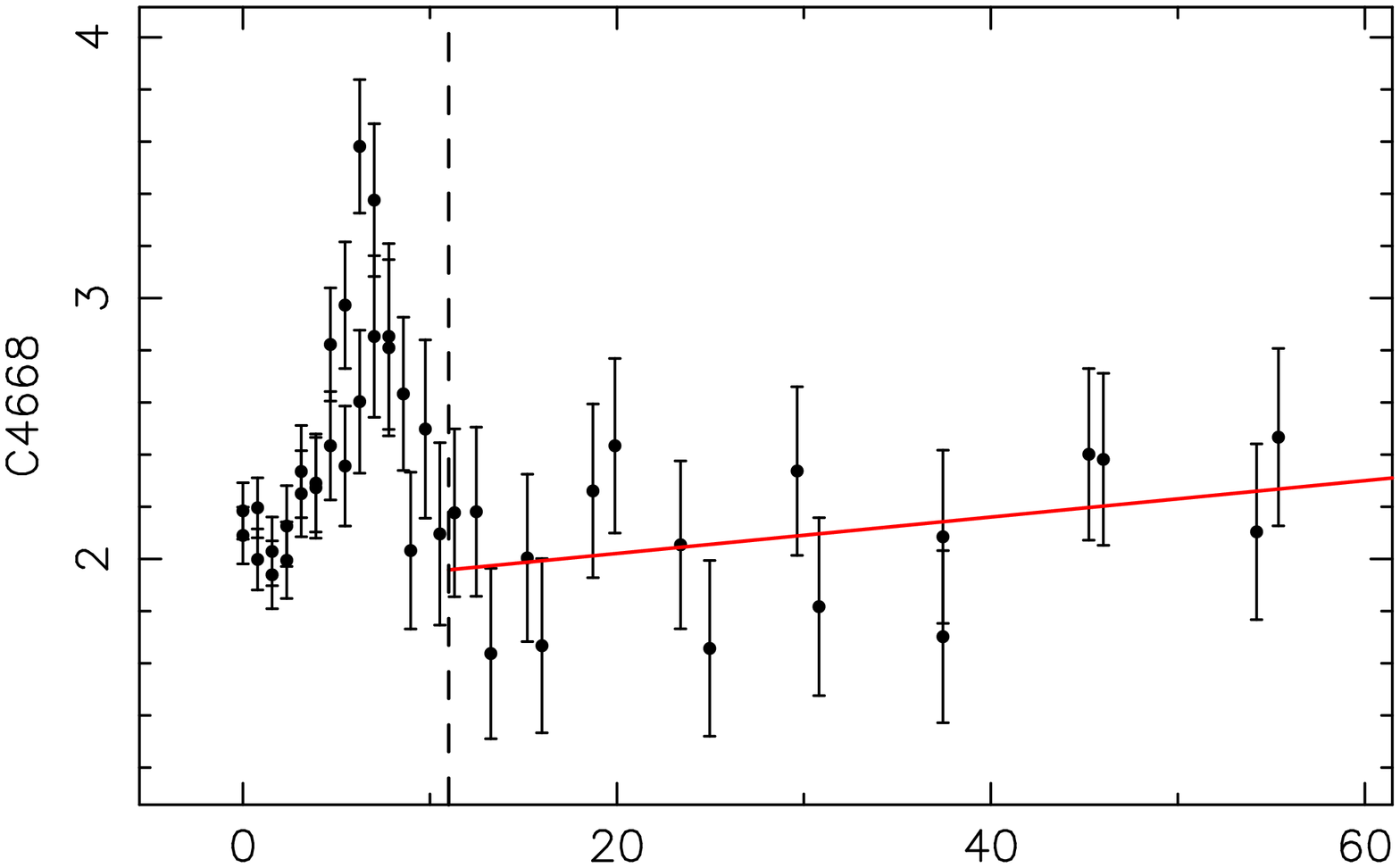}}
\resizebox{0.3\textwidth}{!}{\includegraphics[angle=-90]{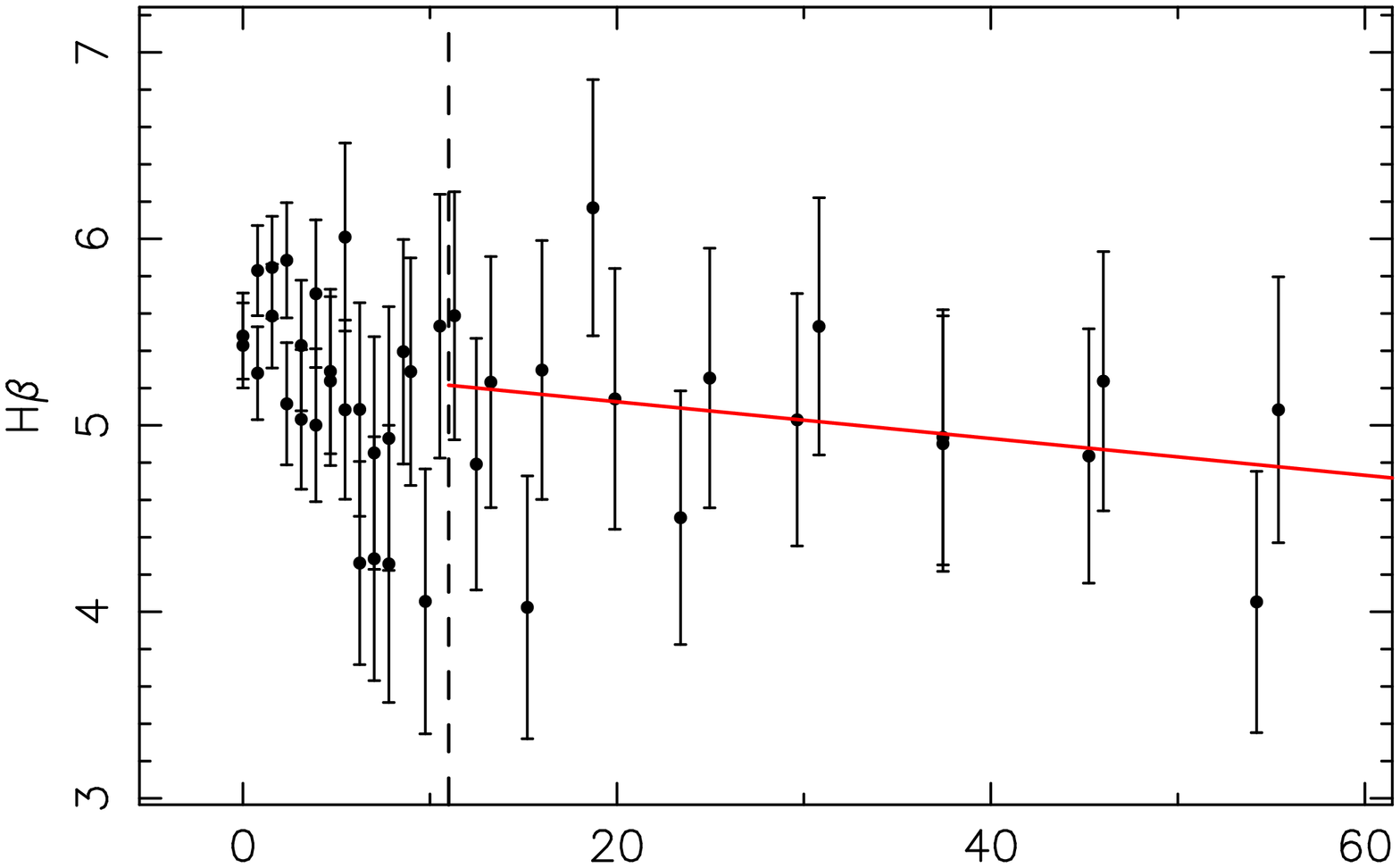}}
\resizebox{0.3\textwidth}{!}{\includegraphics[angle=-90]{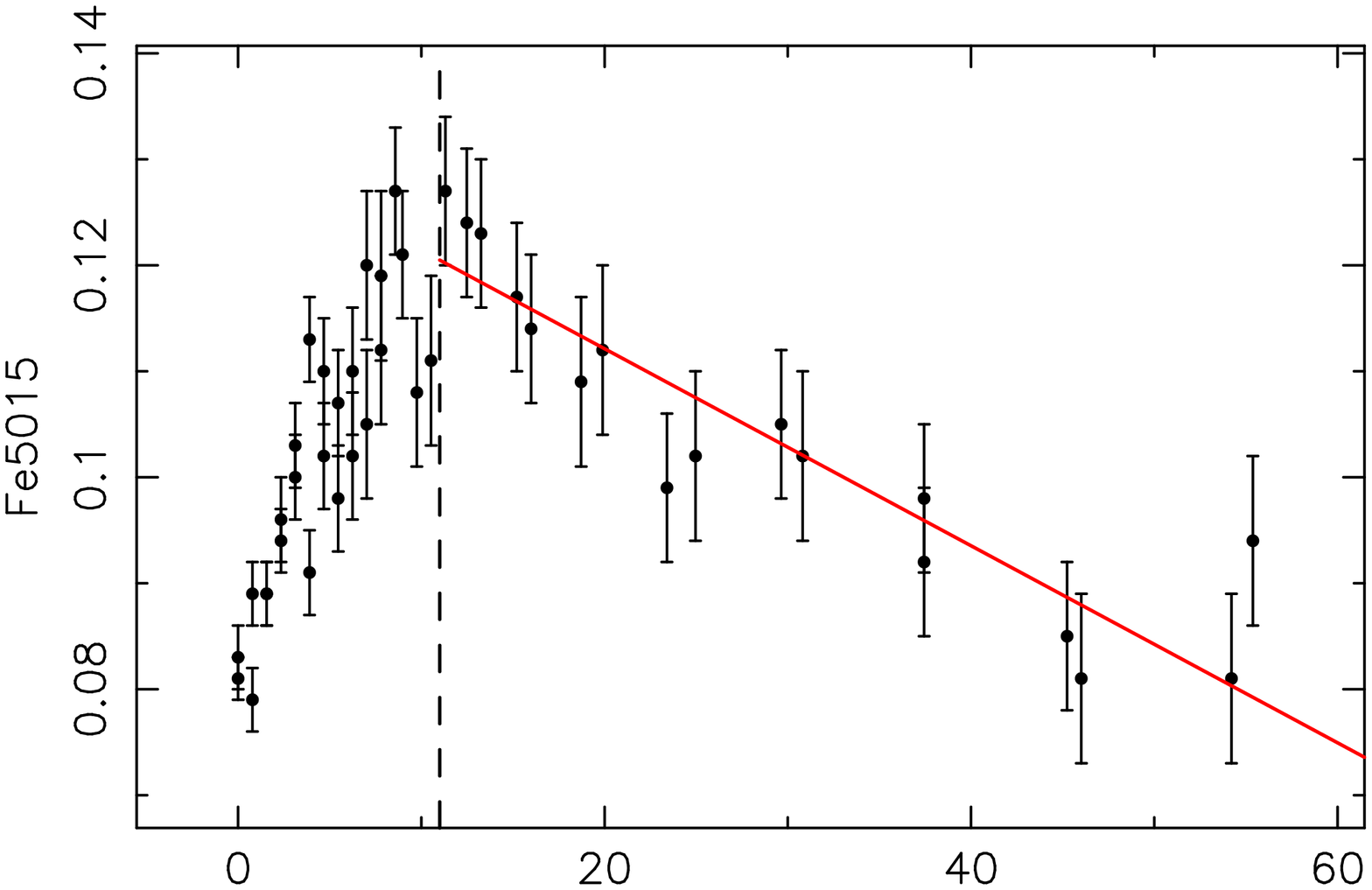}}
\resizebox{0.3\textwidth}{!}{\includegraphics[angle=-90]{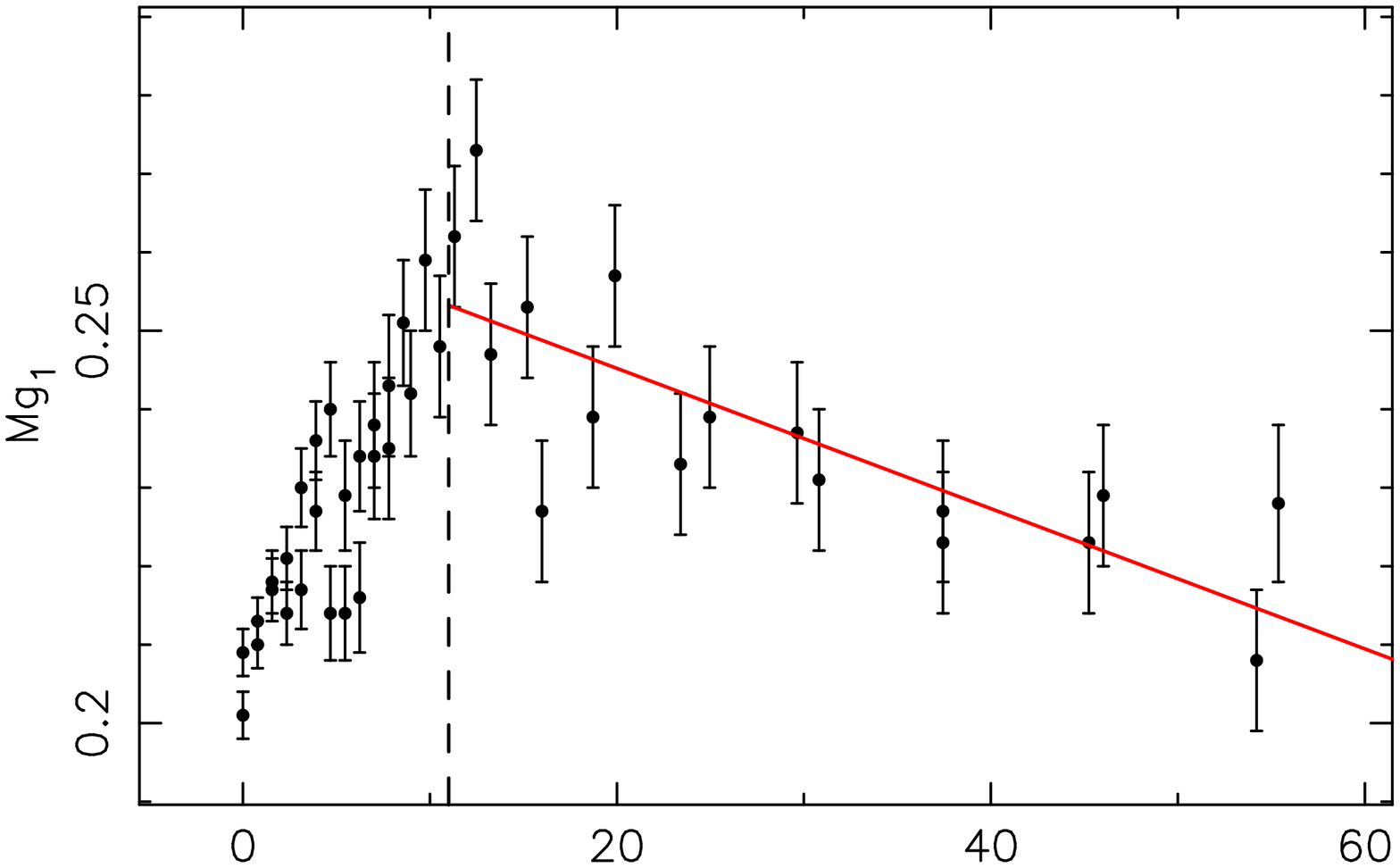}}
\resizebox{0.3\textwidth}{!}{\includegraphics[angle=-90]{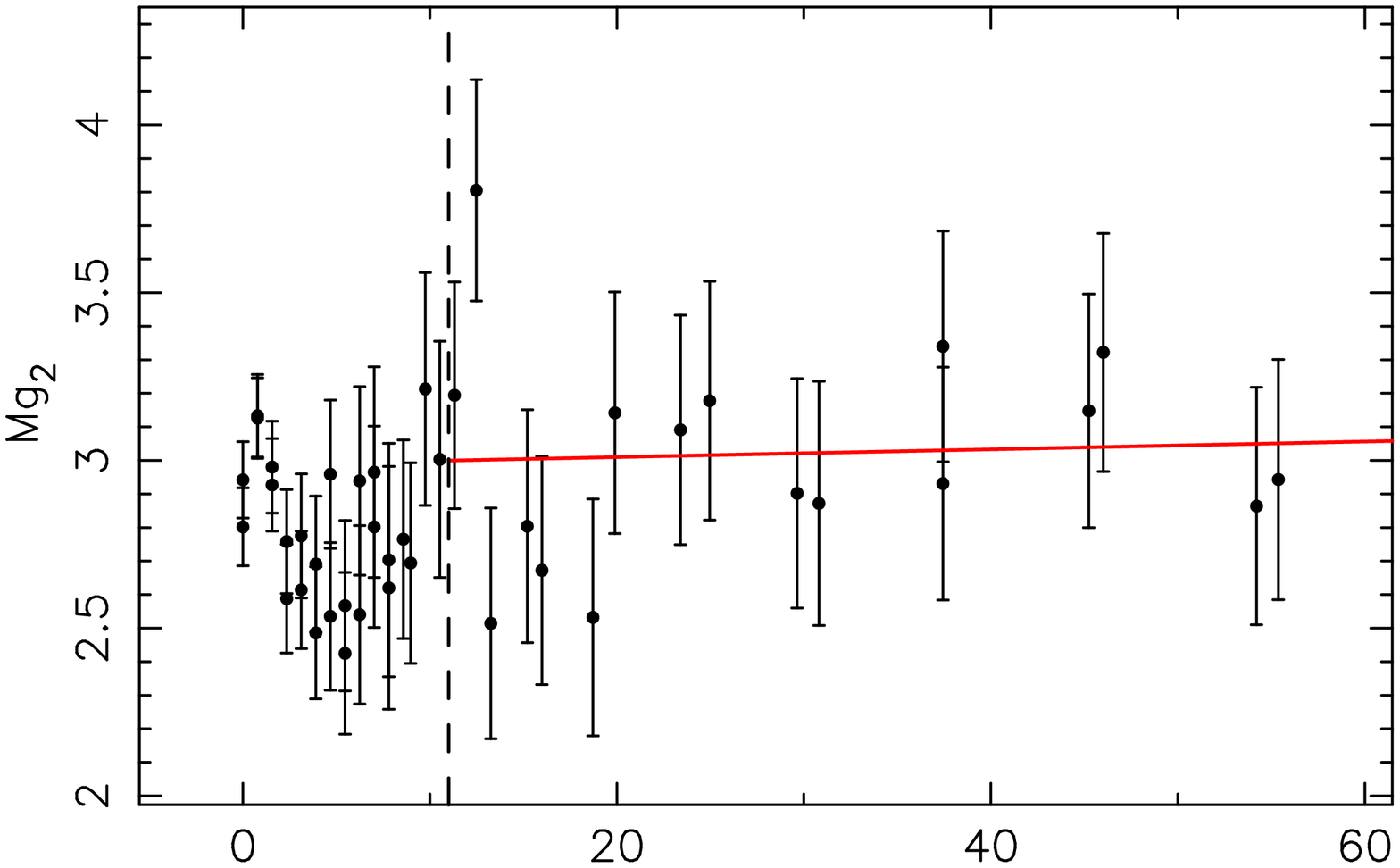}}\hspace{0.85cm}
\resizebox{0.3\textwidth}{!}{\includegraphics[angle=-90]{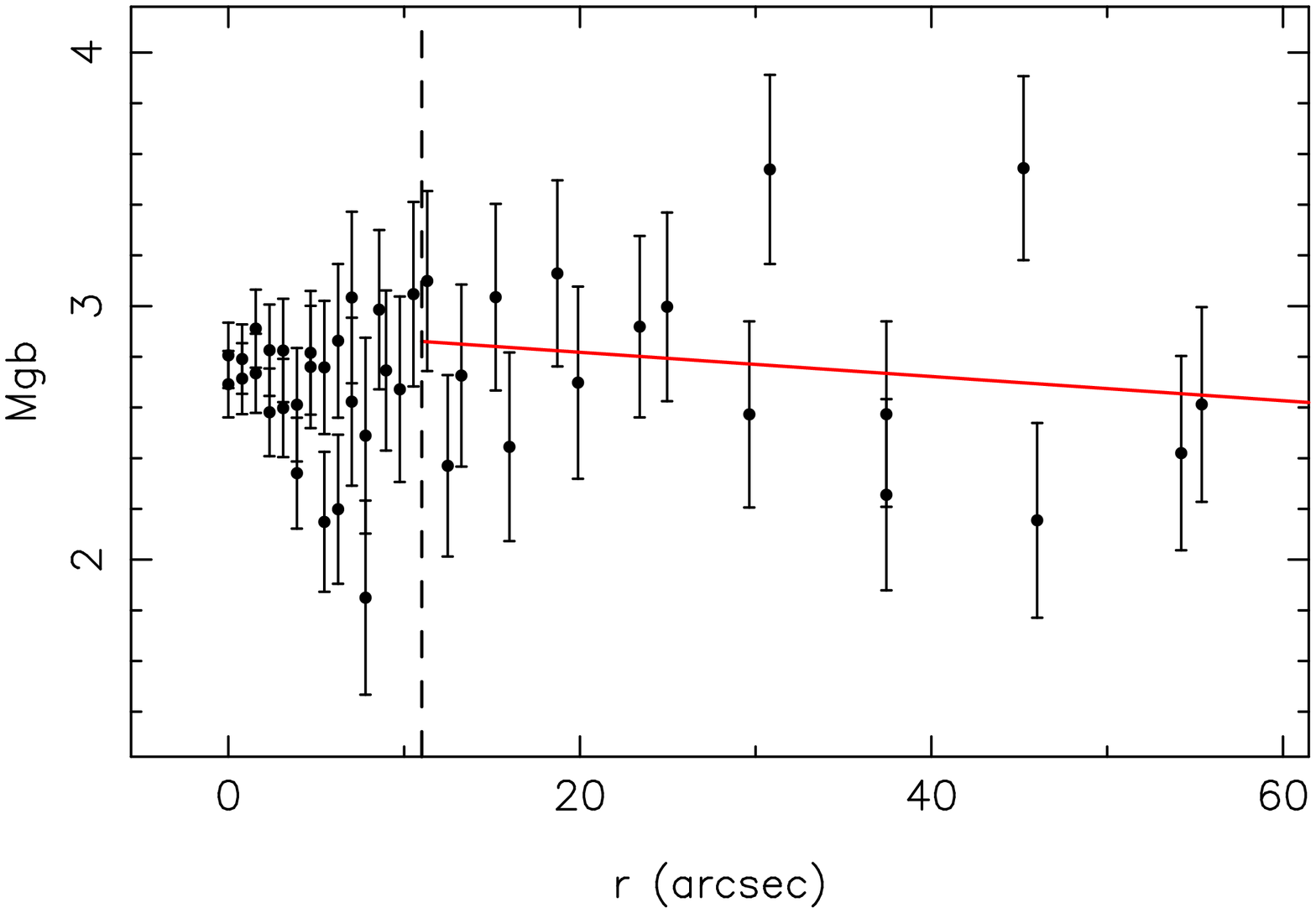}}\hspace{0.85cm}
\resizebox{0.3\textwidth}{!}{\includegraphics[angle=-90]{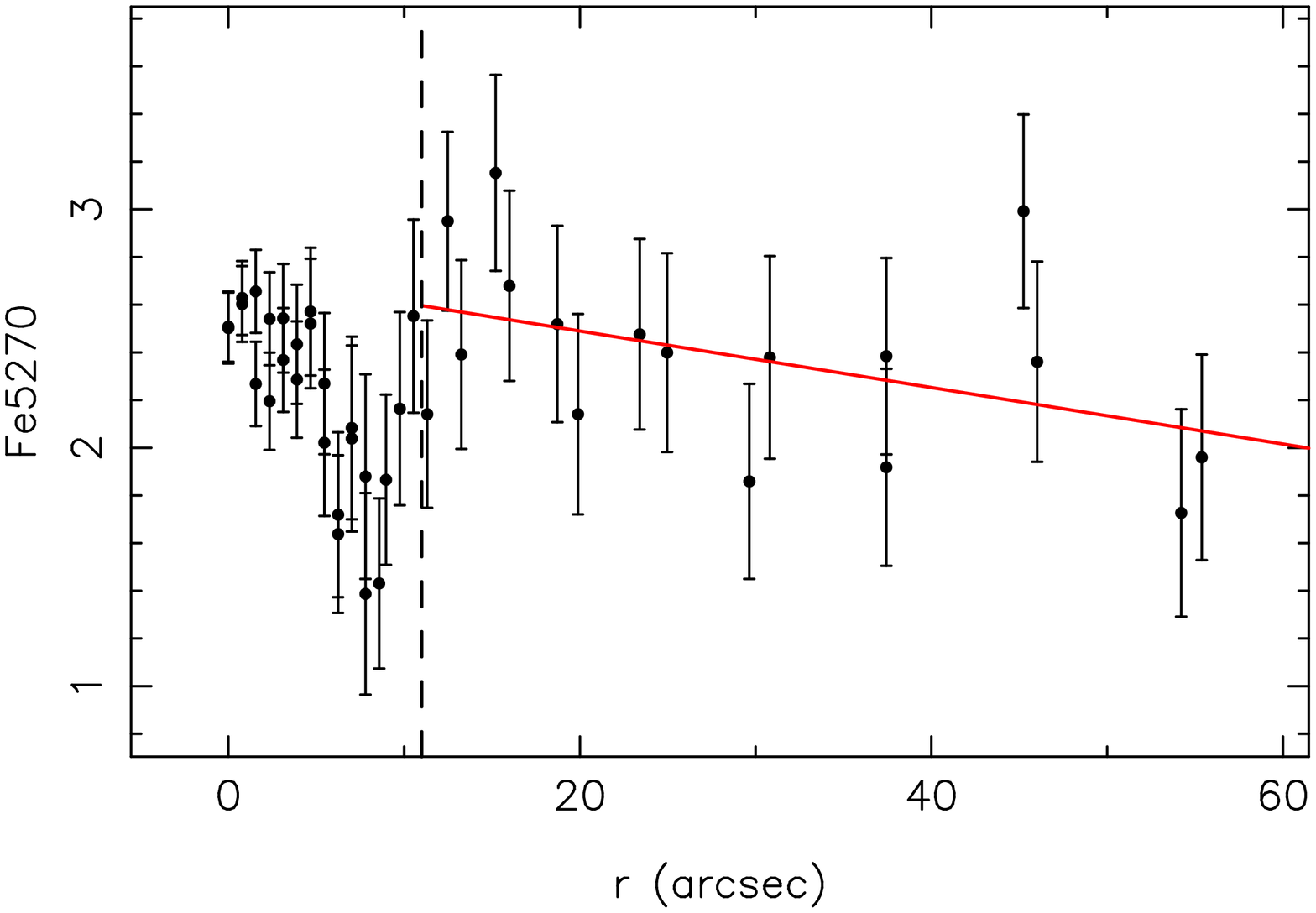}}\hspace{0.85cm}
\resizebox{0.3\textwidth}{!}{\includegraphics[angle=-90]{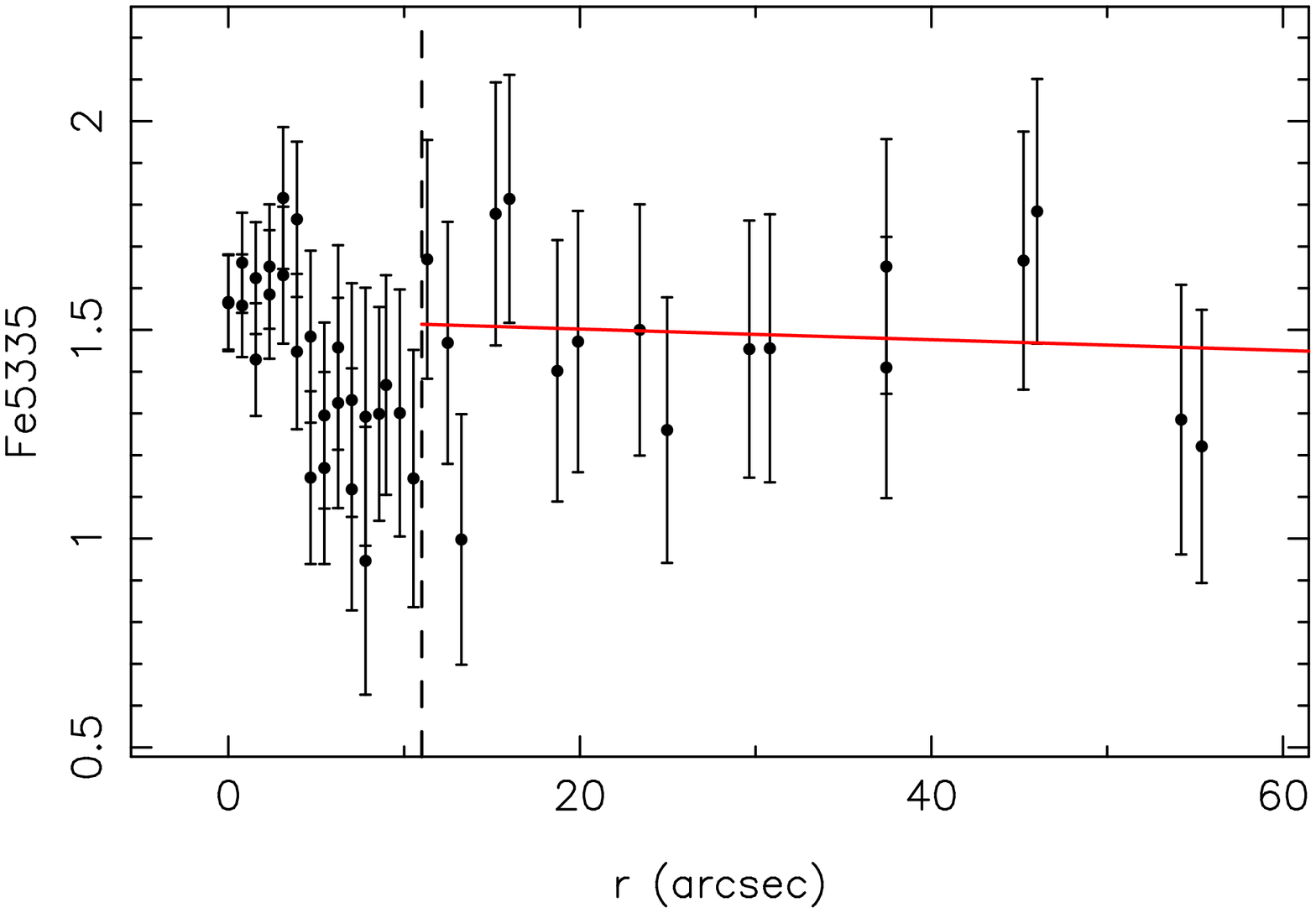}}
\caption{Line-strength distribution in the bar region for all the galaxies}
\end{figure*}
\begin{figure*}
\addtocounter{figure}{-1}
\resizebox{0.3\textwidth}{!}{\includegraphics[angle=-90]{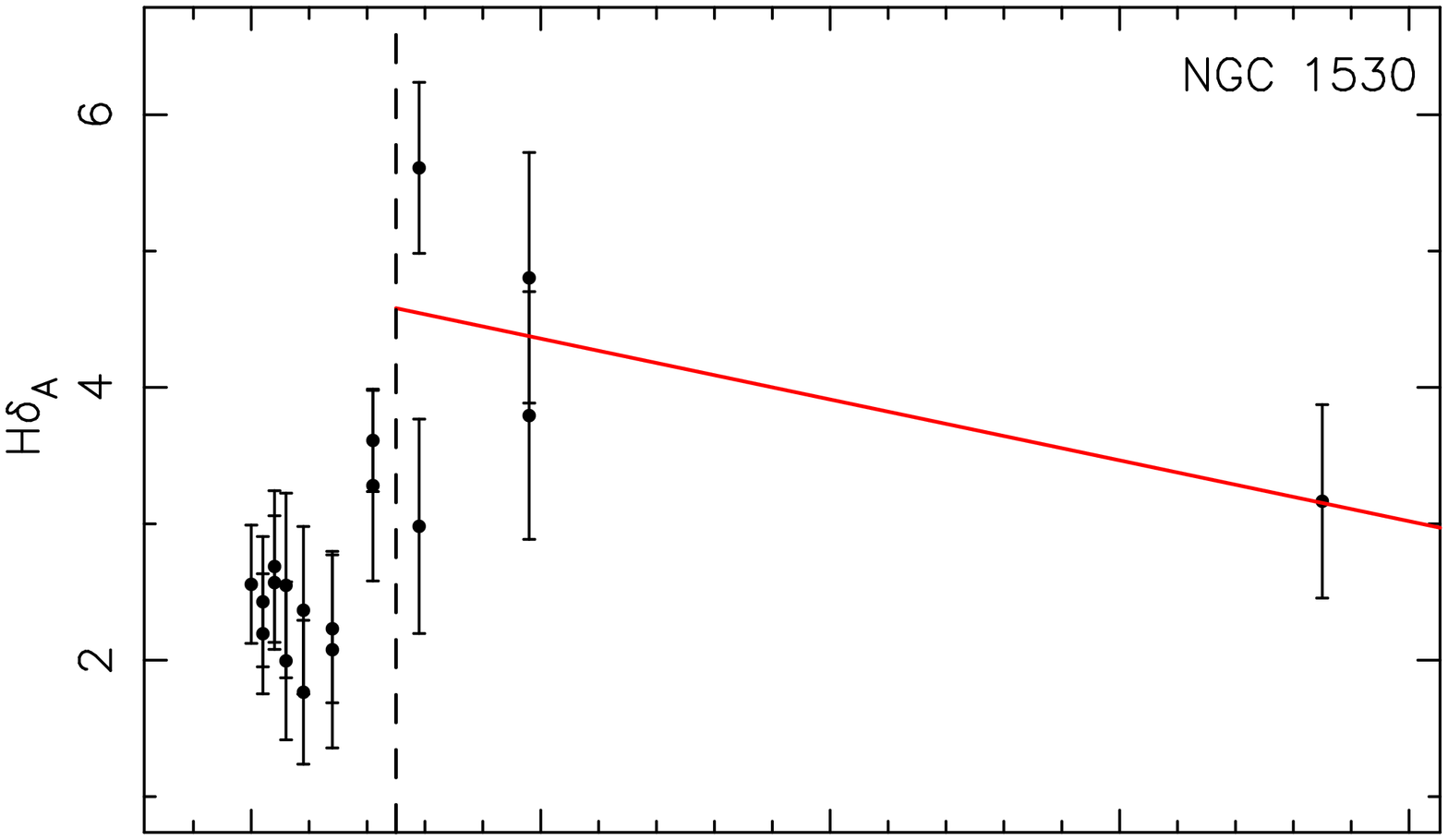}}
\resizebox{0.3\textwidth}{!}{\includegraphics[angle=-90]{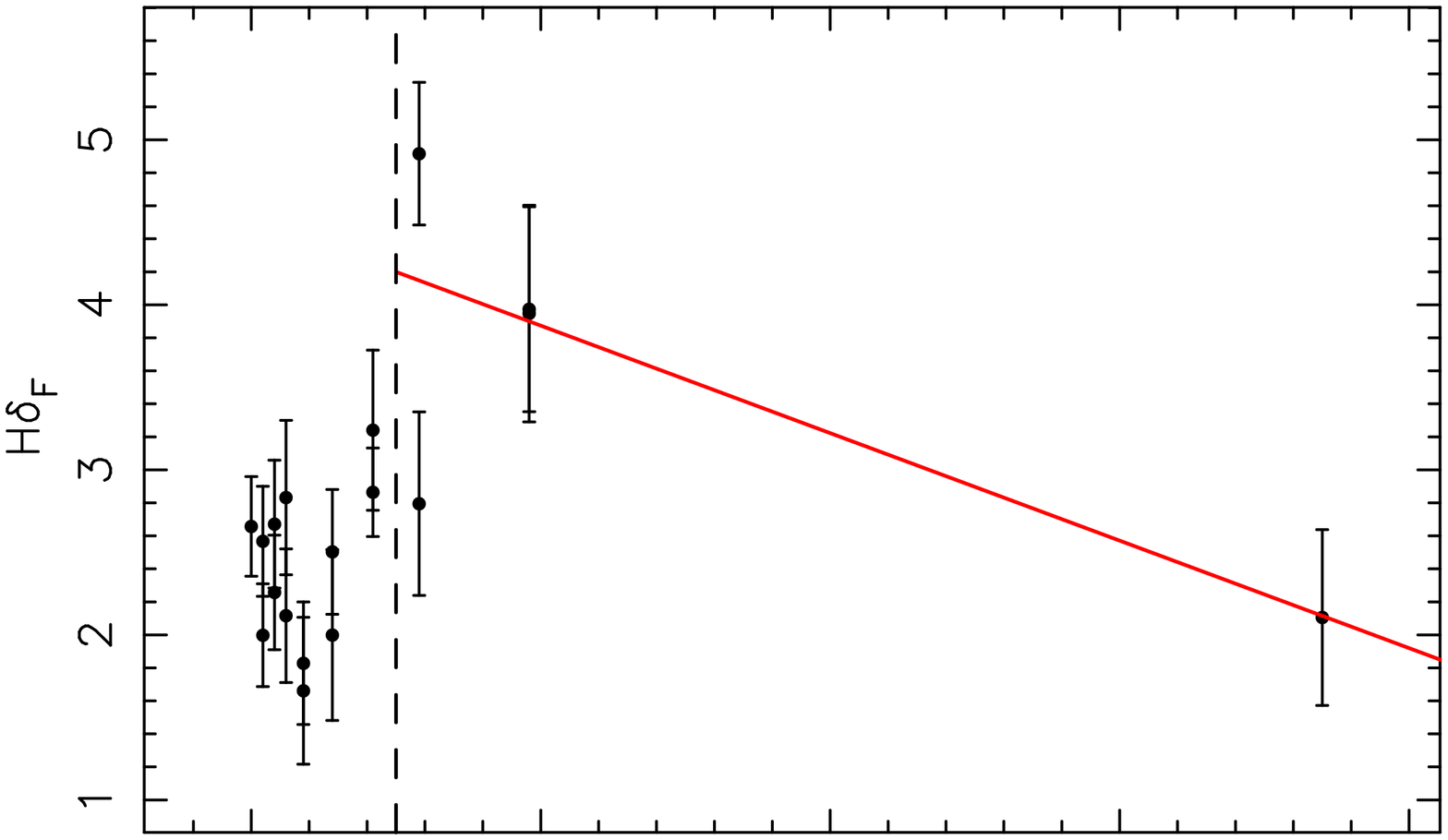}}
\resizebox{0.3\textwidth}{!}{\includegraphics[angle=-90]{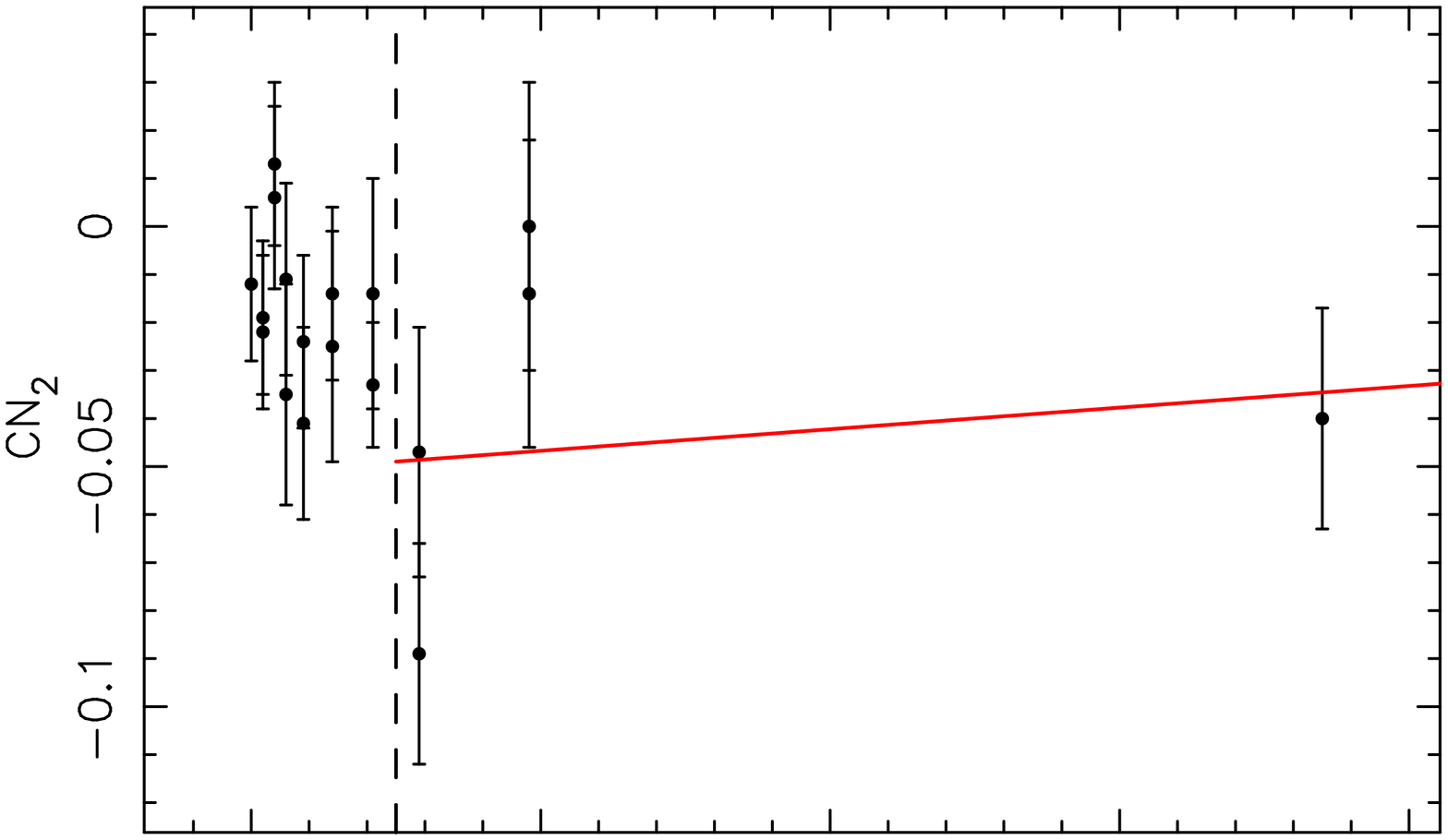}}
\resizebox{0.3\textwidth}{!}{\includegraphics[angle=-90]{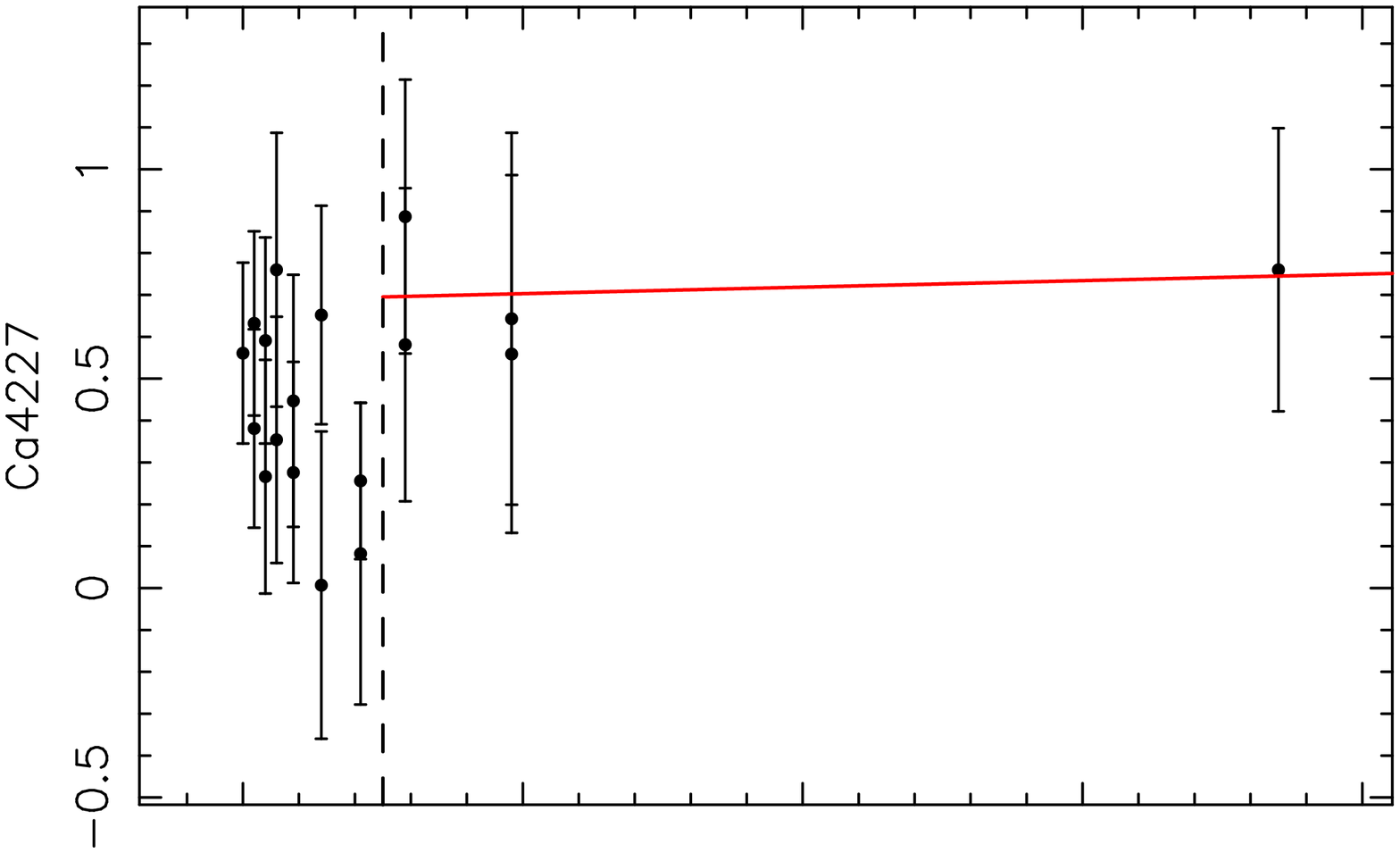}}
\resizebox{0.3\textwidth}{!}{\includegraphics[angle=-90]{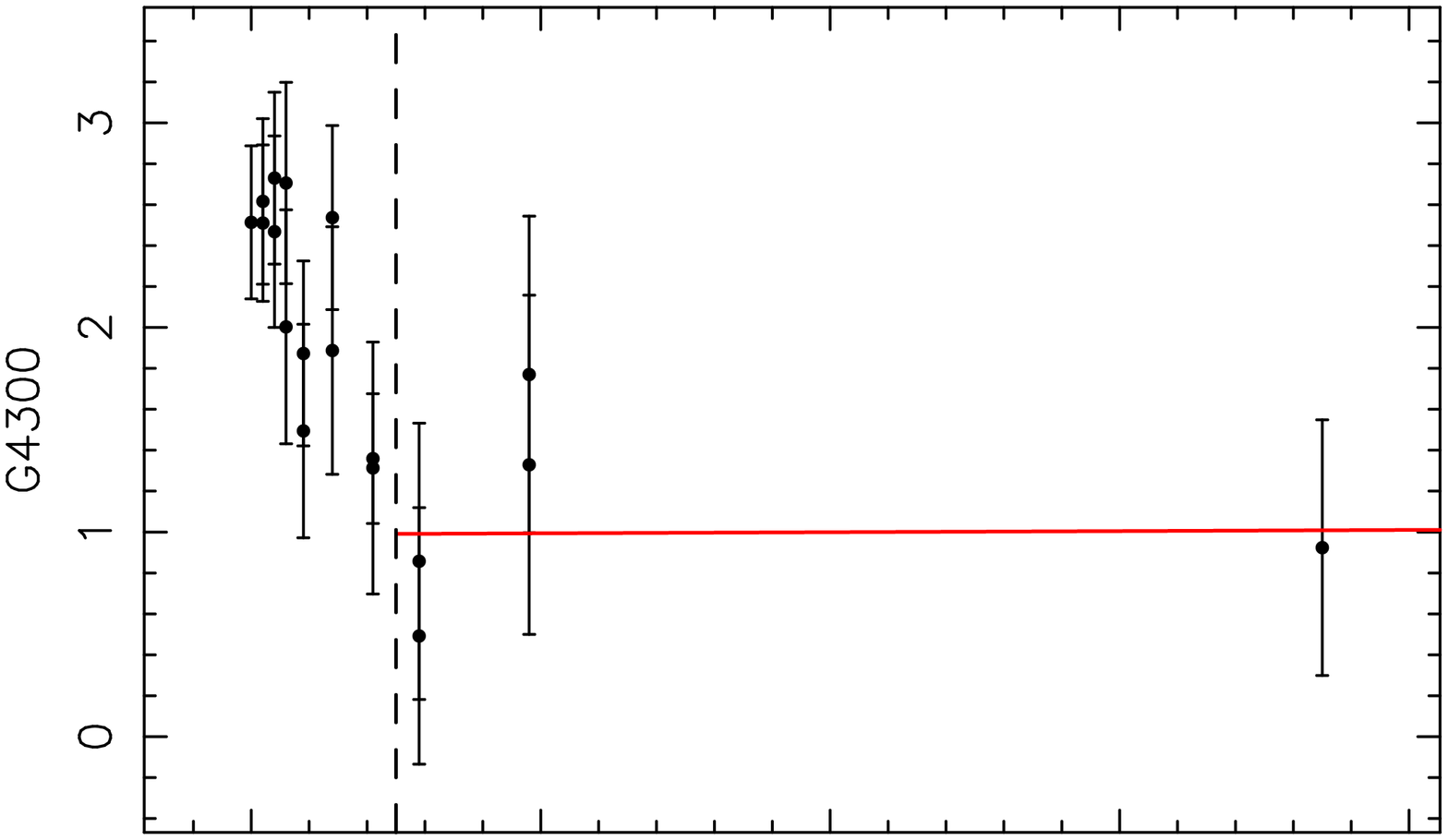}}
\resizebox{0.3\textwidth}{!}{\includegraphics[angle=-90]{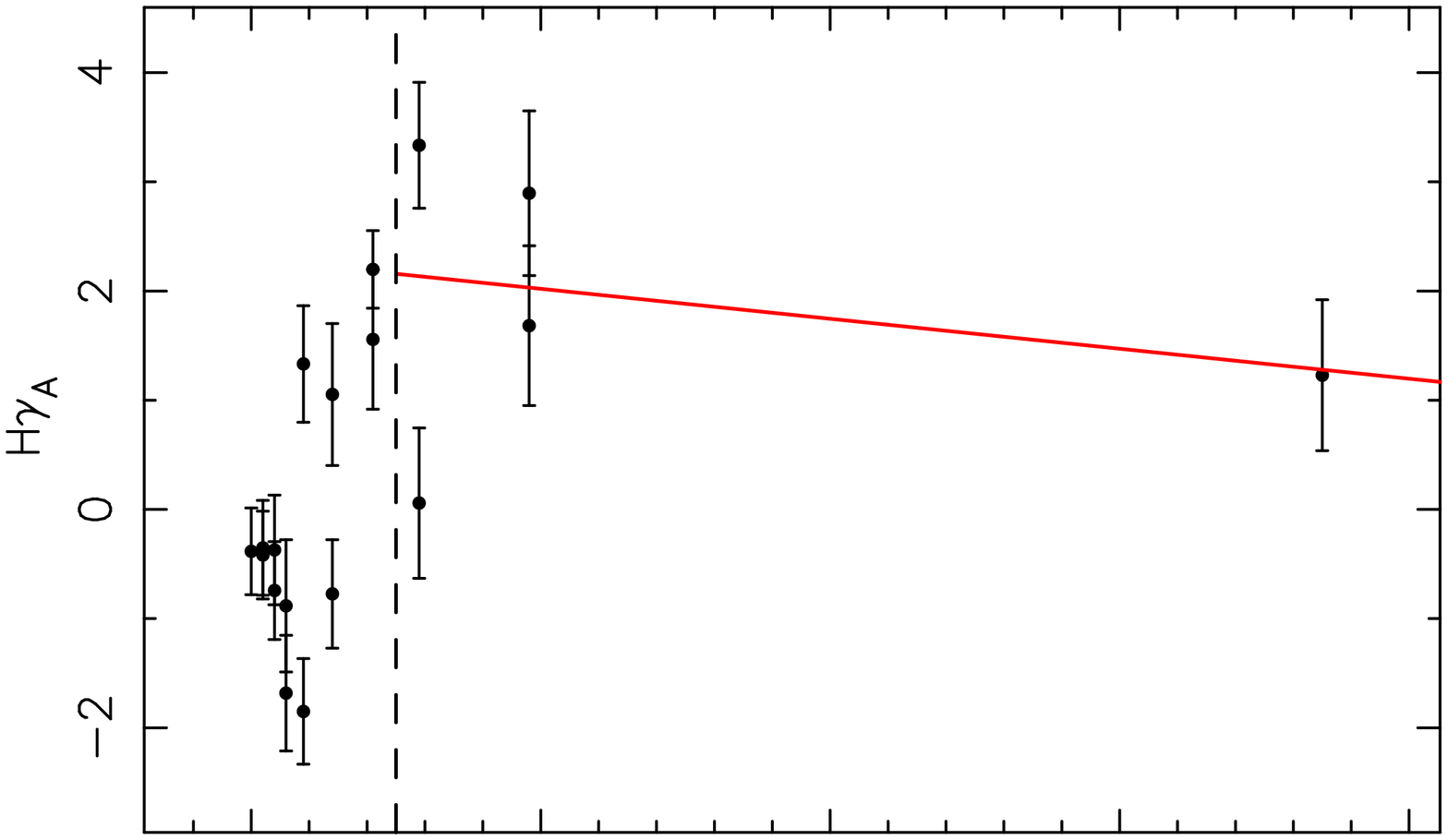}}
\resizebox{0.3\textwidth}{!}{\includegraphics[angle=-90]{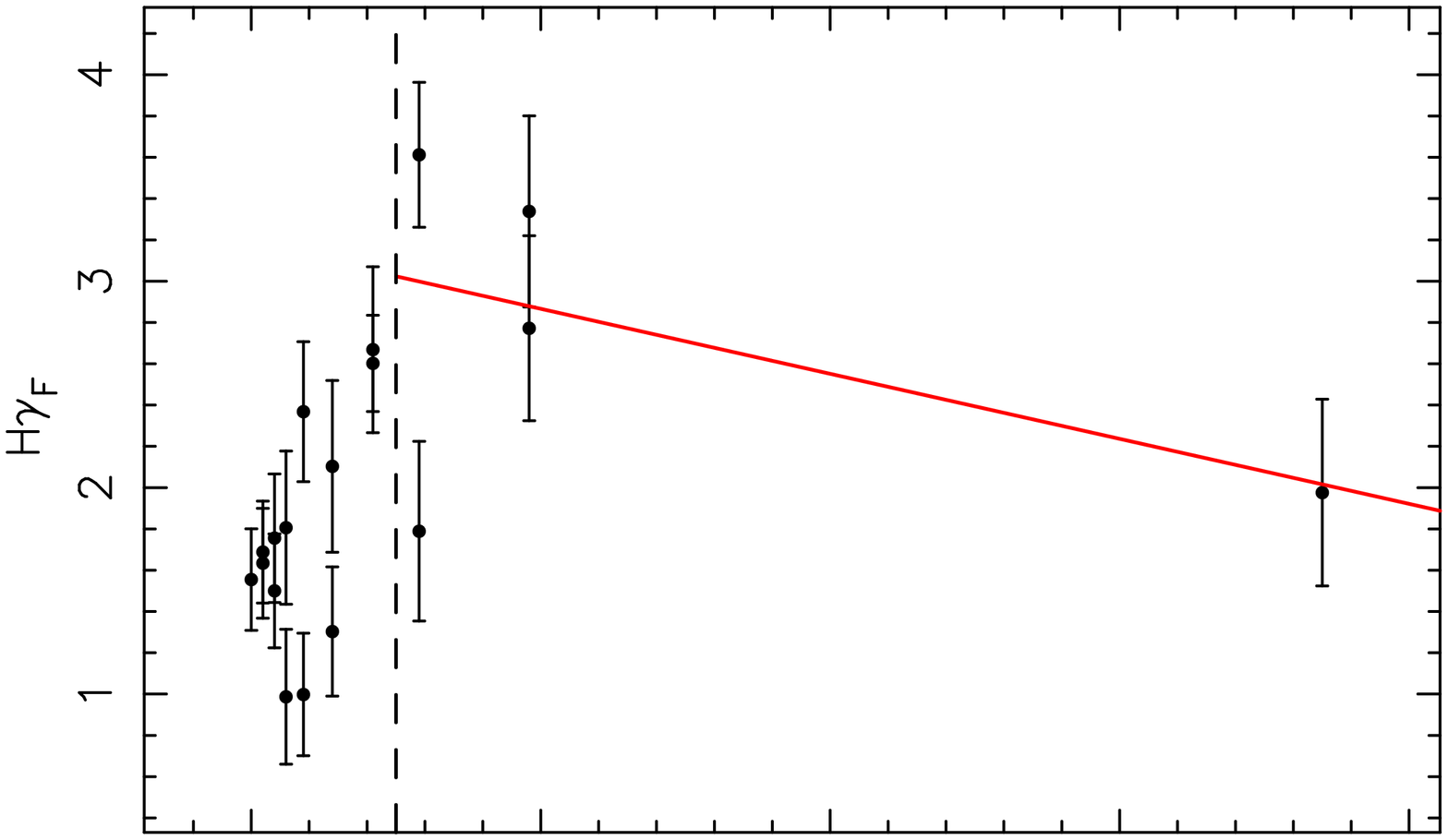}}
\resizebox{0.3\textwidth}{!}{\includegraphics[angle=-90]{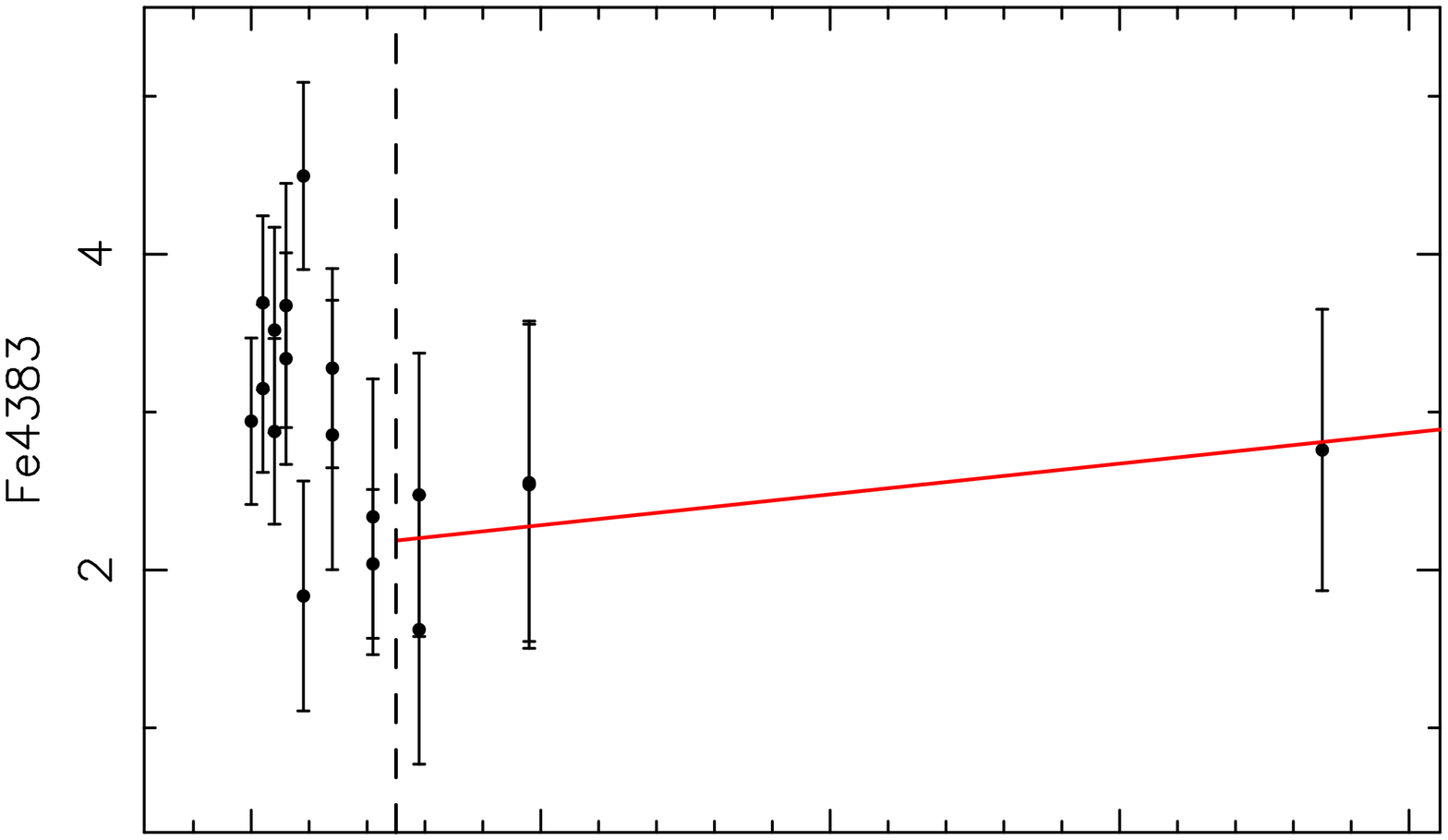}}
\resizebox{0.3\textwidth}{!}{\includegraphics[angle=-90]{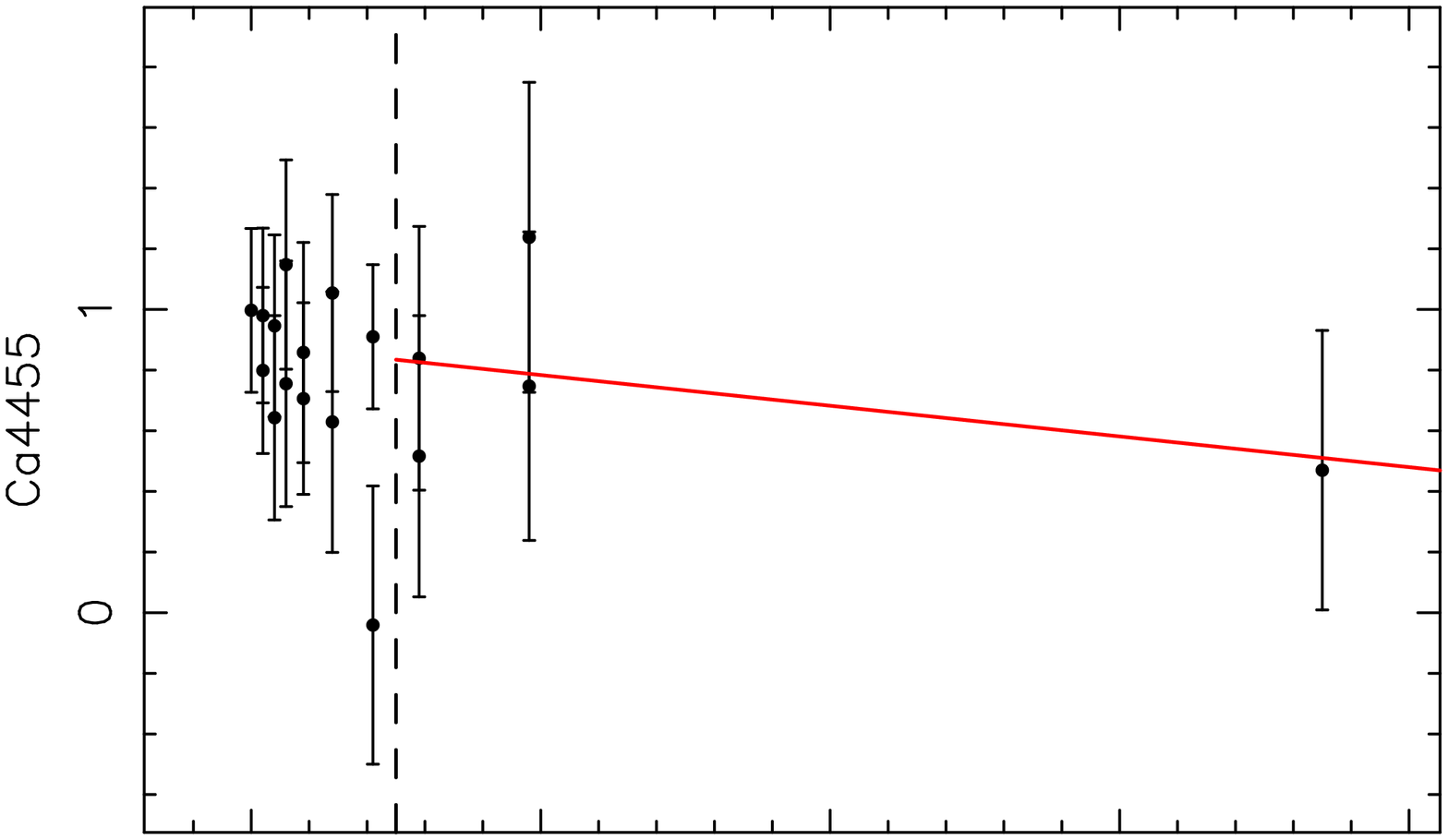}}
\resizebox{0.3\textwidth}{!}{\includegraphics[angle=-90]{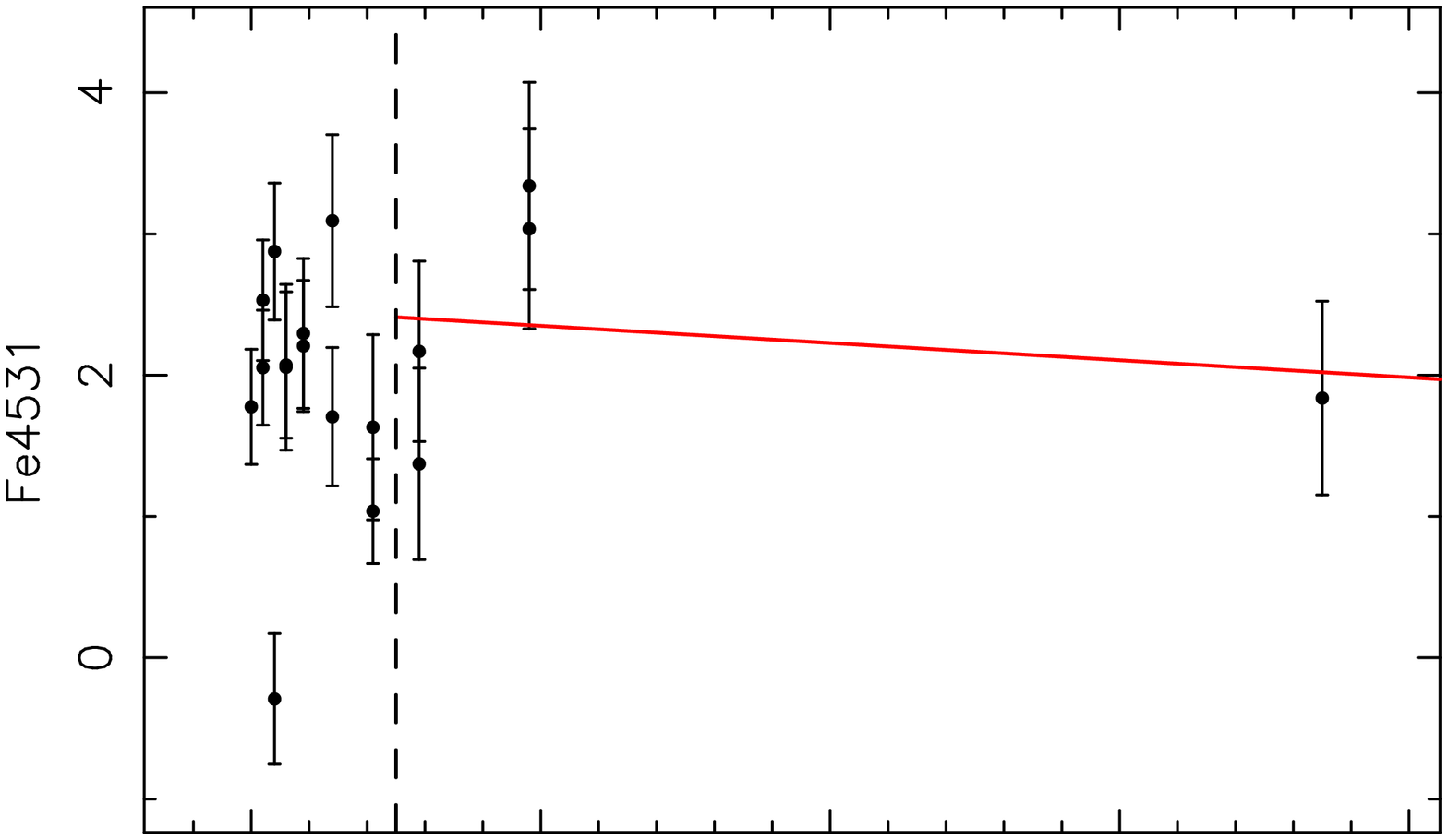}}
\resizebox{0.3\textwidth}{!}{\includegraphics[angle=-90]{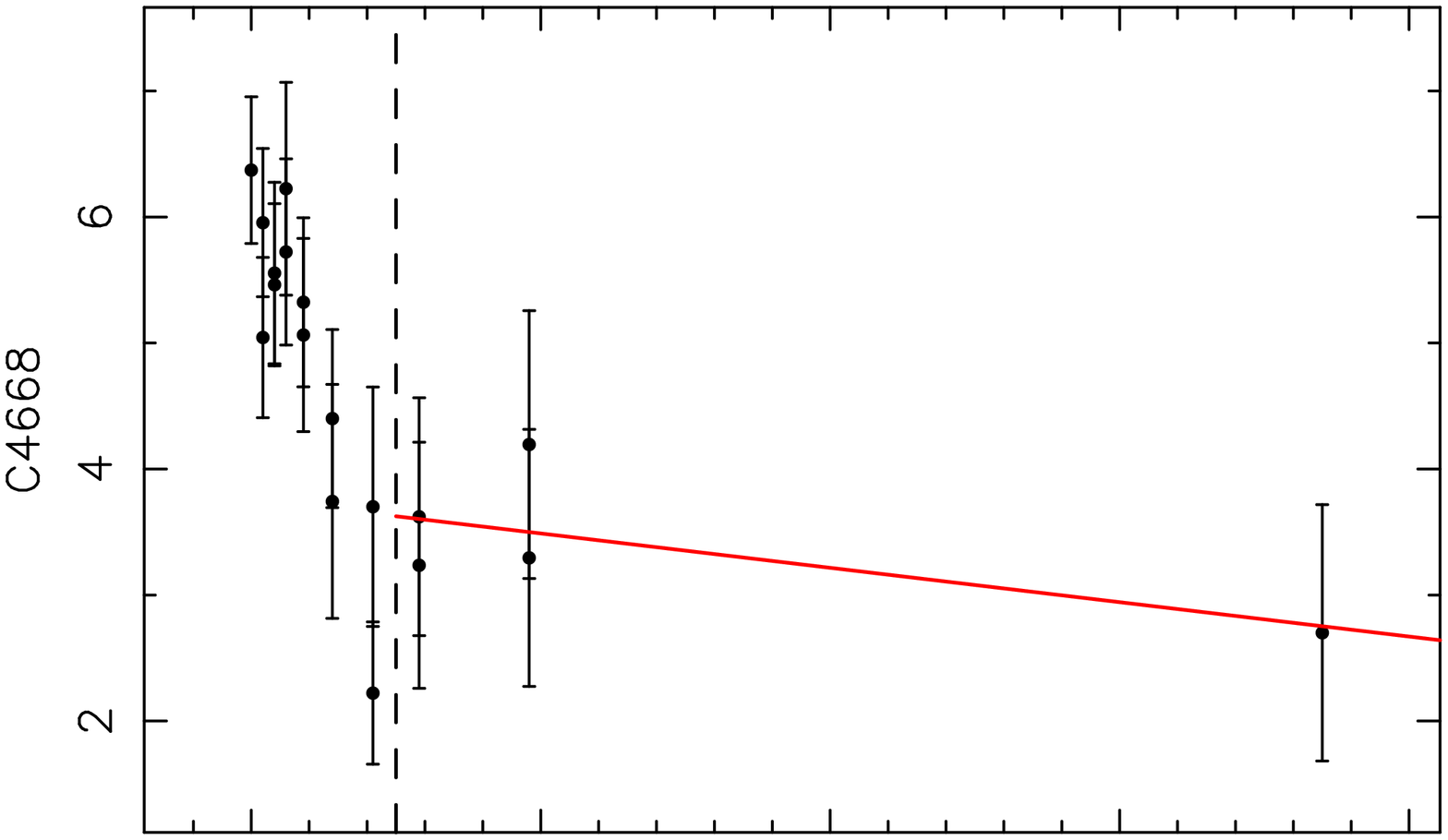}}
\resizebox{0.3\textwidth}{!}{\includegraphics[angle=-90]{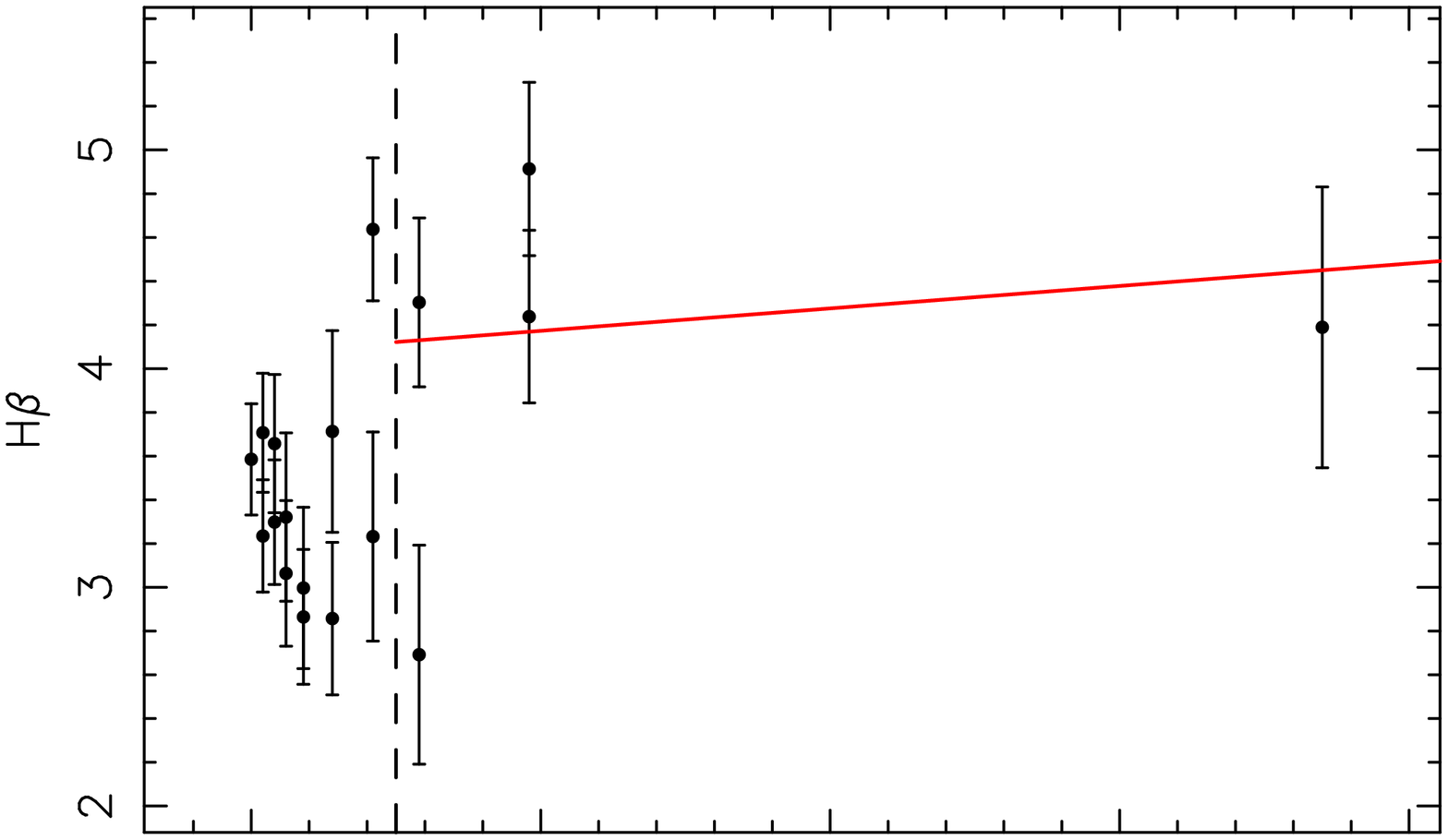}}
\resizebox{0.3\textwidth}{!}{\includegraphics[angle=-90]{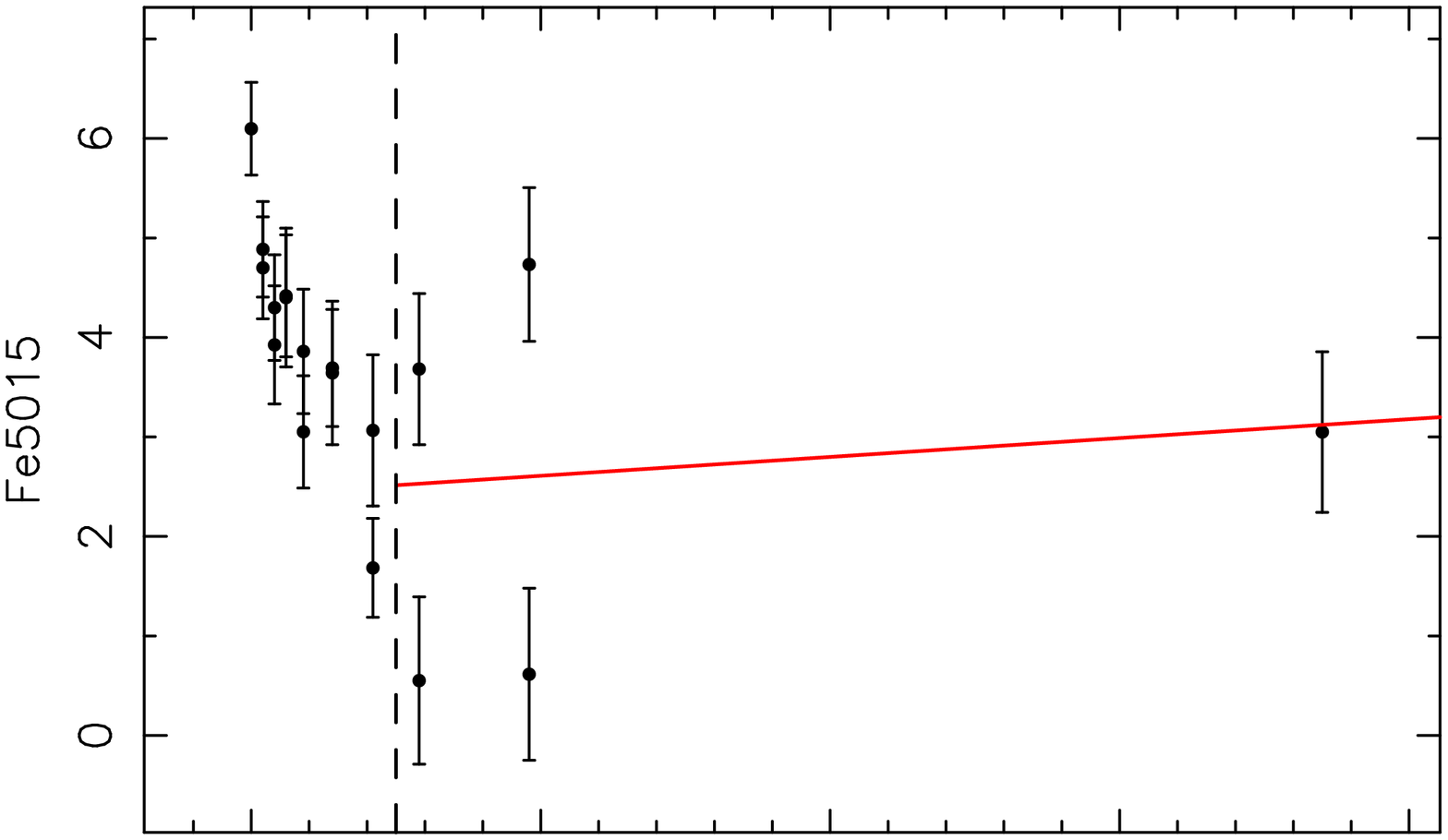}}
\resizebox{0.3\textwidth}{!}{\includegraphics[angle=-90]{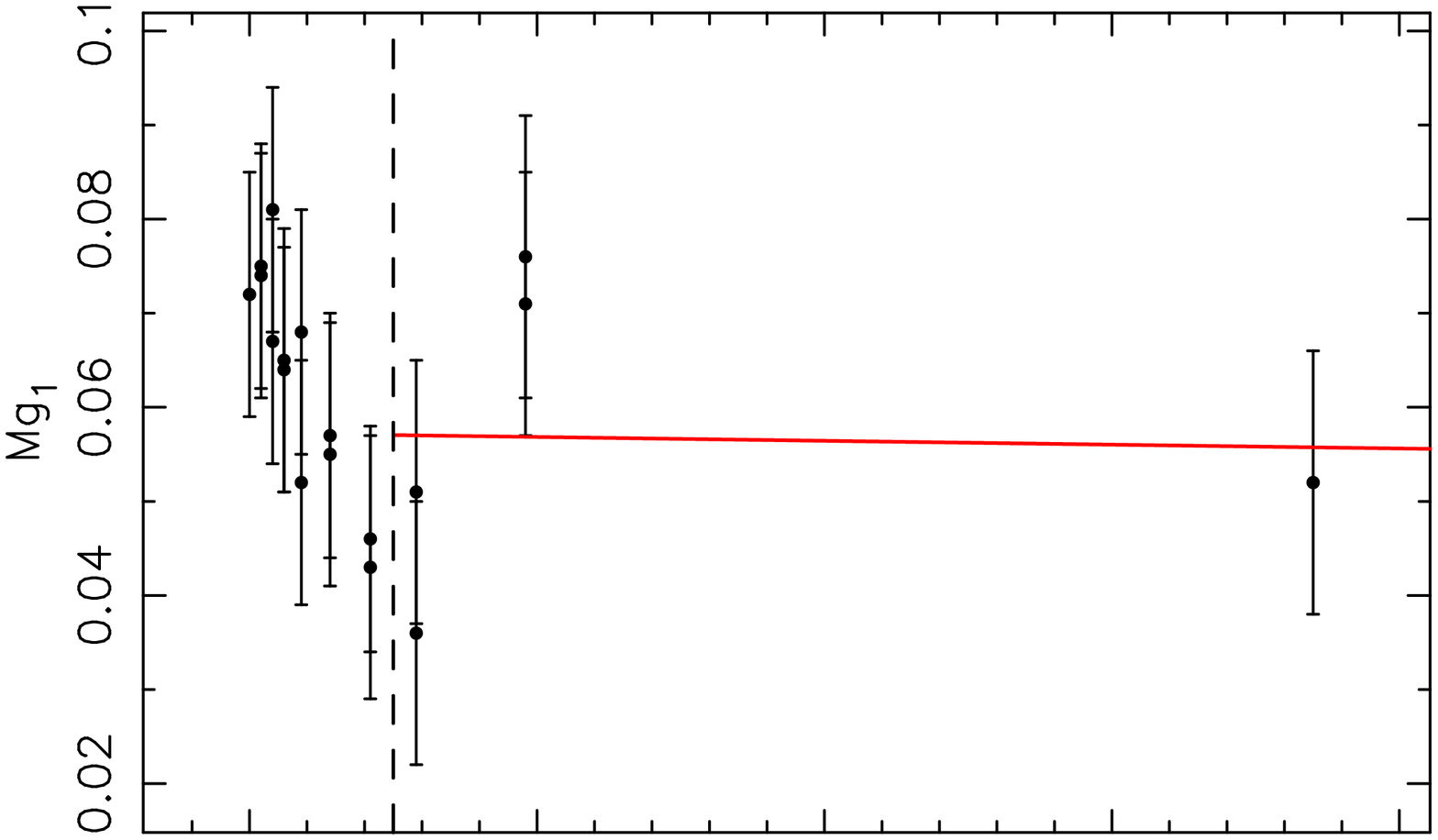}}
\resizebox{0.3\textwidth}{!}{\includegraphics[angle=-90]{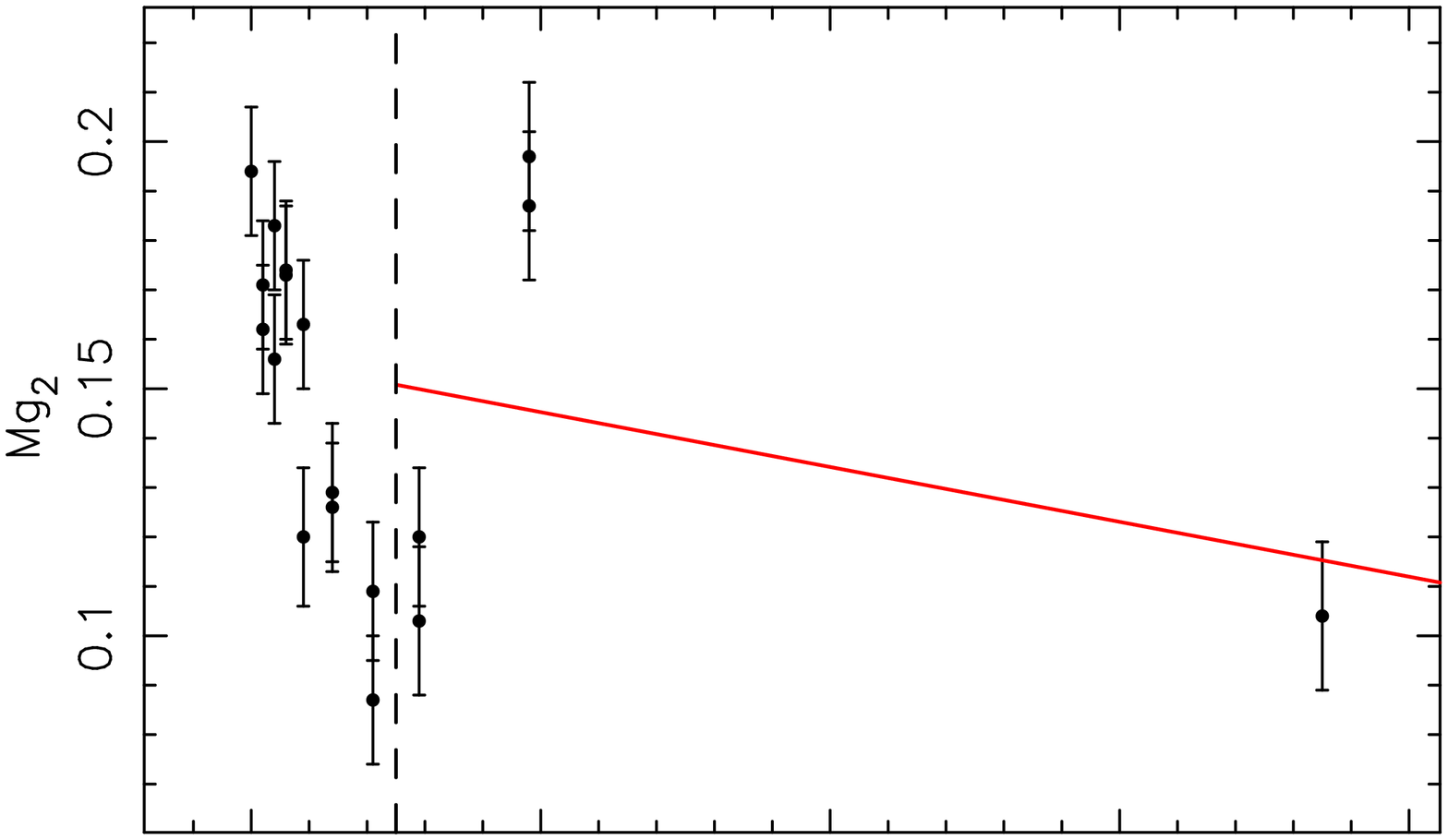}}
\resizebox{0.3\textwidth}{!}{\includegraphics[angle=-90]{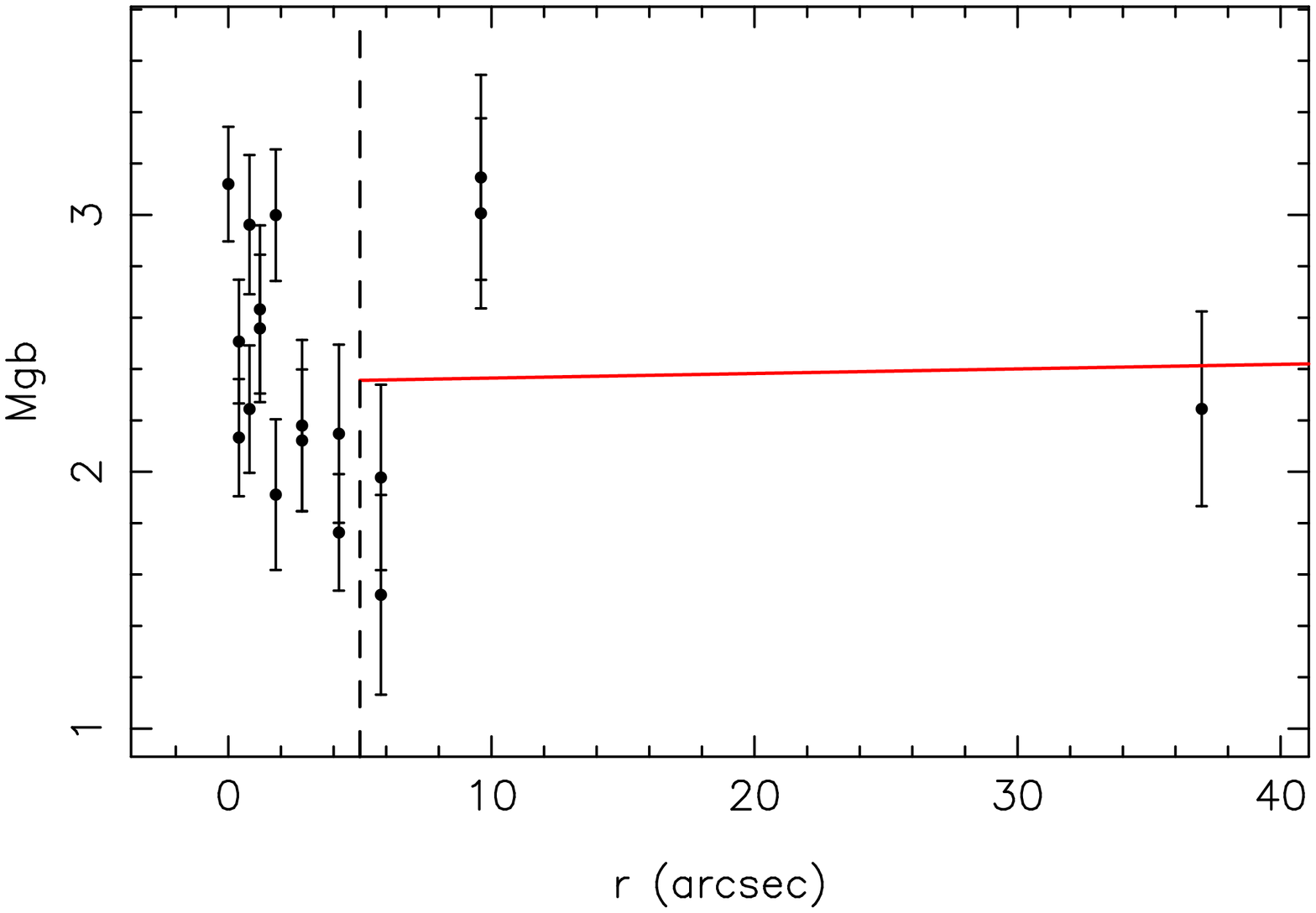}}\hspace{0.85cm}
\resizebox{0.3\textwidth}{!}{\includegraphics[angle=-90]{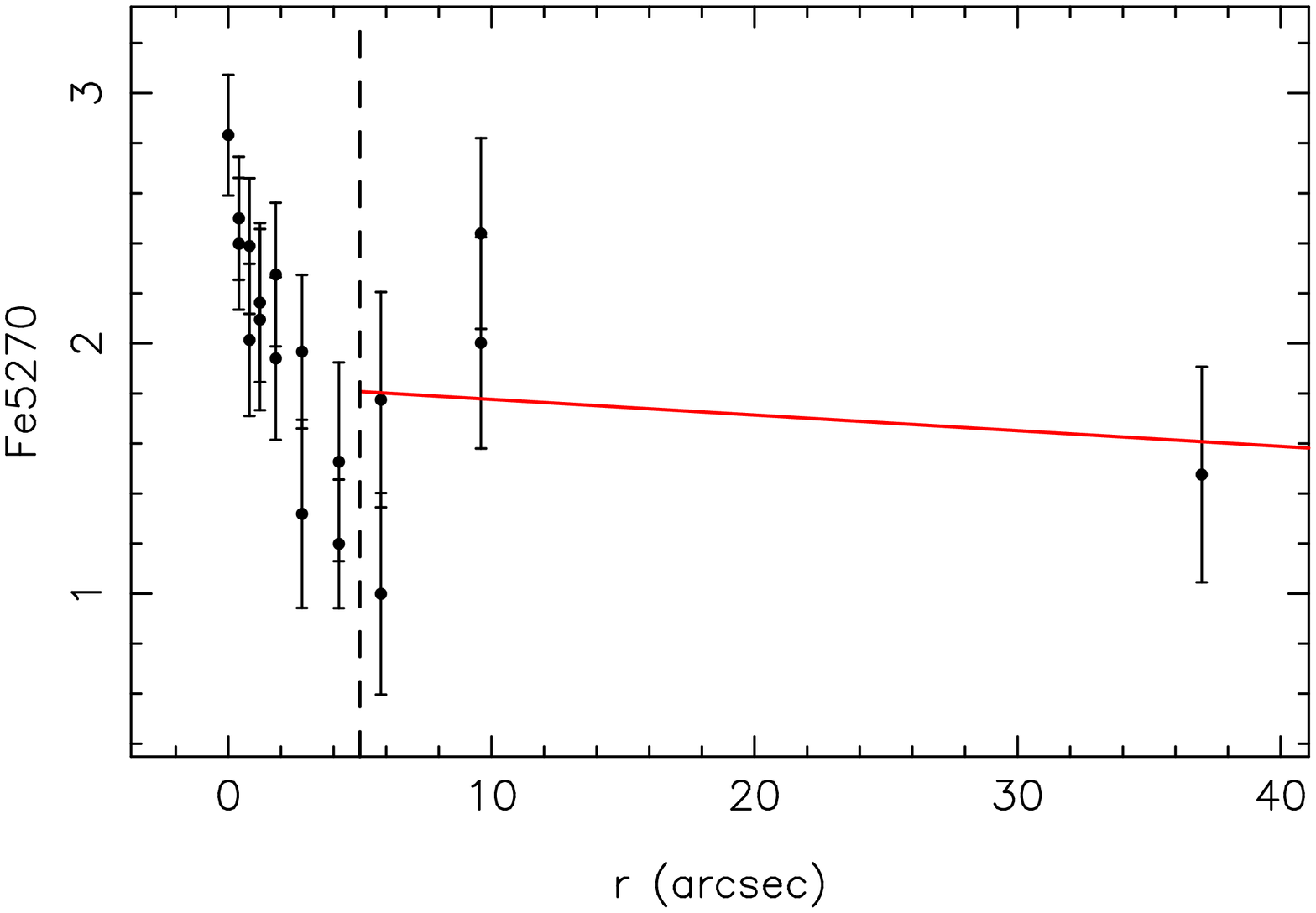}}\hspace{0.85cm}
\resizebox{0.3\textwidth}{!}{\includegraphics[angle=-90]{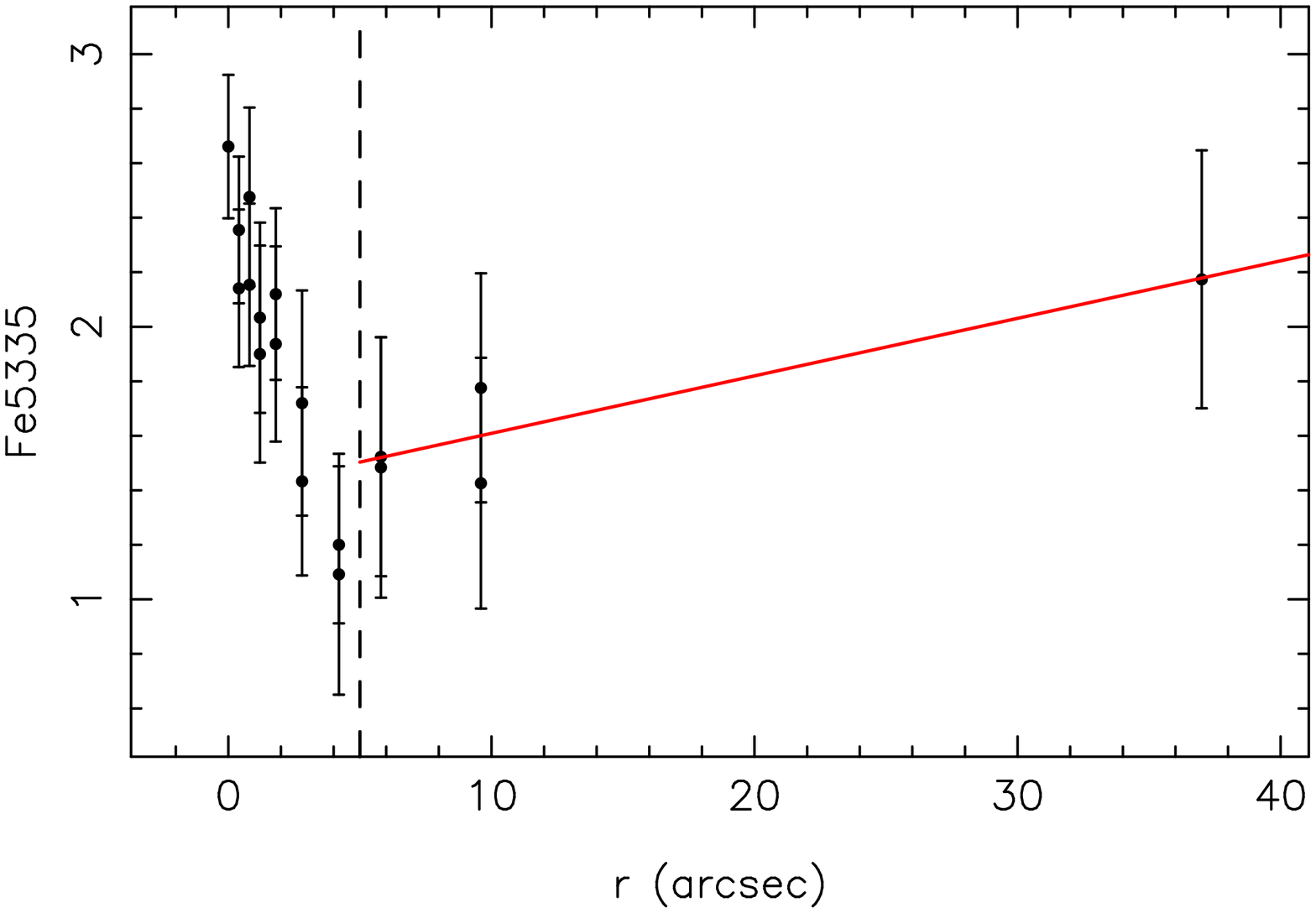}}
\caption{Line-strength distribution in the bar region for all the galaxies}
\end{figure*}
\begin{figure*}
\addtocounter{figure}{-1}
\resizebox{0.3\textwidth}{!}{\includegraphics[angle=-90]{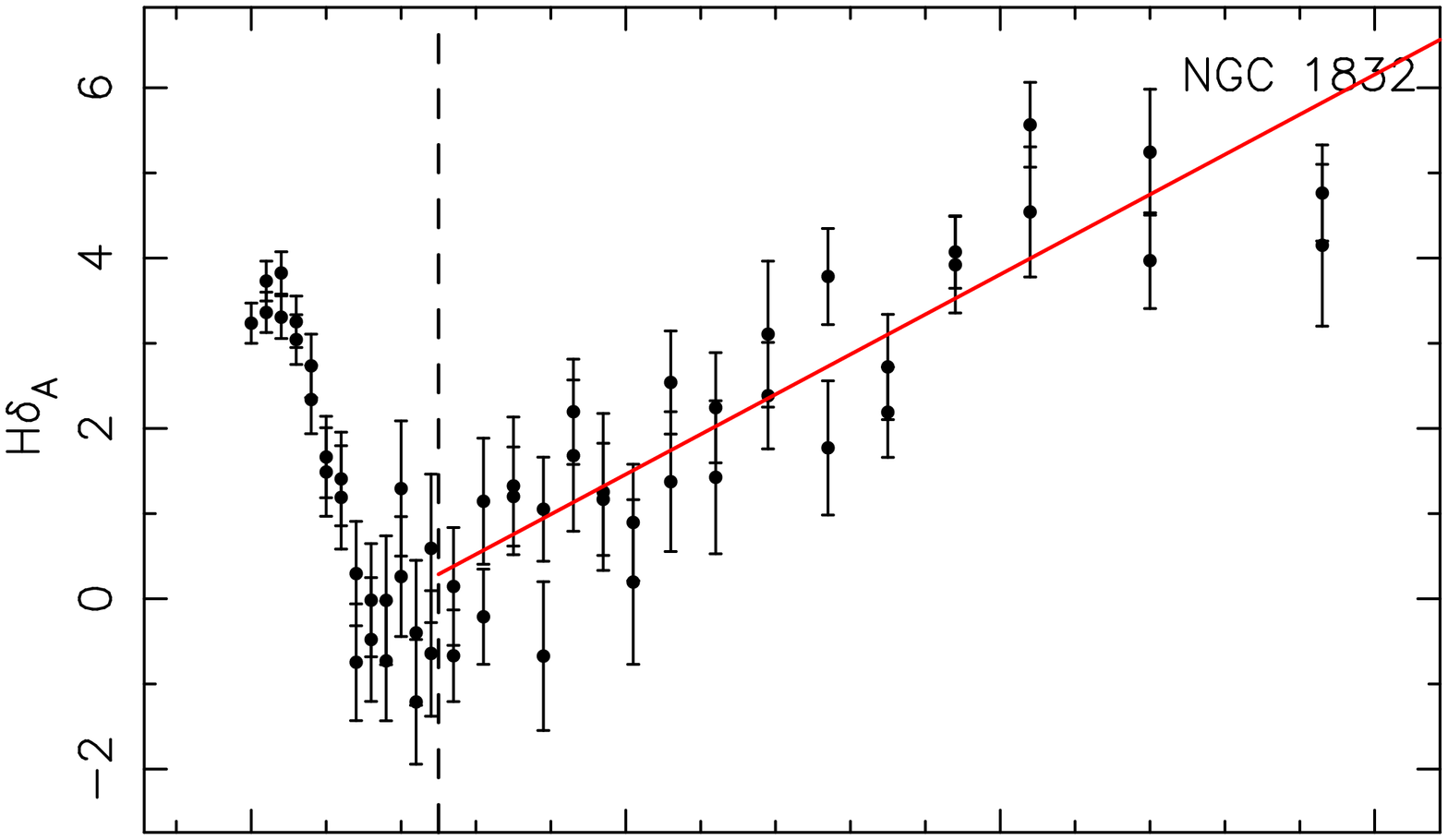}}
\resizebox{0.3\textwidth}{!}{\includegraphics[angle=-90]{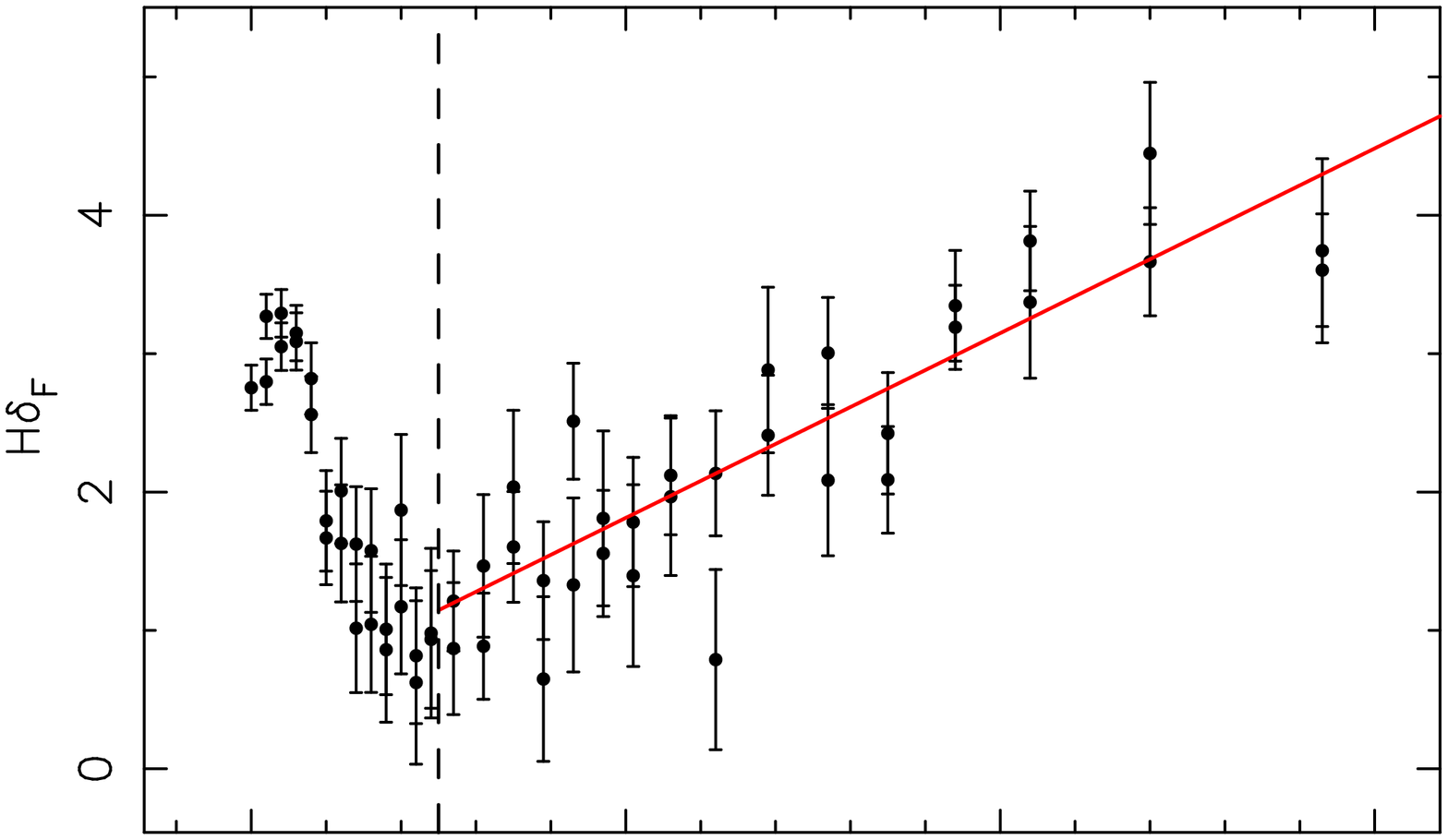}}
\resizebox{0.3\textwidth}{!}{\includegraphics[angle=-90]{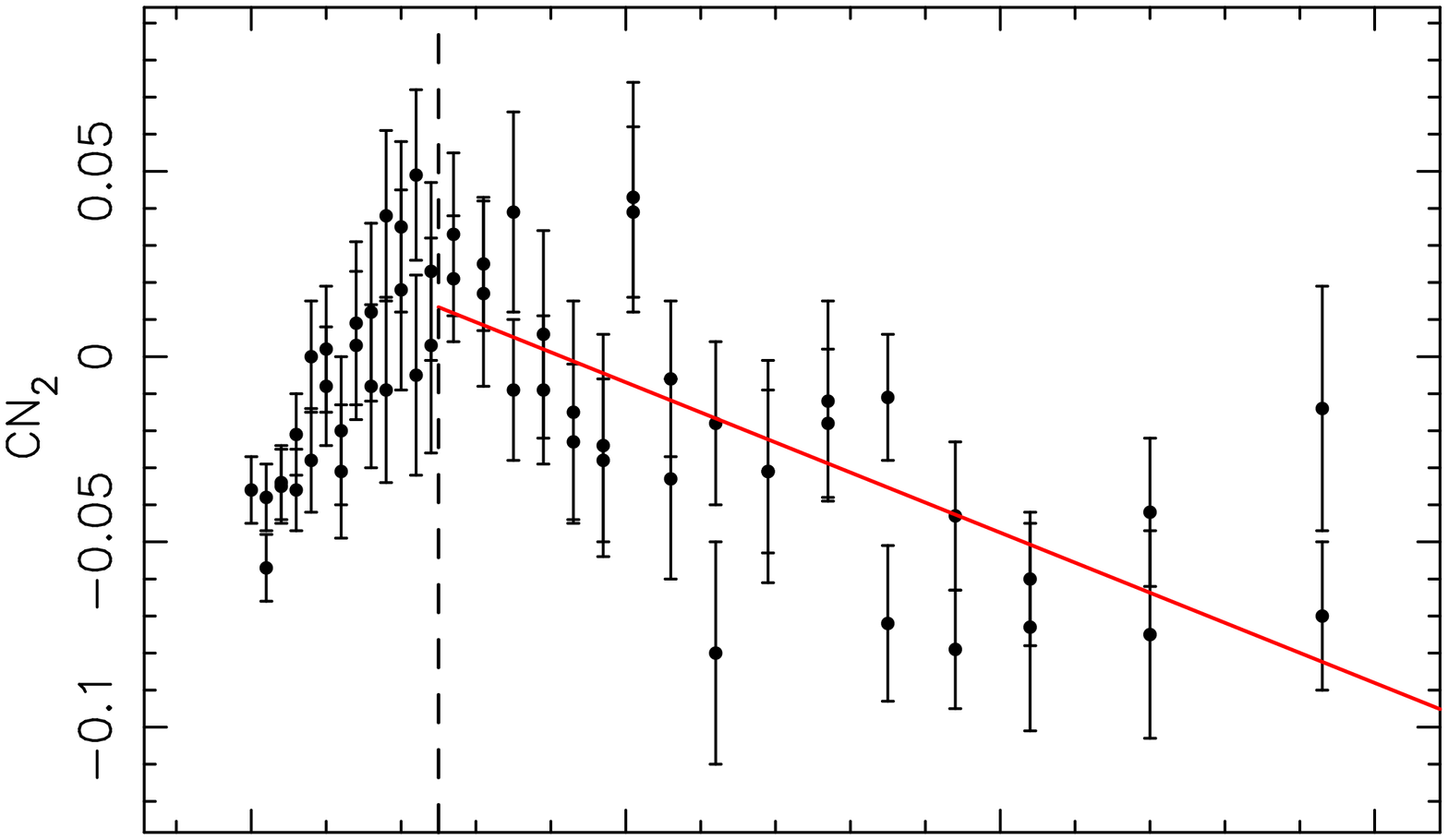}}
\resizebox{0.3\textwidth}{!}{\includegraphics[angle=-90]{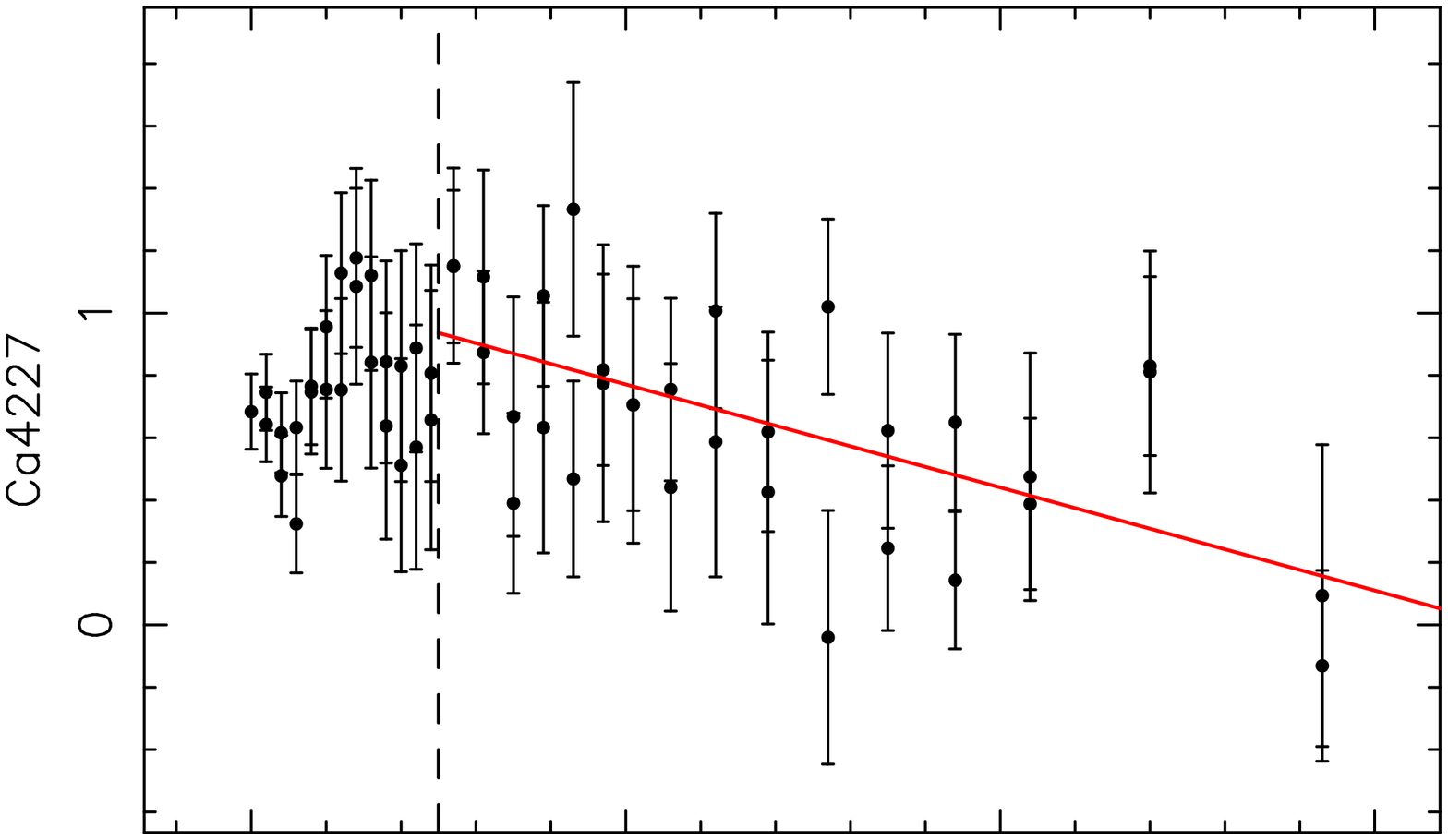}}
\resizebox{0.3\textwidth}{!}{\includegraphics[angle=-90]{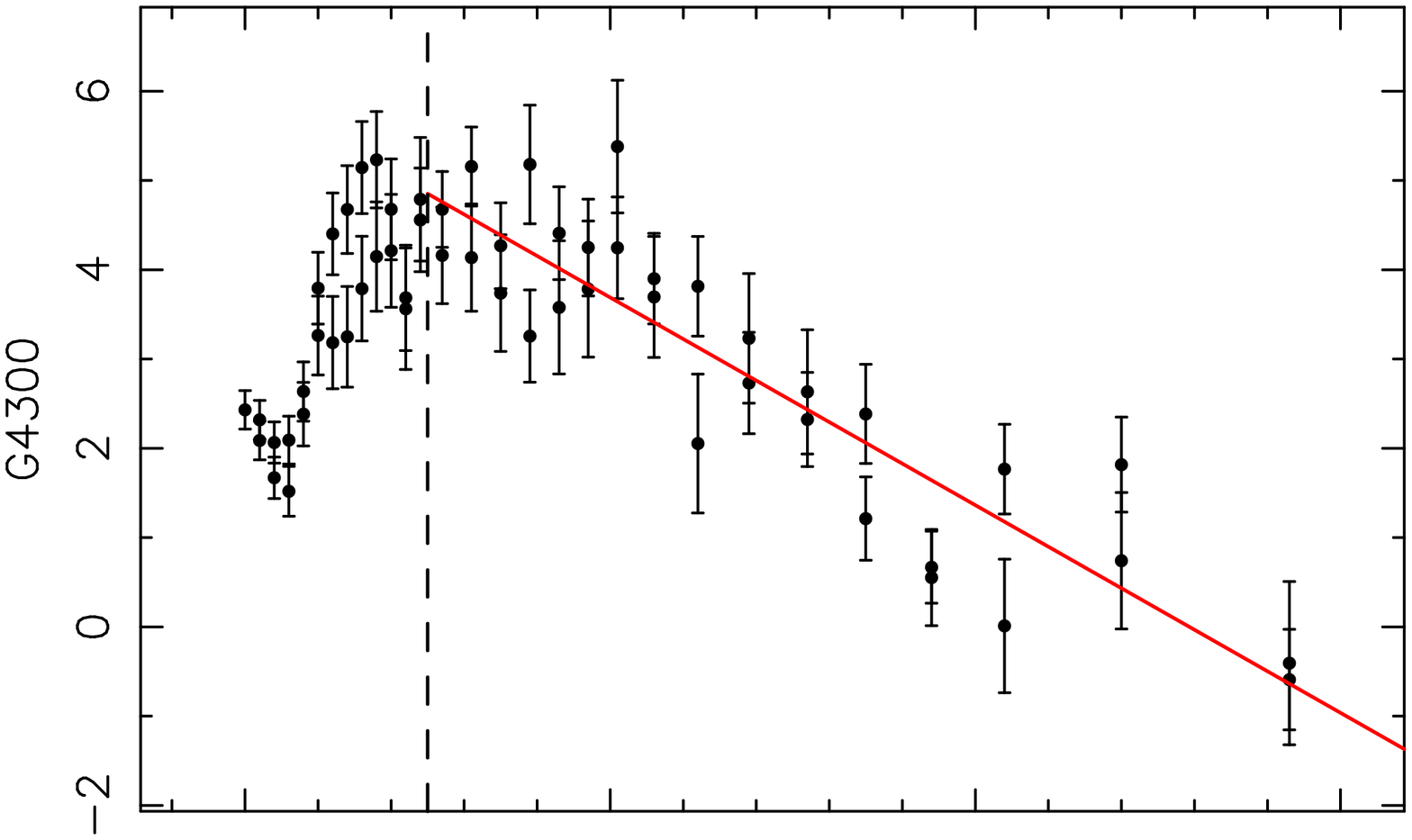}}
\resizebox{0.3\textwidth}{!}{\includegraphics[angle=-90]{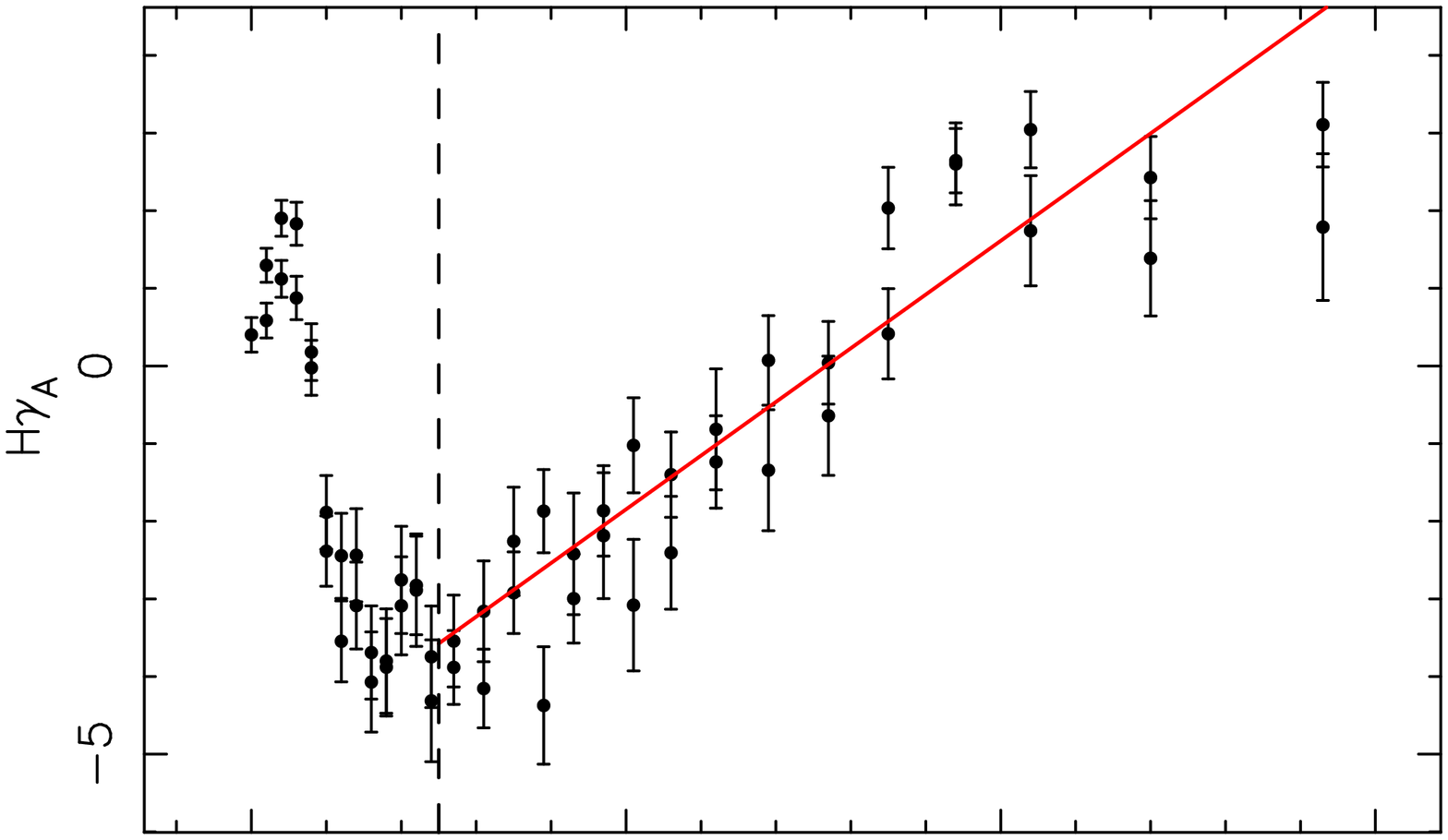}}
\resizebox{0.3\textwidth}{!}{\includegraphics[angle=-90]{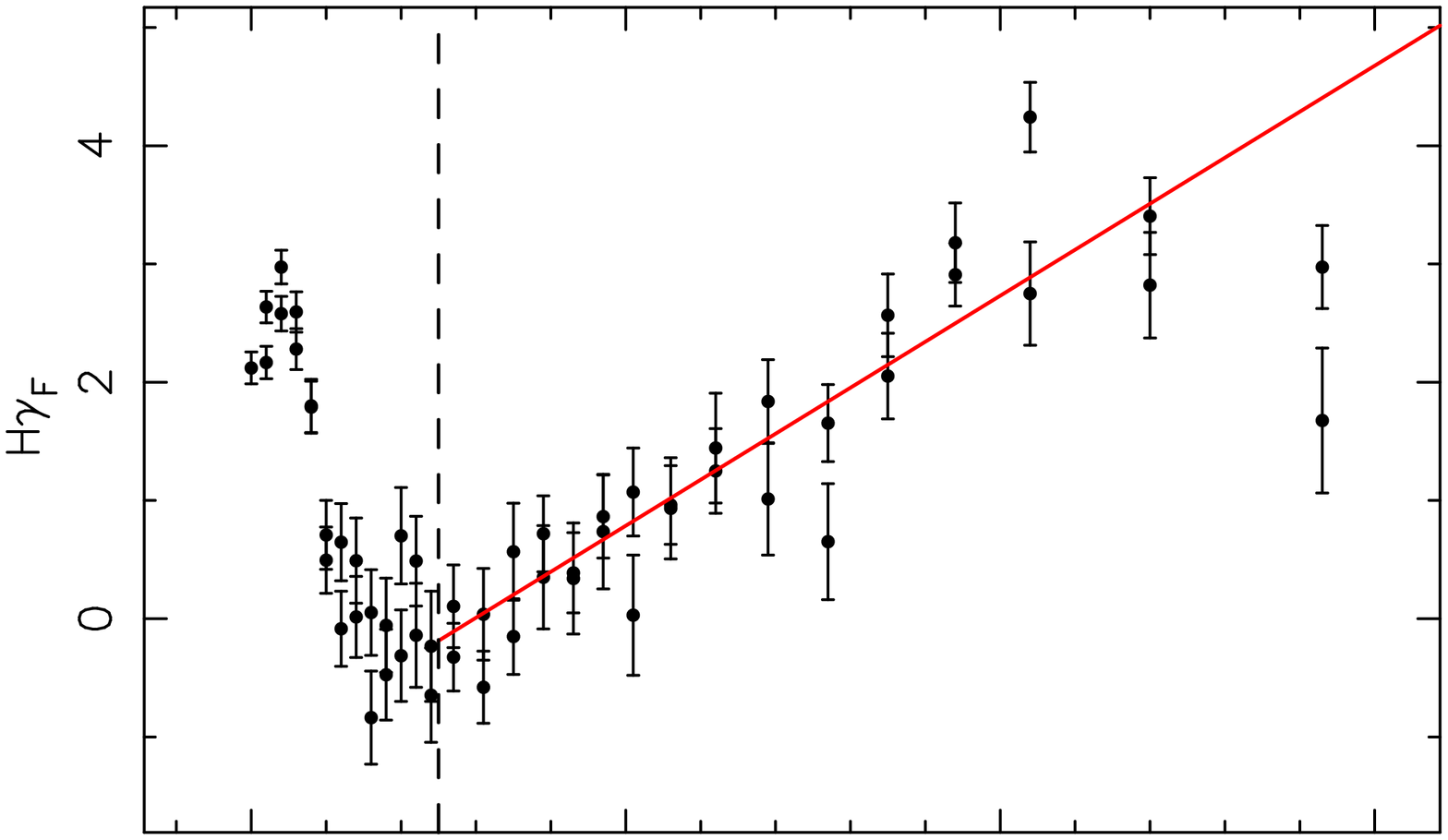}}
\resizebox{0.3\textwidth}{!}{\includegraphics[angle=-90]{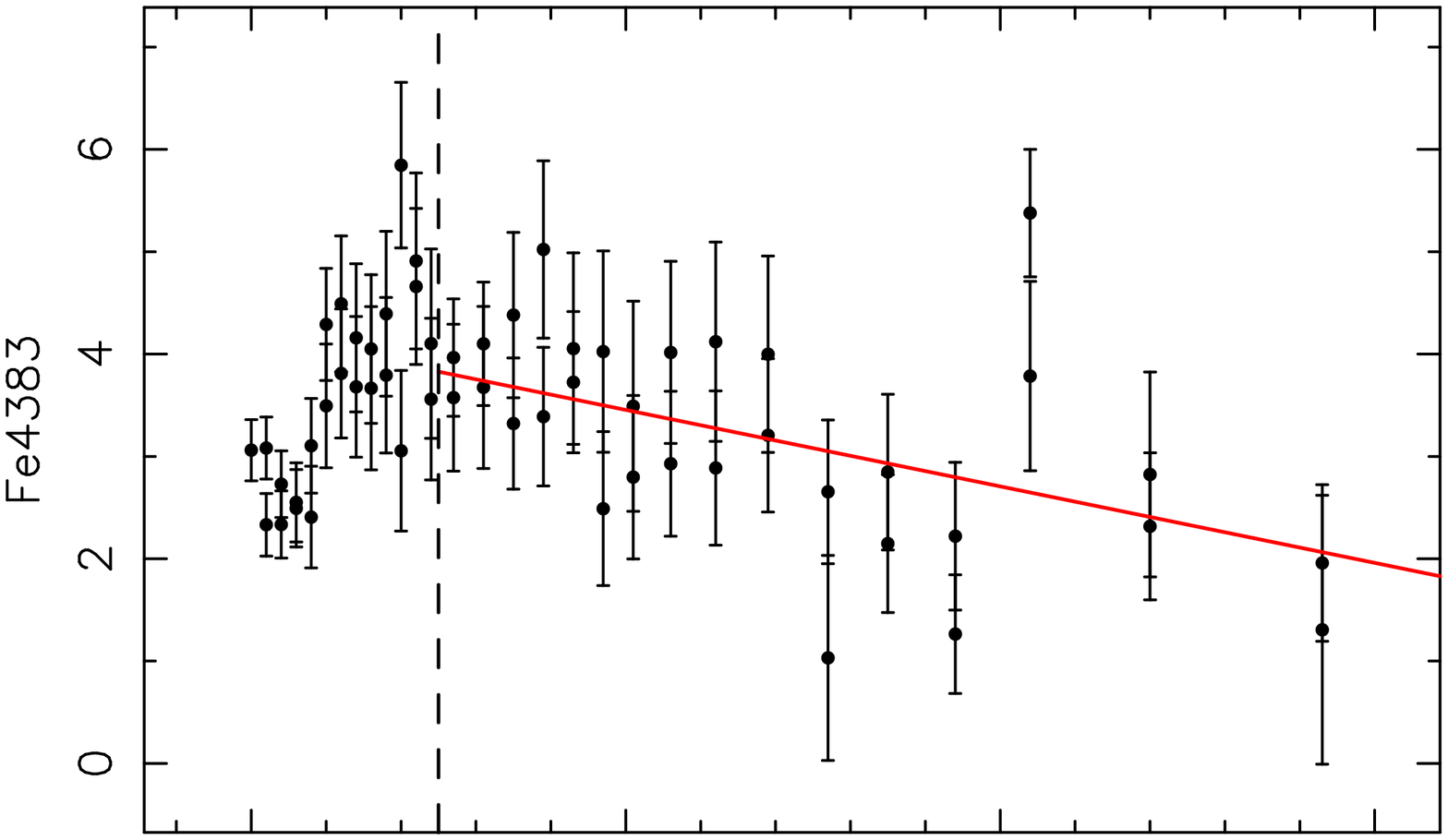}}
\resizebox{0.3\textwidth}{!}{\includegraphics[angle=-90]{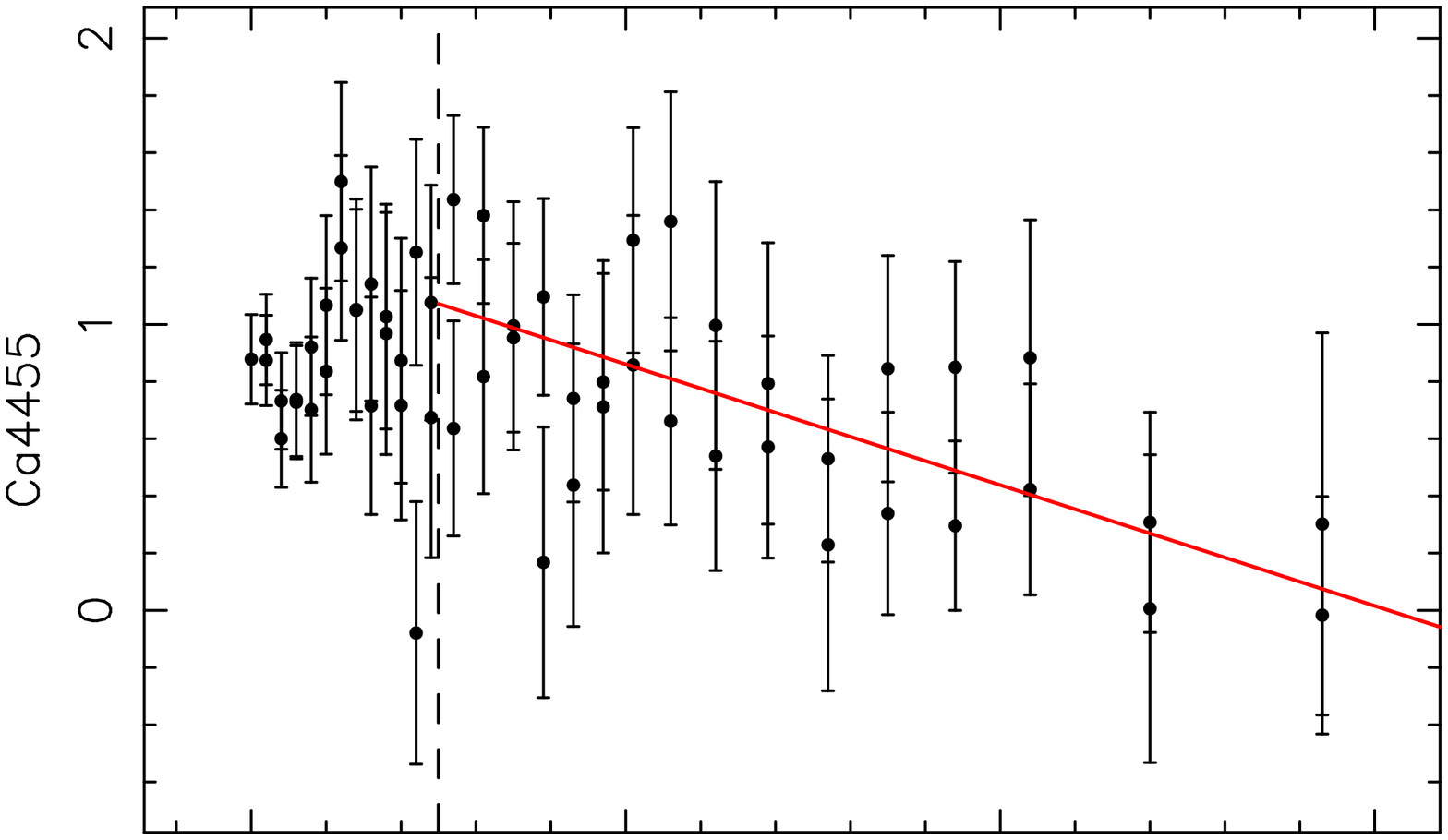}}
\resizebox{0.3\textwidth}{!}{\includegraphics[angle=-90]{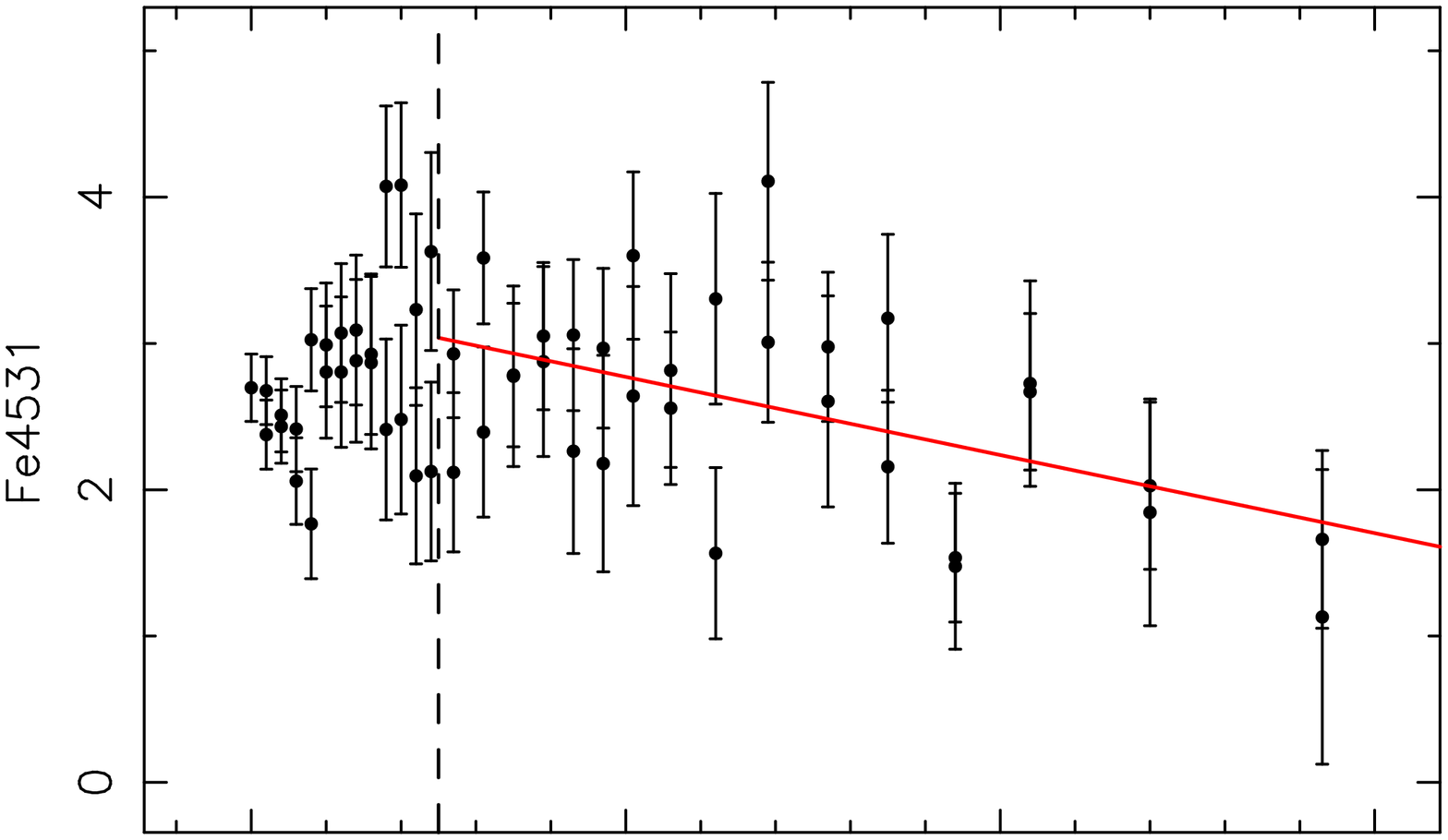}}
\resizebox{0.3\textwidth}{!}{\includegraphics[angle=-90]{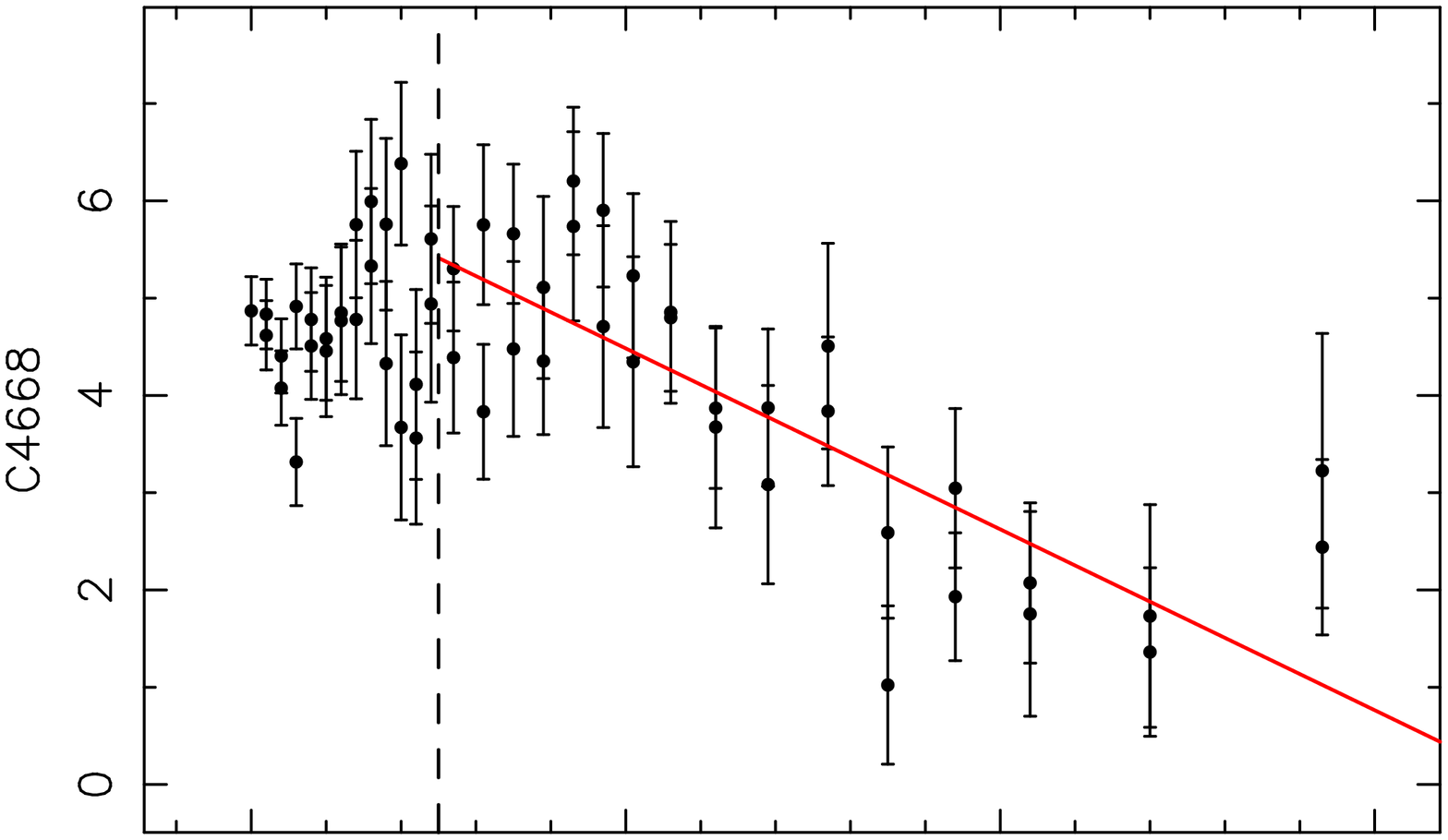}}
\resizebox{0.3\textwidth}{!}{\includegraphics[angle=-90]{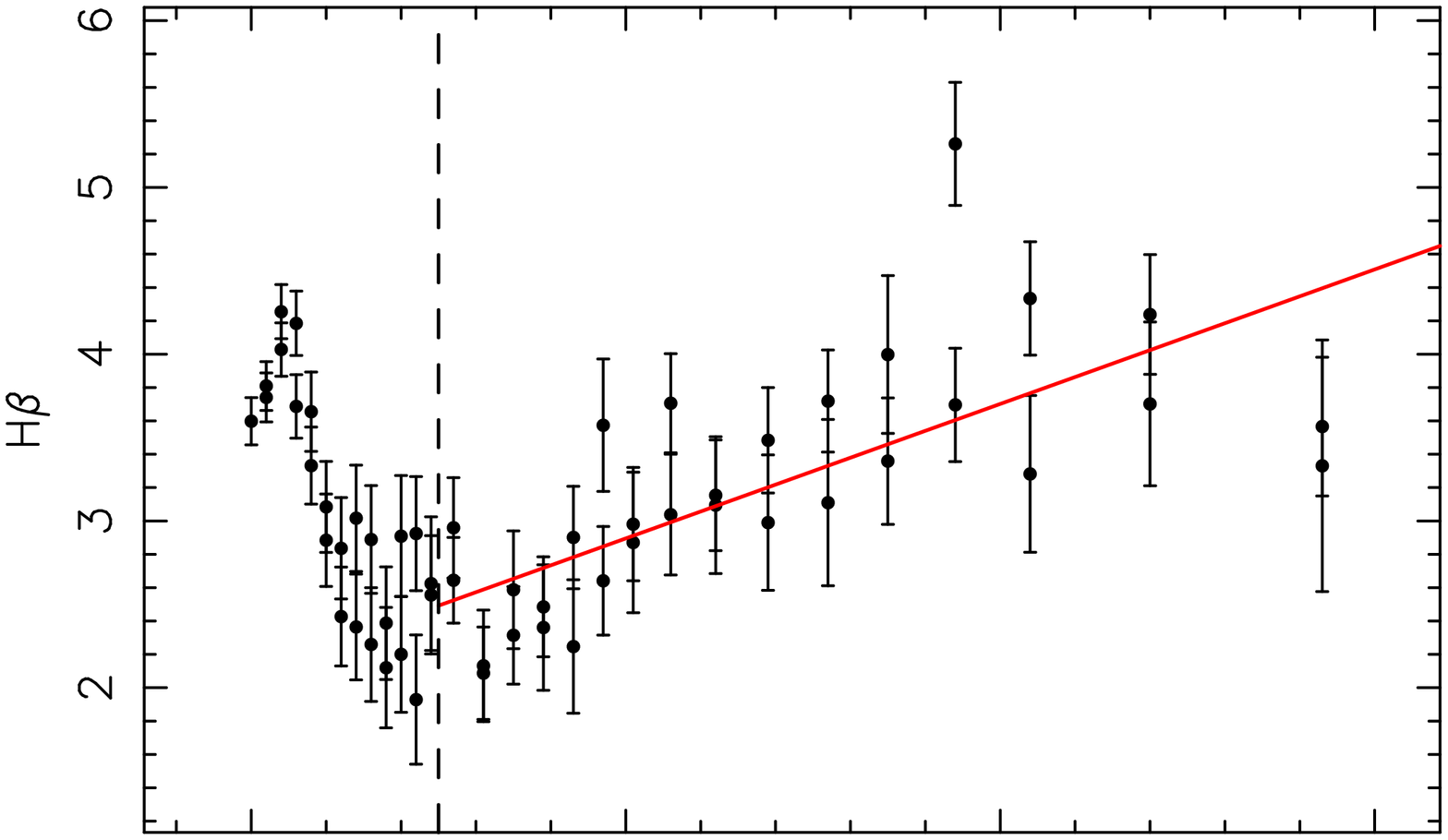}}
\resizebox{0.3\textwidth}{!}{\includegraphics[angle=-90]{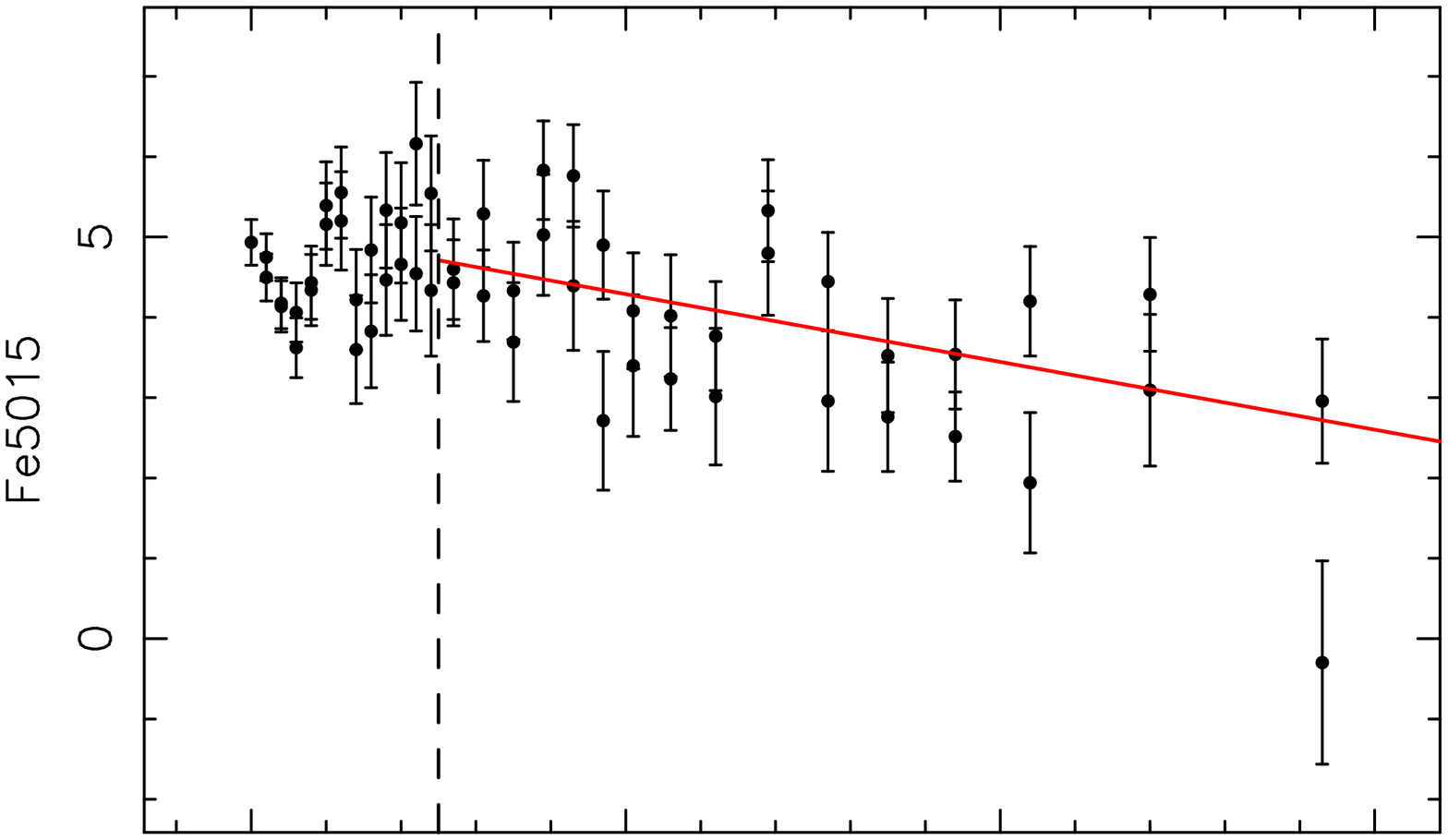}}
\resizebox{0.3\textwidth}{!}{\includegraphics[angle=-90]{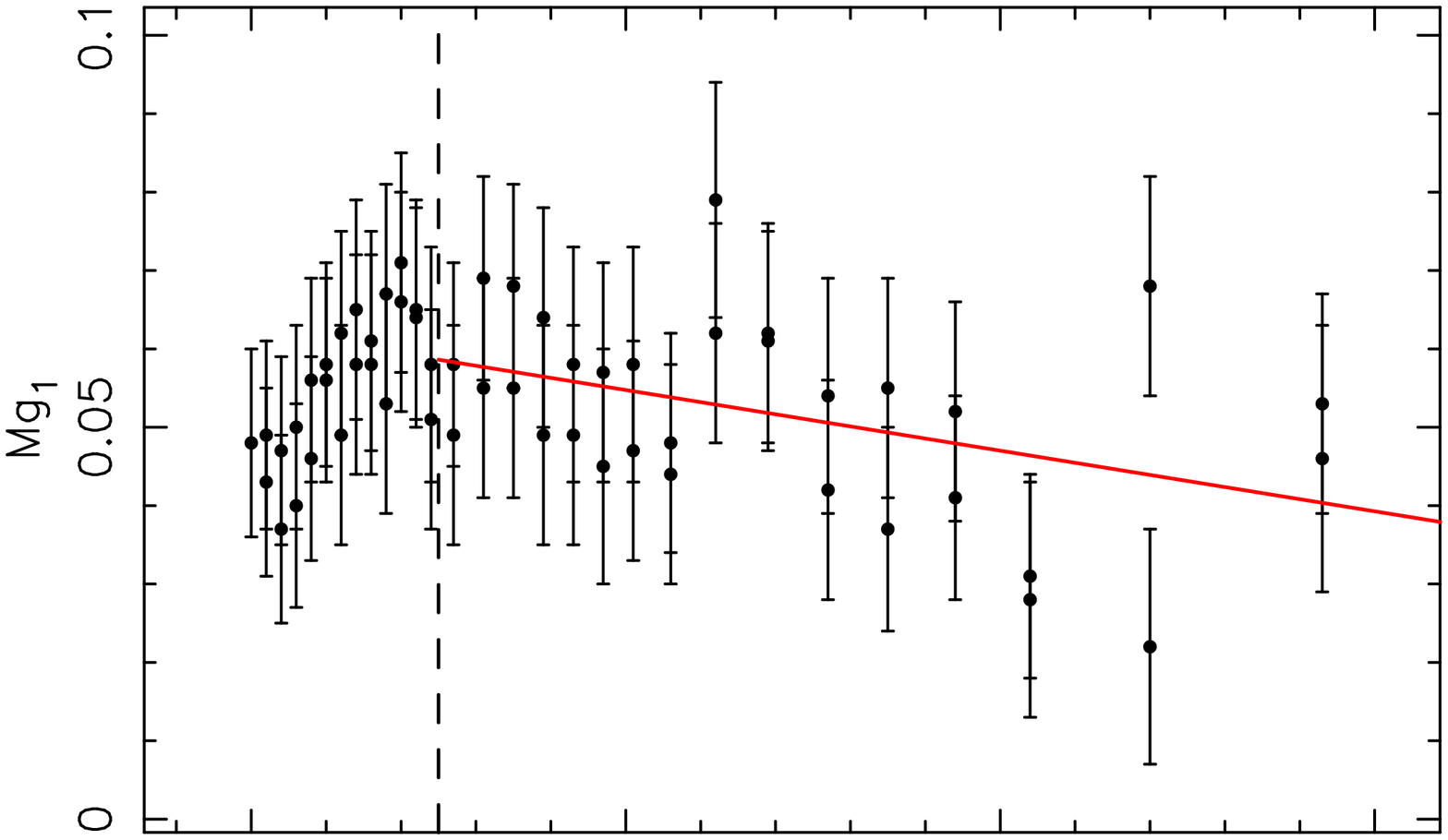}}
\resizebox{0.3\textwidth}{!}{\includegraphics[angle=-90]{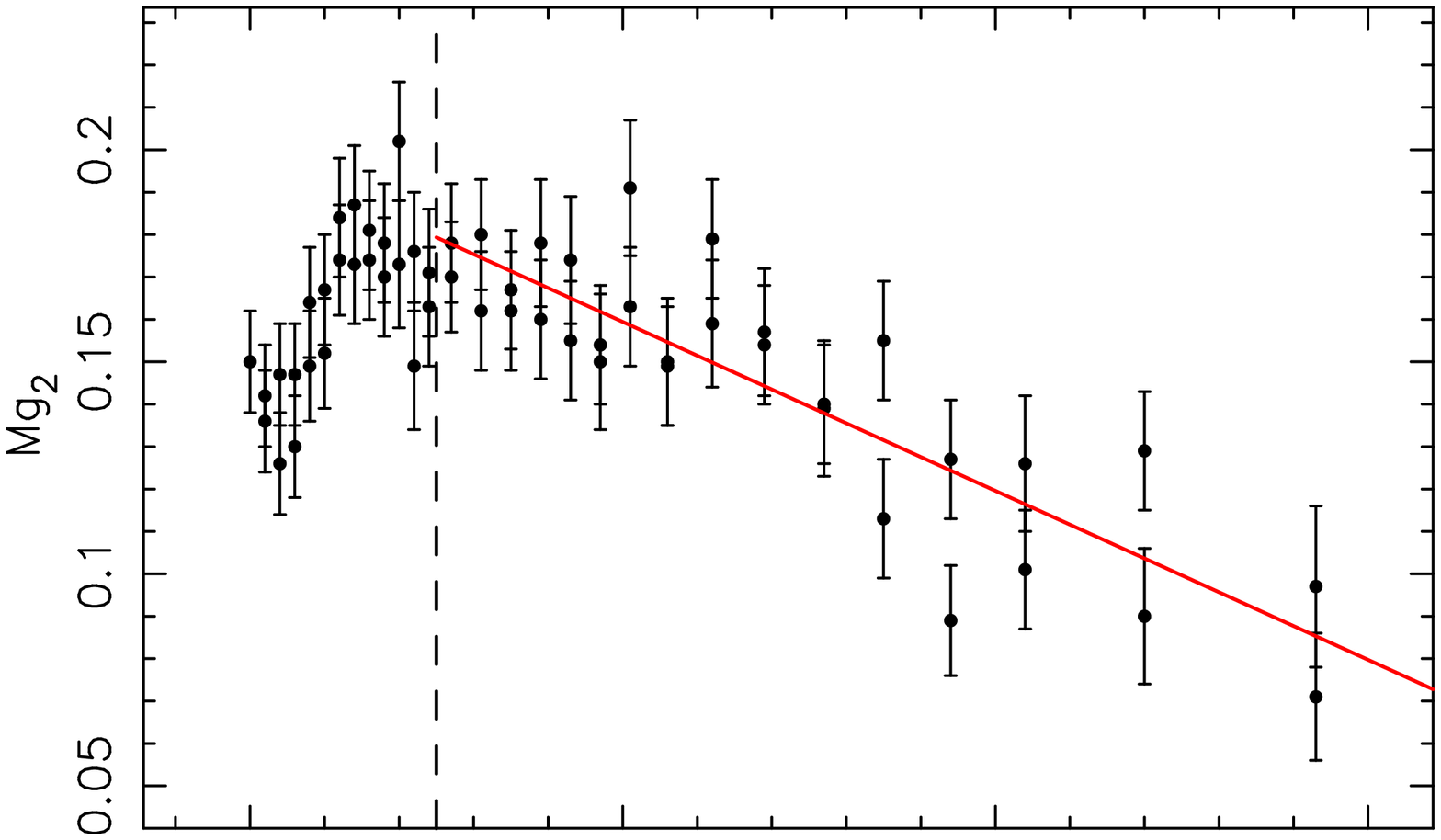}}
\resizebox{0.3\textwidth}{!}{\includegraphics[angle=-90]{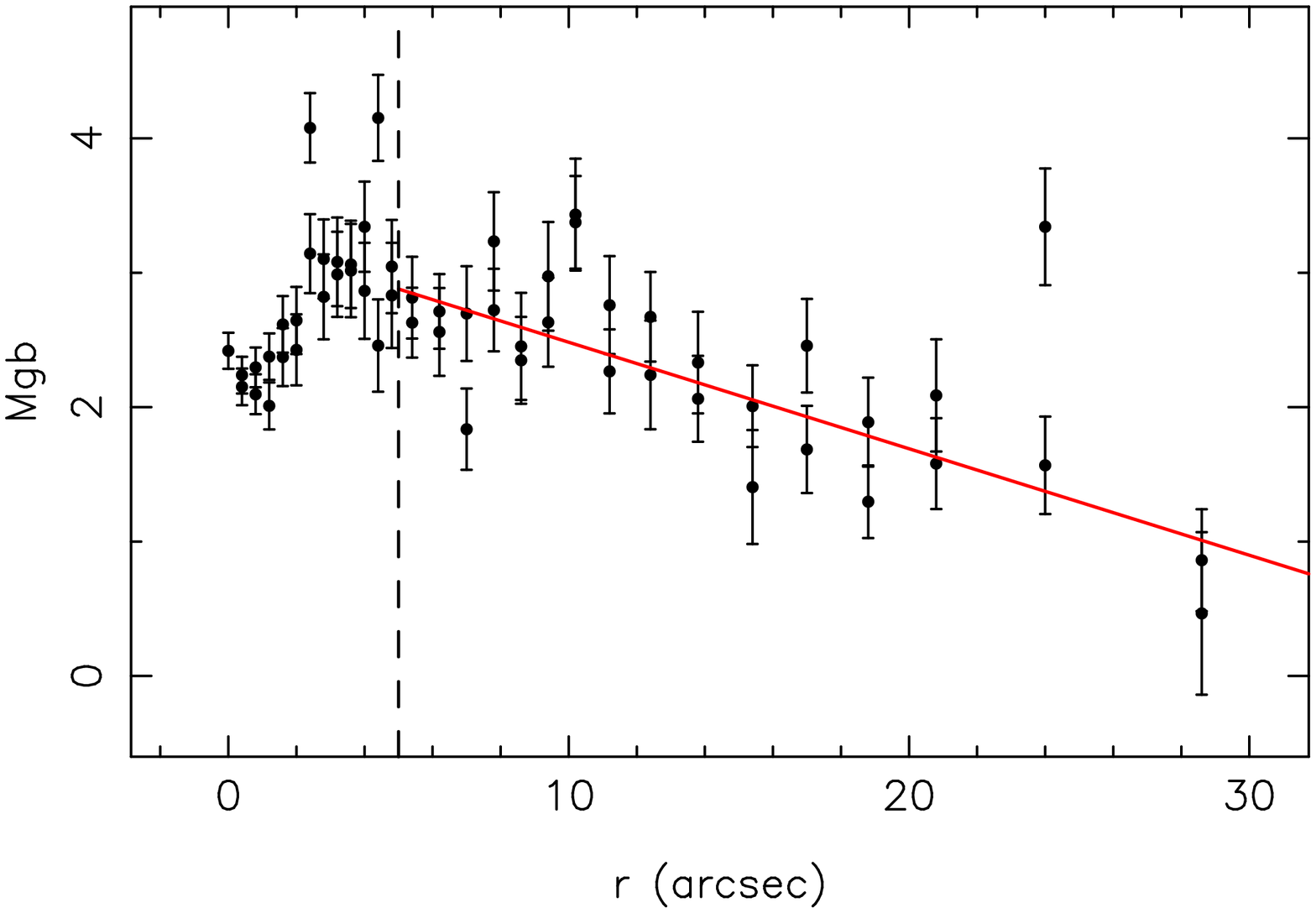}}\hspace{0.85cm}
\resizebox{0.3\textwidth}{!}{\includegraphics[angle=-90]{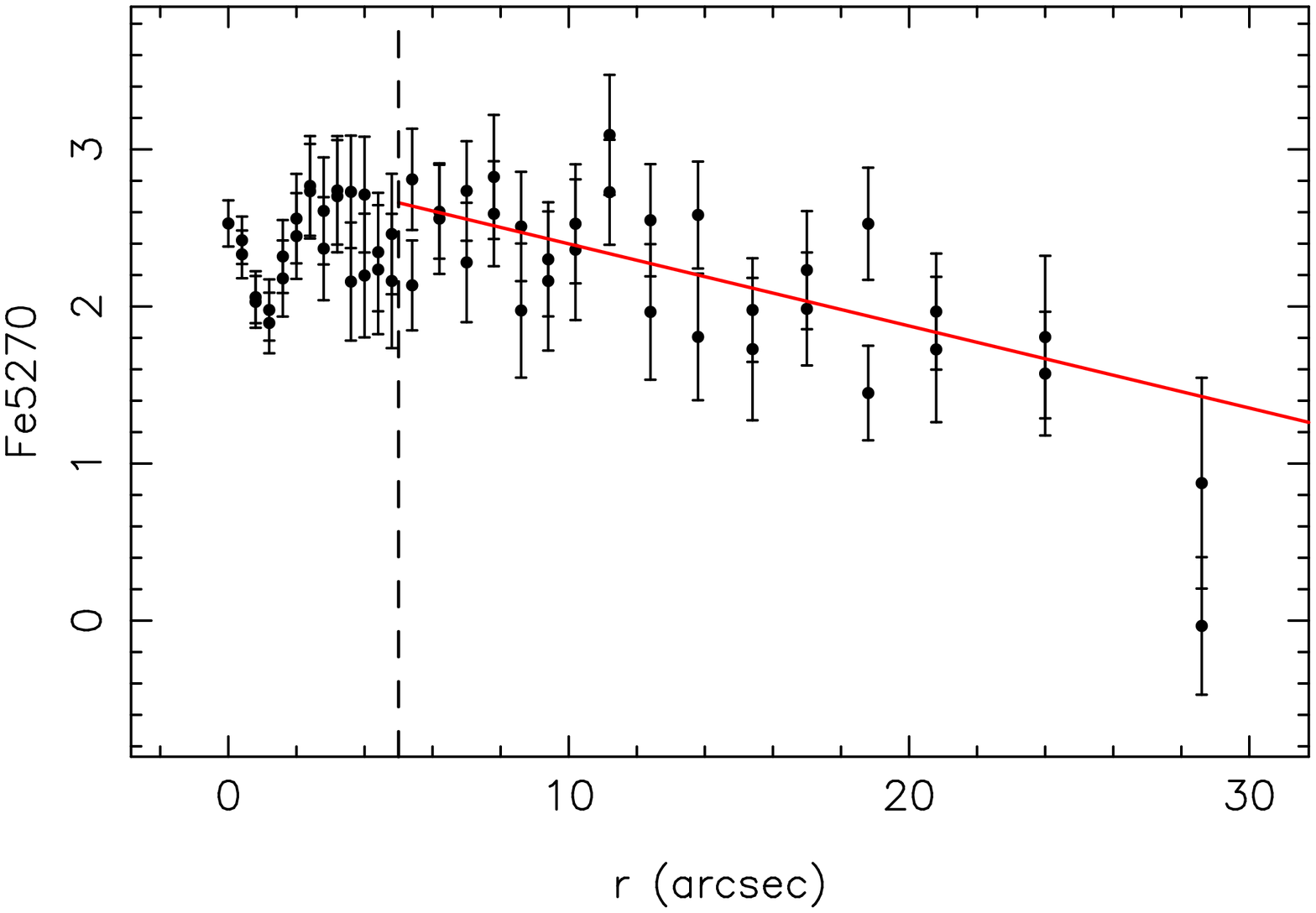}}\hspace{0.85cm}
\resizebox{0.3\textwidth}{!}{\includegraphics[angle=-90]{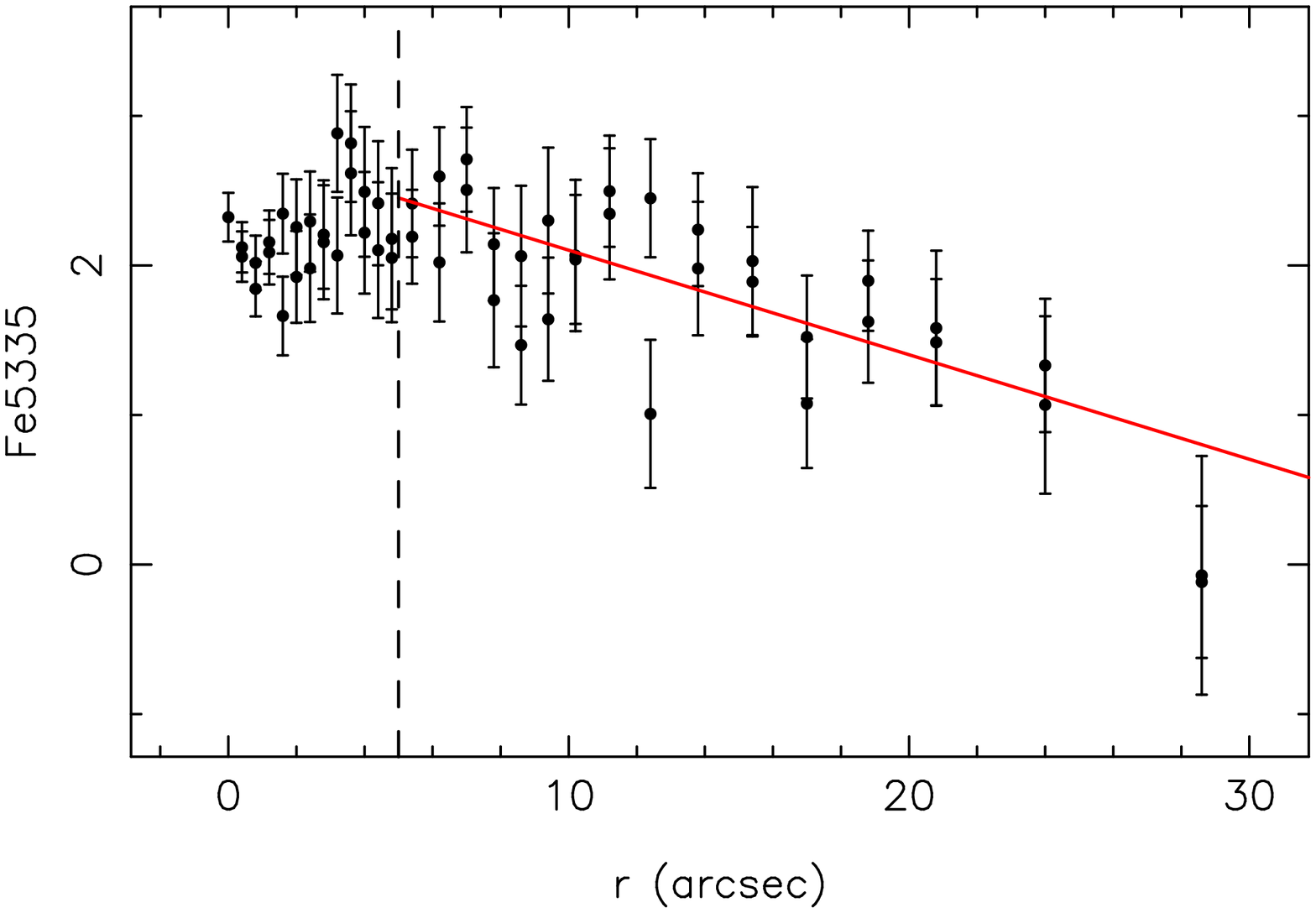}}
\caption{Line-strength distribution in the bar region for all the galaxies}
\end{figure*}

\clearpage
\begin{figure*}
\addtocounter{figure}{-1}
\resizebox{0.3\textwidth}{!}{\includegraphics[angle=-90]{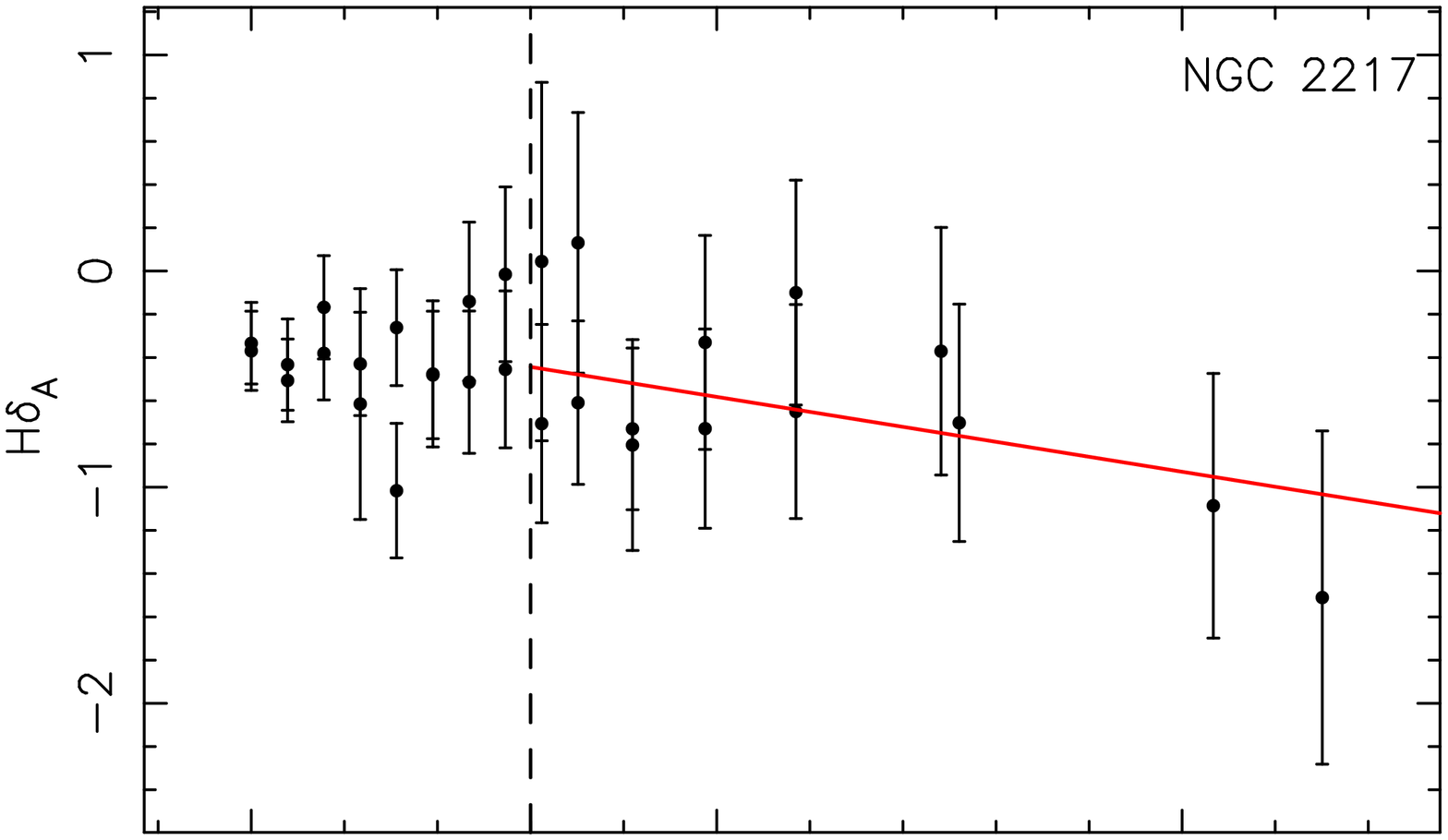}}
\resizebox{0.3\textwidth}{!}{\includegraphics[angle=-90]{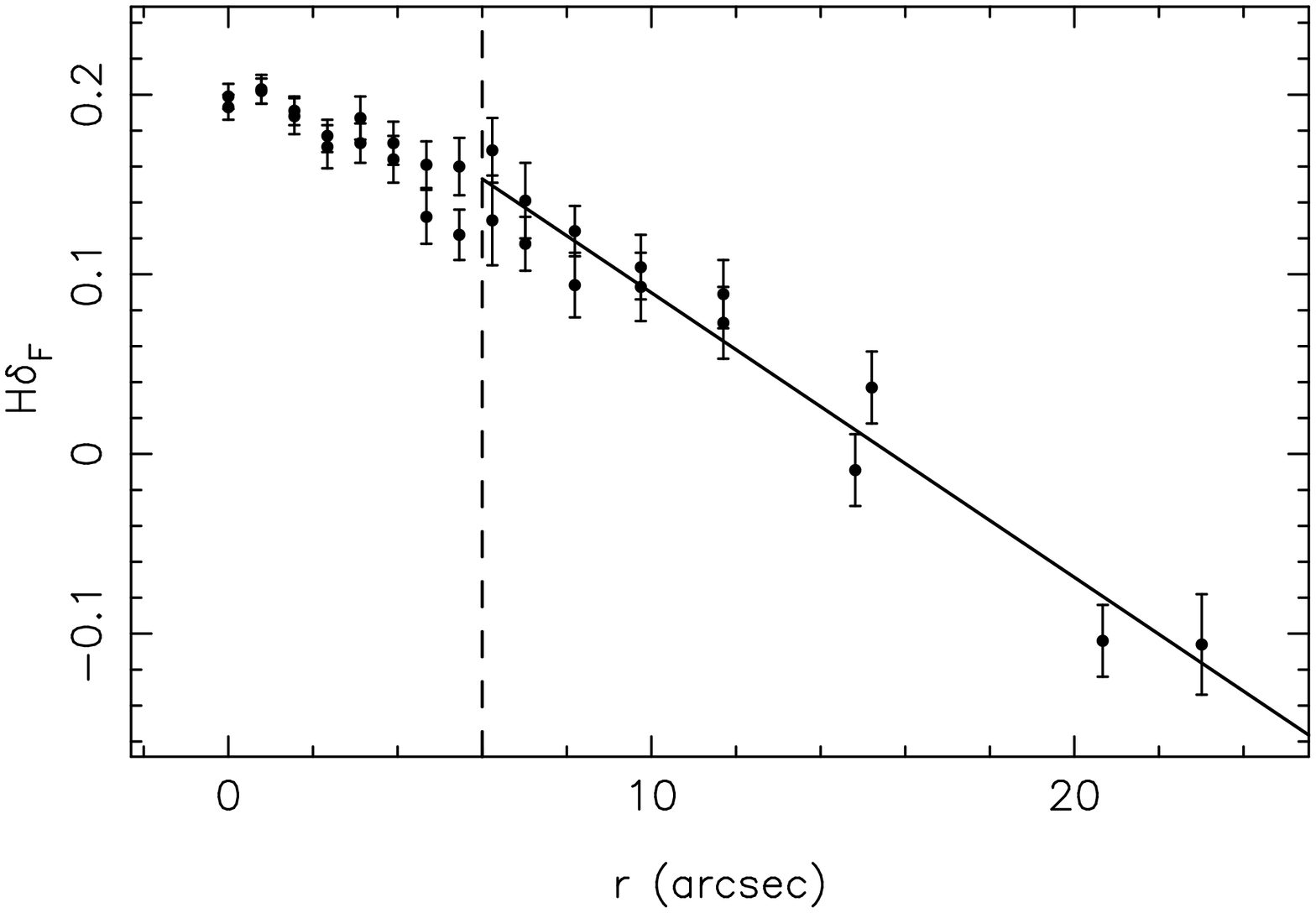}}
\resizebox{0.3\textwidth}{!}{\includegraphics[angle=-90]{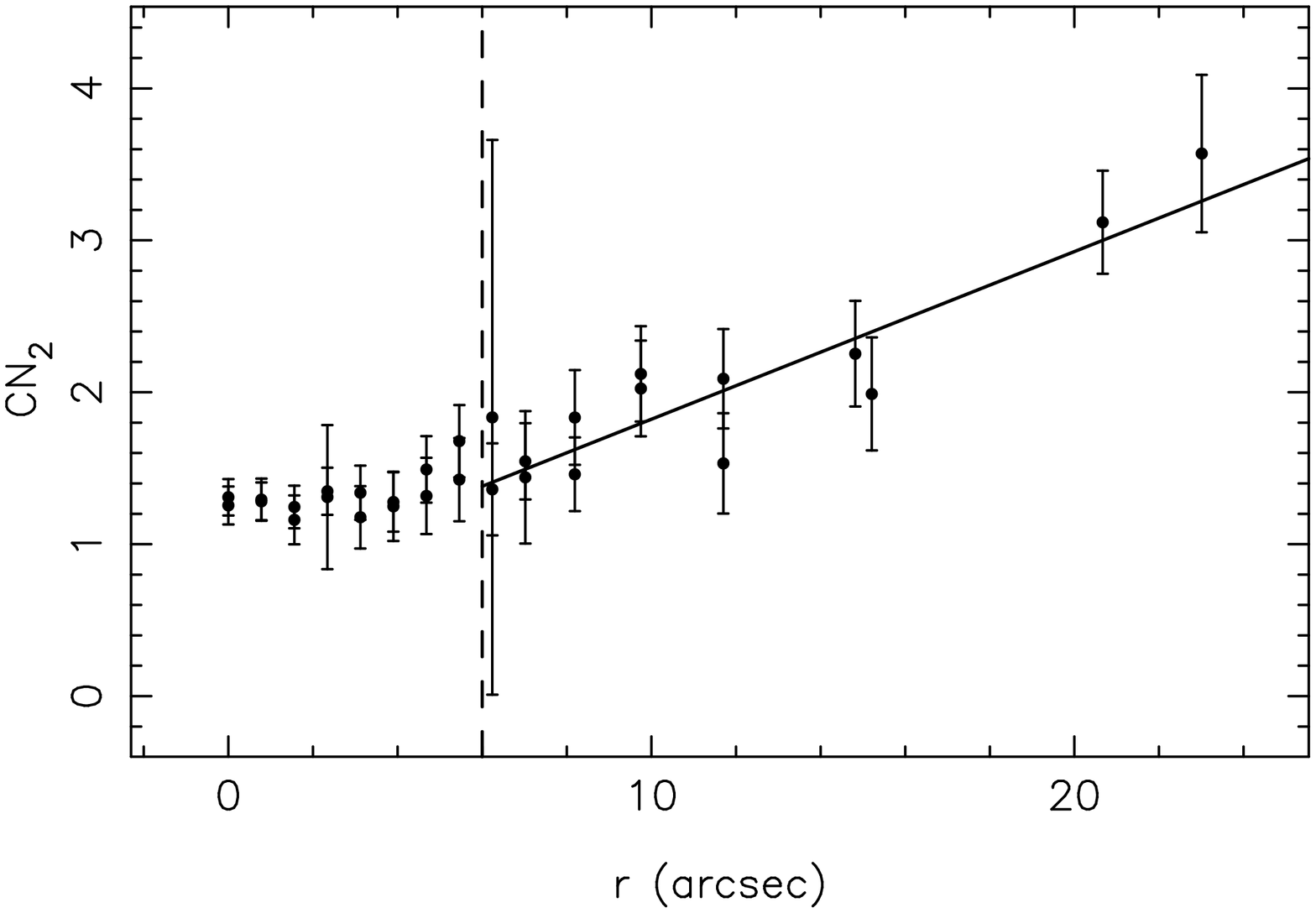}}
\resizebox{0.3\textwidth}{!}{\includegraphics[angle=-90]{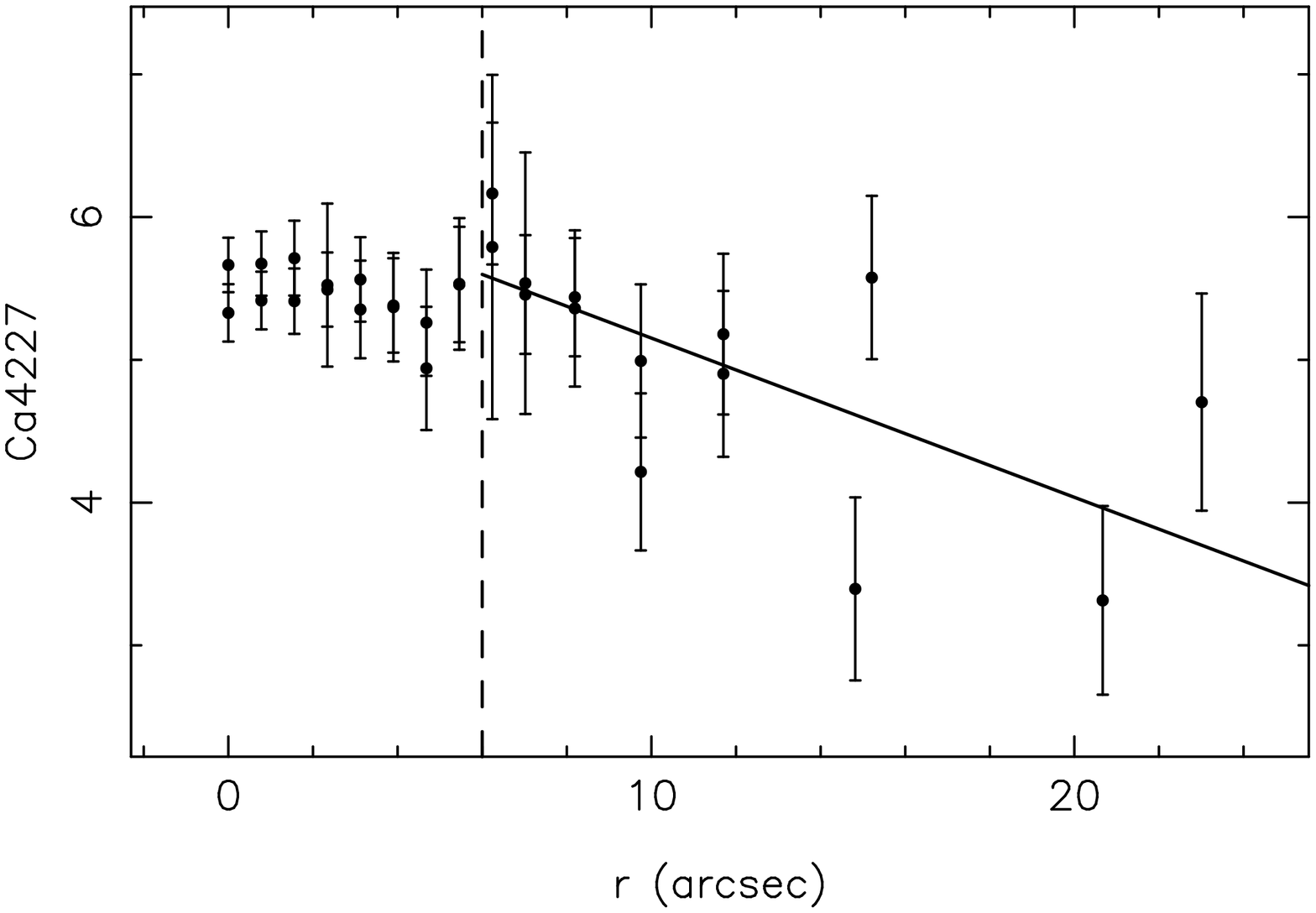}}
\resizebox{0.3\textwidth}{!}{\includegraphics[angle=-90]{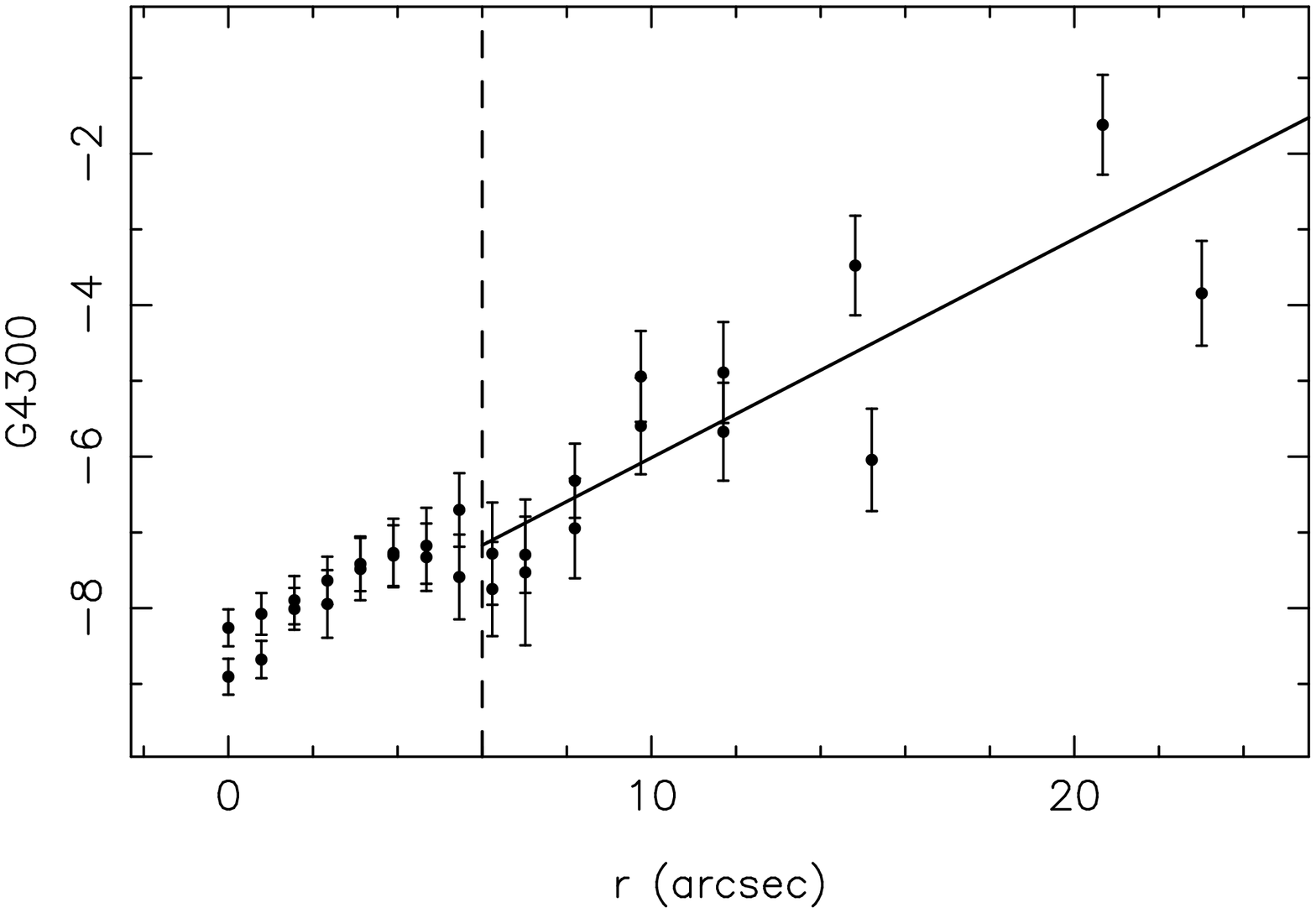}}
\resizebox{0.3\textwidth}{!}{\includegraphics[angle=-90]{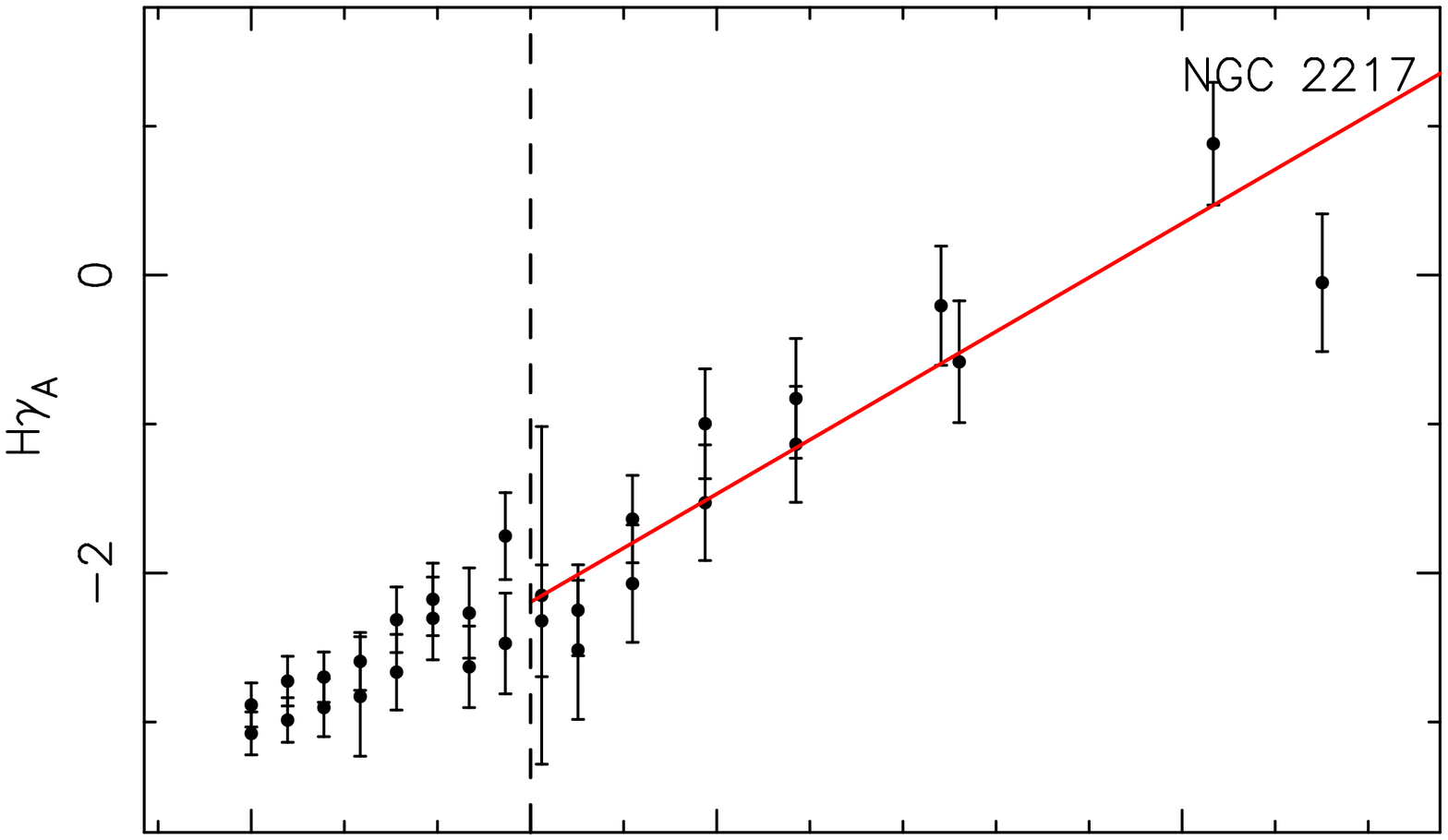}}
\resizebox{0.3\textwidth}{!}{\includegraphics[angle=-90]{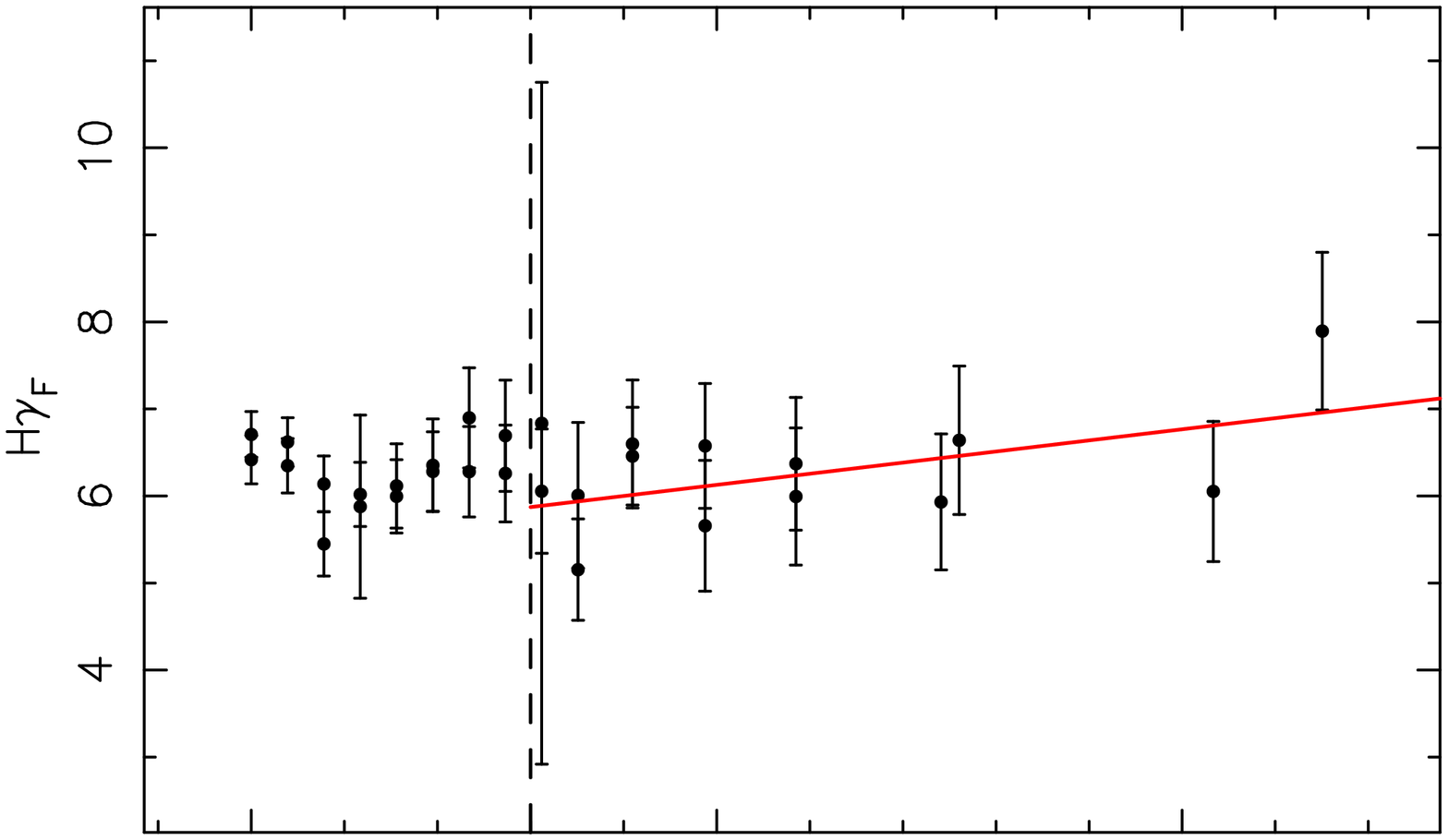}}
\resizebox{0.3\textwidth}{!}{\includegraphics[angle=-90]{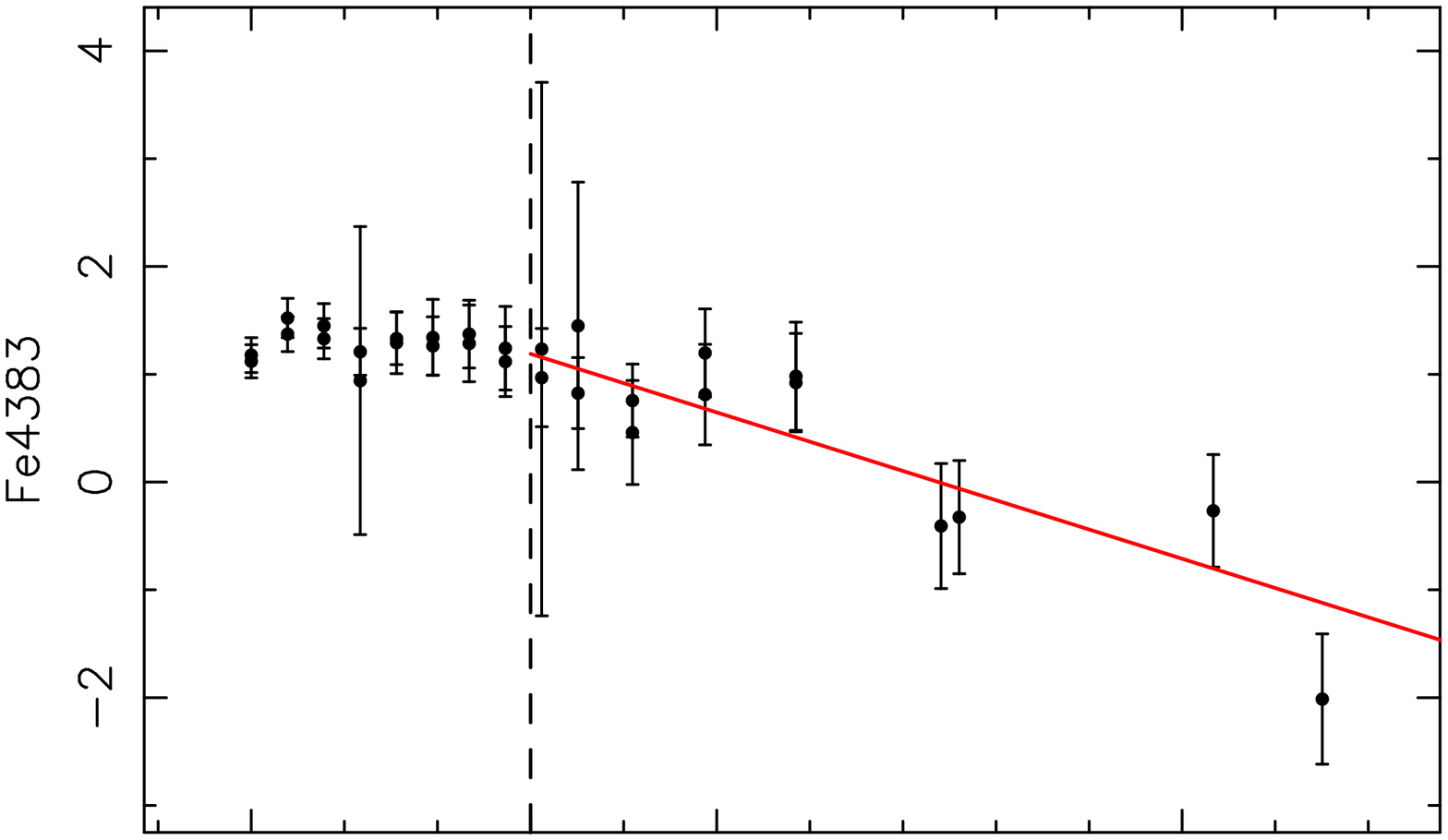}}
\resizebox{0.3\textwidth}{!}{\includegraphics[angle=-90]{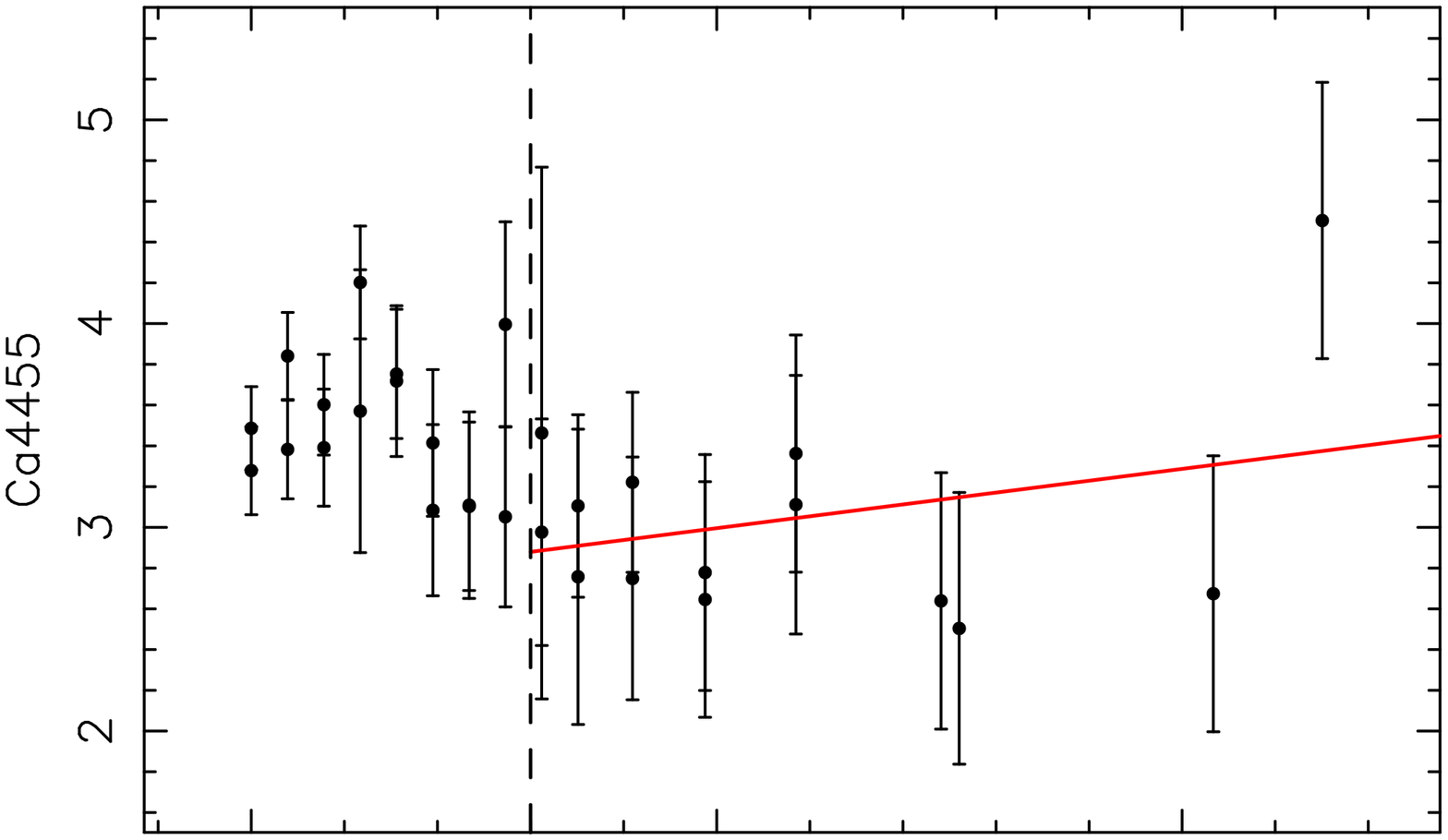}}
\resizebox{0.3\textwidth}{!}{\includegraphics[angle=-90]{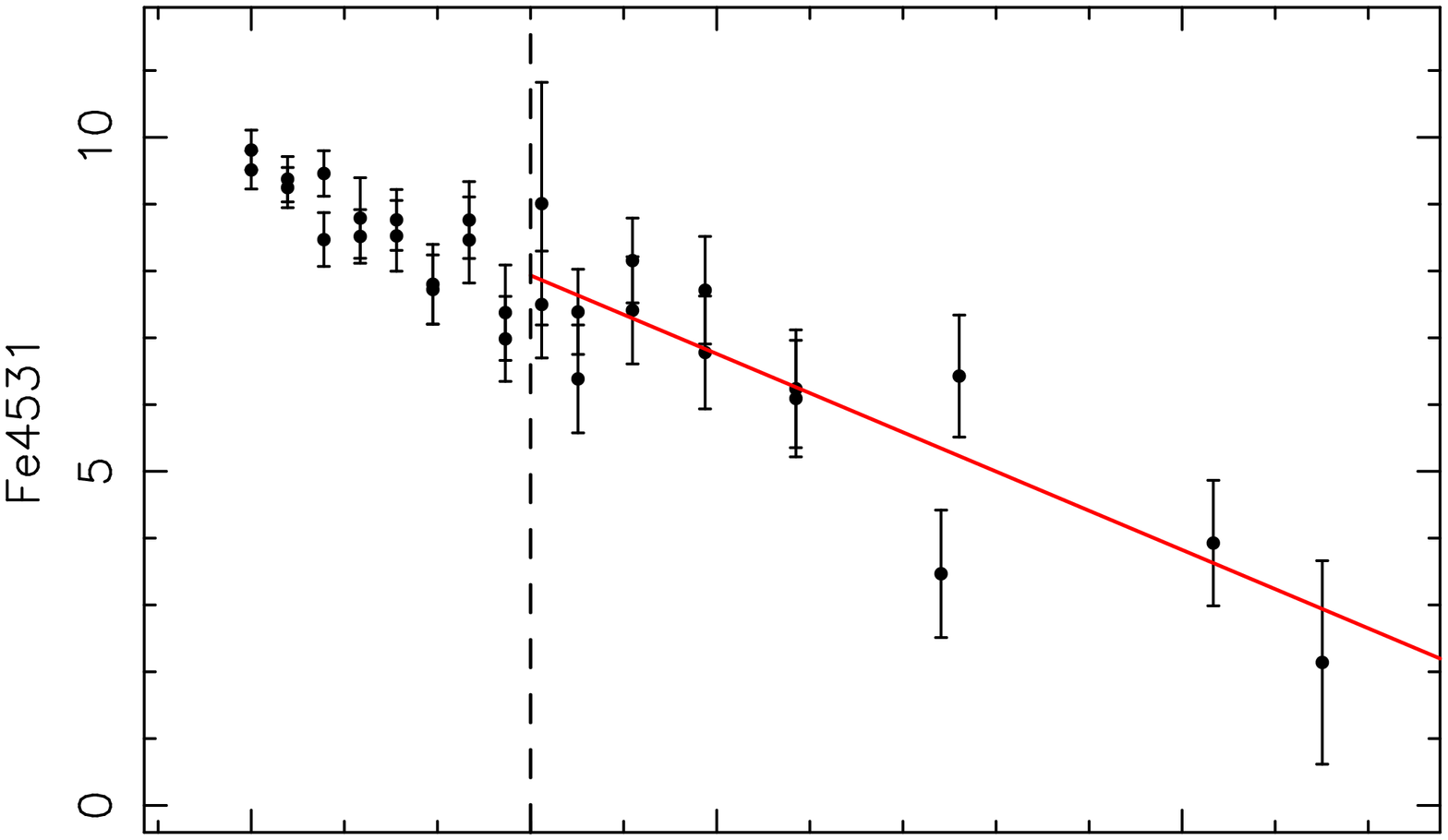}}
\resizebox{0.3\textwidth}{!}{\includegraphics[angle=-90]{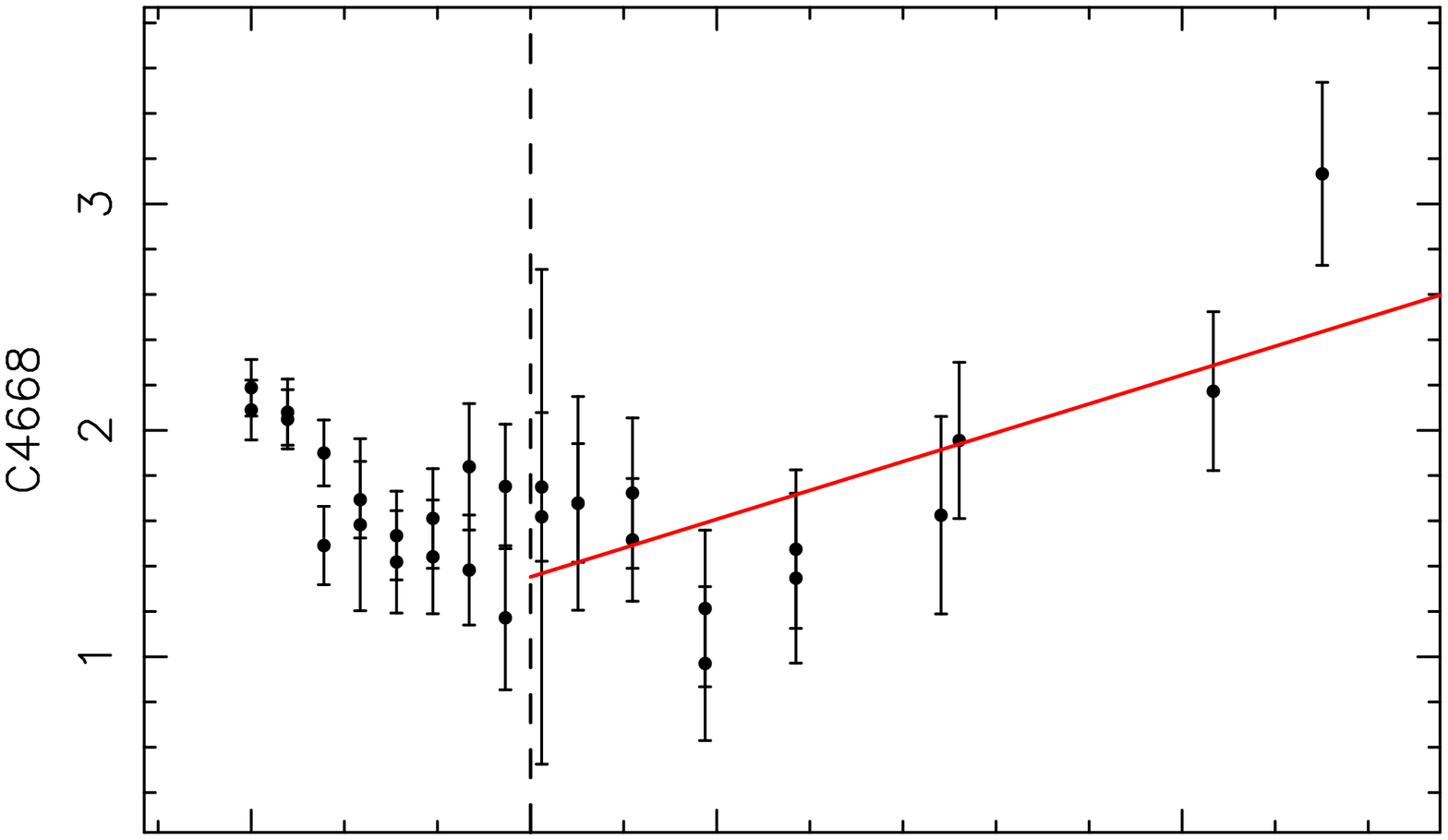}}
\resizebox{0.3\textwidth}{!}{\includegraphics[angle=-90]{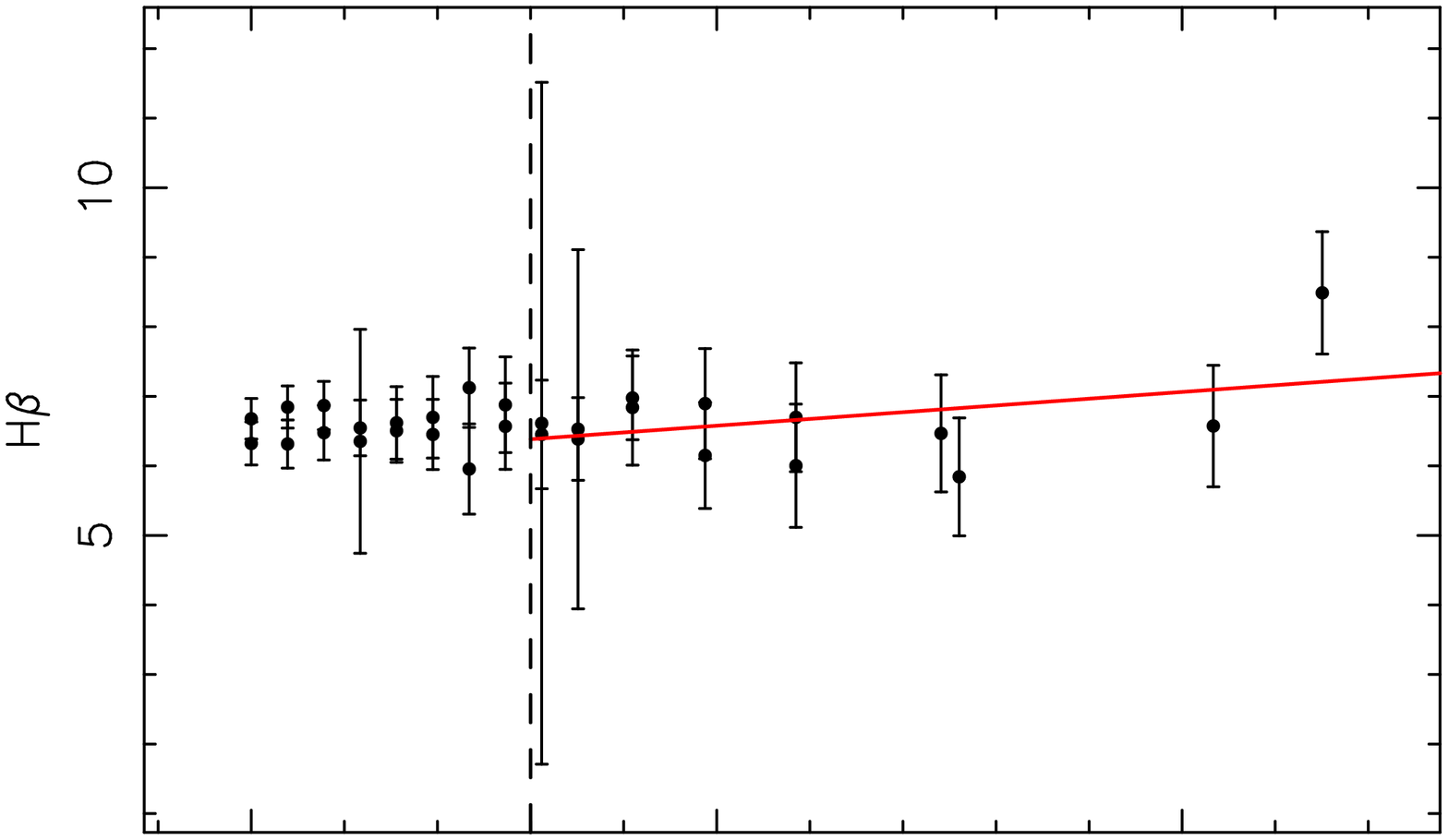}}
\resizebox{0.3\textwidth}{!}{\includegraphics[angle=-90]{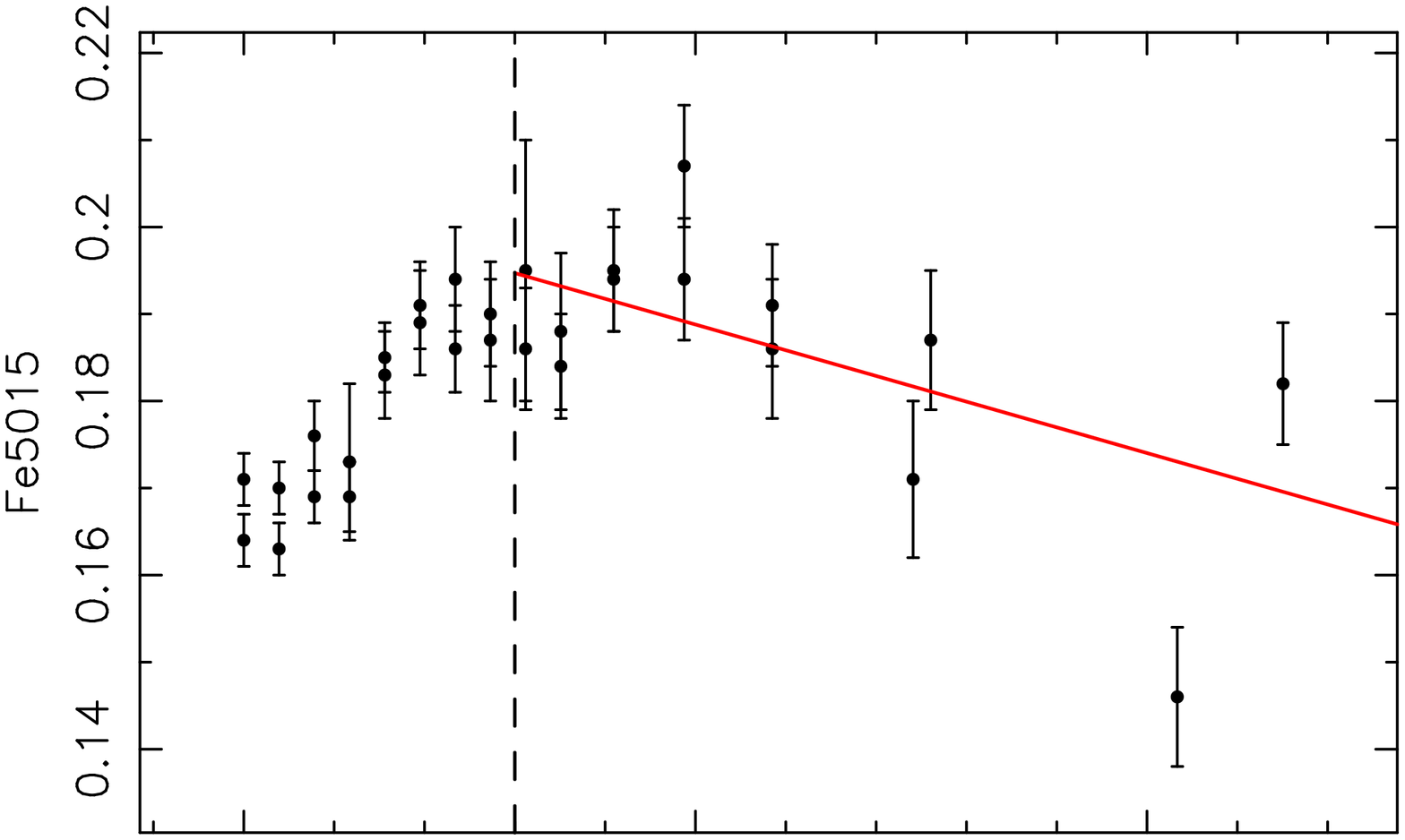}}
\resizebox{0.3\textwidth}{!}{\includegraphics[angle=-90]{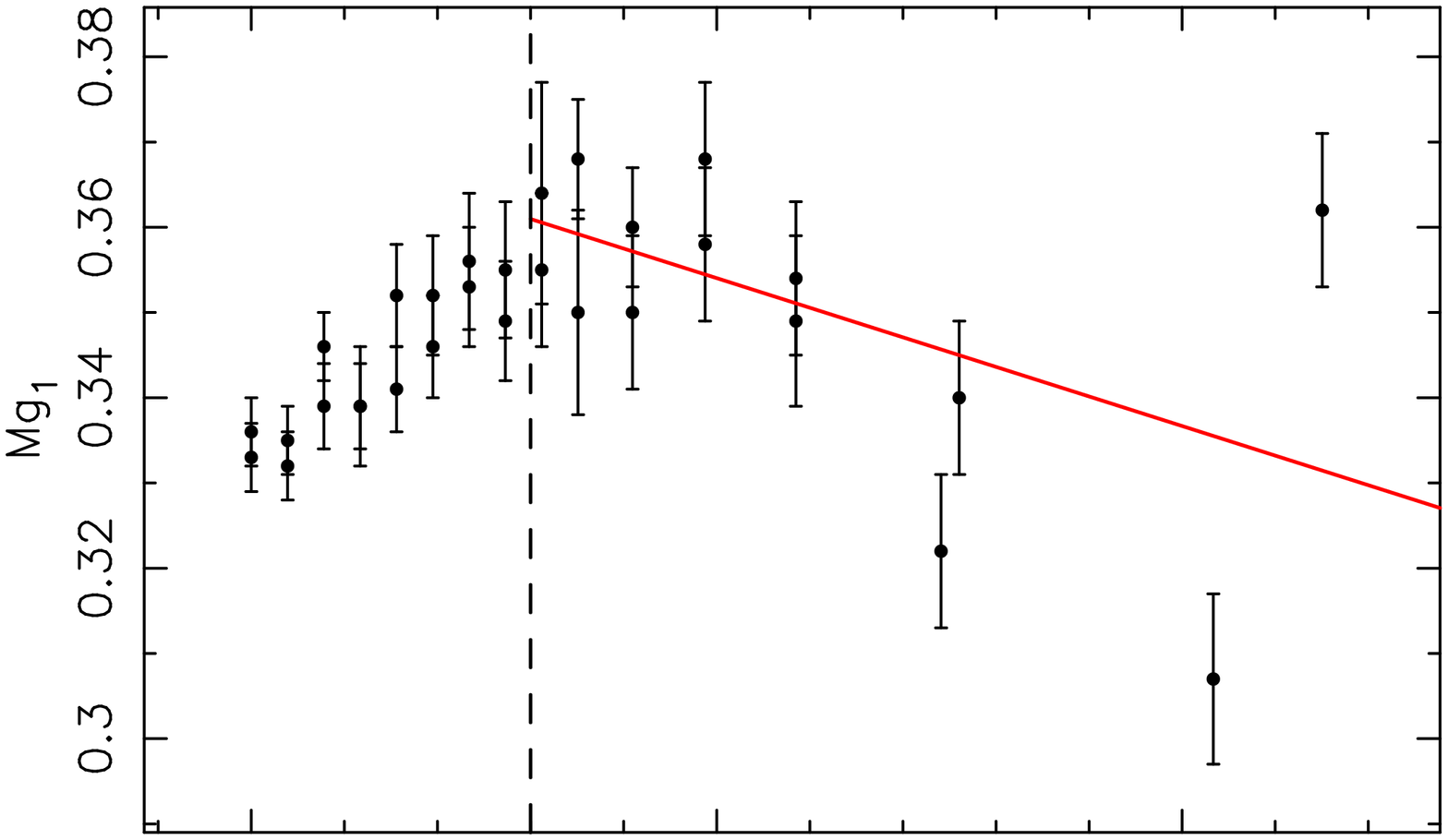}}
\resizebox{0.3\textwidth}{!}{\includegraphics[angle=-90]{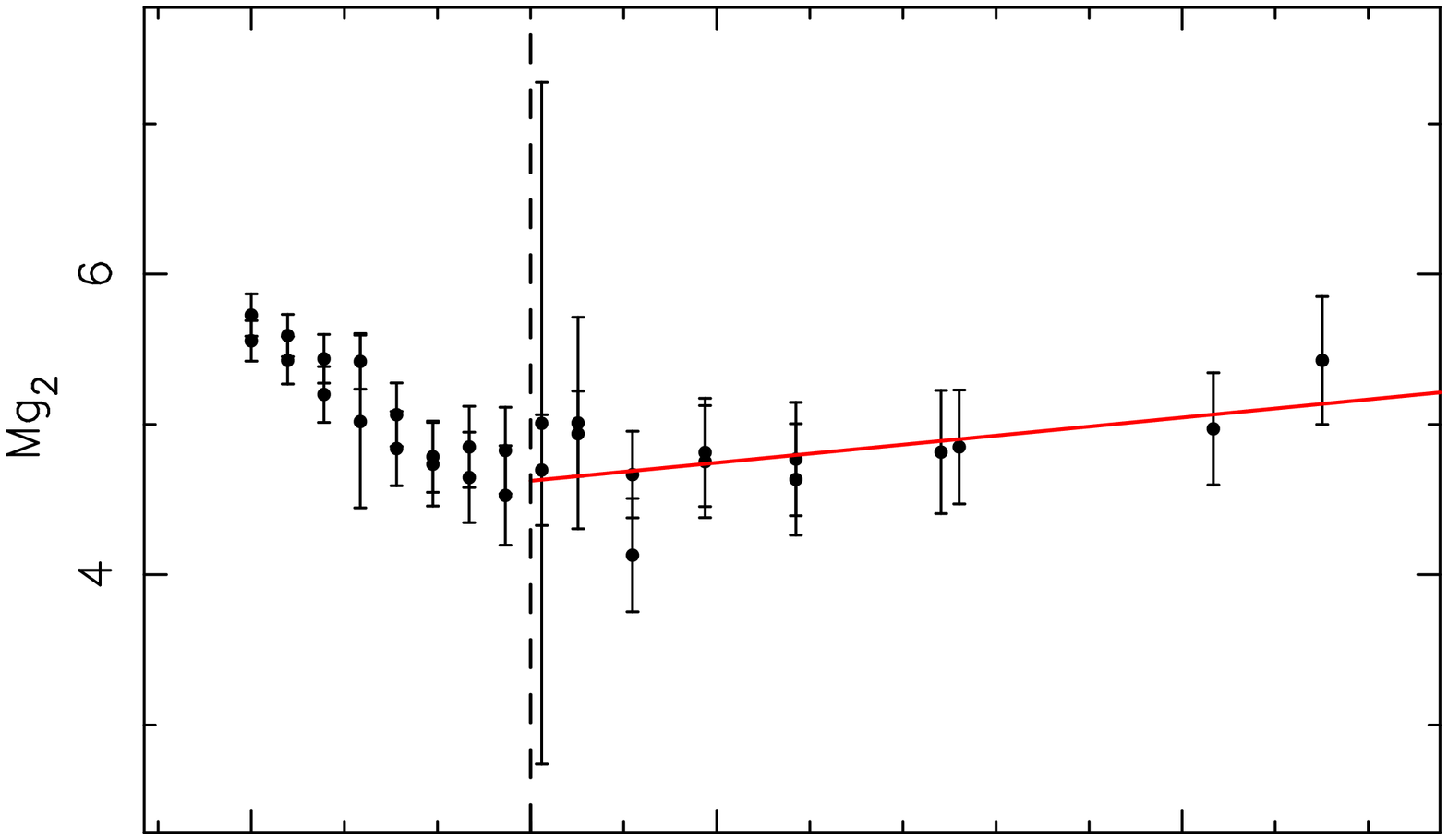}}
\resizebox{0.3\textwidth}{!}{\includegraphics[angle=-90]{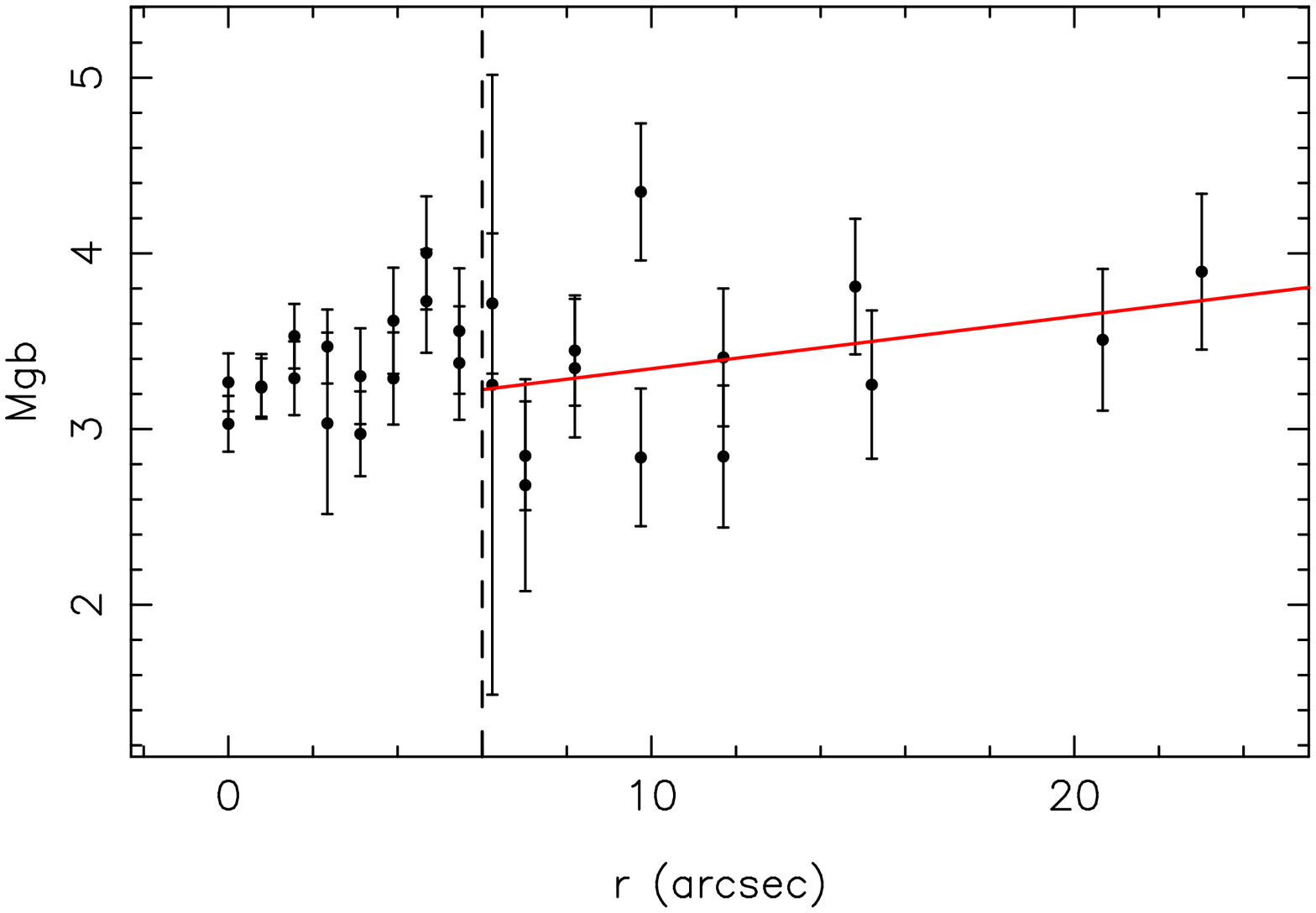}}\hspace{0.85cm}
\resizebox{0.3\textwidth}{!}{\includegraphics[angle=-90]{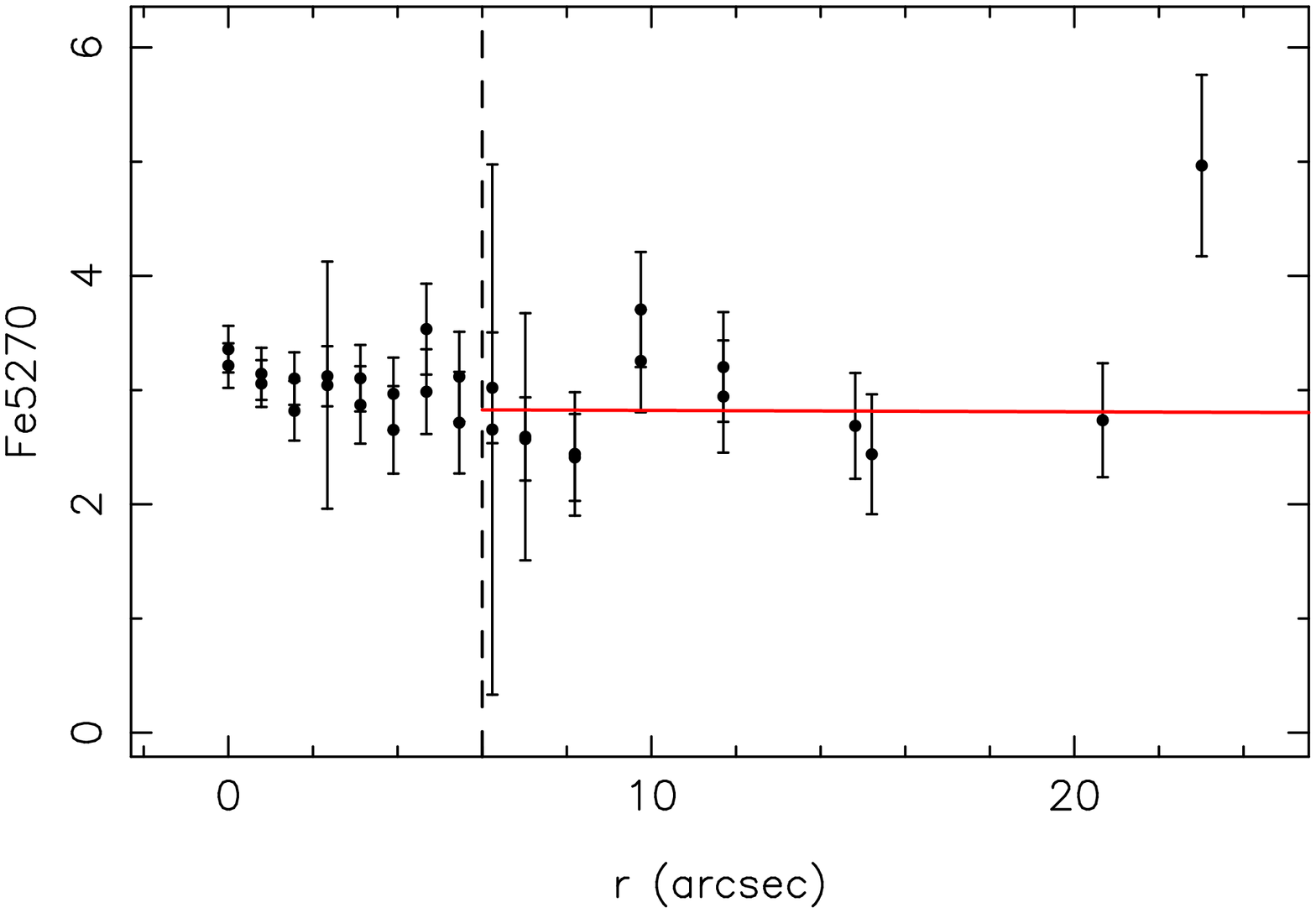}}\hspace{0.85cm} 
\resizebox{0.3\textwidth}{!}{\includegraphics[angle=-90]{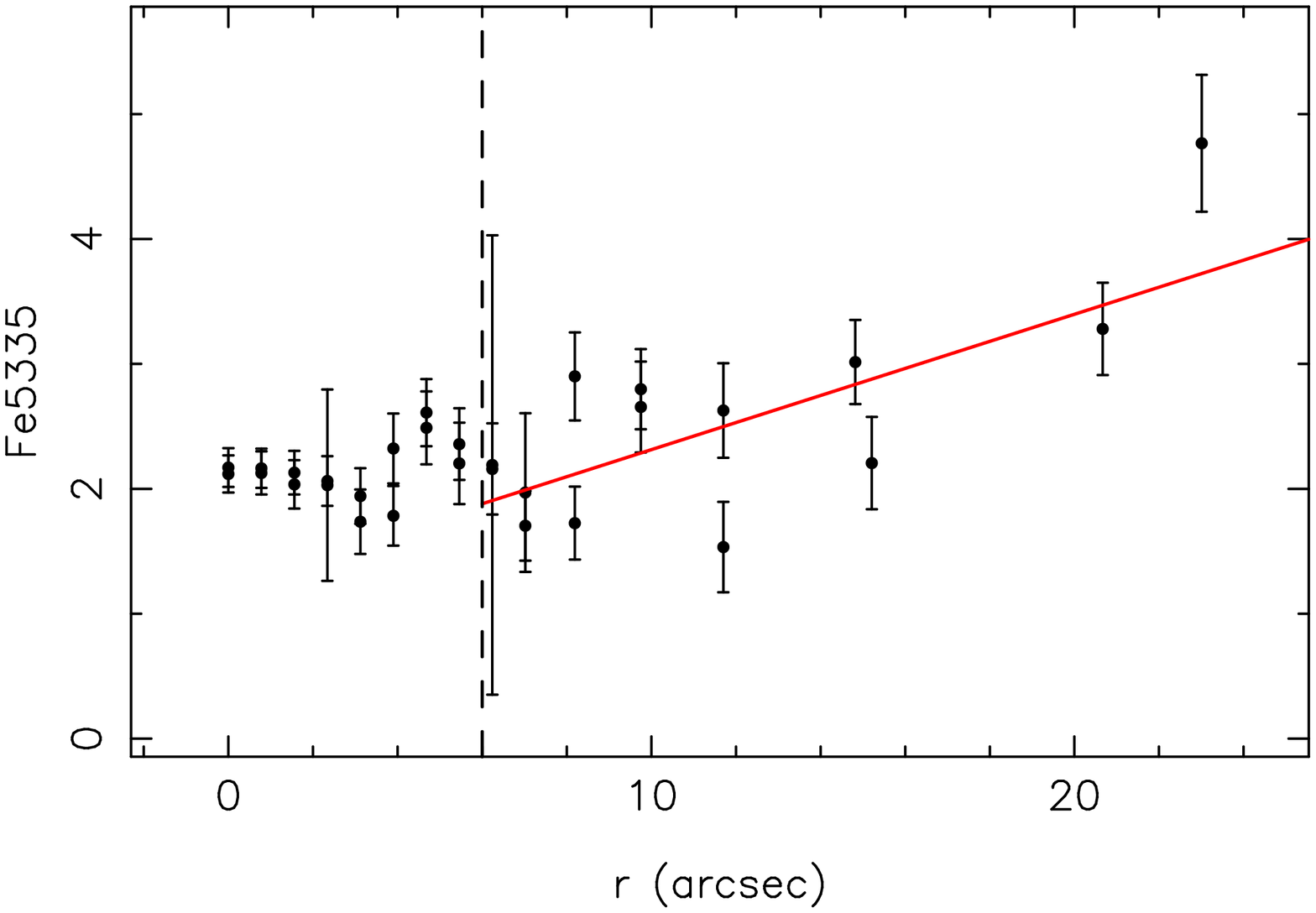}}
\caption{Line-strength distribution in the bar region for all the galaxies}
\end{figure*}

\begin{figure*}
\addtocounter{figure}{-1}
\resizebox{0.3\textwidth}{!}{\includegraphics[angle=-90]{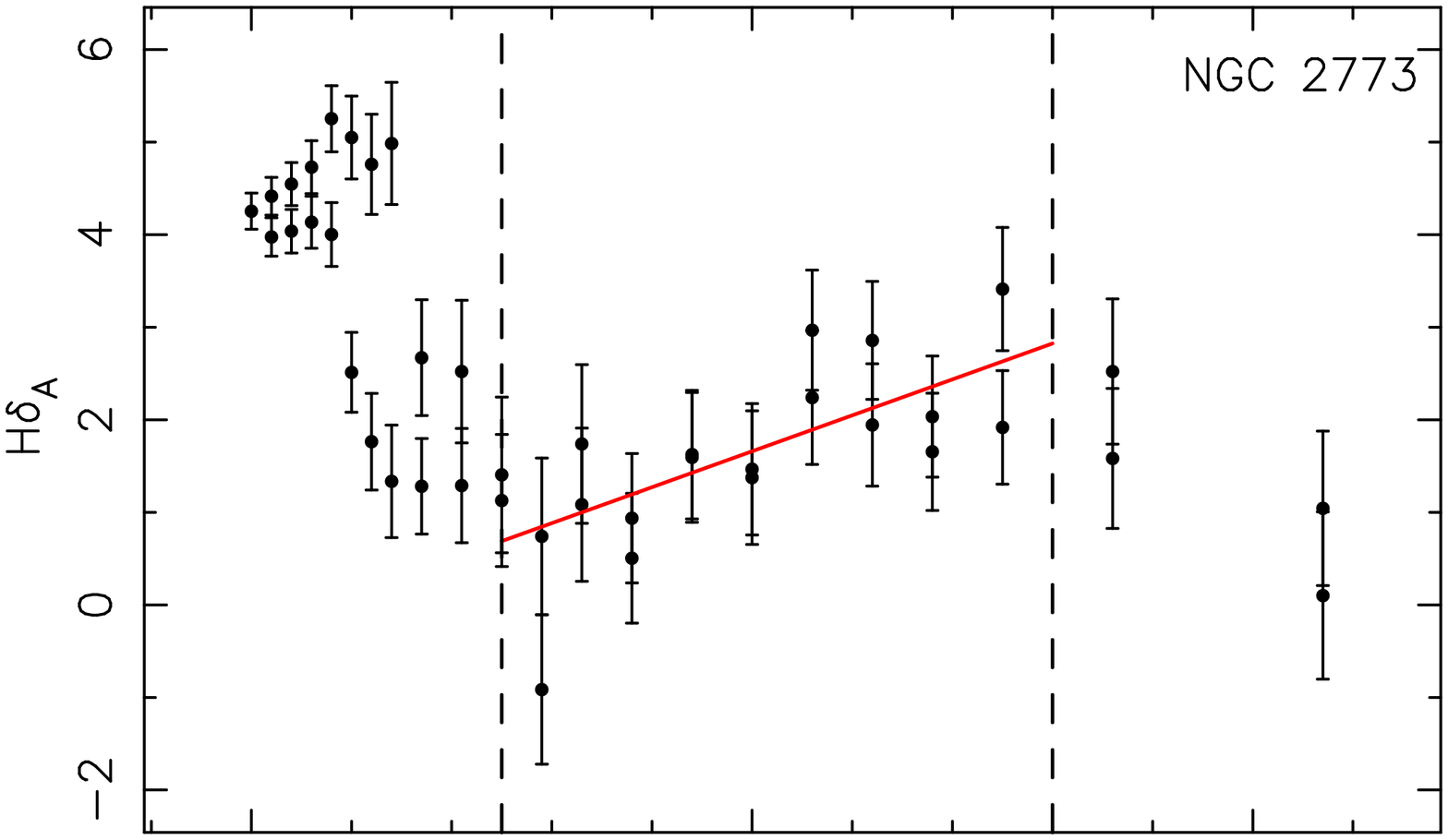}}
\resizebox{0.3\textwidth}{!}{\includegraphics[angle=-90]{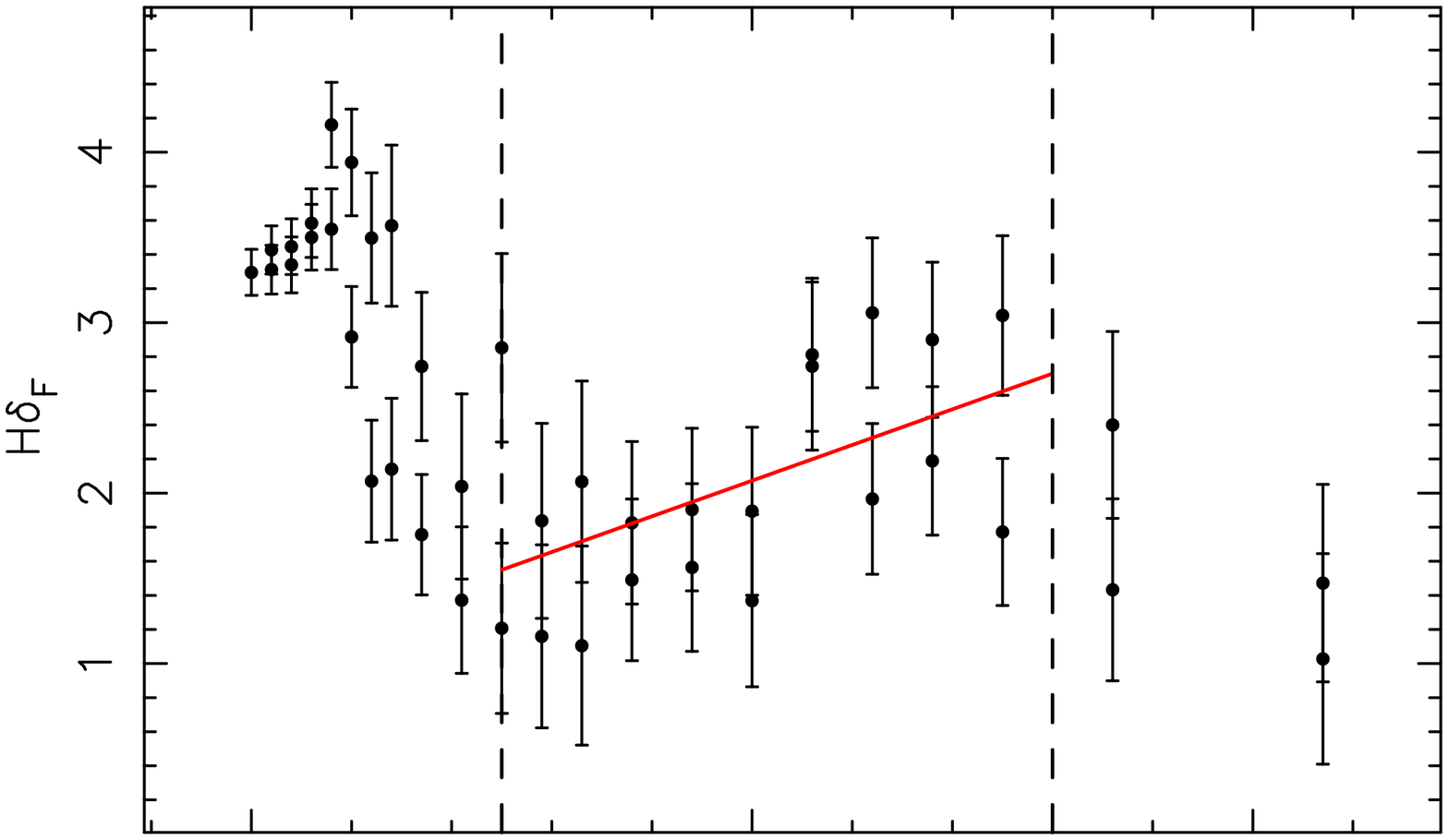}}
\resizebox{0.3\textwidth}{!}{\includegraphics[angle=-90]{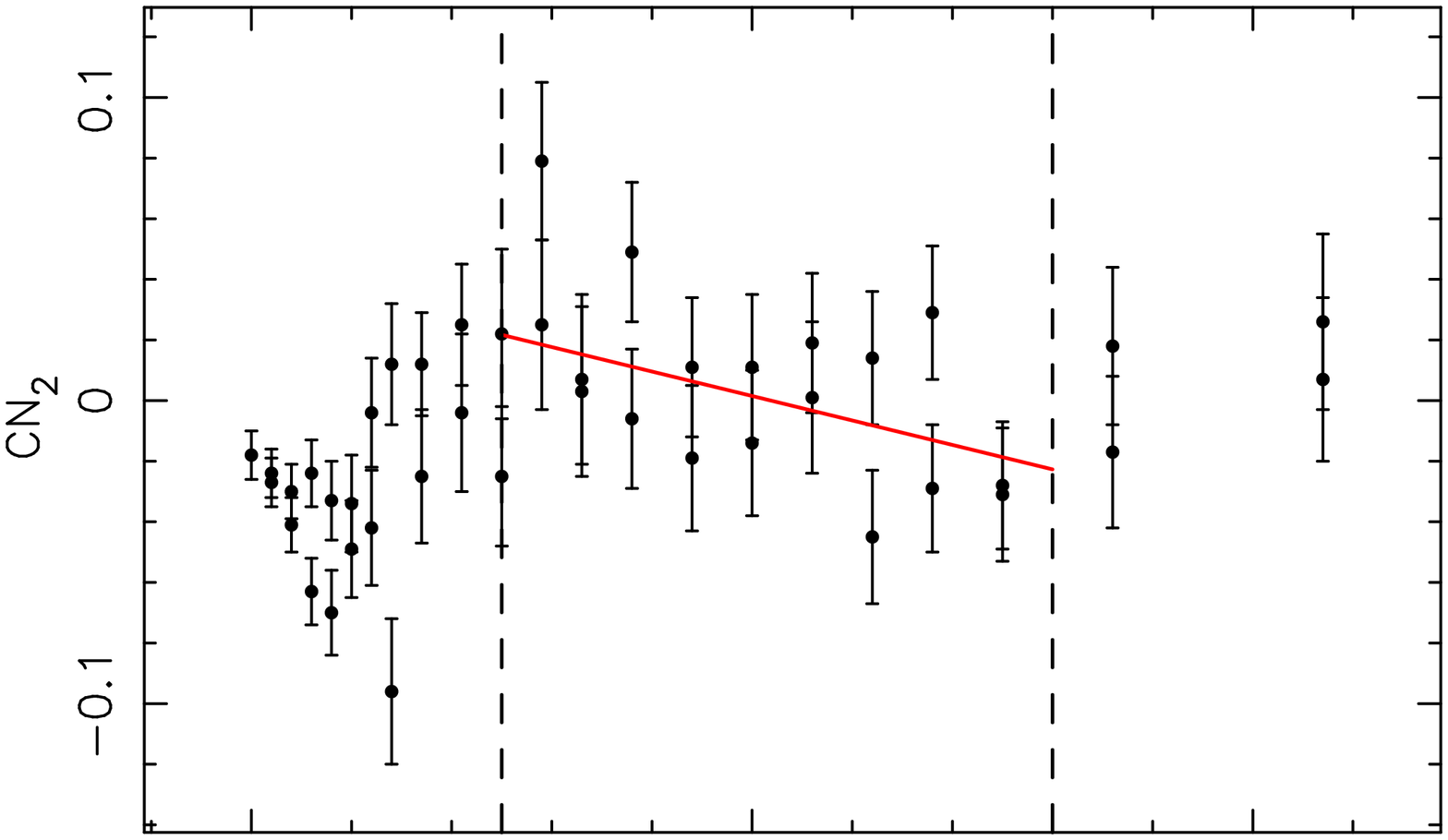}}
\resizebox{0.3\textwidth}{!}{\includegraphics[angle=-90]{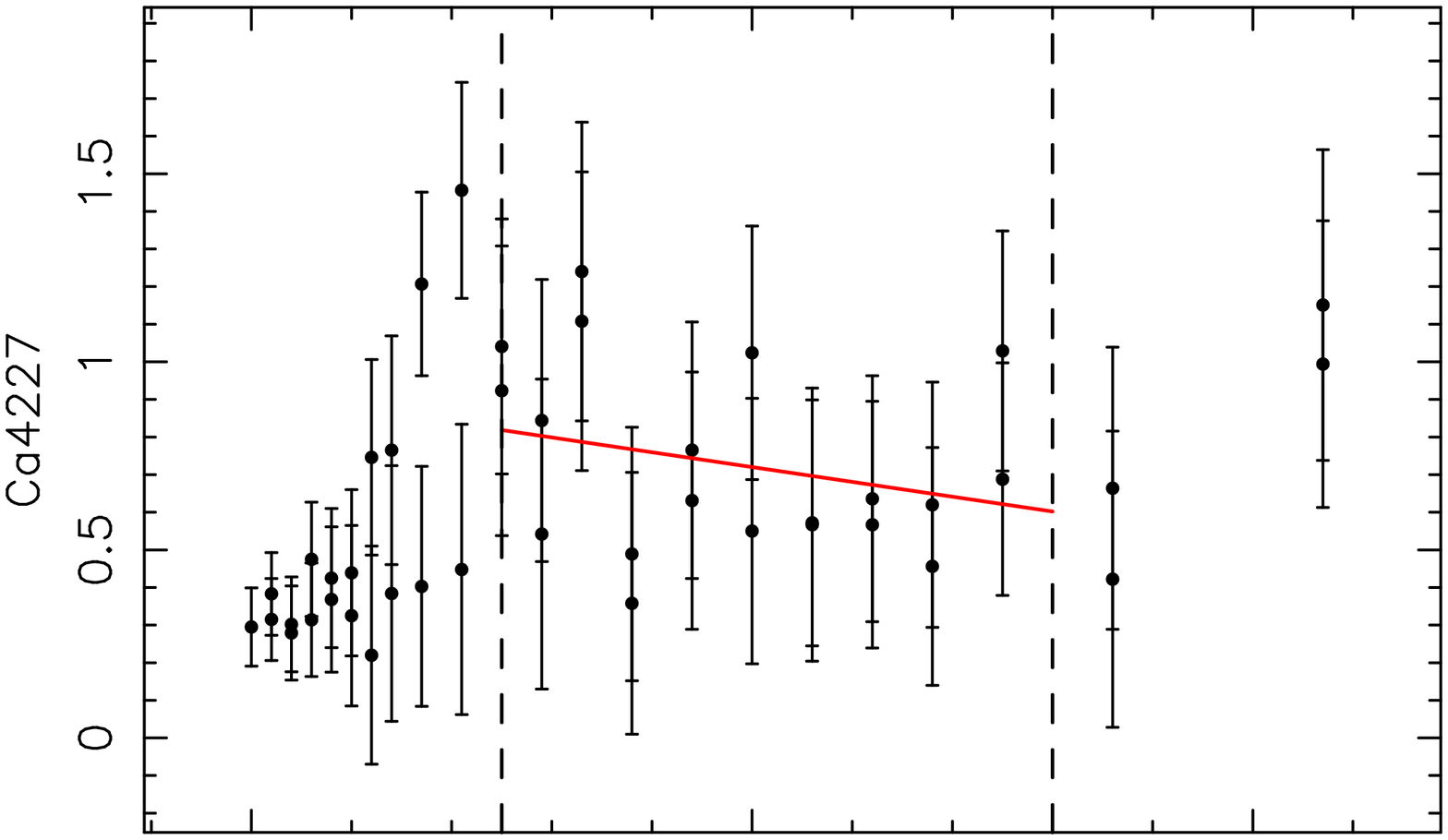}}
\resizebox{0.3\textwidth}{!}{\includegraphics[angle=-90]{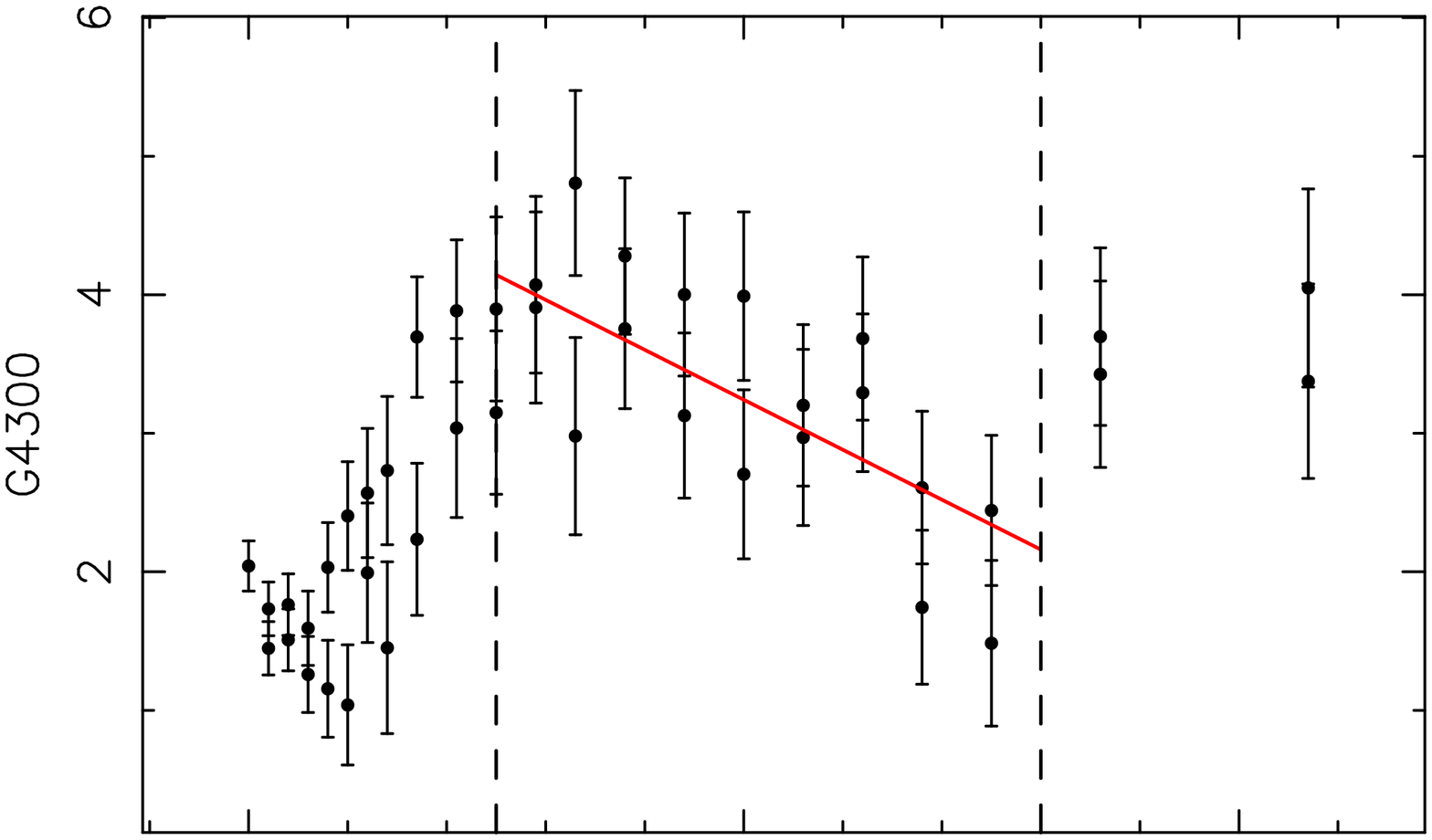}}
\resizebox{0.3\textwidth}{!}{\includegraphics[angle=-90]{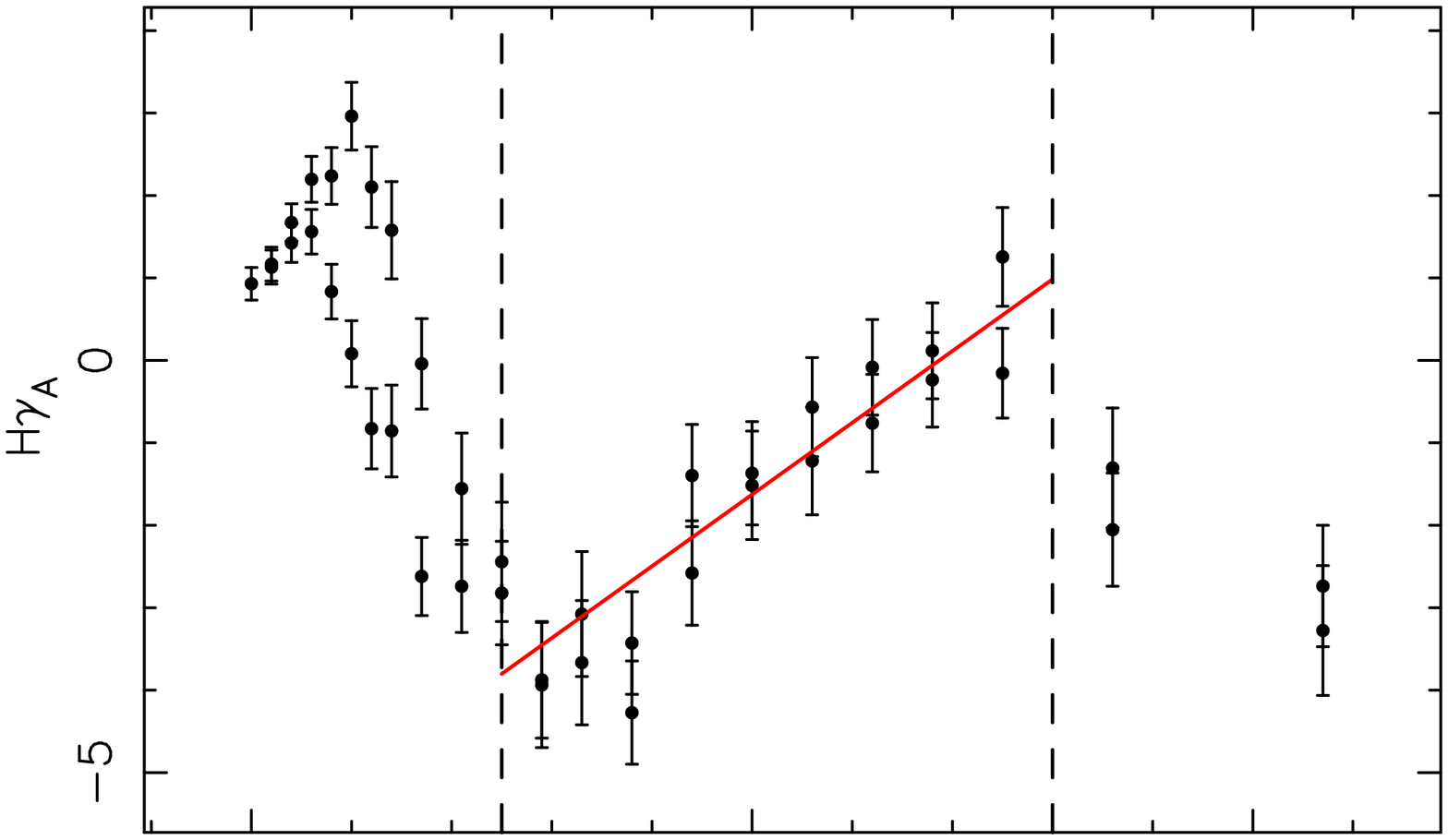}}
\resizebox{0.3\textwidth}{!}{\includegraphics[angle=-90]{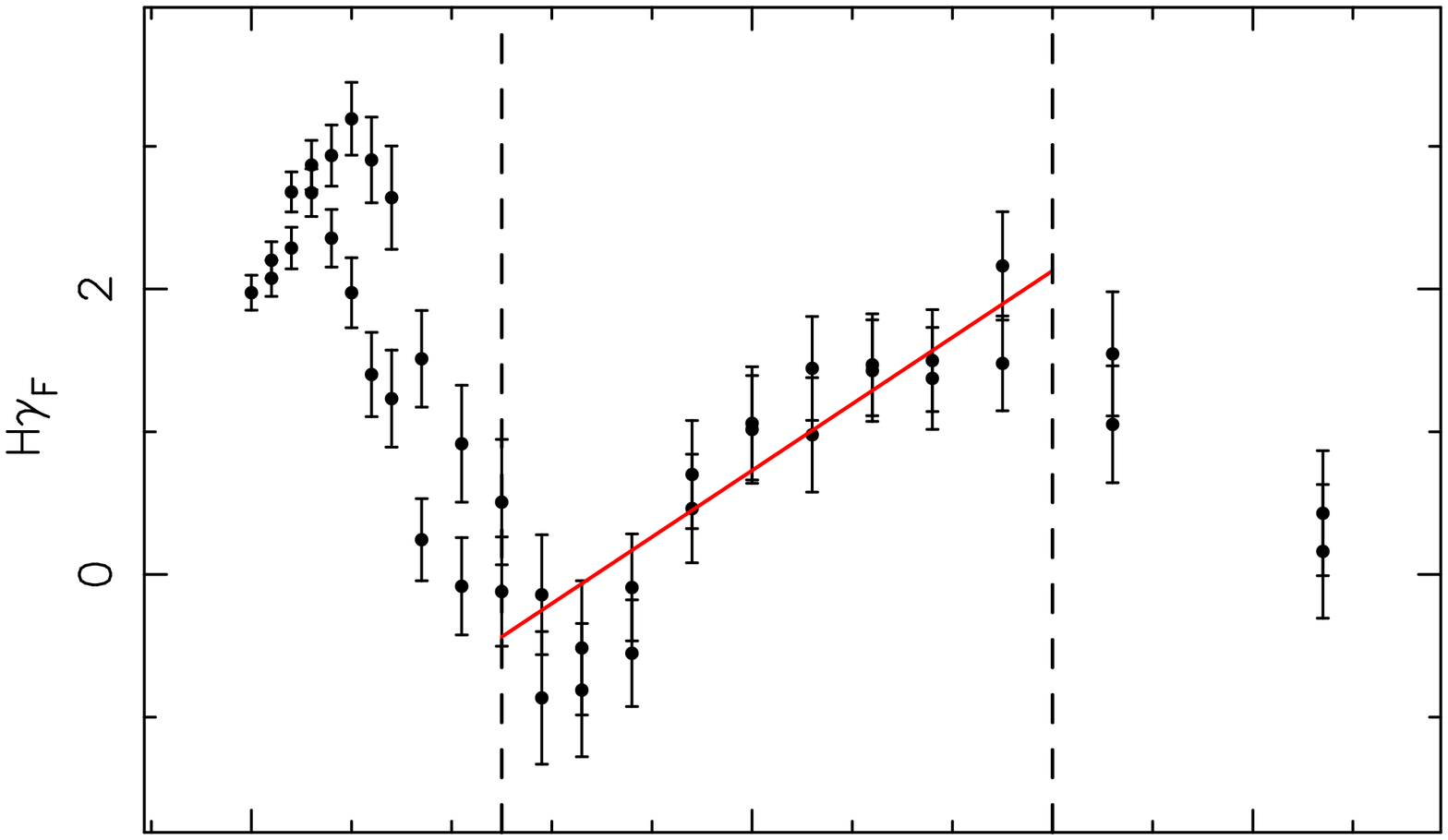}}
\resizebox{0.3\textwidth}{!}{\includegraphics[angle=-90]{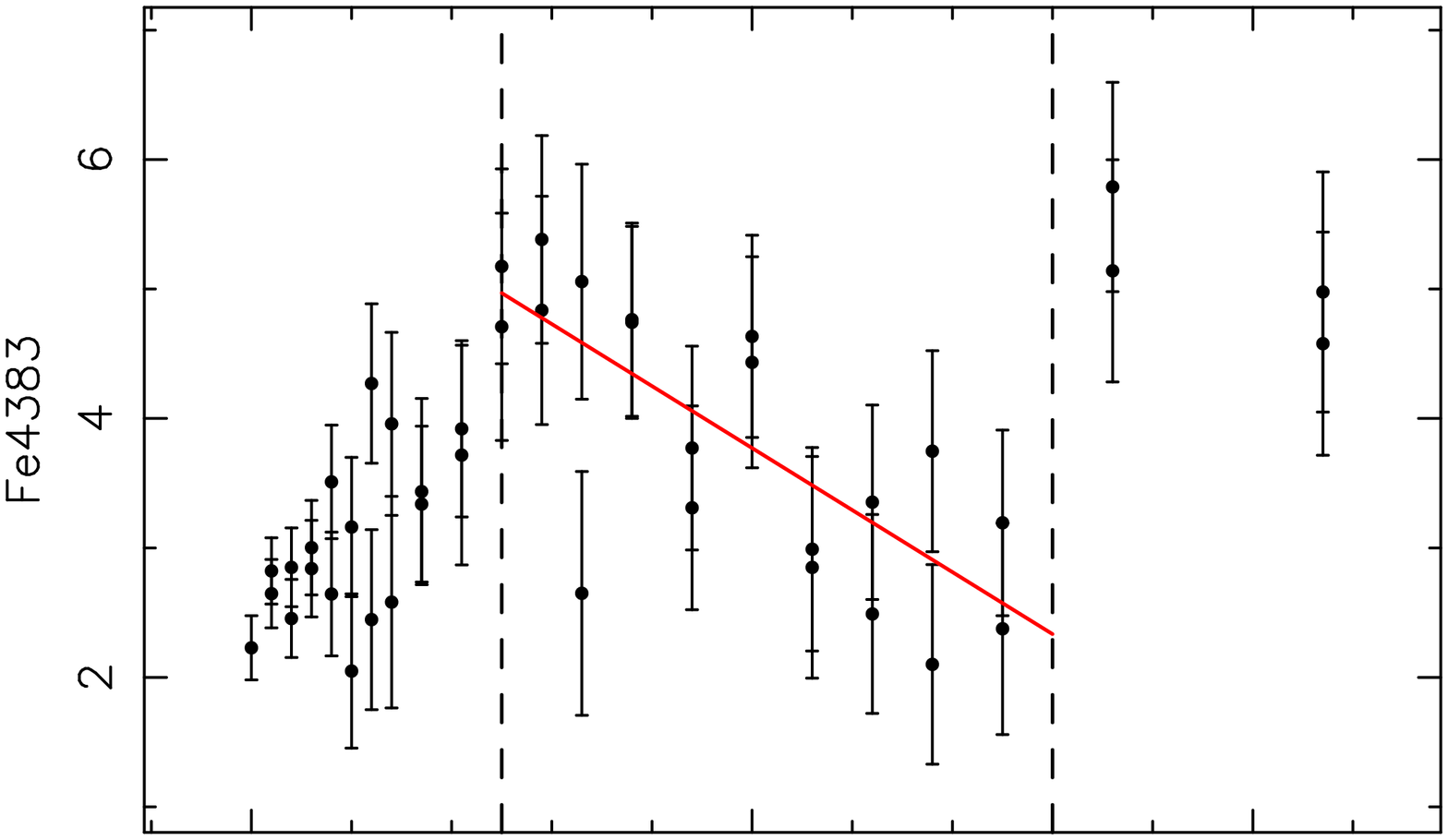}}
\resizebox{0.3\textwidth}{!}{\includegraphics[angle=-90]{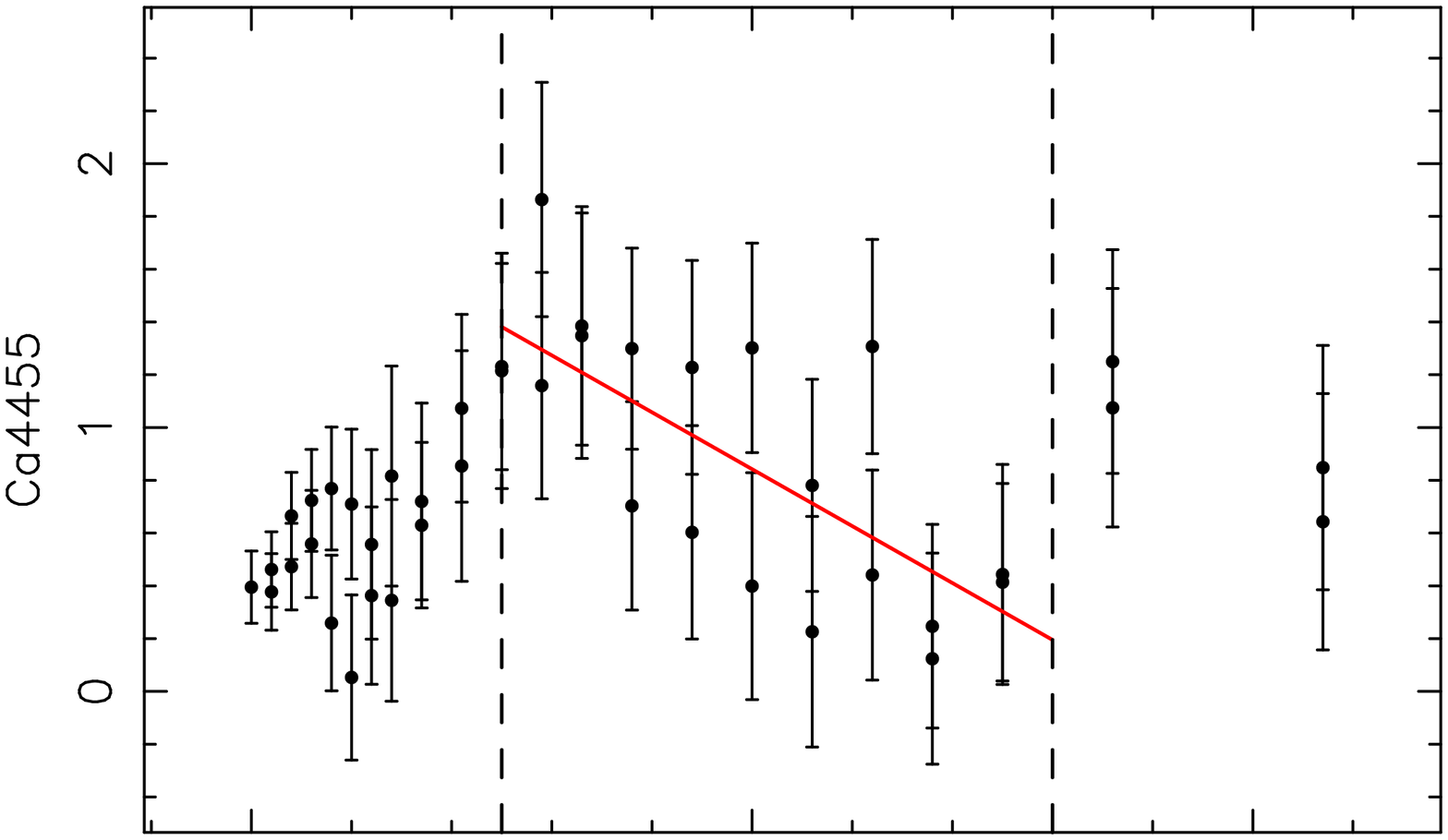}}
\resizebox{0.3\textwidth}{!}{\includegraphics[angle=-90]{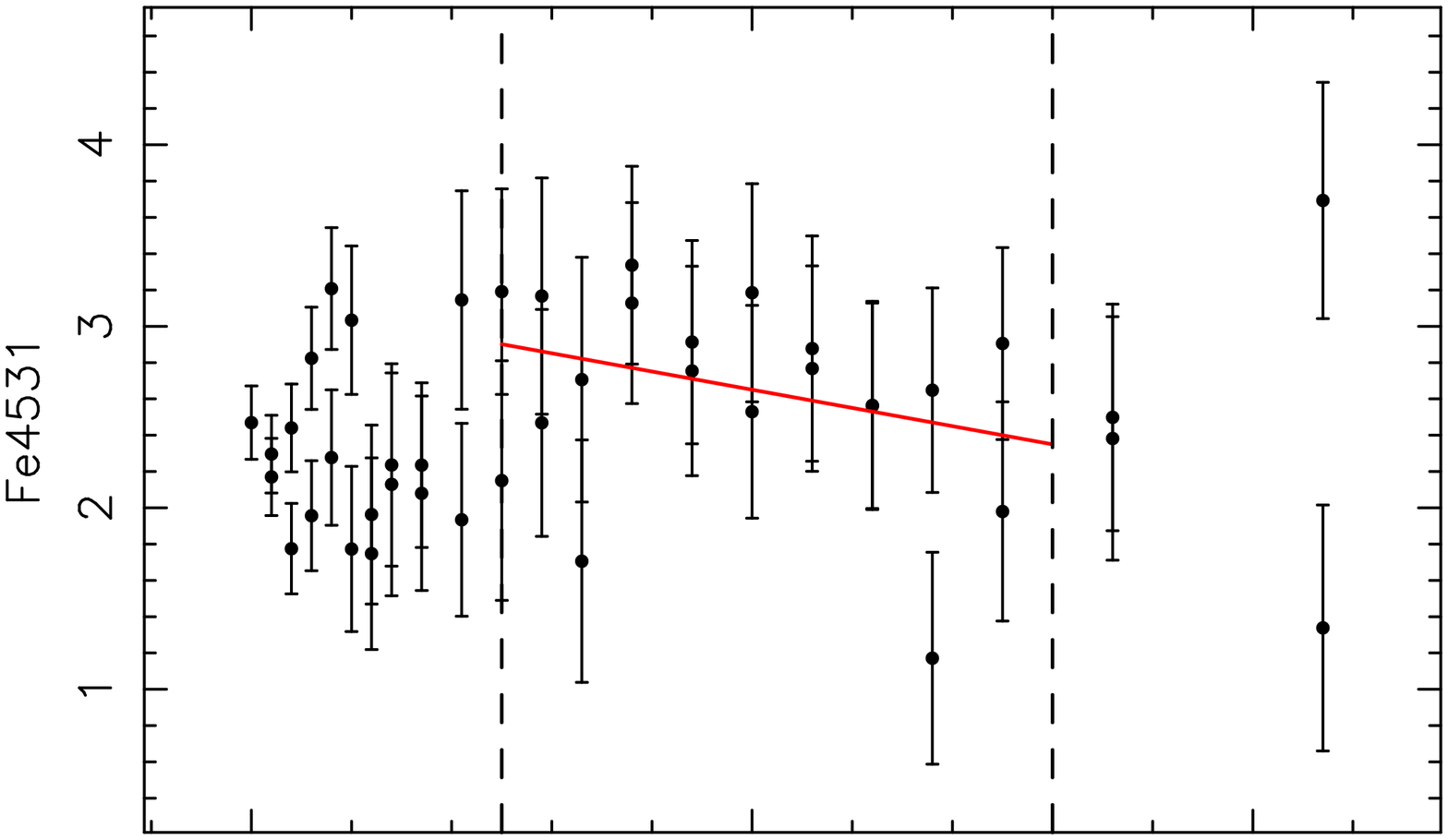}}
\resizebox{0.3\textwidth}{!}{\includegraphics[angle=-90]{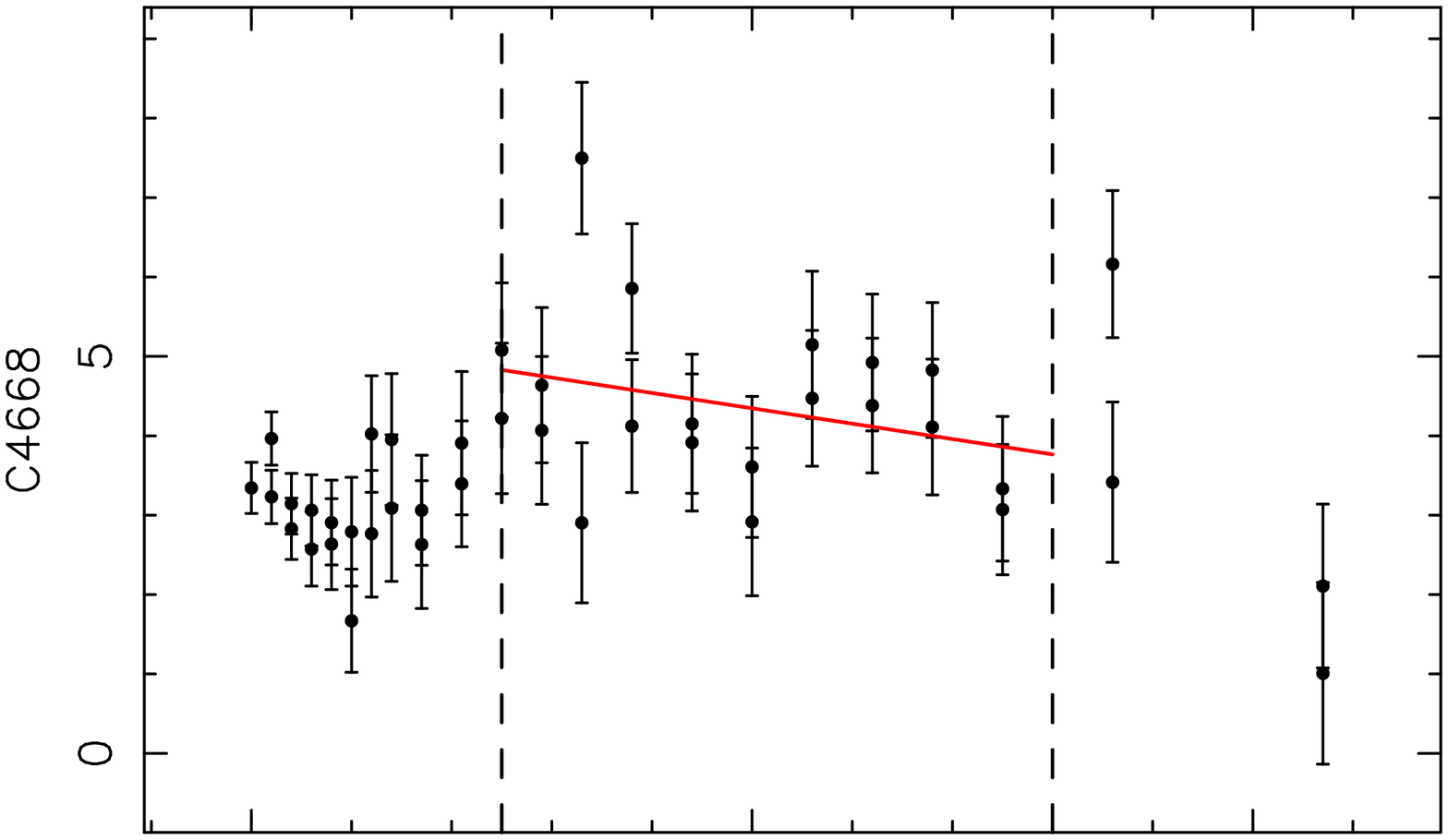}}
\resizebox{0.3\textwidth}{!}{\includegraphics[angle=-90]{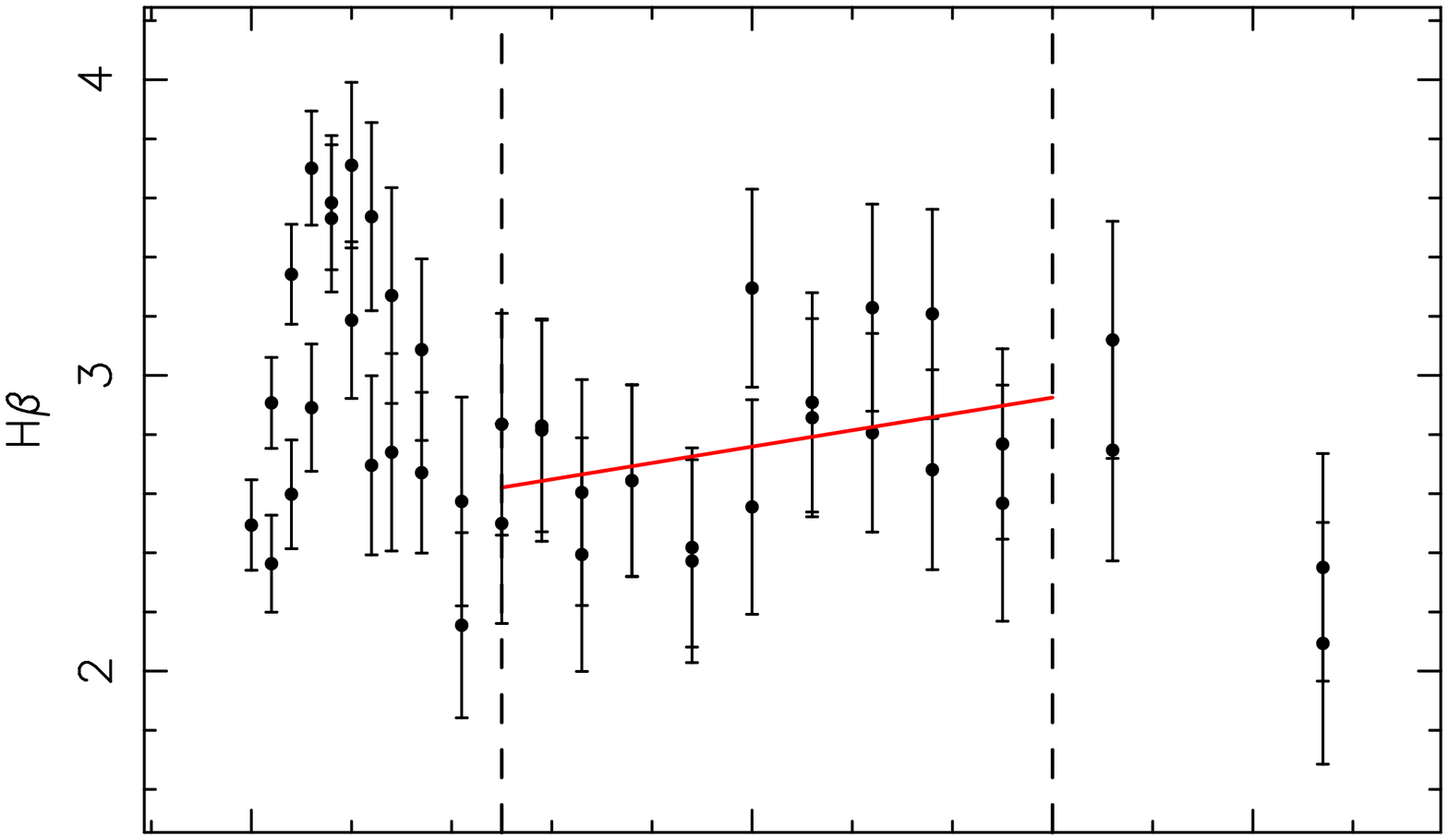}}
\resizebox{0.3\textwidth}{!}{\includegraphics[angle=-90]{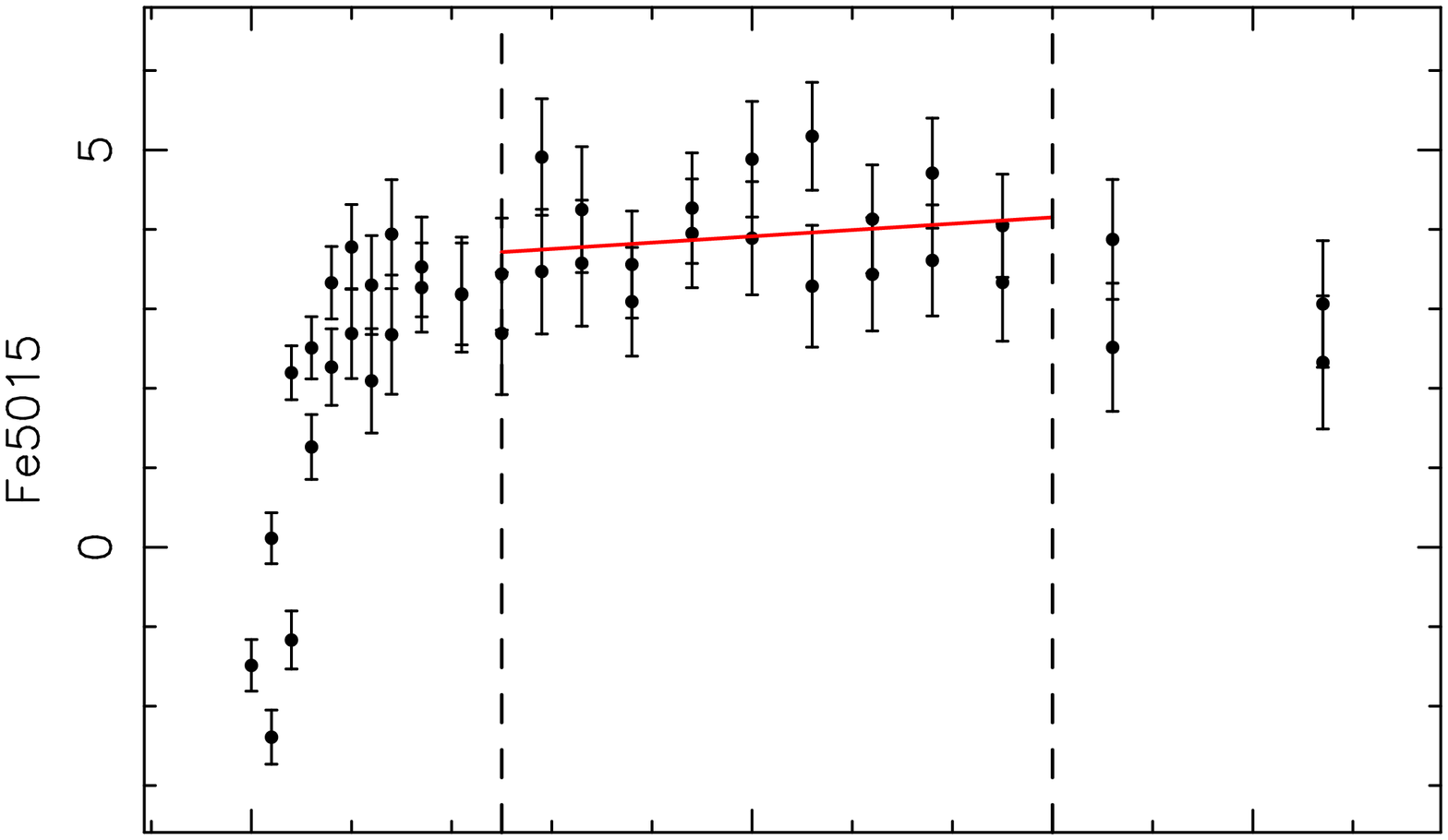}}
\resizebox{0.3\textwidth}{!}{\includegraphics[angle=-90]{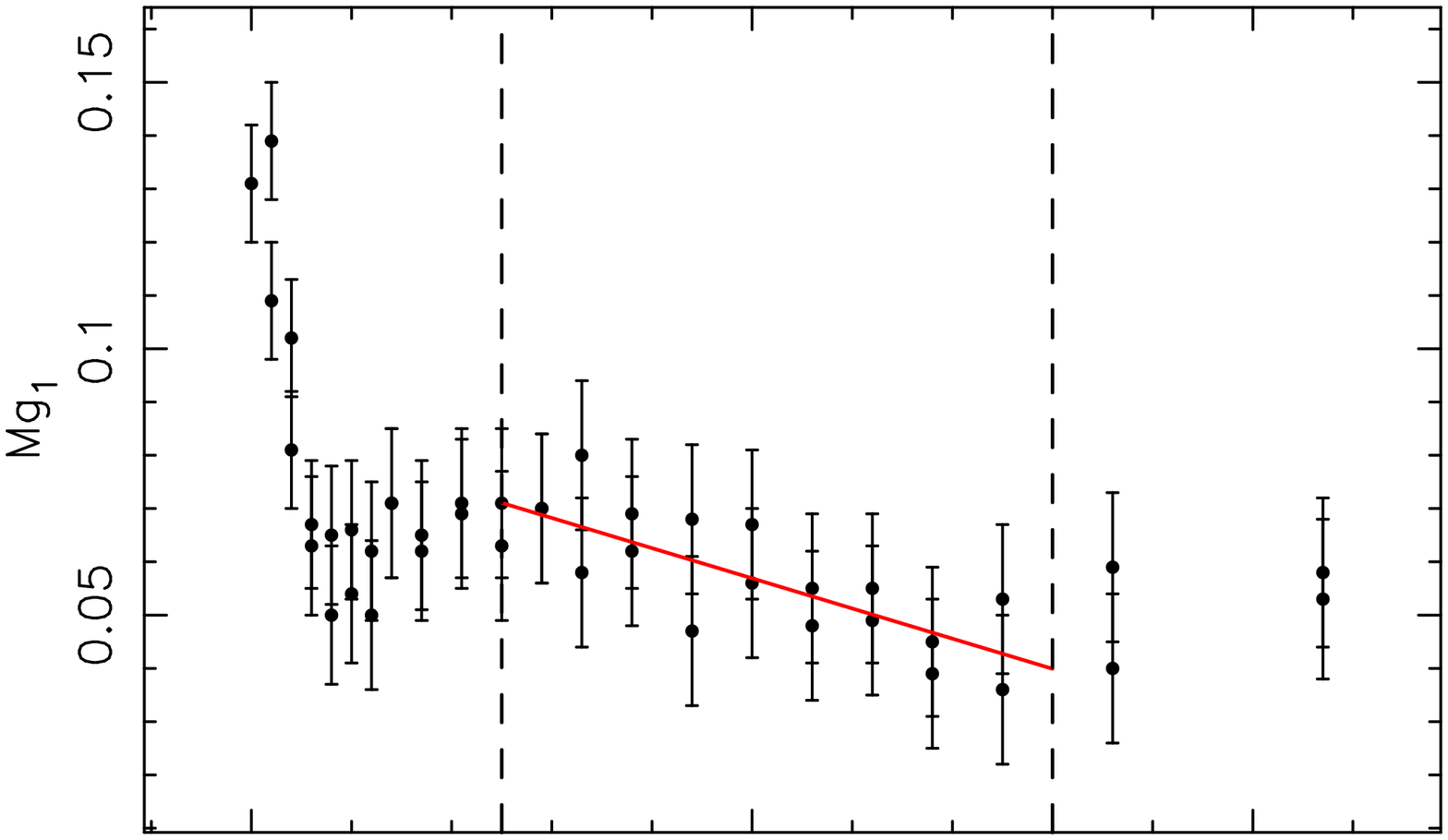}}
\resizebox{0.3\textwidth}{!}{\includegraphics[angle=-90]{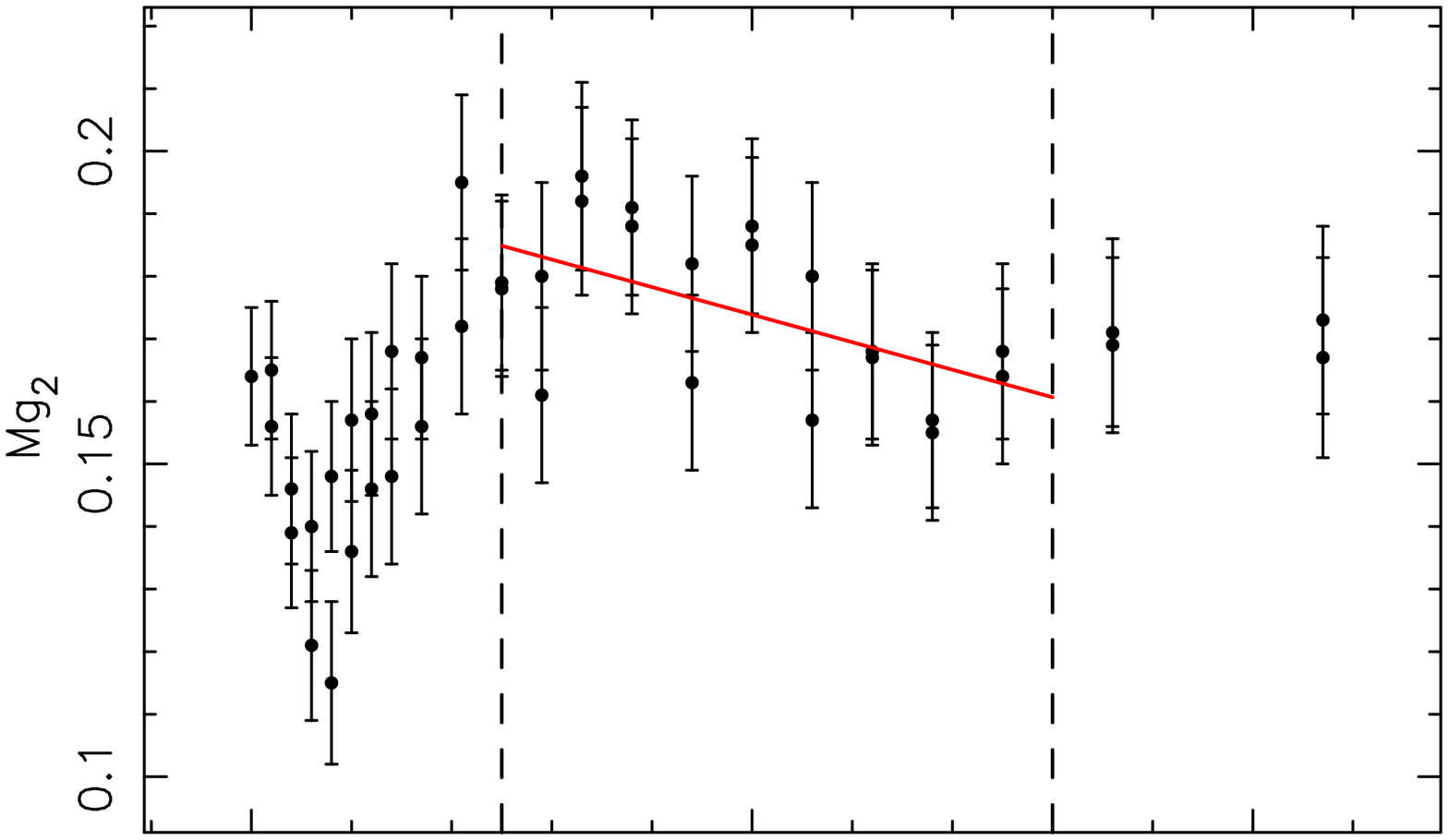}}
\resizebox{0.3\textwidth}{!}{\includegraphics[angle=-90]{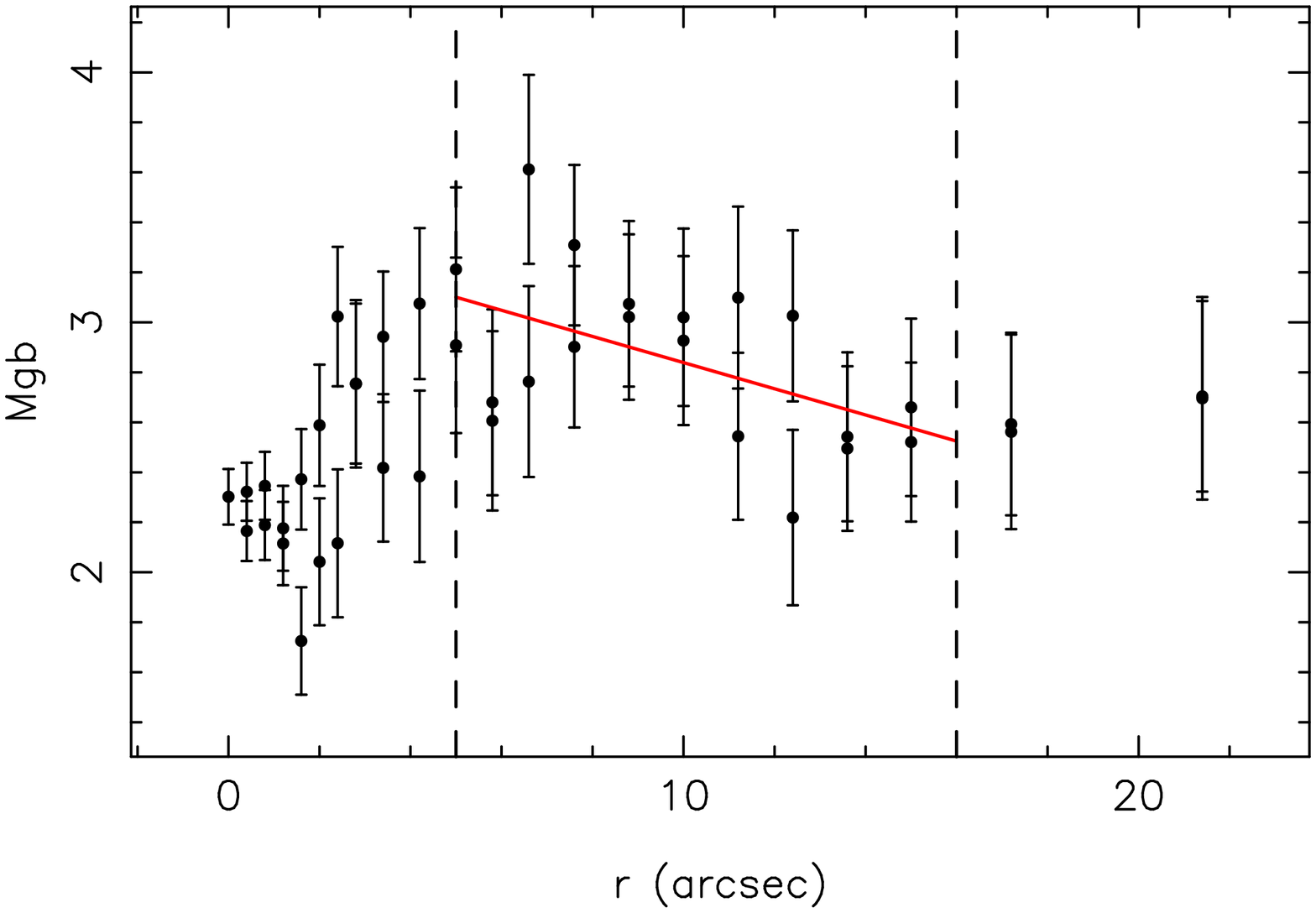}}\hspace{0.85cm}
\resizebox{0.3\textwidth}{!}{\includegraphics[angle=-90]{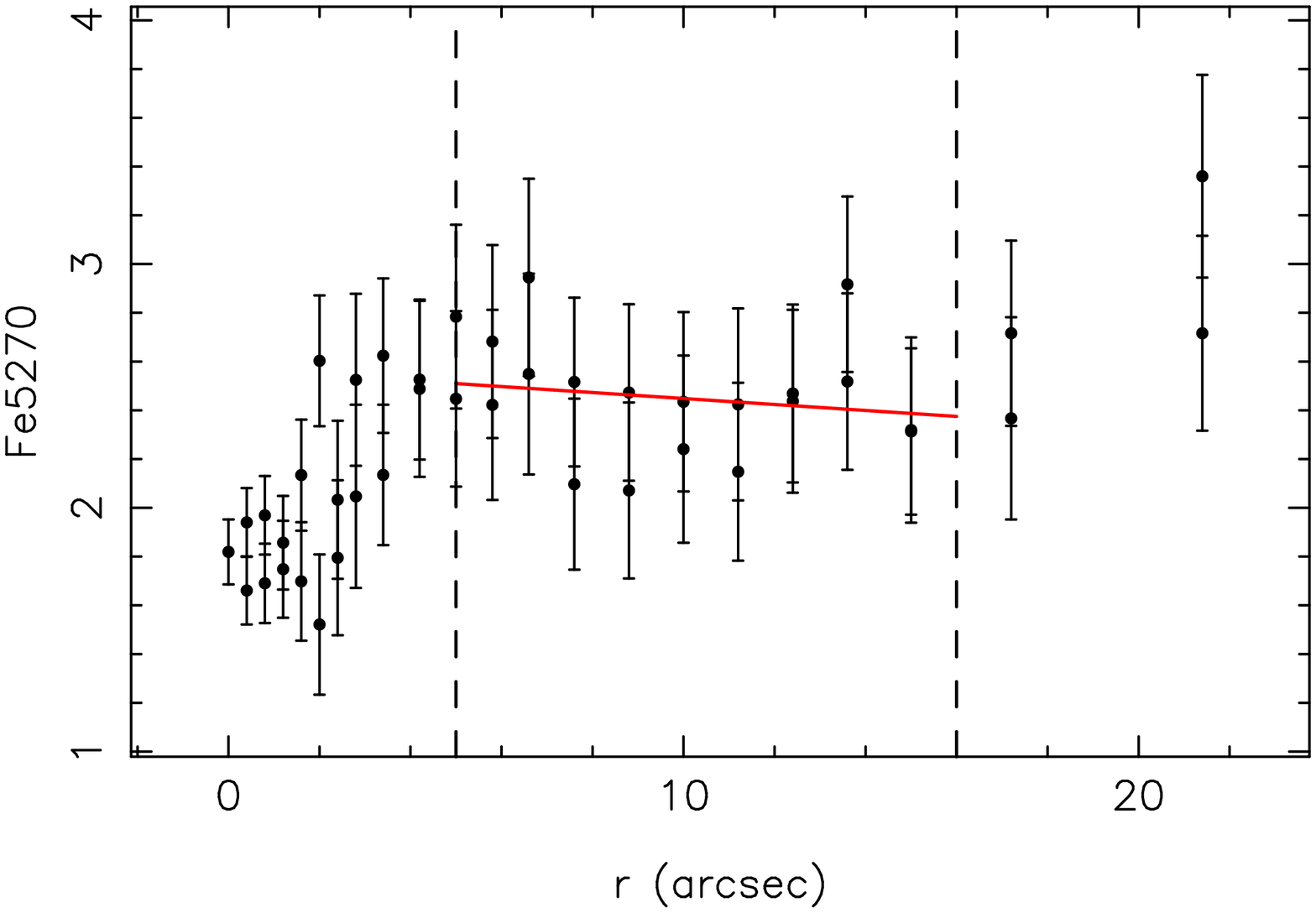}}\hspace{0.85cm}
\resizebox{0.3\textwidth}{!}{\includegraphics[angle=-90]{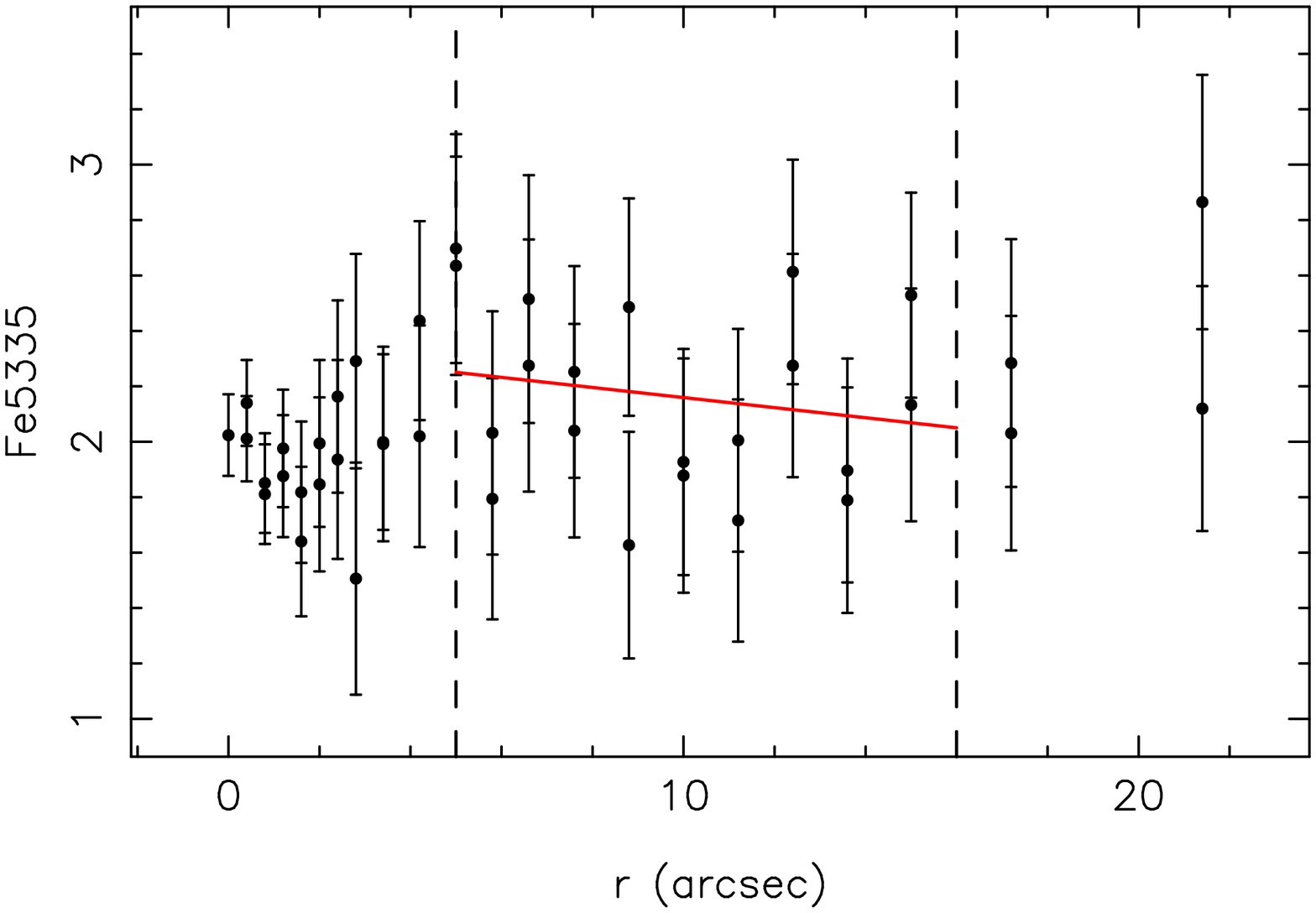}}
\caption{Line-strength distribution in the bar region for all the galaxies}
\end{figure*}

\clearpage
\begin{figure*}
\addtocounter{figure}{-1}
\resizebox{0.3\textwidth}{!}{\includegraphics[angle=-90]{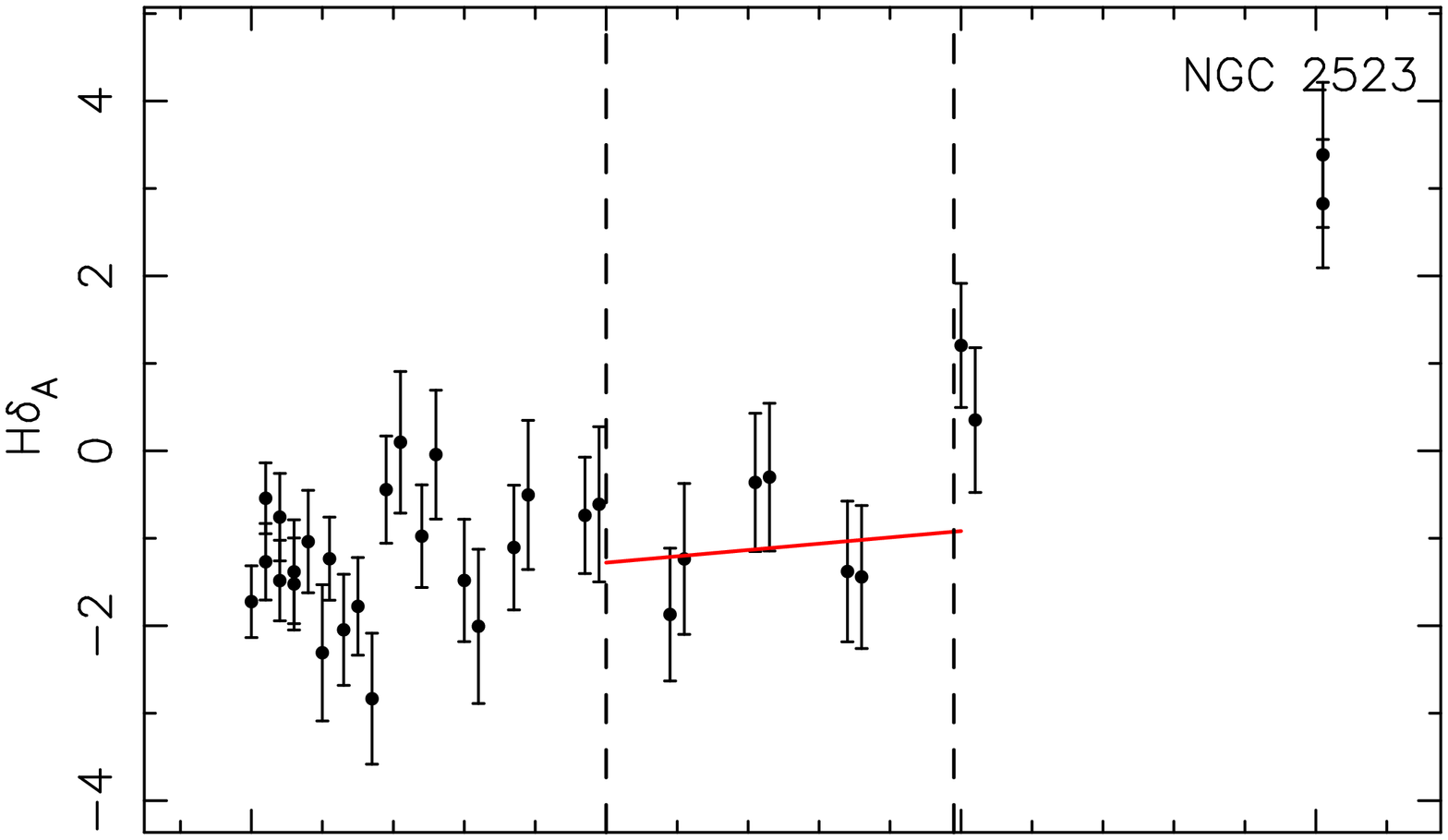}}
\resizebox{0.3\textwidth}{!}{\includegraphics[angle=-90]{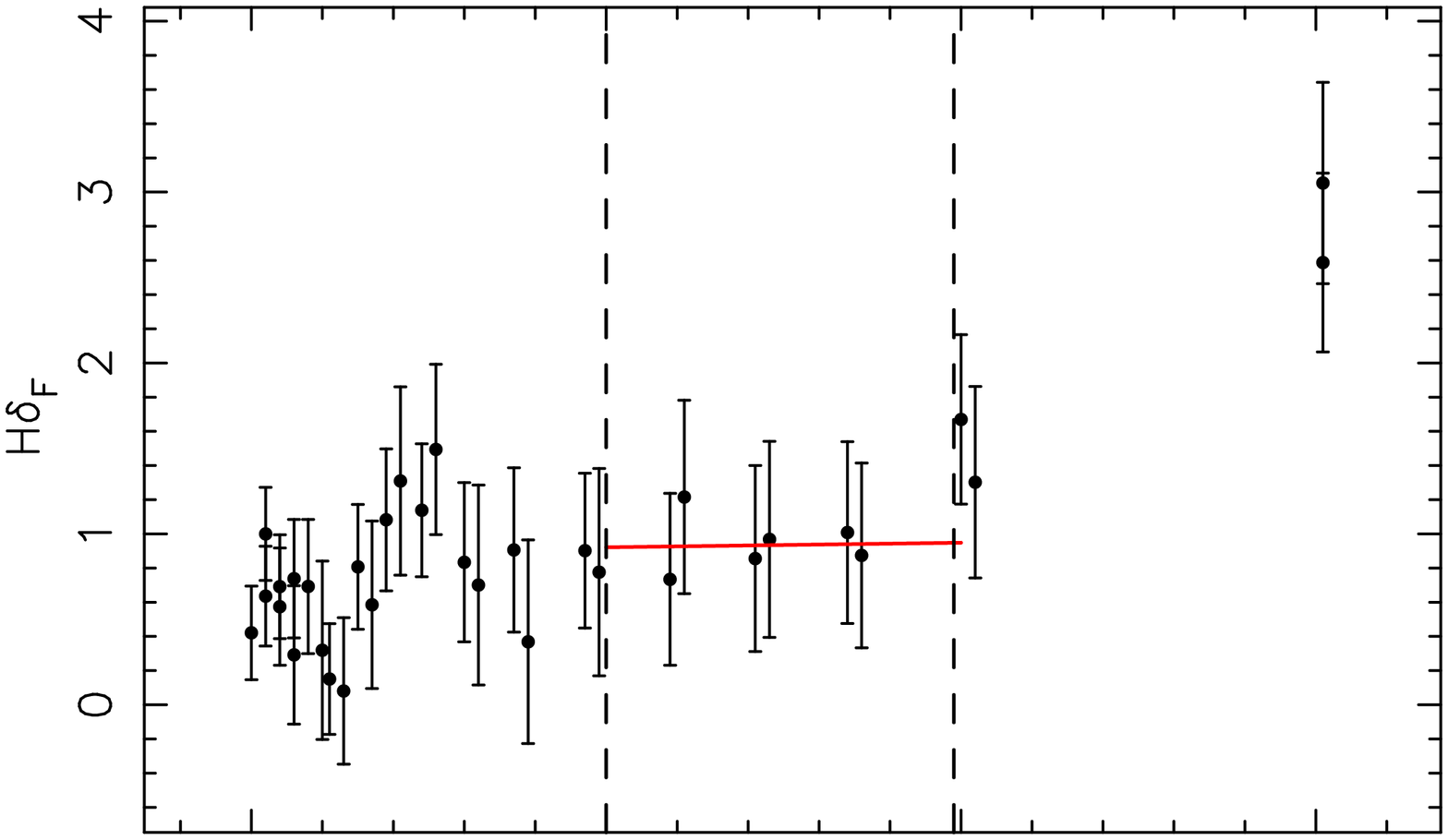}}
\resizebox{0.3\textwidth}{!}{\includegraphics[angle=-90]{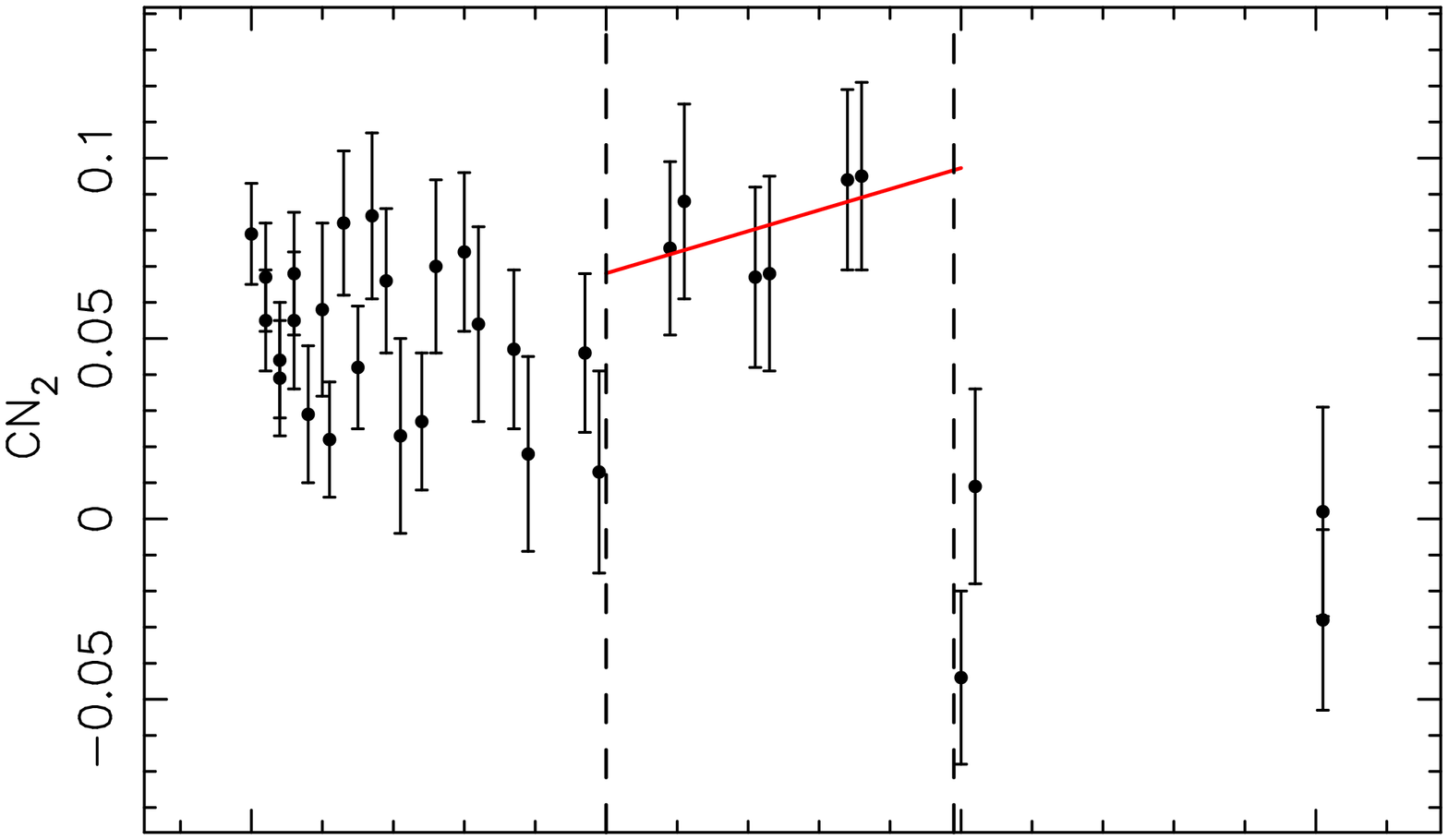}}
\resizebox{0.3\textwidth}{!}{\includegraphics[angle=-90]{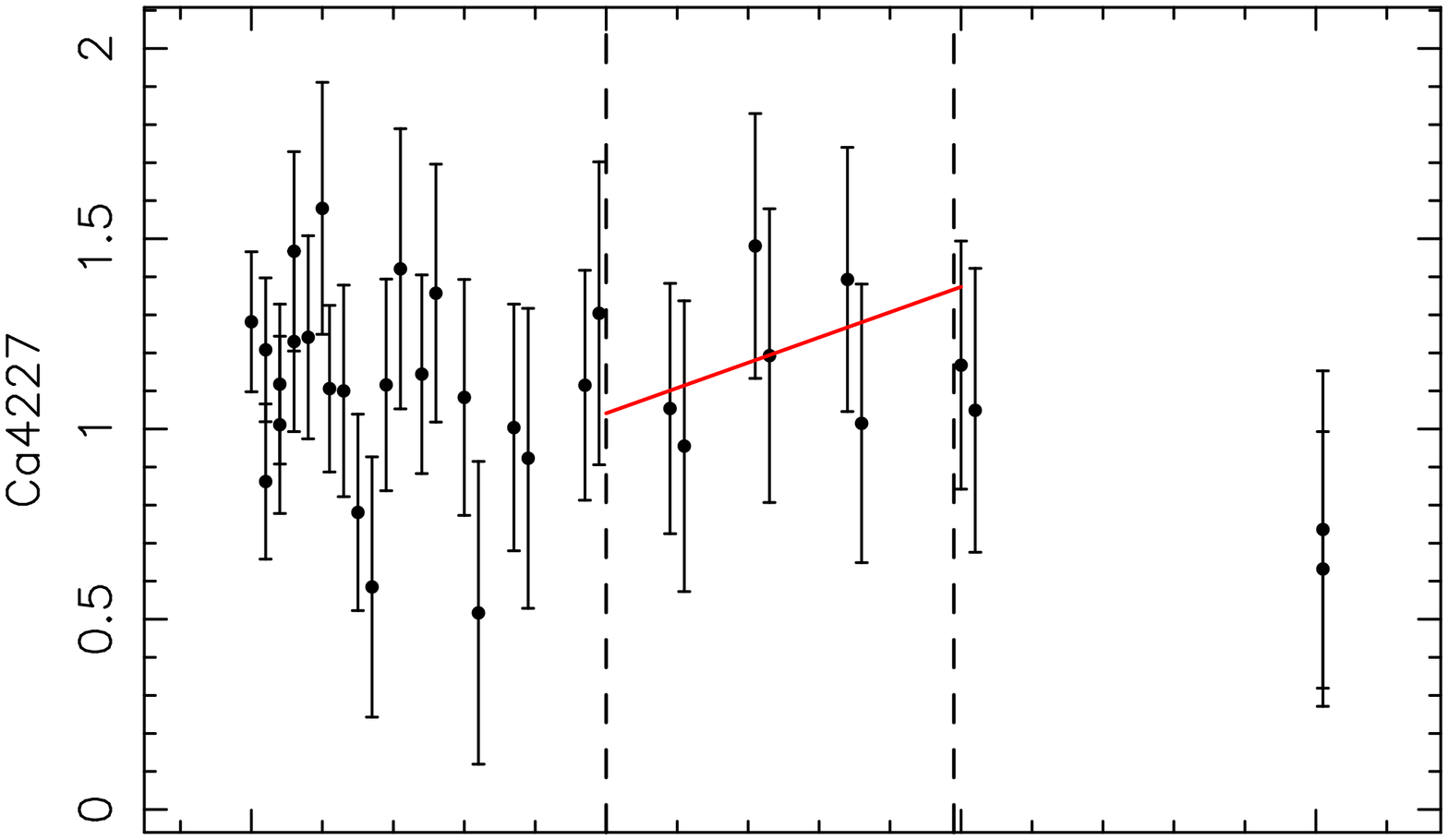}}
\resizebox{0.3\textwidth}{!}{\includegraphics[angle=-90]{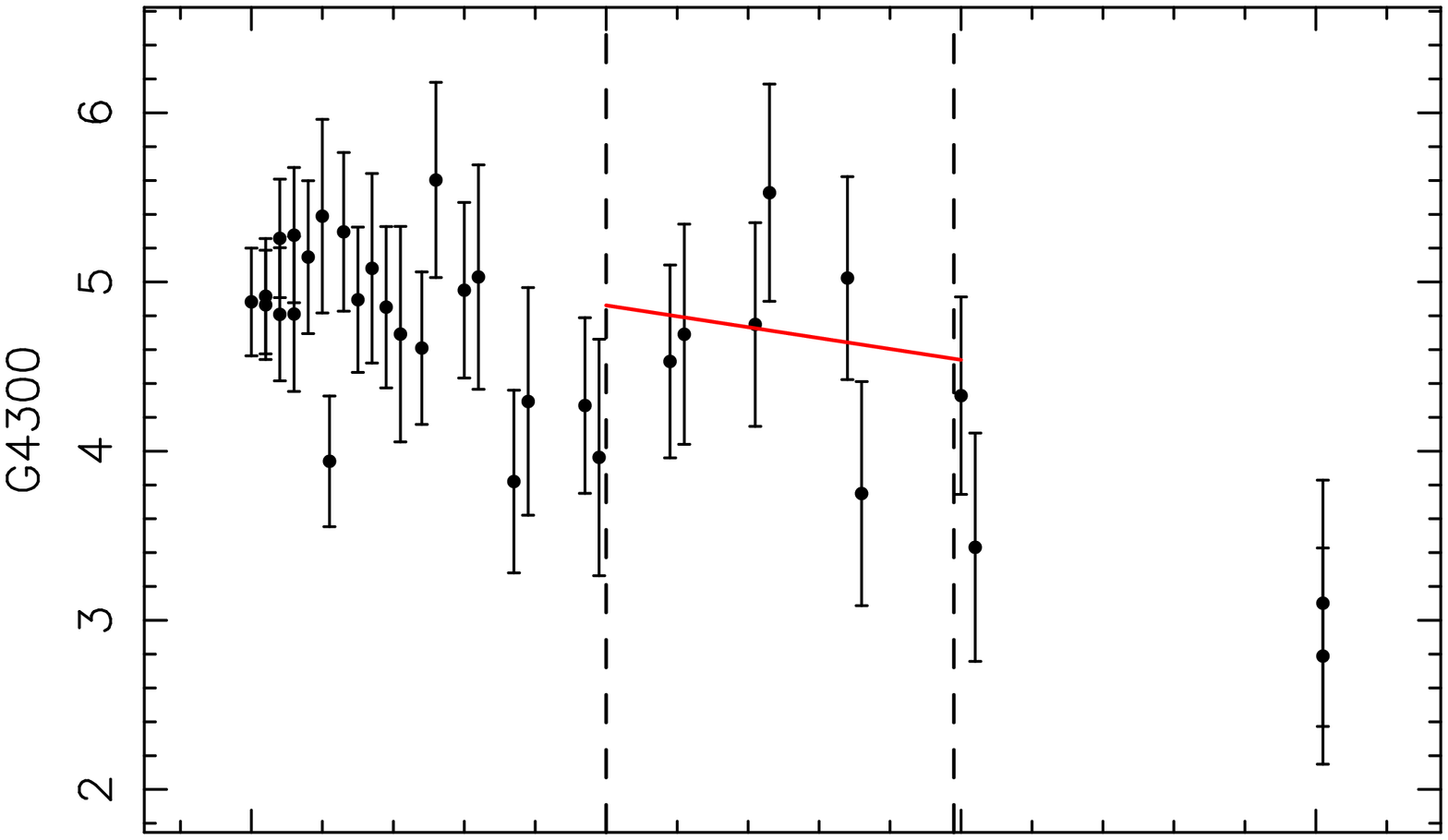}}
\resizebox{0.3\textwidth}{!}{\includegraphics[angle=-90]{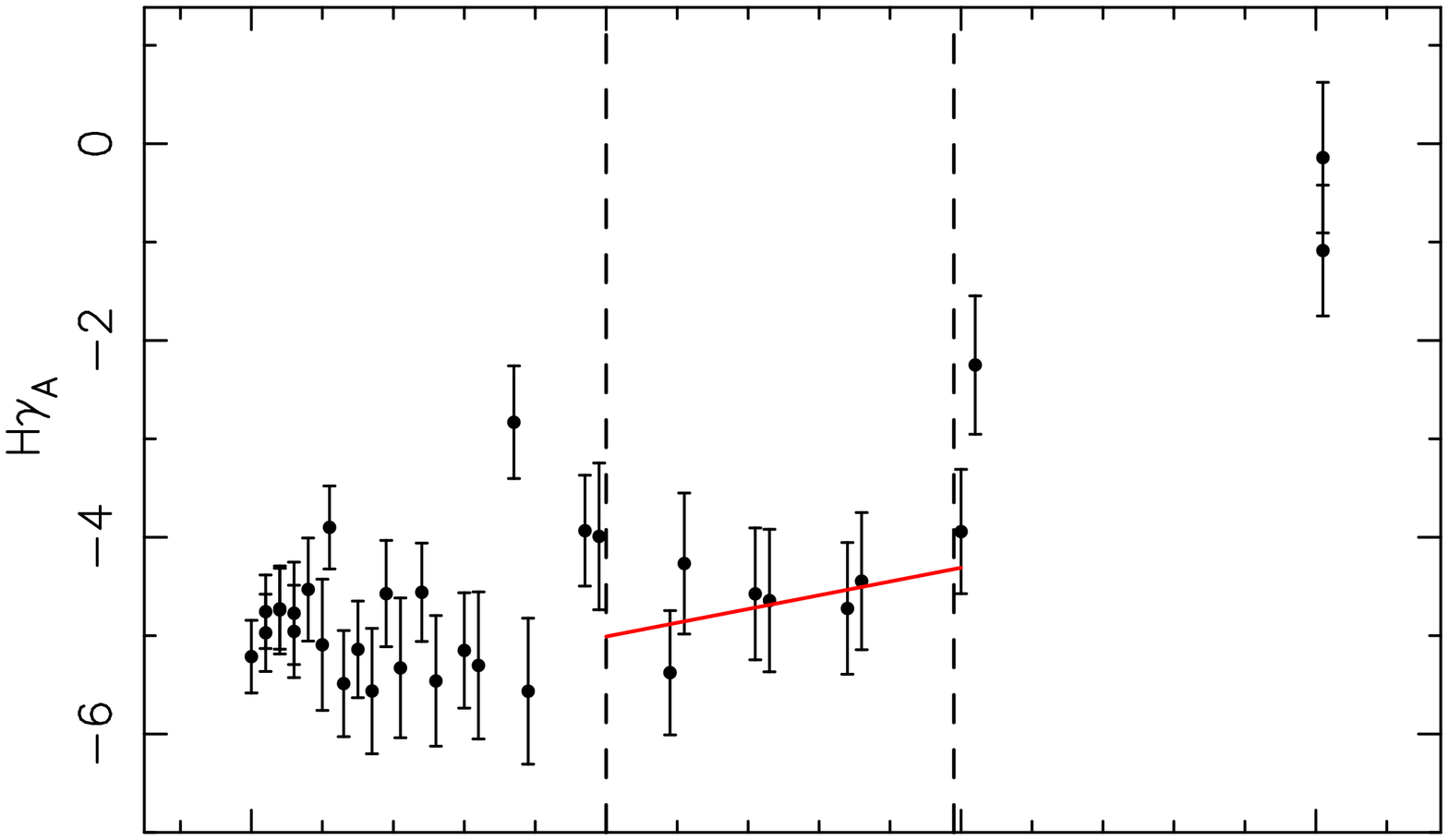}}
\resizebox{0.3\textwidth}{!}{\includegraphics[angle=-90]{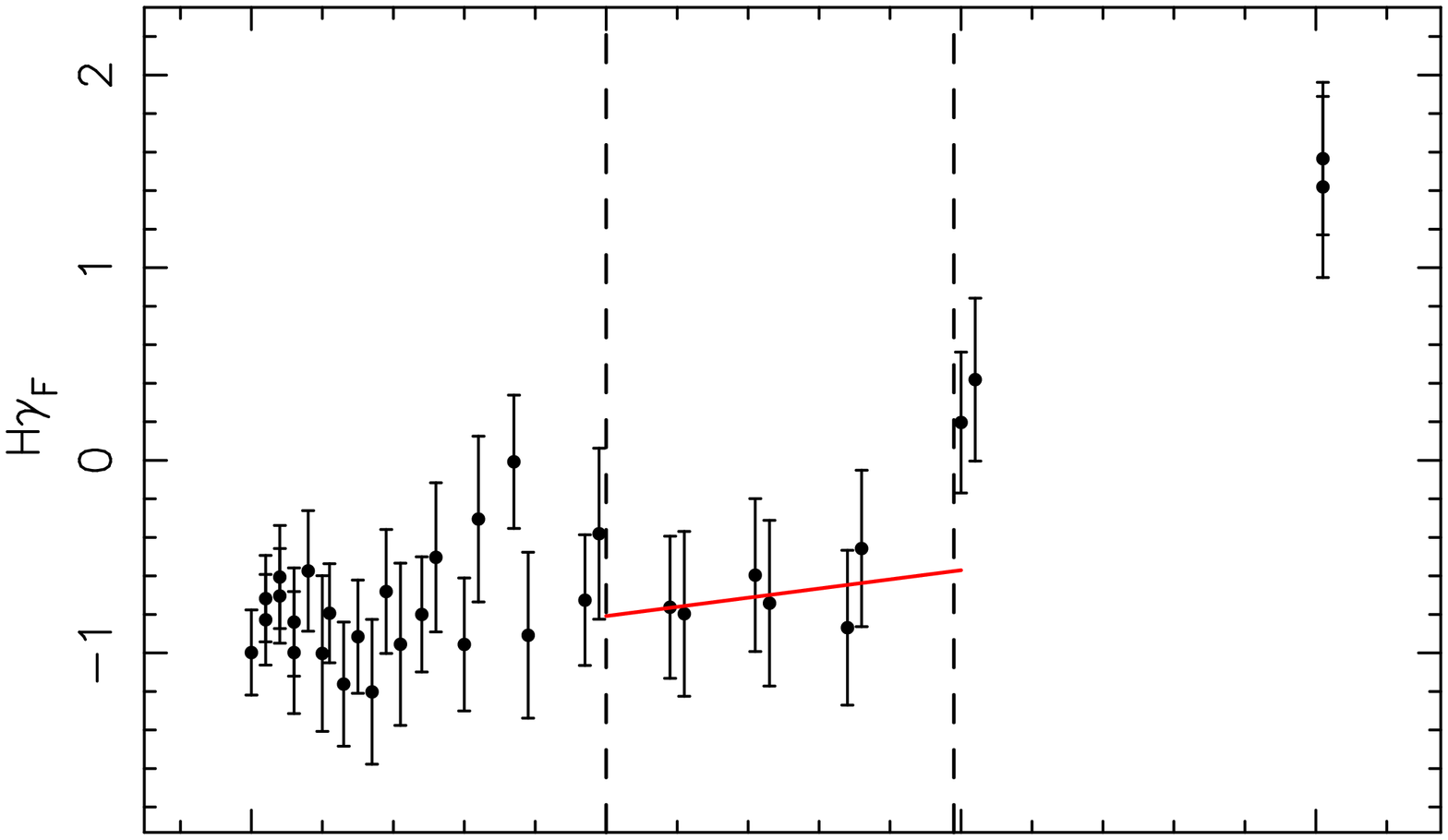}}
\resizebox{0.3\textwidth}{!}{\includegraphics[angle=-90]{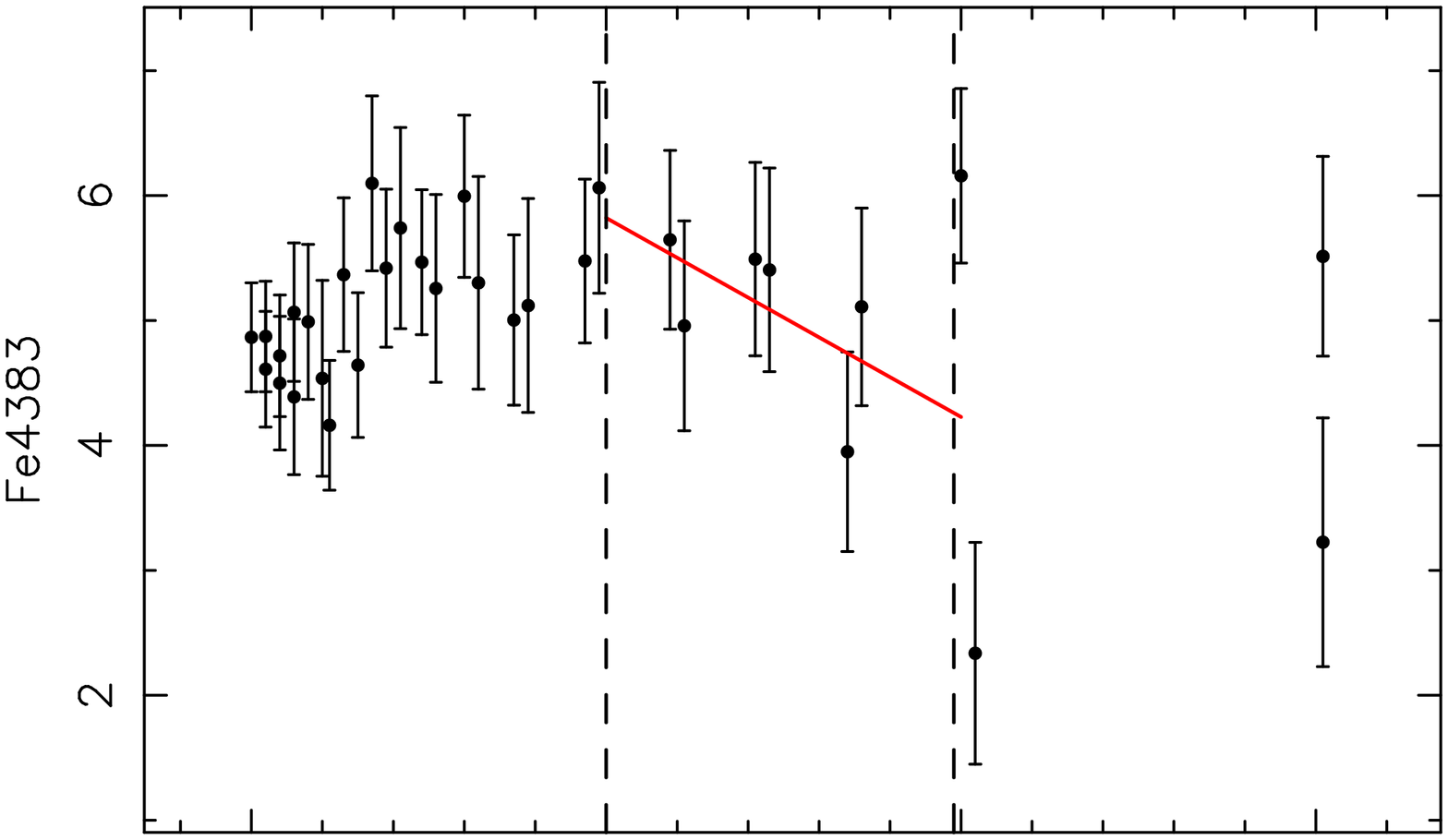}}
\resizebox{0.3\textwidth}{!}{\includegraphics[angle=-90]{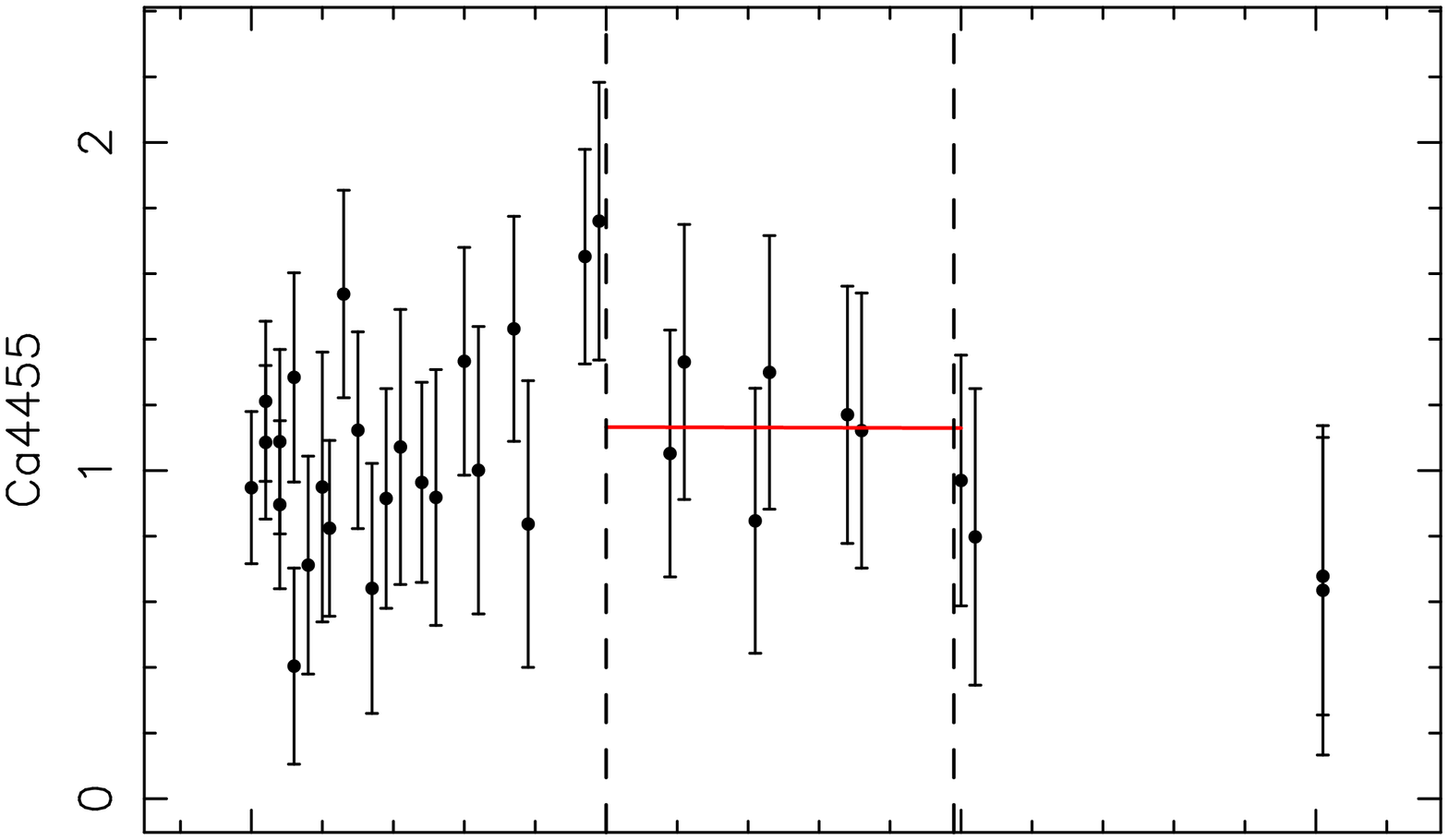}}
\resizebox{0.3\textwidth}{!}{\includegraphics[angle=-90]{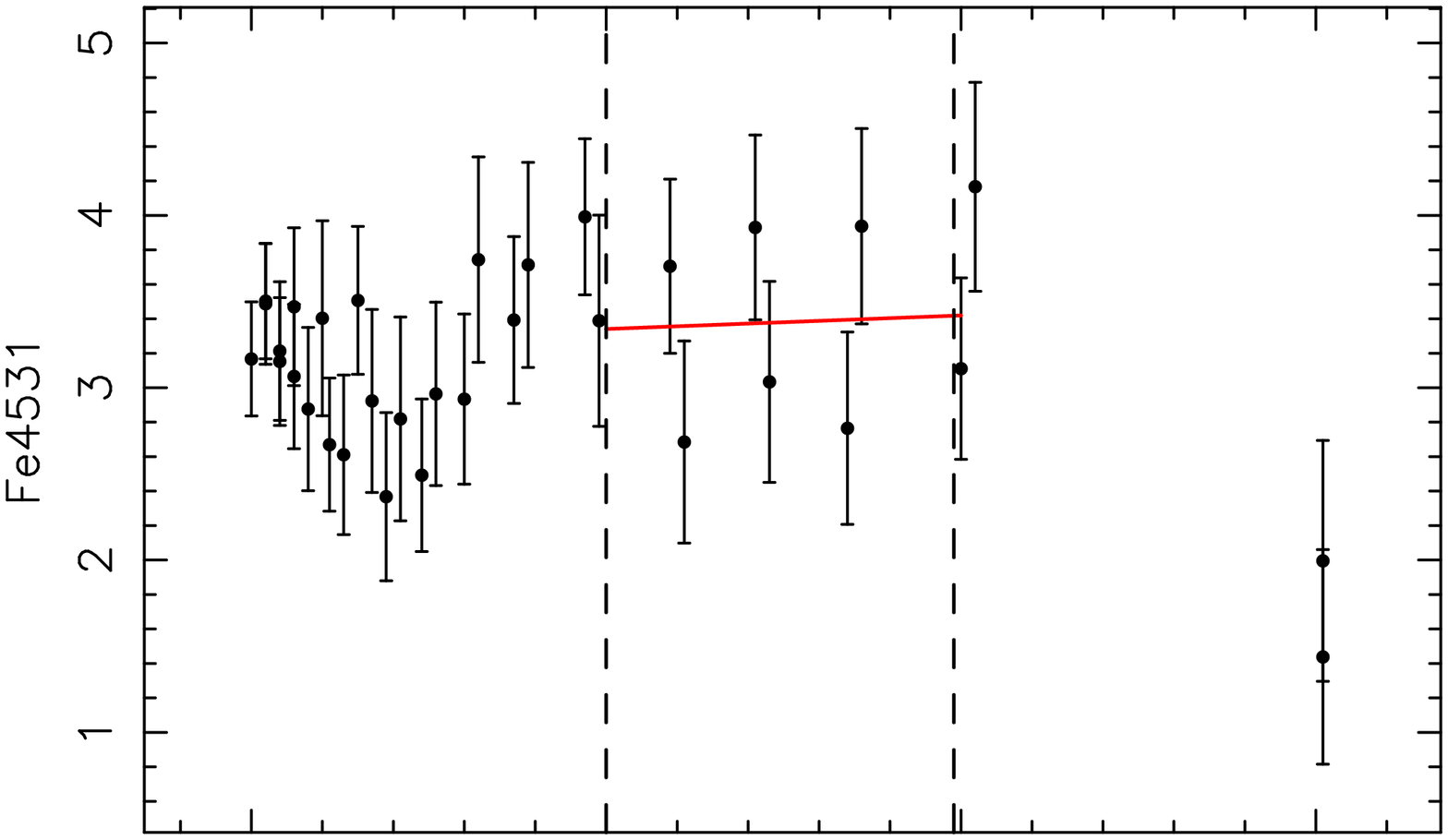}}
\resizebox{0.3\textwidth}{!}{\includegraphics[angle=-90]{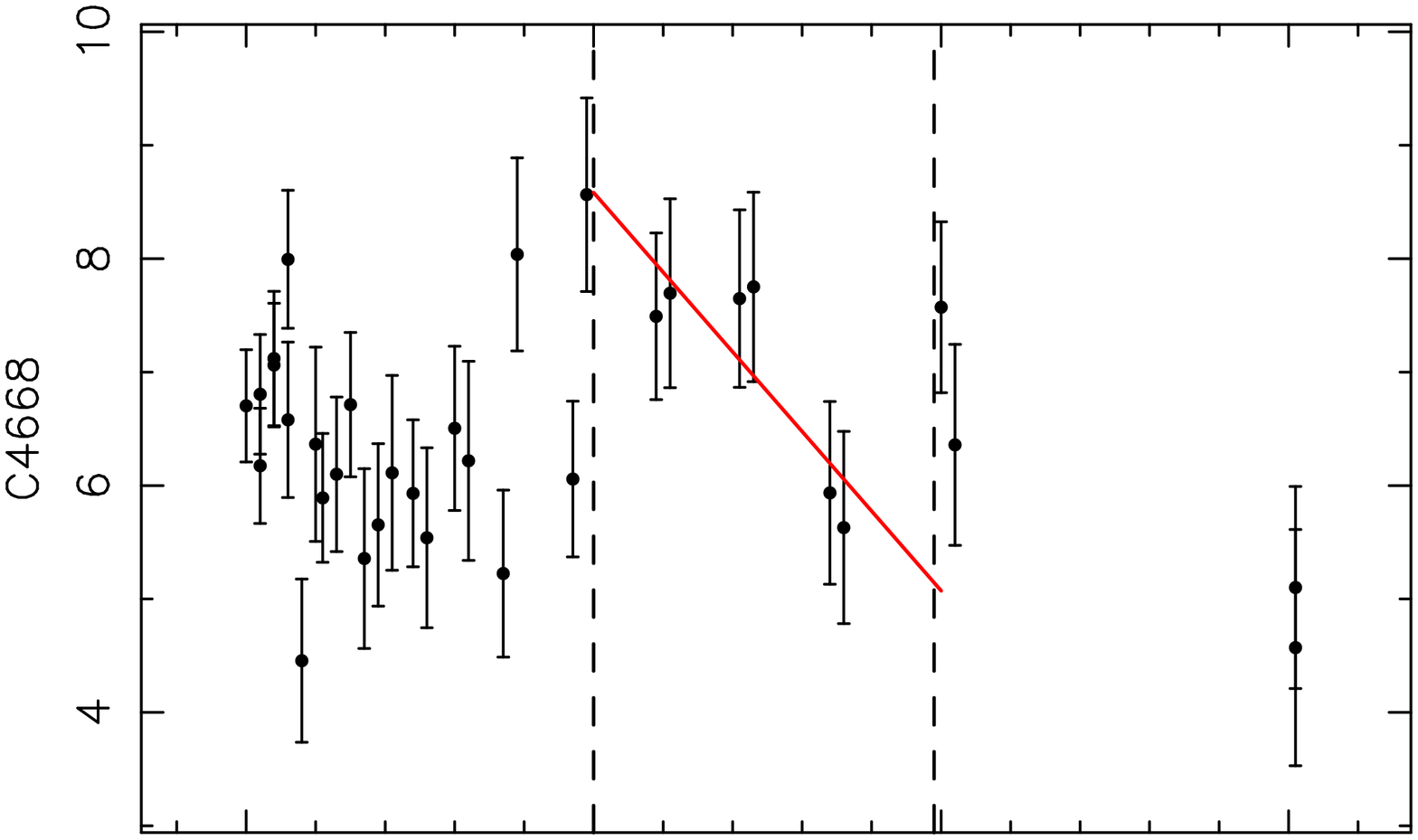}}
\resizebox{0.3\textwidth}{!}{\includegraphics[angle=-90]{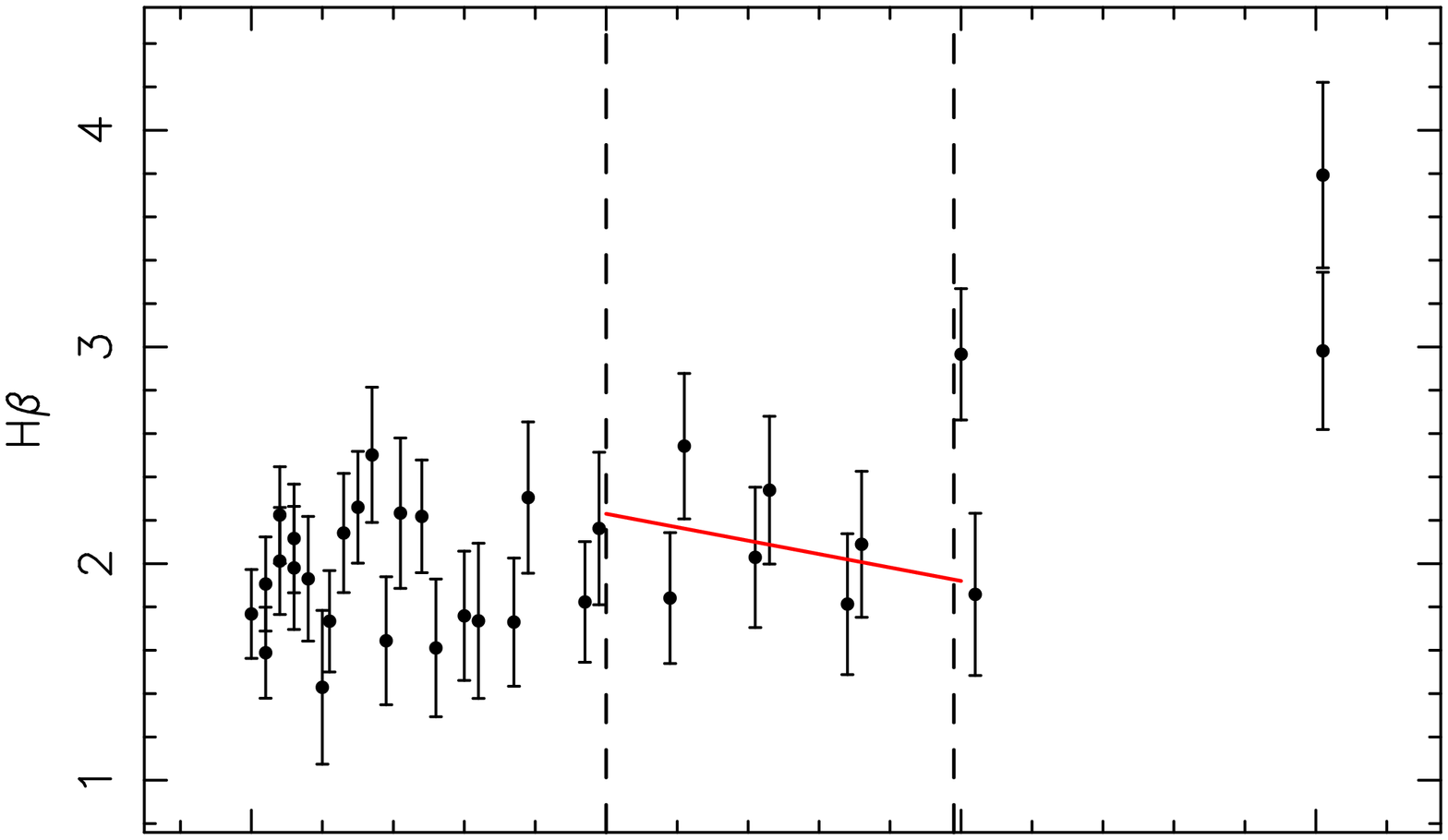}}
\resizebox{0.3\textwidth}{!}{\includegraphics[angle=-90]{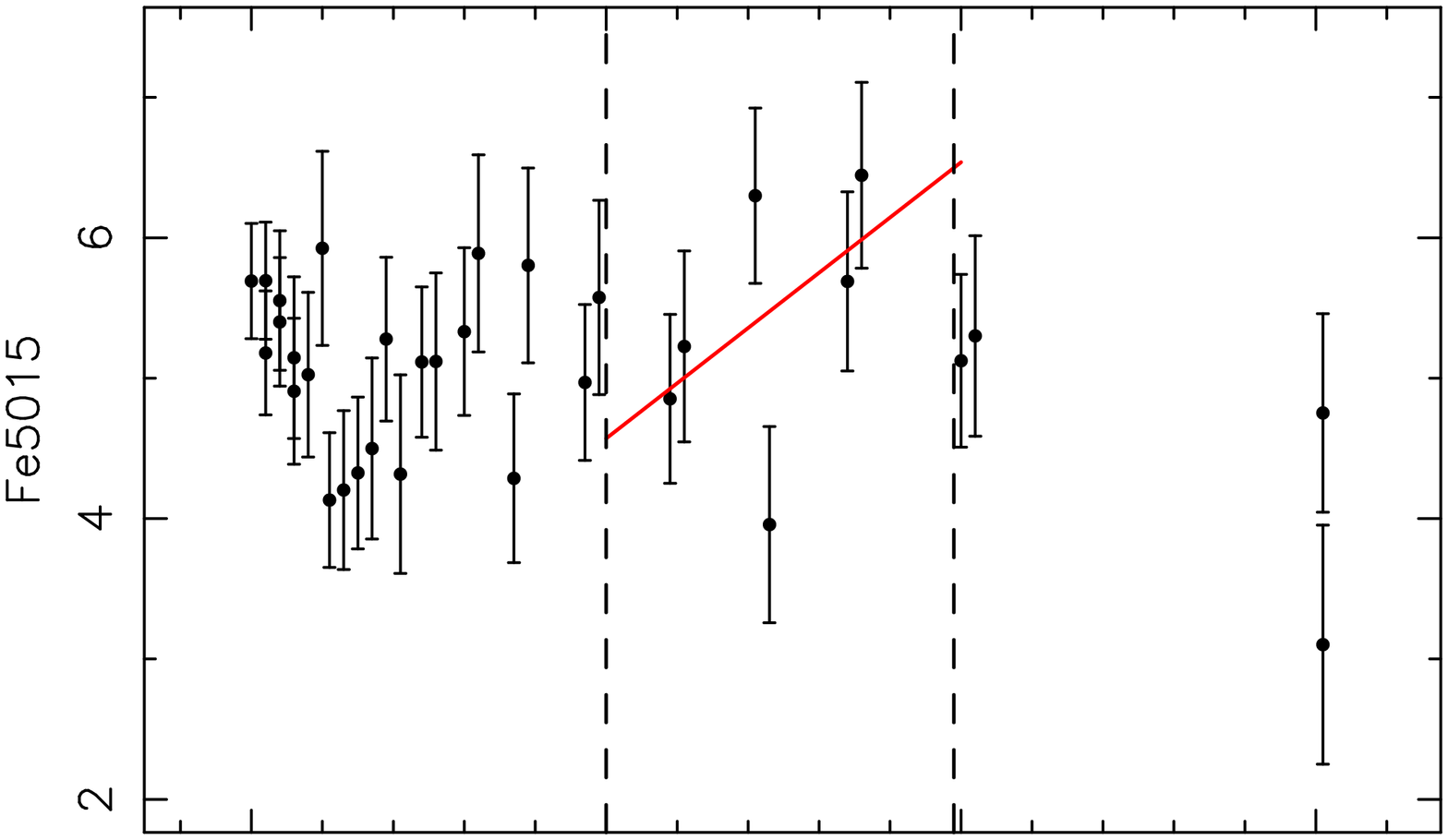}}
\resizebox{0.3\textwidth}{!}{\includegraphics[angle=-90]{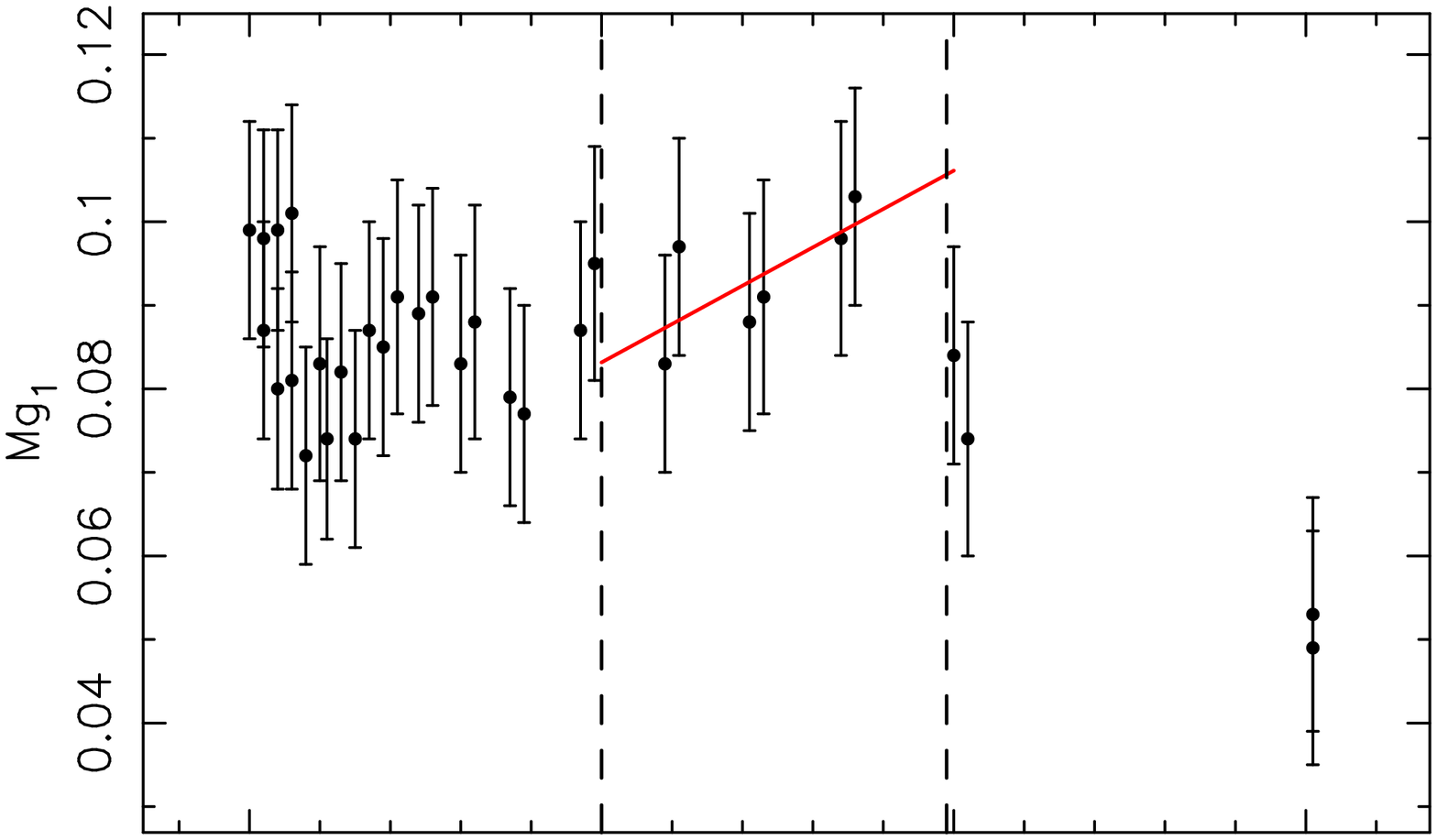}}
\resizebox{0.3\textwidth}{!}{\includegraphics[angle=-90]{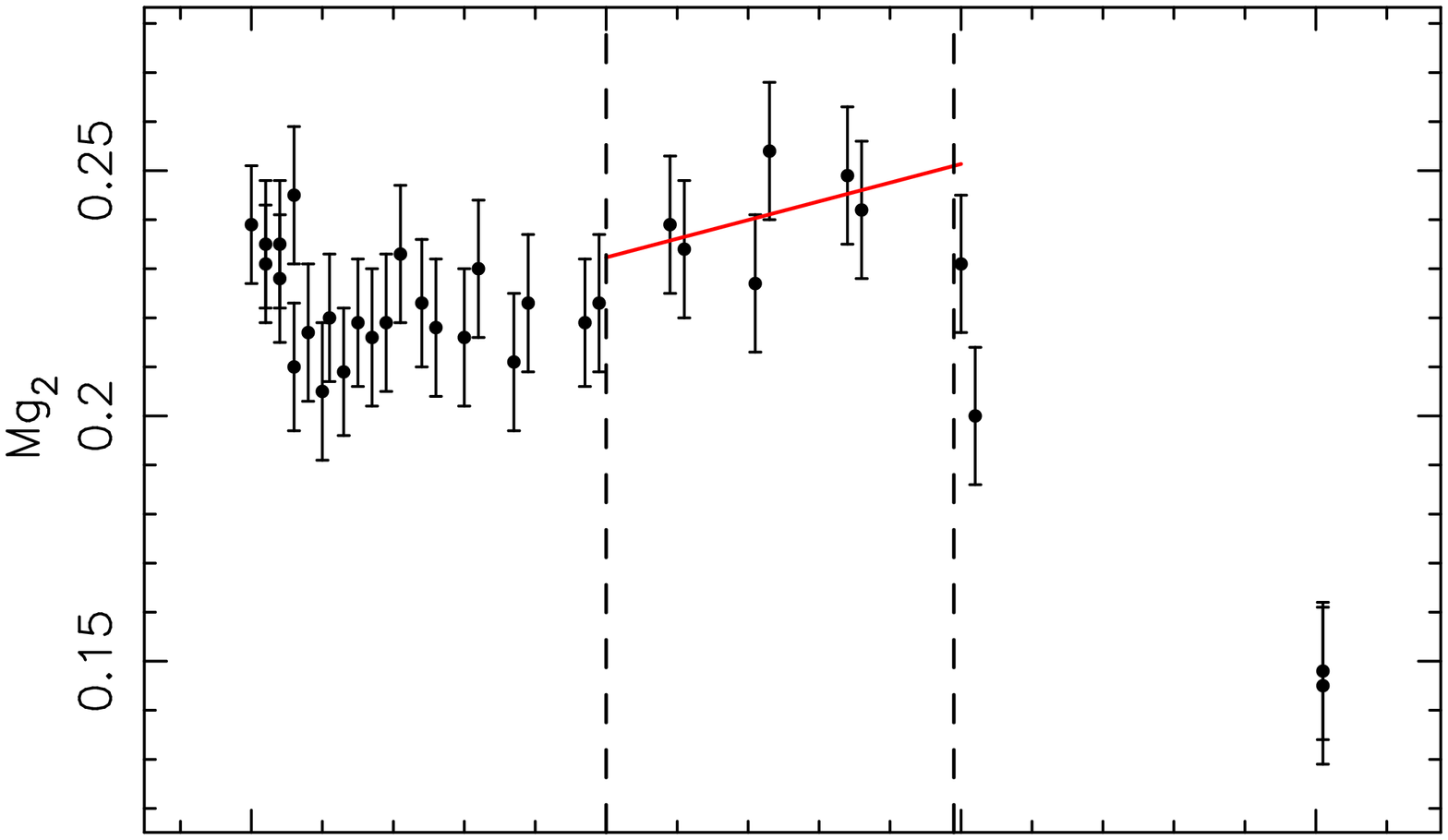}}
\resizebox{0.3\textwidth}{!}{\includegraphics[angle=-90]{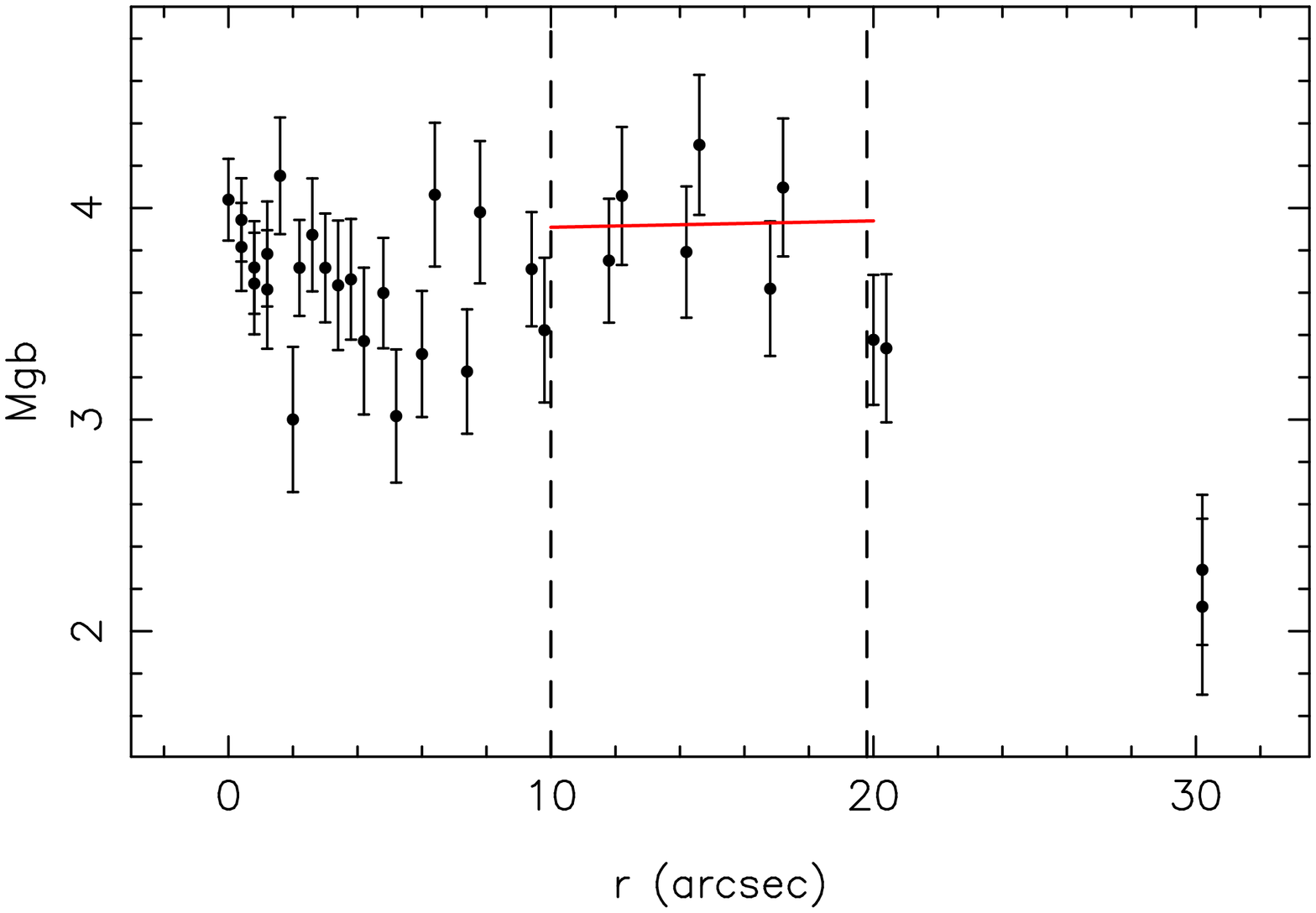}}\hspace{0.85cm}
\resizebox{0.3\textwidth}{!}{\includegraphics[angle=-90]{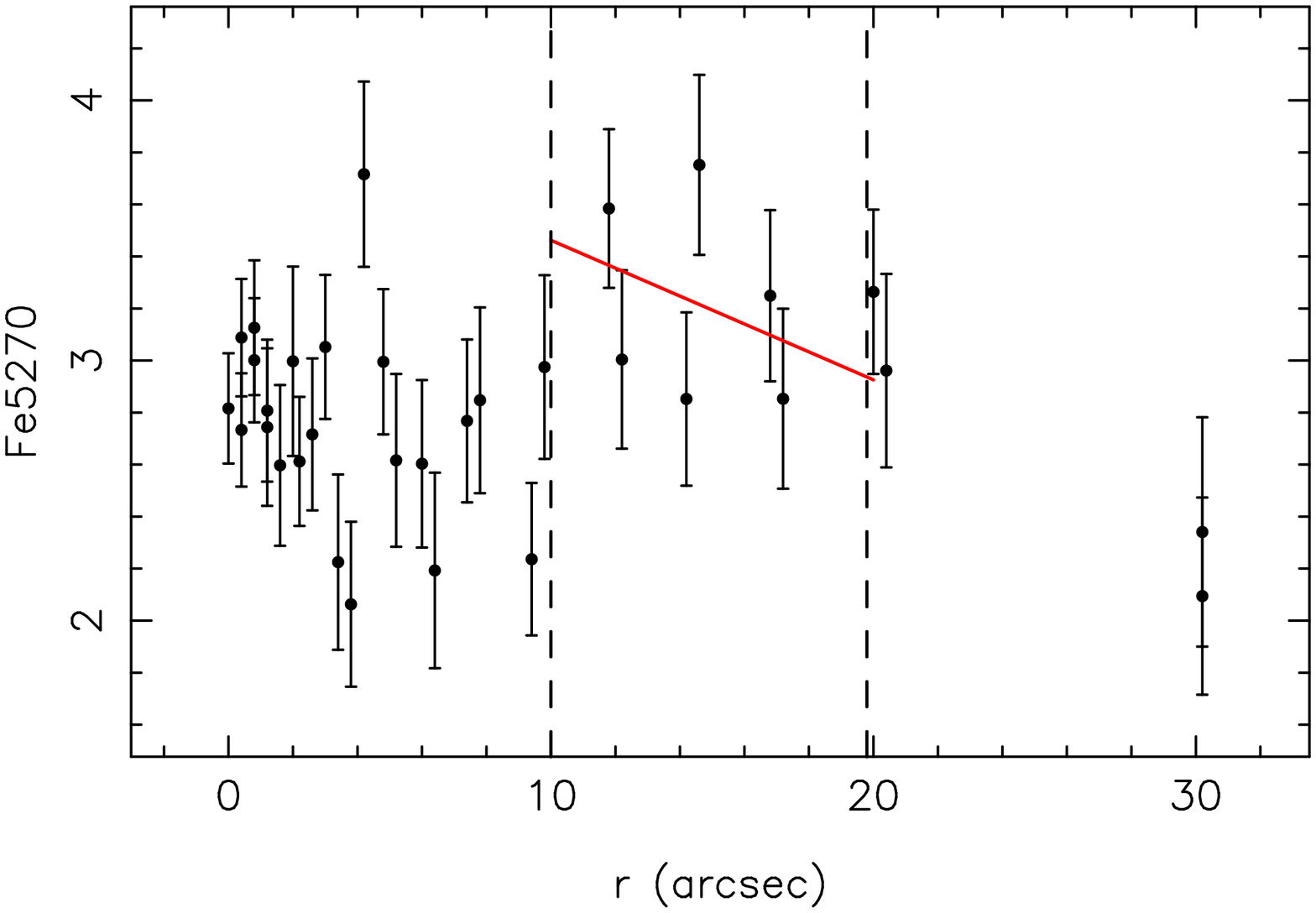}}\hspace{0.85cm}
\resizebox{0.3\textwidth}{!}{\includegraphics[angle=-90]{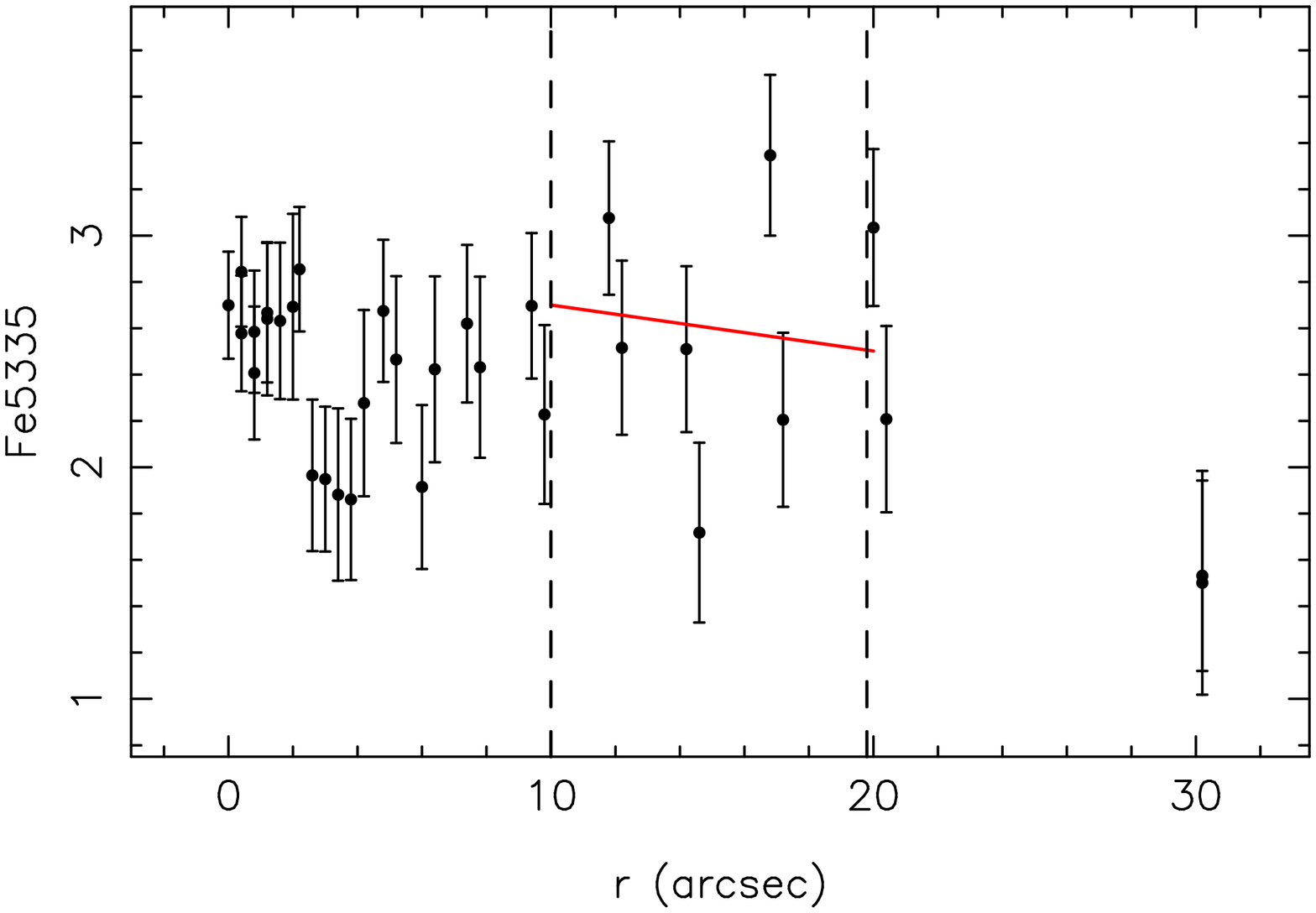}}
\caption{Line-strength distribution in the bar region for all the galaxies}
\end{figure*}

\clearpage
\begin{figure*}
\addtocounter{figure}{-1}
\resizebox{0.3\textwidth}{!}{\includegraphics[angle=-90]{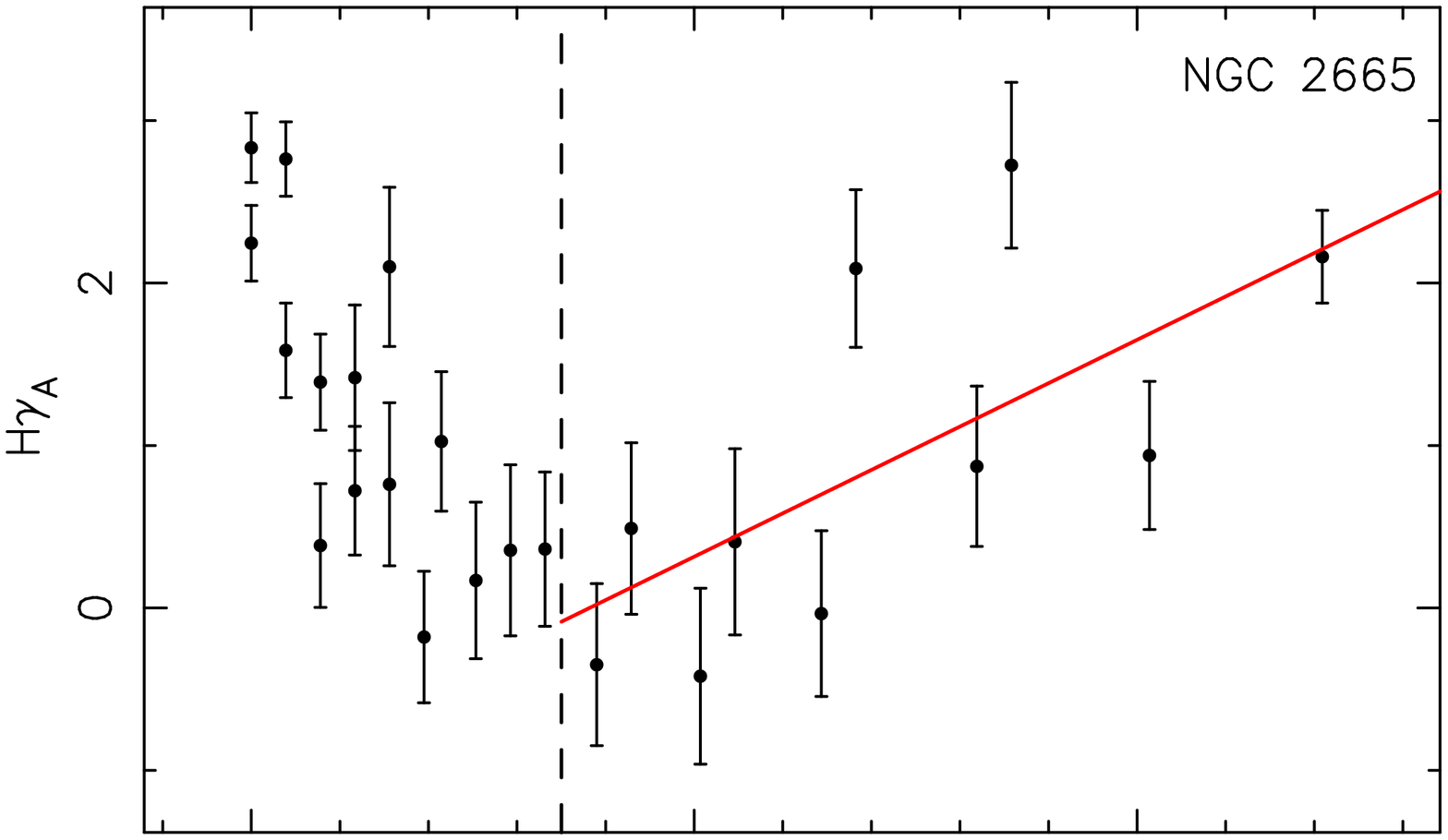}}\hspace{0.85cm}
\resizebox{0.3\textwidth}{!}{\includegraphics[angle=-90]{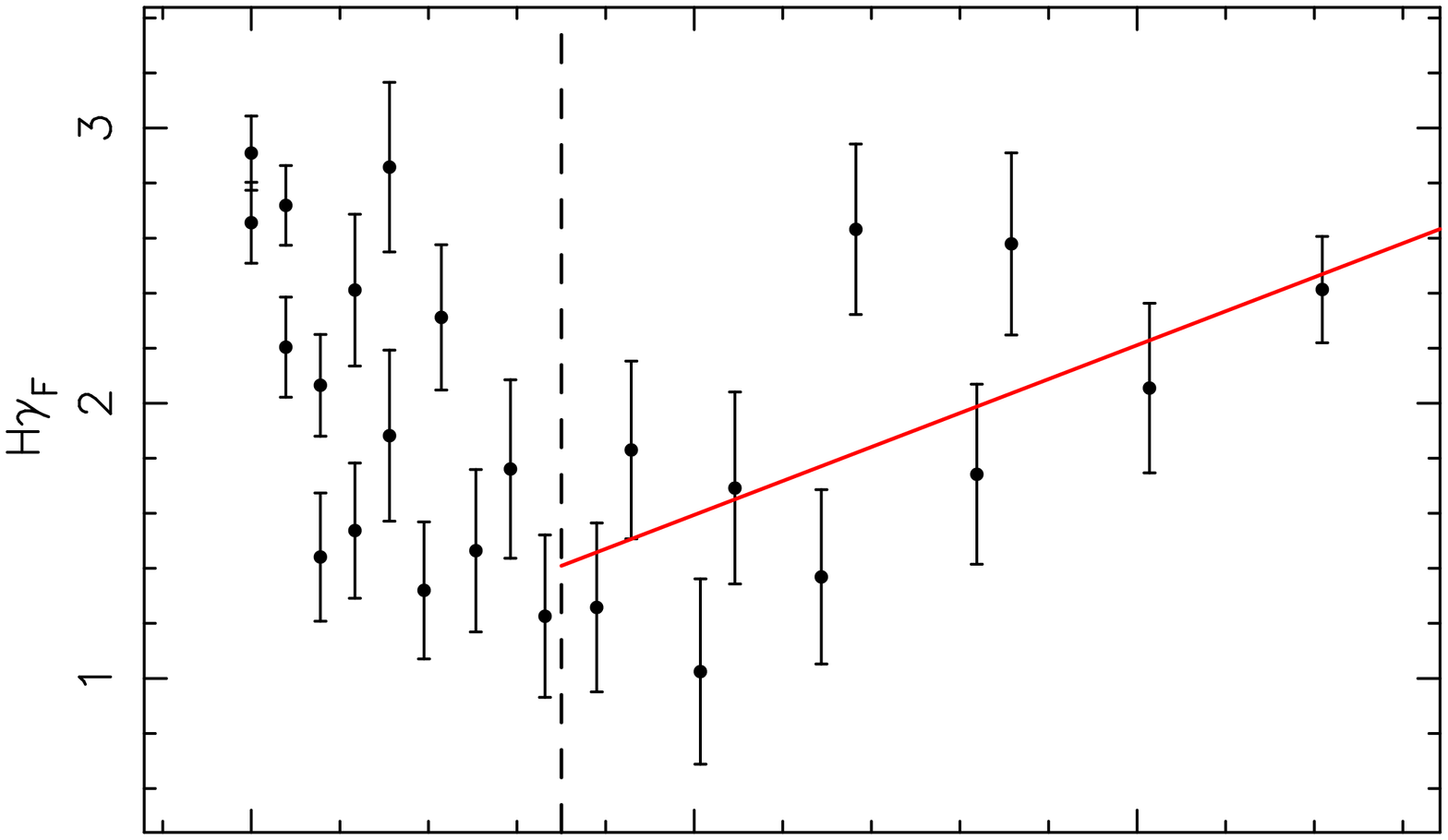}}\\
\resizebox{0.3\textwidth}{!}{\includegraphics[angle=-90]{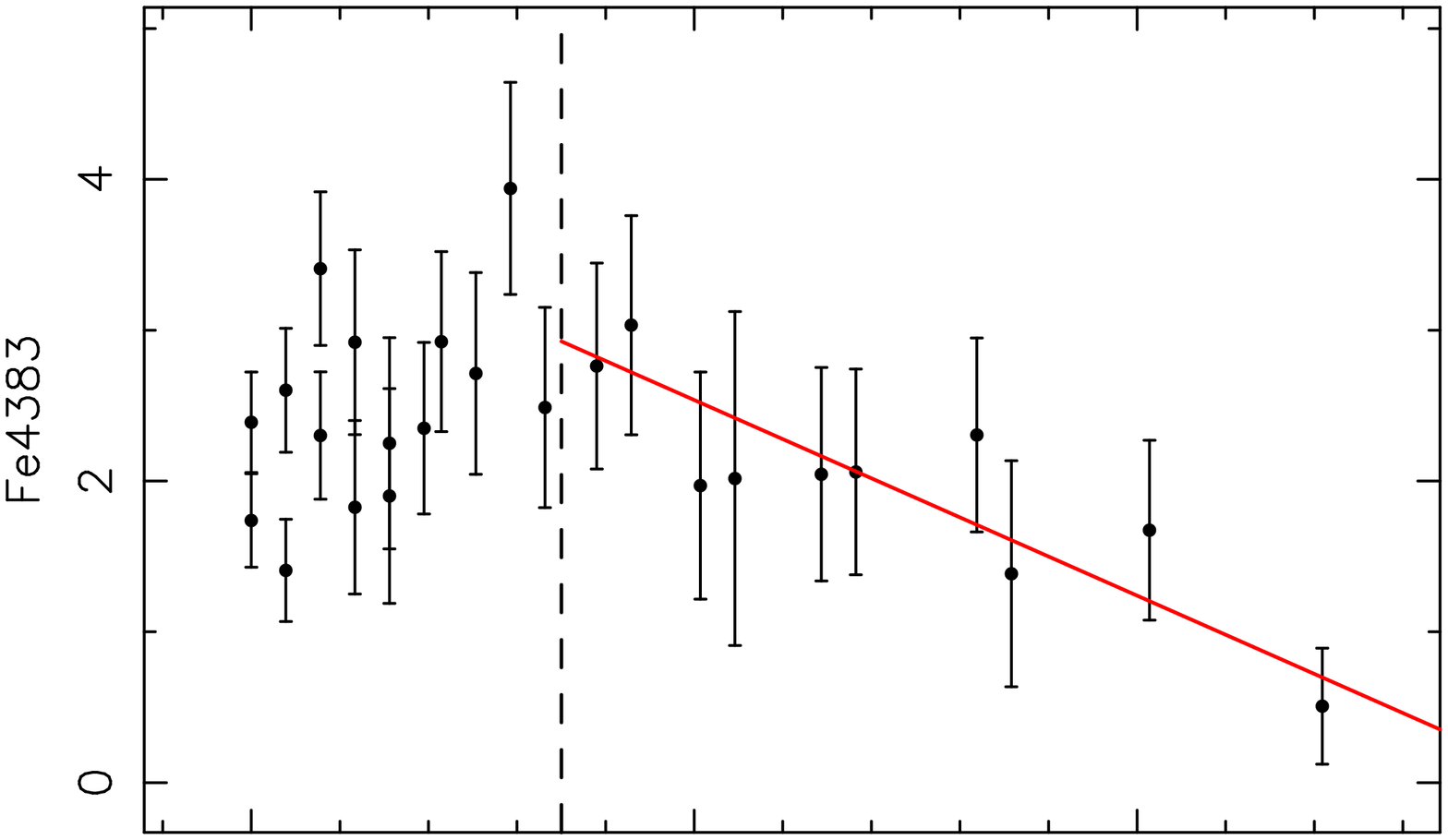}}
\resizebox{0.3\textwidth}{!}{\includegraphics[angle=-90]{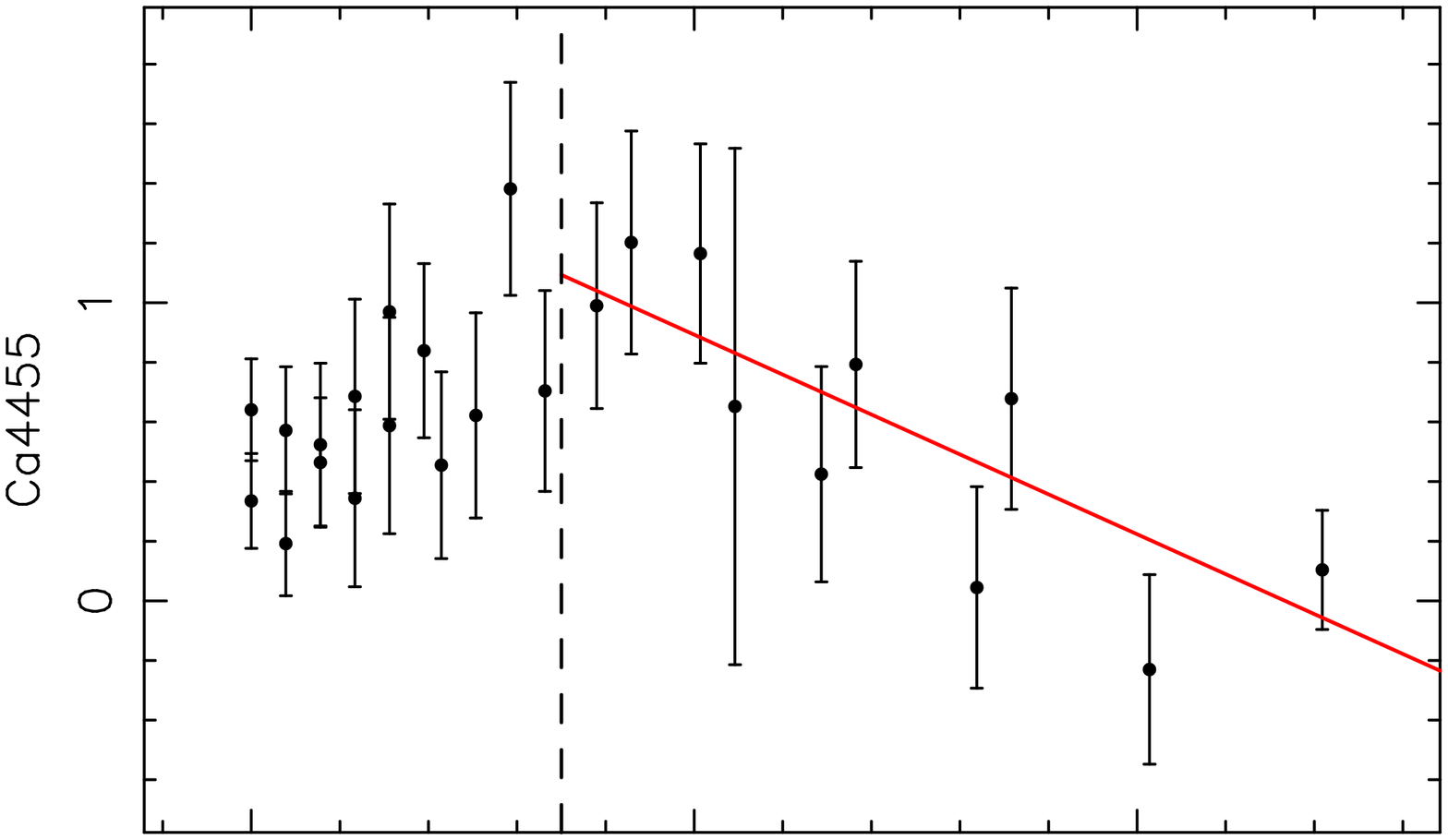}}
\resizebox{0.3\textwidth}{!}{\includegraphics[angle=-90]{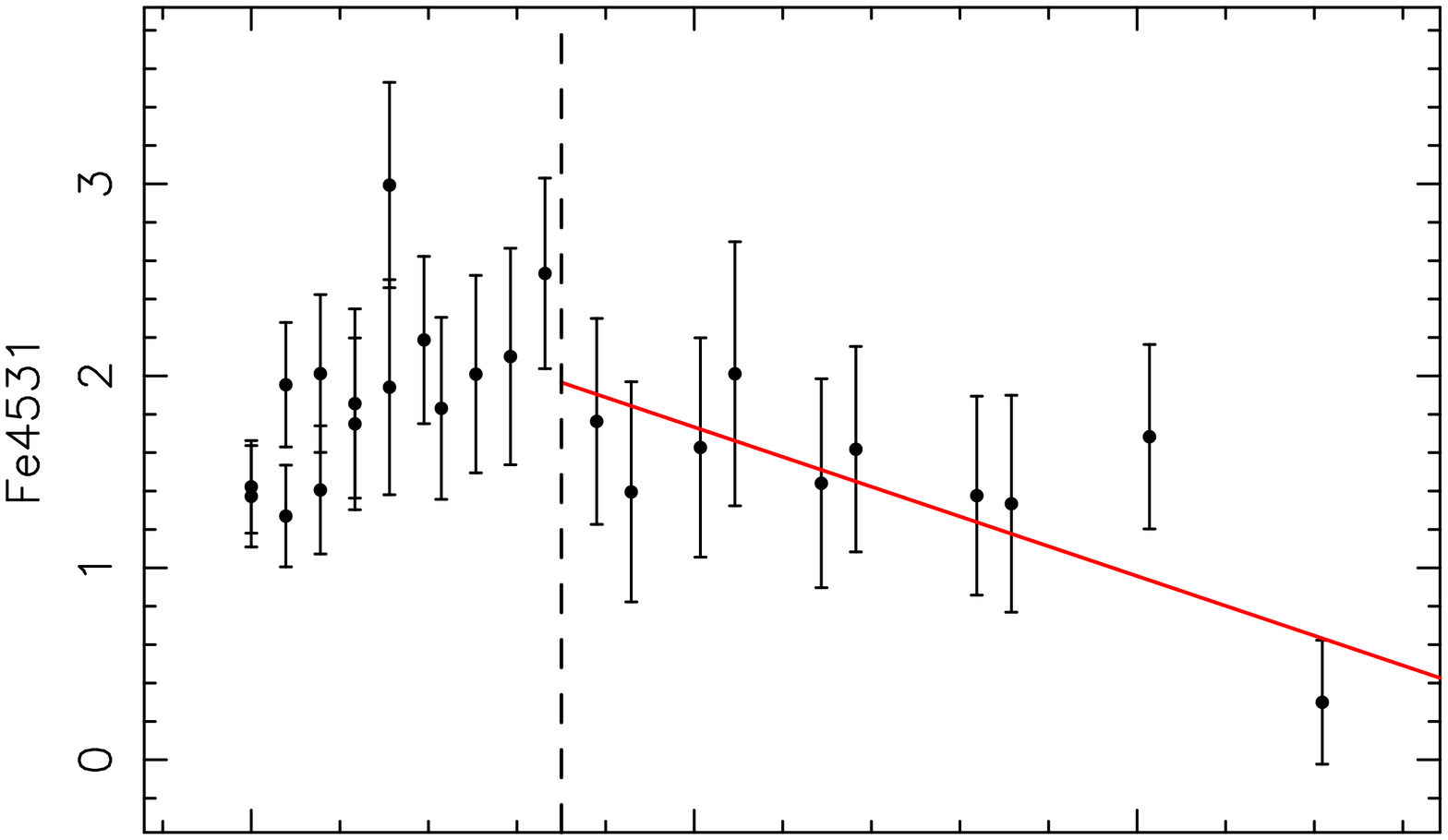}}
\resizebox{0.3\textwidth}{!}{\includegraphics[angle=-90]{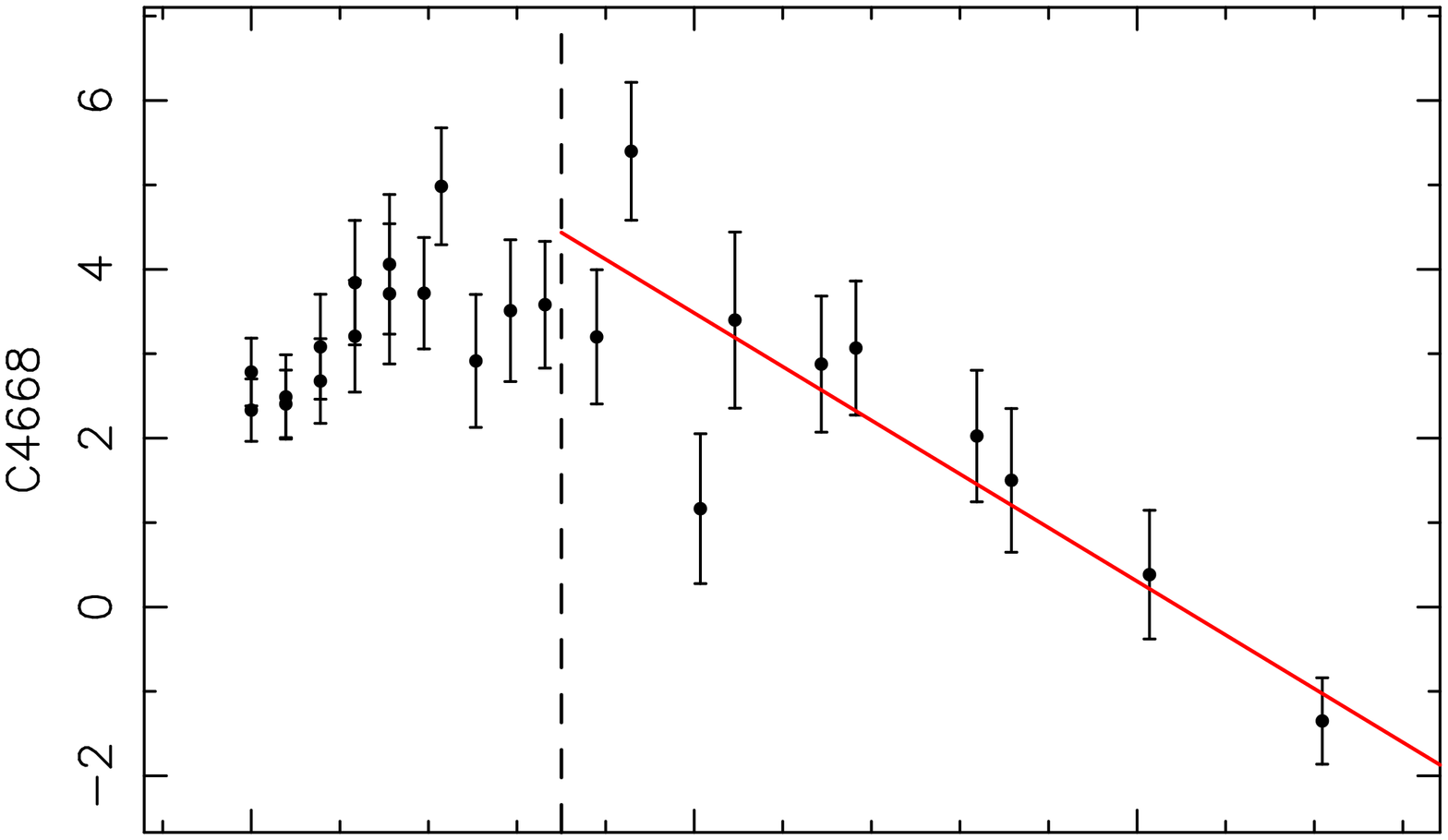}}
\resizebox{0.3\textwidth}{!}{\includegraphics[angle=-90]{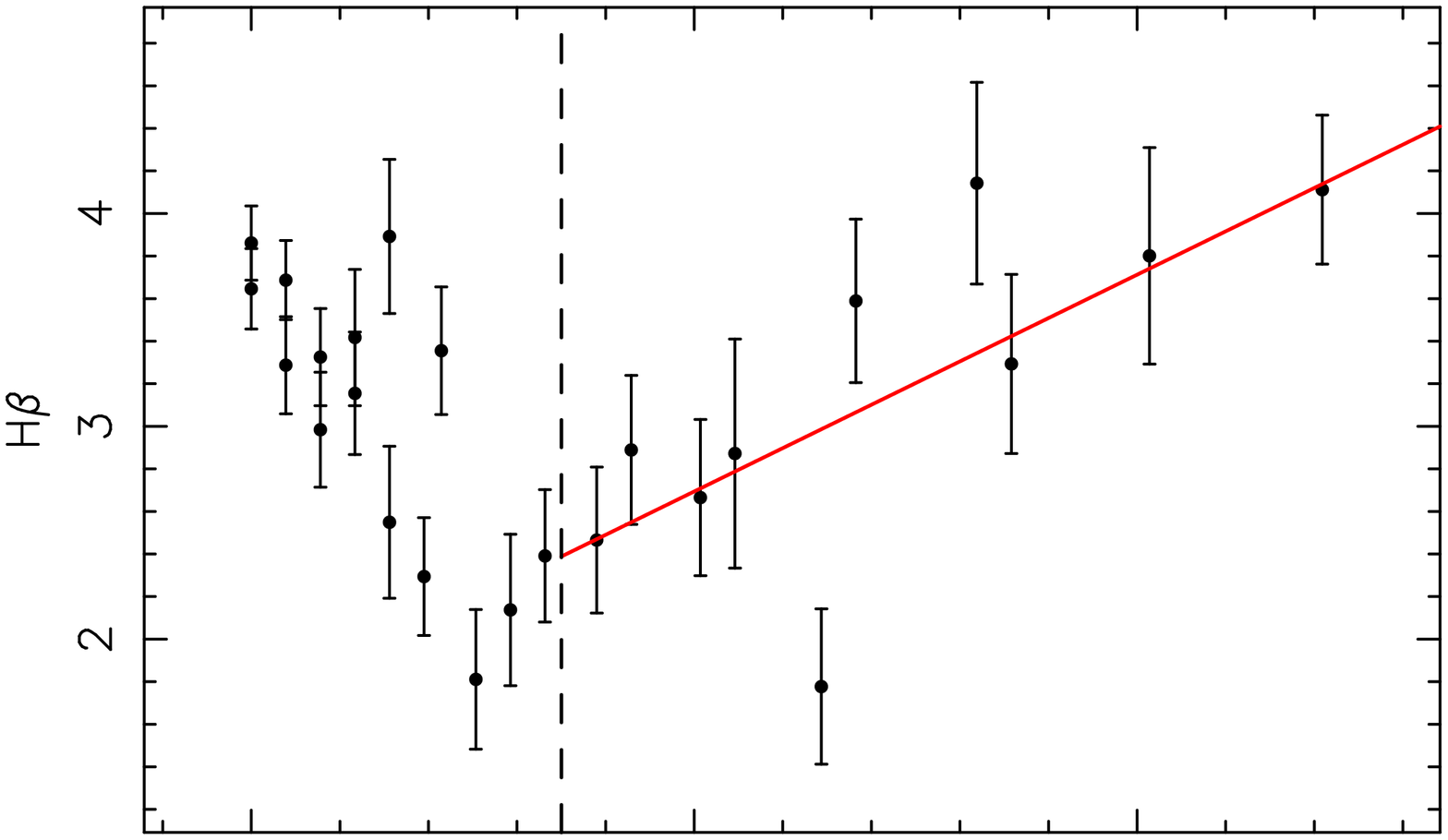}}
\resizebox{0.3\textwidth}{!}{\includegraphics[angle=-90]{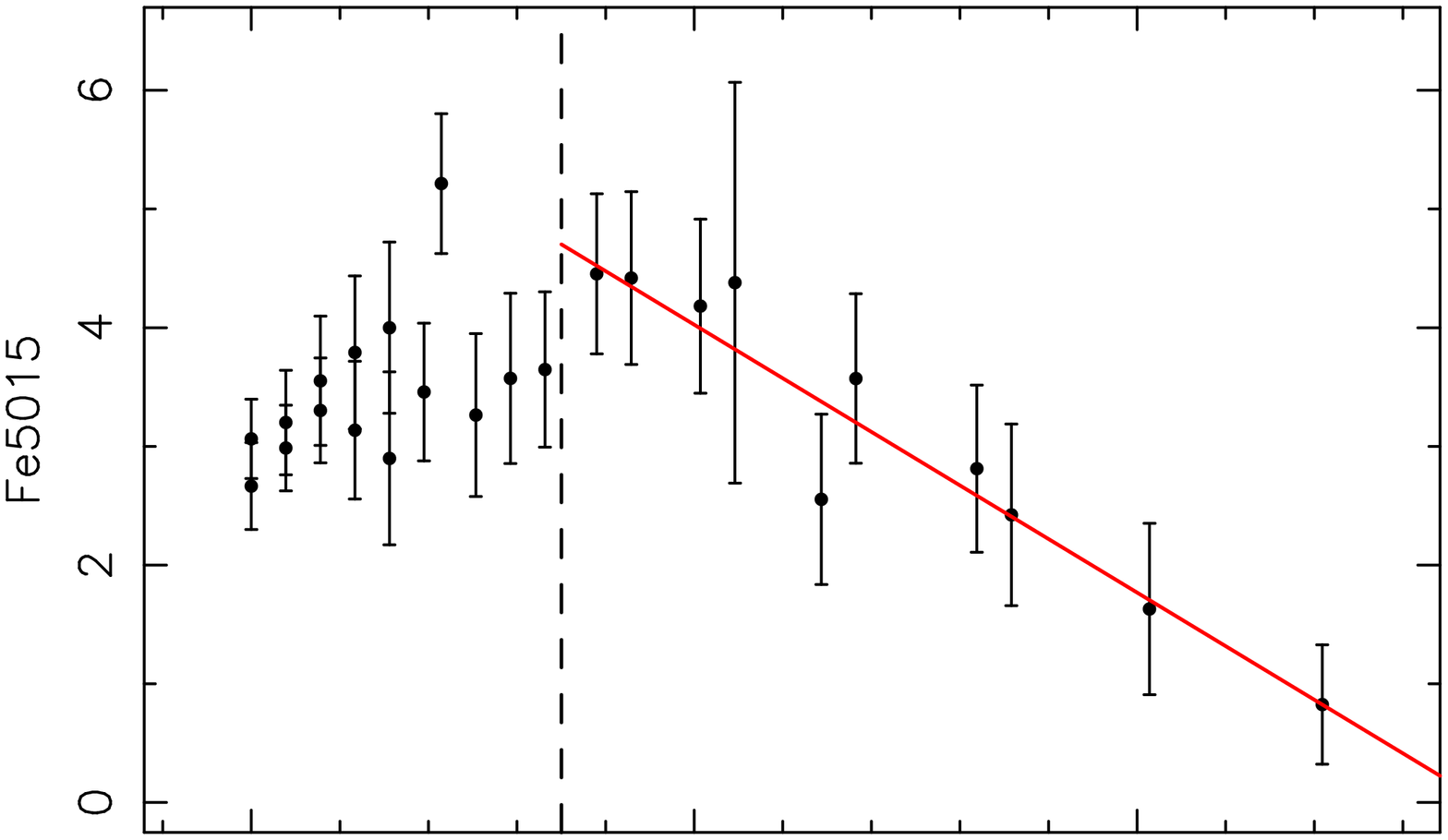}}
\resizebox{0.3\textwidth}{!}{\includegraphics[angle=-90]{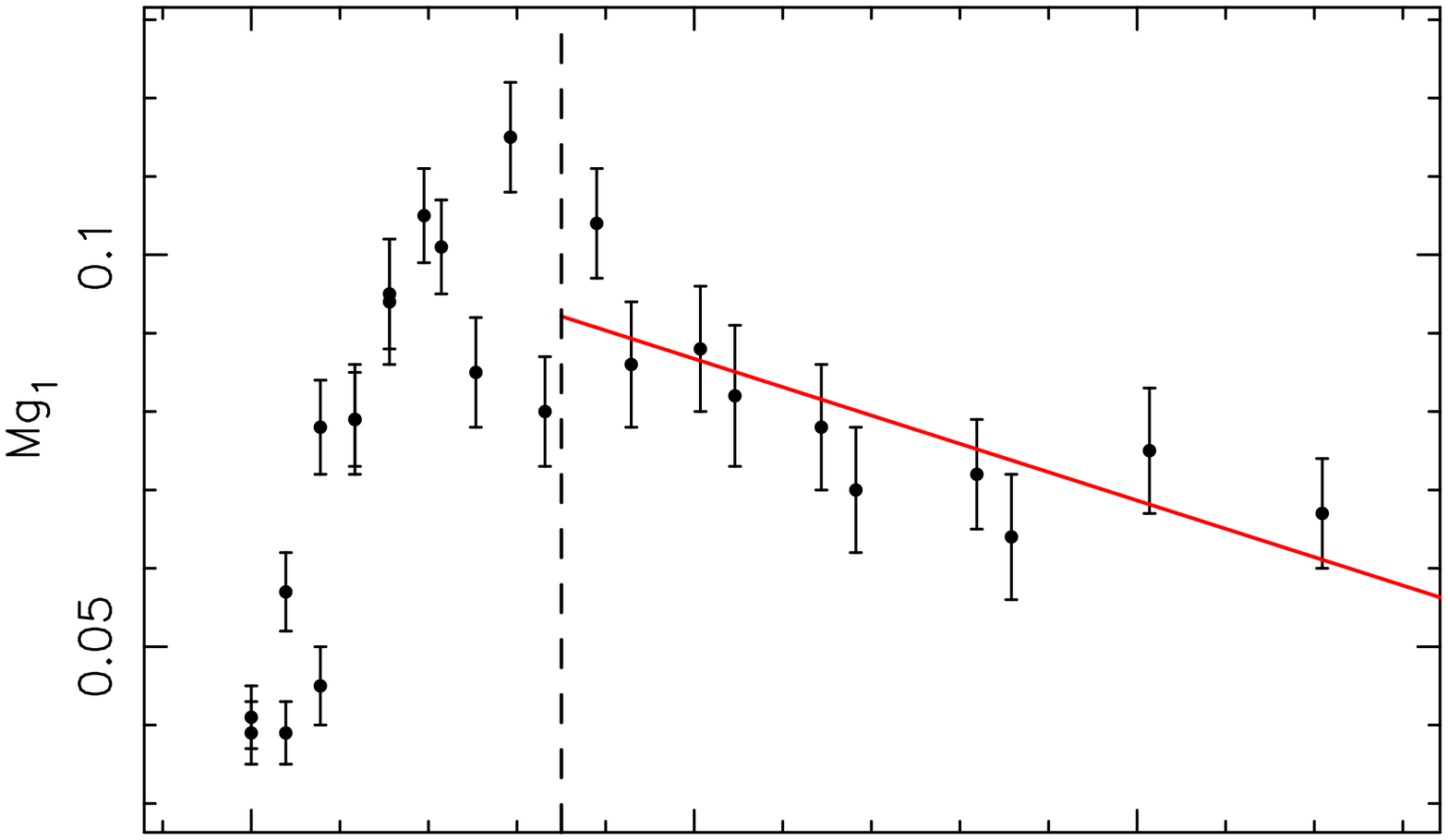}}
\resizebox{0.3\textwidth}{!}{\includegraphics[angle=-90]{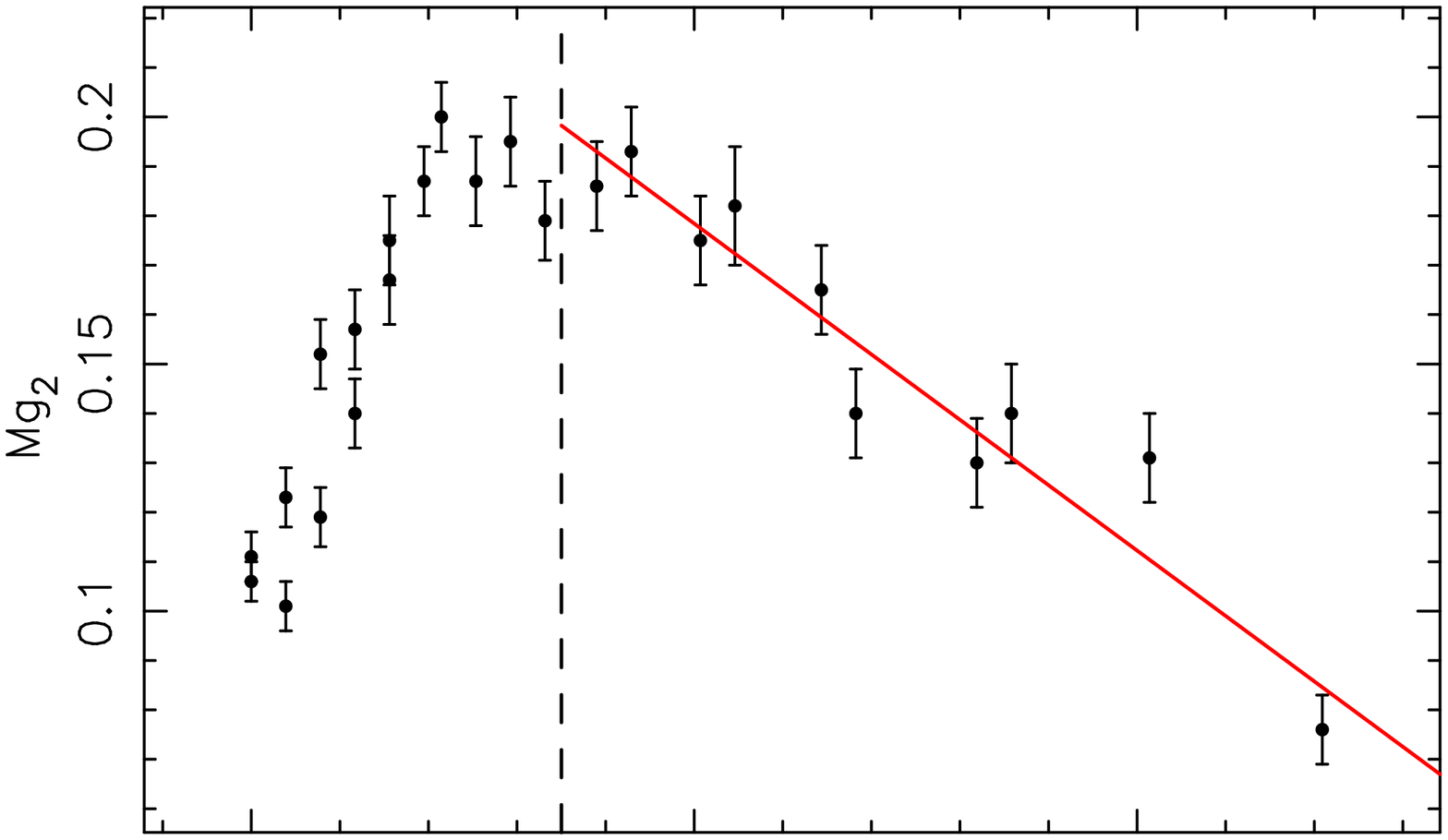}}
\resizebox{0.3\textwidth}{!}{\includegraphics[angle=-90]{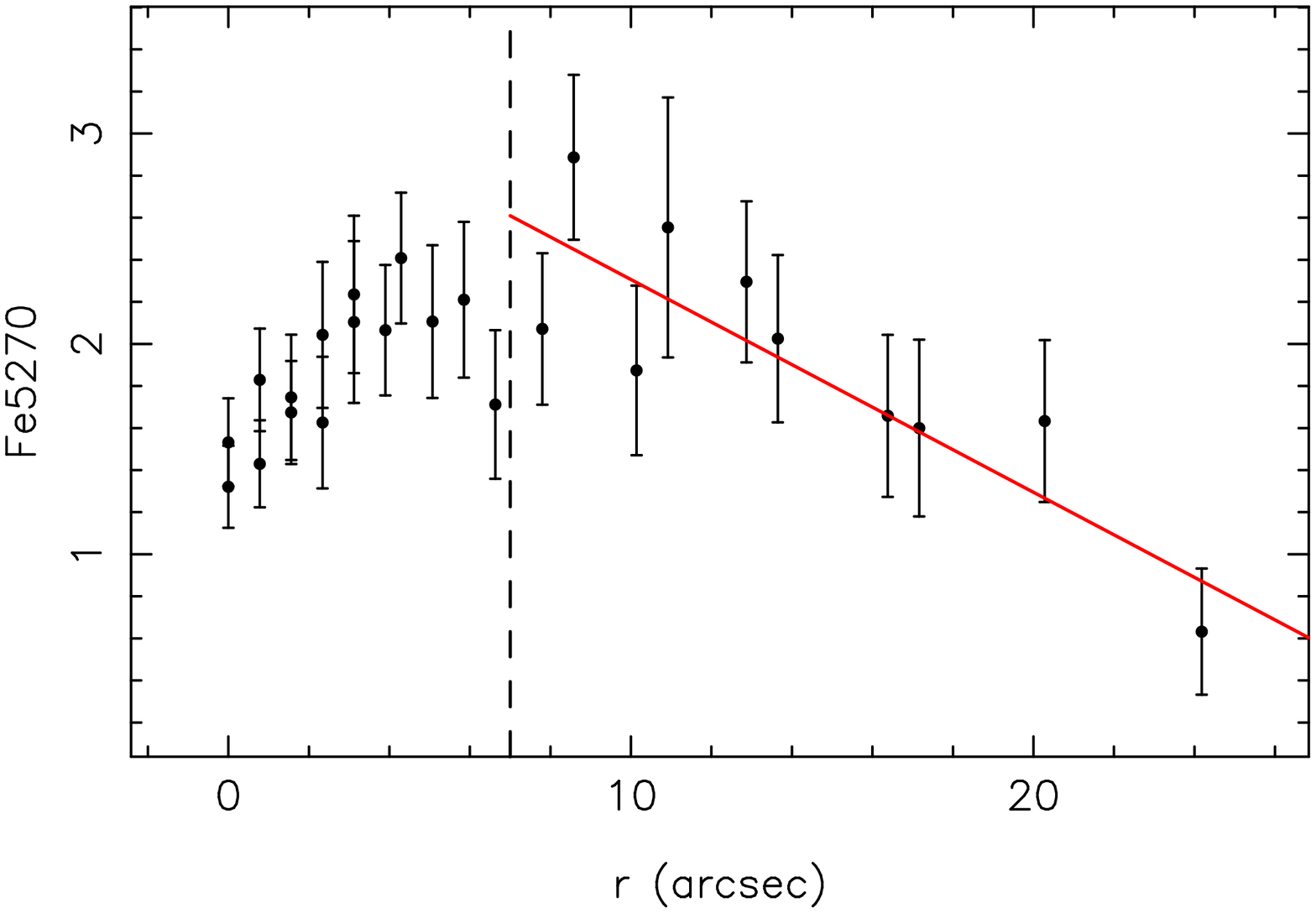}}
\resizebox{0.3\textwidth}{!}{\includegraphics[angle=-90]{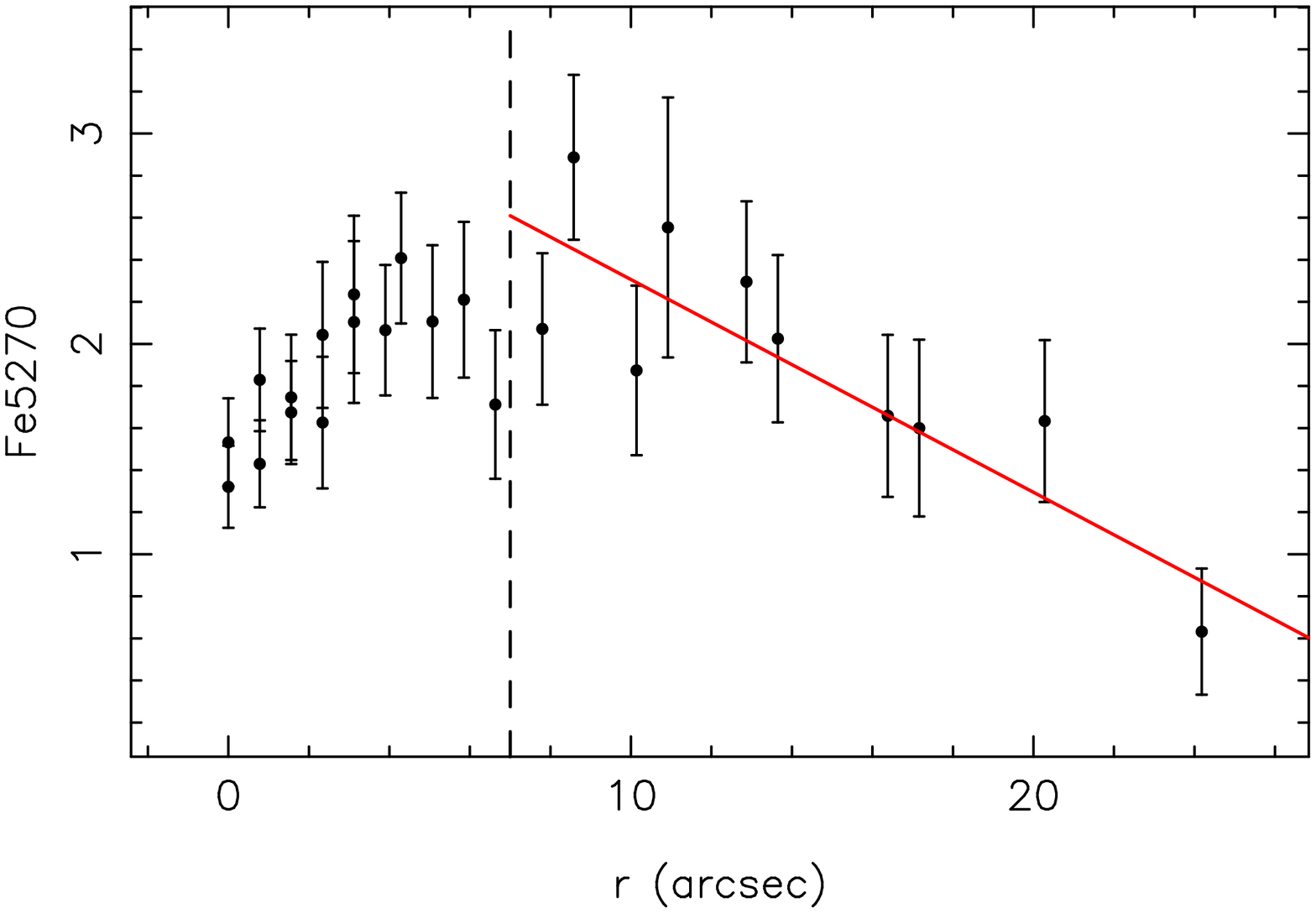}}\hspace{0.85cm}
\resizebox{0.3\textwidth}{!}{\includegraphics[angle=-90]{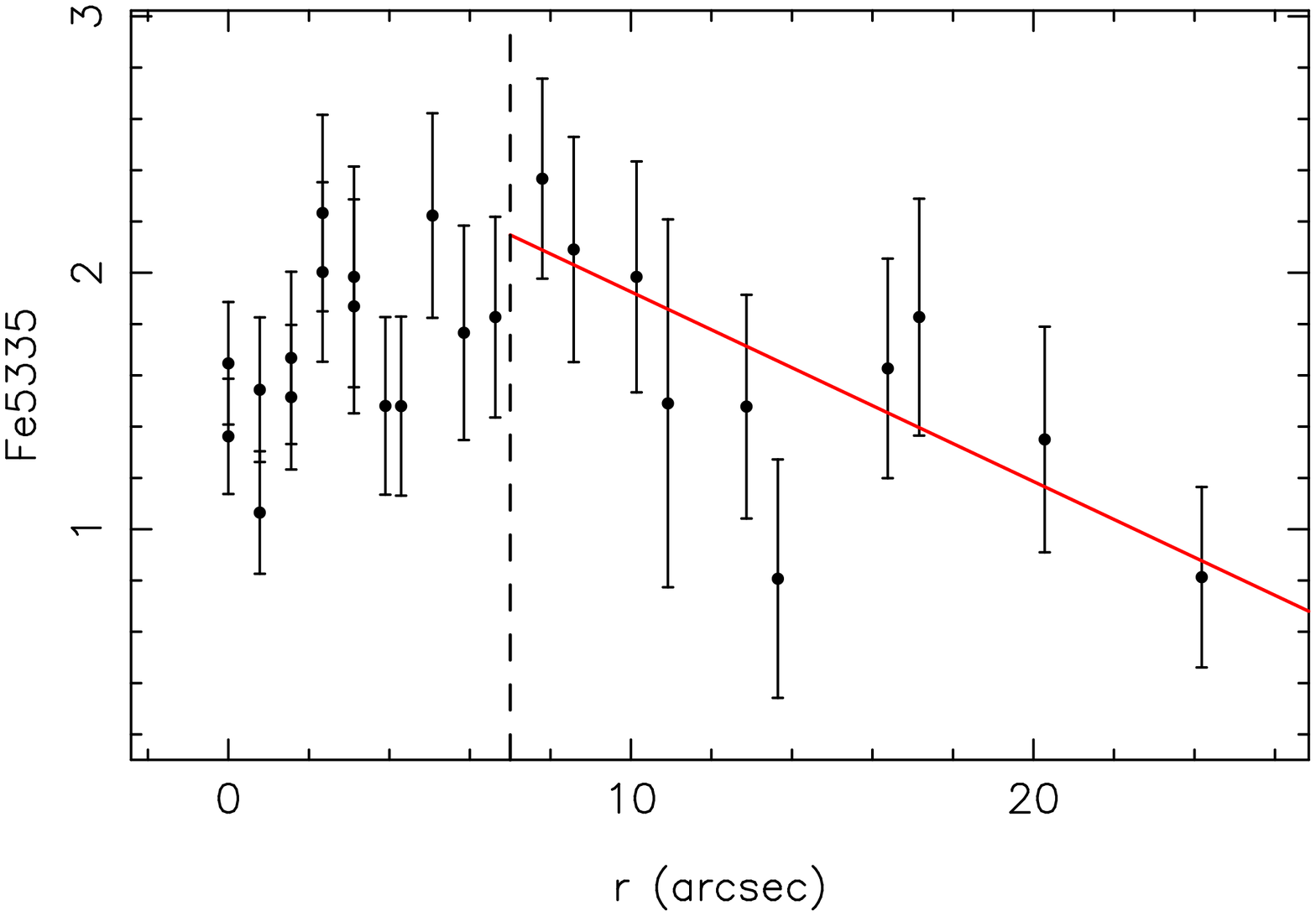}}
\caption{Line-strength distribution in the bar region for all the galaxies}
\end{figure*}

\begin{figure*}
\addtocounter{figure}{-1}
\resizebox{0.3\textwidth}{!}{\includegraphics[angle=-90]{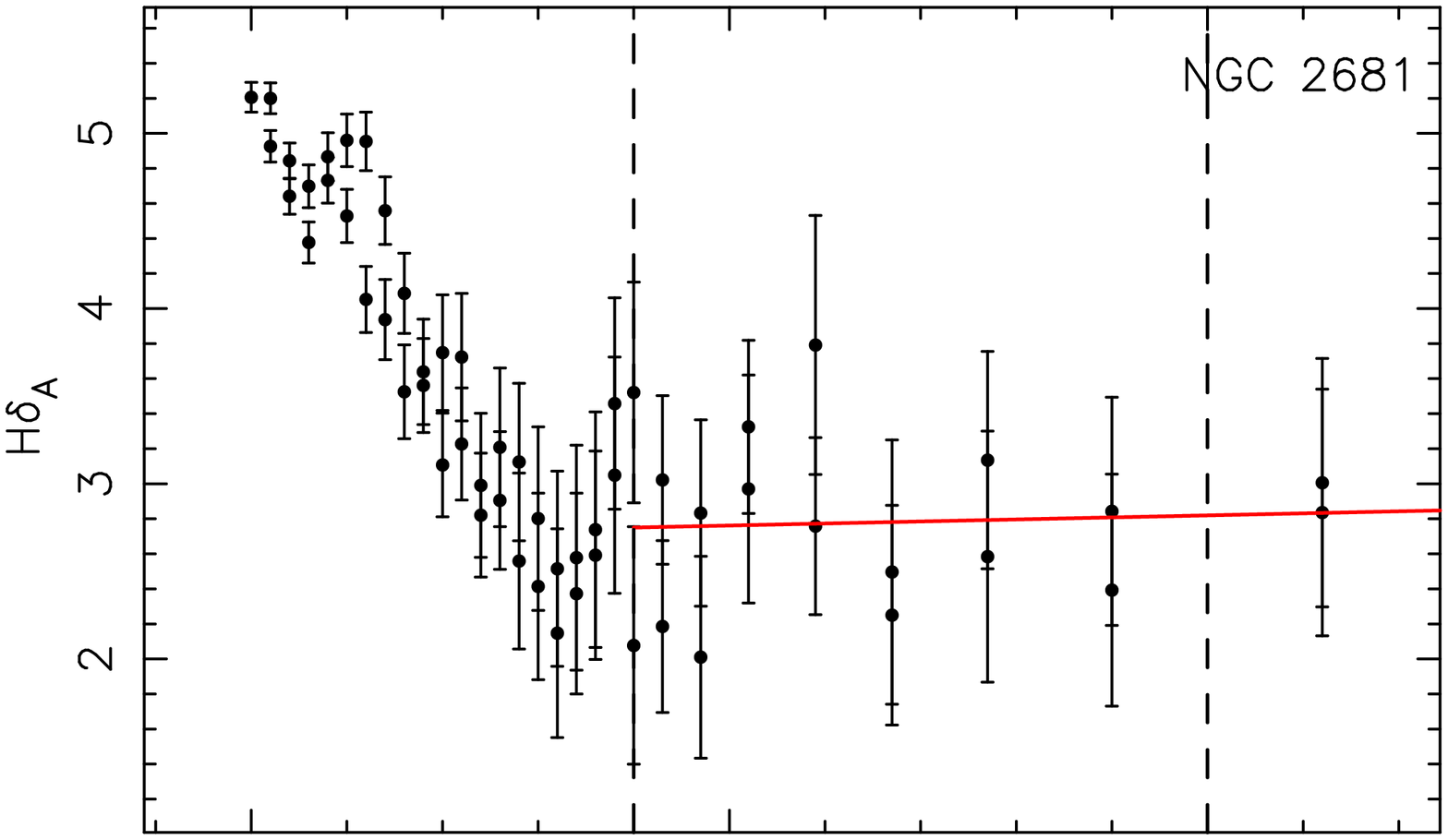}}
\resizebox{0.3\textwidth}{!}{\includegraphics[angle=-90]{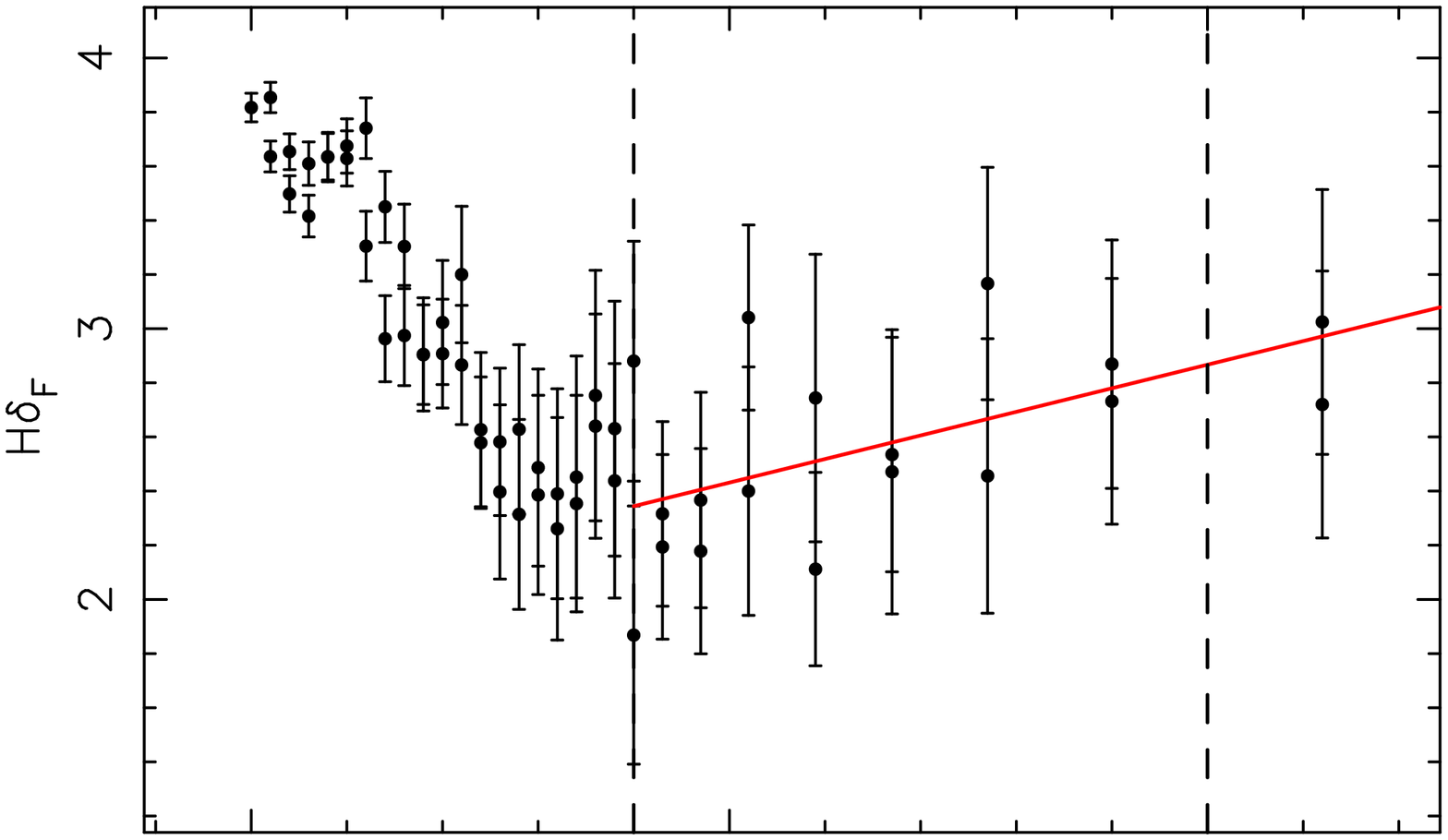}}
\resizebox{0.3\textwidth}{!}{\includegraphics[angle=-90]{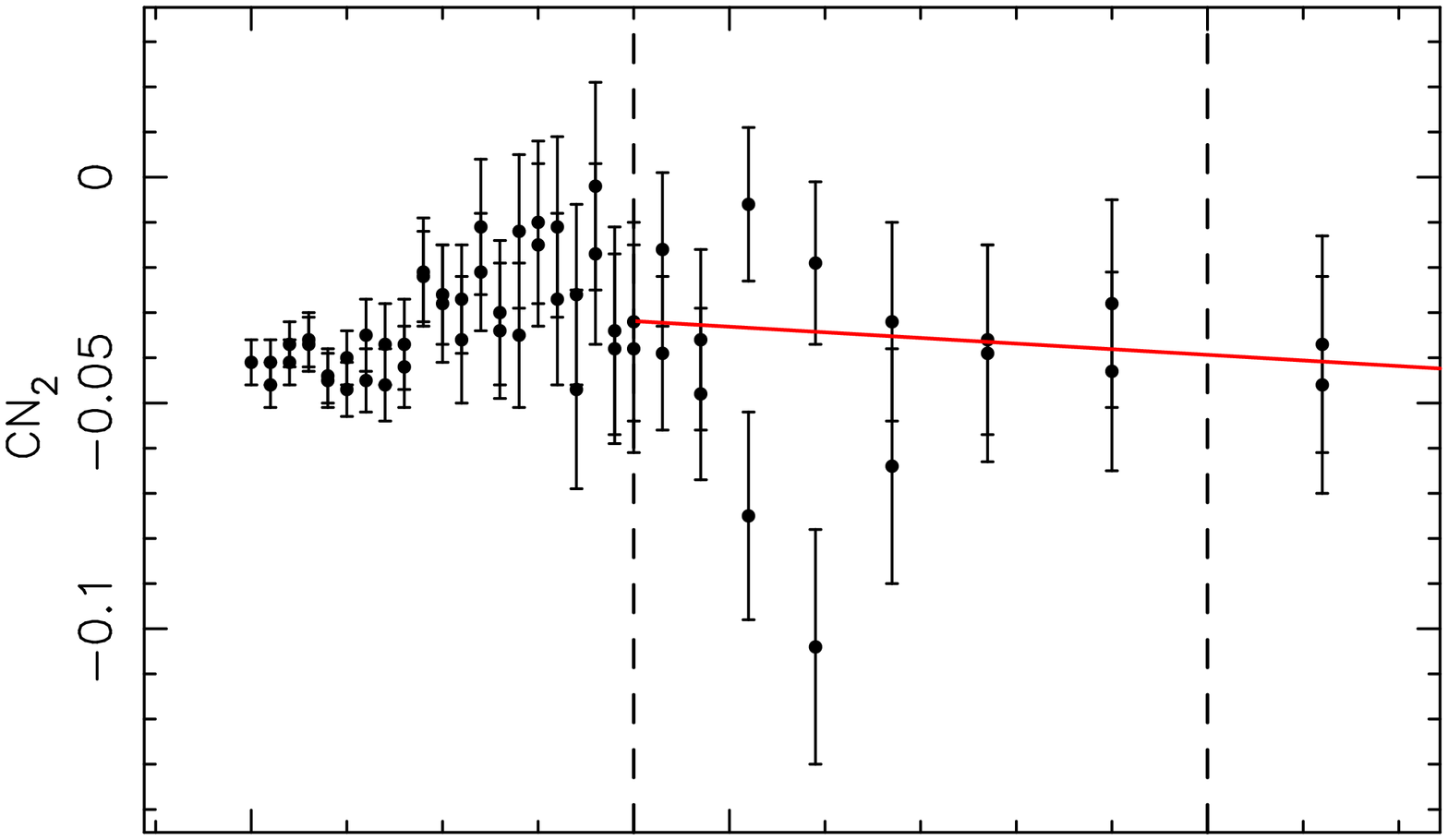}}
\resizebox{0.3\textwidth}{!}{\includegraphics[angle=-90]{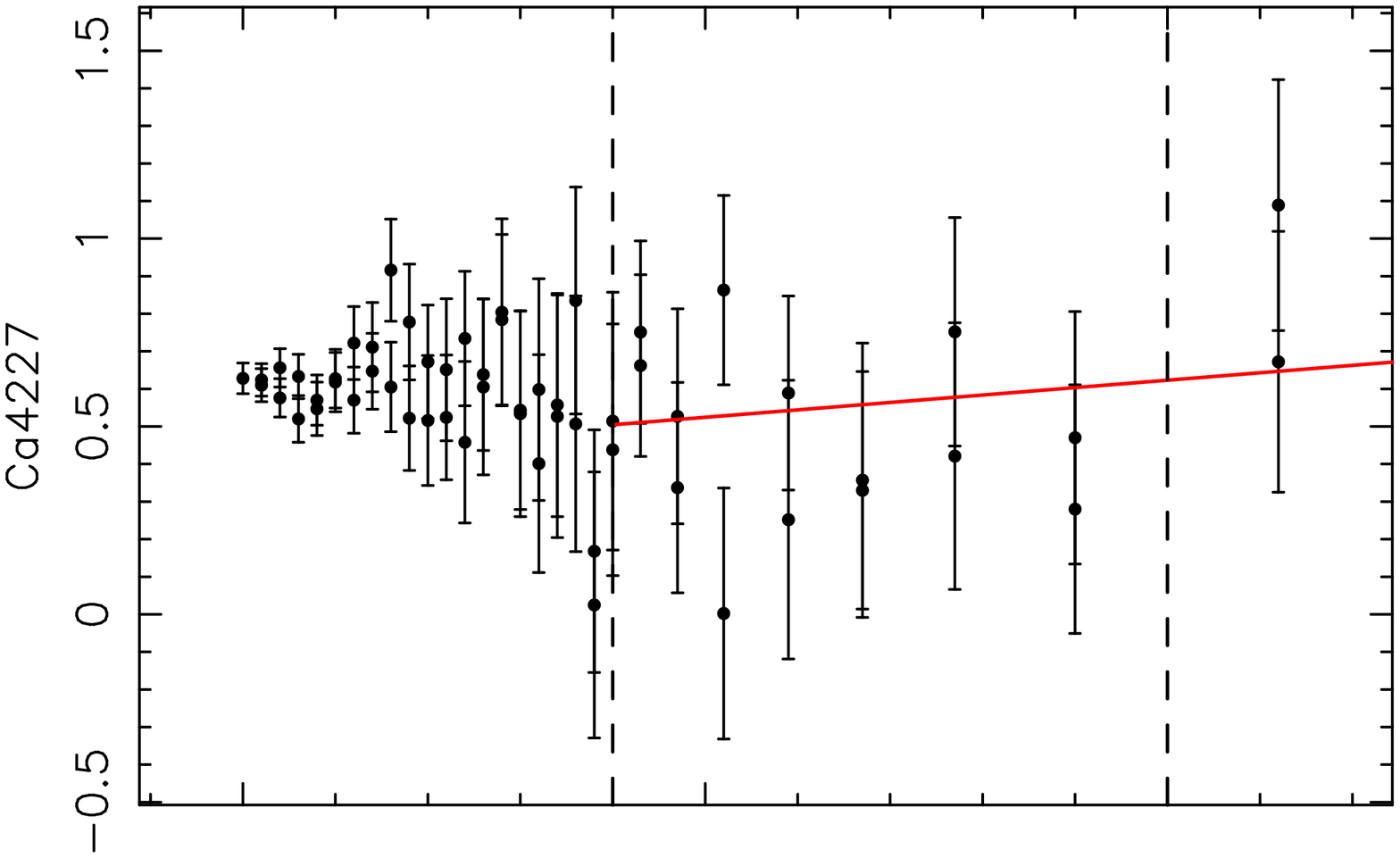}}
\resizebox{0.3\textwidth}{!}{\includegraphics[angle=-90]{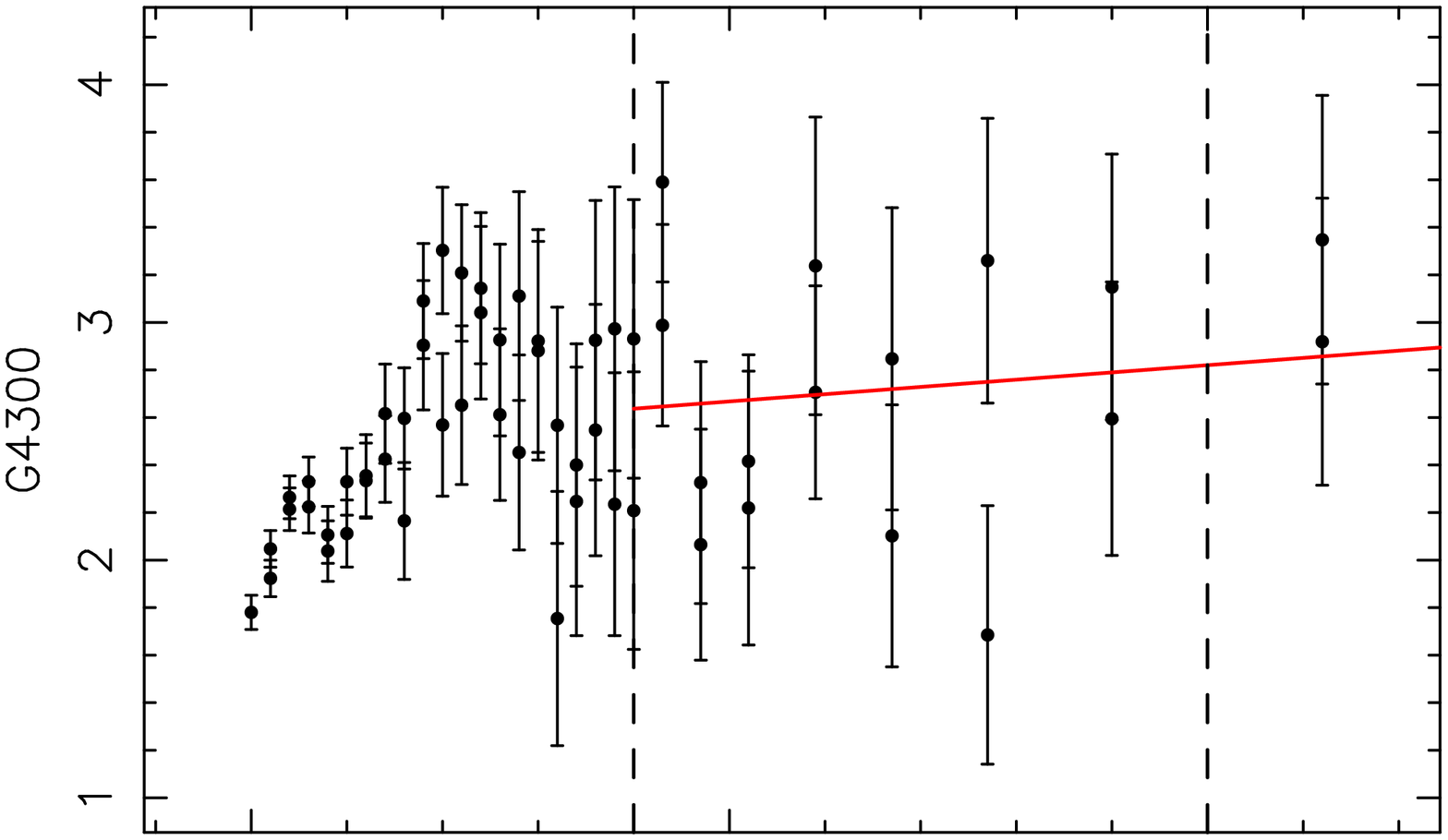}}
\resizebox{0.3\textwidth}{!}{\includegraphics[angle=-90]{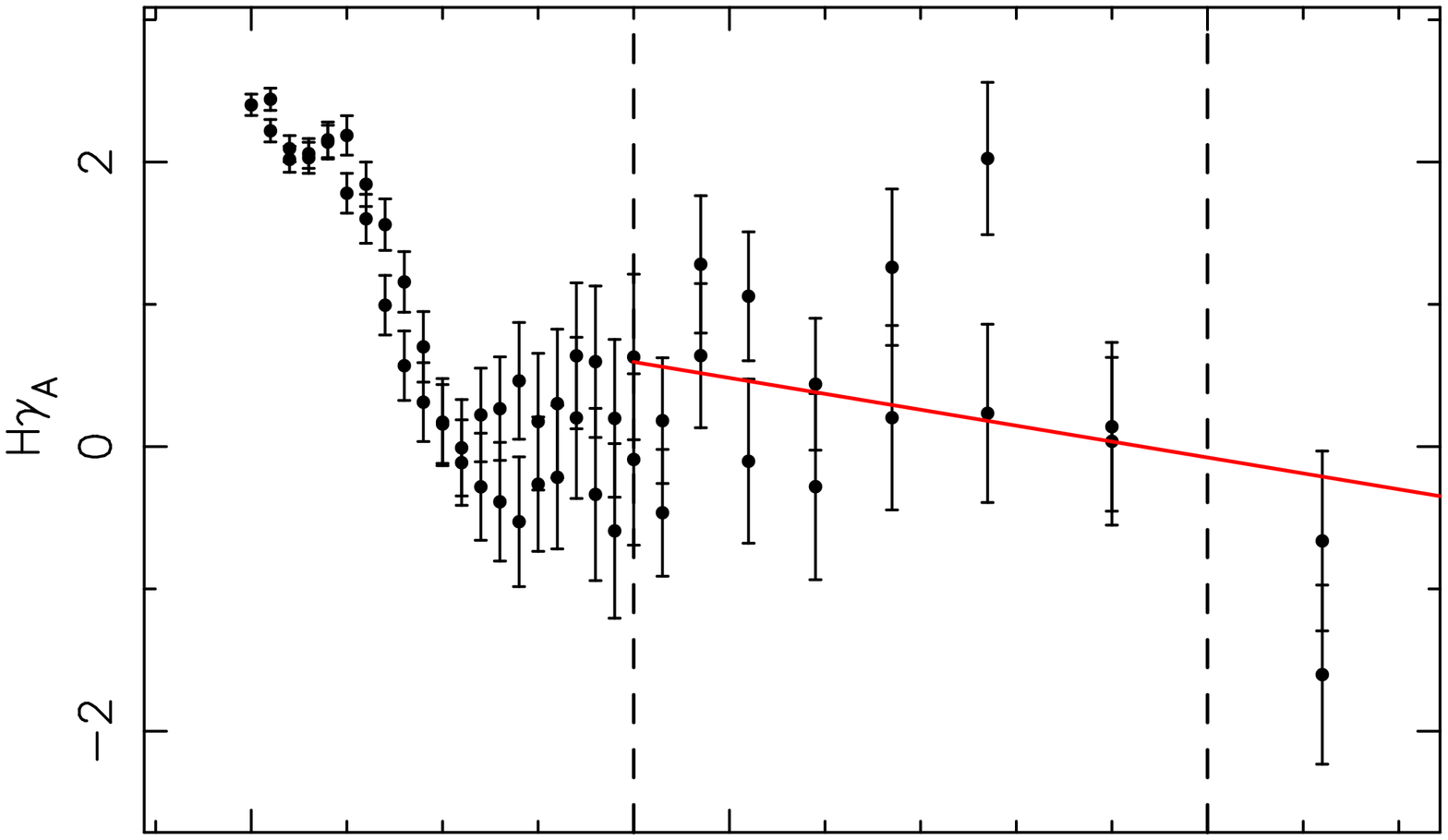}}
\resizebox{0.3\textwidth}{!}{\includegraphics[angle=-90]{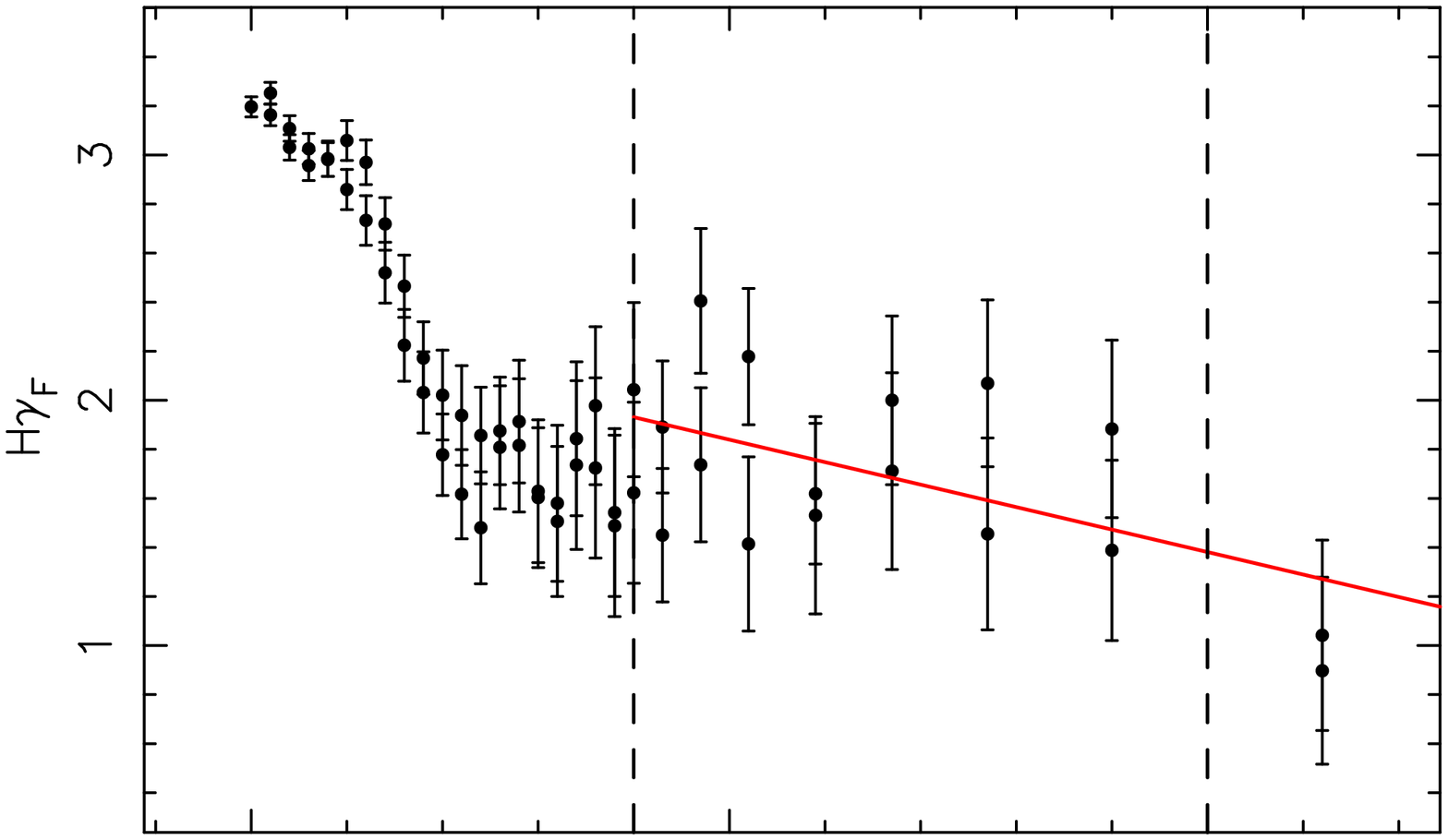}}
\resizebox{0.3\textwidth}{!}{\includegraphics[angle=-90]{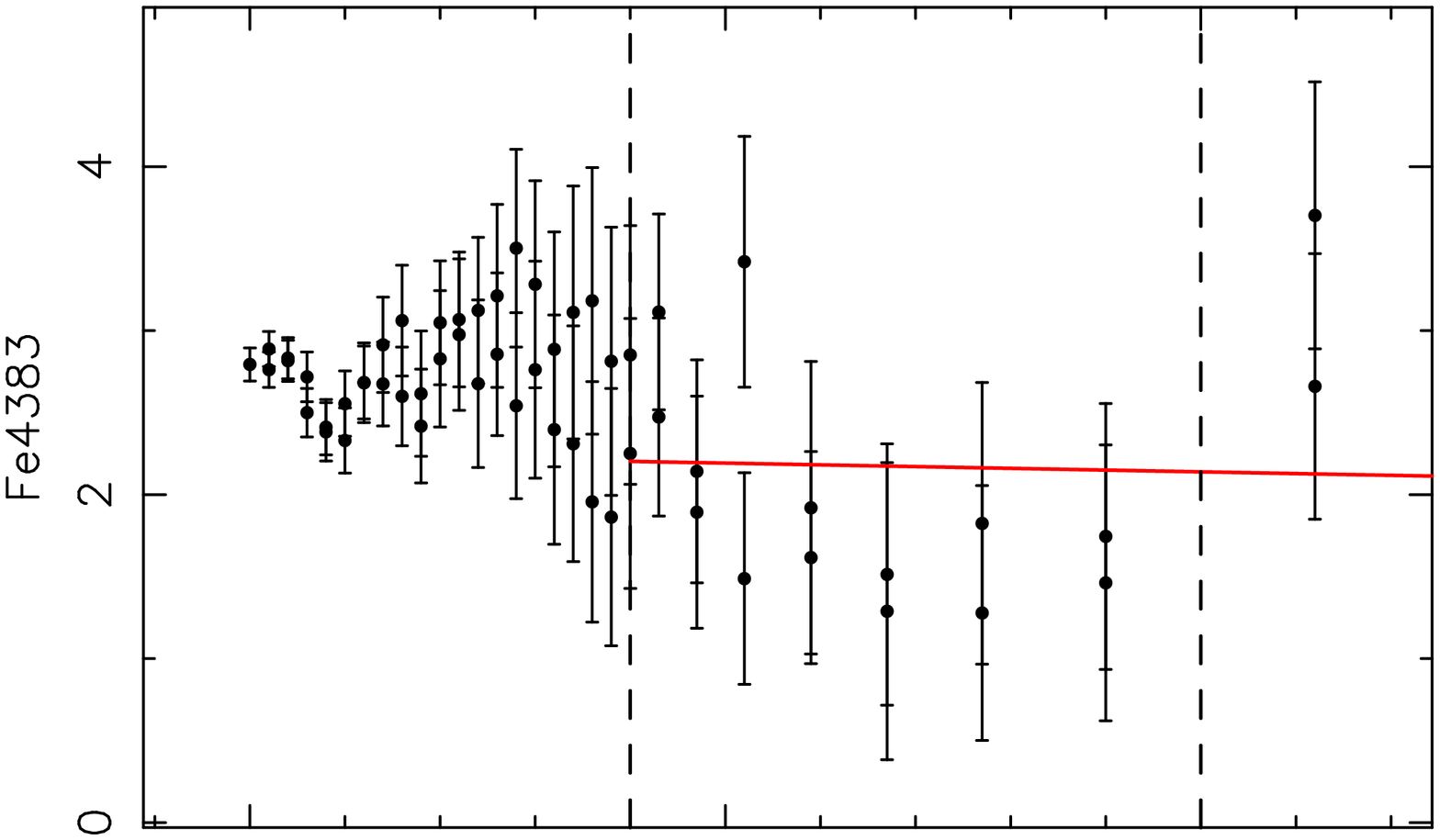}}
\resizebox{0.3\textwidth}{!}{\includegraphics[angle=-90]{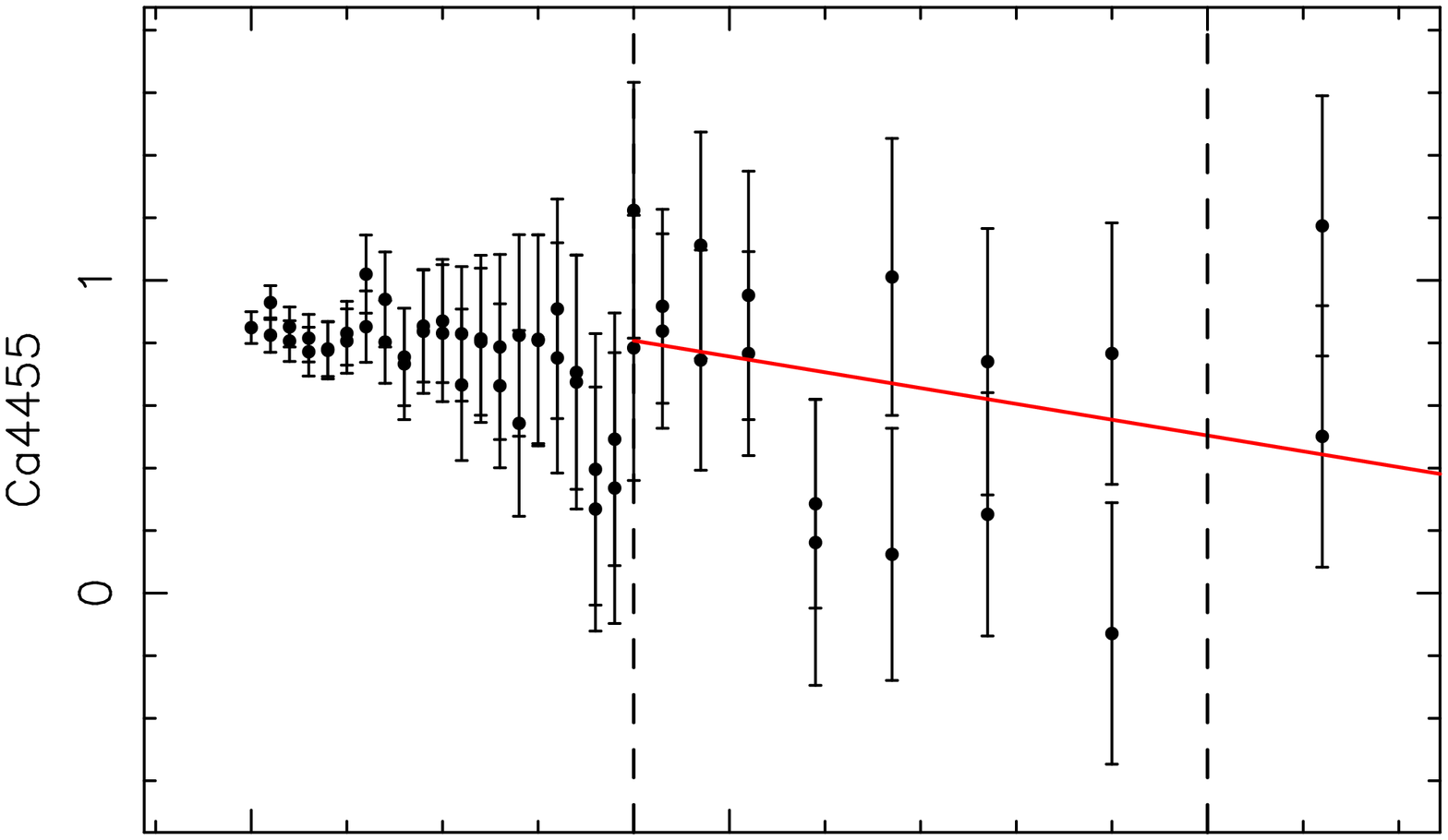}}
\resizebox{0.3\textwidth}{!}{\includegraphics[angle=-90]{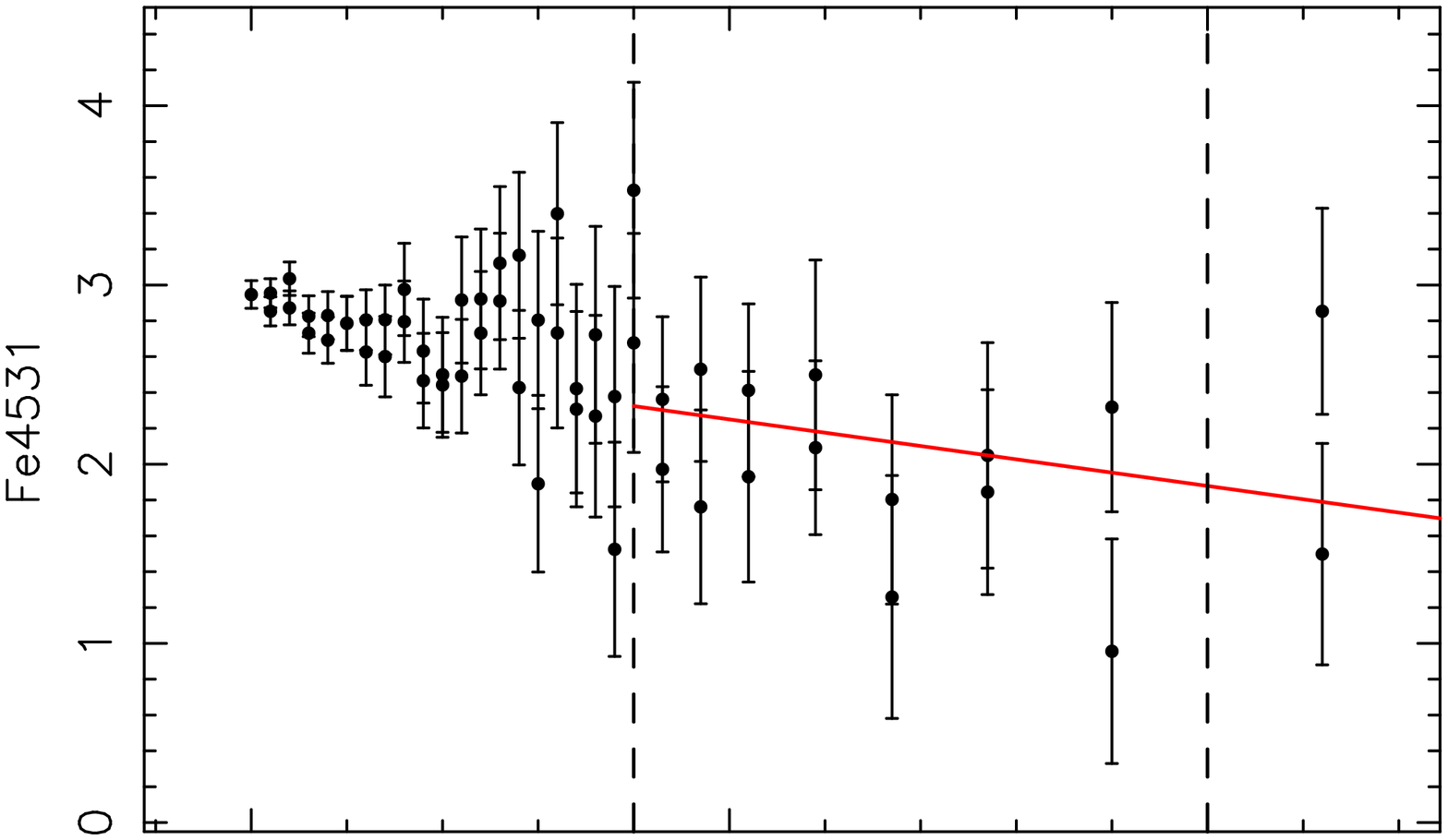}}
\resizebox{0.3\textwidth}{!}{\includegraphics[angle=-90]{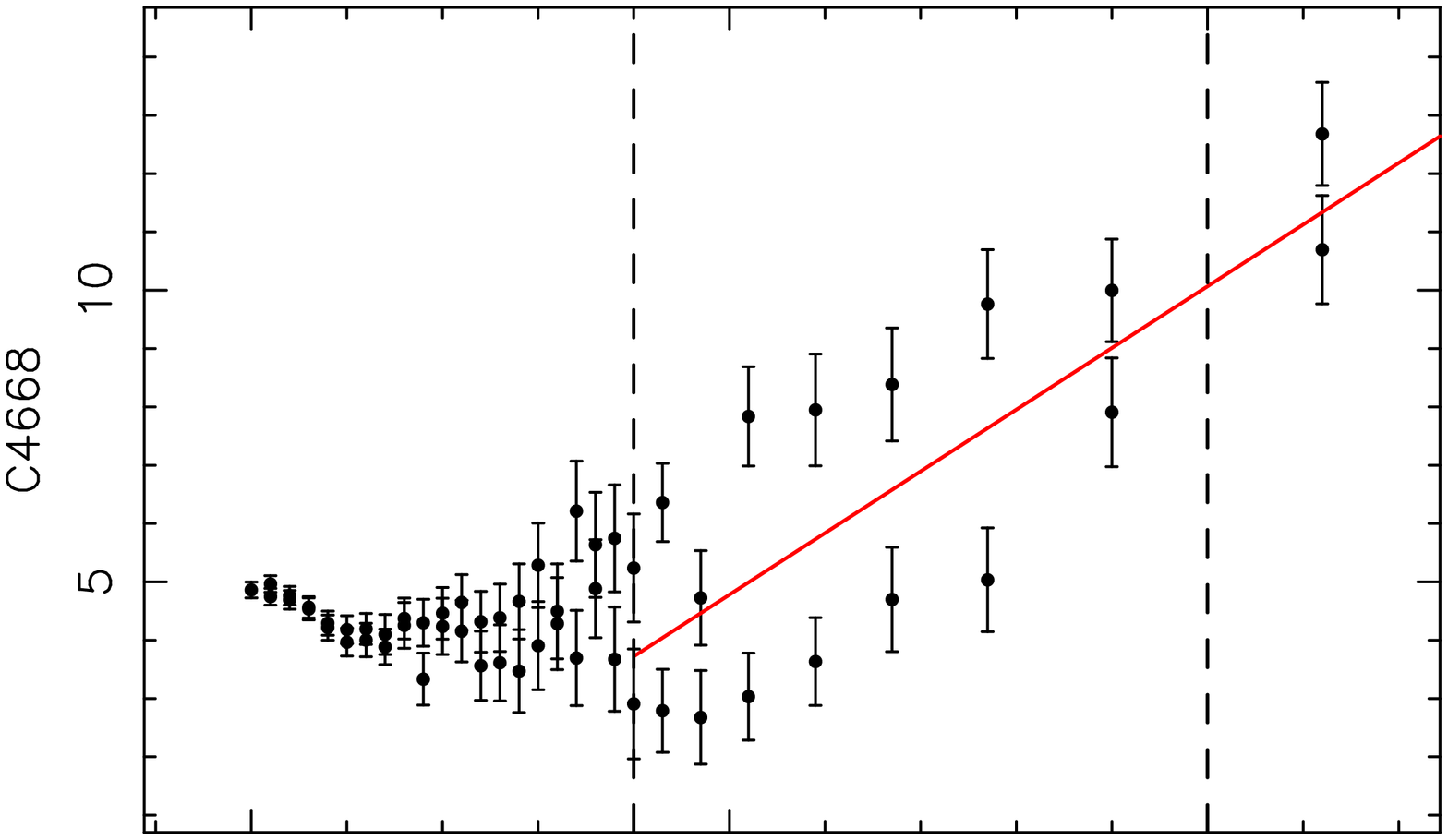}}
\resizebox{0.3\textwidth}{!}{\includegraphics[angle=-90]{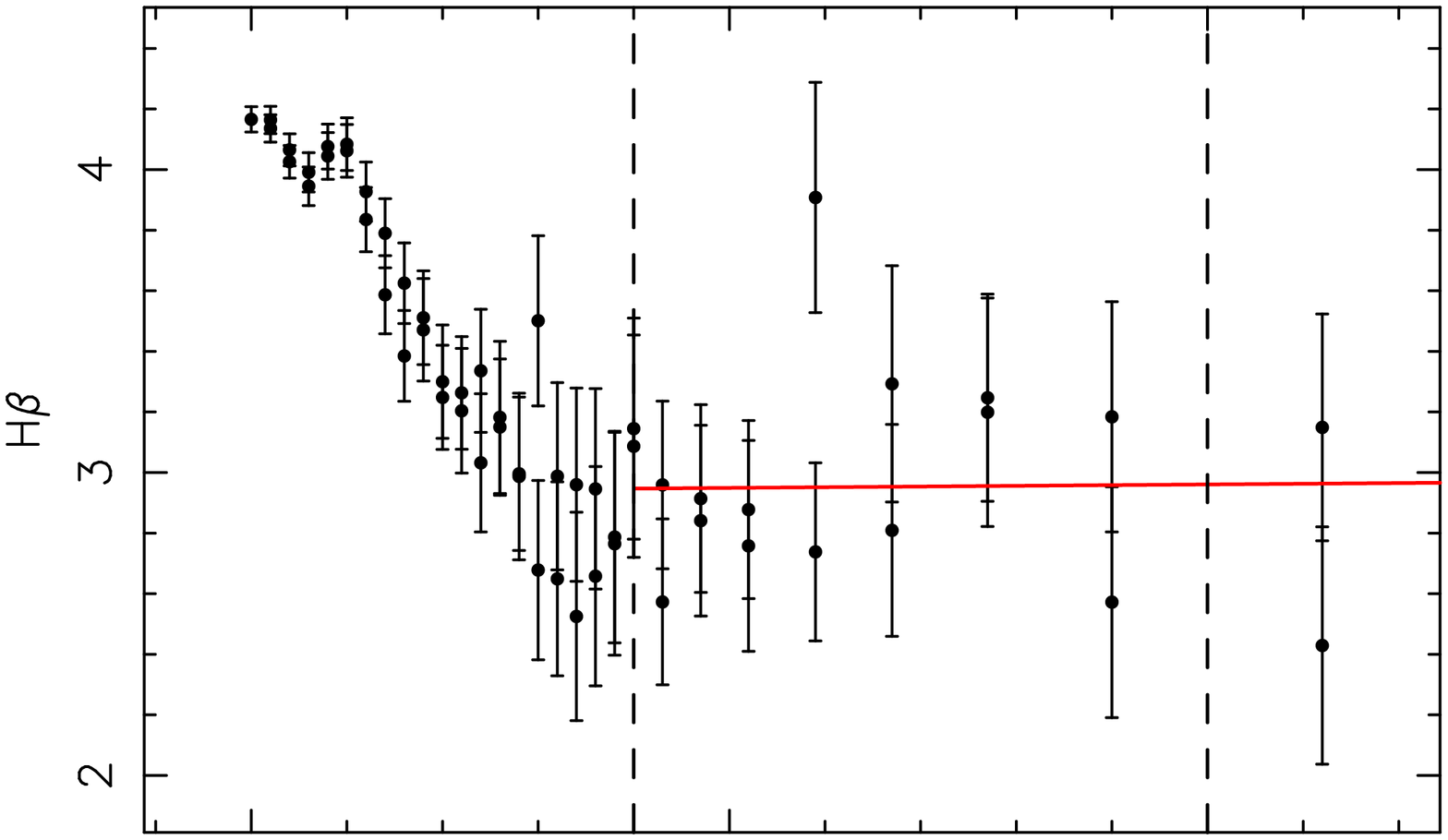}}
\resizebox{0.3\textwidth}{!}{\includegraphics[angle=-90]{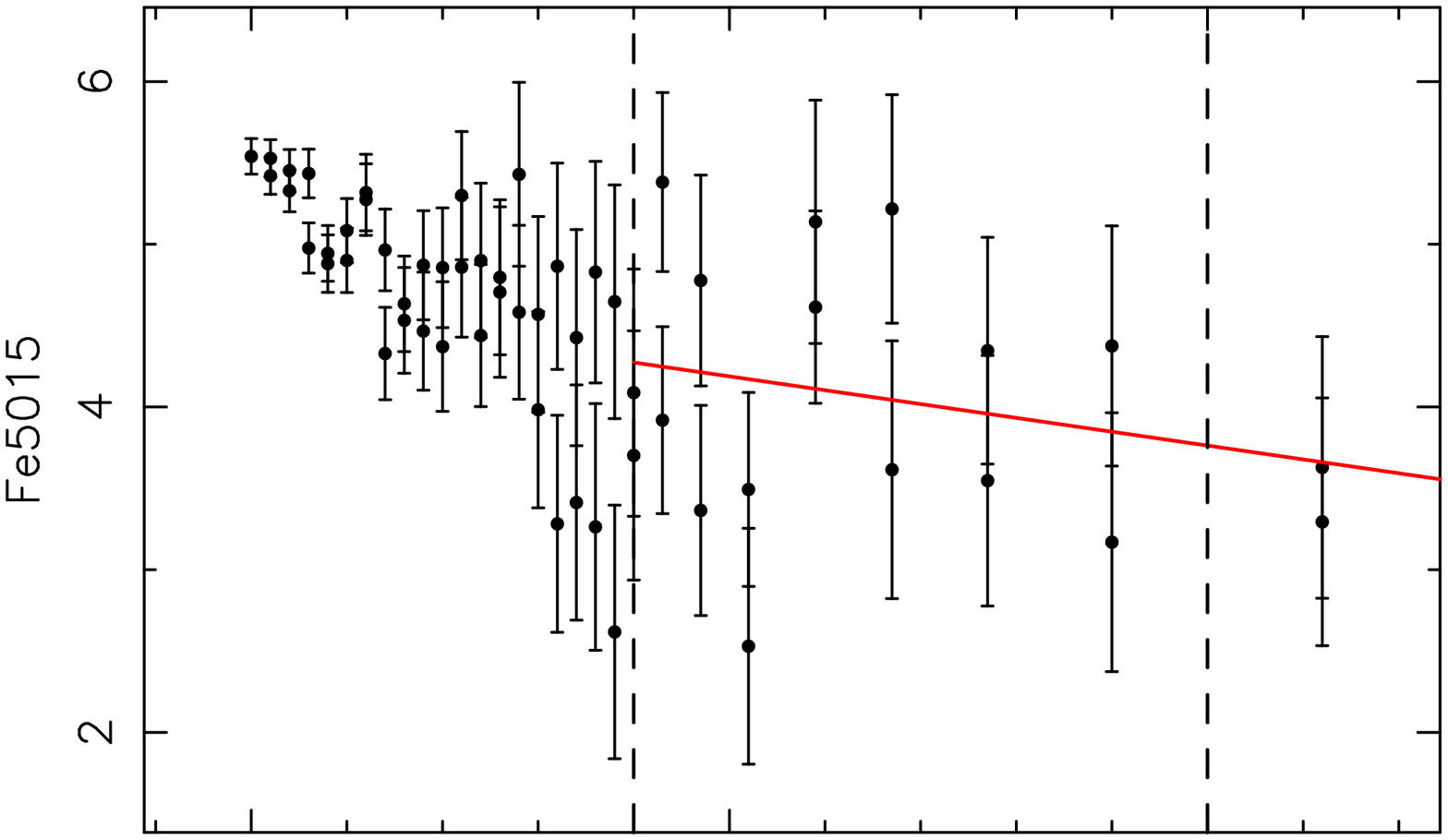}}
\resizebox{0.3\textwidth}{!}{\includegraphics[angle=-90]{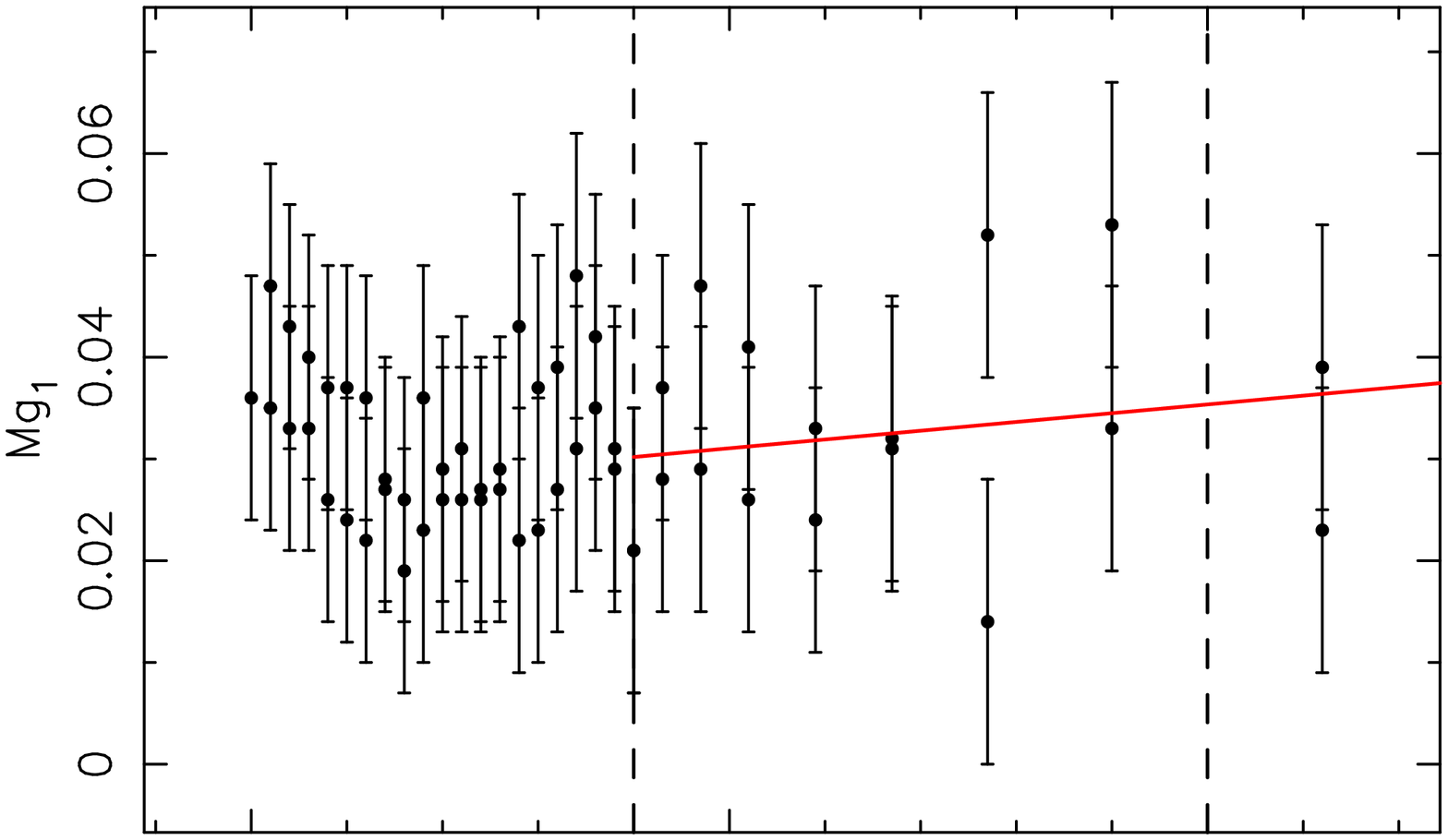}}
\resizebox{0.3\textwidth}{!}{\includegraphics[angle=-90]{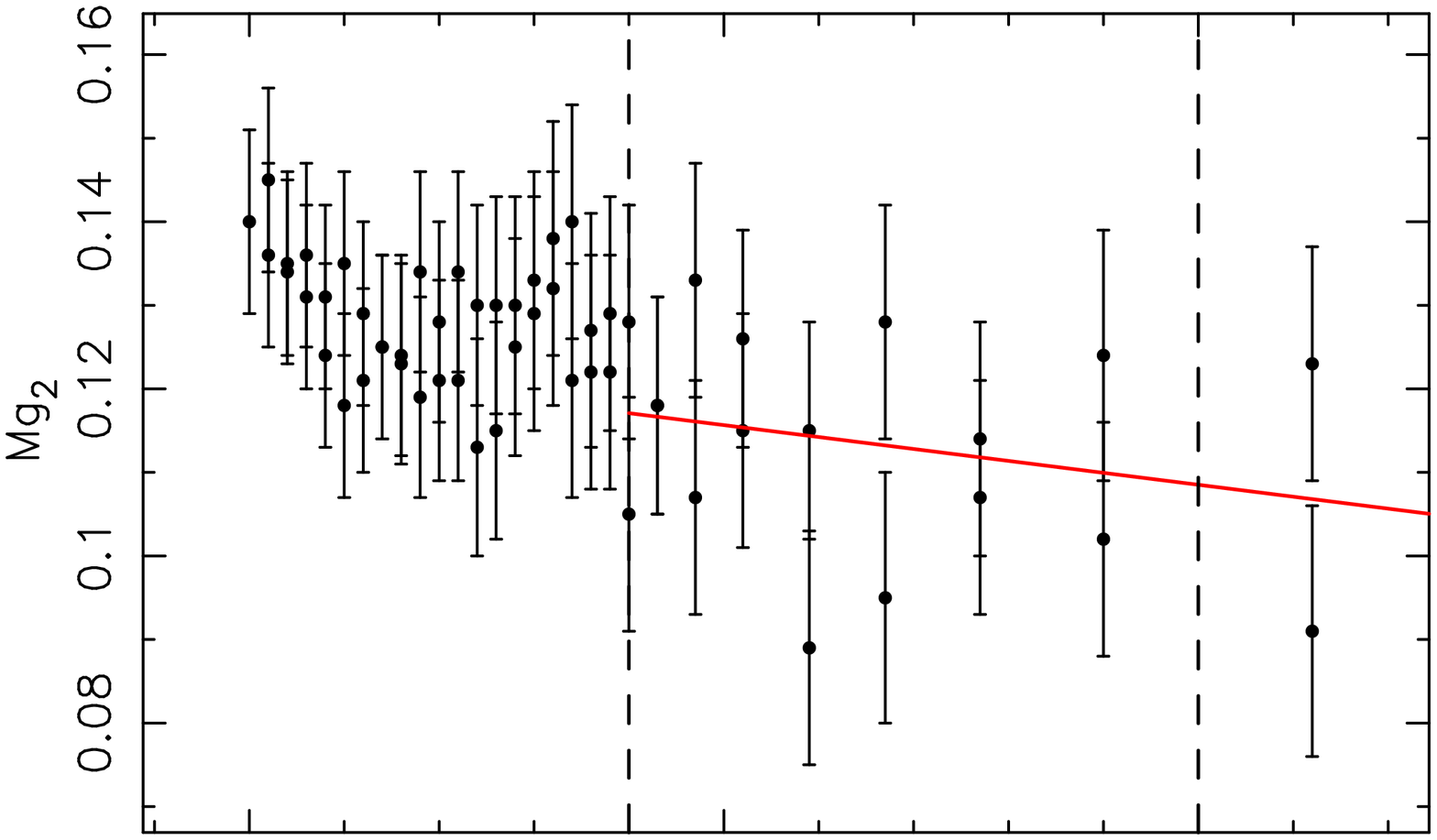}}
\resizebox{0.3\textwidth}{!}{\includegraphics[angle=-90]{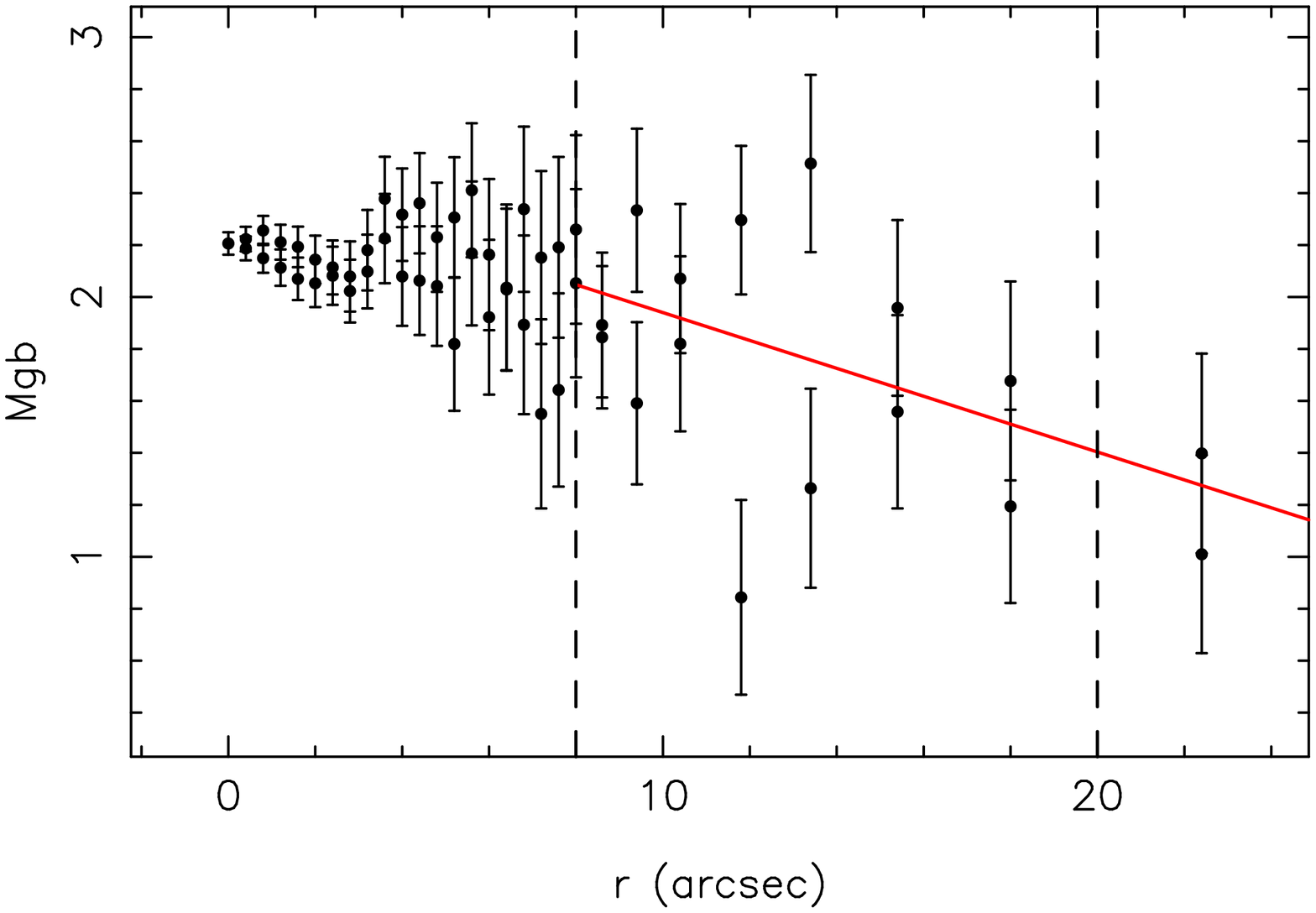}}\hspace{0.85cm}
\resizebox{0.3\textwidth}{!}{\includegraphics[angle=-90]{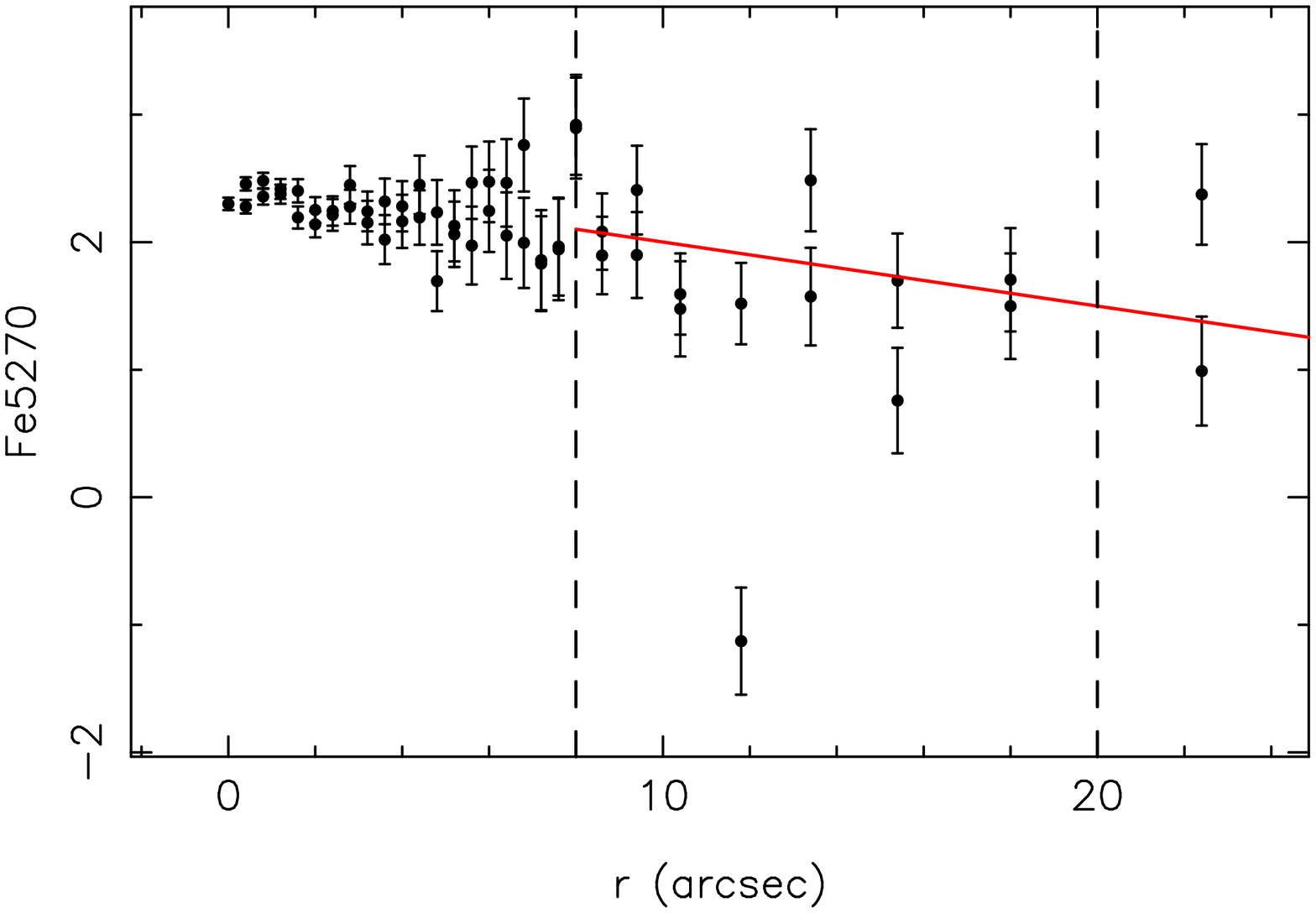}}\hspace{0.85cm}
\resizebox{0.3\textwidth}{!}{\includegraphics[angle=-90]{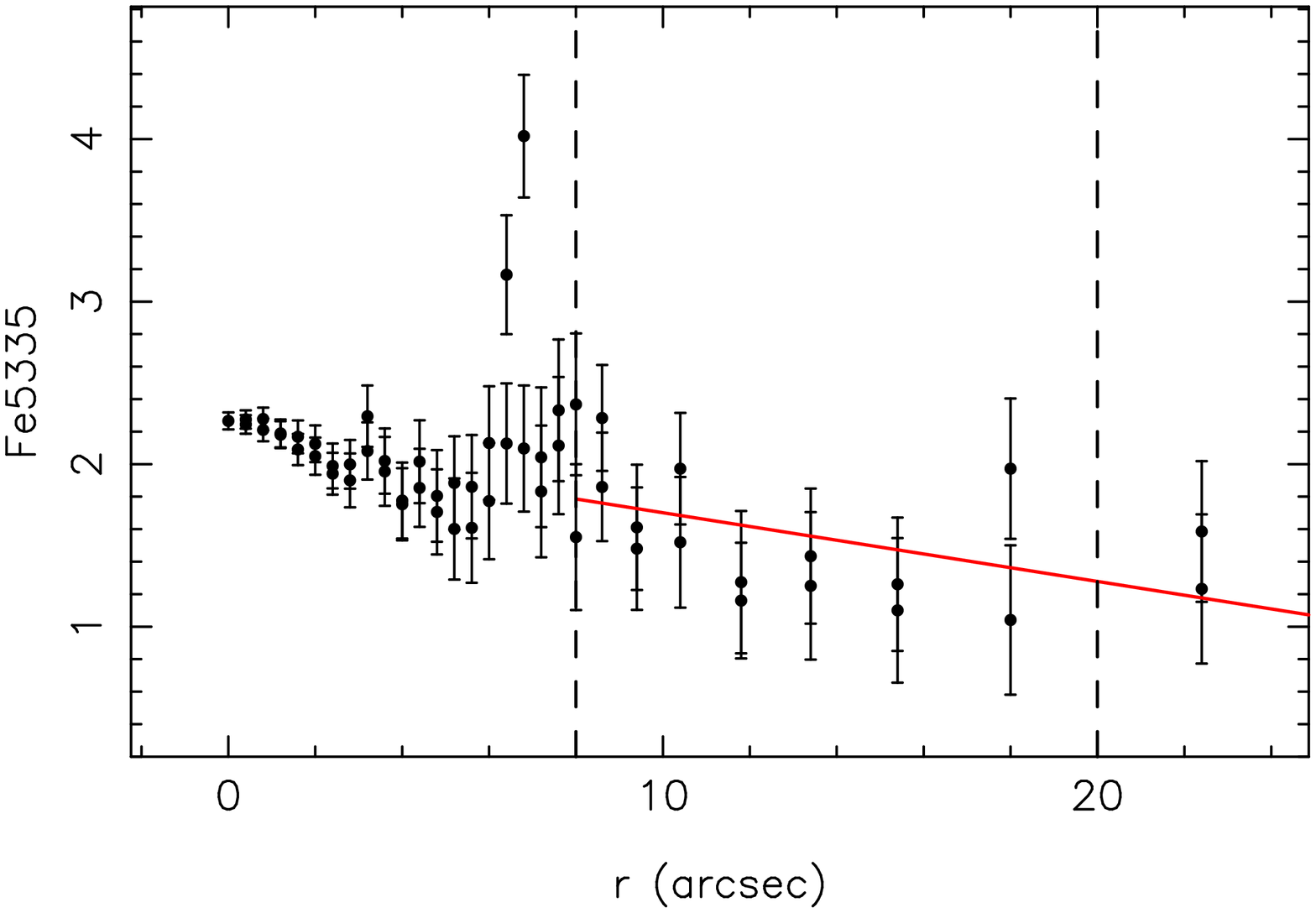}}
\caption{Line-strength distribution in the bar region for all the galaxies}
\end{figure*}
\begin{figure*}
\addtocounter{figure}{-1}
\resizebox{0.3\textwidth}{!}{\includegraphics[angle=-90]{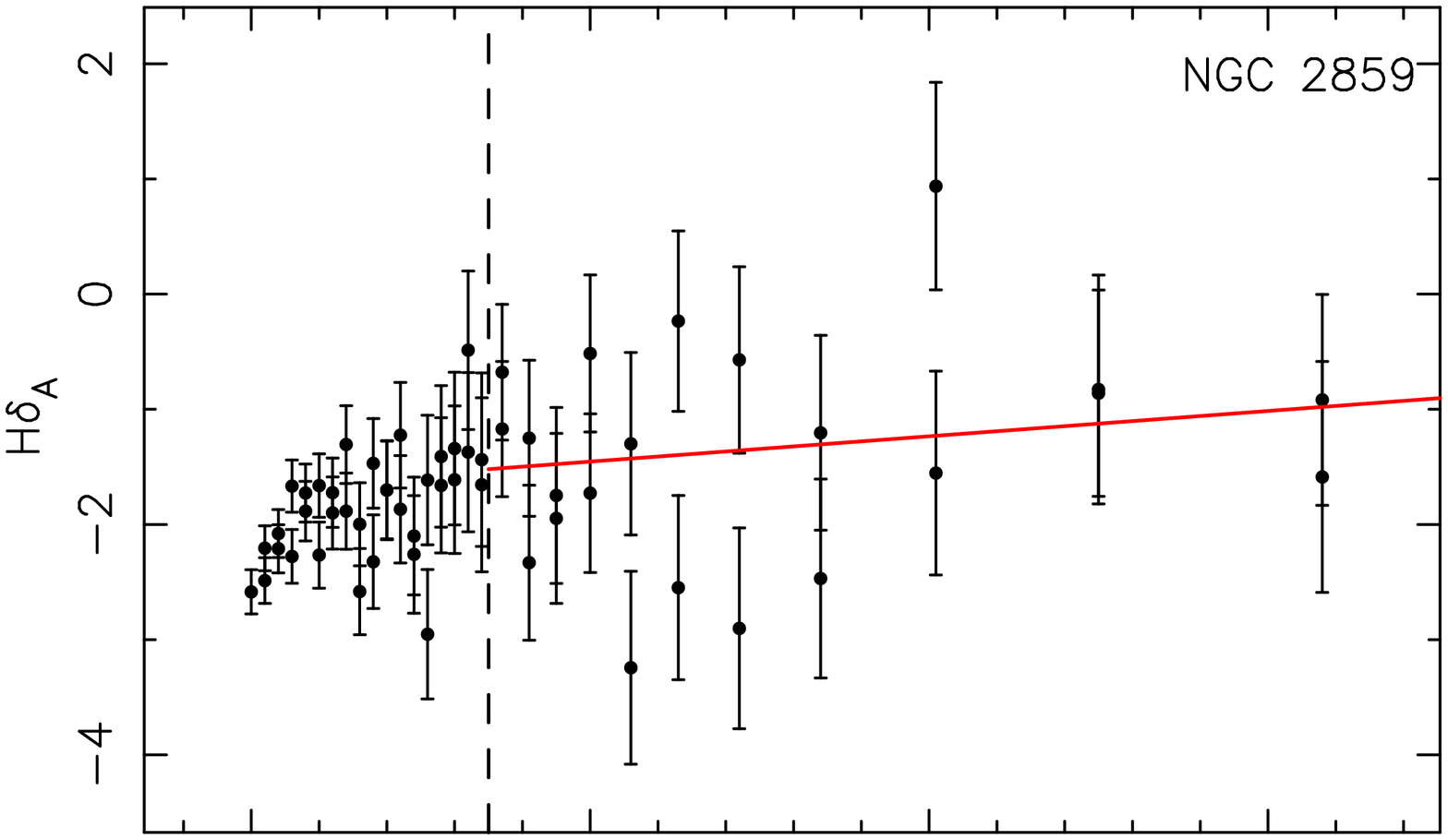}}
\resizebox{0.3\textwidth}{!}{\includegraphics[angle=-90]{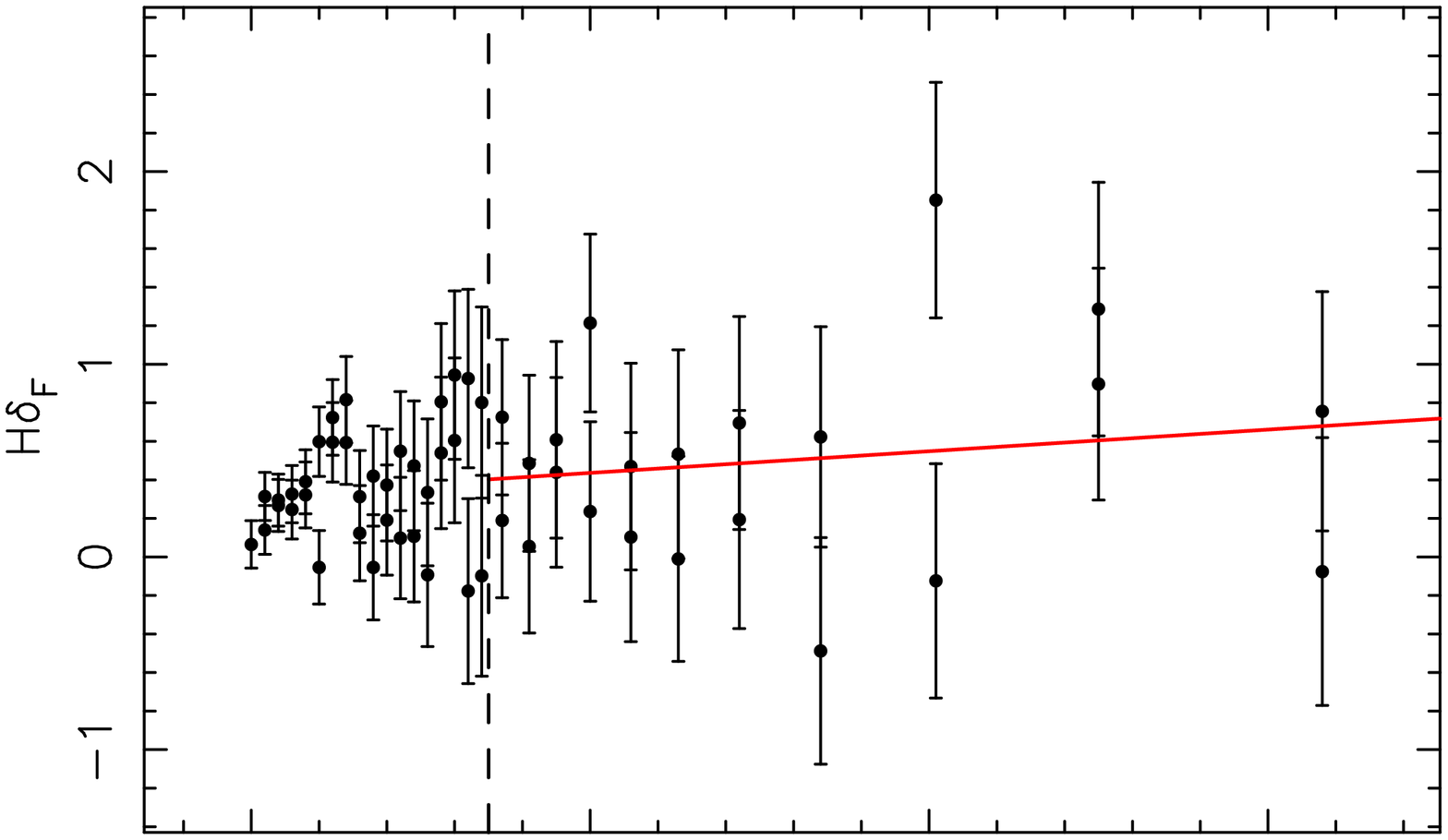}}
\resizebox{0.3\textwidth}{!}{\includegraphics[angle=-90]{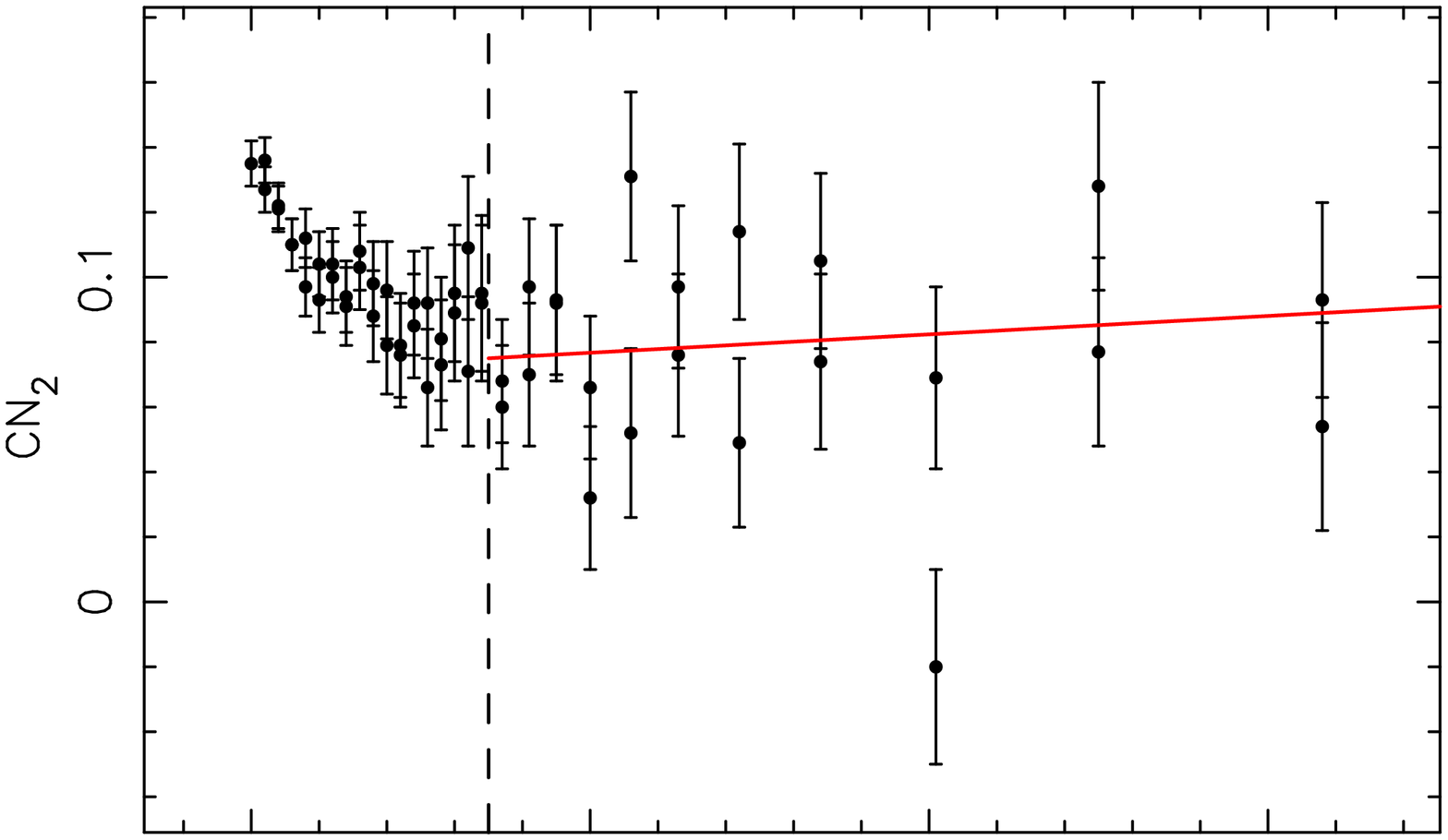}}
\resizebox{0.3\textwidth}{!}{\includegraphics[angle=-90]{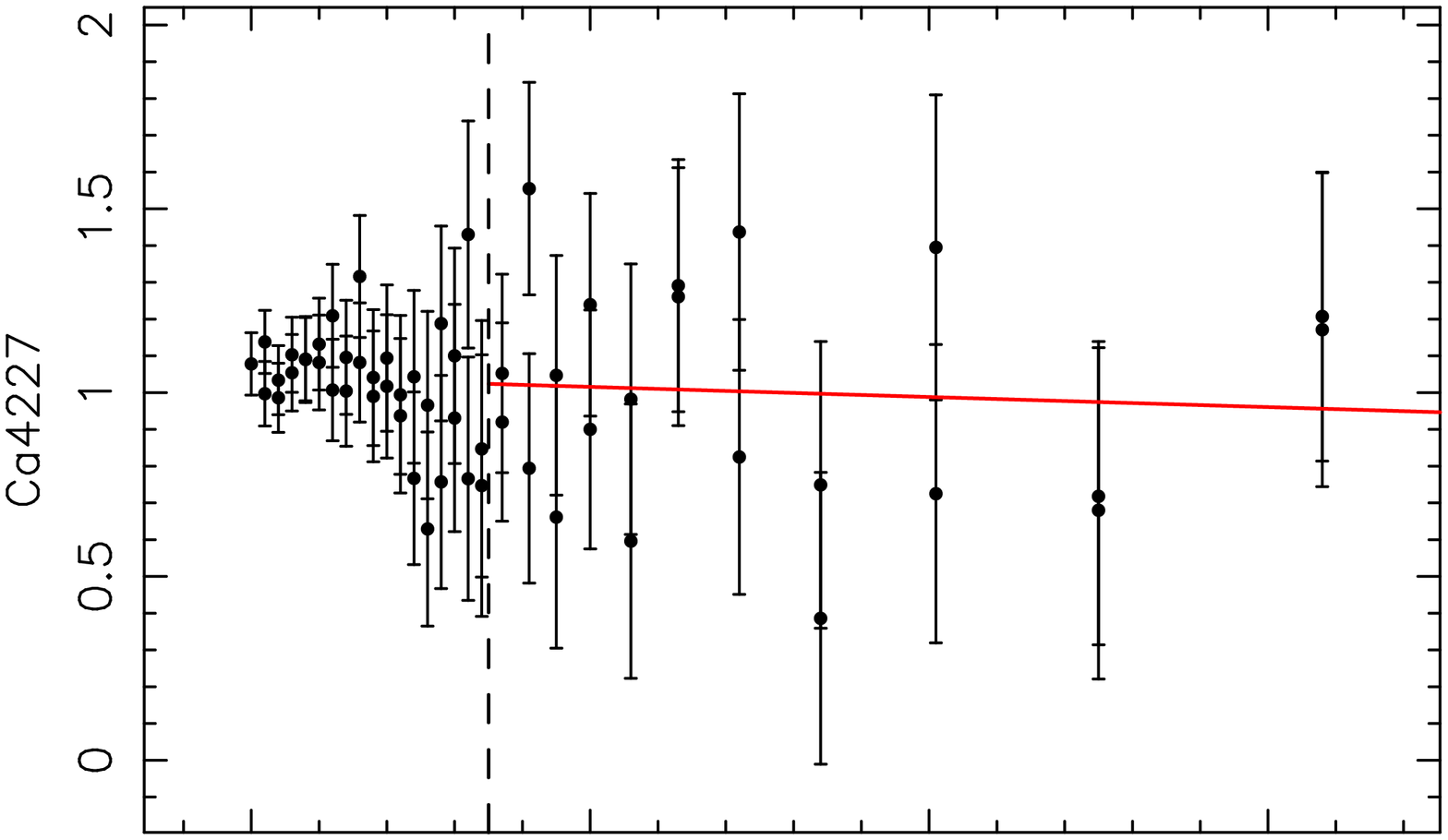}}
\resizebox{0.3\textwidth}{!}{\includegraphics[angle=-90]{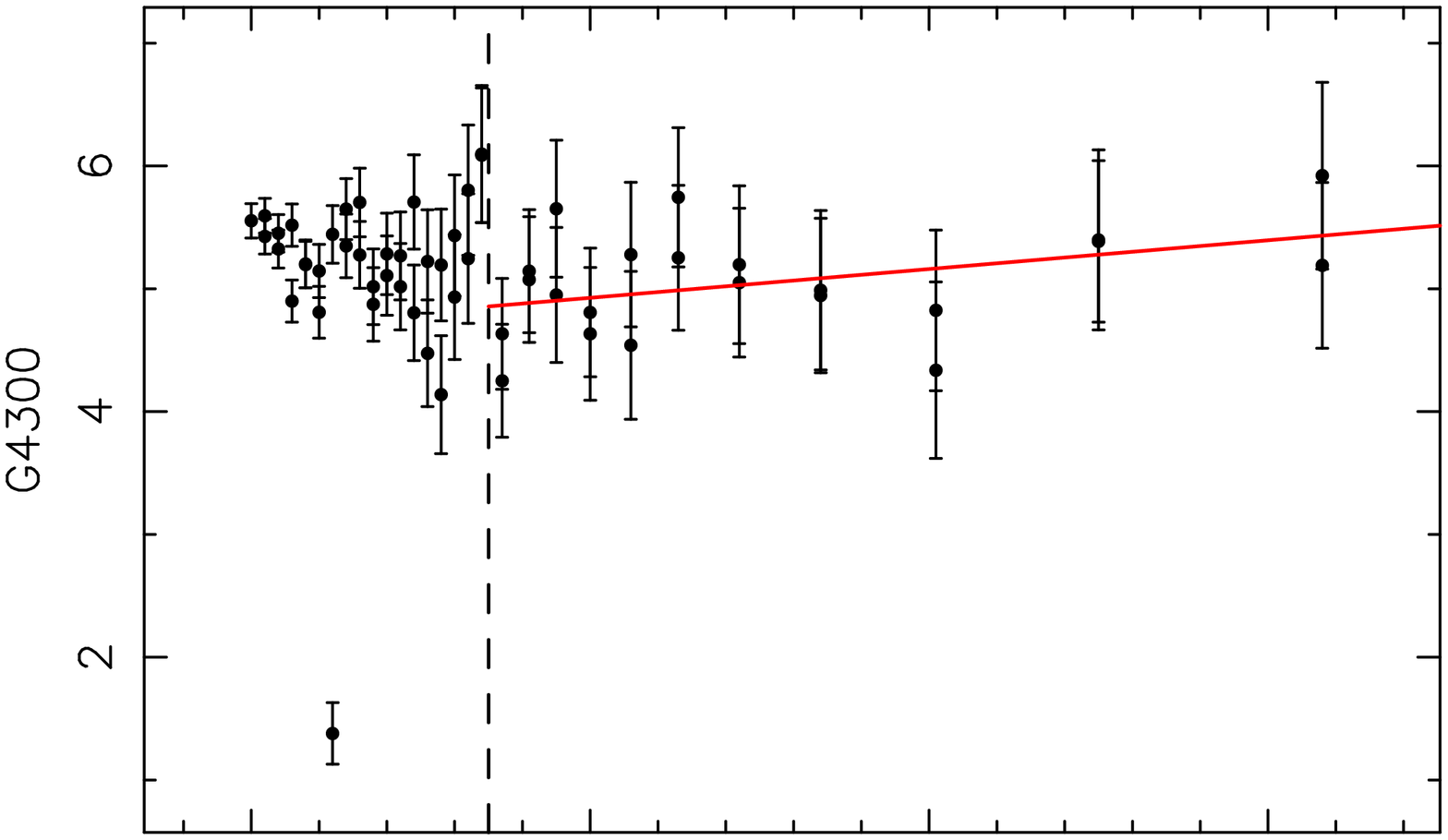}}
\resizebox{0.3\textwidth}{!}{\includegraphics[angle=-90]{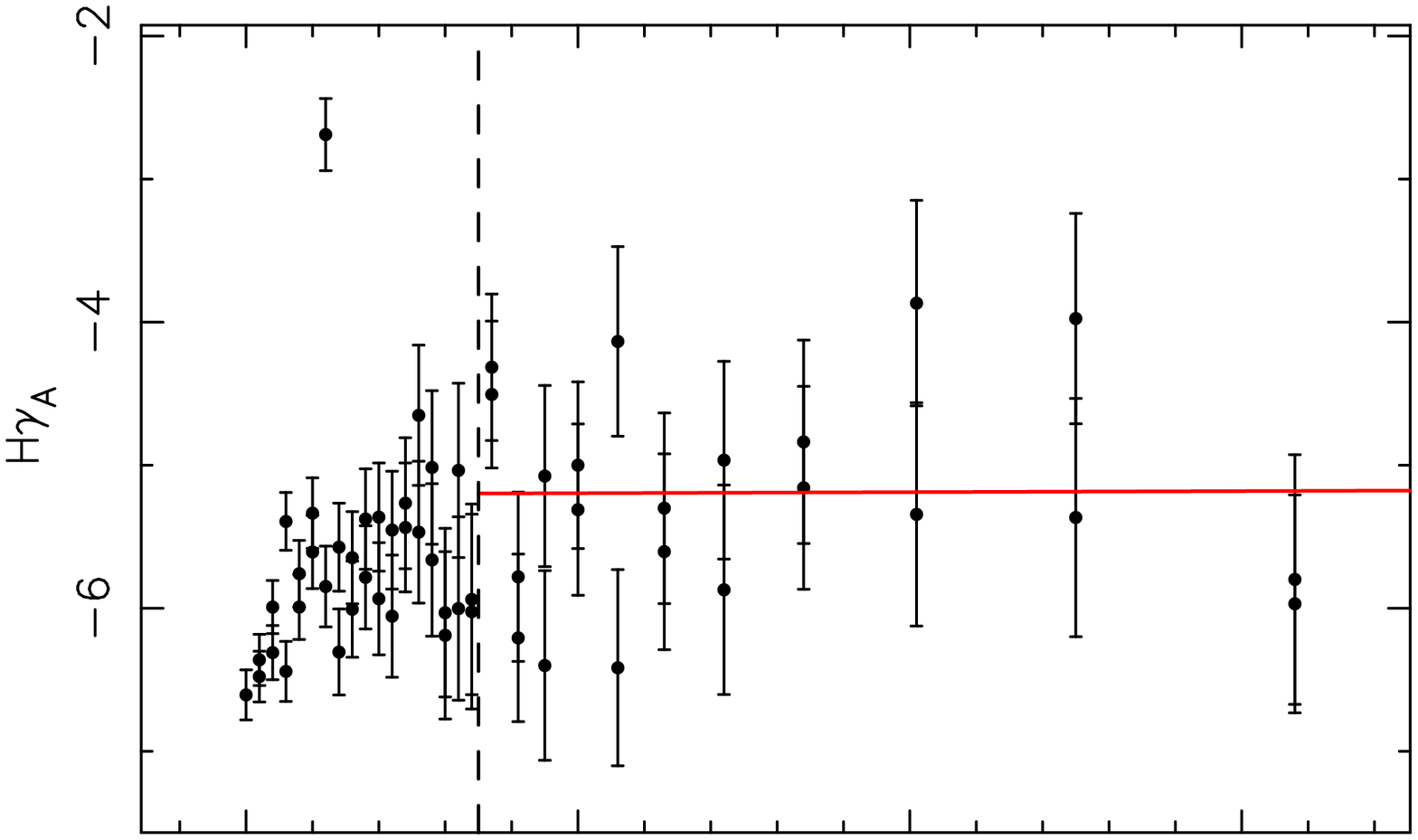}}
\resizebox{0.3\textwidth}{!}{\includegraphics[angle=-90]{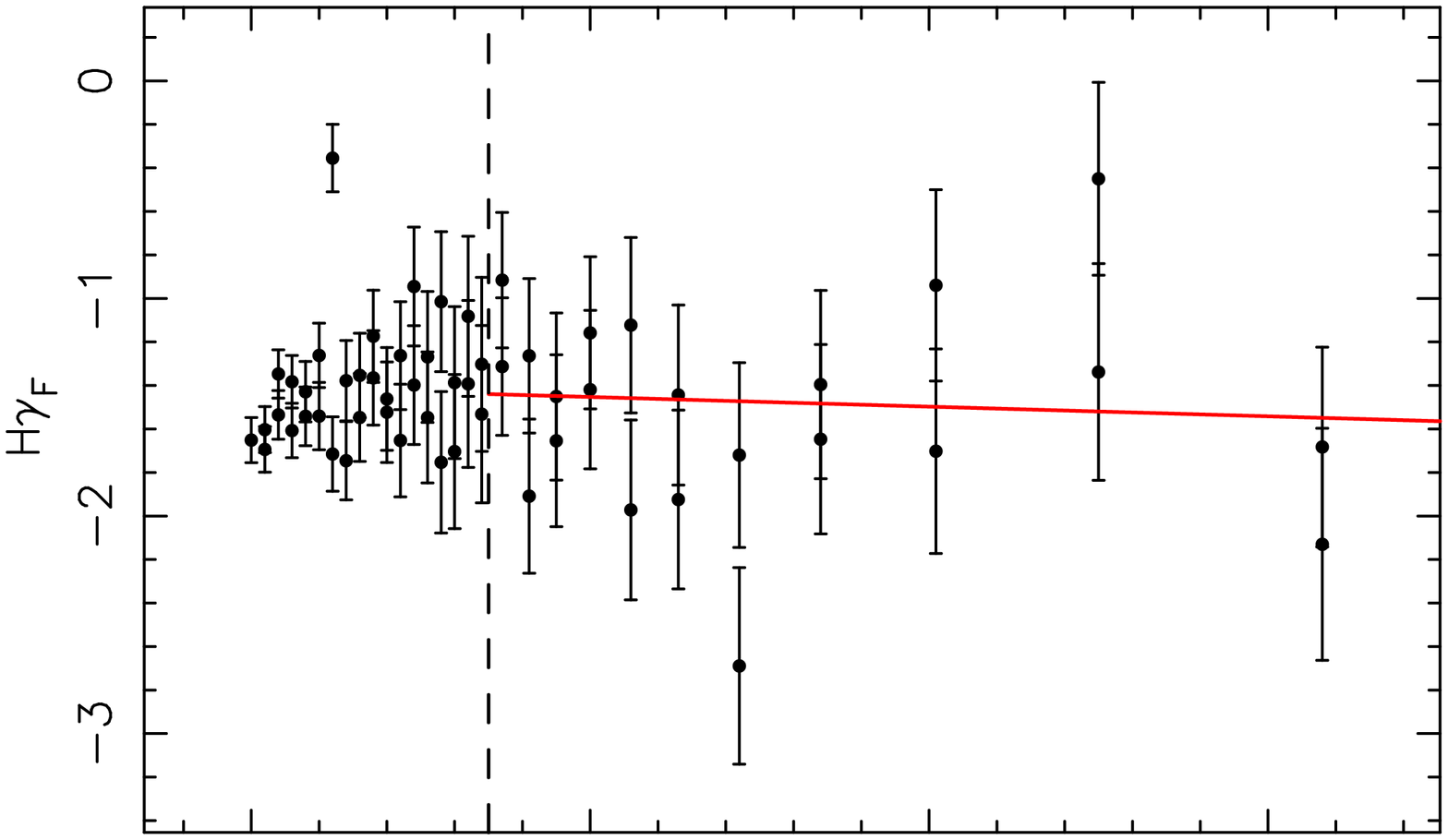}}
\resizebox{0.3\textwidth}{!}{\includegraphics[angle=-90]{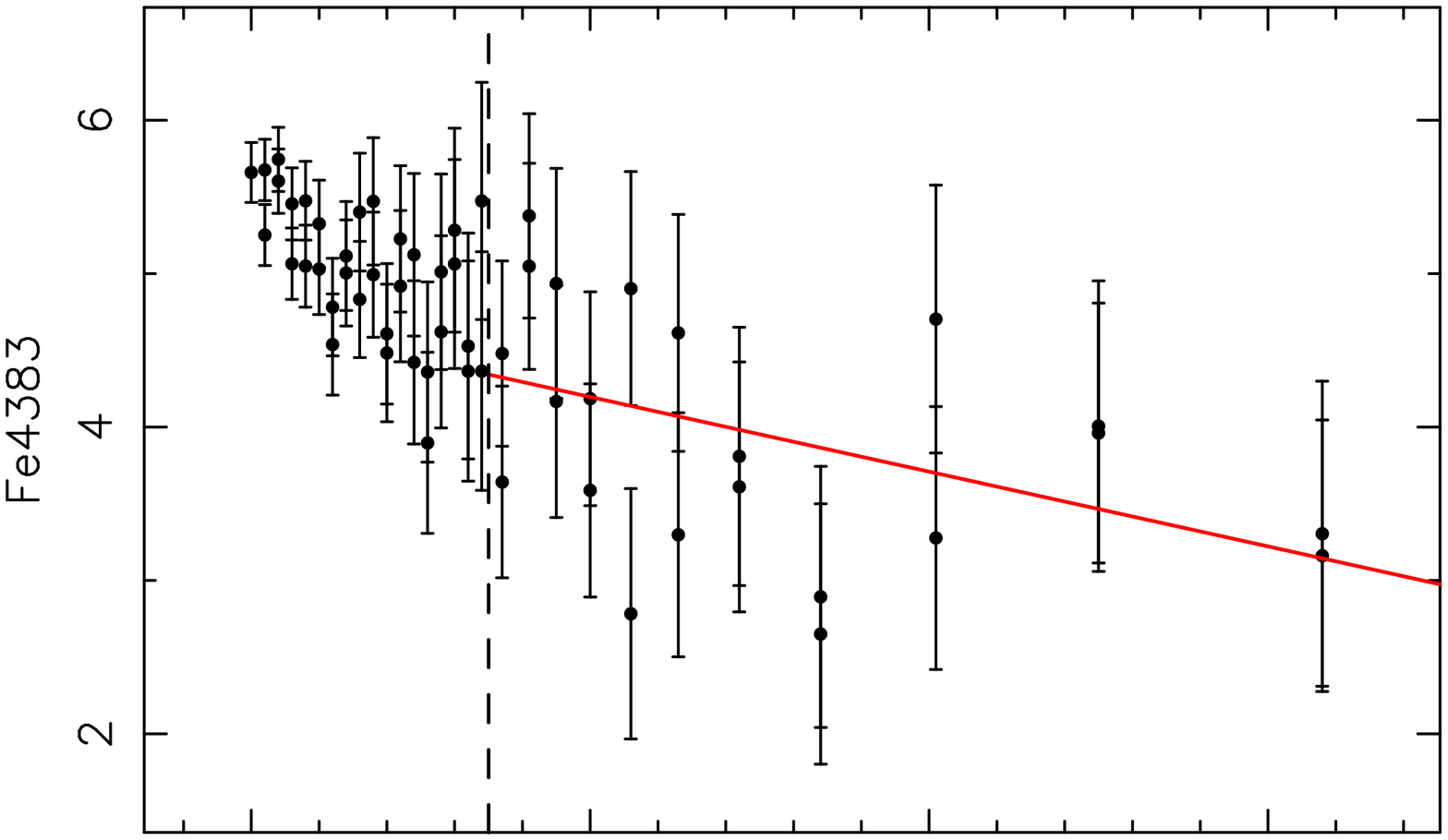}}
\resizebox{0.3\textwidth}{!}{\includegraphics[angle=-90]{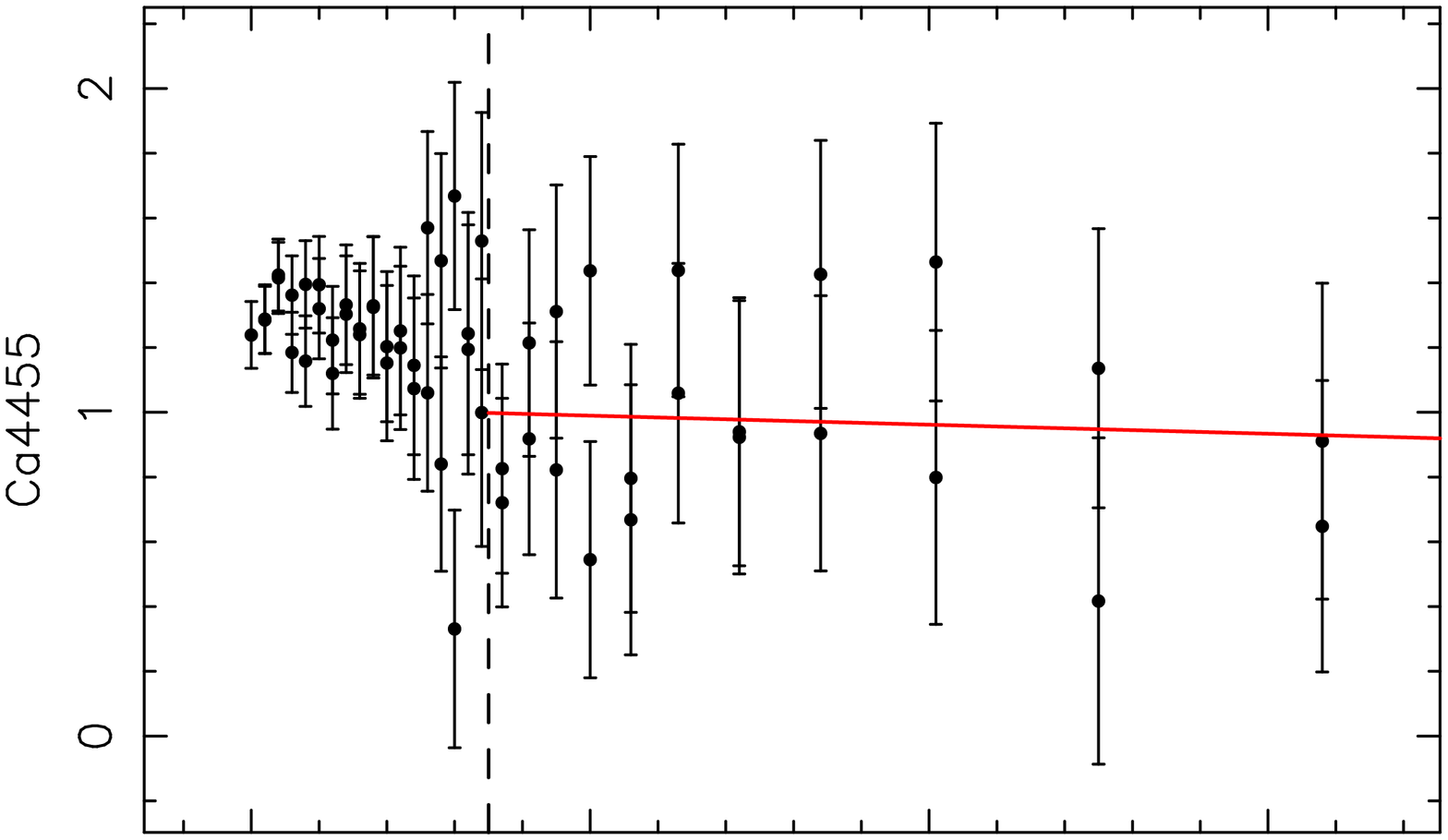}}
\resizebox{0.3\textwidth}{!}{\includegraphics[angle=-90]{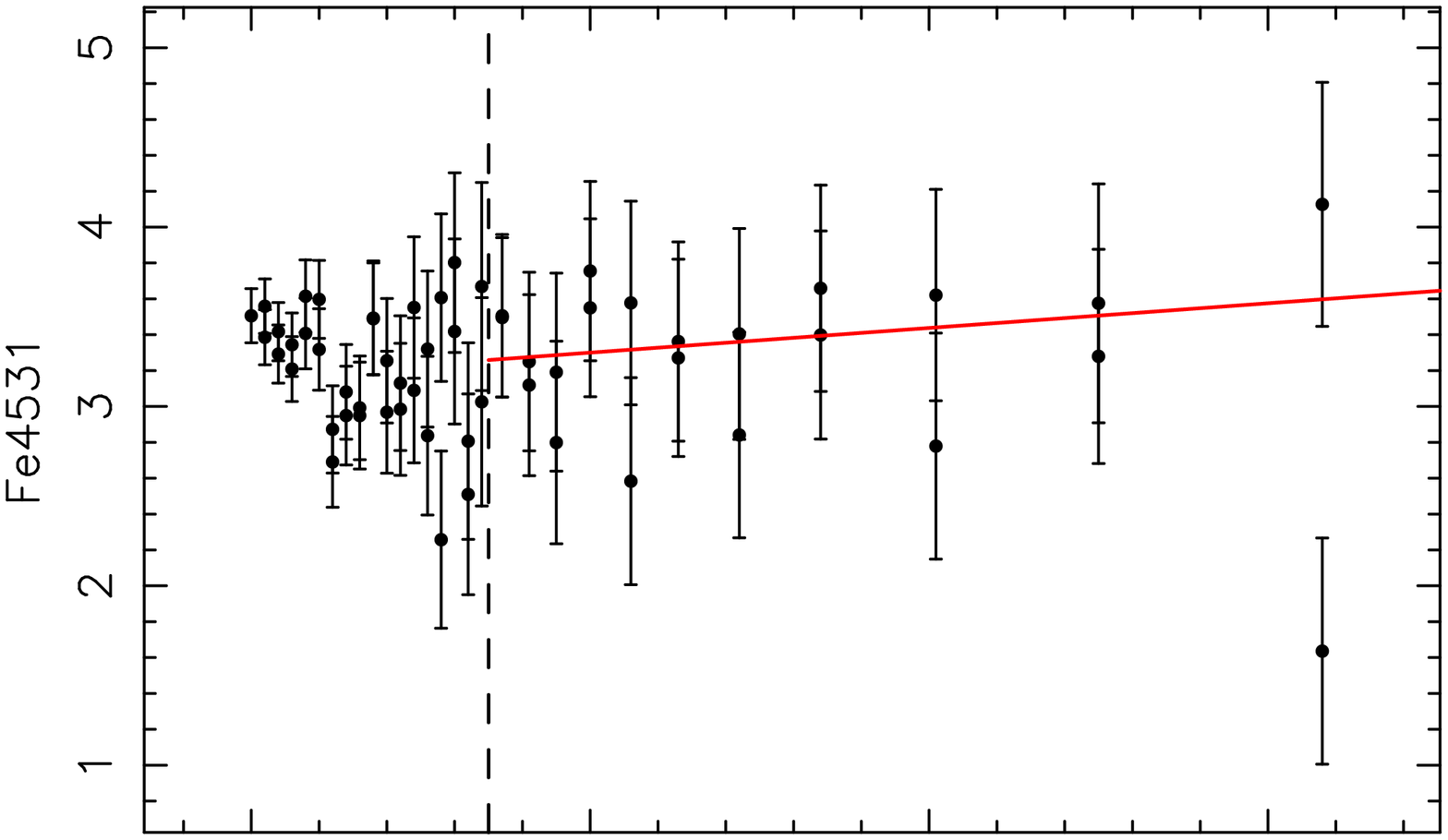}}
\resizebox{0.3\textwidth}{!}{\includegraphics[angle=-90]{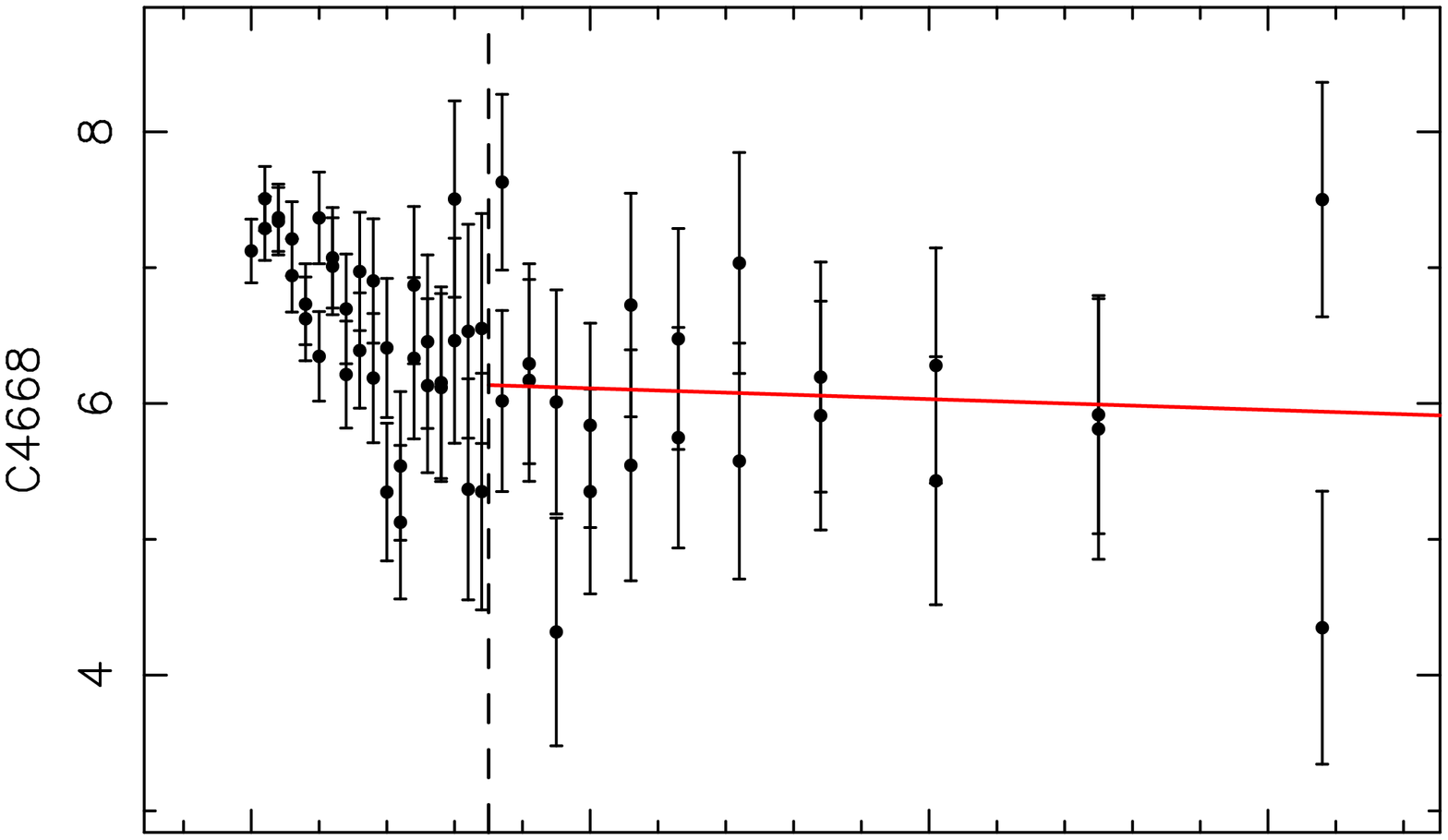}}
\resizebox{0.3\textwidth}{!}{\includegraphics[angle=-90]{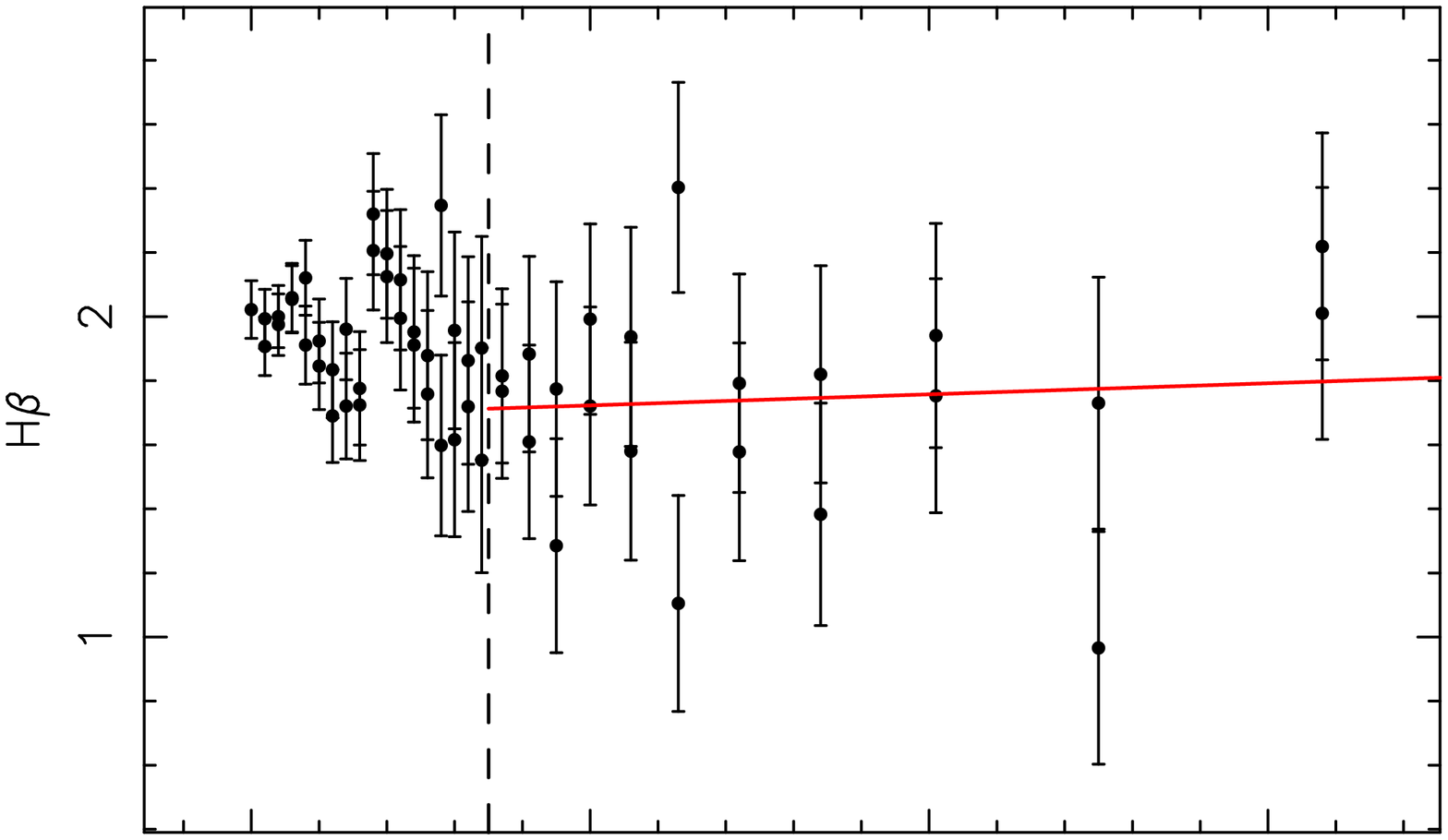}}
\resizebox{0.3\textwidth}{!}{\includegraphics[angle=-90]{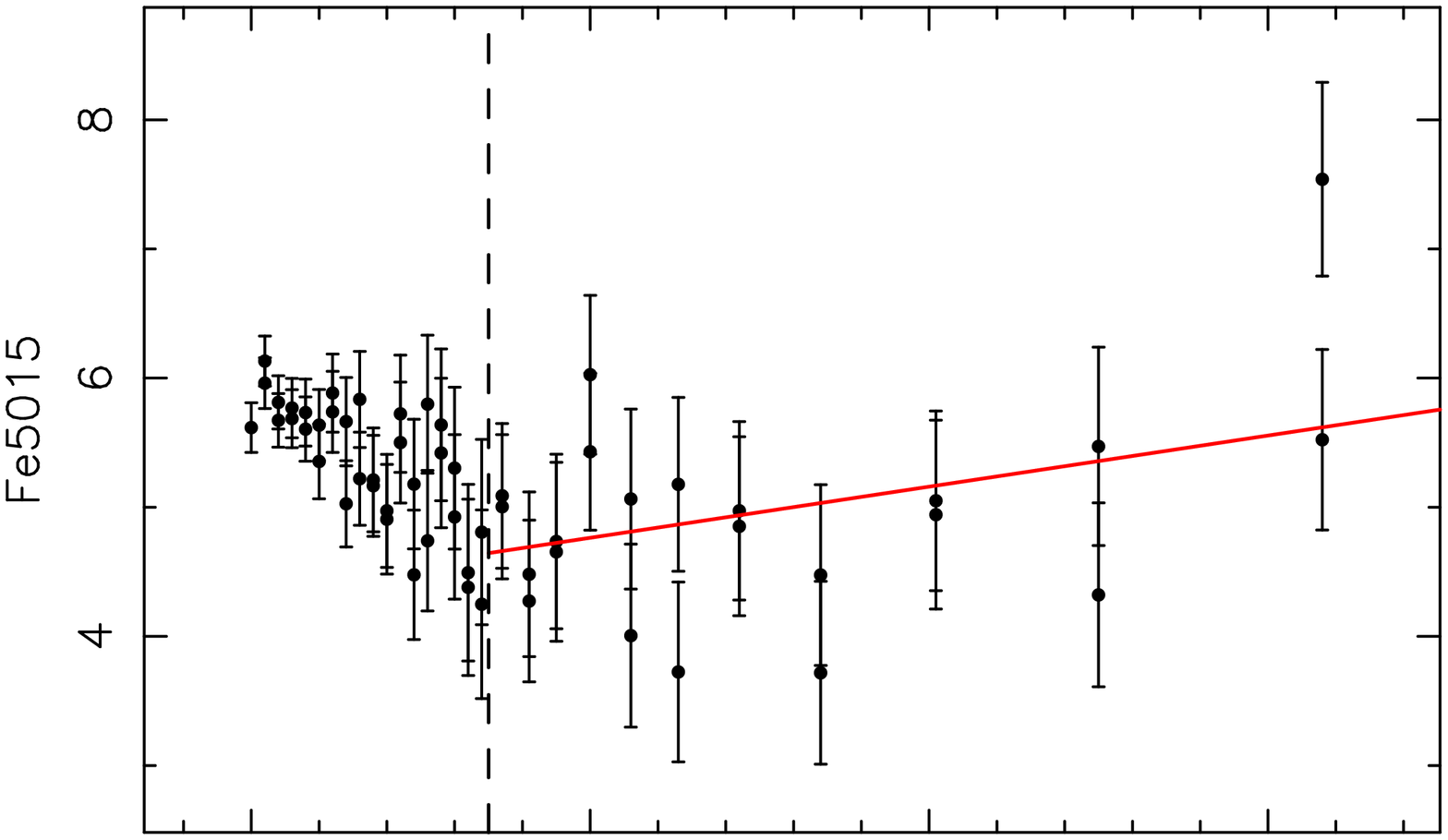}}
\resizebox{0.3\textwidth}{!}{\includegraphics[angle=-90]{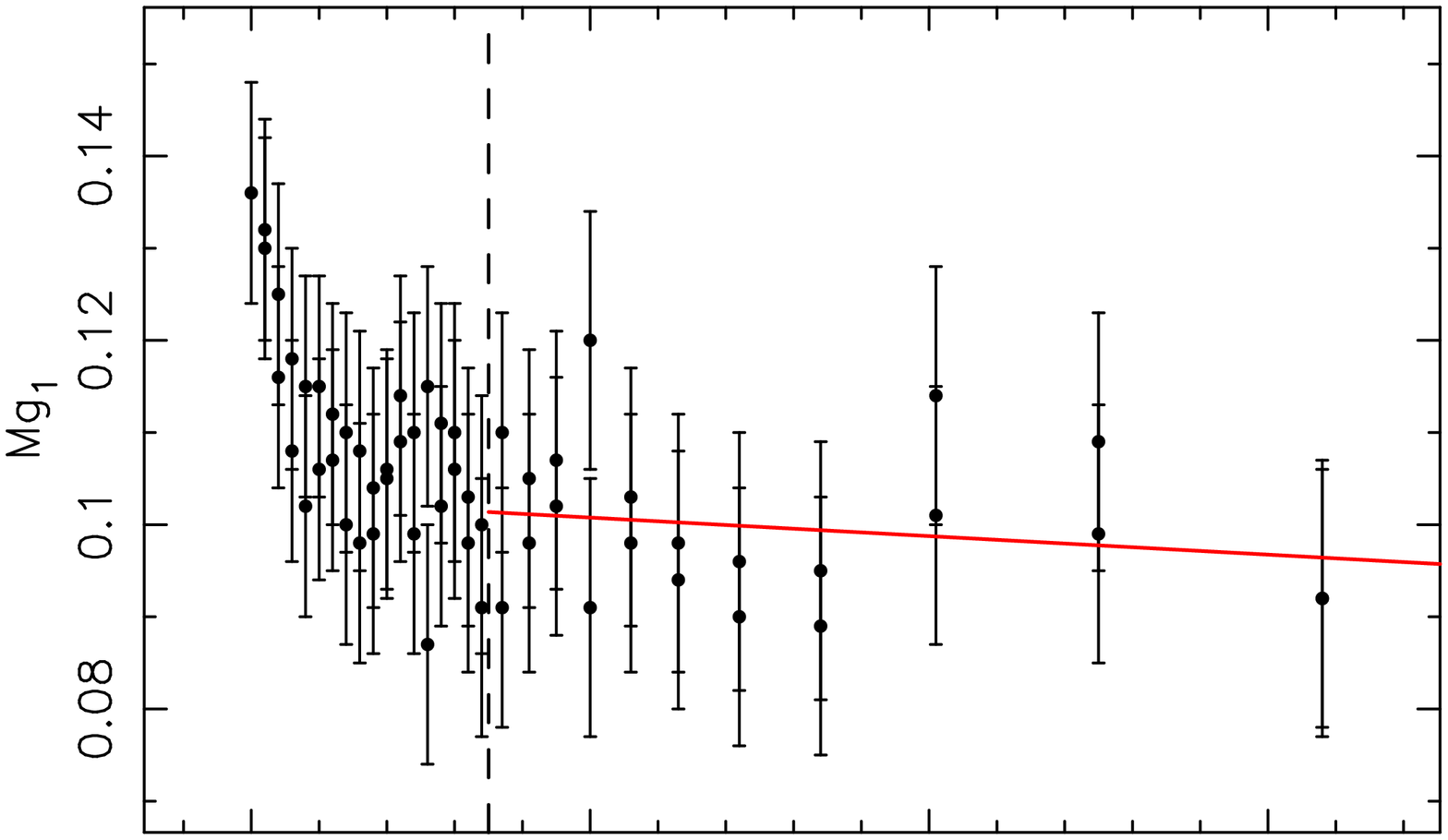}}
\resizebox{0.3\textwidth}{!}{\includegraphics[angle=-90]{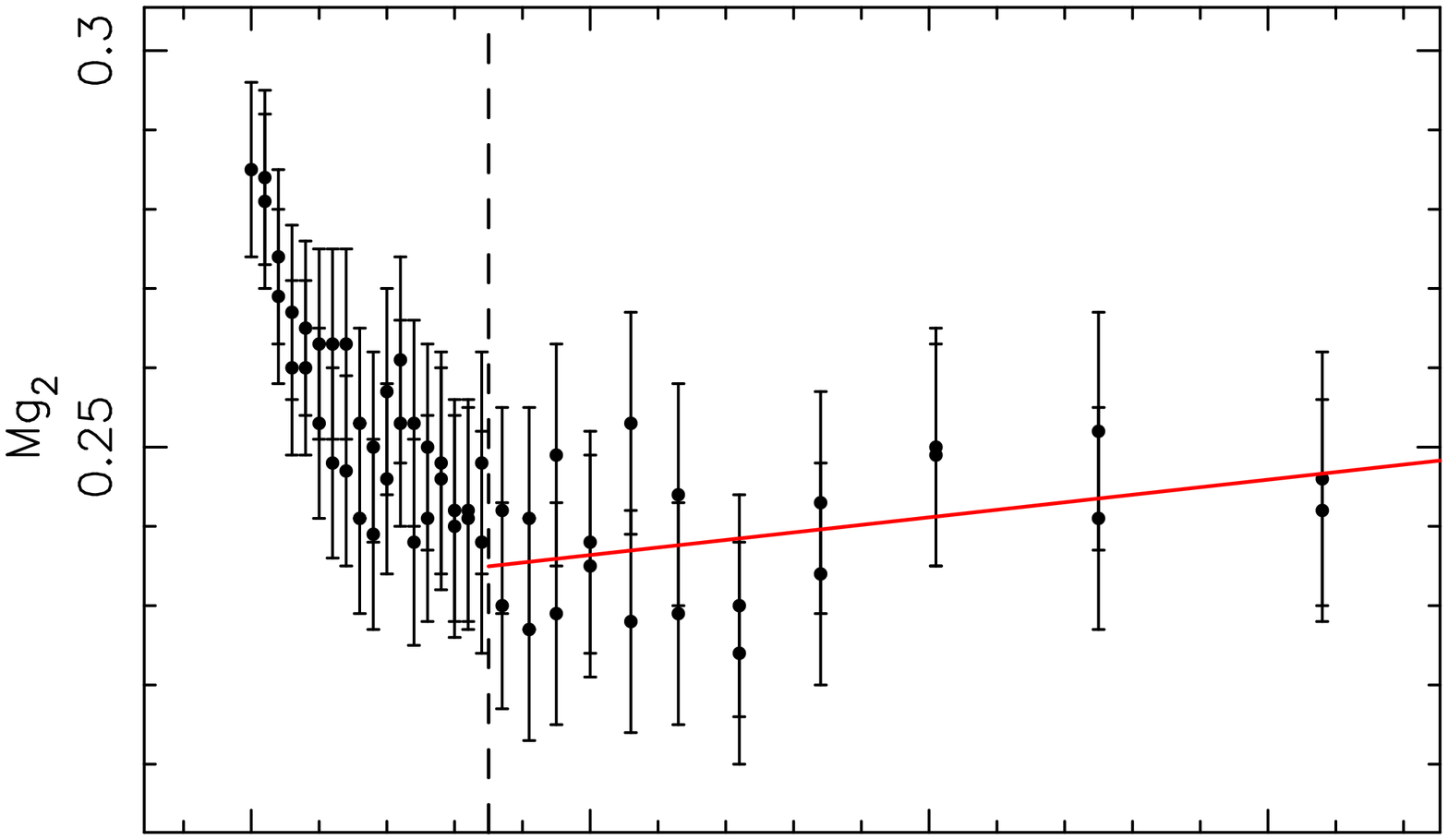}}
\resizebox{0.3\textwidth}{!}{\includegraphics[angle=-90]{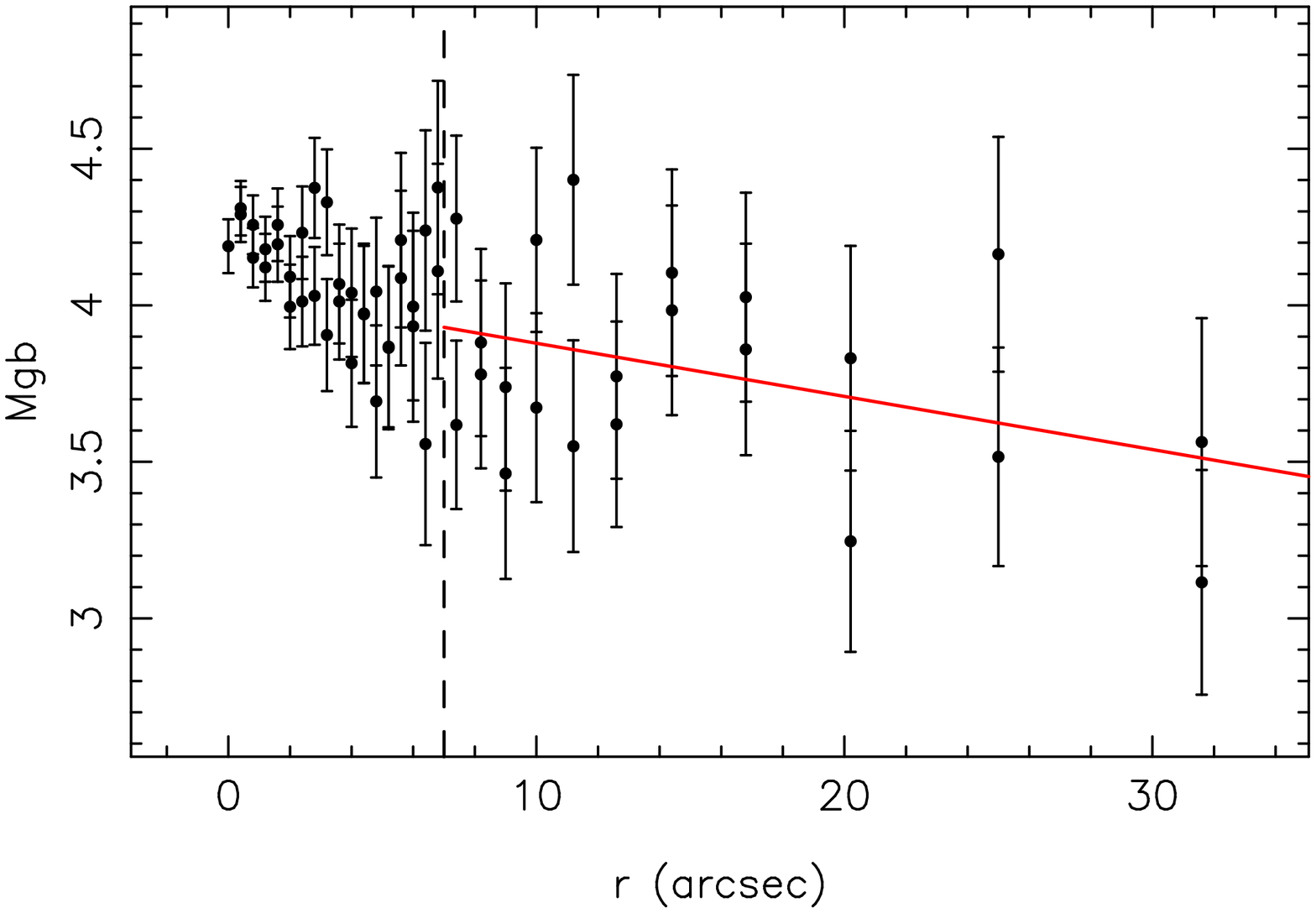}}\hspace{0.85cm}
\resizebox{0.3\textwidth}{!}{\includegraphics[angle=-90]{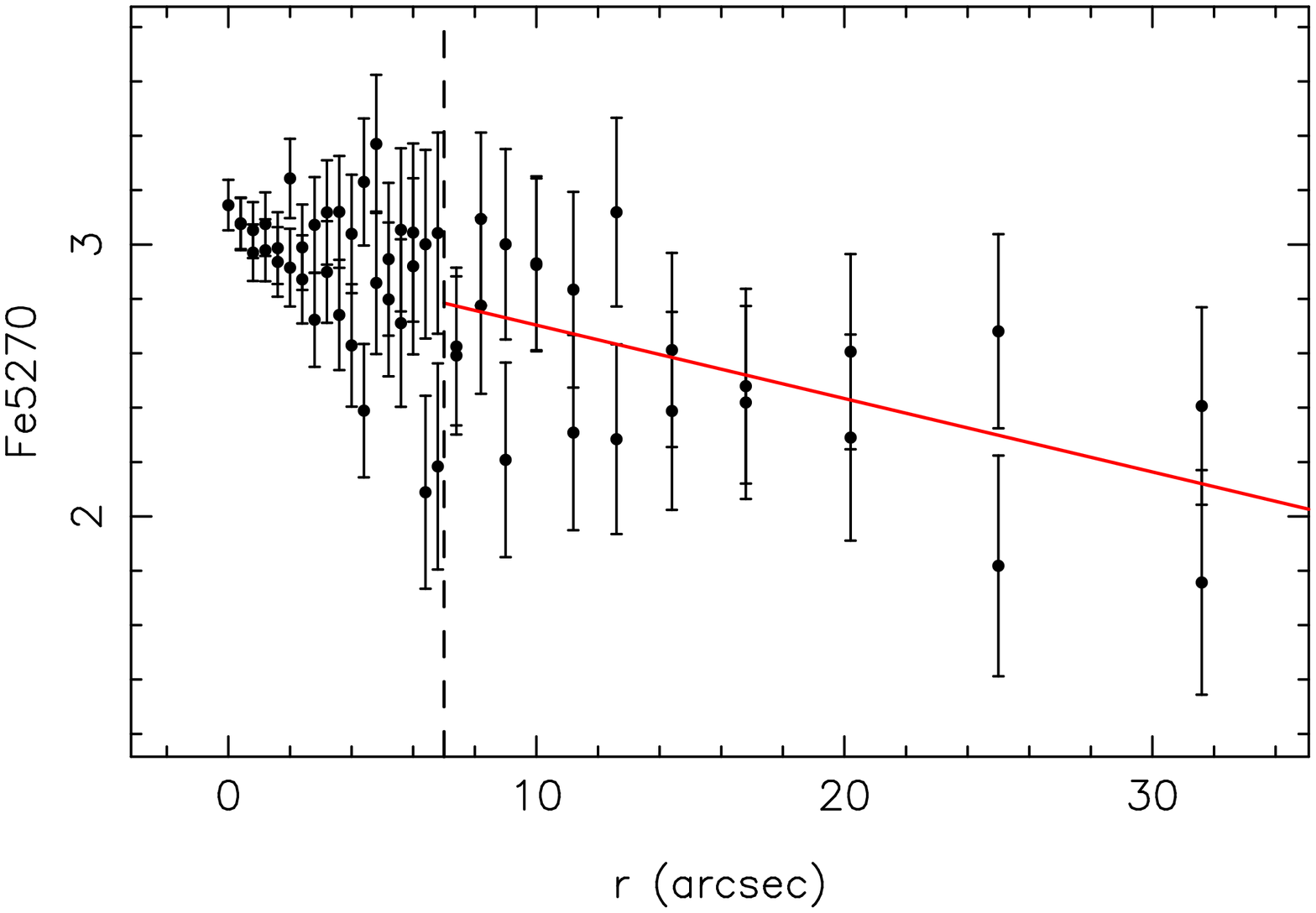}}\hspace{0.85cm}
\resizebox{0.3\textwidth}{!}{\includegraphics[angle=-90]{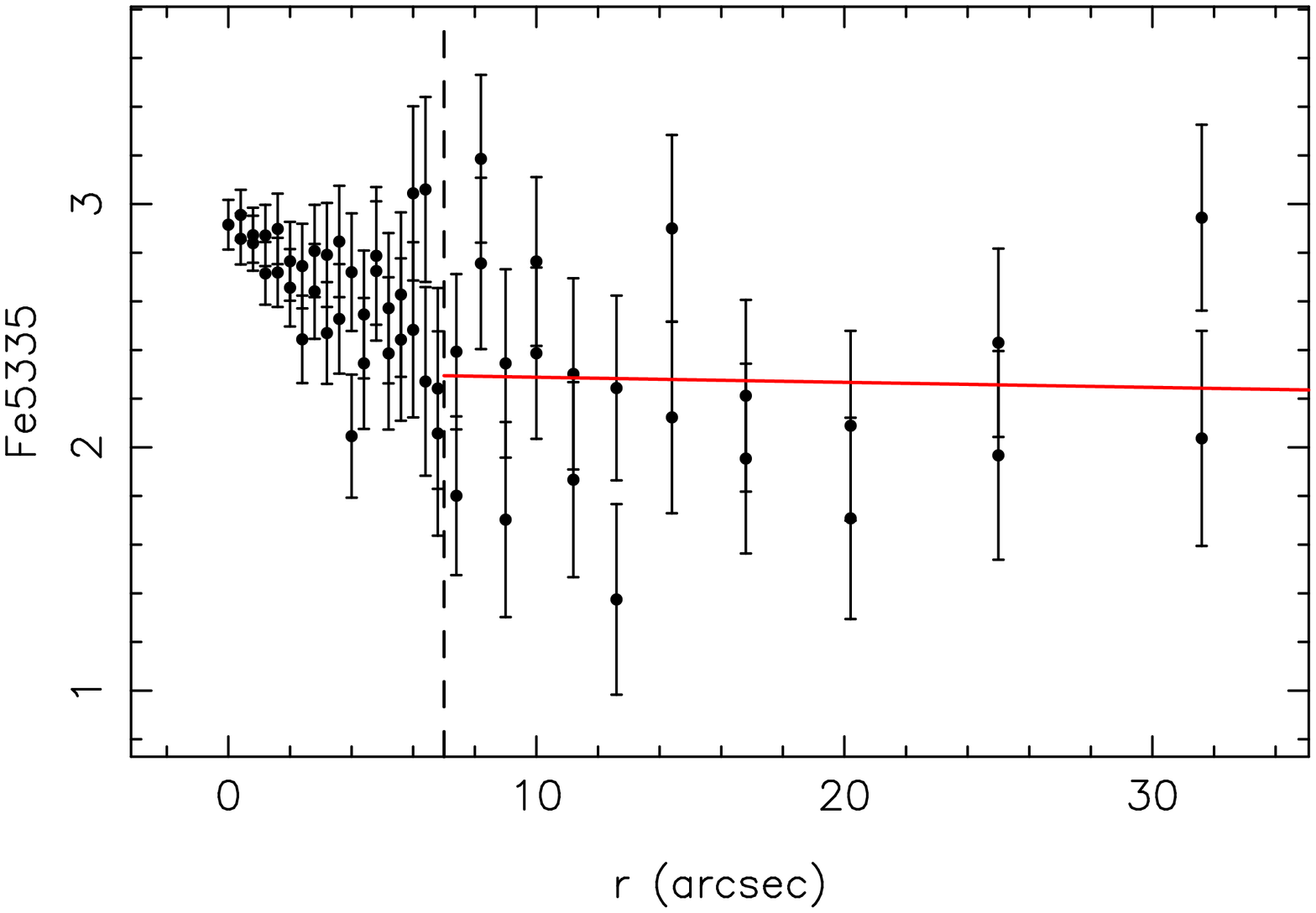}}
\caption{Line-strength distribution in the bar region for all the galaxies}
\end{figure*}

\begin{figure*}
\addtocounter{figure}{-1}
\resizebox{0.3\textwidth}{!}{\includegraphics[angle=-90]{n2859.fol.hda.ps}}
\resizebox{0.3\textwidth}{!}{\includegraphics[angle=-90]{n2859.fol.hdf.ps}}
\resizebox{0.3\textwidth}{!}{\includegraphics[angle=-90]{n2859.fol.cn2.ps}}
\resizebox{0.3\textwidth}{!}{\includegraphics[angle=-90]{n2859.fol.ca4227.ps}}
\resizebox{0.3\textwidth}{!}{\includegraphics[angle=-90]{n2859.fol.g4300.ps}}
\resizebox{0.3\textwidth}{!}{\includegraphics[angle=-90]{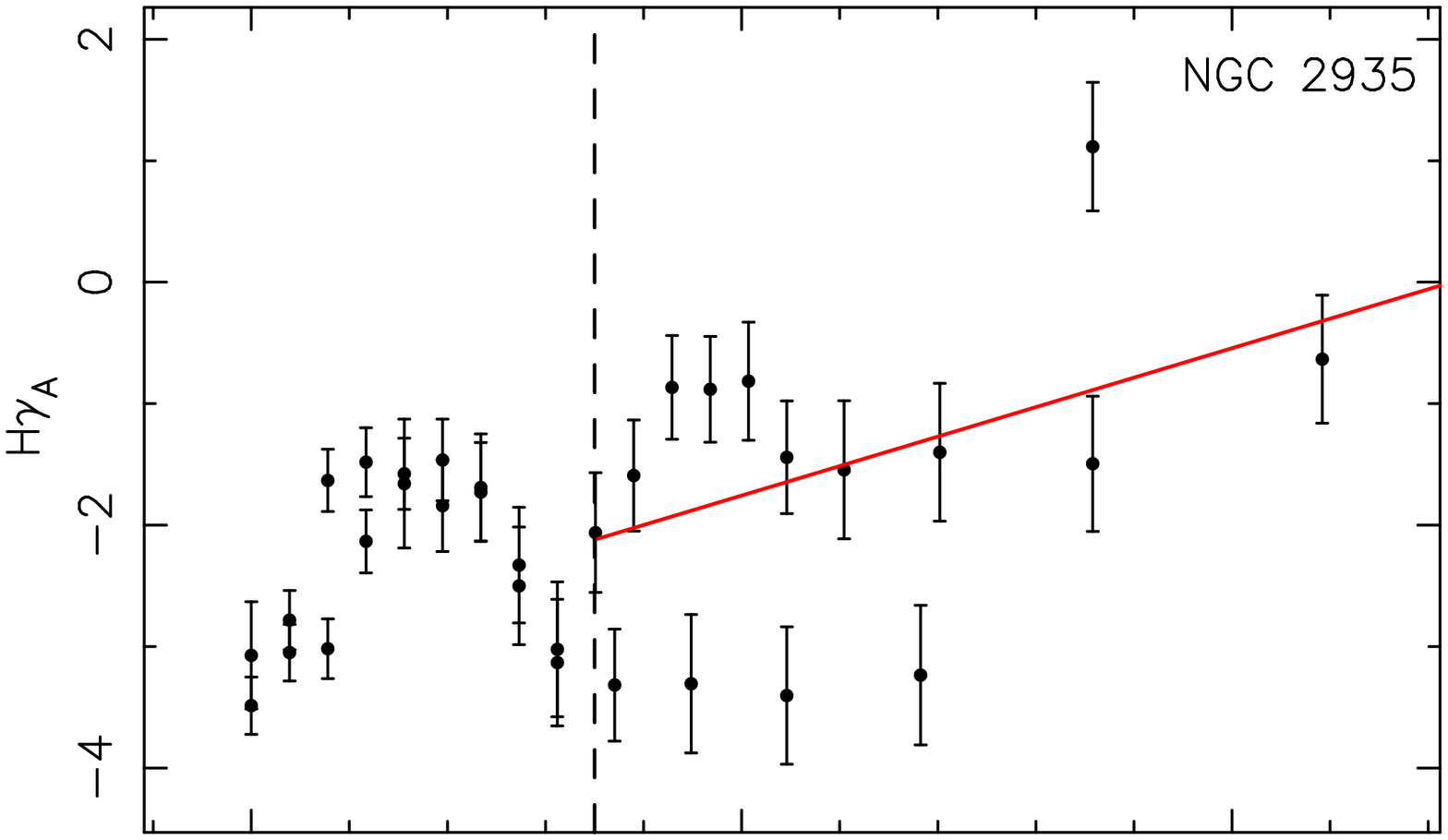}}
\resizebox{0.3\textwidth}{!}{\includegraphics[angle=-90]{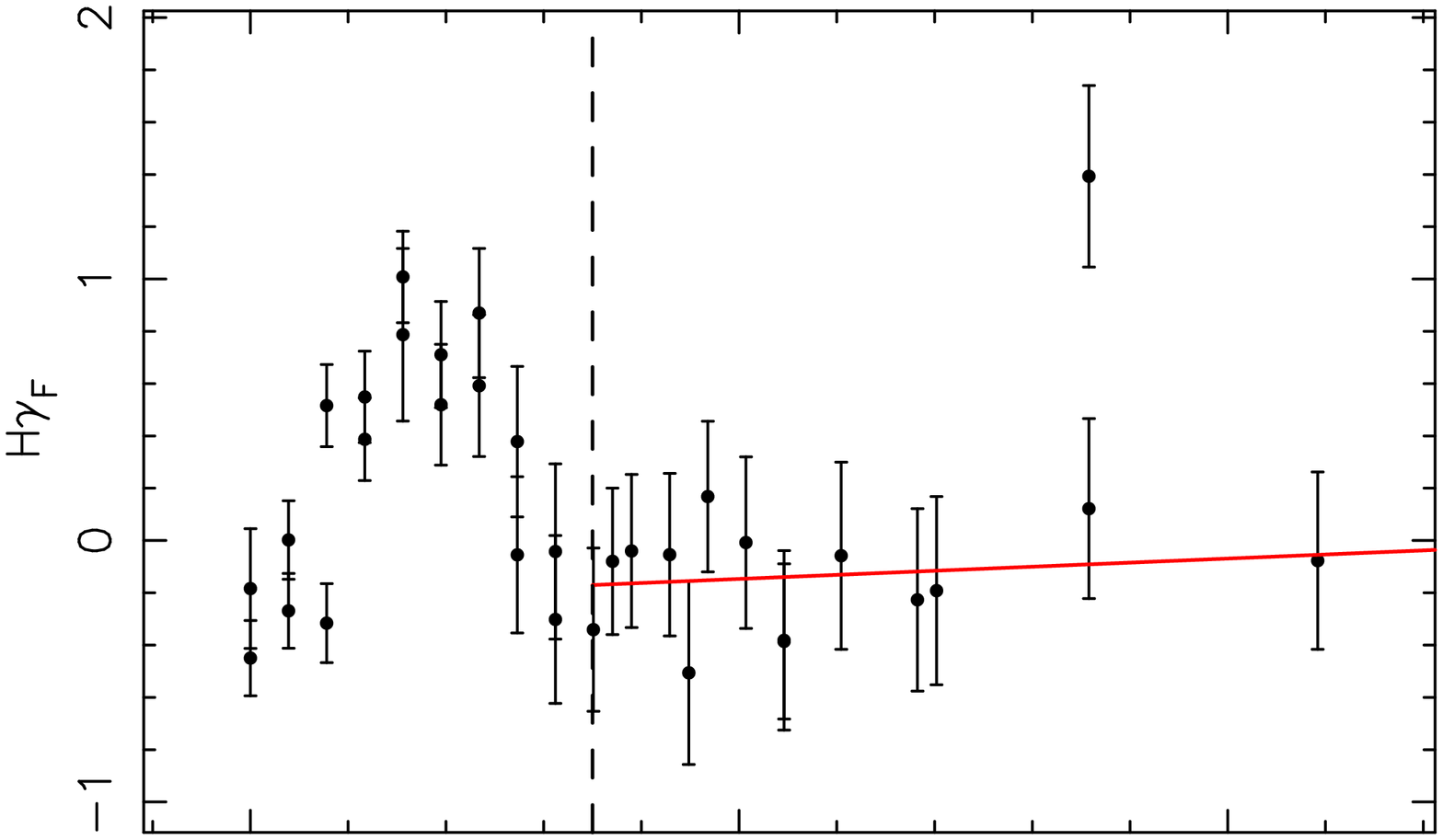}}
\resizebox{0.3\textwidth}{!}{\includegraphics[angle=-90]{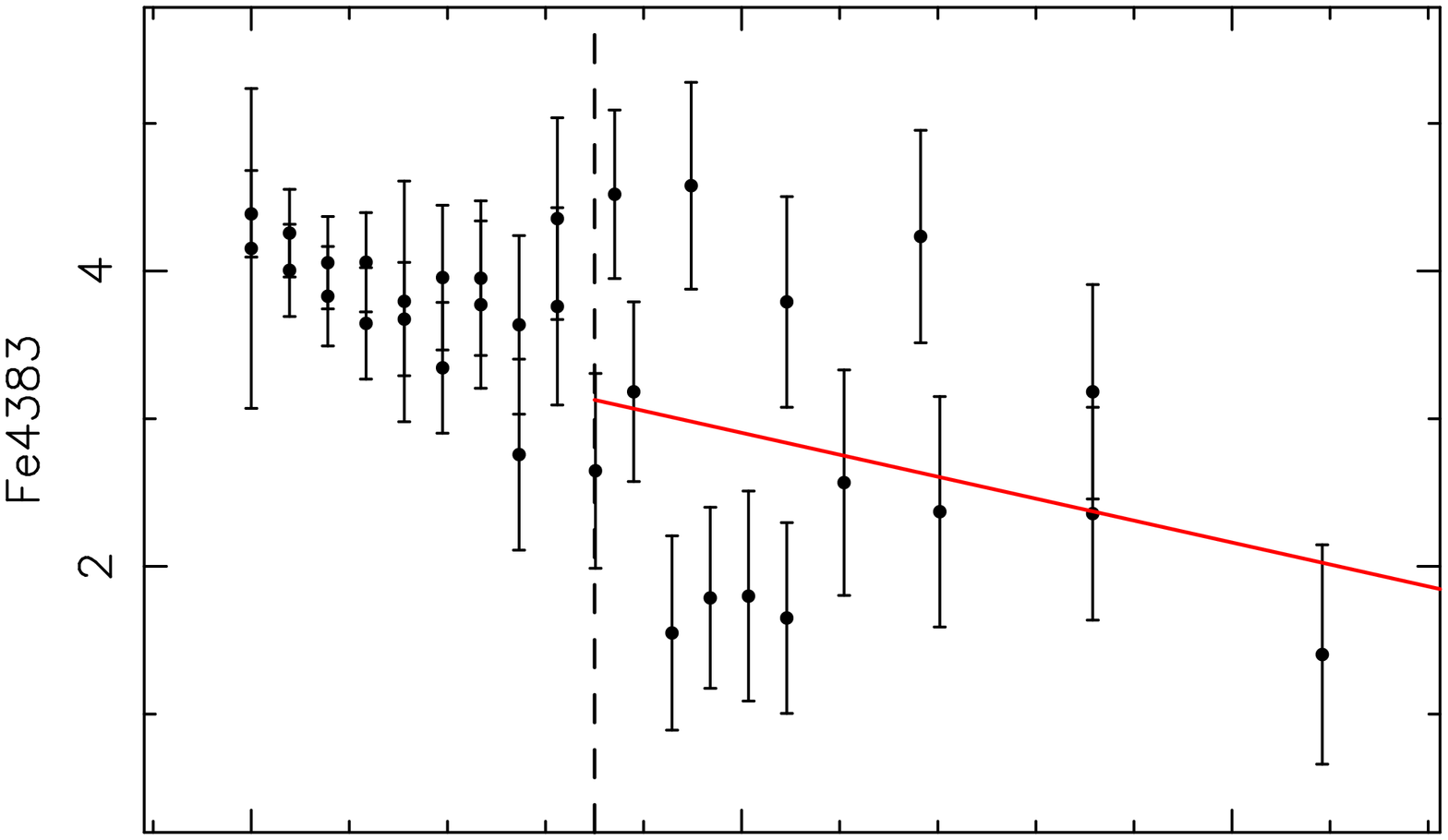}}
\resizebox{0.3\textwidth}{!}{\includegraphics[angle=-90]{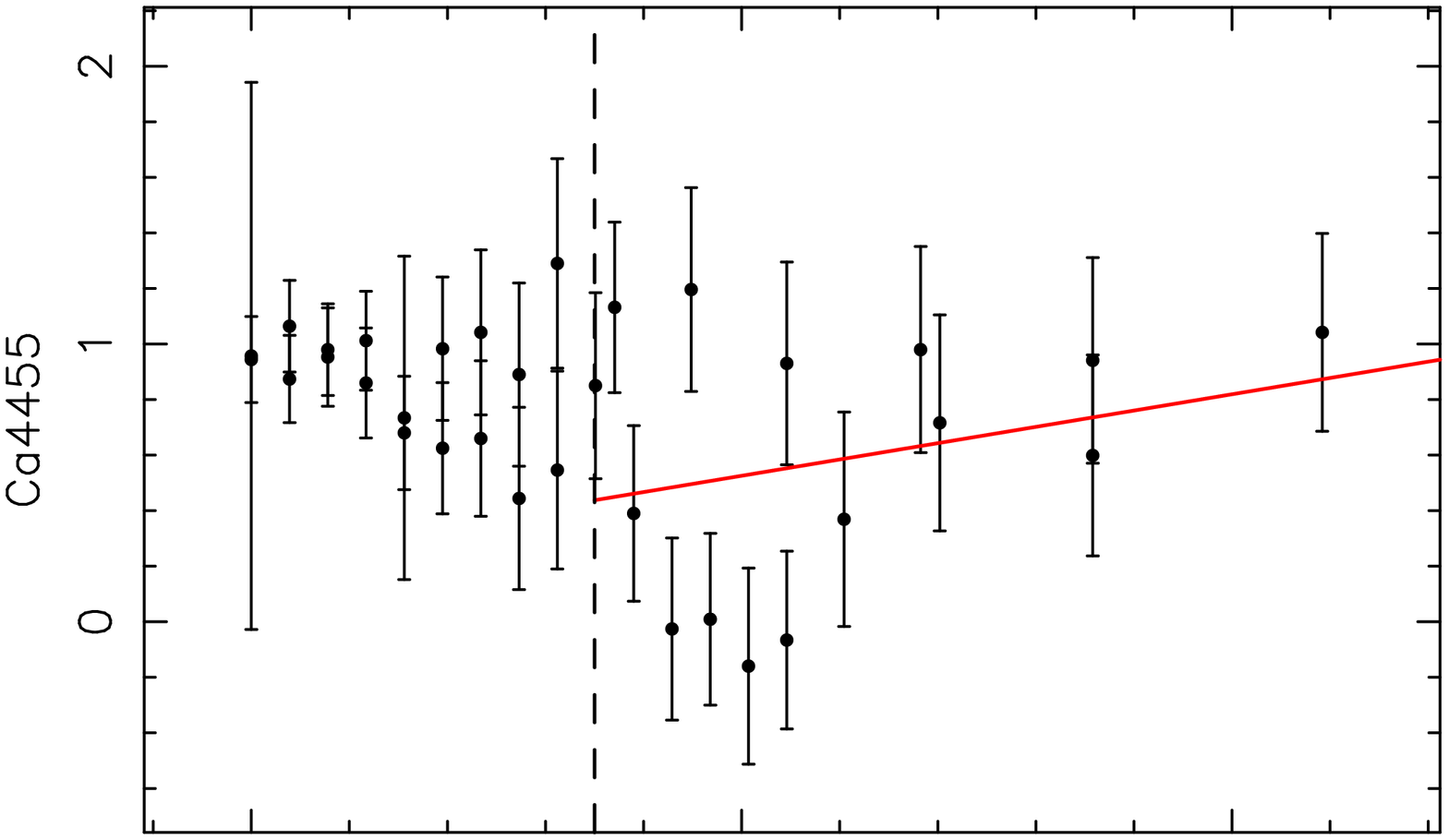}}
\resizebox{0.3\textwidth}{!}{\includegraphics[angle=-90]{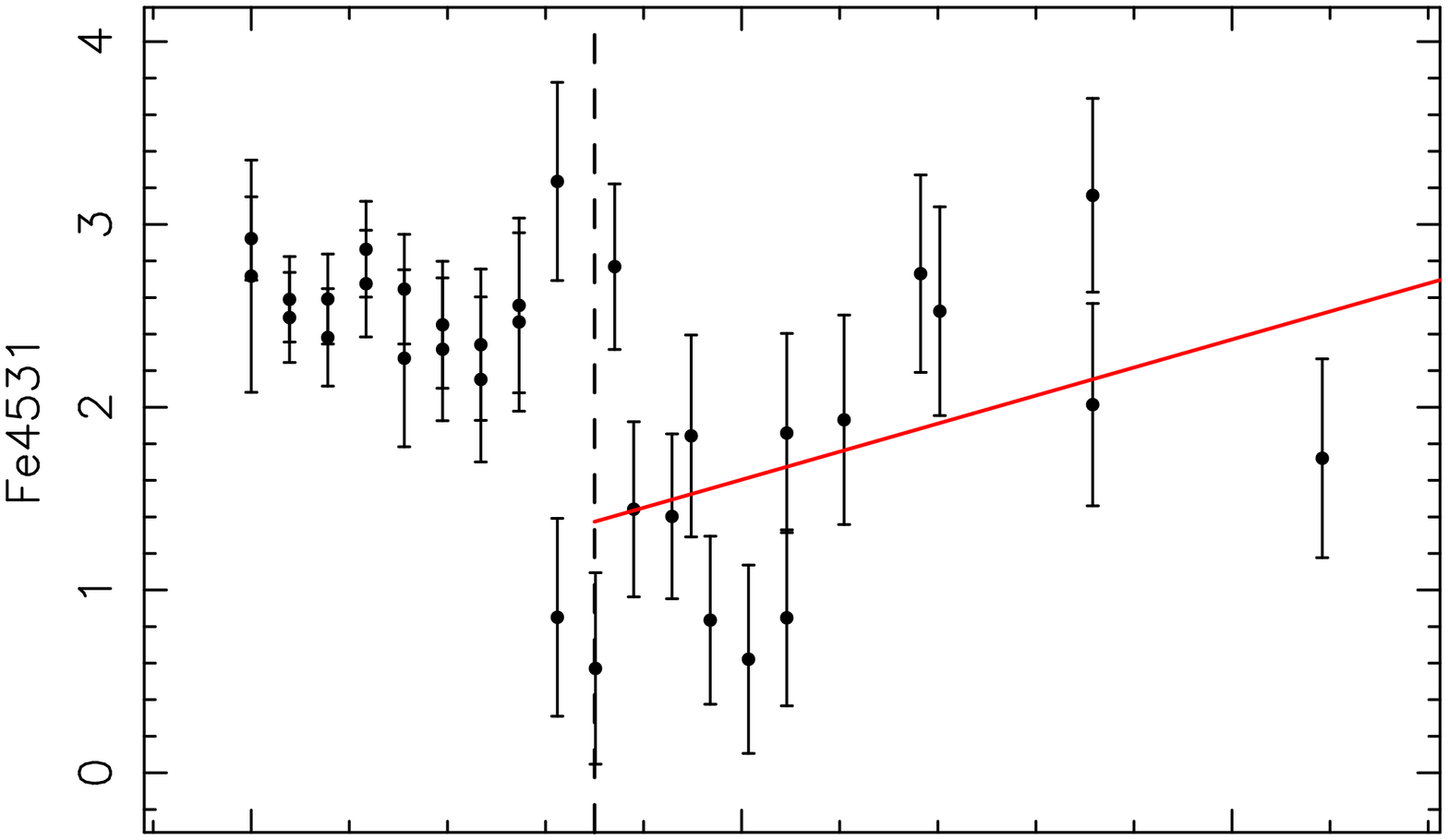}}
\resizebox{0.3\textwidth}{!}{\includegraphics[angle=-90]{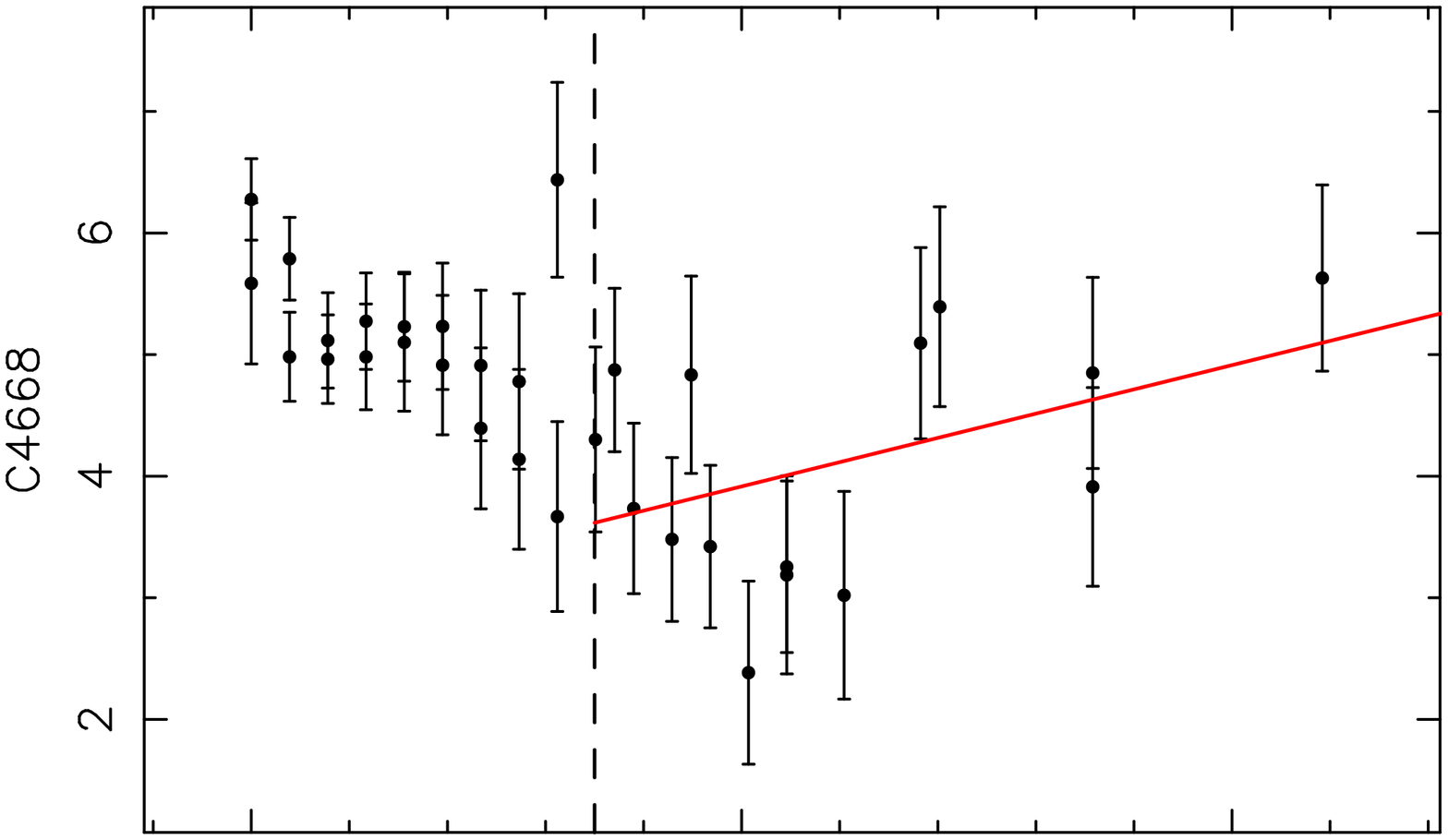}}
\resizebox{0.3\textwidth}{!}{\includegraphics[angle=-90]{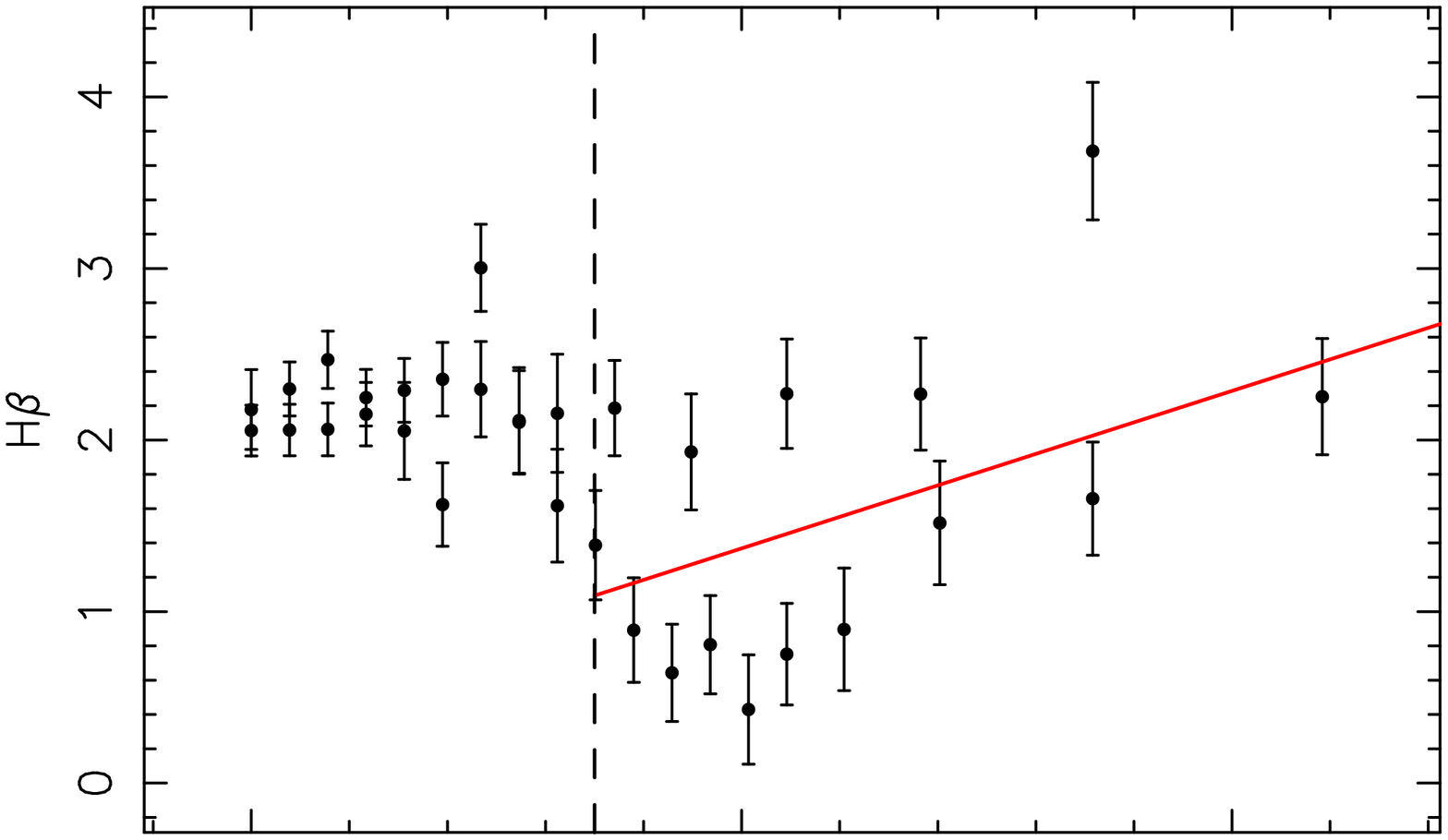}}
\resizebox{0.3\textwidth}{!}{\includegraphics[angle=-90]{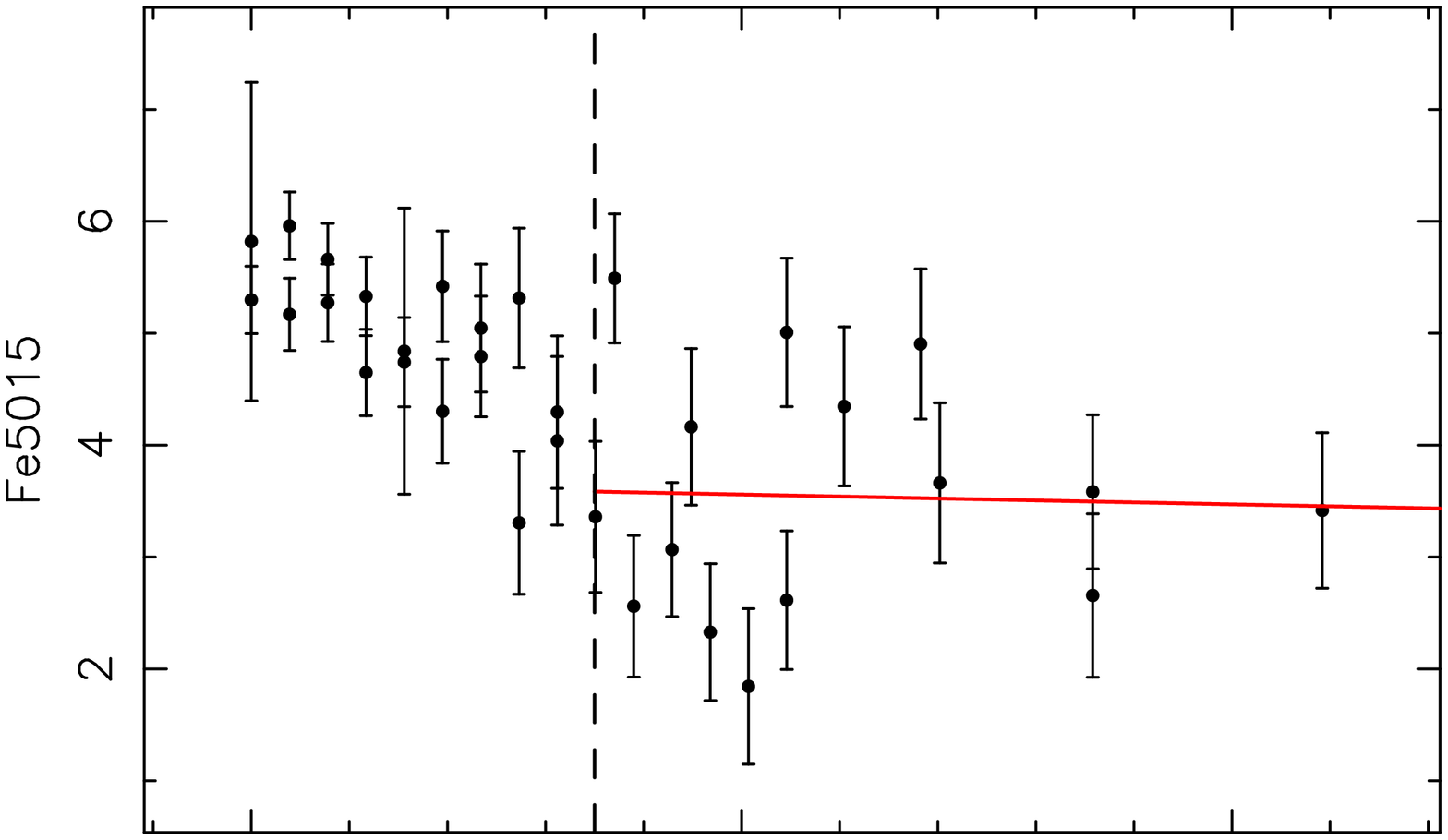}}
\resizebox{0.3\textwidth}{!}{\includegraphics[angle=-90]{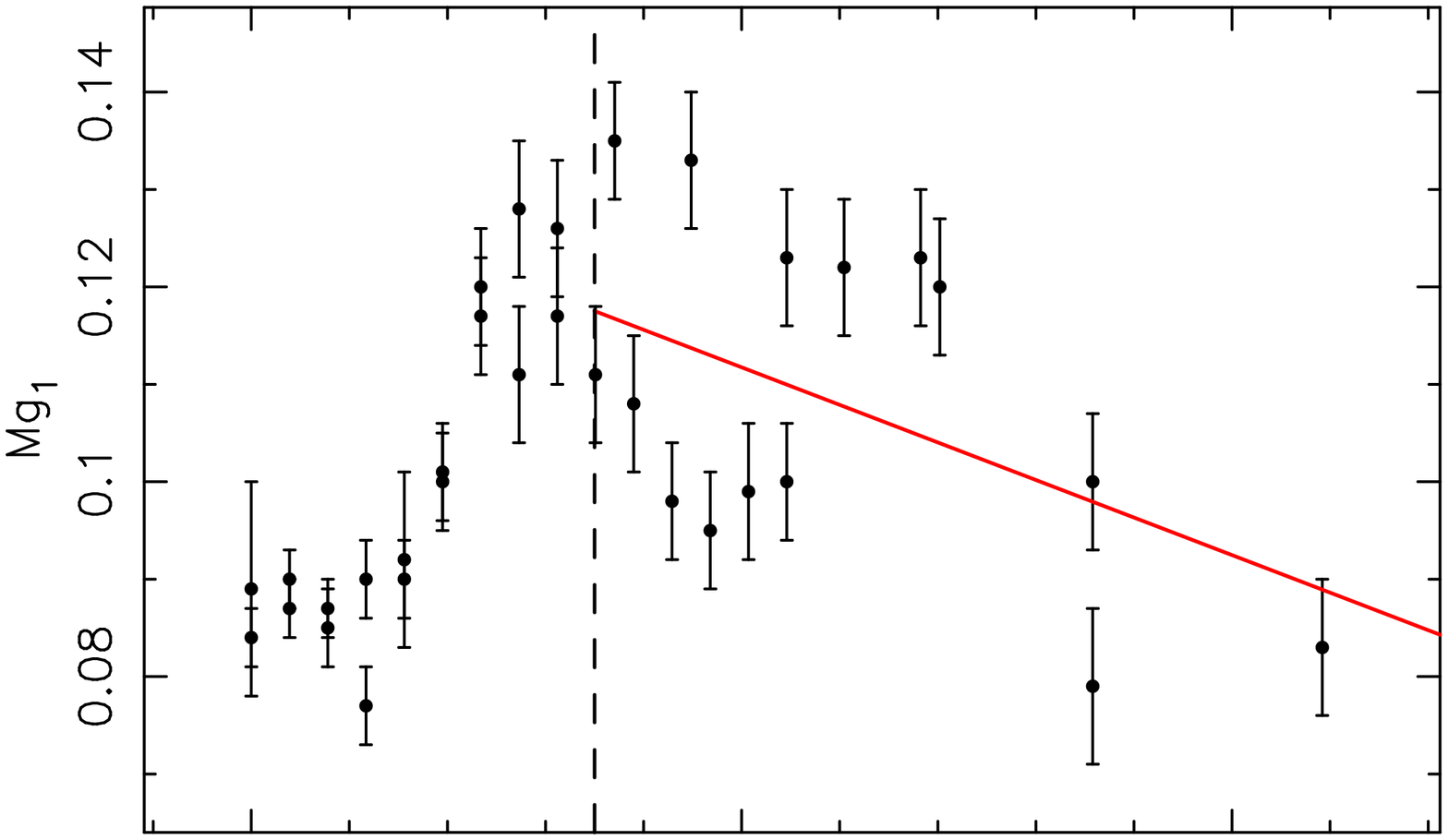}}
\resizebox{0.3\textwidth}{!}{\includegraphics[angle=-90]{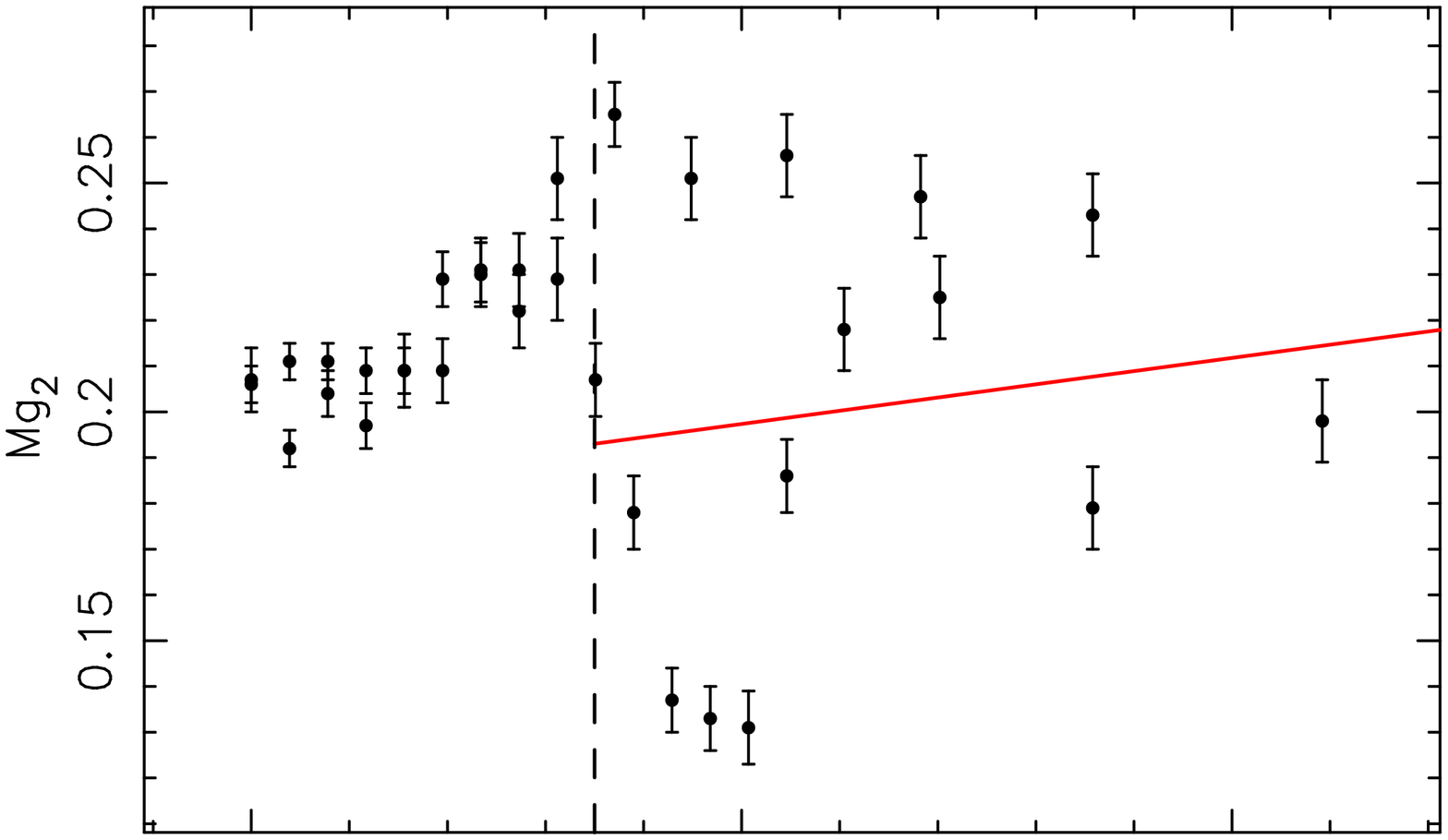}}\hspace{0.85cm}
\resizebox{0.3\textwidth}{!}{\includegraphics[angle=-90]{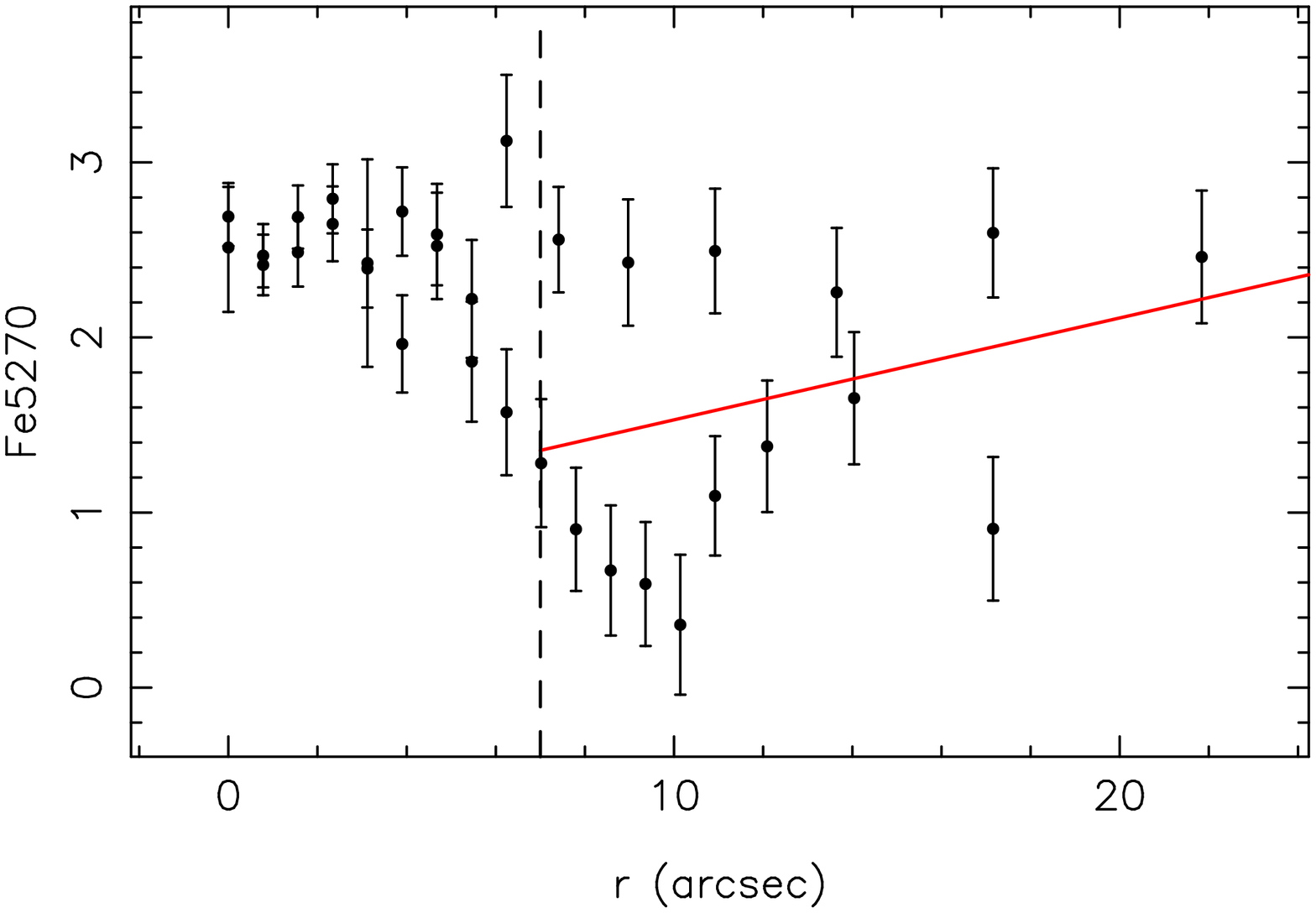}}\hspace{0.85cm}
\resizebox{0.3\textwidth}{!}{\includegraphics[angle=-90]{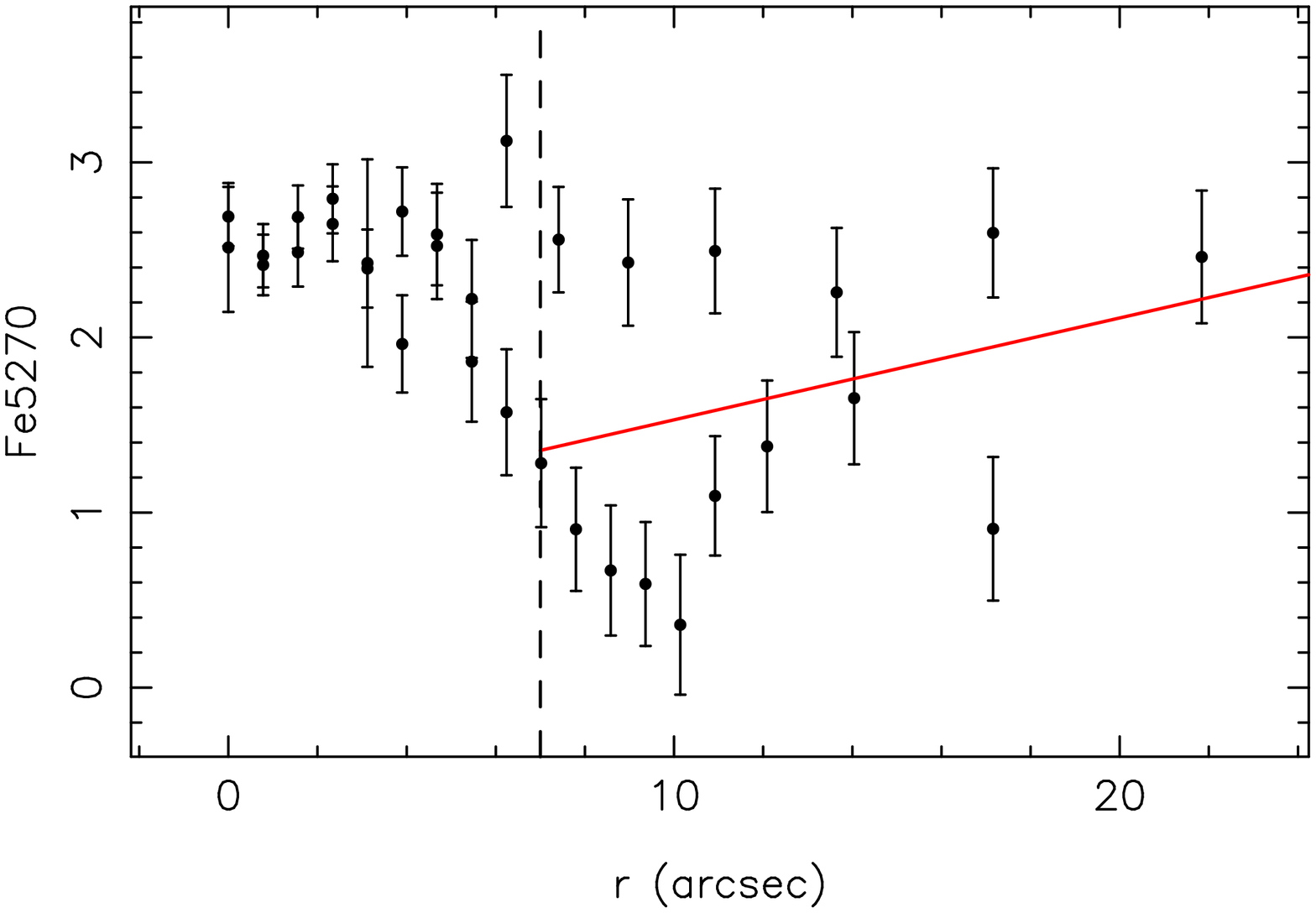}}\hspace{0.85cm}
\resizebox{0.3\textwidth}{!}{\includegraphics[angle=-90]{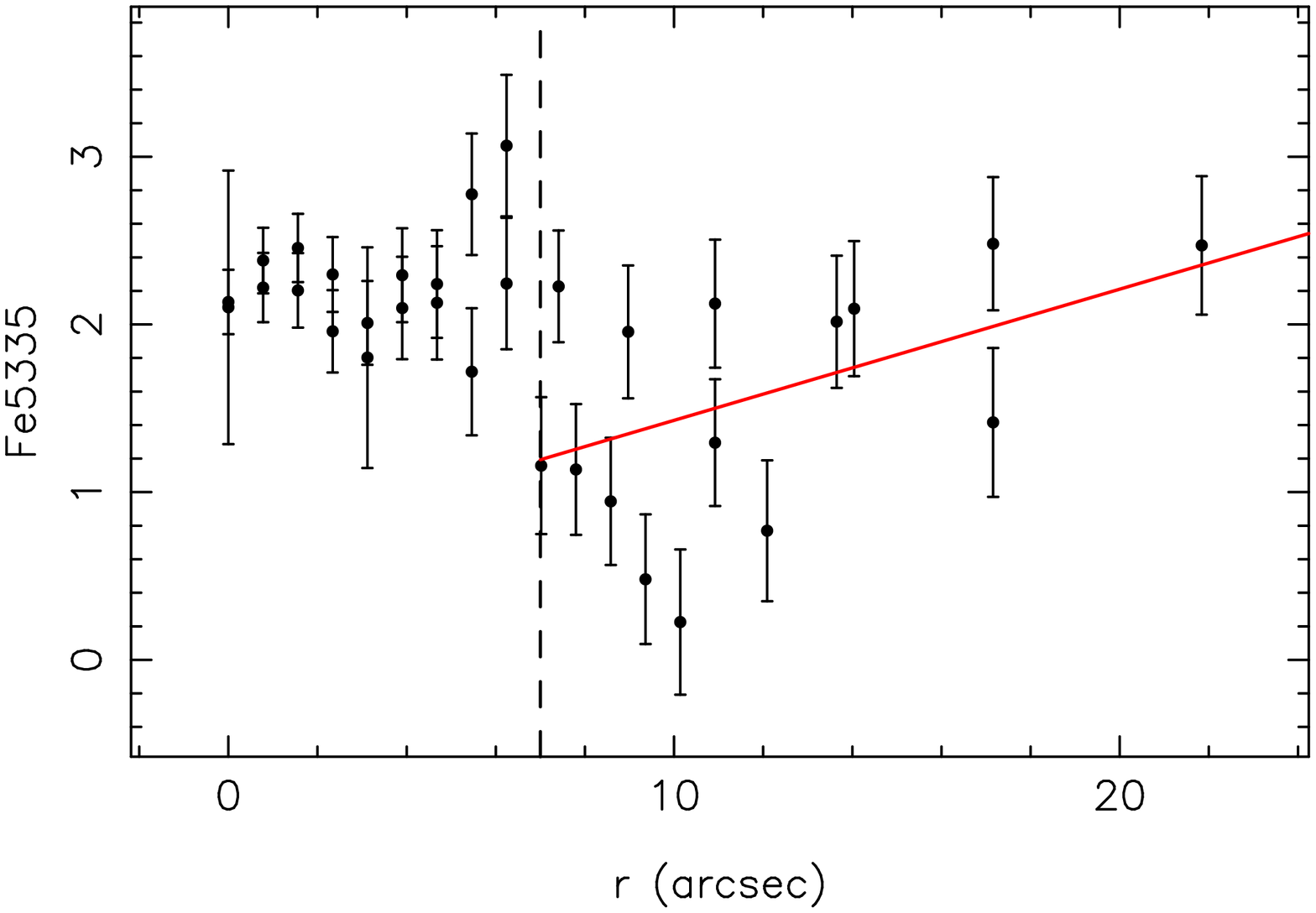}}
\caption{Line-strength distribution in the bar region for all the galaxies}
\end{figure*}

\begin{figure*}
\addtocounter{figure}{-1}
\resizebox{0.3\textwidth}{!}{\includegraphics[angle=-90]{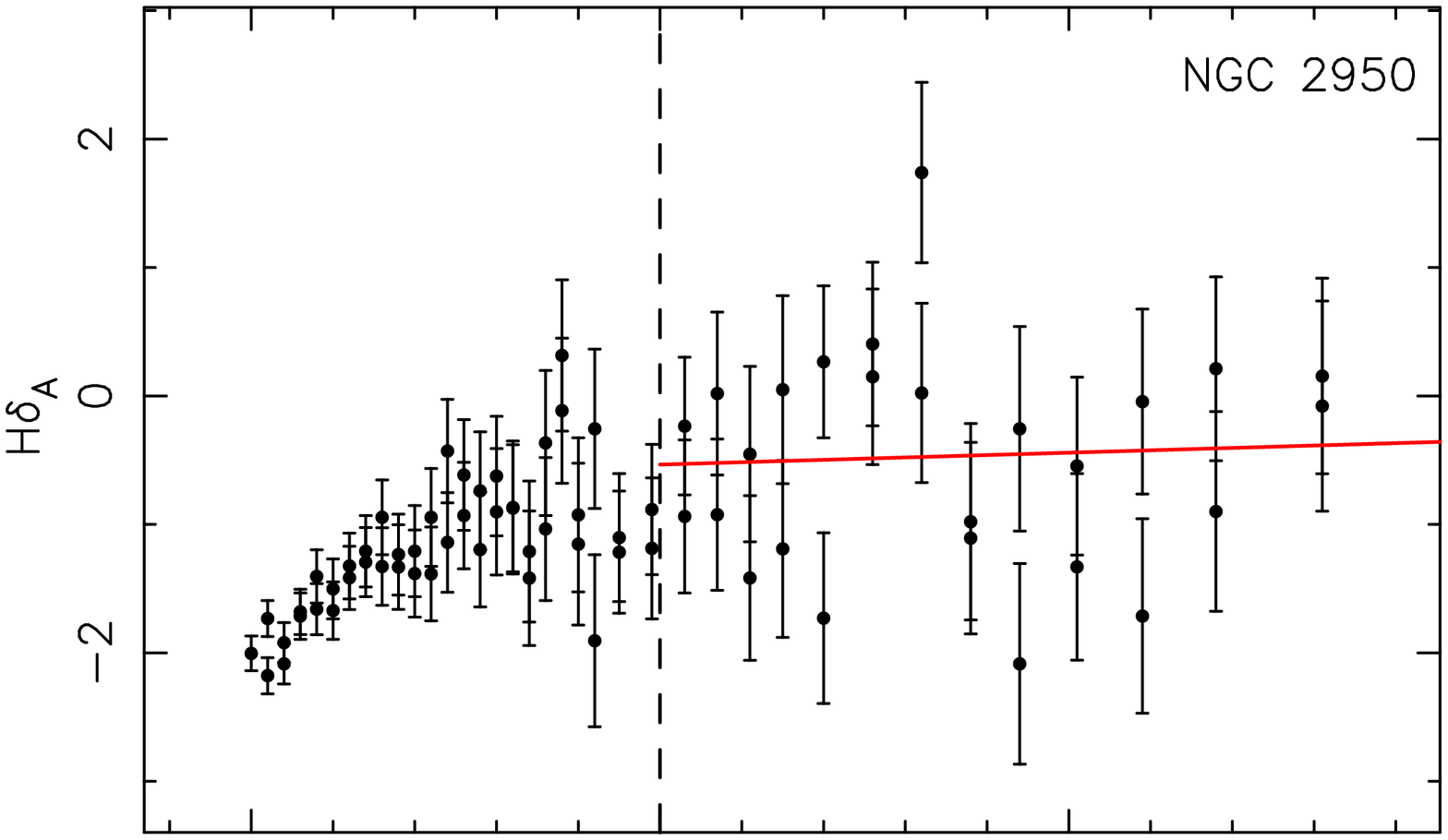}}
\resizebox{0.3\textwidth}{!}{\includegraphics[angle=-90]{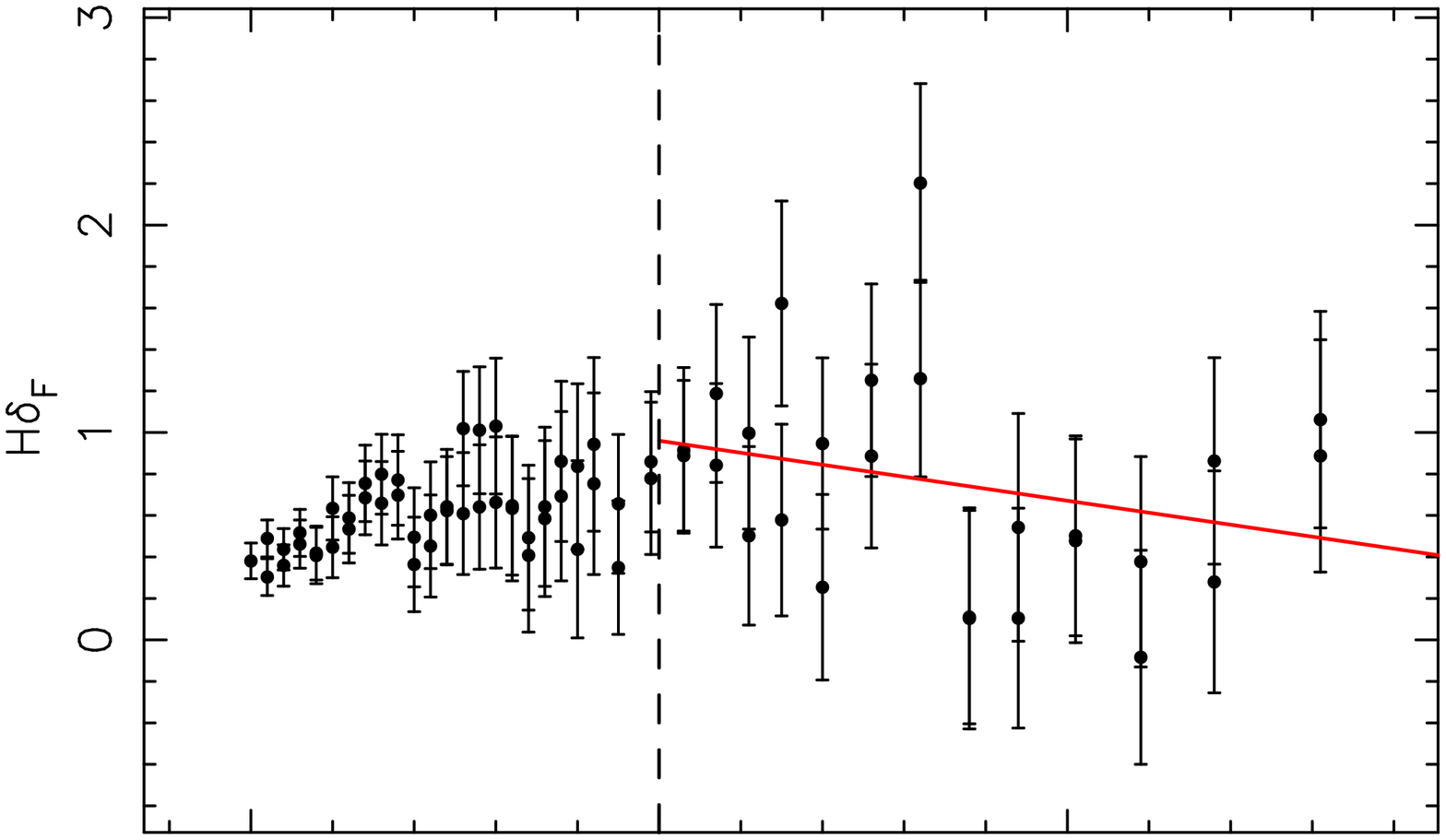}}
\resizebox{0.3\textwidth}{!}{\includegraphics[angle=-90]{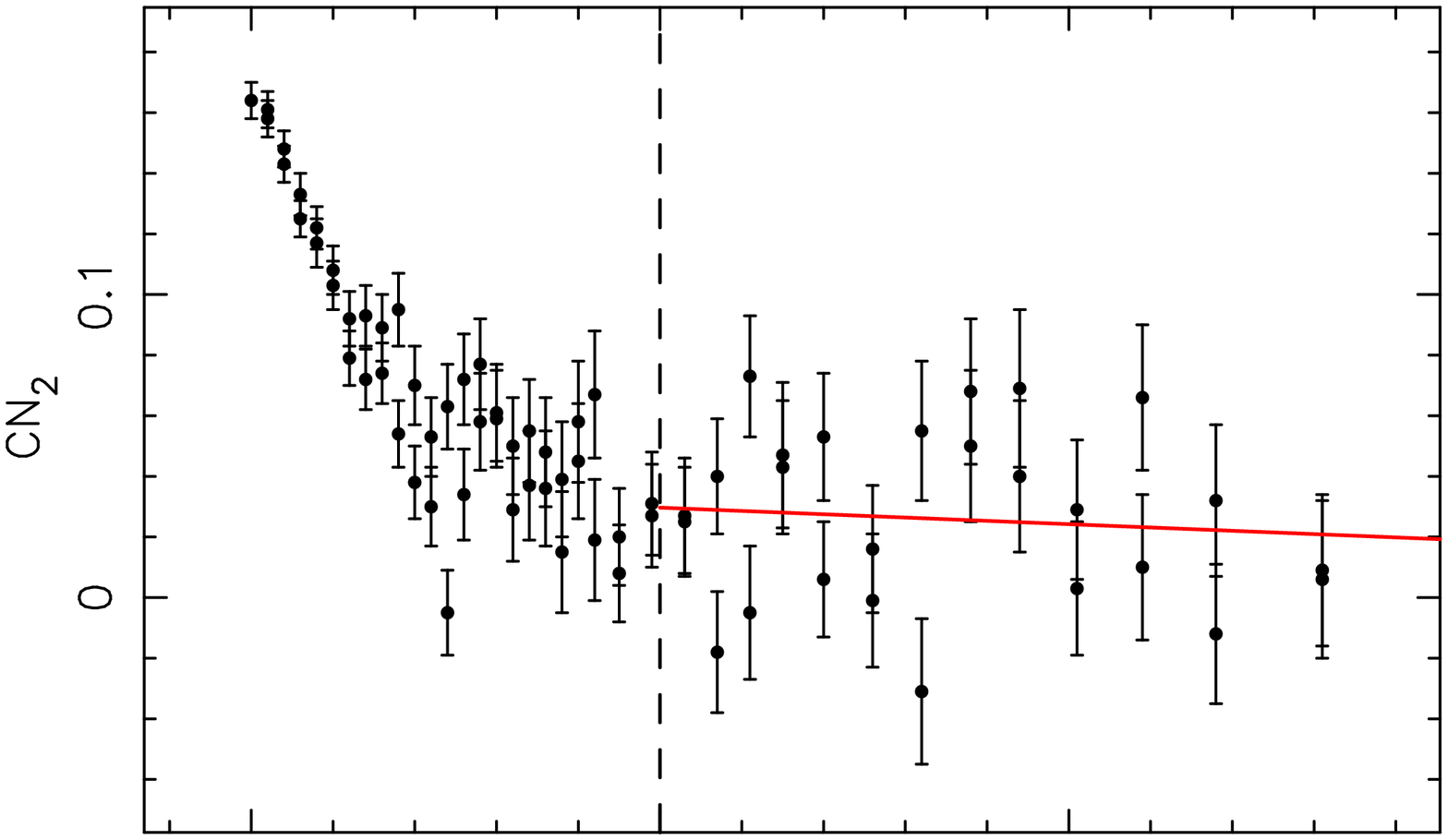}}
\resizebox{0.3\textwidth}{!}{\includegraphics[angle=-90]{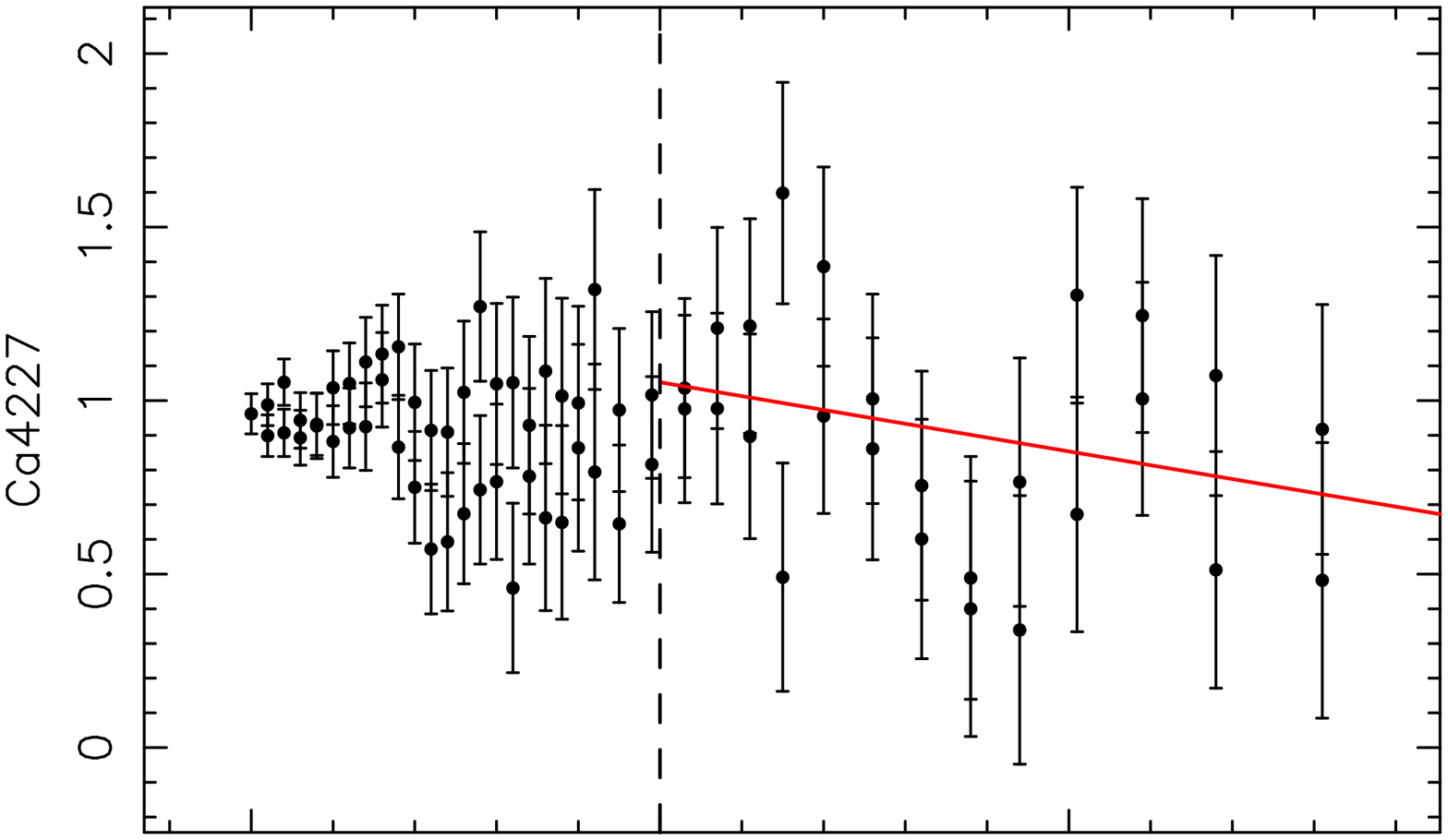}}
\resizebox{0.3\textwidth}{!}{\includegraphics[angle=-90]{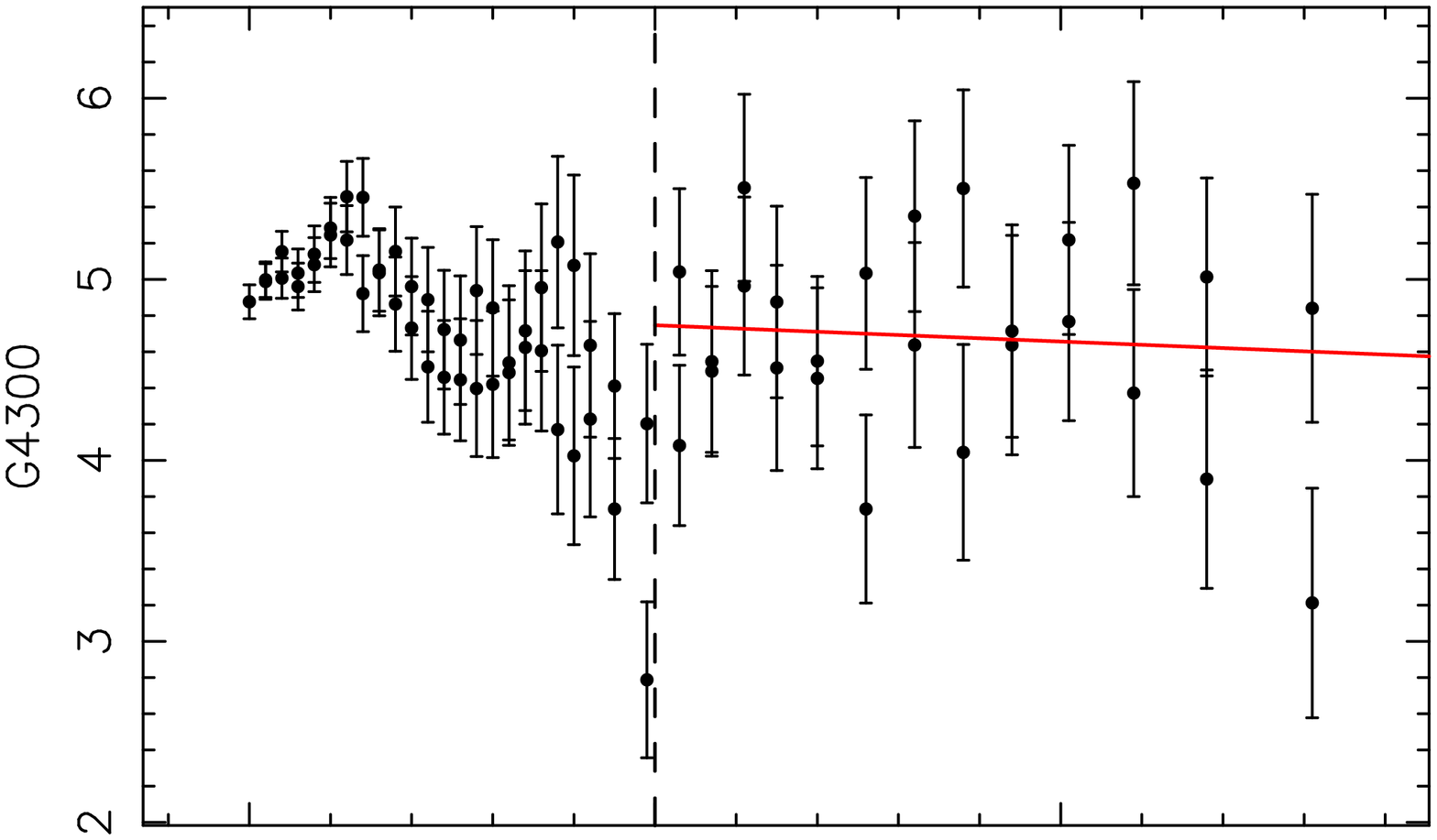}}
\resizebox{0.3\textwidth}{!}{\includegraphics[angle=-90]{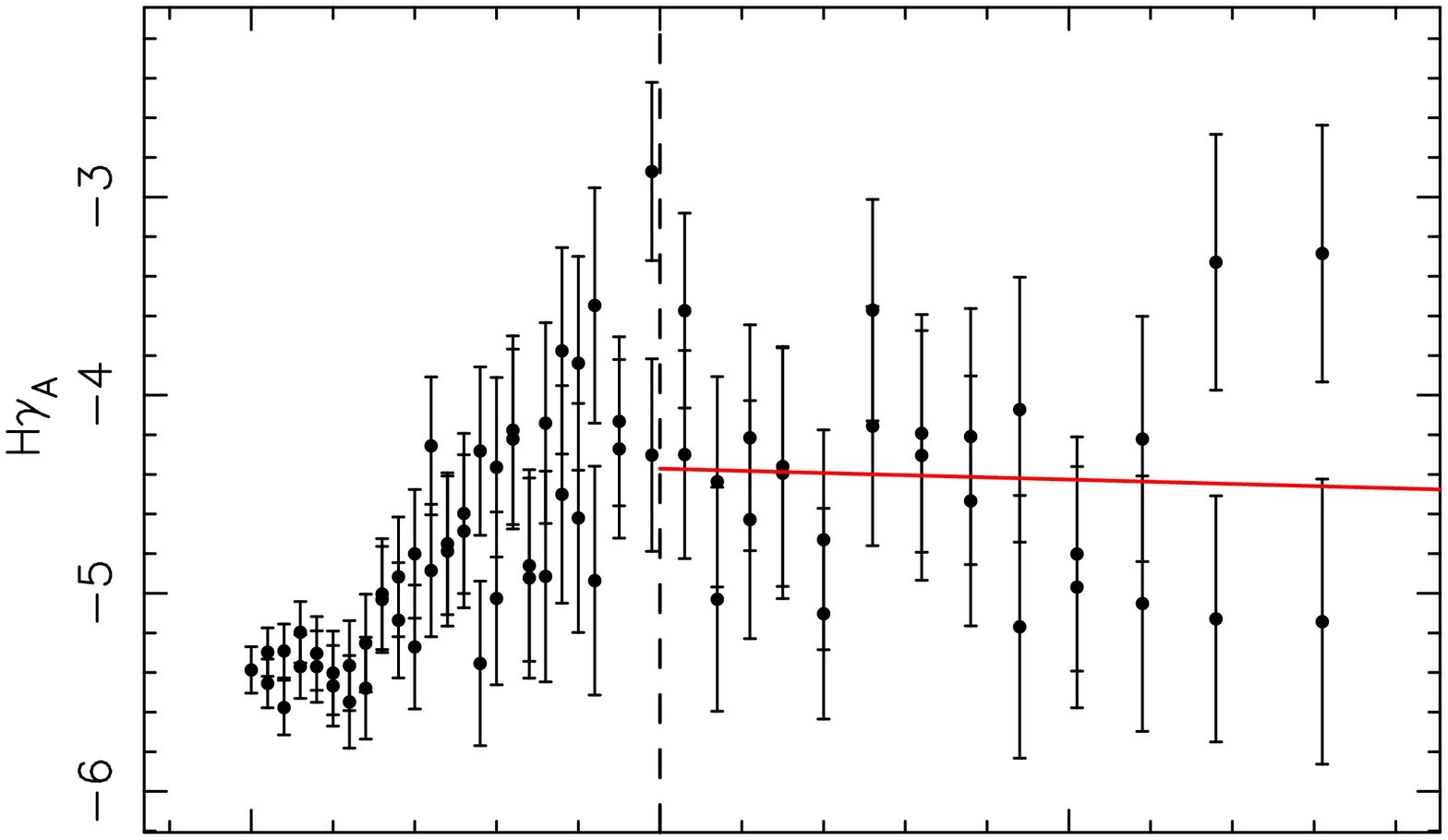}}
\resizebox{0.3\textwidth}{!}{\includegraphics[angle=-90]{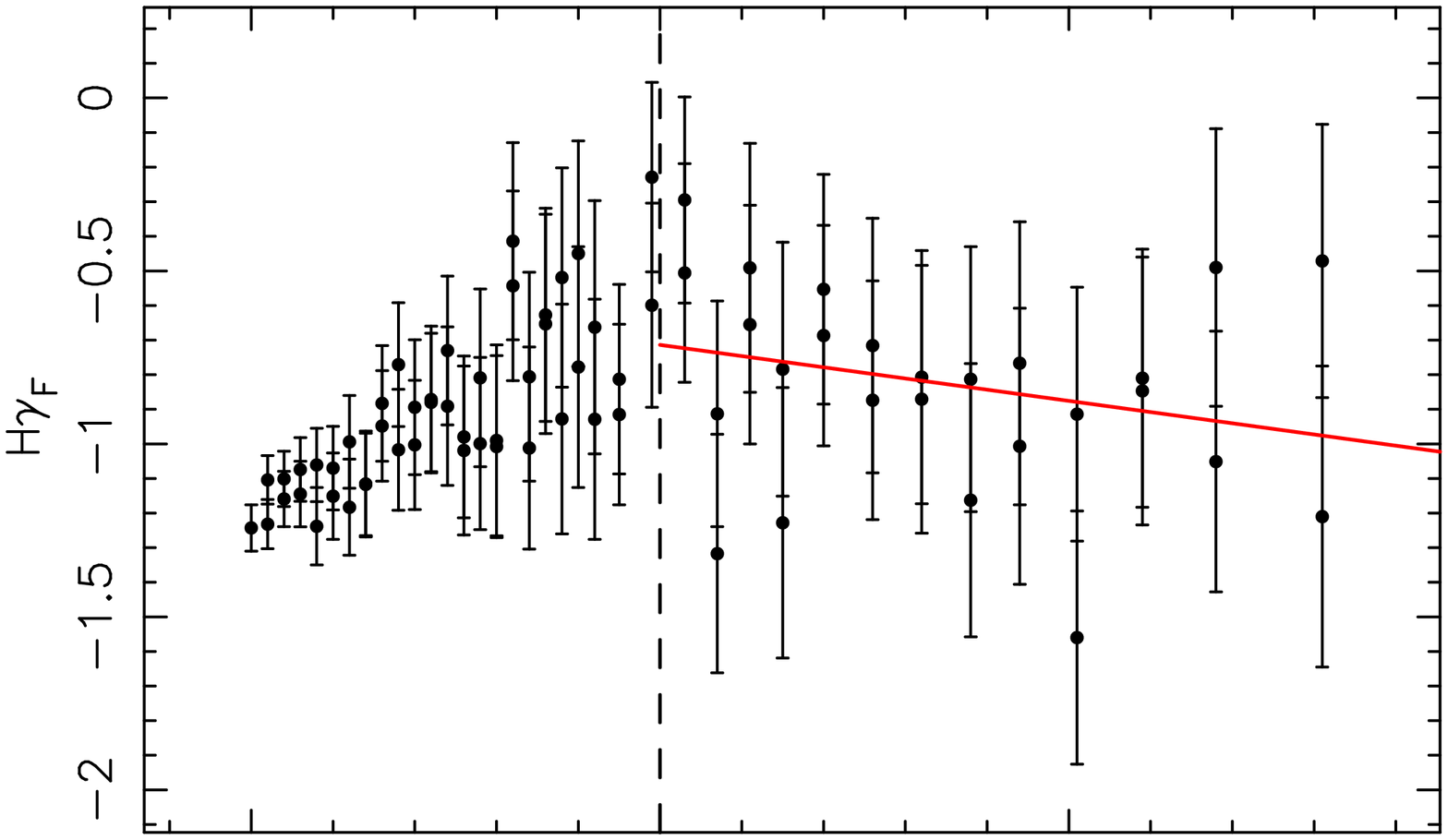}}
\resizebox{0.3\textwidth}{!}{\includegraphics[angle=-90]{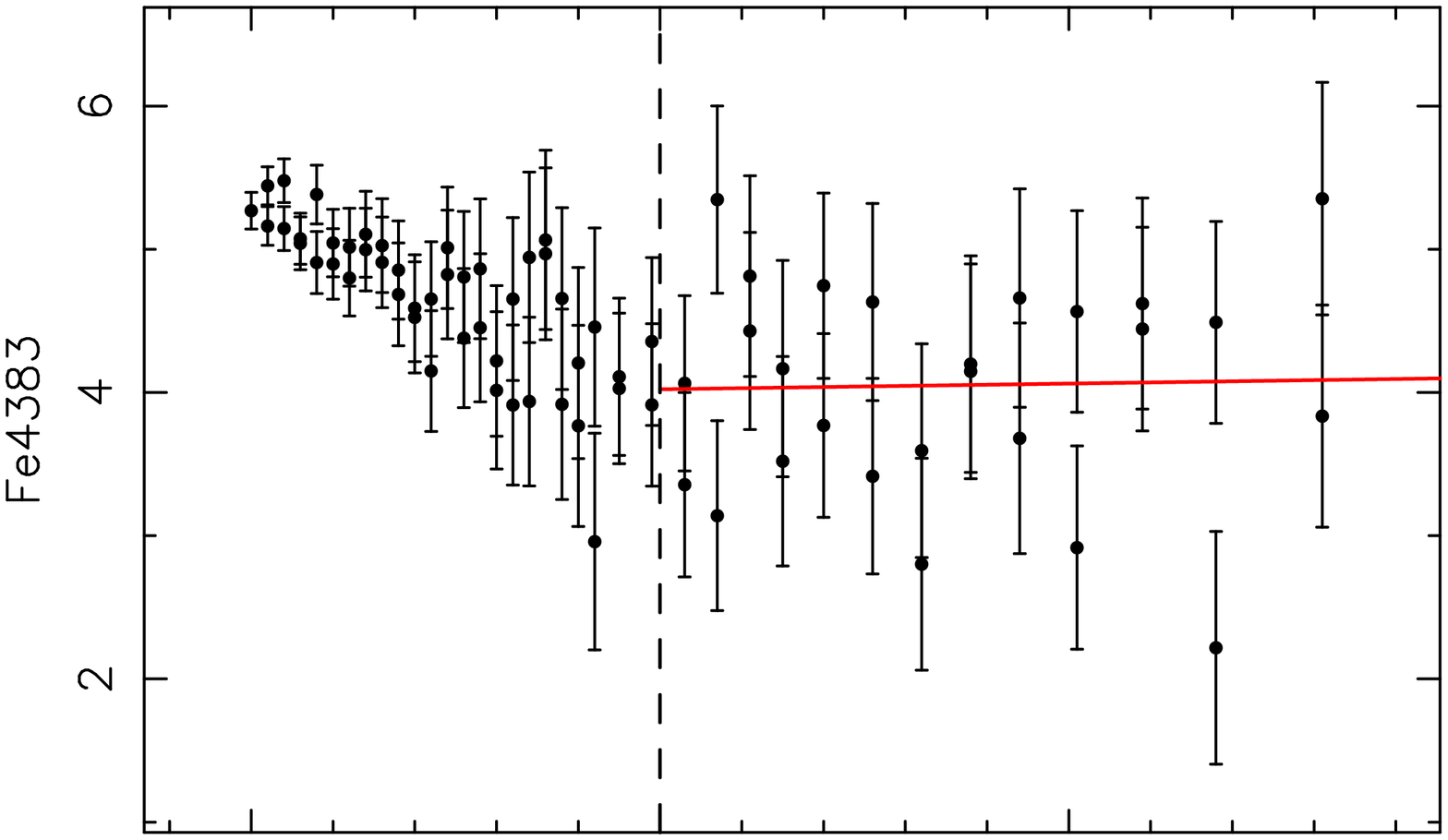}}
\resizebox{0.3\textwidth}{!}{\includegraphics[angle=-90]{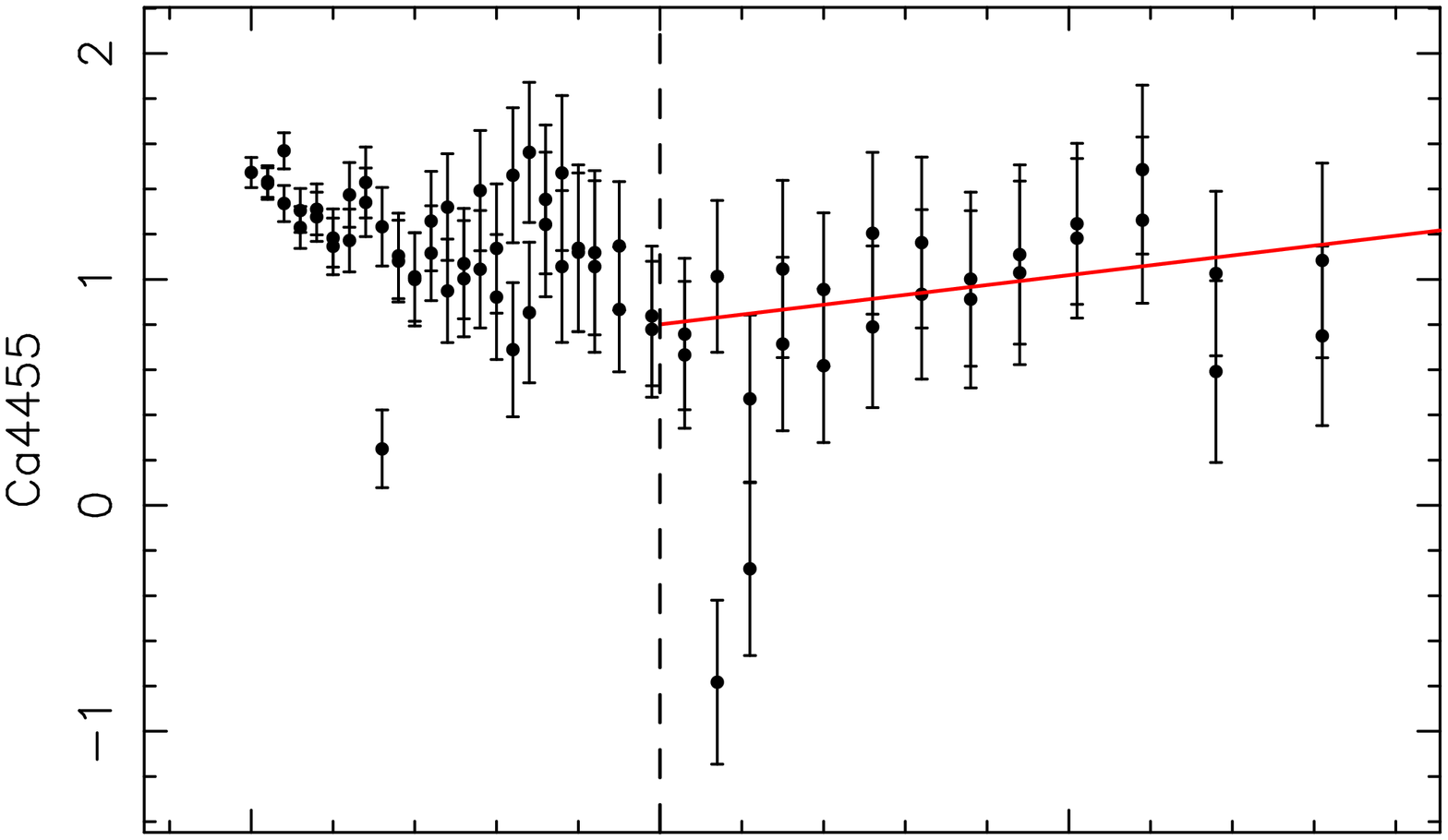}}
\resizebox{0.3\textwidth}{!}{\includegraphics[angle=-90]{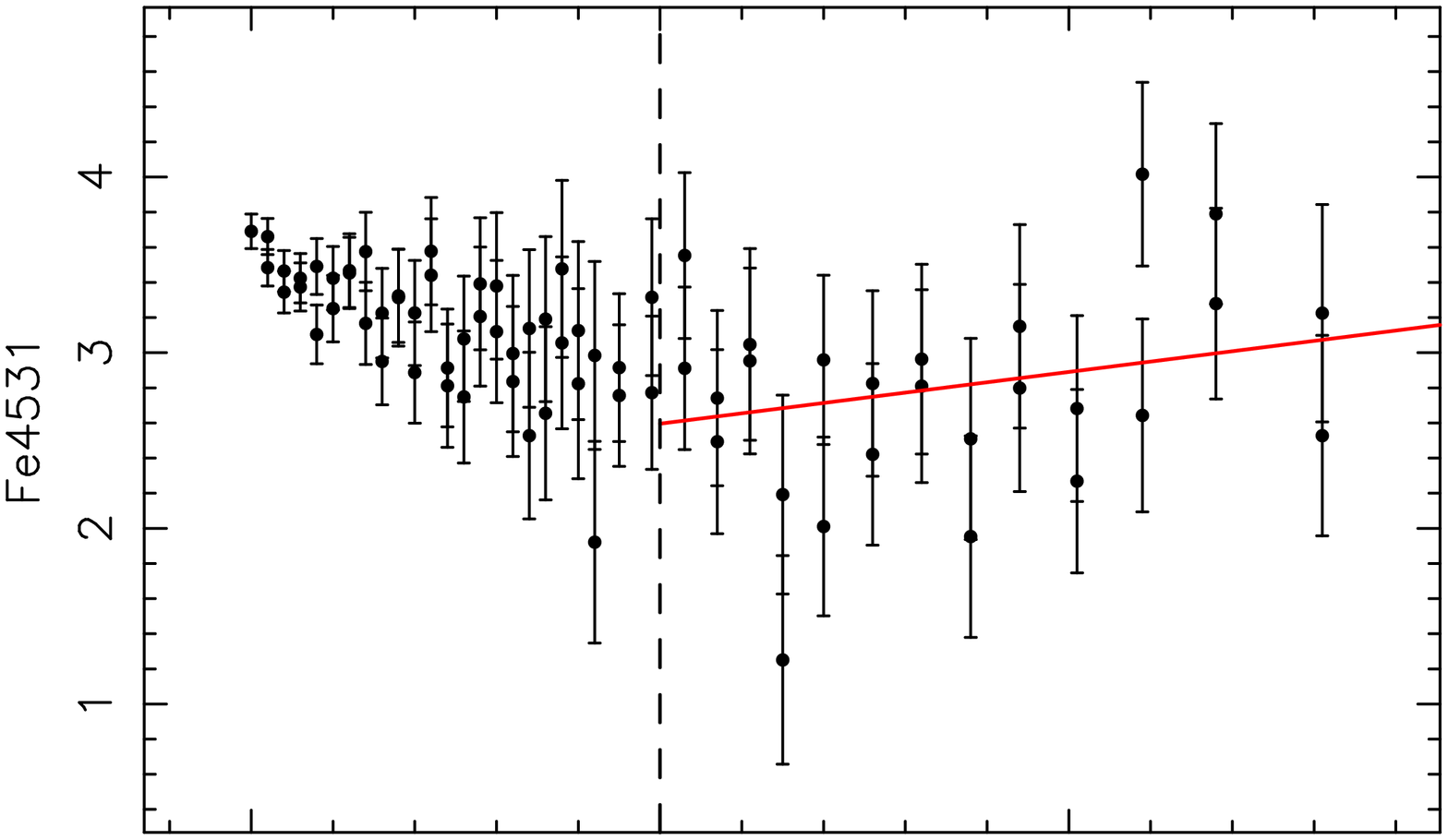}}
\resizebox{0.3\textwidth}{!}{\includegraphics[angle=-90]{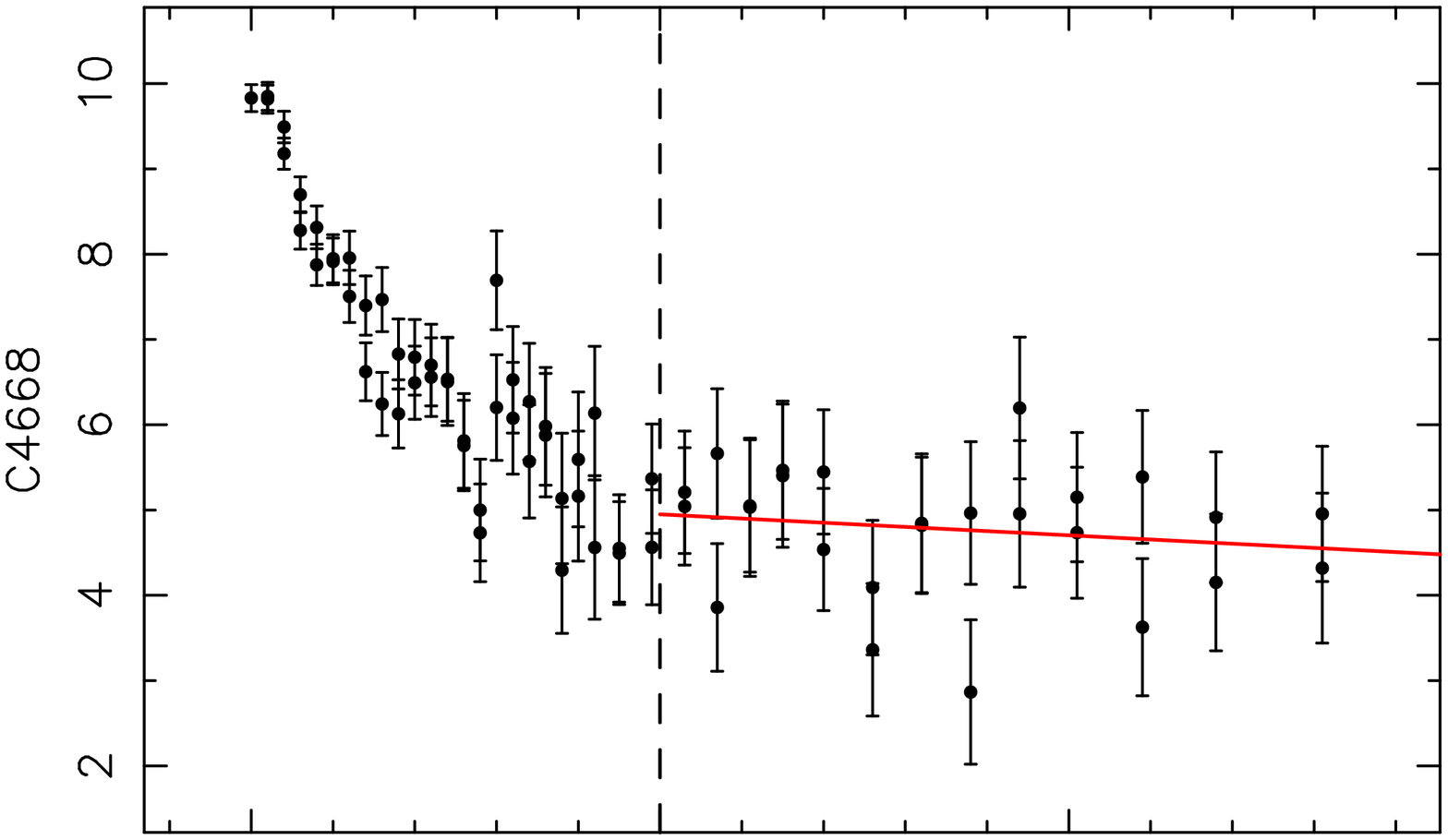}}
\resizebox{0.3\textwidth}{!}{\includegraphics[angle=-90]{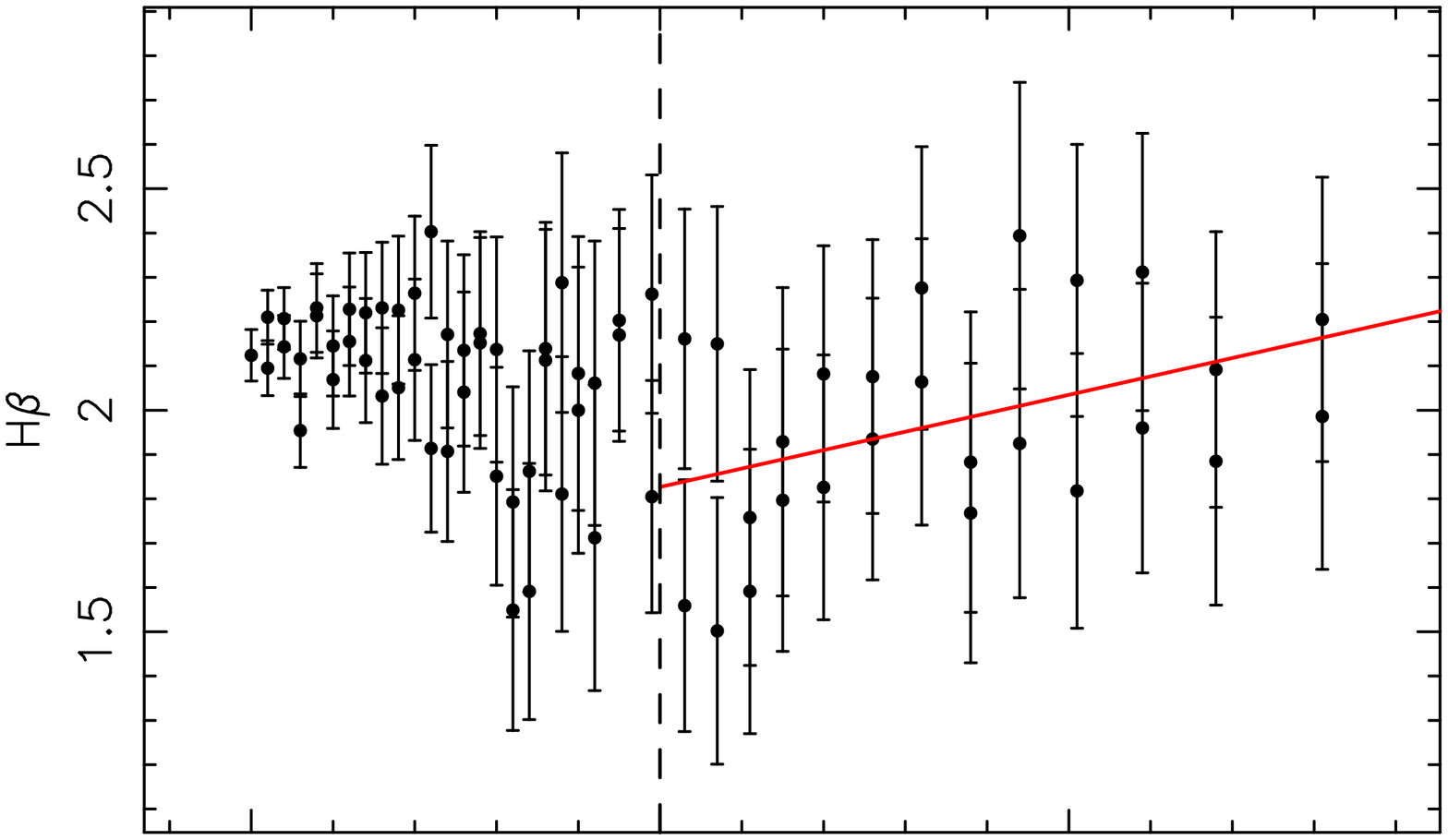}}
\resizebox{0.3\textwidth}{!}{\includegraphics[angle=-90]{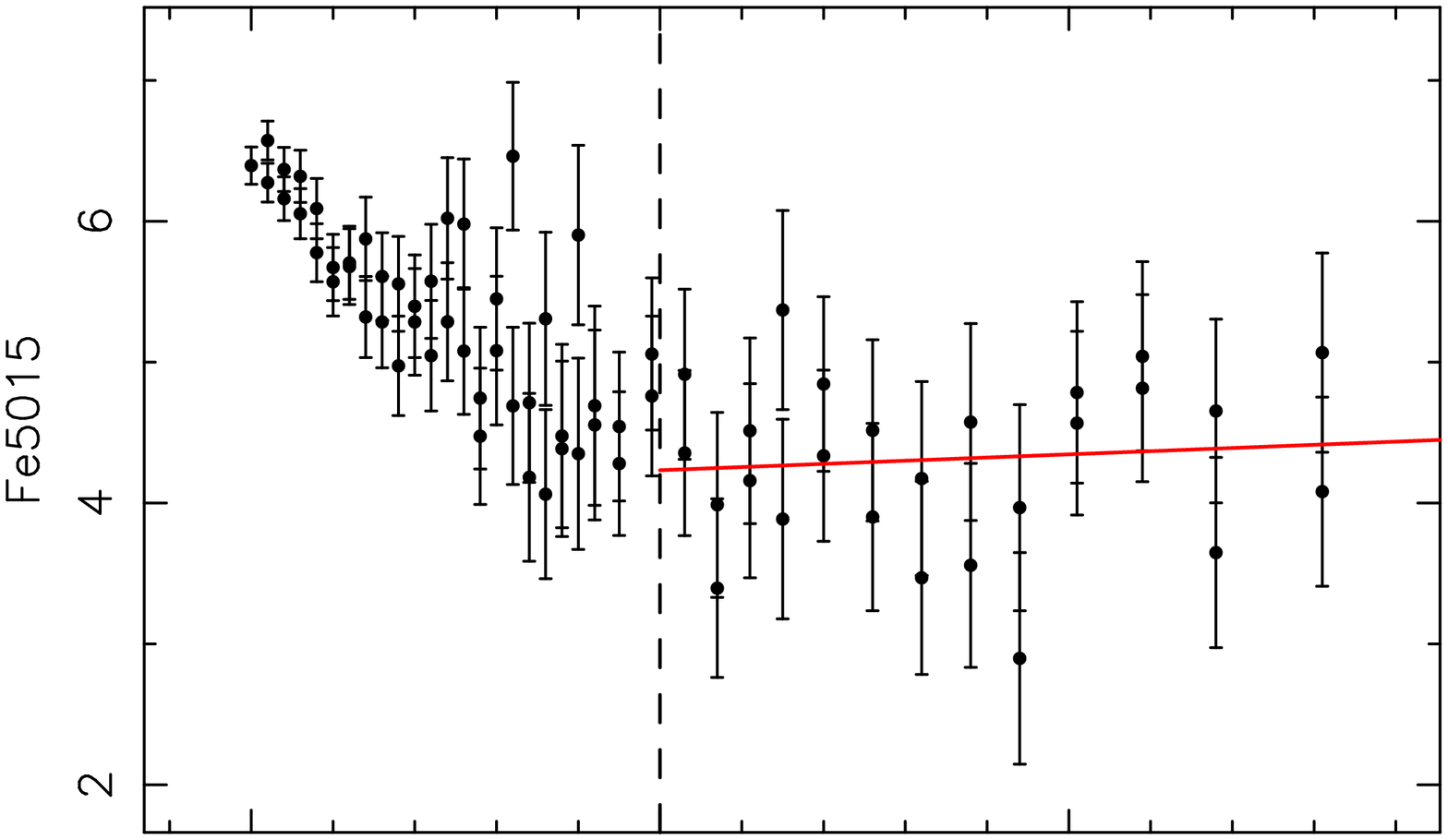}}
\resizebox{0.3\textwidth}{!}{\includegraphics[angle=-90]{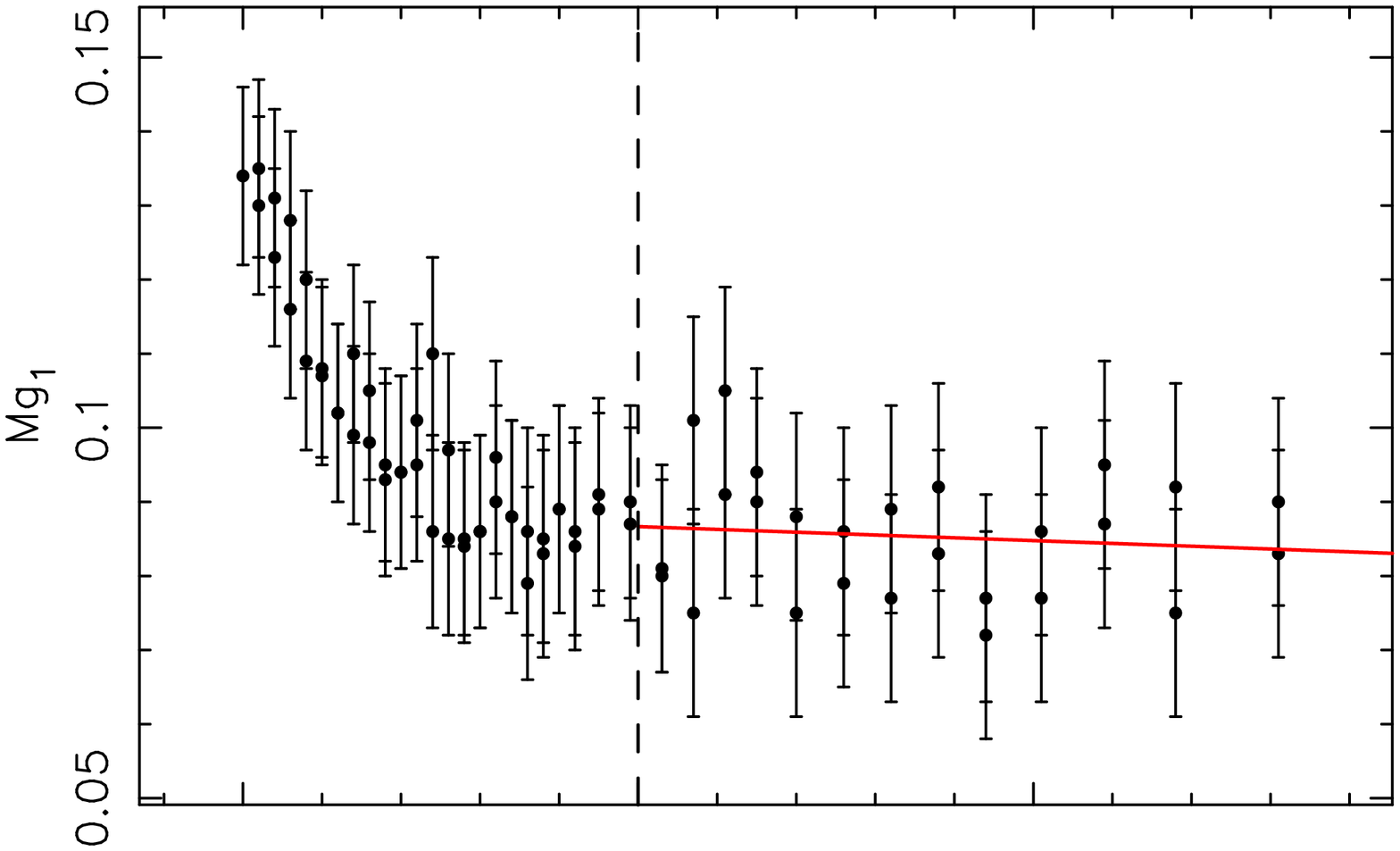}}
\resizebox{0.3\textwidth}{!}{\includegraphics[angle=-90]{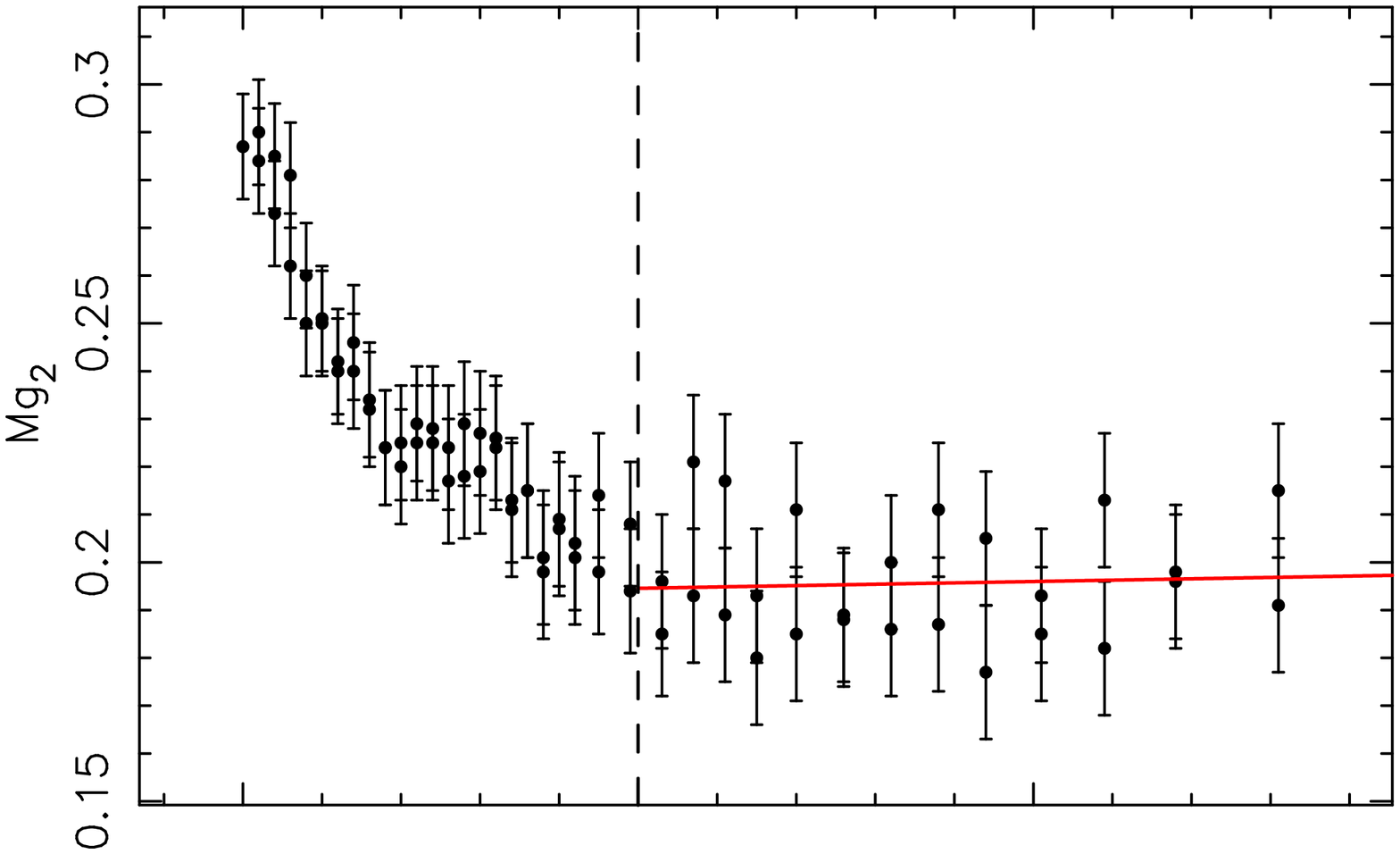}}\hspace{0.85cm}
\resizebox{0.3\textwidth}{!}{\includegraphics[angle=-90]{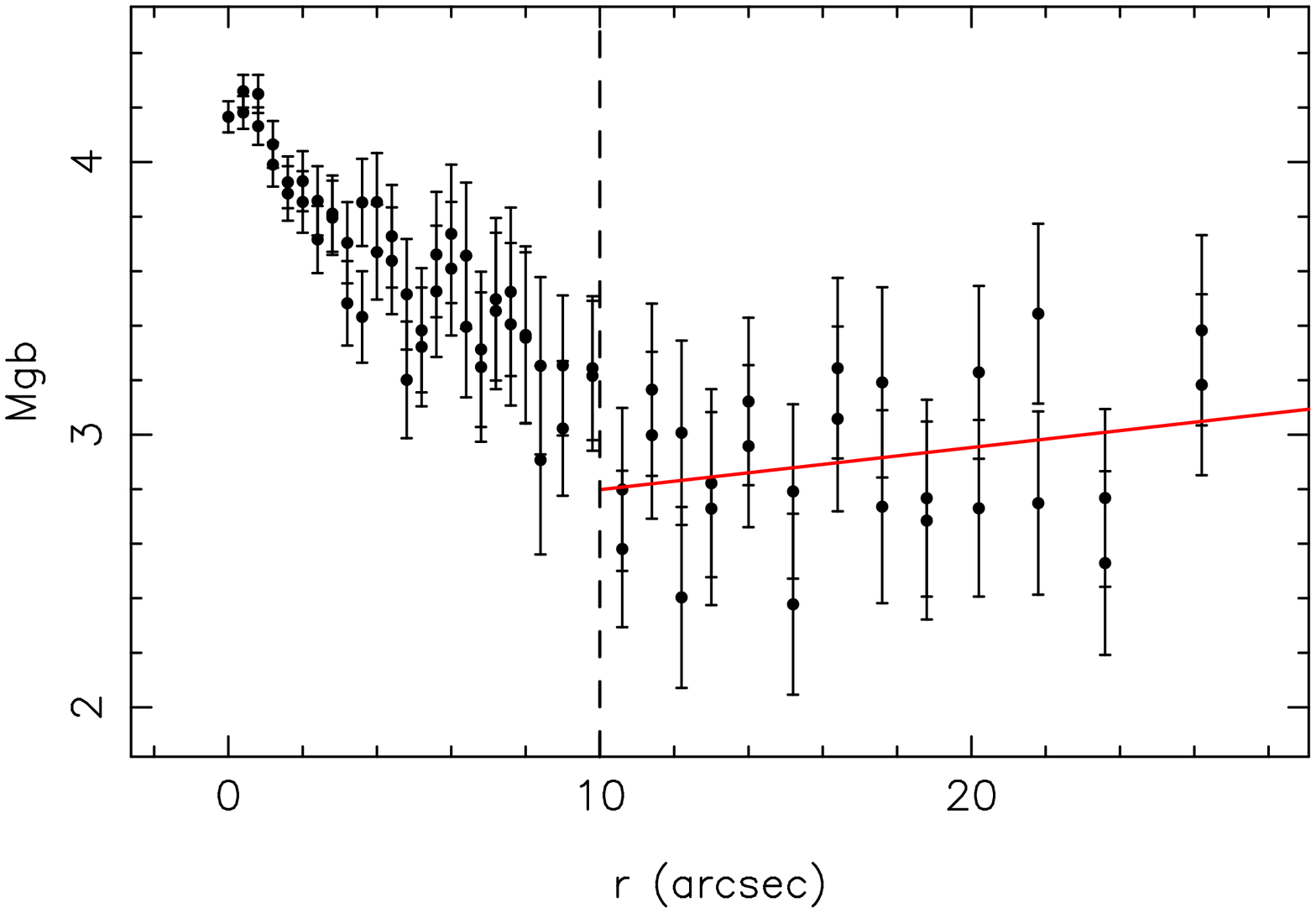}}\hspace{0.85cm}
\resizebox{0.3\textwidth}{!}{\includegraphics[angle=-90]{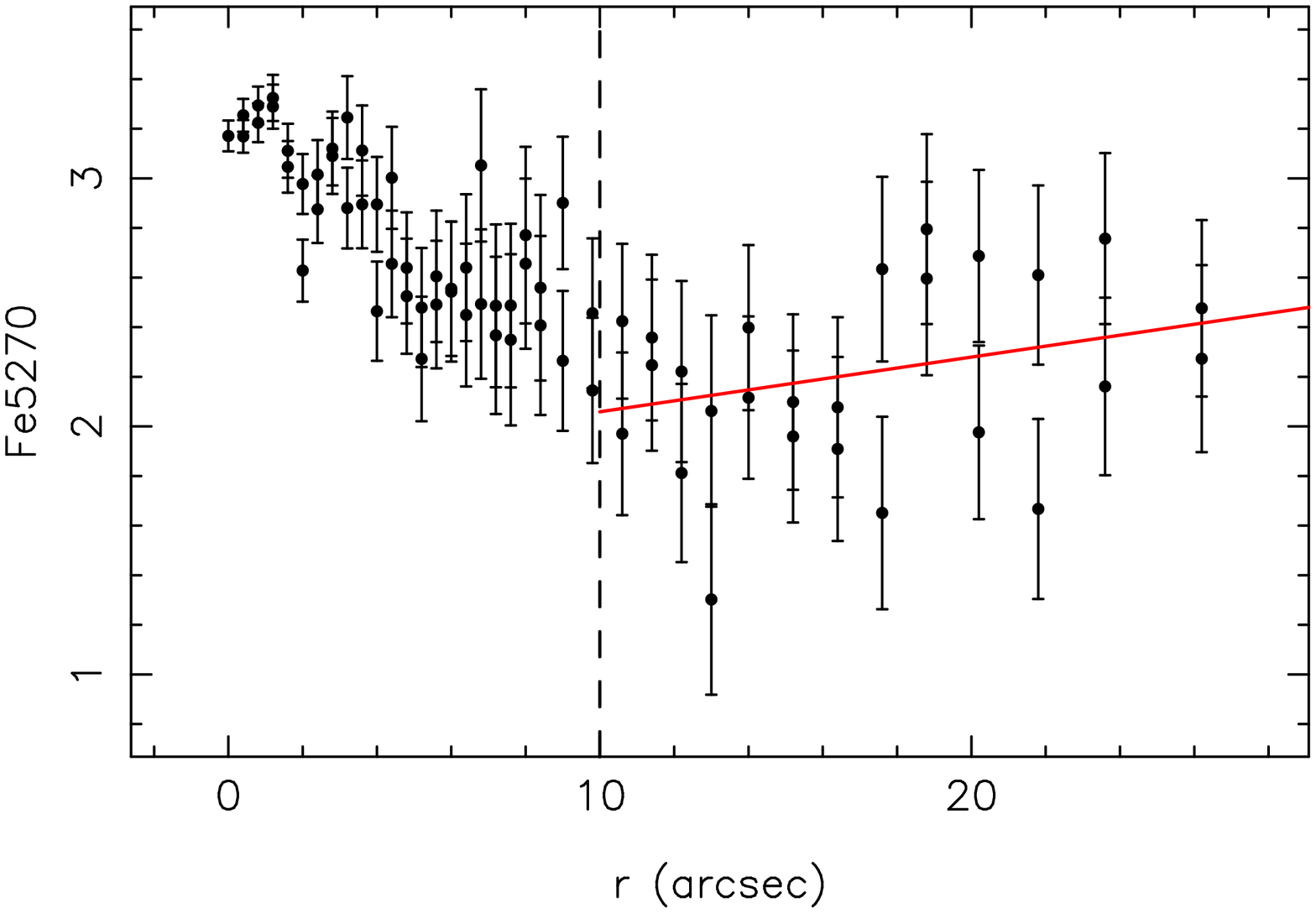}}\hspace{0.85cm}
\resizebox{0.3\textwidth}{!}{\includegraphics[angle=-90]{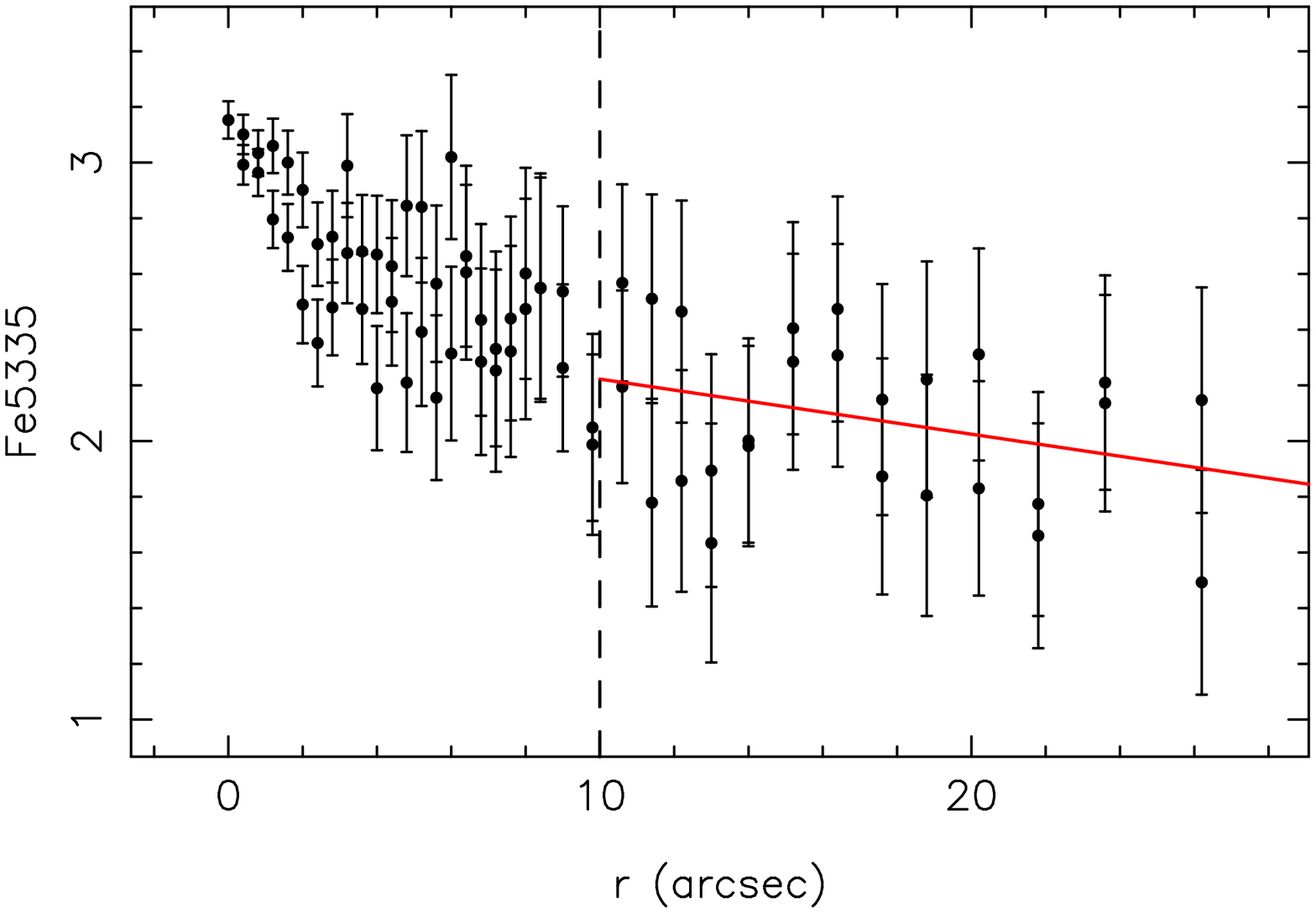}}
\caption{Line-strength distribution in the bar region for all the galaxies}
\end{figure*}

\newpage 
\begin{figure*}
\addtocounter{figure}{-1}
\resizebox{0.3\textwidth}{!}{\includegraphics[angle=-90]{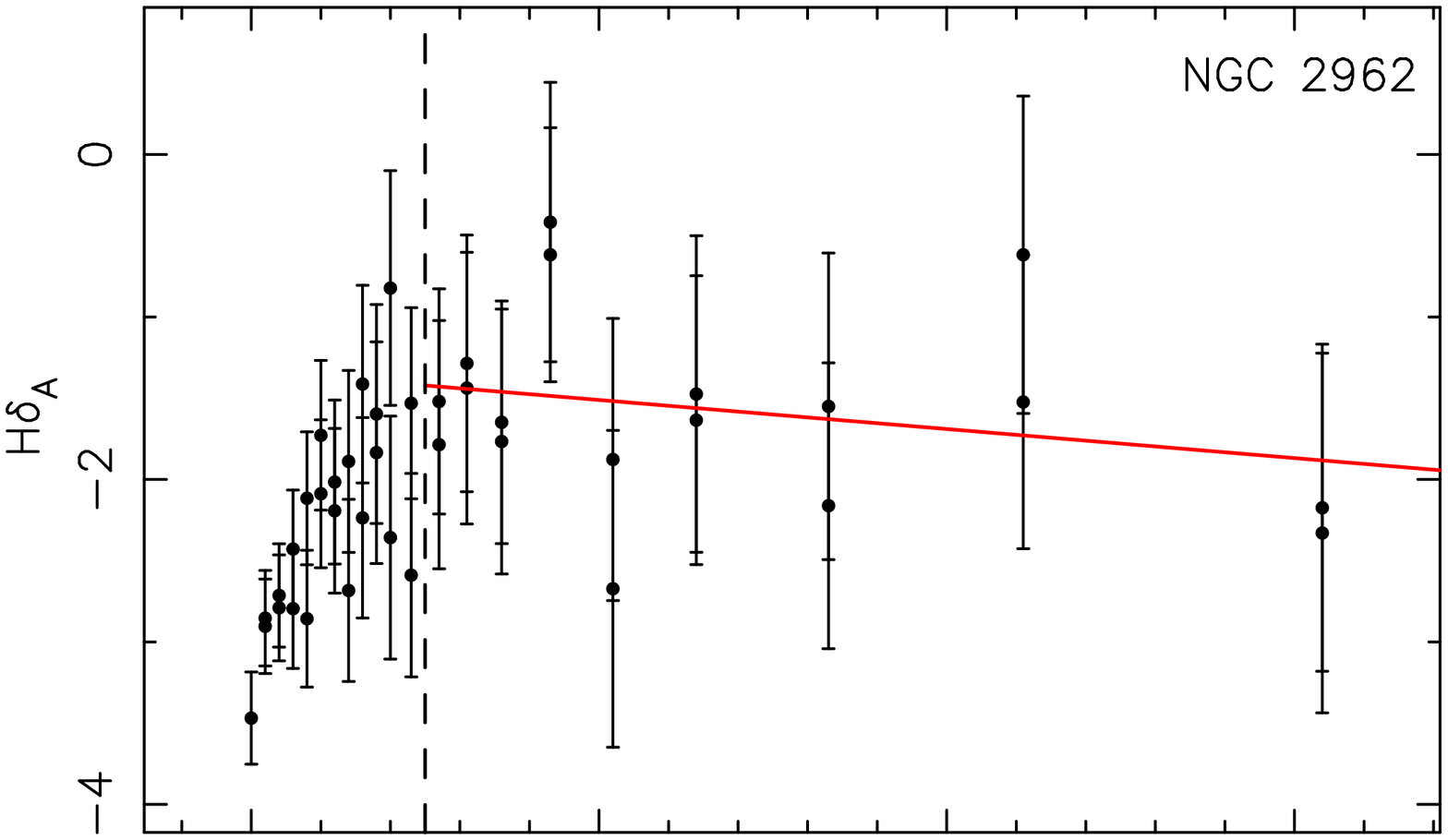}}
\resizebox{0.3\textwidth}{!}{\includegraphics[angle=-90]{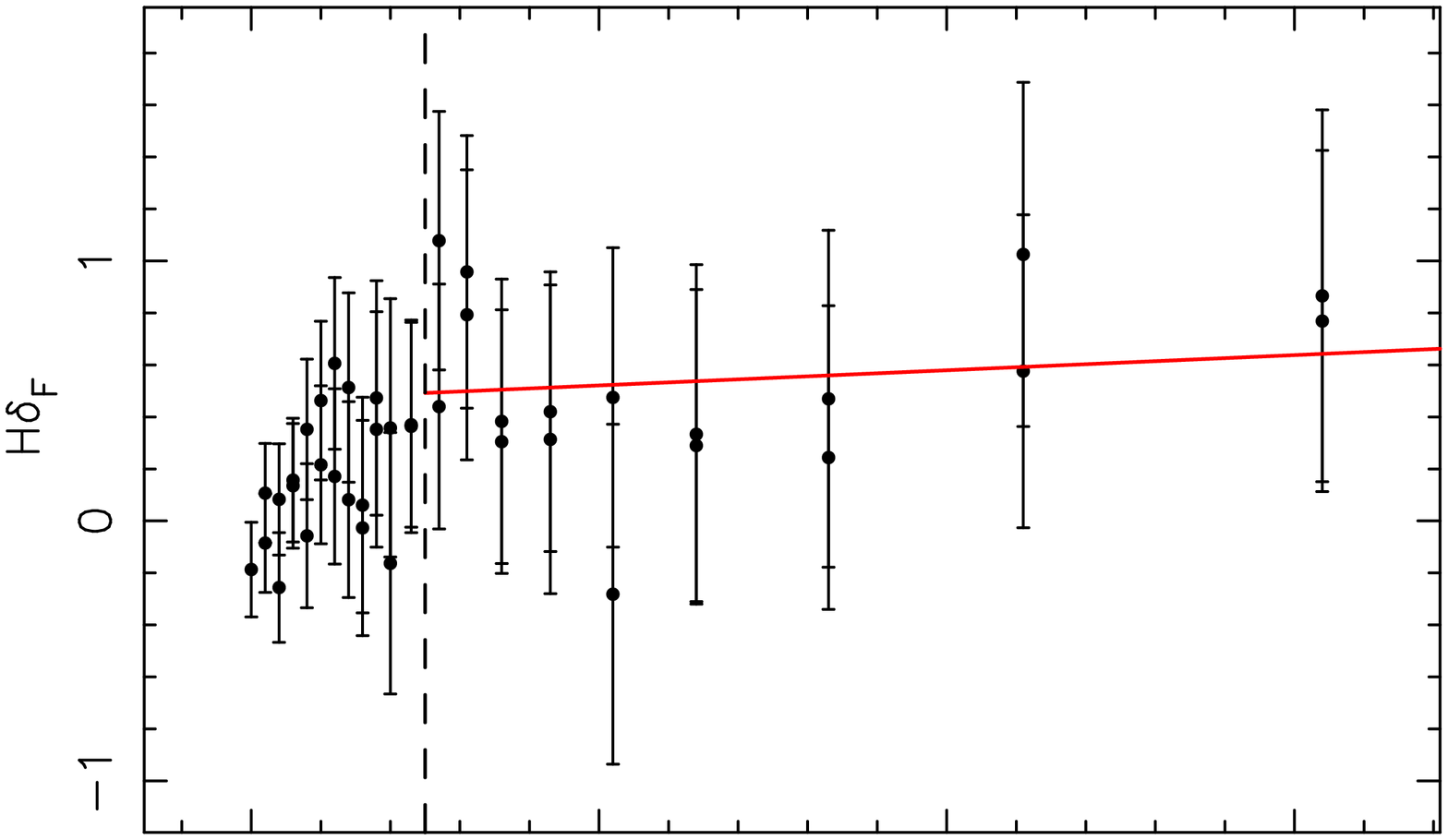}}
\resizebox{0.3\textwidth}{!}{\includegraphics[angle=-90]{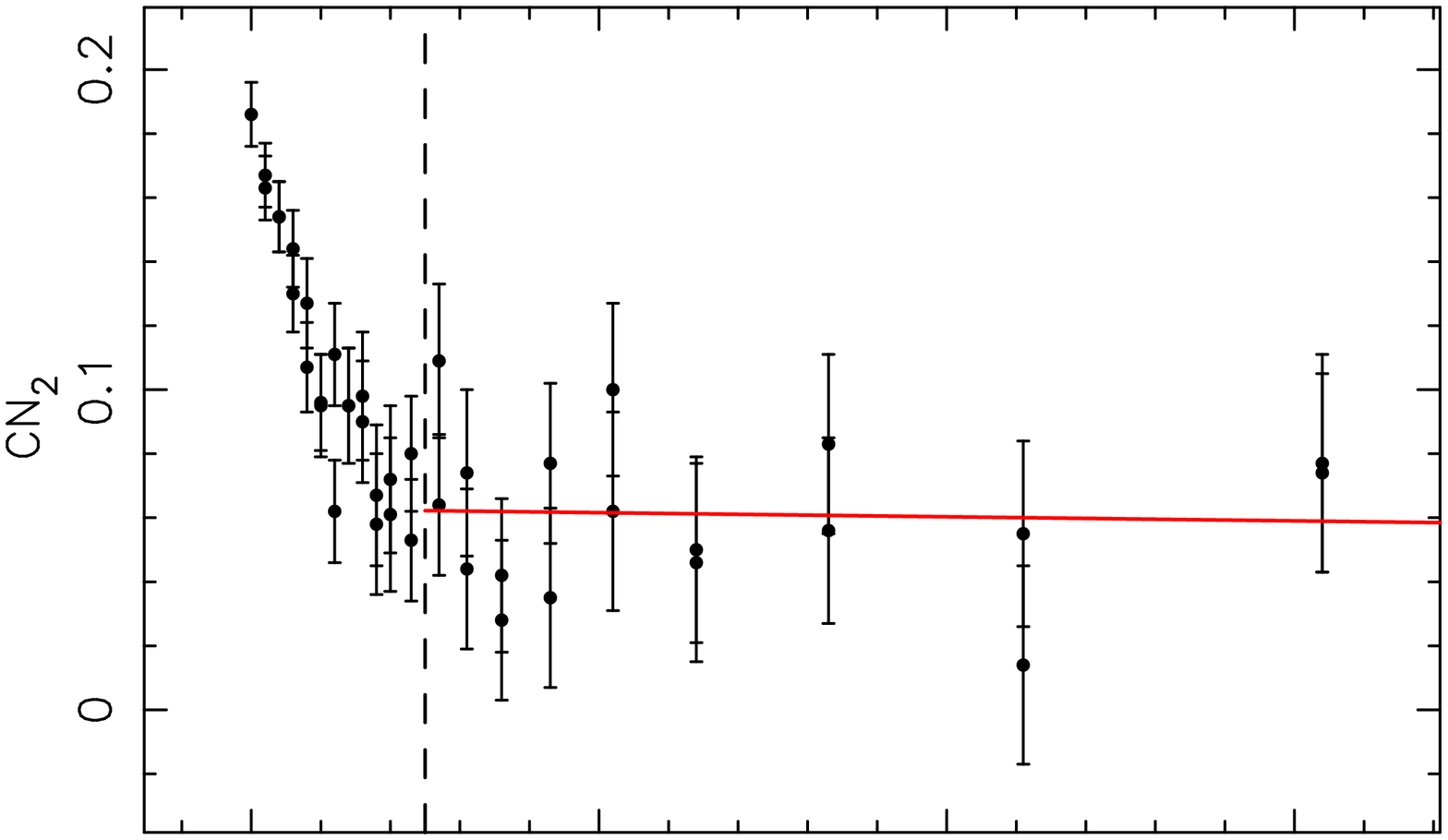}}
\resizebox{0.3\textwidth}{!}{\includegraphics[angle=-90]{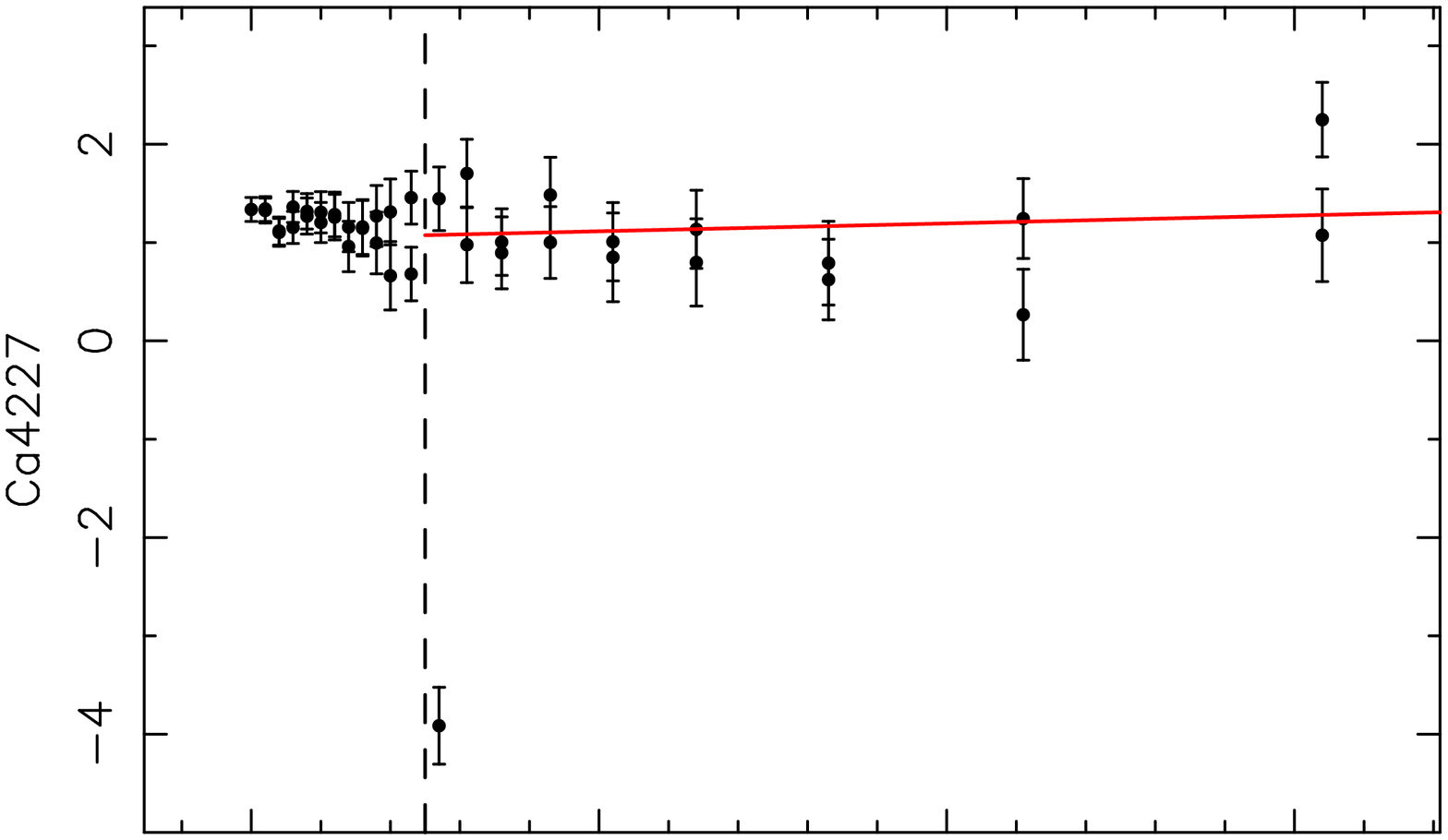}}
\resizebox{0.3\textwidth}{!}{\includegraphics[angle=-90]{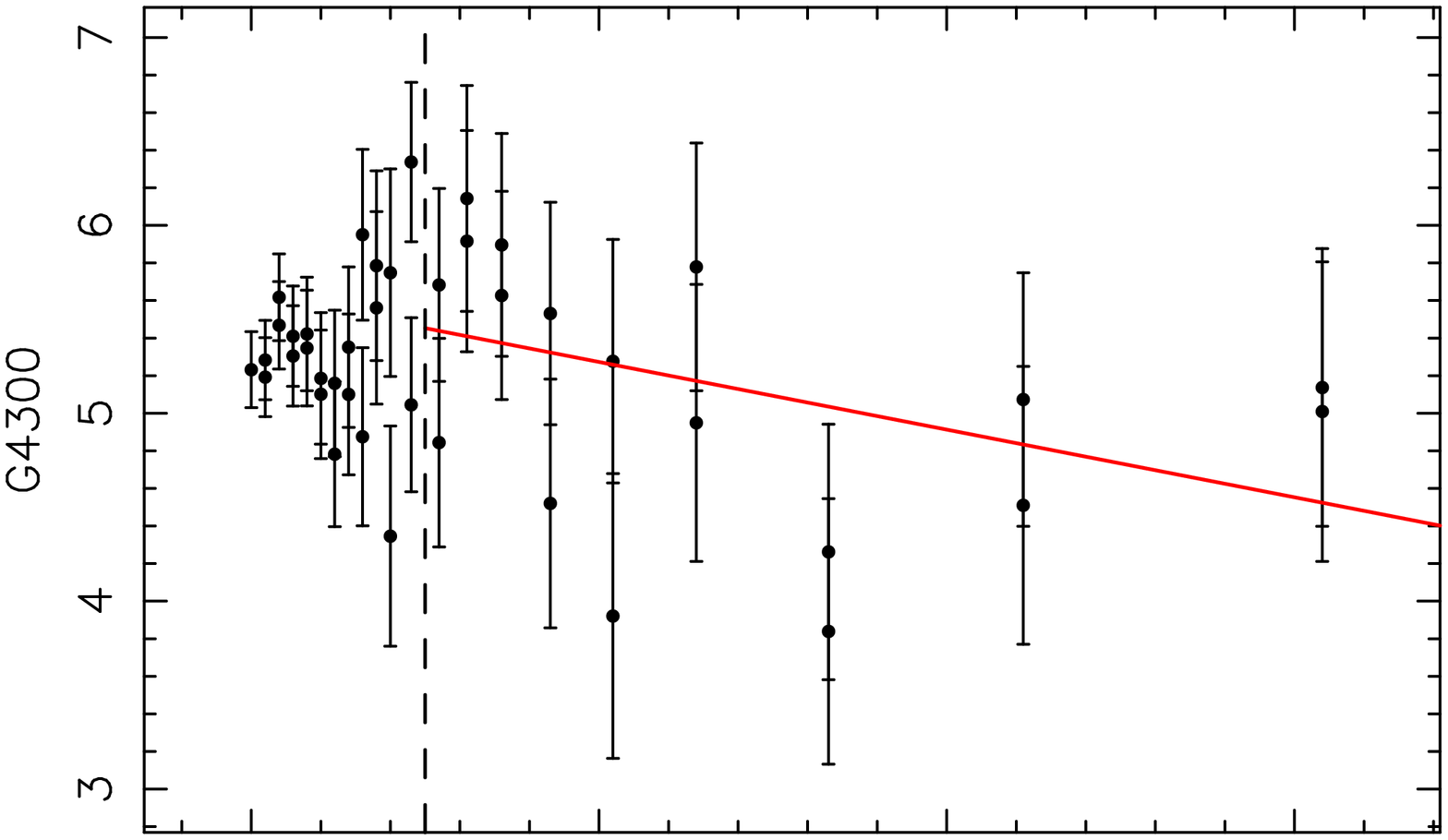}}
\resizebox{0.3\textwidth}{!}{\includegraphics[angle=-90]{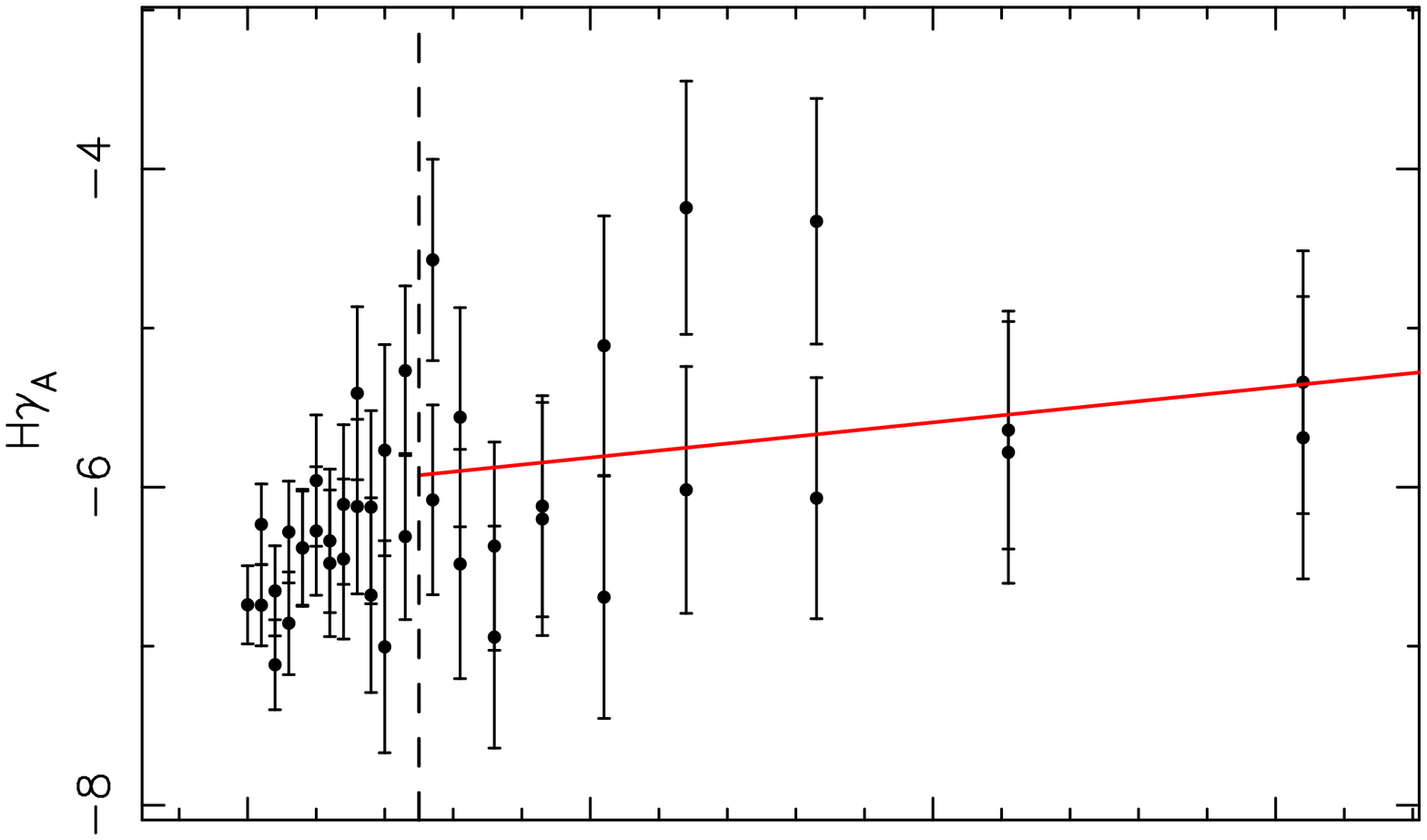}}
\resizebox{0.3\textwidth}{!}{\includegraphics[angle=-90]{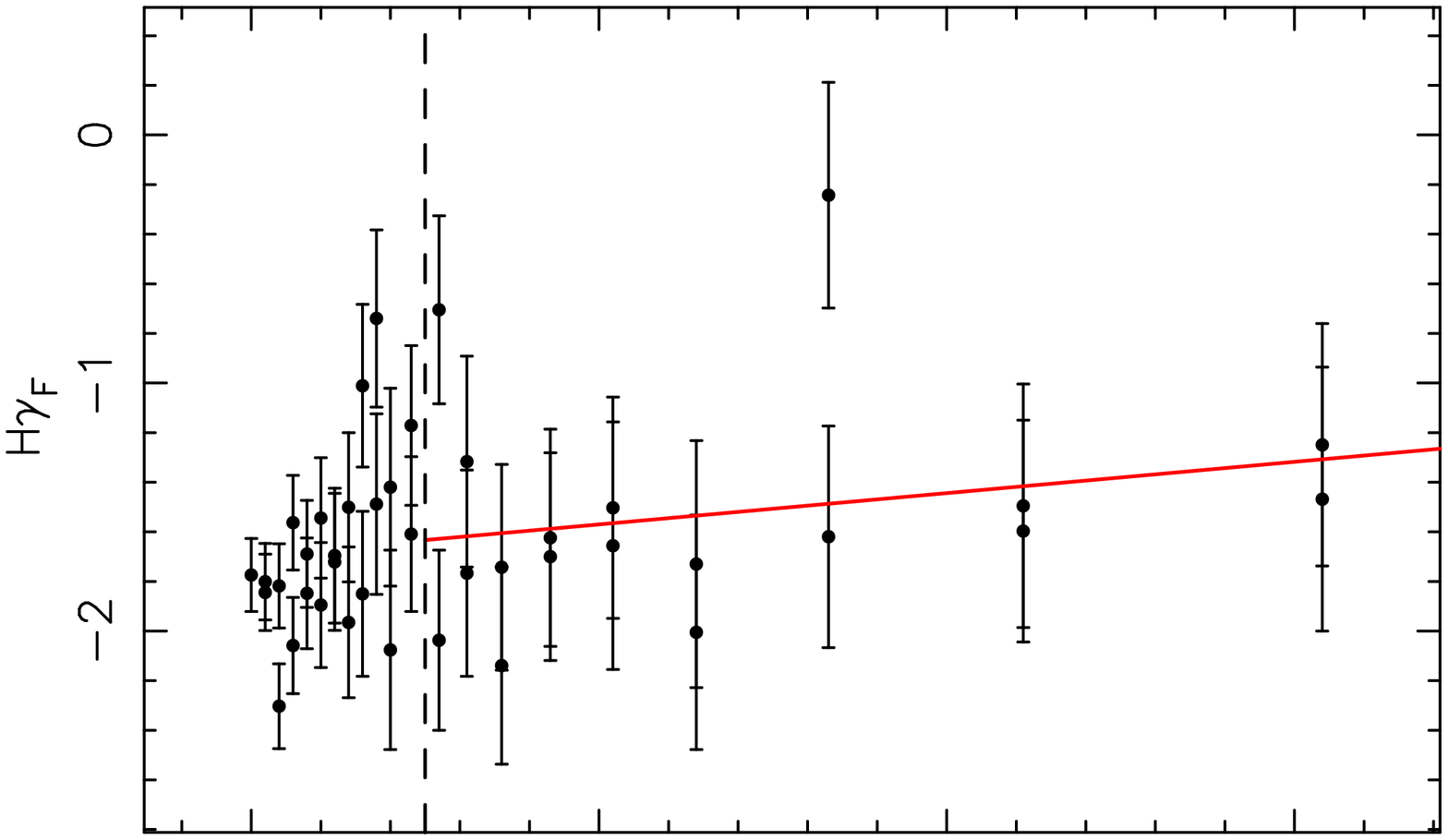}}
\resizebox{0.3\textwidth}{!}{\includegraphics[angle=-90]{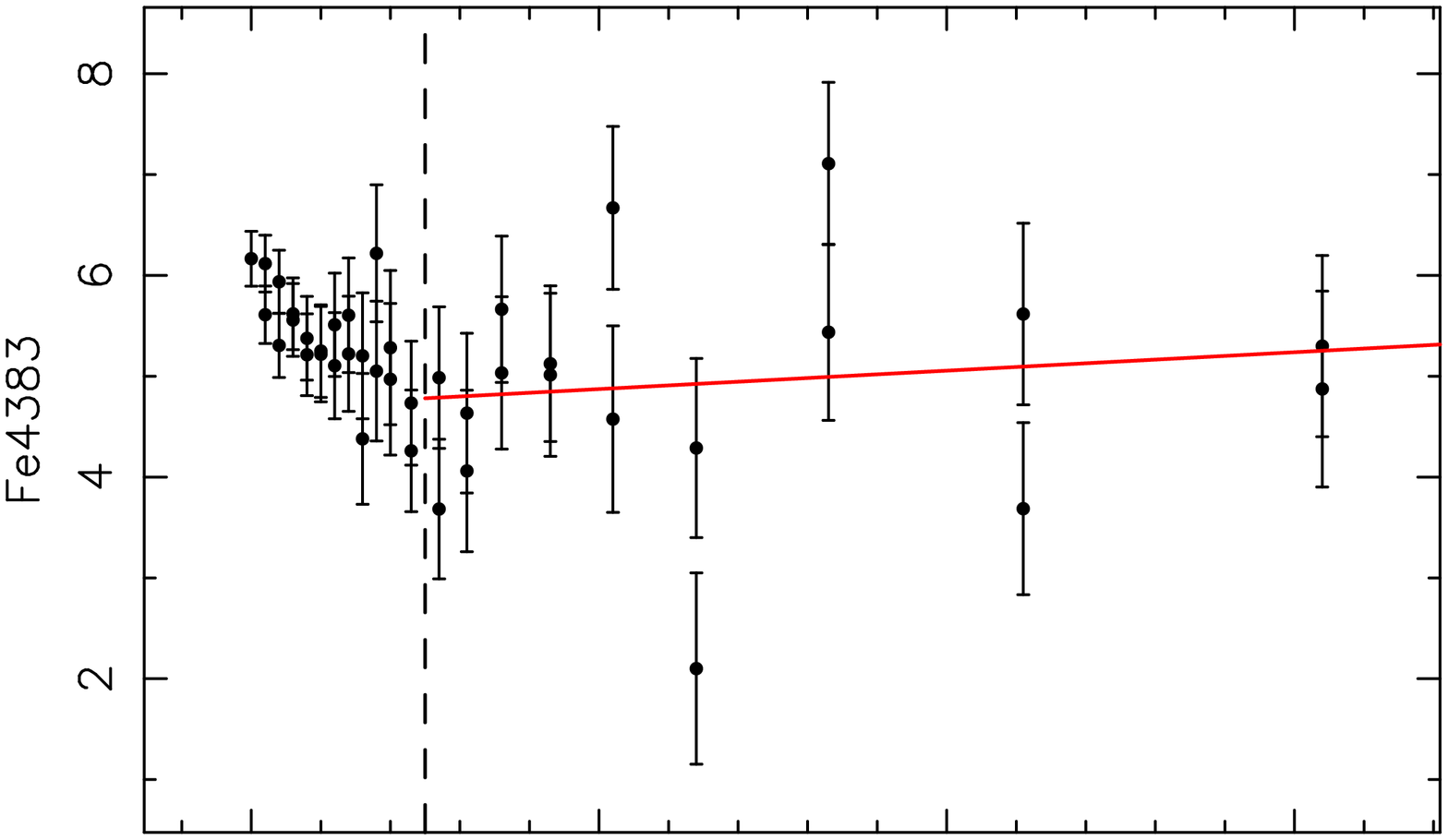}}
\resizebox{0.3\textwidth}{!}{\includegraphics[angle=-90]{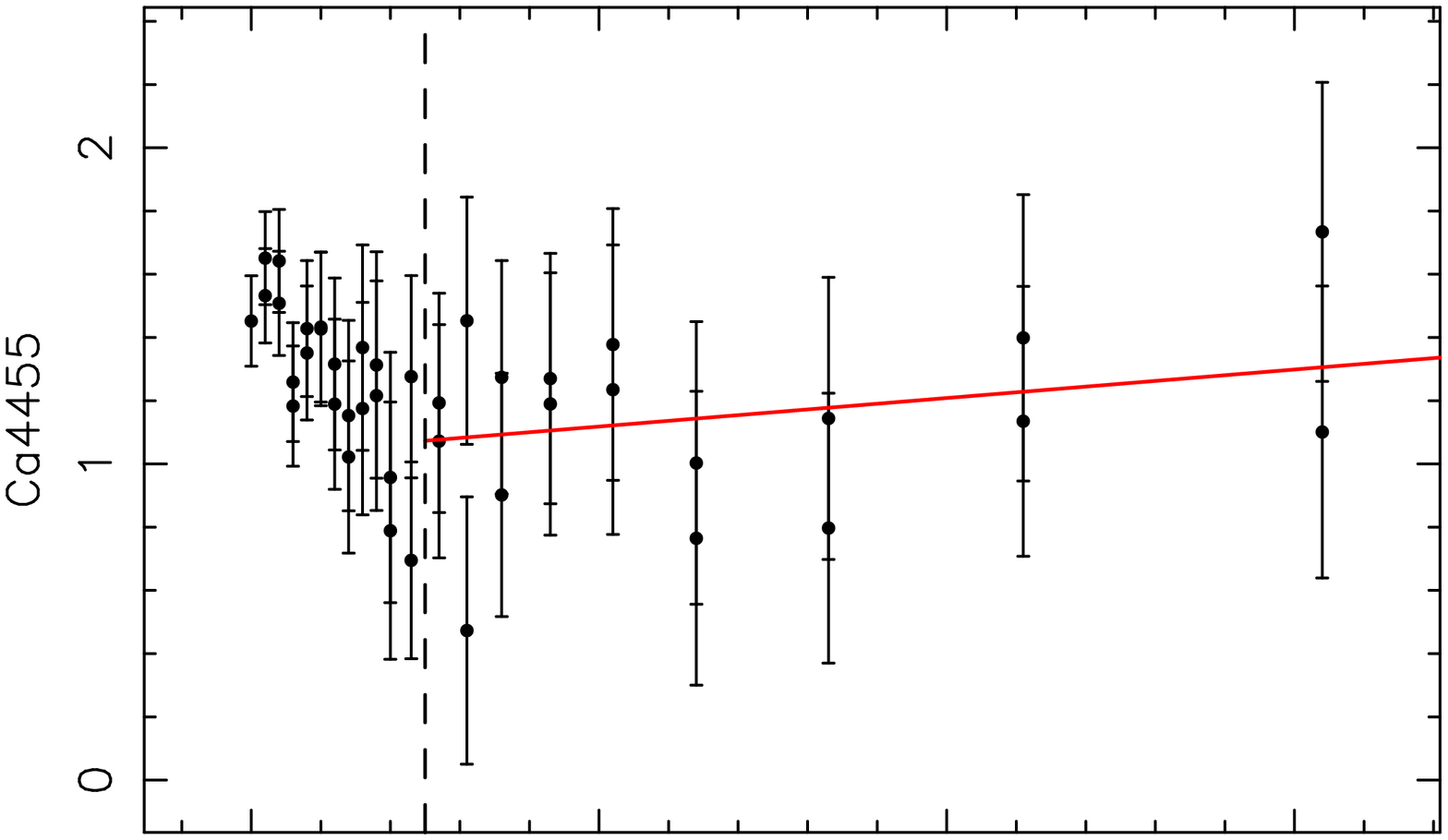}}
\resizebox{0.3\textwidth}{!}{\includegraphics[angle=-90]{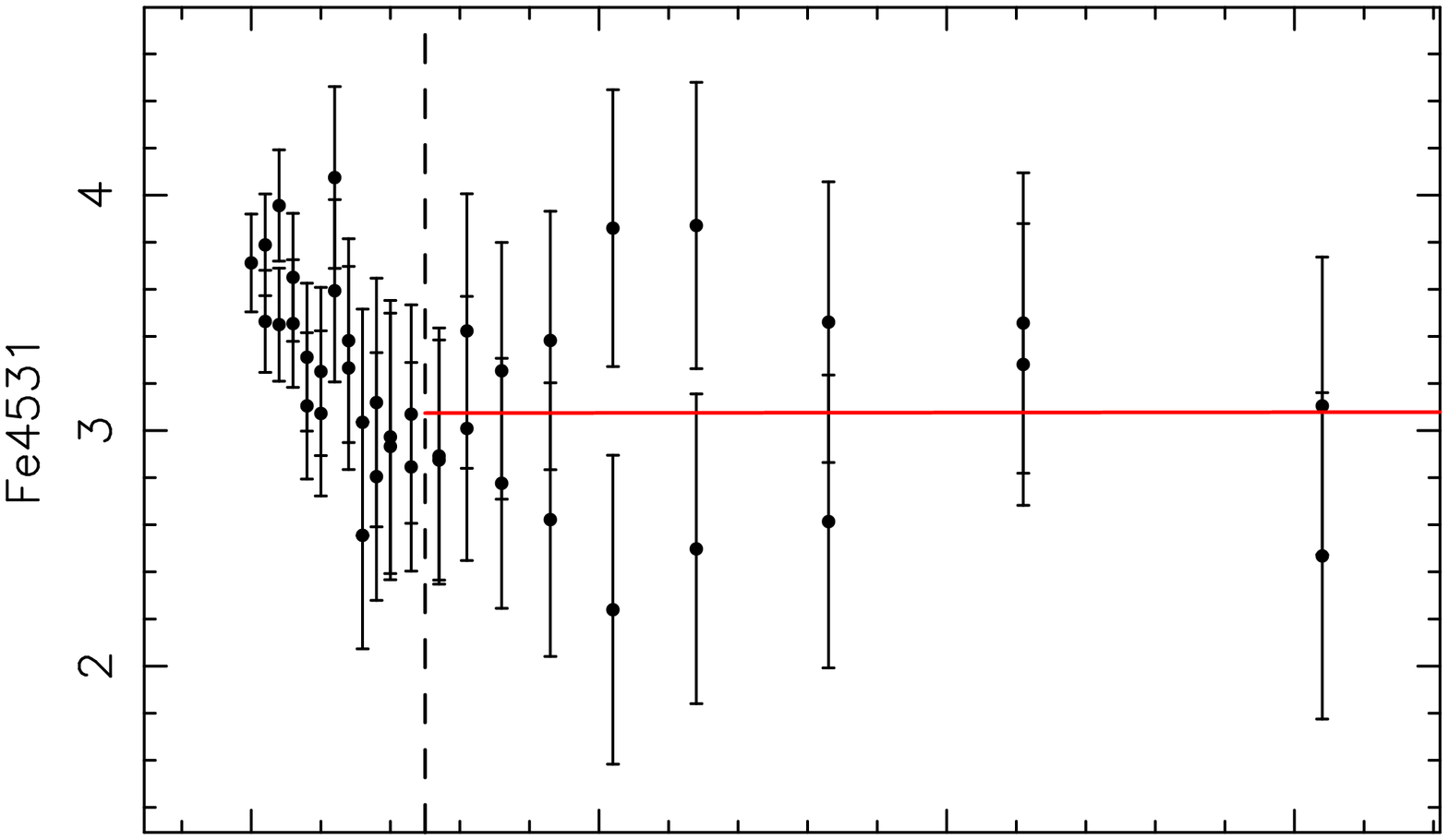}}
\resizebox{0.3\textwidth}{!}{\includegraphics[angle=-90]{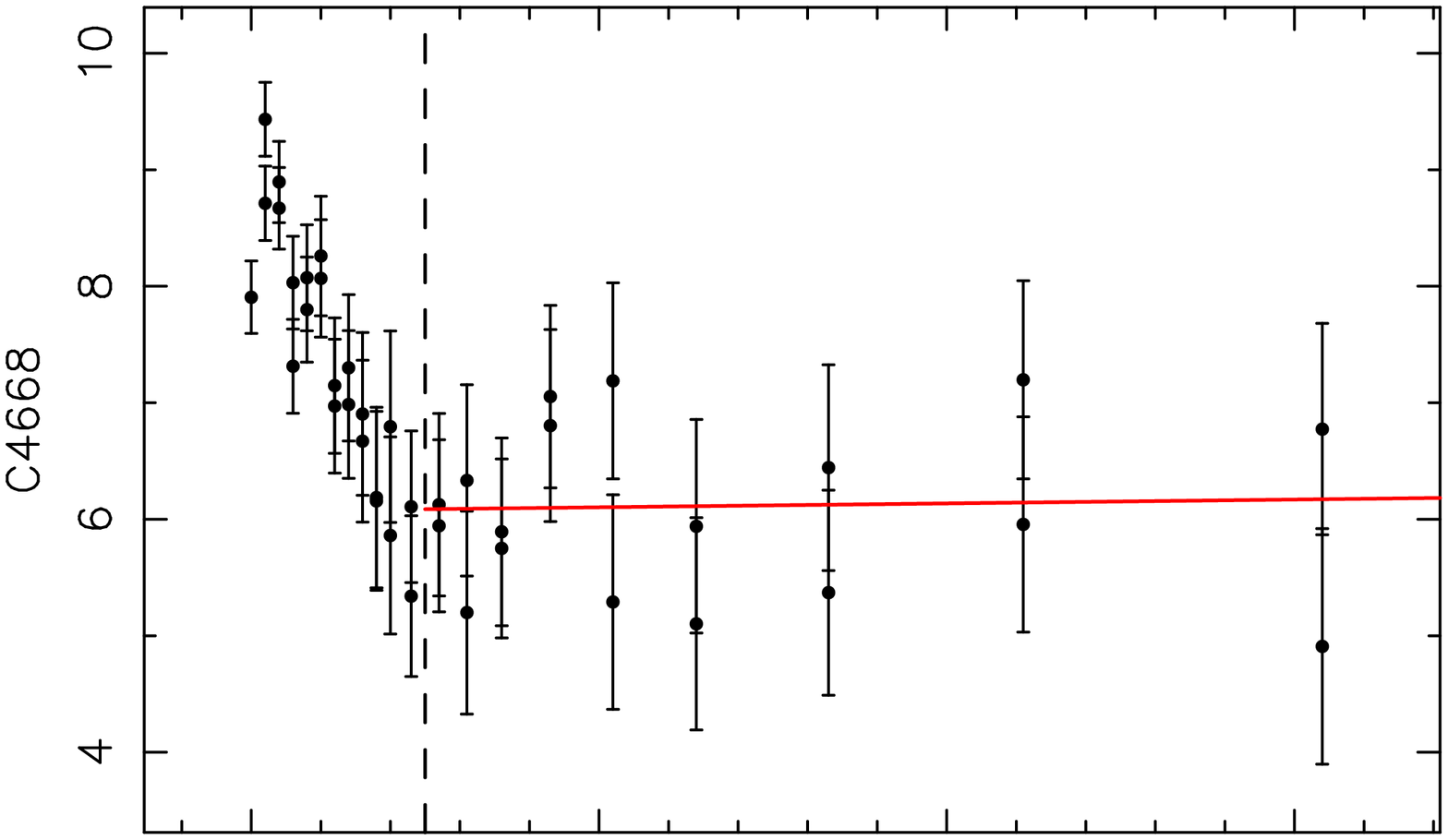}}
\resizebox{0.3\textwidth}{!}{\includegraphics[angle=-90]{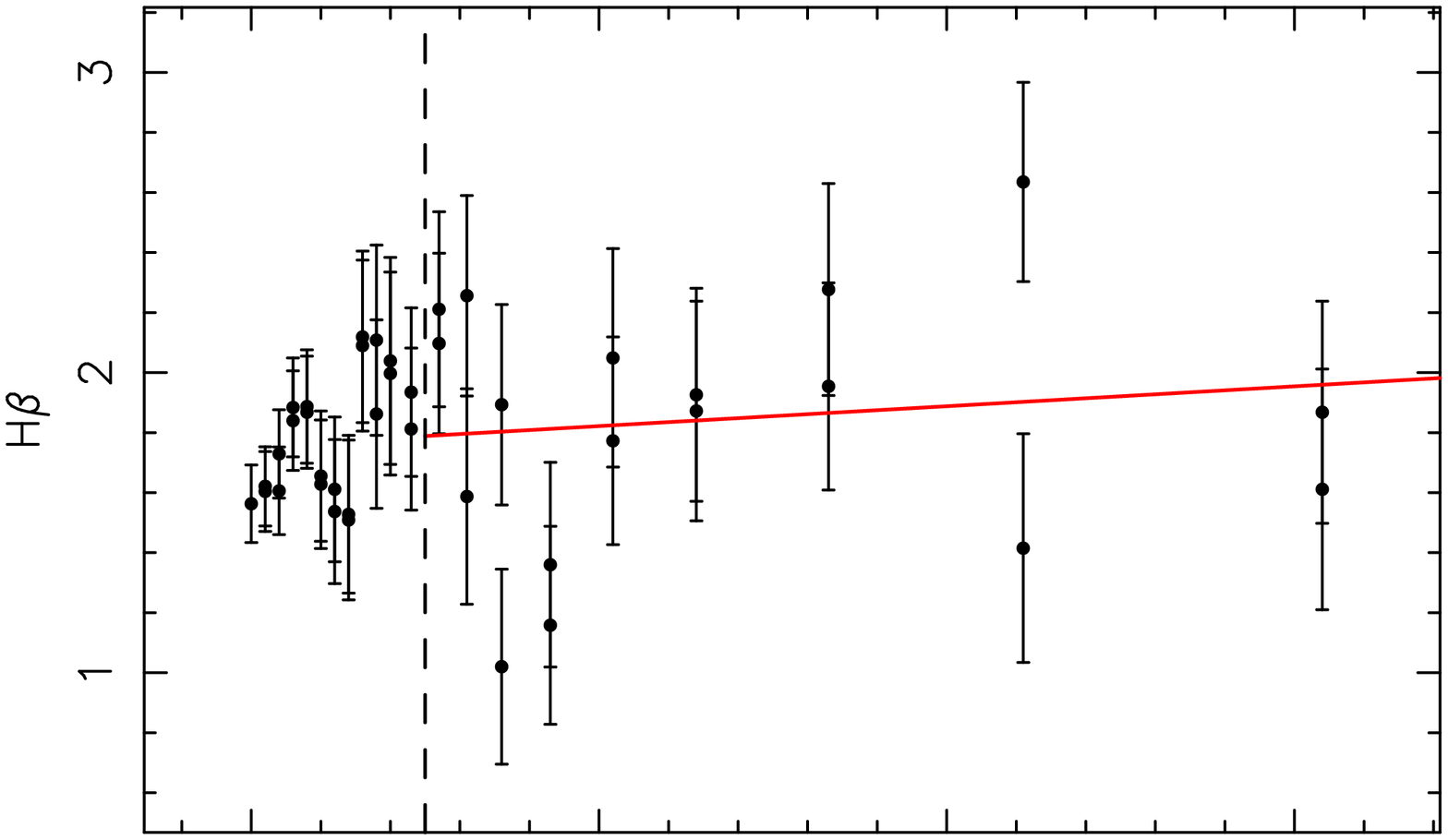}}
\resizebox{0.3\textwidth}{!}{\includegraphics[angle=-90]{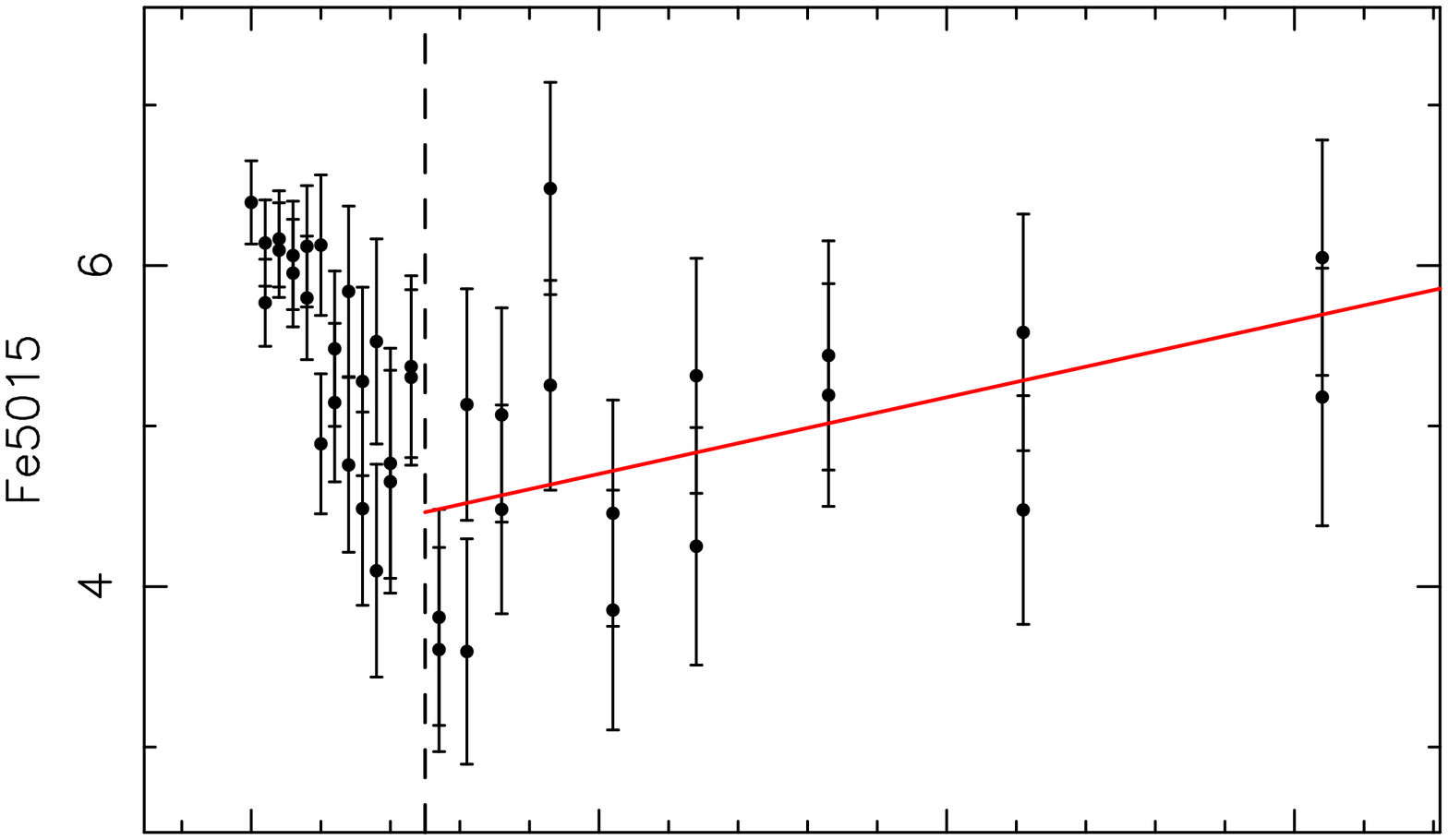}}
\resizebox{0.3\textwidth}{!}{\includegraphics[angle=-90]{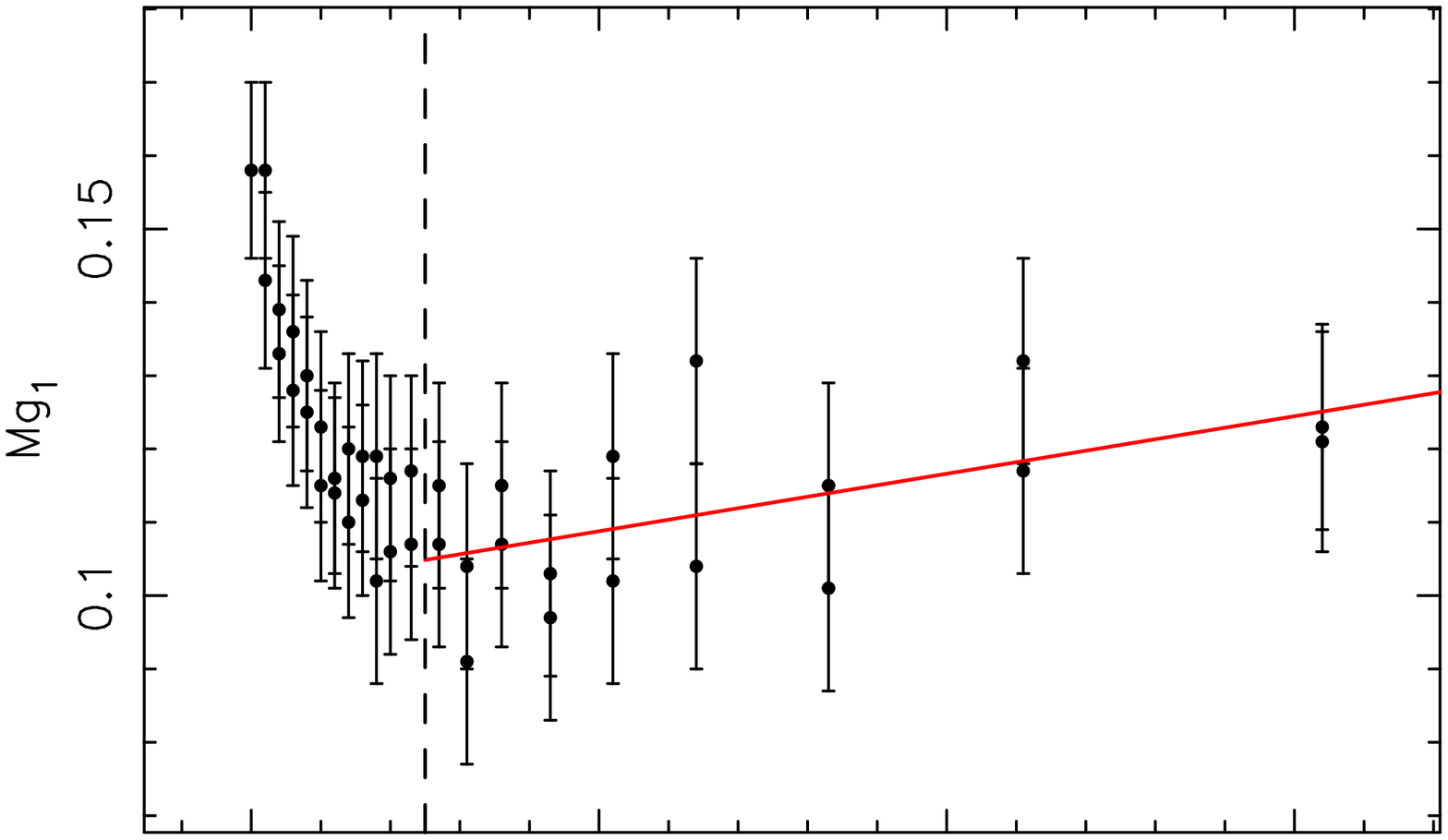}}
\resizebox{0.3\textwidth}{!}{\includegraphics[angle=-90]{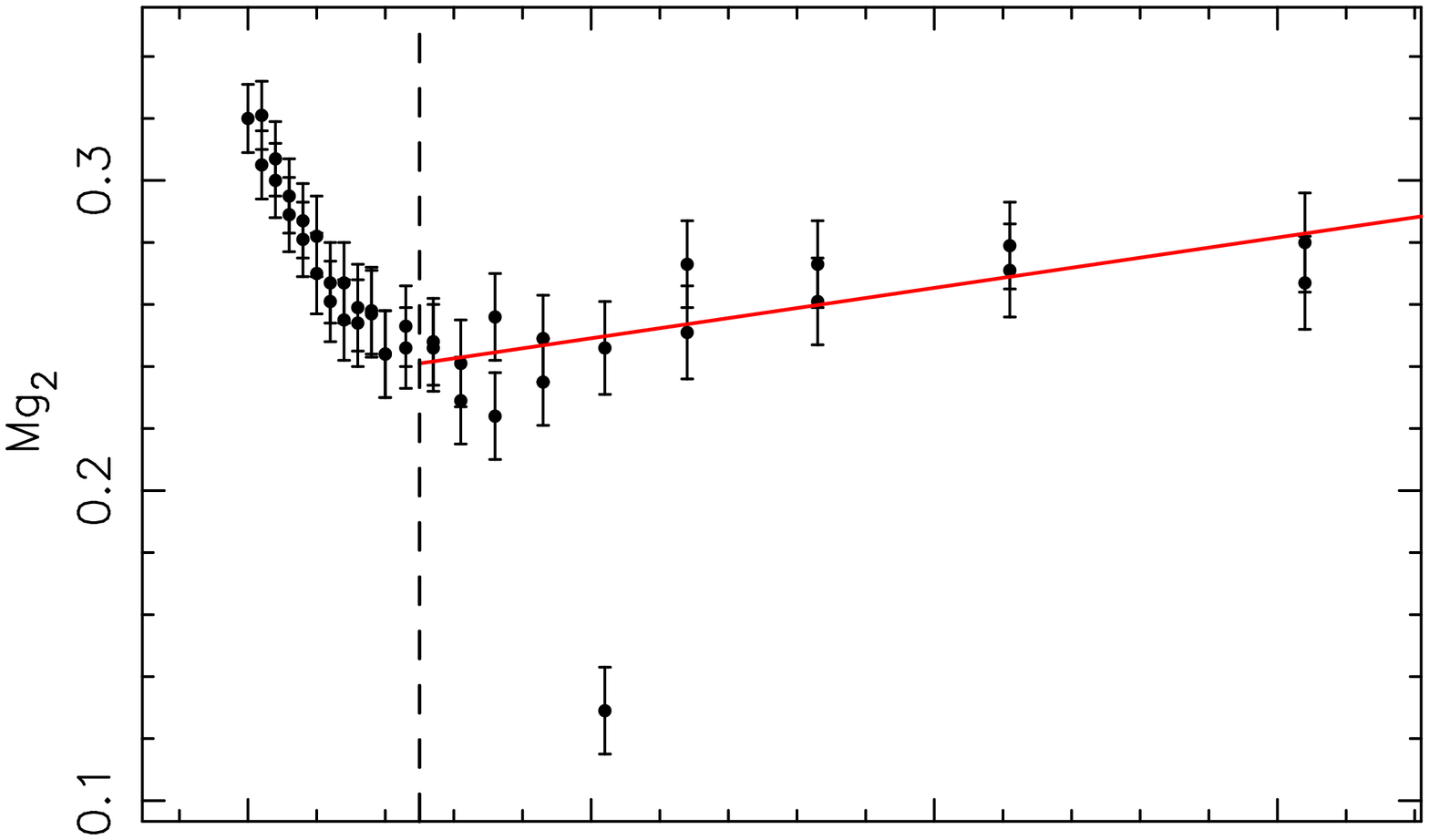}}
\resizebox{0.3\textwidth}{!}{\includegraphics[angle=-90]{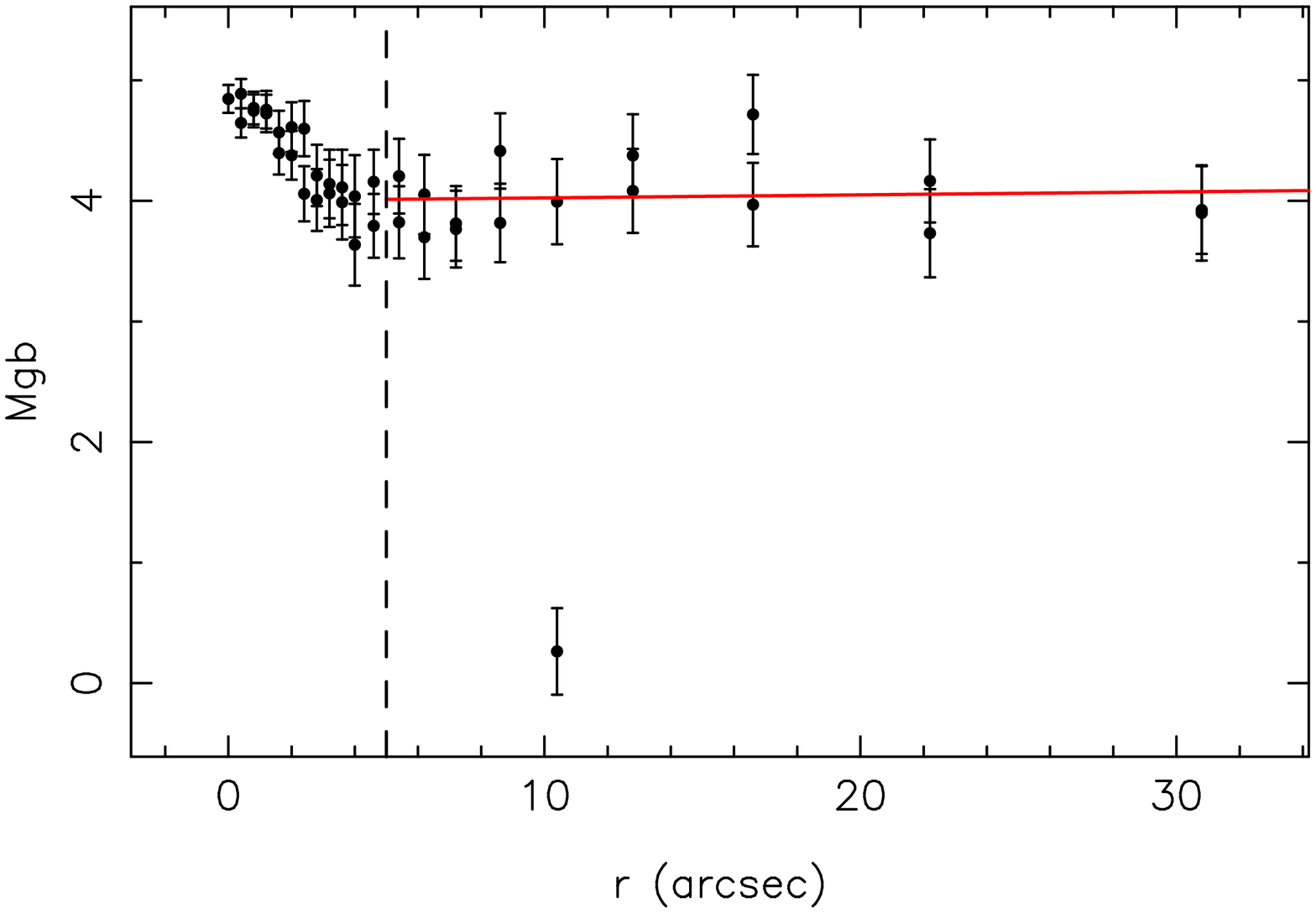}}\hspace{0.85cm}
\resizebox{0.3\textwidth}{!}{\includegraphics[angle=-90]{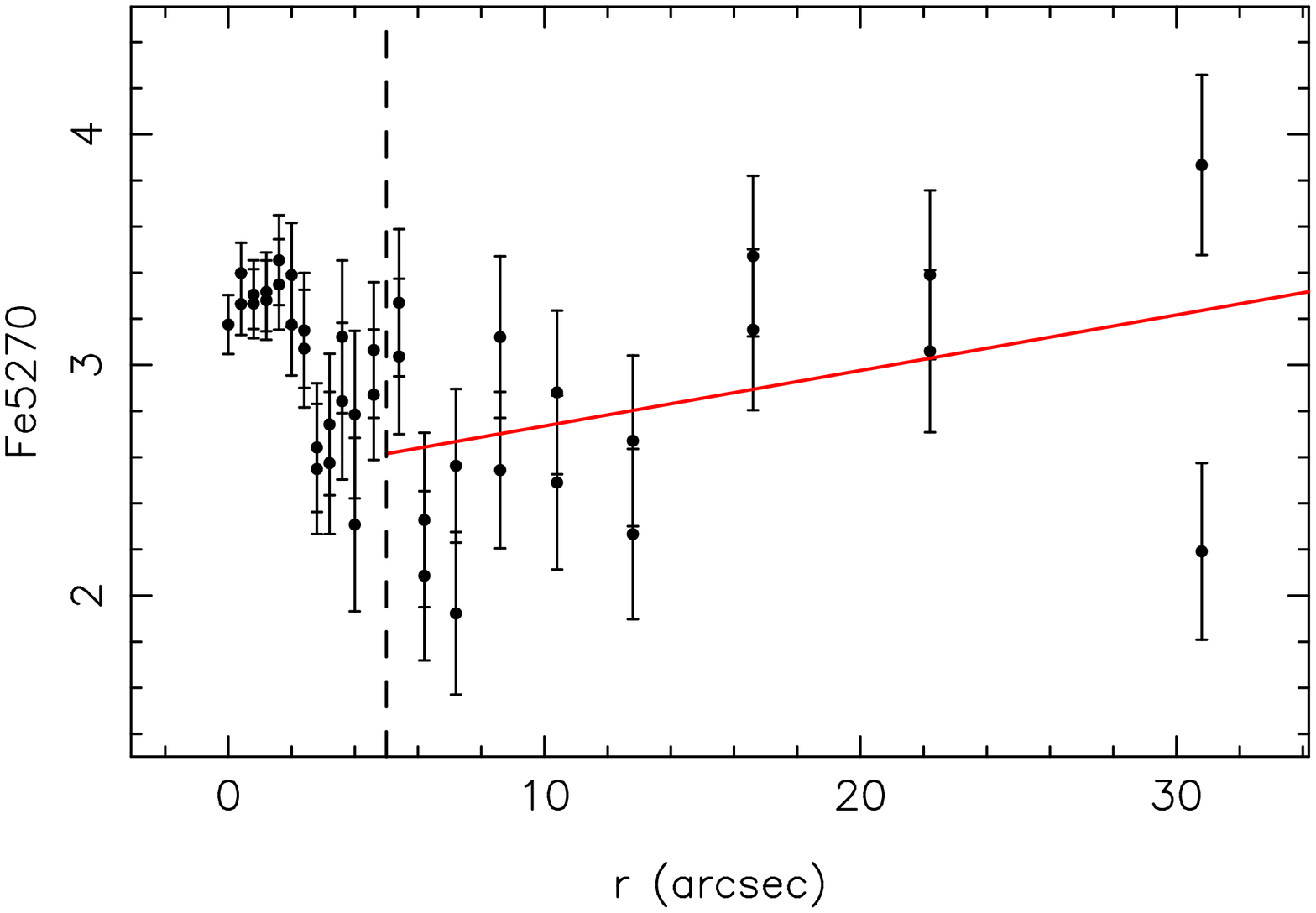}}\hspace{0.85cm}
\resizebox{0.3\textwidth}{!}{\includegraphics[angle=-90]{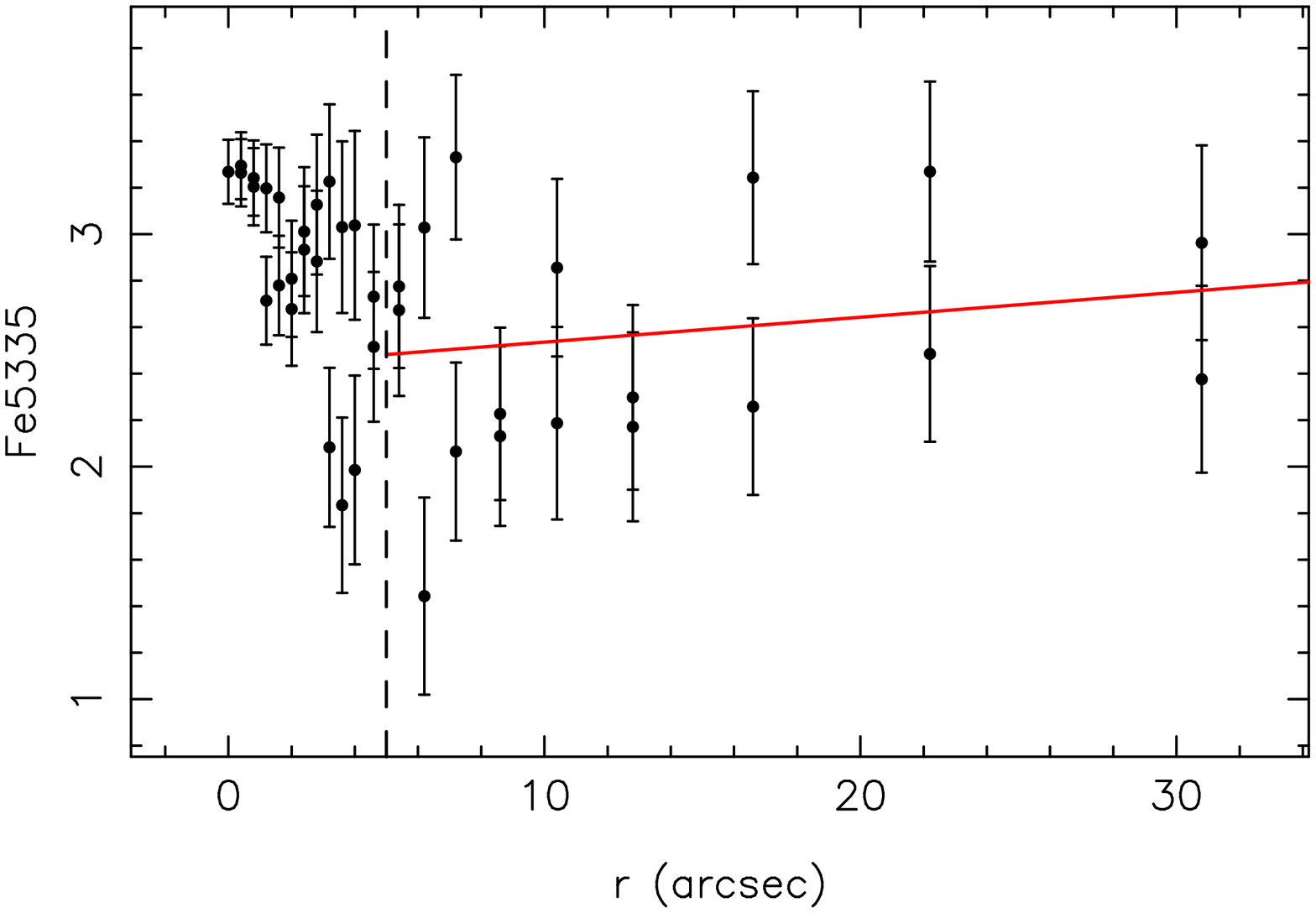}}
\caption{Line-strength distribution in the bar region for all the galaxies}
\end{figure*}

\begin{figure*}
\addtocounter{figure}{-1}
\resizebox{0.3\textwidth}{!}{\includegraphics[angle=-90]{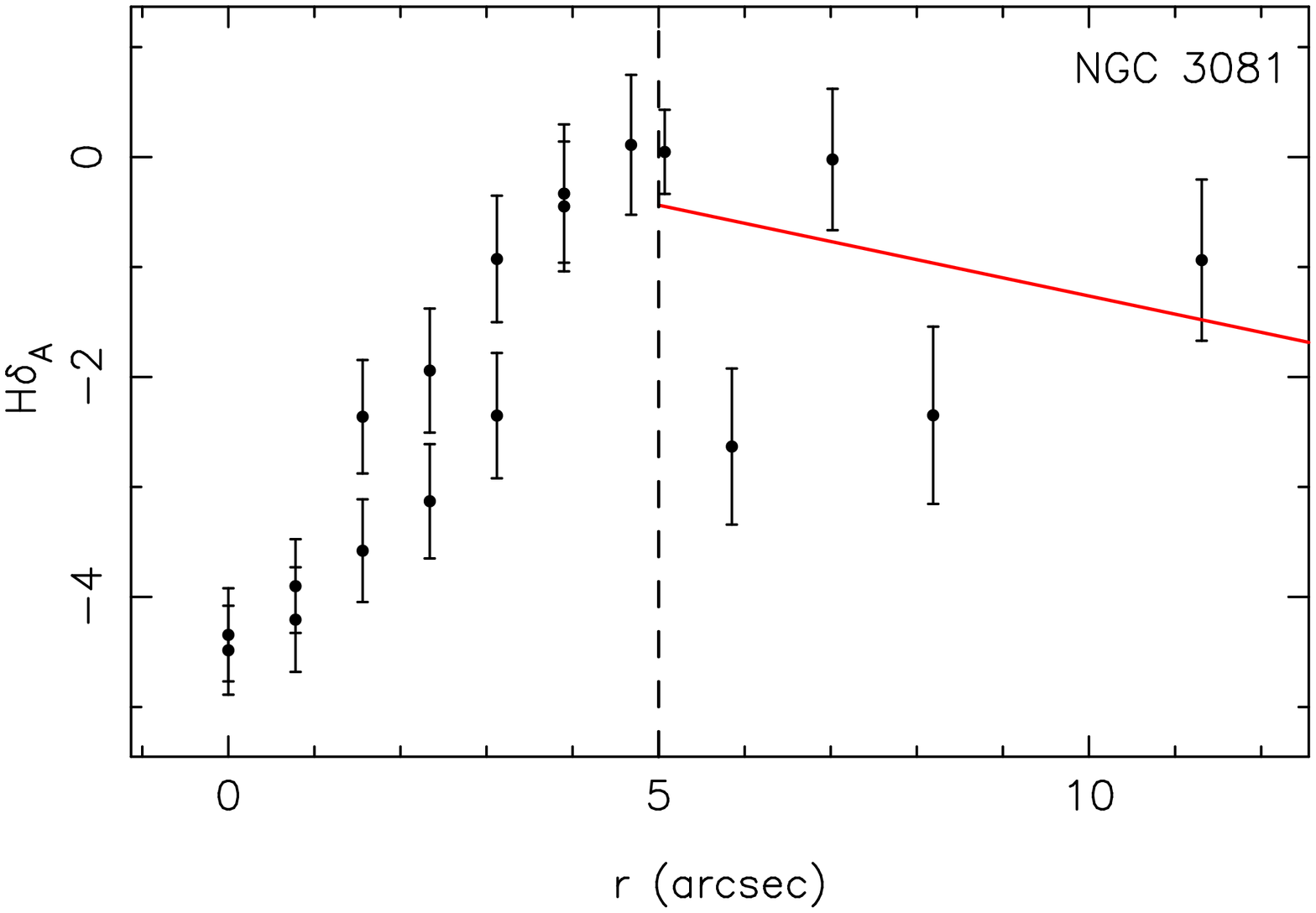}}
\resizebox{0.3\textwidth}{!}{\includegraphics[angle=-90]{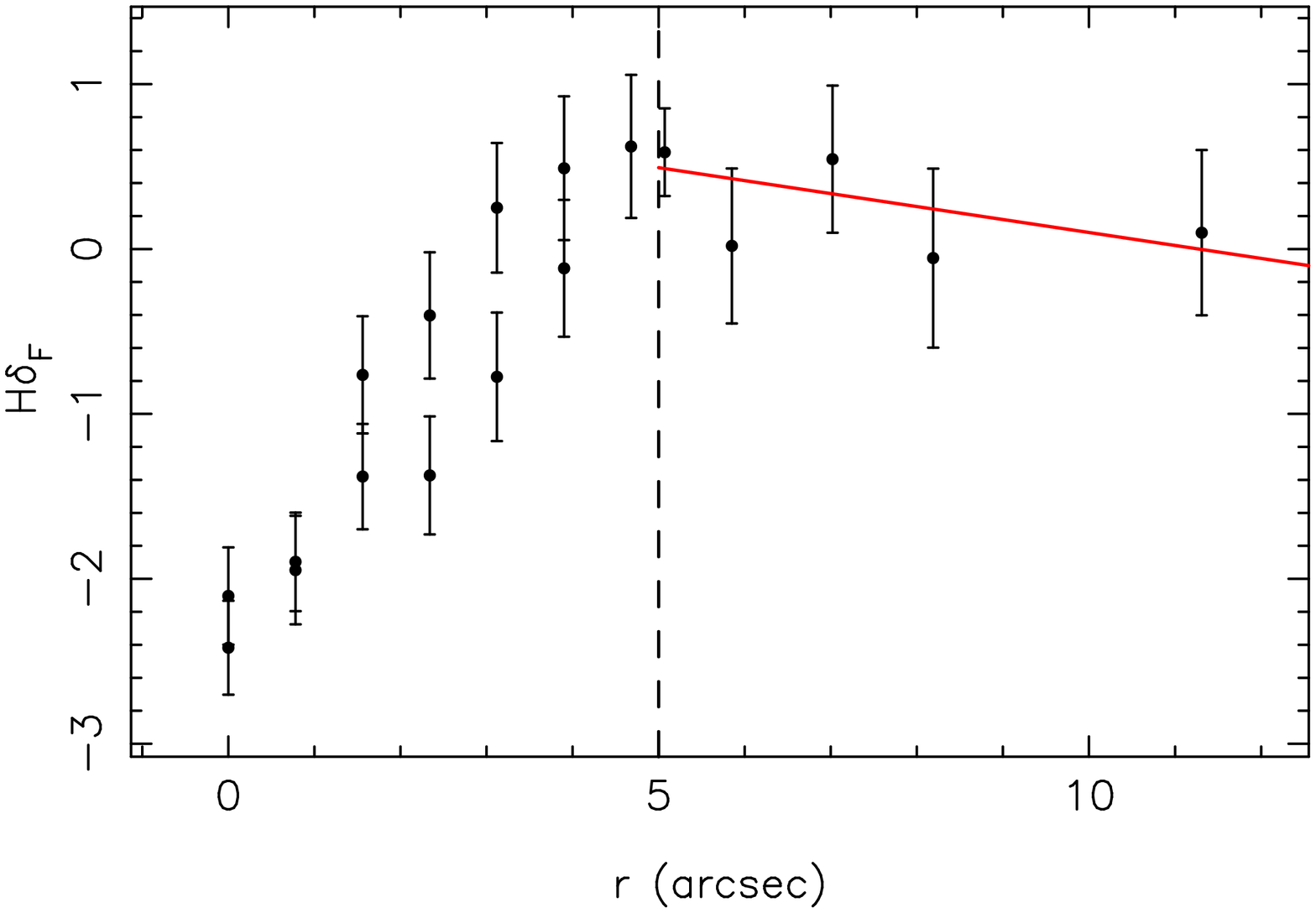}}
\resizebox{0.3\textwidth}{!}{\includegraphics[angle=-90]{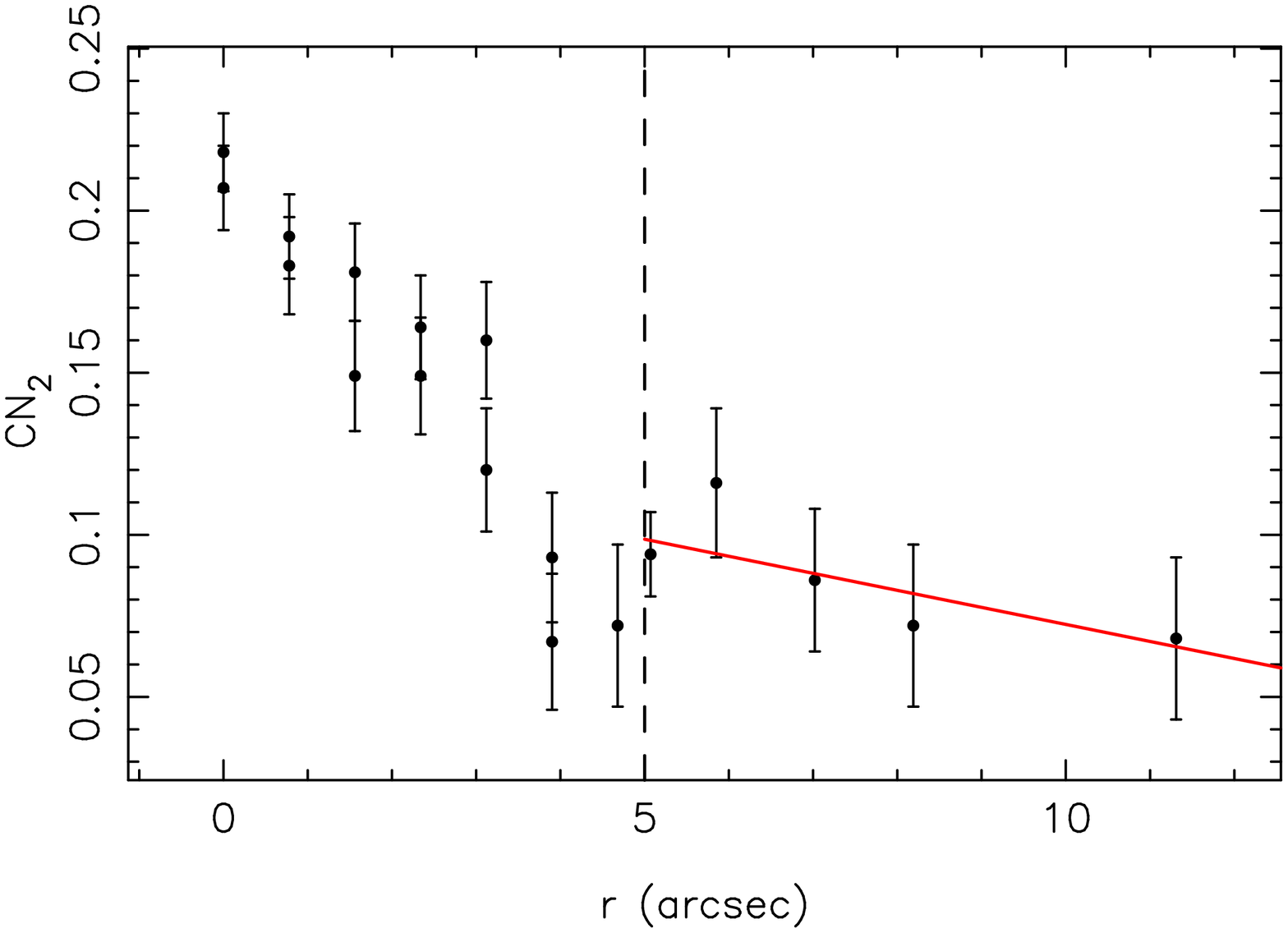}}
\resizebox{0.3\textwidth}{!}{\includegraphics[angle=-90]{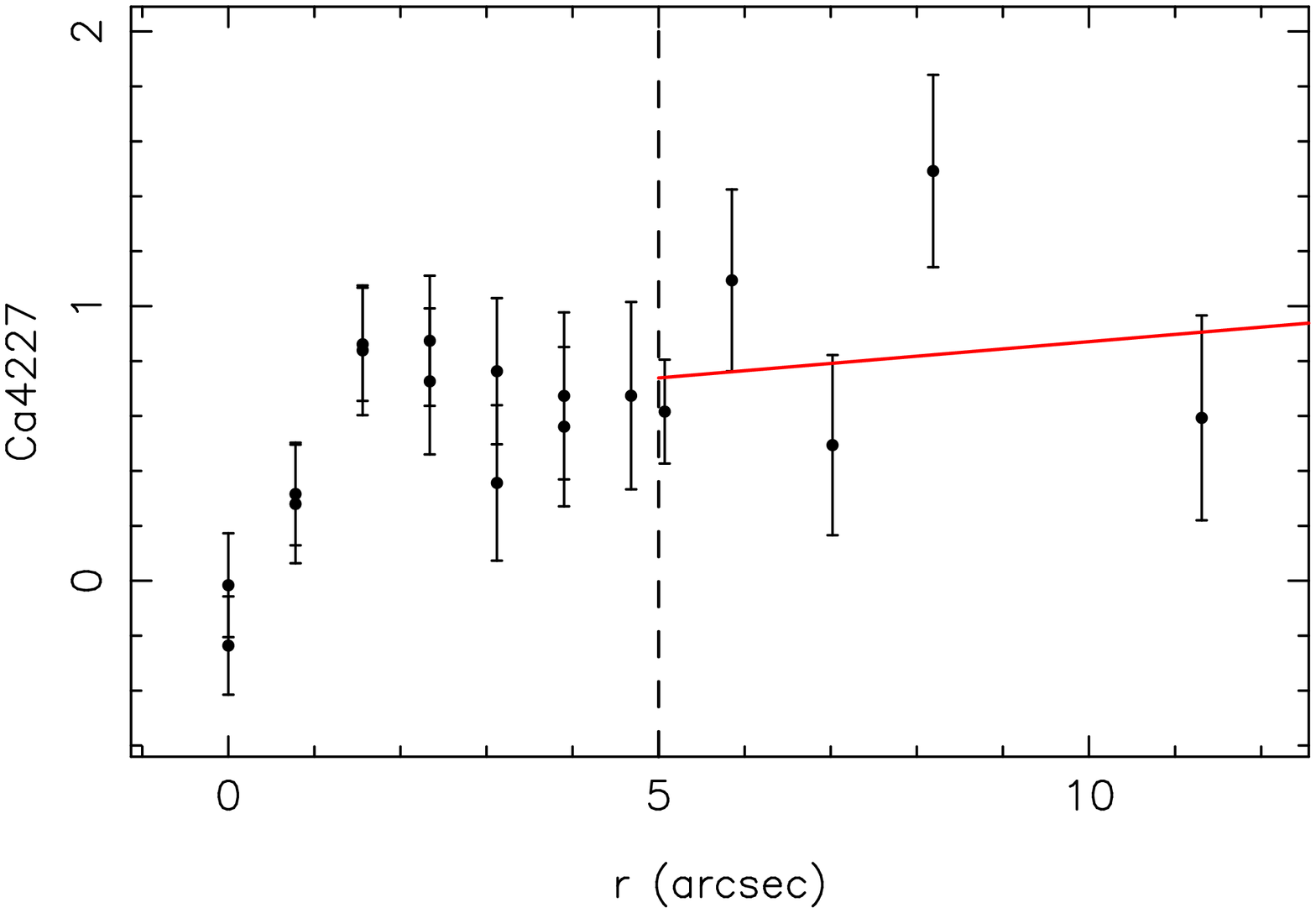}}
\resizebox{0.3\textwidth}{!}{\includegraphics[angle=-90]{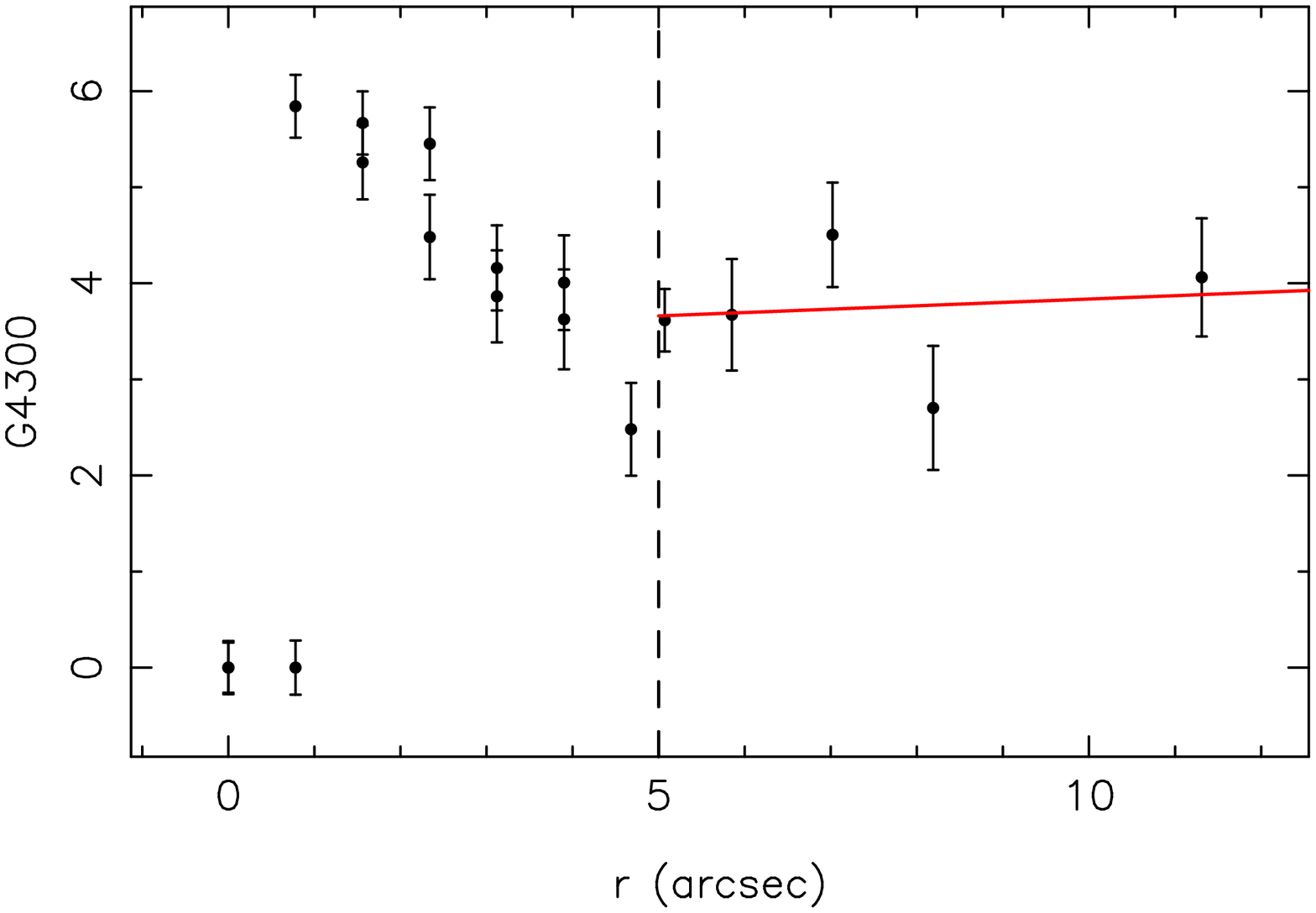}}
\resizebox{0.3\textwidth}{!}{\includegraphics[angle=-90]{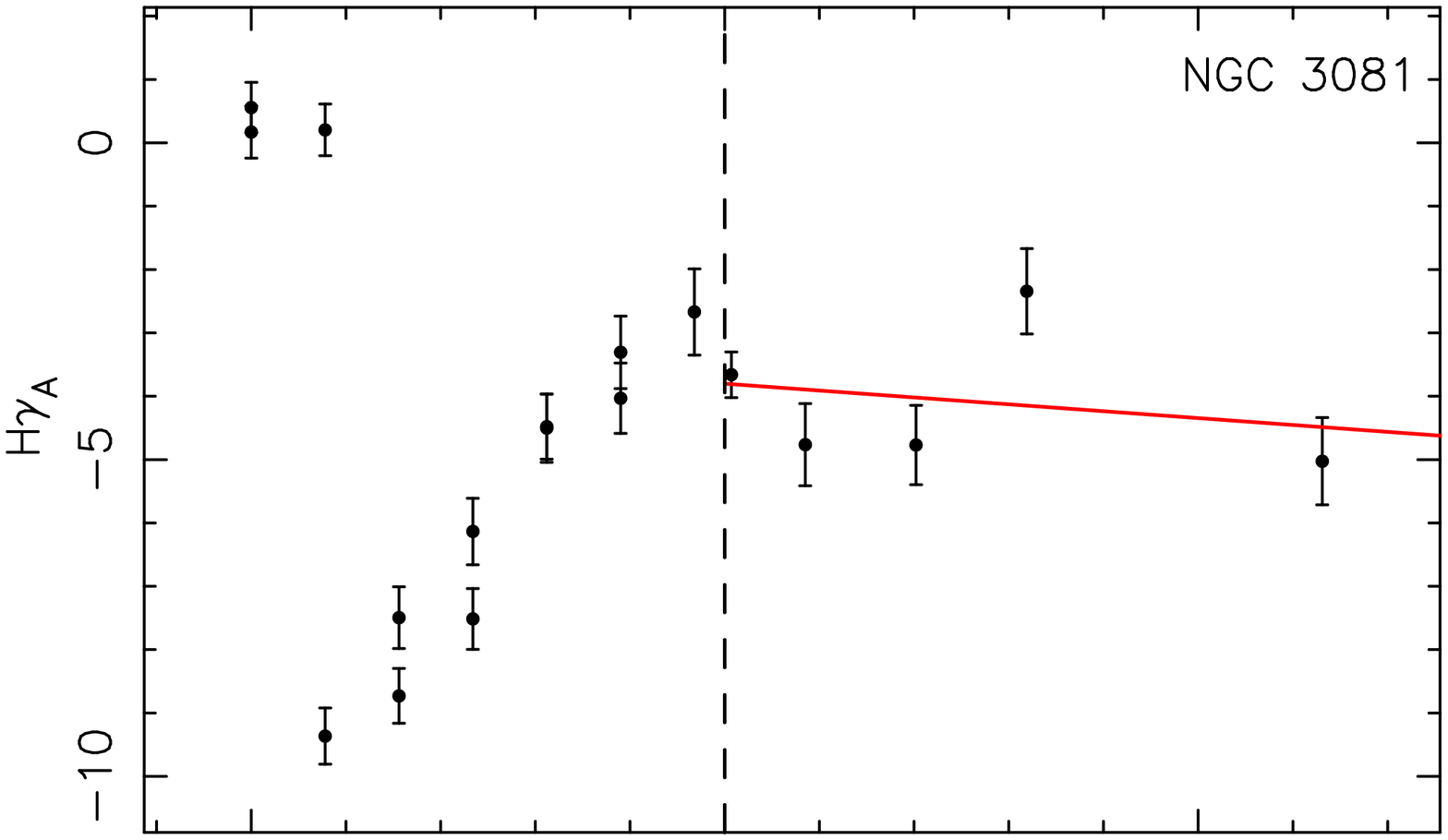}}
\resizebox{0.3\textwidth}{!}{\includegraphics[angle=-90]{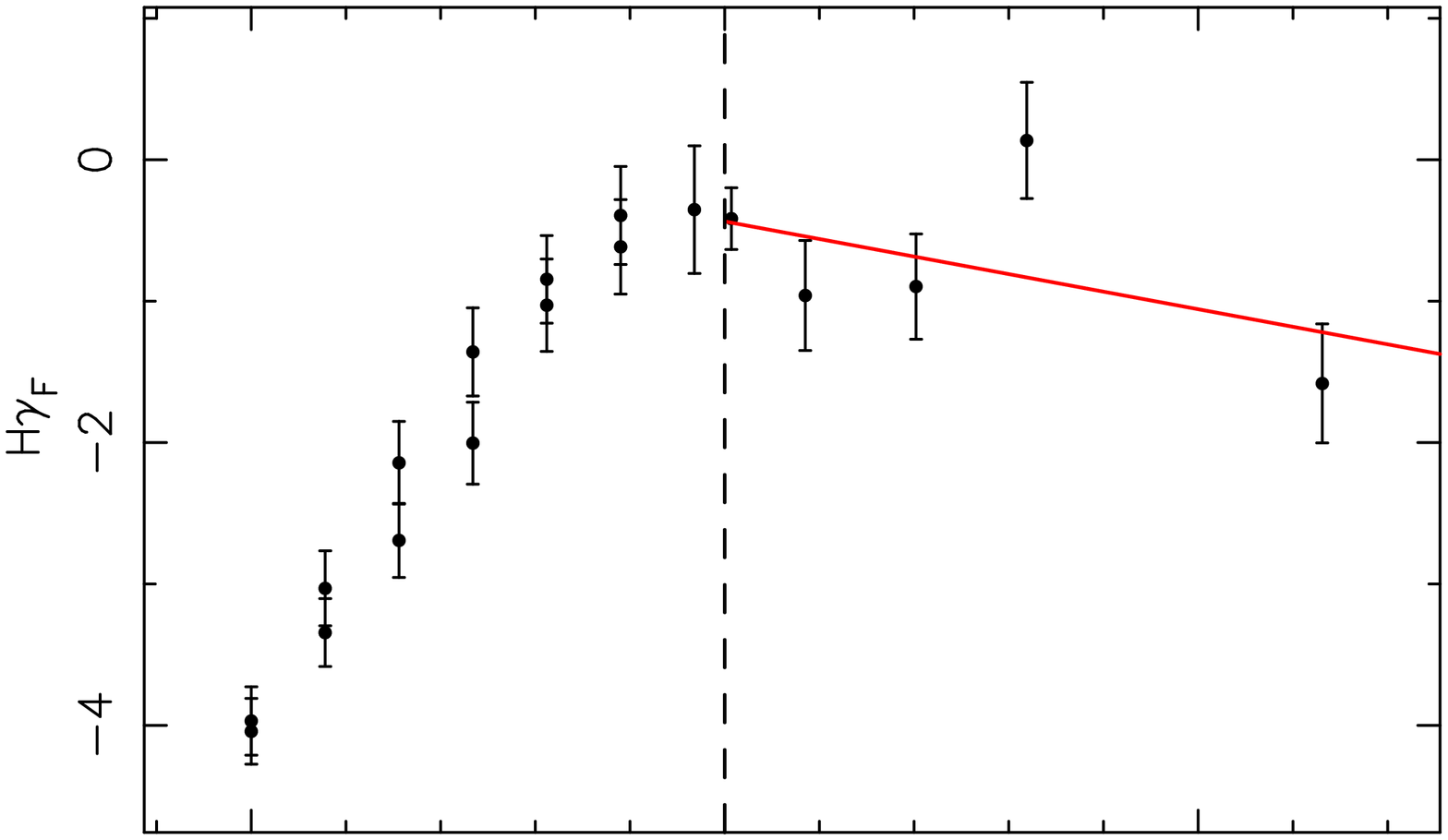}}
\resizebox{0.3\textwidth}{!}{\includegraphics[angle=-90]{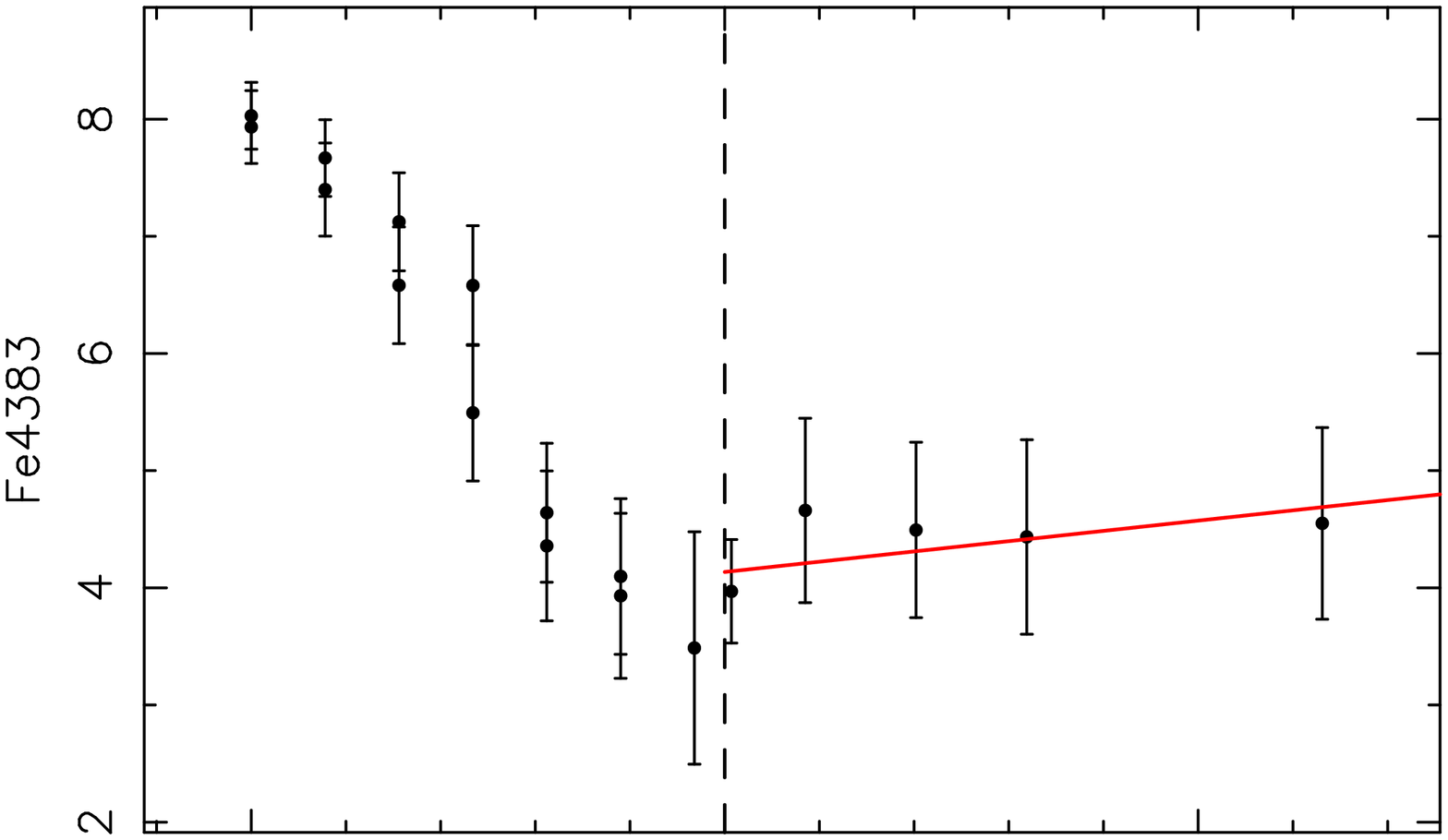}}
\resizebox{0.3\textwidth}{!}{\includegraphics[angle=-90]{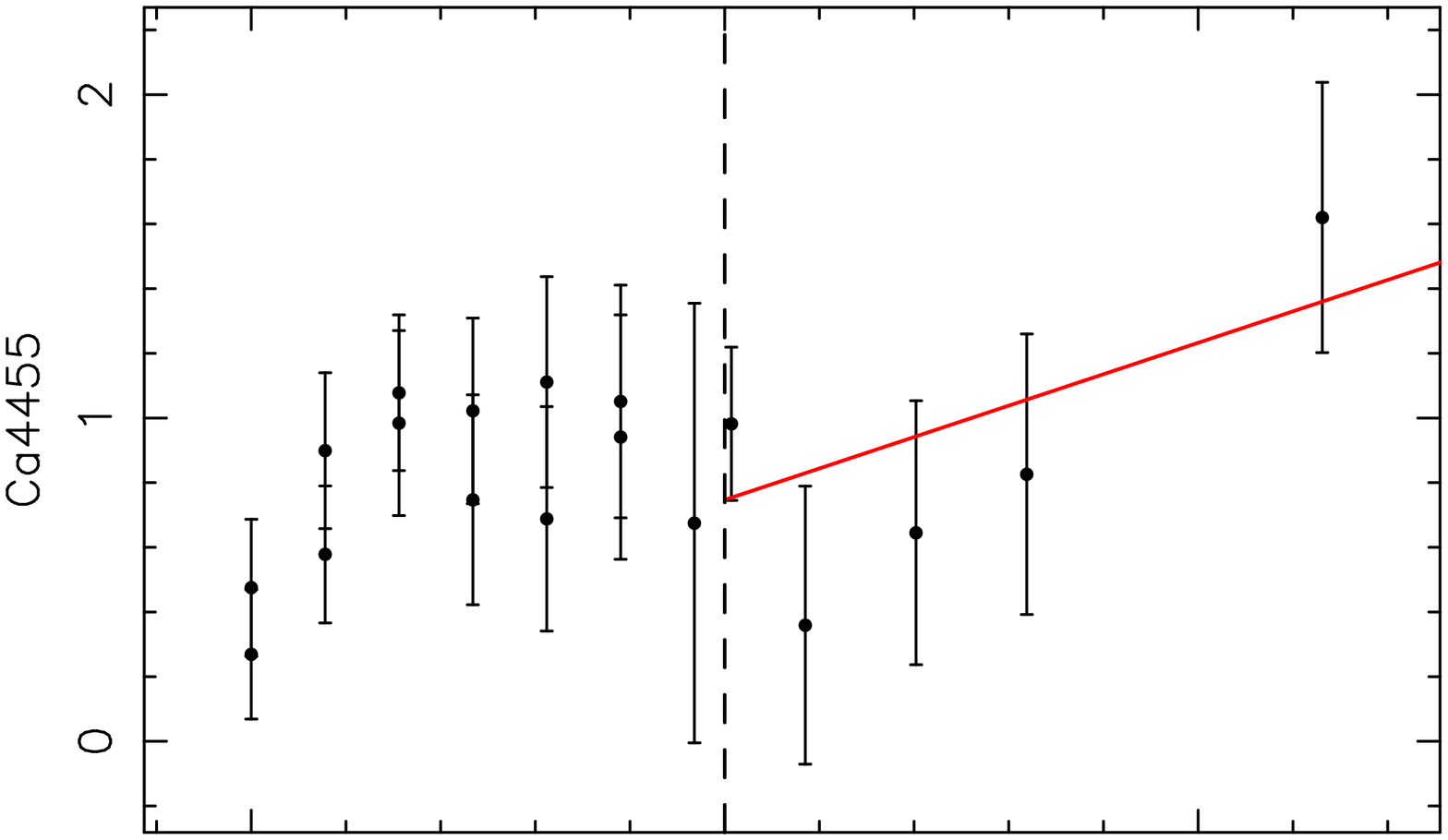}}
\resizebox{0.3\textwidth}{!}{\includegraphics[angle=-90]{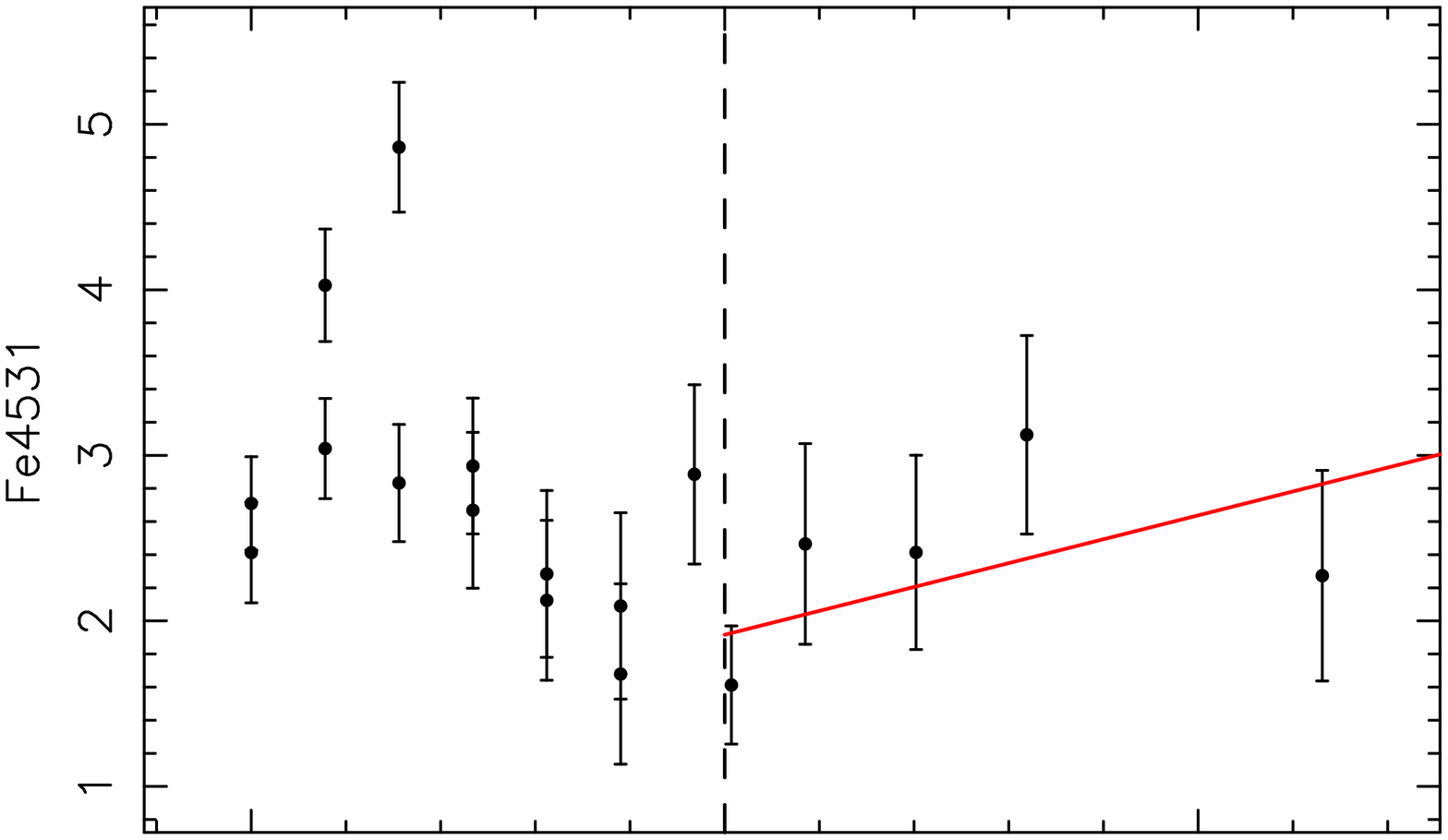}}
\resizebox{0.3\textwidth}{!}{\includegraphics[angle=-90]{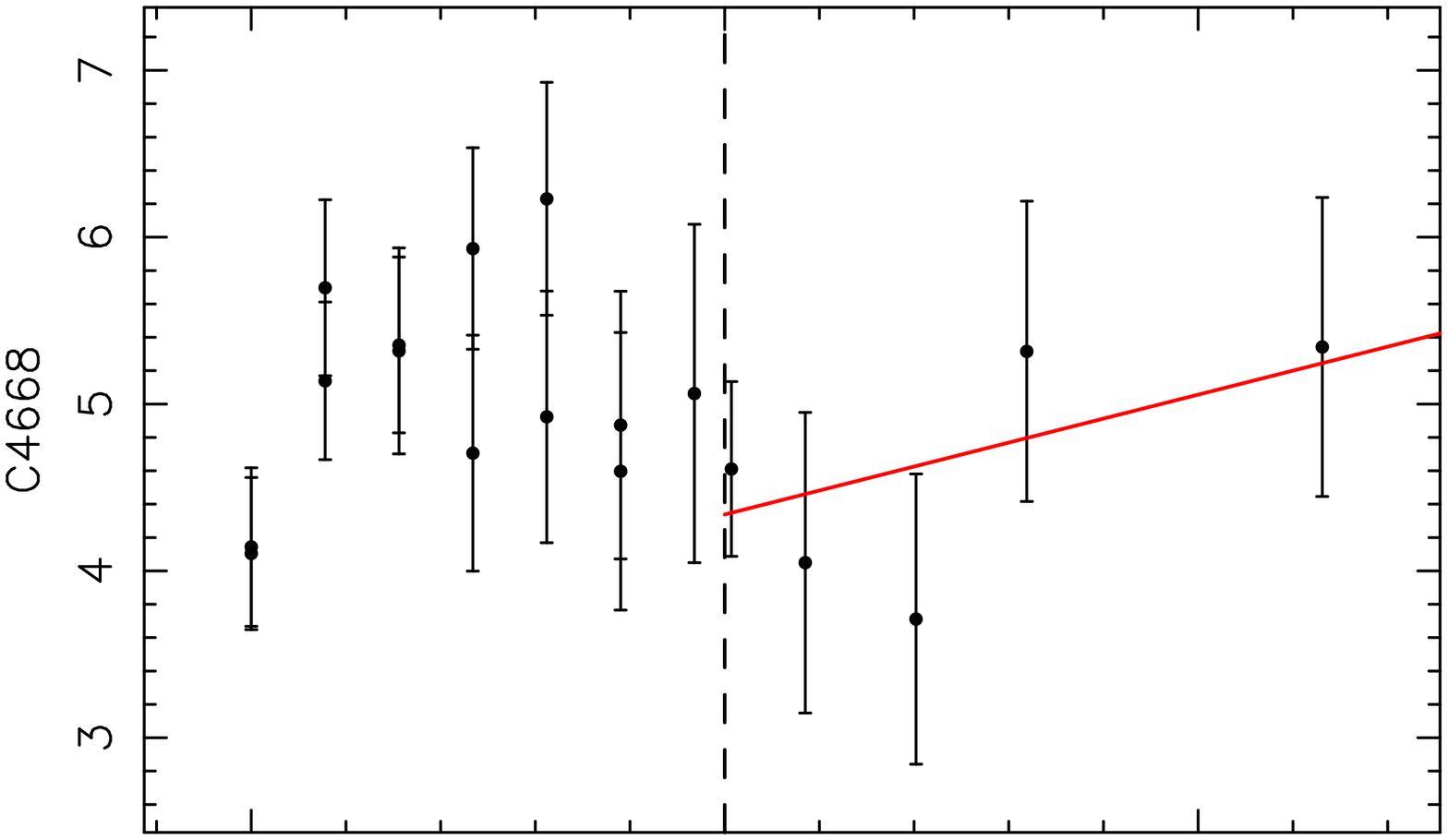}}
\resizebox{0.3\textwidth}{!}{\includegraphics[angle=-90]{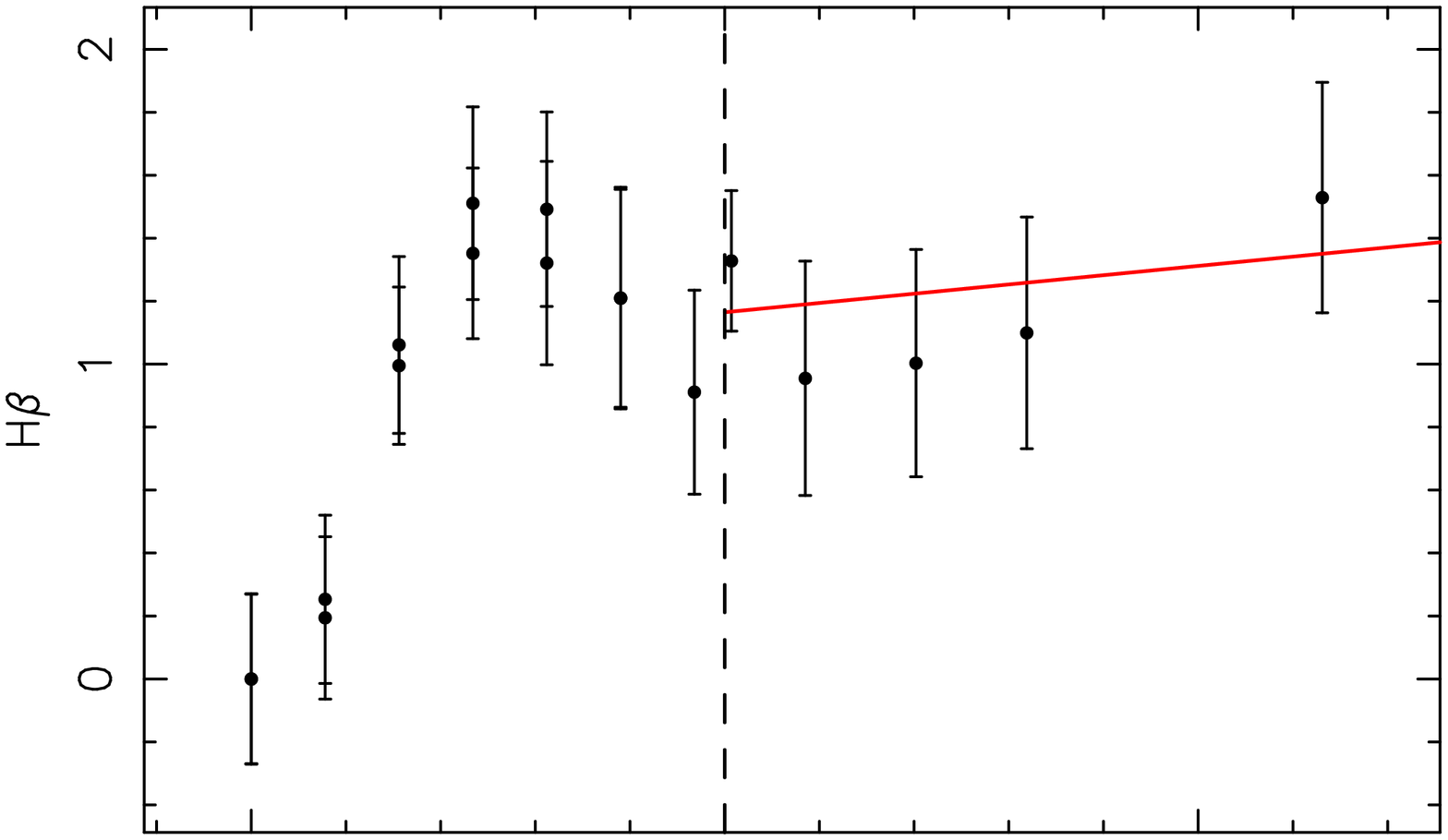}}
\resizebox{0.3\textwidth}{!}{\includegraphics[angle=-90]{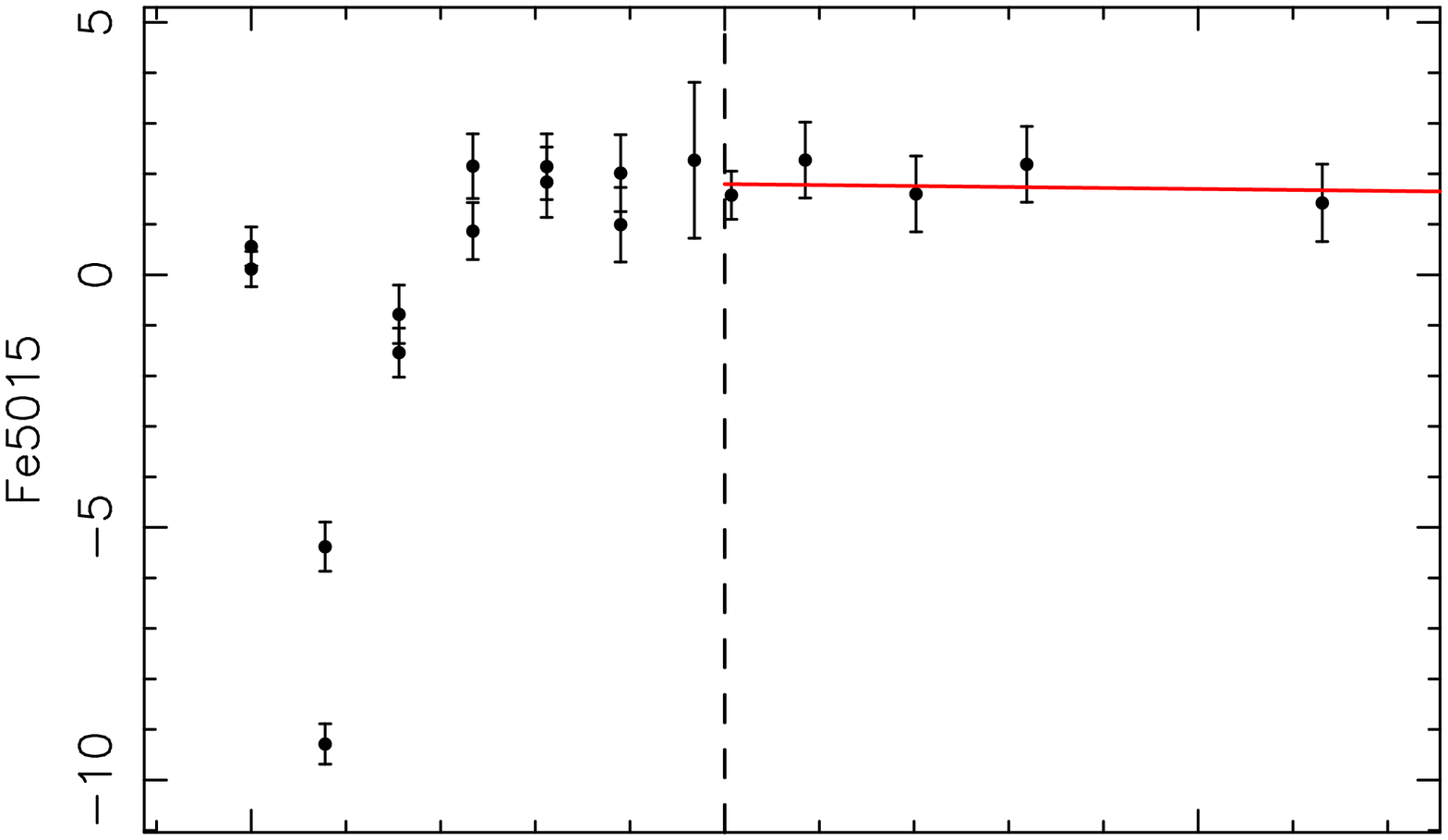}}
\resizebox{0.3\textwidth}{!}{\includegraphics[angle=-90]{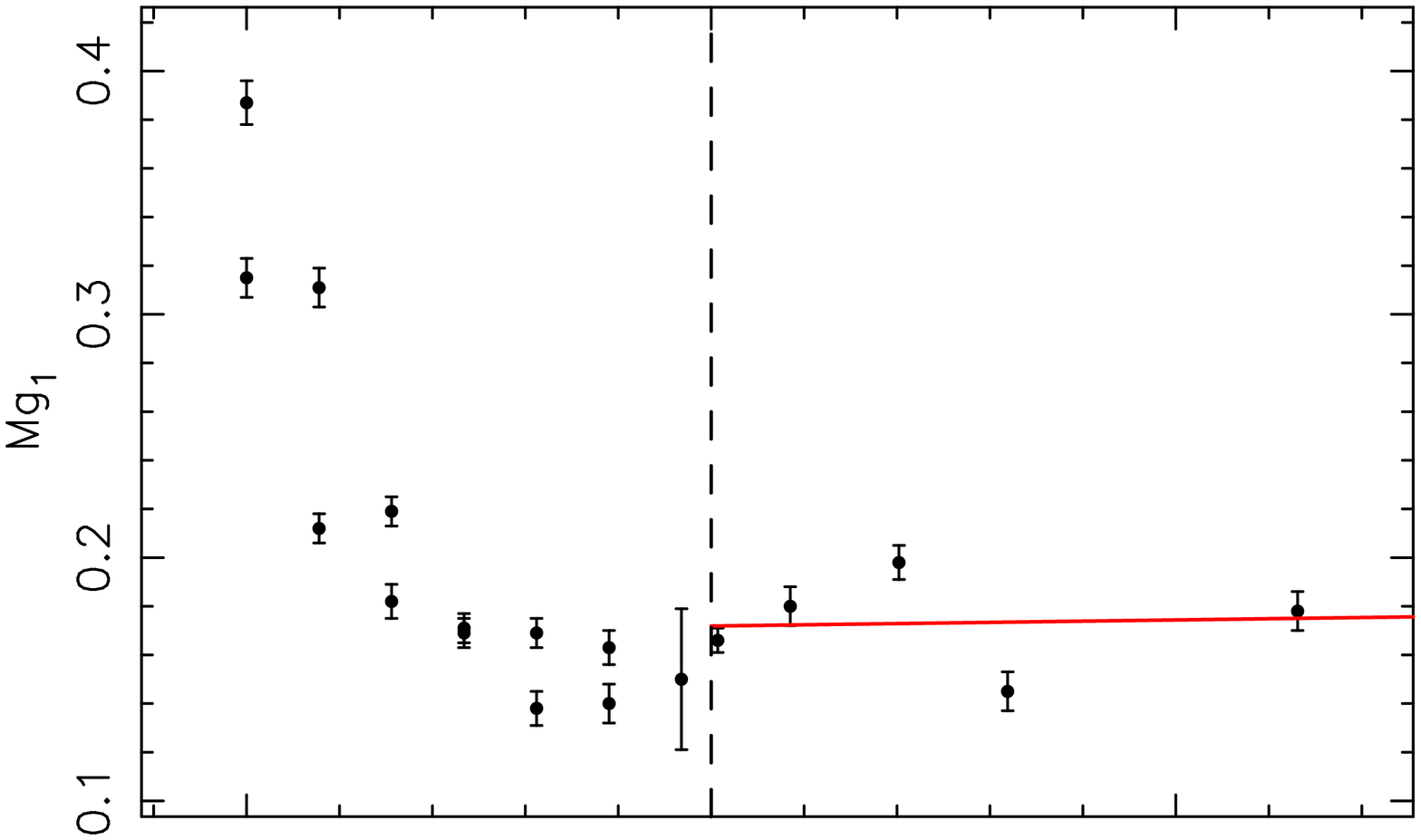}}
\resizebox{0.3\textwidth}{!}{\includegraphics[angle=-90]{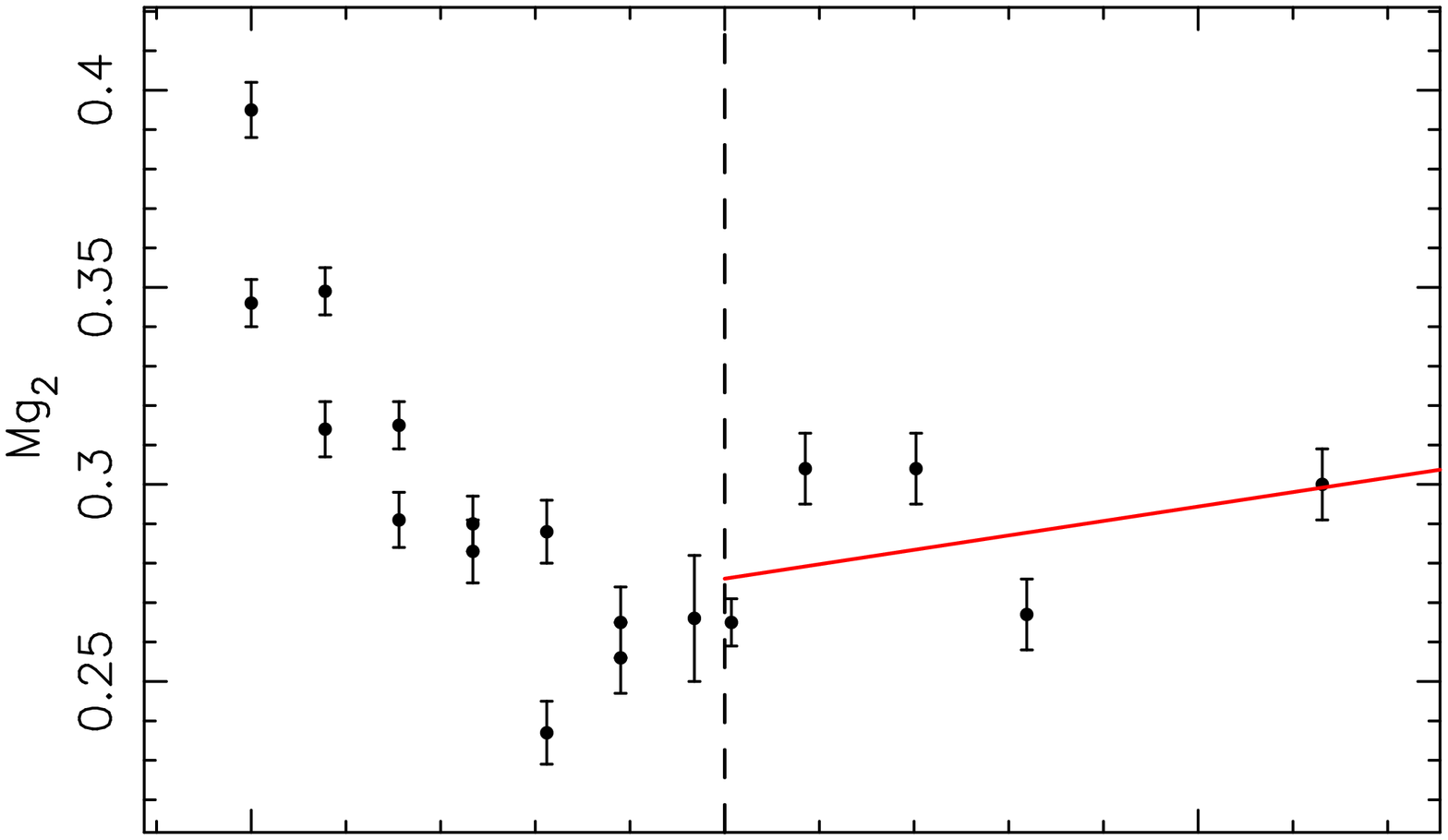}}\hspace{0.8cm}
\resizebox{0.3\textwidth}{!}{\includegraphics[angle=-90]{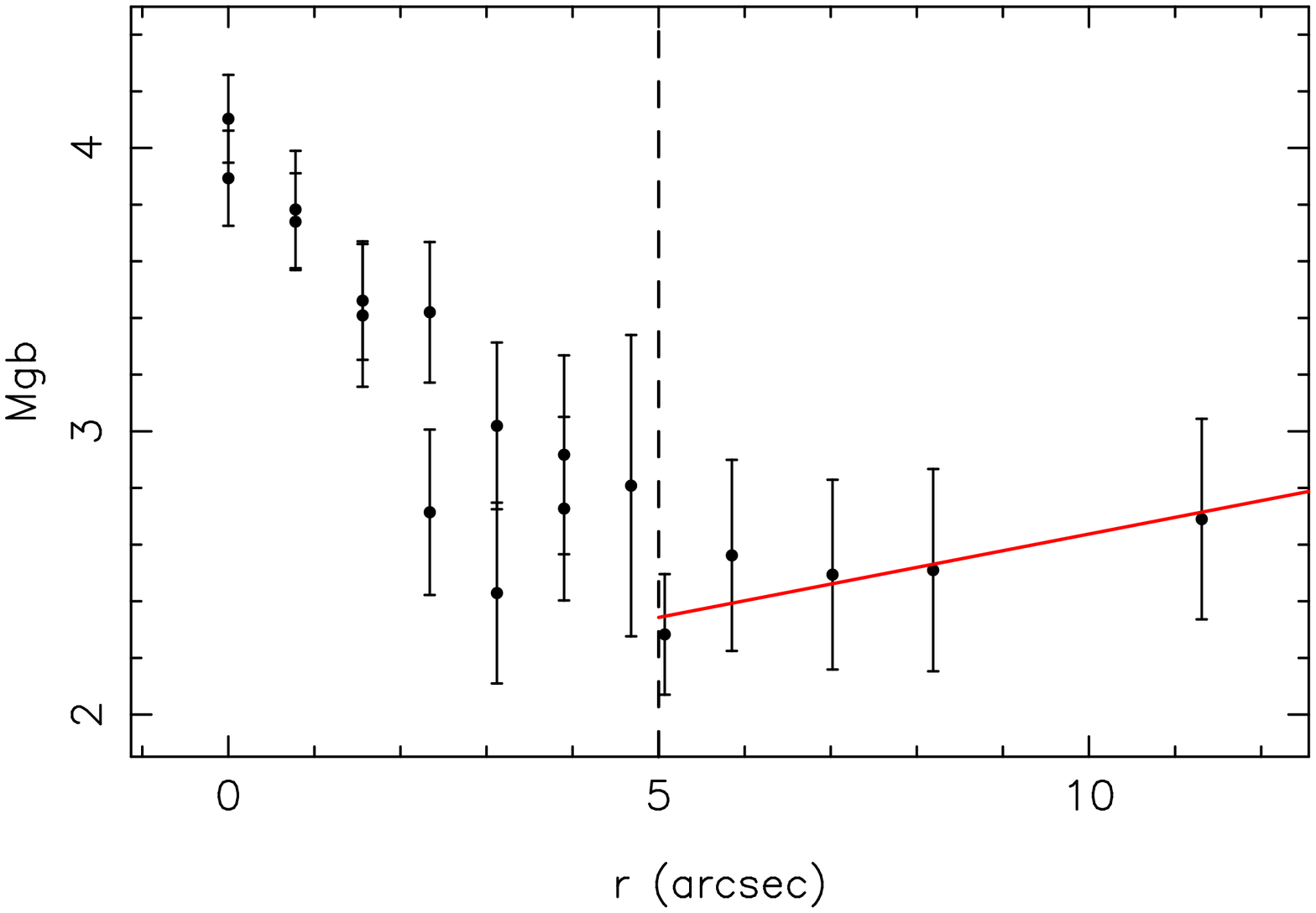}}\hspace{0.8cm}
\resizebox{0.3\textwidth}{!}{\includegraphics[angle=-90]{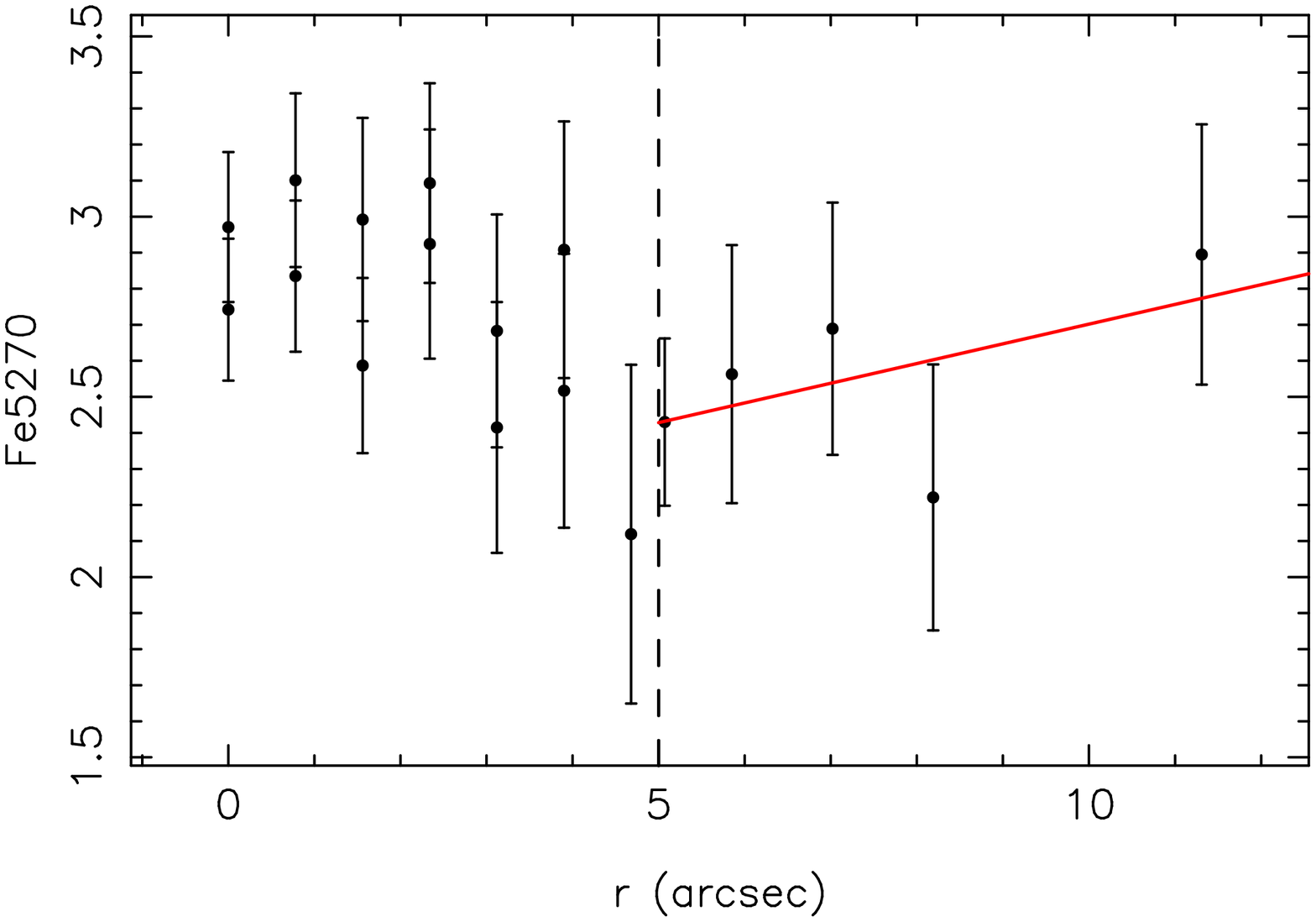}}\hspace{0.8cm}
\resizebox{0.3\textwidth}{!}{\includegraphics[angle=-90]{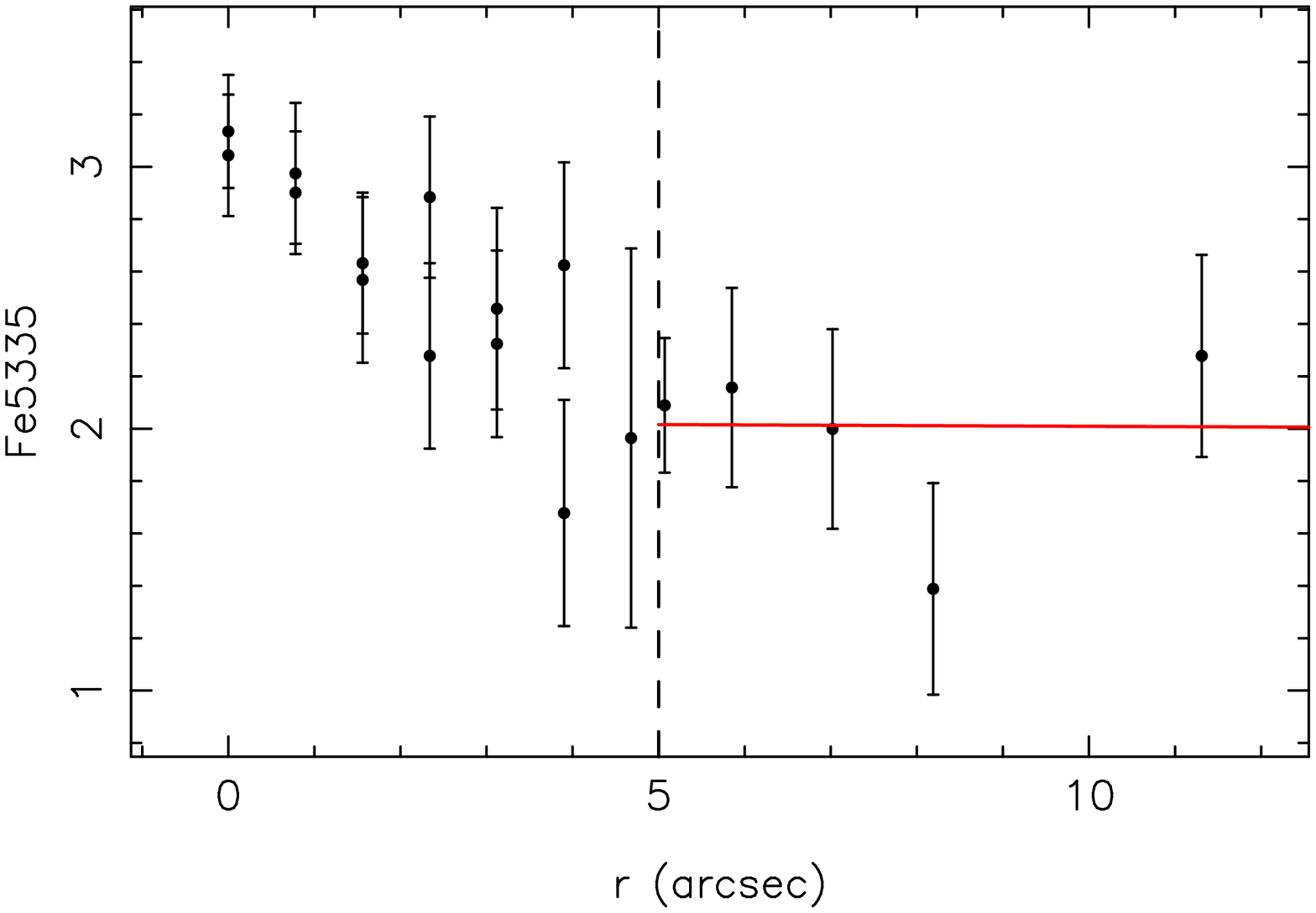}}
\caption{Line-strength distribution in the bar region for all the galaxies}
\end{figure*}

\newpage
\begin{figure*}
\addtocounter{figure}{-1}
\resizebox{0.3\textwidth}{!}{\includegraphics[angle=-90]{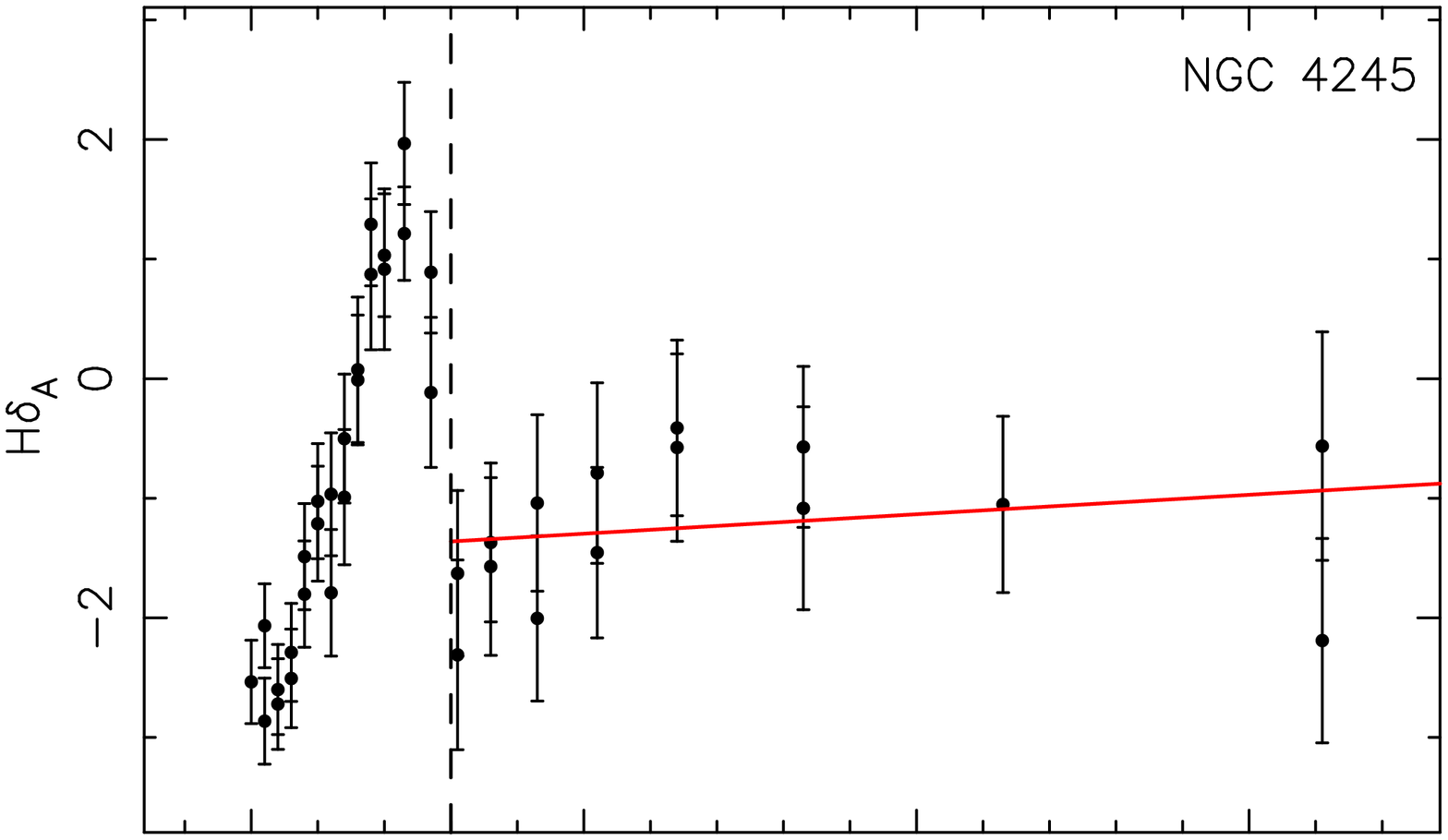}}
\resizebox{0.3\textwidth}{!}{\includegraphics[angle=-90]{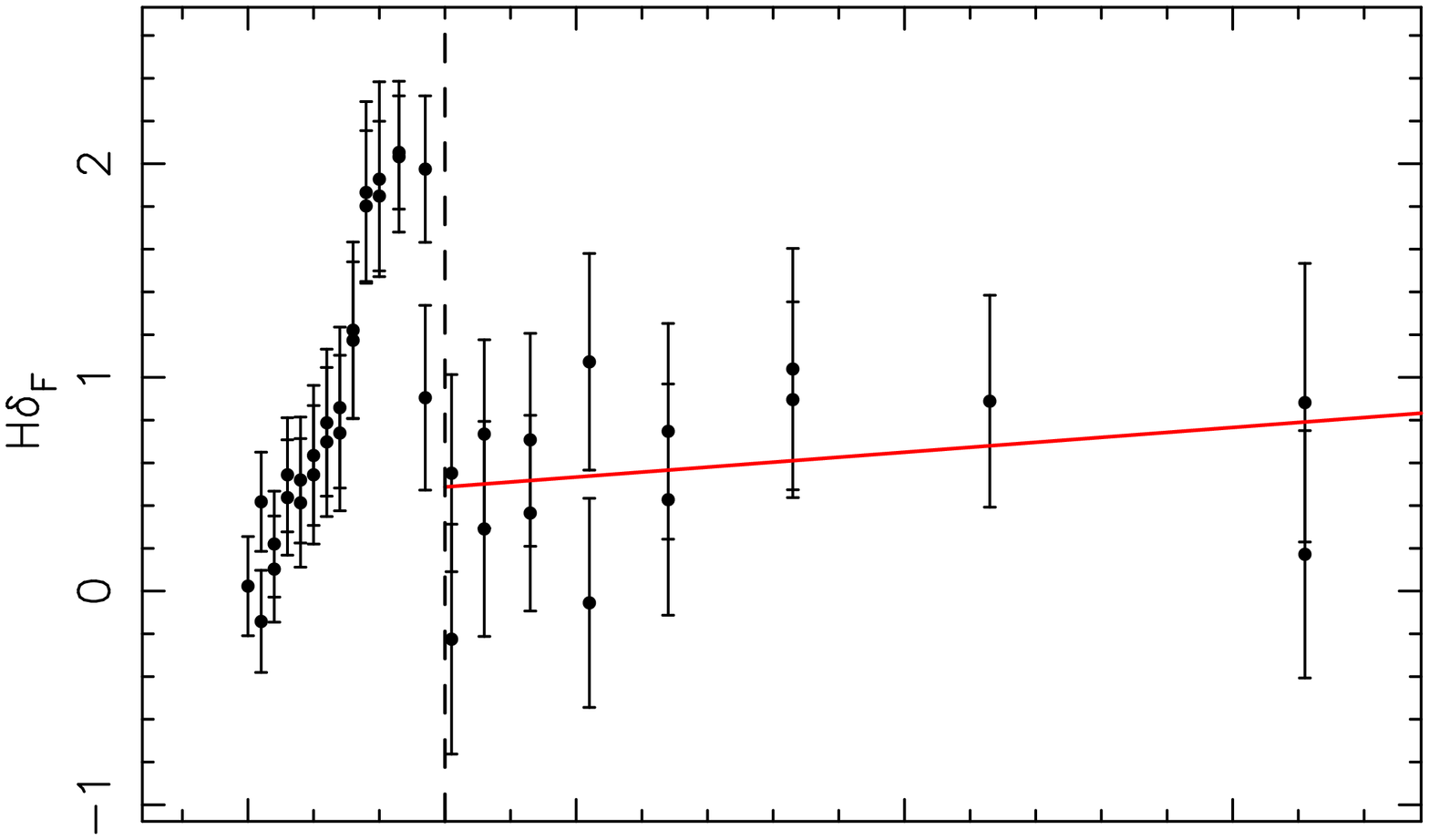}}
\resizebox{0.3\textwidth}{!}{\includegraphics[angle=-90]{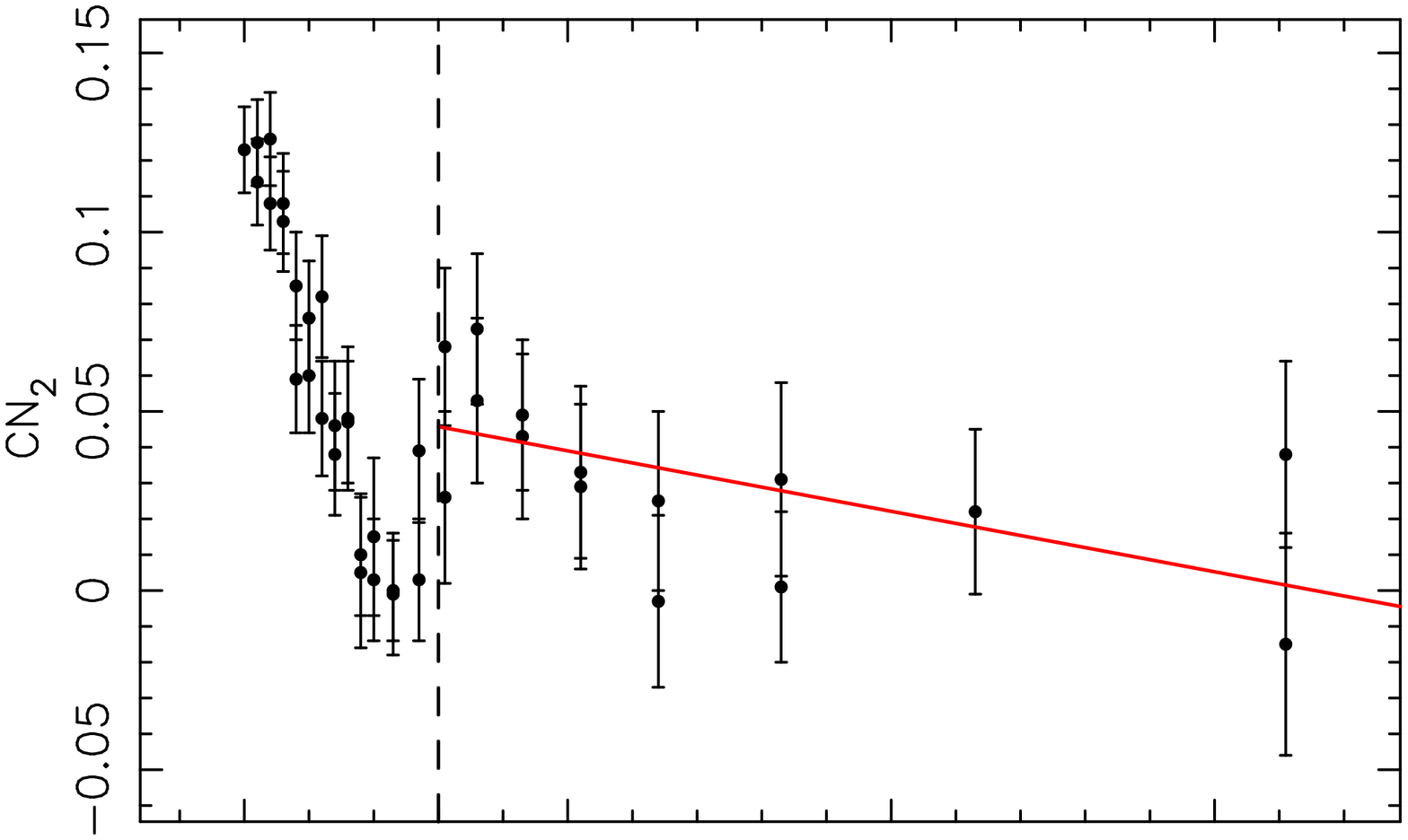}}
\resizebox{0.3\textwidth}{!}{\includegraphics[angle=-90]{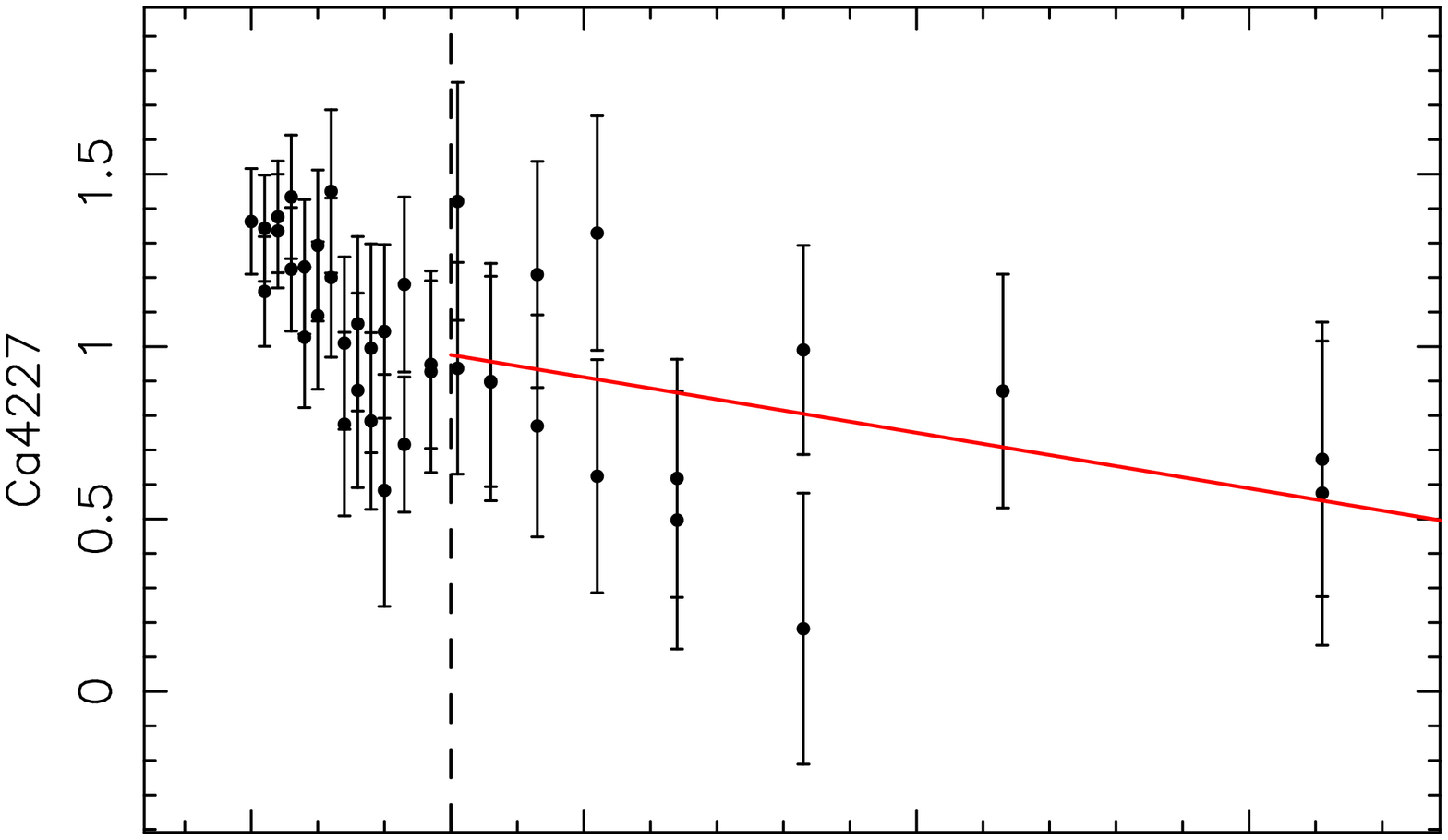}}
\resizebox{0.3\textwidth}{!}{\includegraphics[angle=-90]{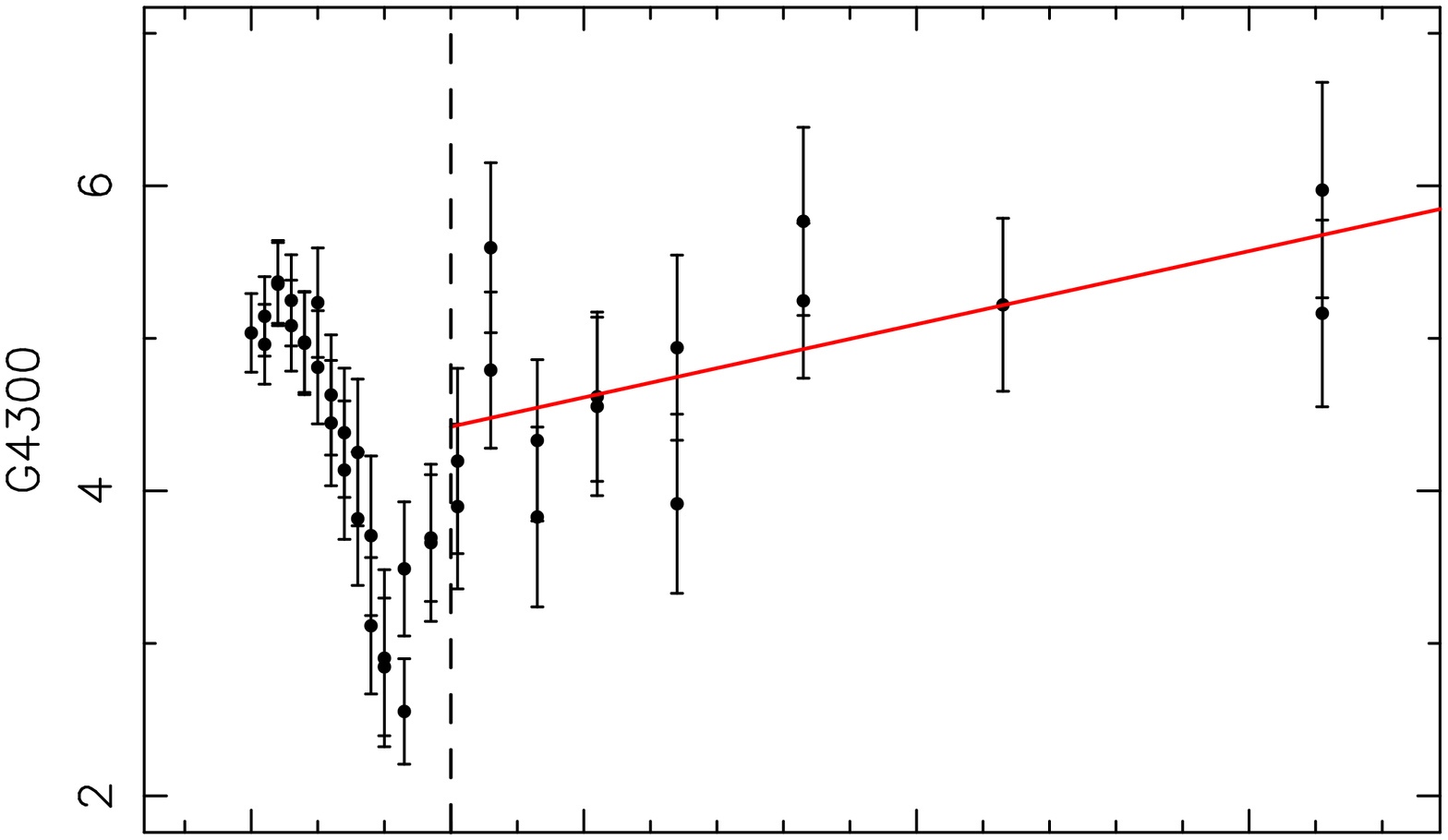}}
\resizebox{0.3\textwidth}{!}{\includegraphics[angle=-90]{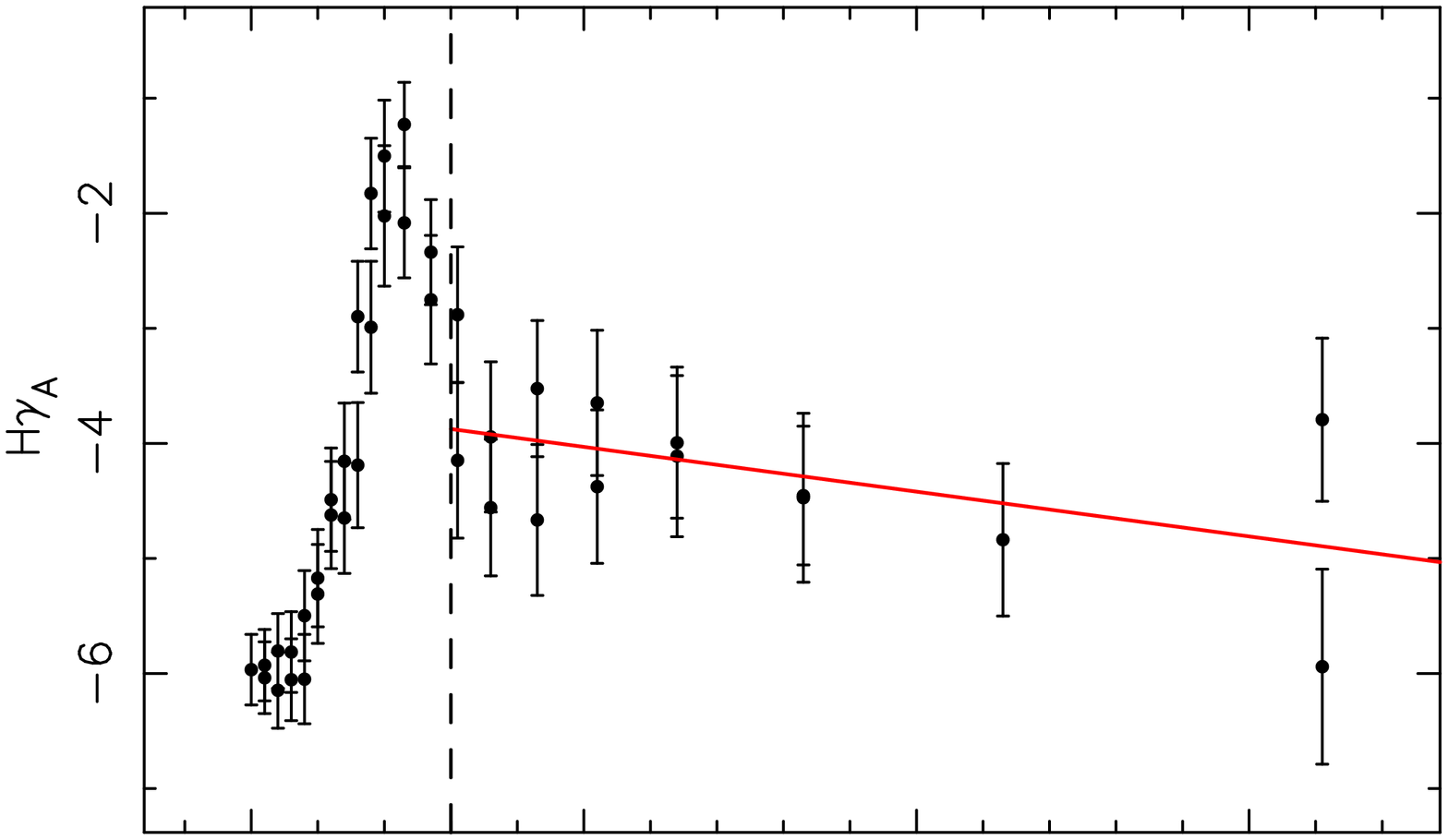}}
\resizebox{0.3\textwidth}{!}{\includegraphics[angle=-90]{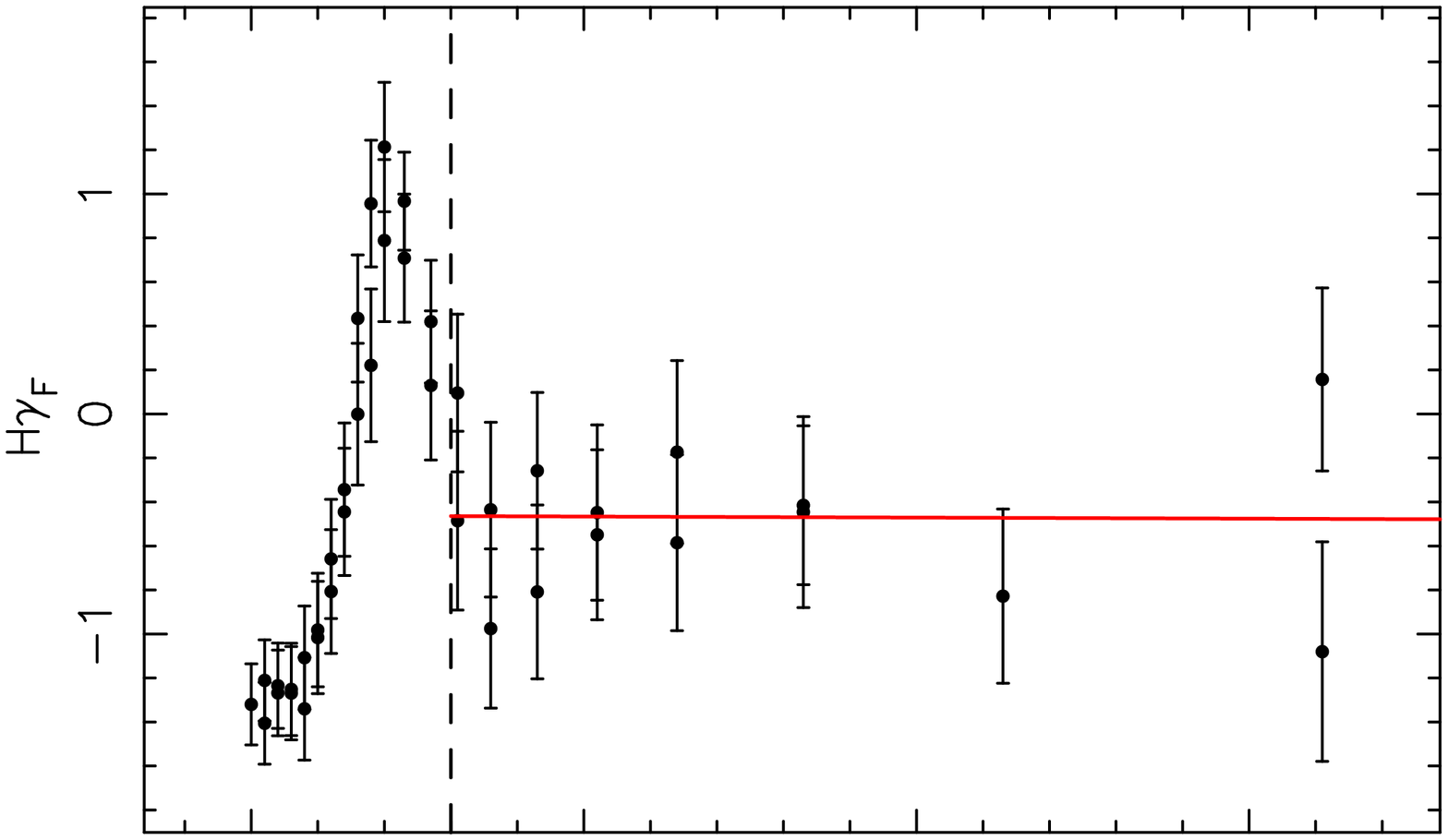}}
\resizebox{0.3\textwidth}{!}{\includegraphics[angle=-90]{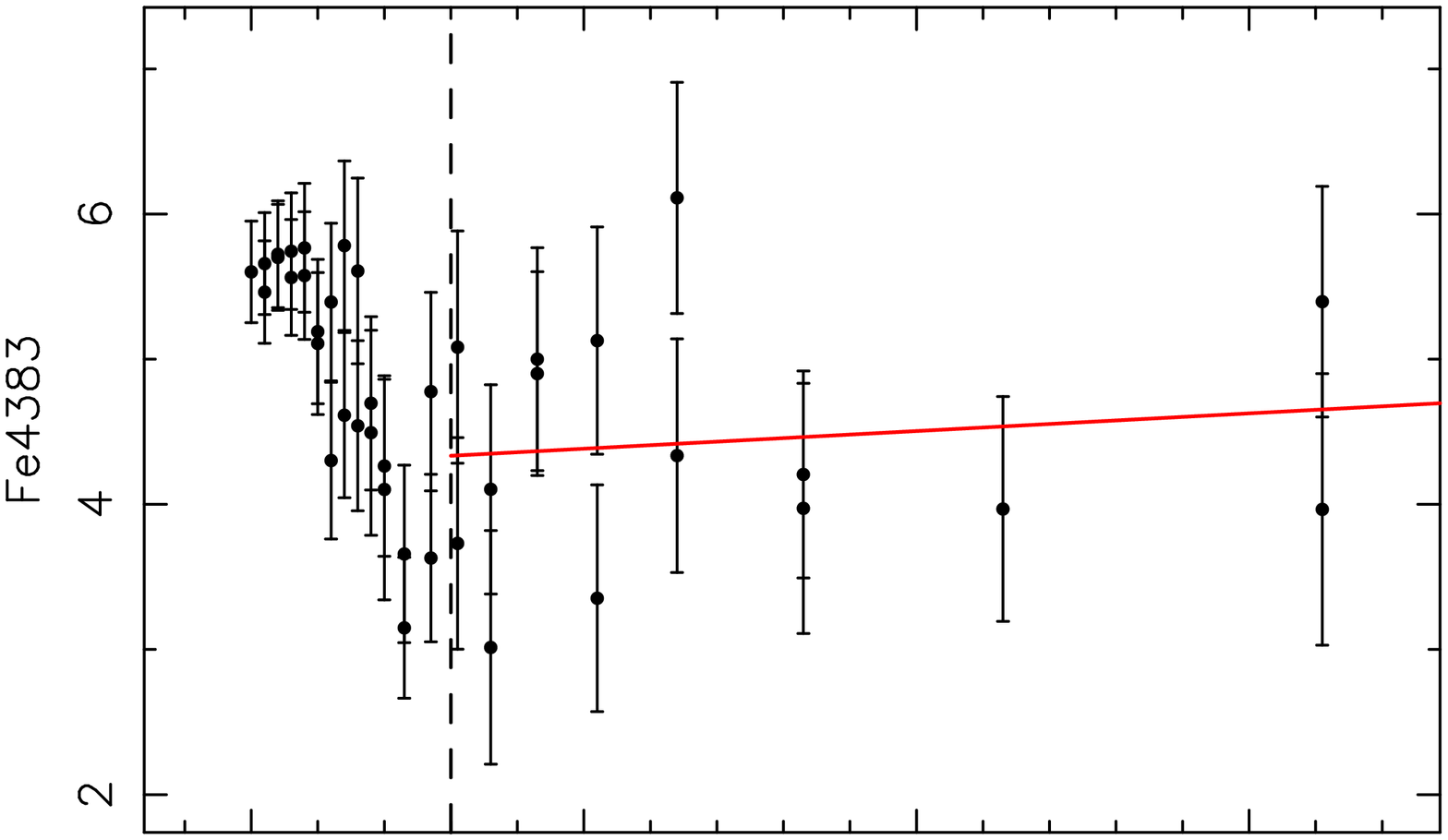}}
\resizebox{0.3\textwidth}{!}{\includegraphics[angle=-90]{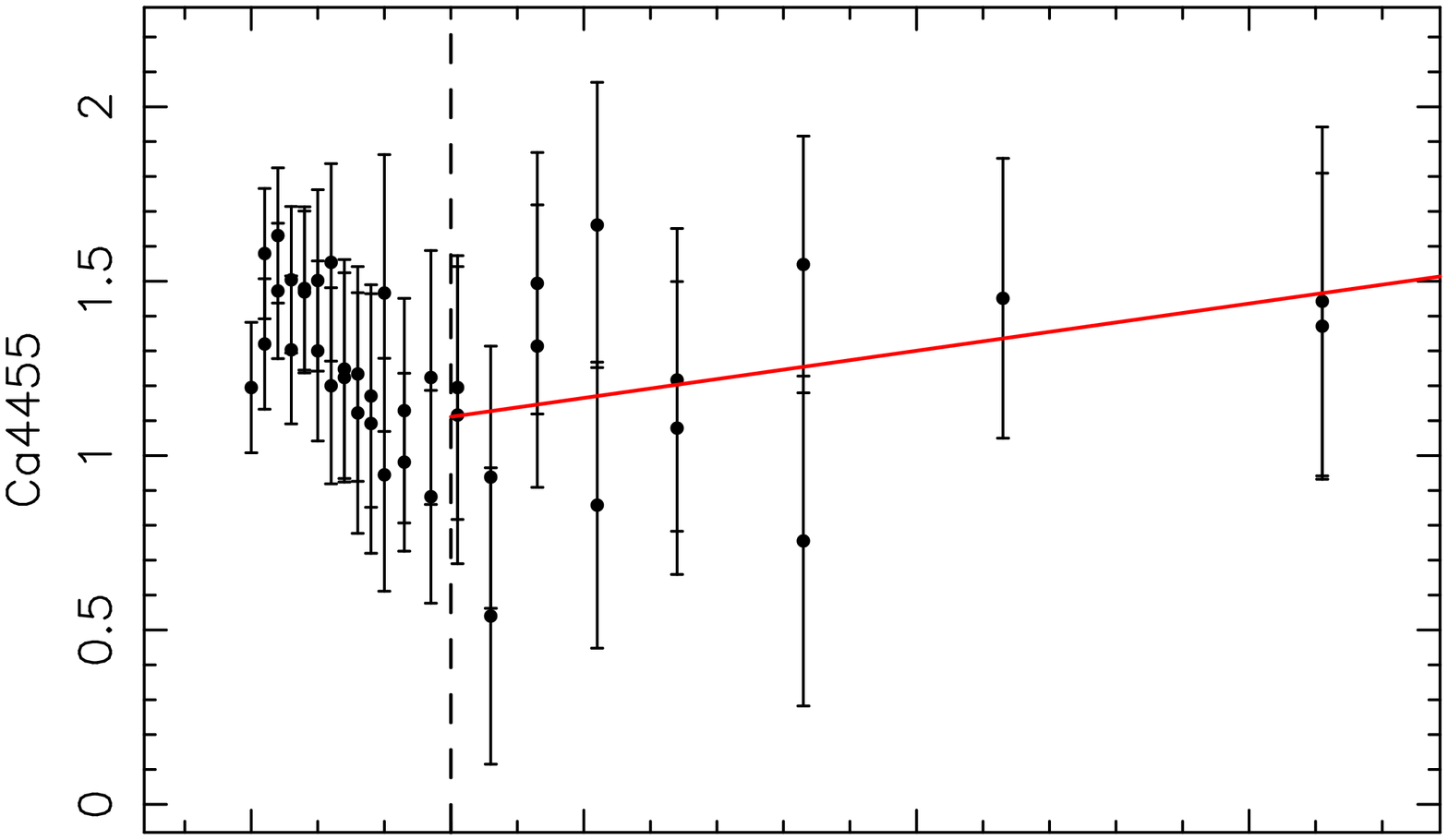}}
\resizebox{0.3\textwidth}{!}{\includegraphics[angle=-90]{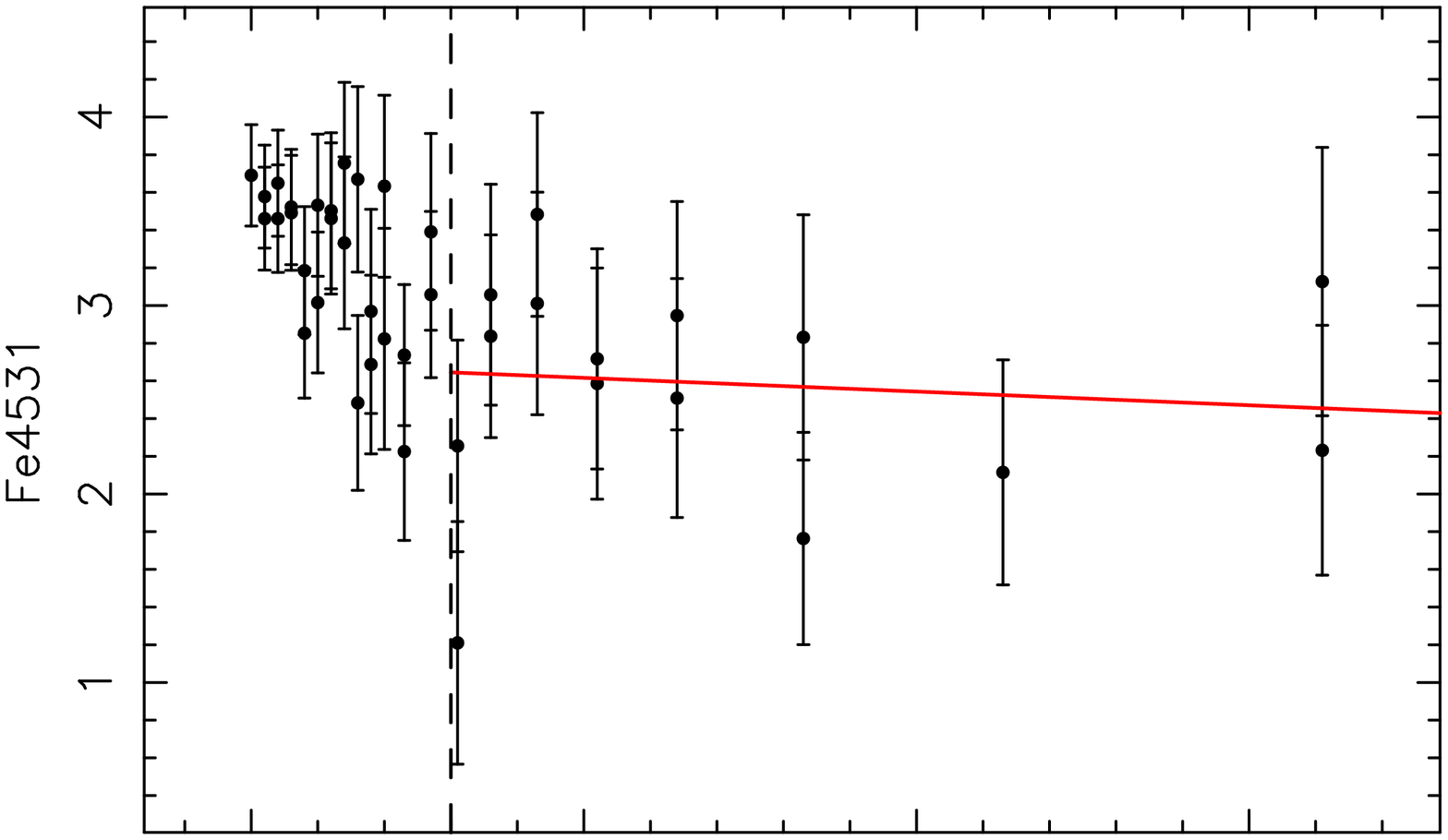}}
\resizebox{0.3\textwidth}{!}{\includegraphics[angle=-90]{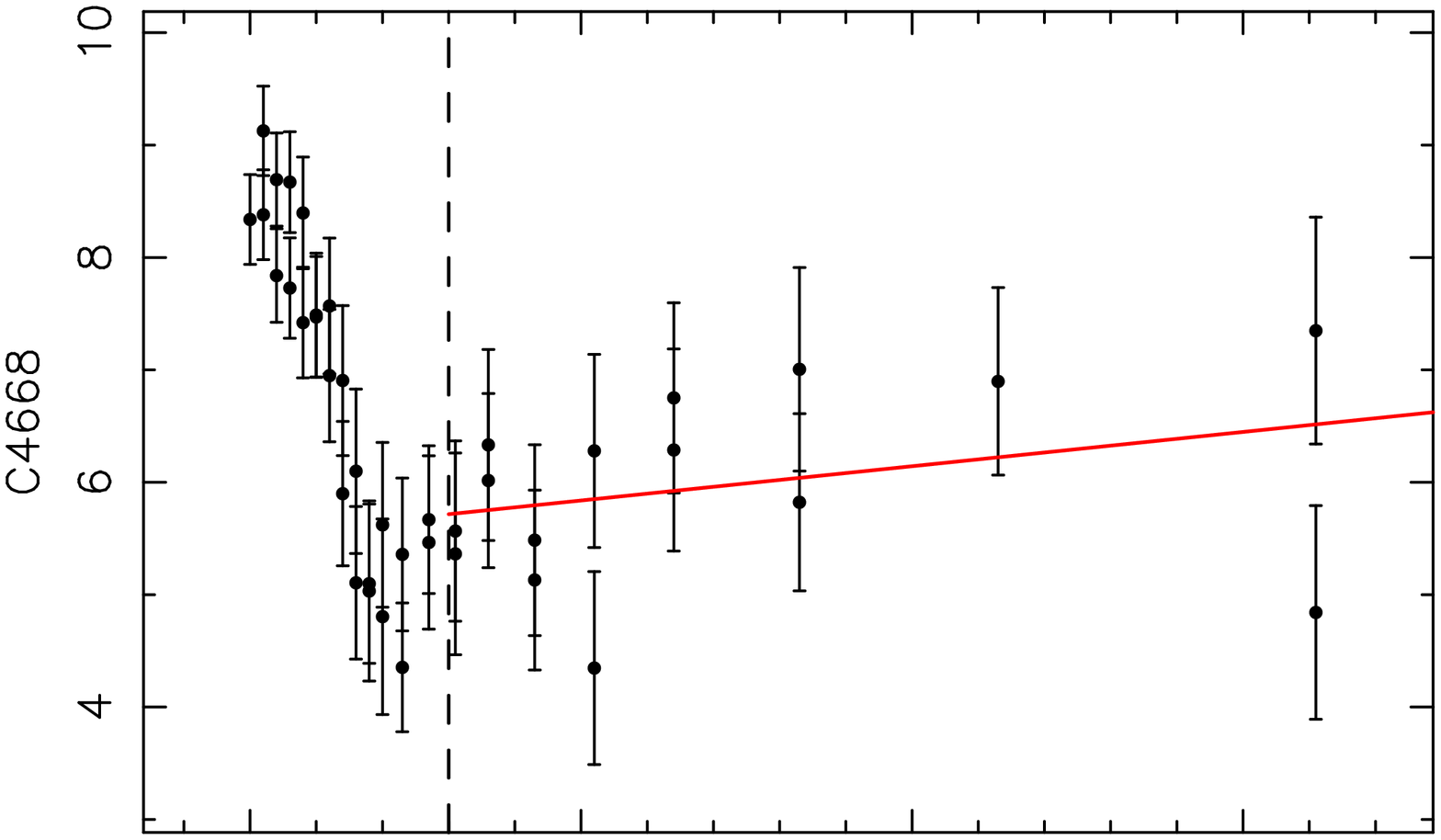}}
\resizebox{0.3\textwidth}{!}{\includegraphics[angle=-90]{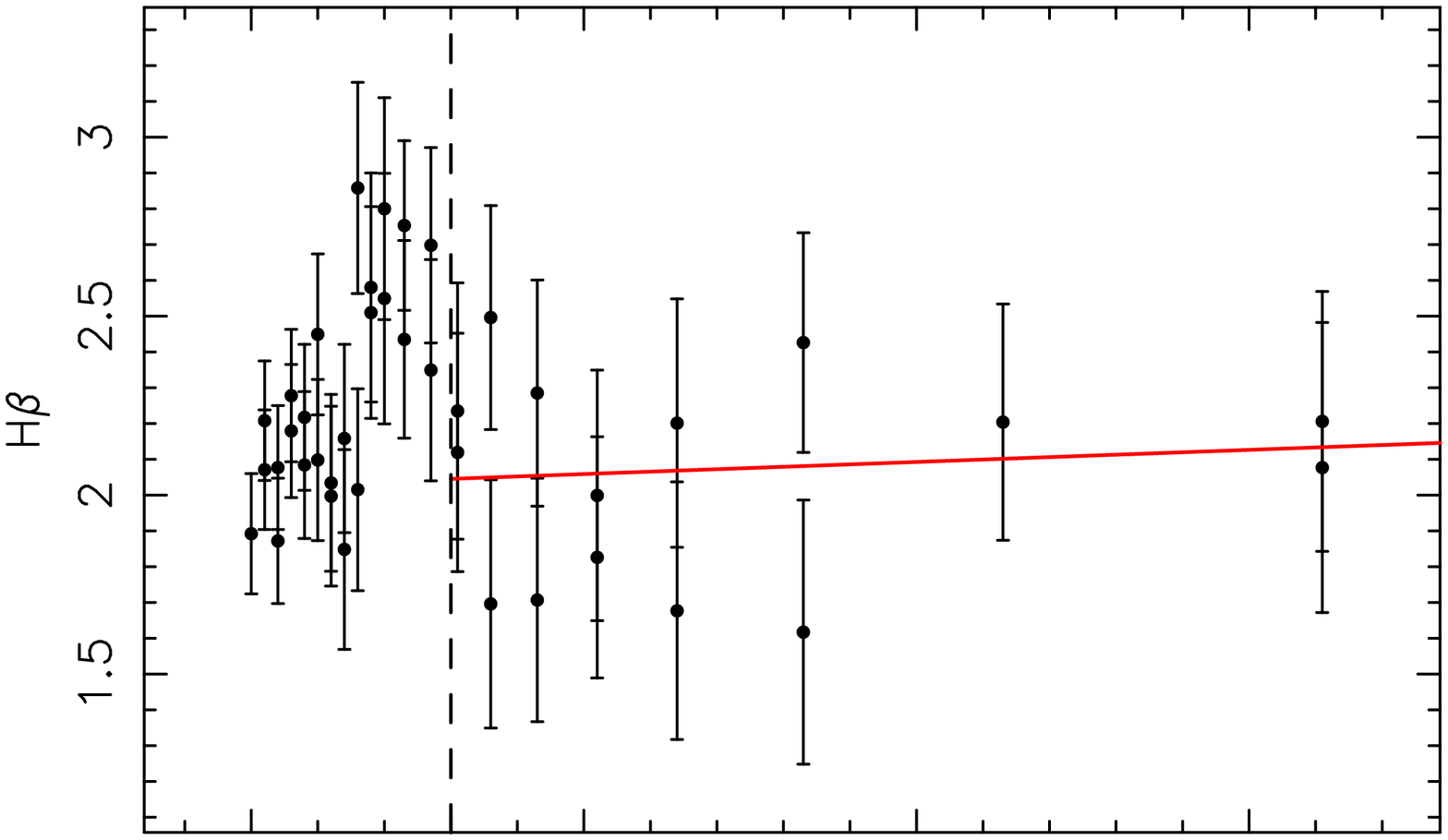}}
\resizebox{0.3\textwidth}{!}{\includegraphics[angle=-90]{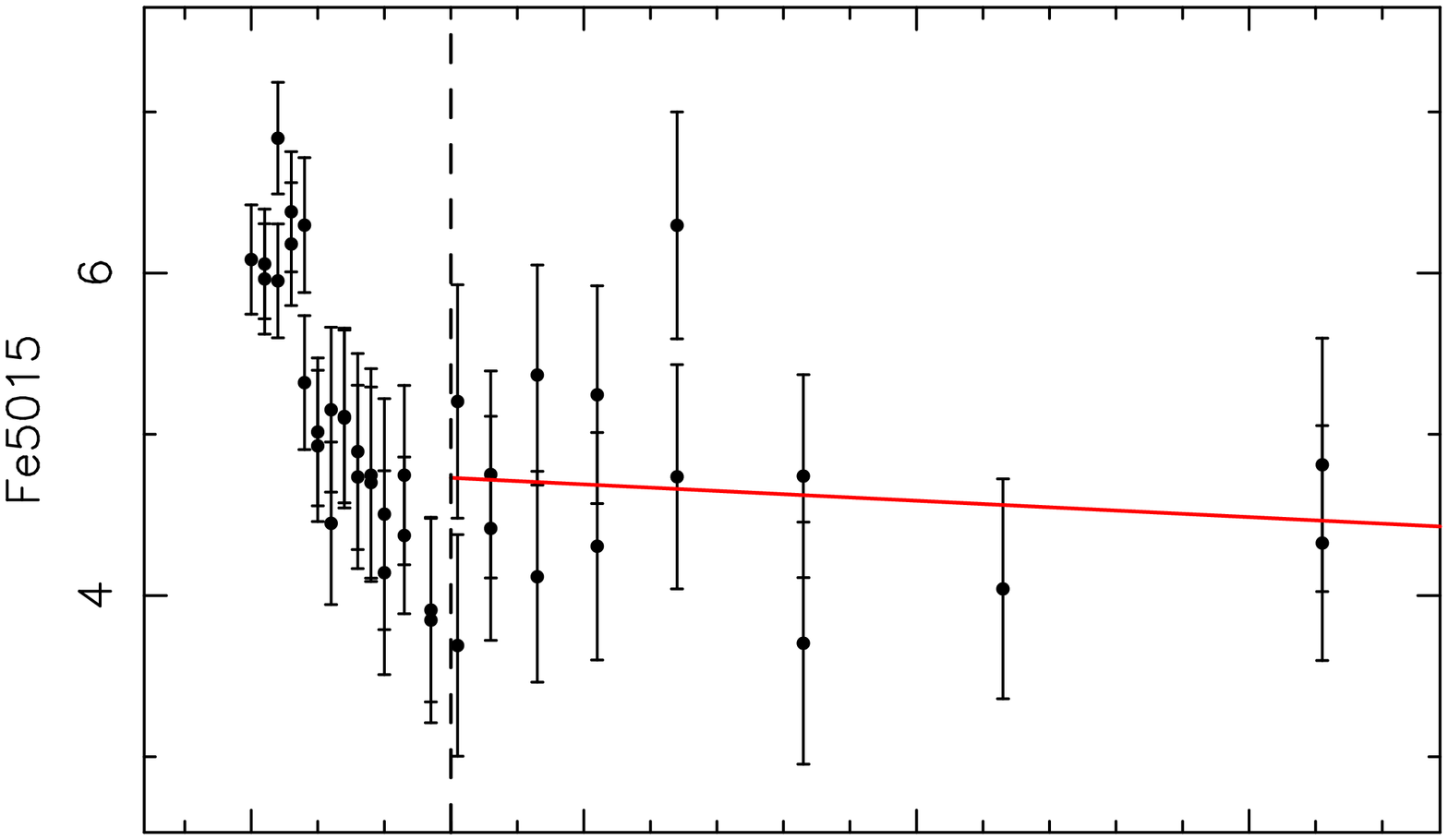}}
\resizebox{0.3\textwidth}{!}{\includegraphics[angle=-90]{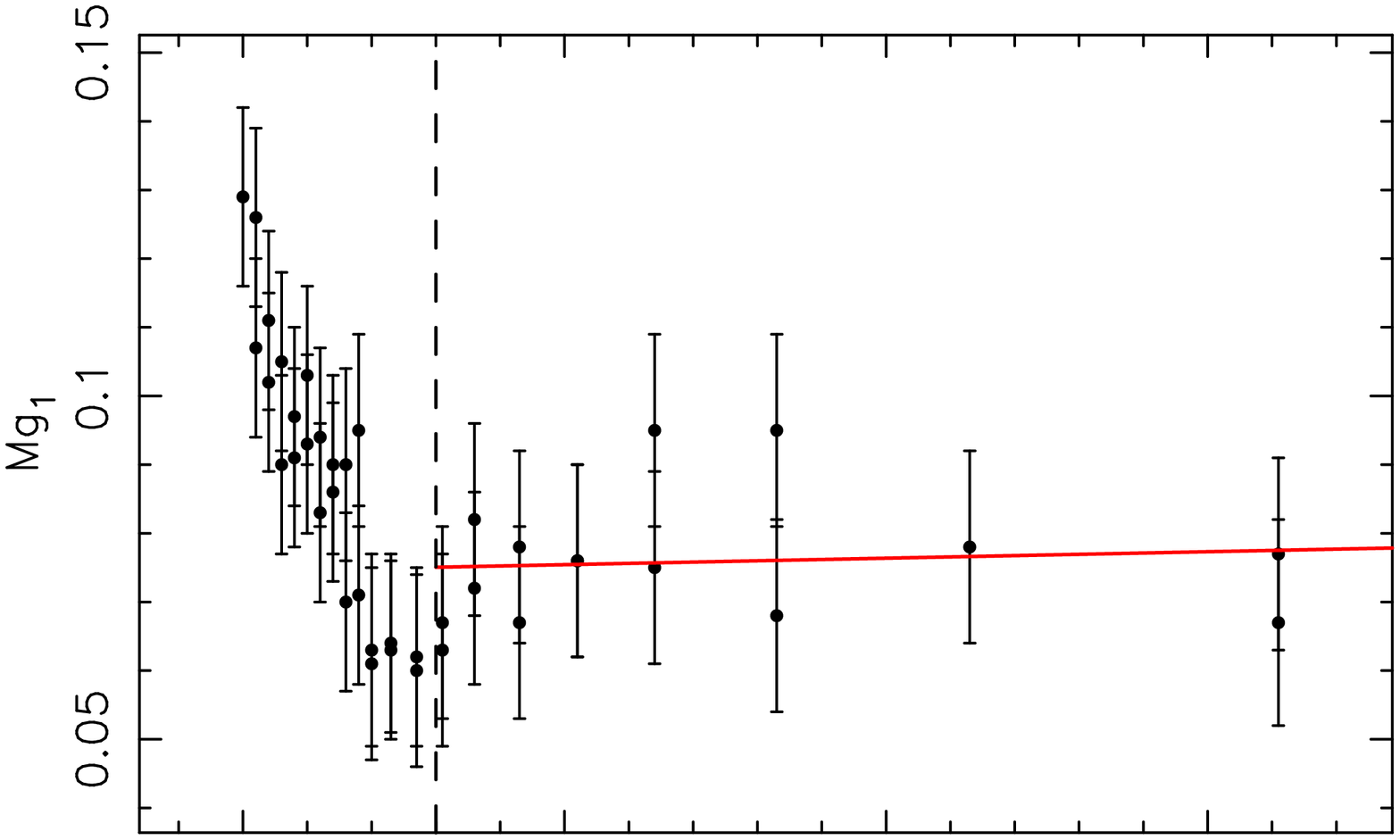}}
\resizebox{0.3\textwidth}{!}{\includegraphics[angle=-90]{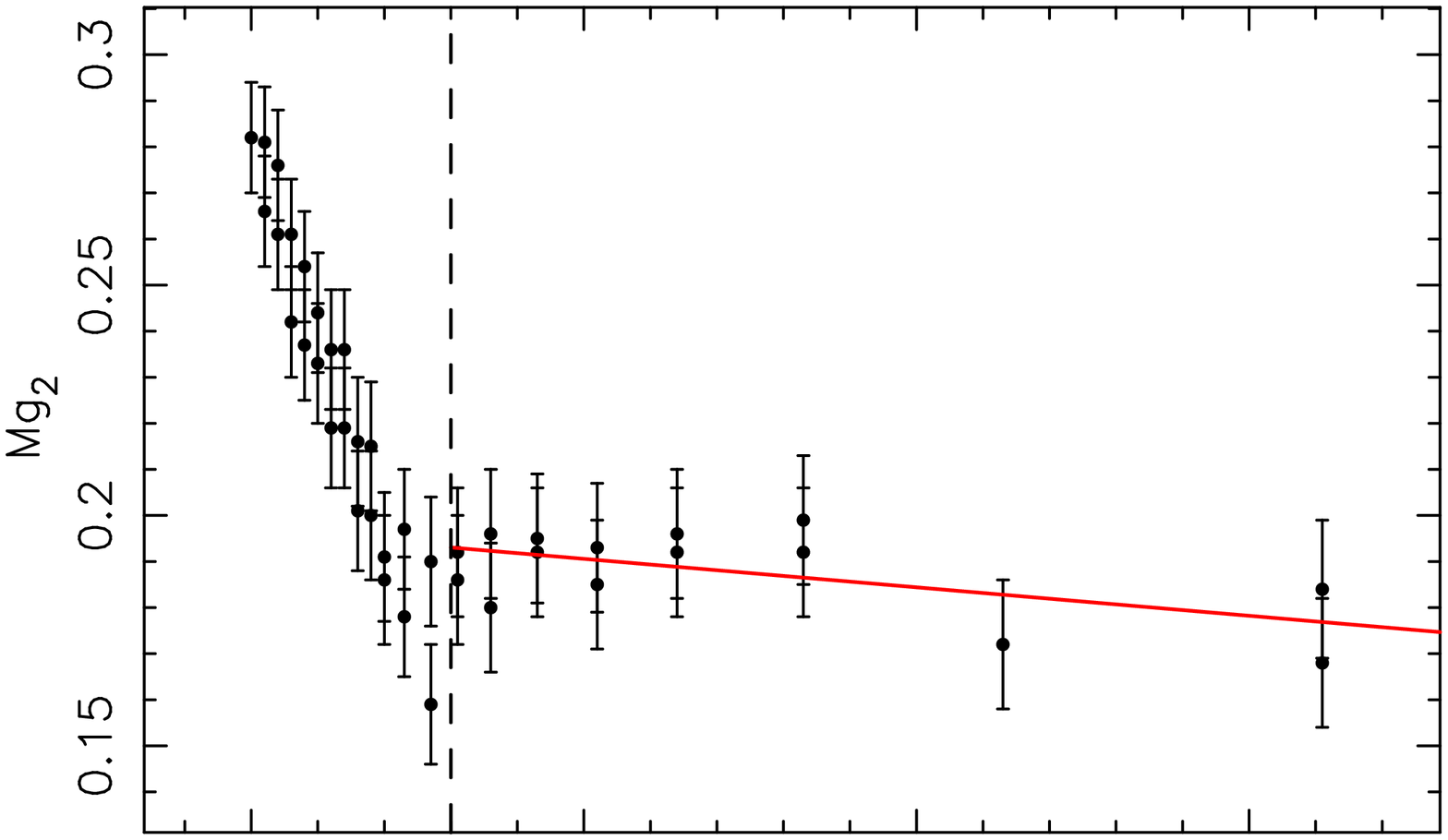}}
\resizebox{0.3\textwidth}{!}{\includegraphics[angle=-90]{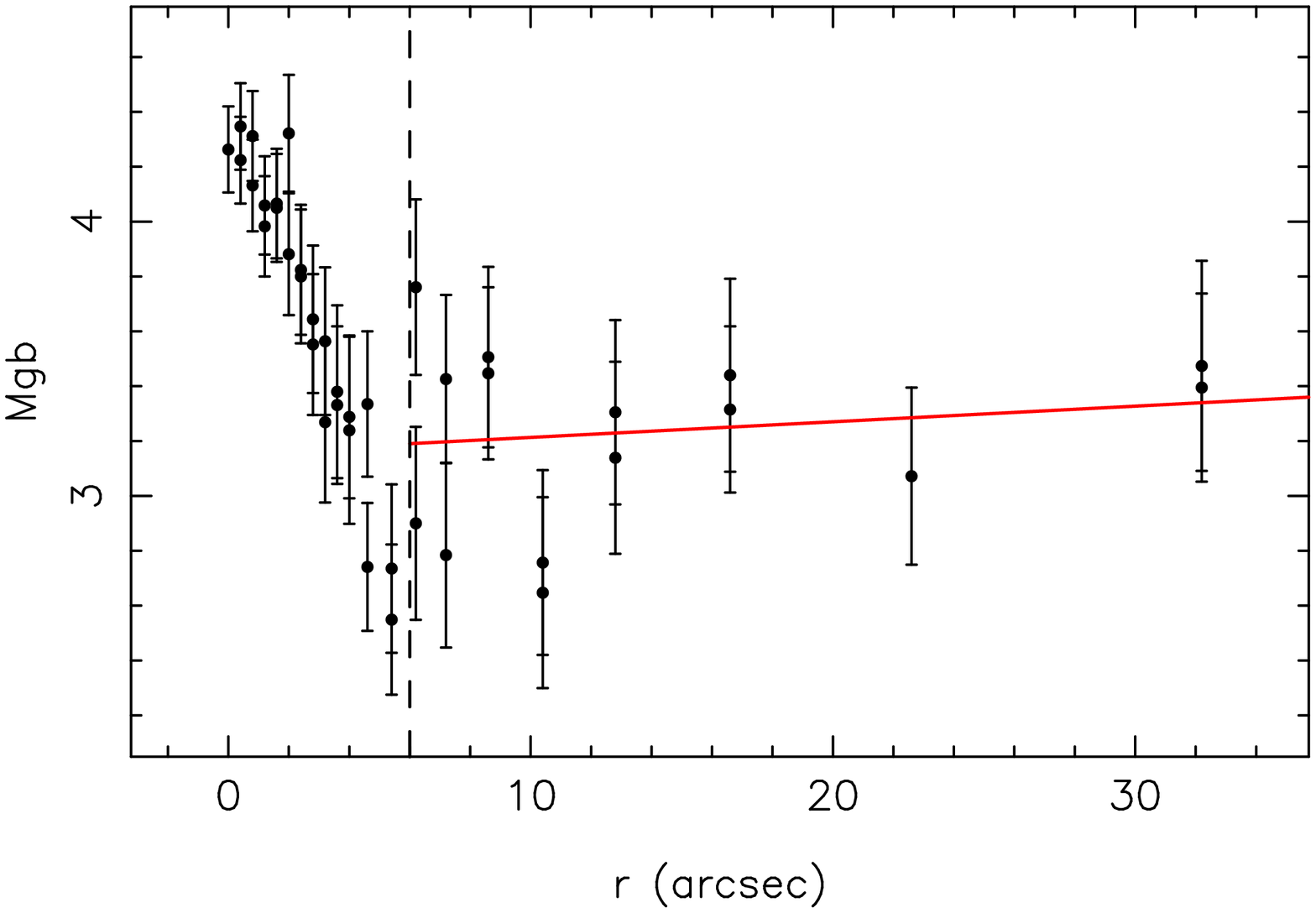}}\hspace{0.85cm}
\resizebox{0.3\textwidth}{!}{\includegraphics[angle=-90]{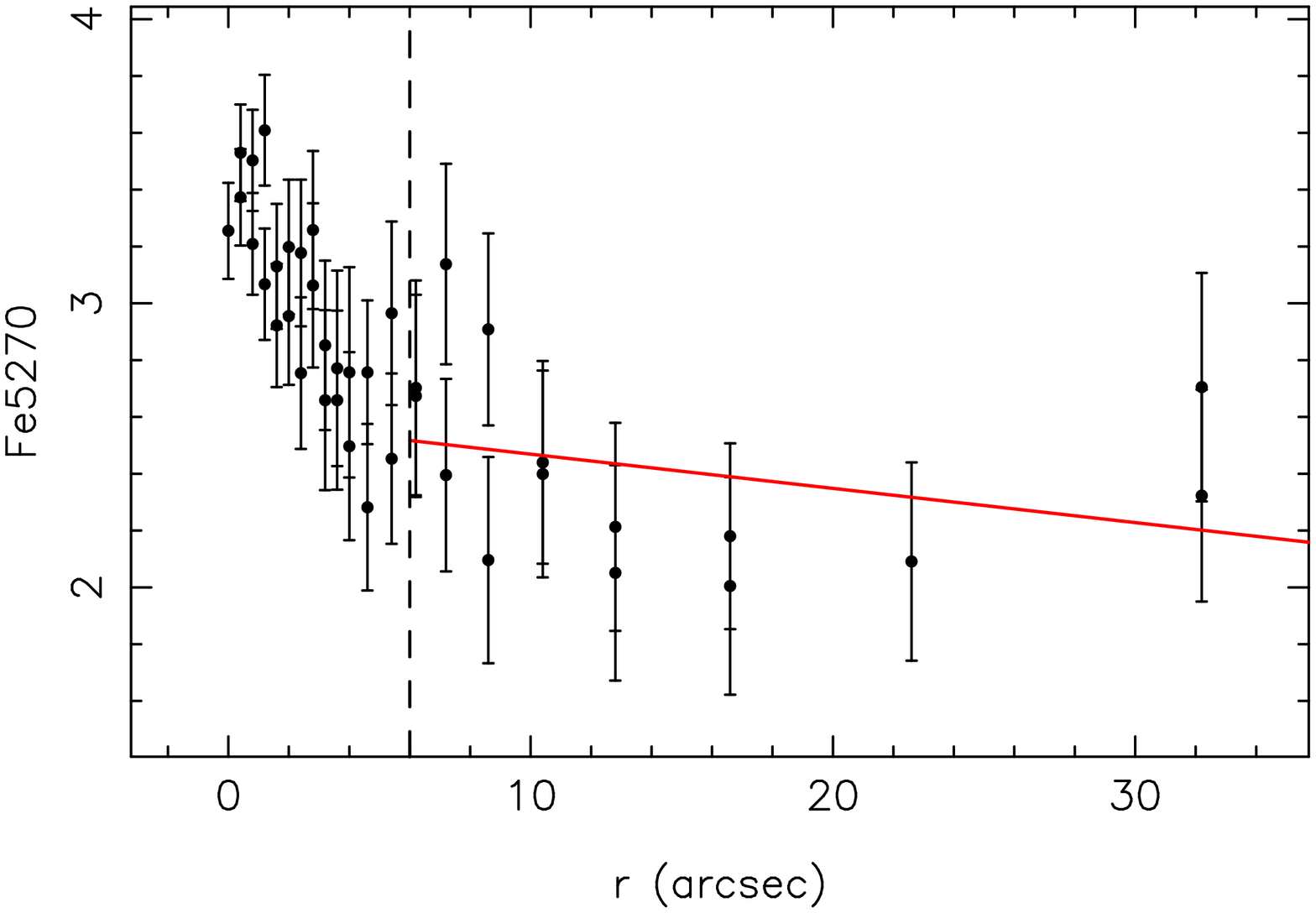}}\hspace{0.85cm}
\resizebox{0.3\textwidth}{!}{\includegraphics[angle=-90]{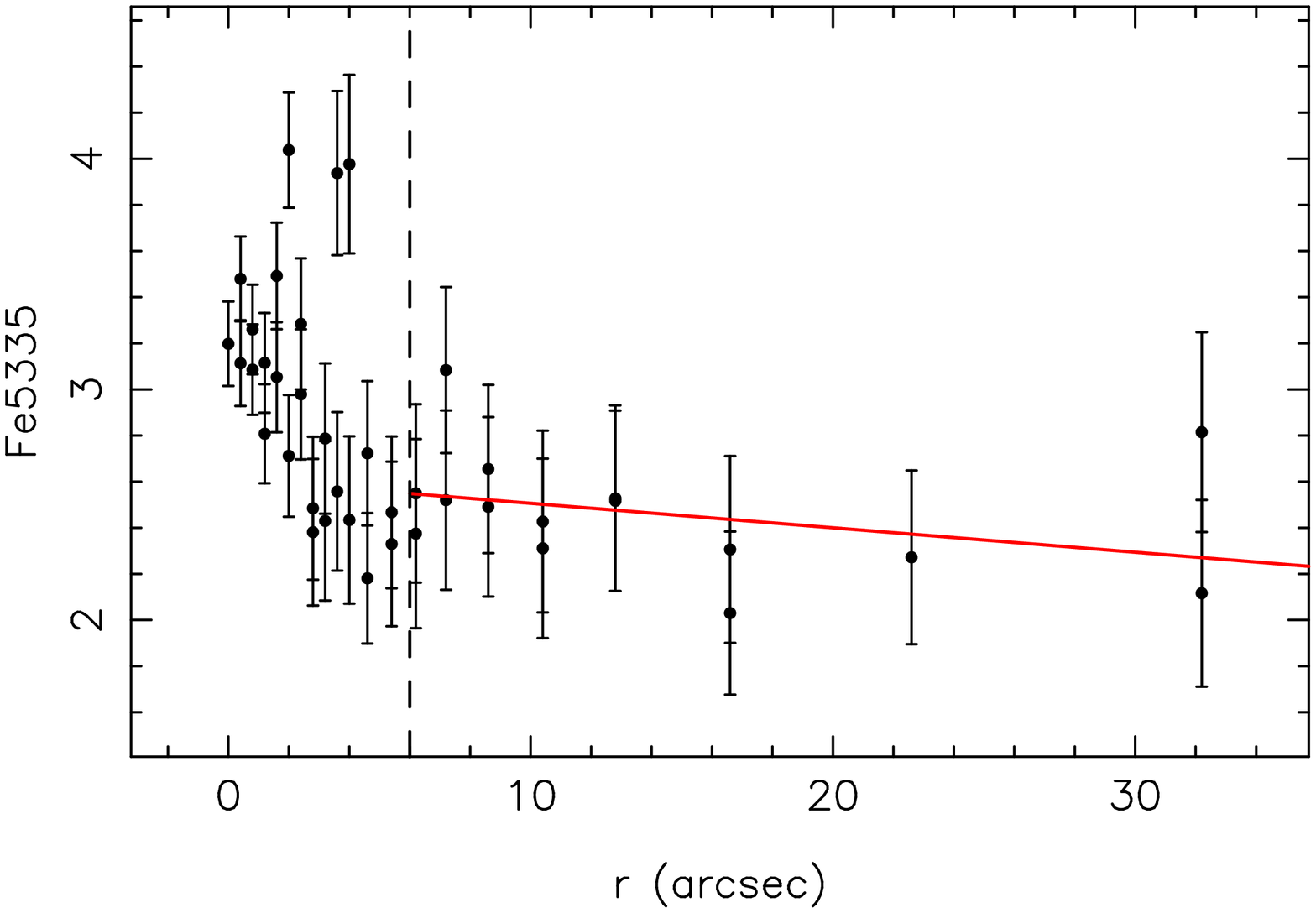}}
\caption{Line-strength distribution in the bar region for all the galaxies}
\end{figure*}

\begin{figure*}
\addtocounter{figure}{-1}
\resizebox{0.3\textwidth}{!}{\includegraphics[angle=-90]{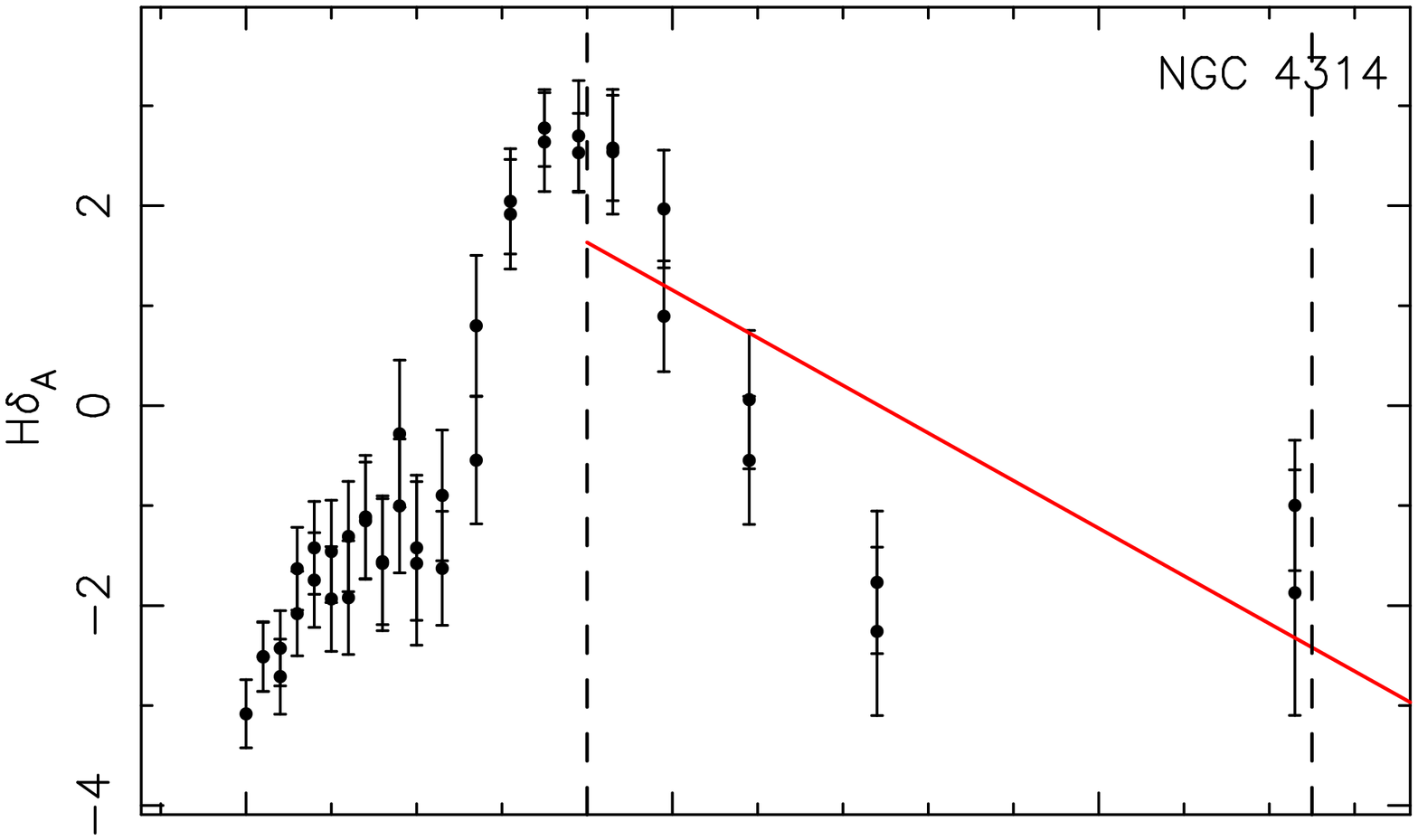}}
\resizebox{0.3\textwidth}{!}{\includegraphics[angle=-90]{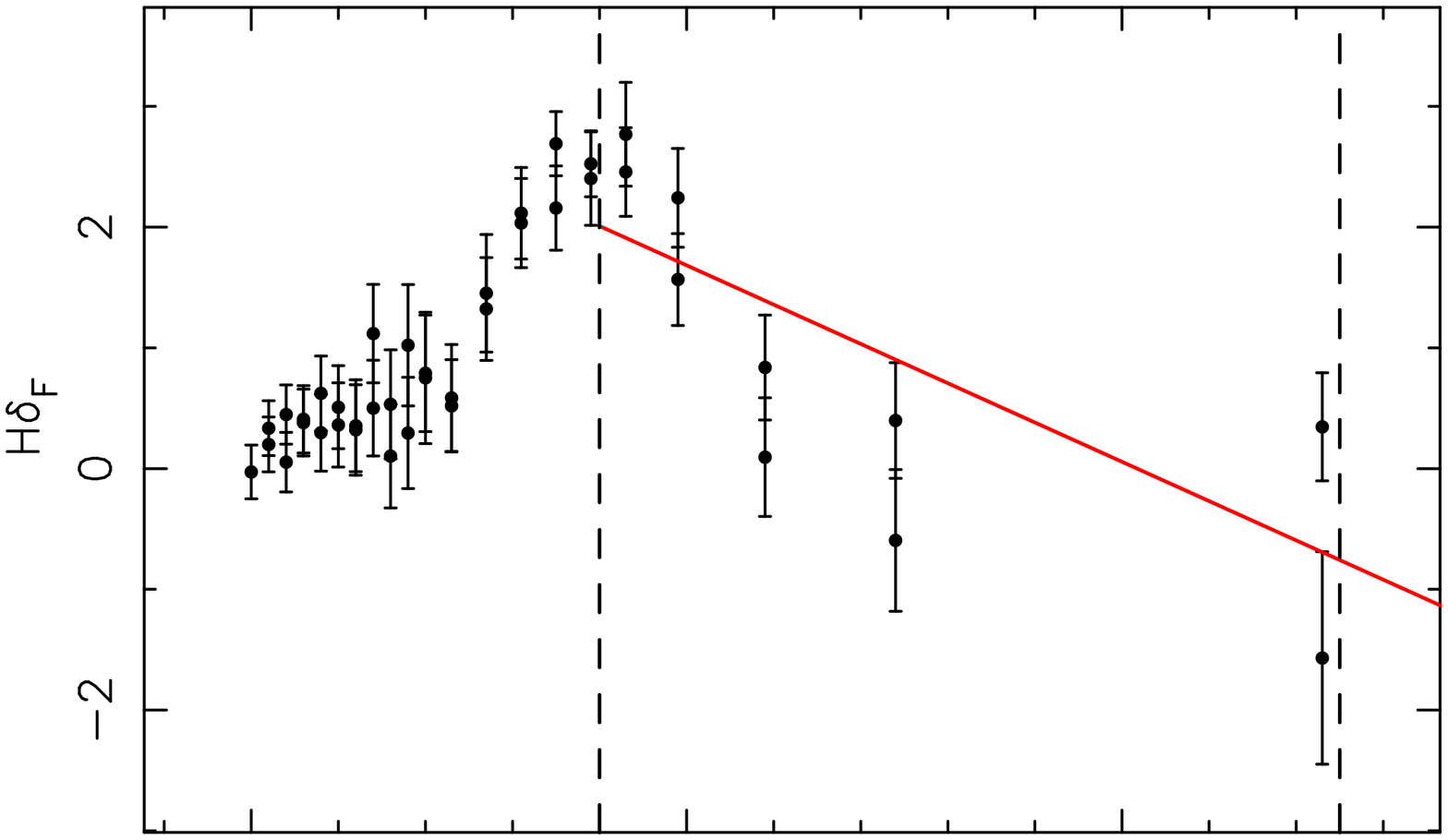}}
\resizebox{0.3\textwidth}{!}{\includegraphics[angle=-90]{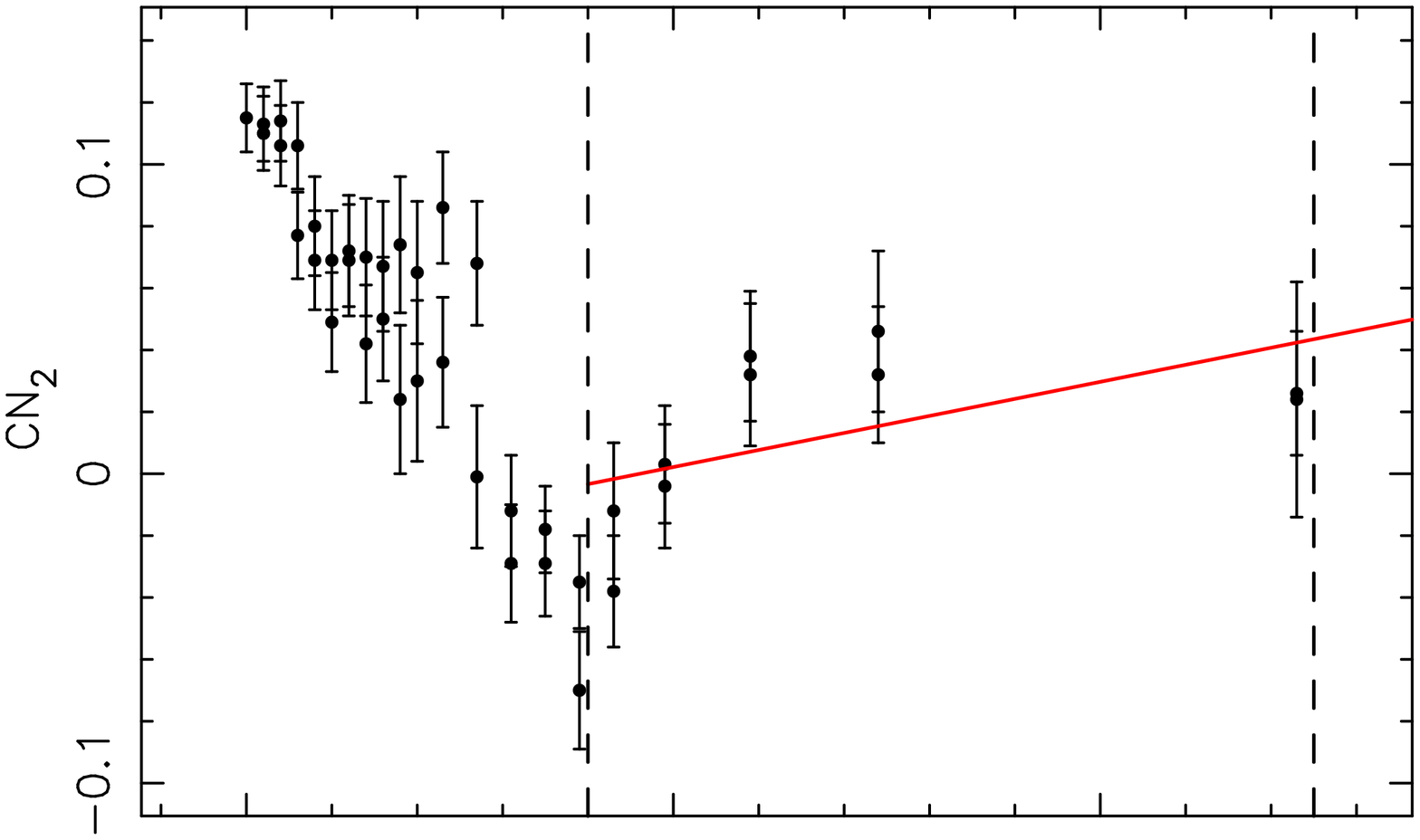}}
\resizebox{0.3\textwidth}{!}{\includegraphics[angle=-90]{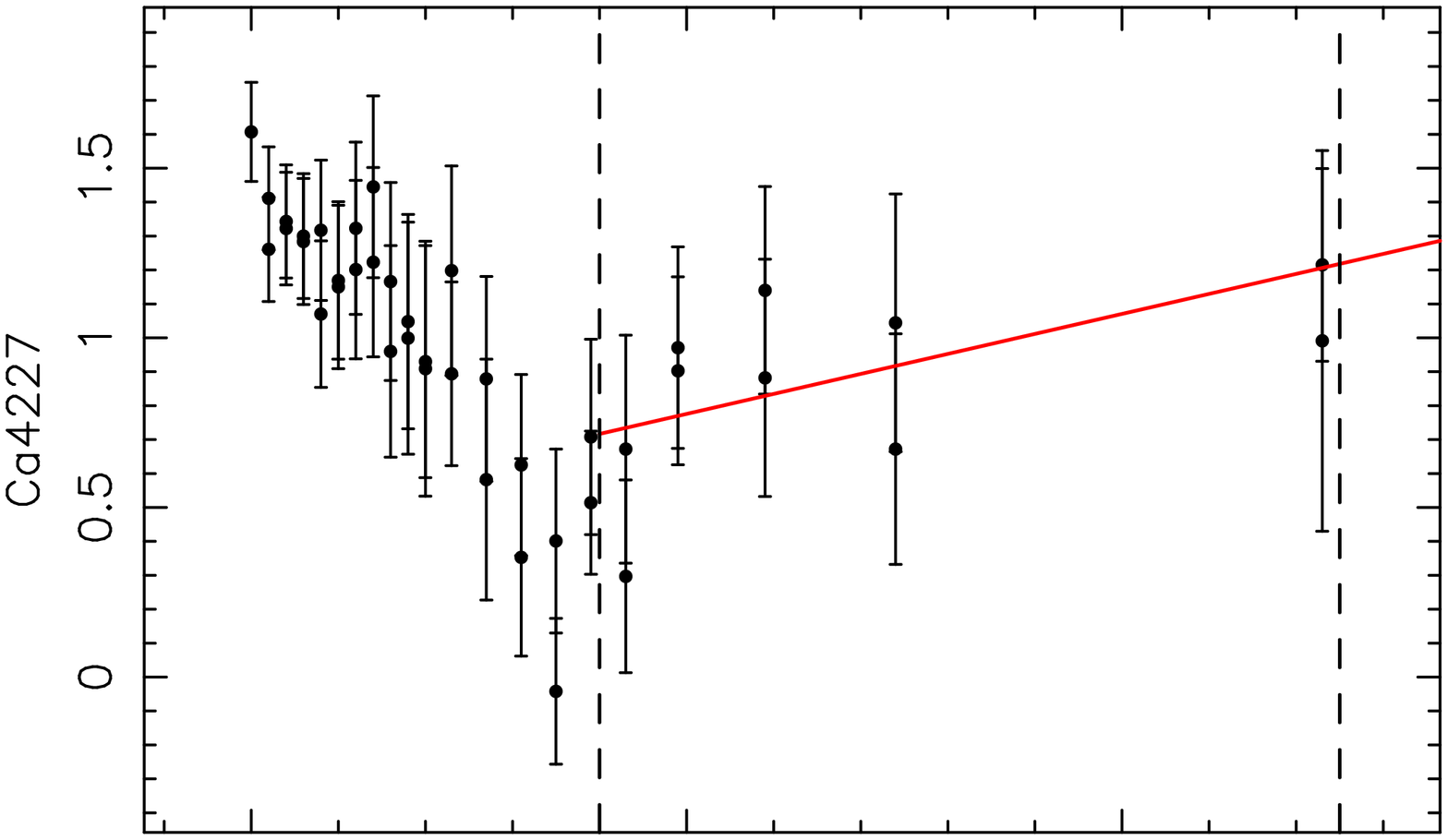}}
\resizebox{0.3\textwidth}{!}{\includegraphics[angle=-90]{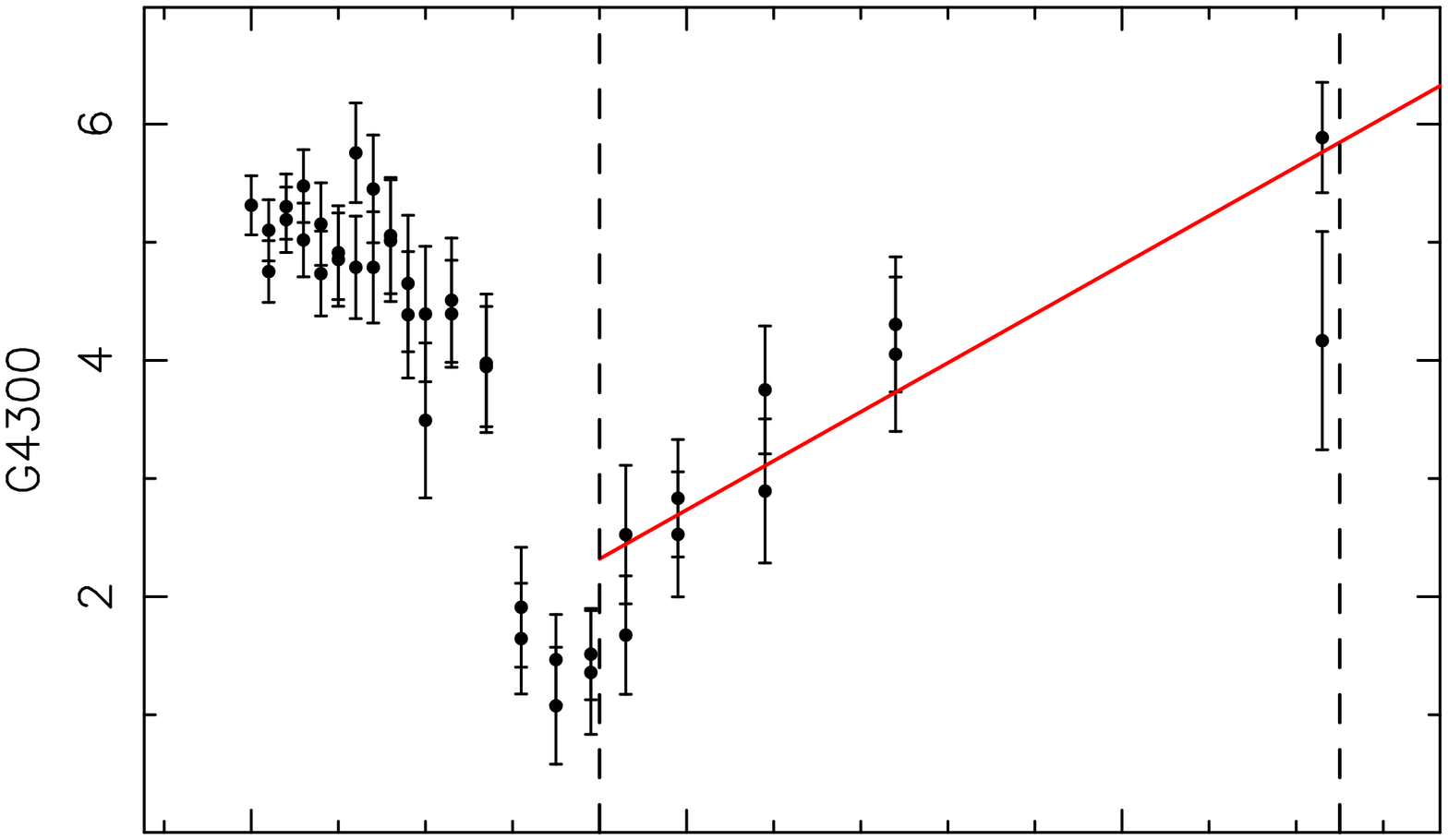}}
\resizebox{0.3\textwidth}{!}{\includegraphics[angle=-90]{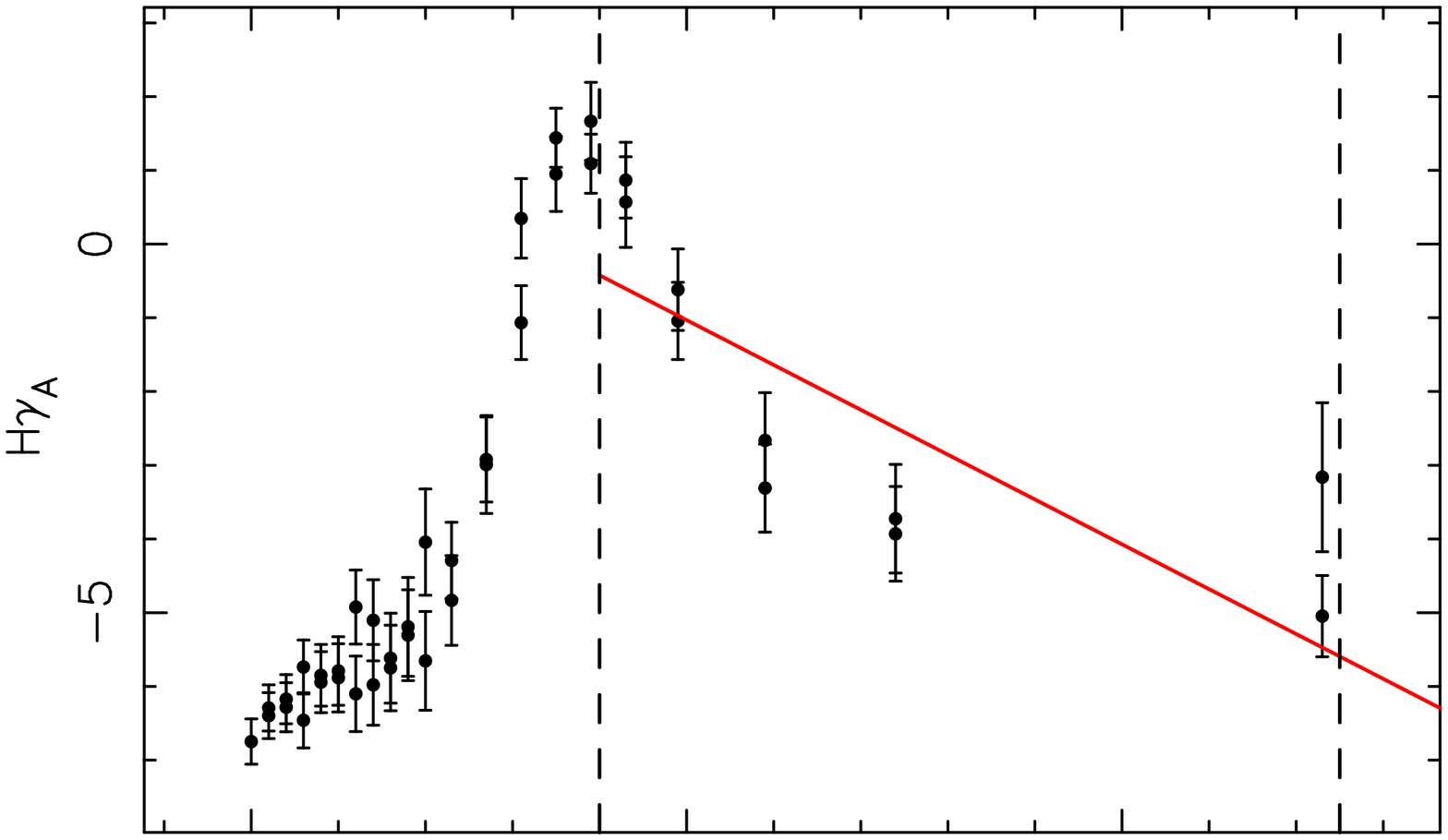}}
\resizebox{0.3\textwidth}{!}{\includegraphics[angle=-90]{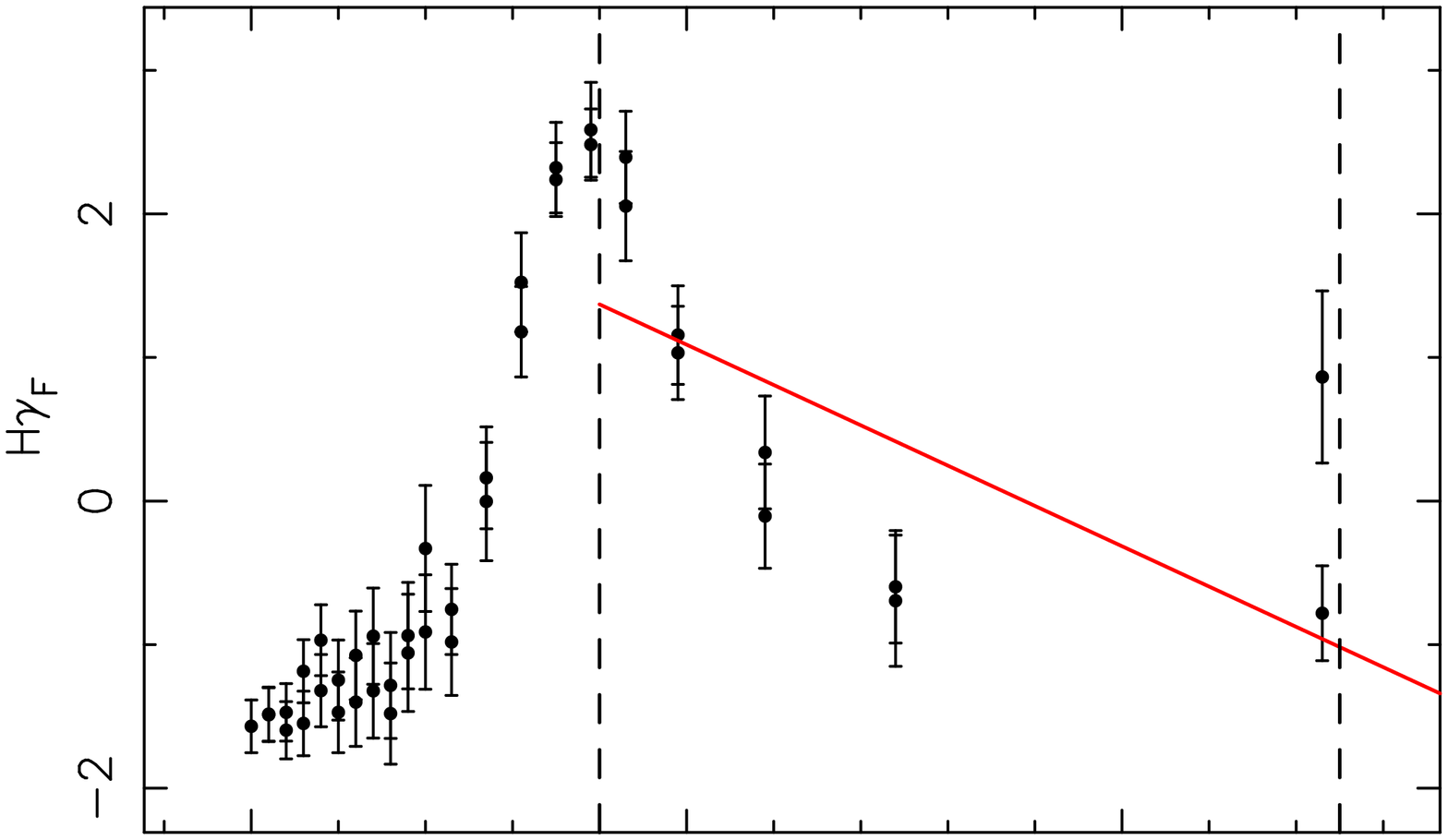}}
\resizebox{0.3\textwidth}{!}{\includegraphics[angle=-90]{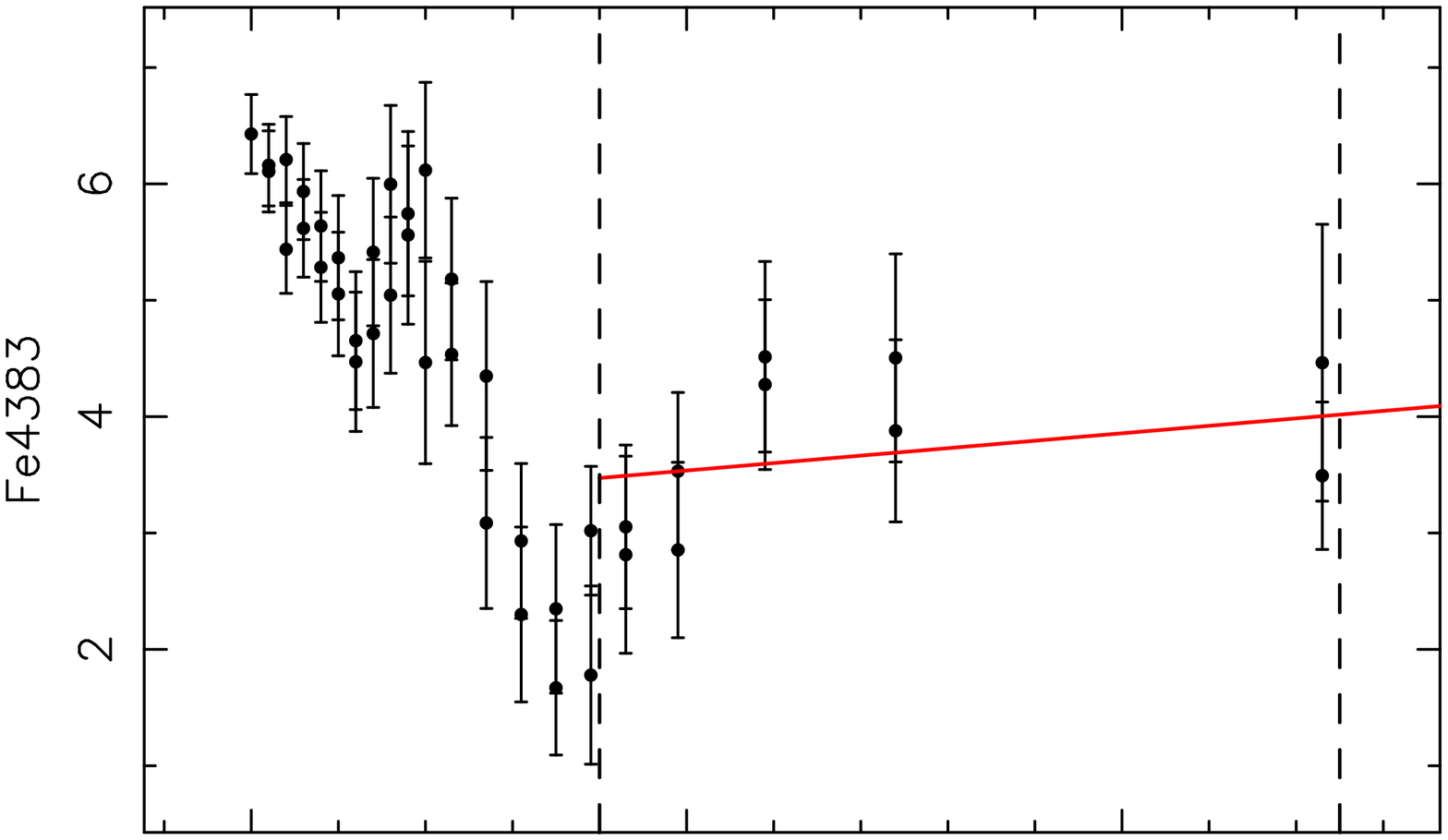}}
\resizebox{0.3\textwidth}{!}{\includegraphics[angle=-90]{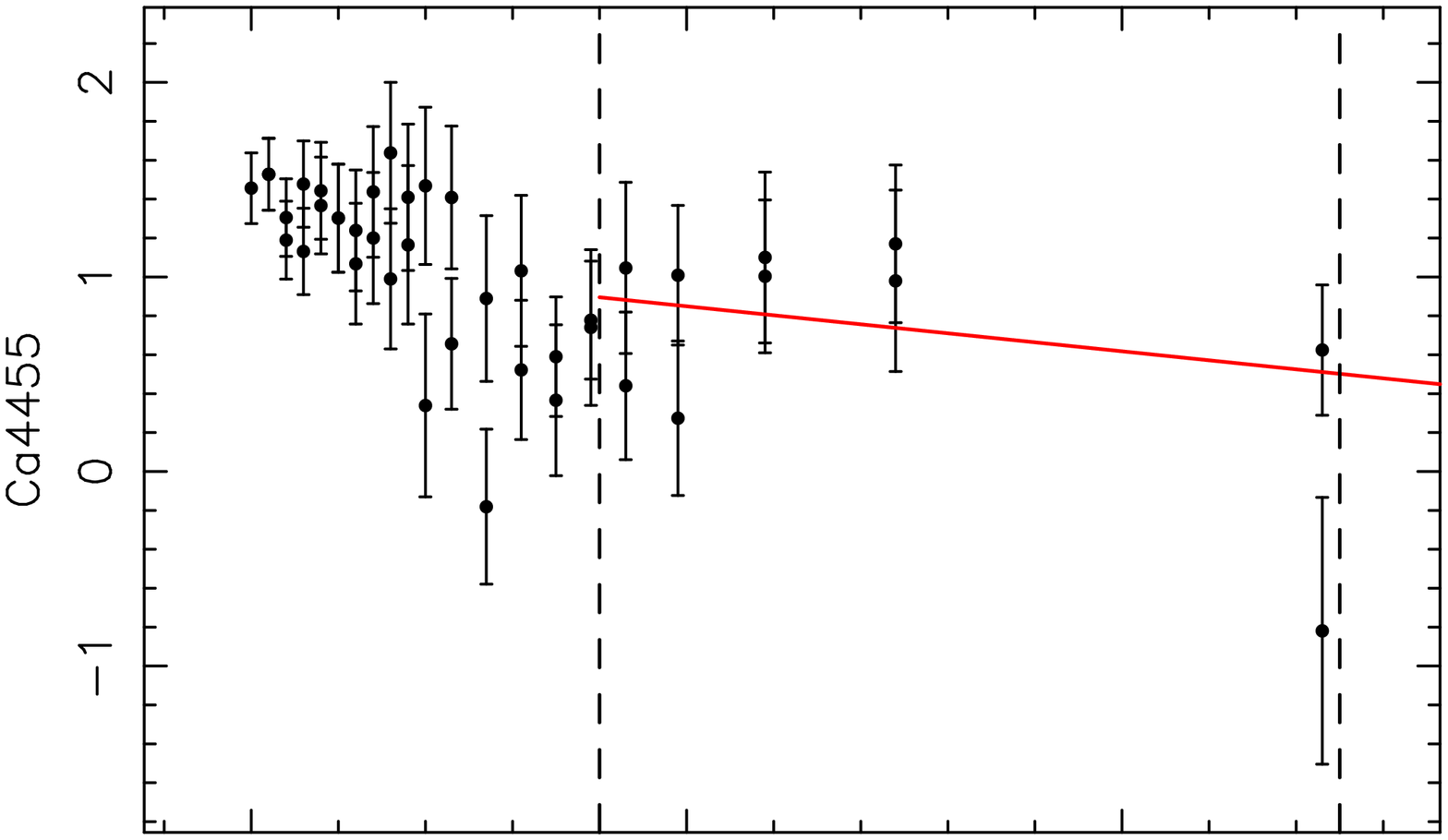}}
\resizebox{0.3\textwidth}{!}{\includegraphics[angle=-90]{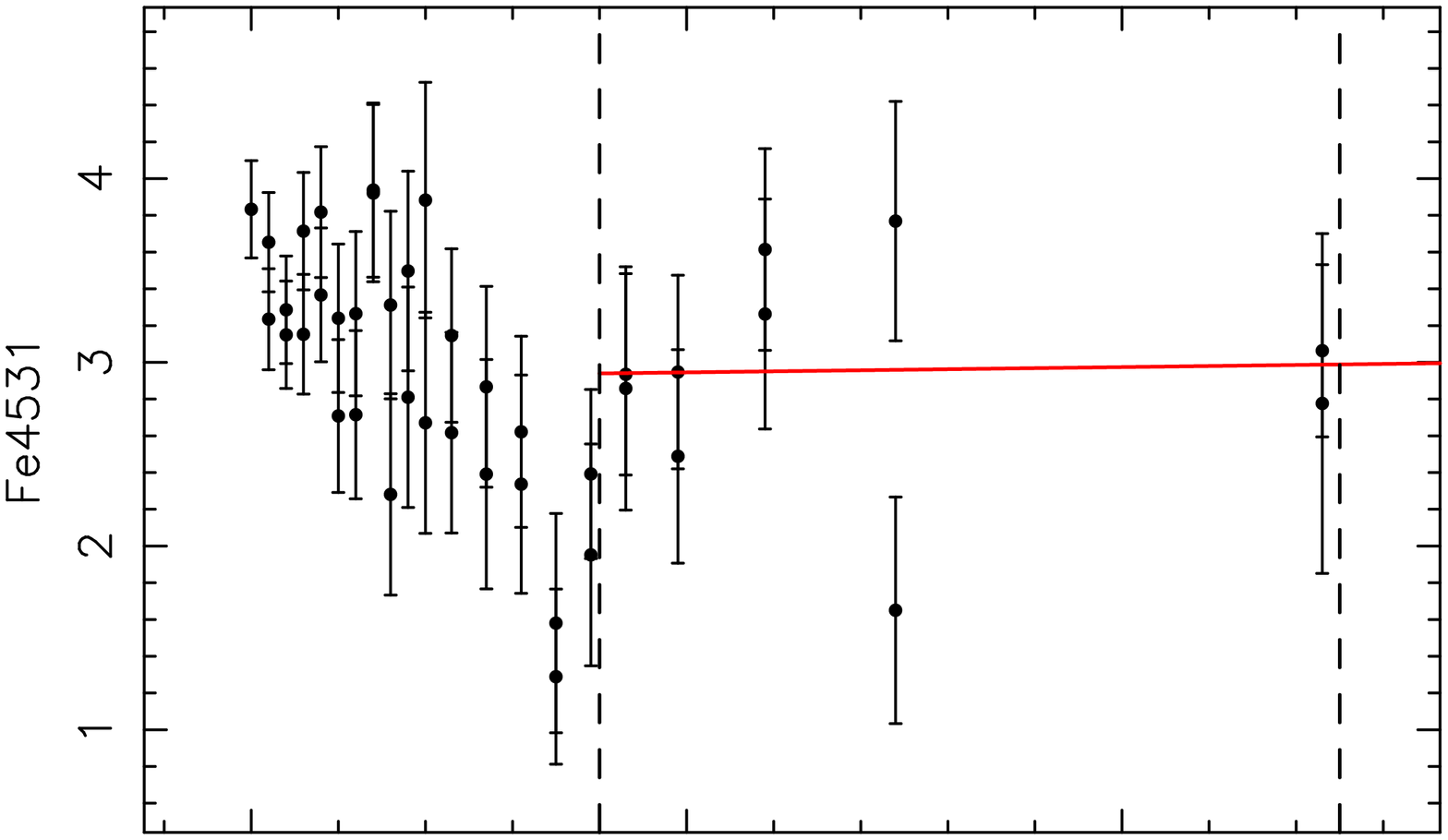}}
\resizebox{0.3\textwidth}{!}{\includegraphics[angle=-90]{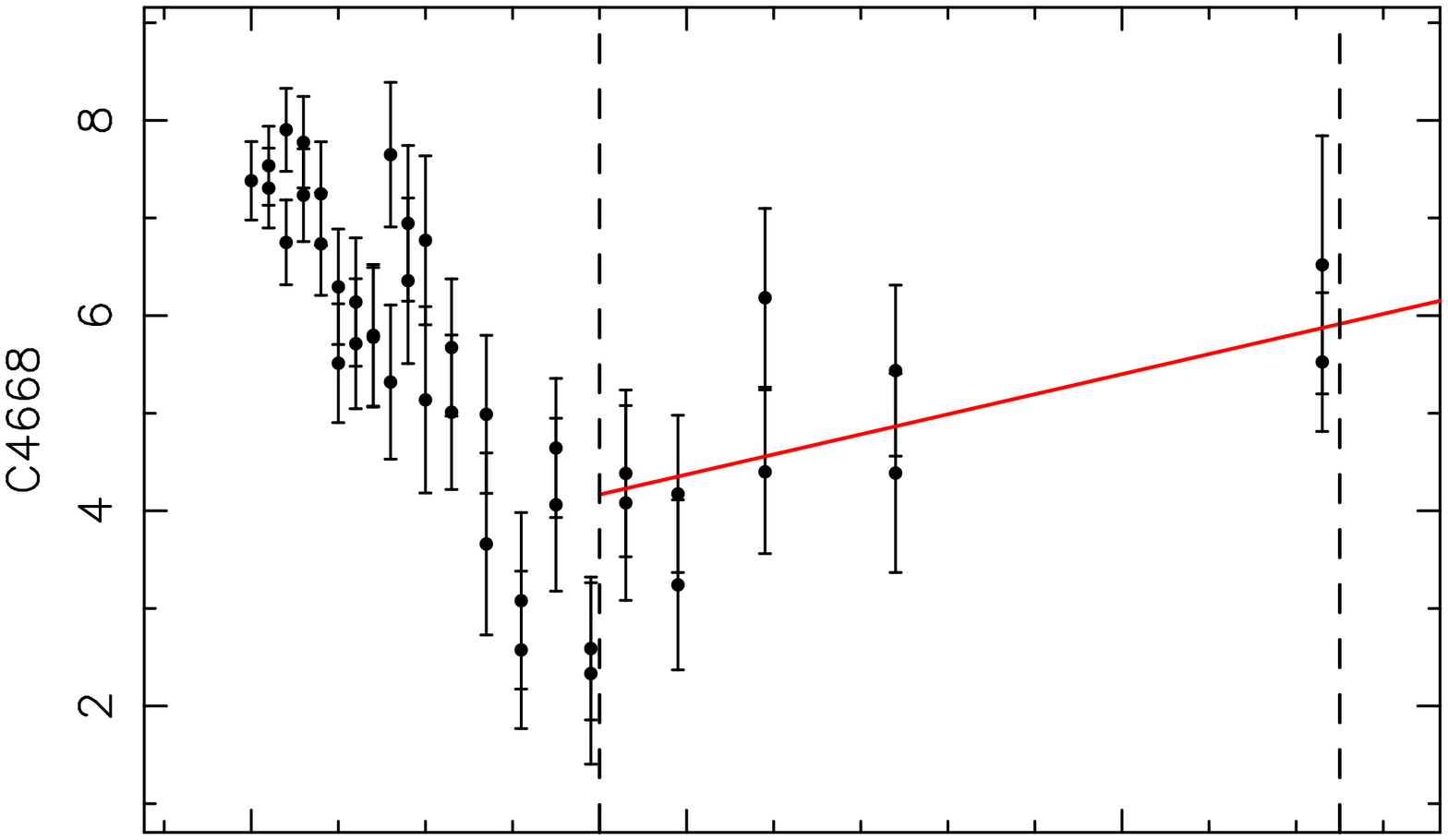}}
\resizebox{0.3\textwidth}{!}{\includegraphics[angle=-90]{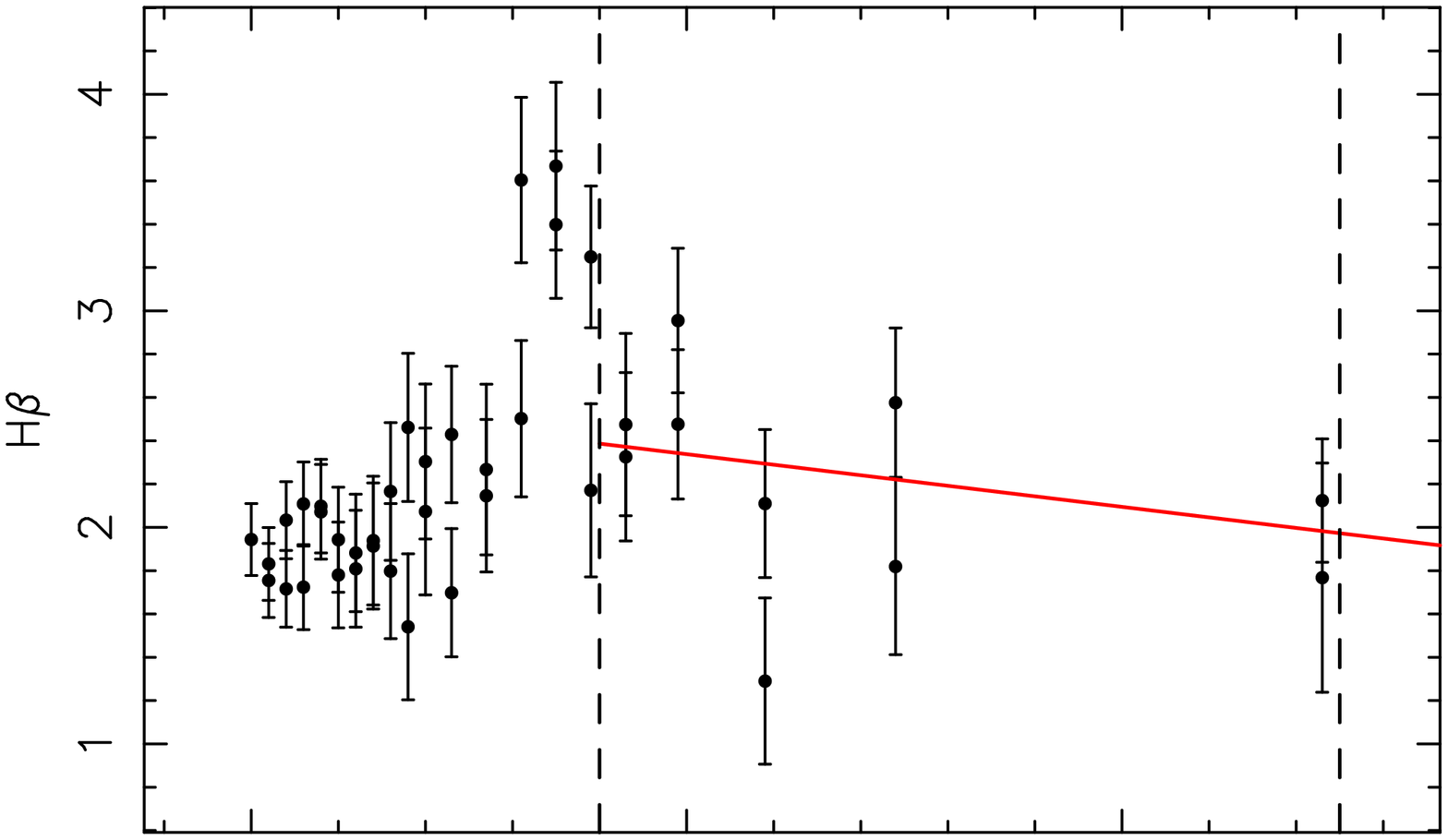}}
\resizebox{0.3\textwidth}{!}{\includegraphics[angle=-90]{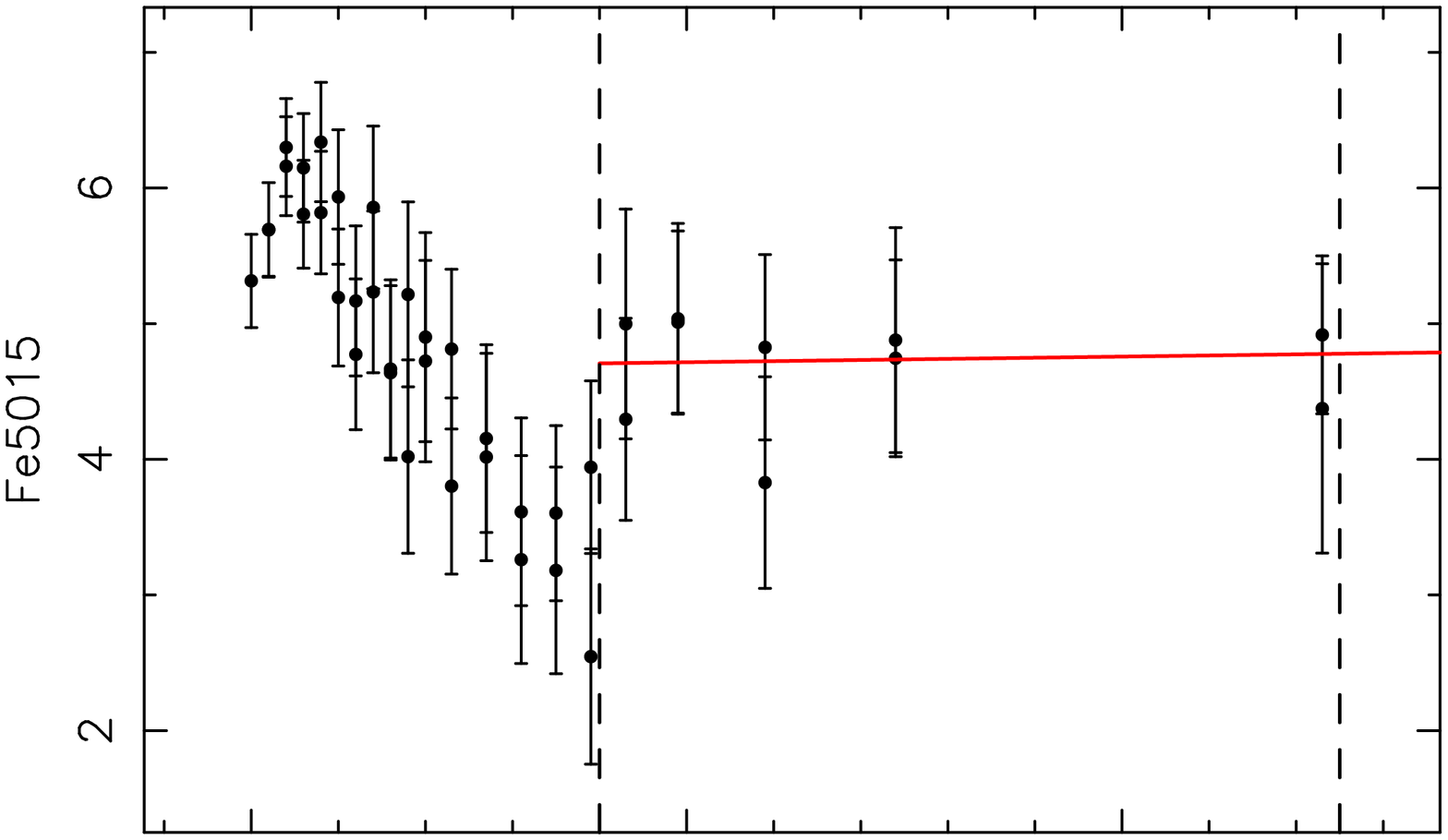}}
\resizebox{0.3\textwidth}{!}{\includegraphics[angle=-90]{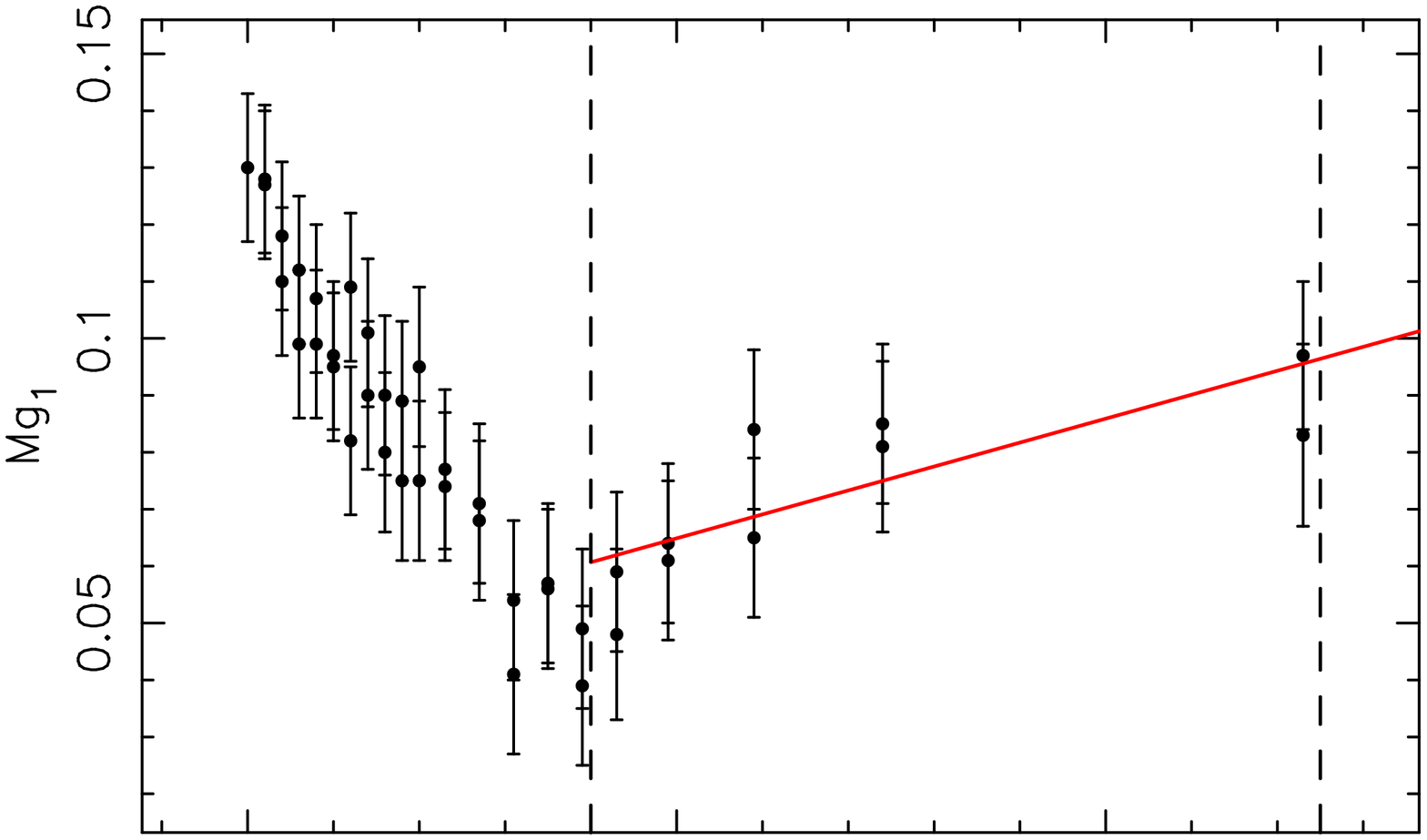}}
\resizebox{0.3\textwidth}{!}{\includegraphics[angle=-90]{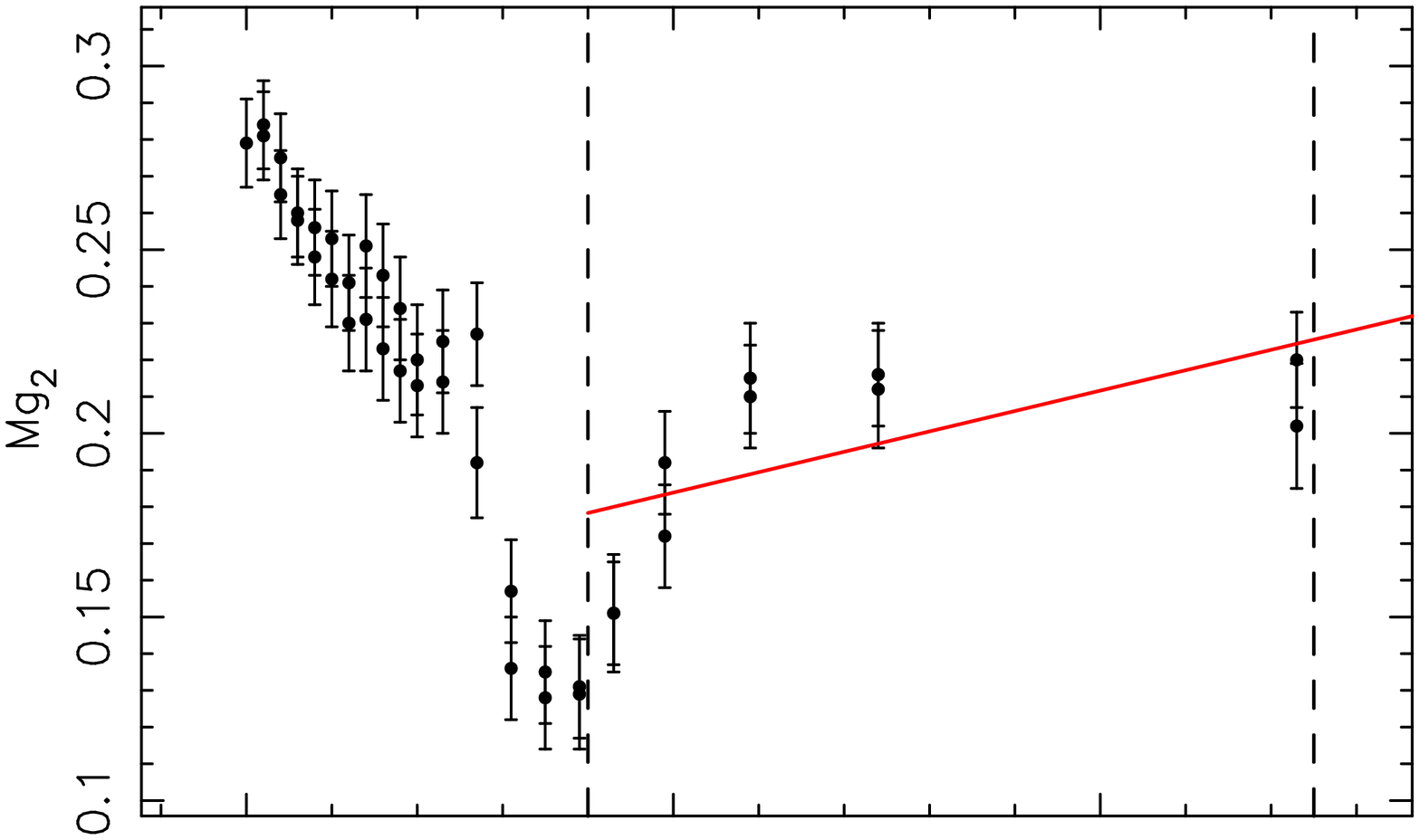}}
\resizebox{0.3\textwidth}{!}{\includegraphics[angle=-90]{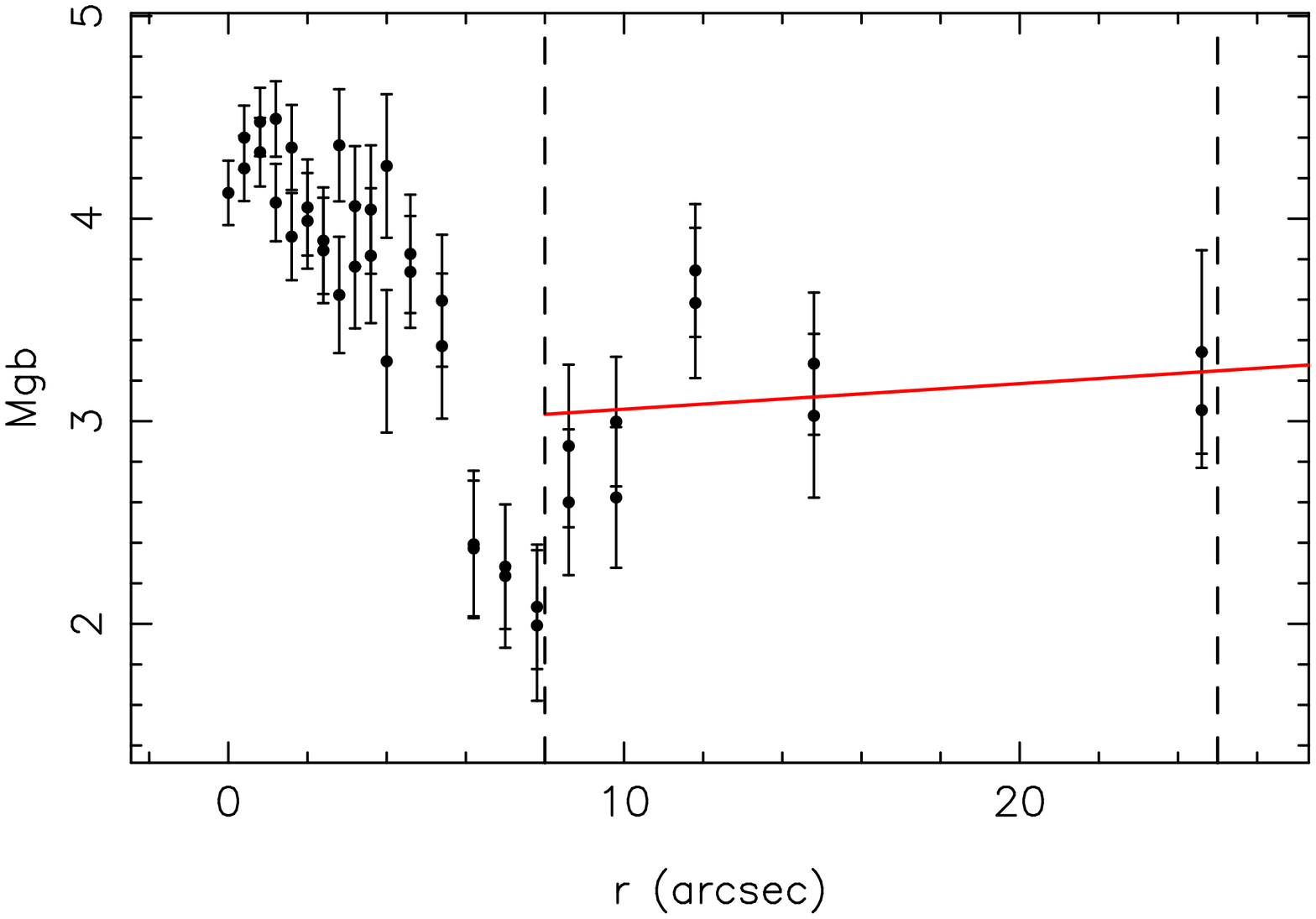}}\hspace{0.85cm}
\resizebox{0.3\textwidth}{!}{\includegraphics[angle=-90]{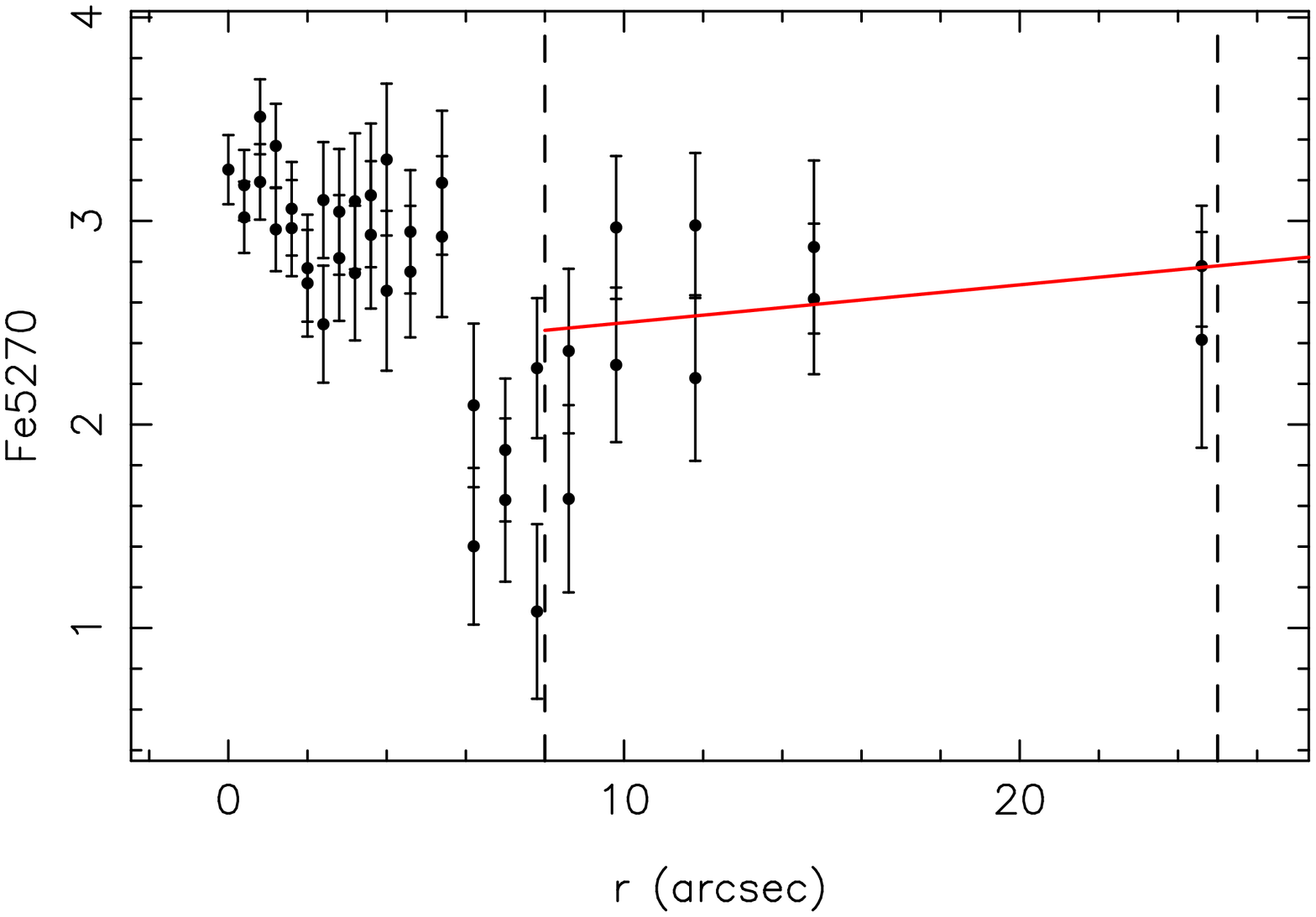}}\hspace{0.85cm}
\resizebox{0.3\textwidth}{!}{\includegraphics[angle=-90]{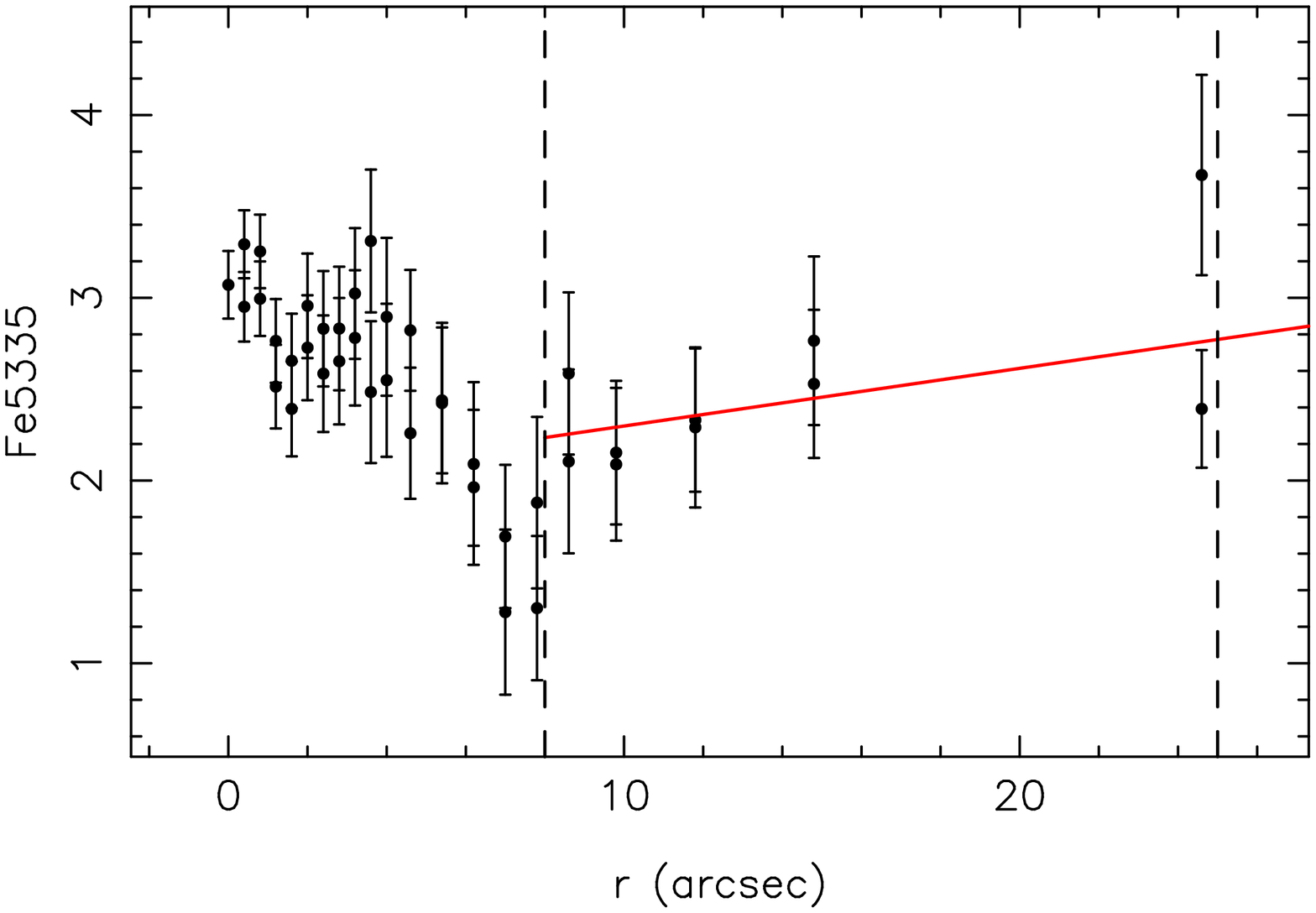}}
\caption{Line-strength distribution in the bar region for all the galaxies}
\end{figure*}
\begin{figure*}
\addtocounter{figure}{-1}
\resizebox{0.3\textwidth}{!}{\includegraphics[angle=-90]{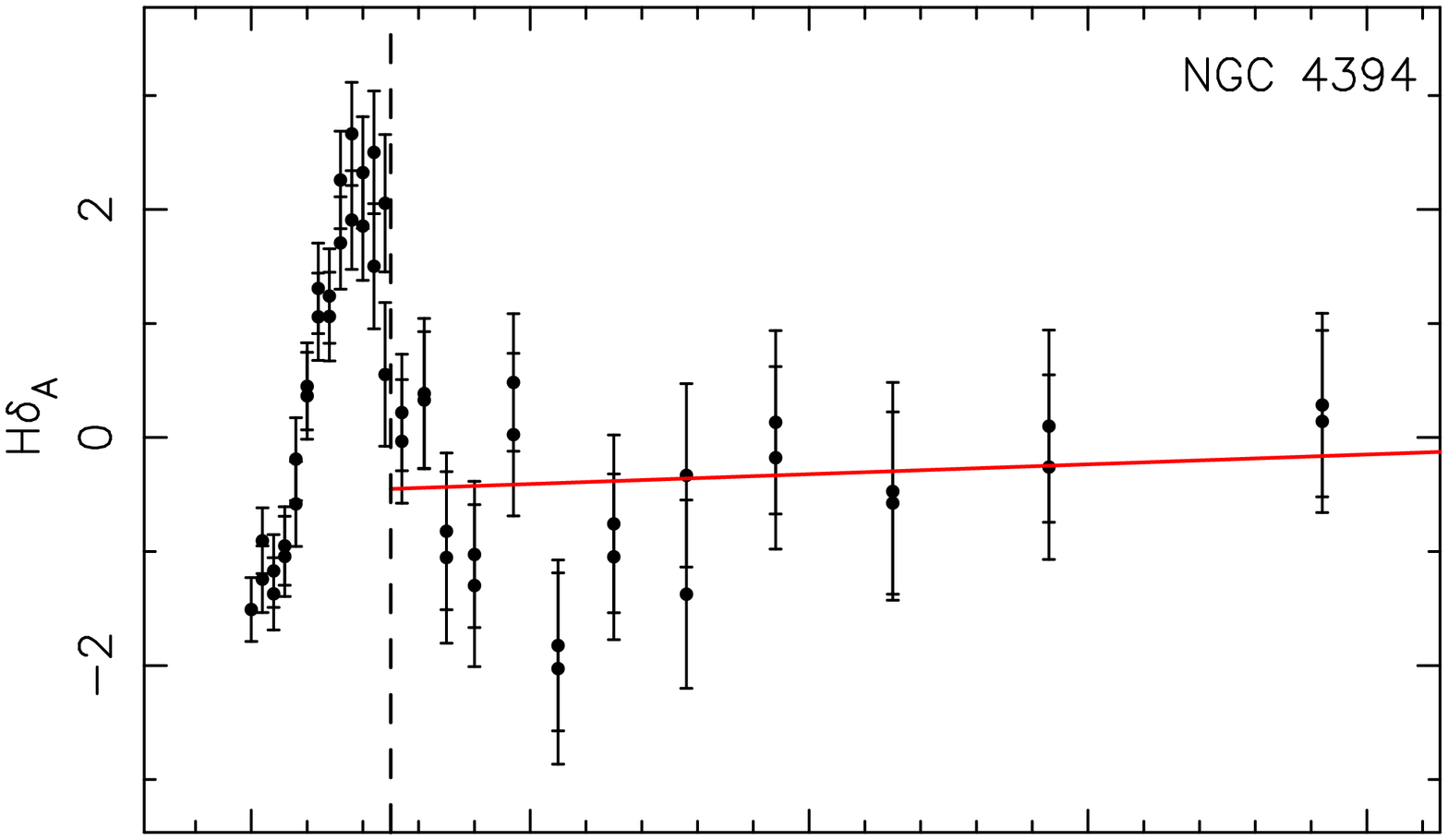}}
\resizebox{0.3\textwidth}{!}{\includegraphics[angle=-90]{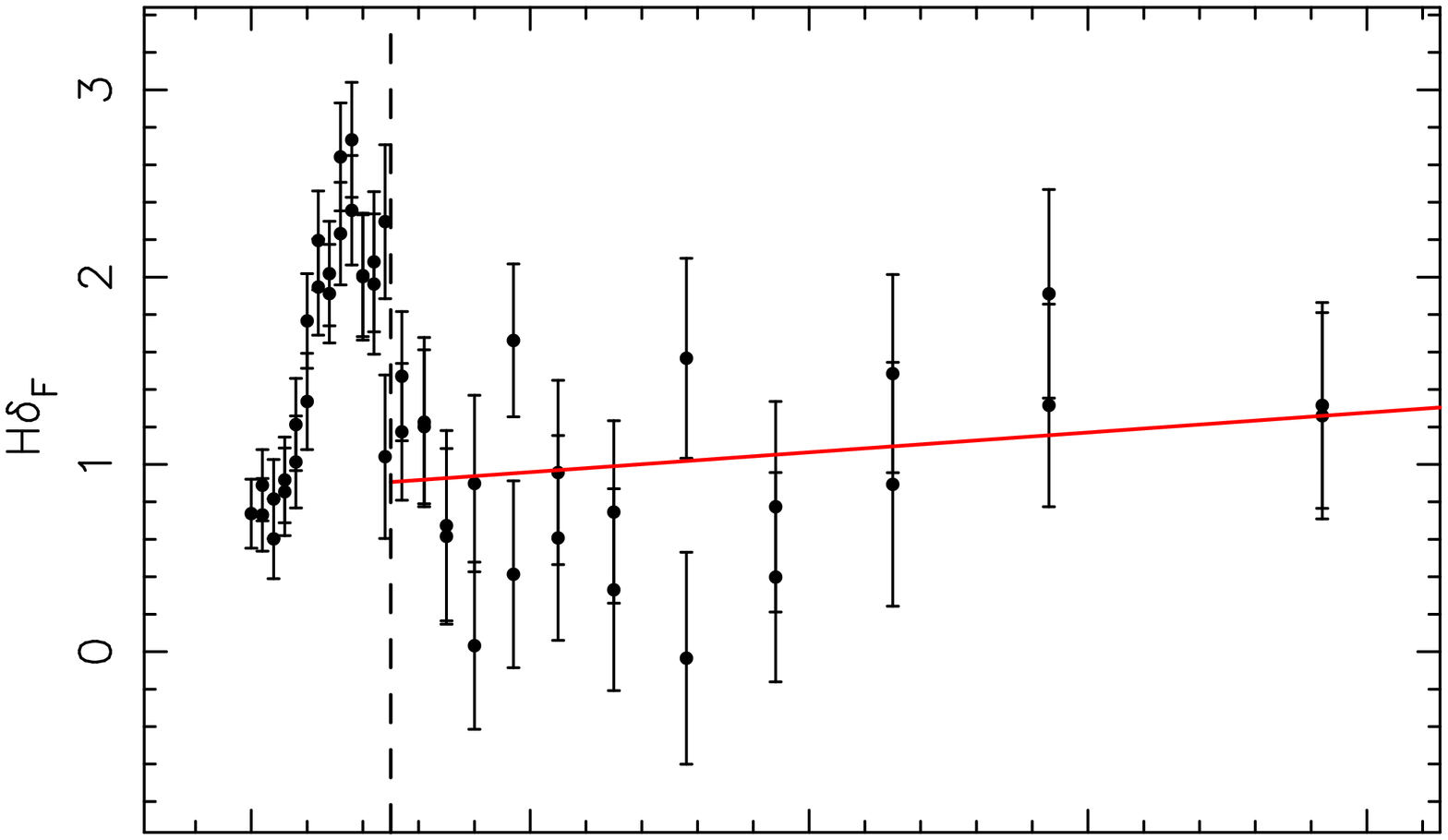}}
\resizebox{0.3\textwidth}{!}{\includegraphics[angle=-90]{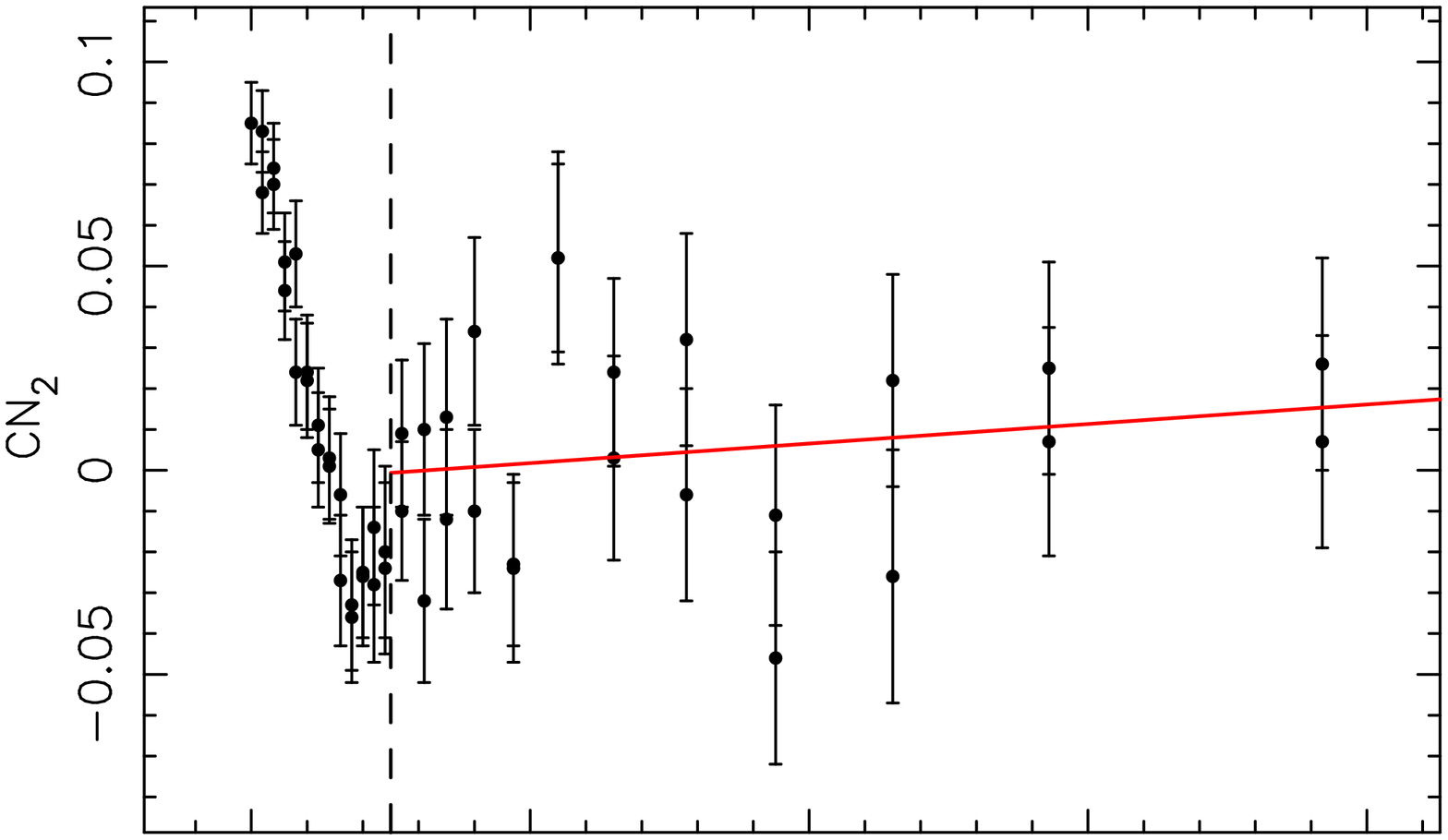}}
\resizebox{0.3\textwidth}{!}{\includegraphics[angle=-90]{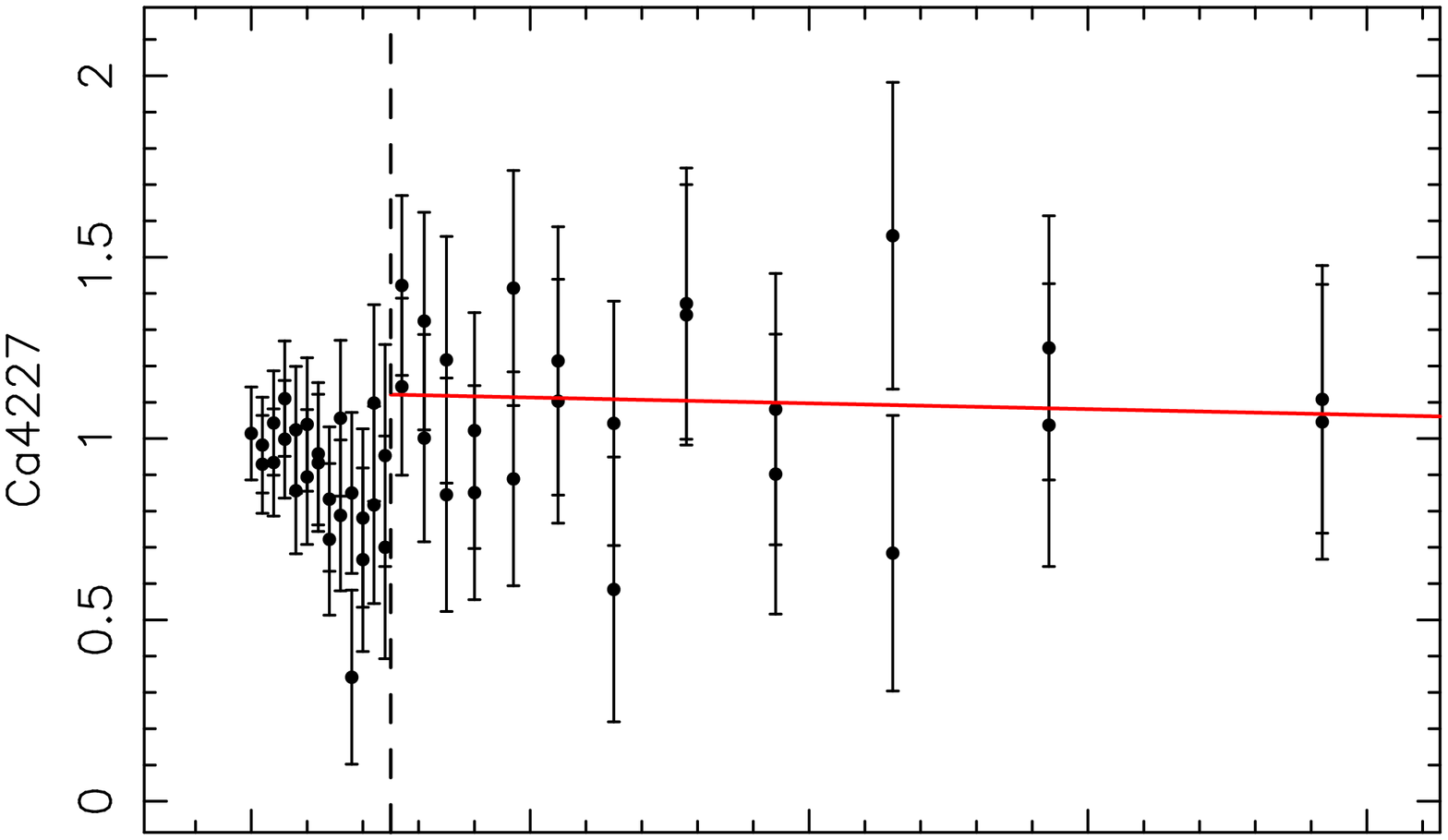}}
\resizebox{0.3\textwidth}{!}{\includegraphics[angle=-90]{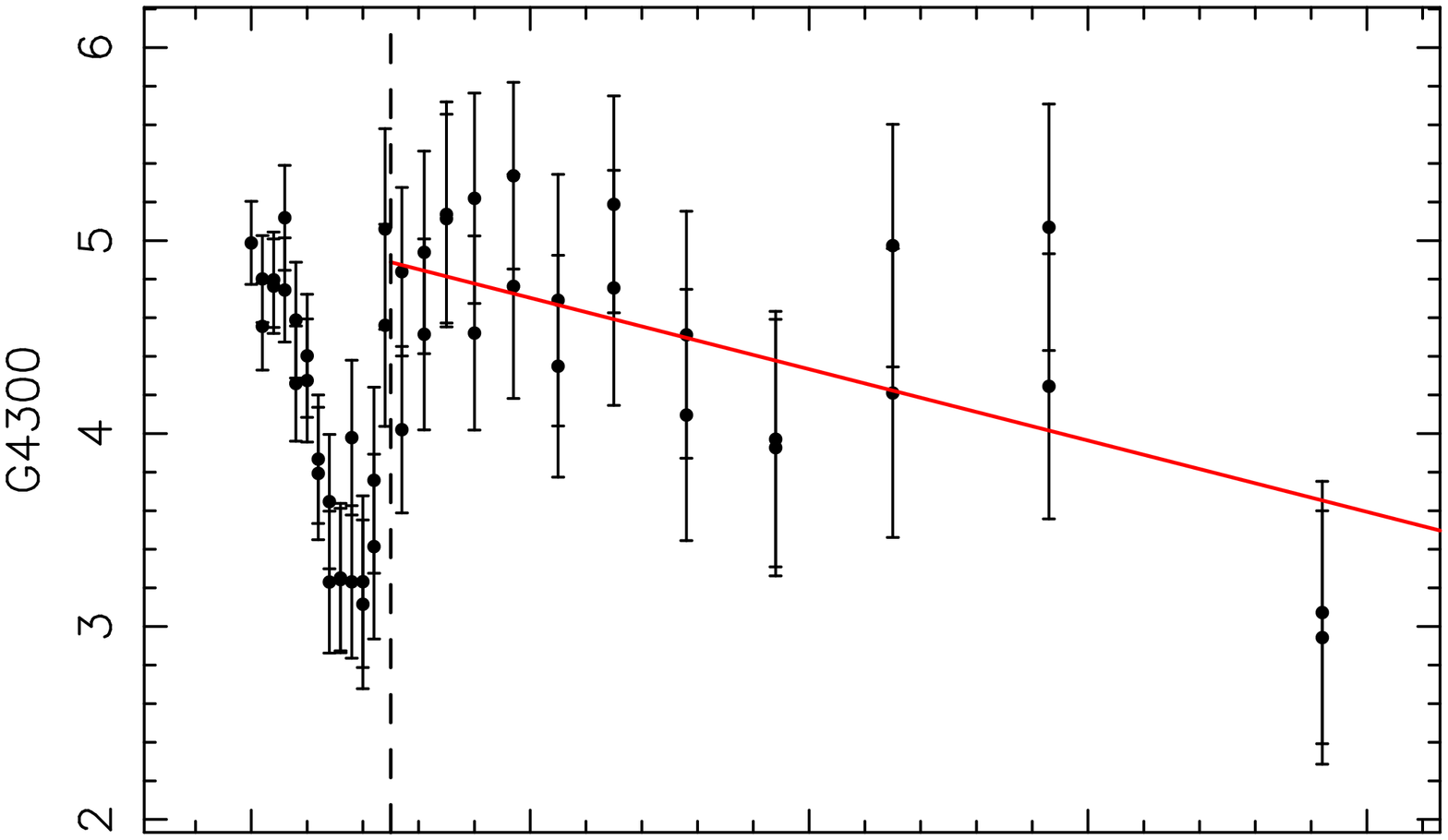}}
\resizebox{0.3\textwidth}{!}{\includegraphics[angle=-90]{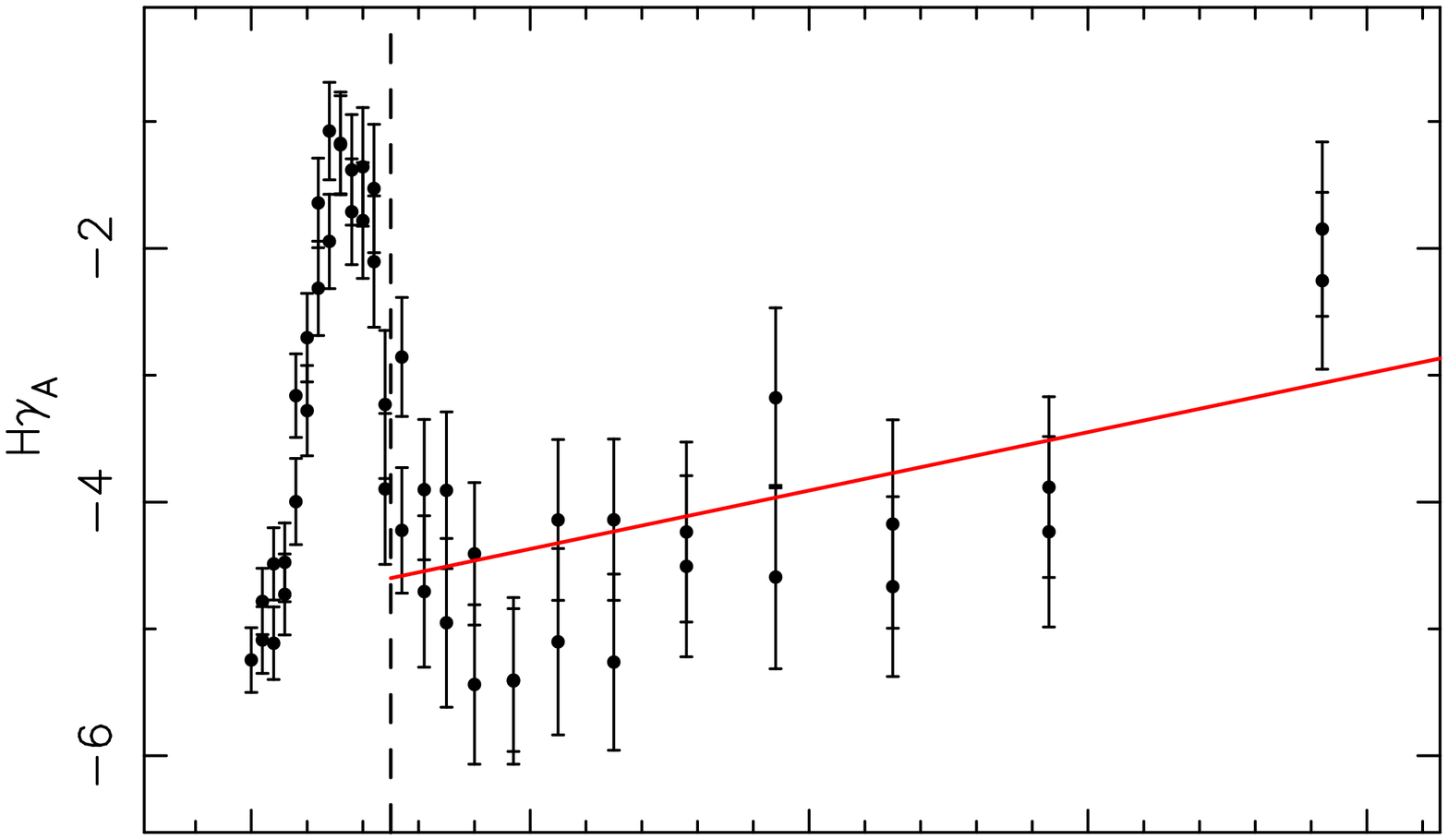}}
\resizebox{0.3\textwidth}{!}{\includegraphics[angle=-90]{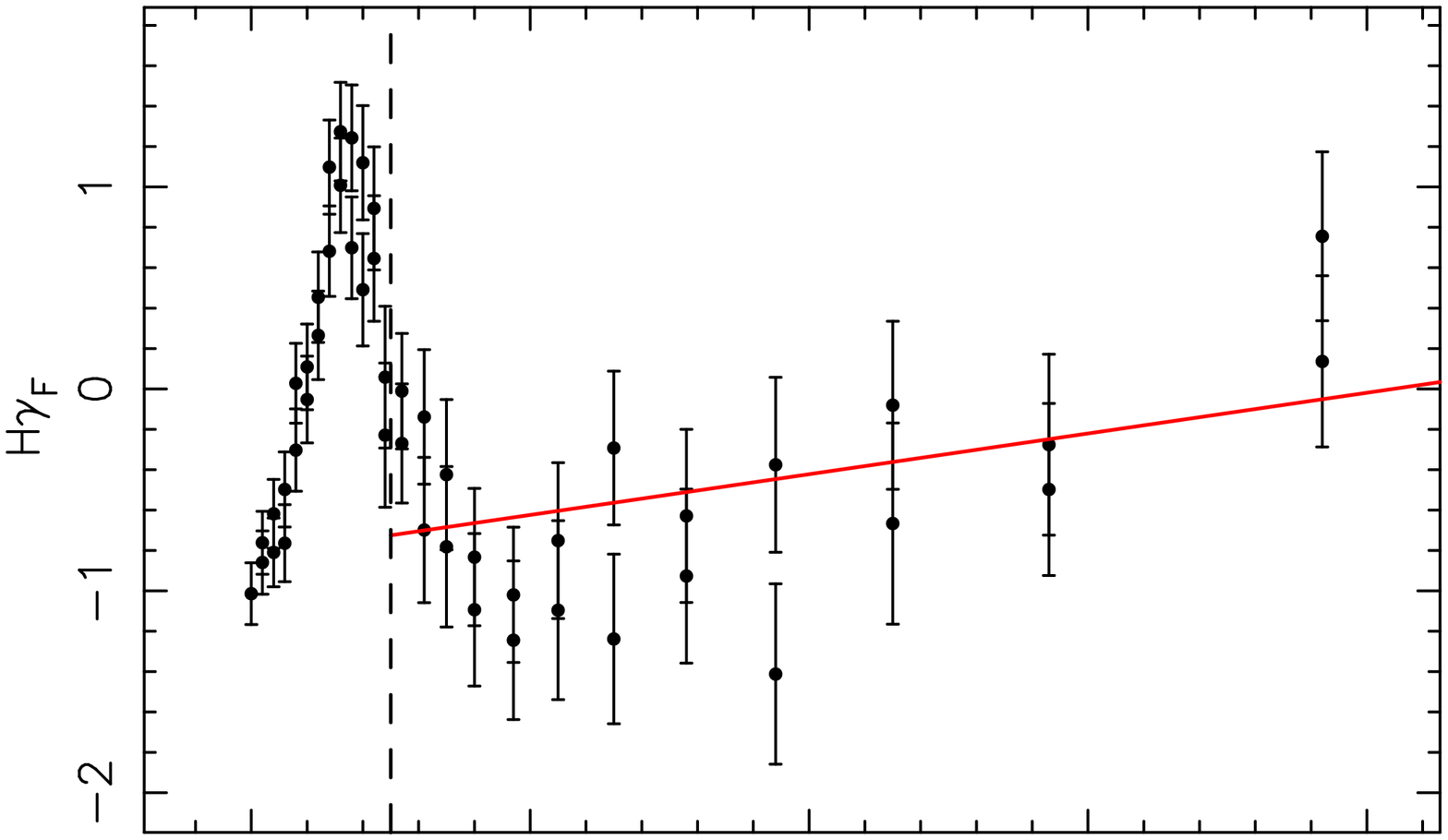}}
\resizebox{0.3\textwidth}{!}{\includegraphics[angle=-90]{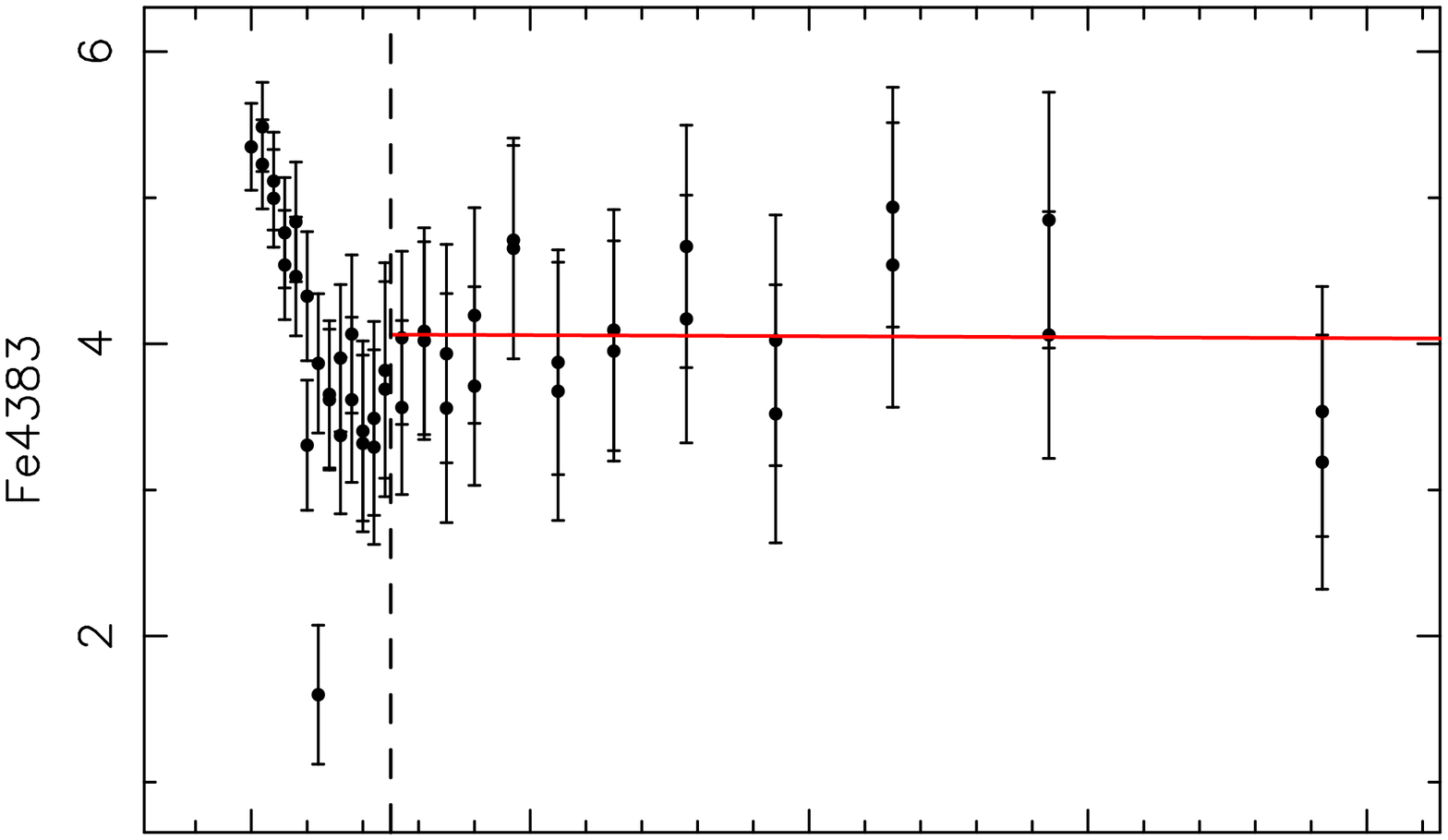}}
\resizebox{0.3\textwidth}{!}{\includegraphics[angle=-90]{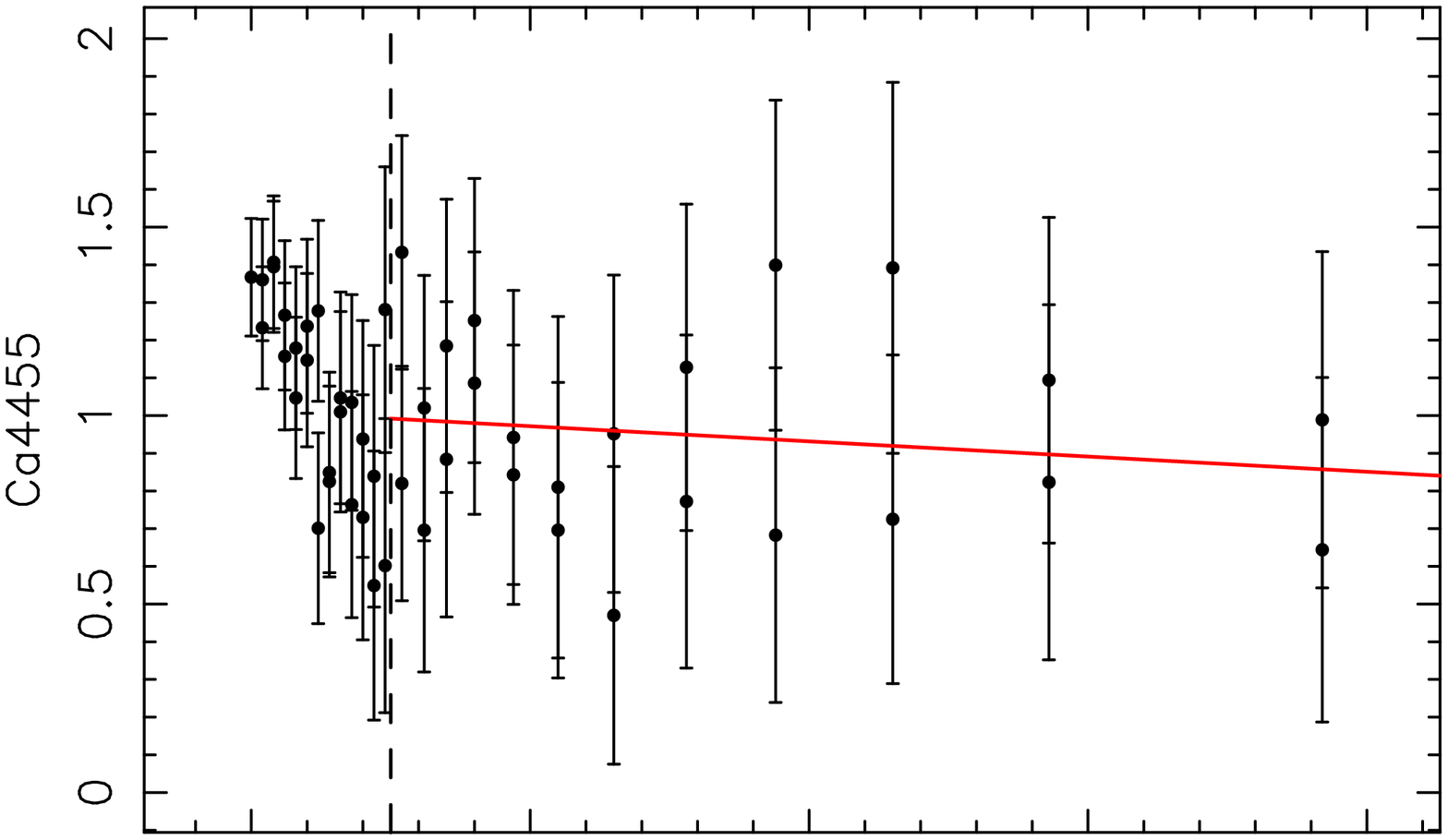}}
\resizebox{0.3\textwidth}{!}{\includegraphics[angle=-90]{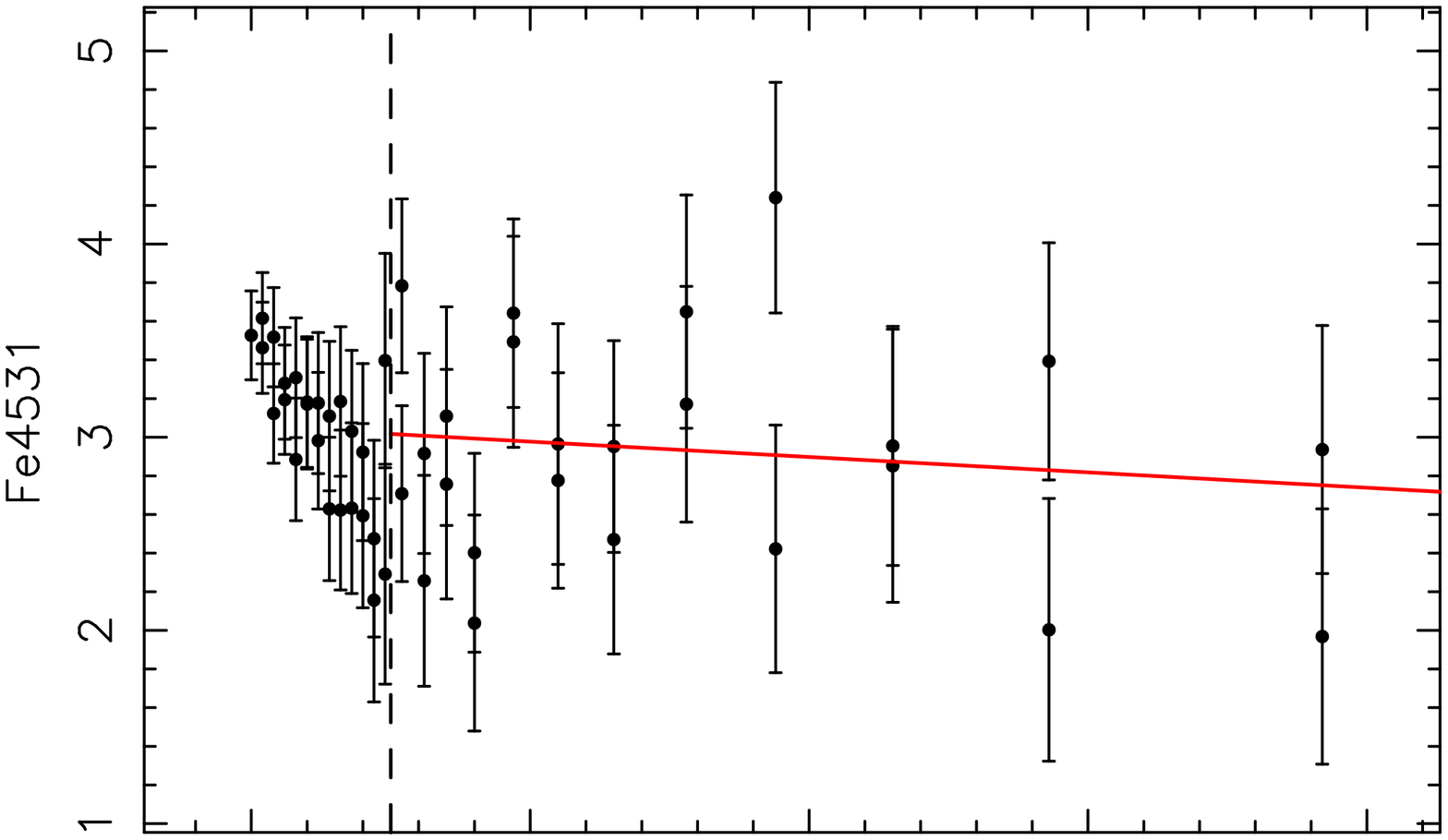}}
\resizebox{0.3\textwidth}{!}{\includegraphics[angle=-90]{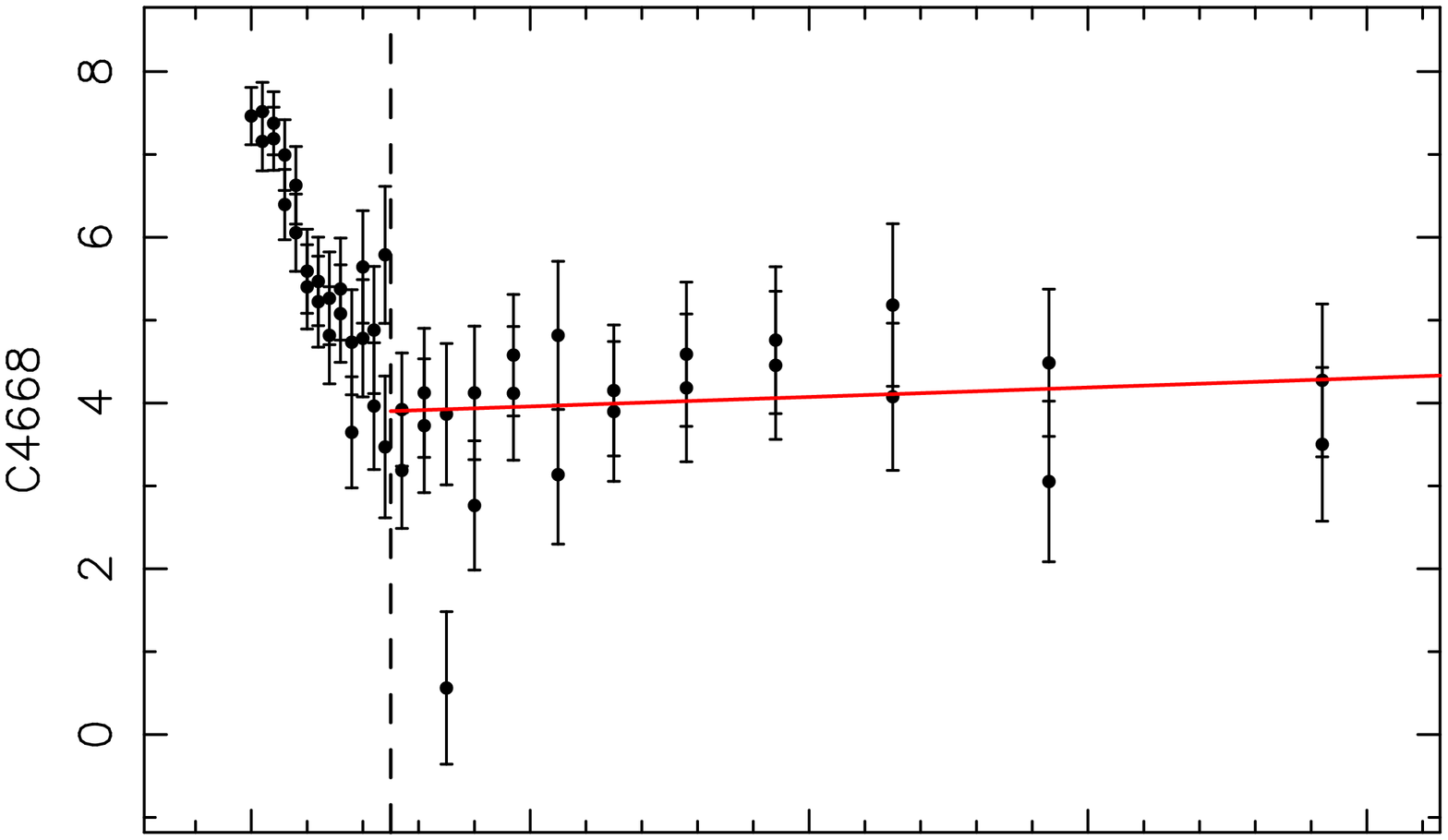}}
\resizebox{0.3\textwidth}{!}{\includegraphics[angle=-90]{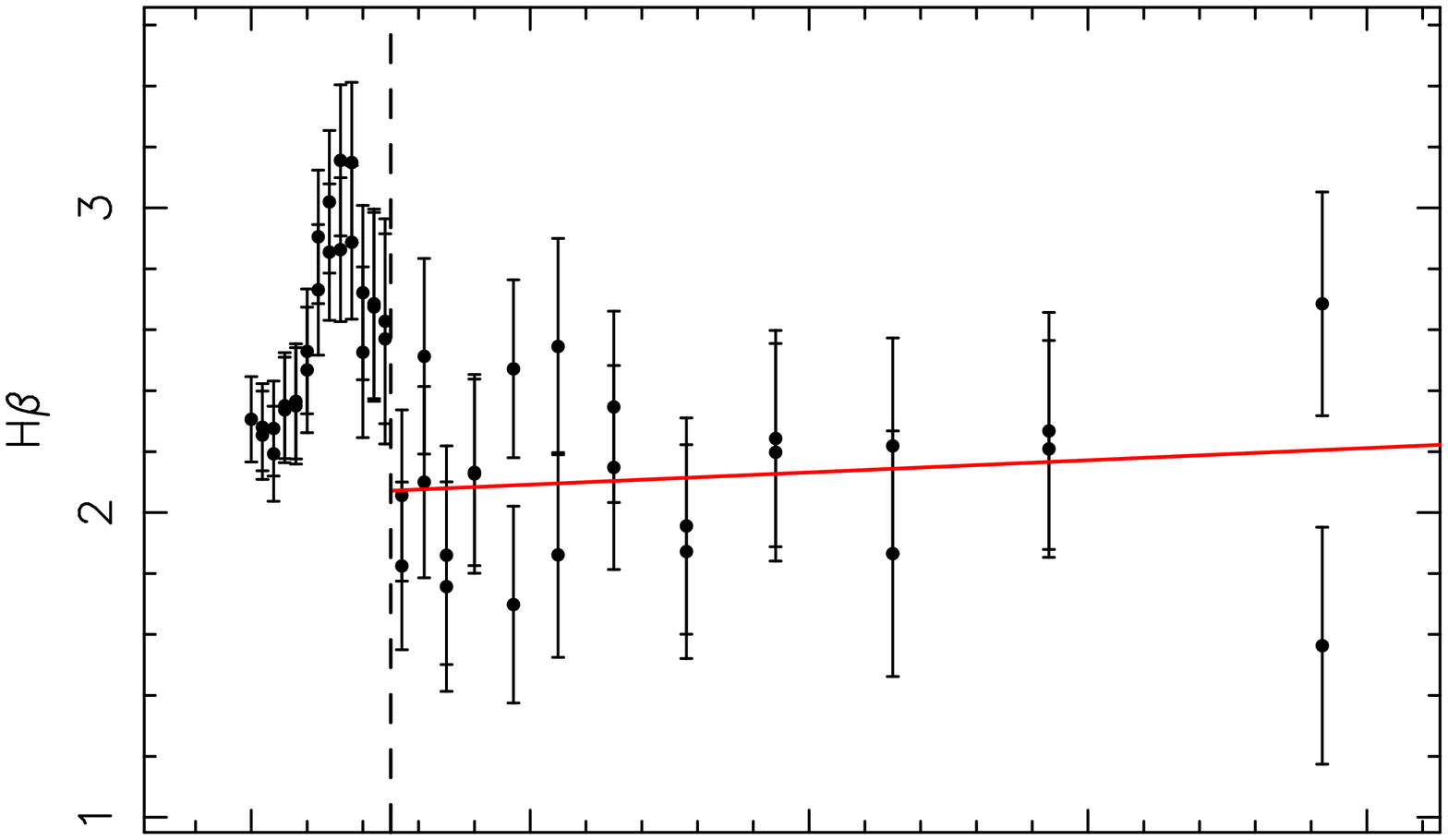}}
\resizebox{0.3\textwidth}{!}{\includegraphics[angle=-90]{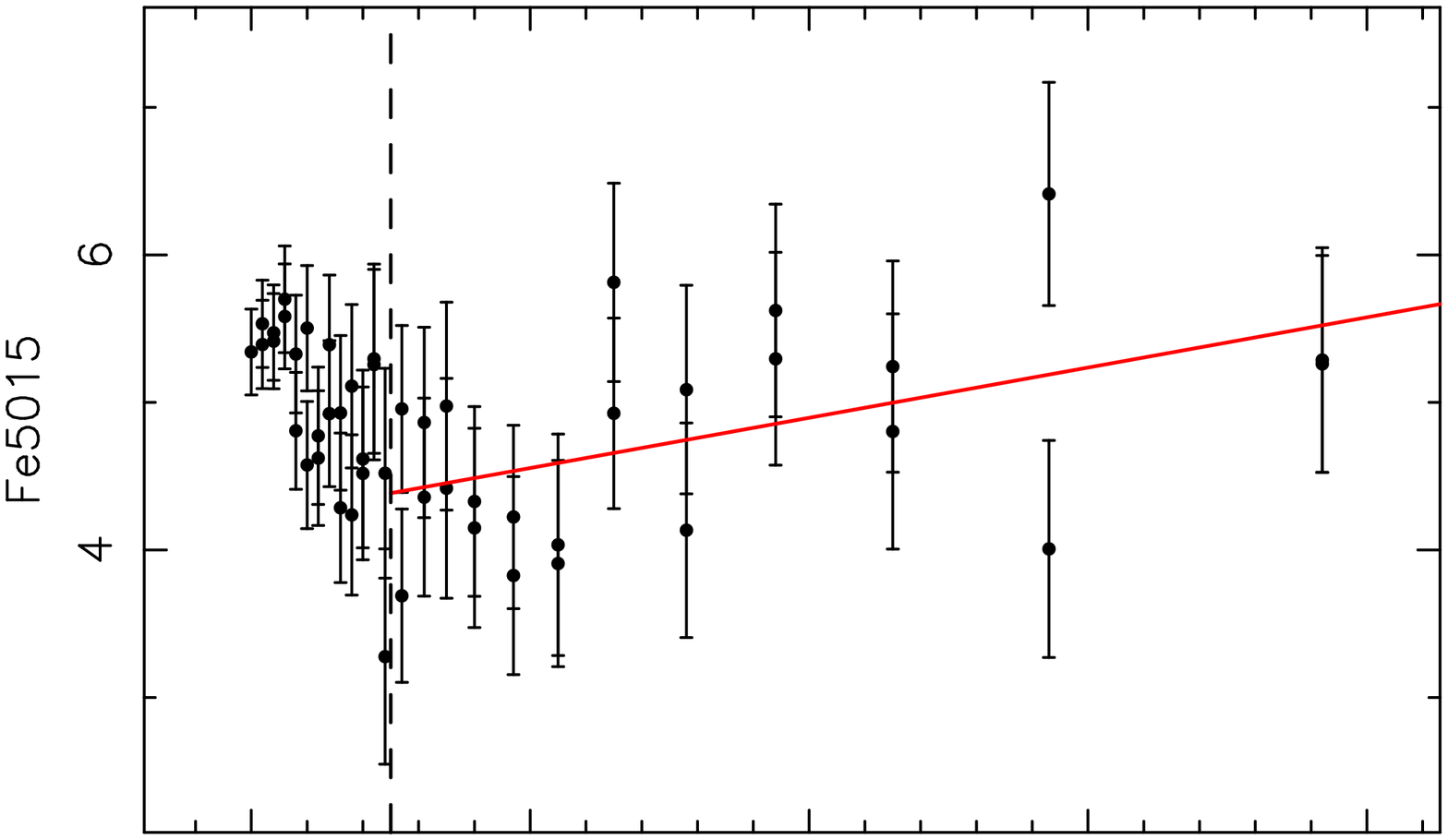}}
\resizebox{0.3\textwidth}{!}{\includegraphics[angle=-90]{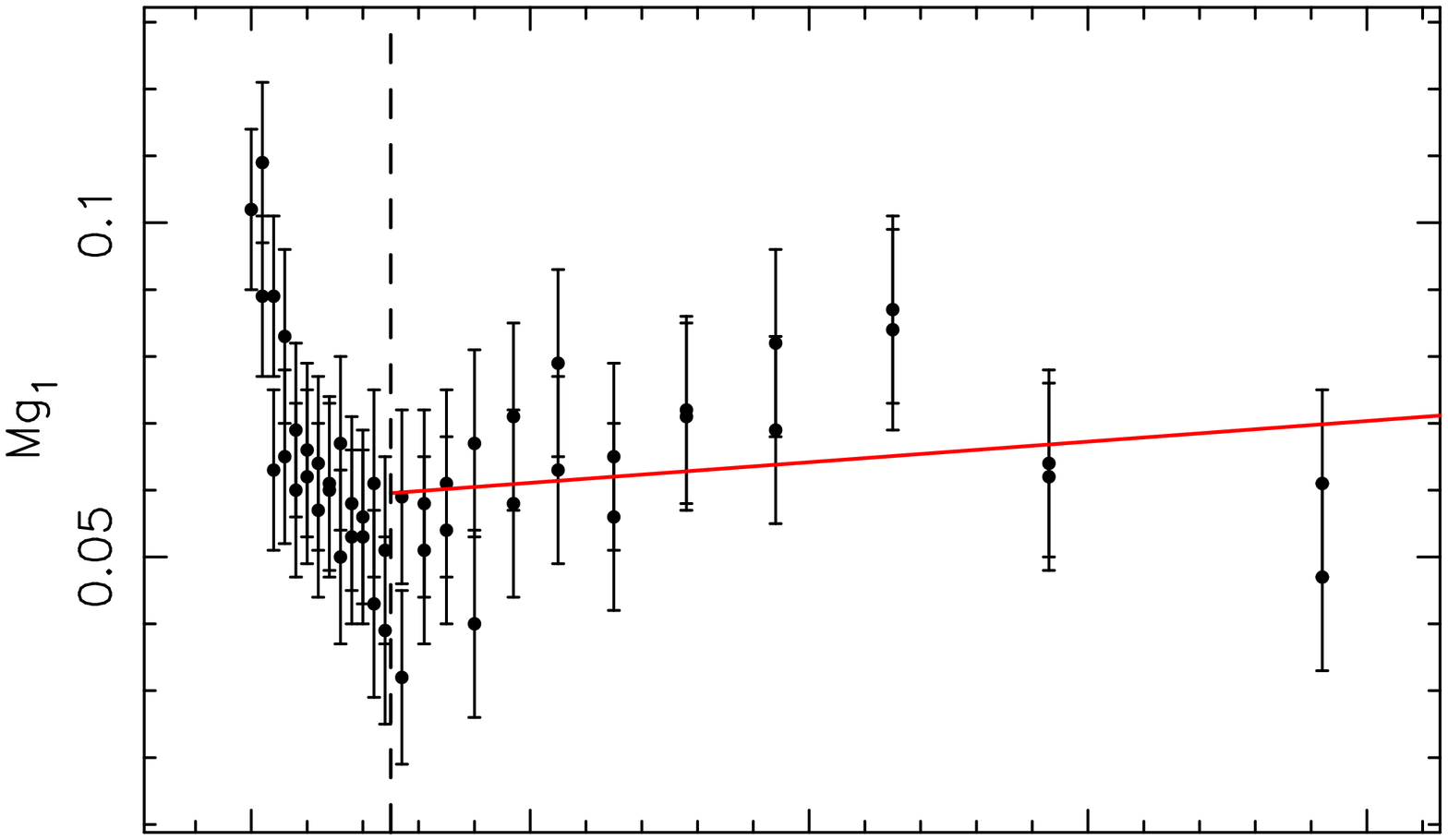}}
\resizebox{0.3\textwidth}{!}{\includegraphics[angle=-90]{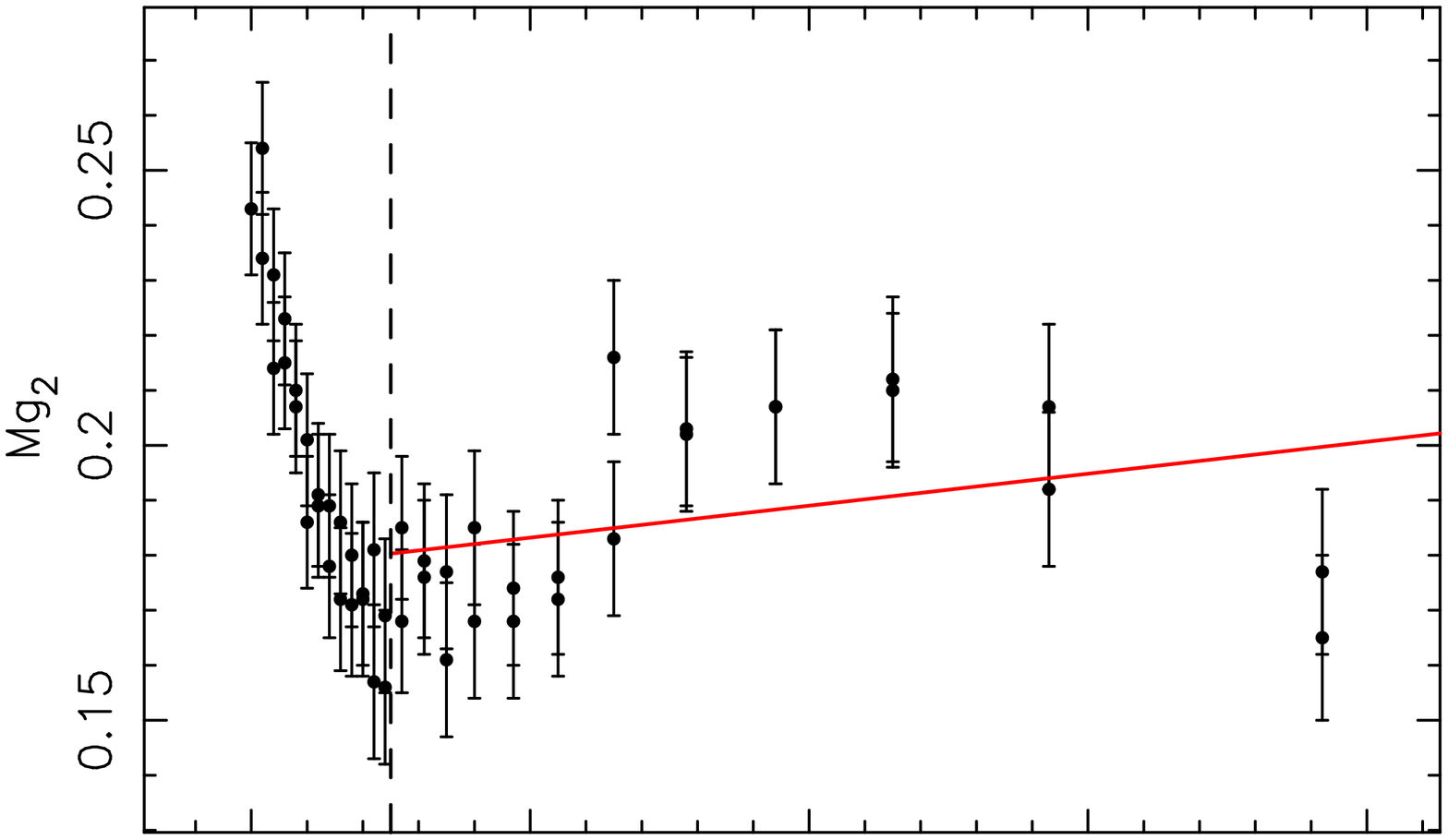}}
\resizebox{0.3\textwidth}{!}{\includegraphics[angle=-90]{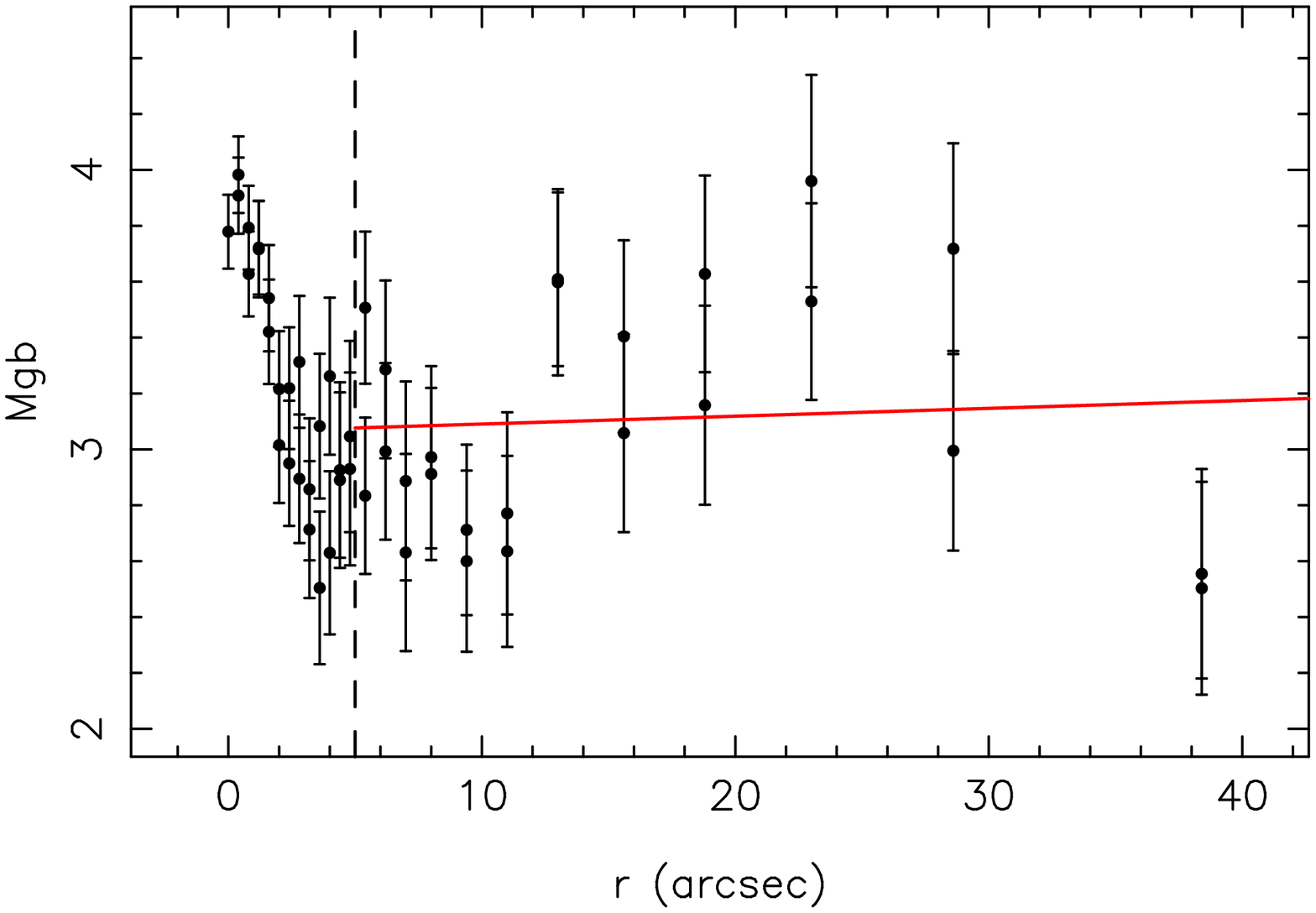}}\hspace{0.85cm}
\resizebox{0.3\textwidth}{!}{\includegraphics[angle=-90]{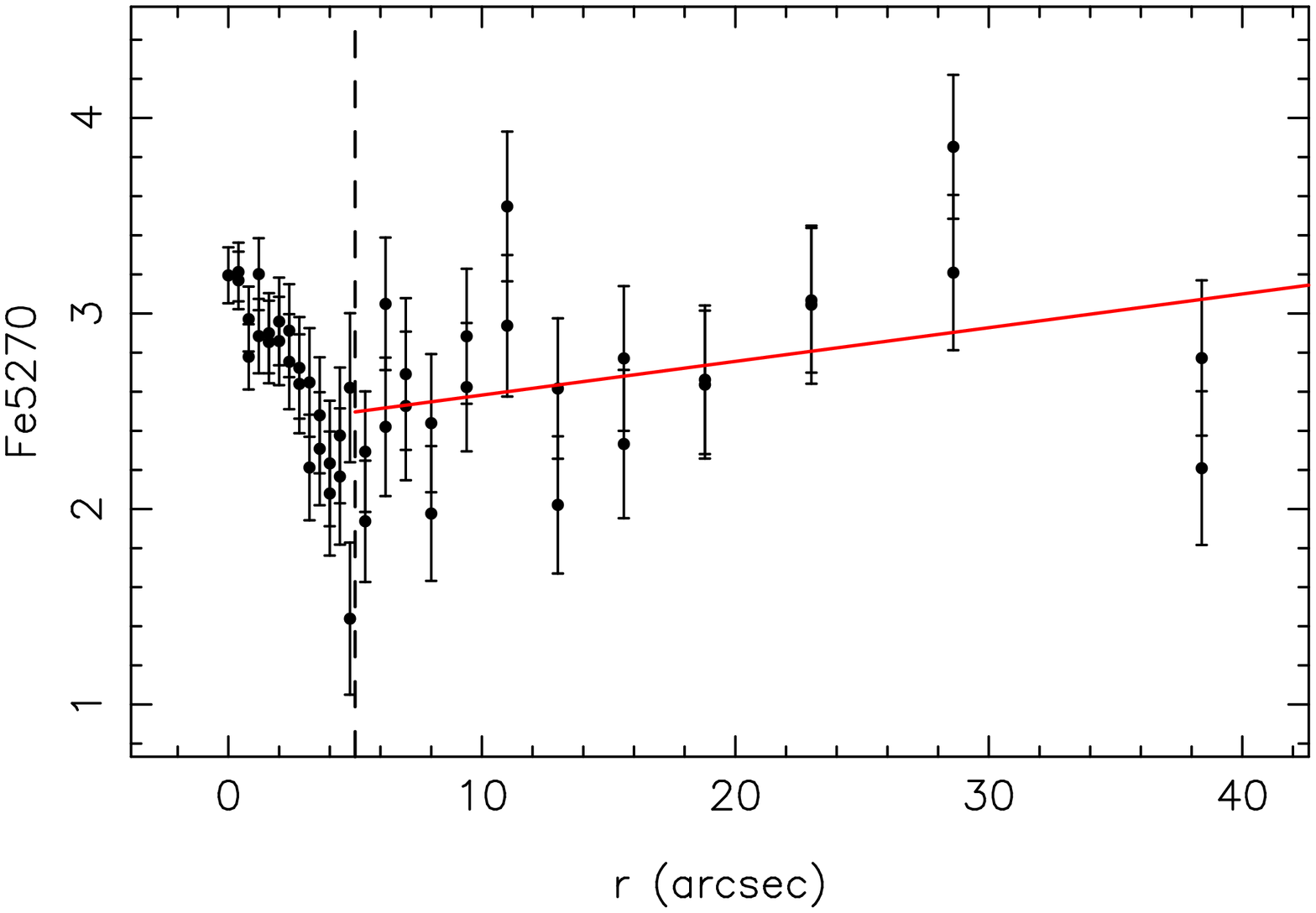}}\hspace{0.85cm}
\resizebox{0.3\textwidth}{!}{\includegraphics[angle=-90]{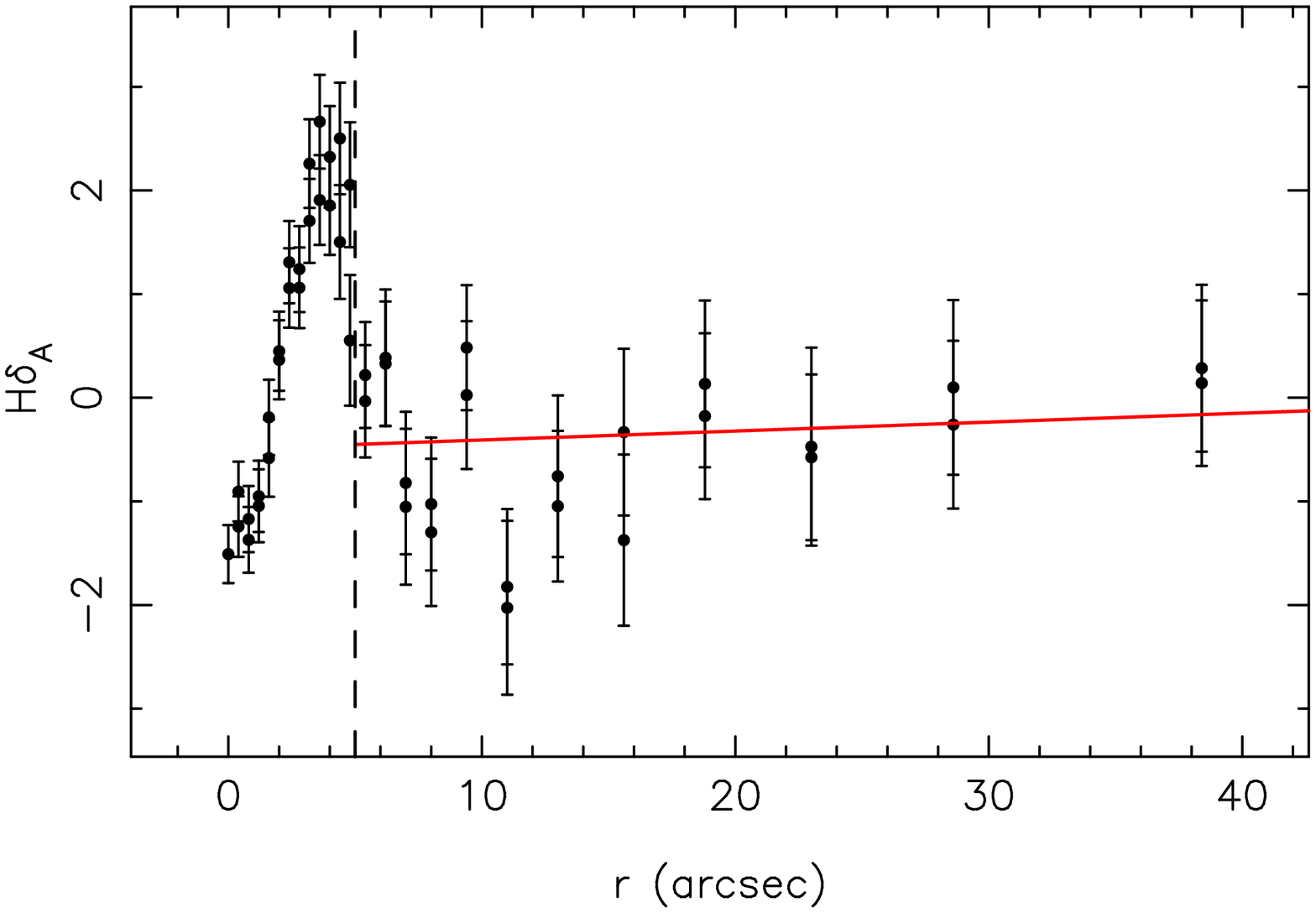}}
\caption{Line-strength distribution in the bar region for all the galaxies}
\end{figure*}


\begin{figure*}
\addtocounter{figure}{-1}
\resizebox{0.3\textwidth}{!}{\includegraphics[angle=-90]{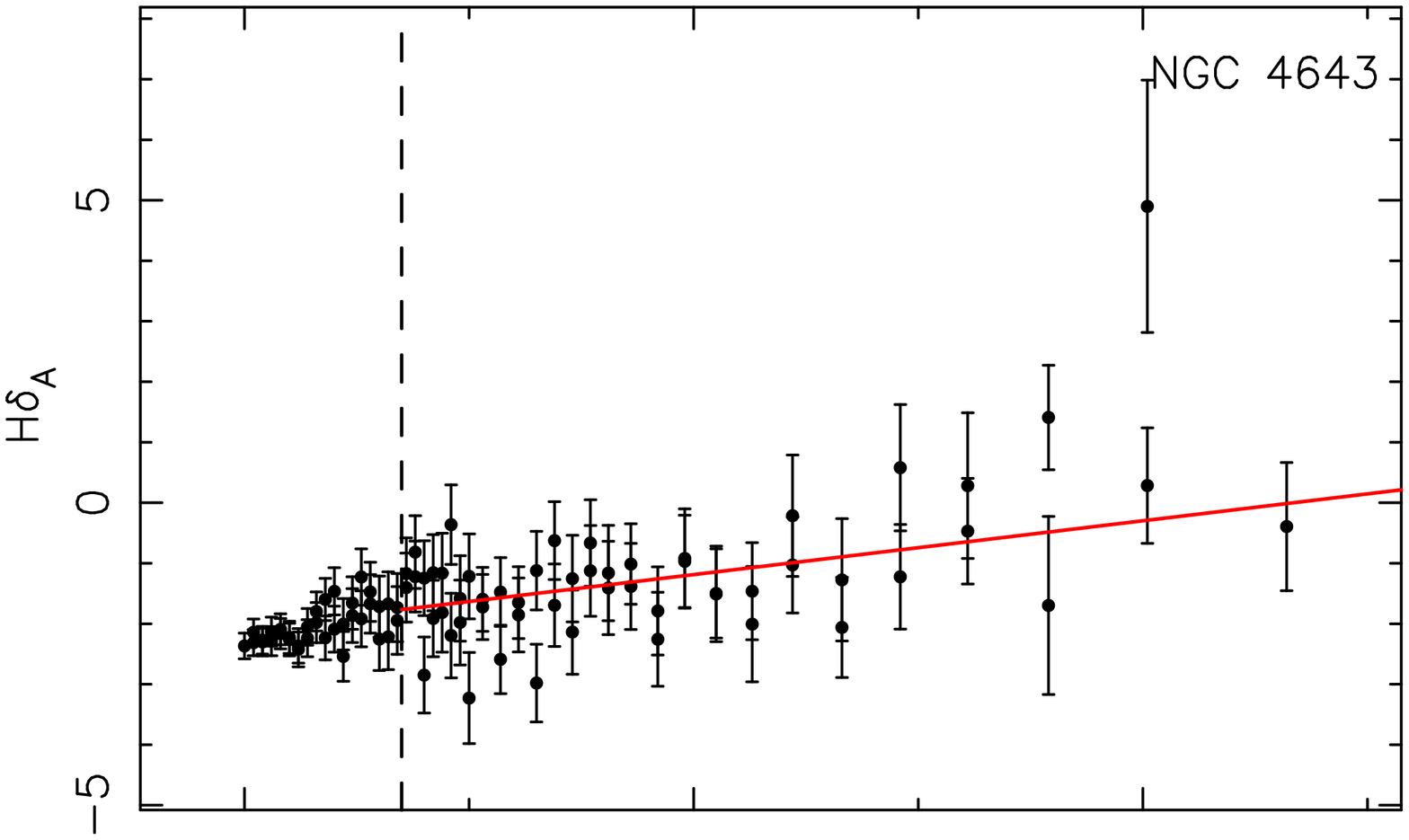}}
\resizebox{0.3\textwidth}{!}{\includegraphics[angle=-90]{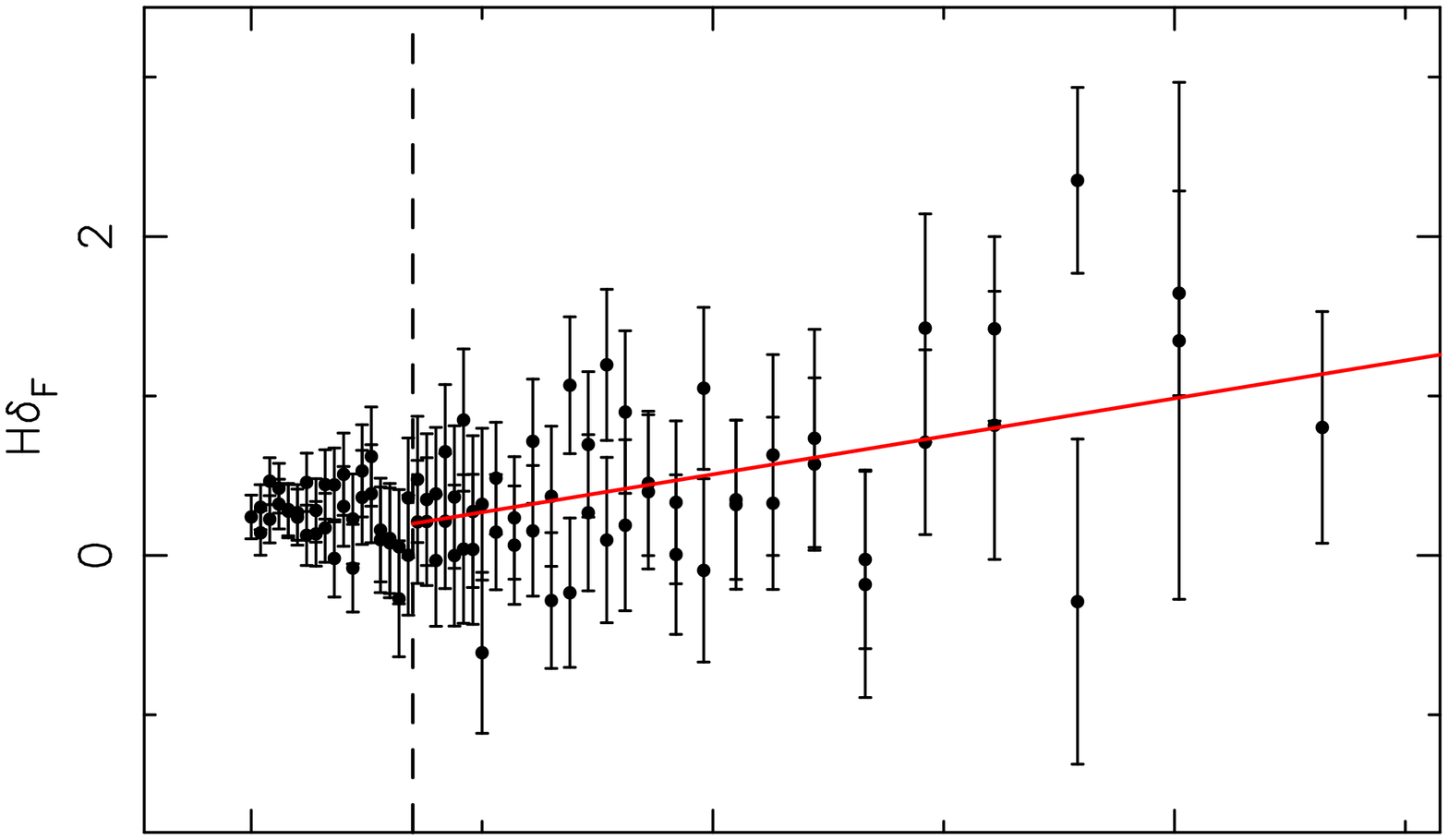}}
\resizebox{0.3\textwidth}{!}{\includegraphics[angle=-90]{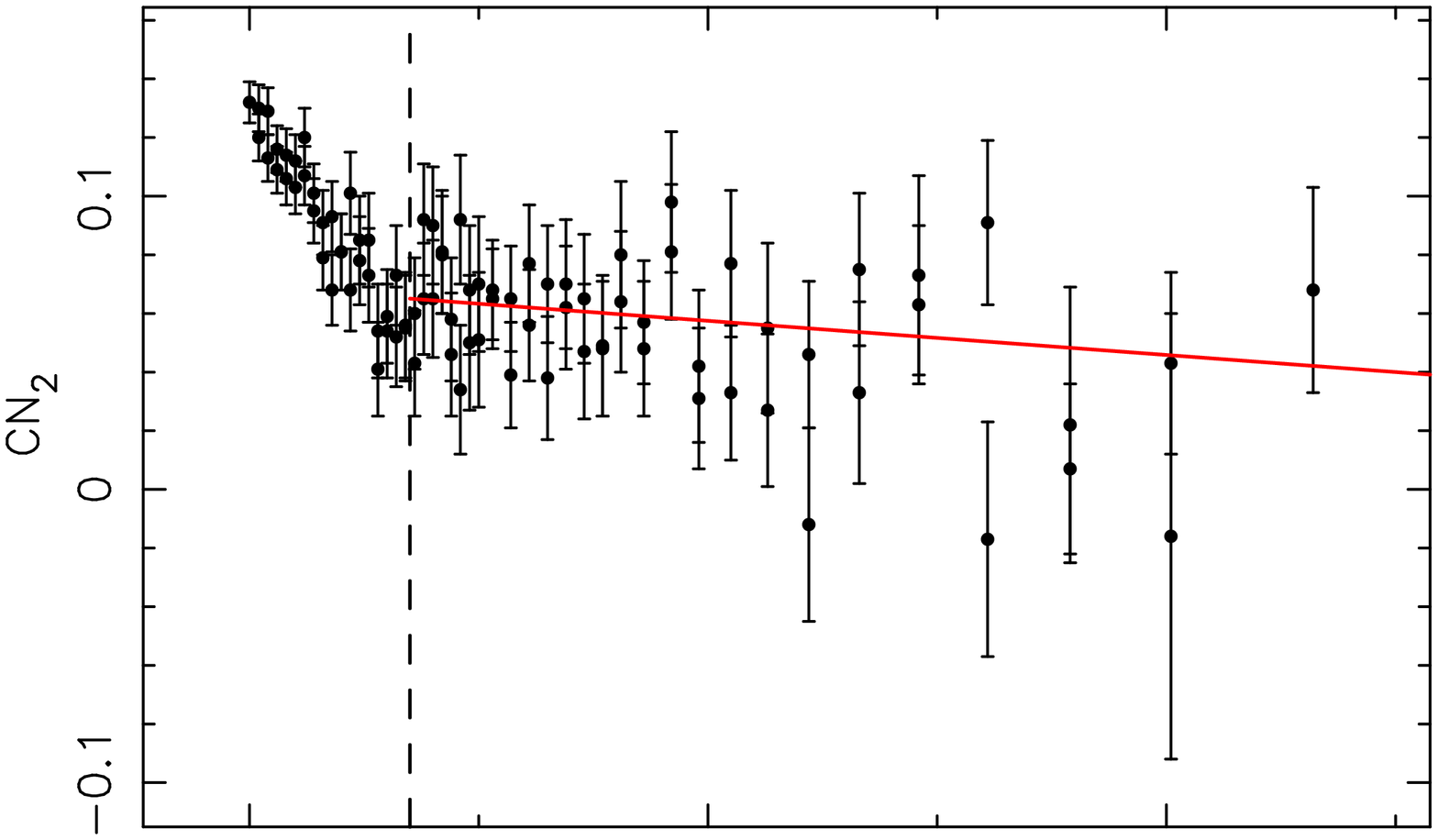}}
\resizebox{0.3\textwidth}{!}{\includegraphics[angle=-90]{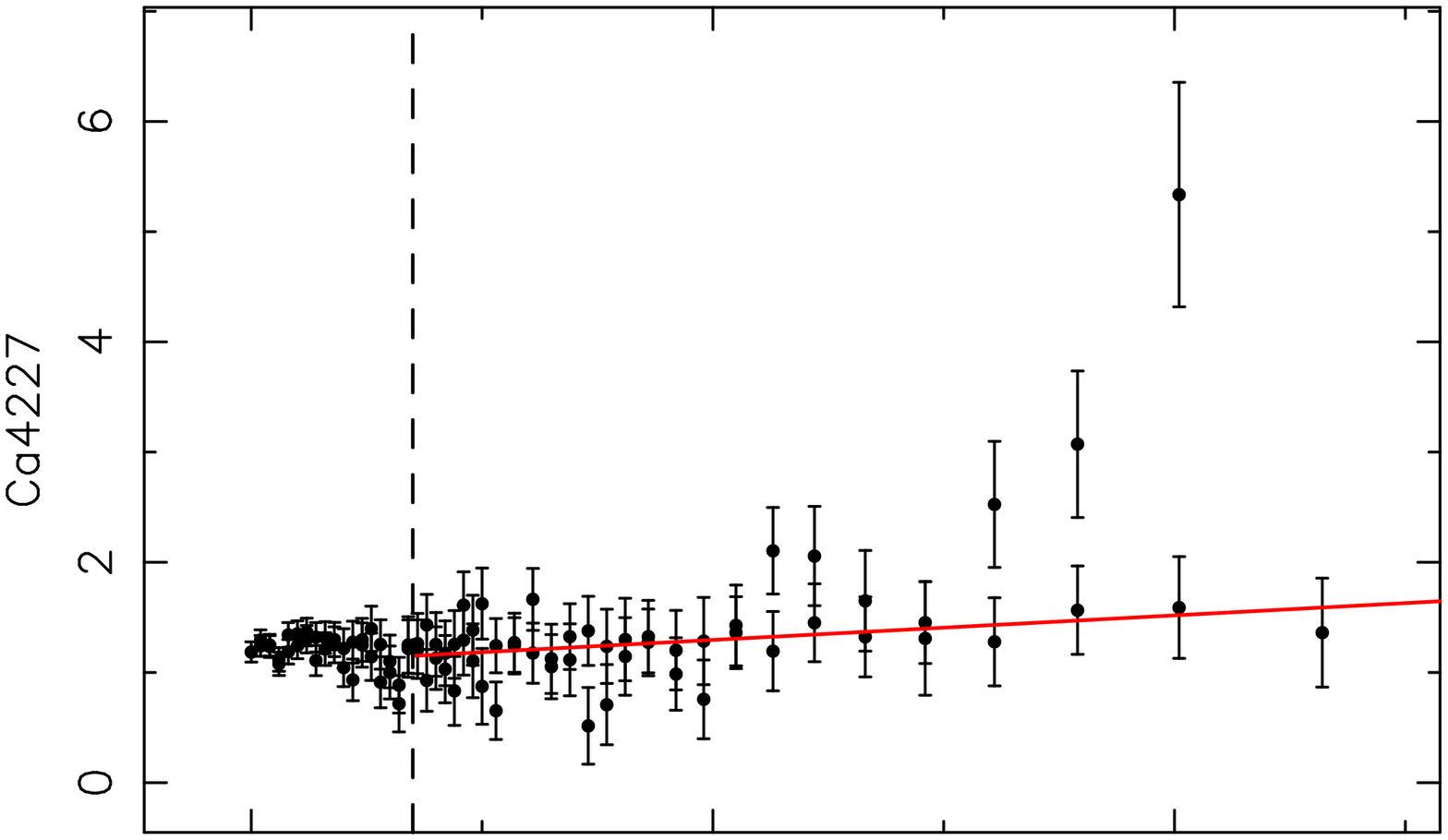}}
\resizebox{0.3\textwidth}{!}{\includegraphics[angle=-90]{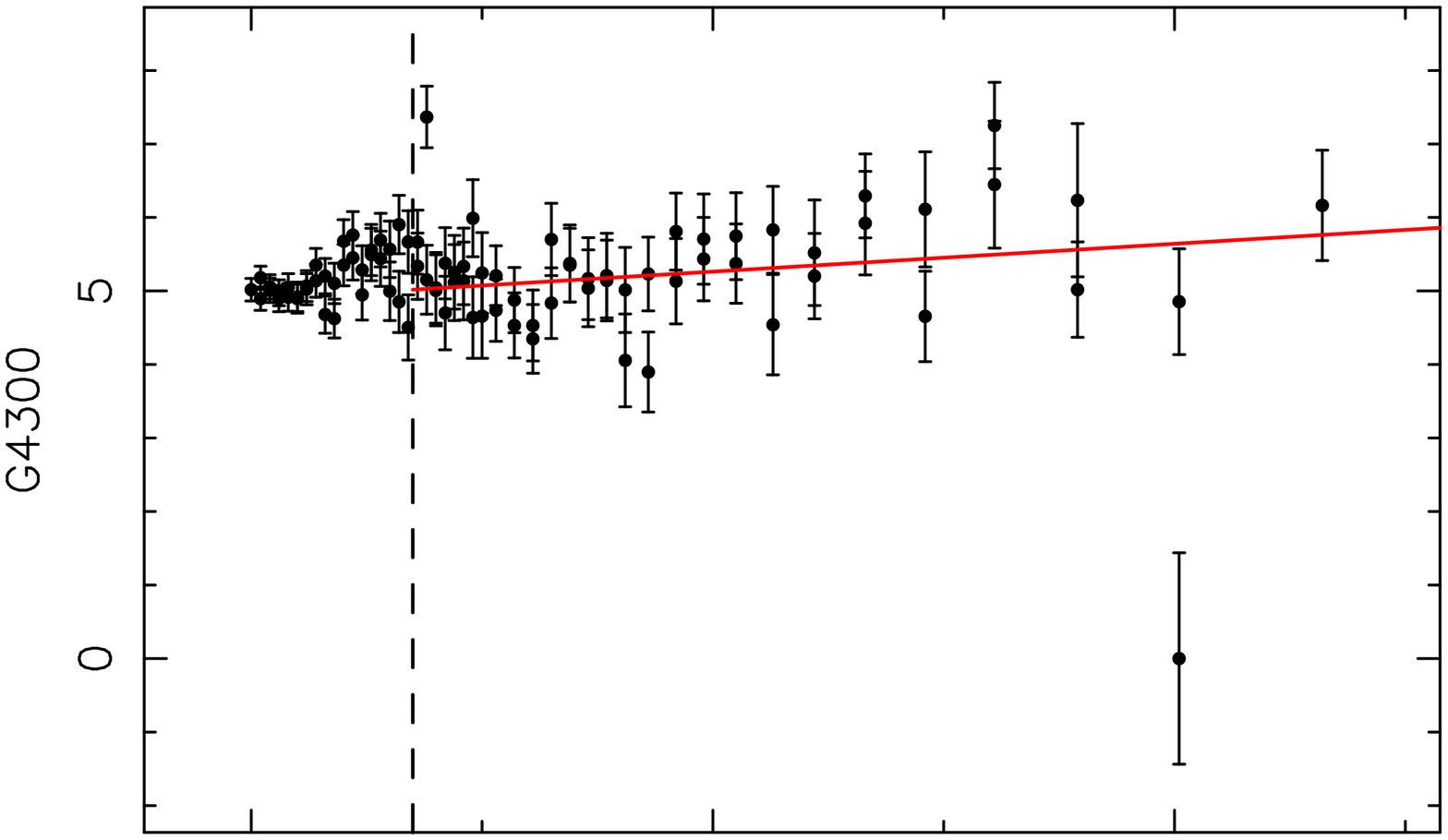}}
\resizebox{0.3\textwidth}{!}{\includegraphics[angle=-90]{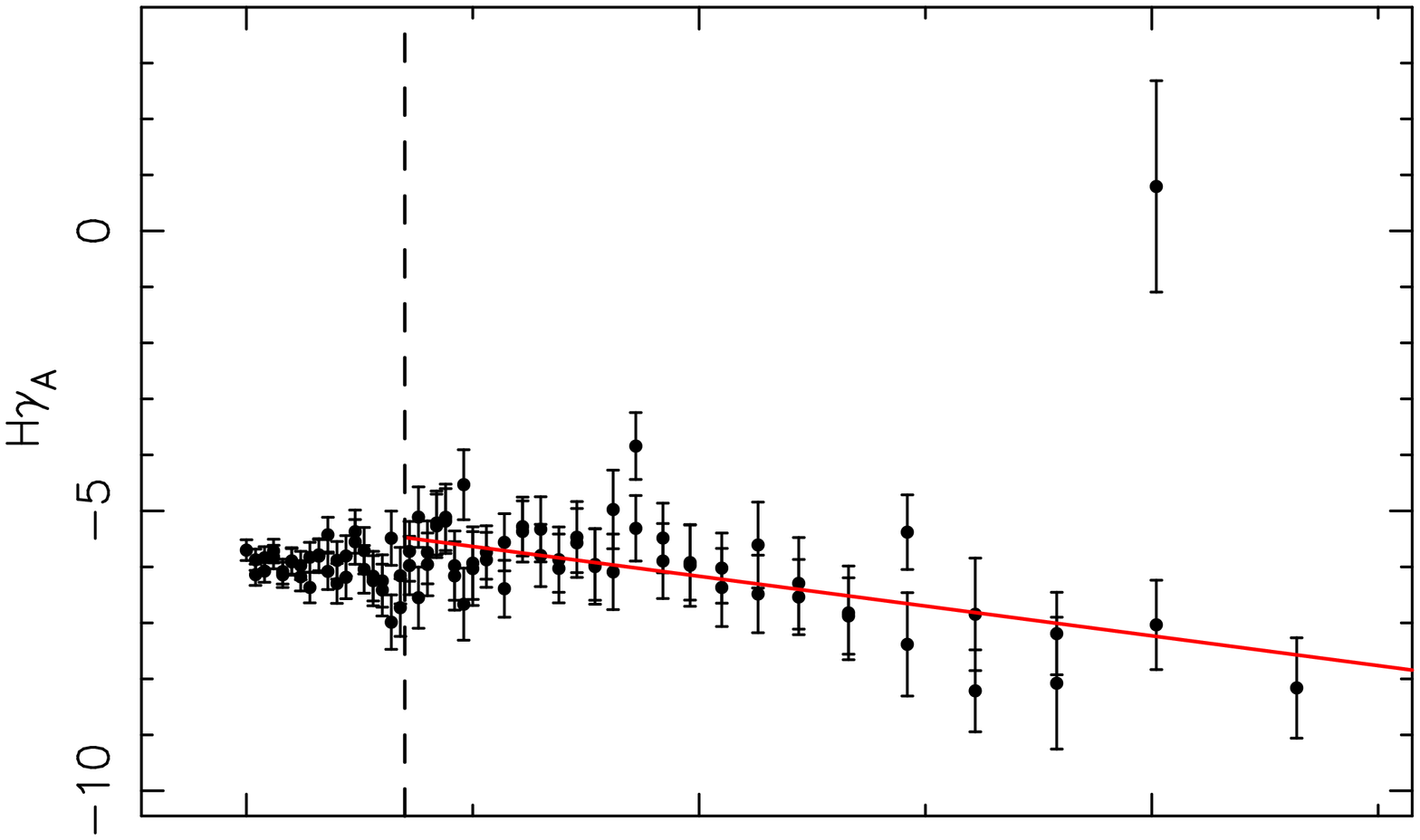}}
\resizebox{0.3\textwidth}{!}{\includegraphics[angle=-90]{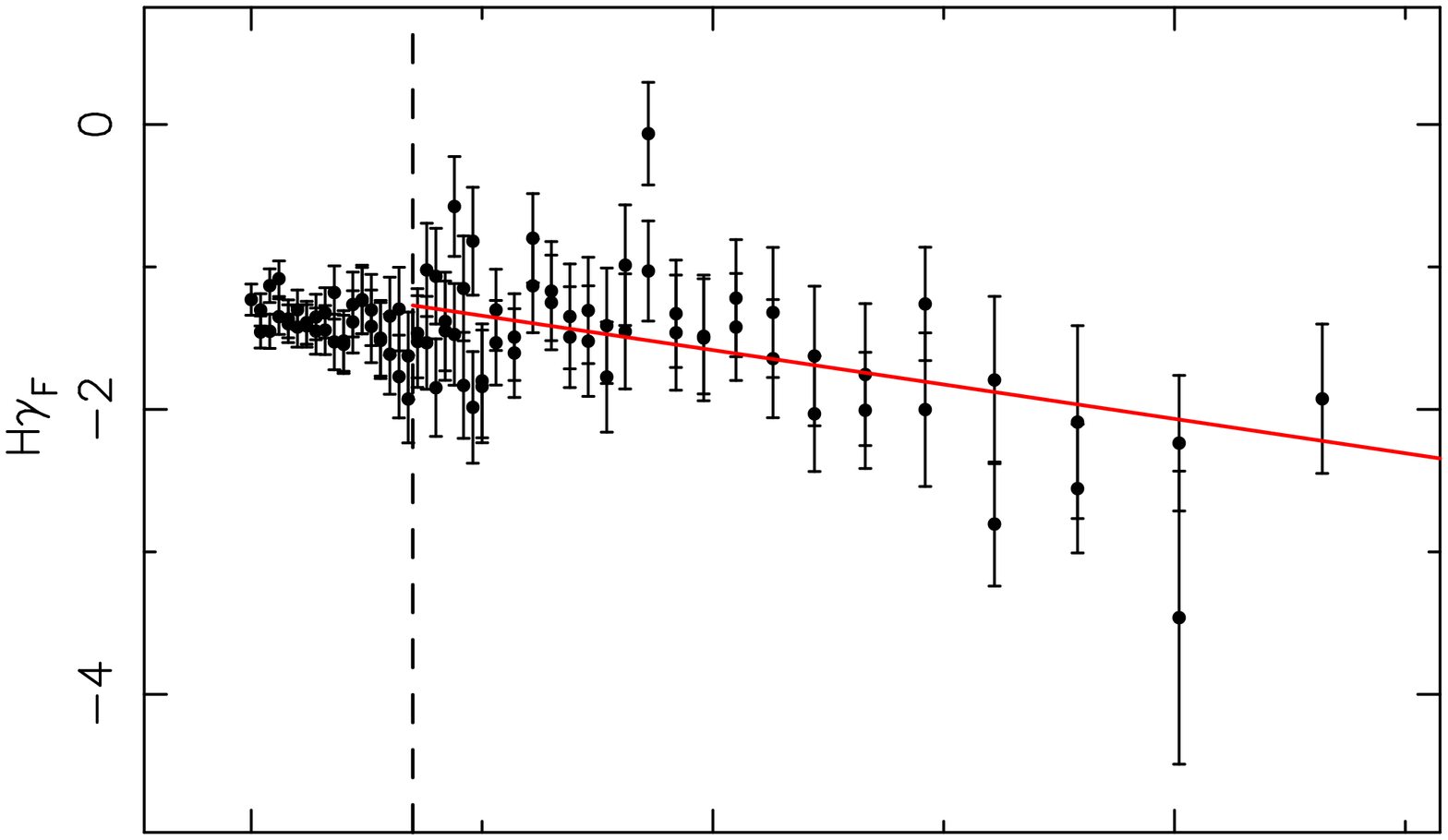}}
\resizebox{0.3\textwidth}{!}{\includegraphics[angle=-90]{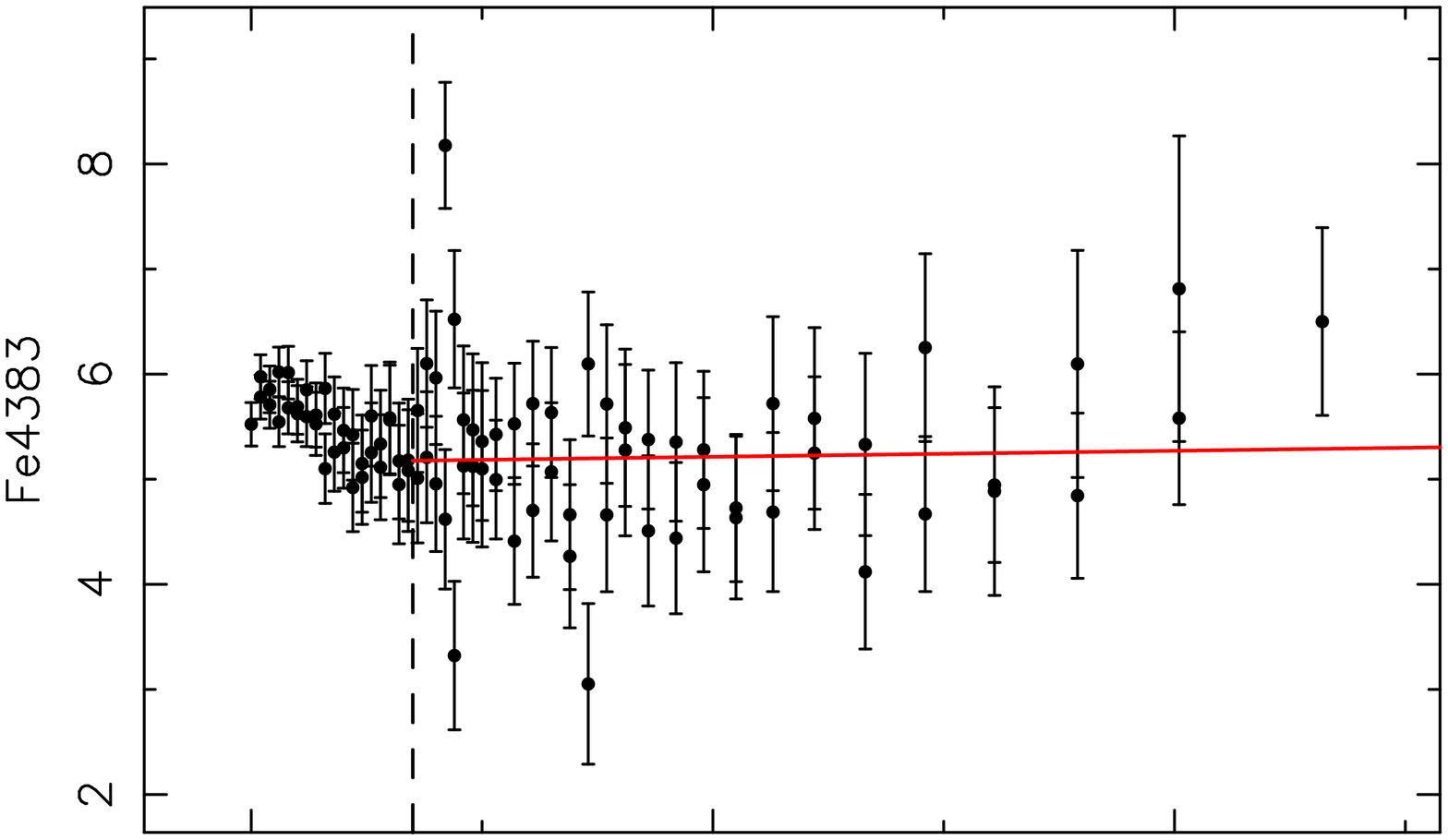}}
\resizebox{0.3\textwidth}{!}{\includegraphics[angle=-90]{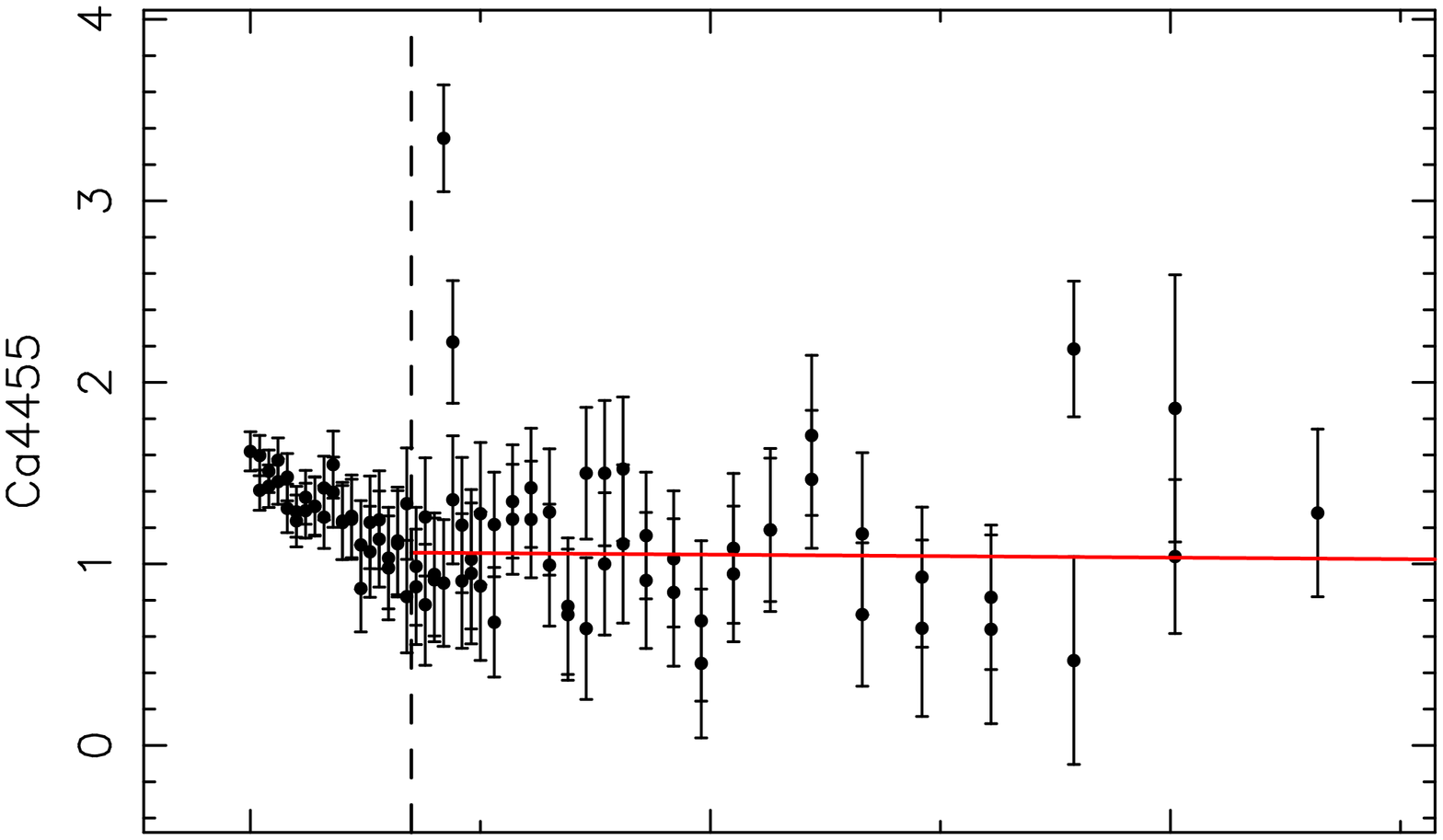}}
\resizebox{0.3\textwidth}{!}{\includegraphics[angle=-90]{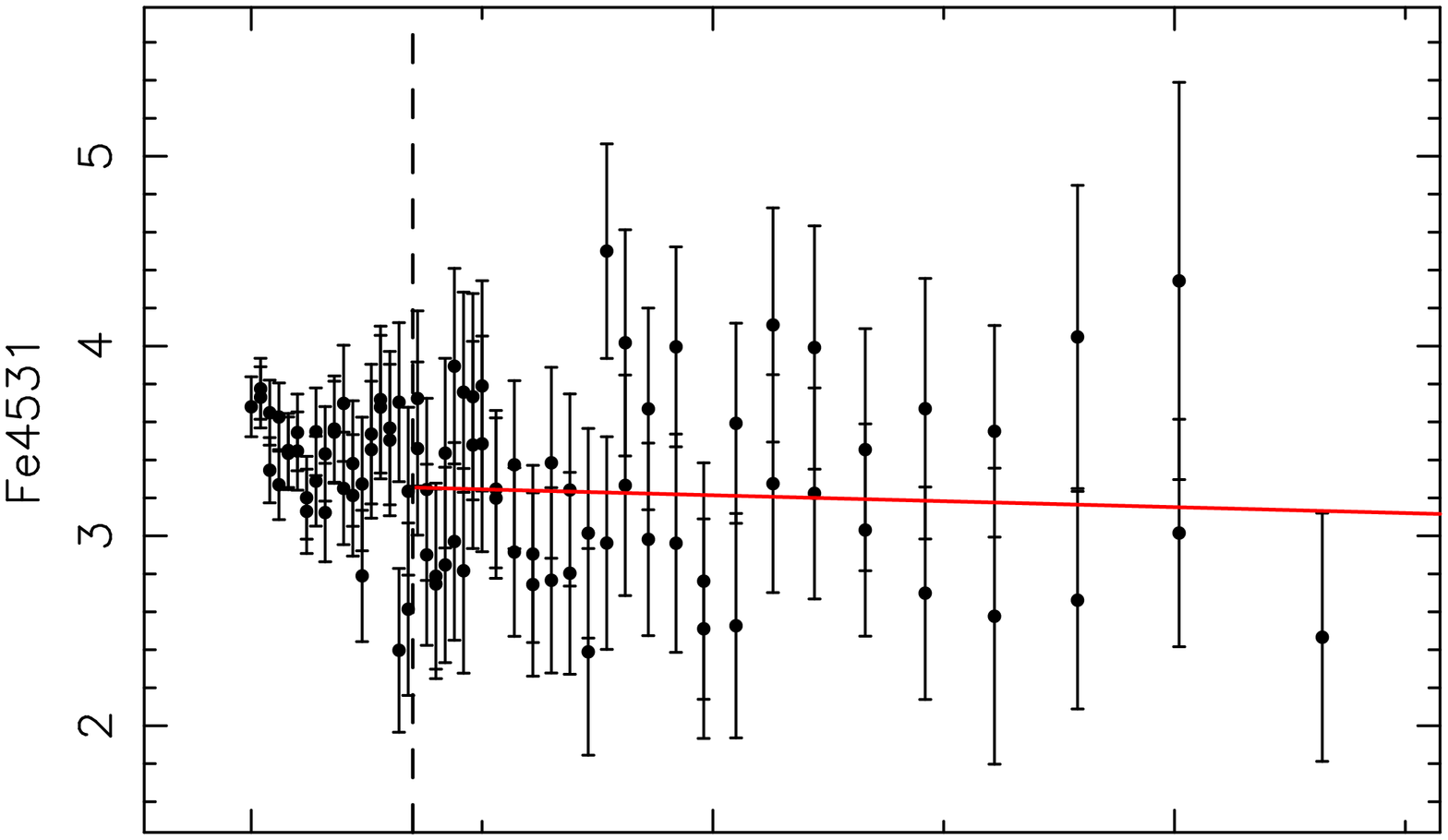}}
\resizebox{0.3\textwidth}{!}{\includegraphics[angle=-90]{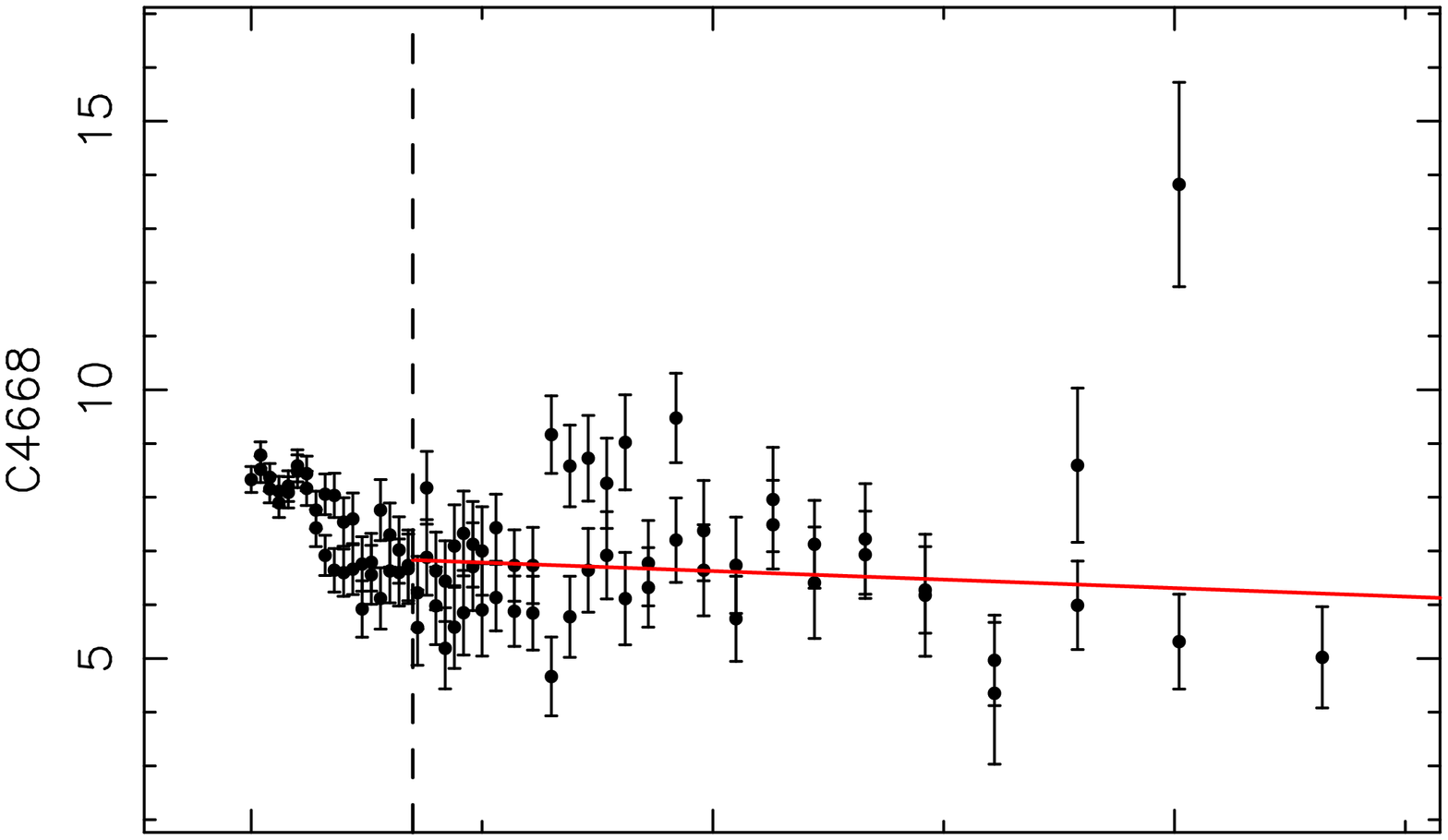}}
\resizebox{0.3\textwidth}{!}{\includegraphics[angle=-90]{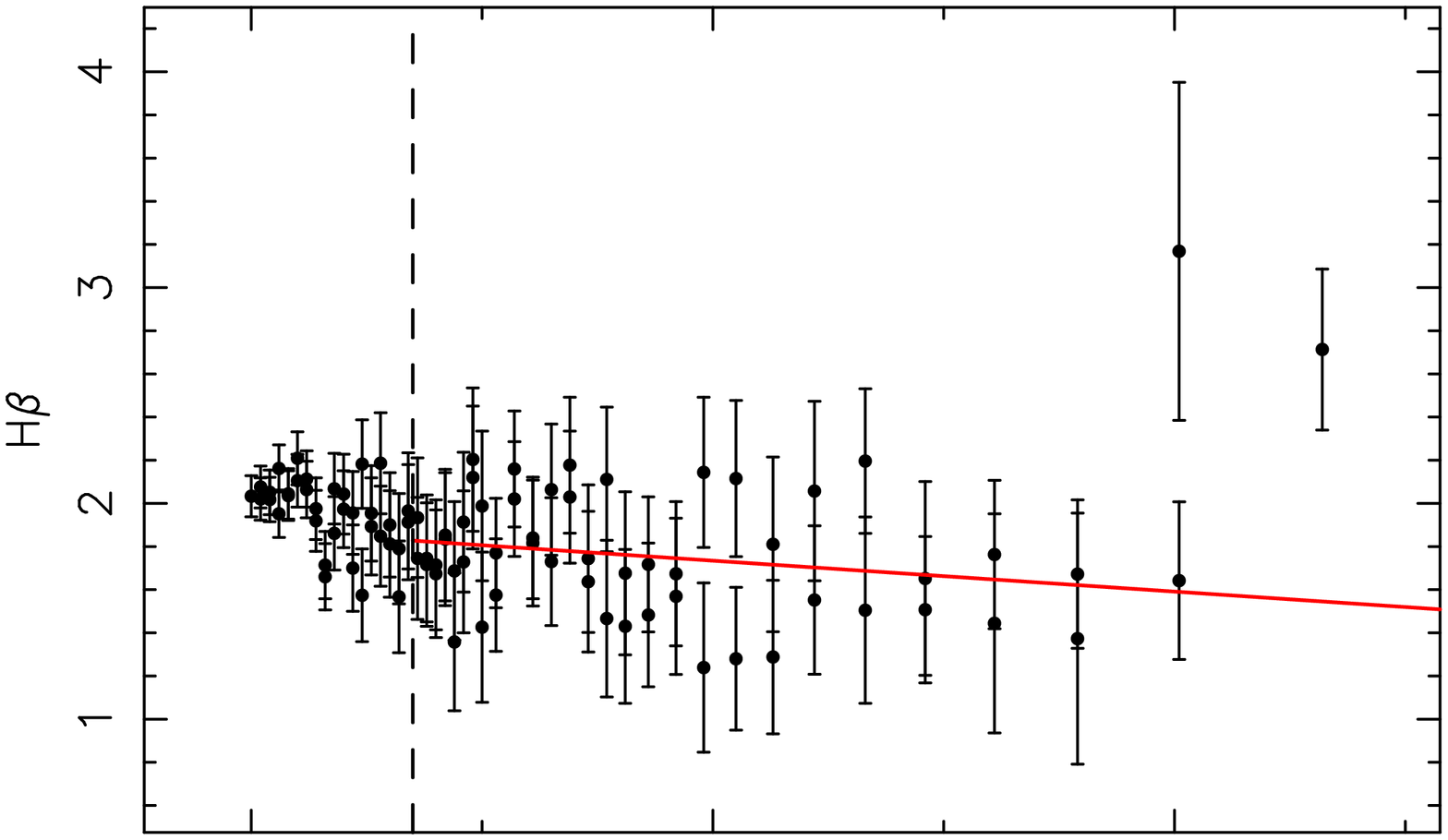}}
\resizebox{0.3\textwidth}{!}{\includegraphics[angle=-90]{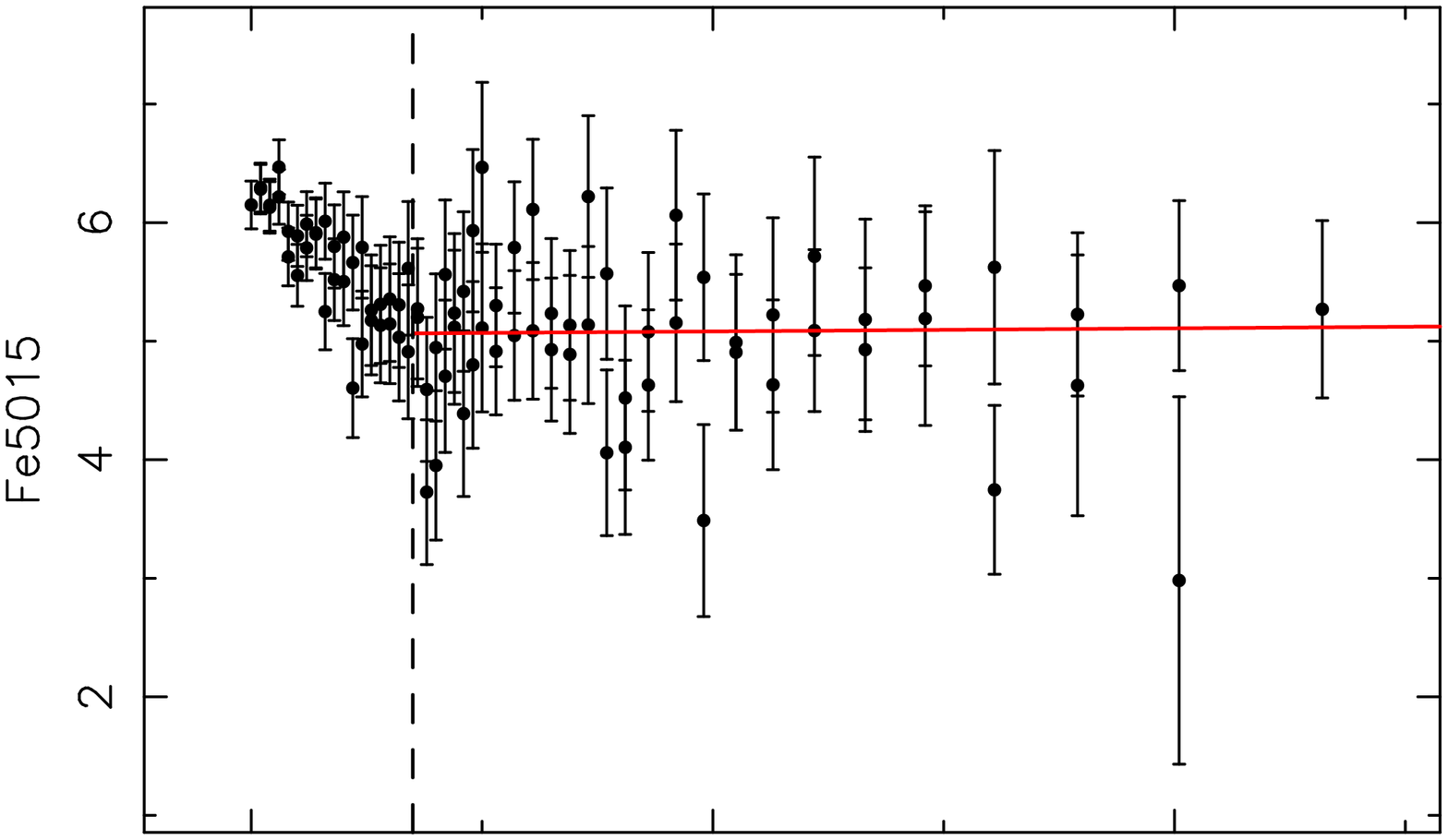}}
\resizebox{0.3\textwidth}{!}{\includegraphics[angle=-90]{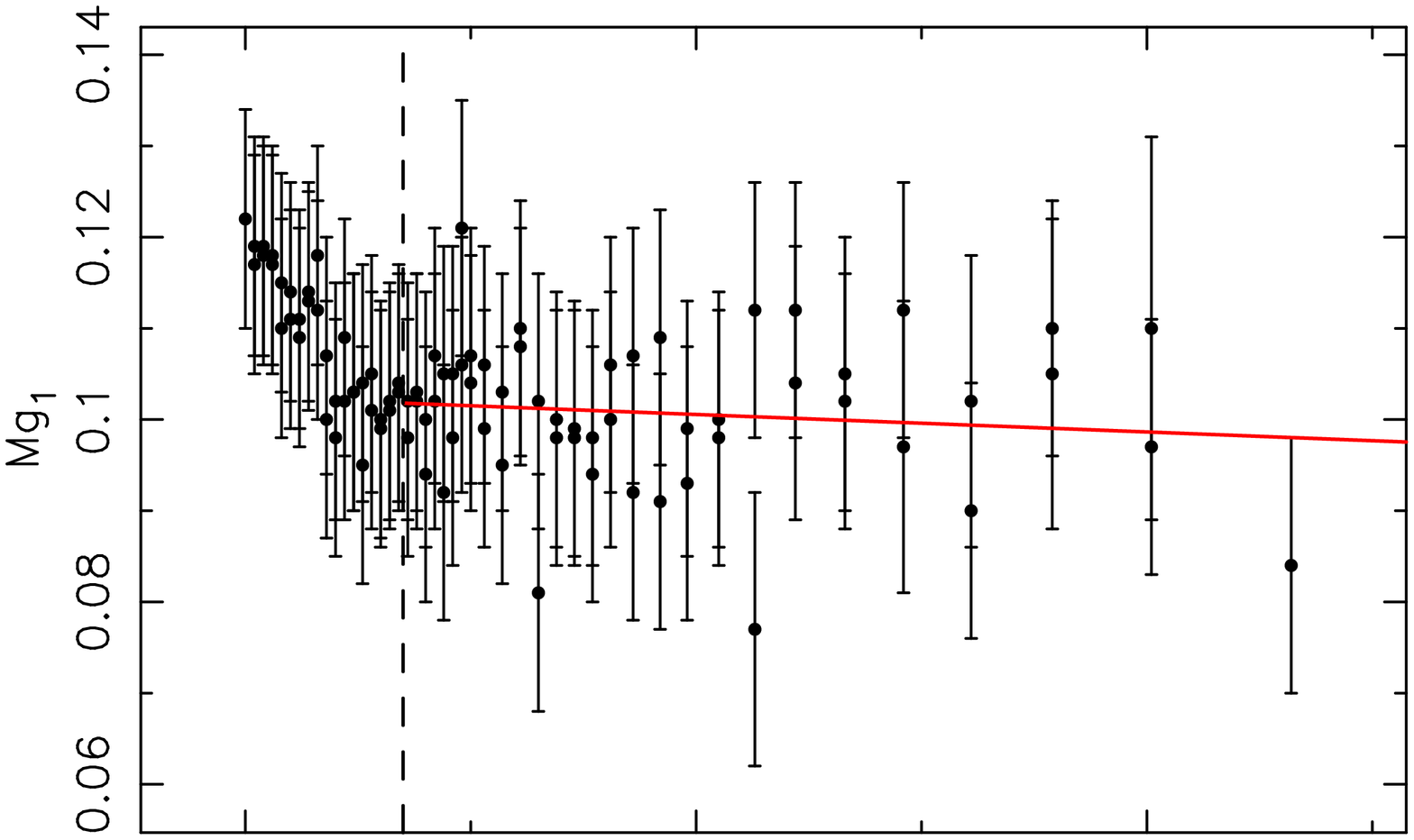}}
\resizebox{0.3\textwidth}{!}{\includegraphics[angle=-90]{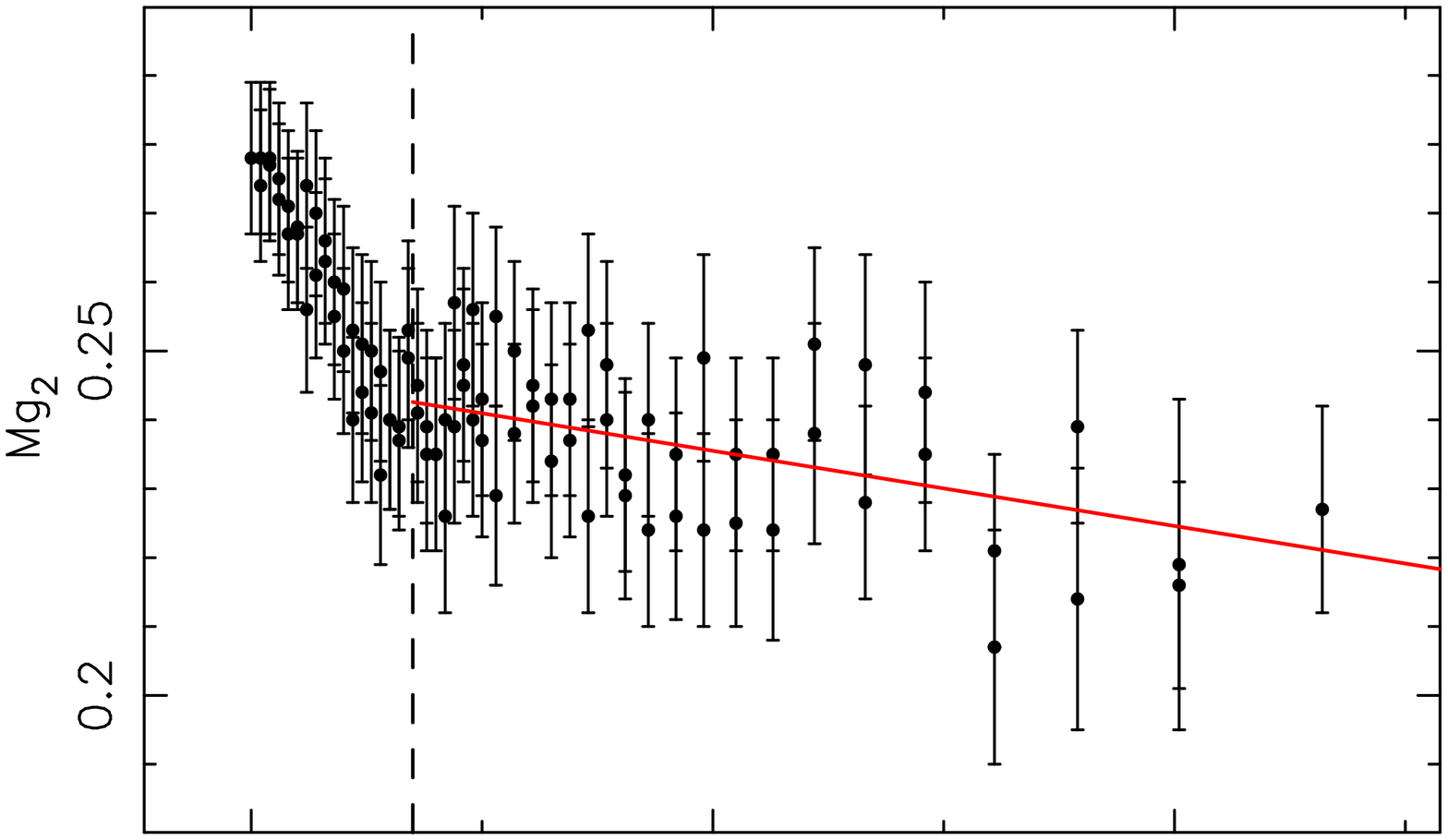}}
\resizebox{0.3\textwidth}{!}{\includegraphics[angle=-90]{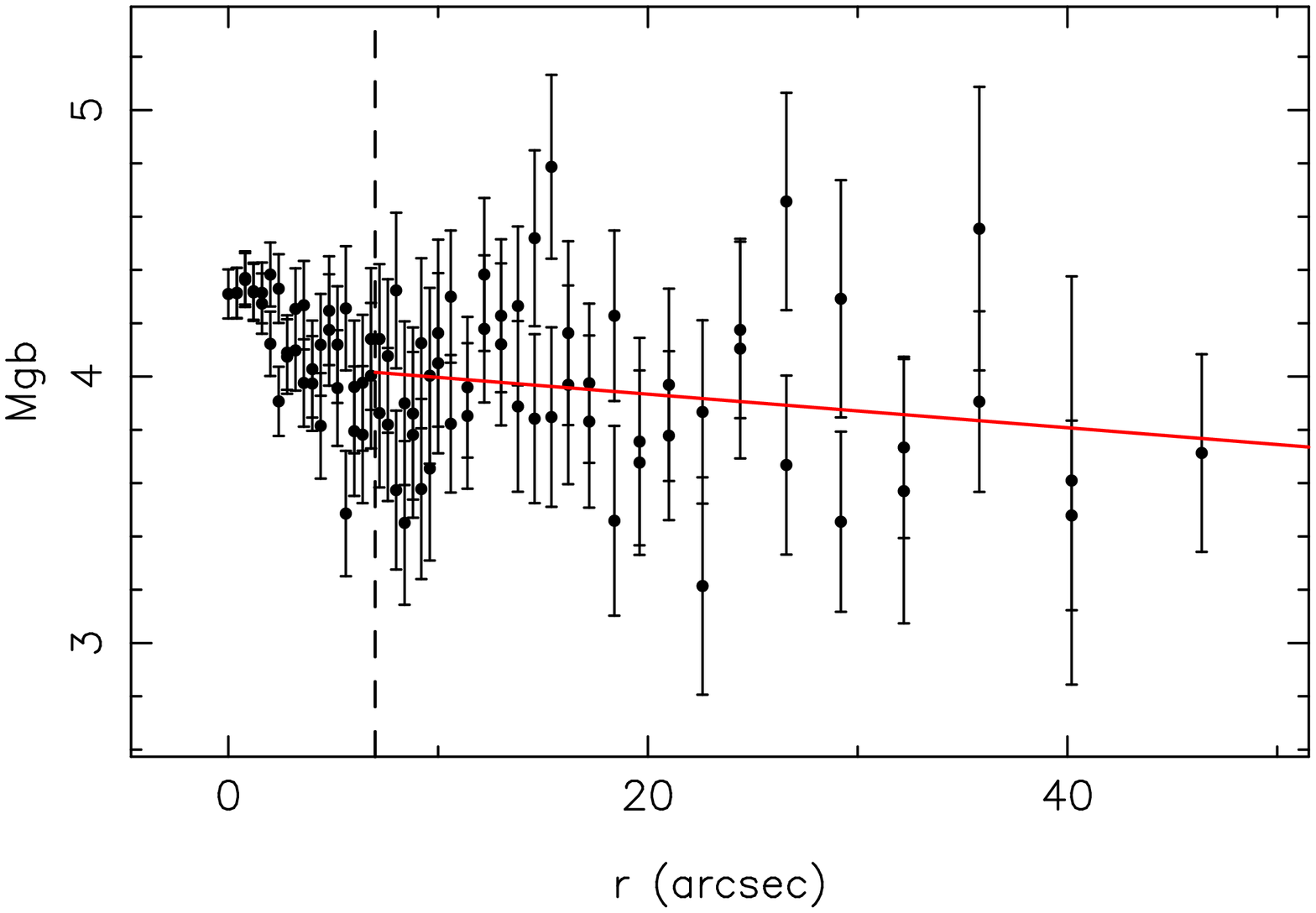}}\hspace{0.85cm}
\resizebox{0.3\textwidth}{!}{\includegraphics[angle=-90]{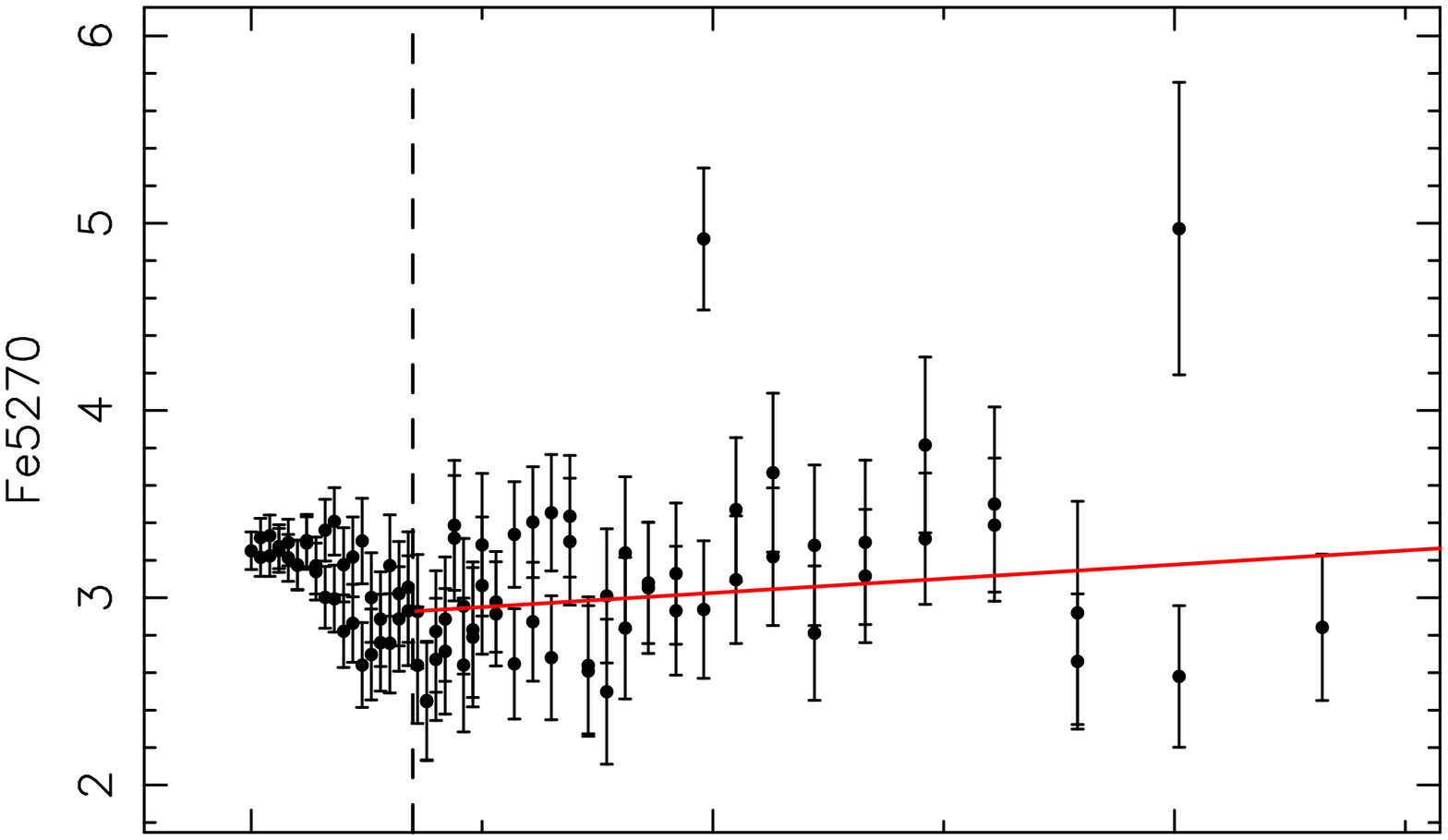}}\hspace{0.85cm}
\resizebox{0.3\textwidth}{!}{\includegraphics[angle=-90]{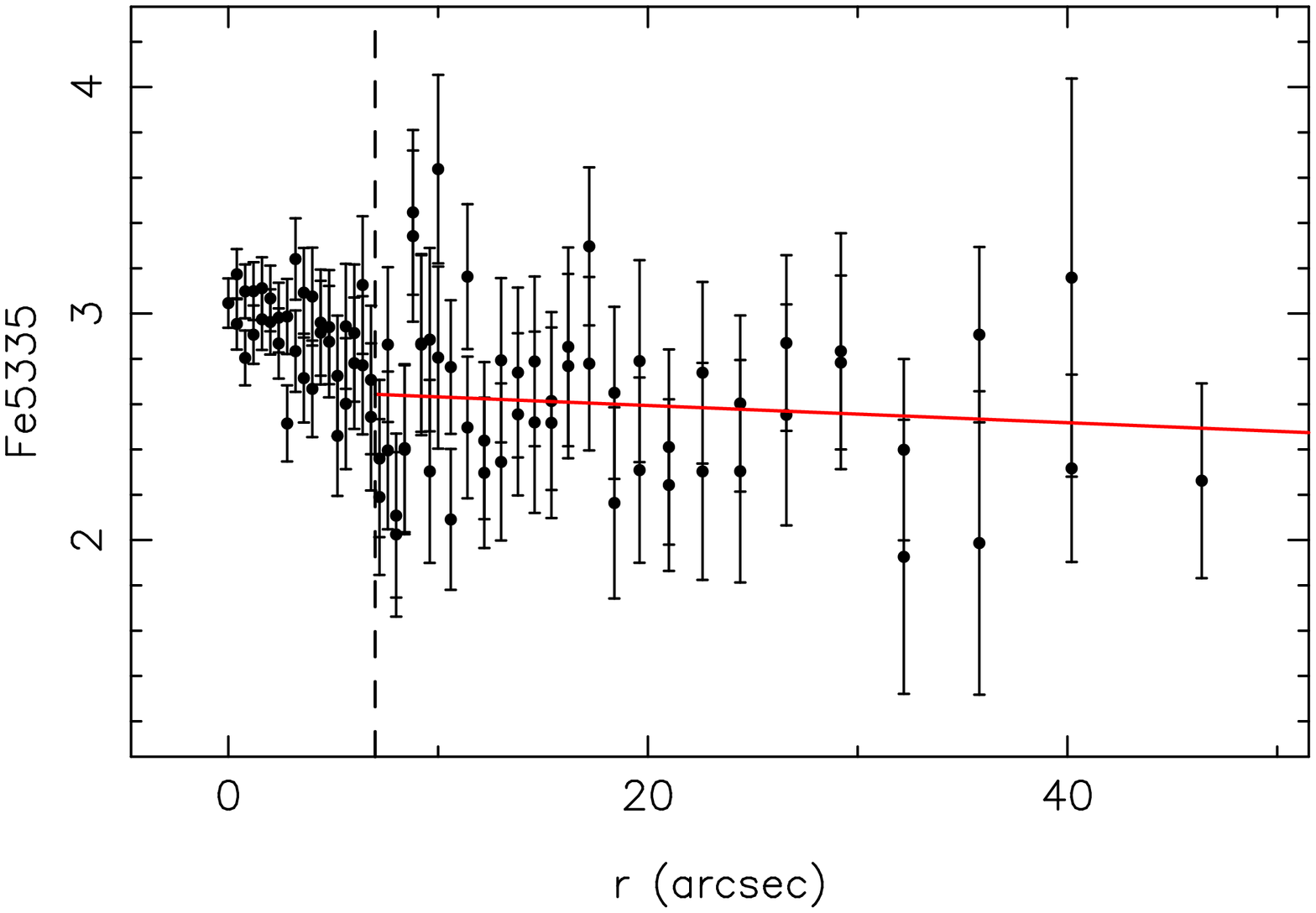}}
\caption{Line-strength distribution in the bar region for all the galaxies}
\end{figure*}

\begin{figure*}
\addtocounter{figure}{-1}
\resizebox{0.3\textwidth}{!}{\includegraphics[angle=-90]{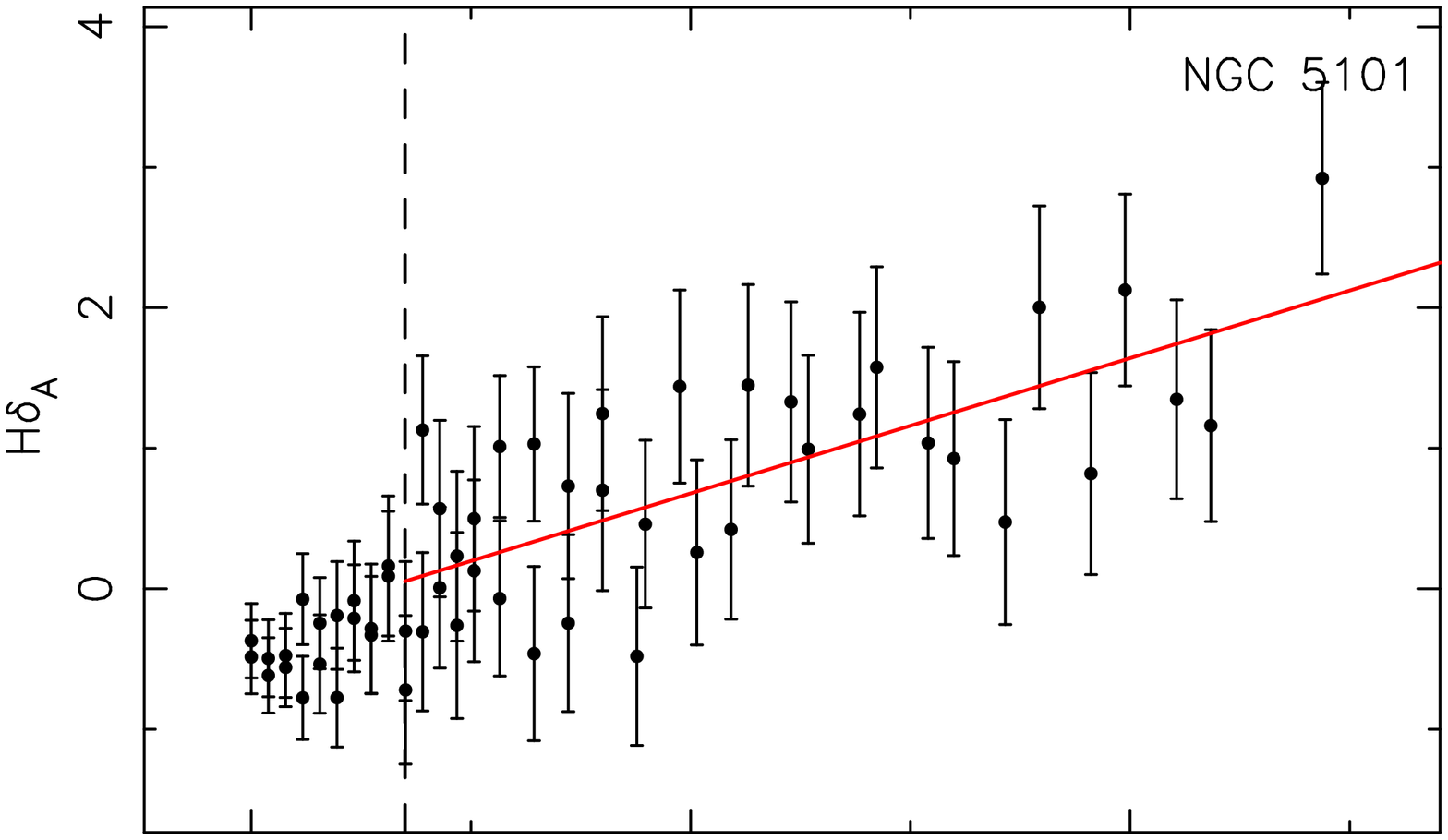}}
\resizebox{0.3\textwidth}{!}{\includegraphics[angle=-90]{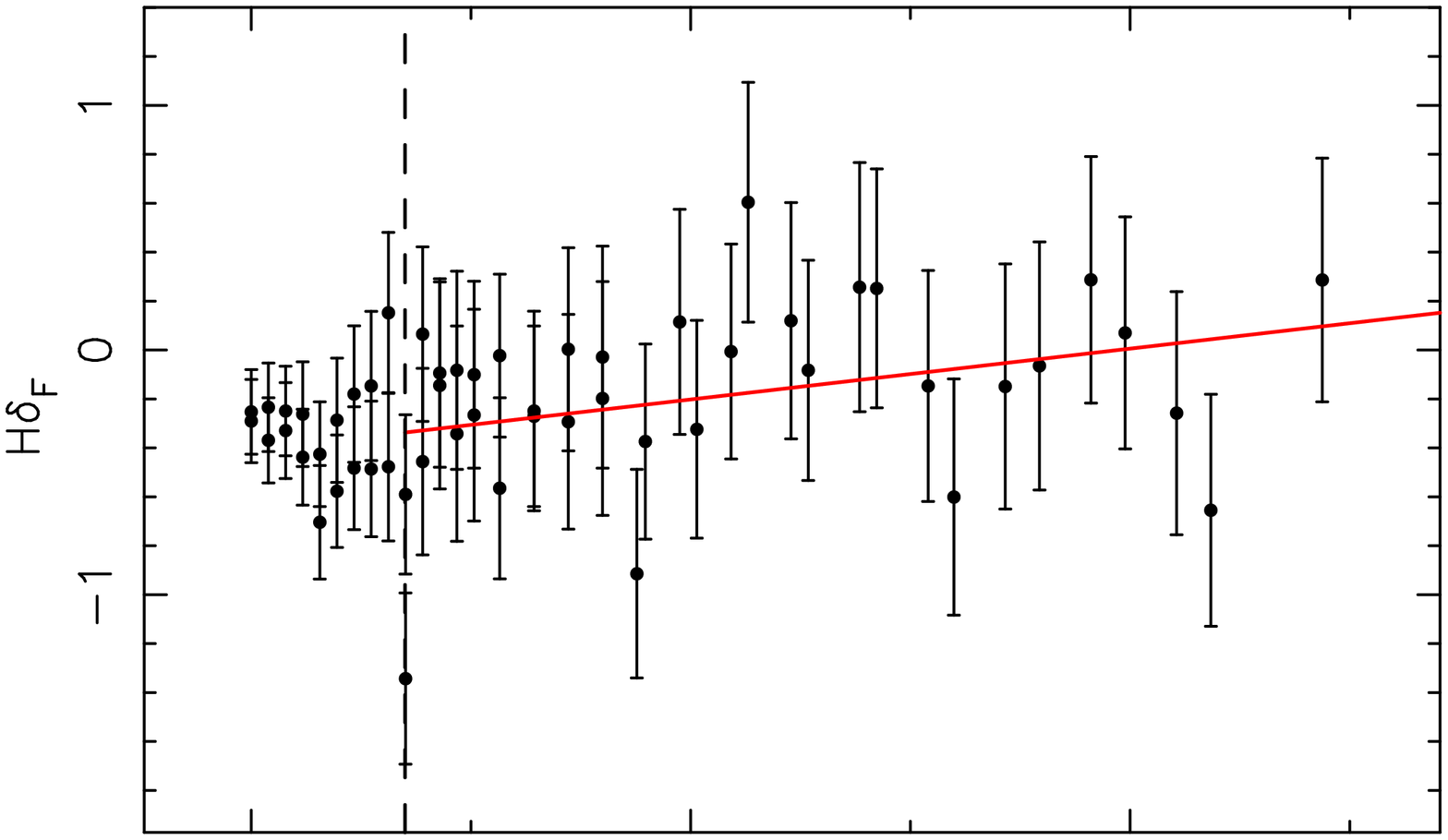}}
\resizebox{0.3\textwidth}{!}{\includegraphics[angle=-90]{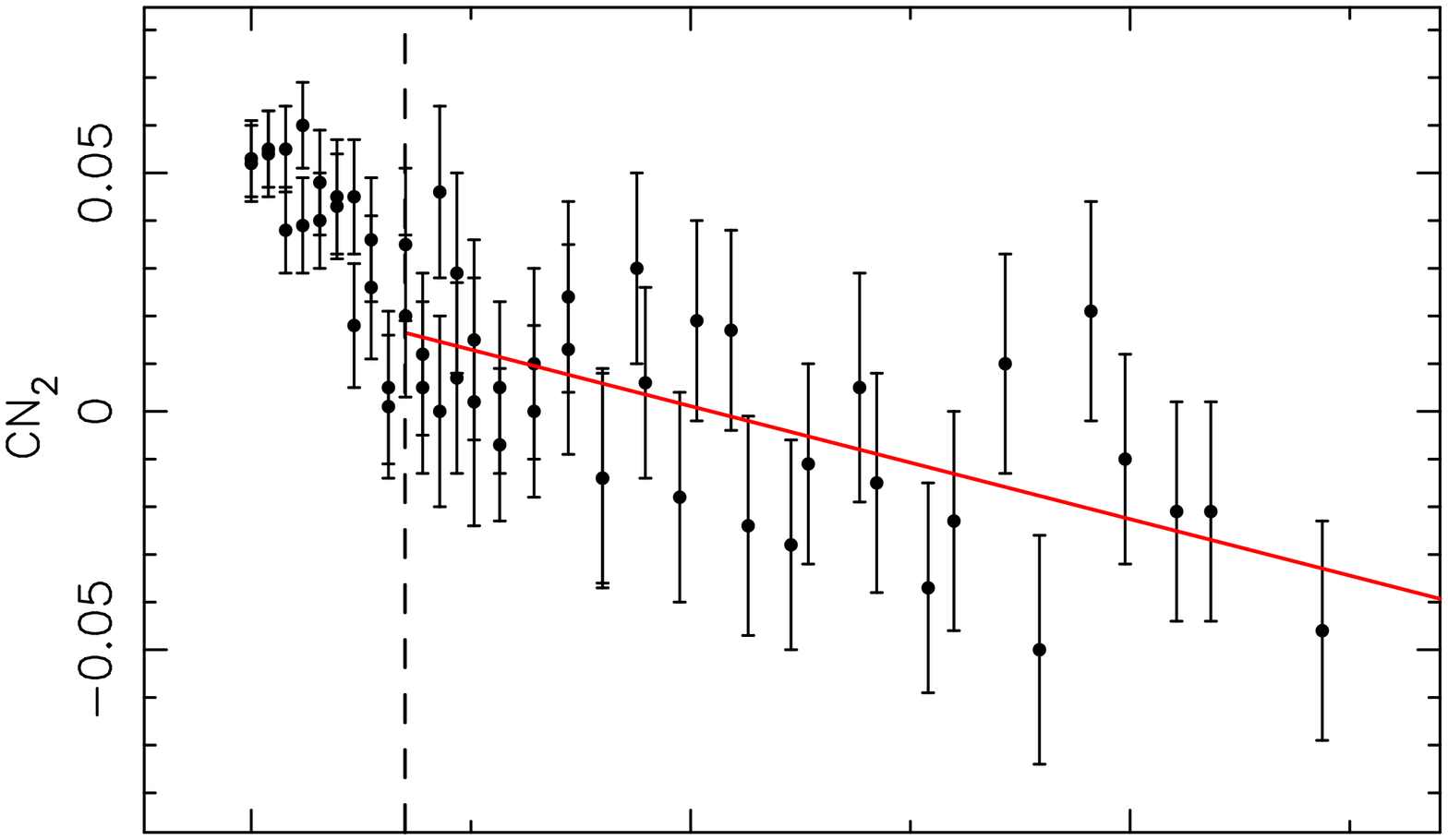}}
\resizebox{0.3\textwidth}{!}{\includegraphics[angle=-90]{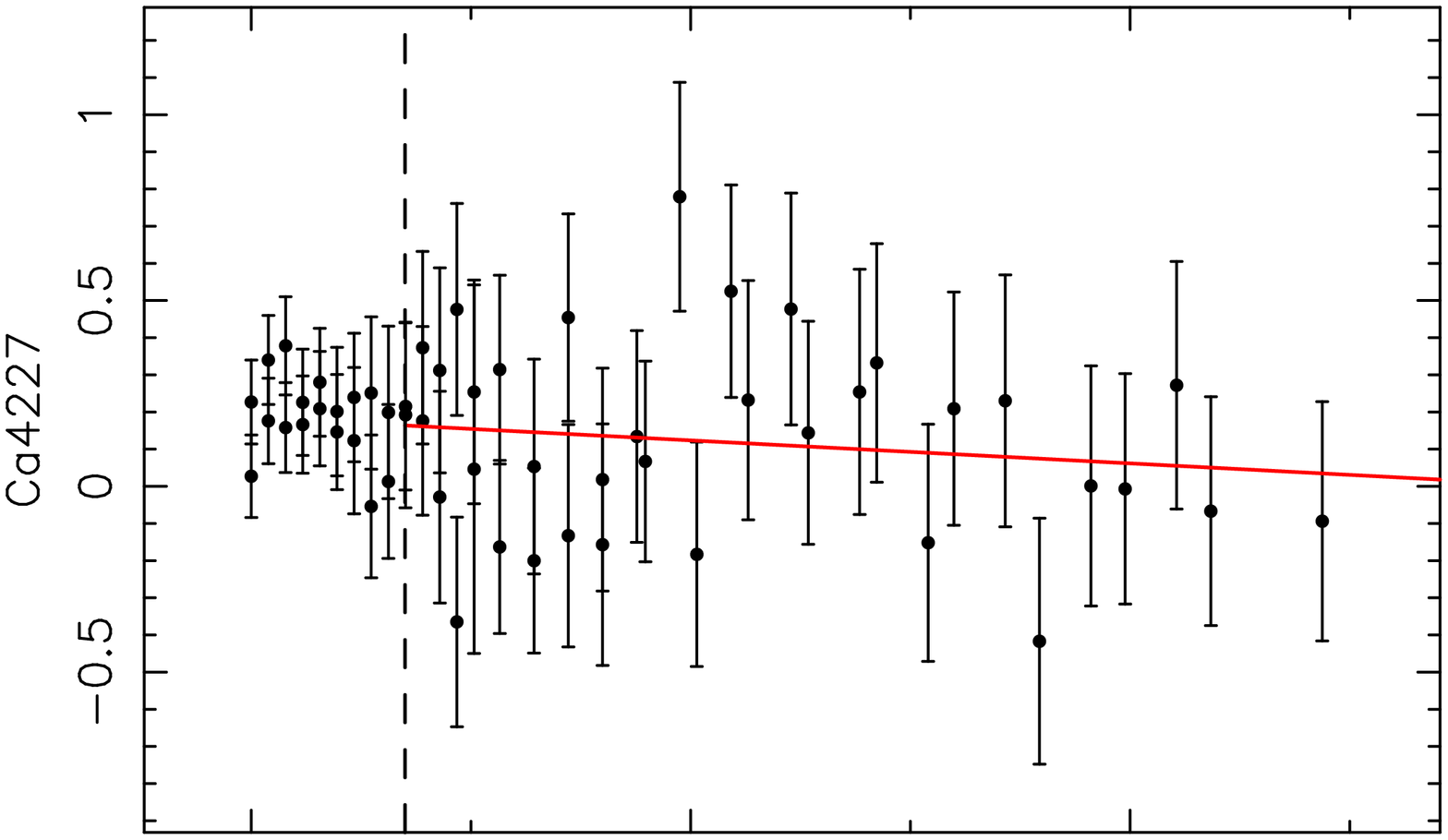}}
\resizebox{0.3\textwidth}{!}{\includegraphics[angle=-90]{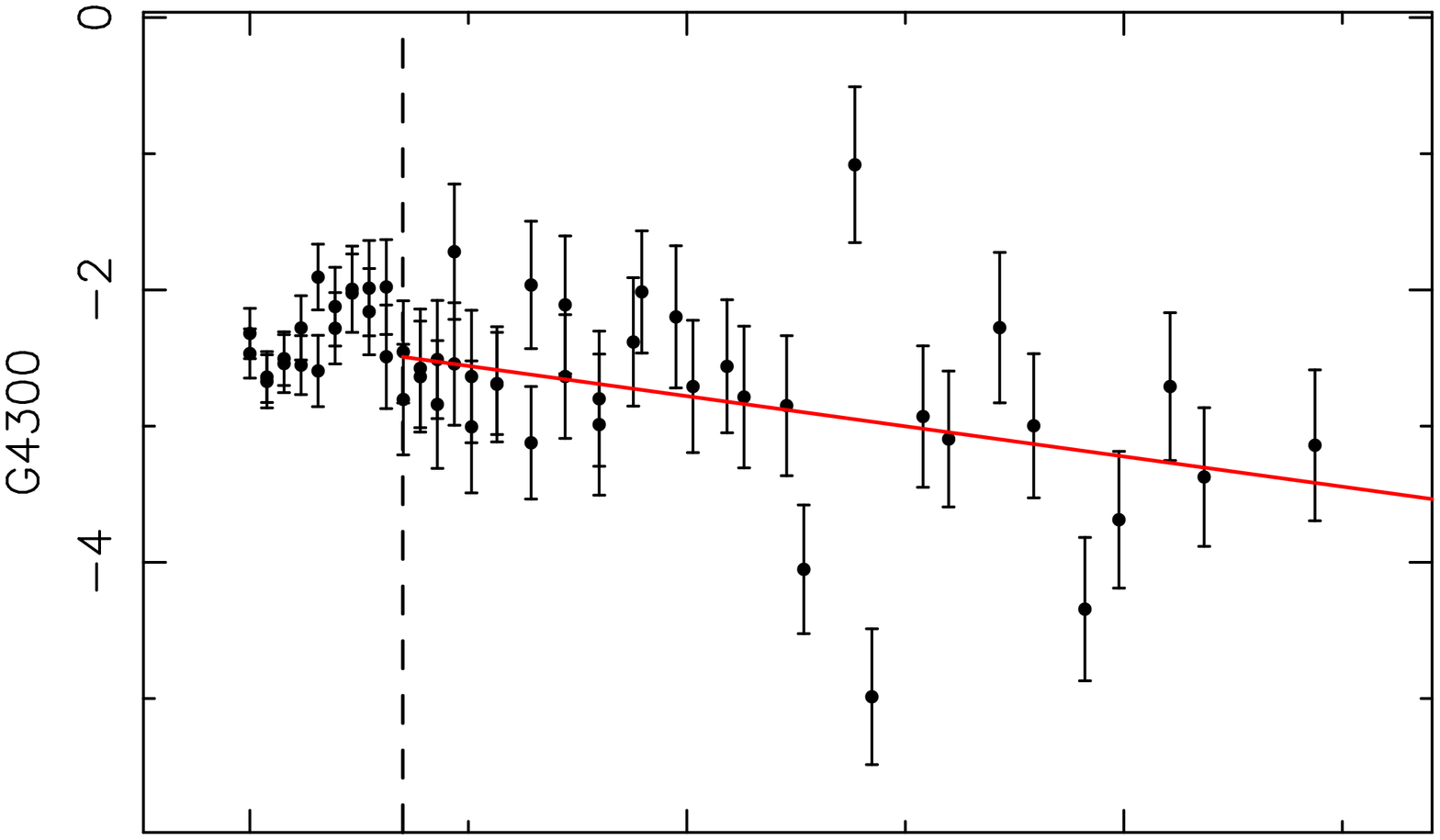}}
\resizebox{0.3\textwidth}{!}{\includegraphics[angle=-90]{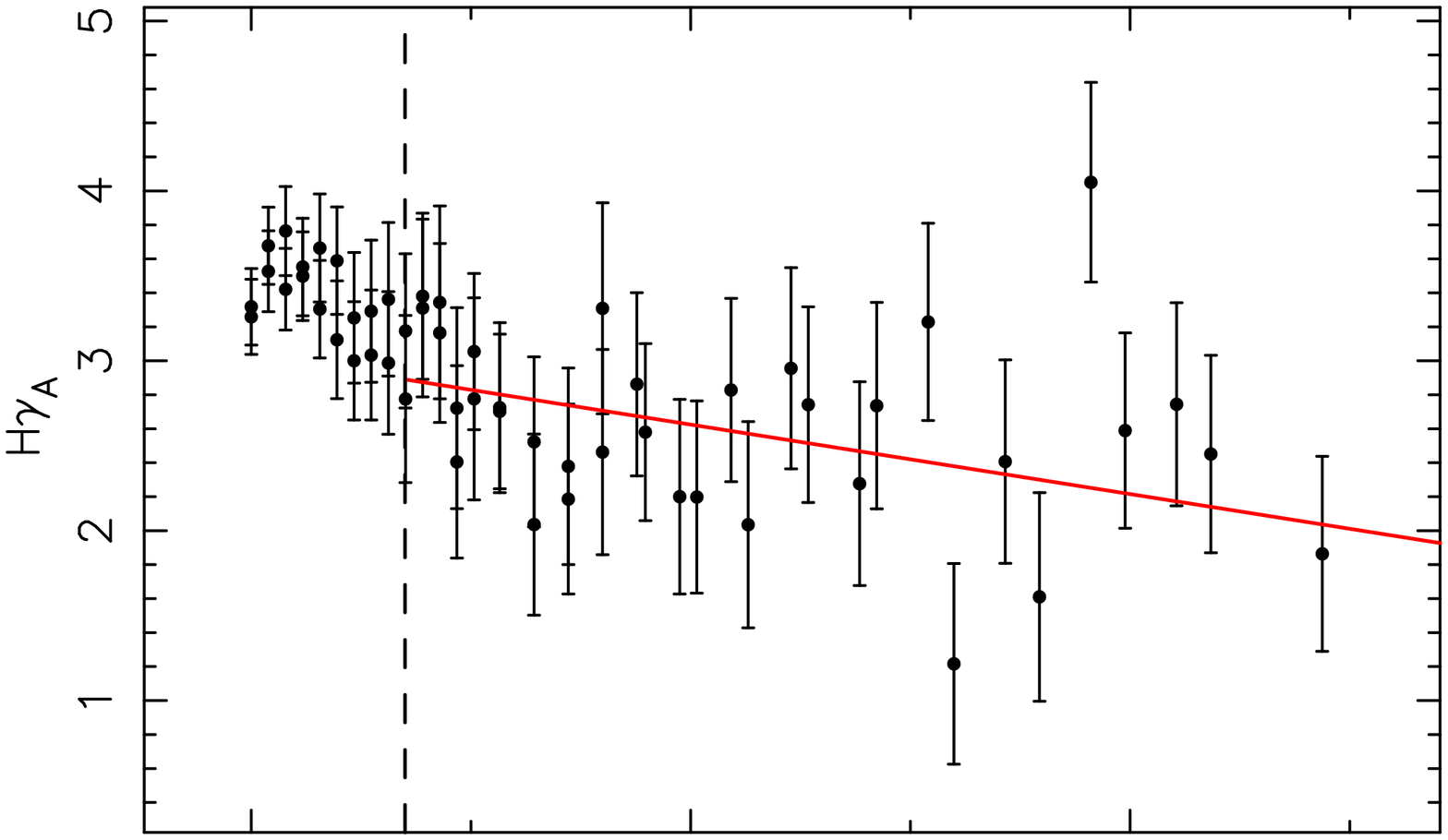}}
\resizebox{0.3\textwidth}{!}{\includegraphics[angle=-90]{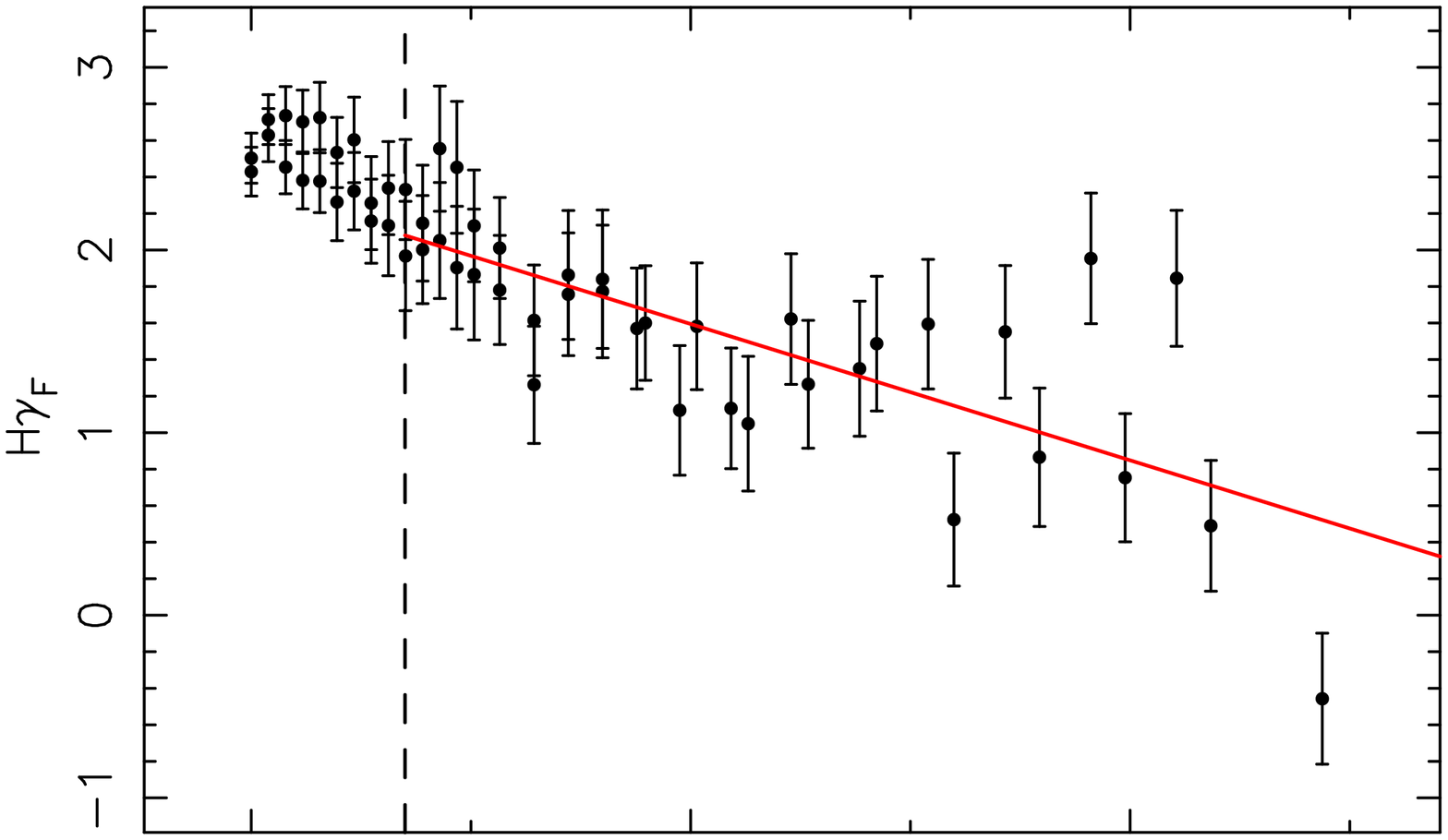}}
\resizebox{0.3\textwidth}{!}{\includegraphics[angle=-90]{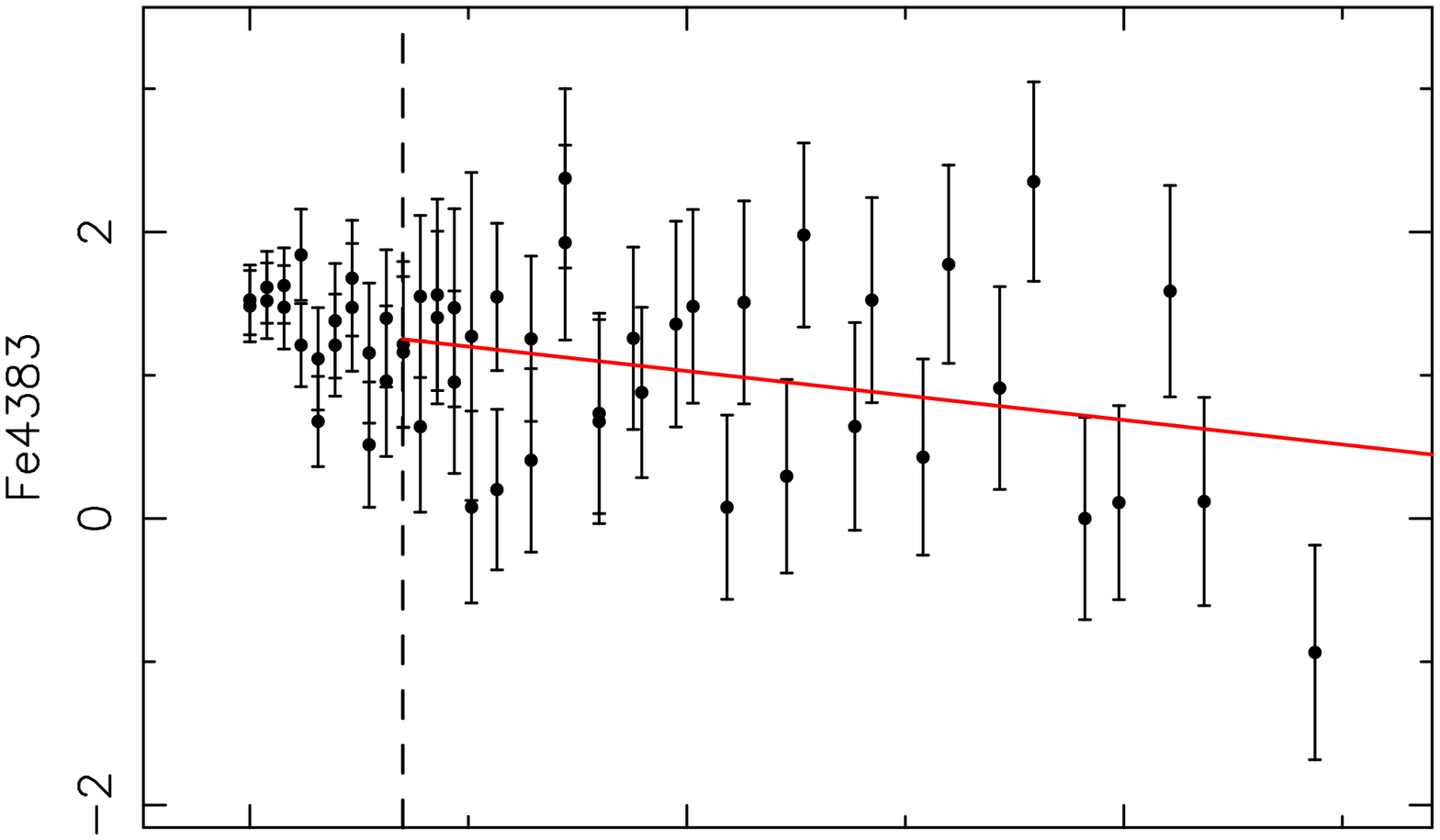}}
\resizebox{0.3\textwidth}{!}{\includegraphics[angle=-90]{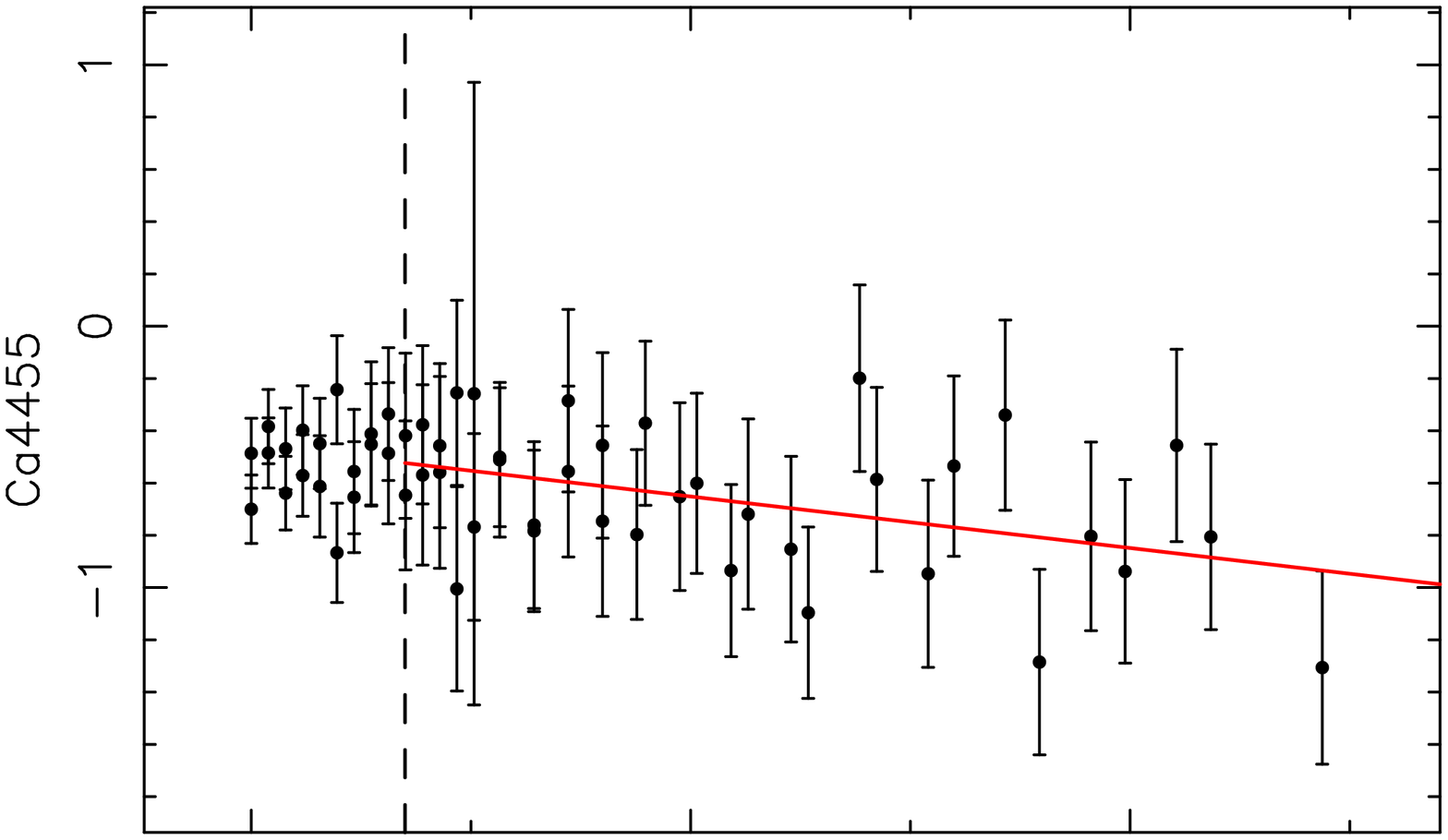}}
\resizebox{0.3\textwidth}{!}{\includegraphics[angle=-90]{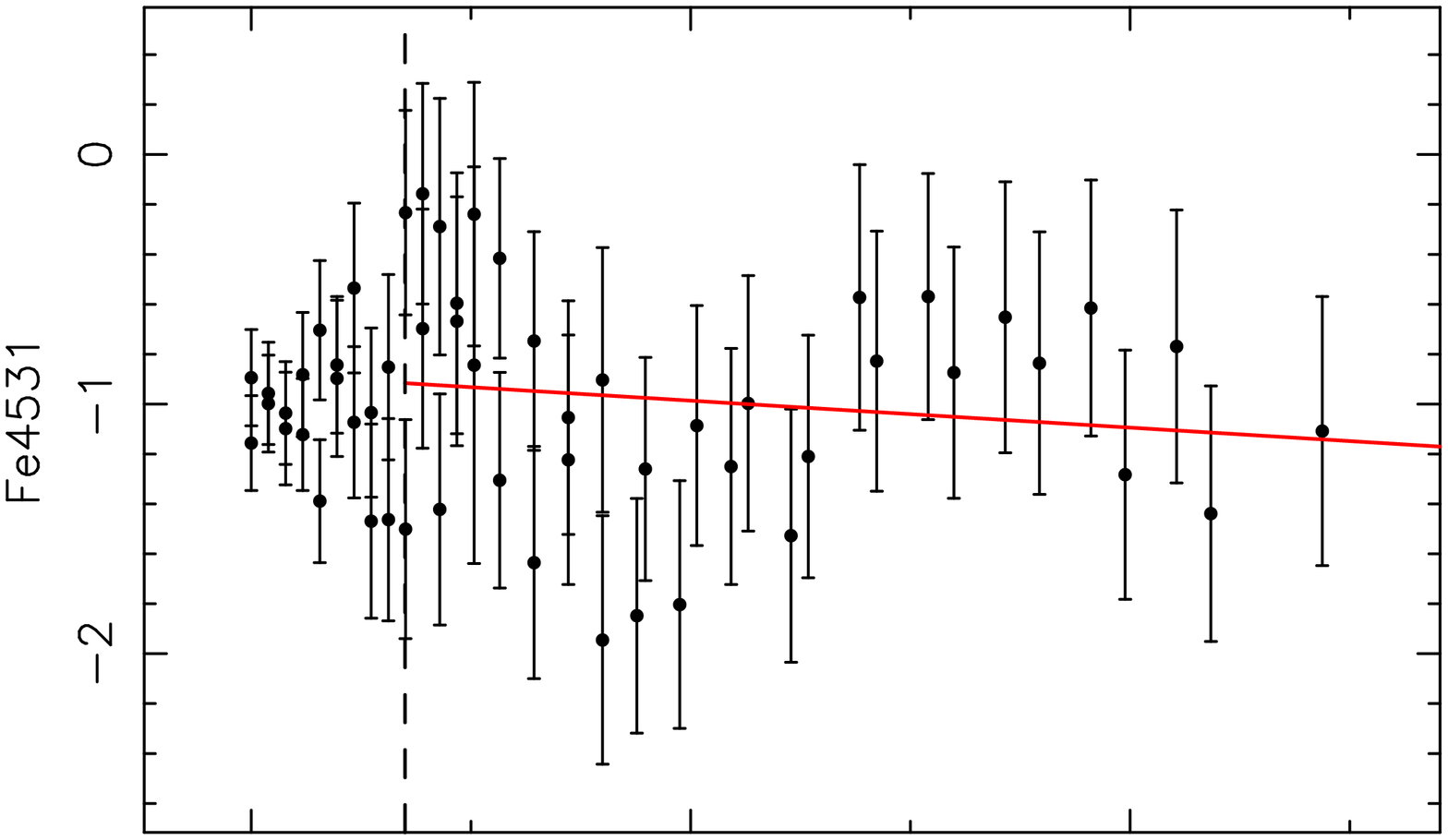}}
\resizebox{0.3\textwidth}{!}{\includegraphics[angle=-90]{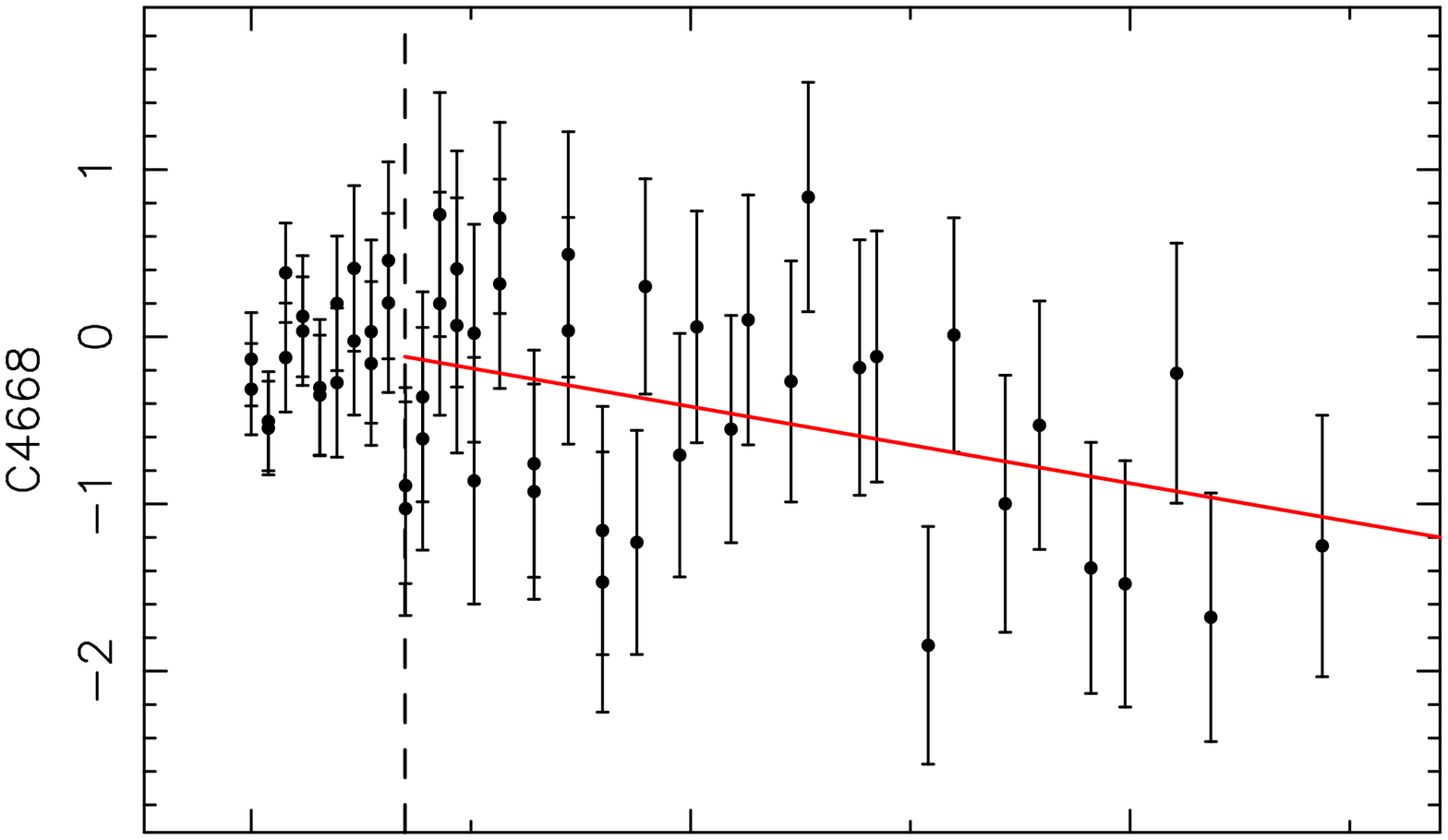}}
\resizebox{0.3\textwidth}{!}{\includegraphics[angle=-90]{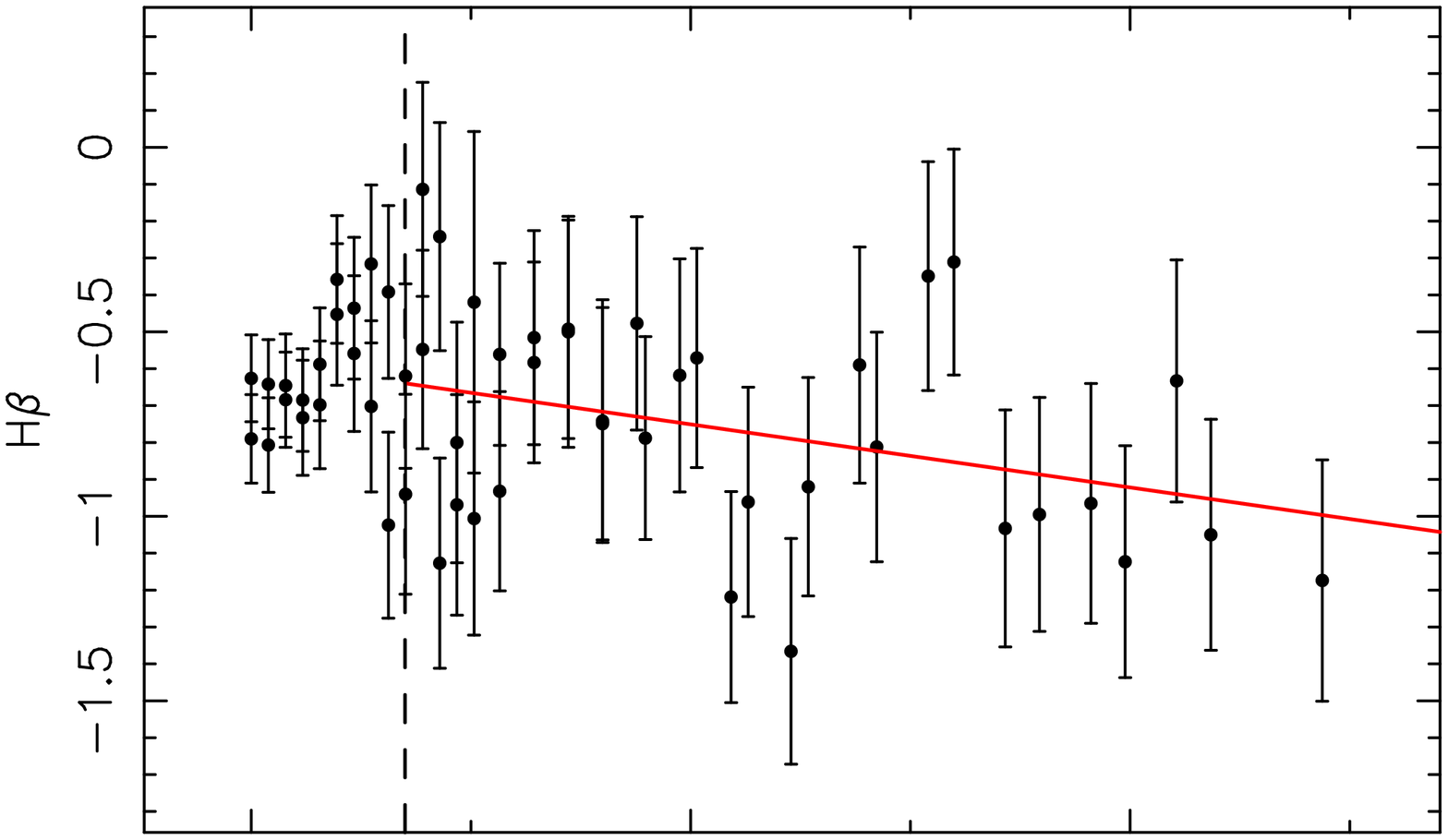}}
\resizebox{0.3\textwidth}{!}{\includegraphics[angle=-90]{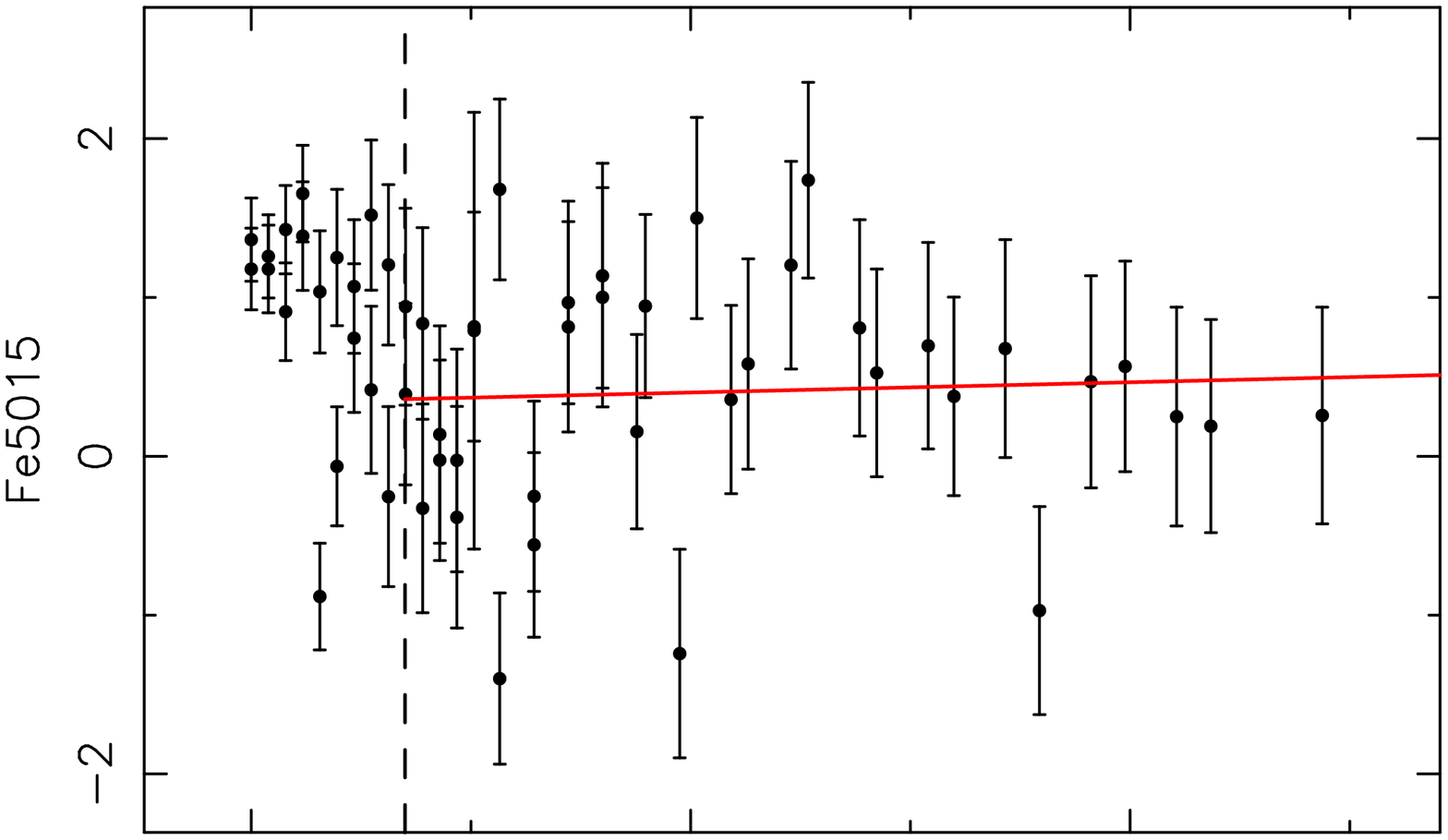}}
\resizebox{0.3\textwidth}{!}{\includegraphics[angle=-90]{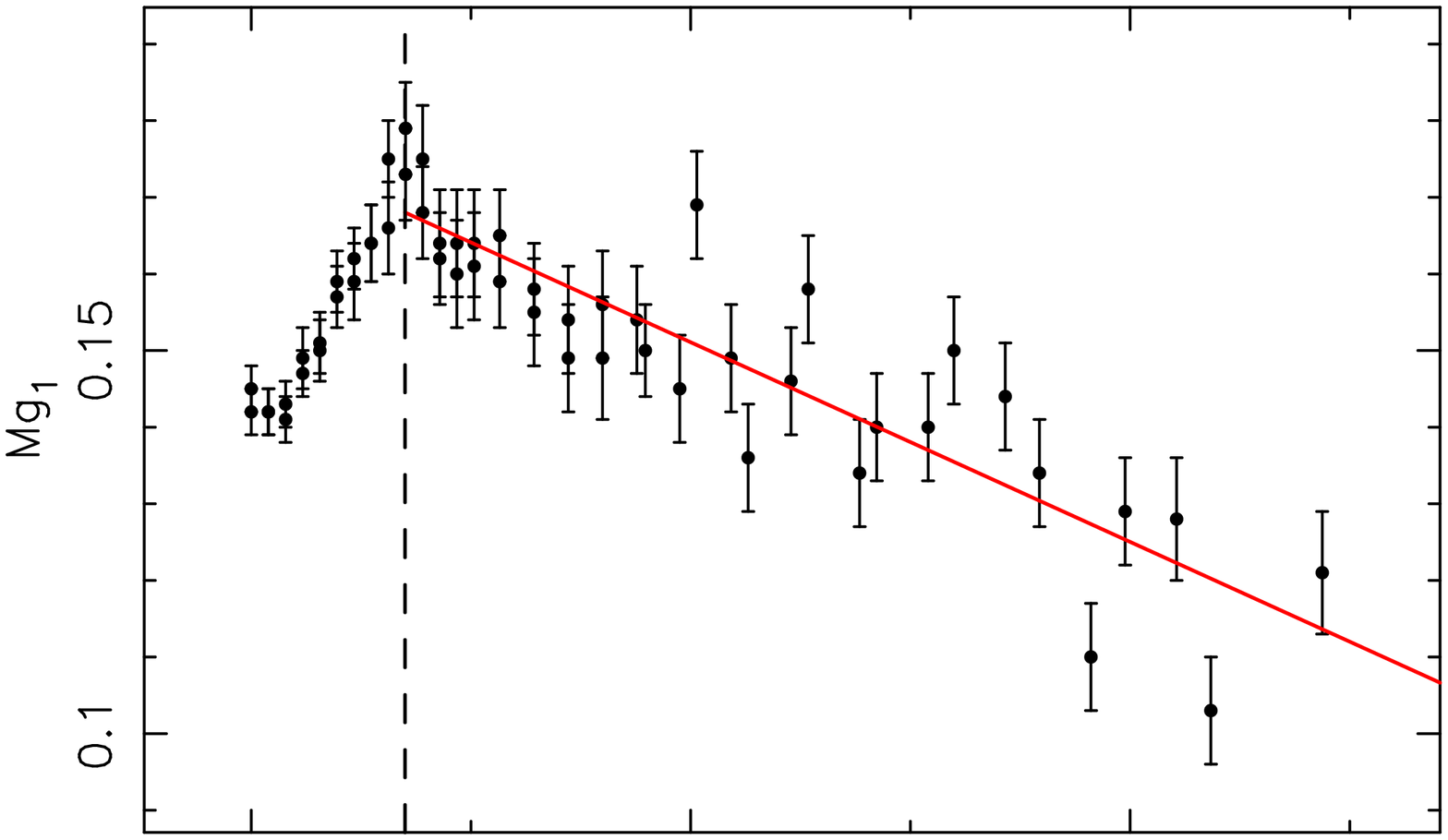}}
\resizebox{0.3\textwidth}{!}{\includegraphics[angle=-90]{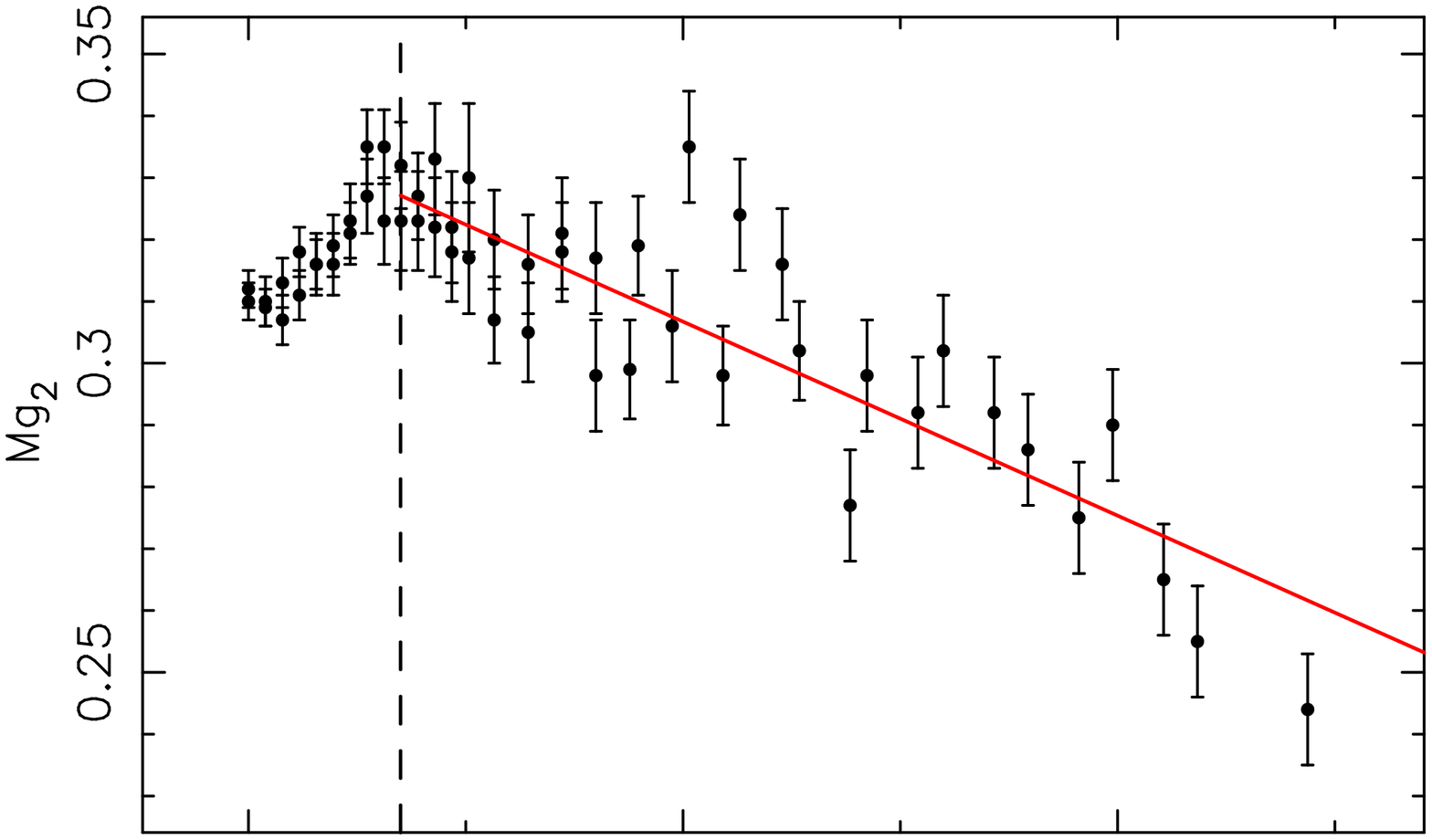}}
\resizebox{0.3\textwidth}{!}{\includegraphics[angle=-90]{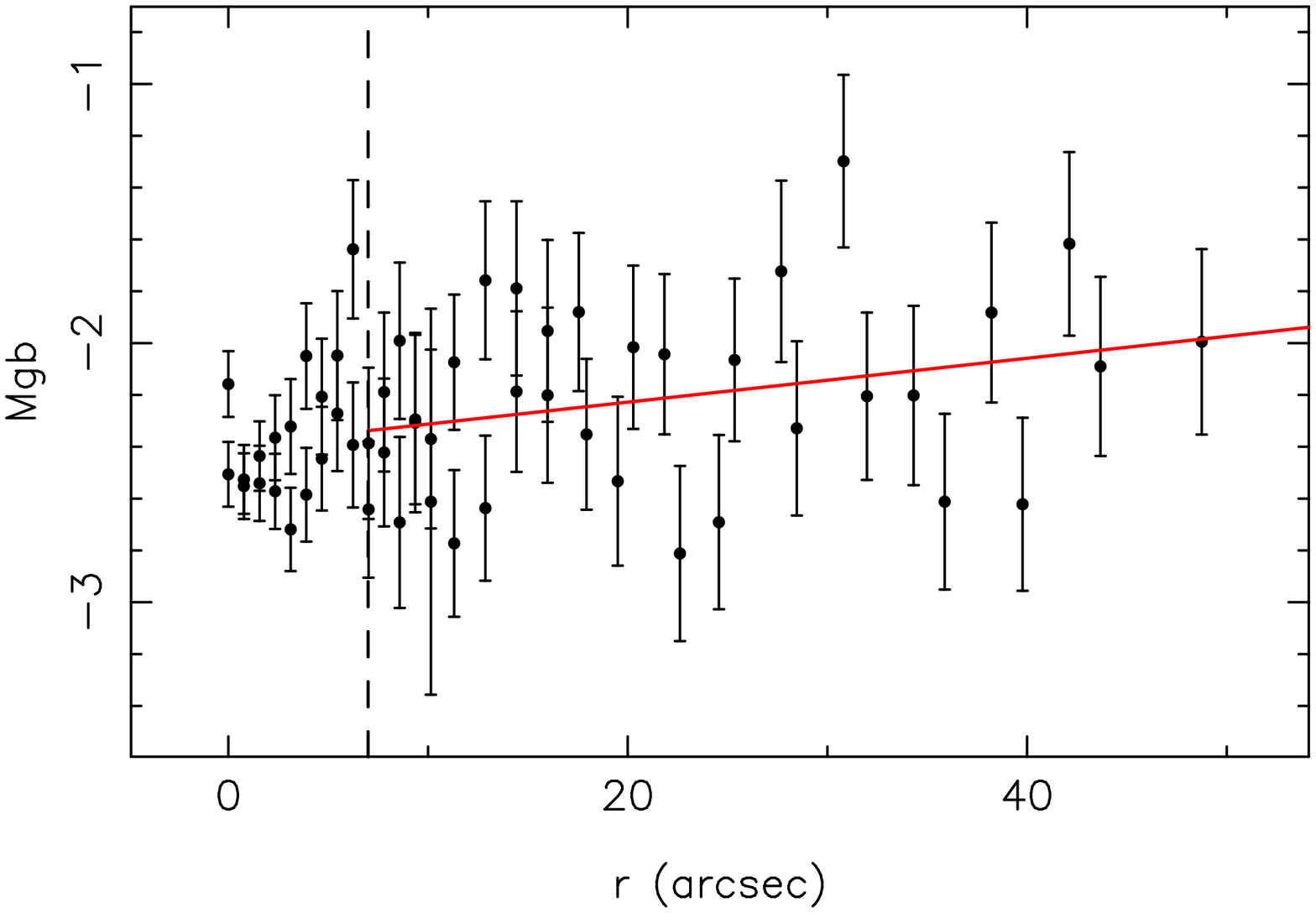}}\hspace{0.85cm}
\resizebox{0.3\textwidth}{!}{\includegraphics[angle=-90]{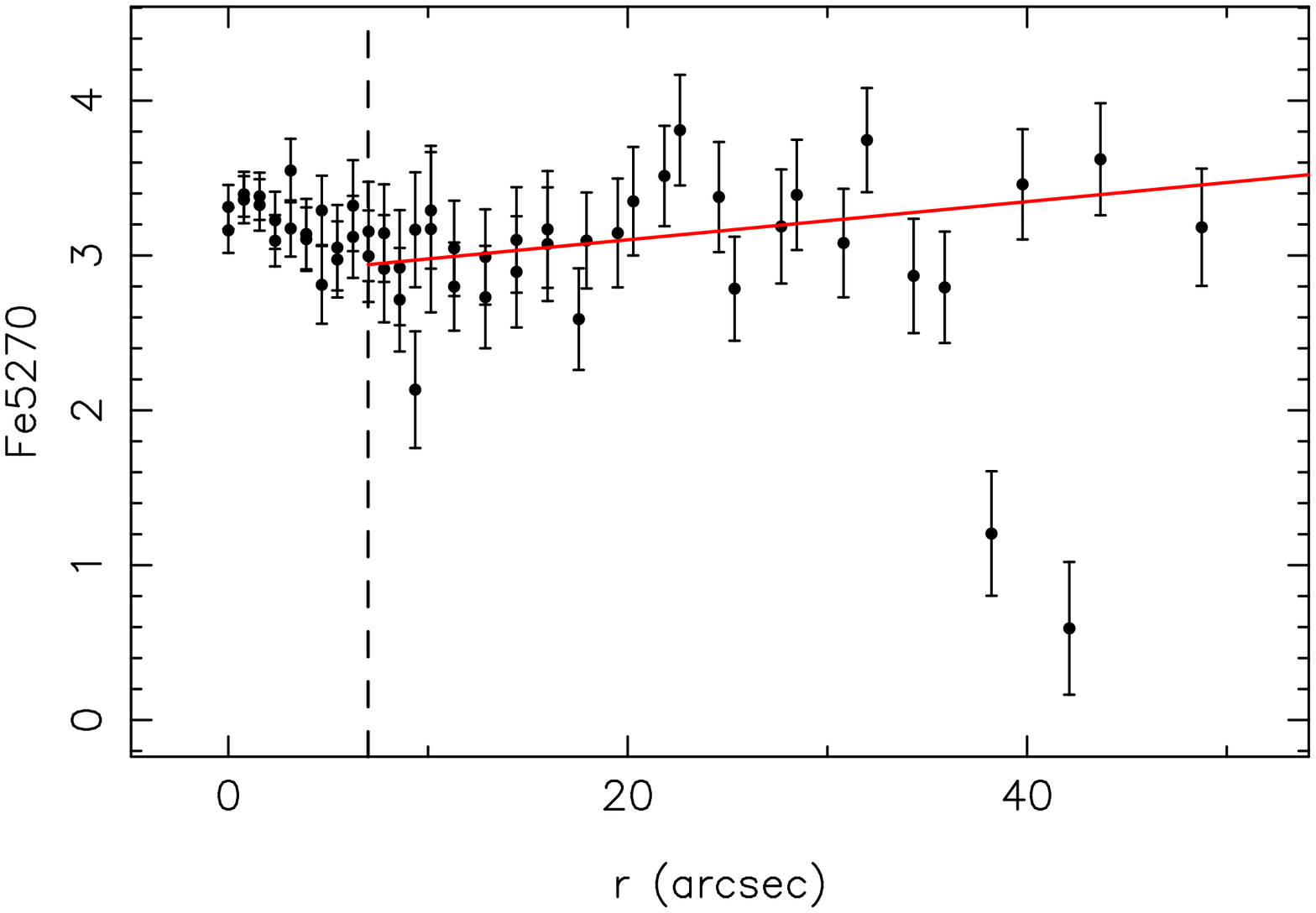}}\hspace{0.85cm}
\resizebox{0.3\textwidth}{!}{\includegraphics[angle=-90]{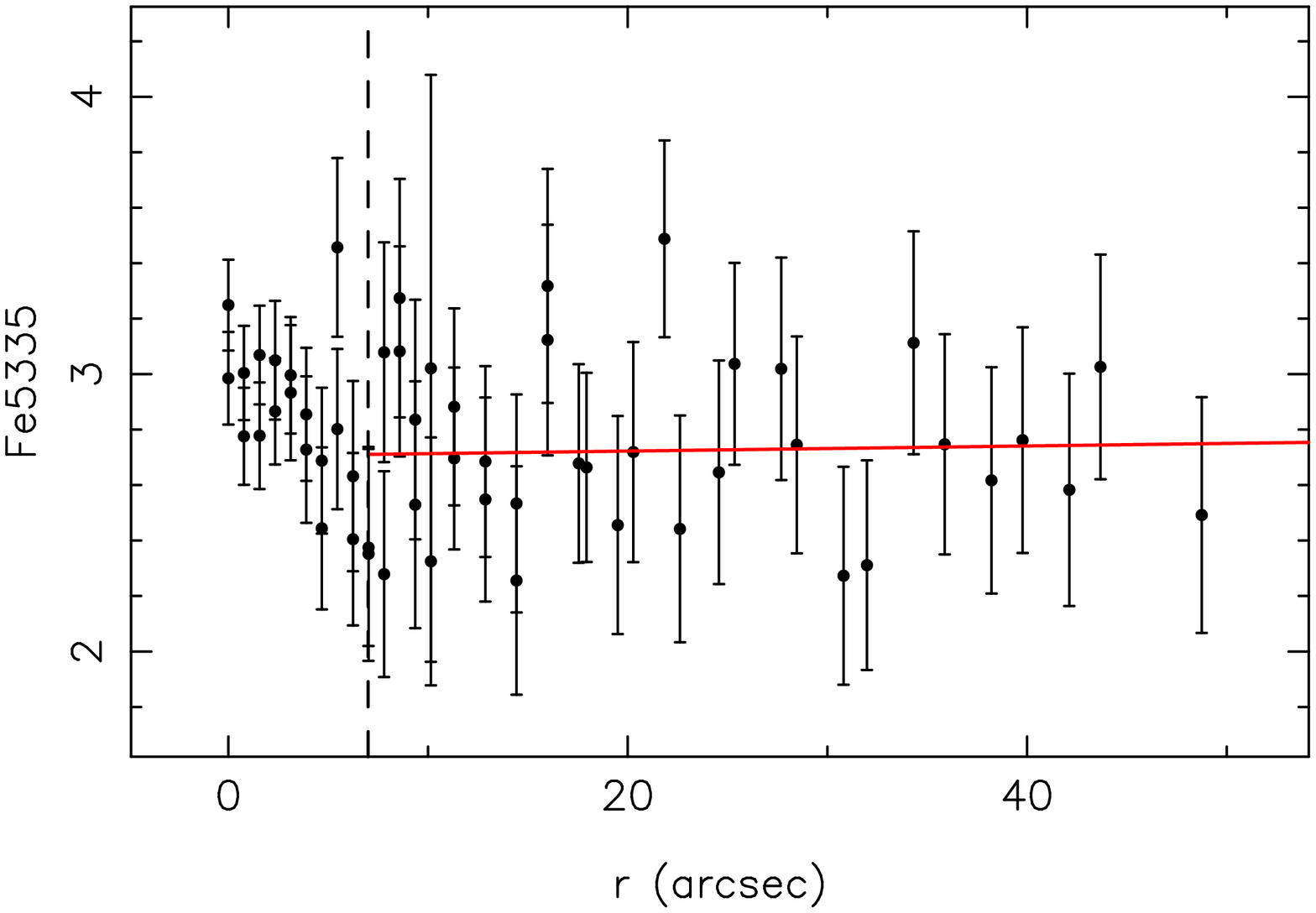}}
\caption{Line-strength distribution in the bar region for all the galaxies}
\end{figure*}

\section{Calibration to the models spectrophotometric systems}
\label{appen.compara.lick}
\begin{figure*}
\resizebox{0.2\textwidth}{!}{\includegraphics[angle=-90]{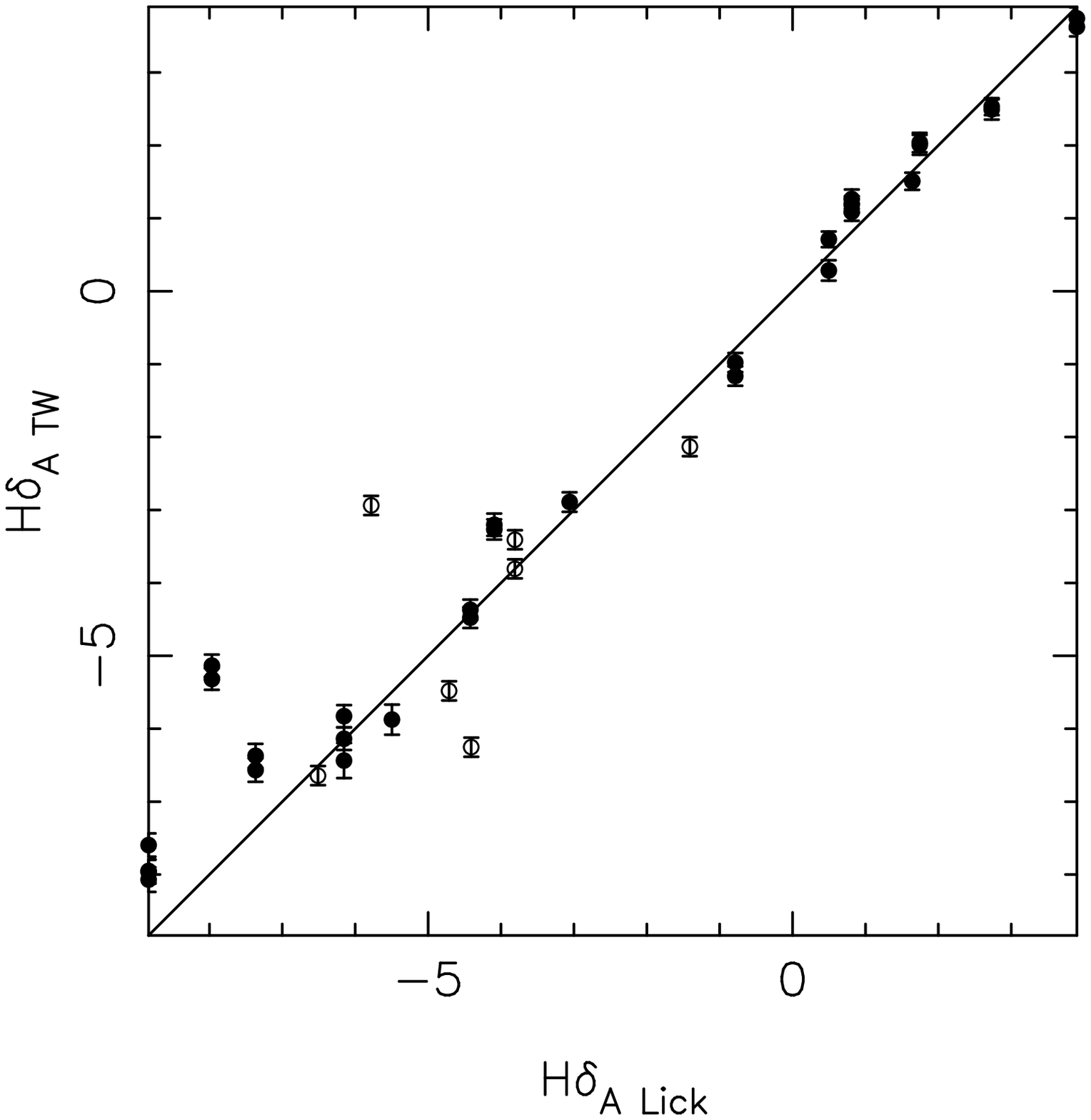}}
\resizebox{0.2\textwidth}{!}{\includegraphics[angle=-90]{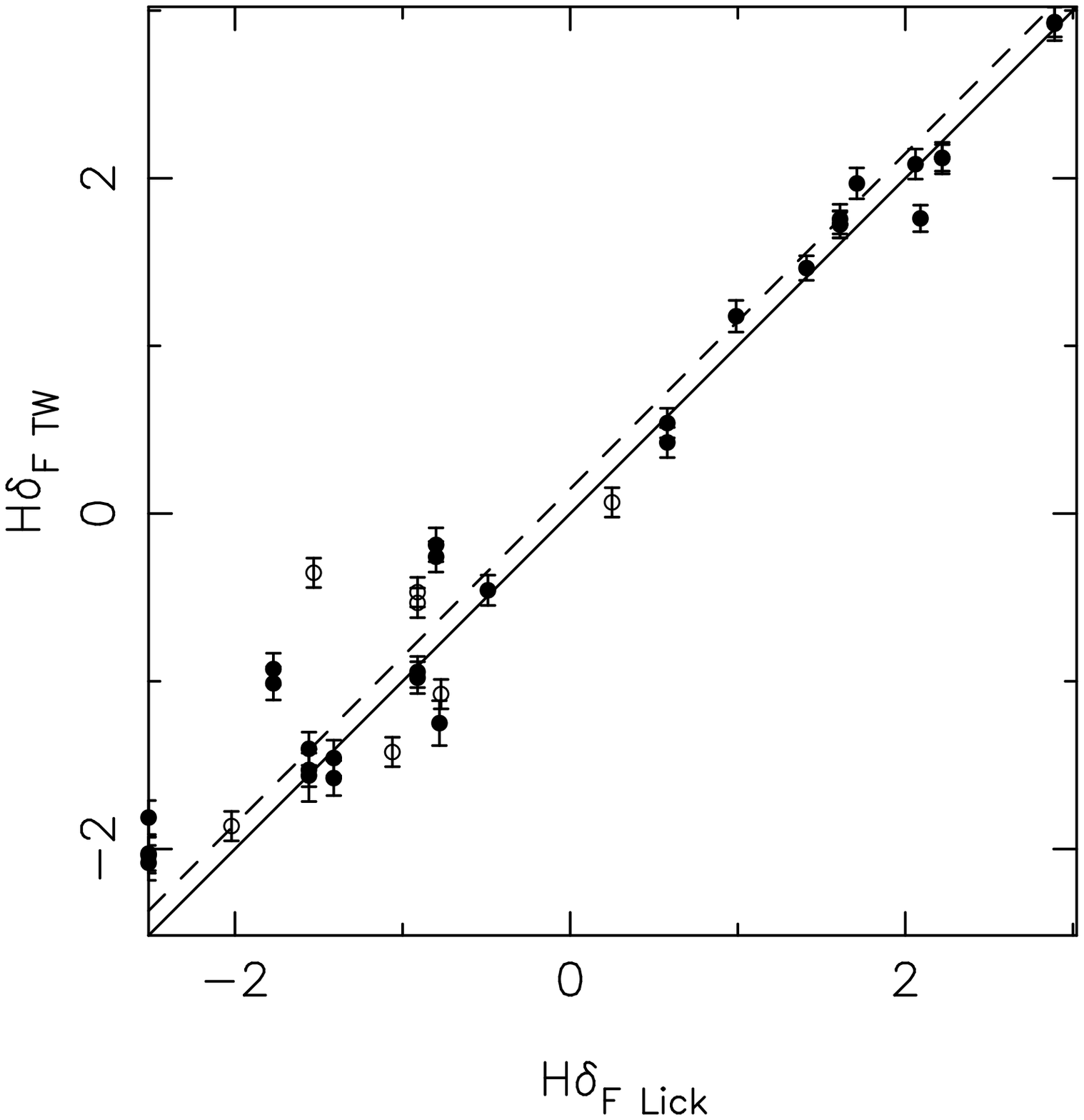}}
\resizebox{0.2\textwidth}{!}{\includegraphics[angle=-90]{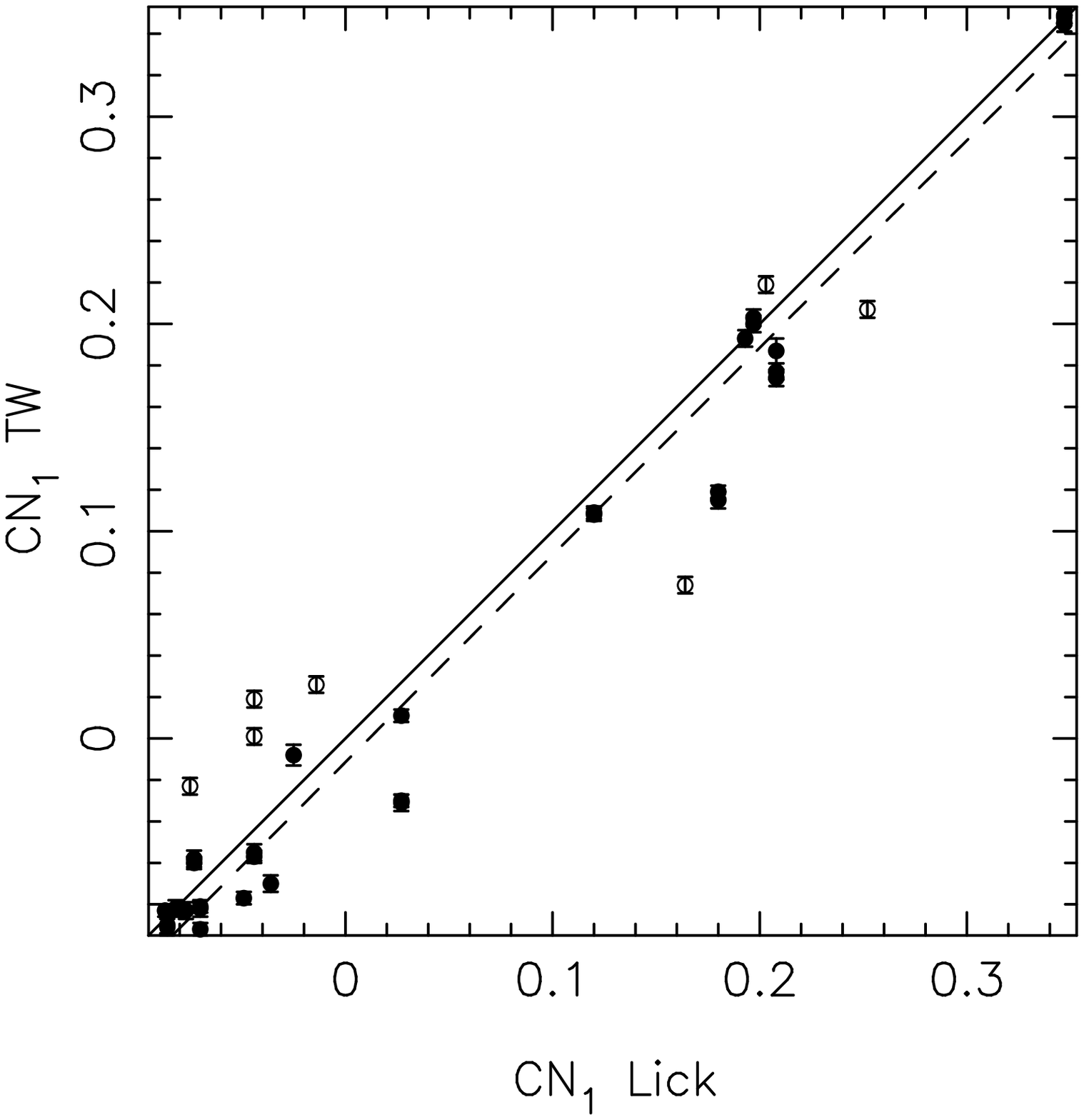}}
\resizebox{0.2\textwidth}{!}{\includegraphics[angle=-90]{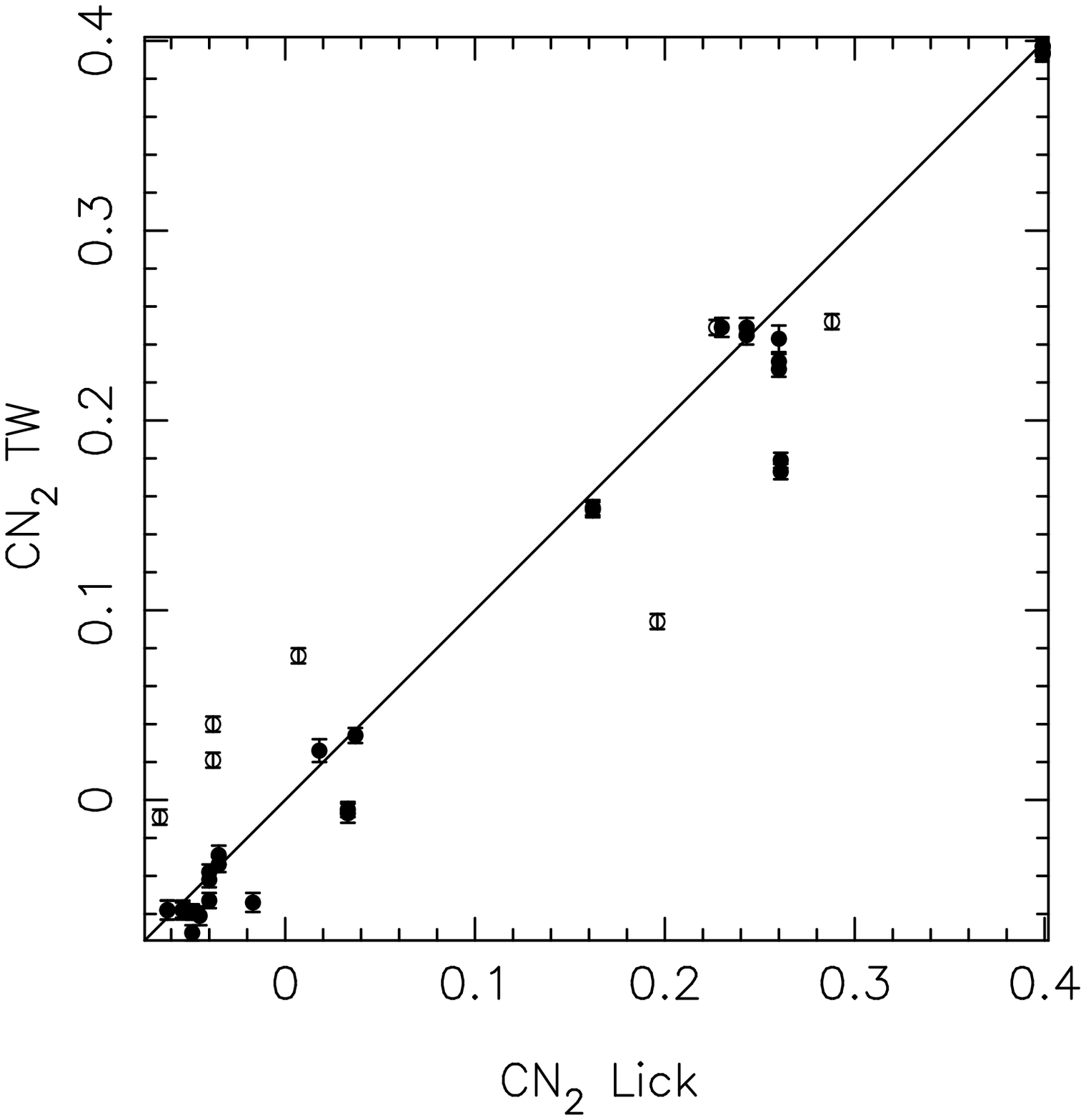}}
\resizebox{0.2\textwidth}{!}{\includegraphics[angle=-90]{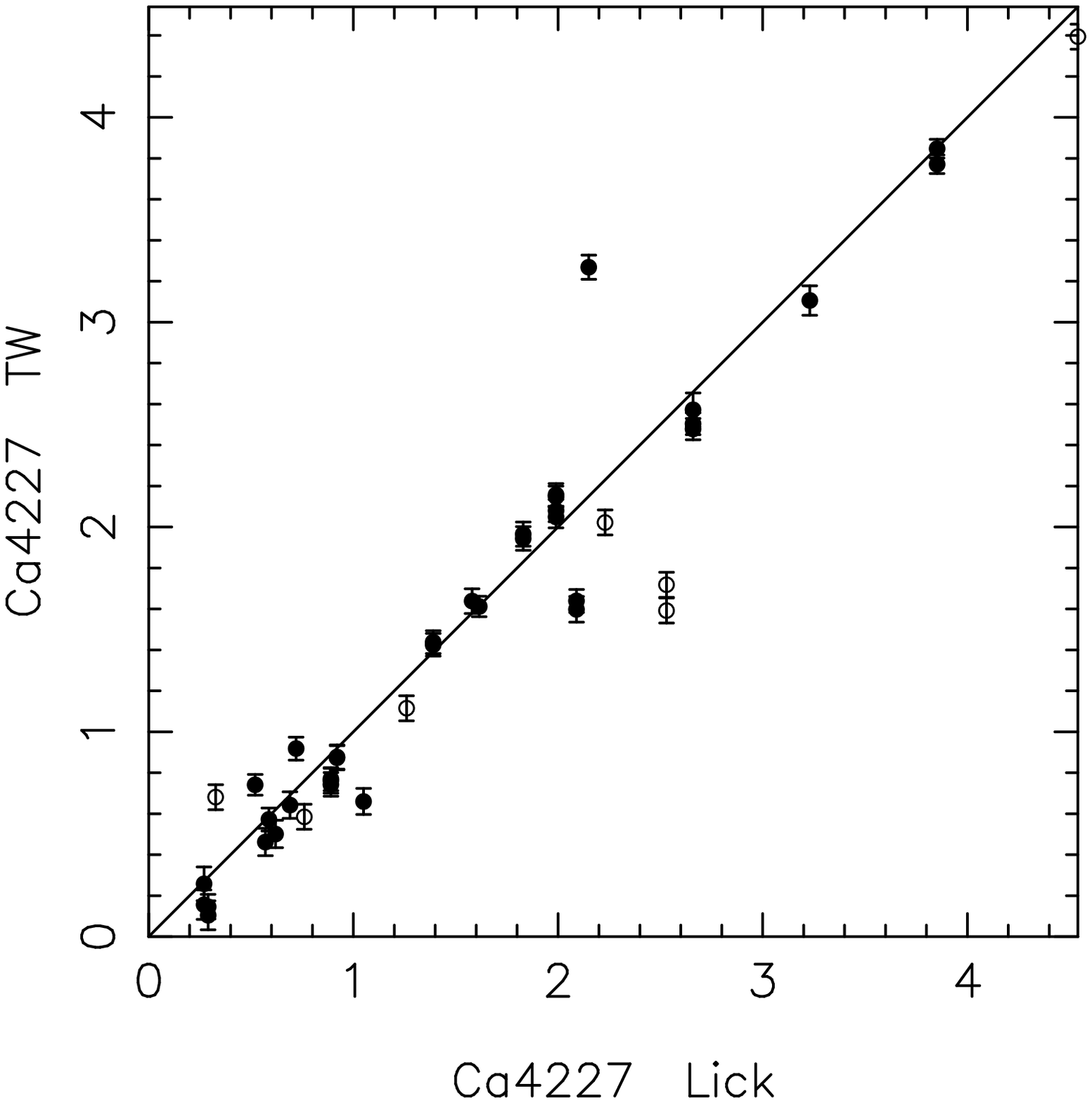}}
\resizebox{0.2\textwidth}{!}{\includegraphics[angle=-90]{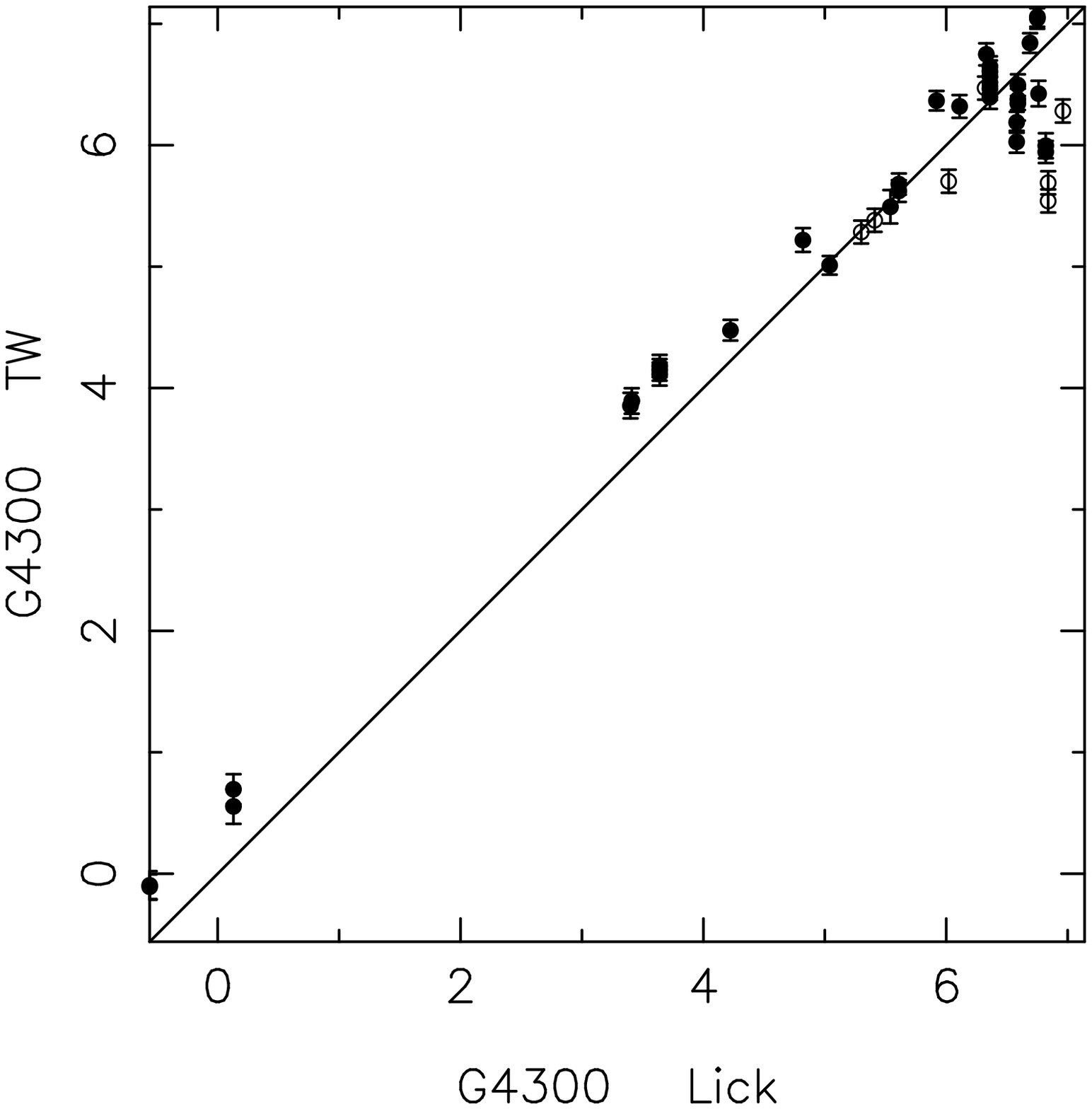}}
\resizebox{0.2\textwidth}{!}{\includegraphics[angle=-90]{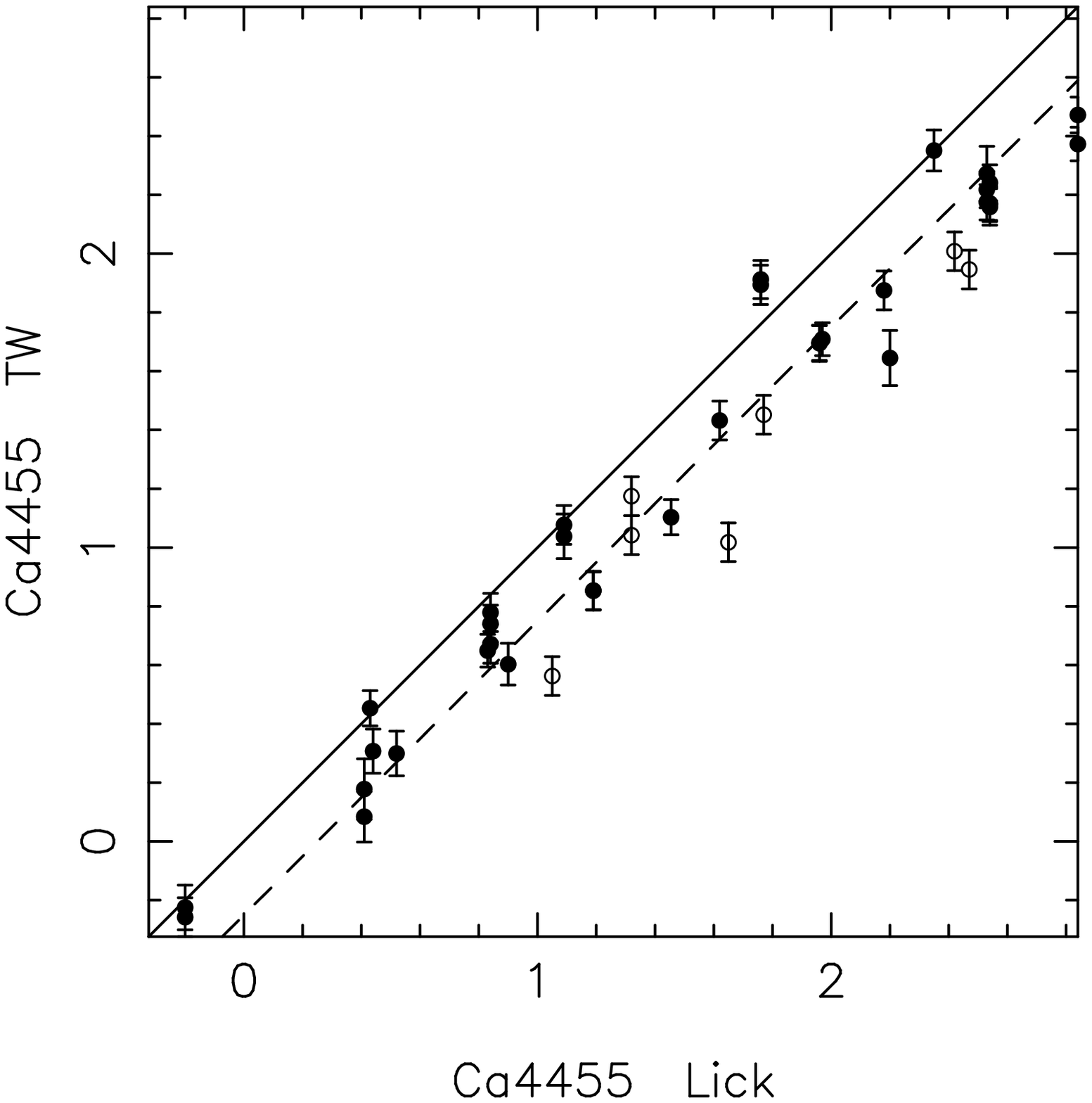}}
\resizebox{0.2\textwidth}{!}{\includegraphics[angle=-90]{hda.compara.lick.ps}}
\resizebox{0.2\textwidth}{!}{\includegraphics[angle=-90]{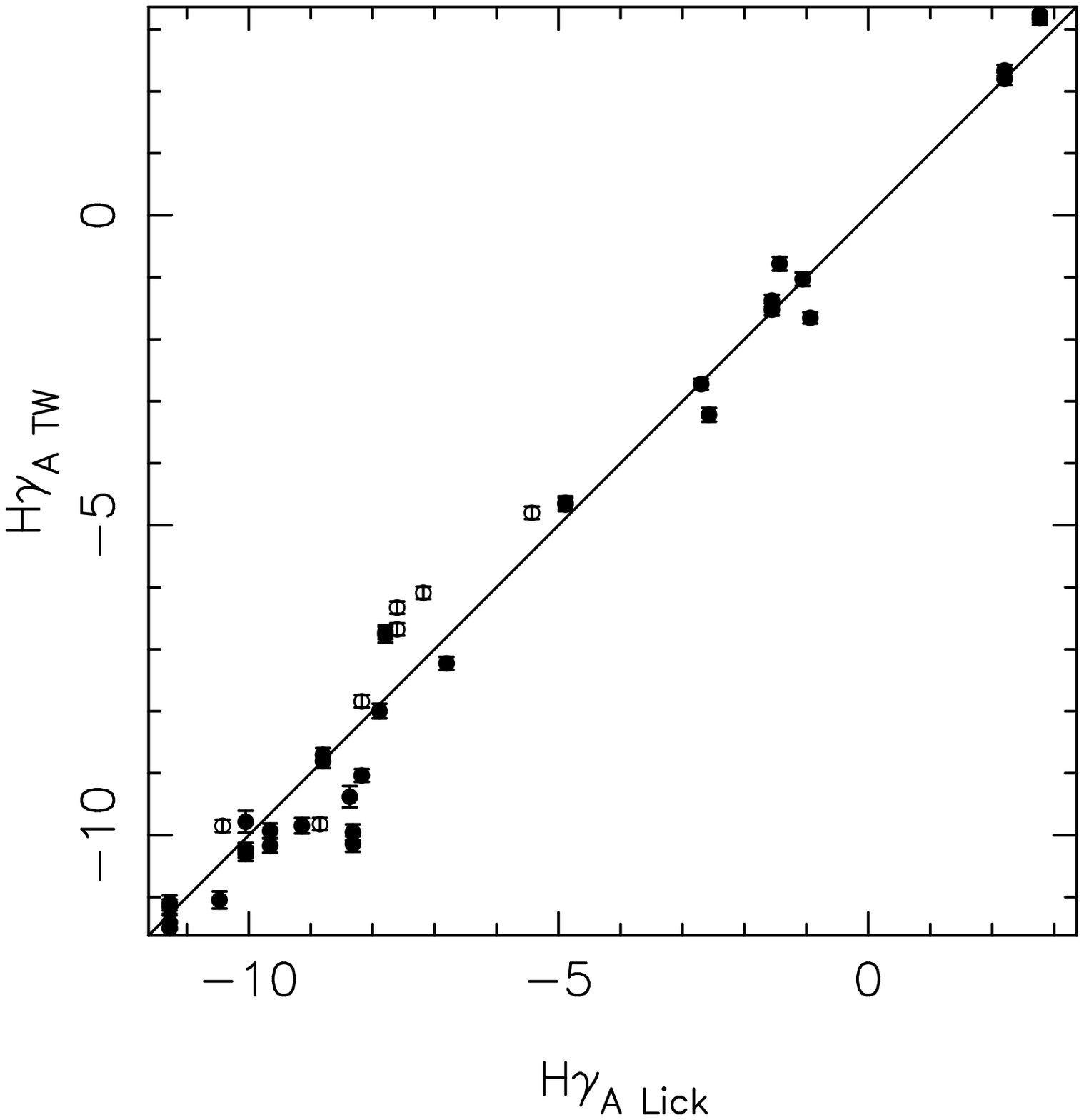}}
\resizebox{0.2\textwidth}{!}{\includegraphics[angle=-90]{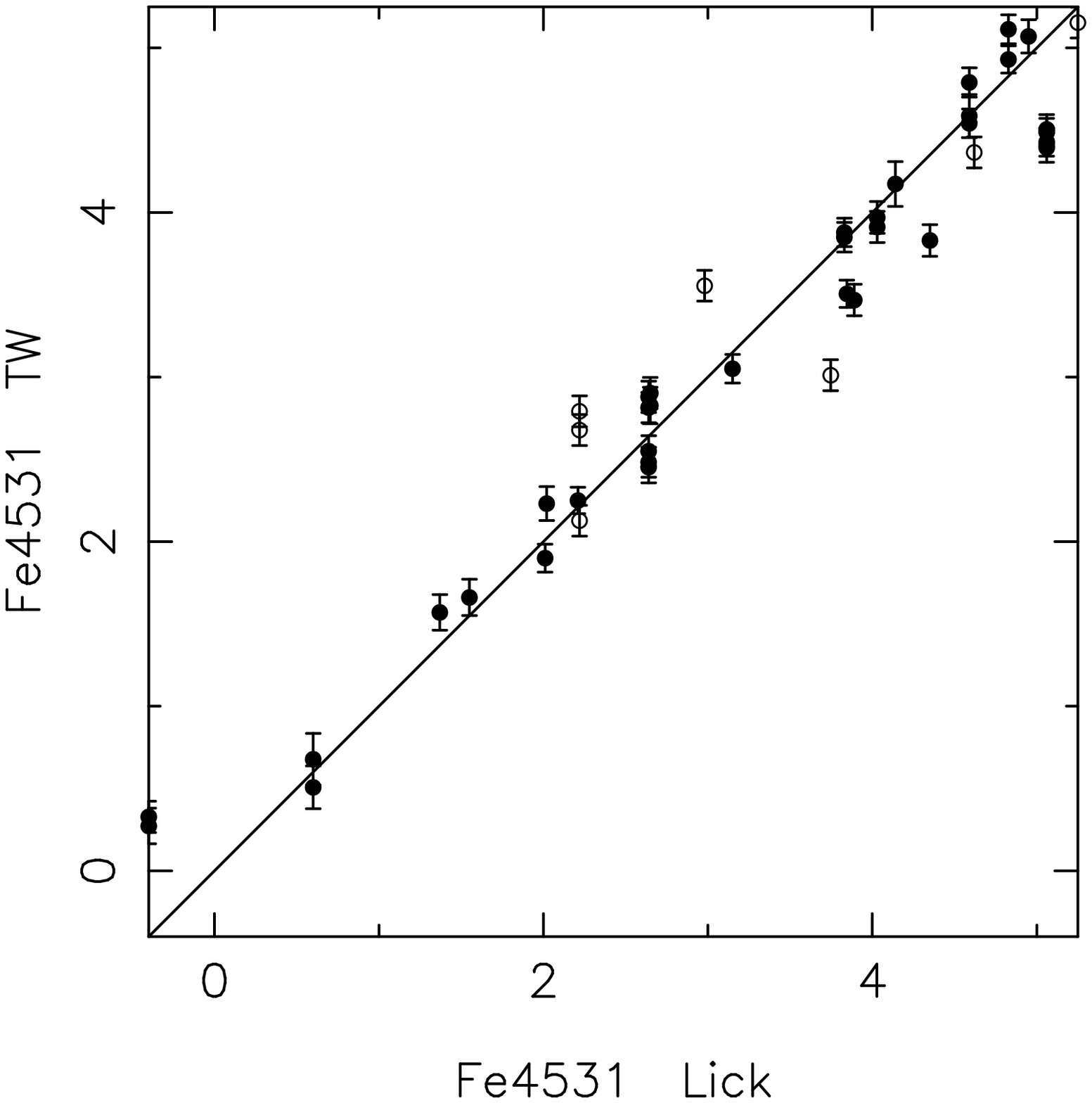}}
\resizebox{0.2\textwidth}{!}{\includegraphics[angle=-90]{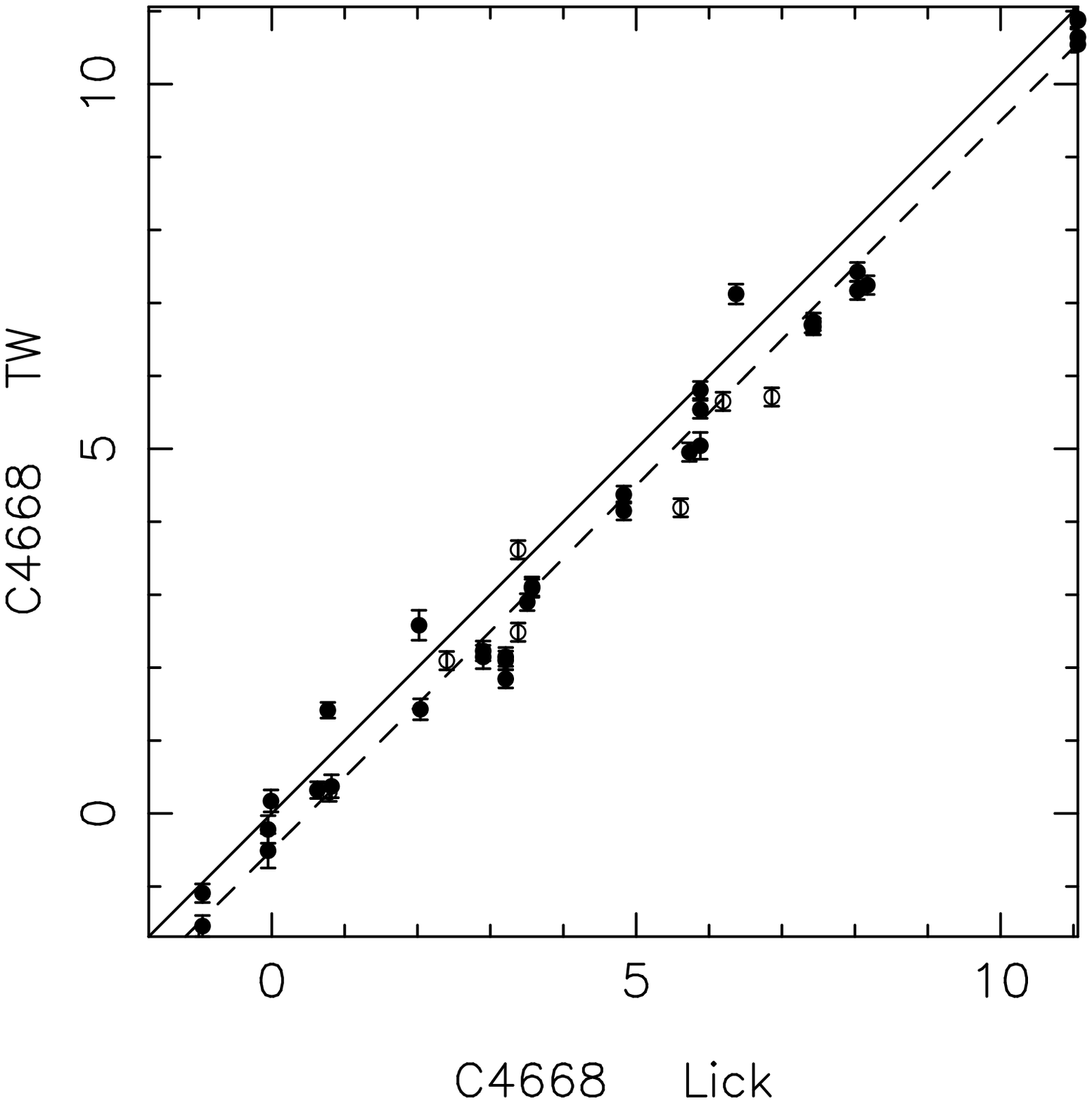}}
\resizebox{0.2\textwidth}{!}{\includegraphics[angle=-90]{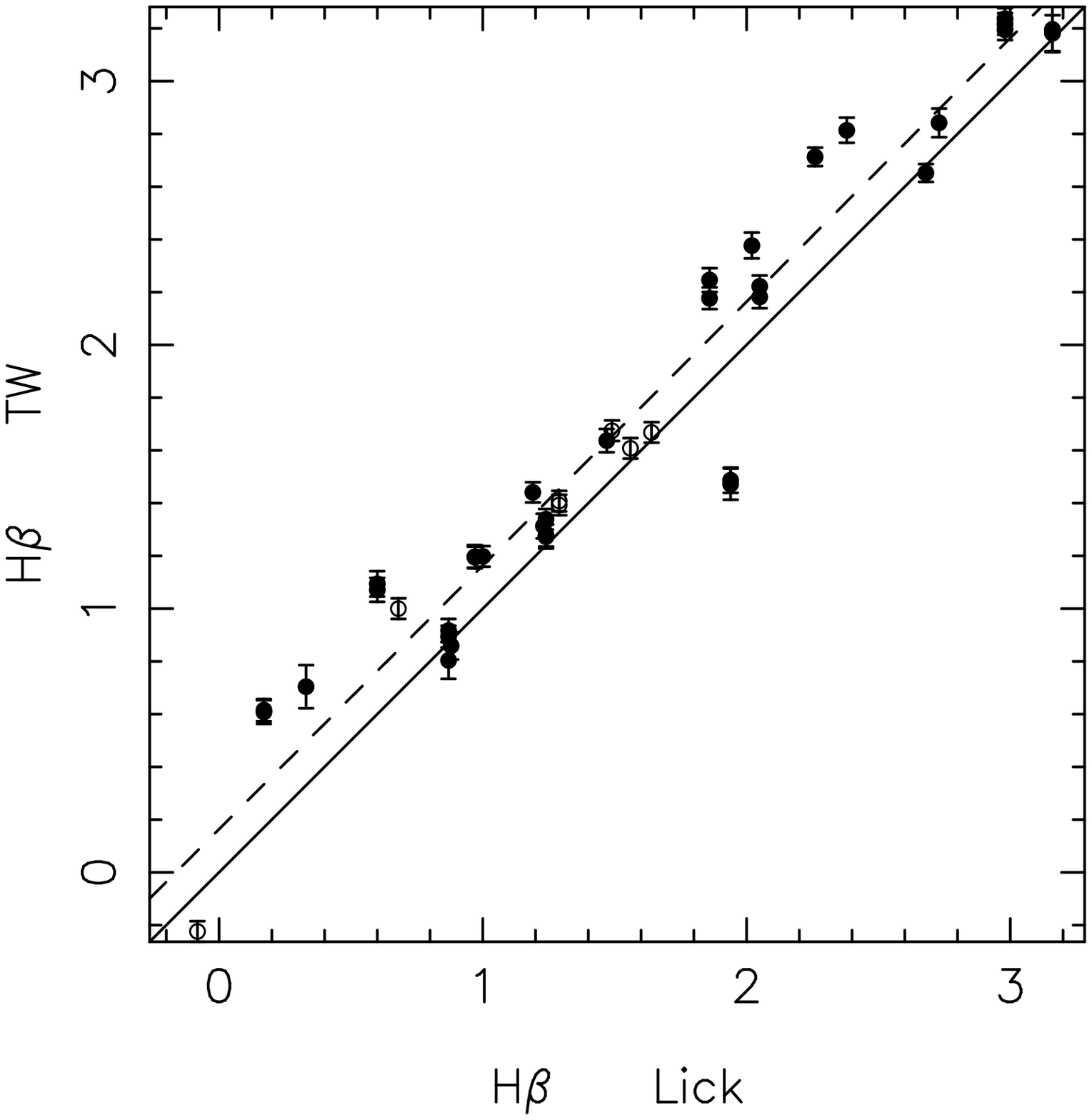}}
\resizebox{0.2\textwidth}{!}{\includegraphics[angle=-90]{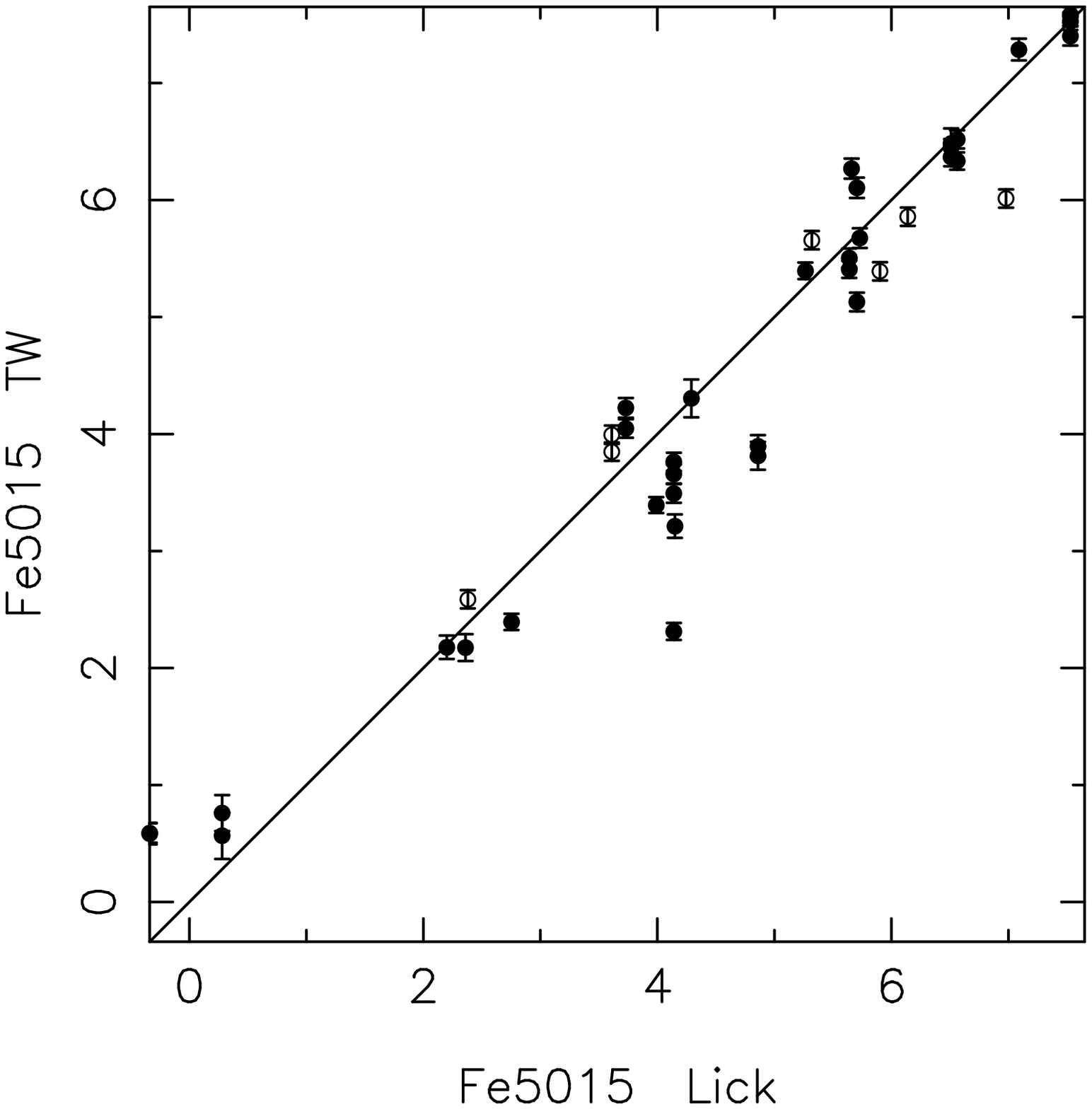}}
\resizebox{0.2\textwidth}{!}{\includegraphics[angle=-90]{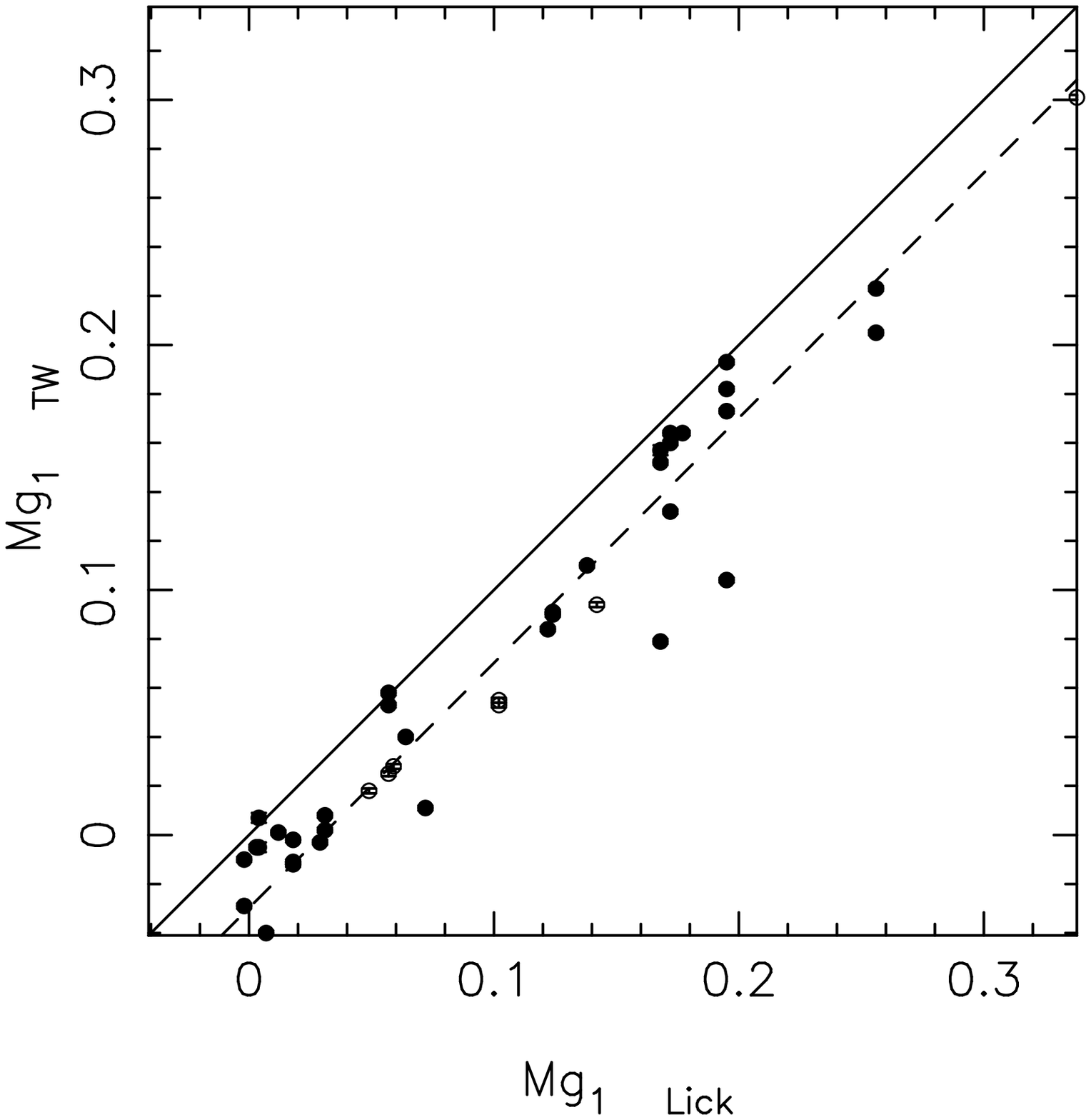}}
\resizebox{0.2\textwidth}{!}{\includegraphics[angle=-90]{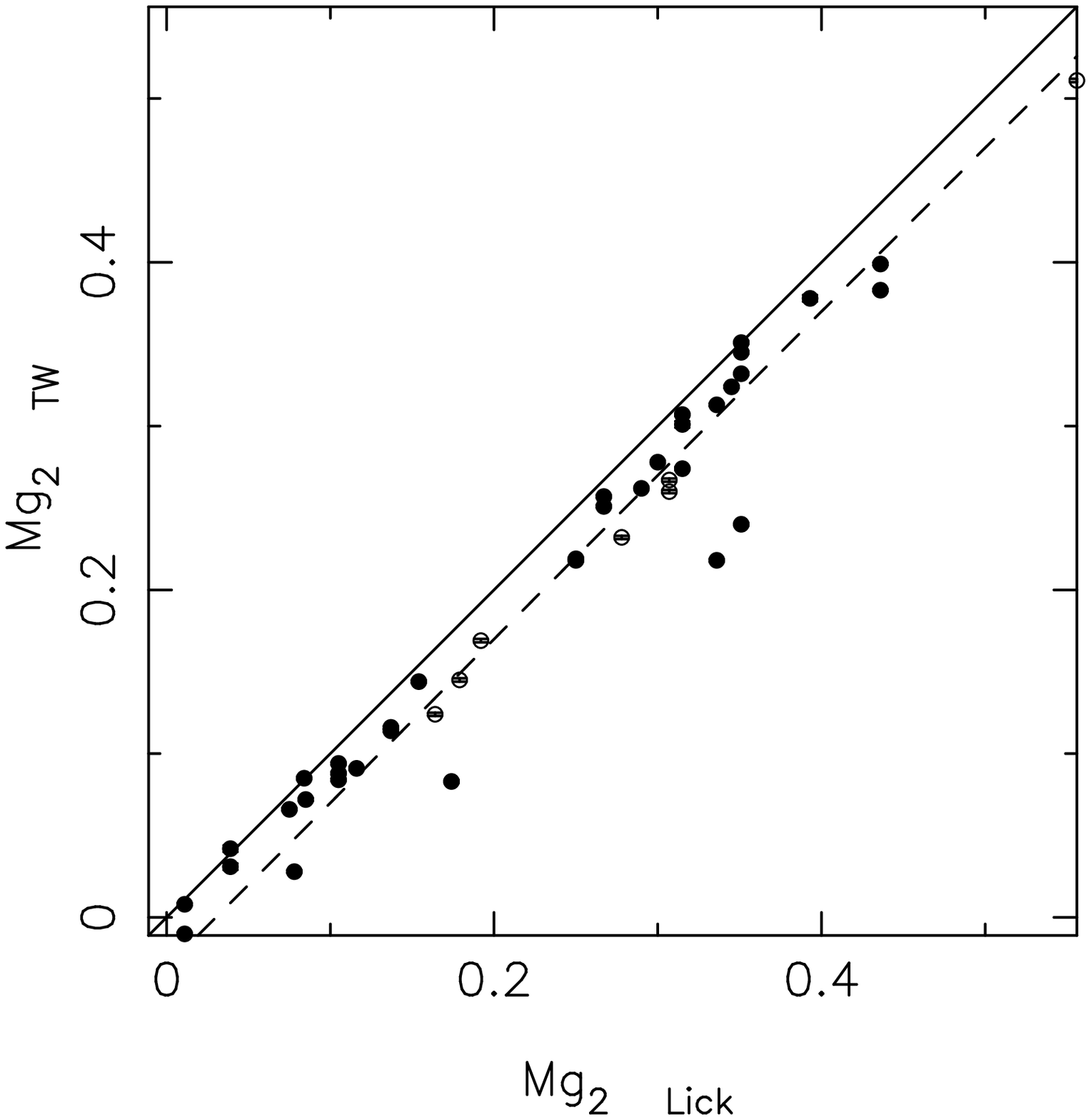}}
\resizebox{0.2\textwidth}{!}{\includegraphics[angle=-90]{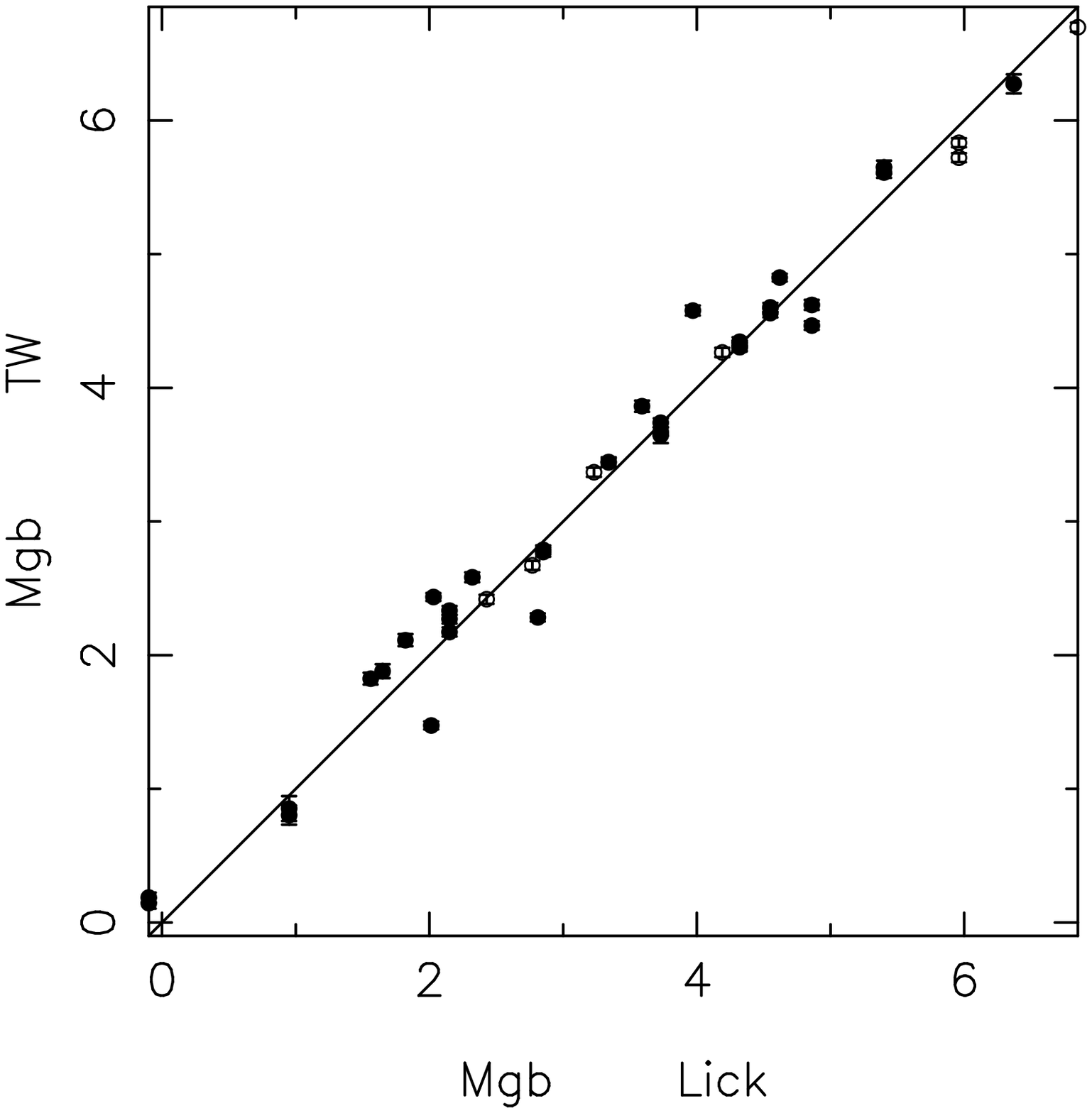}}
\resizebox{0.2\textwidth}{!}{\includegraphics[angle=-90]{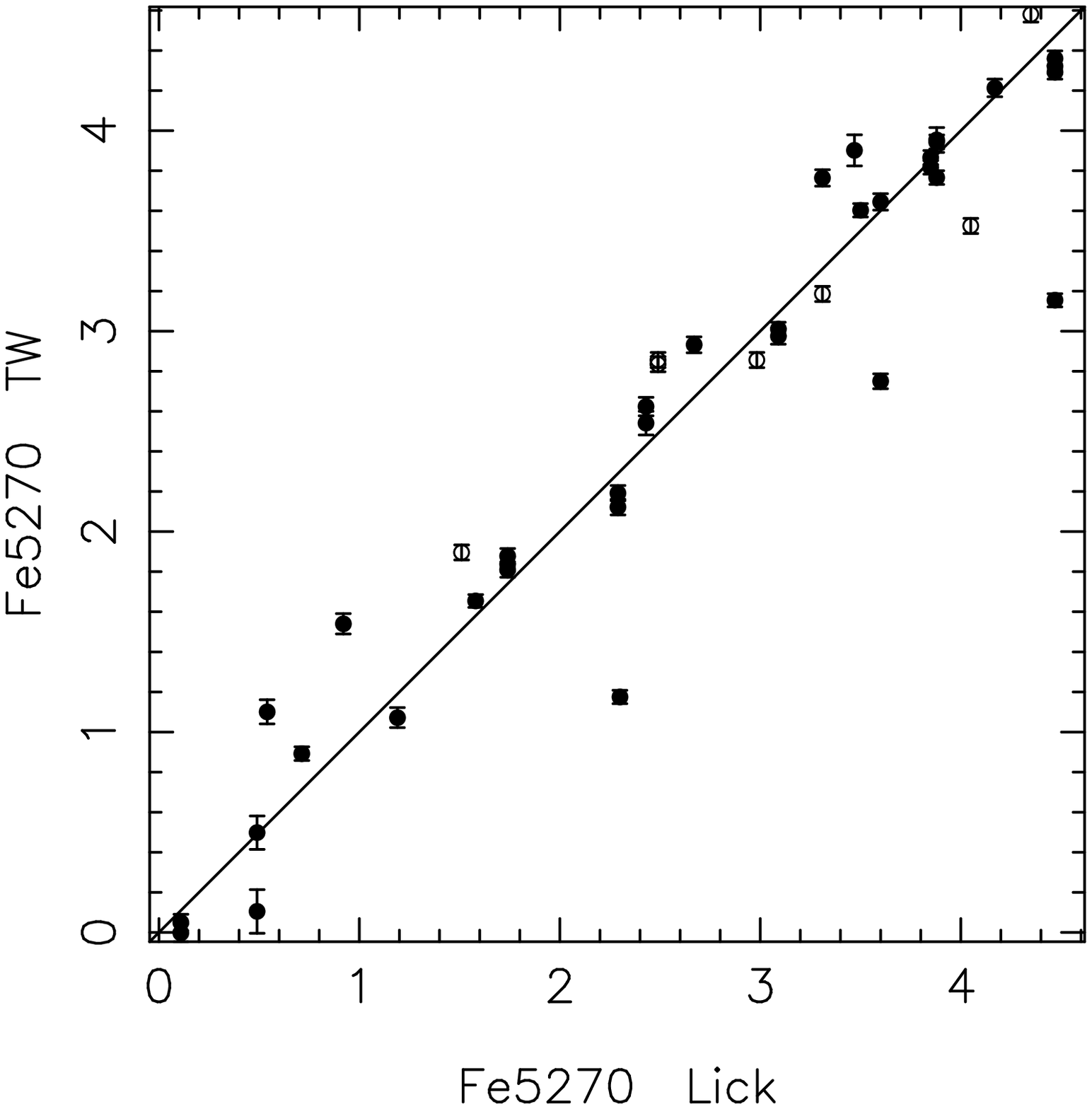}}\hspace{1.1cm}
\resizebox{0.2\textwidth}{!}{\includegraphics[angle=-90]{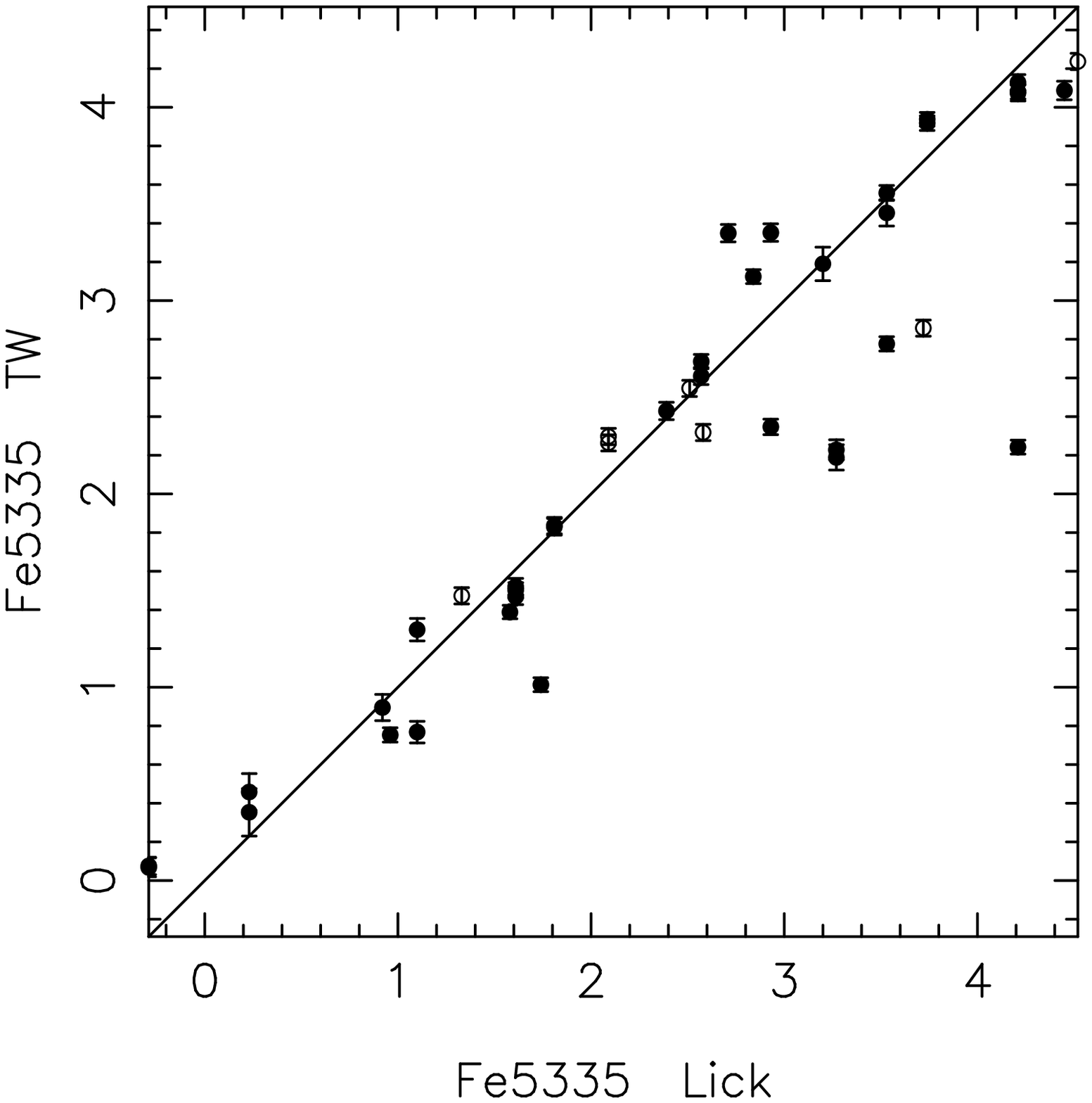}}\hspace{1.1cm}
\resizebox{0.2\textwidth}{!}{\includegraphics[angle=-90]{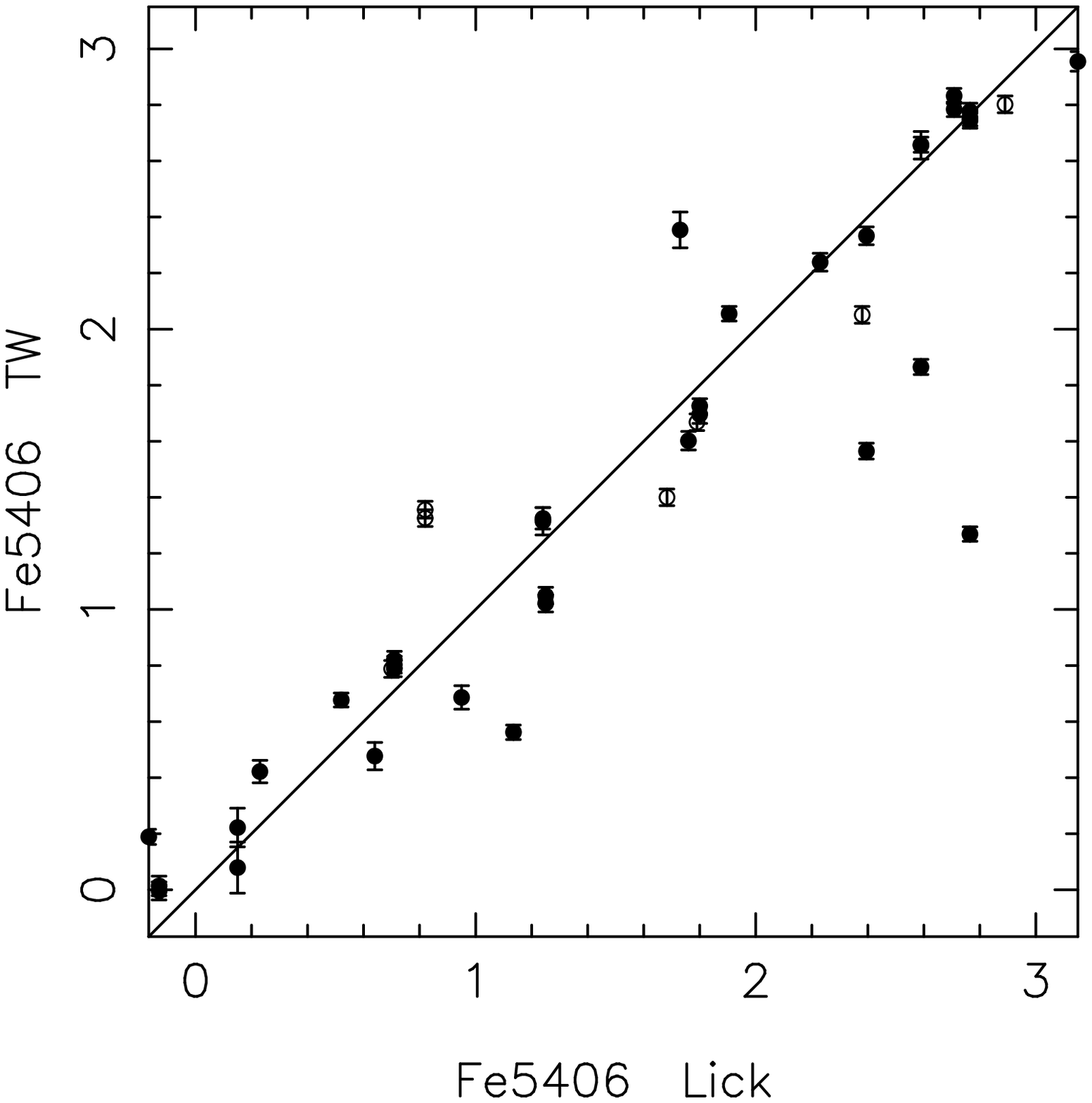}}
\caption{Comparison of the indices measured in the stars in common between 
this work and the Lick/IDS library. \label{fig.lick.off}}
\end{figure*}

The original Lick/IDS spectra were not flux-calibrated by means of a flux-standard
stars but normalised to a quartz-iodide tungsten lamp. The resulting continuum shape
cannot be reproduced and causes small offsets for the indices with a broad wavelength range.
Furthermore, the wavelength dependent resolution of the Lick spectrograph is also 
difficult to reproduce, which cuases differences in other indices with larger dependence
to resolution. In order to compare our measured indices with models based on this library 
we have to calculate these offsets. 
To calculate them, we observed 11 and 35 stars in common with the Lick/IDS library. By comparing 
the indices measured in our stars with those in the Lick/IDS database for the same objects, we 
derived mean offsets for all the indices observed. Fig.~\ref{fig.lick.off} shows this comparison. 
We did not find any systematic difference between the offset obtained with the stars of the first 
and second run and , therefore, 
we calculated just one offset for the two runs. The final offsets are listed in Table~\ref{table.offsets}

\begin{table}
\begin{tabular}{lrr}
index        & Offset (MILES)  &  Offset (Lick)\\
             & this work-miles & this work-Lick\\
\hline
H$\delta_A$  & $-0.18$      &   0.00\\
H$\delta_F$  &   0.000      &  0.15\\
CN$_1$       &   0.000      & -0.011 \\
CN$_2$       &   0.000      &  0.00\\
Ca4227       &   0.0        &   0.00\\
G4300        &   0.2        &   0.0 \\
H$\gamma_A$  &   0.0        &   0.00\\
H$\gamma_F$  &   0.0        &   0.00\\
F4383        &   0.21       &   0.00\\  
Ca4455       &   0.09       &  $-0.25$\\
Fe4531       &   0.12       &   0.00\\
C4668        &  $-0.75$     &  $-0.50$\\
H$\beta$     &   0.00       &  0.163  \\
Fe5015       &   0.00       &  0.00\\
Mg1          &  -0.017      & -0.030\\
Mg2          &   0.0        & -0.030\\
Mgb          &   0.0        &  0.00\\
Fe5270       &   0.0        &  0.0\\
Fe5335       &   0.0        &  0.0\\
Fe5406       &   0.0        &  0.0\\
\hline
\end{tabular}
\caption{Mean offsets measured in the stars in common between our observing runs and MILES (first column) and Lick/IDS
 (second column).\label{table.offsets}}
\end{table}

\section{Stellar Population fits}
\begin{figure*}
\resizebox{0.2\textwidth}{!}{\includegraphics[angle=-90]{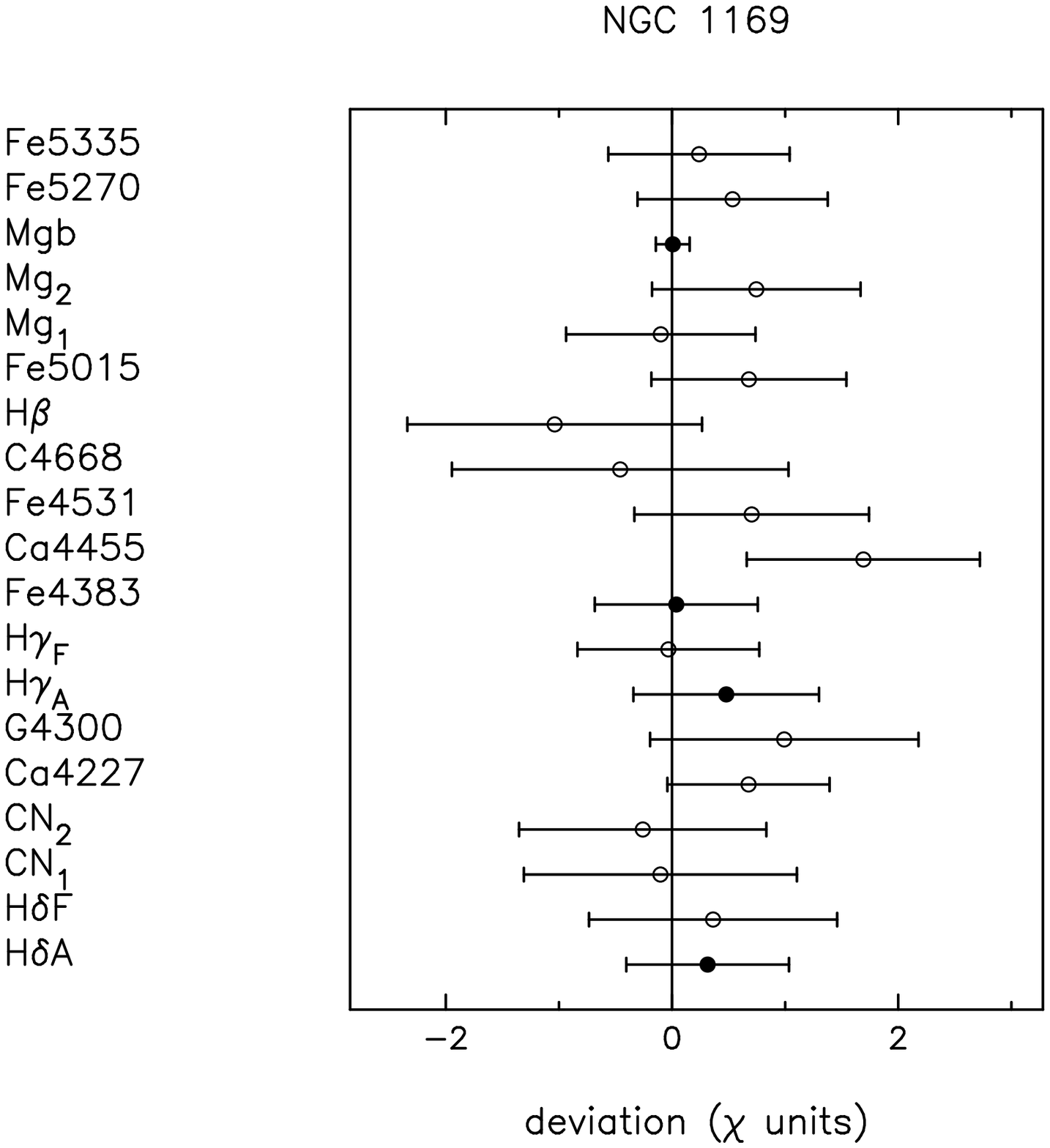}}
\resizebox{0.2\textwidth}{!}{\includegraphics[angle=-90]{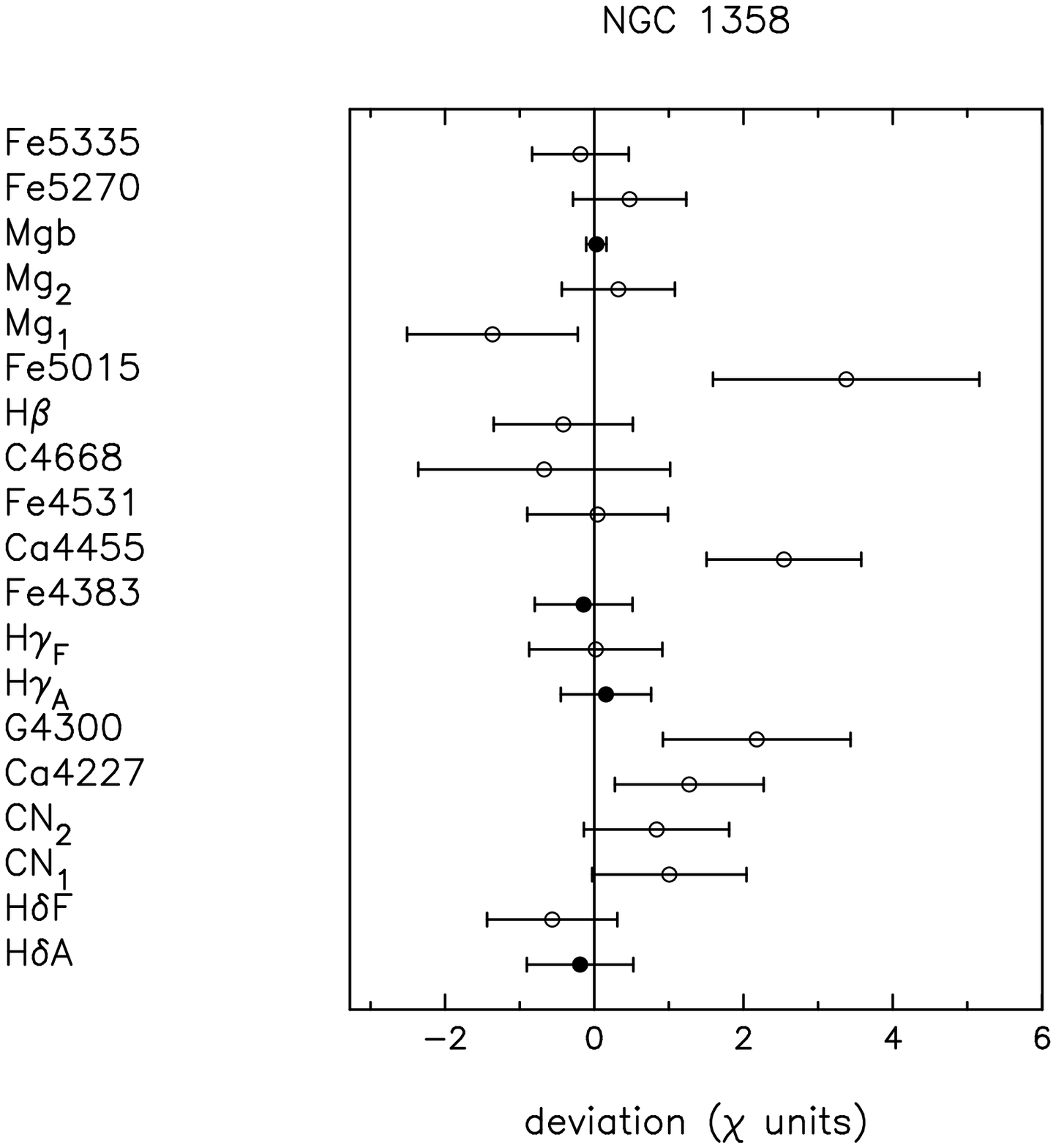}} 
\resizebox{0.2\textwidth}{!}{\includegraphics[angle=-90]{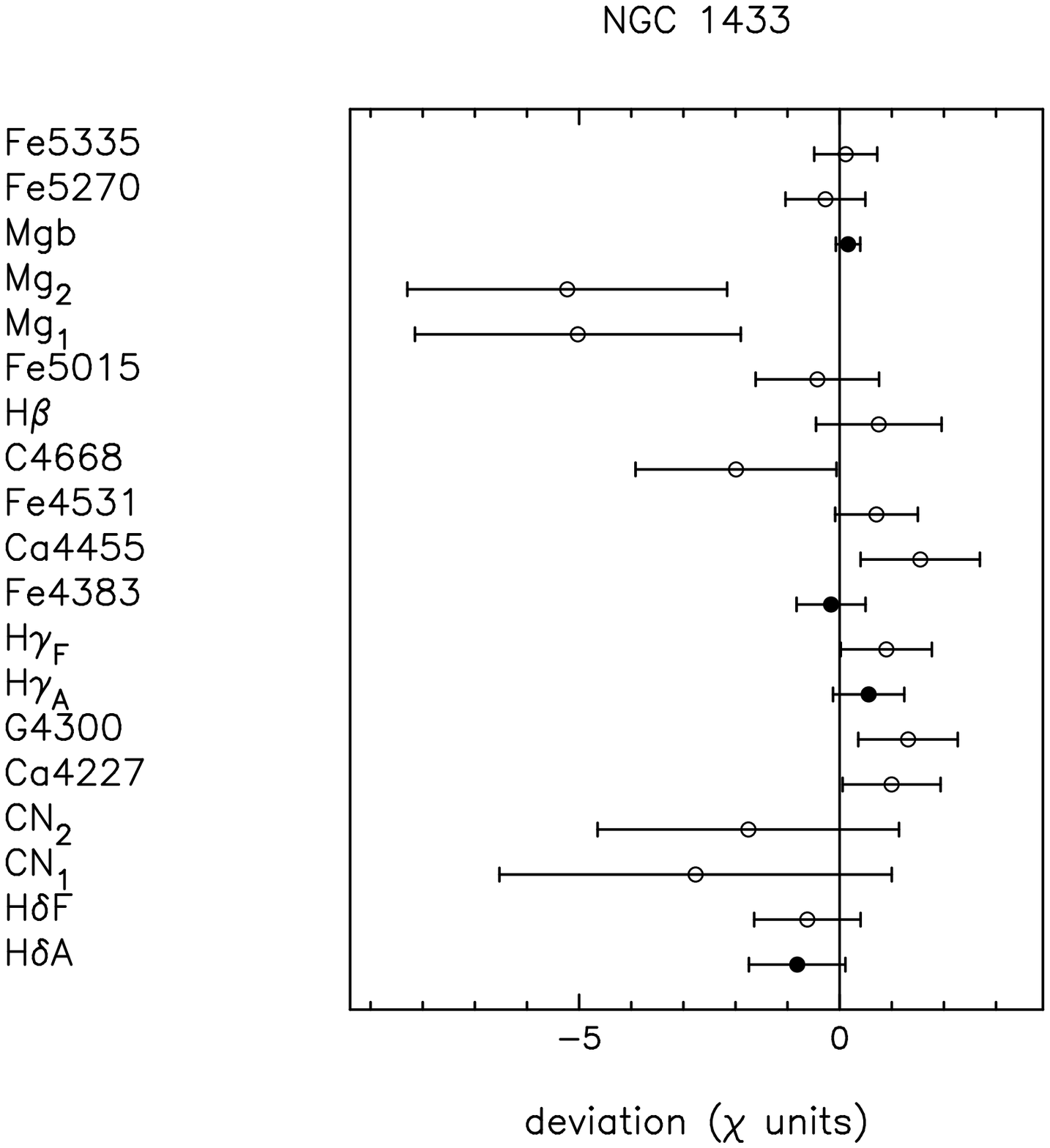}}
\resizebox{0.2\textwidth}{!}{\includegraphics[angle=-90]{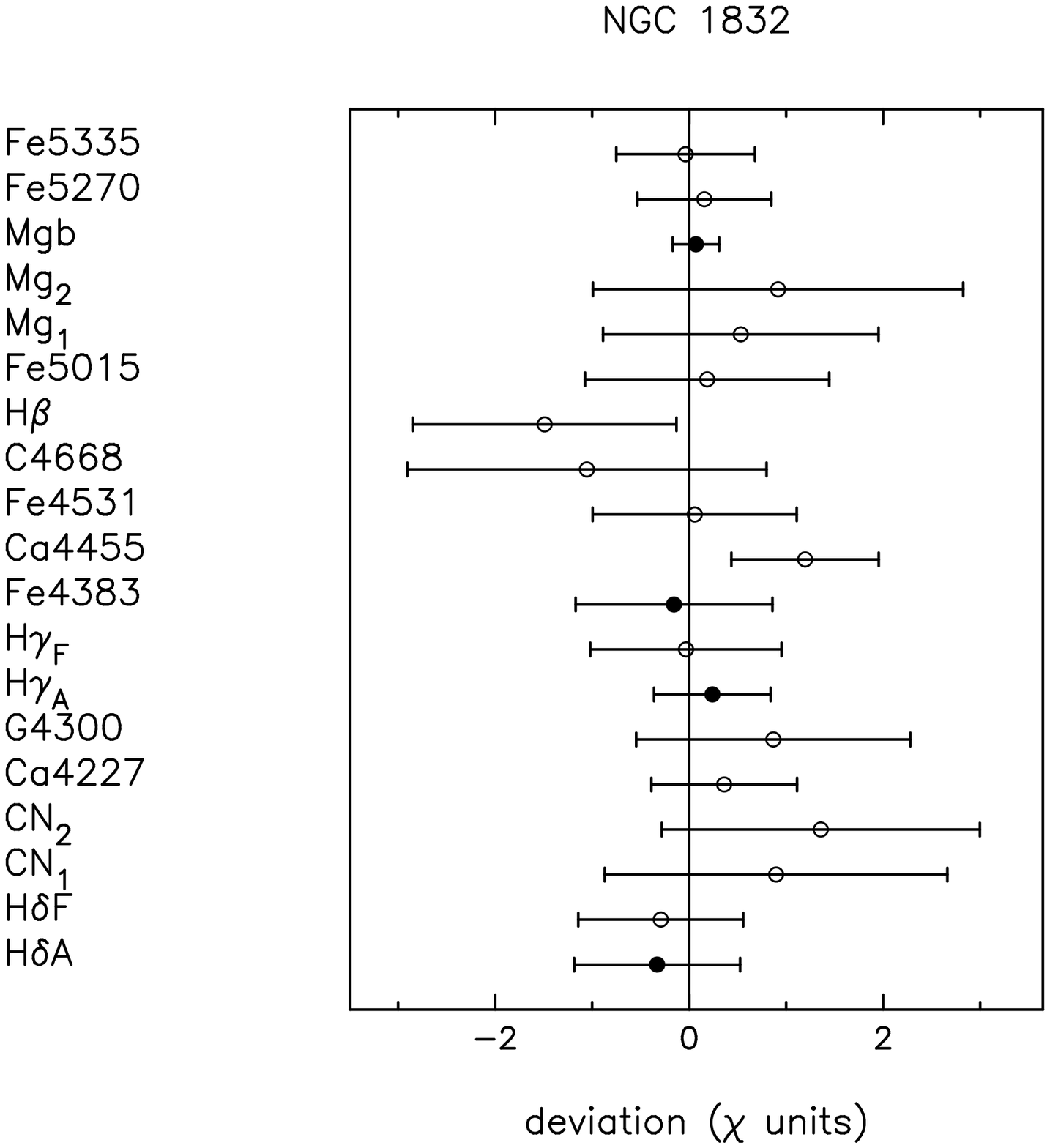}}
\resizebox{0.2\textwidth}{!}{\includegraphics[angle=-90]{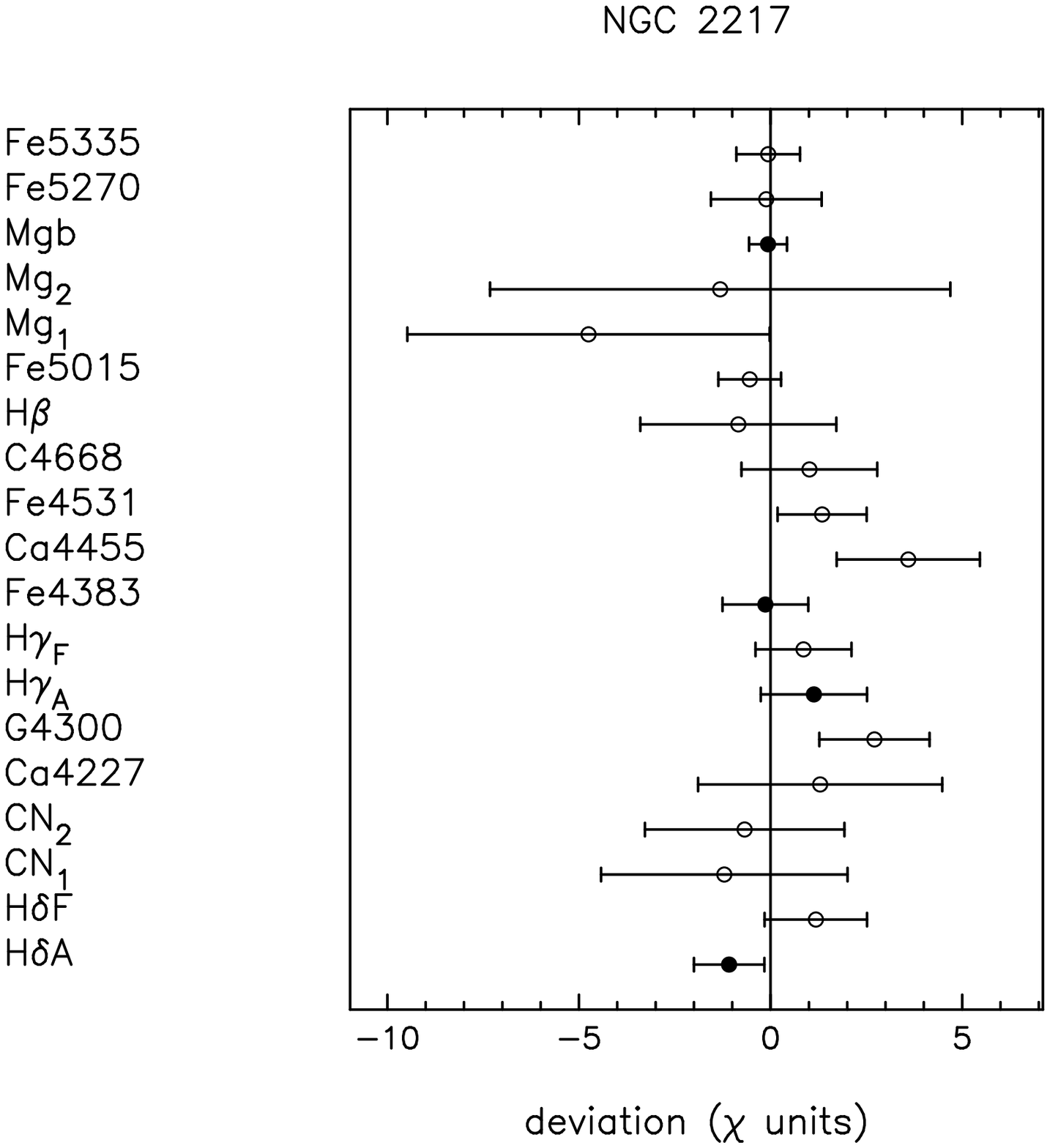}}
\resizebox{0.2\textwidth}{!}{\includegraphics[angle=-90]{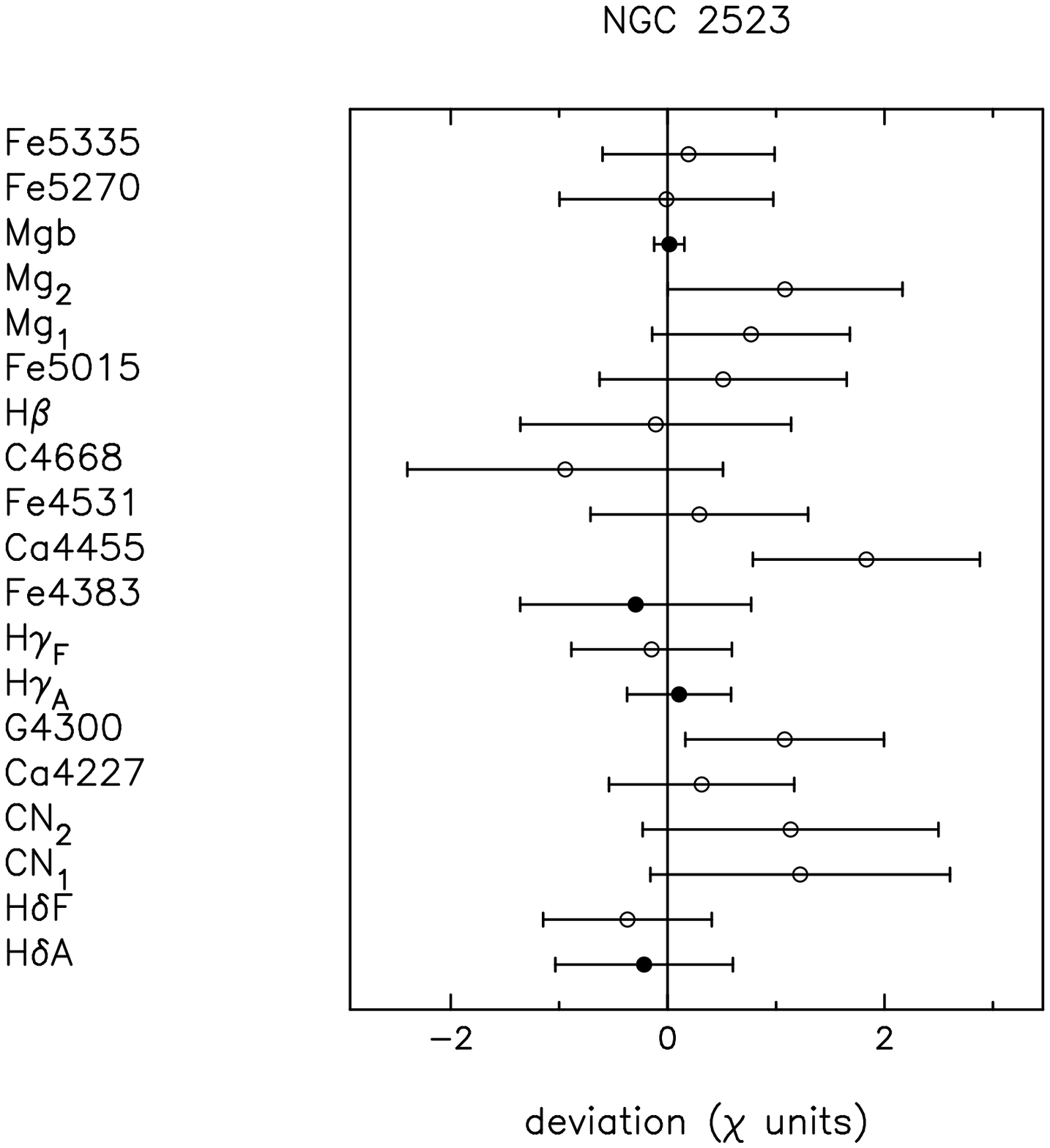}}
\resizebox{0.2\textwidth}{!}{\includegraphics[angle=-90]{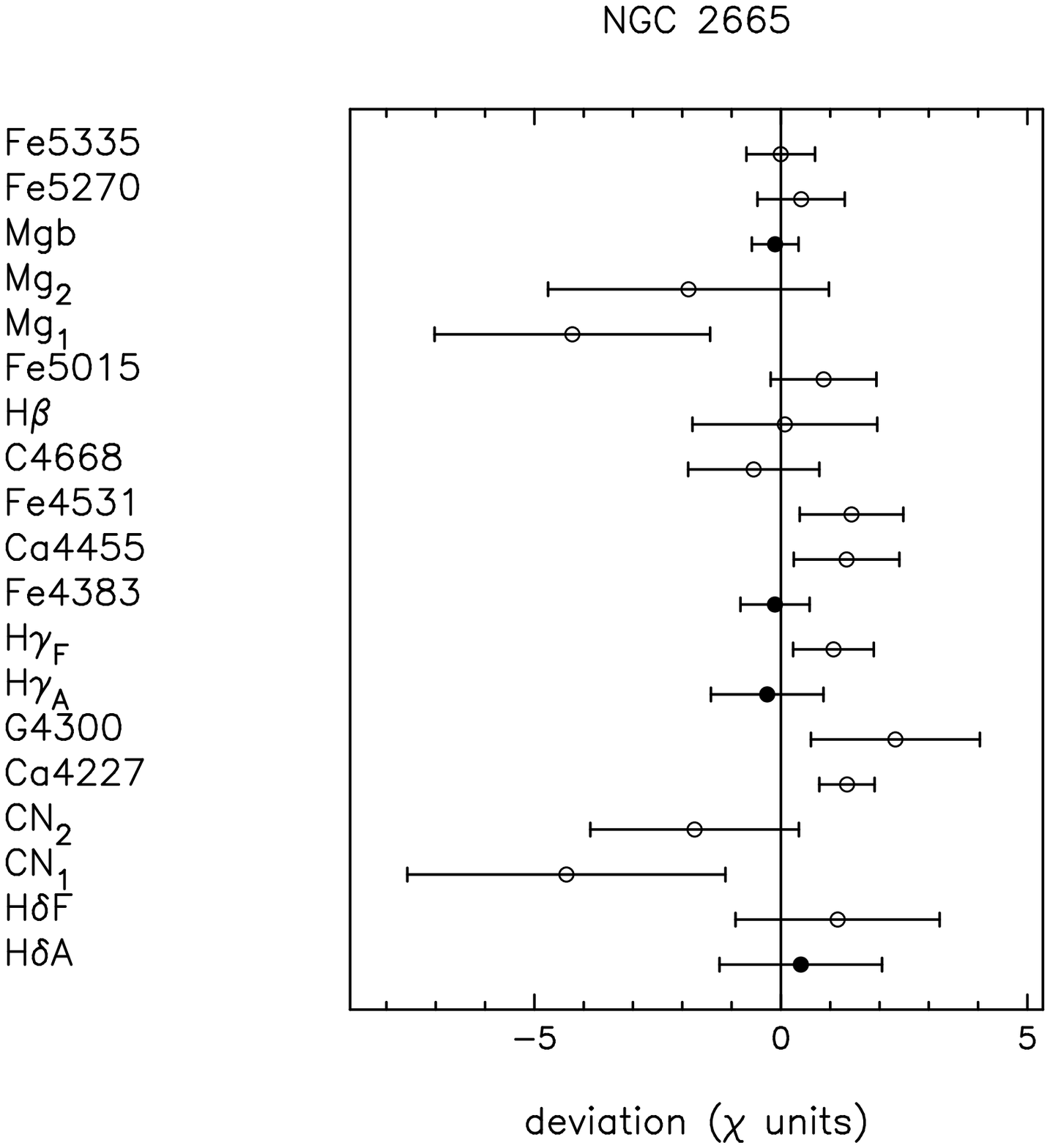}}
\resizebox{0.2\textwidth}{!}{\includegraphics[angle=-90]{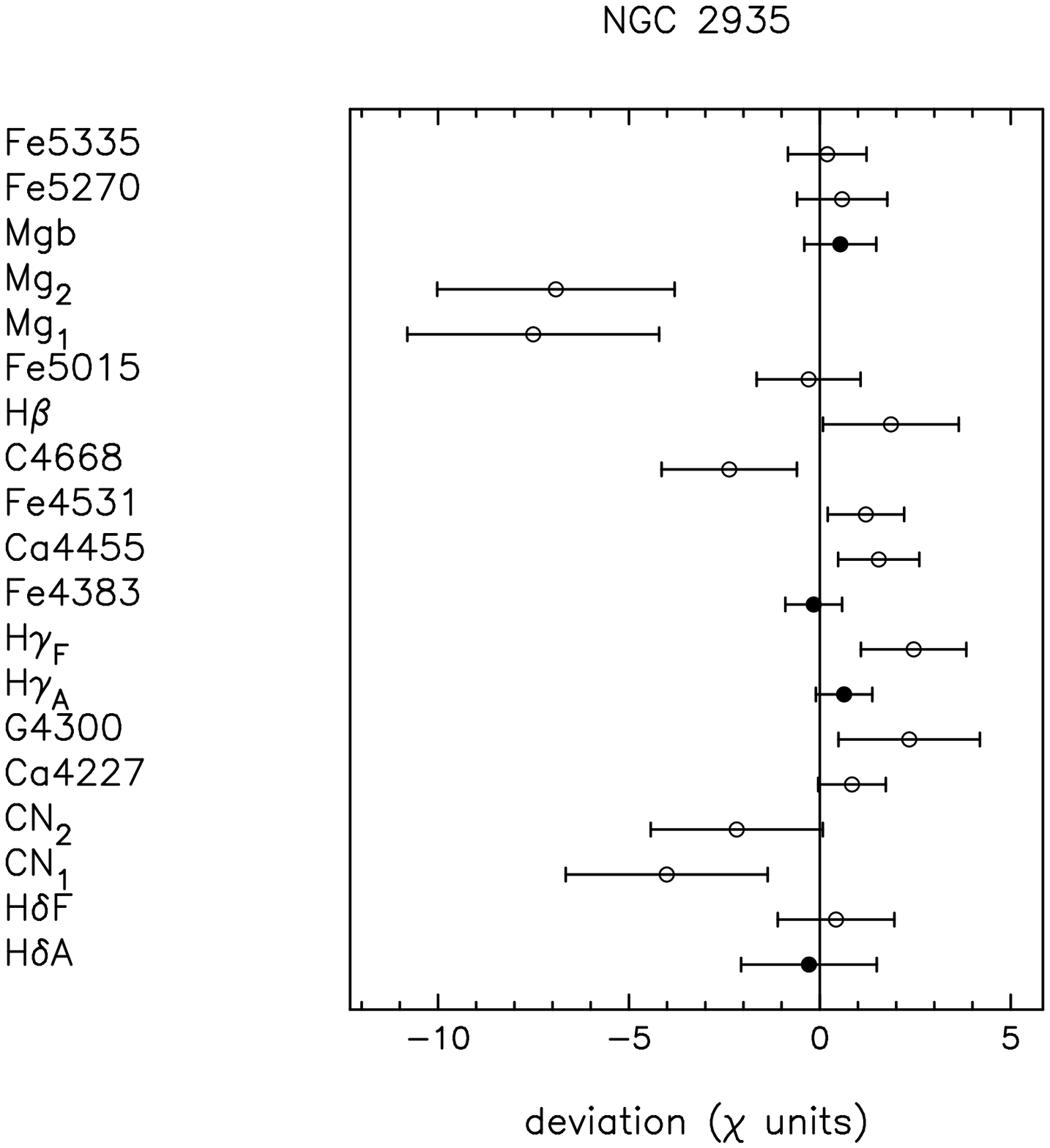}}
\resizebox{0.2\textwidth}{!}{\includegraphics[angle=-90]{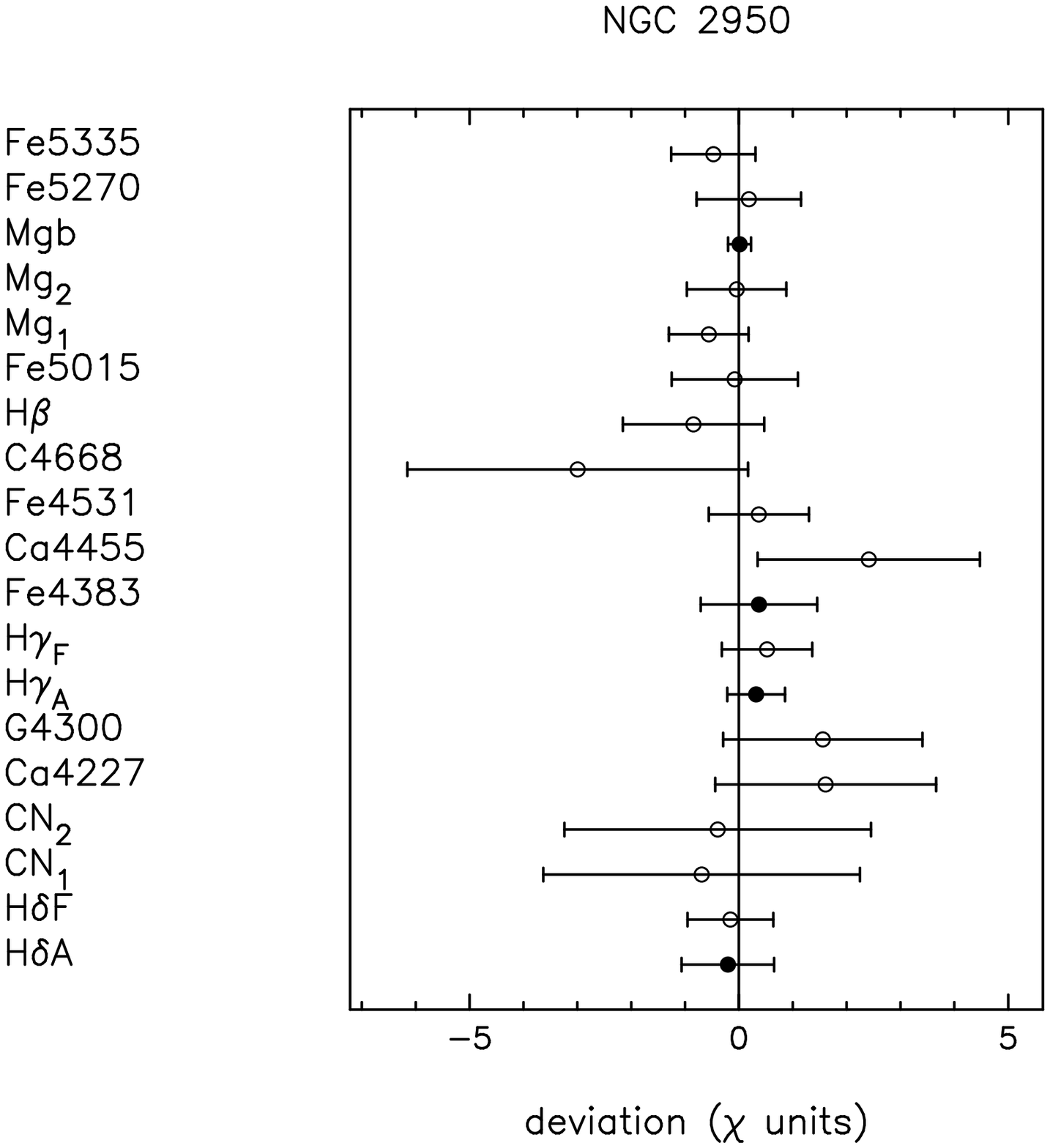}}
\resizebox{0.2\textwidth}{!}{\includegraphics[angle=-90]{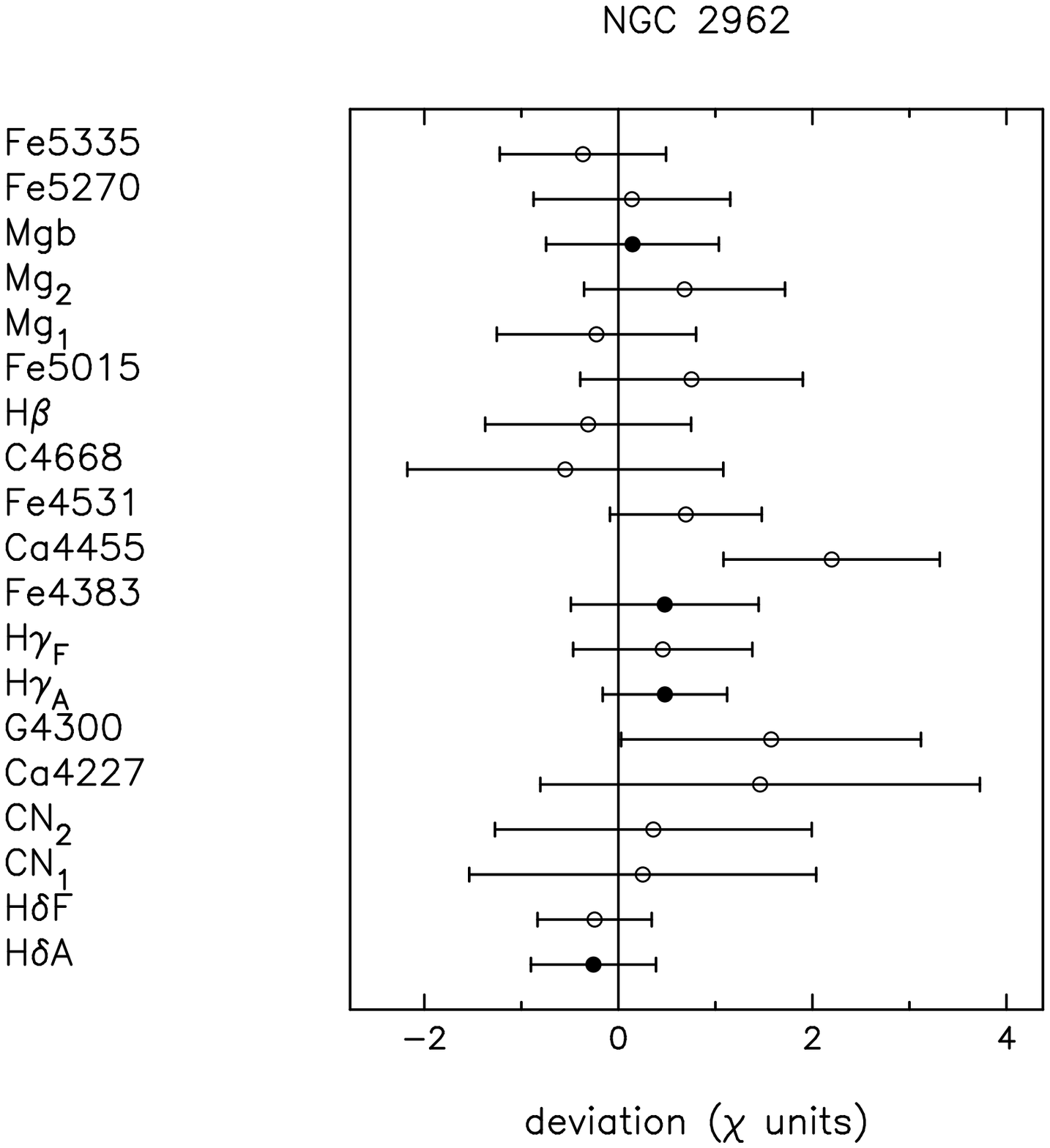}}
\resizebox{0.2\textwidth}{!}{\includegraphics[angle=-90]{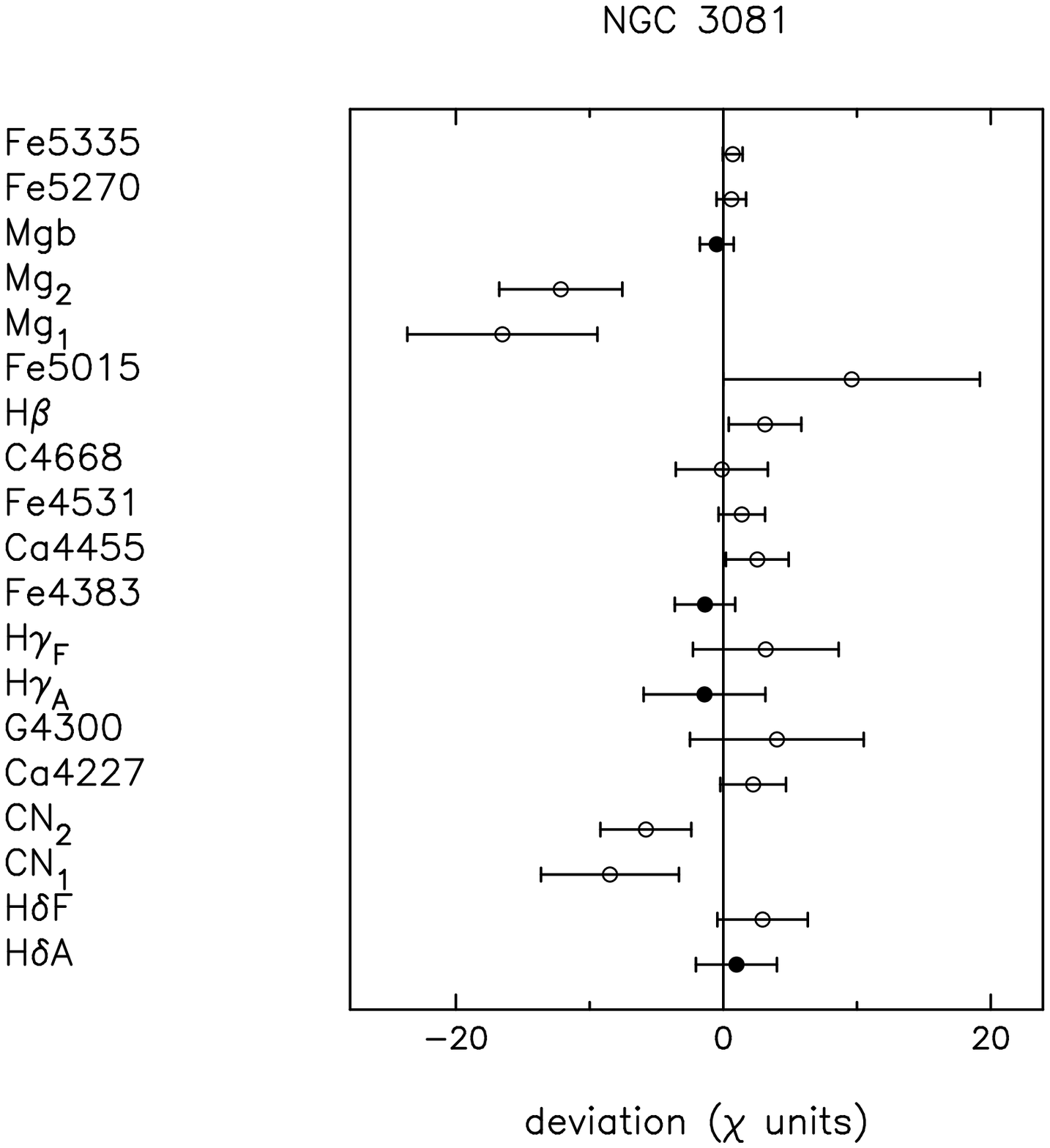}}
\resizebox{0.2\textwidth}{!}{\includegraphics[angle=-90]{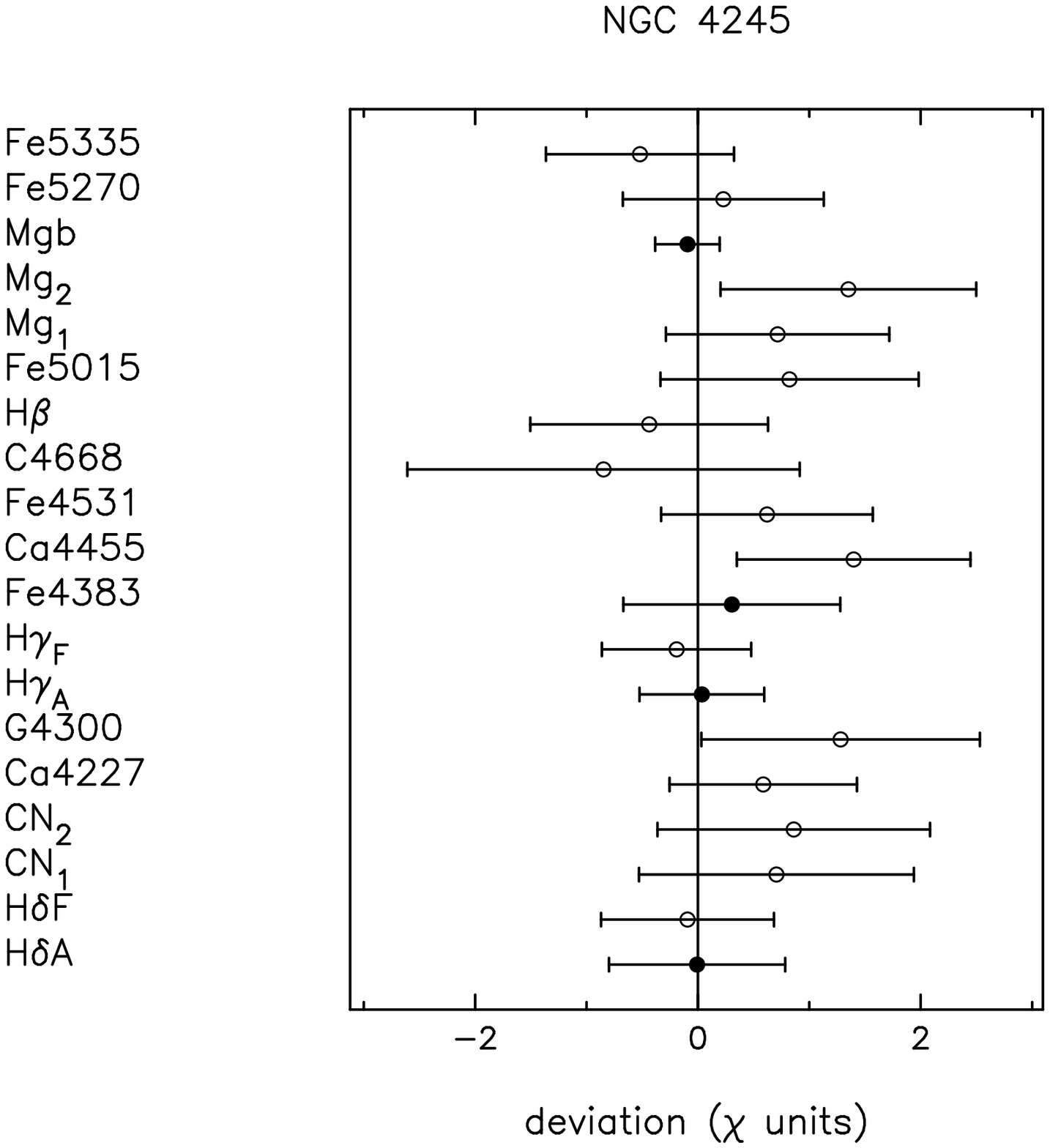}}
\resizebox{0.2\textwidth}{!}{\includegraphics[angle=-90]{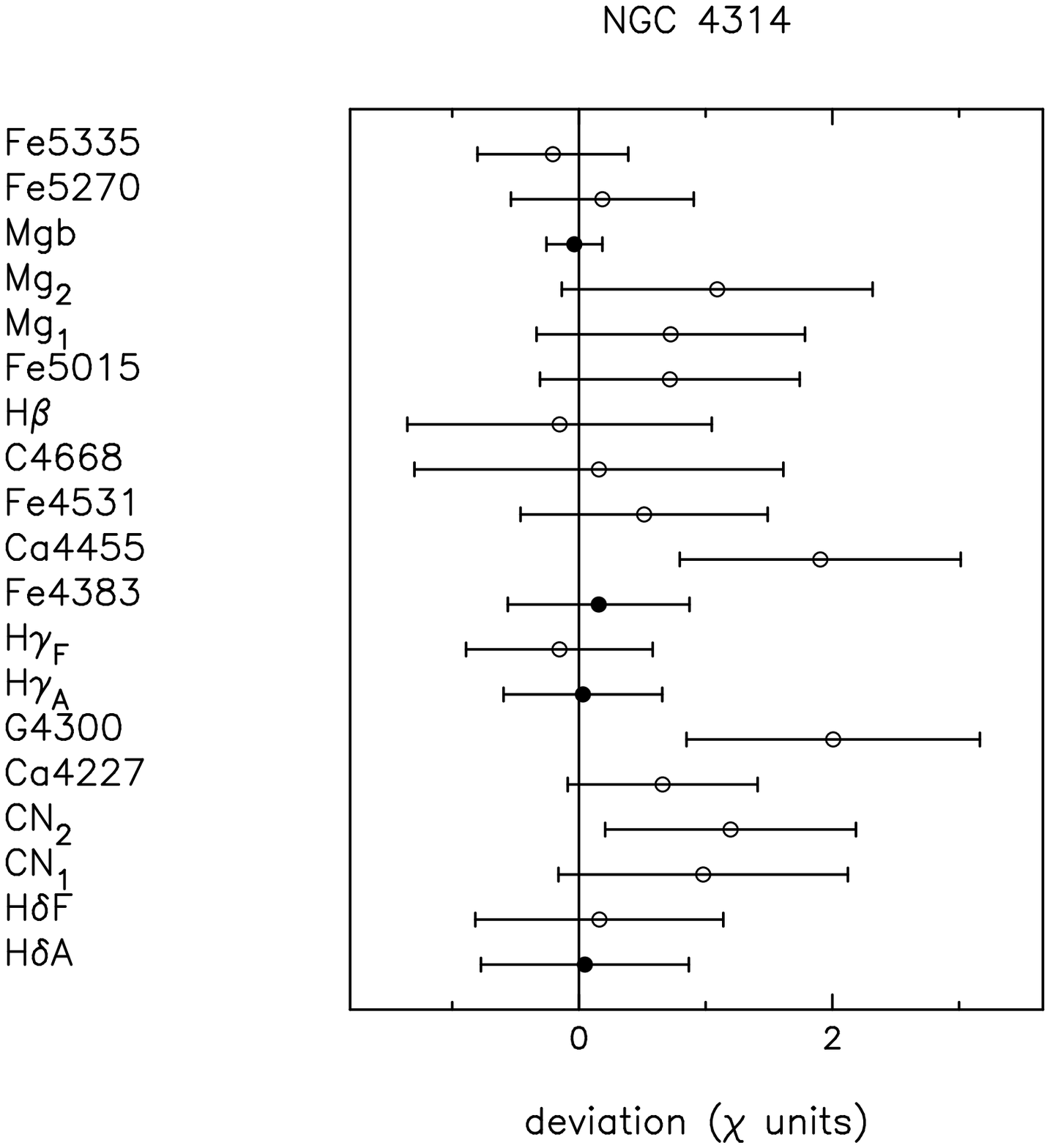}}\hspace{1.1cm}
\resizebox{0.2\textwidth}{!}{\includegraphics[angle=-90]{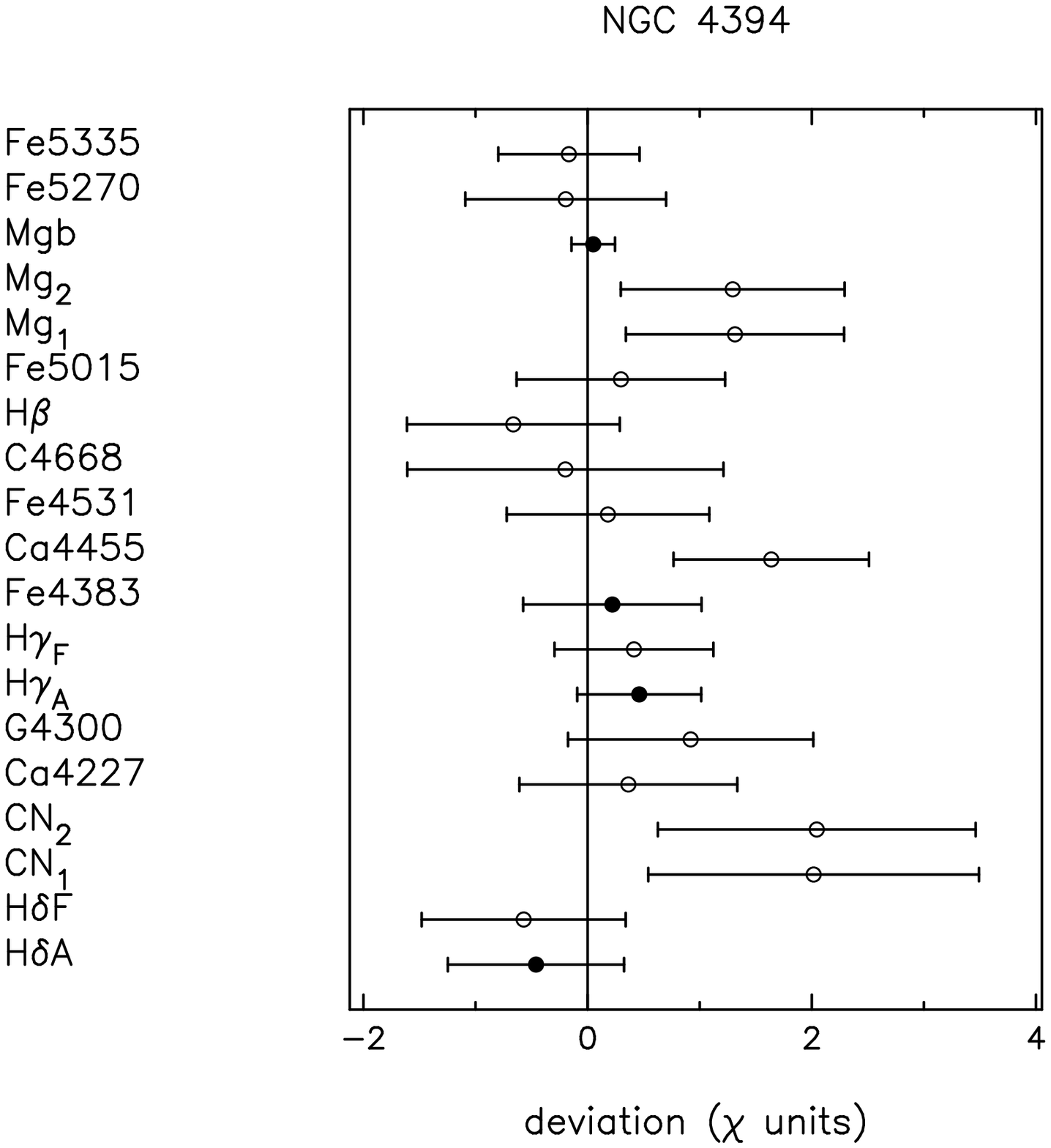}}\hspace{1.1cm}
\resizebox{0.2\textwidth}{!}{\includegraphics[angle=-90]{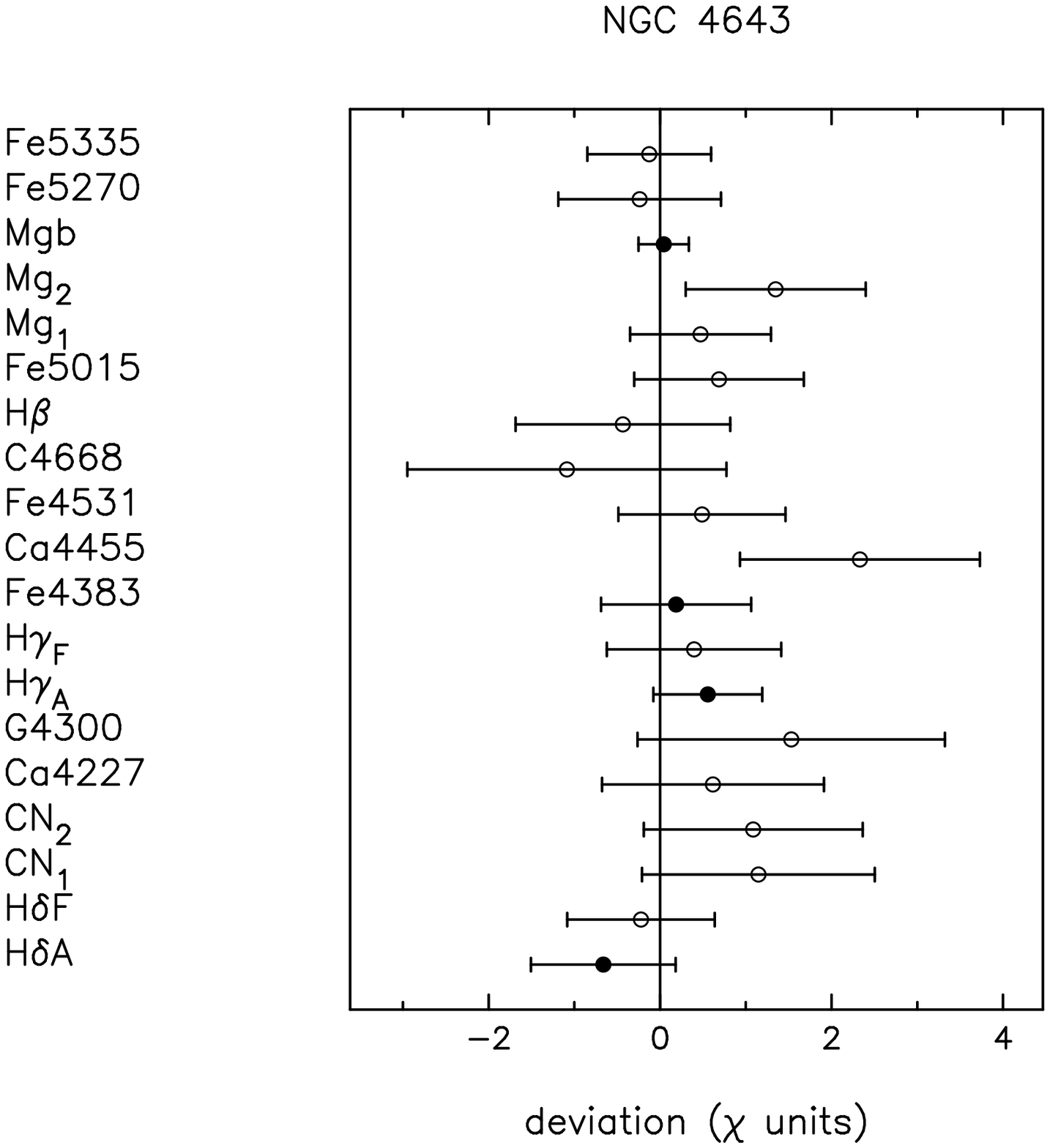}}
\caption{Deviations from best fit values for data compared to the models of 
TMB03. The indices that have been used for the fit are represented with 
filled symbols.\label{fig.chi}}
\end{figure*}

\end{document}